%% file: main.tex
\documentclass[11pt]{book}

\input{header}
\usepackage{datetime2}
\usepackage{lineno}

\usepackage{fancyhdr}
\usepackage{cite} 

\begin{document}

\clearpage
\input{sec_titlepage}

\pagestyle{plain}

\clearpage
\input{authors_all_new}

\pagestyle{headings}

\clearpage
\input{sec_summary}


\pagestyle{fancy}
\fancyhf{}
\fancyhead[L]{\nouppercase\leftmark}
\fancyhead[R]{\thepage}

\clearpage
\renewcommand{\baselinestretch}{0.95}\normalsize
\tableofcontents
\renewcommand{\baselinestretch}{1.0}\normalsize

\chapter{Introduction \label{sec:intro}}

\input{sec_intro}

\chapter{The Facility \label{sec:facility}}
\input{sec_facility}

\chapter{Experiments \label{sec:experiments}}

\input{sec_experiments}

\chapter{Long-Lived Particles \label{sec:bsm1}}
\input{sec_bsm1}

\chapter{Dark Matter and BSM Scattering Signatures \label{sec:bsm2}}
\input{sec_bsm2}


\chapter{Quantum Chromodynamics \label{sec:qcd}}
\input{sec_qcd.tex}


\chapter{Neutrino Physics \label{sec:neutrinos}}
\input{sec_neutrino}


\chapter{Astroparticle Physics \label{sec:astro}}
\input{sec_astro}

\chapter*{Acknowledgments \label{sec:acknowledgements} }
\addcontentsline{toc}{chapter}{Acknowledgements}
\input{acknowledgments_all}


\renewcommand\bibname{References}

{\small \bibliography{references_all,references_added}}

\end{document}

%% file: sec_titlepage.tex

\thispagestyle{plain}	

\vspace*{-2cm}
\snowmass

\vspace*{0.6cm}
\begin{figure*}[h]
\centering
  \includegraphics[width=0.8\textwidth]{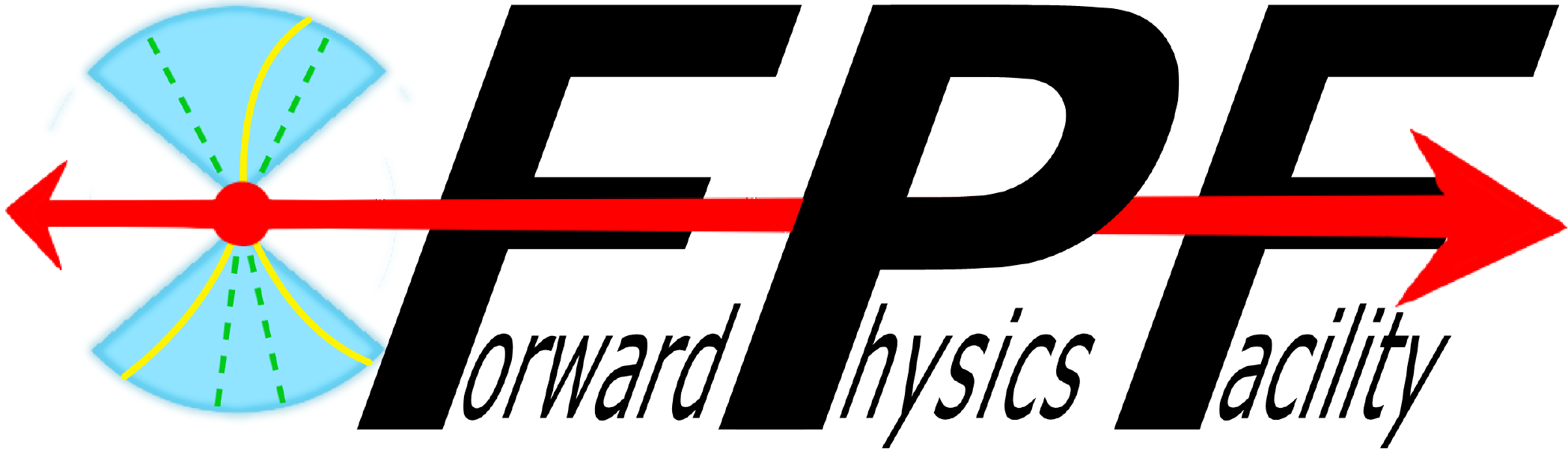}
\end{figure*}

\vspace*{0.8cm}
\begin{center}{\textbf{\huge The Forward Physics Facility \\ \vspace*{0.4cm}
at the High-Luminosity LHC}}\end{center}	

\vspace*{0.8cm}
High energy collisions at the High-Luminosity Large Hadron Collider (LHC) produce a large number of particles along the beam collision axis, outside of the acceptance of existing LHC experiments. The proposed Forward Physics Facility (FPF), to be located several hundred meters from the ATLAS interaction point and shielded by concrete and rock, will host a suite of experiments to probe Standard Model (SM) processes and search for physics beyond the Standard Model (BSM). In this report, we review the status of the civil engineering plans and the experiments to explore the diverse physics signals that can be uniquely probed in the forward region. FPF experiments will be sensitive to a broad range of BSM physics through searches for new particle scattering or decay signatures and deviations from SM expectations in high statistics analyses with TeV neutrinos in this low-background environment. High statistics neutrino detection will also provide valuable data for fundamental topics in perturbative and non-perturbative QCD and in weak interactions. Experiments at the FPF will enable synergies between forward particle production at the LHC and astroparticle physics to be exploited. We report here on these physics topics, on infrastructure, detector, and simulation studies, and on future directions to realize the FPF's physics potential. \\ 

\vspace*{0.2in}
\begin{center}
Snowmass Working Groups \\
EF4,EF5,EF6,EF9,EF10,NF3,NF6,NF8,NF9,NF10,RP6,CF7,TF07,TF09,TF11,AF2,AF5,IF8 
\end{center}

%% file: authors_all_new.tex
\leftline{\footnotesize \hspace*{-0.2in}
UCI-TR-2022-01, CERN-PBC-Notes-2022-001, FERMILAB-PUB-22-094-ND-SCD-T, INT-PUB-22-006, BONN-TH-2022-04}

\renewcommand{\thefootnote}{\fnsymbol{footnote}}
\vspace*{0.6cm}
\begin{center}{{\Large \textsc{Lead Conveners}}\\
\vspace*{0.2cm}
Jonathan~L.~Feng$^{1}$\footnote{Corresponding author: jlf@uci.edu}, 
Felix Kling$^{2}$, 
Mary Hall Reno$^{3}$, 
Juan Rojo$^{4,5}$, 
Dennis~Soldin$^{6}$
}\end{center}

\renewcommand{\thefootnote}{\arabic{footnote}}

\vspace*{-0.3cm}
\begin{center}{{\Large \textsc{Topical Conveners}}\\ 
\vspace*{0.2cm}
Luis~A.~Anchordoqui$^{7}$,
Jamie~Boyd$^{8}$,
Ahmed~Ismail$^{9}$,
Lucian~Harland-Lang$^{10,11}$,
Kevin~J.~Kelly$^{12}$,
Vishvas~Pandey$^{13}$,
Sebastian~Trojanowski$^{14,15}$,
Yu-Dai~Tsai$^{1}$,
}\end{center}

\vspace*{-0.3cm}
\begin{center}{{\Large \textsc{Contributors}}\\
\vspace*{0.2cm}
Jean-Marco~Alameddine$^{16}$,
Takeshi~Araki$^{17}$
}
\end{center}
\begin{center}
{
% ...
}\end{center}

\vspace*{0.2cm}
\begin{center}{\it\footnotesize
$^{1}$ Department of Physics and Astronomy, University of California, Irvine, CA 92697-4575, USA\\
$^{2}$ Deutsches Elektronen-Synchrotron DESY, Notkestr. 85, 22607 Hamburg, Germany\\
$^{3}$ Department of Physics and Astronomy, University of Iowa, Iowa City, IA 52246, USA\\
$^{4}$ Department of Physics and Astronomy, VU Amsterdam, 1081 HV Amsterdam, The Netherlands\\
$^{5}$ Nikhef Theory Group, Science Park 105, 1098 XG Amsterdam, The Netherlands\\
$^{6}$ Bartol Research Institute, Department of Physics and Astronomy,University of Delaware, Newark, DE 19716, USA\\
$^{7}$ Department of Physics and Astronomy, Lehman College, City University of New York, Bronx, NY 10468, USA\\
$^{8}$ CERN, CH-1211 Geneva 23, Switzerland\\
$^{9}$ Department of Physics, Oklahoma State University, Stillwater, OK 74078, USA\\
$^{10}$ Department of Physics, University of Oxford, OX1 3RH, United Kingdom\\
$^{11}$ Rudolf Peierls Centre for Theoretical Physics, University of Oxford, OX1 3PU,  United Kingdom\\
$^{12}$ Theoretical Physics Department, CERN, CH-1211 Geneva 23, Switzerland\\
$^{13}$ Department of Physics, University of Florida, Gainesville, FL 32611, USA\\
$^{14}$ Astrocent, Nicolaus Copernicus Astronomical Center Polish Academy of Sciences, ul. Rektorska 4, 00-614, Warsaw, Poland\\
$^{15}$ National Centre for Nuclear Research, Pasteura 7, 02-093 Warsaw, Poland\\
$^{16}$ Department of Physics, TU Dortmund University, D-44221 Dortmund, Germany\\
$^{17}$ Faculty of Dentistry, Ohu University, 31-1 Sankakudo,Tomita-machi, Koriyama, Fukushima 963-8611, Japan\\
% ...
}\end{center}

%% file: sec_summary.tex
\begin{center}{\LARGE \textsc{Executive Summary}}\end{center}

\begin{figure}[bp]
\centering
\includegraphics[width=0.99\textwidth]{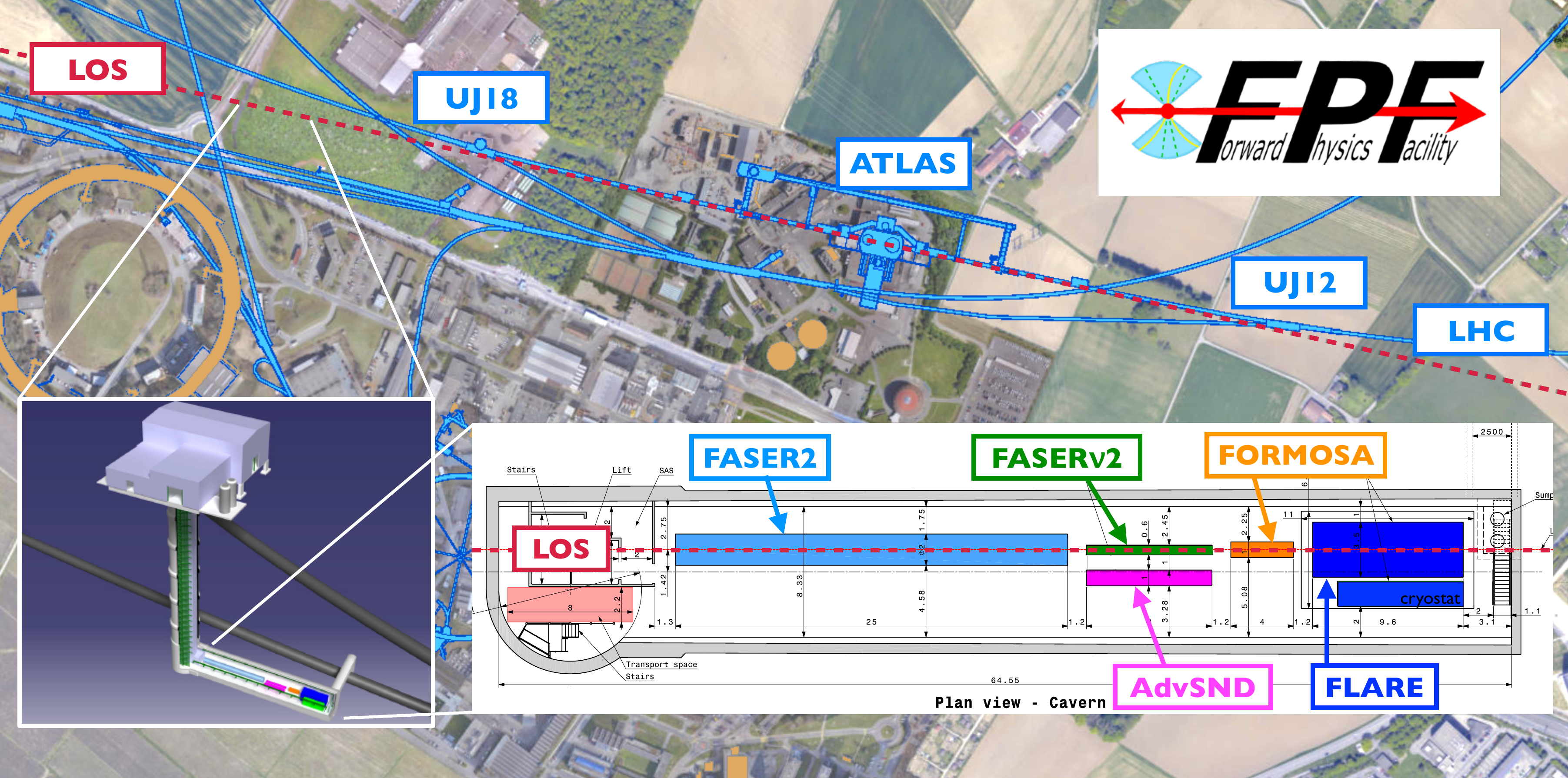}
\caption{The preferred location for the Forward Physics Facility, a proposed new cavern for the High-Luminosity era.  The FPF will be 65~m-long and 8.5~m-wide and will house a diverse set of experiments to explore the many physics opportunities in the far-forward region. 
}
\label{fig:ExecutiveSummaryMap}
\end{figure}

\vspace*{-0.15in} \paragraph*{The Facility}
The Forward Physics Facility (FPF) is a proposal to build a new underground cavern at the Large Hadron Collider (LHC) to host a suite of far-forward experiments during the High-Luminosity LHC era. The existing large LHC detectors have holes along the beam line, and so miss the physics opportunities provided by the enormous flux of particles produced in the far-forward direction.  The FPF will realize this physics potential. A preferred site for the FPF is along the beam collision axis, 617-682 m west of the ATLAS interaction point (IP); see \cref{fig:ExecutiveSummaryMap}. This location is shielded from the ATLAS IP by over 200 m of concrete and rock, providing an ideal location to search for rare processes and very weakly-interacting particles.  FPF experiments will detect $\sim 10^6$ neutrino interactions at the highest human-made energies ever recorded, expand our understanding of proton and nuclear structure and the strong interactions to new regimes, and carry out world-leading searches for a wide range of new phenomena, enhancing the LHC's physics program through to its conclusion in 2040. 

\begin{figure}
  \begin{center}
    \includegraphics[width=0.7\textwidth]{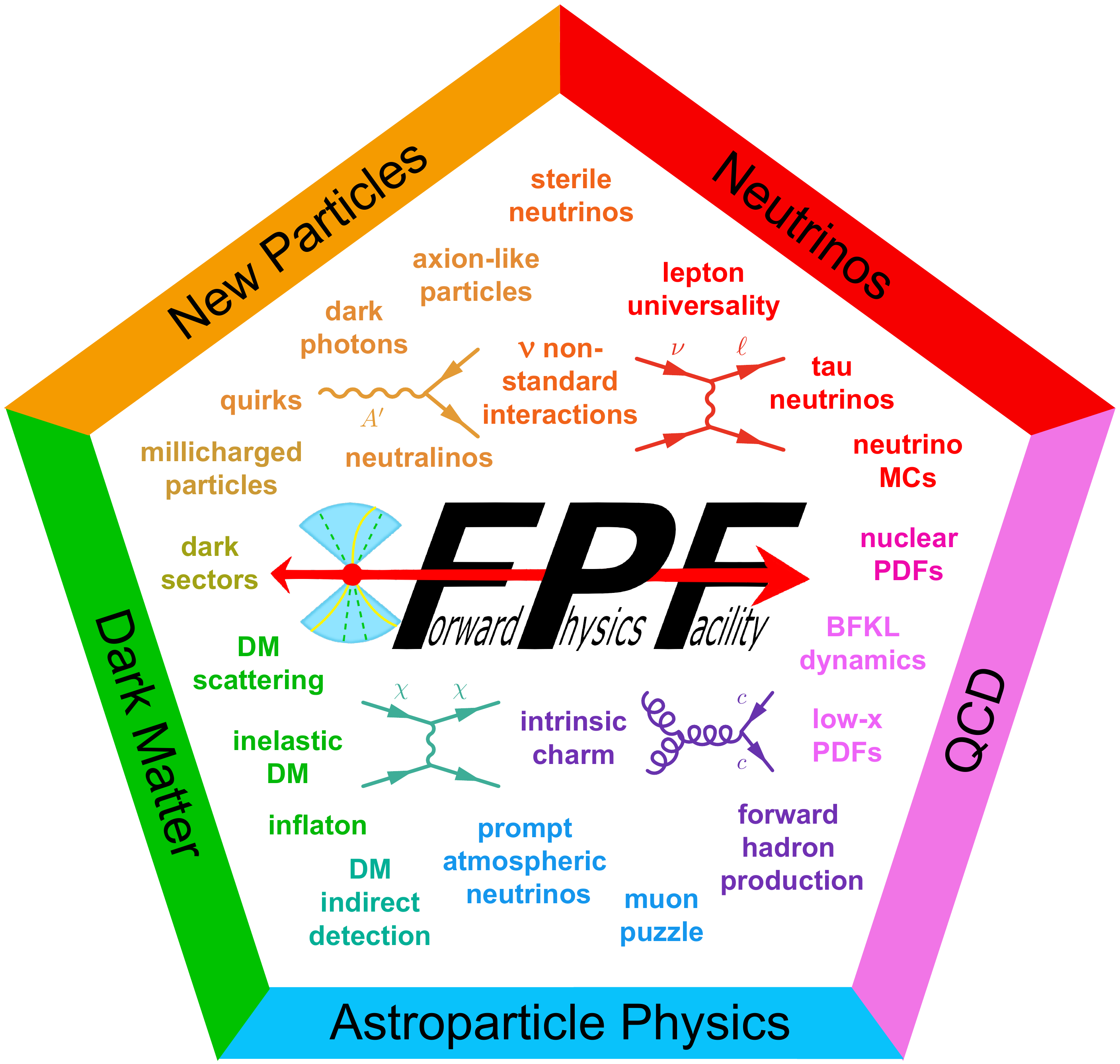}
  \end{center}
\vspace*{-0.1in}
\caption{The Forward Physics Facility will probe topics that span multiple frontiers, including new particles, neutrinos, dark matter, QCD, and astroparticle physics. 
\label{fig:ExecutiveSummaryPhysics}
\vspace*{-0.1in}
}
\end{figure}

\vspace*{-0.08in} \paragraph*{Experiments}
The FPF is uniquely suited to exploit physics opportunities in the far-forward region, because it will house a diverse set of experiments, each optimized for particular physics goals. The envisioned experiments and their physics targets are shown in \cref{fig:ExecutiveSummaryPhysics}. FASER2, a magnetic spectrometer and tracker, will search for light and weakly-interacting states, including long-lived particles, new force carriers, axion-like particles, light neutralinos, and dark sector particles.  FASER$\nu$2 and Advanced SND, proposed emulsion and electronic detectors, respectively, will detect $\sim 10^6$ neutrinos and anti-neutrinos at TeV energies, including $\sim 10^3$ tau neutrinos, the least well-understood of all known particles.  FLArE, a proposed 10-tonne-scale noble liquid detector, will detect neutrinos and also search for light dark matter.  And FORMOSA, a detector composed of scintillating bars, will provide world-leading sensitivity to millicharged particles and other very weakly-interacting particles across a large range of masses.

\vspace*{-0.08in} \paragraph*{Physics Beyond the Standard Model}
The FPF will allow tests of a wide variety of theories of physics Beyond the Standard Model (BSM), explaining outstanding questions such as the hierarchy problem, neutrino masses, the nature of dark matter, inflation, and the matter-antimatter asymmetry of the universe. New particles and forces can be detected at the FPF in different ways. Many theories feature light, weakly-coupled particles that have lifetimes long enough to be produced at the ATLAS IP and subsequently decay within FPF detectors, like FASER2. Alternatively, DM particles may scatter inside a dense detector like Advanced SND, FASER$\nu$2, or FLArE and produce visible signatures. Both electron and nuclear scattering are possible. Finally, some models have new states, for example, millicharged particles, which would leave non-standard energy deposits in detectors. Such particles could be observed at FORMOSA and possibly other detectors. 

Searches for new heavy particles benefit from the unparalleled energies at the LHC, and the FPF will provide leading sensitivities if such states are preferentially produced in the forward direction, as in the case of quirks. Searches for light states may also be enhanced at the energy frontier, as they are produced with very high boosts, allowing probes of shorter lifetimes, or through rare $B$ decays and similar processes that are much less common at other facilities. These capabilities result in unique projected sensitivities in many BSM models that surpass current, or even future expected, limits.

\vspace*{-0.08in} \paragraph*{Quantum Chromodynamics}
The FPF has the promising potential to probe our understanding of the strong interactions as well as of proton and nuclear structure.  It will be sensitive to the very forward production of light hadrons and charmed mesons, providing access to both the very low-$x$ and the very high-$x$ regions of the colliding protons. The former regime is sensitive to novel QCD production mechanisms, such as BFKL effects and non-linear dynamics, as well as the gluon parton distribution function (PDF) down to $x\sim 10^{-7}$, well beyond the coverage of other experiments and providing key inputs for astroparticle physics. The latter regime provides information on open questions relating to the high-$x$ PDFs, and in particular intrinsic charm. In addition, the FPF acts as a neutrino-induced deep-inelastic scattering (DIS) experiment with TeV-scale neutrino beams. The resulting measurements of neutrino DIS structure functions represent a valuable handle on the partonic structure of nucleons and nuclei, particularly their quark flavour separation, that is fully complementary to the charged-lepton DIS measurements expected at the upcoming Electron-Ion Collider (EIC).

\vspace*{-0.08in} \paragraph*{Neutrino Physics}
The LHC produces high energy and intense fluxes of all flavors of neutrinos and anti-neutrinos in the forward region.  Ten-tonne-scale experiments at the FPF are being designed to detect $\sim 10^5$ $\nu_e$, $\sim 10^6$ $\nu_{\mu}$, and $\sim 10^3$ $\nu_{\tau}$ interactions with energies between several hundreds of GeV and a few TeV, an energy range that has not been directly probed for any neutrino flavor.  In addition, by measuring the charge of the resulting muons in charged-current interactions, muon and tau neutrinos and anti-neutrinos will be distinguished.  These neutrino event will significantly extend accelerator cross section measurements and provide the first opportunity for detailed studies of tau neutrinos and anti-neutrinos.  They will also open up new avenues to discover or constrain BSM physics effects in neutrino production, propagation, and interactions, with important implications for QCD and astroparticle physics.

\vspace*{-0.08in} \paragraph*{Astroparticle Physics}
The FPF provides opportunities for interdisciplinary studies at the intersection of high-energy particle physics and modern astroparticle physics. Cosmic rays enter the atmosphere with energies up to $10^{11}\,\mathrm{GeV}$ and beyond, where they produce large cascades of high-energy particles. The development of these extensive air showers is driven by hadron-ion collisions under low momentum transfer in the non-perturbative regime of QCD.  Measurements at the FPF will improve the modeling of high-energy hadronic interactions in the atmosphere, reduce the associated uncertainties of air shower measurements, and thereby help to understand the properties of cosmic rays, such as their energy and mass, which is crucial to discover their origin.  Moreover, atmospheric muons and neutrinos produced in these extensive air showers in the far-forward region are the main background for searches of high-energy astrophysical neutrinos with large-scale neutrino telescopes.  The FPF will help to understand the atmospheric neutrino flux and reduce the uncertainties for astrophysical neutrino searches in the context of multi-messenger astrophysics. 

\vspace*{-0.08in} \paragraph{Timeline and Cost}
The FPF is well aligned with the 2020 European Strategy Update's first recommendation that ``the full physics potential of the LHC and the HL-LHC...should be exploited.''  To fully exploit the far-forward physics opportunities, many of which will disappear for several decades if not explored at the FPF, the FPF should be available for as much of the HL-LHC era as possible.  The FPF requires no modifications to the LHC, and all of the planned experiments are relatively small, inexpensive, and fast to construct. A very preliminary costing for the FPF has yielded estimates of 25 MCHF for the construction of the new shaft and cavern and 15 MCHF for all necessary services. To this must be added the cost of the individual experiments.  A possible timeline is for the FPF to be built during Long Shutdown 3 from 2026-28, the support services and experiments to be installed starting in 2029, and the experiments to begin taking data not long after the beginning of Run 4. Such a timeline is guaranteed to produce exciting physics results through studies of very high energy neutrinos, QCD, and other SM topics, and will additionally enhance the LHC's potential for groundbreaking discoveries that will clarify the path forward for decades to come.

%% file: sec_intro.tex
Particle colliders have been used for decades to discover the fundamental building blocks of the universe and study their properties.  The high energy frontier is now at the Large Hadron Collider (LHC) at CERN, which began colliding protons with protons in 2010 and is expected to run until 2040.  Soon after the LHC started, the Higgs boson was discovered~\cite{ATLAS:2012yve,CMS:2012qbp}, but so far no other new fundamental particles have been found.  At the same time, deep mysteries remain, including the origin of neutrino masses, the identity of dark matter, and many others.  These problems provide overwhelming evidence that we are far from a complete understanding of the universe, and they strongly motivate new experiments that will deepen our understanding of the Standard Model (SM) and maximize our potential for discovering new physics in the years to come.

An important question is whether opportunities for groundbreaking discoveries are currently being missed at the LHC.  History provides a cautionary tale.  In 1971, the first proton-proton collider, the Intersecting Storage Rings (ISR), began operating at CERN.  As recounted in numerous talks and articles celebrating the ISR's 50th anniversary last year~\cite{ISR50,DiscoveryMachines}, when the ISR began operating, physicists believed that new discoveries would be made by observing particles emitted in the forward region, that is, roughly parallel to the beamline.  Detectors therefore focused on this region.  For this reason, the ISR missed the discovery of the charm quark, which was discovered at Brookhaven and SLAC in 1974 in what is now recognized as one of the most important breakthroughs in the history of physics.  

Recently, it has been recognized that we may be missing opportunities at the LHC for a similar, but opposite, reason.  Having absorbed the lessons of the 1970's, the current large detectors at the LHC are well instrumented at large angles relative to the beamline.  Unfortunately, they have holes in the far-forward direction.  Particles produced parallel to the beamline are therefore undetected, leading to the possibility that new discoveries may have simply escaped the LHC through these holes in the far-forward region.  

In fact, it is now known that interesting physics opportunities have indeed been missed in the far-forward region.  In May 2021, the FASER Collaboration announced the detection of neutrino candidates using an 11 kg pilot detector placed in the far-forward region for a month in 2018~\cite{FASER:2021mtu}. These were the first neutrino candidates ever detected at a collider, and the highest energy neutrino candidates ever seen from a terrestrial source.  To date, an entire program of neutrino physics has been missed at the LHC, and it is natural to wonder if even greater discoveries could be made with dedicated experiments placed in the far-forward region.

Motivated by such considerations, in the last 3 years, three new far-forward detectors have been approved and constructed at the LHC: FASER~\cite{Feng:2017uoz, FASER:2018ceo, FASER:2018bac, FASER:2018eoc}, FASER$\nu$~\cite{FASER:2020gpr, Abreu:2019yak, FASER:2021mtu}, and SND@LHC~\cite{SHiP:2020sos, Ahdida:2750060}.  Despite their small size (meter-scale) and inexpensive and rapid construction, these detectors will significantly extend the LHC's physics program when Run 3 begins in mid-2022.  Exploiting the enormous fluxes of particles in the far-forward direction, and shielded from the ATLAS interaction point (IP) by 100 m of rock and concrete, FASER, FASER$\nu$, and SND@LHC will together detect $\sim 10,000$ TeV-energy neutrinos and search for signs of new particles, with important implications for models of new physics, dark sectors, QCD, neutrino physics, and astroparticle experiments.

FASER, FASER$\nu$, and SND@LHC are currently located in previously abandoned service tunnels constructed for the Large Electron-Positron Collider in the 1980’s.  These locations were never intended to house experiments and the necessary services, and they cannot accommodate larger detectors or additional experiments.  At the same time, it has become abundantly clear that these detectors do not fully exploit the possibilities offered by the far-forward region. 

The Forward Physics Facility (FPF) is a proposal to construct a dedicated facility to house a suite of far-forward experiments during the High-Luminosity LHC era.  Studies of potential sites for the FPF have now converged on two preferred options.  In the first, a purpose-built facility is excavated with a new shaft and new cavern providing 65~m of space along the beam collision axis or line of sight (LOS), 617-682~m west of the ATLAS interaction point (IP).  An alternative option is to expand the current location of FASER and FASER$\nu$ with alcoves to provide space along the LOS 480-521 m to the east of the ATLAS IP.  Based on the expected costs, and a number of important benefits for the experiments, the new purpose-built facility is currently considered the baseline option for the FPF.  Experiments currently planned for the FPF include upgraded versions of the existing detectors (FASER2, FASER$\nu$2, and Advanced SND), as well as new experiments, including FORMOSA, which will search for millicharged particles and related signals, and FLArE, a noble liquid TPC, which will detect neutrinos and also search for light dark matter produced by the LHC. 

The FPF's special location makes its experiments uniquely sensitive to many SM and BSM phenomena, and its physics capabilities are complementary to those of other current and proposed experiments at the LHC. Besides the large LHC experiments probing high-$p_T$ physics, these include a number of smaller detectors performing SM measurements in the forward region, including ALFA~\cite{AbdelKhalek:2016tiv}, AFP~\cite{Grinstein:2016sen},  CASTOR~\cite{CMS:2020ldm}, CT-PPS~\cite{CMS:2014sdw}, LHCf~\cite{LHCf:2008lfy}, and TOTEM~\cite{Anelli:2008zza}.  These are located in or around the LHC beam pipe close to either the ATLAS or CMS IP, but, in contrast to the FPF, are not shielded from these IPs by hundreds of meters of concrete and rock. Such shielding is essential for searches for extremely rare phenomena, where the goal is to reduce backgrounds so that even a few events over the course of the entire HL-LHC era will be sufficient to claim a signal.  The FPF is also complementary to the existing experiments MilliQan~\cite{Haas:2014dda,Ball:2016zrp} and MoEDAL~\cite{Acharya:2014nyr}, as well as proposed experiments, such as ANUBIS~\cite{Bauer:2019vqk}, CODEX-b~\cite{Gligorov:2017nwh, Aielli:2019ivi}, and MATHUSLA~\cite{Chou:2016lxi, MATHUSLA:2018bqv, MATHUSLA:2020uve}, which also aim to search for long-lived particles and other new physics at the LHC, but, in contrast to FPF experiments, are located at large angles relative to the beamline. Last, there are also important synergies of FPF physics with experiments running or proposed at other facilities, including, for example, BSM searches at beam dump experiments, such as SHiP~\cite{Alekhin:2015byh} at the SPS.  Compared to fixed target experiments, FPF experiments have lower $pp$ interaction rates, but higher center-of-mass energies.  Of course, practically, the FPF also has the virtue of being completely parasitic, requiring no dedicated beam time and no modifications to the existing accelerator structures.

The FPF was first proposed in May 2020 and has been the subject of 4 dedicated meetings held in November 2020~\cite{FPF1Meeting}, May 2021~\cite{FPF2Meeting}, October 2021~\cite{FPF3Meeting}, and February 2022~\cite{FPF4Meeting}. In parallel, the FPF activities have been strongly supported by CERN's Physics Beyond Colliders Study Group~\cite{PBCWebpage} and through the activities of numerous Snowmass 2021 working groups~\cite{SnowmassWebpage}. A brief Letter of Interest was submitted to Snowmass in August 2020~\cite{SnowmassFPF}, and the status of the FPF was summarized by 80 authors in a 75-page document in October 2021~\cite{Anchordoqui:2021ghd}.  This current document contains a more comprehensive summary of the status of the FPF, including studies of the facility and its environment in \cref{sec:facility}, the proposed far-forward experiments in \cref{sec:experiments}, and its unique potential to discover long-lived particles, detect DM and other scattering signatures, and study QCD, neutrinos physics, and astroparticle physics in Chapters~\ref{sec:bsm1}, \ref{sec:bsm2}, \ref{sec:qcd}, \ref{sec:neutrinos}, and \ref{sec:astro}, respectively.

The FPF is well aligned with the 2020 European Strategy Update's first recommendation that ``the full physics potential of the LHC and the HL-LHC...should be exploited''~\cite{EuropeanStrategyGroup:2020pow}.
To realize this goal, the FPF and its experiments should be available for as much of the HL-LHC era as possible.  A possible timeline is for the FPF to be built during Long Shutdown 3 from 2026-28, the support services and experiments to be installed starting in 2029, and the experiments to begin taking data not long after the beginning of Run 4.  To realize this timeline, Conceptual Design Reports for the FPF and all experiments must be prepared in the near future, to be followed by Technical Design Reports, approvals, and funding.  The timeline benefits from the fact that the purpose-built facility can be mostly constructed even while the LHC is running and requires no modifications to the LHC, while all of the planned experiments are small, fast, and inexpensive, relative to most collider detectors.  Of course, the driving force is the FPF's potential to enrich the physics program of the LHC.  If not constructed for the HL-LHC era, many of the FPF's physics opportunities will be lost for at least several decades.  On the other hand, if prepared for the HL-LHC era, the FPF is guaranteed to produce exciting physics results through studies of neutrinos, QCD, and other SM topics, and it will enhance the LHC's potential for groundbreaking discoveries that will clarify the path forward for decades to come.

%% file: sec_facility.tex
\vspace*{-.2in}

\contributors{Jamie Boyd, Jonathan L.~Feng (conveners), 
Jean-Marco Alameddine, 
Kincso~Balazs, 
Michele Battistin, 
Caterina Bertone, 
Stewart T.~Boogert, 
Francesco Cerutti, 
Jean-Pierre Corso, 
Lucie Elie, 
Stephen Gibson, 
Silvia Grau, 
Timo Hakulinen, 
Angelo Infantino, 
Helena Lefebvre, 
Mickael Lonjon, 
Angel Navascues Cornago, 
Pierre Ninin, 
Laurence J.~Nevay, 
Rui Nunes, 
John Osborne, 
Guillermo Peon, 
Wolfgang Rhode, 
Tim Ruhe, 
Marta Sabate-Gilarte, 
Alexander Sandrock, 
Pierre Thonet, 
and Heinz Vincke}

The physics goals of the FPF require that it be located on the beam collision axis or line of sight (LOS) near an LHC interaction point (IP).  The location should also be sufficiently shielded from the IP to provide a very low-background environment for studies of neutrinos and searches for other very weakly-interacting particles.  In this Chapter, we present the results of studies to identify suitable locations for the FPF and to understand the particle fluxes and backgrounds at these locations.

\begin{figure}[tbp]
  \centering
  \includegraphics[width=0.95\textwidth]{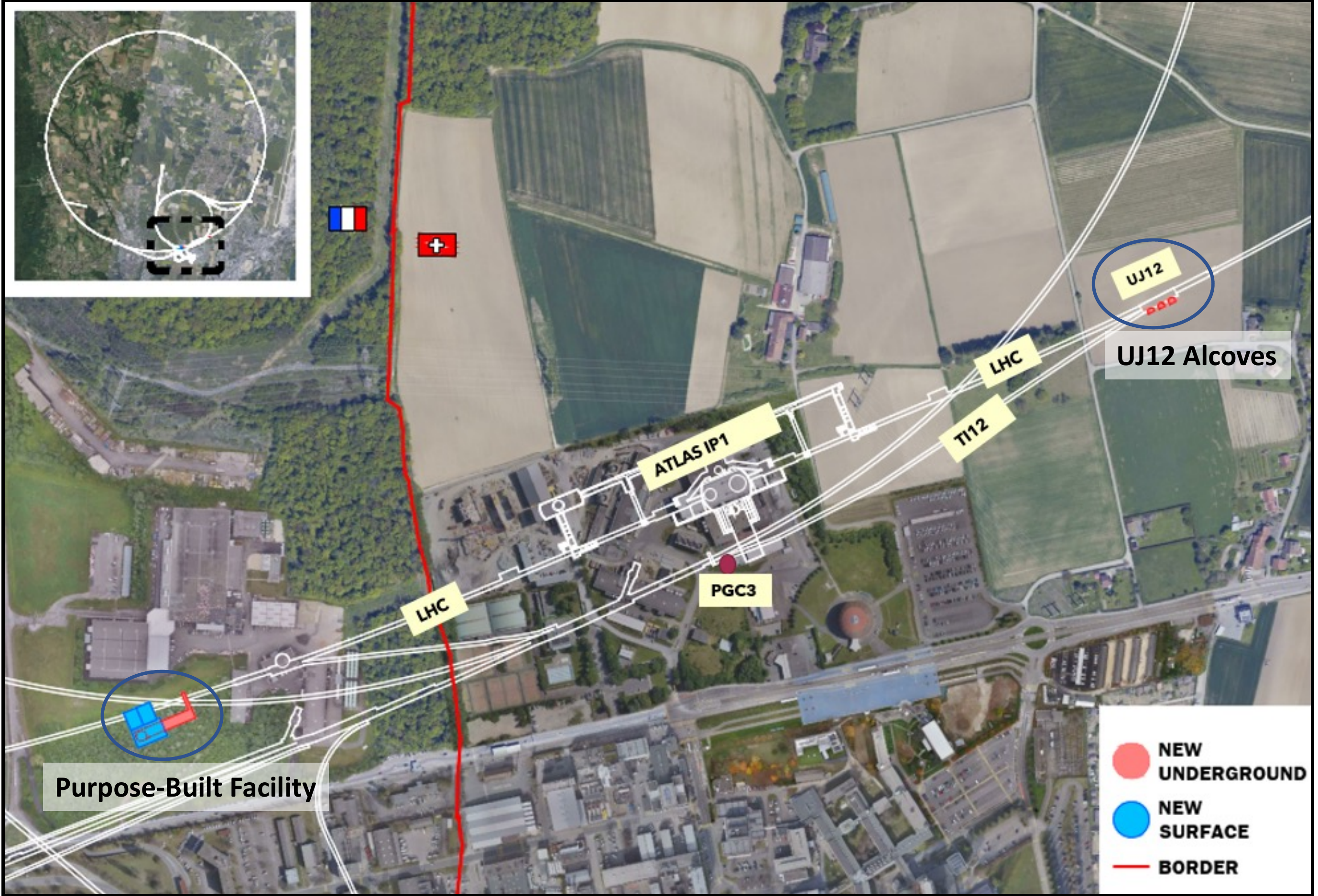}
  \caption{The locations of the two preferred FPF sites currently under consideration.  In the baseline option, a new cavern and shaft are excavated to create a new purpose-built facility, providing a detector hall along the LOS roughly 617–682~m west of the ATLAS IP (IP1) in France. An alternative option is to build alcoves to extend the existing UJ12 cavern, which is located 480–521~m east of the ATLAS IP in Switzerland.}
  \label{fig:CE:GeneralLayout}
\end{figure}

The civil engineering (CE) studies have been based on the requirement that the FPF be approximately 500-600~m away from a high-luminosity LHC IP on the LOS. Following an initial study of the existing LHC infrastructure and geological conditions, several options were considered to accommodate the facility around both the ATLAS IP (IP1) and the CMS IP (IP5). The options considered included constructing a new facility, which could be built around the needs of the experiments, and widening or expanding the existing LHC infrastructure, with the potential benefit of minimizing the cost and the disruption to LHC operations and reducing the overall schedule of the required CE works. The many possibilities were then narrowed down to two preferred options: (1) a new purpose-built facility, approximately 617–682~m west of the ATLAS IP, and (2) alcoves extending the existing UJ12 cavern, which is 480–521~m east of the ATLAS IP. The locations of these two options are shown in \cref{fig:CE:GeneralLayout}. Based on the expected costs and a number of important benefits for the experiments, the new purpose-built facility is currently considered the baseline option for the FPF.

This Chapter is organized as follows.  In \cref{sec:dedicated} the purpose-built facility is described, including the experimental cavern, access shaft, safety gallery, and support buildings and infrastructure. The necessary services for this facility are discussed in \cref{sec:services}.  We then present the UJ12 alcoves option in \cref{sec:alcoves}.  Preliminary estimates of the engineering costs for both options are given in \cref{sec:costs}, and the benefits of the purpose-built facility are summarized in \cref{sec:dedicatedfacilitybenefits}.  

We then describe the results of studies to evaluate the particle fluxes and backgrounds in the FPF.  \texttool{FLUKA}  simulation studies are described in \cref{sec:FLUKA}, and the implications of these studies for radiation protection studies are discussed in \cref{sec:RP}.  Complementary studies with the \texttool{BDSIM} and \texttool{PROPOSAL} simulation programs are described in \cref{sec:BDSIM} and \cref{sec:proposal}, respectively.  Finally, the location and design of a sweeper magnet to reduce the dominant muon background are summarized in \cref{sec:sweepermagnet}.

\section{Purpose-Built Facility  \label{sec:dedicated}}


Cvili engineering generally represents a signiﬁcant portion of the eﬀort for physics projects like the FPF. For this reason, CE studies are of critical importance to ensure a viable and cost-eﬃcient conceptual design. This section provides an overview of FPF CE studies for the purpose-built facility, including key considerations and the current design being studied.  As noted above, the purpose-built facility is now considered the baseline implementation of the FPF at CERN. The main advantage of having such a new facility is not being limited in size and length. In comparison to options extending the existing LHC infrastructure, the facility would be designed around the needs of the experiments. 

Studies for the purpose-built facility benefit from many similar projects carried out at CERN. A recent example carried out from 2018-21 is the CE works at Point 1 for the HL-LHC (the so called UPR), which involved the digging of similar shaft and tunnel/cavern structures and the installation of the needed services. Studies, designs, and lessons learned during the UPR construction helped to make rapid progress in the conceptual design of the FPF purpose-built facility.

\begin{figure}[tbp]
  \centering
  \includegraphics[width=0.9\textwidth]{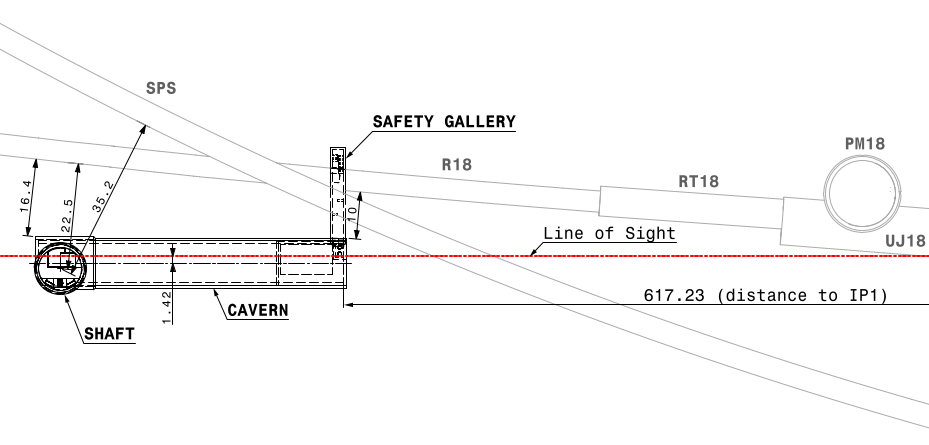}
  \caption{Situation plan of the purpose-built facility, located approximately 617-682~m to the west of IP1 on the French side of CERN land, 10~m away from the LHC tunnel. The SPS beamline is 30 m above the FPF cavern.}
  \label{fig:CE:NewFac1}
\end{figure}

The proposed location begins approximately 617~m from IP1 on the French side of CERN land, 10~m away from the LHC tunnel, as shown in \cref{fig:CE:GeneralLayout}. A more detailed view is given in \cref{fig:CE:NewFac1}. 

The major CE elements required to implement the FPF are:
\vspace*{-0.5em}
\begin{itemize}
\setlength\itemsep{-0.4em}
\item An 88~m-deep access shaft.
\item A 65~m-long experimental cavern.
\item A safety gallery connecting the FPF cavern to the LHC tunnel.
\item Support buildings and infrastructure.
\end{itemize}
\vspace*{-0.5em}
These are described in turn in the following subsections.

Vibrations during the digging of the shaft and cavern could have a detrimental effect on the performance of the LHC. A study is ongoing to understand what part of the FPF CE works could be carried out during LHC operations. Based on observations during the UPR works, it is expected that a significant part of the FPF works could be done during LHC running. Of course, the excavation of the safety gallery and the connection to the LHC tunnel will have to be done during an LHC shutdown.

\subsection{Experimental Cavern}

\begin{figure}[tbp]
  \centering
  \includegraphics[width=1.0\textwidth]{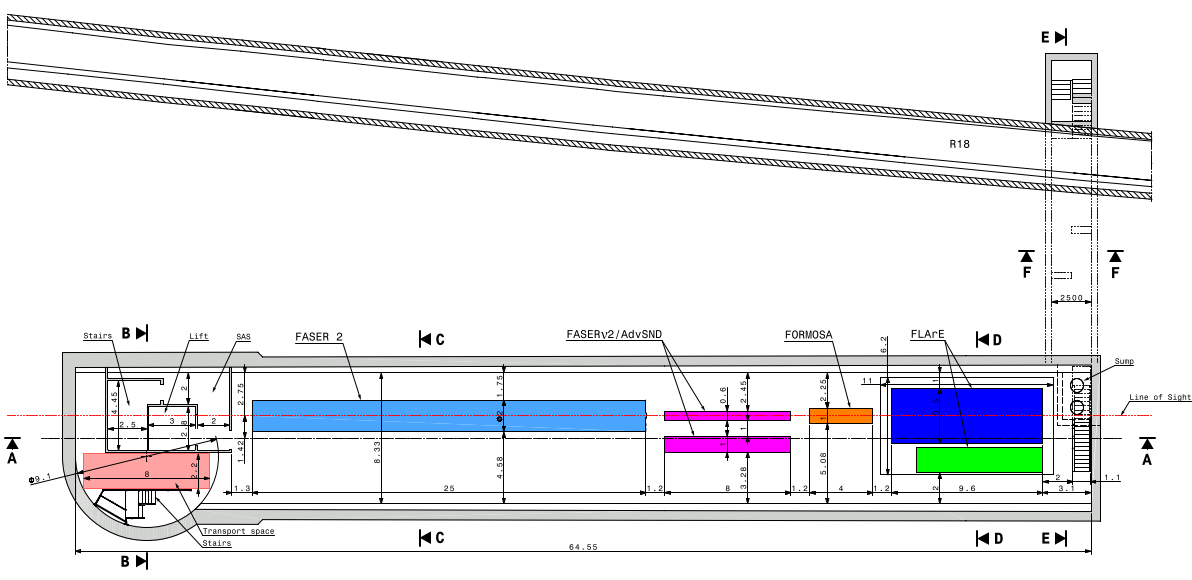}
  \caption{General layout of the FPF experimental cavern. The colored boxes indicate the possible experiments (and their dimensions) that could be installed in this option, including FASER2 to search for long-lived particles, FASER$\nu$2 and AdvSND to study neutrinos and search for new particles, FORMOSA to search for mCPs, and FLArE to detect neutrinos and search for DM. The green box is a possible cooling unit for FLArE.}
  \label{fig:CE:NewFac2}
\end{figure}

\begin{figure}[tbp]
  \centering
  \includegraphics[width=1.0\textwidth]{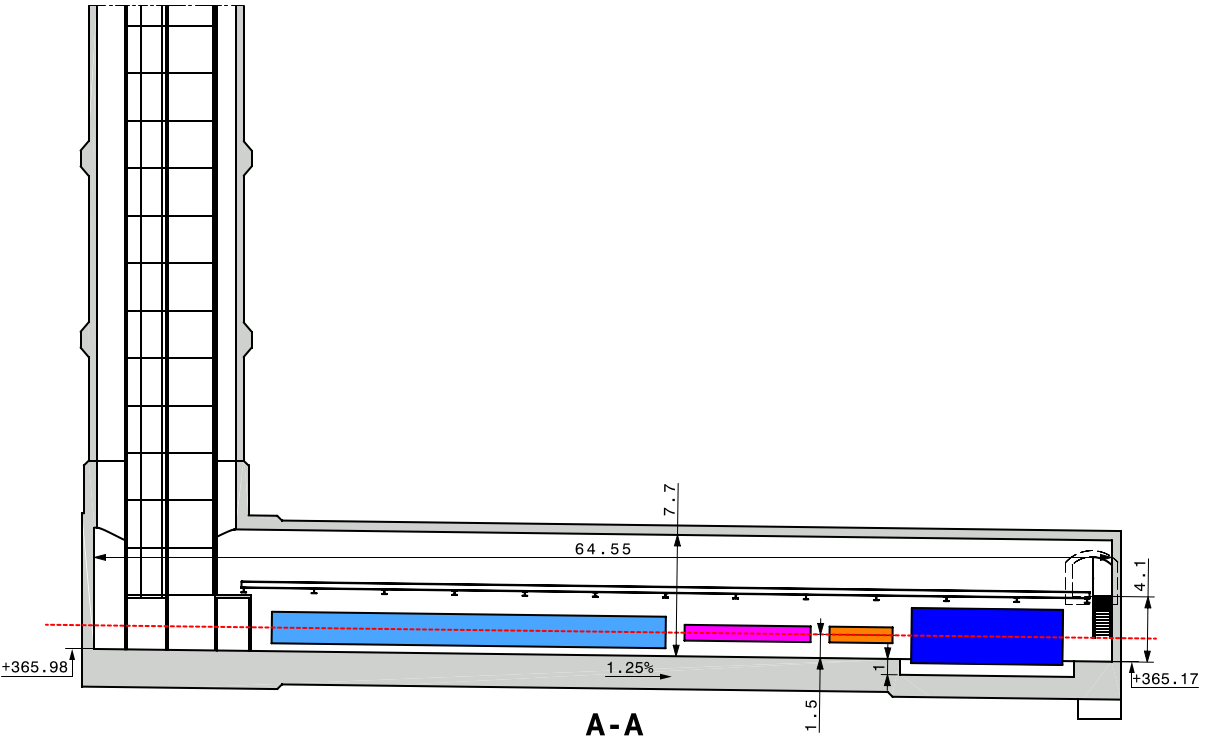}
  \caption{Section through the cavern with the proposed experiments and crane system.}
  \label{fig:CE:NewFac3}
\end{figure}

\begin{figure}[tbp]
  \centering
  \includegraphics[width=0.9\textwidth]{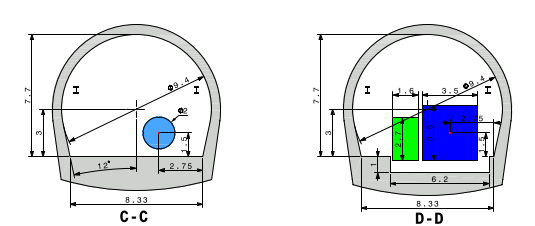}
  \caption{Cross sections through the cavern. Section C-C shows the FASER2 detector, and section D-D shows the FLArE detector in blue, a possible cooling unit in green, and a 1~m-deep trench for safety in the case of cold gas leakage.}
  \label{fig:CE:NewFac4}
\end{figure}

To meet the physics requirements, the experimental cavern is designed to be located on the LOS, beginning approximately 617~m from the ATLAS IP1 and 10~m away from the LHC tunnel. The cavern will be 65~m-long and 8.5~m-wide, leaving enough space around the experiments for easy access for transport and installation of the required services, as shown in \cref{fig:CE:NewFac2}. The ﬂoor level is set at 1.5~m under the LOS, with a 1.25\% fall towards IP1, following the inclination of the LOS. The experiments are centralized on the LOS and are served by a crane system along the cavern, as shown in \cref{fig:CE:NewFac3} and \cref{fig:CE:NewFac4}. For safety reasons, given the potential of cold gas leakage, a 1~m-deep trench is foreseen under the LAr detector (FLArE), as shown in \cref{fig:CE:NewFac4}.

\subsection{Access Shaft}

\begin{figure}[tbhp]
  \centering \hspace*{0.6in}
  \includegraphics[width=0.15\textwidth]{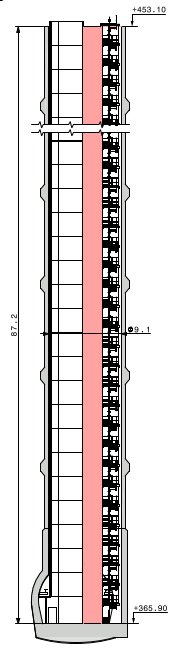} \hspace*{0.3in}
    \includegraphics[width=0.6\textwidth]{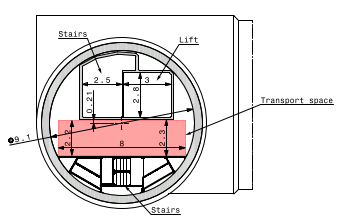}
  \caption{Shaft equipped with lift and staircase. The red area showing the space reserved for transport.}
  \label{fig:CE:NewFac5}
\end{figure}

The cavern is connected to the surface through an 88~m-deep and 9.1~m-diameter access shaft located on the top of the cavern. It will be equipped with a lift and staircase for access with enough space reserved for transport, as shown in the \cref{fig:CE:NewFac5}.

\subsection{Safety Gallery}

\begin{figure}[tbhp]
  \centering
  \includegraphics[width=1.0\textwidth]{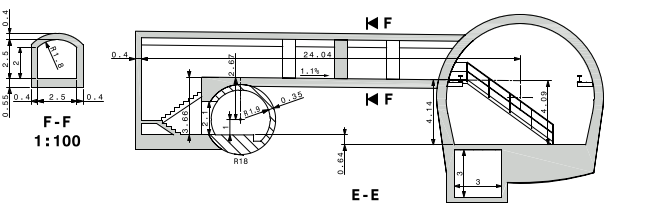}
  \caption{Section through the LHC tunnel, safety gallery, and experimental cavern.}
  \label{fig:CE:NewFac6}
\end{figure}

To comply with CERN’s safety requirements and avoid any possible dead ends, a safety gallery will connect the experimental cavern to the LHC and serve as a secondary emergency exit, as shown in \cref{fig:CE:NewFac6}. 

A key virtue of the dedicated facility is that access to the cavern will be possible during LHC operations. This would allow the installation of services and experiments, as well as maintenance and upgrades of experiments, to be possible at any time. A radioprotection (RP) study (discussed in detail in \cref{sec:RP}) has been carried out to assess the feasibility of this due to the radiation level in the cavern, which shows this is sensitive to the design of the safety gallery connecting the cavern to the LHC tunnel. 

\begin{figure}[tbhp]
  \centering
  \includegraphics[width=1.0\textwidth]{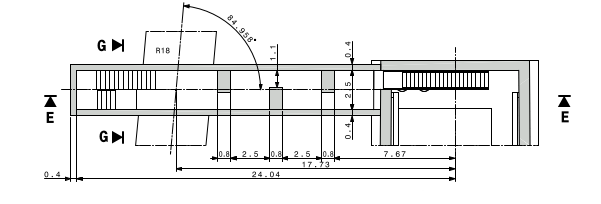}
  \caption{Plan view of the safety gallery.}
  \label{fig:CE:NewFac7}
\end{figure}

Based on the first RP study, the initial layout of the gallery has been modified to further reduce the dose levels in the cavern. As part of the modification, a third chicane wall was added, the thickness of the walls was increased from 40 cm to 80 cm, and a change was made in the location of the walls, as shown in \cref{fig:CE:NewFac6} and \cref{fig:CE:NewFac7}. A new RP study will be made to verify the effectiveness of the above-mentioned modifications. 

The safety gallery will only be used as an emergency escape route from the FPF cavern into the LHC tunnel, with an interlocked access door between the two. In the case that this door is opened, LHC operation will be automatically stopped for safety reasons.

\subsection{Support Buildings and Infrastructure}

\begin{figure}[hp]
  \centering
    \includegraphics[width=0.95\textwidth]{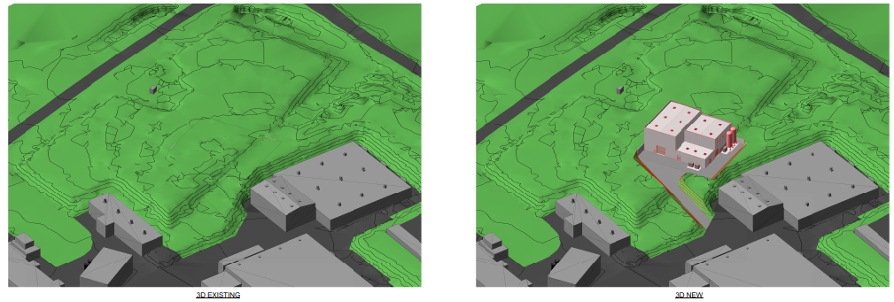}
  \caption{Proposed surface buildings for the facility (left shows the existing situation, right with the proposed surface buildings included).}
  \label{fig:CE:SB_1}
\end{figure}

\begin{figure}[bp]
  \centering
    \includegraphics[width=0.95\textwidth]{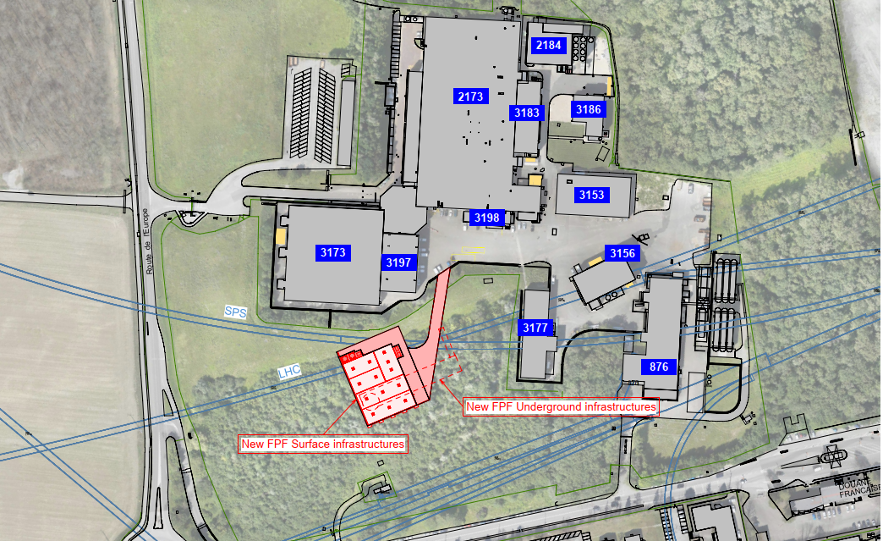}
  \caption{General layout of the proposed surface buildings.}
  \label{fig:CE:SB_2}
\end{figure}

Above ground, an access building and two auxiliary buildings are proposed to house the necessary infrastructure and utilities for the experiments, with their size being based on similar projects at CERN. The proposed layout of the buildings is shown in Figs.~\ref{fig:CE:SB_1} and \ref{fig:CE:SB_2}.

The 33~m-long and 21~m-wide access building located over the shaft will provide access from the ground level to the experiments for both personnel and equipment. It is designed as a basic steel portal frame structure with an internal height of 15~m, however the walls on the south and southwest will be part-formed from a retaining wall to support the excavation. The hall will be equipped with a 25~t overhead crane to lower the experiments into the cavern.

\begin{figure}[tbp]
  \centering
    \includegraphics[width=0.49\textwidth]{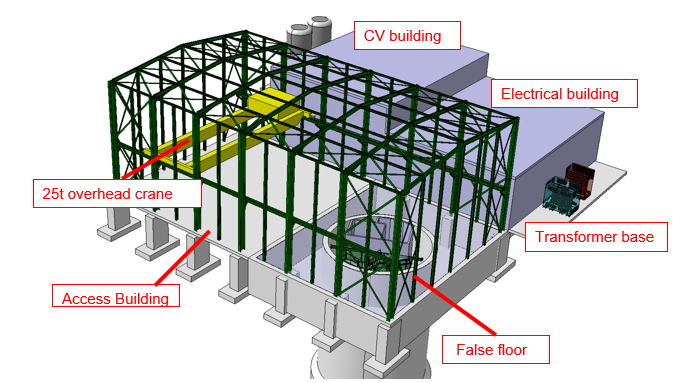}
  \includegraphics[width=0.49\textwidth]{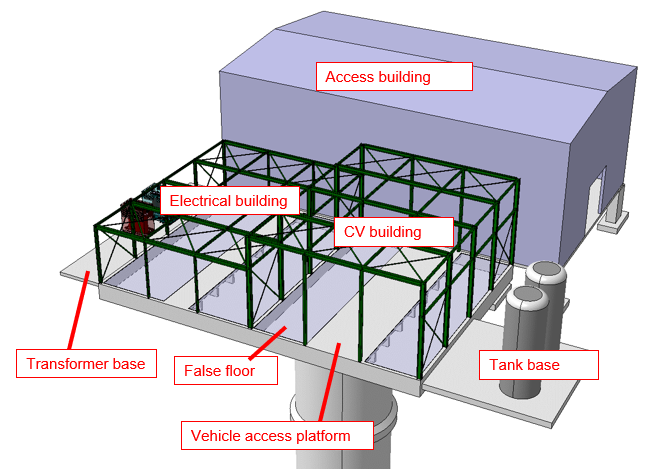}
  \caption{3D model of the proposed surface buildings.}
  \label{fig:CE:SB_3}
\end{figure}

The service buildings for the electrical, cooling, and ventilation infrastructure are also characterized as steel portal frame structures, similar to the access building. The cooling and ventilation building will be 20.5~m long, 21~m wide, and 13.5~m high, with the wall on the west side part-formed from a retaining wall to support the excavation. The electrical building will be adjacent to the cooling and ventilation building and will be 20.5~m long, 12~m wide, and 5.5~m high. Both buildings will have a 1.2~m-deep false floor to allow the services to be distributed into the shaft. The proposed design of the surface buildings is shown in \cref{fig:CE:SB_3}. 

\begin{figure}[tbp]
  \centering
    \includegraphics[width=0.80\textwidth]{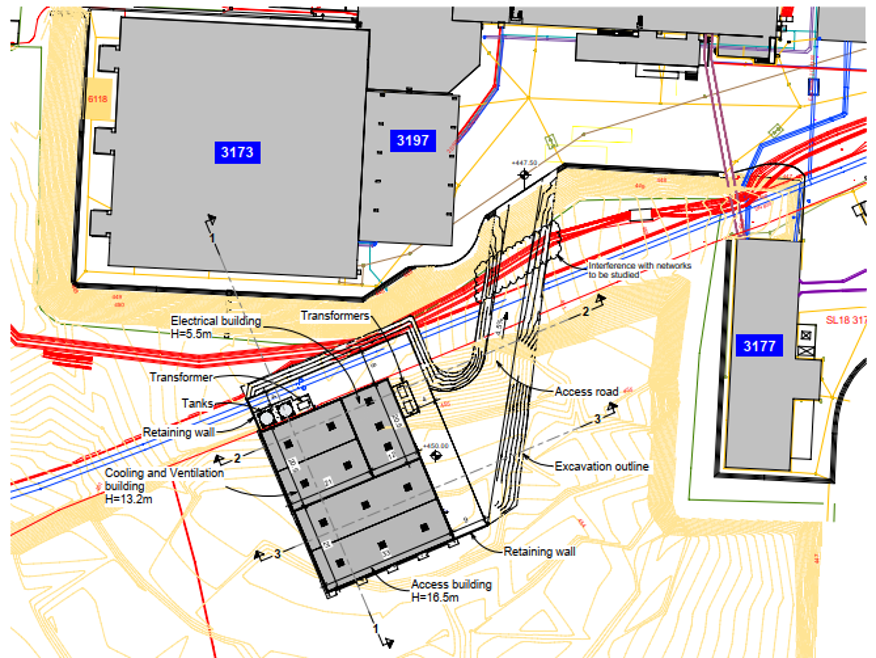}
  \caption{Section through the surface buildings showing the proposed excavated volume and existing site levels.}
  \label{fig:CE:SB_4}
\end{figure}

An access road will be provided, linking to the existing roads and infrastructure of the SM18 buildings at the northeast, as shown in \cref{fig:CE:SB_4}. A requirement for a maximum gradient of 6\% has been respected, in line with the requirements of CERN’s transportation services. 

\begin{figure}[tbp]
  \centering
      \includegraphics[width=1.\textwidth]{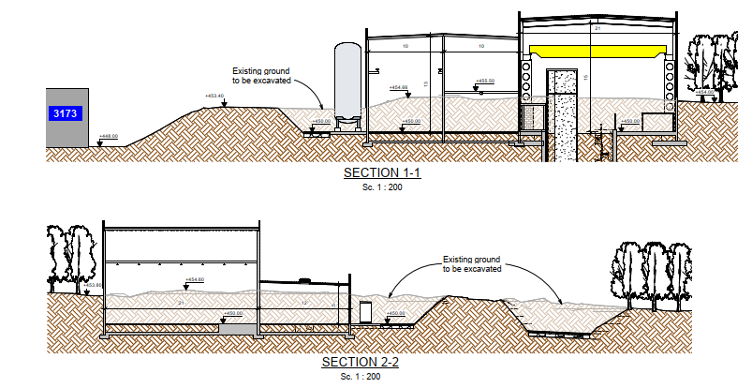}
    \includegraphics[width=1.\textwidth]{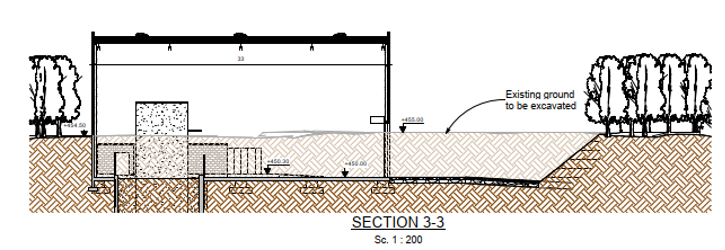}
  \caption{Section through the surface buildings showing the proposed excavated volume and existing site levels. The section lines are shown in \cref{fig:CE:SB_4}.}
  \label{fig:CE:SB_5}
\end{figure}

The volume of earthworks arising from the project is significant, predominantly because of the existing site levels and ground conditions. The proposed location has been previously used as a spoil disposal area, and the ground levels vary between 453-457~m above sea level, 5-8~m above the existing infrastructures in the surrounding area. As a result, to reduce the volume of the excavated material as much as possible, and taking into consideration the ground condition, the finished ground level of the buildings is proposed to be at 450 m, as shown in \cref{fig:CE:SB_5}.

\section{Services \label{sec:services}}


Given the early stage of the project and the lack of detailed designs and requirements for the proposed experiments in the FPF, further work is required for a detailed understanding of the needed services.

Based on similar underground facilities at CERN, it is clear that the purpose-built facility would need dedicated services, including electrical distribution, ventilation system, transport/handling infrastructure, communication infrastructure, access and alarm systems, and safety systems.  This contrasts with the case of the UJ12 alcoves option, to be discussed in \cref{sec:alcoves}, where most of the needed services would be available from close by within the LHC infrastructure. 

The details and costs for some of the needed infrastructure and services are fairly well known, whereas in other cases there are very large uncertainties, mostly stemming from the lack of detailed requirements. A very preliminary costing of the main services is summarized in~\cref{tab:services}.\footnote{Note that, for a large LAr TPC detector, an additional gas extraction system would be needed for safety reasons, and this has not been included in this costing.}  For the approximate overall costing of the facility to be presented in \cref{sec:costs}, a total of 15 MCHF for services has been assumed to also account for items not included in \cref{tab:services}. 

\begin{table}[tbp]
\setlength{\tabcolsep}{3pt}
\centering
\begin{tabular}{l||l|c} 
    \hline\hline
       {\bf Item} & {\bf Details} & {\bf Approximate cost} \\
        & & {\bf (MCHF)} \\
        \hline
        Electrical Installation & 2MVA electrical power & 1.5 \\
            \hline
        Ventillation & Based on HL-LHC underground installation & 7.0 \\
            \hline
        Access/Safety Systems & Access system & 2.5 \\
 & Oxygen deficiency hazard & \\ 
 & Fire safety & \\
 & Evacuation & \\
     \hline
     Transport/Handling & Shaft crane (25 t) & 1.9 \\
 Infrastructure & Cavern crane (25 t)  & \\
 & Lift & \\
            \hline
            {\bf Total}  & & {\bf 12.9} \\
\hline\hline
\end{tabular}
\caption{Breakdown of the main services and infrastructure for the dedicated FPF facility, with a very preliminary costing. This costing was done in mid-2021 and so, where applicable, reflects prices at that time.}
\label{tab:services}
\end{table}

\section{UJ12 Alcoves Option \label{sec:alcoves}}

An alternative to the purpose-built facility is the UJ12 alcoves option, in which the existing UJ12 cavern is expanded on one side with separate alcoves to accommodate the experiments and to provide the space needed around them. UJ12 is part of the LHC tunnel system and is 480-521~m west of the ATLAS IP1 at CERN’s site in Switzerland, as shown in \cref{fig:CE:GeneralLayout}.

\begin{figure}[tbhp]
  \centering
  \includegraphics[width=0.85\textwidth]{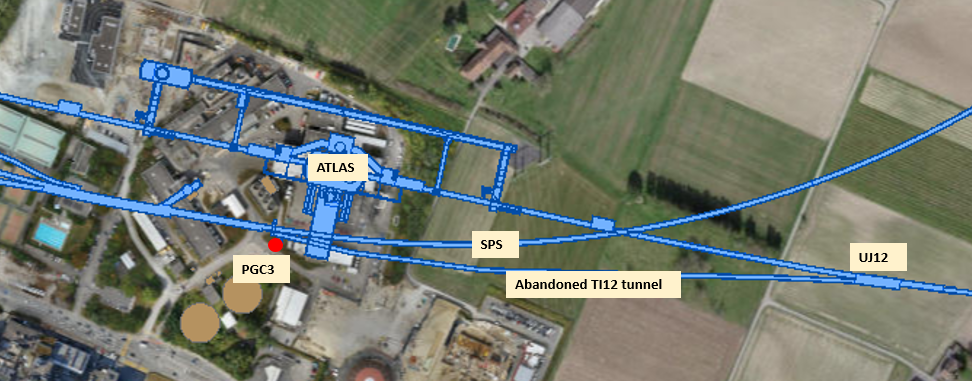}
  \caption{Access to the UJ12 cavern using the existing PGC3 shaft and passing through the 536~m-long TI12 tunnel.}
  \label{fig:CE:PGC3}
\end{figure}

\begin{figure}[tbhp]
  \centering
  \includegraphics[width=0.40\textwidth]{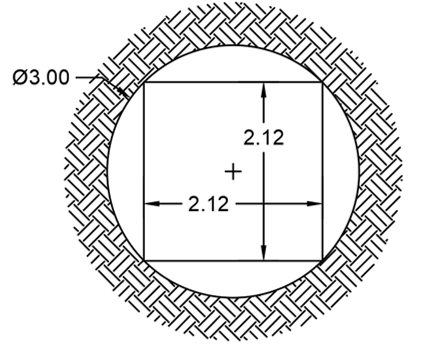}
  \caption{Internal dimensions of the PGC3 shaft.}
  \label{fig:CE:PGC3dimensions}
\end{figure}

A significant drawback of the UJ12 option is the difficulty of accessing the work site. Transport to UJ12 along the LHC from ATLAS is limited to small equipment that is approximately 1 m wide, prohibiting using this route for an excavating machine for any works in UJ12. As an access point, it is therefore envisaged to use the existing 40~m-deep PGC3 shaft located on the top of the abandoned TI12 tunnel and then pass through the 536~m-long TI12 tunnel, which currently houses the FASER experiment, as shown in \cref{fig:CE:PGC3}. The PGC3 shaft has an internal diameter of 3 m, as shown in \cref{fig:CE:PGC3dimensions}, which imposes significant space constraints, and the works need to be designed around what can be achieved with only small equipment.

\begin{figure}[tbhp]
  \centering
  \includegraphics[width=1.0\textwidth]{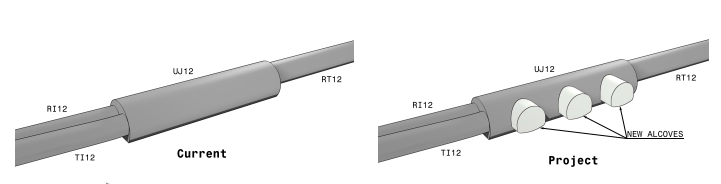}
  \caption{Proposed CE works for the UJ12 alcoves option.}
  \label{fig:CE:UJ12works1}
\end{figure}

\begin{figure}[tbhp]
  \centering
  \includegraphics[width=0.7\textwidth]{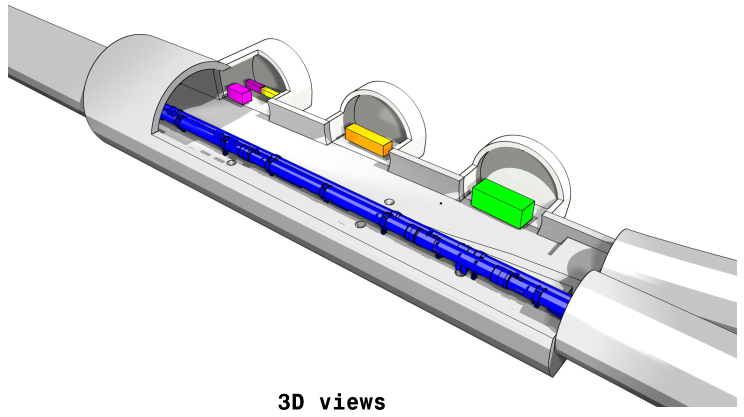}
  \caption{3D model of the UJ12 alcoves option with possible experiments shown.}
  \label{fig:CE:UJ12works2}
\end{figure}

\begin{figure}[tbhp]
  \centering
  \includegraphics[width=0.8\textwidth]{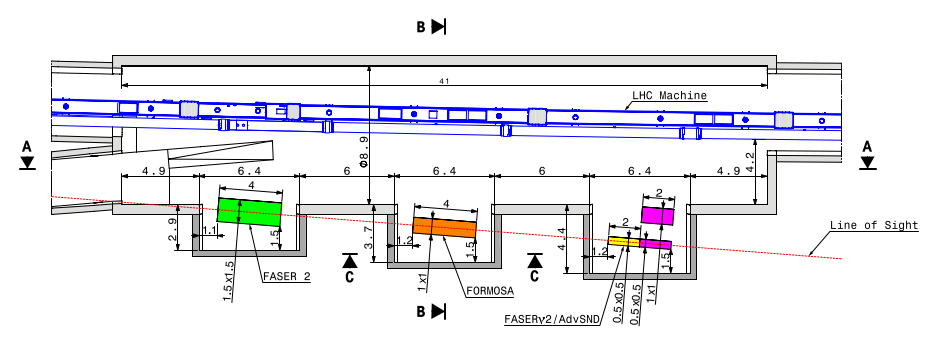}
  \caption{Plan view of the UJ12 alcoves option. The coloured boxes indicate the possible experiments (and their dimensions) that could be installed in the alcoves, including FASER2 to search for long-lived particles, FORMOSA to search for millicharged particles, and FASER$\nu$2 and AdvSND to detect neutrinos and search for DM.}
  \label{fig:CE:UJ12works3}
\end{figure}

\begin{figure}[tbhp]
  \centering
  \includegraphics[width=0.8\textwidth]{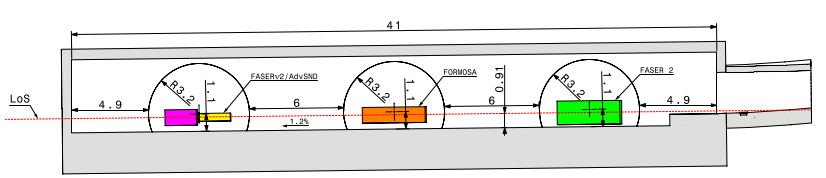}
  \caption{Longitudinal section through the UJ12 cavern with alcoves and possible experiments.}
  \label{fig:CE:UJ12works4}
\end{figure}

Following the conceptual design studies, the baseline layout includes three alcoves of 6.4~m width, but with different lengths of 2.9 m, 3.7 m, and 4.4 m, as shown in Figs.~\ref{fig:CE:UJ12works1}, \ref{fig:CE:UJ12works2}, \ref{fig:CE:UJ12works3}, and \ref{fig:CE:UJ12works4}. It must be noted that the impact of the foreseen works on the existing wall of the cavern and the cavern itself has yet to be fully assessed. All the works must be carried out in a way that minimises the impact on the existing facility. It is assumed that all the existing services and equipment will be removed from the cavern prior to the works. This would include temporarily removing 4 LHC dipole magnets, a 60~m-long section of the QRL cryogenic line, and also electrical and ventilation equipment. Initial studies suggest that this would be possible during a Long Shutdown between LHC runs, but it would entail significant work for many CERN teams.

\section{Engineering Costs \label{sec:costs}}

The cost of construction of the FPF is difficult to estimate at such an early stage of the study. The variability of ground conditions, inflation, change of scope, and lack of detailed design means that developing a high level of confidence is not possible. For FPF costing purposes, a comparative costing was adopted, based on the presented layouts. 

A very preliminary cost estimate suggests that the cost of the dedicated purpose-built facility, including 15 MCHF for the needed services, as discussed in \cref{sec:services}, would be about 40 MCHF. The cost of the UJ12 alcoves option would be about 15 MCHF. The accuracy of the estimates is considered Class 4 -- Study or Feasibility, with the actual cost possibly 15–30\% lower or 20–50\% higher~\cite{CEcosting}. Until the project requirements are further developed, it is suggested that a suitable band to adopt would be 20\% lower to 40\% higher for CE costs.

\section{Choice of Baseline Facility \label{sec:dedicatedfacilitybenefits}}

Given the preliminary costing of the two options studied, with only a factor of 2.7 difference in cost, there is a strong preference from the physics community to make the purpose-built facility the baseline FPF option. The main reasons for this are:
\vspace*{-0.5em}
\begin{itemize}
\setlength\itemsep{-0.4em}
\item Much more space for the experiments, allowing for them to be designed/optimized for physics reach rather than around the available space. This also allows for experiments to be placed somewhat off-axis,\footnote{It is worth noting that as the detector is moved off-axis, a larger detector is needed to fully cover a given rapidity range. For example, a 1~m$^2$ detector centred 1~m away from the LOS, will cover only approximately 16\% of the full solid angle corresponding to this rapidity range ($6.7 < |y| < 7.8$).} as is motivated by some of the physics goals;
\item Much better access for equipment to be transported into the experimental area, since there are strict size and weight limitations for transporting items to UJ12 in the LHC complex;
\item A LAr TPC detector (FLArE) with a strong physics case, could be installed in the dedicated facility, but not the UJ12 alcoves option, given safety requirements;
\item For the dedicated facility, it is expected that people can access the experiments during beam operations (pending the RP study detailed in~\cref{sec:RP}), which allows improved flexibility in terms of scheduling of detector installation and maintenance;
\item Beam backgrounds will be minimized for experiments in the dedicated facility, which may be important for experiments searching for rare and low-energy signatures (such as neutrino interactions or dark matter scattering);
\item The dedicated facility's location would allow a factor of two larger lever arm for a sweeper magnet to deflect background muons away from the LOS, as discussed in~\cref{sec:sweepermagnet}. 
\end{itemize}

\section{FLUKA Studies of the FPF Environment and Backgrounds \label{sec:FLUKA}}


The physics goals of the FPF require a low-background environment so that weakly-interacting particles and very rare processes may be observed.  In addition, an important potential benefit of the purpose-built facility discussed above is that the radiation backgrounds in the cavern may be low enough to allow access to the FPF even during LHC operations.  

For all of these aspects, it is important to have an accurate understanding of the particle fluxes and radiation environment in the FPF cavern.  In this section, preliminary results from \texttool{FLUKA}  are presented, with a particular focus on the fluxes of high-energy muons that are the dominant particle physics background for many FPF signals.  \texttool{FLUKA}  studies may also be used to determine the low-energy radiation backgrounds, and their implications for radiation processes, access to the cavern, and safety will be discussed in the following \cref{sec:RP}.

\subsection{Introduction to FLUKA}

At particle colliders, it is essential to characterize the radiation field to cope with the multiple effects of the interaction of regular and accidental beam losses on machine and detector components. To quantify these effects starting from the relevant loss terms, multipurpose Monte Carlo codes are a critical tool, enabling the evaluation of macroscopic quantities through the microscopic description of particle transport and interactions in matter. This requires tracking through magnetic fields, as well as accounting for all applicable electromagnetic and nuclear processes over an extremely wide energy range. The code's reliability is verified through individual benchmarking of physics models against exclusive data. A profitable calculation requires modeling the machine and detector geometry, including material information, to a challenging degree of accuracy. This allows in turn for an inclusive validation against measurements from extended monitor systems (see below).

At CERN, \texttool{FLUKA}~\cite{FLUKA:web,FLUKA:new,Battistoni:2015epi} is the reference tool to assess the machine protection aspects and the complementary radiation protection scope, as well as the machine-induced background to experiments. It is regularly and extensively used for the whole accelerator chain, from the beam dump design of low-energy injectors as Linac4, up to LHC collimation and the High Luminosity (HL) upgrade of the LHC. For future colliders, \texttool{FLUKA}  also plays a crucial role, starting in the early stages of planning, for both accelerator and detector design and for both hadron and lepton machines. This means that simulations have to deal with protons up to some million TeV (the beam energy with a target at rest required to reach 100 TeV center-of-mass collisions) down to the lowest transport limit of 100 eV photons (for the study of lepton ring synchrotron radiation). Such a task calls for the continuous improvement of the different interaction models, having in mind that, despite the very high energy of beams of interest, several quantities, such as those related to radionuclide inventory, are extremely sensitive to the low-energy nuclear physics ingredients ruling the reaction fragment de-excitation.

Together with this physics-oriented effort, notable technical developments have made it possible to automatize the construction of consistent geometry models of several-hundred-meter accelerator portions~\cite{Mereghetti:2012zz,Vlachoudis:2009qga}.

The operation of the LHC has provided the opportunity to probe the degree of reliability of simulations performed over the long course of the LHC design phase. In particular, the Beam Loss Monitor (BLM) system, consisting of a few thousand ionization chambers all along the 27 km beam line, provides on-line measurements of the energy released by the particle shower originated by beam particle interactions, and it triggers beam aborting if the detected values exceed pre-defined thresholds. Particle shower calculations make it possible to predict BLM signals for different loss scenarios, correlating them at the same time with the energy deposition levels in the most exposed or sensitive elements and the radiation levels, namely differential particle fluences, in areas of interest. Various representative examples of comparisons between BLM measurements and \texttool{FLUKA}  predictions are presented in Ref.~\cite{Lechner:2019vgh}.

\subsection{The FLUKA Model of the ATLAS Insertion}

For the FPF, it is particularly important to understand the particle debris generated by colliding beams at the ATLAS IP during HL-LHC running. Given an inelastic cross section of about 80 mb (including diffractive events), the collision products carry almost 7 kW towards each side of IP1 at the ultimate instantaneous luminosity of $7.5 \times 10^{34}~\cm^{-2}~\text{s}^{-1}$ for 7 TeV proton operation.  Only a limited fraction of this power ($<5$\%) impacts the detector itself, while 80\% of it is transported out of the experiment’s cavern by high-energy, forward-angle particles travelling inside the beam vacuum chamber through the accelerator elements. These particles amount on average to 6 per single $pp$ collision out of the 155 particles emerging on average from IP1 after neutral pion decay, and are mostly photons, charged pions, protons, and neutrons.  The accelerator elements will be protected by a 1.8 m-long copper absorber, called the TAXS,  located near the interface between the cavern and the LHC tunnel about 20~m from IP1, and featuring a 60~mm-diameter cylindrical aperture, considerably larger than the present one. The TAXS absorbs another 8\%, while the following 60~m-long string composed of single-bore superconducting quadrupoles (the final focus triplet), corrector magnets and separation dipole (D1) take 22\% (as part of the above 80\%, which does not include the TAXS fraction, though). In fact, the majority of the energetic pions matching the TAXS aperture are then bent by the magnetic field onto the massive beam screen structure traversing the aforementioned string and embedding tungsten layers to protect the 150~mm-diameter aperture of the magnets.

\begin{figure}[tbp]
  \centering
  \includegraphics[width=0.8\textwidth]{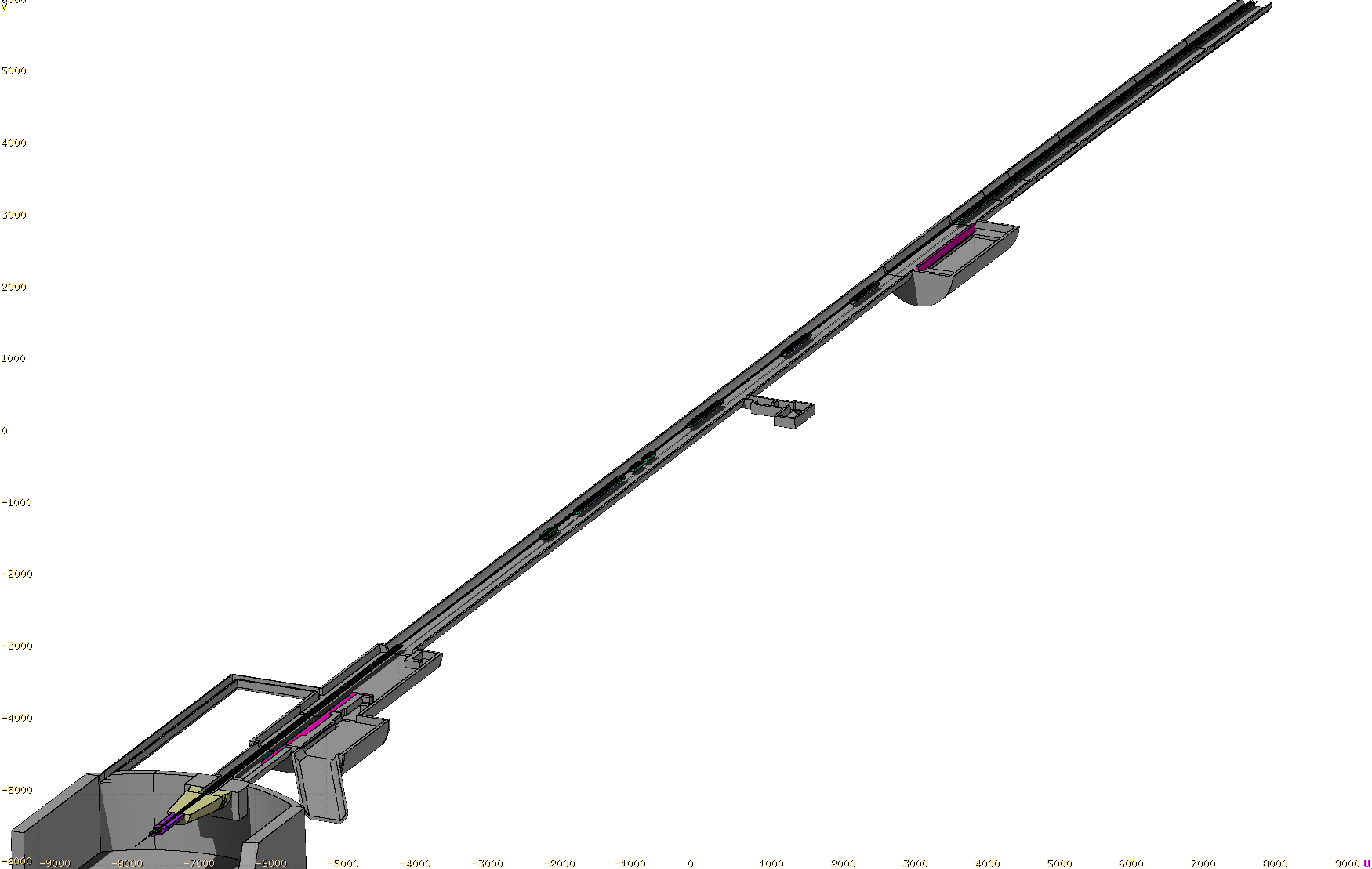}
  \caption{View of a portion of the \texttool{FLUKA}  model of the HL-LHC. On the bottom left, one can see the ATLAS forward shielding surrounding the TAXS. The LHC tunnel extends to the west of the ATLAS IP up to the first DS cells (top right), with the violet wall shielding the RR alcove that sits at 250~m from the ATLAS IP.}
  \label{fig:FLUKA:IR1}
\end{figure}

A new version of the other main absorber (TAXN) will sit 45~m after the D1, incorporating the transition between one single central aperture and two separate symmetrical apertures of 88~mm-diameter and intercepting the LOS of neutral particles coming from IP1. Its effectiveness in protecting downstream elements will be slightly weakened by the adoption of horizontal beam crossing in IP1, since, in this case, the axis of the debris cone, instead of hitting exactly in-between the two TAXN twin apertures (as for vertical crossing), moves closer to the external aperture, as a function of the crossing angle. For the baseline half crossing angle of 250 $\mu$rad, the ATLAS TAXN is expected to absorb 20\% of the one-side debris power.

An 8~m-long superconducting twin bore recombination dipole will open to the matching section, taking 100~m of the experimental insertion up to the Dispersion Suppressor (DS). The matching section features four main superconducting quadrupole assemblies (numbered Q4 to Q7) and ends 270~m from IP1. Less than 10\% of the one-side debris power, mostly carried by photons, neutrons, and protons, is absorbed in this machine segment, concentrated in the first (TCLPX4) of the three metallic horizontal collimators to be installed on the outgoing beam aperture to further shield the cold magnet coils. Most of the remaining debris power goes to the preceding tunnel walls. 

The third TCL collimator, in front of Q6, can also provide an effective cleaning of the initial part of the DS, where the beam lines are bent by the LHC main dipoles and no layout modification is planned for the HL-LHC era. Nevertheless, beyond the TCL6 range, losses take place in the DS odd half-cells, according to the periodicity of the single turn dispersion, as already regularly observed. They consist of protons that underwent diffraction at IP1 and are affected by a magnetic rigidity deficit of the order of 1\%, leading them to touch the beam screen of the outgoing beam chamber in the horizontal plane towards the center of the ring.

\subsection{Radiation Characterization in the Dispersion Suppressor}

The implementation in \texttool{FLUKA}  of the ATLAS insertion model described above has enabled multiple studies, mainly oriented to HL-LHC design~\cite{ZurbanoFernandez:2020cco}, but also serving other purposes, for instance, the evaluation of background and neutrino fluxes for SND@LHC~\cite{Ahdida:2750060}. We present here preliminary results for the calculation of the flux of high-energy muons reaching FPF along the ATLAS LOS, which originate in both primary IP1 collisions and downstream shower development. 

\begin{figure}[tbp]
  \centering
  \includegraphics[width=0.45\textwidth]{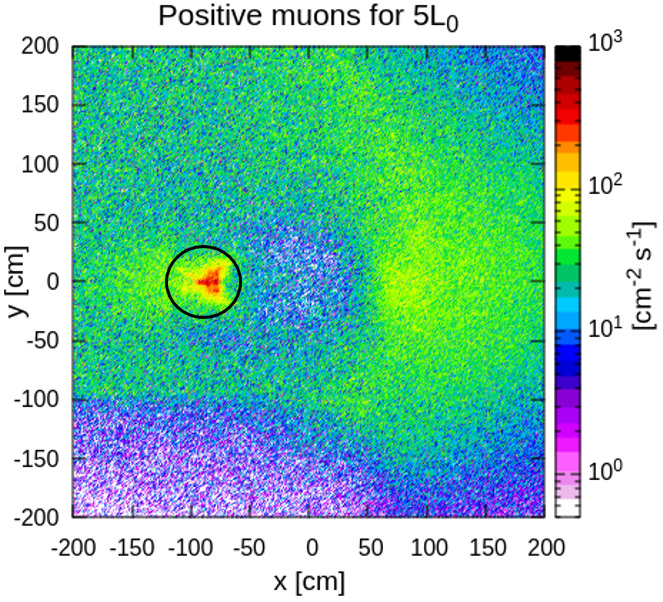}
  \includegraphics[width=0.45\textwidth]{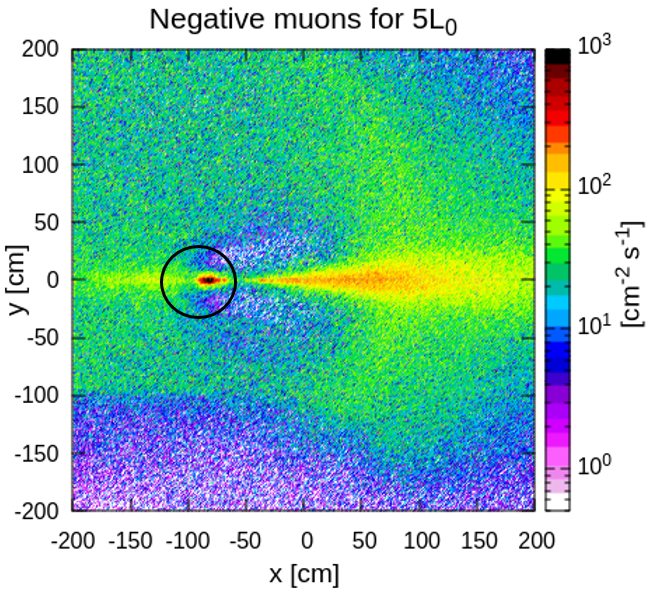}
  \caption{Positive (left) and negative (right) muon fluence rate distributions over a 16 m$^{2}$ square centered on the ATLAS LOS at 348.5~m from IP1. Values are normalized to the HL-LHC \textit{nominal} luminosity of $5 L_0 = 5 \times 10^{34}$ cm$^{-2}$ s$^{-1}$. The black circle represents the superconducting magnet transverse section, indicating the accelerator line position at the end of half-cell 9. }
  \label{fig:FLUKA:muons2D}
\end{figure}

\begin{figure}[tbp]
\centering
\includegraphics[width=0.85\textwidth]{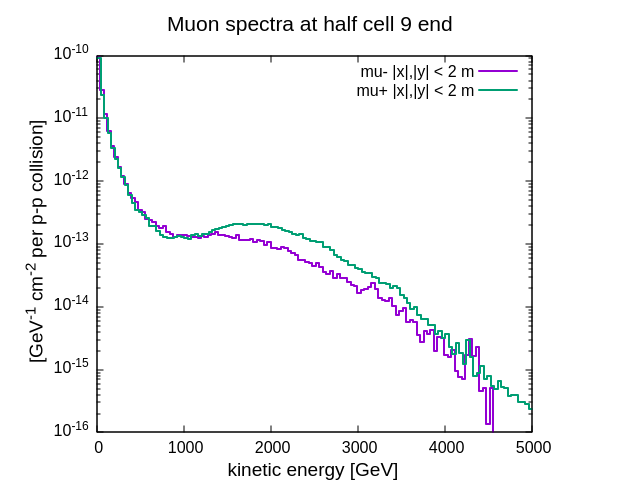}
\caption{Positive (green) and negative (violet) muon fluence spectra averaged over a 16 m$^{2}$ square centered on the ATLAS LOS at 348.5~m from IP1. Values are normalized to one $pp$ inelastic collision; the ultimate HL-LHC luminosity of $7.5 \times 10^{34}$ cm$^{-2}$ s$^{-1}$ corresponds to $6 \times 10^{9}$ inelastic $pp$ collisions per second.}
  \label{fig:FLUKA:muonspe}
\end{figure}

Thanks to a dedicated optimization, featuring a suitable transport threshold adjustment and combined biasing techniques, including artificially increasing the decay probability of parent mesons and controlling statistical weight fluctuations, it was possible to produce a meaningful muon sample just upstream of the proposed sweeper magnet location, as shown in \cref{fig:FLUKA:muons2D}.
For both positive and negative muons, the fluence is maximal on the accelerator line, a bit less than 1~m from the ATLAS LOS at $x=y=0$. In contrast, a fluence minimum is found at the LOS for positive muons, while negative muons display a concentration on the horizontal plane ($y=0$) on the external side of the ring, compatible with the bending action of the DS dipole field. The positive and negative muon energy spectra, averaged over the square of \cref{fig:FLUKA:muons2D}, are reported in \cref{fig:FLUKA:muonspe}, indicating that positive muons are predominant above 1 TeV.

The muon samples shown constitute the source term for a second step simulation implementing the sweeper magnet (see \cref{sec:sweepermagnet}) and the further propagation of muons to the FPF through the rock.  The goal of this second step will be to quantify the muon background, as well as the related dose equivalent contribution (see \cref{sec:RP}).

\subsection{Validation of FLUKA Estimates \label{sec:validation}}

In preparation for the FASER experiment, {\em in situ} measurements were made during 2018 LHC running in the TI12 and TI18 tunnels in the LHC, 480~m from IP1 and along or close to the LOS. Measurements were made using emulsion-based detectors, which can be used to determine the number of charged particles that traverse the detector while it is installed and very accurately measure the track angles.  There is, however, no knowledge of the time the particles cross the detector. In addition to this, measurements were made with a TimePix3 beam loss monitor (BLM), which is able to measure the rate of charged particles with excellent time resolution, but, given a lack of calibration, could not give an absolute rate. The measurements are discussed in detail in the FASER Technical Proposal~\cite{FASER:2020gpr}. The measurements were used to validate the \texttool{FLUKA} estimate of the muon flux for the 2018 LHC running conditions on the LOS.

The emulsion detector observed a clear peak of charged particles entering the detector with an angle consistent with the direction from IP1.  The number of particles observed in this angular peak was $(1.2 - 1.9) \times 10^4~\text{fb}~\text{cm}^{-2}$, when normalised by the luminosity that the detector was exposed to.  This can be compared to the estimate from \texttool{FLUKA} of $2.0 \times 10^4~\text{fb}~\text{cm}^{-2}$ with an uncertainty of ${\cal O}(50\%)$. The \texttool{FLUKA} simulation result is consistent with the measurement within the expected uncertainties from \texttool{FLUKA} . The measurements with the BLM showed that the observed rate is correlated with the instantaneous luminosity in IP1 during an LHC fill, with the rate falling off during the fill as the luminosity decreases. Again, this is consistent with the \texttool{FLUKA}  expectation that the background muon rate is originating from collision debris. With two circulating beams that were not colliding, the measured rate was very close to zero.  The agreement between the \texttool{FLUKA}  simulations and {\em in situ} measurements for the 2018 setup of the LHC gives confidence that \texttool{FLUKA}  estimates for the HL-LHC will describe the background with reasonable accuracy. To further validate the simulations of the muon backgrounds in this region, when Run 3 begins in 2022, the FASER Collaboration plans to make further measurements using small emulsion detectors placed at various distances from the LOS.

\section{Radiation Protection Studies \label{sec:RP}}


\subsection{Radiation Protection at CERN}

The CERN Radiation Protection (RP) rules are provided in the so-called ``Safety Code F''~\cite{SafetyCodeF, Forkel-Wirth:1533023}. The objective of Safety Code F is to define the rules for the protection of personnel, the population, and the environment from ionising radiation produced at CERN. Safety Code F is based on and updated to the most advanced standards of European and other relevant international legislations, including the legislation of CERN host states France and Switzerland. 

With regard to the design of new facilities, different RP aspects must be taken into account at the design level, including shielding requirements, radiation levels during operation (prompt) and technical stops (residual), area classification, radiation monitoring, and the activation for future disposal of radioactive wastes. Among these, area classification and dose limits are discussed here. These aspects are particularly relevant, as they determine whether access to the new experimental cavern will be possible during LHC beam operation. 

Areas inside CERN’s perimeter are classified as a function of the effective dose a person would receive during his stay in the area under normal working conditions and routine operation. The potential external, as well as internal, exposures must be taken into account when assessing the effective dose researchers may receive when working in the area considered. The exposure limitation in terms of effective dose is ensured by limiting correspondingly the operational quantity ambient dose equivalent rate $\dot{H}^{*}(10)$ for exposure from external radiation, and the action levels of specific airborne radioactive material (airborne radioactivity) and specific surface contamination at the corresponding workplaces for exposure from incorporated radionuclides. In addition, exposure of people working on the CERN site, the public, and the environment must remain below the dose limits under normal, as well as abnormal, conditions of operation. \cref{tab:RP:area_classificaiton} shows the limits for area classification of Non-designated and Supervised Radiation Areas at CERN.  These limits are relevant for the experimental FPF cavern, which will be discussed in the following.

\begin{table}[tbp]
\centering
\begin{tabular}{c||c|c} 
\hline\hline
\multirow{2}{*}{Area} & Annual dose limit & Ambient dose equivalent rate  \\
& (year) [mSv] & (low-occupancy) [$\mu$Sv/h] \\ 
\hline
Non-designated & 1 & 2.5 \\
Supervised & 6 & 15 \\ 
\hline\hline
\end{tabular}
\caption{The effective dose limits for area classification at CERN for Non-Designated and Supervised Radiation Areas. Dose limits for Controlled Radiation Area not reported since not relevant for this study. FPF is considered a low-occupancy area, i.e. $<20\%$ working time.}
    \label{tab:RP:area_classificaiton}
\end{table}

\subsection{Radiation Protection FLUKA Simulations}

The RP-\texttool{FLUKA}  simulations aim to determine the prompt radiation levels in the new purpose-built FPF cavern and in the shaft for different scenarios (normal/abnormal LHC operation); verify the accessibility of the experimental cavern during LHC and SPS operation; and check the effectiveness of the chicane in the safety tunnel.

The \texttool{FLUKA}  model of the LHC tunnel (\cref{fig:FLUKA:IR1}) presented in \cref{sec:FLUKA} has been modified to include a detailed model of the new experimental cavern. This model contains $\sim500$~m of LHC beam line elements (from 250~m to 750~m to the west of IP1) and the experimental cavern and its access shaft, based on technical drawings provided by CERN CE. The safety tunnel connecting the LHC tunnel to FPF includes a chicane made of $2\times40$~cm concrete walls, representing the baseline layout at the time of computation. Simulations were conducted with and without the chicane walls to verify its effectiveness.

Several source terms were considered to simulate the operational and accidental scenarios relevant for the RP calculations: 
\vspace*{-0.5em}
\begin{itemize}
\setlength\itemsep{-0.4em}
    \item Beam-gas interactions: this source term is relevant for normal LHC operation, as it originates from inelastic interactions of the 7 TeV proton beam with residual gas within the beam pipes.
    \item Direct muon component from IP1: this source term is also relevant for normal LHC operation, and originates from muons directly streaming from ATLAS's $pp$ collisions and from the particle showers produced in the Long Straight Section (LSS) by the collision debris.
    \item Loss of the LHC beam: this source is relevant only for abnormal LHC operation (accidental scenario). It considers the loss of the full 7 TeV proton beam on the MB.B15R1, the superconductive dipole placed in front of the safety tunnel connection to the LHC tunnel.
    \item Loss of the SPS beam: this source term is relevant for the access of the FPF shaft, due to the vicinity of the two infrastructures, and it considers the loss of the 450 GeV proton beam in the SPS tunnel.
\end{itemize}

In all LHC simulations, Beam 1, the clockwise beam traveling from the ATLAS IP toward the FPF, was simulated. In addition, the HL-LHC beam intensity was simulated using a scaling/normalization factor to consider 2748 bunches and $2.3\times10^{11}$ proton per bunch. With regard to beam-gas interactions, a conservative residual gas-density of $1.0\times10^{15}$ H$_{2}$~m$^{-3}$ was used.  As reported in Ref.~\cite{Bruning:782076}, this value ensures a 100 hour beam lifetime. Recent studies conducted at CERN~\cite{Bilko:2725326} indirectly determined lower residual gas densities during Run 2 operations ($2.25-0.25\times10^{13}$ H$_{2}$~m$^{-3}$), with higher values registered at the beginning of the physic run.

Finally, the prompt ambient dose equivalent was scored in the LHC tunnel, the safety tunnel, and the new experimental cavern by using a Cartesian XYZ mesh.

\subsection{Radiation Protection Aspects and Constraints}

\begin{figure}[tbhp]
\centering
\includegraphics[width=0.87\textwidth]{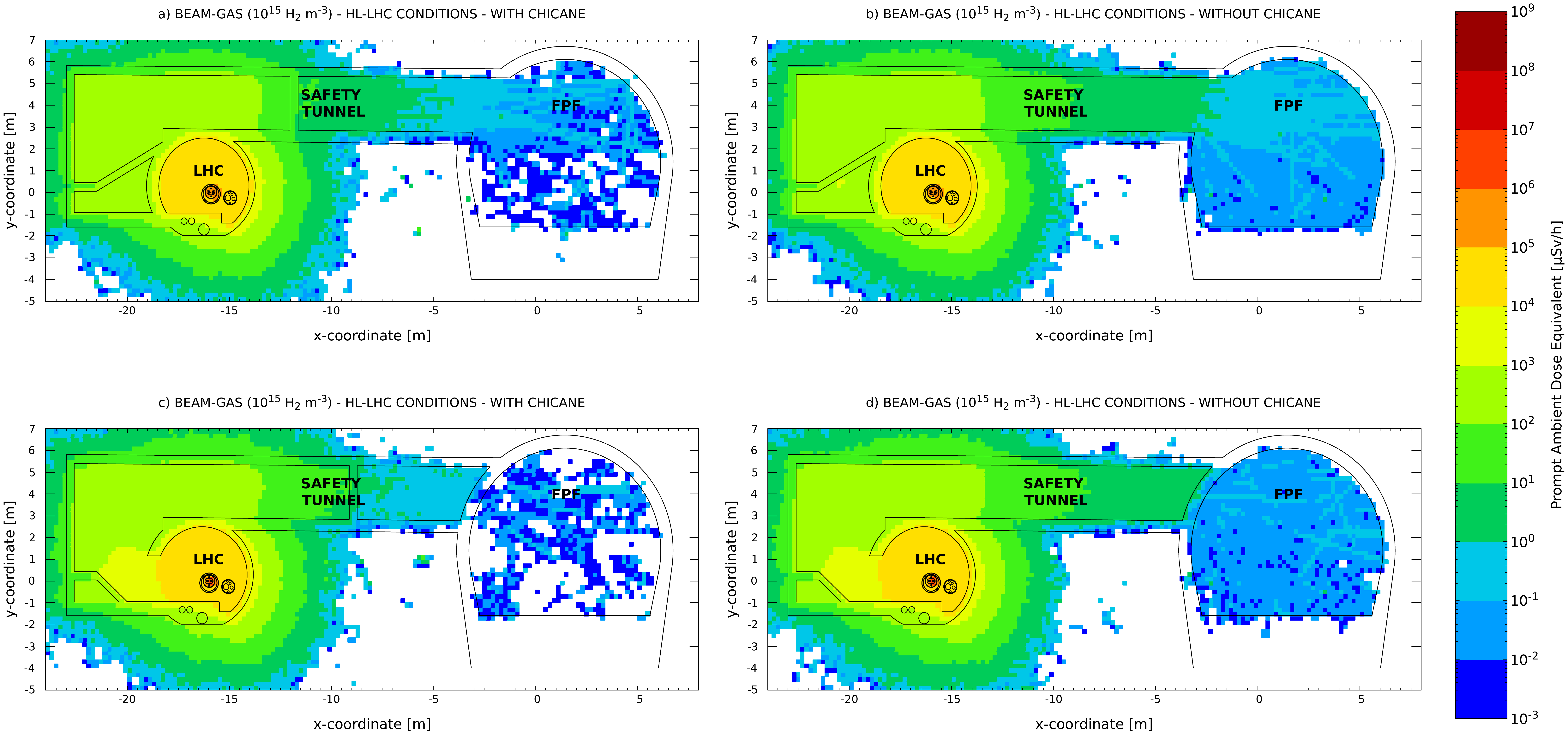}
\caption{Prompt ambient dose equivalent rate during LHC operation (beam-gas interaction -- HL-LHC beam conditions). Panels a/c (left) are with the chicane and panels b/d (right) are without the chicane. Panels a/b (top) and c/d (bottom) are generated at different distances from the IP1 ($z$-coordinates) to show the connection of the safety tunnel with the FPF and the LHC tunnel, respectively.}
  \label{fig:RP:2D_XY_beamgas}
\end{figure}

\begin{figure}[htbp]
\centering
\includegraphics[width=0.87\textwidth]{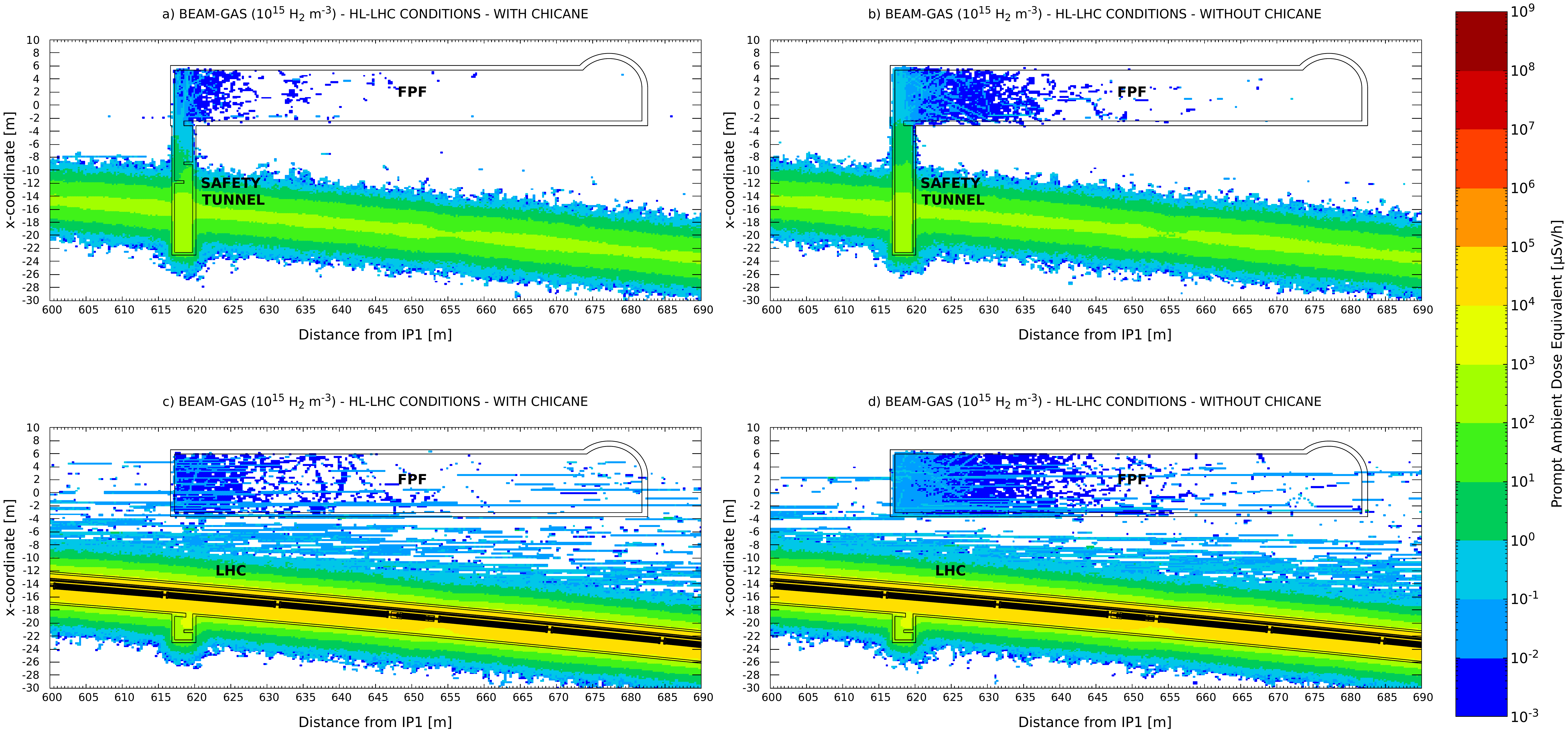}
\caption{Prompt ambient dose equivalent rate during LHC operation (beam-gas interaction -- HL-LHC beam conditions). Panels a/c (left) are with the chicane and panels b/d (right) are without the chicane.  Panels a/b (top) and c/d (bottom) are generated at different heights ($y$-coordinates) to show the connection of the safety tunnel with the FPF and the LHC tunnel, respectively.}
  \label{fig:RP:2D_ZX_beamgas}
\end{figure}

\cref{fig:RP:2D_XY_beamgas,fig:RP:2D_ZX_beamgas} show the prompt ambient dose equivalent rates, in $\mu$Sv/h, during normal LHC operation. The particle shower due to beam-gas interactions streams through the safety tunnel into the FPF cavern, but the presence of the chicane lowers the $\dot{H}^{*}(10)$ at the FPF entrance. 

\begin{figure}[tbhp]
\centering
\includegraphics[width=0.87\textwidth]{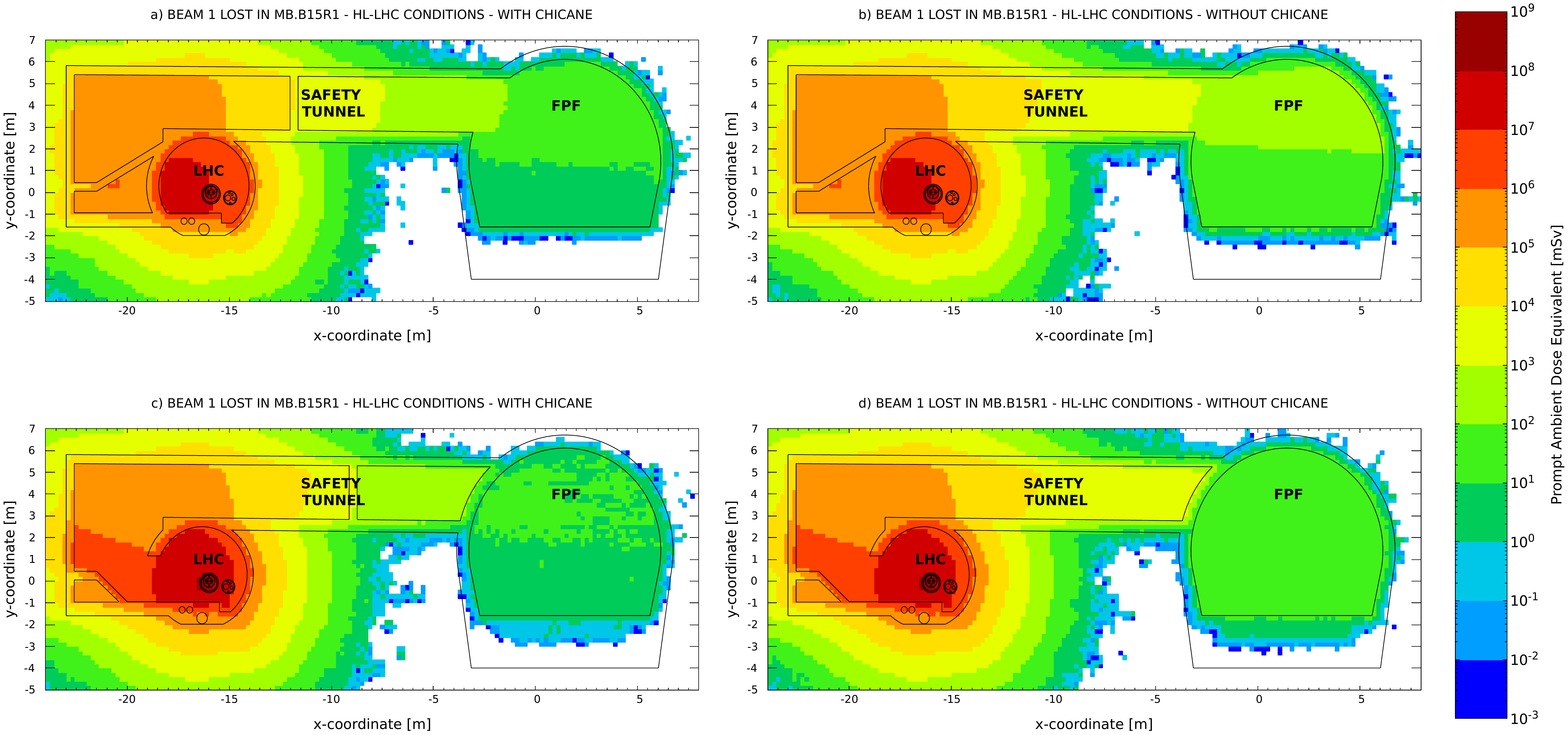}
\caption{Prompt ambient dose equivalent when the full beam is lost on the MB.B15R1 dipole (accidental scenario -- HL-LHC beam conditions). Panels a/c (left) are with the chicane and panels b/d (right) are without the chicane.  Panels a/b (top) and c/d (bottom) are generated at different distances from the IP1 ($z$-coordinates) to show the connection of the safety tunnel with the FPF and the LHC tunnel, respectively.}
  \label{fig:RP:2D_XY_accident}
\end{figure}

\begin{figure}[tbhp]
\centering
\includegraphics[width=0.87\textwidth]{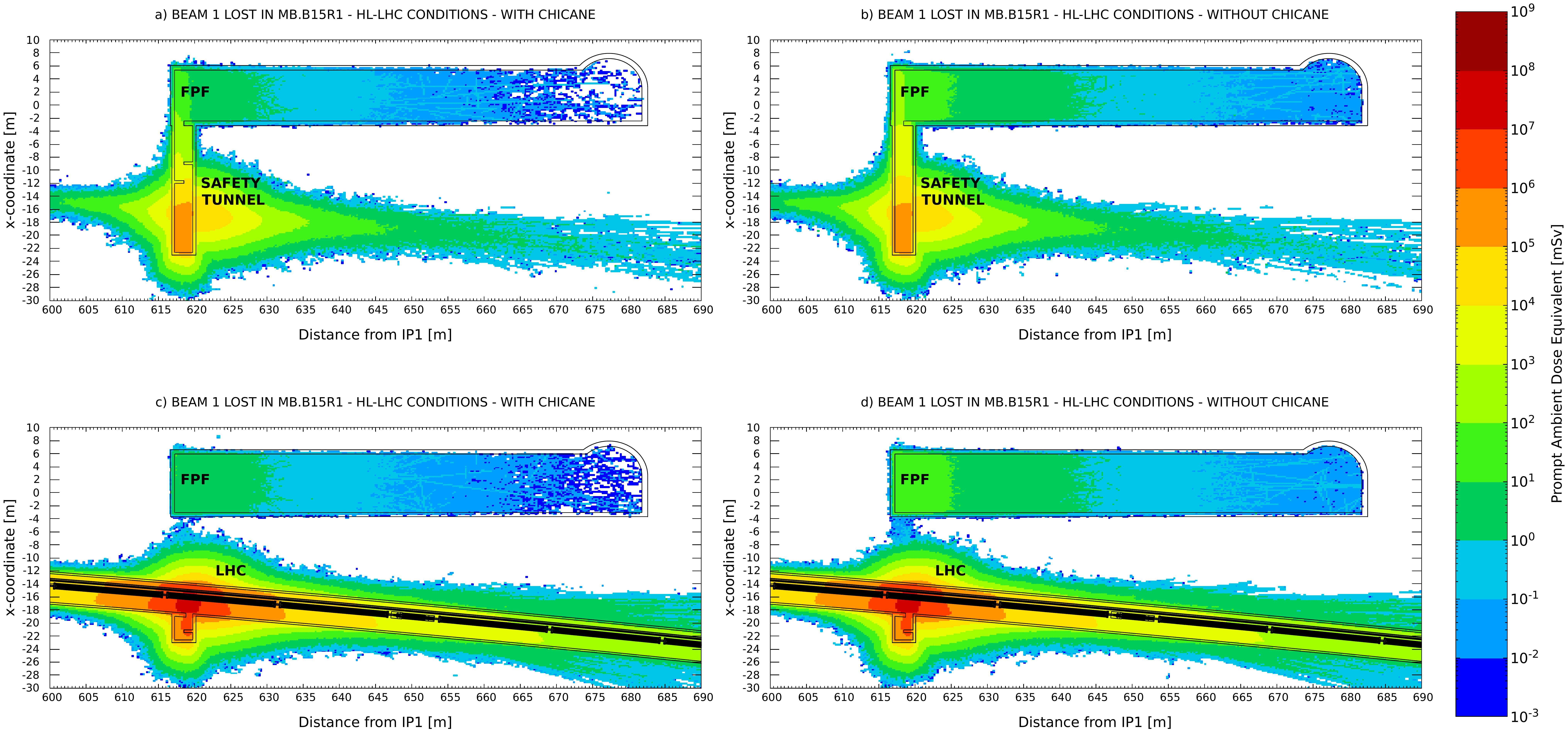}
\caption{Prompt ambient dose equivalent when the full beam is lost on the MB.B15R1 dipole (accidental scenario -- HL-LHC beam conditions). Panels a/c (left) are with the chicane and panels b/d (right) are without the chicane.  Panels a/b (top) and c/d (bottom) are generated at different heights ($y$-coordinates) to show the connection of the safety tunnel with the FPF and the LHC tunnel, respectively.}
  \label{fig:RP:2D_ZX_accident}
\end{figure}

\begin{figure}[tbhp]
  \centering
  \includegraphics[width=0.55\textwidth]{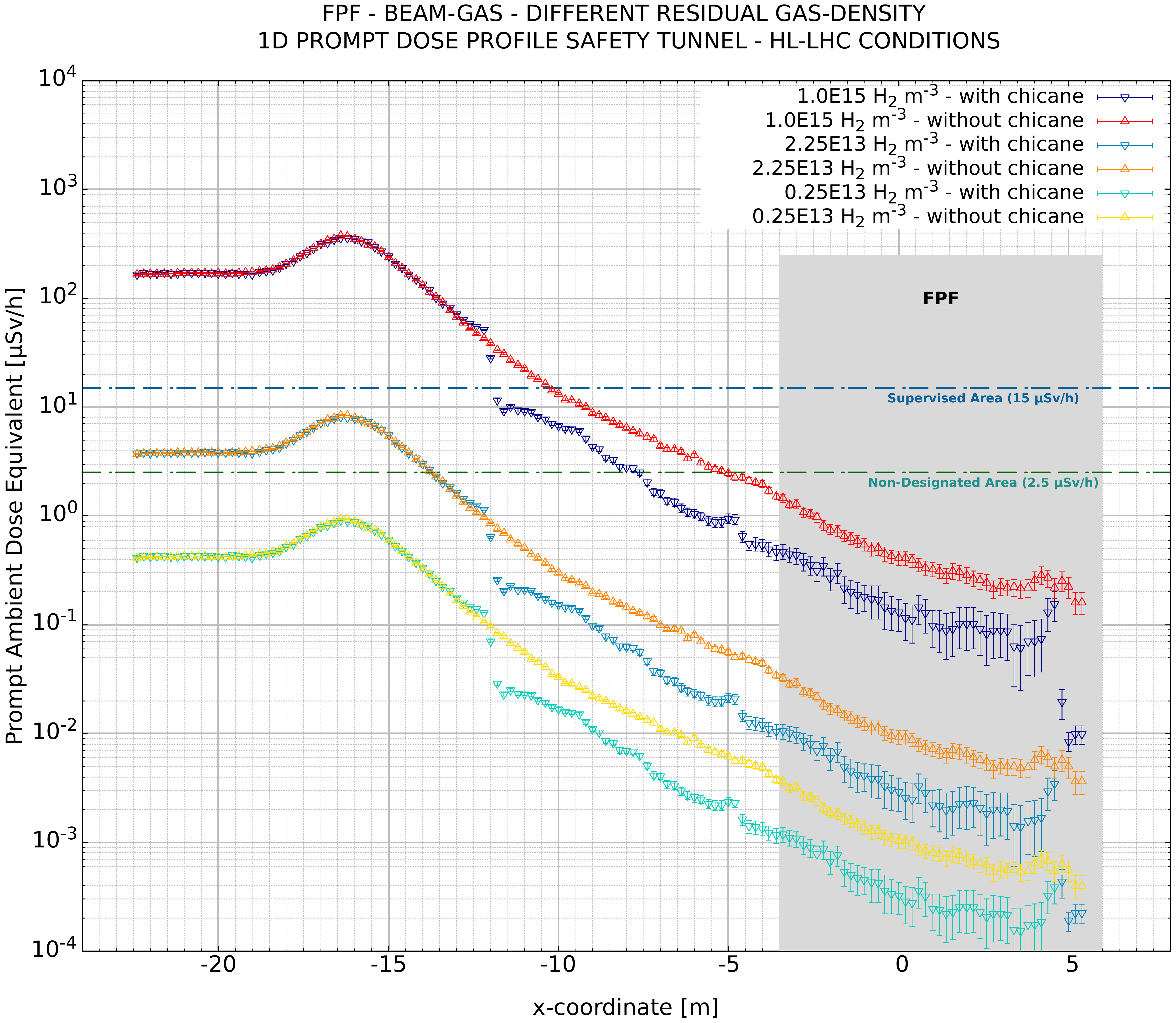}
  \caption{Prompt ambient dose equivalent rate profile in the safety tunnel during LHC operation (beam-gas interaction -- HL-LHC beam conditions). Different residual gas densities have been considered to take into account the machine conditioning, and results are presented both with and without the chicane. The footprint (width) of the FPF has been highlighted with a gray box.}
  \label{fig:RP:1D_X_beamgas}
\end{figure}

\begin{figure}[tbhp]
  \centering
  \includegraphics[width=0.55\textwidth]{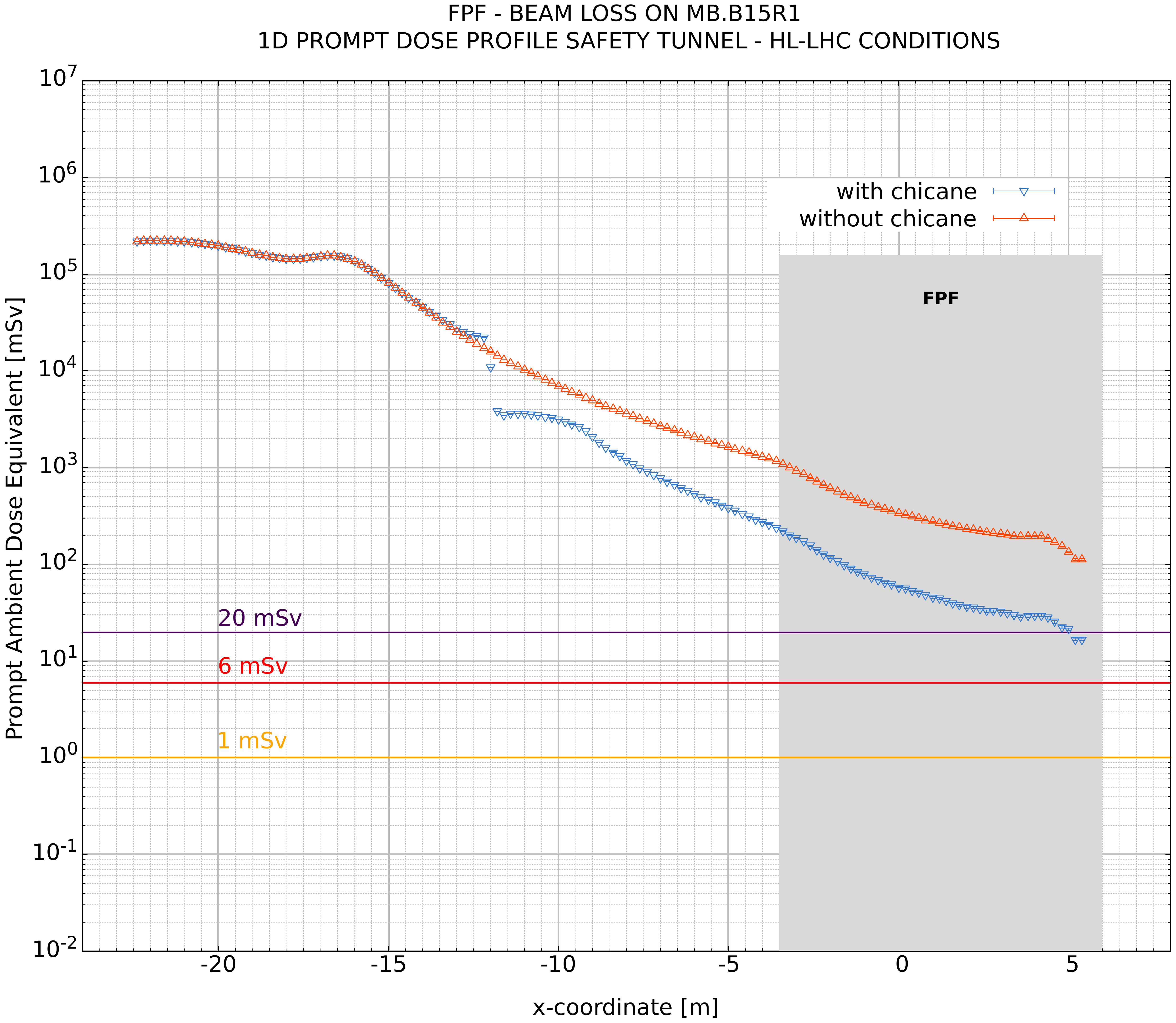}
  \caption{Prompt ambient dose equivalent profile in the safety tunnel when the full beam is lost on the MB.B15R1 dipole (accidental scenario -- HL-LHC beam conditions). Different residual gas densities have been considered to take into account the machine conditioning, and results are presented both with and without the chicane. The footprint (width) of the FPF has been highlighted with a gray box.}
  \label{fig:RP:1D_X_accident}
\end{figure}

\cref{fig:RP:2D_XY_accident,fig:RP:2D_ZX_accident} show the prompt ambient dose equivalent in mSv if the full LHC 7 TeV proton beam is accidentally lost on MB.B15R1, the superconductive dipole placed in front of the safety tunnel connection to the LHC tunnel. Similar to the normal operation case, the particle shower generated during this undesired event streams through the safety tunnel and could be potentially harmful for people standing at the entrance of the FPF cavern.

A quantitative assessment of the radiation levels along the safety tunnel, for LHC normal and abnormal operation, is provided through the 1D profile shown in \cref{fig:RP:1D_X_beamgas,fig:RP:1D_X_accident}. \cref{fig:RP:1D_X_beamgas} shows the $\dot{H}^{*}(10)$ considering different residual gas densities in the LHC beam screen: independently from the shielding configuration (with or without the chicane), even considering a conservative residual gas density of $1.0\times10^{15}$ H$_{2}$~m$^{-3}$, the $\dot{H}^{*}(10)$ remains below the limit for Non-Designated Radiation Areas. On the other hand, the results shown for the accidental scenario in \cref{fig:RP:1D_X_accident} show that the cumulative $H^{*}(10)$ exceeds the annual dose limit for classified personnel (20 mSv), even including the positive effect of the chicane.

With regard to the other source terms to be considered, the loss of the full 450 GeV proton beam in the SPS tunnel, which is most relevant for the FPF shaft, produces a negligible dose, since the distance between the shaft and the SPS tunnel is $>35$ m. The direct muon contribution coming from IP1/LSS1 to the prompt ambient dose equivalent rate needs to be further investigated and simulations are currently ongoing.

The verification of the accessibility of the experimental cavern during LHC operation requires the evaluation of different scenarios/source terms, considering both normal and abnormal operation. At present, the limiting scenario is the possible loss of the full LHC beam on the MB.B15R1 dipole:~possible mitigation actions such as adding a turn in the safety tunnel, adding another wall (``triple chicane'') and thickening the chicane walls, might be evaluated and integrated into the CE model. However, the missing direct muon contribution might have an impact on the accessibility of the FPF cavern during operation, which needs to be addressed.

\section{BDSIM Studies of the FPF Environment and Backgrounds \label{sec:BDSIM}}


\subsection{Introduction}

For any particle physics experiment in proximity to an accelerator, an understanding of the background sources, their origin, the particle types, and their spectra is crucial at both the design stage as well as during operation. To predict and understand these backgrounds requires the use of Monte Carlo techniques, as the particle fluence close to an accelerator originates from many different and indirect sources.  For the FPF specifically, we focus on the region of the LHC close to IP1, where the ATLAS experiment is located.  This entails predicting the particle flux due to:
\vspace*{-0.4em}
\begin{enumerate}
\setlength\itemsep{-0.4em}
\item $pp$ collisions at IP1 (and potentially also Pb-Pb and Pb-$p$)
\item Inelastic proton interactions with residual vacuum gas in the LHC arcs
\item Other beam losses in the arcs due to other-IP physics debris, collimation losses, and other beam losses
\end{enumerate}

The first is expected to be the dominant contribution, as the FPF is on the LOS of the IP where the collisions occur, and they will produce high-energy, penetrating particles, such as muons and neutrinos. The second and third sources are from a significantly different direction, but are expected to be small contributions, since the distance of between 10\,m and 16\,m from the FPF inside wall to the LHC tunnel is expected to be sufficient to absorb the majority of background particles from these sources.

To simulate the particle fluence, a 3D radiation transport model is required, along with the transport and description of many subatomic particles. Both electromagnetic and hadronic interactions with the material of the accelerator, the tunnel, and the surrounding rock must also be simulated, as well as the deflection of charged particles by the many unique magnetic fields of the accelerator magnets.

In contrast to conventional transverse-orientated detectors, the far-forward location of the FPF requires simulation of the accelerator complex including its varied conditions throughout operation. The magnet strengths (the \emph{optics}) are varied throughout each fill and generally throughout the operation (the \emph{Run}) for various purposes. The crossing-angle of the colliding beams as well as the beam size (and therefore divergence angle) are varied to maintain, or level, the luminosity at the collision point to best serve the experiments. Throughout the Run, the optics may generally be improved upon depending on the machine performance, machine protection, the upstream injector-chain performance, and collimation performance.

\texttool{BDSIM}~\cite{Nevay:2018zhp} is a Monte Carlo tool based on \texttool{Geant4}~\cite{GEANT4:2002zbu, Allison:2016lfl, Allison:2006ve}, ROOT~\cite{Antcheva:2011zz}, and CLHEP that creates \texttool{Geant4} models of accelerators from an optical description (one concerned with the magnetic strengths), such as MADX~\cite{MADX-website}, used as the magnetic description of the LHC. It includes a library of geometries for many accelerator components in many styles including those of the LHC. It also includes parameterised magnetic fields for all of the conventional accelerator magnets, as well as the ability to load and interpolate field maps. Crucially, being based on \texttool{Geant4}, it allows the production and tracking of all subatomic particles provided by \texttool{Geant4}, as well as the extensive physics processes for them. \texttool{Geant4} is a particle physics library commonly used in detector simulation and underpins the Monte Carlo for countless experiments in particle physics as well as being used in the space and medical physics domains. It is regularly updated with the latest developments from the field. Similar codes also used for this purpose include \texttool{FLUKA}~\cite{Battistoni:2015epi, FLUKA:new}, described above in \cref{sec:FLUKA}, and MARS~\cite{osti_1282121,Mokhov:2017klc}.

Although \texttool{BDSIM} provides a library of approximate and scalable geometries suitable for most applications, it is required in many cases to provide a more accurate geometry for a specific installation, where relevant. In this case, \texttool{BDSIM} provides the ability to load GDML format geometry, which is the geometry persistence format of \texttool{Geant4} based on XML. The Python library \texttool{pyg4ometry}~\cite{Walker:2020lpq} is used to create, convert, and composite geometry. This allows detailed descriptions in other formats, such as the IR1 tunnel complex at CERN described in \texttool{FLUKA}  geometry format, to be converted and incorporated into the model.

An important feature of \texttool{BDSIM} is the ability to filter and store select trajectories of particles in a linked data structure in ROOT format. Therefore, a \texttool{BDSIM} model allows not only fluences to be estimated, but also may be used to provide insight into the origin of background sources and their production mechanisms.

As an input, event generator output in \texttool{HepMC}~\cite{Buckley:2019xhk} format can be used for each event.  \texttool{BDSIM} is also usable as a C++ class to simulate individual events for a given model.  Therefore, it can be integrated into an analysis framework if required, or used to generate a static Monte Carlo sample. The output of \texttool{BDSIM} is a ROOT-format file that is standard in high-energy physics, and either ROOT itself or the included tools with \texttool{BDSIM} can be used to perform a large scalable analysis, including skimming.  \texttool{BDSIM} has been used for studying accelerator beam losses in a variety of machines; the models described here were originally developed for LHC collimation studies~\cite{Abramov:2020nos, Walker:2020lpq}.

\subsection{BDSIM Model of the LHC IP1}

A model of the LHC accelerator from IP1 towards the FPF was created using \texttool{BDSIM} . This model was originally developed for the FASER experiment~\cite{Lefebvre:2021yzn} at both the pilot detector location (``C-side'' of IP1 at the TI18 LOS location) and the FASER experiment location (``A-side'' of IP1 at the TI12 LOS location). Several data sources are used to prepare the \texttool{BDSIM} input automatically using its associated \textit{pybdsim} Python library. These include:
\begin{enumerate}
\setlength\itemsep{-0.4em}
\item A MADX ``Twiss Table'' file providing an optical description of the magnets
\item A detailed aperture model from the LHC Collimation Group (BE-ABP-NDC)
\item Corresponding collimator settings for the optical configuration
\item Tunnel geometry converted from \texttool{FLUKA}  format
\item Tunnel geometry from \texttool{BDSIM} and \texttool{pyg4ometry}
\item Select geometry pieces for various components (e.g., TAN, JSCA(A,B) shielding).
\end{enumerate}

Items (1) and (3) are expected to vary throughout the Run and the model can be easily adjusted to reflect these through automated preparation. The model was originally  created for LHC Run 2 (2015-18) and Run 3 (2022-25). However, for the FPF, during HL-LHC~\cite{ZurbanoFernandez:2020cco}, a new model is required as the accelerator layout will have changed. There are several key differences from the LHC-era machine, namely:
\vspace*{-0.5em}
\begin{itemize}
\setlength\itemsep{-0.4em}
\item New TAXS absorber replaces the TAS absorber with increased aperture
\item New TAXN absorber replaces the TAN absorber
\item The `D1' separation dipole is now superconducting
\item New collimators for incoming and outgoing beams in the layout
\end{itemize}

The TAXS absorber (currently TAS in the LHC) is a cylindrical copper absorber approximately 19\,m from the IP. The TAXN (currently TAN in the LHC) is an absorber between the two separation dipoles to protect the machine from predominantly neutral physics debris. The layout and geometry in the \texttool{BDSIM} model has been updated according to the optics configuration \textit{HL-LHC V1.5}.

\begin{figure}[tbp]
\centering
\includegraphics[width=0.95\textwidth]{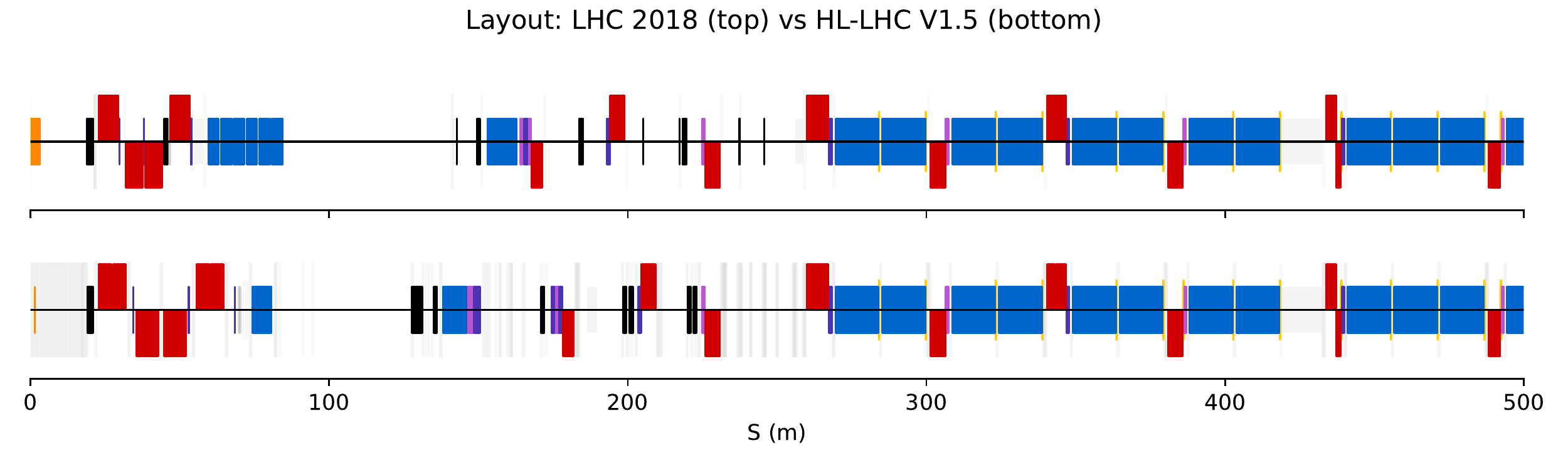}
\caption{Comparison of LHC (top) and HL-LHC (bottom) machine layouts from IP1 at $S = 0$\,m along the direction of Beam 1 (\textit{outgoing}). Dipoles are shown in blue, quadrupoles in red, sextupoles in yellow, and collimators in black.}
\label{fig:bdsim:hllhclhccomparison}
\end{figure}

\begin{figure}[tbp]
\centering
\includegraphics[width=0.75\textwidth]{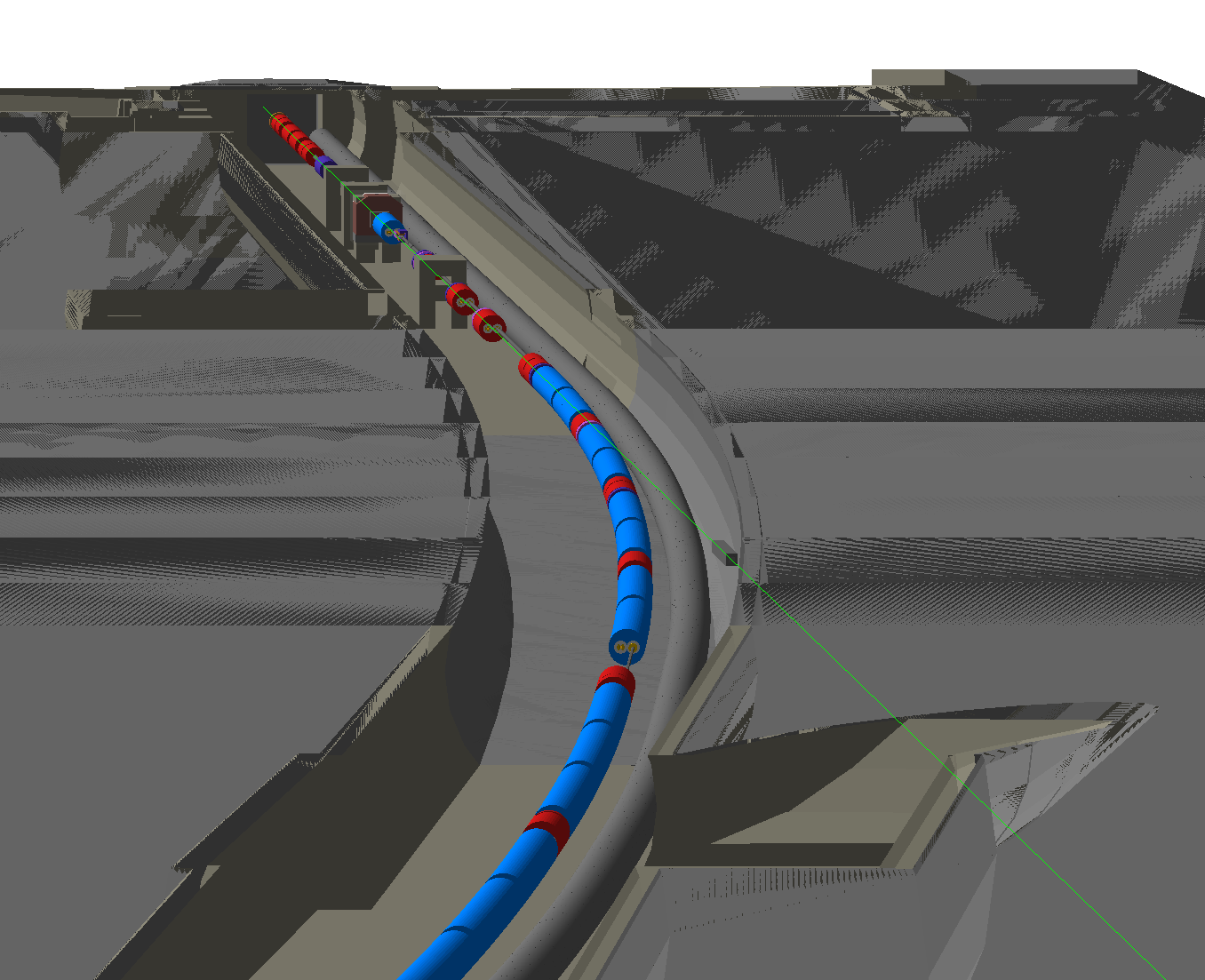}
\caption{Visualisation of the \texttool{BDSIM} model of the HL-LHC from ATLAS IP1 along the Beam 1 direction to the TI18 tunnel and the FPF. The geometry above $Y = 1$~m has been removed for the visualisation. A green line shows the LOS from the IP (top left), through the accelerator (blue and red magnets), the cryogenic coolant line (grey tube beside it), a sweeper magnet (small grey box), and a partial view of the ascending ramp of the TI18 tunnel (lower right).}
\label{fig:bdsim:modelview1}
\end{figure}

A 3D view of the complete \texttool{BDSIM} model is shown in \cref{fig:bdsim:modelview1}. The proposed FPF cavern is not simulated, as a particle flux on its entrance is desired in this study and will have no effect on that result. In the future, the geometry for the cavern will be added.

\subsection{Simulation Procedure}

To generate a Monte Carlo sample, the \texttool{CRMC}~\cite{CRMC} event generator tool is used to generate the $\sqrt{s} = 14$\,TeV $pp$ collisions. \texttool{CRMC} provides several underlying generators including \texttool{EPOS-LHC}~\cite{Pierog:2013ria}, \texttool{QGSJet}~\cite{Ostapchenko:2010vb}, and \texttool{Sibyll}~\cite{Ahn:2009wx}. For this study, \texttool{Sibyll} was used. \texttool{CRMC} writes a \texttool{HepMC} format file that is subsequently loaded by \texttool{BDSIM} . \texttool{BDSIM} is set to simulate only loaded particles in the forward direction of the model and with a pseudorapidity $\eta \geq 2.3$ for simulation efficiency. A minimum kinetic energy cut of 10\,GeV was used. \texttool{Geant4} V10.7.p03 was used with reference physics list \texttool{FTFP\_BERT} (a complete physics list including electromagnetic, decay, and hadronic processes).  \textit{HL-LHC V1.5} MADX optics were used with a primary proton beam energy of 7\,TeV. The crossing angle was $250\,\mu$rad. An aperture model was used from the BE-ABP-NDC collimation group.

Cross section biasing is used to reduce the computational time required to estimate muon fluxes by increasing the cross section of several particles' decay processes. Given the small forward area at a great distance, this biasing scheme is required to estimate the relevant quantities in a reasonable amount of time on an available computer farm. The numerical biasing factors used per particle are shown in \cref{tab:bdsim:biasingfactors}.

\begin{table}[bp]
\centering
    \begin{tabular}{c||c|c}
    \hline \hline
    Particle & Factor in Vacuum \& Air  & Factor in Material \\
    \hline
    $\pi^{\pm}$  & $50$  & $10^4$ \\
    K$^{\pm}$    & $50$  & $10^3$ \\
    \hline \hline
    \end{tabular}
\caption{Numerical scaling factors applied to the \texttool{Geant4} \texttt{Decay} physics process for charged pions and charged kaons. Two factors were generally applied, firstly for inside the beam pipe vacuum and any air, and secondly for other material.}
    \label{tab:bdsim:biasingfactors}
\end{table}

\begin{figure}[tbp]
\centering
\includegraphics[width=\textwidth]{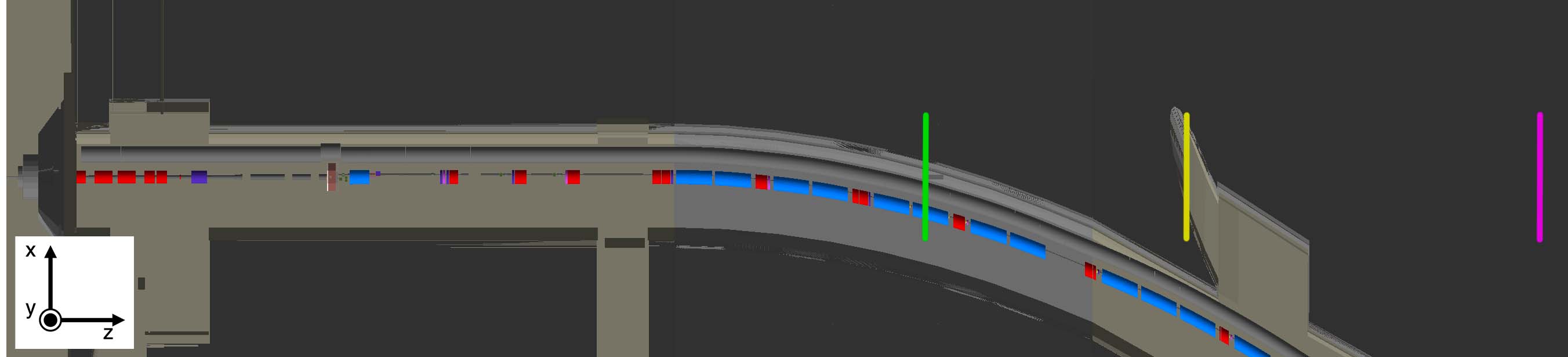}
\caption{Visualisation of the \texttool{BDSIM} model of the HL-LHC from IP1 at ATLAS (left) along the Beam 1 direction (to the right) in plan view with the geometry above $Y = 1$\,m removed. Three sample planes are shown, all centred on the LOS from IP1. At $Z= 370$\,m (green), a plane is placed in front of the potential sweeper magnet location as the tunnel starts to bend. At $Z = 475$\,m (yellow), a plane is located in the TI18 tunnel as it connects to the main LHC tunnel; this was the location of the FASER pilot detector and the current location of SND@LHC.  Finally, a plane at $Z = 617$\,m (pink) is placed before the entrance to the proposed FPF cavern. The $Z$ dimension has been compressed by a factor of 10. }
\label{fig:bdsim:modelview2}
\end{figure}

Passive and invisible planes, called ``samplers'' in \texttool{BDSIM}, were placed at several positions in the model to record muons and neutrinos only, as it is foreseen that these are the most relevant Standard Model particles at the FPF for this study. These samplers record the kinematic variables of particles in a 2D plane at a given position with a given square size. Samplers were placed at 3 locations in the LOS of IP1, as shown in \cref{fig:bdsim:modelview2}. These were at distances 370\,m, 475\,m, and 617\,m, which correspond to just into the tunnel after the accelerator starts to bend; the opening into the TI18 tunnel; and in front of the FPF. The data from the samplers can be analysed individually or together, as well as be re-launched into the model or another for further study.

The model preparation was validated by tracking 10,000 primary protons from the IP to the end of the model, analysing the beam size, mean offset, and calculated optical Twiss functions from the tracks recorded in samplers after every magnet. The tracking showed excellent agreement with the MADX model.

\begin{figure}[tbp]
  \centering
  \includegraphics[width=0.4\textwidth]{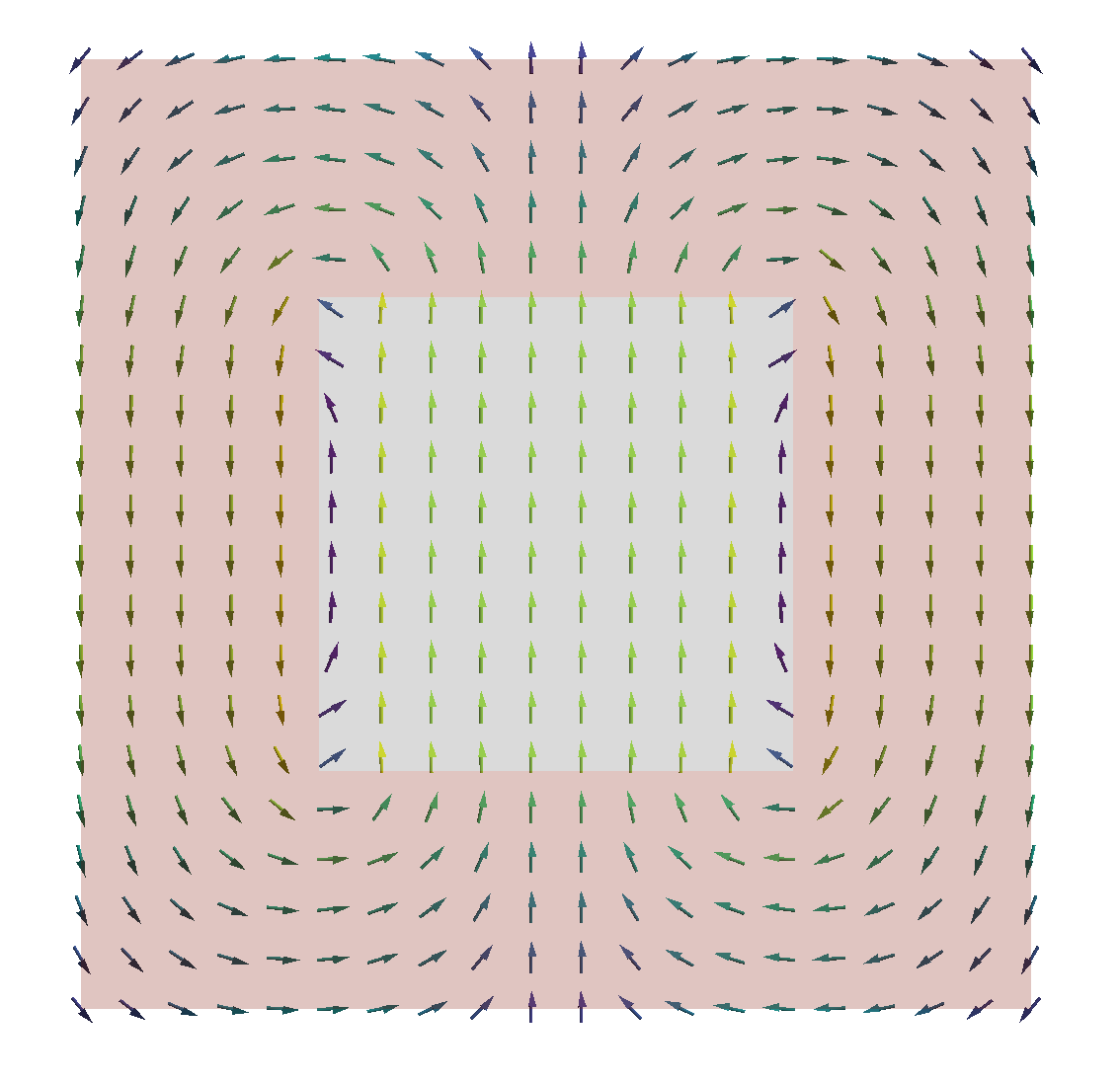}
  \caption{A 2D view of the proposed SmCo sweeper magnet with an overlaid magnetic field map. The full width is 20\,cm, and the outer layer (red) is iron.}
  \label{fig:bdsim:sweeperfield}
\end{figure}

For this study, a proposed sweeper magnet was included. It was $20 \times 20 $\,cm in area, with a peak field of 1.4\,T, and placed in the LOS ($X,Y = 0,0$\,m) at $Z = 370$\,m.  The field was orientated vertically, with the return flux in an iron layer 10\,cm wide. The core of the magnet was samarium cobalt. A cross section of it with overlaid magnetic field map is shown in \cref{fig:bdsim:sweeperfield}. The field map is based on previous permanent magnet designs and was provided by the SY-STI-BMI group at CERN. The geometry was created in GDML using \texttool{pyg4ometry}.

A Monte Carlo sample of 100M $pp$ events was generated using the Royal Holloway computing cluster.  This sample includes the trajectories of (exclusively) muons and neutrinos that reach the 2\,$\times$\,2\,m sample plane at the entrance to the FPF. Additionally, the select parent track trajectories linking each particle back to the primary vertex are stored. This will enable a detailed analysis of origins and production mechanisms in the future.

\subsection{Muon and Neutrino Fluxes} \label{sec:bdsim_fluxes}

\begin{figure}[tbp]
\centering
\includegraphics[width=0.48\textwidth]{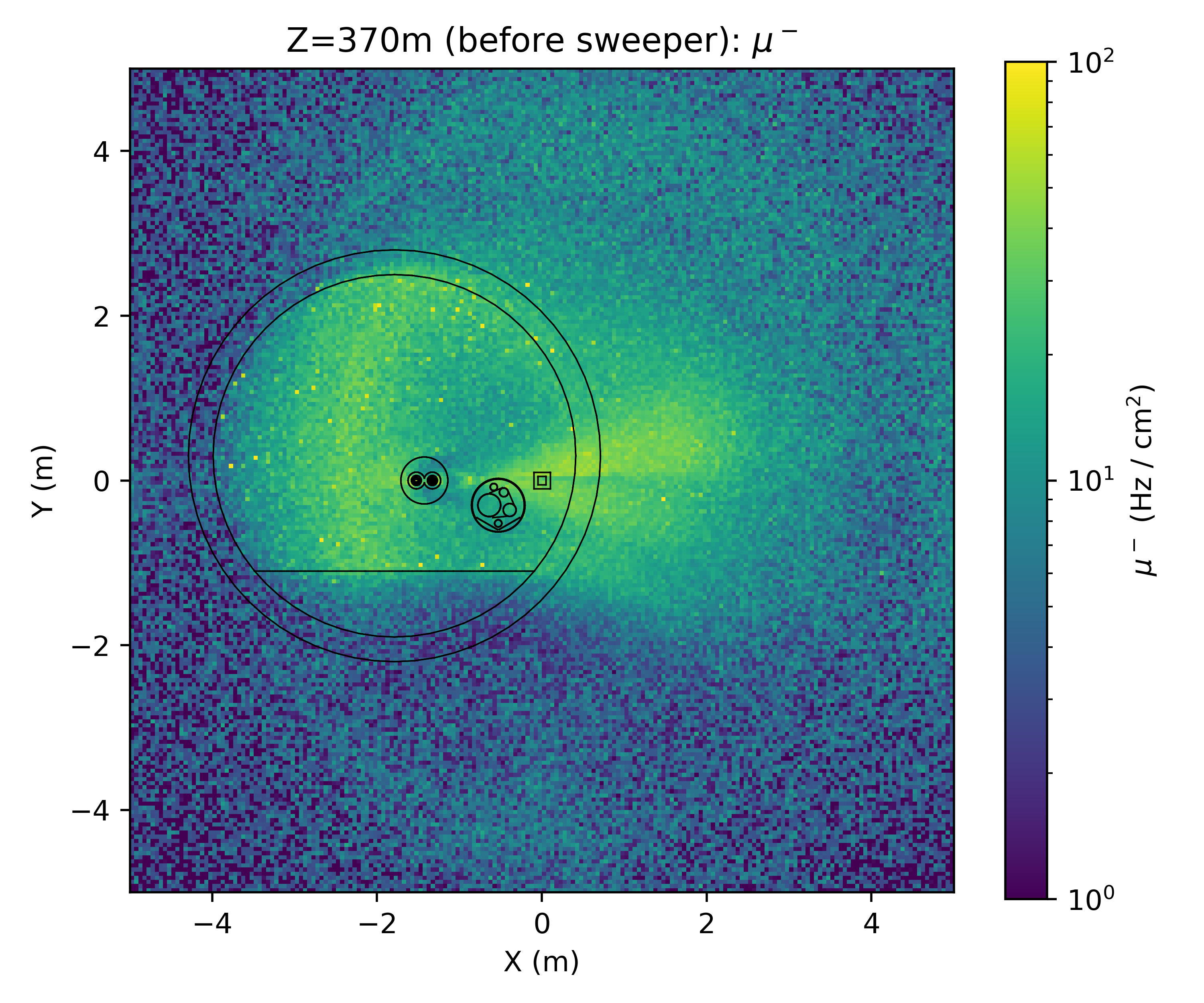}
\includegraphics[width=0.48\textwidth]{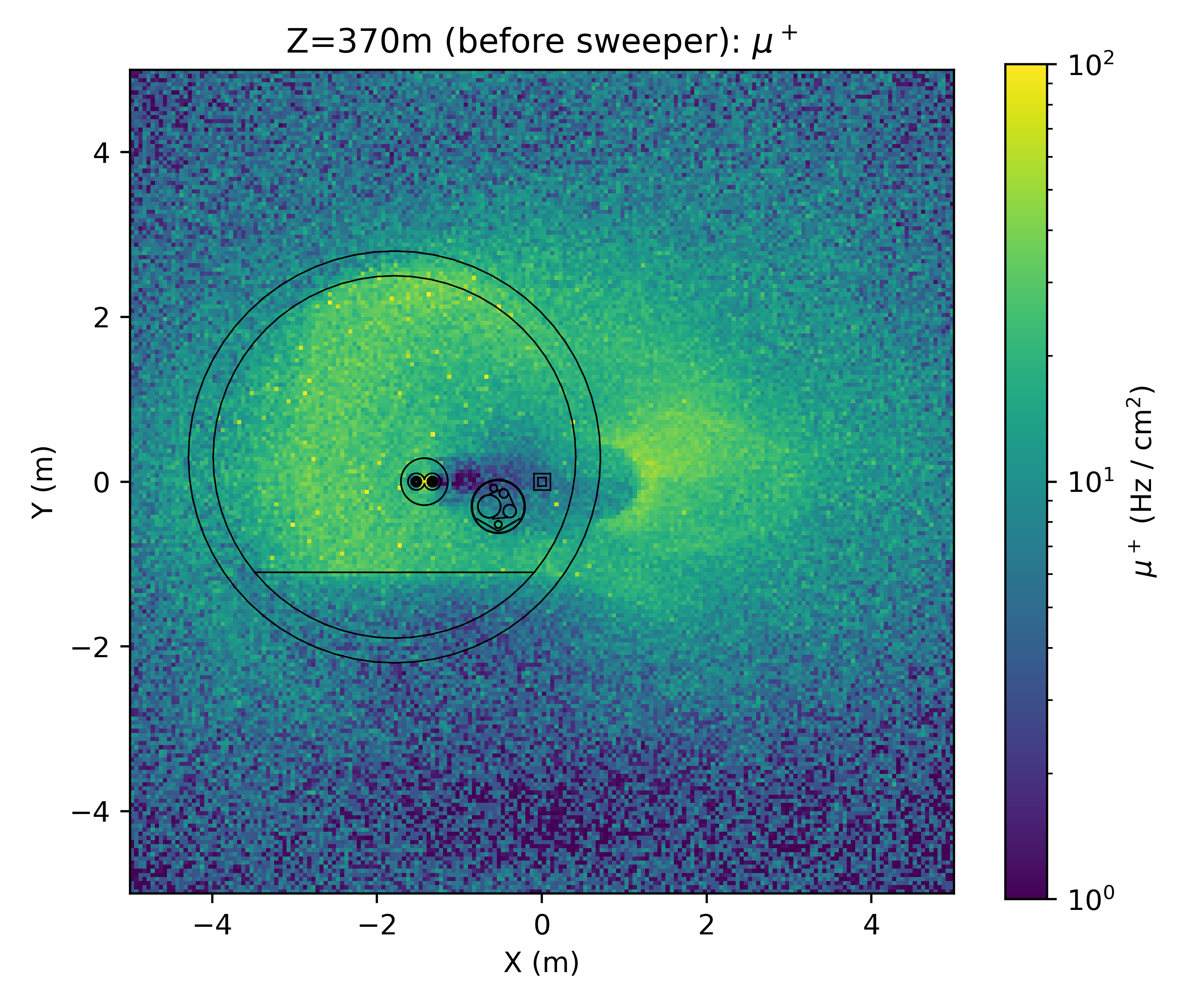}
\caption{The 2D muon distribution at $Z = 370$\,m from IP1 in front of the proposed sweeper magnet location. Outlines of the LHC magnets (left tube) cold-mass, the cryogenic coolant line (right tube), and the tunnel are shown in black. The proposed sweeper magnet is shown as a square outline also in black. The data is scaled to the nominal HL-LHC luminosity of $5 \times 10^{34}$~cm$^{-2}$\,s$^{-1}$.}
\label{fig:bdsim:muon2dsweeper}
\end{figure}

\begin{figure}[tbp]
\centering
\includegraphics[width=0.48\textwidth]{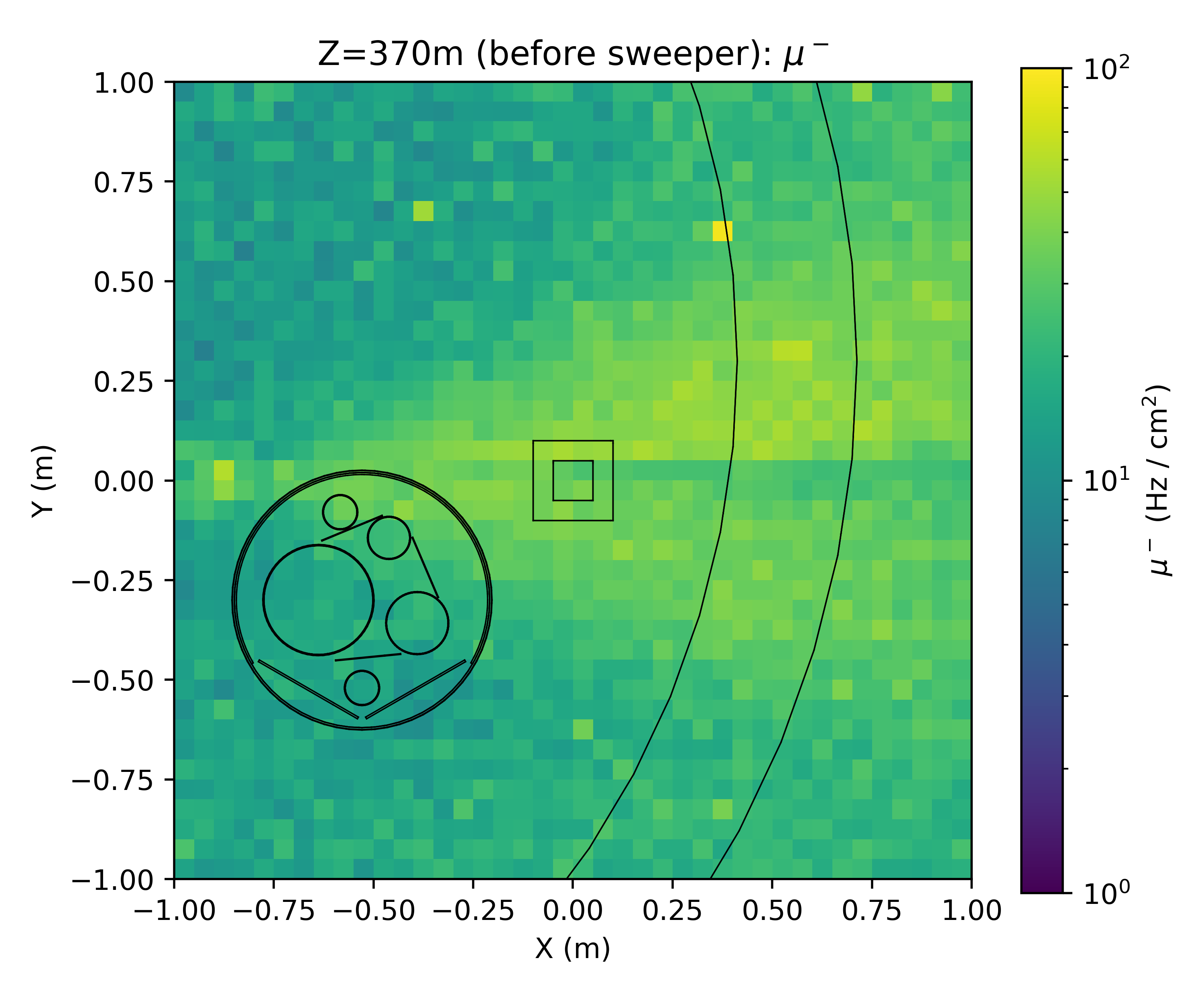}
\includegraphics[width=0.48\textwidth]{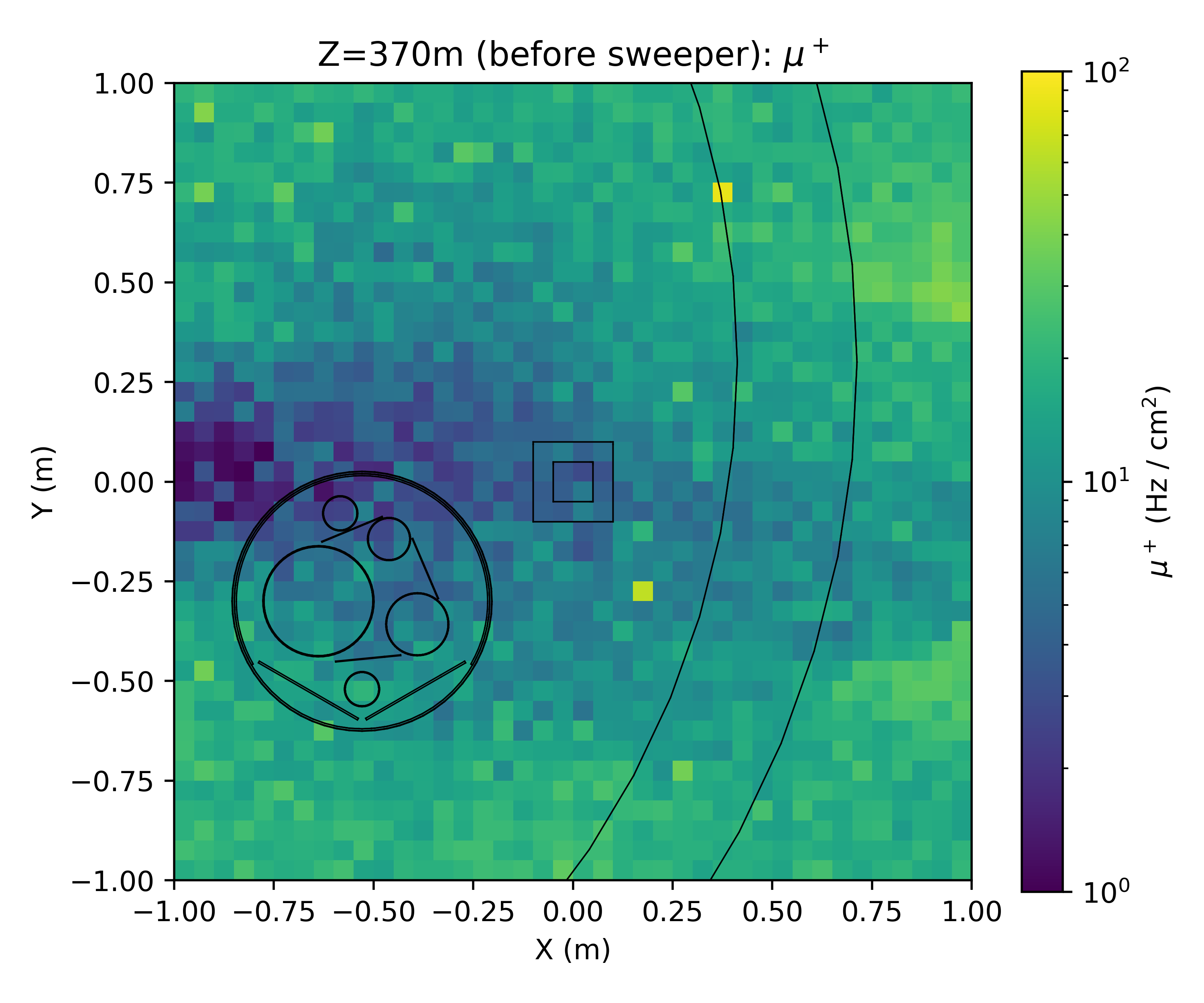}
\caption{The 2D muon distribution at $Z = 370$\,m, viewed more closely than in \cref{fig:bdsim:muon2dsweeper} and showing the muon flux at the location of a sweeper magnet. The data is scaled to the nominal HL-LHC luminosity of $5 \times 10^{34}$~cm$^{-2}$\,s$^{-1}$.}
  \label{fig:bdsim:muon2dsweeperzoom}
\end{figure}

\begin{figure}[tbp]
  \centering
  \includegraphics[width=0.48\textwidth]{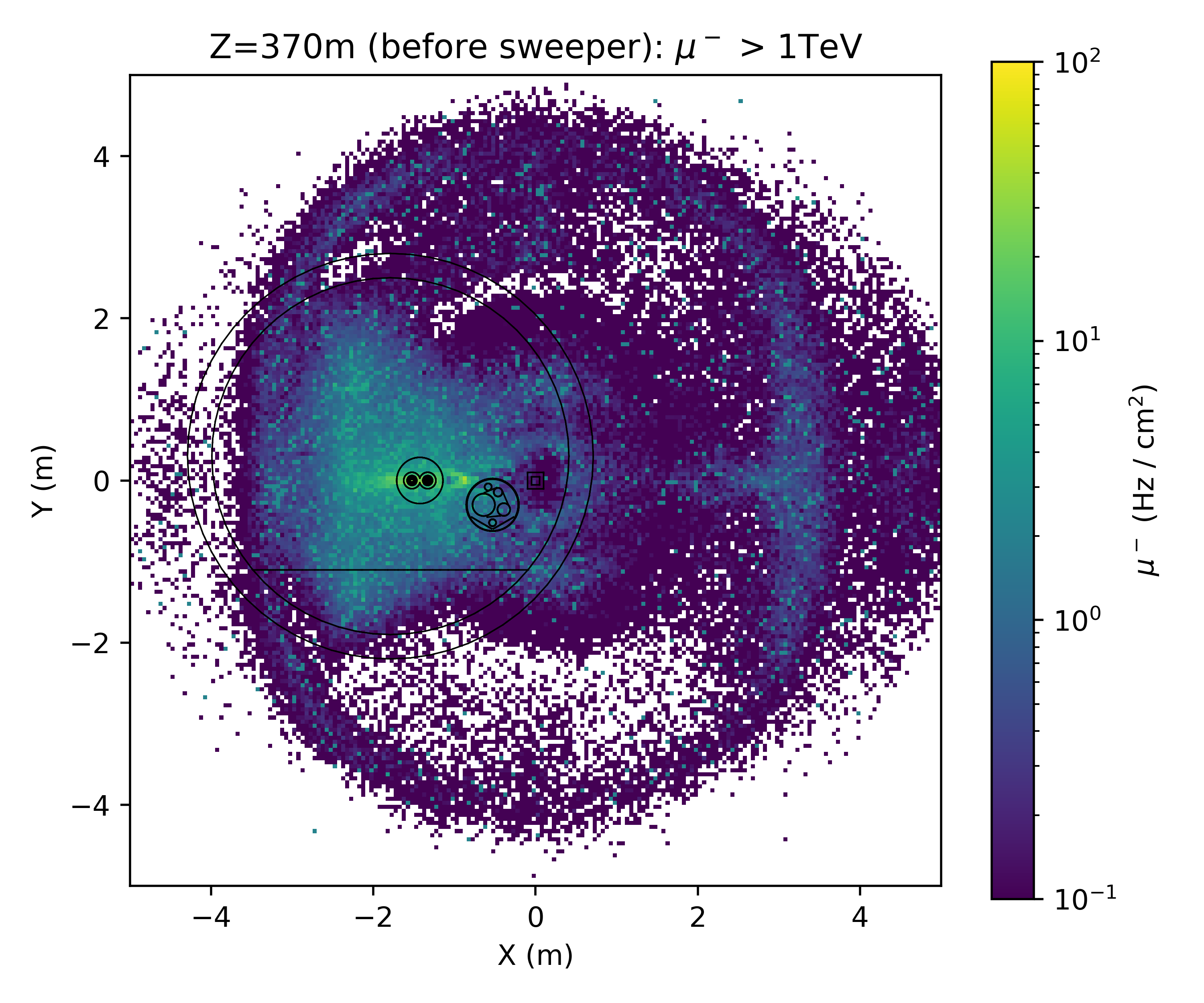}
  \includegraphics[width=0.48\textwidth]{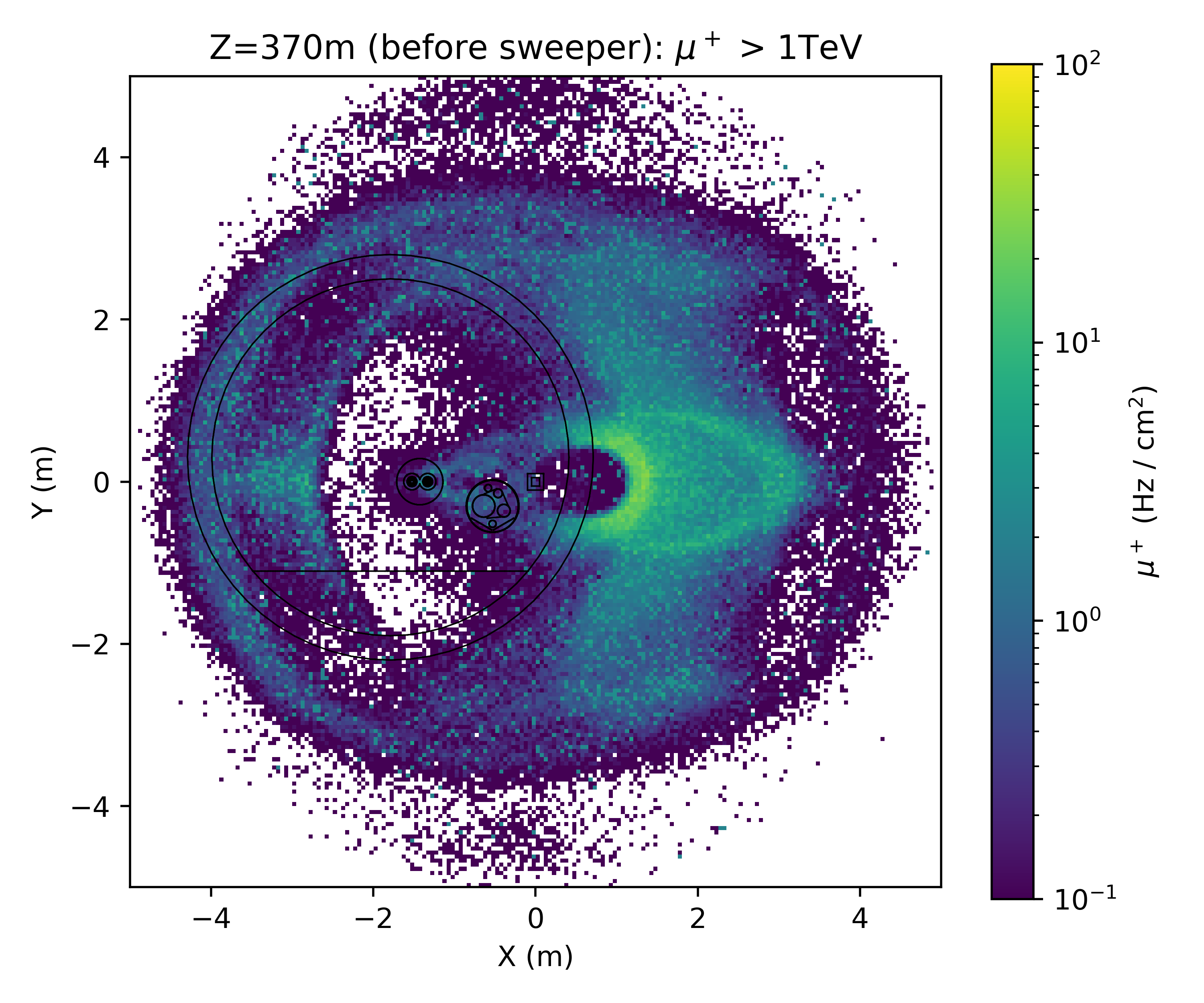}
  \caption{The 2D muon distribution at $Z = 370$\,m, as in \cref{fig:bdsim:muon2dsweeper}, but only for muons with $E_k > 1$\,TeV. The data is scaled to the nominal HL-LHC luminosity of $5 \times 10^{34}$~cm$^{-2}$\,s$^{-1}$.}
  \label{fig:bdsim:muon2dsweeper1tev}
\end{figure}

The \texttool{BDSIM} studies provide results for the distribution of muons along the LOS at several distances from the ATLAS IP.  The results can be displayed using the coordinate system shown in \cref{fig:bdsim:modelview2}. Firstly, the 2D distribution of muons at $Z = 370$\,m (the green line in \cref{fig:bdsim:modelview2}) is shown in \cref{fig:bdsim:muon2dsweeper}, with a magnified view in \cref{fig:bdsim:muon2dsweeperzoom}.  As the primary proton energy is 7\,TeV, very high-energy, and therefore penetrating, muons can be produced. A subset of the muons shown in \cref{fig:bdsim:muon2dsweeper} is displayed in \cref{fig:bdsim:muon2dsweeper1tev}, where only muons with $E_k > 1$\,TeV are shown.  These are less frequent but are more likely to reach the FPF through the accelerator complex and the surrounding rock.

\begin{figure}[tbp]
  \centering
  \includegraphics[width=0.45\textwidth]{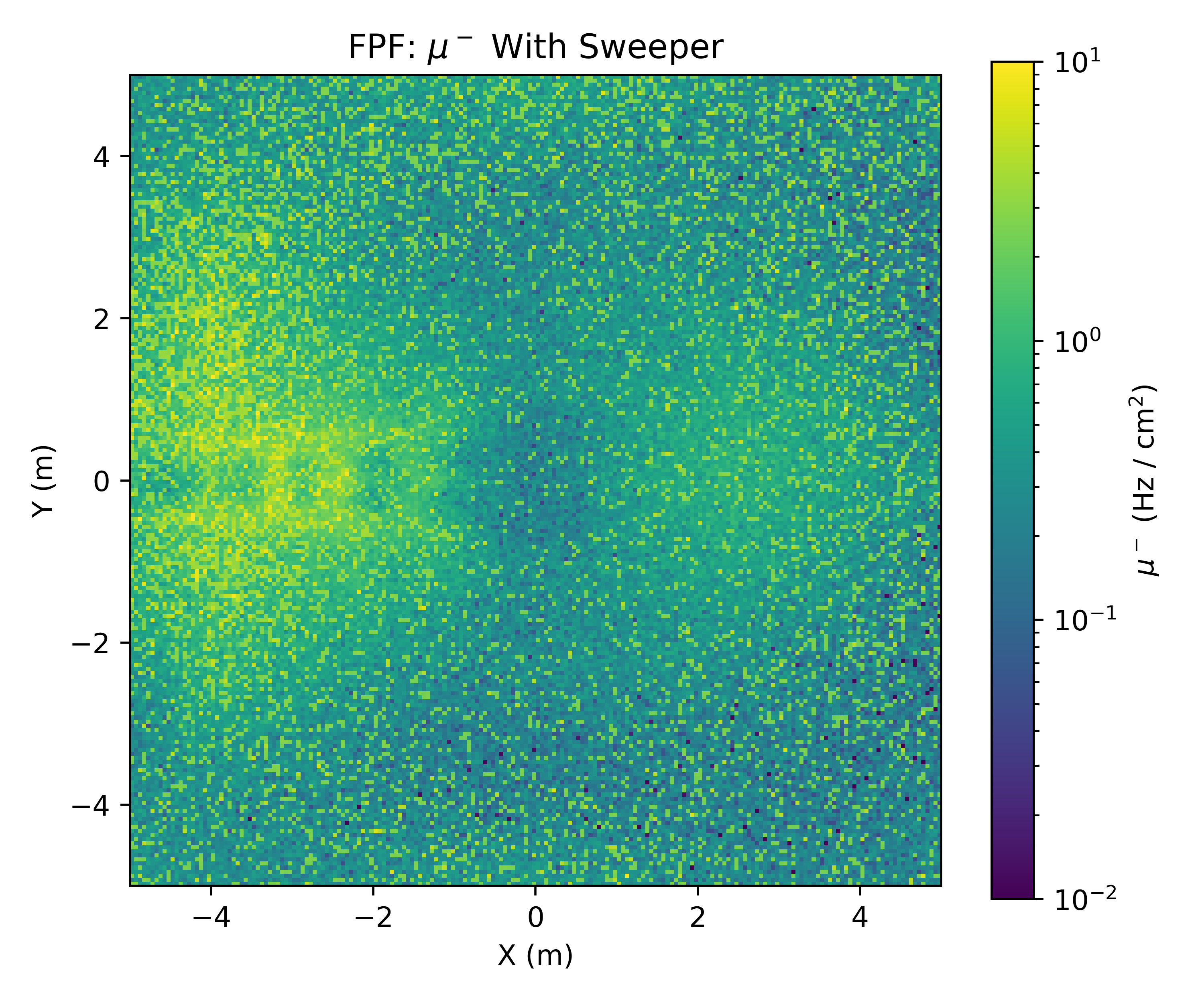}
  \includegraphics[width=0.45\textwidth]{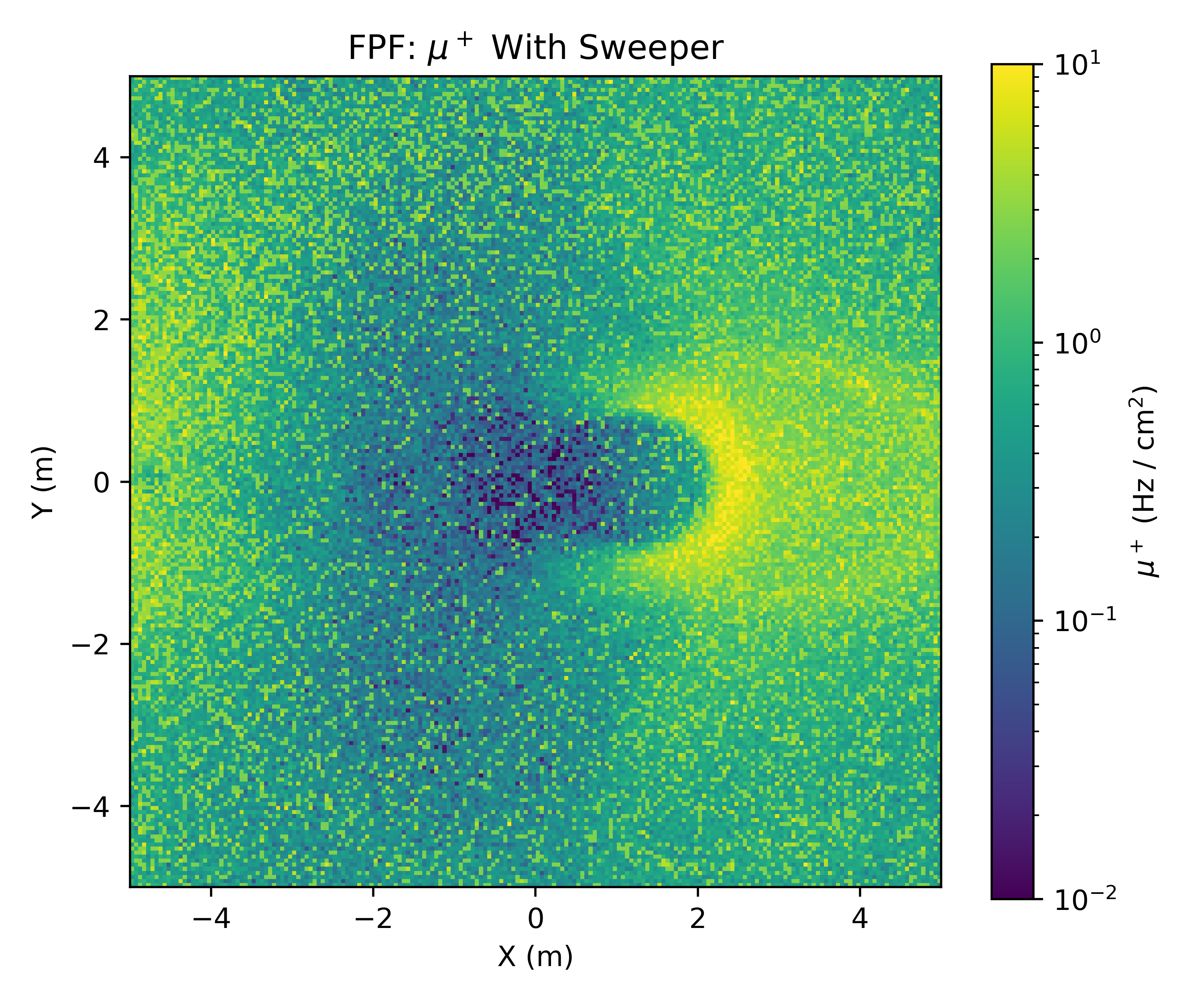}
  \caption{2D Muon distribution at $Z = 617$\,m from IP1 in front of the proposed dedicated 
  FPF location. The data is scaled to the nominal HL-LHC luminosity of 
  $5 \times 10^{34}$~cm$^{-2}$\,s$^{-1}$.}
  \label{fig:bdsim:fpfmuondistribution}
\end{figure}

\begin{figure}[tbp]
\centering
\includegraphics[width=0.48\textwidth]{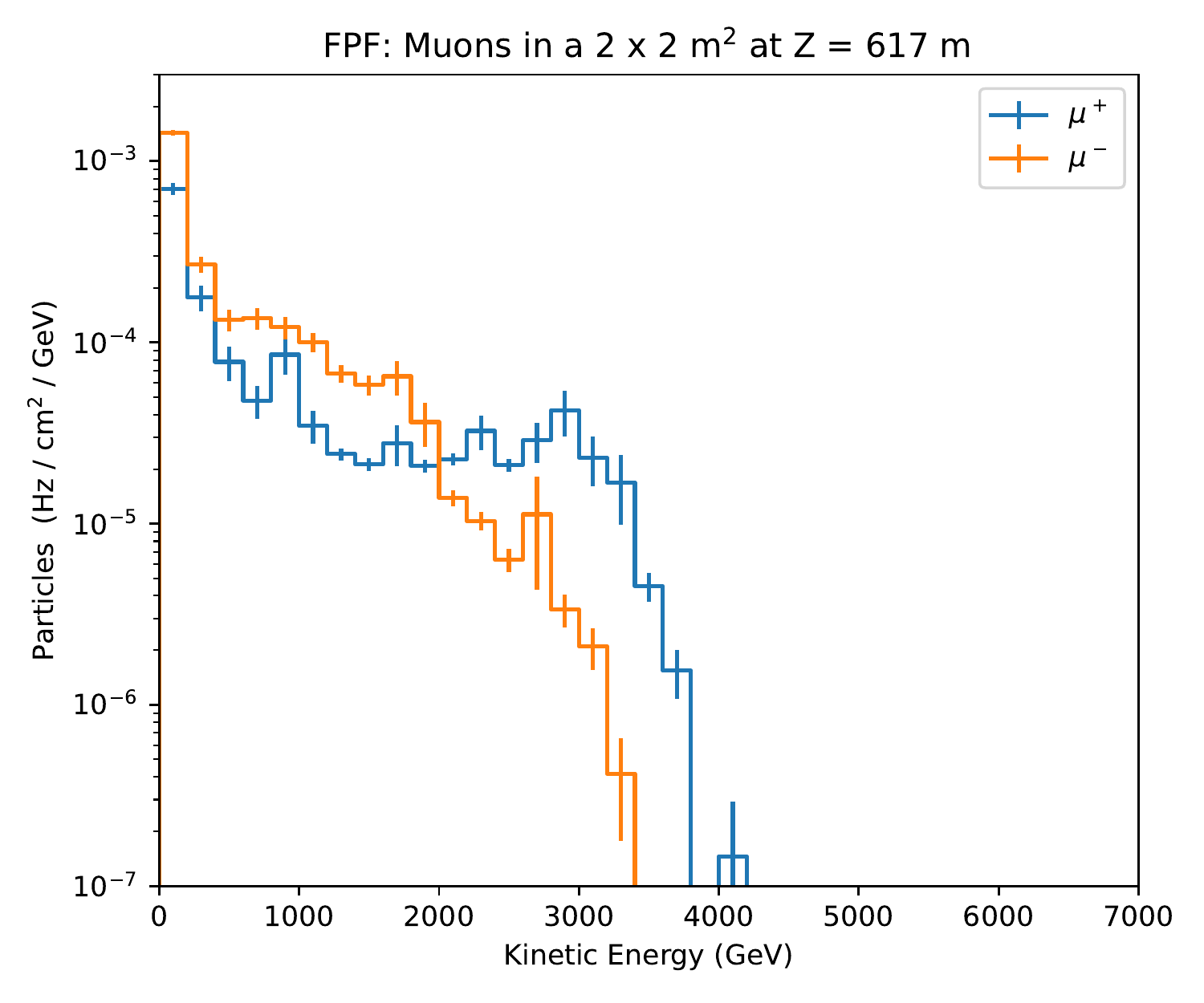}
\includegraphics[width=0.48\textwidth]{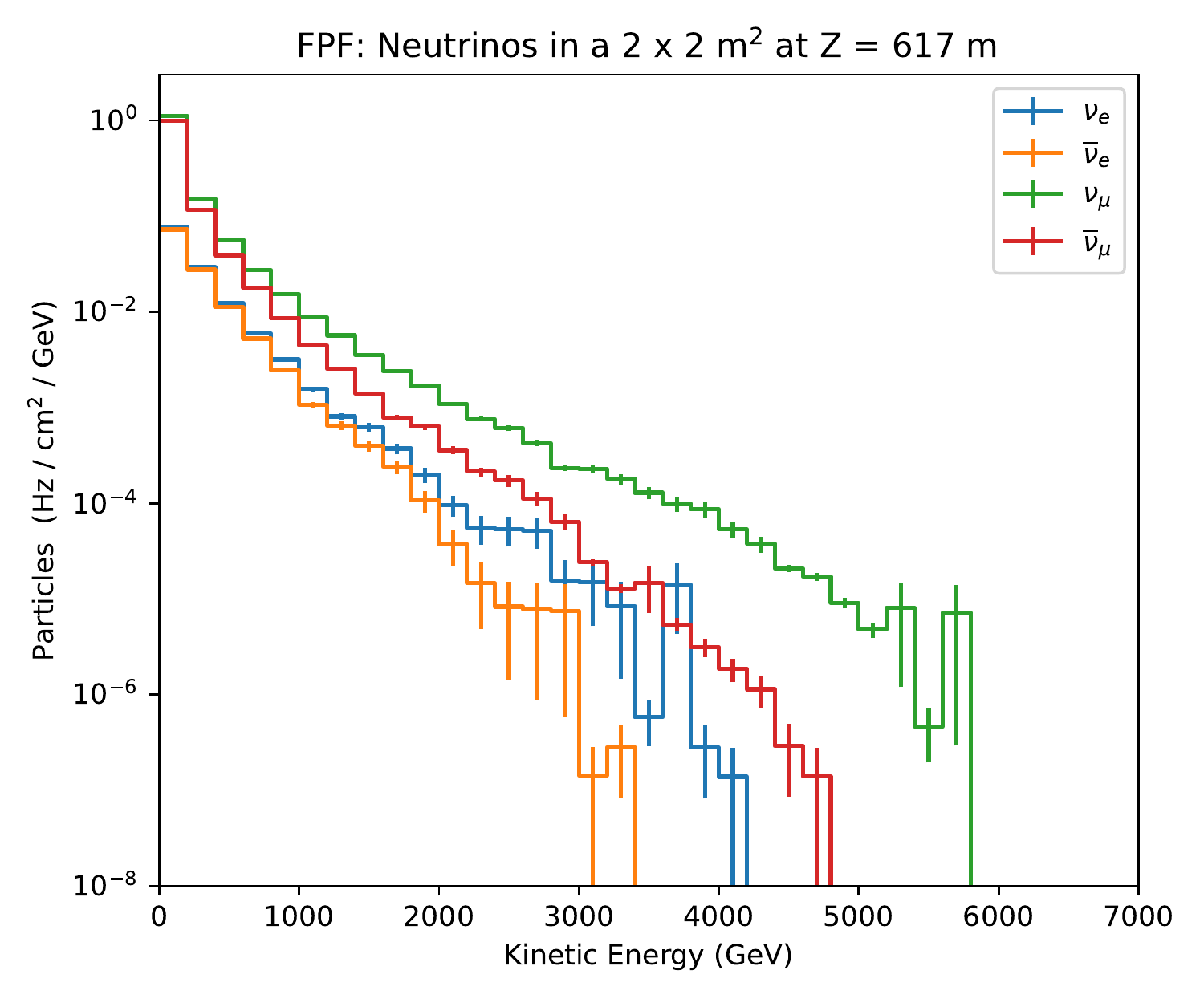}
\caption{Muon (left) and neutrino (right) spectra at the FPF location at $Z = 617$\,m integrated across an area of $2~\text{m} \times 2~\text{m}$ and scaled to the nominal HL-LHC luminosity of $5 \times 10^{34}$~cm$^{-2}$\,s$^{-1}$.}
\label{fig:bdsim:fpfspectra}
\end{figure}

The 2D muon distribution in front of the entrance to the FPF, $Z = 617$\,m from IP1 (the pink line in \cref{fig:bdsim:modelview2}), is shown in~\cref{fig:bdsim:fpfmuondistribution}. The spectrum of the muons and neutrinos at the FPF entrance plane is shown in \cref{fig:bdsim:fpfspectra}. It should be noted that the production of $\tau^{\pm}$ is not implemented in the version of \texttool{Geant4} used (10.7.p03), but the recently released \texttool{Geant4} V11.0 can produce them through positron annihilation. Here, we also expect only electron and muon neutrinos to be present in the simulation. The high end of the kinetic energy spectra is still statistically limited, but is expected to continue downwards.

The predicted muon fluxes based on averaging the $2~\text{m} \times 2~\text{m}$ sample plane at the FPF location are given in \cref{tab:bdsim:muonfluxes}. These are integrated across all kinetic energies in the simulation ($E_k \geq 10$\,GeV).

\begin{table}[tbp]
\centering
\begin{tabular}{c||c|c}
\hline \hline
Scenario & $\mu^{-}$ (Hz/cm$^{2}$) & $\mu^{+}$ (Hz/cm$^{2}$) \\
\hline
With sweeper  & $0.4937 \pm 0.0145$  & $0.2827 \pm 0.0139$  \\
No sweeper    & $0.5342 \pm 0.0163$  & $0.2881 \pm 0.0139$  \\
\hline \hline
\end{tabular}
\caption{Predicted muon fluxes at the entrance to the FPF at $Z = 617$\,m from IP1 along the LOS, based on the nominal HL-LHC luminosity of $5 \times 10^{34}$~cm$^{-2}$\,s$^{-1}$. The rate is calculated from a sample area of $2~\text{m} \times 2~\text{m}$ across all kinetic energies from the minimum simulated kinetic energy of 10\,GeV.}
\label{tab:bdsim:muonfluxes}
\end{table}

\begin{figure}[tbp]
\centering
\includegraphics[width=0.85\textwidth]{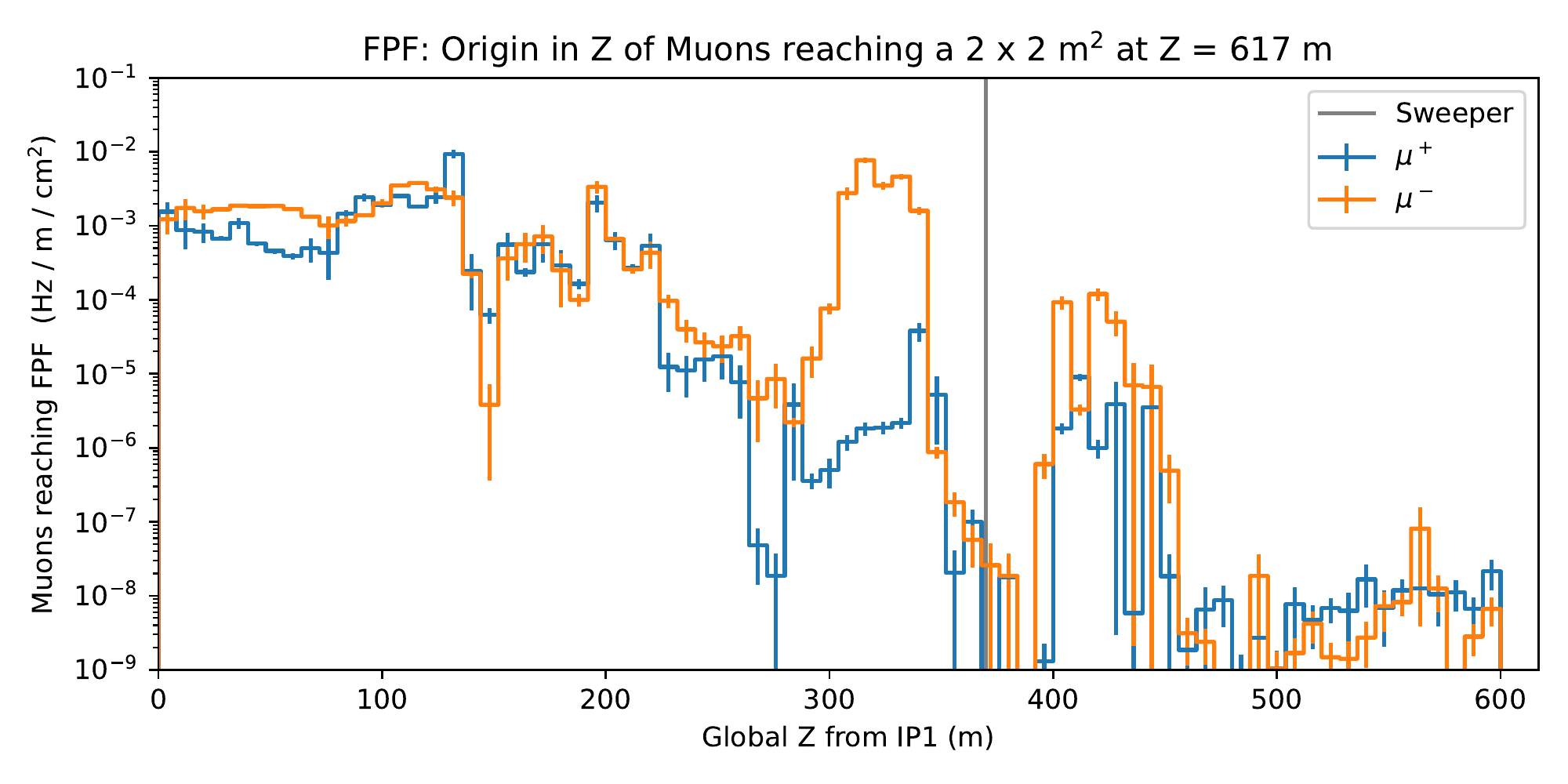}
\caption{The origin of muons reaching the $2~\text{m} \times 2~\text{m}$ sample plane at the $Z = 617$\,m FPF location. This figure shows where such muons were created as a function of the global Cartesian $Z$ coordinate, that is, the distance from IP1.  The results are scaled to the nominal HL-LHC luminosity of $5 \times 10^{34}$~cm$^{-2}$\,s$^{-1}$.}
\label{fig:bdsim:fpfmuonorigin}
\end{figure}

Using the trajectory information stored in ROOT-format \texttool{BDSIM} output files, we can visualise the origin of muons that reach the FPF $2~\text{m} \times 2~\text{m}$ sample plane.  The origins are shown in \cref{fig:bdsim:fpfmuonorigin} as a function of the global Cartesian coordinate $Z$. The distribution shows that the majority of muons originate from before the accelerator begins to curve at approximately 350\,m, and there appears to be significant structure. Many of the peaks (e.g., from $Z = 100$\,m to $Z = 250$\,m) are explained by the TAXN (the absorber at the 2-beam separation point), as well as by subsequent collimators that are designed to protect the machine from the high-power forward physics debris. Secondary particles can in turn produce muons that may reach the FPF from these locations. The two broad peaks are initially believed to be from the increasing dispersion (transverse position-energy correlation) in the accelerator as we enter the arcs. The increased dispersion causes greater transverse excursions for off-energy particles resulting in their loss and interaction with the accelerator, eventually producing muons.
These results provide an initial insight into where muons originate from and how they can be be mitigated with the choice of sweeper magnet location.

\subsection{Outlook}

The model presented is a first step in modelling the FPF muon backgrounds and further improvements are required. Firstly, a simplified geometry based on the aperture can be exported for a \texttool{RIVET} module suitable for testing new and existing generators and their models. This method and tool is described in Ref.~\cite{Kling:2021gos}. Such a technique allows rapid iteration to test key quantities and responses of potential  experiments to different physics models. The Monte Carlo sample here required $\sim 80000$~cpu-hours to be generated, despite the decay cross section biasing.  Whilst the simulation data provided is highly useful in understanding backgrounds, the approach using \texttool{RIVET} by Ref.~\cite{Kling:2021gos} using a simplified geometry from this model will allow many more variations to be studied.

Secondly, the geometry can be improved further to include the collimator vacuum tanks and cooling apparatus. The \texttool{BDSIM}-provided LHC magnet geometries will also be improved.  The magnetic fields used in the yokes of the magnets are parameterised multipole fields according to the Biot–Savart law. However, more accurate field maps will be used in the future. Additional systematic errors can be investigated, including the varied proton beam optical parameters (e.g., the crossing angle) and the effect of the rock density uncertainty.

Here, only the contribution from collisions at IP1 is presented. The contribution from the inelastic interaction of protons with residual gas in the vacuum can be simulated, as can any physics debris from other IPs that can cause proton losses, albeit at a low rate, throughout the whole accelerator.

The model presented and \texttool{BDSIM} in general present a method that can be used as part of a physics analyses and Monte Carlo chain to predict quantities at the FPF, as well as estimate the impact of design choices including a sweeper magnet.

\section{The PROPOSAL Framework For Simulating Particles Fluxes\label{sec:proposal}}


In addition to the \texttool{FLUKA}  and \texttool{BDSIM} frameworks discussed in \cref{sec:FLUKA} and \cref{sec:BDSIM}, respectively, the simulation framework \texttool{PROPOSAL} may be used to estimate particle fluxes at the FPF. \texttool{PROPOSAL} provides 3D Monte Carlo simulations of high-energy leptons and photons~\cite{Koehne:2013gpa}.  It is used in several experiments, for example, in the simulation chain of the IceCube Neutrino observatory, to propagate muons and taus~\cite{IceCube:2021uhz}.  However, its design is focused on flexibility, so that the framework can be used for a wide range of applications.

One key advantage of \texttool{PROPOSAL} is that the framework provides the options to find an optimal tradeoff between simulation precision and simulation performance for each individual use case. This is realized by introducing a combination of energy cuts: an absolute energy cut $e_\text{cut}$, as well as a relative energy cut $v_\text{cut}$. Energy losses above these energy cuts are treated by sampling each interaction individually, while all energy losses below these energy cuts are treated continuously. Furthermore, the settings for the energy cuts can be varied for the different parts of the simulation environment. This is especially useful for setups where specific areas may have to be simulated with a higher focus on precision, while other areas can be simulated with a focus on performance to save computing time.

To precisely simulate the energy loss processes of muons, an accurate knowledge of the underlying cross sections is necessary. For this purpose, \texttool{PROPOSAL} provides the possibility to choose from a set of different theoretical models, including up-to-date parametrizations of muon energy losses. The effects of these modern parametrizations, in comparison to commonly used parametrizations in other simulation tools, can be of the order of up to two percent, depending on the energy range~\cite{Sandrock:2018ivj}. Especially for muon propagation over long distances, these differences may have an observable impact on the simulation results. Furthermore, the flexible structure of \texttool{PROPOSAL} allows for a straightforward implementation of additional parametrizations into the framework, for example, including BSM physics, if necessary. 

To simulate the effects of magnetic fields on charged particles, which is relevant to estimate the effects of sweeper magnets on the muon background in the FPF, magnetic field deflection may be implemented directly in \texttool{PROPOSAL} in the future. As an alternative, \texttool{PROPOSAL} can be used together with the particle cascade simulation tool \texttool{CORSIKA~8}~\cite{Engel:2018akg}. In this case, the magnetic field deflection would be provided by \texttool{CORSIKA~8}, while the physics of muon interactions are provided by \texttool{PROPOSAL} via an existing interface.

\section{Sweeper Magnet \label{sec:sweepermagnet}}


For many of the physics goals of the FPF, reducing the rate of background particles traversing the FPF experiments would be very beneficial. For example, for emulsion-based neutrino detectors such as FASER$\nu$2, reducing the rate of background charged particles would allow the emulsion films to be exchanged less often, significantly reducing the cost. For a LAr-based neutrino/dark matter detector, reducing the background rate will allow improved physics sensitivity, and may allow one to loosen some of the requirements, resulting in a cheaper and simpler detector. In this section, we discuss the use of a sweeper magnet, to be installed on the LOS significantly in front of the FPF, to reduce the background rate from muons, is discussed. The effectiveness and feasibility of such a sweeper magnet are under study, and this section gives a snapshot of the current studies.

\subsection{Sweeper Magnet Location}

\begin{figure}[tbp]
  \centering
  \includegraphics[trim={0cm 0cm 0cm 1.5cm}, clip, width=0.95\linewidth]{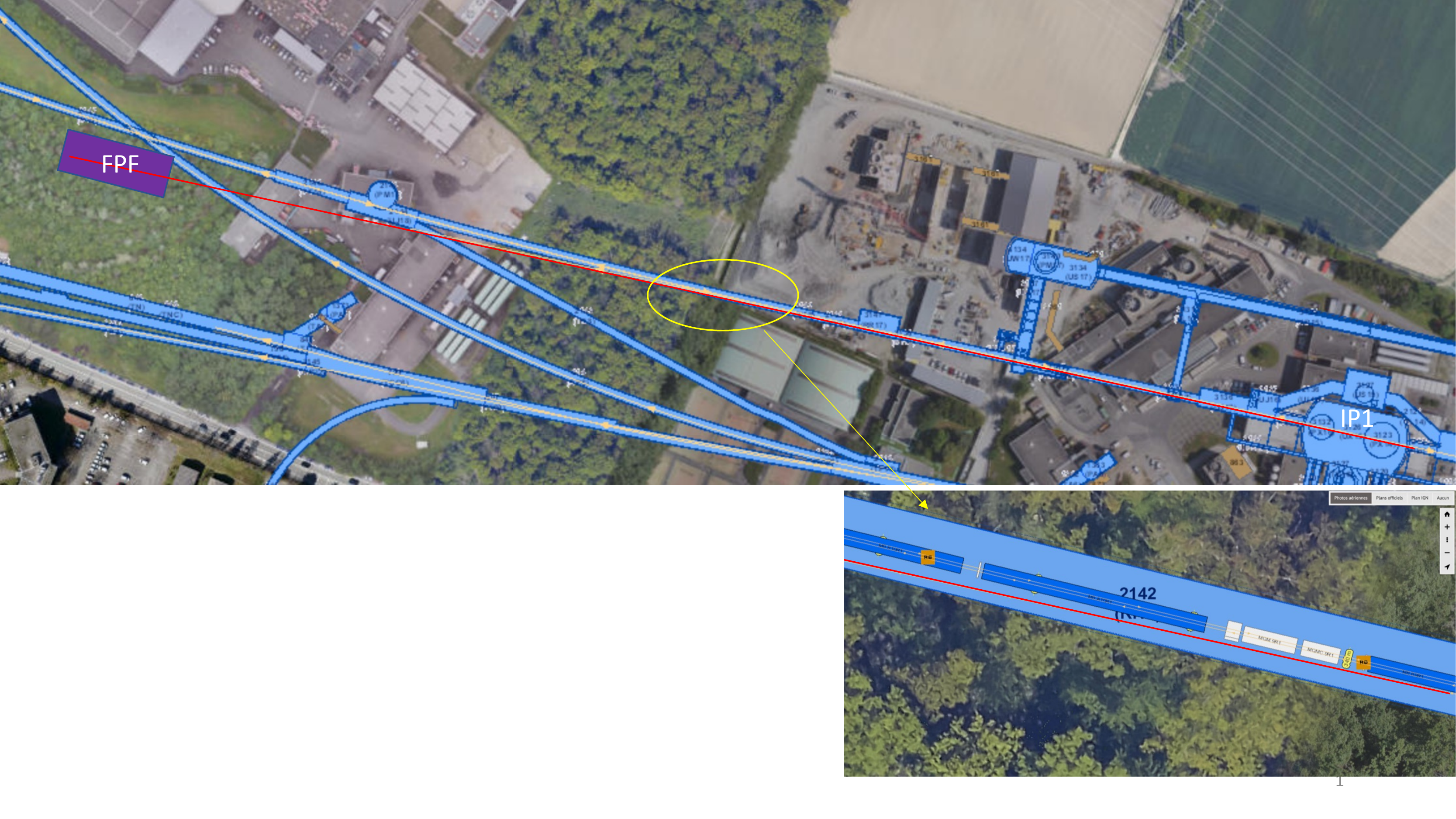}
  \caption{Plan view of the LHC complex between the ATLAS IP and the proposed FPF location. The LOS is shown as a red line, and the region where it has left the LHC magnet cryostats, but remains in the LHC tunnel, is highlighted in the zoomed in region. (Taken from the CERN GIS portal.)}
  \label{fig:sweeperMagnetLocation}
\end{figure}

Early feasibility studies carried out for the FASER experiment showed that we expect a significant rate of high-energy muons travelling along the LOS; see \cref{sec:validation} and Ref.~\cite{FASER:2020gpr}. The rate on the LOS is reduced due to the LHC magnets, particularly the separation/re-combination dipoles, D1 and D2. \texttool{FLUKA}  simulations and {\em in situ} measurements show that the direction of the muons is consistent with following the LOS from the IP, which means that a sweeper magnet should be placed on the LOS in between the IP and the FPF. The LOS is inside the LHC beam pipe for the long straight section of the LHC machine, and then inside the LHC magnet cryostat volume for the first part of the arc.  However, after about 350m from the IP, the LOS leaves the cryostat and remains in the LHC tunnel for about a further 50m. This is the location where a sweeper magnet could be installed, and it is shown in~\cref{fig:sweeperMagnetLocation}. 

\begin{figure}[tbp]
\centering
\includegraphics[width=0.85\textwidth]{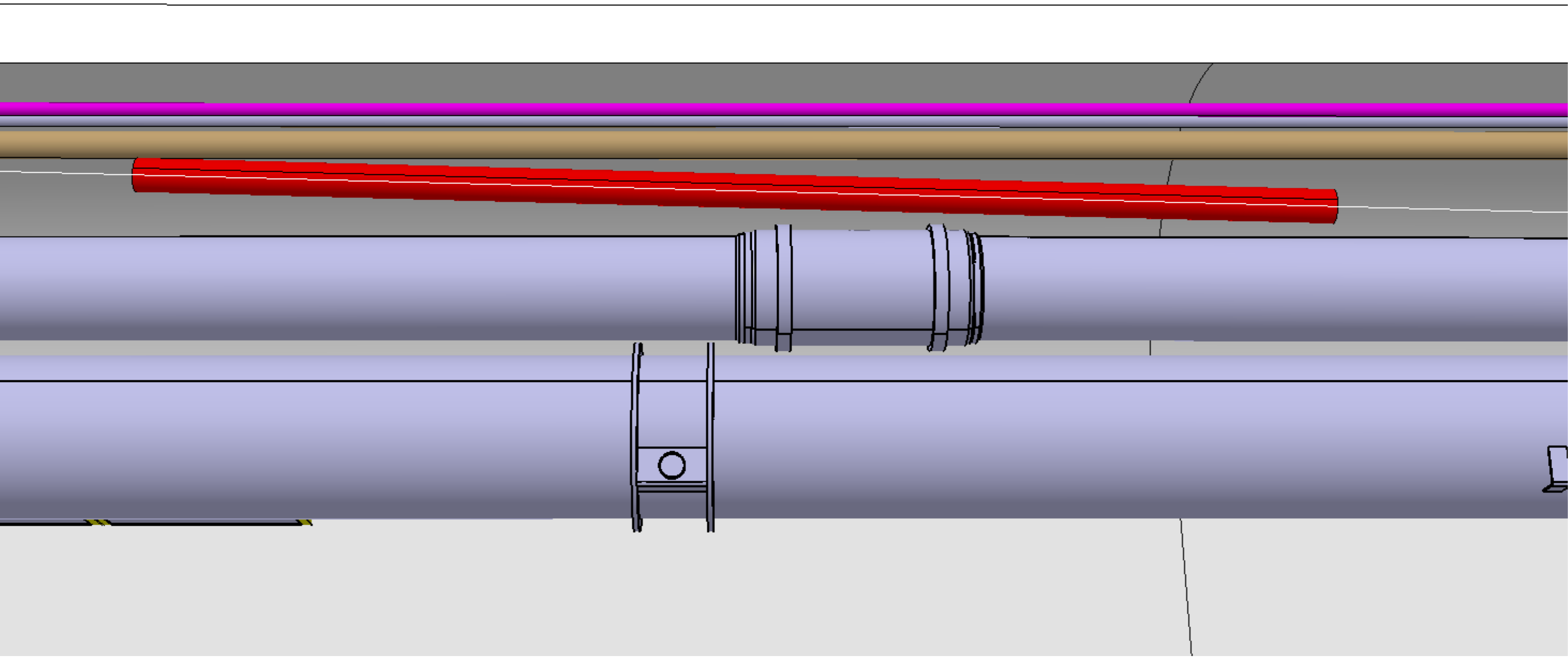}
\includegraphics[width=0.85\textwidth]{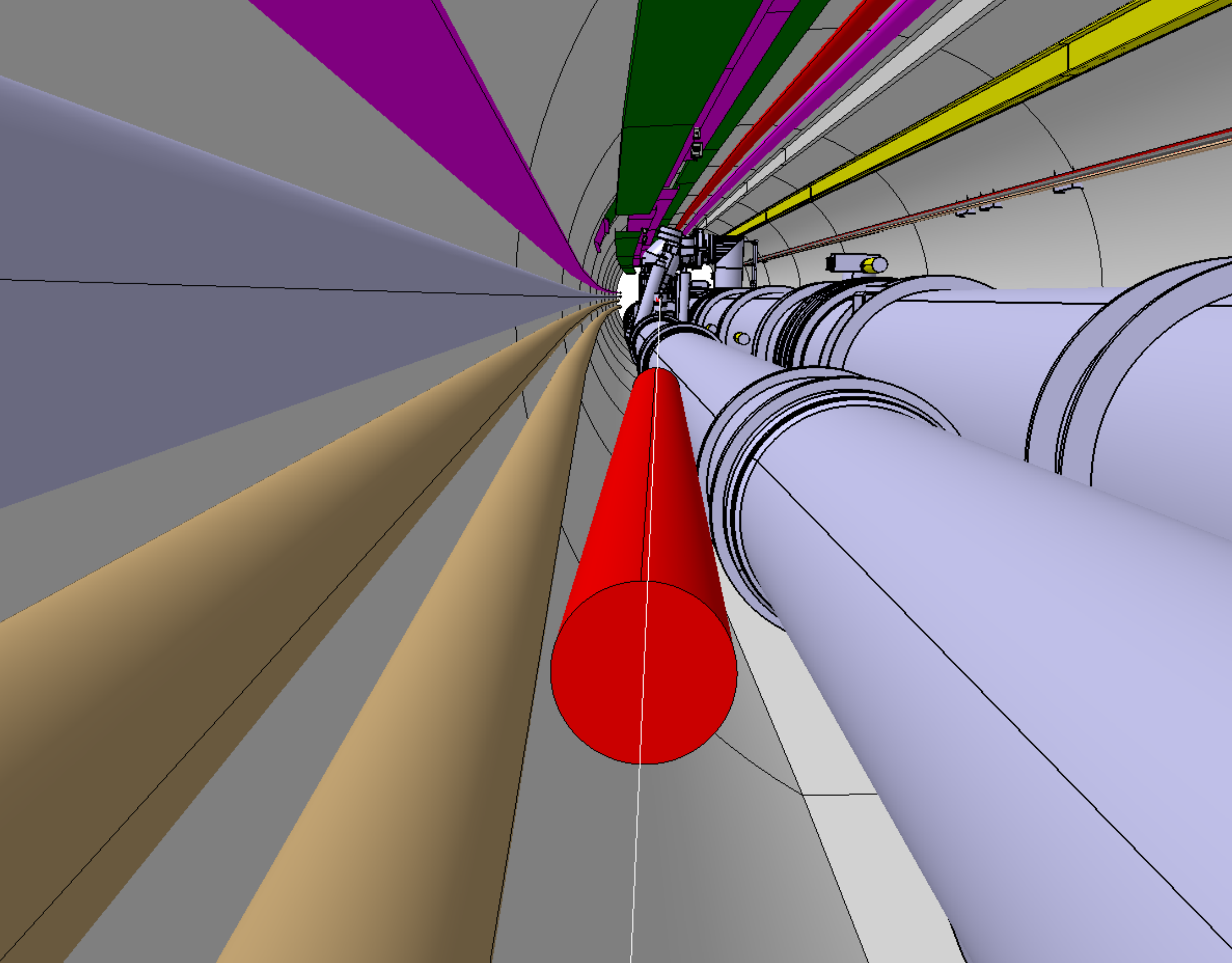}
\caption{Plan view (top) and side view (bottom) from the initial integration study to place a 7~m-long, 20~cm-diameter sweeper magnet on the LOS. The magnet is shown as a red cylinder. (The location shown is where the LOS leaves the beampipe to the west of ATLAS, but the symmetric location to the east of ATLAS and in the direction of the FPF is expected to be essentially identical.)  }
\label{fig:sweeperMagnetInitIntStudy}
\end{figure}

A preliminary integration study, using the 3D model of the LHC machine in this region, was carried out. This suggested that there would be space for a 7~m-long magnet, of transverse size 20~cm diameter, to be  placed on the LOS. This magnet would be about 200~m from the FPF, providing a significant lever arm for the deflected muons to move away from the LOS before they get to the FPF.  \cref{fig:sweeperMagnetInitIntStudy} shows the plan view and side view of the magnet compared to the existing LHC infrastructure from this integration study. The magnet is very close to the QRL cryogenic line, which is a delicate piece of equipment in the LHC tunnel. The study does not take into account the support of the magnet, or the handling equipment needed to install and remove it from this location.

\begin{figure}[tbp]
  \centering
    \includegraphics[width=0.9\textwidth]{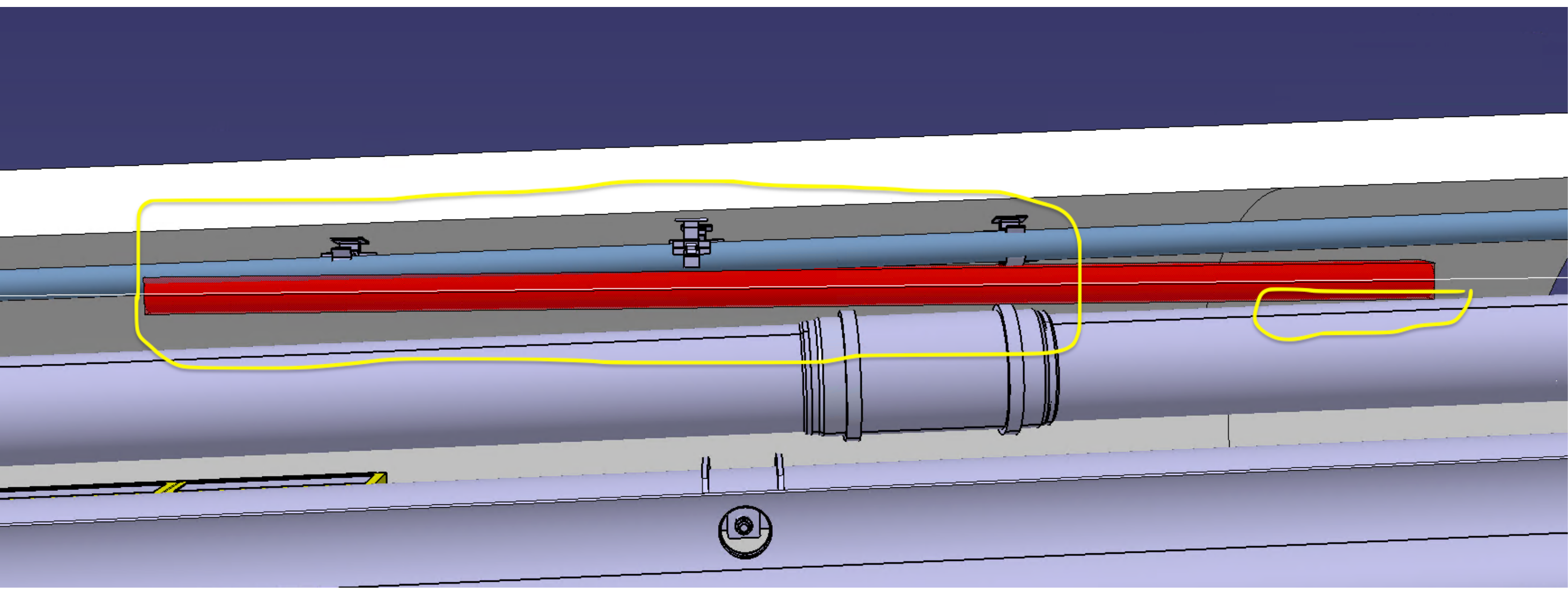}
    \includegraphics[width=0.9\textwidth]{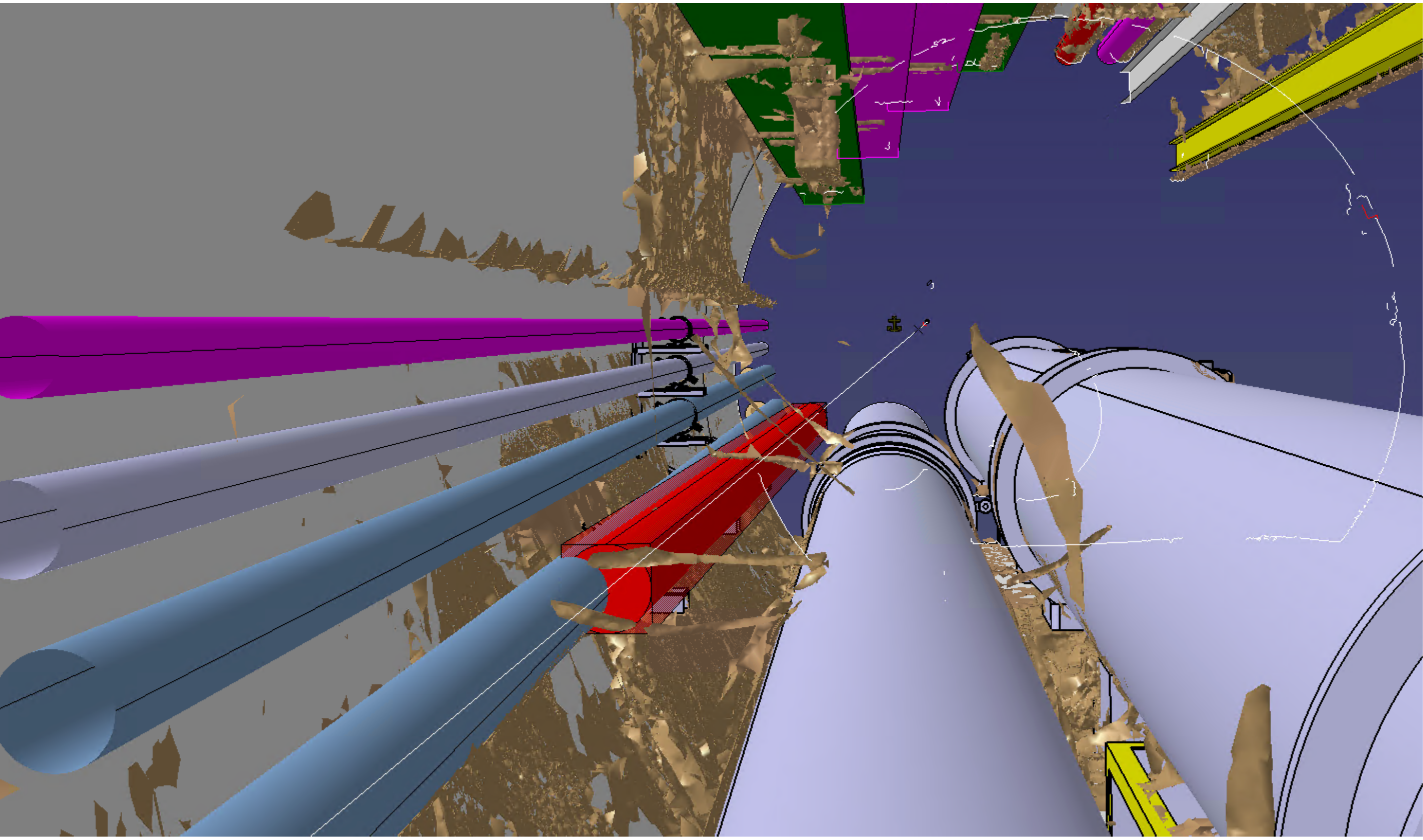}
  \caption{Integration study of the FPF sweeper magnet, using the LHC geometry updated after a laser scan in the relevant LHC area. The top panel shows the plan view, with the clash between the infrastructure, particularly the cryogenic warm return line (WRL), and the proposed sweeper magnet envelope highlighted in yellow. The bottom panel shows the view from in front of the sweeper magnet.  (The location shown is where the LOS leaves the beampipe to the west of ATLAS, but the symmetric location to the east of ATLAS and in the direction of the FPF is expected to be essentially identical.)}
  \label{fig:sweeperMagnetSecondIntStudy}
\end{figure}

Following this study, a laser scan was carried out in this area of the LHC tunnel to validate the 3D integration model used in the integration study. Unfortunately, this revealed a number of components that were not included in the original 3D model used.  These components significantly reduce the available space for the sweeper magnet, assuming no modifications to the infrastructure are made. In this case the magnet length would be limited to about 2~m, as shown in~\cref{fig:sweeperMagnetSecondIntStudy}. We are currently investigating if some of the infrastructure in this area of the LHC could be minimally modified, to allow the 7~m magnet to be installed there.

\subsection{Conceptual Magnet Design}

\begin{figure}[tbp]
  \centering
  \includegraphics[width=0.8\linewidth]{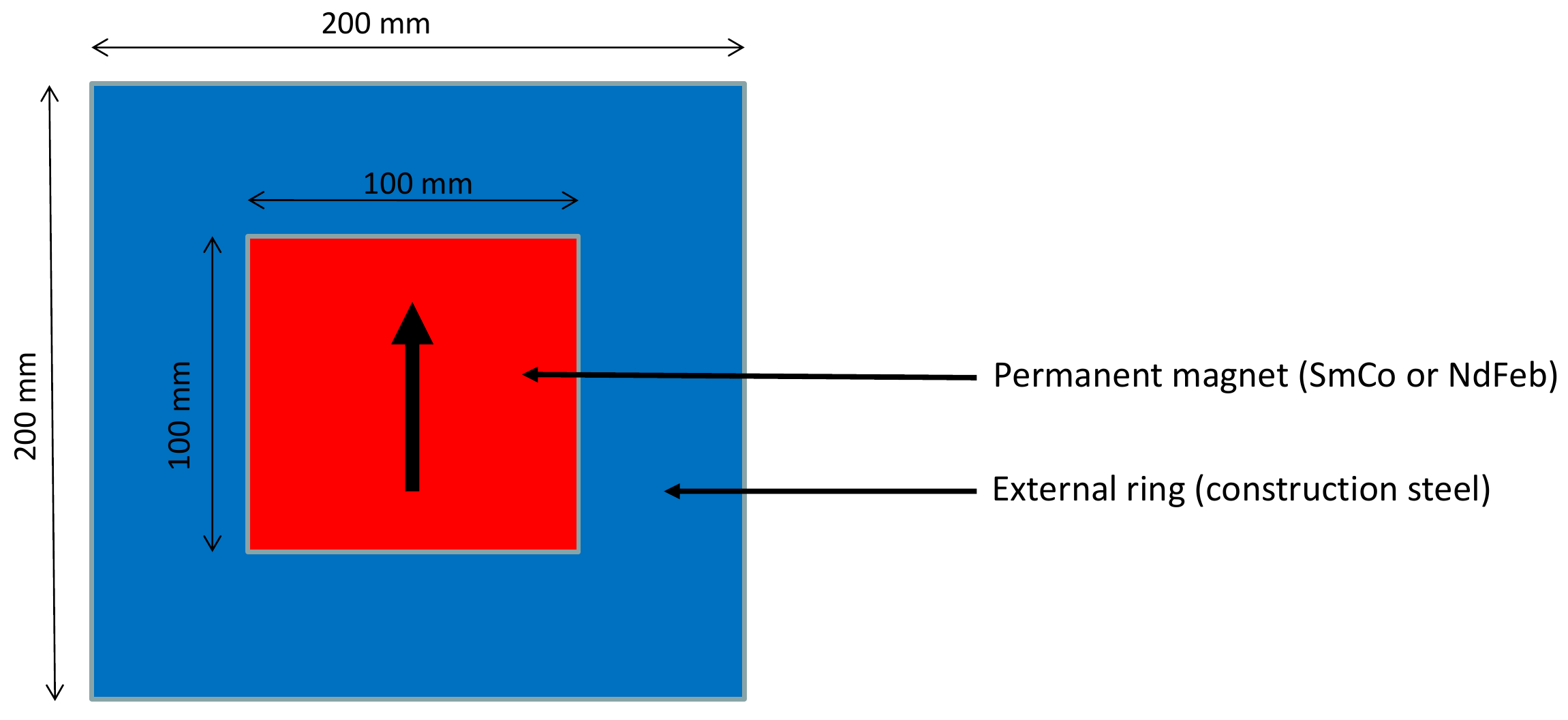}
  \caption{Figure showing the conceptual design of the proposed FPF sweeper magnet.}
  \label{fig:sweeperMagnetDesign}
\end{figure}

\begin{figure}[tbp]
  \centering
  \includegraphics[trim={0cm 1cm 0cm 1.5cm}, clip, width=1.0\linewidth]{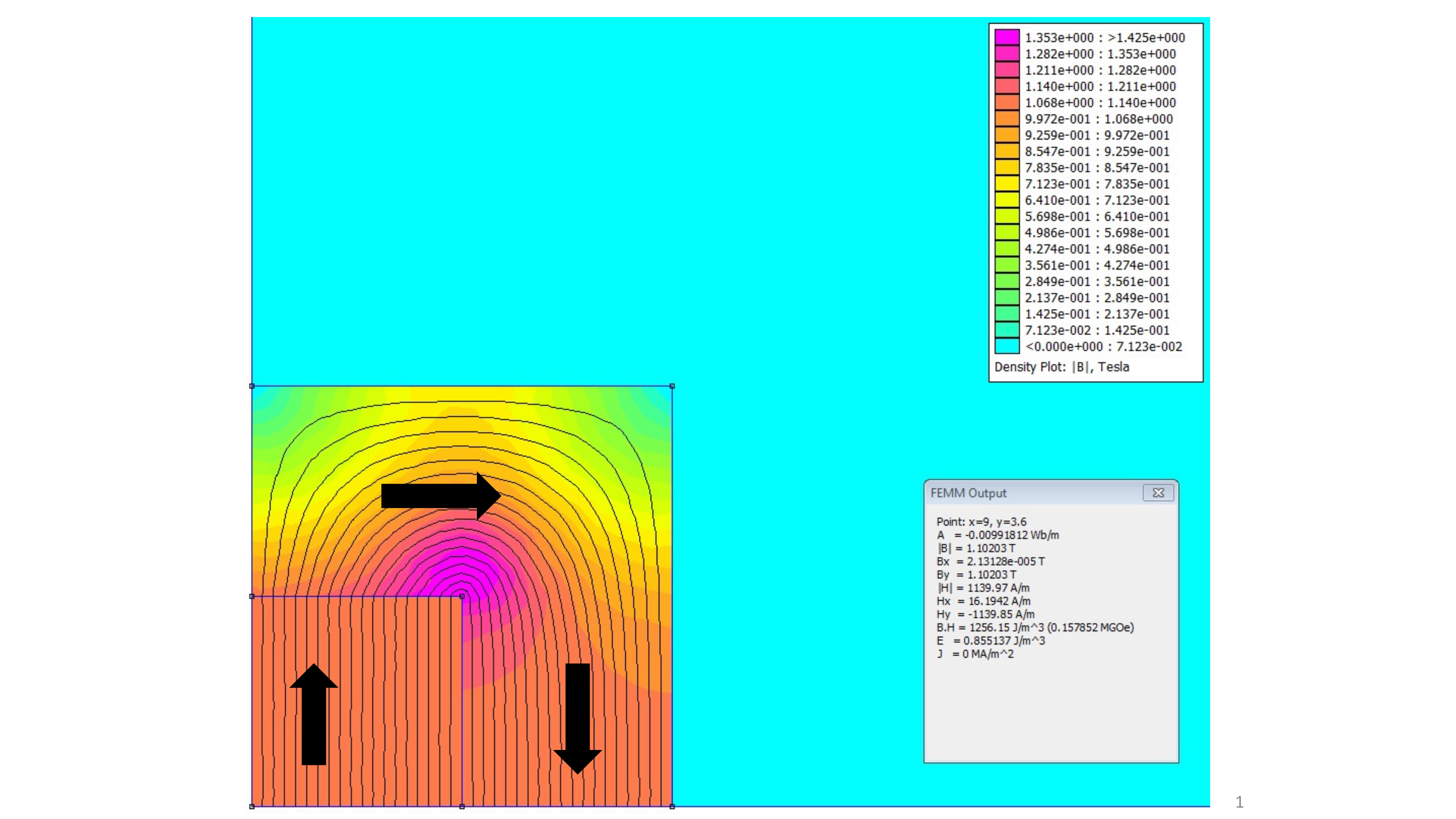}
  \caption{Figure showing the calculated magnetic field distribution inside the proposed sweeper magnet. Only a quarter of the magnet transverse area is shown, but the field is symmetric in the horizontal and vertical directions. The field is calculated using FEMM~\cite{Magnet:FEM}.}
  \label{fig:sweeperMagnetFieldMap}
\end{figure}

The location of the sweeper magnet in the LHC arc has significant radiation when the LHC is operating. It is therefore difficult to reliably operate a power converter in this region, which suggests that a permanent magnet could be a desirable solution for the sweeper magnet. A conceptual design for a possible permanent sweeper magnet is shown in~\cref{fig:sweeperMagnetDesign}. The magnet is made up of a central block of permanent magnetic material (e.g., NdFeb or SmCo) of 10 cm by 10 cm transverse dimensions, surrounded by an outer ring of construction steel (making up a 20~cm by 20~cm transverse size magnet). The magnetic field distribution for this configuration is shown in~\cref{fig:sweeperMagnetFieldMap} with a central field of 1.1~T for a SmCo magnet or 1.4~T for a NdFeb magnet.\footnote{There is a risk that the magnet could become de-magnitized by radiation for the NdFeB magnetic material.  This will need to be studied further before deciding on what material to use.} The radial stray magnetic field is negligible at the level of $<$0.002~T at 10 cm from the magnet, which is not expected to be problematic for the LHC beam or any equipment in the LHC tunnel.  Assuming a field integral of 7 Tm for the sweeper magnet, this would sweep a 100 GeV (500 GeV) muon 4.2~m (0.8~m) from the LOS at the FPF (200~m away). However, it still needs to be studied with simulation if the magnet would also sweep muons into the LOS, and if so what the overall reduction of the background along the LOS would be.

For the conceptual design mentioned above, a 7~m-long magnet would weigh 2.3 tonnes and would likely be constructed and installed in 1~m-long sections. The cost of such a magnet (not taking into account the support and handling infrastructure, nor any modifications that would need to be made to the tunnel infrastructure to create room for the magnet to be installed) is about 150 kCHF. 

%% file: sec_experiments.tex
\contributors{Jonathan L.~Feng (convener), 
Akitaka Ariga, 
Tomoko Ariga, 
Alan J.~Barr, 
Jianming Bian, 
Jamie Boyd,
David~W.~Casper, 
Matthew~Citron, 
Giovanni~De~Lellis, 
Albert~De~Roeck, 
Antonia~Di~Crecsenzo, 
Milind~V.~Diwan, 
Christopher S.~Hill, 
Tomohiro Inada, 
Richard Jacobsson, 
Masahiro Komatsu,
Josh~McFayden, 
Toshiyuki Nakano, 
Hidetoshi Otono, 
Filippo~Resnati, 
Hiroki Rokujo, 
Osamu Sato, 
Yosuke Takubo, 
Yu-Dai Tsai, 
and Wenjie Wu}

The FPF will provide space along the LOS for a suite of experiments to explore the diverse physics signals that can be uniquely probed in the forward region.  At present, there are five experiments that have been proposed for the FPF.  These include FASER2 to search for long-lived particles; FASER$\nu$2 and AdvSND to study neutrinos and search for new particles; FLArE to detect neutrinos and search for DM; and FORMOSA to search for mCPs and related particles. The detector design for these experiments is a work in progress.  In this Chapter, we describe the current status of each of these experiments and detectors in turn.

\section{FASER2 \label{sec:faser2}}


\subsection{Physics Goals and Design Considerations \label{sec:faser2physics}}

The existing FASER experiment is already set to probe new parameter space in the search for BSM physics. However, the overall size of FASER, and therefore its possible decay volume, has been heavily constrained, ever since the initial stages of planning, by the available space underground. This directly affects the sensitivity and reach obtainable by FASER, as, for many representative BSM models, the sensitivity is directly related to the length and radius of the decay volume. This strongly motivates the case for an enlarged detector, FASER2, which was already explored in the original FASER proposal~\cite{Feng:2017uoz}, the FASER Letter of Intent~\cite{FASER:2018ceo}, Technical Proposal~\cite{FASER:2020gpr}, and physics reach~\cite{FASER:2018eoc} documents. 

In previous studies, the nominal FASER2 design is comprised of a decay volume 5m in length and 2m in diameter. This results in an angular acceptance of neutral pions that increases from 0.6\% in FASER to 10\% in FASER2, as shown in Fig.~5 (left) of Ref.~\cite{FASER:2018eoc}.  In addition, there is a significant improvement in sensitivities to LLPs produced in the decays of heavy mesons, due to the additional acceptance to $B$-meson production, as shown in Fig.~5 (right) of Ref.~\cite{FASER:2018eoc}.  The larger decay volume also improves sensitivity to larger LLP masses and longer LLP lifetimes. The combined effect of all these factors is an improvement in reach of 4 orders of magnitude for some models.

There are several key design considerations for FASER2. The larger radius puts less emphasis on the importance of being directly on-axis. The significant improvement in sensitivity to higher mass LLPs has the consequence of exposing FASER2 to a more complicated mixture of decay channels; this strongly motivates the need for particle identification capabilities, for example, between electrons, muons, pions, and kaons. In addition, the factor of 10 increase in decay volume radius corresponds to a factor of 100 increase in area that needs to be instrumented. It therefore becomes much more challenging to accommodate an extended version of the ATLAS SCT tracker module configuration, due to cost considerations and the services required. However, the marked increase in detector length of FASER2 creates the potential to achieve larger decay product separations with different and possibly cheaper technology. The overall increase in detector size will also lead to a larger background rate, which is likely to require more complicated trigger and data analysis techniques.

\subsection{Detector Configurations \label{sec:faser2detector}}

Given these considerations there is much to be studied in terms of possible detector configurations and technologies. So far studies have focused on general size/layout optimisations. Several possibilities for decay volume sizes and locations have been considered, as shown in~\cref{tab:FASER2_configs}.  These are based on the constraints imposed from the FPF facility scenarios, which we will refer as the ``Alcoves'' and ``New Cavern'' options. 

\begin{table}[tbp]
    \centering
    \begin{tabular}{l|p{1.5cm}|p{2cm}|p{2.5cm}|p{2.5cm}|p{2.5cm}}
    \hline\hline
Scenario & Distance to IP [m]	& Available Length [m] &	Decay Volume Length [m]	& Available  Diameter [m] & Decay Volume Diameter [m] \\
\hline
Original (F2) &	480 & 15 & 5 & 2 & 2 (/ 1 / 0.5) \\
Alcoves (S1) &	500 &	5	& 1.5 (/ 2) & 1.5 (/ 2) & 2 / 1 (/ 0.5) \\
New Cavern (S2) & 620 &	25	& 10 (/ 15 / 20) &	2 &	2 / 1 (/ 0.5) \\
\hline\hline
\end{tabular}
\caption{Possible FASER2 design parameters, including the FASER2 parameters of Ref.~\cite{FASER:2018eoc} (``Original'') and configurations appropriate for the two preferred FPF scenarios: the UJ12 alcoves option (``Alcoves'') and the baseline purpose-built facility option (``New Cavern''). Numbers in brackets show configurations that have been considered, but are not shown in the figures or discussed in the text.}
\label{tab:FASER2_configs}
\end{table}

\begin{figure}[tbhp]
  \centering
  \includegraphics[width=0.49\textwidth]{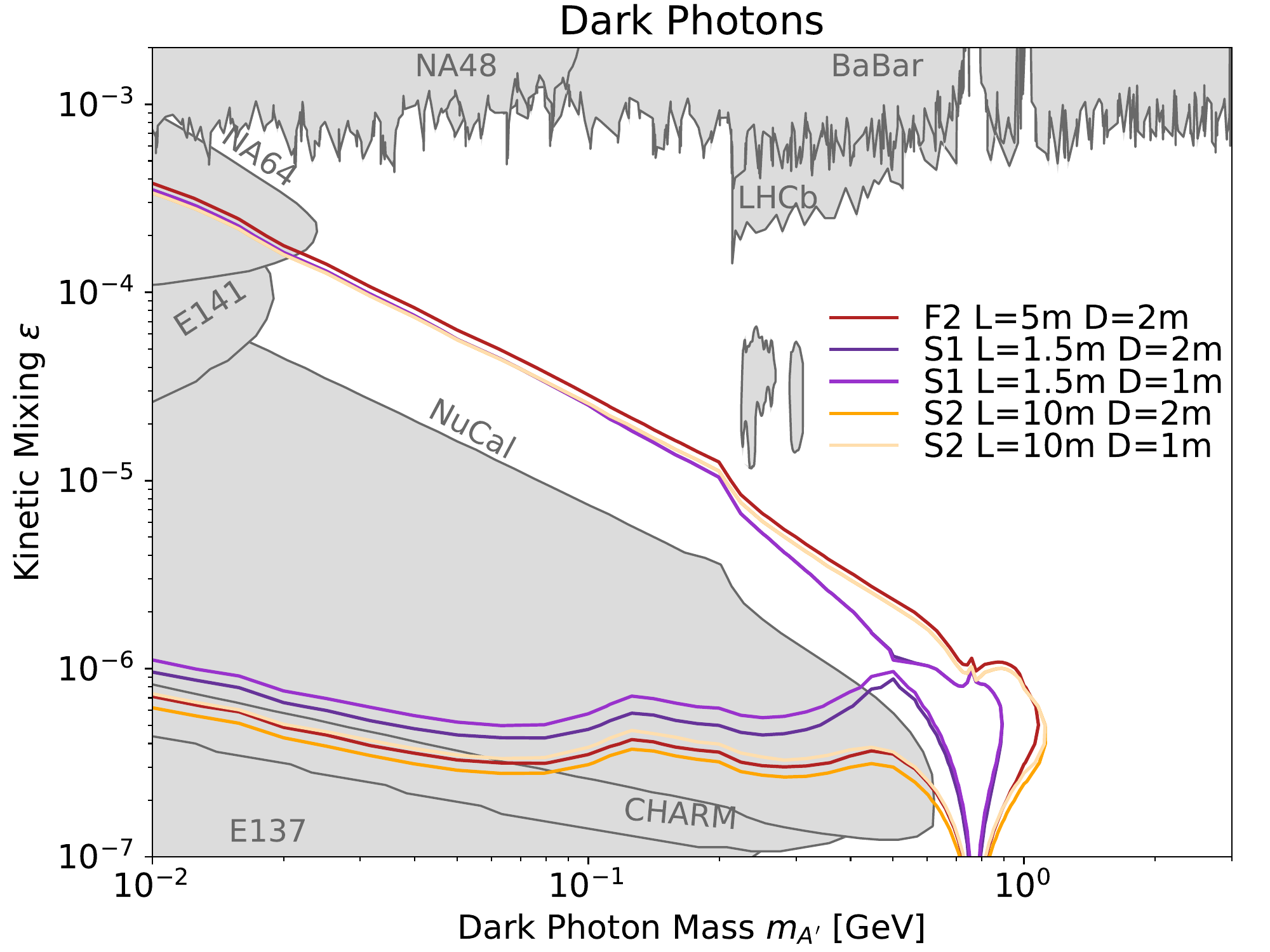}
  \includegraphics[width=0.49\textwidth]{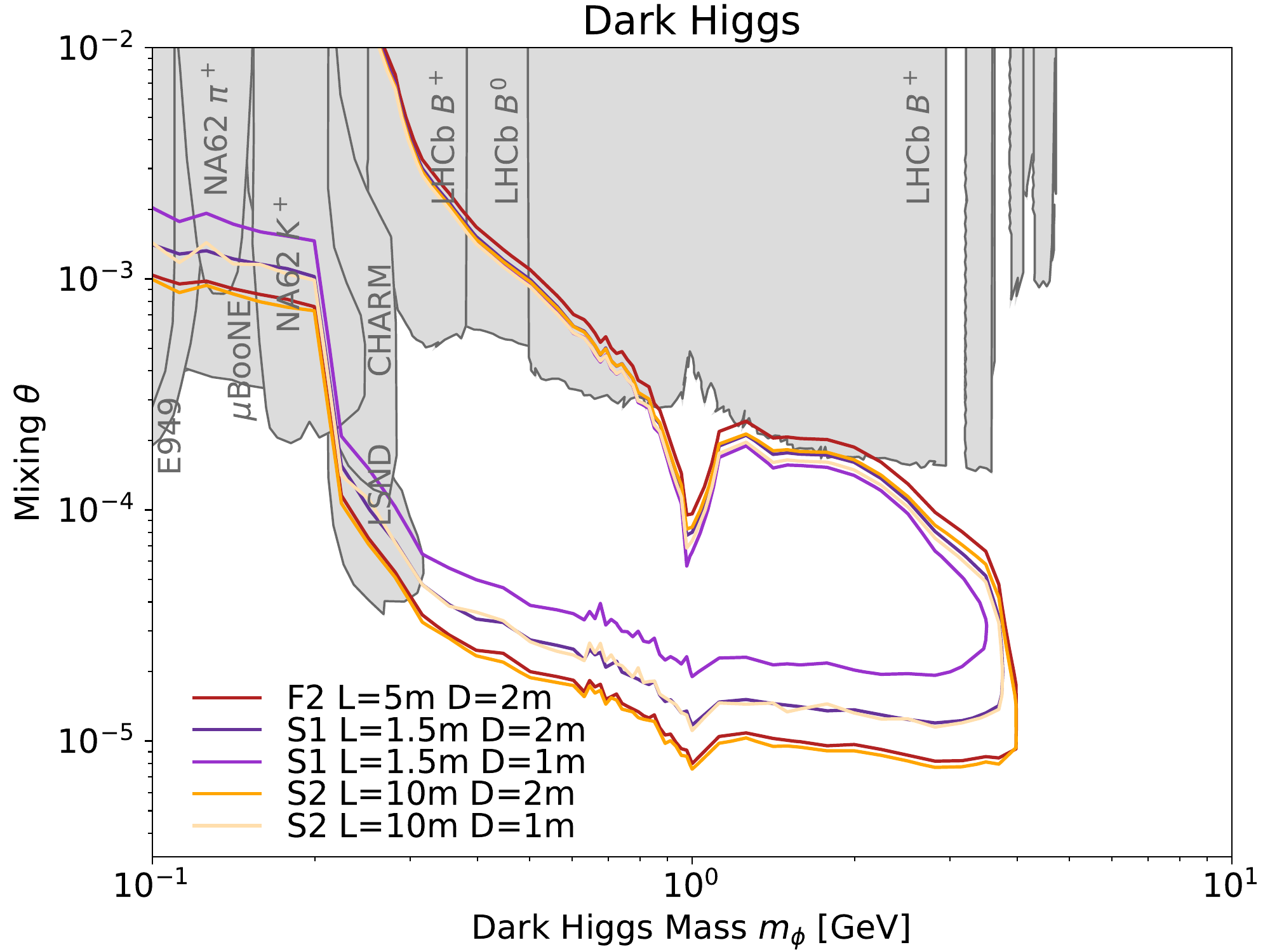}  
  \caption{Projected discovery sensitivities for various FASER2 scenarios in the dark photon (left) and dark Higgs boson (right) models.}
  \label{fig:FASER2-Scenarios}
\end{figure}

\cref{fig:FASER2-Scenarios} shows the sensitivity to dark photon (left) and dark Higgs boson (right) models for a selection of the possible FASER2 scenarios shown in~\cref{tab:FASER2_configs}.  The sensitivity contours have been determined using the \texttool{FORESEE} tool~\cite{Kling:2021fwx}, described in \cref{sec:foresee}.  As already discussed, the sensitivity to dark photons in the original FASER2 configuration is significantly improved with respect to FASER. However, the Alcoves option does not allow for such a large detector, and the figure shows a significant loss of sensitivity with respect to the original configuration. On the other hand, the New Cavern option is able to accommodate a detector that can recover and even improve upon the original FASER2 sensitivity, making it the strongly preferred scenario. The only downside to the New Cavern scenario is the slight shift in sensitivity to lower couplings, resulting from its increased distance from the ATLAS IP, but this is a rather small effect.  Similar conclusions can be drawn for the dark Higgs boson sensitivity, where the effect of the increased radius is even stronger, because of the enhancement in acceptance of $B$-meson decays, as already discussed.

\begin{figure}[tbp]
  \centering
  \includegraphics[width=0.95\textwidth]{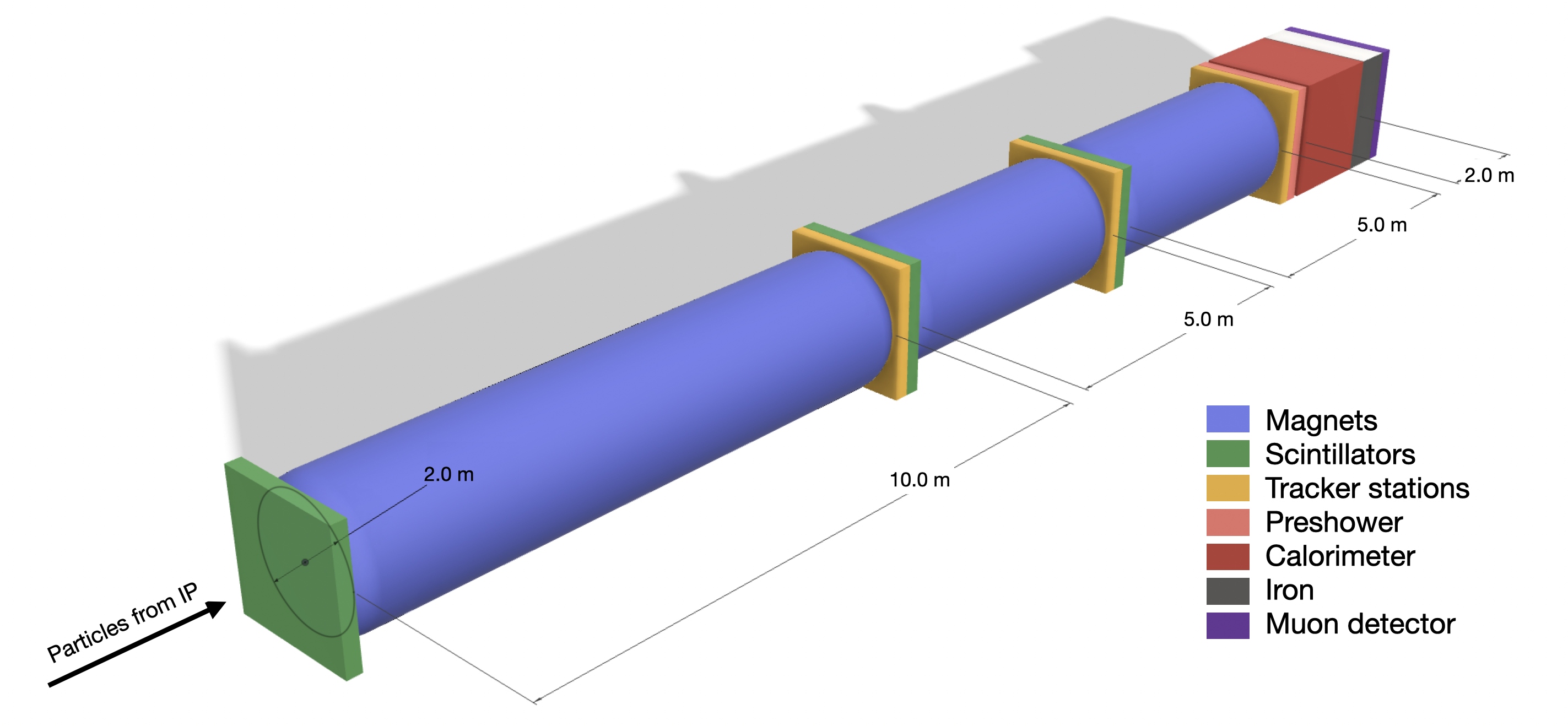}
  \caption{Schematic diagram of the proposed FASER2 detector.}
  \label{fig:FASER2-Design}
\end{figure}

The FASER2 design can be optimised for either the New Cavern or Alcoves option. The design is not yet strictly defined, but it will be similar to FASER in the general philosophy, modulo changes needed to ameliorate some of the additional challenges already described. A schematic layout of the FASER2 detector, assuming the baseline New Cavern option, is given in~\cref{fig:FASER2-Design}. The veto system will be scintillator-based, similar to FASER. The significantly increased area of the active volume makes it impractical to use silicon tracker technology. A SiPM and scintillating fibre tracker technology, such as LHCb's SciFi detector~\cite{Hopchev:2017tee}, is a strong candidate to replace the ATLAS SCT modules used in FASER. In addition, Monitored Drift Tube (MDT) technology, similar to that used in the ATLAS New Small Wheel~\cite{Kawamoto:1552862}, is also being considered, although this option requires the use of gases in the LHC tunnel that could be problematic for the UJ12 alcoves scenario. 

For the magnets, superconducting technology would be required to maintain sufficient field strength across the much larger aperture.  Suitable technology for this already exists and can be built for FASER2. There are several possibilities for the cooling of such magnets; the use of cryocoolers and the possibility to share a single cryostat across several magnets are being considered. The calorimeter needs to have sufficient spatial resolution to be able to identify particles at $\sim1-10$~mm separation; good energy resolution; improved longitudinal separation with respect to FASER; and the capability to perform particle identification, separating, for example, electron and pions. Dual readout calorimetry~\cite{Lee:2017xss,Antonello:2018sna} is a good candidate to satisfy all these requirements. Finally, the ability to identify separately electrons and muons would be very important for signal characterisation, background suppression, and for the interface with FASER$\nu$2 (see \cref{sec:fasernu2}) and other detectors. To achieve this, a mass of iron will be placed after the calorimeter, with sufficient depth to absorb pions and other hadrons, followed by a detector for muon identification.

\subsection{Magnet and Tracker Requirements \label{sec:faser2geant4}}

\begin{figure}[tbp]
  \centering
  \includegraphics[width=0.49\textwidth]{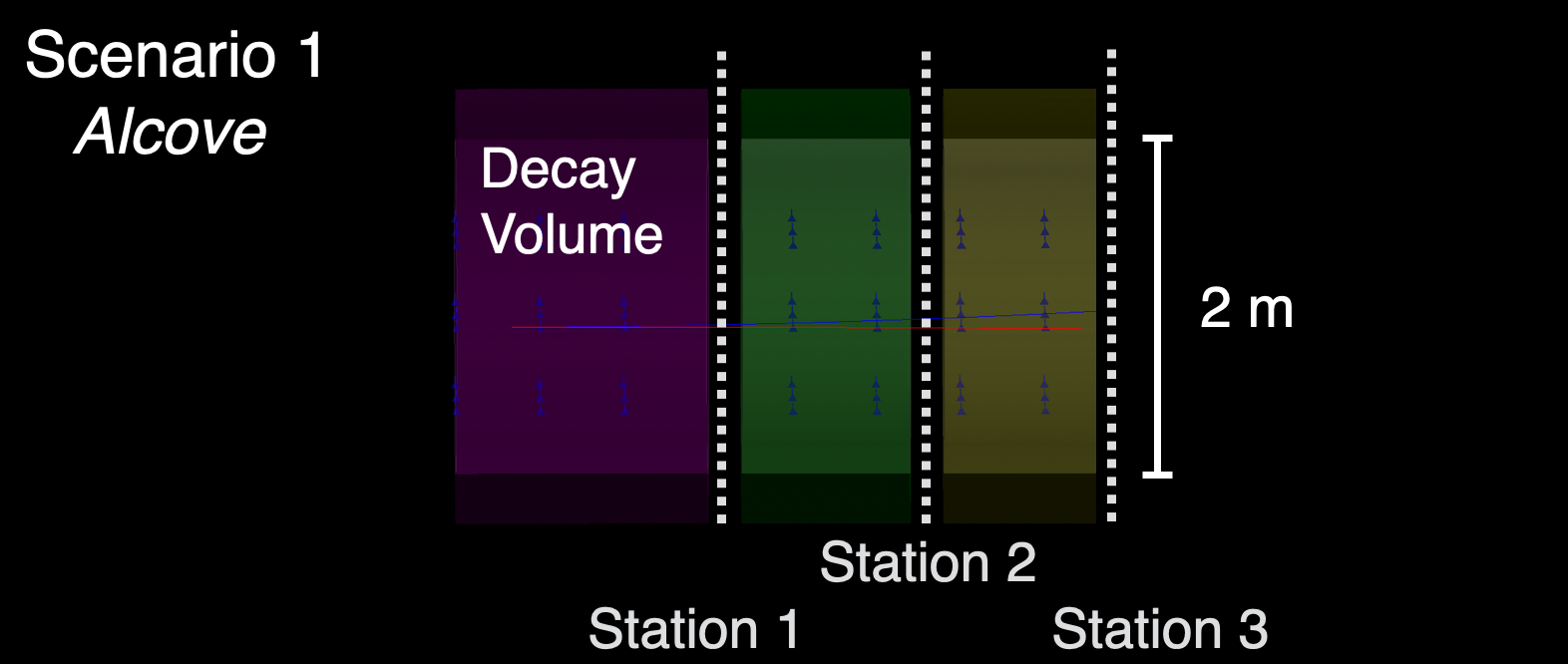}
  \includegraphics[width=0.49\textwidth]{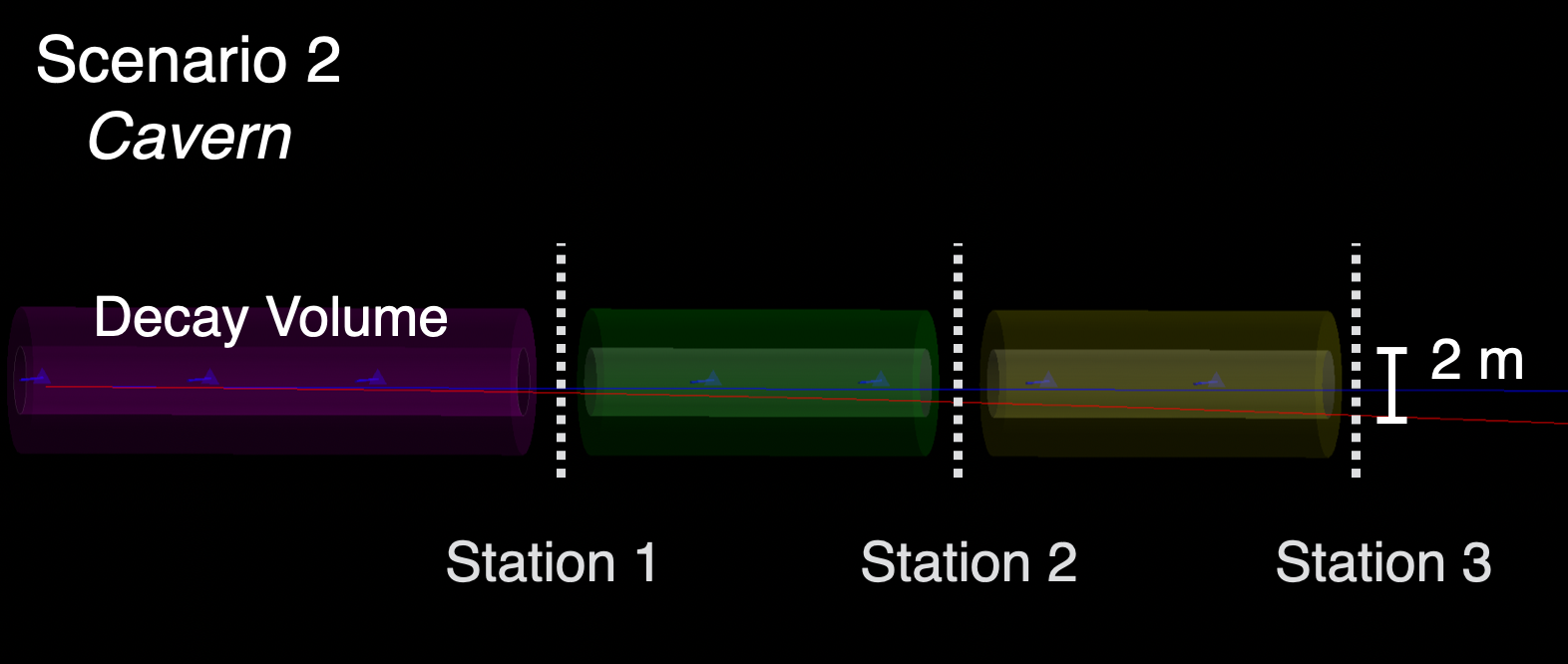}  
  \caption{Schematic views of the FASER2 \texttool{Geant4} simulations for the FPF Alcoves (left) and New Cavern (right) options.  The tracker stations are indicated by vertical dashed lines. For the Alcoves design, the tracker stations are 1.5, 2.7, and 3.9 m from the front of the decay volume. For the New Cavern design, the stations are 10.0, 15.5, and 21.0 m from the front of the decay volume. }
  \label{fig:FASER2-sims}
\end{figure}

The FASER2 design requirements that must be defined with highest priority are the needs of the spectrometer, both in terms of the magnetic field strength and the tracker resolution. Studies are underway to investigate the characteristic particle separations, starting with dark photons that decay to $e^+ e^-$ pairs. \cref{fig:FASER2-sims} shows a schematic of the \texttool{Geant4}~\cite{GEANT4:2002zbu} simulations for the Alcoves and New Cavern scenarios. In each case, tracker station positions are indicated by vertical dashed lines. 

\begin{figure}[tbhp]
\centering
\includegraphics[width=0.49\textwidth]{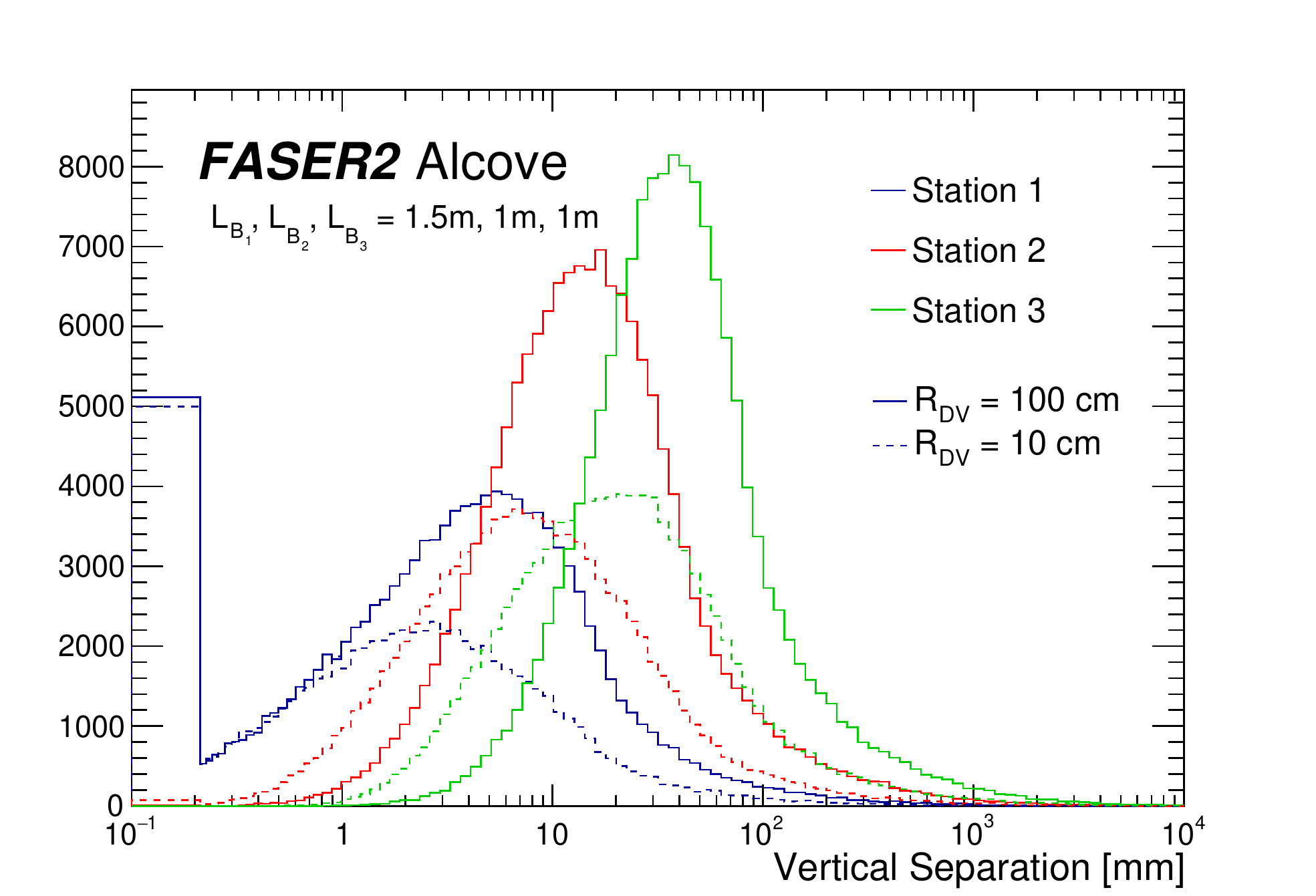}
\includegraphics[width=0.49\textwidth]{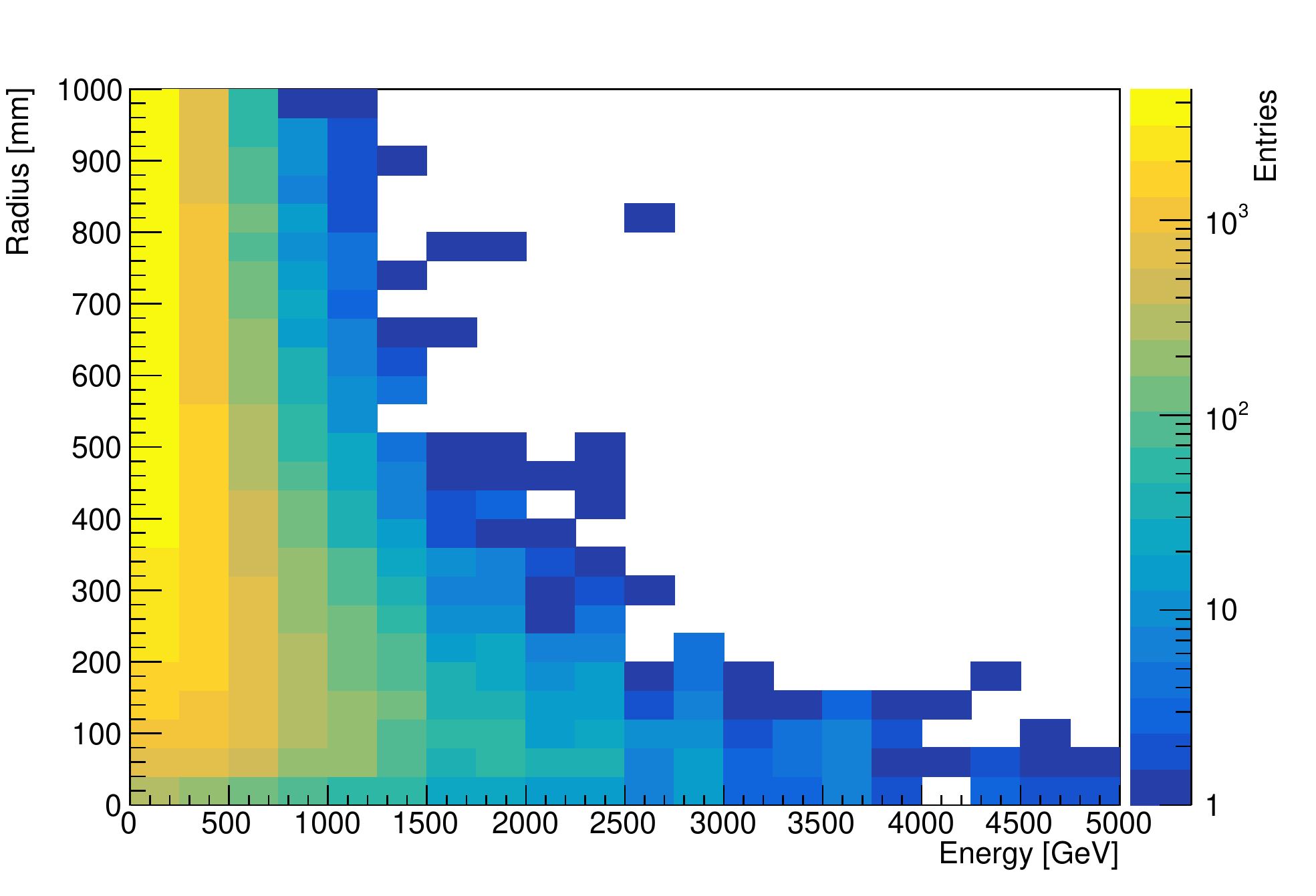}  
\caption{Left: Results from \texttool{Geant4} simulations of the FASER2 Alcoves detector for the distribution of transverse separations of $e^+$ and $e^-$ tracks resulting from 100 MeV dark photons decaying to $e^+ e^-$ pairs in a 1 T magnetic field.  The different colors are the distributions at the various tracker station locations shown in~\cref{fig:FASER2-sims}.  The solid contours are for all dark photons decaying in the decay volume.  The dashed contours are for the subset of dark photons that decay within a decay volume radius of $R_{\text{DV}} = 10~\text{cm}$ from the LOS. Right: The distribution of dark photon decays in the $(\text{Energy}, \text{Radius})$ plane, where the Energy is the energy of the dark photon, and Radius is the distance from the LOS at which it decays.  Highly energetic dark photons decay closer to the LOS.}
\label{fig:FASER2-alcove-g4}
\end{figure}

In \cref{fig:FASER2-alcove-g4} (left), the solid contours show the distribution of transverse spatial separation between the $e^+$ and $e^-$ tracks resulting from a 100 MeV dark photon decaying to $e^+ e^-$ pairs in a 1 T magnetic field.  With this magnetic field strength, the track separations are approximately 5~mm at Station 1 and 20~mm at Station 3 (and at the calorimeter).  These lengths determine the spatial resolutions that would be needed to reliably identify separate decay products.

Considering the same signal model, and taking into account the almost twice as strong magnetic field assumed here, the particle separations are much larger than might naively be expected based on what was observed in FASER~\cite{FASER:2018ceo} for a similar longitudinal detector layout. This is explained by \cref{fig:FASER2-alcove-g4} (right), which shows that with a larger-radius decay volume, there is much more acceptance for lower energy dark photons, whose decay products are then easier to separate with a given magnetic field. The dashed contours in~\cref{fig:FASER2-alcove-g4} (left) show the distribution of spatial separations for the $e^+$ and $e^-$ tracks for the subset of dark photons that are produced in the decay volume of the same transverse size as FASER, that is, within a decay volume radius of $R_{\text{DV}} = 10~\text{cm}$ from the LOS. In this case, it can be seen that the particle separations are significantly reduced and more in line with what might be expected from a naive extrapolation from FASER.

\begin{figure}[tbp]
  \centering
  \includegraphics[width=0.49\textwidth]{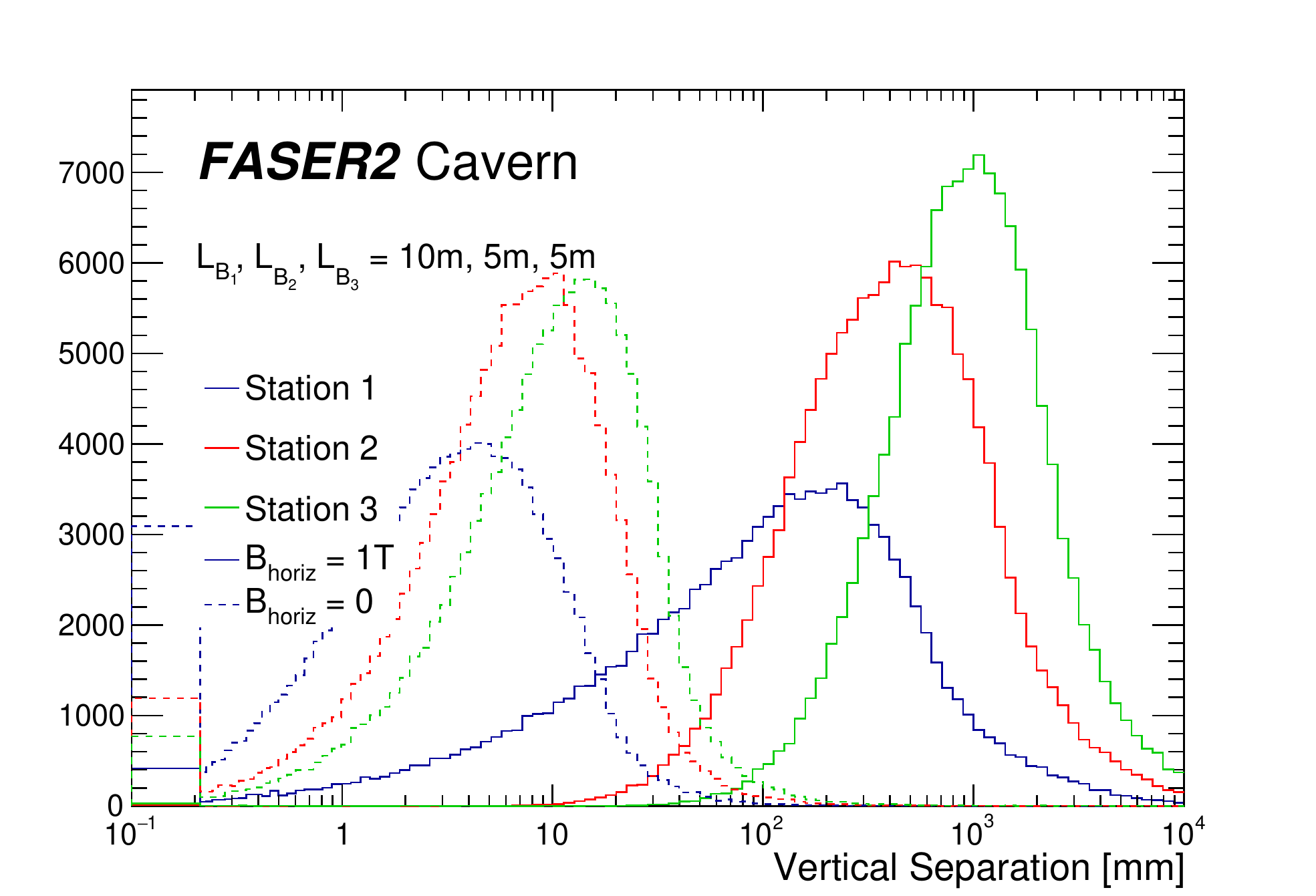}
  \includegraphics[width=0.49\textwidth]{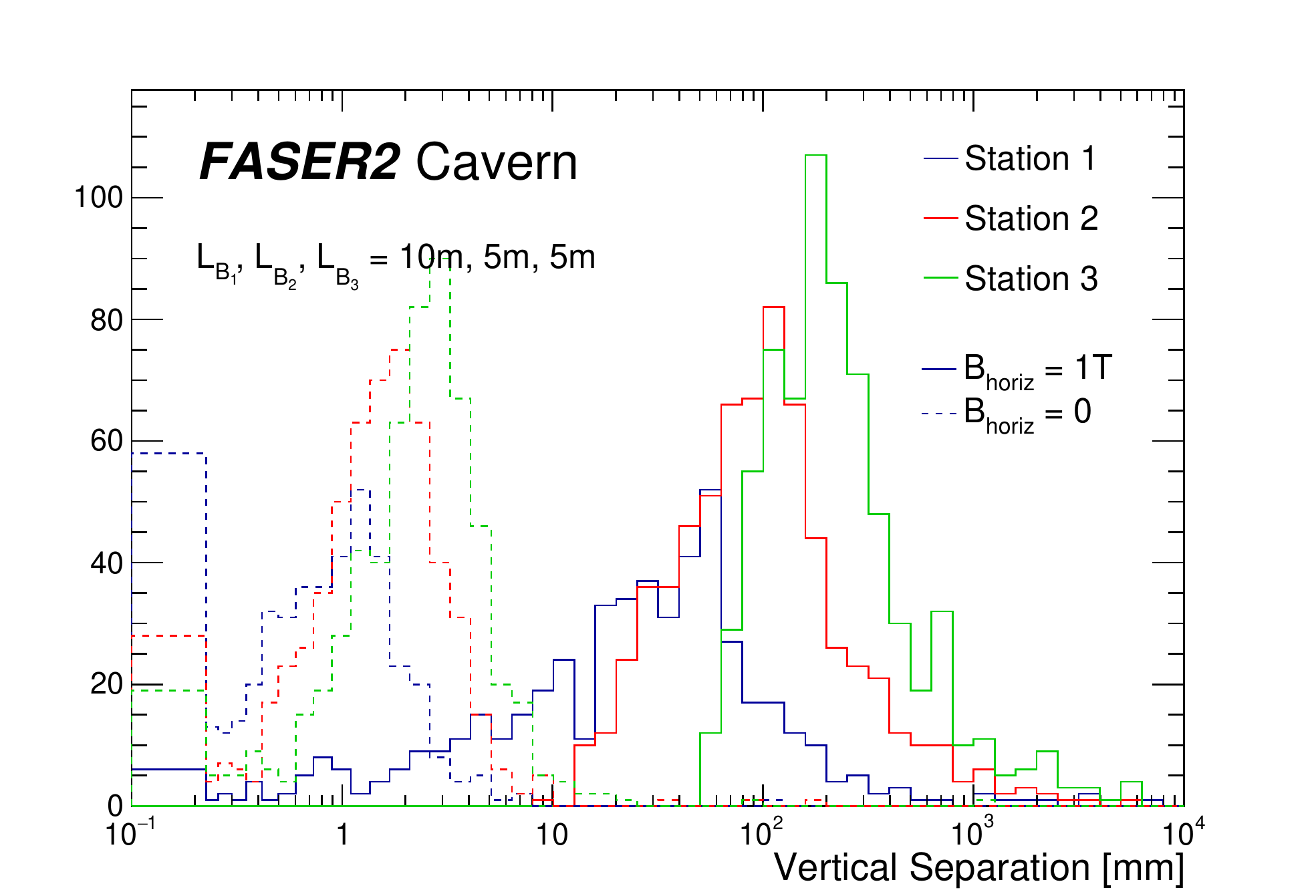}  
\caption{Left: Results from \texttool{Geant4} simulations of the FASER2 New Cavern detector for the distribution of transverse separations of $e^+$ and $e^-$ tracks resulting from 100 MeV dark photons decaying to $e^+ e^-$ pairs in a 1 T magnetic field (solid) or no magnetic field (dashed).  The different colors are the distributions at the various tracker station locations shown in~\cref{fig:FASER2-sims}.  Right: The same as in the left panel, but for dark photons with a fixed energy of 1 TeV.}
  \label{fig:FASER2-cavern-g4}
\end{figure}

In \cref{fig:FASER2-cavern-g4} (left), the same distributions of transverse track separations are shown, but for the New Cavern detector.  For the same 1 T magnetic field, the significantly increased detector length results in large separations even at the first station. A tracker resolution of 50~mm would be very efficient for Station 1 and much more coarse resolutions would required for Stations 2 and 3 and for the calorimeter. With such large separations it is worth investigating whether a magnet is needed to obtain sufficiently large separations. In the same figure, the dashed line show the separations without any magnetic field and even then they are comparable to the Alcove scenario. However, this is obviously not representative of all energy ranges. \cref{fig:FASER2-cavern-g4} (right) shows the same information but for dark photons with an energy of 1~TeV. Here, even with a 1 T magnetic field, the resolutions needed for good track separation are approximately 10~mm at Station 1 and approximately 100~mm at Station 3 and at the calorimeter. 

However, the most important consideration is the impact of the different particle separation efficiencies on the sensitivity of FASER2. There is a non-trivial interplay between the LLP mass, coupling, decay product separation, and reach, and so this requires dedicated study. 

\begin{figure}[tbp]
  \centering
  \includegraphics[width=0.49\textwidth]{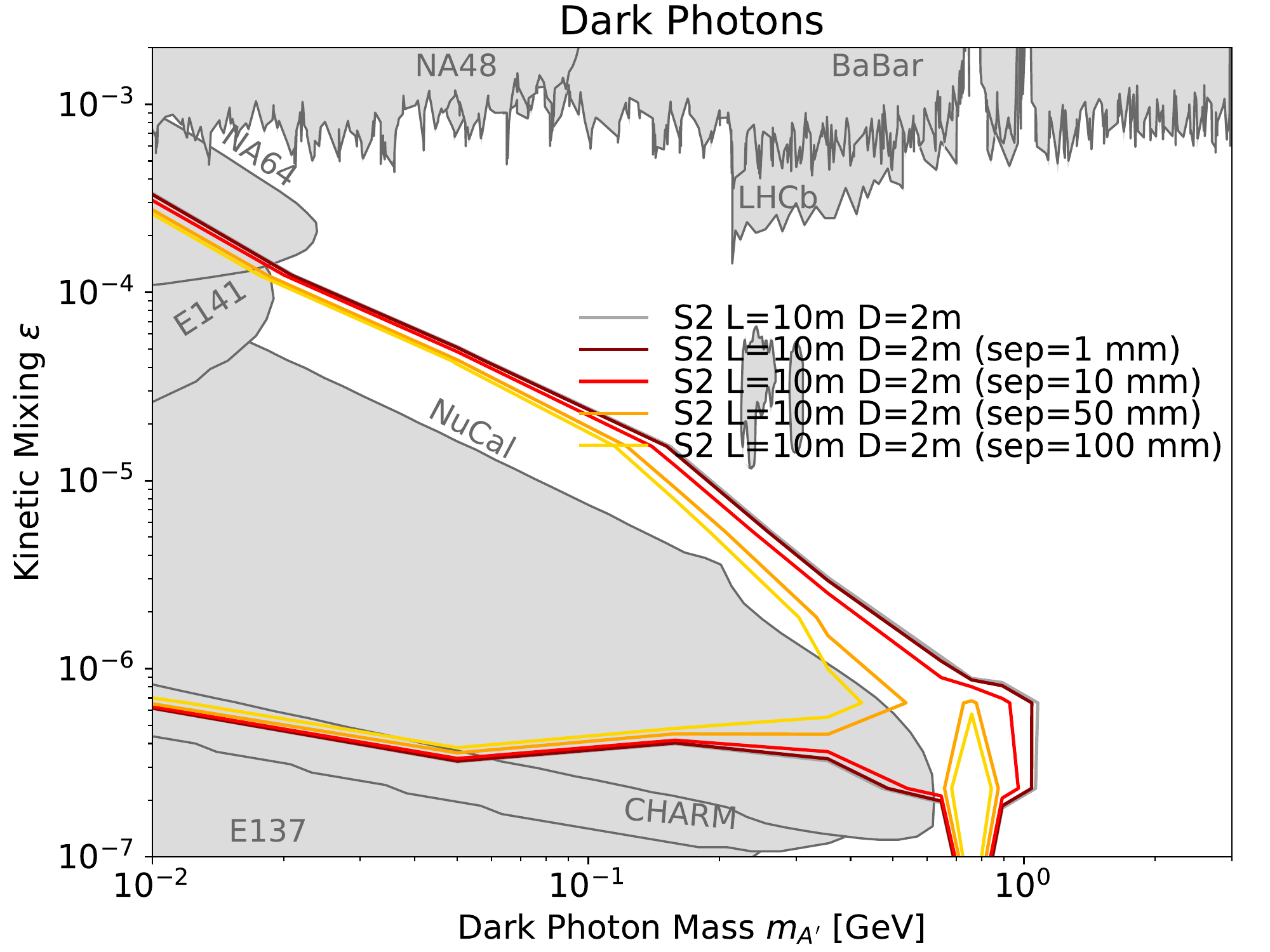}
  \includegraphics[width=0.49\textwidth]{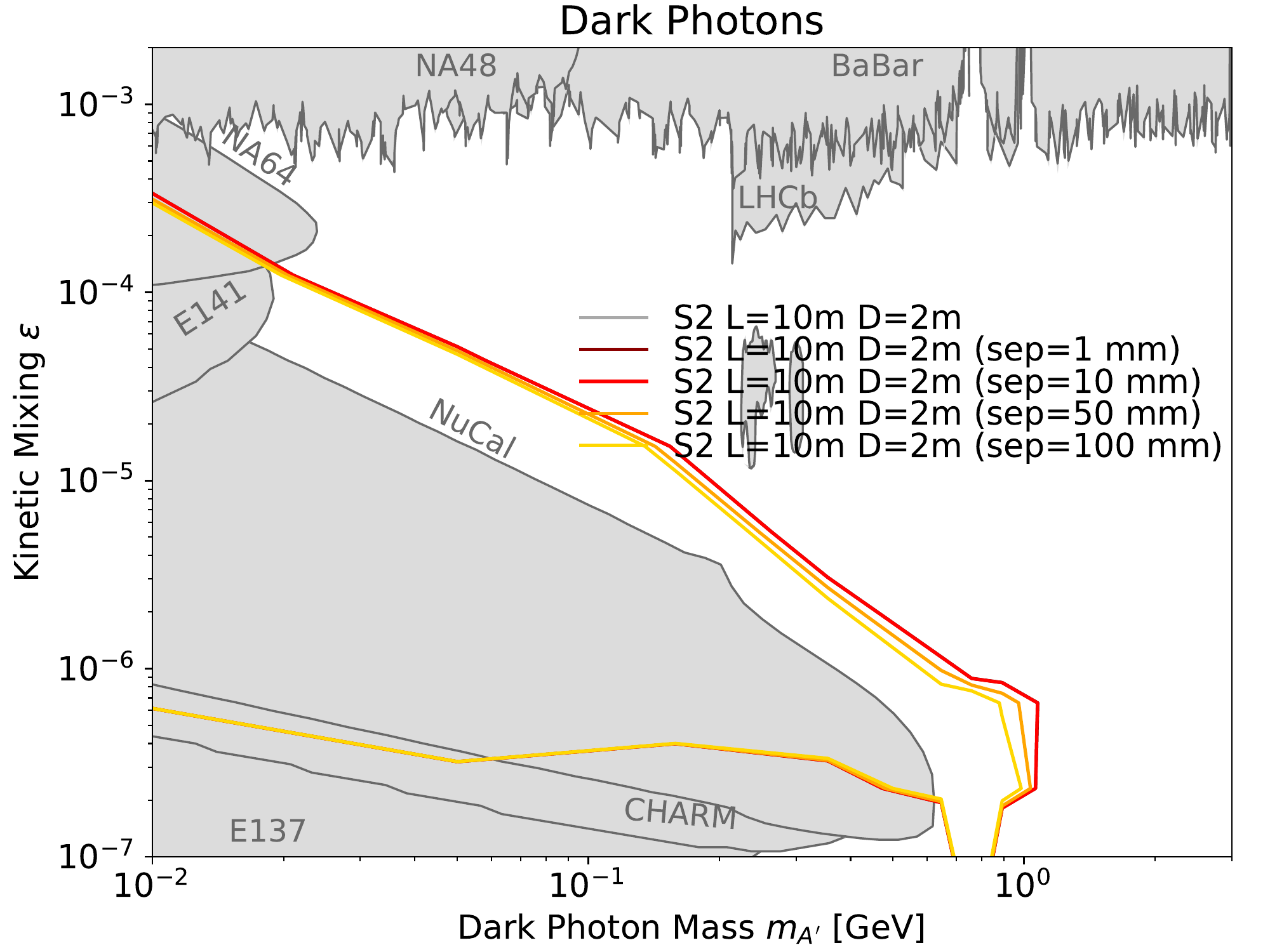}
  \caption{Effect on the sensitivity to dark photon decays $A' \to e^+ e^-$ for different requirements on the electron separations. These results are for the FASER2 New Cavern configurations indicated, with the separation requirement applied at tracker Station 1 (left) and at tracker Station 3/calorimeter (right).}
  \label{fig:FASER2-separation-reach}
\end{figure}

The effect of different separation cuts on the reach for the New Cavern scenario is shown in~\cref{fig:FASER2-separation-reach} for Station 1 (left) and Station 3/Calorimeter (right). For a requirement of 10~mm separation, the reduction in sensitivity is small enough that the tracker technologies under consideration would be more than sufficient. The particle separations are also large enough at the calorimeter that even a relatively coarse granularity could be sufficient, given the field strength of the proposed magnets.

To conclude, the physics potential of a larger-scale successor to FASER is clear. Possible scenarios for this larger detector are being explored and initial studies strongly indicate a preference for a FPF with a dedicated new cavern. Much progress has been made on defining the possible FASER detector designs and identifying detector technologies. Several studies are still required to finalise the design boundaries of FASER2, such as understanding the physics needs and possible detector performance capabilities for LLP mass and pointing reconstruction and particle identification.

\section{FASER$\nu$2 \label{sec:fasernu2}}


\subsection{Physics Goals \label{sec:FASERnu2physics}}

As a component of the existing FASER experiment, the FASER$\nu$ neutrino detector~\cite{FASER:2019dxq, FASER:2020gpr} was designed to detect collider neutrinos for the first time and examine their properties at TeV energies. In 2021, the FASER Collaboration reported the detection of the first neutrino interaction candidates at the LHC in a 29 kg pilot detector exposed in 2018~\cite{FASER:2021mtu}. Starting in 2022 at LHC Run 3, FASER$\nu$, with a total tungsten target mass of 1.1 tonnes, will measure about 10,000 flavor-tagged, charged-current (CC) neutrino interactions, opening up the new field of neutrino physics at colliders.

So far, no cross section data are available at the TeV-energy scale. For muon neutrinos, the FASER$\nu$ measurements will probe the gap between current accelerator measurements ($E_\nu < 360~\gev$)~\cite{ParticleDataGroup:2018ovx} and existing IceCube data ($E_\nu > 6.3~\tev$)~\cite{Aartsen:2017kpd}. For electron and tau neutrinos, the FASER$\nu$ cross section measurements will extend existing data to much higher energies. In addition to CC interactions, neutral-current (NC) interactions can be measured. Such measurements can provide a new limit on the nonstandard interactions of neutrinos to complement existing limits~\cite{Ismail:2020yqc}. Furthermore, as discussed in \cref{sec:qcd}, forward hadron production, which is poorly constrained by other LHC experiments, can be investigated using FASER$\nu$. In addition, uncertainties in forward charm production limit the clarification of the atmospheric neutrino background to astrophysical neutrino observations using neutrino telescopes; see \cref{sec:astro}. FASER$\nu$ measurements of high-energy electron neutrinos, which mainly originate from charm decays, can provide the first data on high-energy and large-rapidity charm production, providing vital input from a controlled environment for astrophysical neutrino observations. 

In LHC Run 3, FASER$\nu$ will detect of $\sim {\cal O}(10)$ tau neutrinos and anti-tau neutrinos. This will be a welcome supplement to the small number of tau neutrinos that have been detected so far, but will be insufficient to probe tau neutrino properties in detail.  Given the status of the tau neutrino as the least well-studied SM particle, there is strong motivation to study it more thoroughly with measurements that may be included among the precision probes of the general flavor structure of quarks and leptons.

The FASER$\nu$2 detector is designed as a much larger successor to FASER$\nu$ to greatly extend the physics potential for tau neutrino studies. It will be an emulsion-based detector able to identify heavy flavor particles produced in neutrino interactions, including $\tau$ leptons and charm and beauty particles. 
In the HL-LHC era, FASER$\nu$2 will be able to perform precision tau neutrino measurements and heavy flavor physics studies, eventually testing lepton universality in neutrino scattering and new physics effects. Furthermore, FASER$\nu$2 can provide extraordinary opportunities for a broad range of physics studies, with additional and important implications for QCD, neutrino physics, and astroparticle physics, as described in \cref{sec:qcd,sec:neutrinos,sec:astro}.

\subsection{Detector Requirements \label{sec:FASERnu2detector}}

To detect and distinguish the various neutrino flavors, the requirements for the FASER$\nu$2 detector are as follows: 
\vspace*{-0.2cm}
\begin{description}
\setlength{\itemsep}{0pt}
\item [On-axis] The detector should be centered along the beam collision axis to maximize the neutrino interaction event rate of all three flavors ($\nu_e, \nu_\mu, \nu_\tau$), and also to maximize the detected neutrino energies.
\item [Large mass] The detector should have a high-density and large-mass target to maximize the neutrino interaction event rate within the space constraints of the underground cavern. 
\item [Flavor sensitivity] The detector should be able to identify different lepton flavors: sufficient target material to identify muons; finely sampled detection layers to identify electrons and to distinguish them from gamma rays; and good position and angular resolutions to detect tau and charm decays. 
\item [Energy reconstruction] The detector should be able to measure muon and hadron momenta and the energy of electromagnetic showers, and be able to estimate neutrino energy. 
\item [Tau-neutrino/anti-tau neutrino separation] To distinguish $\nu_{\tau}$ and $\bar{\nu}_{\tau}$ interactions in the case of tau decays to muons, charge measurement of muons is required. 
A global analysis that combines information from FASER$\nu$2 with the FASER2 spectrometer, described in \cref{sec:faser2}, with the help of an interface detector, is required.
\end{description}

\begin{figure}[tbp]
\centering
\includegraphics[width=0.95\linewidth]{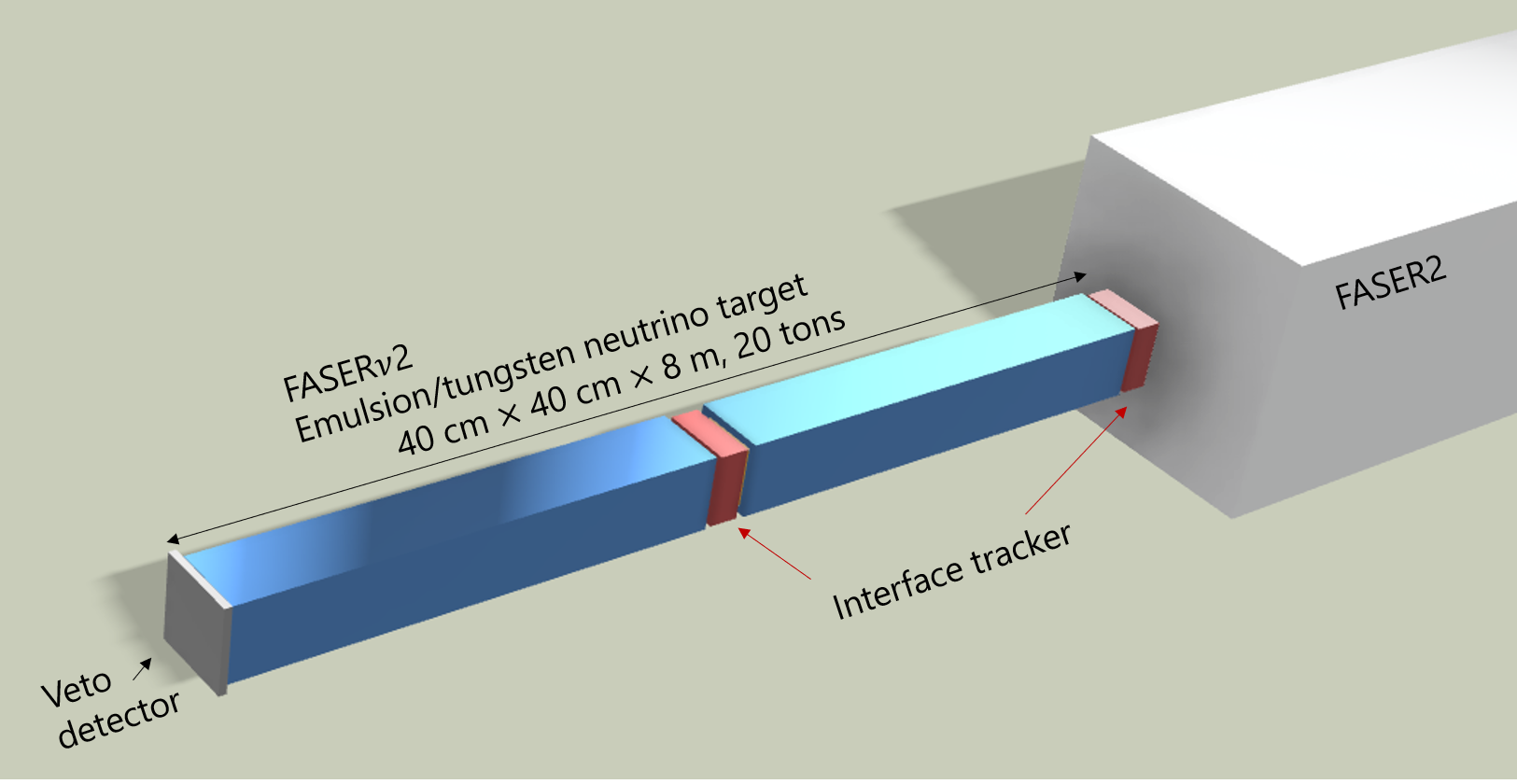}
\caption{Conceptual design of the FASER$\nu$2 detector~\cite{Anchordoqui:2021ghd}.}
\label{fig:FASERnu2}
\end{figure}

\cref{fig:FASERnu2} shows a view of the FASER$\nu$2 detector. Its ideal location is in front of the FASER2 spectrometer along the beam collision axis. The FASER$\nu$2 detector is envisioned to be composed of 3300 emulsion layers interleaved with 2 mm-thick tungsten plates. The total volume of the tungsten target is 40 cm $\times$ 40 cm $\times$ 6.6 m, and the mass is 20 tonnes. The FASER$\nu$2 detector will also include a veto detector and interface detectors to the FASER2 spectrometer, with one interface detector in the middle of the emulsion modules and the other detector downstream of the emulsion modules to make the global analysis and muon charge measurement possible. Similar to FASER2, the veto system will be scintillator-based, and the interface detectors could be based on the SiPM and scintillating fibre tracker technology. The detector length, including the emulsion films and interface detectors, will be $\sim$8~m. Both the emulsion modules and interface detectors will be situated in a cooling system (not drawn). 

\begin{figure}[tbp]
\centering
\includegraphics[width=0.65\linewidth]{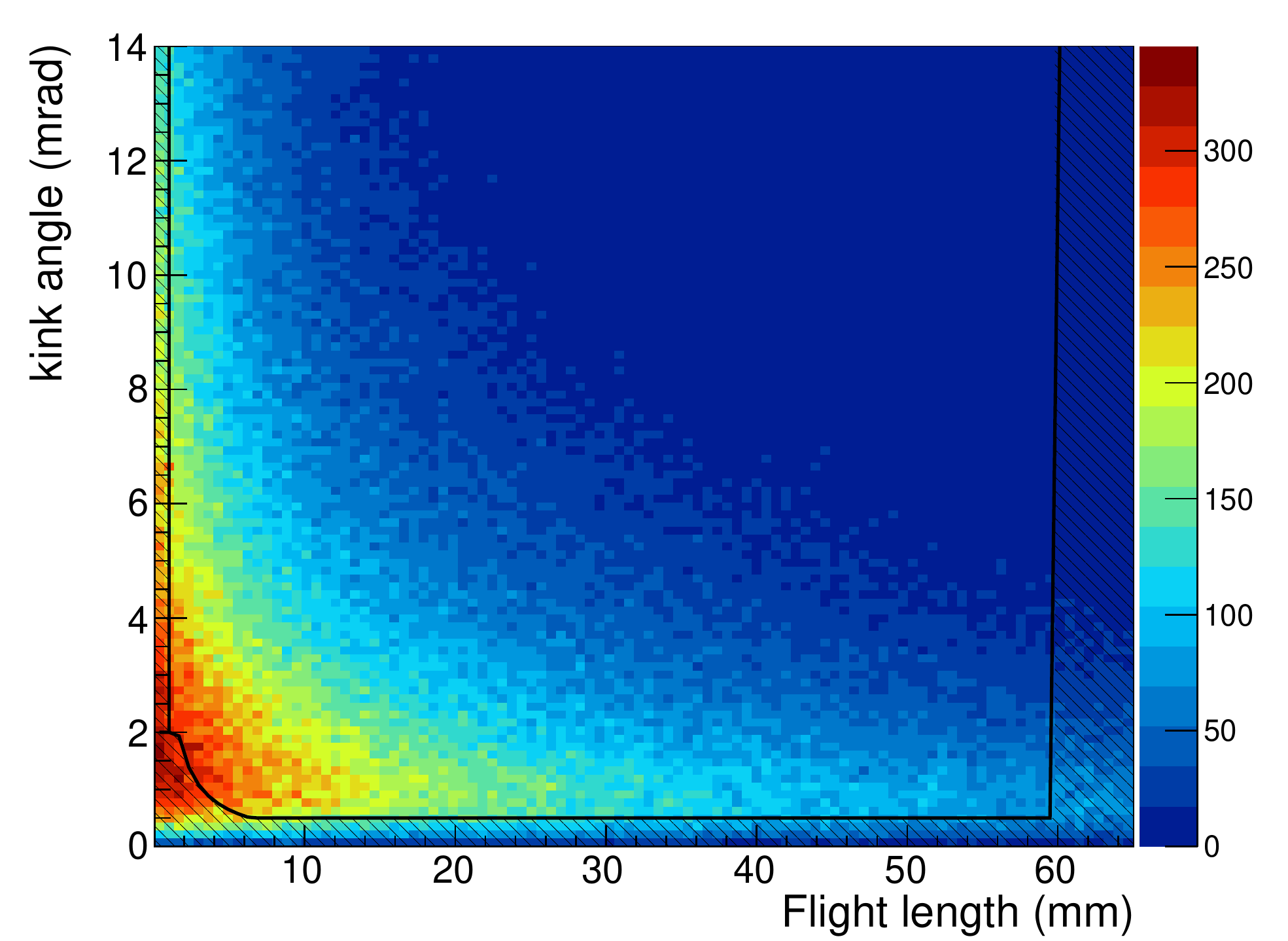}
\caption{The distribution of events in the $(\tau~\text{flight length}, \text{kink angle})$ plane for $\nu_{\tau}$ CC interaction events that produce tau leptons that decay with 1-prong topology. }
\label{fig:taukinkangle}
\end{figure}

\begin{figure}[tbp]
\centering
\includegraphics[width=0.75\linewidth]{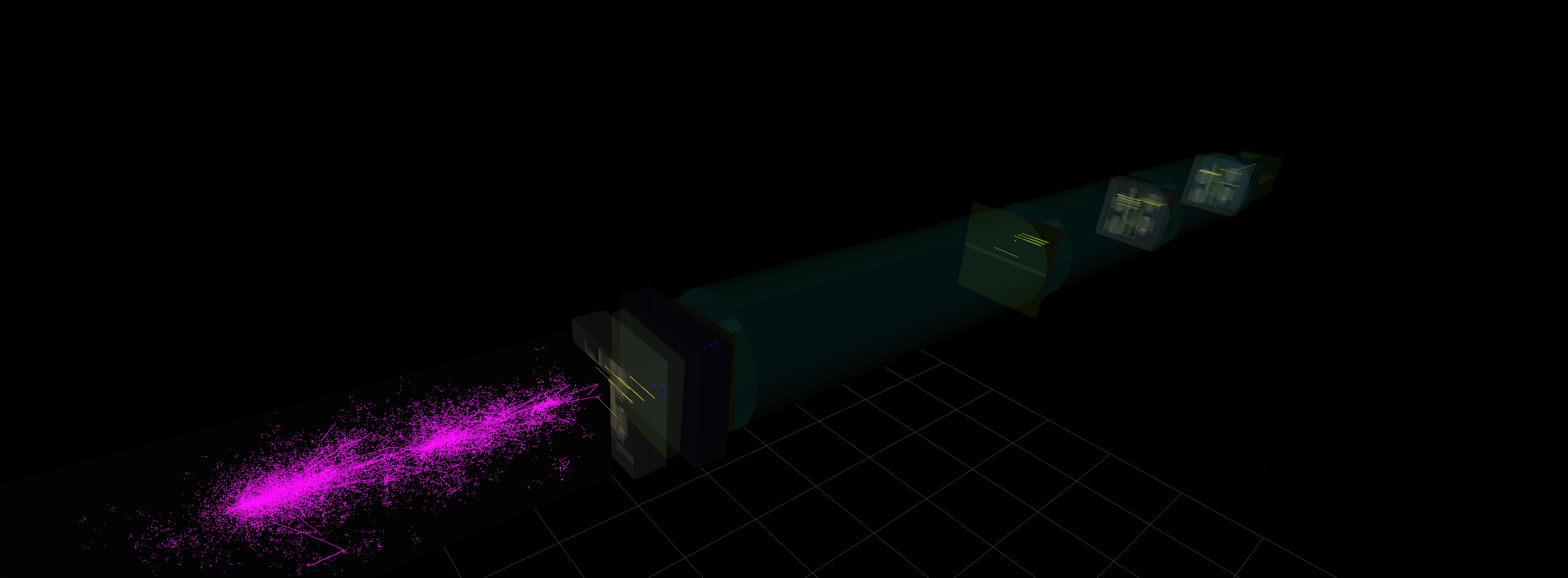} \\ 
\vspace*{0.1in}
\includegraphics[width=0.75\linewidth]{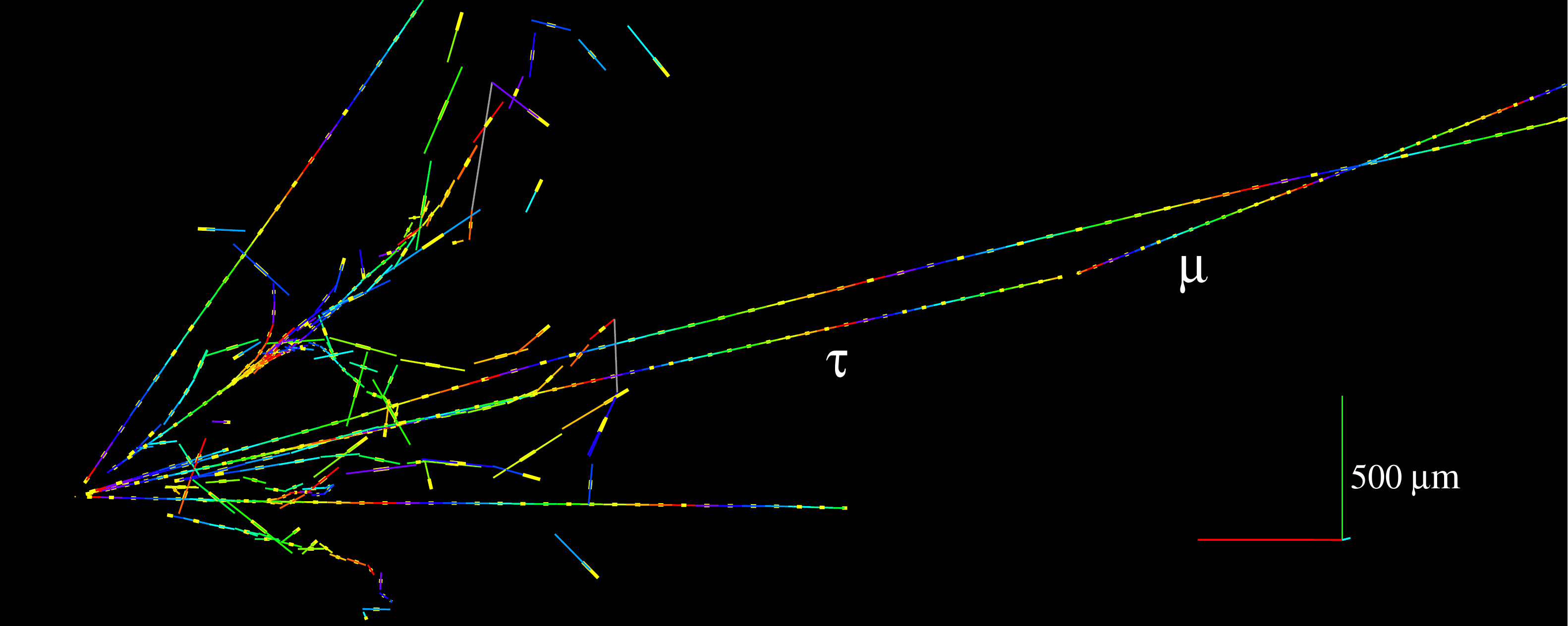}
\caption{Top: Event display of a simulated $\nu_{\tau}$ event with the $\tau^{-}$ decaying into a $\mu^{-}$ in FASER$\nu$ and the muon passing through the FASER spectrometer. Bottom: A magnified view of the part of the event in the FASER$\nu$ emulsion detector.
}
\label{fig:nutau}
\end{figure}

Tau neutrino CC interactions produce $\tau$ leptons, which have 1-prong decays 85\% of the time. \cref{fig:taukinkangle} shows the distribution of these events in the $(\tau~\text{flight length}, \text{kink angle})$-plane~\cite{FASER:2019dxq}. The mean $\tau$ flight length is 3~cm. To detect a kink, the $\tau$ must cross at least one emulsion layer.  In addition, the kink angle should be larger than four times the angular resolution and more than $0.5~\mrad$, and the flight length should be less than 6~cm, where the last requirement is implemented to reduce hadronic backgrounds. \cref{fig:nutau} shows event displays of a simulated $\nu_{\tau}$ event in FASER$\nu$ and FASER. The $\nu_{\tau}$ interacts and the tau decays into a muon in FASER$\nu$, and the muon then passes through the FASER spectrometer. Similar events are expected in FASER$\nu$2 and FASER2.

The high muon background in the LHC tunnel might be an experimental limitation. The emulsion detector readout and reconstruction work for track densities up to $\sim 10^6$ tracks/cm$^2$. To keep the detector occupancy low, the possibility of sweeping away such muons with a magnetic field placed upstream of the detector is currently being explored, as described in~\cref{sec:sweepermagnet}. Considering the expected performance, the emulsion films will be replaced every year during the winter stops.

\subsection{Emulsion Film Production \label{sec:FASERnu2emulsion}}

The emulsion sensitive layers consist of silver bromide micro-crystals, which are semiconductors with a band gap of 2.684~eV, dispersed in a gelatin substrate. The diameter of the crystals which will be used for FASER$\nu$2 will be approximately 200~nm. An emulsion detector with 200~nm crystals has a spatial resolution of 50~nm.  The two-dimensional intrinsic angular resolution of a double-sided emulsion film with 200 nm-diameter crystals and a base thickness of 210~$\mu$m is therefore 0.35 mrad. More details on the emulsion technology are summarized in~\cite{Ariga:2020lbq}.

\begin{figure}[tbp]
\centering
\includegraphics[width=0.6\linewidth]{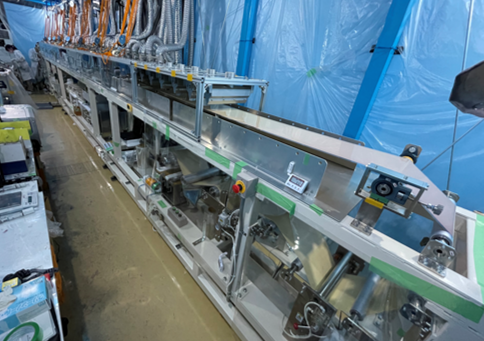}
\caption{The emulsion film coating system.}
\label{fig:coating_system}
\end{figure}

The emulsion gel and film production will be performed at a large-scale production facility established at Nagoya University. The sensitivity of the emulsion layers was examined by exposing the produced emulsion to several tens of MeV electrons, measuring $\sim$45 grains per 100~$\mu$m along with the trajectory of particles.  This sensitivity is sufficient for detecting minimum ionizing particles by setting an emulsion thickness of larger than 50~$\mu$m.  The emulsion gel produced can then be used to produce films with two 65~$\mu$m emulsion layers deposited on both sides of a 210~$\mu$m plastic base by using the coating system shown in~\cref{fig:coating_system}. 
The capability of the facility to produce emulsion films can reach $\sim 2000~\m^2$ per year. It is possible to produce 3300 emulsion films, corresponding to a total of $\sim 530~\m^2$ of emulsion films, for FASER$\nu$2 every year. The production of emulsion gel and films will be scheduled half a year before each installation.

\subsection{Readout and Analysis \label{sec:fasernu2readout}}

\begin{figure}[htbp]
\centering
\includegraphics[width=0.6\linewidth]{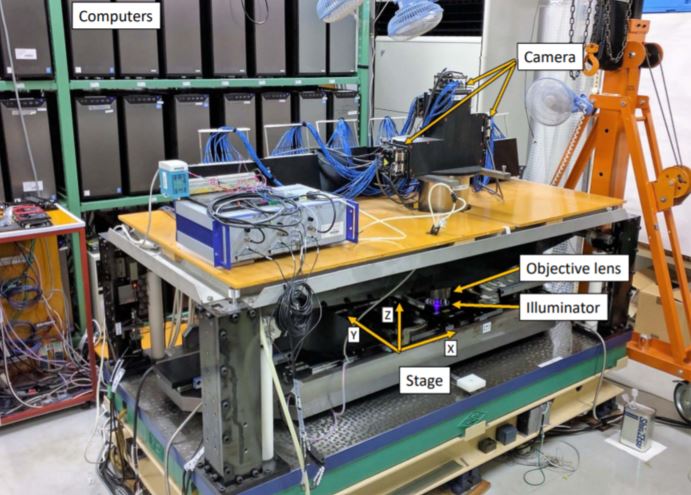}
\caption{
The fast emulsion readout system HTS~\cite{Yoshimoto:2017ufm}.
}
\label{fig:hts}
\end{figure}

Analyses of the data collected in the emulsion modules will be based on the readout of the full emulsion volume using the Hyper Track Selector (HTS) system~\cite{Yoshimoto:2017ufm}. The readout speed of the HTS system is 0.45~m$^2$/hour/layer, which is a big leap from previous generations. Currently, an upgraded system HTS2, which is about five times faster, is under commissioning and a further upgraded system HTS3 with about 10~m$^2$/hour/layer is being developed. The readout speeds of these systems are summarized in~\cref{table:scanning_speed}, and a photo of the HTS system is shown in \cref{fig:hts}. 
The total emulsion film surface to be analyzed in FASER$\nu$2 is $\sim$530 m$^2$/year, implying a readout time of $\sim$2400 hours/year with HTS or $\sim$420 hours/year with HTS2. It will be possible to finish reading the data taken in each year within a year using either of the above systems.

\begin{table}[tbhp]
\centering
\begin{tabular}{l|c|c}
\hline\hline
\                          & Field of view [mm$^2$]      & Readout speed [cm$^2$/hour/layer] \\
\hline 
SUTS (used in OPERA)       & 0.04                        & 72            \\ \hline 
HTS \ (running)            & 25                          & 4500          \\ \hline 
HTS2 (under commissioning) & 50                          & 25000         \\ \hline\hline
\end{tabular}
\caption{Comparison of emulsion scanning systems and their performance properties.
}
\label{table:scanning_speed}
\end{table}

Reconstruction of the emulsion data will enable the localization of neutrino interaction vertices, identification of muons, the measurement of charged particle momenta by the multiple Coulomb scattering method, and the energy measurement of electromagnetic showers. Additionally, with the FASER2 spectrometer using the interface detectors, the charges of muons will be identified. 

In the HL-LHC, given the 20 times luminosity and 20 times target mass of FASER$\nu$, FASER$\nu$2 will collect two orders of magnitude higher statistics than FASER$\nu$, allowing precision measurements of neutrino properties for all three flavors. For tau neutrinos, $\sim 2,300$ ($\sim 20,000$) $\nu_{\tau}$ CC interactions are expected, using the event generator \texttool{Sibyll-2.3d} (\texttool{DPMJet~3.2017}), as shown in~\cref{table:event_rate}. As for the uncertainty on the tau neutrino flux, forward charm production is poorly constrained by other experiments, but it can be studied by measuring electron neutrino interactions in FASER$\nu$2. Roughly 178k (668k) $\nu_e$ CC interactions are expected in FASER$\nu$2, using the event generator \texttool{Sibyll~2.3d} (\texttool{DPMJet~3.2017}). Electron neutrinos at high energies above $\sim$500 GeV, which mainly originate from charm decays, can constrain forward charm production. The major remaining uncertainty could be at the 20\% level from the dependency on the charm species, which other experiments or theoretical predictions might further constrain.

\begin{table}[htbp]
\centering
\begin{tabular}{l||l|c|c|c}
\hline \hline
\             & Generator & $\nu_e$+$\bar{\nu}_e$ CC  & $\nu_\mu$+$\bar{\nu}_\mu$ CC  & $\nu_\tau$+$\bar{\nu}_\tau$ CC  \\
\hline 
FASER$\nu$    & \texttool{Sibyll}    & 0.9k                 & 4.8k                     & 15   \\
\             & \texttool{DPMJet}    & 3.5k                 & 7.1k                     & 97   \\ \hline
FASER$\nu$2   & \texttool{Sibyll}    & 178k                 & 943k                     & 2.3k \\
\             & \texttool{DPMJet}    & 668k                 & 1400k                    & 20k  \\ \hline \hline
\end{tabular}
\caption{The expected number of neutrino interactions obtained using two different event generators, \texttool{Sibyll-2.3d} and \texttool{DPMJet~3.2017}, for FASER$\nu$~\cite{Kling:2021gos} and FASER$\nu$2~\cite{Anchordoqui:2021ghd}.
}
\label{table:event_rate}
\end{table}

\section{AdvSND \label{sec:advsnd}}


\subsection{Physics Goals \label{sec:AdvSNDphysics}}

The Advanced SND project is meant to extend the physics case of the SND@LHC experiment~\cite{Ahdida:2750060}. It will consist of two detectors: one placed in the same $\eta$ region as  SND@LHC, i.e.~$7.2 < \eta < 8.4$, hereafter called FAR, and the other one in the region $4 < \eta < 5$, hereafter denoted NEAR.  In the first part of this section, we review the way the physics case would be extended, and in the second part, we describe the detector design and layout in more detail. These two detectors are meant to operate during Run 4 of the LHC (i.e., HL-LHC) and beyond. The FPF would host the FAR detector. The NEAR detector, given the higher average angle, would have to be placed more upstream to get a sizeable azimuth angle coverage. Note that the extension of the physics case covered here is related to neutrinos and, more generally, to SM physics, while the BSM physics case is described in \cref{sec:bsm1}.   

{\bf QCD Measurements.} 
Electron neutrinos in the pseudorapidity range of SND@LHC, $7.2 < \eta < 8.4$, are mostly produced by charm decays. Therefore, $\nu_e$s can be used as a probe of charm production in an angular range where the charm yield has a large uncertainty, to a large extent coming from the gluon parton distribution function (PDF); see \cref{sec:qcd}. Electron neutrino measurements can thus constrain the uncertainty on the gluon PDF in the very small (below $10^{-5}$) $x$ region. The interest therein is two-fold: first, the gluon PDF in this $x$ domain will be relevant for Future Circular Collider (FCC) detectors; and second, the measurement will reduce the uncertainty on the flux of very-high-energy (100 PeV) atmospheric neutrinos produced in charm decays, an essential input to the study of neutrinos from astrophysical sources, as discussed in \cref{sec:astro}. The charm measurement by SND@LHC in Run 3 will be affected by a systematic uncertainty at the level of 30\% and by a statistical uncertainty of 5\%. The large systematic uncertainty mostly comes from the procedure linking neutrinos to charm. To reduce this uncertainty, the NEAR detector of AdvSND comes into play, since the charm yield was measured with high precision by LHCb in the $4.0 < \eta < 4.5$ region~\cite{Aaij:2015bpa}. The comparison between neutrino measurements and LHCb direct charm measurements will reduce the systematic uncertainties, with the goal of bringing them down to the level of the statistical one. 

\begin{figure}[tbp]
\centering
\includegraphics[width=0.65\textwidth]{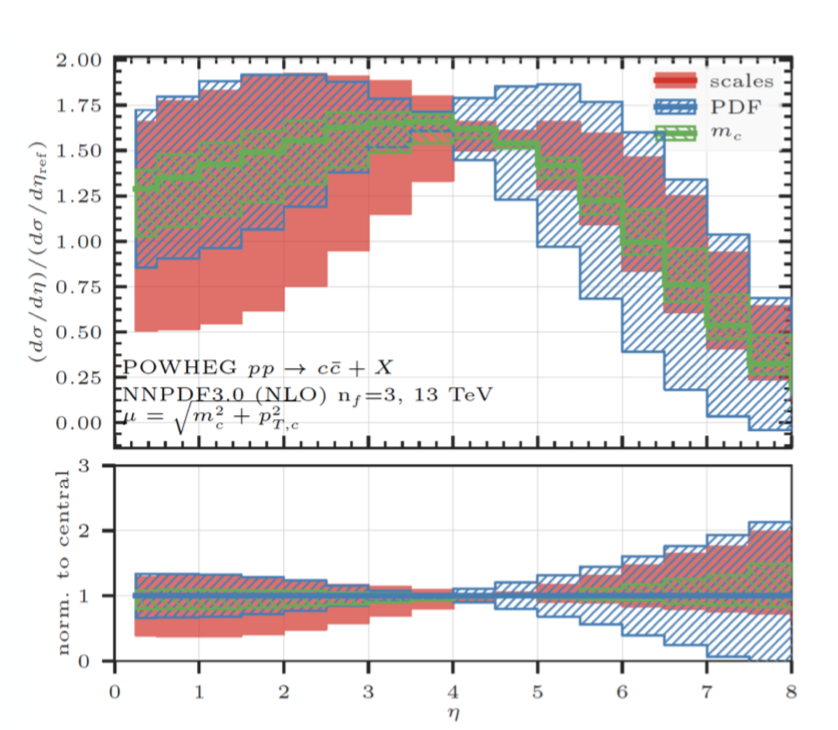}
\caption{Ratio between the differential cross section for charm production at 13 TeV and the differential cross section at 7 TeV, with the latter evaluated in the pseudorapidity range $ 4 < \eta < 4.5$~\cite{Ahdida:2750060}. }
\label{fig:ratiocharm}
\end{figure}

The operation in Run 4 of the FAR detector will reduce the statistical uncertainty below 1\%, as it is clear from~\cref{tab:NuFlux_Far}. \cref{fig:ratiocharm} shows the ratio between charm measurements in different $\eta$ regions normalised to the LHCb measurement: the gluon PDF uncertainty provides the largest contribution.  AdvSND will measure charm in two $\eta$ regions: in the $7.2 < \eta < 8.4$ region with the FAR detector and in the $4.5 < \eta < 5$ range with the NEAR one, where the relative contribution of PDF uncertainty is even larger. 

\begin{table}[hbtp]
\centering
\begin{tabular}{ c | c c | c c }
\hline \hline
 \multicolumn{5}{c}{AdvSND - FAR} \\
\hline
       & \multicolumn{2}{c|}{$\nu$ in acceptance} & \multicolumn{2}{c}{CC DIS} \\
Flavour &  hardQCD: $c\overline{c}$& hardQCD: $b\overline{b}$  &hardQCD: $c\overline{c}$& hardQCD: $b\overline{b}$  \\
\hline
$\nu_\mu$ +   $\bar{\nu}_\mu$   & $6.3 \times 10^{12}$ & $1.5 \times 10^{11}$ &  $1.2 \times 10^{4}$ &  200 \\
$\nu_e$  +  $\bar{\nu}_e$ & $6.7       \times 10^{12}$ & $1.7 \times 10^{11}$ & $1.2 \times 10^{4}$  & 220 \\
$\nu_\tau$  +  $\bar{\nu}_\tau$   & $7.1 \times 10^{11}$ & $4.7 \times 10^{10}$  &  880 & 40 \\
\hline
 Tot &\multicolumn{2}{c|}{ $1.4 \times 10^{13}$ }& \multicolumn{2}{c}{ $2.5 \times 10^{4}$ } \\
\hline \hline
\end{tabular}
\caption{The number of neutrinos passing through the FAR detector of AdvSND and the number of CC neutrino interactions in the detector target, assuming 3000\,fb$^{-1}$, as estimated with the \texttool{Pythia~8} generator. \label{tab:NuFlux_Far}}
 \end{table}

\begin{figure}[tbp]
\centering
\includegraphics[width=0.98\textwidth]{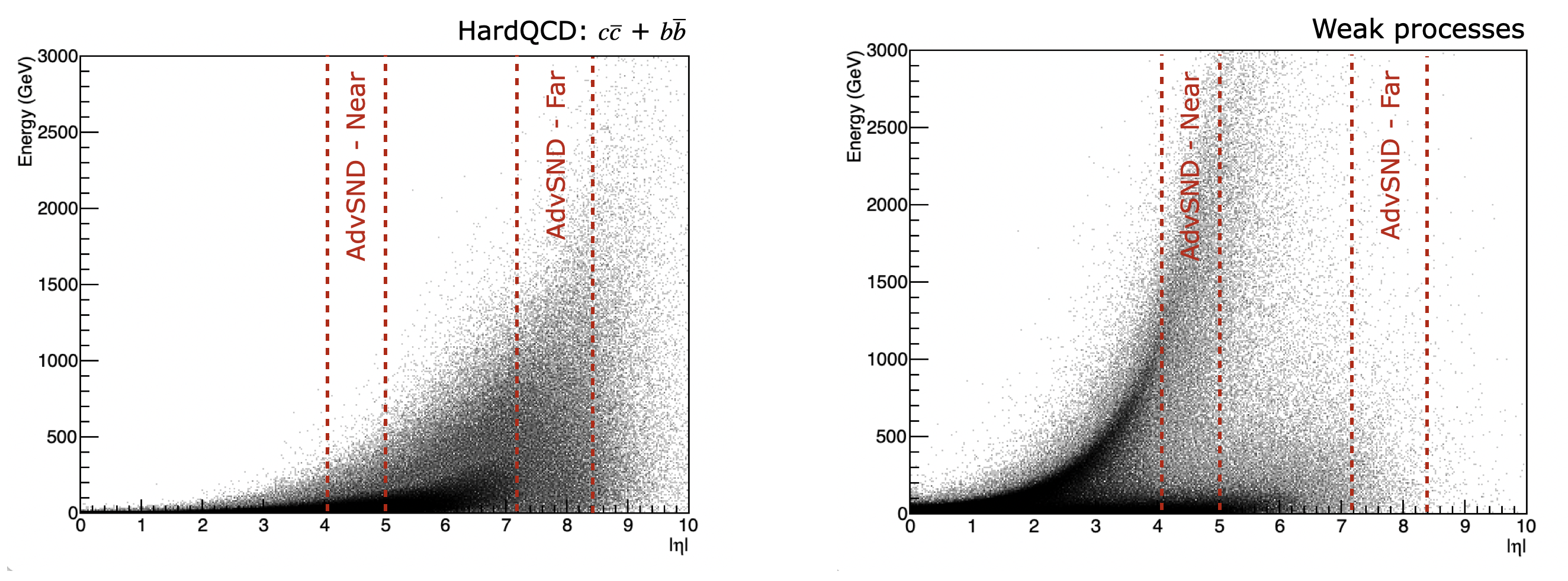}
\caption{Scatter plot of neutrino energy versus pseudorapidity $\eta$ in $b$, $c$ (left) and $W$ (right) decays. All neutrino flavours are included~\cite{Beni:2019gxv}. The AdvSND regions are highlighted.}
\label{fig:eta-energy}
\end{figure}

{\bf Neutrino Cross Section Measurements.} \cref{fig:eta-energy} shows the scatter plots of neutrino energy versus $\eta$ for neutrinos originating from $b$ and $c$ and from $W$ decays. Neutrinos from leptonic $W$ decays are seen to be kinematically well separated~\cite{Beni:2019gxv}. Note that LHCb has measured charm, beauty, and $W$ production cross sections in the $2 < \eta < 5$ range: 1.5 nb for $W$, 144 $\mu$b for beauty, and 8.6 mb for charm~\cite{Aaij:2015bpa}. The $W$ measurement was carried out at 7 TeV while the other two were done at 13 TeV.  Accounting for all that, and considering the case of tau neutrinos, which shows a low branching ratio in charm decays ($c \rightarrow \nu_\tau \sim 5 \times 10^{-3}$), we expect a factor $10^5$ more charm-induced than $W$ and $Z$-induced $\nu_\tau$s. The role of $W$ and $Z$ decays is therefore marginal in this context and we focus on charm and beauty in the following. 

\begin{figure}[tbp]
\centering
\includegraphics[width=0.98\columnwidth]{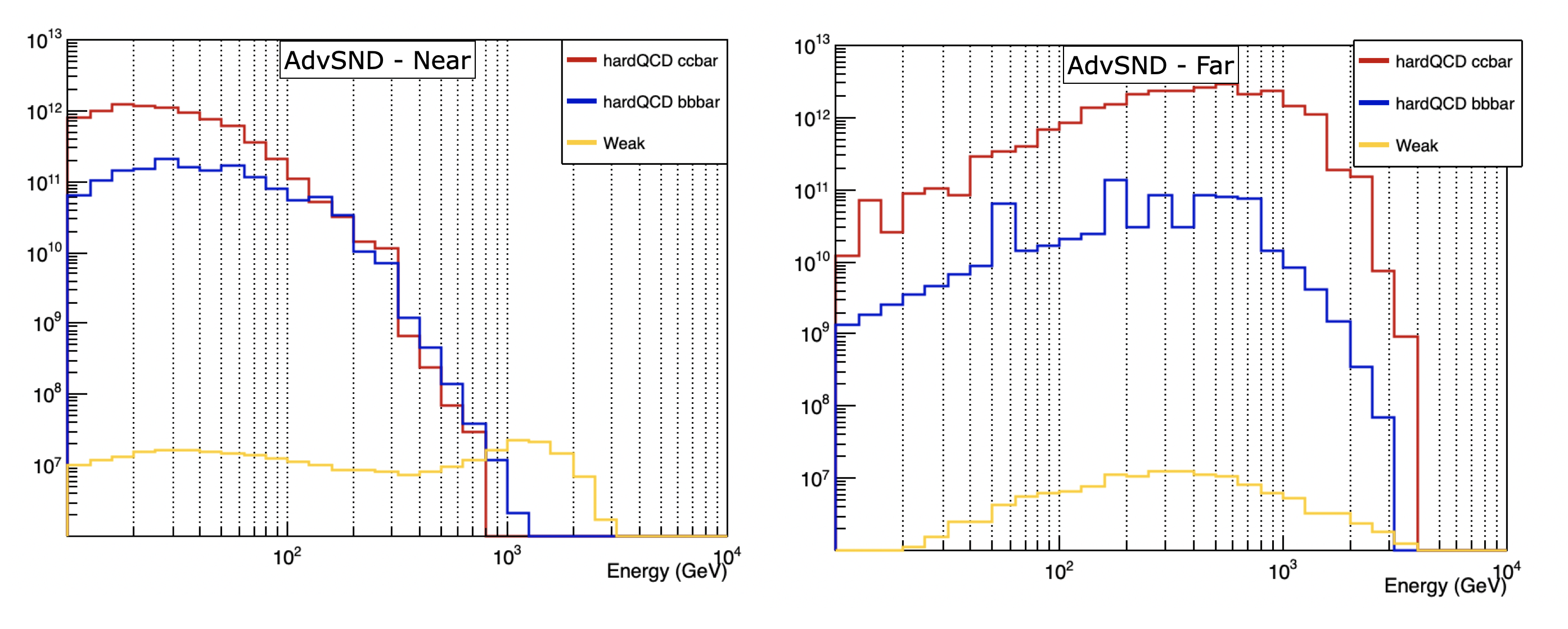}
\caption{Neutrino energy spectra for three different sources: charm, beauty and $W$, $Z$ bosons. The NEAR (left) and FAR (right) locations are considered. }
\label{fig:energy_all}
\end{figure}

\cref{fig:energy_all} shows the neutrino energy spectra for the two $\eta$ regions, separately for the different neutrino parents. The energy spectra of charm and beauty-induced neutrinos is much softer in the NEAR location, as expected. 

The large uncertainty on the charm-induced neutrino flux in the large $\eta$ region prevents SND@LHC from making a neutrino cross section measurement. AdvSND will instead be able to perform this measurement with the NEAR detector, since the neutrino flux from charm and beauty in the $4.0 < \eta < 4.5$ region is very reliable, given the measurements performed by LHCb~\cite{Aaij:2015bpa}. This will lead to a neutrino cross section measurement with very small systematic uncertainties of all three neutrino flavours, including tau neutrinos. The expected number of events in the NEAR detector is given in~\cref{tab:NuFlux_Near}. The lower average energy of neutrinos in the NEAR location results in a lower neutrino cross section, which explains the differences between the neutrino yields in the two detectors, despite of the similar flux. 

\begin{table}[hbtp]
\centering
\begin{tabular}{c | c c | c c}
\hline\hline
 \multicolumn{5}{c}{AdvSND - NEAR} \\
\hline
       & \multicolumn{2}{c|}{$\nu$ in acceptance} & \multicolumn{2}{c}{CC DIS} \\
Flavour &  hardQCD: $c\overline{c}$& hardQCD: $b\overline{b}$  &hardQCD: $c\overline{c}$& hardQCD: $b\overline{b}$  \\
\hline
$\nu_\mu$ +   $\bar{\nu}_\mu$   & $2.1 \times 10^{12}$ & $3.3 \times 10^{11}$ & 980 &  200 \\
$\nu_e$  +  $\bar{\nu}_e$ & $2.2       \times 10^{12}$ & $3.3 \times 10^{11}$ & 1000  & 200 \\
$\nu_\tau$  +  $\bar{\nu}_\tau$   & $2.7 \times 10^{11}$ & $1.4 \times 10^{11}$  &  80 & 50 \\
\hline
 Tot &\multicolumn{2}{c|}{ $5.4 \times 10^{12}$ }& \multicolumn{2}{c}{ $2.5 \times 10^{3}$ } \\
\hline\hline
 \end{tabular}
  \caption{The number of neutrinos passing through the NEAR detector of AdvSND and the number of CC neutrino interactions in the detector target, assuming 3000\,fb$^{-1}$, as estimated with the \texttool{Pythia~8} generator.
  \label{tab:NuFlux_Near} }
 \end{table}

Thus, one expects the leading uncertainty to be the statistical one: a few percent for electron and muon neutrinos and about 10\% for tau neutrinos as one can derive from~\cref{tab:NuFlux_Near}. Notice that the yield of muon neutrinos from $\pi$ and $K$ decays is not included in this table.

{\bf Lepton Flavour Universality with Neutrino Interactions.}
In the $7. 2 < \eta < 8.4$ region, electron and tau neutrinos come essentially only from charm decays. Therefore, the uncertainty on the flux cancels out in the ratio, which can then be used to test lepton flavour universality with neutrino interactions. The corresponding measurement by SND@LHC is dominated by a 30\% statistical uncertainty due to the poor $\nu_\tau$ statistics. AdvSND will reduce the statistical uncertainty down to less than 5\%; see~\cref{tab:NuFlux_Far}. At this point, the systematic uncertainty due to the charm quark hadronization fraction into $D_s$ mesons, $f_{D_s}$, would be leading. This would turn into a measurement of lepton flavour universality at the 20\% level.

More constraints on this ratio could come from the NEAR detector, where all charmed hadron species, including $D_s$, have been identified by the LHCb Collaboration. Given the expected number of electron and tau neutrino interactions reported in~\cref{tab:NuFlux_Near}, lepton flavour universality with electron and tau neutrinos could be tested with an accuracy of 10\%. 

Lepton flavour universality can also be tested with the electron to muon neutrino ratio. In this case, charm can be considered also as the source of muon neutrinos if an energy cut is applied. A tentative value of 600 GeV is assumed as the energy threshold in the forward location with $7.2 < \eta < 8.4$, and SND@LHC will have a 10\% accuracy in this ratio for both systematic and statistical uncertainty. With the FAR detector operating at the FPF, AdvSND can reduce the statistical uncertainty down to a few percent, and the accuracy will be limited by the systematic uncertainty. It has to be underlined that the systematic uncertainty will have to be re-evaluated in this region in the high luminosity regime.    

The NEAR detector would have a much smaller systematic uncertainty and therefore the reach is driven by the statistical uncertainty, at the level of a few percent.

\subsection{Detector Layout \label{sec:AdvSNDdetector}}

Both detectors will be made of three elements. The upstream one is the target region for the vertex reconstruction and the electromagnetic energy measurement with a calorimetric approach. It will be followed downstream by a hadronic calorimeter and a muon identification system. The third and most downstream element will be a magnet for the muon charge and momentum measurement, thus allowing for neutrino/anti-neutrino separation for muon neutrinos and for tau neutrinos in the muonic decay channel of the $\tau$ lepton.

\begin{figure}[tbp]
    \centering
    \includegraphics[width=0.9\textwidth]{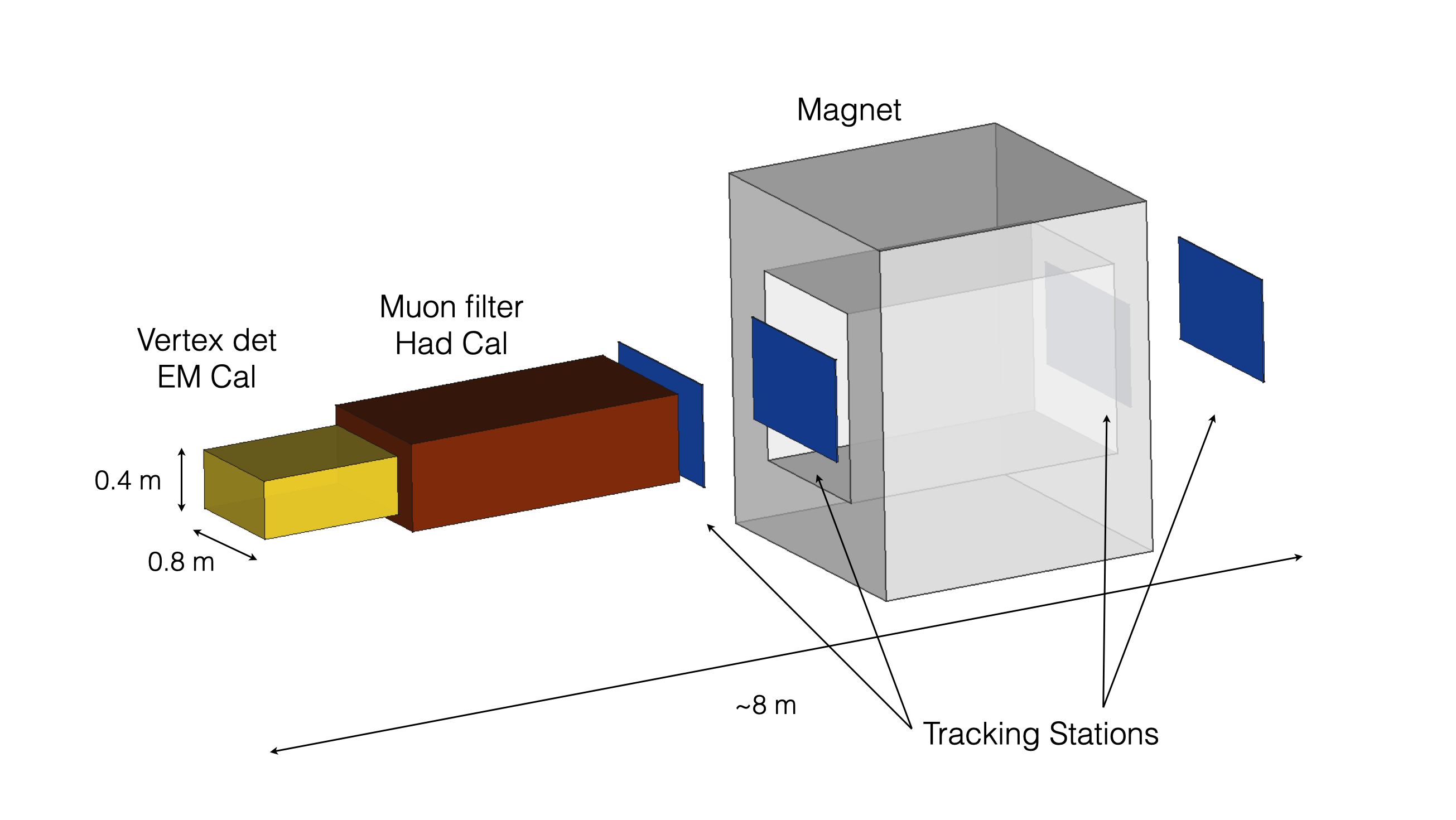}
    \caption{Layout of the AdvSND detector. }
    \label{fig:advsnd}
\end{figure}

The target will be made of thin sensitive layers interleaved with tungsten plates, for a total mass of $\sim 5$ tons. The use of nuclear emulsion at the HL-LHC is prohibitive due to the very high luminosity that would make the replacement rate of the target incompatible with technical stops. As discussed in \cref{sec:sweepermagnet}, this motivates the study of a sweeper magnet to reduce the muon background.  As an alternative approach, the Collaboration is investigating the use of compact electronic trackers with high spatial resolution fulfilling both tasks of vertex reconstruction with micrometer accuracy and electromagnetic energy measurement. The hadronic calorimeter and the muon identification system will be about 10~$\lambda$, which will bring the average length of the hadronic calorimeter to about 12~$\lambda$, thus improving the muon identification efficiency and energy resolution.  The magnetic field strength is assumed to be about 1 T over about 2~m length. A schematic view of the detector is shown in~\cref{fig:advsnd}. 

\begin{figure}[tbp]
    \centering
    \includegraphics[width=0.6\textwidth]{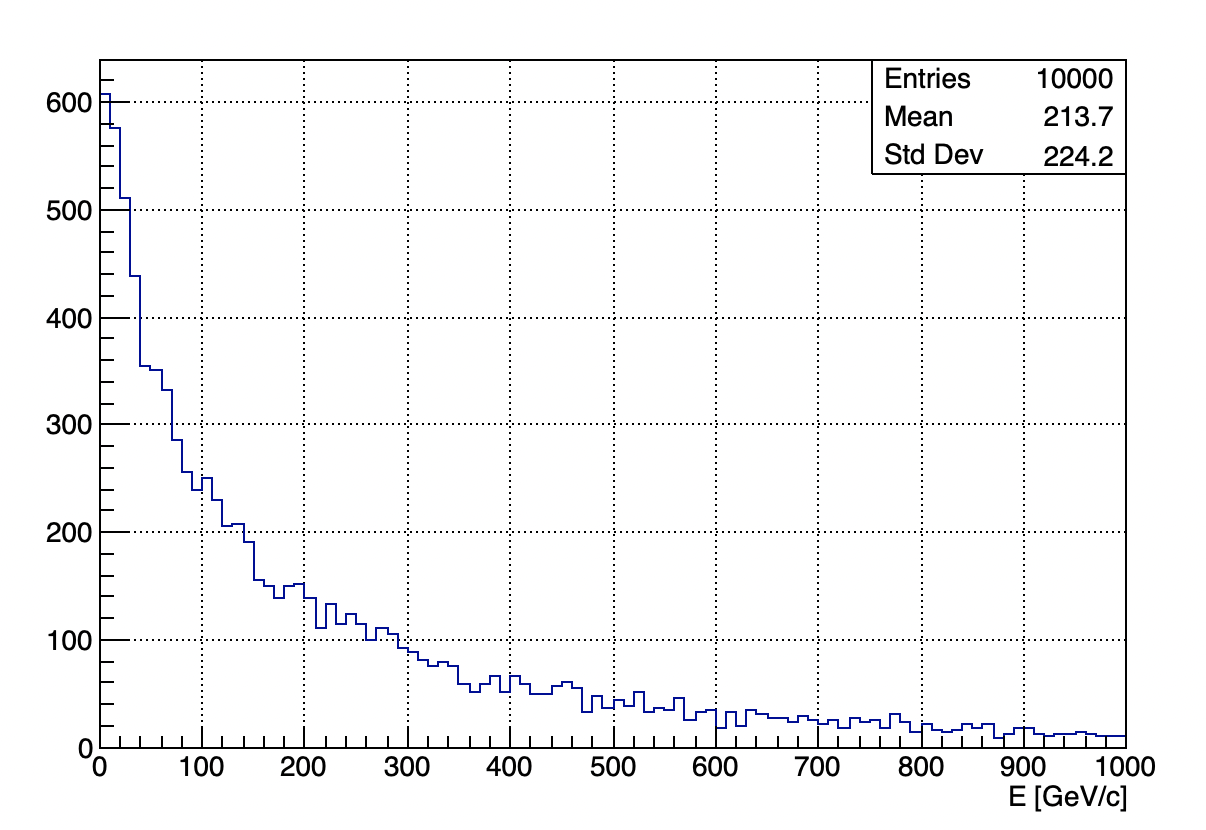}
    \caption{Momentum spectrum of muons in $\nu_\mu$ CC interactions in the $7.2 < \eta < 8.4$ region. }
    \label{fig:muon_spectrum}
\end{figure}

\begin{figure}[tbp]
    \centering
    \includegraphics[width=0.8\textwidth]{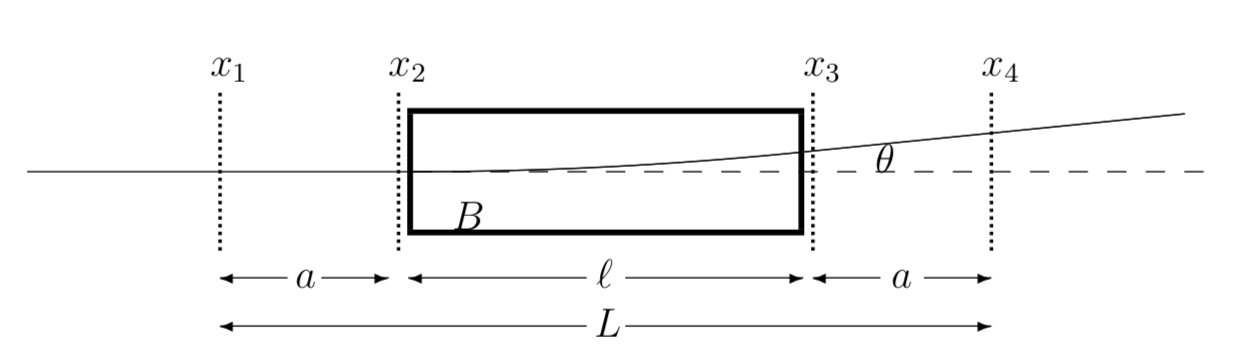}
    \caption{Schematic drawing of a magnetic spectrometer measuring the bending angle of a particle track. }
    \label{fig:magnet_scheme}
\end{figure}

The magnet is a key element in the detector design, because it makes it possible to distinguish $\nu_{\mu}$ from $\bar{\nu}_{\mu}$ and $\nu_{\tau}$ from $\bar{\nu}_{\tau}$ when the resulting tau lepton decays to a muon.  \cref{fig:muon_spectrum} shows the momentum spectrum of muons induced by the CC interactions of muon neutrinos in the $7.2 < \eta < 8.4$ region.  The layout of a spectrometer measuring the bending angle of a track is shown in~\cref{fig:magnet_scheme} with all the relevant parameters. We describe in the following the design of this kind of spectrometer for AdvSND. 

After traversing a magnet with field strength $B$ and length $\ell$, the track of a charged particle with momentum $p$ is bent by an angle $\theta = \dfrac{\ell}{r} = \dfrac{\mbox{e} B \ell}{p}$. The bending angle $\theta$ is determined by two tracking planes that are located before the magnet and separated by the lever arm $a$, which measure the track coordinates $x_1$ and $x_2$, and two tracking planes that are located behind the magnet and also separated by the lever arm $a$, measuring the track coordinates $x_3$ and $x_4$. From the track coordinates the bending angle is determined to be
\begin{equation} 
\theta = \frac{x_4 - x_3}{a} - \frac{x_2 - x_1}{a} \ .
\end{equation}
Denoting the measurement error of a track coordinate for a tracking station by $\epsilon$, the error on the bending angle measurement is $\Delta \theta = \dfrac{2 \epsilon}{a}$. Hence, the momentum resolution of the spectrometer is
\begin{equation} \frac{\Delta p}{p} \approx \frac{\Delta \theta}{\theta} = \frac{2 \epsilon p}{\mbox{e} B \ell a} \ .
\end{equation}
For a given total length $L$ of the spectrometer the choice of the length $\ell$ of the magnet and of the lever arm $a$, which results in the best momentum resolution, is defined by $a = \dfrac{L}{4} = \dfrac{\ell}{2}$.  For the muon charge assignment, the maximum muon momentum, $p_{\text{max}}$, for which a charge assignment is possible, can be defined by requiring at least a $4 \sigma$ separation from infinite momentum, corresponding to a momentum resolution of
\begin{equation} 
\frac{\Delta p}{p} = \frac{1}{4} \qquad \mbox{for} \qquad p = p_{\text{max}} \ ,
\end{equation}  
Thus, the maximum momentum, up to which a muon charge assignment is possible, is obtained: 
\begin{equation} p_{\text{max}} = \frac{\mbox{e} B \ell a}{8 \epsilon} \ .
\end{equation}
Assuming a magnetic field of $B = 1$\,T, a spatial resolution of the tracking chambers of $\epsilon = 100\, \mu$m, $\ell = 2\,$m, and $a = 1\,$m, the spectrometer allows for a charge assignment up to 750 GeV/$c$, thus covering 95\% of the momentum spectrum.  The overall length of the spectrometer is 4~m. 

\cref{tab:geometry} summarises the main parameters of the two locations and the corresponding detectors. 
\begin{table}[hbtp]
\centering
\begin{tabular}{l | c | c }
\hline\hline
&  AdvSND - NEAR & AdvSND - FAR \\
\hline
$\eta$ & [4.0, 5.0] &  [7.2, 8.4] \\ 
target mass (tonne) & 5 & 5  \\
front surface (cm$^2$) & $120 \times 120$ & $100 \times 55$ \\
distance from IP (m) & 55 & 630 \\
\hline\hline
 \end{tabular}
\caption{Parameters of the two AdvSND detectors in the NEAR and FAR locations. \label{tab:geometry} }
 \end{table}

\section{FLArE \label{sec:FLArE}}


A liquid argon time projection chamber (LArTPC) is considered for the suite of detectors for the FPF. Such a detector offers the possibility to precisely determine particle identification, track angle, and kinetic energy over the large dynamic range from 10~MeV to many hundreds of GeV\cite{Rubbia:1977zz,MicroBooNE:2021rmx}. Such a detector is therefore well motivated by the requirements of neutrino detection~\cite{Anchordoqui:2021ghd} and light dark matter searches~\cite{Batell:2021blf, Batell:2021aja,Batell:2021snh}.  A key motivation is simply the detection and measurement of TeV-scale neutrino events from a laboratory-generated, well-characterized source.  The LHC also represents the only possible terrestrial source of high-energy intense tau neutrinos.  The data set that will be gathered at the FPF with highly capable detectors will be unique and broadly valuable to particle physics and astrophysics.  

The key requirement for both measurements, neutrinos and light dark matter, is the ability to trigger and collect particles that come from an LHC IP and produce an event within the fiducial volume of the detector, in the presence of large muon backgrounds from the high luminosity running of the LHC. The detector must also be able to contain the events, reconstruct the kinematics, and identify the neutrino type.  Identification of tau neutrinos presents a particular challenge, requiring both high spatial and kinematic resolution.  In the case of dark matter events, an energetic, isolated, forward-going electron must be identified and its energy measured. A liquid argon (or noble liquid) TPC provides the opportunity to have high (mm-scale) spatial resolution in 3 dimensions, along with excellent electromagnetic calorimetry.  However, the issue of trigger using scintillation light and event reconstruction needs considerable R\&D.  Much of this R\&D can benefit from the investments made in the DUNE and protoDUNE~\cite{DUNE:2021hwx} experiments.

\subsection{Physics Requirements} 

Because of the spatial, energy, and time resolution of the LArTPC, it is a well motivated by the requirements of neutrino detection and the light DM search~\cite{Batell:2021blf, Batell:2021aja,Batell:2021snh}.  In particular, the TPC is an excellent choice for the detection and measurement of energetic electromagnetic showers.  Single muon tracks as well as showers of hadronic tracks also benefit from the superb spatial and charge resolution of this detector. The detector has no insensitive mass and therefore the energy loss and scattering can be measured along a long track. This capability leads to  superb particle identification at momenta of $\sim$ 1 GeV and also to excellent momentum resolution for high energy muons.  The kinematic resolutions in angle and momentum and how they affect various backgrounds for neutrino physics at the TeV scale needs further study. 

The detector is expected to measure millions of neutrino interactions, including tau neutrinos. The detector should have sufficient capability to measure these very high energy ($>100~\gev$) events, so that the cross section for each flavor can be measured. Identification of tau neutrinos with low backgrounds needs detailed simulations and reconstruction studies. As an approximate estimate, there will be about 50 high-energy neutrino events per ton per ${\rm fb^{-1}}$; this is approximately the integrated luminosity expected to be collected every day during the high-luminosity running of the LHC.  The majority of this flux will be muon neutrinos, with electron neutrinos forming about 1/5 of the event rate.  The tau neutrino rate is expected to be $\sim 0.1$ event${\rm /ton/fb^{-1}}$ with a very large uncertainty due to QCD modeling in the forward direction.  The high energy electron neutrino and the tau neutrino fluxes come from charm meson decays in the forward region, and therefore careful measurements of  these event types has broad implications for particle physics, as described in \cref{sec:qcd}, \cref{sec:neutrinos} and \cref{sec:astro}. 

\cref{lartab} summarizes the main parameters of a LArTPC for the FPF. A detector with a fiducial mass of approximately 10~tonnes of liquid argon is envisioned. We are also considering this same detector with a filling of liquid krypton. The dimensions of the TPC are extremely preliminary and could depend considerably on event containment and the energy loss  properties of LAr or LKr.   For both the LAr and LKr options, the efficiency for well-measured events that happen in the downstream portion of the detector can be enhanced by the addition of a hadron calorimeter, which is included in the table. 

For $3~{\rm ab^{-1}}$, such a detector will collect as many as millions of muon neutrino/antineutrino CC events, about a hundred thousand electron neutrino events, and thousands of tau neutrino events. These numbers have large uncertainties due to the poorly understood production cross section in the forward region~\cite{Bai:2020ukz,Bai:2021ira}. It is also important to note that this flux of events will have the same time structure as the LHC accelerator with a bunch spacing of 25 ns. At the same time, muons from interactions at the IP will produce a background flux of about $\sim 1$~muon/cm$^2$/s at the nominal maximum luminosity of $5\times 10^{34}~\text{cm}^{-2}~\text{s}^{-1}$ at the HL-LHC. As the luminosity is increased to $\sim 7.5\times 10^{34}~\text{cm}^{-2}~\text{s}^{-1}$ in later years of the HL-LHC, the background muon flux will correspondingly increase. As discussed in \cref{sec:sweepermagnet}, this muons flux may be reduced by the addition of a sweeper magnet placed between the ATLAS IP and the FPF.  

If the TPC can be operated with liquid krypton, several advantages are expected. The radiation length of LKr (4.7 cm) is much shorted than LAr (14 cm) leading to much more compact electromagnetic showers~\cite{miller:arkrxe, schinzel:krcalor, constantini:na48}. This performance naturally leads to much higher event containment for neutrino events. The higher density of LKr should also allow higher event rate. The overall increase of useful event rate is expected to be almost a factor of 2 at energies above 1 TeV.  Detailed simulations of event reconstruction need to be performed, but the better resolution from LKr is expected to lead to much better performance for tau neutrinos.  The choice of krypton for the TPC is, however, not simple because of intrinsic radioactivity due to $^{85}$Kr. 

\begin{table}[tbp]
\setlength{\tabcolsep}{3pt}
\centering
\begin{tabular}{ p{0.24\linewidth} || p{0.23\linewidth} | p{0.45\linewidth} }
 \hline
 \hline
 & Value & Remarks \\
 \hline 
 Detector length &  7~m & Not including cryostat \\ 
 TPC drift length &  $>0.5$ ~m & 2 TPC volumes with HV cathode in center \\ 
 TPC height &  $>1.3~m$ &  Dimension dependant on containment \\ 
 Total LAr mass & $\sim 16$~tonnes & Volume in the cryostat \\ 
 Total LKr mass & $\sim 27.5$~tonnes &  As an option \\ 
 Fiducial mass LAr/LKr & $10/17$~tons &  \\ 
 Charge Readout & wires or pixels & Hybrid approach is possible \\ 
 Light readout & SiPM array & Needed for neutrino trigger \\ 
 Background muon rate & $\sim 1/$cm$^2/$s & Maximum luminosity of $5\times10^{34}/$cm$^2/$s \\
 Neutrino event rate & $\sim 50/$ton/fb$^{-1}$ & For all flavors of neutrinos \\ 
 Cryostat type & Membrane 0.5 m & Thickness of membrane \\
 \hline 
 Hadronic calorimeter & $\sim 6-10 \lambda$ & Interactions lengths \\ 
 Dimension &  2m$\times$2m$\times$1.6m(depth)  & Fe/Scint sandwich \\ 
 Muon Range & To be determined &  \\ 
 \hline
 \hline
\end{tabular}
\caption{Detector parameters for a LArTPC for the FPF. The top part of the table shows the nominal geometric parameters for the time projection chamber to be considered for the FPF, and the bottom part shows the additional hadron and muon detectors. \label{lartab}}  
\end{table} 

\subsection{Detector Design Considerations}  

The nominal configuration for the LArTPC detector would include a central cathode operating at a large high voltage and two anode planes on two sides of the detector parallel to the beam from the ATLAS IP. The electric field between the cathode and the anode will be at $\sim500$~V/cm, providing a drift field for ionization electrons; the drift time for a 0.5~m-long drift will be about 0.3~ms. For a detector with an approximate cross section of $1$~m$^2$, there will be about 3 muon tracks within a single drift time. Neutrino and dark matter events must be selected out of these overlaying background particle trajectories. For the TPC, a readout using wires or pixels is possible\cite{Qian:2018qbv}.   A readout of the scintillation light is crucial to allow the measurement of the distance along the drift. It is also important for the selection of events that originate in the detector (such as a neutrino or a dark matter event), as well as generating the trigger necessary for acquiring the data.  Neutrino events need to be identified at the trigger level as events with tracks or showers that originate from a common vertex within the detector volume. As stated above, the expected event rate is $\sim 50$ per day per ton; the detector scintillation system and the trigger selection must be designed to retain high efficiency for these real events, while maintaining a low $\ll 1$Hz trigger rate.  

For the detection of neutrino and dark matter events  at the 10-ton fiducial mass scale, further simulation work is needed to understand event reconstruction and background rejection, especially for tau neutrino events.  For detector design, in particular, simulation work is needed to understand neutrino event  containment and energy resolution in a 7 m-long detector.  Study of kinematic resolution in the case of wire readout versus pixel readout is needed. And finally, the design and performance of the photon detector system needs to be investigated and demonstrated by R\&D.  Lastly, a LkrTPC would have remarkable resolution for electromagnetic showers and the event containment is expected to excellent.  In either the LAr or LKr case, a downstream hadronic calorimeter and a muon range detector is needed to contain particles escaping from the TPC for energetic neutrino events.  Looking further to the future, the addition of magnetic field and momentum measurement either with a downstream magnet or as part of the TPC needs to be explored.

\subsection{Cryogenics and Noble Liquid Circulation System}  

FLArE will be deployed in an underground cavern, and so it is necessary to develop compact liquid cryogenic facilities to condense, recirculate, and filter liquid argon or krypton. The CERN cryogenics group with their enormous experience is expected to take the major responsibility for this facility at CERN. Targeted R\&D to understand the differences between liquid argon and liquid krypton recirculation and purification systems is also important.  In particular, the liquid krypton system will have additional requirements because of the higher mass and the need to keep losses very low.  There could also be differences due to the needed purity levels.  

The LArTPC will be installed in a membrane cryostat with passive insulation and with nominal inner dimensions of $2.1 \times 2.1 \times 8.2 $~m$^3$. These cryostat dimensions allow excellent high voltage safety for the TPC dimensions in~\cref{lartab} using experience from previous designs. However, the cryostat design will need to be closely coordinated with the TPC as the design changes.  The membrane cryostat technology allows the cryostat to be constructed underground.  The insulation, being passive, ensures reliable and safe long-term performance.  The cryogenic system must re-condense the boil-off, keeping the ullage absolute pressure stable to better than 1 mbar, and purify the liquid argon bath. A standard approach is to re-condense the argon with a heat exchanger with liquid nitrogen. A LAr flow of 500~kg/h through the purification circuit is considered sufficient to reach and maintain the required LAr purity.

The total heat input due to the cryostat and the cryogenics system is estimated to be of the order of few~kW: $<~$few~kW from the cryostat depending on the size, 1~kW from the GAr circuit, 1~kW from the LAr purification and 1~kW from other inefficiencies.  To this, the detector electronics should be added.  A Turbo-Brayton (~8 m x 1.6 m x 2.7 m) TBF-80 unit from Air Liquid installed in the vicinity of the cryostat provides approximately 10~kW cooling power from $\approx$100~kW electrical power and 5 kg/s of water at ambient temperature. Liquid argon and nitrogen storage tanks are required above ground and connected via piping to the underground cryogenics.  Exhaust of gasses will be done to the atmosphere on the surface. For safety reasons, as noted in \cref{sec:facility}, the cryostat in the cavern is placed in a trench 1.5~m deep, 6.9~m wide, and 12.6~m long, which collects the argon in case of a leak.  Oxygen deficiency is the main risk associated with the LArTPC. A properly designed ventilation system will constantly extract air in the proximity of the cryostat/cryogenics.  A detection of low oxygen content by the ODH system in the cavern and in the trench will trigger  increased air extraction. The design and technology for the cryosystem is well understood because of the experience from protoDUNE. If Lkr is considered as a fill, then the requirements for the cryosystem need to be further examined in detail, but there is also experience with a LKr system at CERN from the NA48 experiment, which has run for a long time~\cite{constantini:na48}.

\subsection{Research and Development} 

The FLArE detector concept requires a few key R\&D efforts.  Neutrino events at energies of 1 TeV and above from an accelerator source have not been detected, and simulations of these events require much tuning of software codes and comparison with existing data.  This effort will lead to a much better understanding of kinematic resolutions and the design of a hadron calorimeter and a muon rangefinder.  In particular, detailed simulations are needed to understand the signal to background ratio achievable for tau neutrinos with event shape and kinematic cuts and other software techniques. The design of the TPC and the cathode and anode readout depends strongly on the spatial and charge resolution that will be needed for particle identification. The three options are: a wire readout, a pixel readout, or a hybrid readout.  The wire readout has been analyzed thoroughly~\cite{Qian:2018qbv} and is best suited for very large detectors with very low channel occupancy.  On the other hand, a pixel readout could be quite expensive in terms of channel counts.  A new type of readout plane is in the process of development for the DUNE vertical drift detector; this design could be appropriate for FLArE. Finally, the design and performance of the photon detector system needs to be investigated and demonstrated by R\&D.  In particular, the photon system has to serve three functions in order of increasing difficulty:  separation of beam related muon and neutrino events, accurate timing performance to measure the location  of the neutrino or dark matter events in the TPC, and association of the neutrino or dark matter events with a bunch crossing in the collider detector.

\vspace*{0.1in}
\noindent {\bf Time Projection Chamber Design}
\vspace*{0.05in}

\begin{figure}[tbp]
\centering
\includegraphics[width=0.99\textwidth]{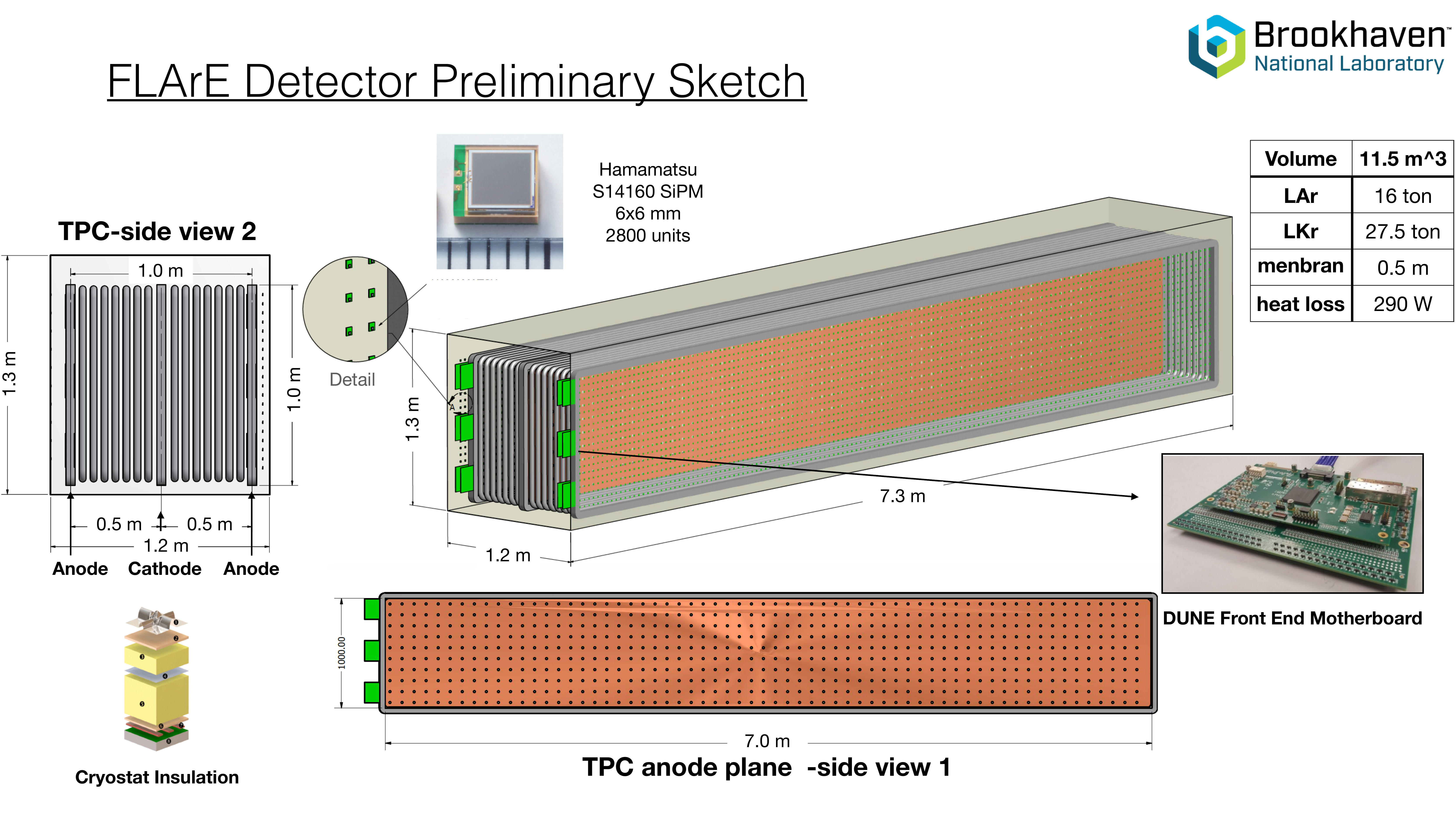} 
\caption{A preliminary sketch of the FLArE detector. The concept must be further engineered for integration and high-voltage safety.  The cryostat technology from GTT (Gaztransport and Technigaz) has corrugations that allow expansion/contraction of the metallic surfaces. For a small cryostat such as for FLArE, the cryostat must be redesigned with flat walls.  The space between the field cage and the cryostat is for high-voltage safety. Possible electronics readout is indicated, but must be designed in detail for this detector. } 
\label{fig:layout}
\end{figure}

\cref{fig:layout} shows the preliminary conceptual design for the FLArE detector, and \cref{lartab} summarizes the main parameters.  This nominal design is being implemented with \texttool{Geant4} for detector response and physics reach simulations. A detector with a fiducial mass of approximately 10~tonnes (24 tonnes) of liquid argon (liquid krypton) is envisioned. During the HL-LHC era, more than 500 neutrino events per day are expected in the detector. 

The TPC serves to measure the ionization tracks in the liquid with close-to-ideal spatial resolution for an electronic detector.  The ionization electrons are drifted over a length of 0.5 to 1 m away from the cathode plane in the center and detected by their induced currents in electrodes on the anode plane.  The electrodes (wires or pixels) can be arranged according to the needed spatial resolution down to a few mm.  The TPC has three essential components in the design:~the high voltage cathode and the field cage, the anode and the electrode design, and the readout electronics.  

The cathode and the field cage define the static electric field.  High voltage must be provided using a penetration (feed-through) in the cryostat. The design of this system is now very well understood from experience from protoDUNE and ICARUS detectors. Fields of 500 volts/cm can now easily be sustained over several meters of drift.  Two important challenges that must be confronted for the design of FLArE are the availability of space above the detector for the installation of the HV feedthrough, and providing a cathode that permits scintillation light  to reach both sides of the TPC.  A particular issue arises for liquid krypton due to the high level of intrinsic radioactivity ($\sim 500 {\rm bq/cm^3} $) which creates a cloud of charge that reduces the applied electric field (space charge).  This problem needs careful examination and may require smaller TPC gaps or high fields.  Both the liquid argon and krypton options require careful evaluation for space charge due to high rates of muon tracks from the LHC.  

The anode and the electrode design determines the spatial resolution for measurement of ionization in 3D.  A design using several (more than 3) planes of wires has been used in the protoDUNE test detector. It is also the design for ICARUS, MicroBooNE, and the first module of DUNE. The wires in each plane allow measurement of a single plane projection of the ionization image. Drifting electrons induce currents in each of the planes of wires before being collected on the final wire plane. Combining three 2D images electronically  produces a 3D pattern. This technique, however, requires low track multiplicity in the events to reduce reconstruction ambiguities~\cite{Qian:2018qbv} and is otherwise subject to inefficient event recognition. Another option is being examined in the context of the DUNE second module, in which a rigid plane is employed with small holes to allow electrons to pass.  Induced currents on metalized electrodes on both sides of the rigid  plane (or electronic board) are used to create a projection of the image. This technique could lead to an inexpensive pixelized detector that is not subject to reconstruction ambiguities.    

Typical amount of charge from a minimum ionizing particle through liquid argon (or krypton) is about $\sim 1-2\times 10^4$ electrons for a few mm of track length.  The collected charge depends on the electric field strength, purity of the liquid, and losses due to attachment over the drift distance.  The liquid must have very high purity ($\sim 0.1 $ppb) to allow charge drift over 1 meter without significant attenuation.  Detection of induced current pulses from this small charge requires low noise  amplification of the pulses as close to the electrode as possible.  Such low noise cryogenic electronics has now been fully developed and tested in the protoDUNE detector.  The entire readout chain of amplification, digitization, and data transmission has been designed to work in liquid argon.  The FLArE detector can fully utilize this technology with minimal changes regardless of the detailed design of the electrode.  

\vspace*{0.1in}
\noindent {\bf Detector Response and Physics Reach Simulation}
\vspace*{0.05in}

\begin{figure}[tbp]
\centering
\includegraphics[width=0.47\textwidth]{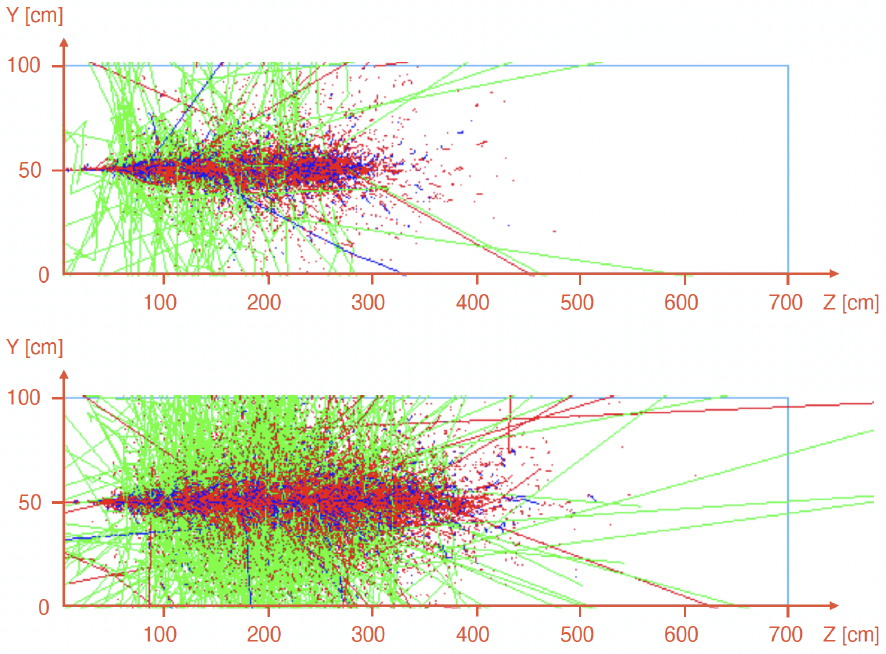} 
\hfill
\includegraphics[width=0.47\textwidth]{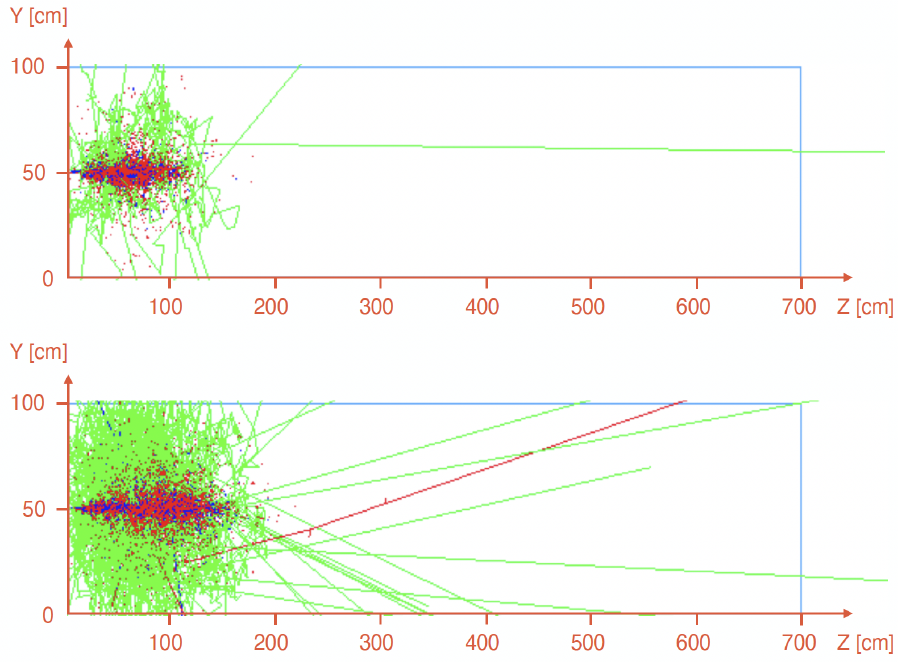} 
\\
\caption{\texttool{Geant4} simulations of electromagnetic showers in LAr (left) and LKr (right) produced by 200 GeV (top) and 1 TeV (bottom) electrons. The showers start at $z=0$, and all particles, except photons, are shown, with the color indicating the particle's charge: negative (red), neutral (green), and positive (blue). The shower depth in LKr is shorter due to it’s smaller radiation length.
\label{fig:containment}}
\end{figure}

Simulations based on various neutrino and DM event generators and \texttool{Geant4} detector modeling play a critical role in understanding FLArE's response to all relevant physics processes and evaluating its physics research for neutrino studies and DM searches.  Preliminary simulations have been performed to compare the development of electromagnetic showers in LAr and LKr.  These results are shown in \cref{fig:containment} and demonstrate LKr's remarkable resolution for electromagnetic showers and excellent event containment.  Further studies will be used to optimize the experiment design and understand event reconstruction and event selection; study the kinematic resolution in the case of wire readout versus pixel readout; study the performance of the proposed photon detection system as well as alternatives; and investigate the potential benefits of a magnetic field, created either by a downstream magnet or as part of the TPC. 

\vspace*{0.1in}
\noindent {\bf Photon Sensor System}
\vspace*{0.05in}

A new photon sensor system must be designed for FLArE.  This system provides three key functions: (1) The system will provide an accurate measurement of the time of the ionization event in the detector so that the event vertex can be located in the drift direction in the TPC. The location is essential for reconstruction of the kinematics of the event. (2) The time is also needed to isolate the several particle tracks that are expected to be within a single drift time. The tracks that are background muons can then be identified and removed  from the data. (3) The photon system must have sufficient granularity or pixelization to allow basic fast pattern recognition to select interesting events at the trigger level. Interesting neutrino or DM events are those that originate in the active volume of the detector.  The decision to read out the detector must be carried out quickly so that the TPC data that are in the pipeline can be recorded without dead time.  It will also be interesting to examine the physics and the technical possibility of triggering the ATLAS data acquisition based on a neutrino trigger from FLArE.  This requires the examination of the time resolution that can be achieved and the time available for a trigger decision and transmission to the ATLAS level 1 system.

The photon sensors detect the ample ultra-violet scintillation light that is produced by ionization events in liquid argon (128 nm) or liquid krypton (150 nm).  This scintillation light has two components with short (few ns) and long (1600 and 90 ns for LAr and LKr, respectively) time constants. It is common in such detectors to convert these photons to longer wavelengths by using a wavelength shifter, such as tetraphenyl butadiene (TPB), so that the shifted light can be detected in photomultiplier tubes or in silicon-photomultiplier (SiPM) sensors.  There is a preference for using SiPM technology, as it is rapidly advancing and becoming more affordable~\cite{DUNE:2021hwx}.  SiPMs also take much less space inside the cryostat, thus saving the precious sensitive volume for physics.  Both technologies need to be assessed because of the expected count rates of signal and background events in the detector. In particular, the high-energy muons and the intrinsic radioactivity ($^{39}$Ar decays for liquid argon and $^{85}$Kr decays for liquid krypton) are expected to contribute large instantaneous and average currents for these sensors. The low energy threshold for a trigger decision will determine the physics reach for a number of topics, such as low mass dark matter scattering, and the evaluation will require detailed simulations, as well as detector evaluation and measurements in the laboratory.

\vspace*{0.1in}
\noindent {\bf AI/Machine Learning-Based Trigger Studies}
\vspace*{0.05in}

A key challenge for FLArE is triggering and identifying DM and neutrino signals with low backgrounds.  As noted above, approximately 500 high-energy neutrino events will be seen each day in a 10-tonne detector at the HL-LHC. The majority of this flux will be muon neutrinos, with electron neutrinos forming about 1/5 of the event rate.  The rates for neutrino and DM signals are much lower than the muon background rate of $\sim 1 \rm{/cm^2/sec}$.  Therefore, traditional simple trigger schemes, such as multiple detector signal coincidence in some time window, will be overwhelmed. 

To solve this problem, machine learning (ML) techniques may be used early on in the trigger to reduce the trigger time and the rate of saved events. These techniques may also reduce the trigger energy threshold if they are applied to the task of event reconstruction itself. Convolutional neural networks (CNNs) in real-time event processing and triggering may also be implemented. The CNNs will use event-level PDS pixel maps to classify events directly for the fast trigger. The TPC information can also be investigated for slower trigger decisions.  Novel Transformer-based~\cite{2017arXiv170603762V, 2018arXiv181004805D} deep neural networks may also be used to combine charge, timing, and location information in an interpretable way for FLArE trigger and event identification.

\section{FORMOSA \label{sec:FORMOSA}}


The FPF provides an ideal location for a next generation experiment to search for BSM particles that have an electrical charge that is a small fraction of that of the electron. Although the value of this fraction can vary over several orders of magnitude, we generically refer to these new states as millicharged particles (mCPs). Since these new fermions are typically not charged under QCD, and because their electromagnetic interactions are suppressed by a factor of $(Q/e)^2$, they are ``feebly" interacting and naturally arise in many BSM scenarios that involve dark or otherwise hidden sectors. General purpose detectors at the LHC are not sensitive to the deposits of such particles and so experimental observation of mCPs requires a dedicated detector. 

As proposed in Ref.~\cite{Foroughi-Abari:2020qar}, where it is referred to as FORMOSA-II\footnote{FORMOSA-I refers to a demonstrator prototype that could be installed in UJ12/TI12 experimental areas near the current FASER experiment.}, an experiment to search for mCPs at the FPF would consist of a scintillator-based detector~\cite{Haas:2014dda, Ball:2016zrp} technically similar to what the milliQan collaboration will install in the PX56 drainage gallery near the CMS IP at LHC Point 5 for LHC Run~3~\cite{milliQan:2021lne}, but with a significantly larger active area and a more optimal location with respect to the expected mCP flux. During Run 2 of the LHC, the milliQan collaboration installed a prototype scintillator-based detector (the milliQan ``demonstrator") in the PX56 draining gallery at LHC P5 near the CMS IP. This device was used successfully to search for mCPs, proving the feasibility of such a detector~\cite{Ball:2020dnx}. The results of the milliQan demonstrator provide valuable insights into the design and operation of the FORMOSA detector. 

The FORMOSA detector is primarily targeted at discovering low charge signals (described in detail in \cref{sec:bsm2}), however, sensitivity to alternative signatures, such as a heavy neutrino electric dipole moment, can be expected~\cite{Sher:2017wya}. In addition, more involved mCP signatures can be considered. This can include multiple mCPs traversing the detector~\cite{Izaguirre:2015eya}. In the case of the discovery of mCPs, an expanded detector could be constructed that can distinguish the details of the signal signature.

\subsection{Detector Design}

To be sensitive to the small $dE/dx$ of a particle with $Q \lesssim 0.1e$, an mCP detector must contain a sufficient amount of sensitive material in the longitudinal direction pointing to the IP. Plastic scintillator, such as, for example, Eljen EJ-200~\cite{Eljen} or Saint-Gobain BC-408~\cite{SG}, provides a detection medium with the best combination of photon yield per unit length, response time, and cost. The FORMOSA detector is planned to be a $1~\mathrm{m} \times 1~\mathrm{m} \times 5~\mathrm{m}$ array of plastic scintillator. The length of the bars is determined by the desire to efficiently reconstruct a particle of charge ${\cal O}(10^{-3})$, however, this may be optimised further. No gain in sensitivity is expected using a larger area thin scintillator detector (as planned for the Run 3 milliQan detector~\cite{milliQan:2021lne}) as the reach in charge is typically below that which such a detector could efficiently reconstruct.

\begin{figure}[tbp]
  \centering
  \includegraphics[width=0.7\textwidth]{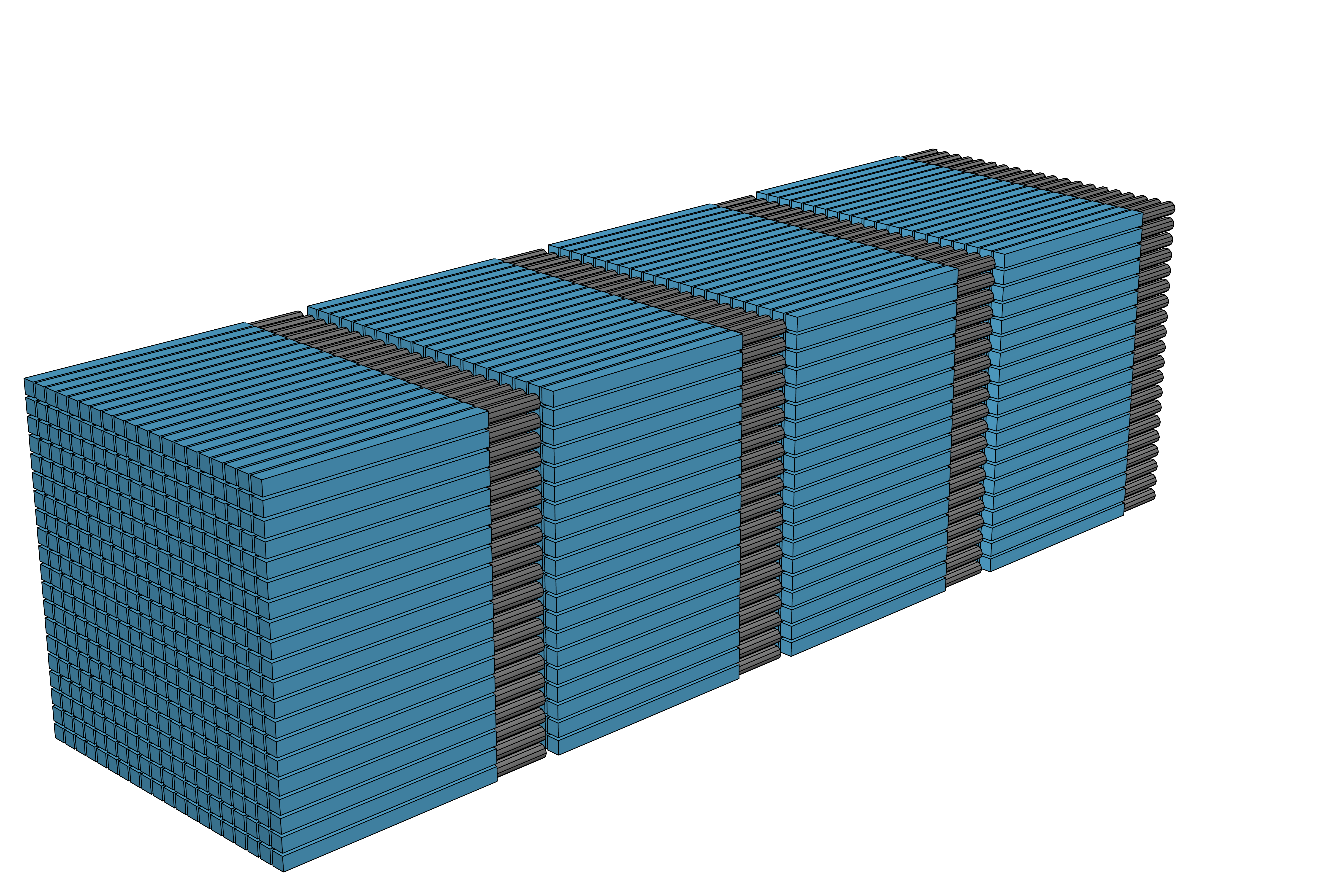}
  \caption{A diagram of the FORMOSA detector components. The scintillator bars are shown in blue connected to PMTs in black.}
  \label{fig:formosa-bars}
\end{figure}

The array will be oriented such that the long axis points at the ATLAS IP and it will be located on the beam collision axis. The array contains four longitudinal ``layers" arranged to facilitate a 4-fold coincident signal for feebly interacting particles originating from the ATLAS IP. Each layer in turn contains one hundred $5~\mathrm{cm}\times5~\mathrm{cm}\times100~\mathrm{cm}$ scintillator ``bars" in a $20\times20$ array. The bars will be held in place by a steel frame. A conceptual design of the FORMOSA detector is shown in \cref{fig:formosa-bars}. 

Although omitted for clarity in \cref{fig:formosa-bars}, three additional scintillator ``panels" of $5~\mathrm{cm}\times 100~\mathrm{cm}\times400~\mathrm{cm}$, placed on each side of the detector, will be used to actively veto cosmic muon shower and beam halo particles. Finally, scintillator panels of $5~\mathrm{cm}\times 100~\mathrm{cm}\times100~\mathrm{cm}$ will be placed on the front and back of the detector to aid in the identification of muons resulting from LHC proton collisions. 
 
To maximize sensitivity to the smallest charges, each scintillator volume will be coupled to a high-gain photomultiplier tube (PMT) capable of efficiently reconstructing the waveform produced by a single photoelectron (PE).  To reduce random backgrounds, mCP signal candidates will be required to have a quadruple coincidence of hits with $N_{\rm PE}\ge 1$ within a small time window. The PMTs must therefore measure the timing of the scintillator photon pulse with a resolution of $\le5$ ns.  Three different species of PMT were deployed as part of the milliQan demonstrator: Hamamatsu R878, Hamamatsu R7725~\cite{Hamamatsu}, and Electron Tube 9814B~\cite{ElectronTubes}. While all three species meet the minimal requirements for the FORMOSA detector, the R7725 and Electron Tube PMTs were found to have the best dark rate, timing, and response performance.  The afterpulsing properties of the PMT species used will also have significant impact on the background faced by the detector. To avoid sensitivity to residual magnetic fields in the cavern, each PMT will be wrapped with mu-metal shielding. This shielding consists of two parts:~one is directly around the PMT within the mount, and another thin layer is wrapped around the outside of a completed bar covering a region 2 cm on either side of the photo-cathode position.

Waveforms from each PMT will be digitized, read out, and stored for offline analysis. As the pulse rate per PMT is large, a trigger will be used to record only those waveforms during interesting time windows when signal-like activity in the detector is observed with at least 3 layers in a $2\times2\times4$ bar region having a pulse above the single PE threshold. To avoid the rate being dominated by through-going muons, large pulses in the front and end panels will be vetoed in the trigger. The digitisation can be performed using 25 16-channel CAEN V1743 digitizers~\cite{CAENV1743}, operating at $1.6\times10^{9}$ samples per second with 12-bit resolution, providing 1024 samples within a 640 ns acquisition window. The 16 channels are arranged into 8 trigger groups, each of which can output a trigger bit via LVDS. These trigger bits can then be combined by dedicated electronics to form the trigger decision. 

Commercial CAEN HV power supply modules connected to fan-out boards will provide power for up to 12 groups of up to 12 PMTs (144 PMTs total) for each HV supply module. Therefore, three CAEN A1535DN power supply modules can be utilised to power all scintillator bars and panels.

\subsection{Backgrounds and Sensitivity}

Even though the pointing, 4-layered, design will be very effective at reducing background processes, small residual contributions from sources of background that mimic the signal-like quadruple coincidence signature are expected. These include overlapping dark rate pulses, cosmic muon shower particles and beam muon afterpulses. In Ref.~\cite{milliQan:2021lne}, data from the milliQan prototype was used to predict backgrounds from dark rate pulses and cosmic muon shower particles for a closely related detector design and location. Based on these studies, such backgrounds are expected to be negligible for FORMOSA. An additional background is expected from the large through-going muon rate of $\sim1\ \rm{Hz}/\rm{cm}^{2}$. Although the large pulses from the muons can easily be distinguished from the signal, the afterpulses caused by ionising in the PMTs can mimic the signal signature. As detailed in Ref.~\cite{Foroughi-Abari:2020qar}, it is expected that these can be rejected by vetoing a 10 $\mu$s time window in the detector following through-going beam muons. To more effectively track the paths of beam muons through the detector, it is possible to segment the veto panels at the front and back of the detector to form hodoscopes. This may allow areas impacted by beam muons to be identified, rather than necessitating deadtime for the full detector. In addition, input on the beam muon paths could be provided to other detectors within the FPF to allow improved background rejection. Finally, the use of a sweeper magnet (detailed in \cref{sec:sweepermagnet}) to lessen the rate of beam muons incident on the detector could significantly reduce the deadtime and backgrounds in the detector.

\begin{figure}[tbp]
  \centering
  \includegraphics[width=0.7\textwidth]{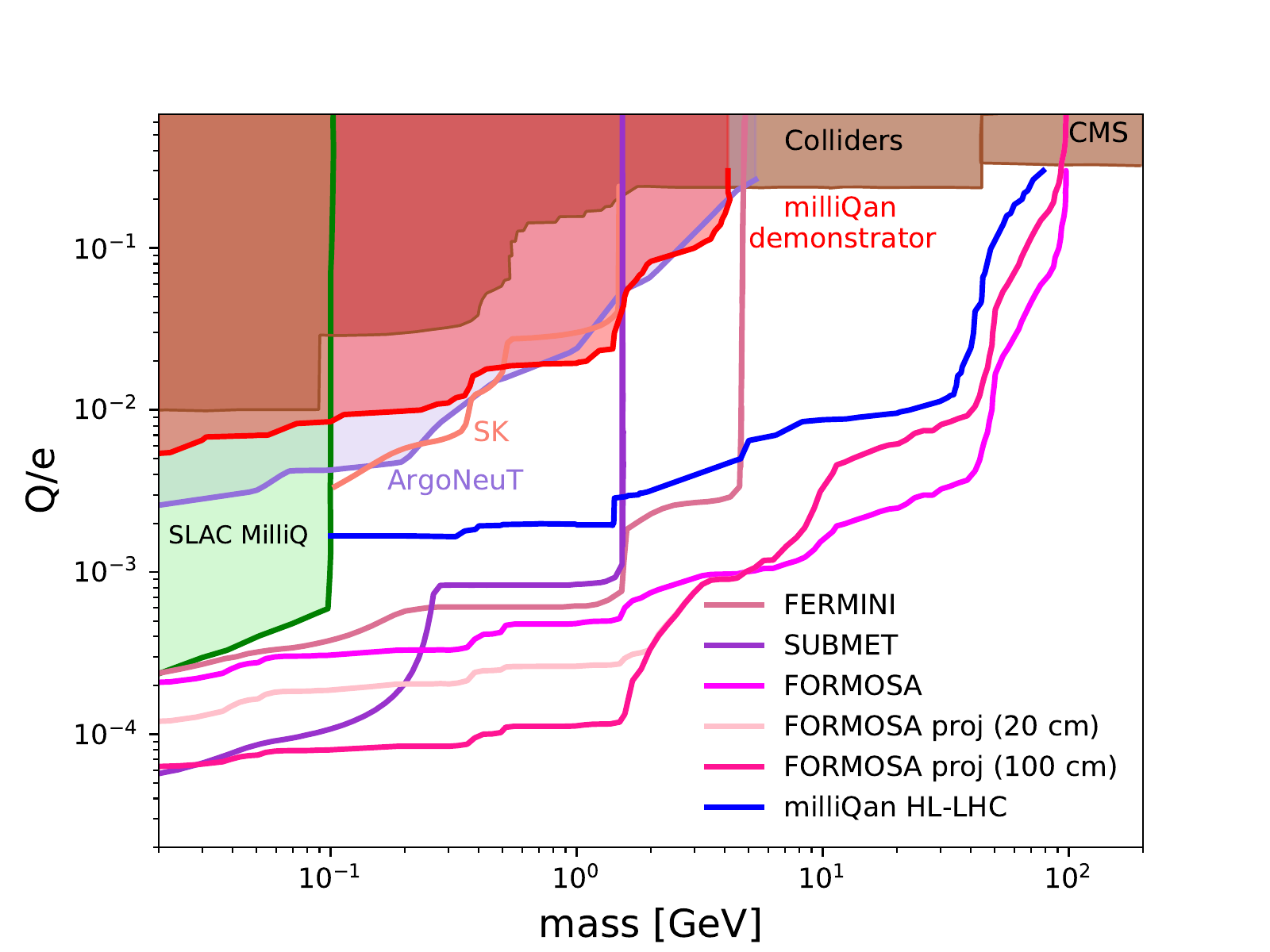}
  \caption{Expected sensitivity of FORMOSA compared to other constraints from previous and proposed experiments.}
  \label{fig:formosa-sens}
\end{figure}

The signal process is simulated from a range of production modes, as detailed in Ref.~\cite{Foroughi-Abari:2020qar}. This provides the expected flux for each mass and charge of the signal. This can then be used to determine the expected limits as shown in \cref{fig:formosa-sens}. For much of the parameter space, the sensitivity is limited by the efficiency of the scintillator bars to detect through-going mCPs. In this regime, the mCP flux is very high, and so only a small area of higher performance scintillator can allow substantial gains in sensitivity. One possibility is an upgraded design in which an additional $2 \times 2 \times 4$ bars are installed using a higher performance scintillator, such as LaBr3(Ce).  By placing these bars close to the larger plastic scintillator, the active veto background rejection capabilities of the array for sources such as cosmic showers are maintained. Shielding of these would mitigate backgrounds for the larger detector caused by the radioactivity of the LaBr3(Ce).  In this scenario, the optimal charge reach could be lowered by as much as a factor of 5.

Finally, the FLArE detector may also have sensitivity to mCPs. It could be possible for FORMOSA to prove $\sim$ 10 cm (40 cm) resolution tracking of mCP paths at the front (back) of FLArE . This could help FLArE  provide an independent measurement of the properties of the mCP signal. The combination of measurements from multiple detectors in the FPF can provide powerful insights into both background and signal processes.

%% file: sec_bsm1.tex
\contributors{Ahmed Ismail, Felix Kling, Sebastian Trojanowski, Yu-Dai Tsai (conveners), Takeshi Araki, Kento Asai, Martin Bauer, Brian Batell, Mathias Becker, Asher Berlin, Enrico Bertuzzo, Joseph Bramante, Adrian Carmona, Garv Chauhan, Emanuelle Copello, Luc Darm\'e,  Neda Darvishi, Sergey Demidov, Patrick deNiverville, Frank F. Deppisch, Bhupal Dev, Jordy de Vries, Keith R. Dienes, Herbi K. Dreiner, Bhaskar Dutta,  Sebastian Ellis, Yasaman Farzan, Jonathan L.~Feng, Max Fieg, Ana Luisa Foguel, Patrick Foldenauer, Saeid Foroughi-Abari, Jean-Fran\c cois Fortin, Elina Fuchs, Sumit Ghosh, Dmitry Gorbunov, Julian Y. Günther, Steven~P. Harris, Julia Harz, Fei Huang, Juan Carlos Helo Herrera, Martin Hirsch, Ameen Ismail, Yongsoo Jho, Krzysztof Jodlowski, Dmitry Kalashnikov, Timo J.  K{\"a}rkk{\"a}inen, Jongkuk Kim, Pyungwon Ko, Dominik Köhler, Suchita Kulkarni, Jason Kumar, Hye-Sung Lee, Seung J. Lee, Jinmian Li, Shuailong Li, Wei Liu, Zhen Liu, Kunfeng Lyu, Mohammad~R. Masouminia, Kirtimaan Mohan, Martin Mosny, Saurabh Nangia, Takaaki Nomura, Nobuchika Okada, Hidetoshi Otono, Gilad Perez, Simon Pl\"atzer, Digesh Raut, Peter Richardson, Leszek Roszkowski,  Peter Reimitz, Adam Ritz, Christiane Scherb, Pedro Schwaller, Dipan Sengupta, Takashi Shimomura, Kuver Sinha, Huayang Song, Shufang Su, Wei Su, Yosuke Takubo, Marco Taoso,  Brooks Thomas, Zeren Simon Wang, Martin Winkler, Xun-Jie Xu, Tevong You, Yongchao Zhang, Guanghui Zhou, Renata Zukanovic Funchal}\\

Light new particles appear generically in BSM models that motivate the electroweak scale, explain the origin of dark matter, describe the mechanism underlying neutrino masses, solve the strong CP problem, and explain the baryon asymmetry of the universe. Some of the most common signatures of these models arise from new long-lived particles that could decay in detectors such as FASER2. This section describes such models in the context of the FPF and their potential long-lived particle signatures; see also contribution to Snowmass 2021 \textsl{Big Idea: light long-lived particles}~\cite{RF06WP_1} and \textsl{RF6 - Overview of Facilities and Experiments}~\cite{RF06WP_4} for further general discussion. Separately, theories predicting BSM particles that scatter within the FPF are described in the following section. Models that feature new long-lived states are organized below by the type of new particle (vector, scalar, fermion, or pseudoscalar) that they would predict at the FPF. Some models have more than one new particle that plays a relevant role in FPF phenomenology as well. Within the relevant subsections, we compile the theories, final states, and expected FPF sensitivities for these models. 

\begin{figure*}[th]
\centering
\includegraphics[width=0.99\textwidth]{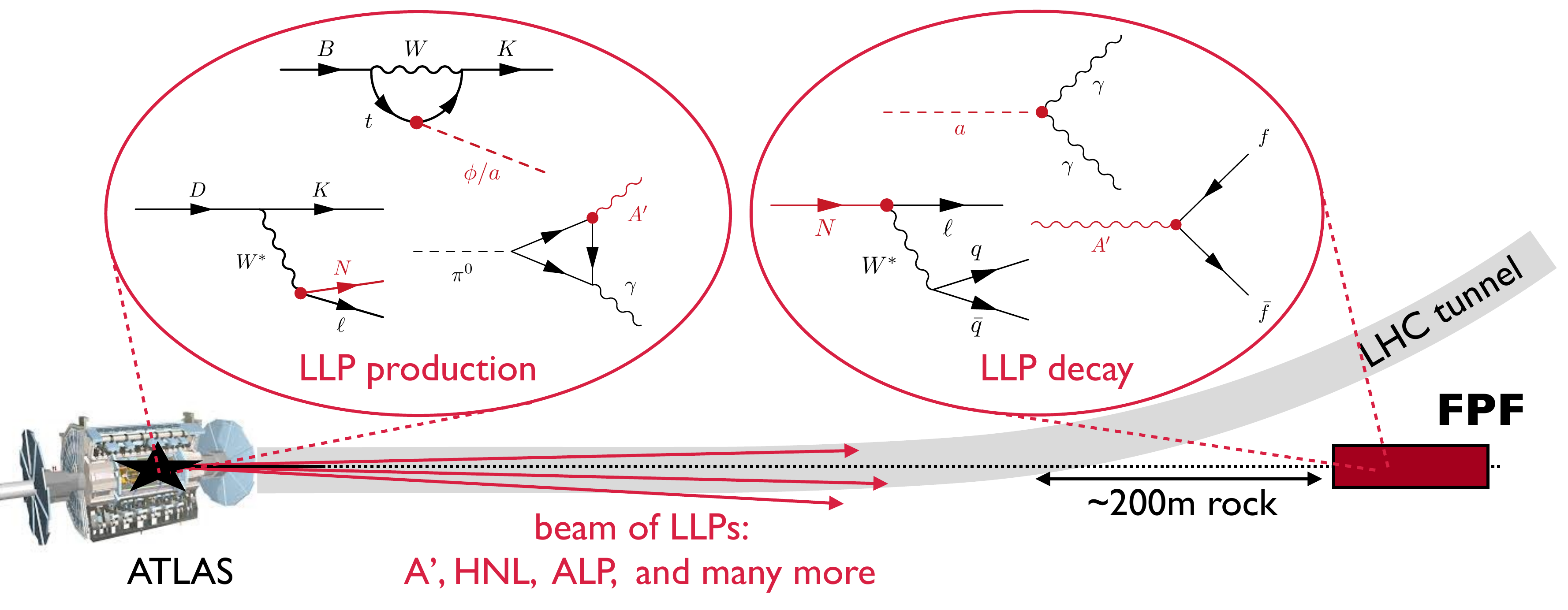}
\caption{Examples of long-lived particle signatures at the FPF. New states can be produced at the ATLAS IP through meson decay (pictured), bremsstrahlung, Primakoff production, etc. When the relevant interaction scales are much smaller than the typical LHC parton energy, a well-collimated beam of the BSM particles is created. Then, if these particles interact very weakly with the SM, they can propagate downstream and decay to SM final states inside the FPF. Both the production and decay mechanisms are model-dependent.}
\label{fig:bsm_llp_sum1}
\end{figure*}

The ultimate source of BSM states at the FPF is the proton collisions at the LHC, and it is useful to briefly describe how these collisions could give rise to light new particles. First, a large flux of mesons in the forward direction is expected due to the GeV scale of QCD interactions and the typical TeV parton energy. In fact, over the lifetime of the HL-LHC, there will be $4 \times 10^{17}$ neutral pions, $6 \times 10^{16}$ $\eta$ mesons, $2 \times 10^{15}$ $D$ mesons, and $10^{13}$ $B$ mesons produced in the direction of the FPF. In many models, meson decay is a major source of new particles, resulting in high-energy beams of these particles at large rapidity. Depending on the theory, BSM particles can also be produced through proton bremsstrahlung or even through Drell-Yan production. The following contributions describe the main production mechanisms of BSM in a range of different models, as well as potential final states that can be achieved through scattering or decay. \cref{fig:bsm_llp_sum1} illustrates the production of long-lived BSM particles and their FPF decay signatures. When the scale of the interactions responsible for new particle production is much smaller than the TeV energies typical at the LHC, collimated beams of long-lived particles are naturally produced in the direction of the FPF.

\begin{figure*}[th]
\centering
\includegraphics[width=0.70\textwidth]{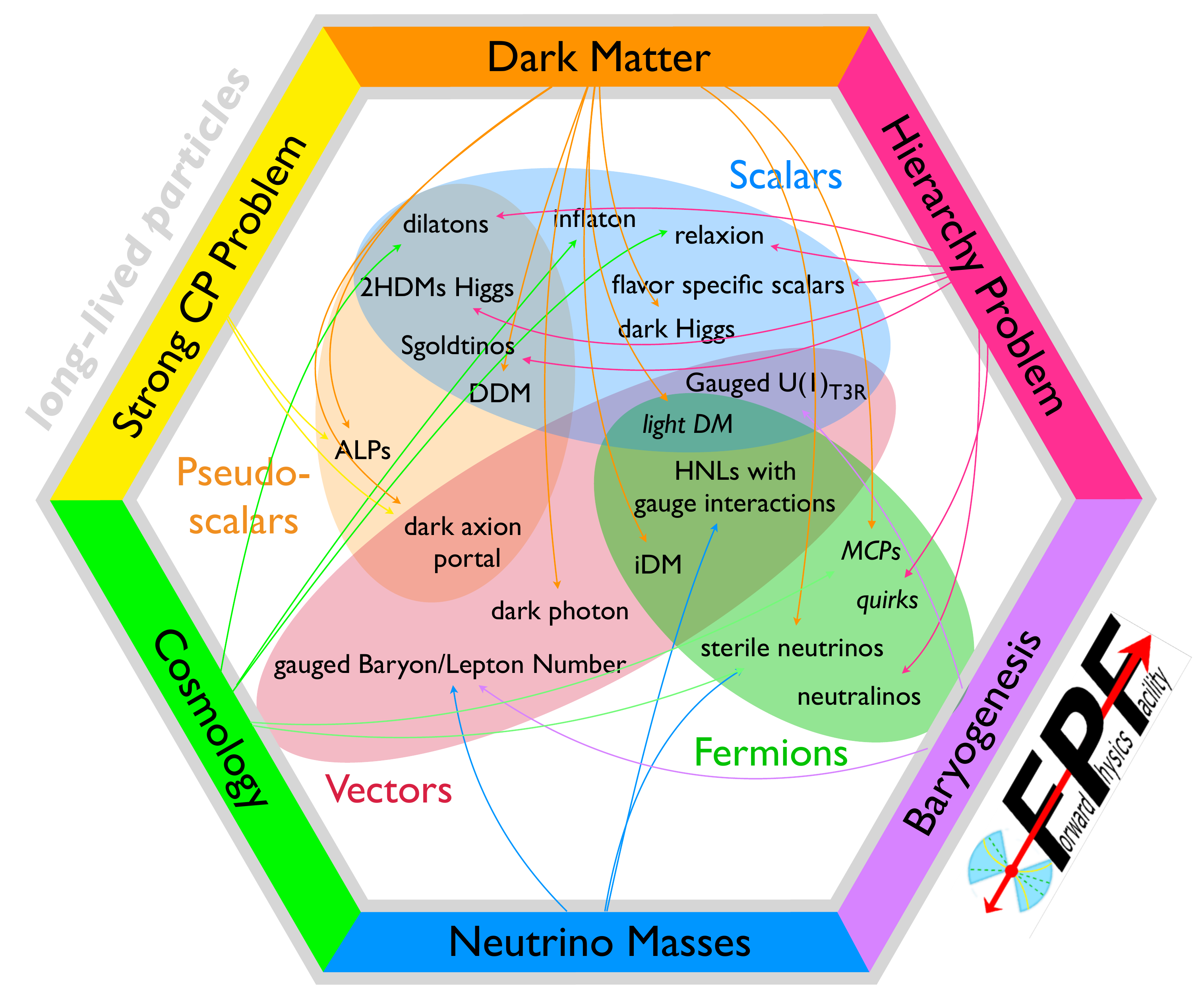}
\caption{New physics models with FPF signatures considered in this report. The models are grouped by the type of BSM particle they contain that is typically most relevant for the FPF. Connections are shown to the multitude of new physics motivations on the outside of the figure. Models in italics give rise to FPF signatures involving the scattering of BSM states rather than long-lived particle decays, and are treated in the following section.}
\label{fig:bsm_llp_sum2}
\end{figure*}

In many theories, these long-lived particles can be identified with states that are necessary to resolve fundamental outstanding questions. For instance, long-lived scalars appear in relaxion theories as well as in models of inflation. In other cases, light new states are obligatory features of models which resolve outstanding experimental anomalies, such as novel gauge symmetries to explain the measured anomalous magnetic moment of the muon. \cref{fig:bsm_llp_sum2} shows the multitude of models with long-lived particle signatures at the FPF and how they are connected to different needs for BSM physics. Theories predicting scattering signatures, as discussed in the following section, are also shown.

The rest of this section is organized as follows. We begin in \cref{sec:foresee} by describing the \texttool{FORESEE} package for use in simulating and analyzing BSM events at the FPF. Then, we turn to long-lived particles, with the next several sections dealing with new particles decaying to visible final states in FPF detectors which are vectors in \cref{sec:bsm_llp_vec}, scalars in \cref{sec:bsm_llp_scalar}, fermions in \cref{sec:bsm_llp_fermion}, and ALPs in \cref{sec:bsm_llp_alp}. We also discuss models in which more than one new particle could give a viable FPF signature in \cref{sec:bsm_llp_nonminmal}. Further ways in which new physics can indirectly be detected at the FPF are described in other sections. Taken together, the suite of possible models of light new physics will be exceedingly well-tested at the FPF.

\section{Monte Carlo Tools for BSM: FORESEE}
\label{sec:foresee}

An important part of the successful proposal of each new physics program is the development of convenient numerical tools that can be used in phenomenological simulations and can support related experimental efforts. To facilitate such BSM studies in the far-forward region of the LHC, we introduce a numerical package, namely the \textbf{FOR}ward \textbf{E}xperiment \textbf{SE}nsitivity \textbf{E}stimator, or \texttool{FORESEE}~\cite{Kling:2021fwx}.\footnote{The package is available at \href{https://github.com/KlingFelix/FORESEE}{ https://github.com/KlingFelix/FORESEE}. Detailed instructions of how to run the package are provided in tutorial jupyter notebooks therein.} The package can be used to analyze the expected sensitivity reach for new physics models predicting the existence of unstable LLPs, light DM, and mCPs in various experiments in the FPF and beyond, including the far-forward region of the future, high-energy hadron colliders. 

The main features of \texttool{FORESEE} can be divided into two categories: $1)$ it can be used as a standalone simulation tool to produce sensitivity reach plots for a growing list of popular BSM scenarios, $2)$ it provides the user with a set of useful numerical data for separate simulations of far-forward physics. In the former case, the package allows for a flexible definition of the detector geometries and cuts used in the analysis, as well as for modifying the LLP properties under study. These include, i.a., the relative importance of different LLP production and detection modes. On the other hand, the package provides a number of far-forward spectra of light mesons and other SM species that can straightforwardly be employed in independent simulations.

\begin{figure*}[t]
\centering
\includegraphics[width=0.510\textwidth]{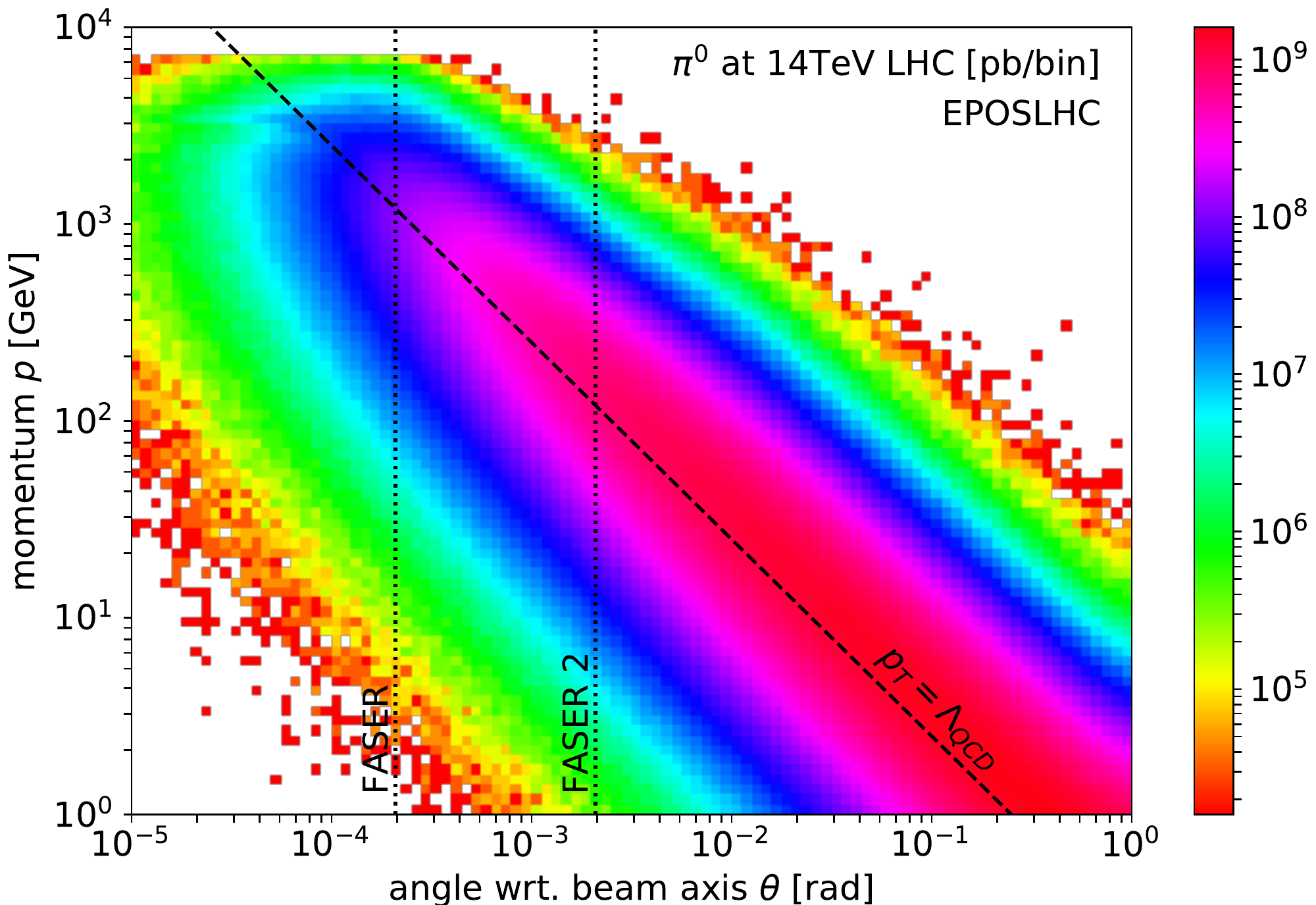}
\hfill
\includegraphics[width=0.477\textwidth]{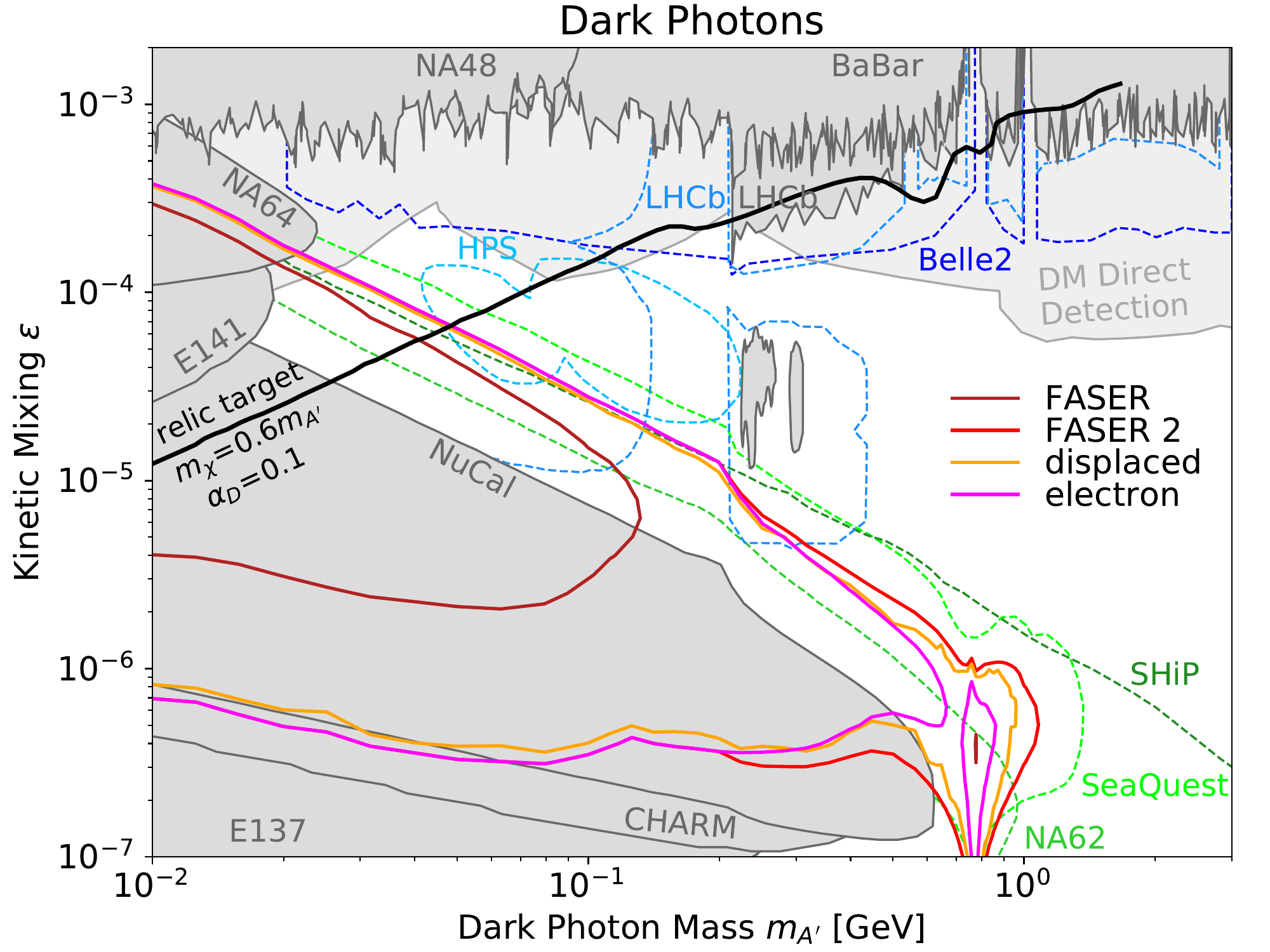}
\caption{Left: Distribution of $\pi^0$ mesons in the forward hemisphere for $14~\tev$ collision energy at the LHC obtained with \texttool{EPOS-LHC}~\cite{Pierog:2013ria}. The distribution is shown in the ($\theta$,$p$) plane, where $\theta$ is the meson’s angle with respect to the beam axis and $p$ is its momentum. The characteristic transverse momentum scale $p_T\sim\Lambda_{\textrm{QCD}}\sim 250~\mev$ is indicated with the diagonal black dashed line. The angular acceptances for the FASER detector to take data during LHC Run 3 and for the proposed FASER2 experiment in the FPF are highlighted by the vertical black dotted lines. Right: Dark photon sensitivity reach lines in $(m_{A^\prime},\epsilon)$ plane obtained with the \texttool{FORESEE} package for the FASER (brown solid lines) and FASER2 (red) detectors. We also show the expected FASER2 sensitivity to only $A^\prime\to e^+e^-$ decay channel (magenta) and to all the possible visible decays happening in a detector displaced by a $1~\m$ distance off the beam collision axis (orange). The dark gray-shaded regions correspond to previous bounds, while colorful dashed lines represent future sensitivity for selected other searches, as indicated in the plot. We also show current constraints on the dark photon parameter space from dark matter direct detection searches (light gray-shaded region). Here, we assume complex scalar DM $\chi$, the fixed mass ratio $m_\chi/m_{A^\prime}=0.6$, and the dark coupling constant equal to $\alpha_D=0.1$. Taken from Ref.~\cite{Kling:2021fwx}.}
\label{fig:foresee}
\end{figure*}

Light new physics particles traveling towards the FPF can be produced in the $pp$ collisions at the LHC due to at least several different production mechanisms. One such possible efficient production mode is due to rare decays of light mesons. Notably, one expects e.g. about $\mathcal{O}(10^{18})$ neutral pions to be produced during the HL-LHC era, while this number is one to two orders of magnitude lower for heavier mesons, depending on their mass. The far-forward meson spectra can be predicted using dedicated hadronic interaction models that have been greatly improved in recent years, cf. Ref.~\cite{Akiba:2016ofq} for review of related efforts for the LHC. In the package, an extensive set of such spectra for both mesons and other SM species is provided. These are listed in \cref{tab:spectra} along with the respective Monte Carlo (MC) generator tools used to obtain them. For the lighter mesons and far-forward photons, the user has the freedom to choose between different MC tools employed in the analysis (\texttool{EPOS-LHC}~\cite{Pierog:2013ria}, \texttool{QGSJet II-04}~\cite{Ostapchenko:2010vb}, \texttool{SIBYLL~2.3c}~\cite{Ahn:2009wx, Riehn:2015oba}). We follow their implementation in the \texttool{CRMC} package~\cite{CRMC}. The heavier species have been modeled with \texttool{Pythia 8.2}~\cite{Sjostrand:2006za, Sjostrand:2014zea} with the Monash tune~\cite{Skands:2014pea}. The spectra of $J/\psi$, $\psi(2S)$, and $\Upsilon(1,2,3S)$ mesons are provided for 14 TeV LHC following the discussion in Ref.~\cite{Foroughi-Abari:2020qar}.

In the left panel of \cref{fig:foresee}, we present exemplary such spectrum of neutral pions in the $(\theta,p)$ plane, where $p$ is the meson momentum and $\theta$ is its angle with respect to the beam collision axis. The spectrum peaks along the $p_{T}\sim m_\pi\sim \Lambda_{\textrm{QCD}}$ line. As a result, the most energetic mesons are produced with $\theta\sim p_T/p\ll 1$, i.e., along the beam collision axis. The meson spectra files can be found in the \texttt{files/hadrons} directory in the package. The spectrum is given in tables with the first two columns corresponding to the bin position in ($\log_{10}{\theta}$, $\log_{10}{(p/\gev)}$) variables and the third column providing the weights of each of the bins in units of pb/bin. To this end,  the angle $\theta$ is given in radians and the values of the production cross sections in the bins correspond to the forward hemisphere only. The weights can be multiplied by the relevant integrated luminosity (e.g., $3~\iab$ for the HL-LHC phase) to obtain the number of mesons in each of the bins. The filenames corresponding to different spectra are the meson ID in the MC particle numbering scheme~\cite{ParticleDataGroup:2020ssz}. 

\begin{table}[t]
\setlength{\tabcolsep}{1pt}
\centering
    \begin{tabular}{c|c||c|c|c|c}
      \hline\hline
     \multirow{2}{*}{Particle category} & \multirow{2}{*}{Particles} & \multicolumn{4}{c}{Generators}\\
    \cline{3-6}
     & & EPOS-LHC & QGSJET II-04 & SIBYLL 2.3c & \texttool{Pythia~8.2}\\
     \hline
 Photons & $\gamma$ & \checkmark & \checkmark & \checkmark & \\
 \multirow{2}{*}{Light hadrons} & $\pi^0$, $\pi^+$, $\eta$, $\eta^\prime$, $\omega$, $\rho$, $\phi$, $n$, $p$ & \multirow{2}{*}{\checkmark} & \multirow{2}{*}{\checkmark} & \multirow{2}{*}{\checkmark} & \multirow{2}{*}{} \\
 & $K^+$, $K_L$, $K_S$, $K_0^*$, $K^{*+}$, $\Lambda$& & & & \\
 Charm hadrons & $D^+$, $D^0$, $D_s^+$, $\Lambda_c$ &  &  & \checkmark & \checkmark \\
 Beauty hadrons & $B^0$, $B^+$, $B_s$, $B_c^+$, $\Lambda_b$ &  &  &  & \checkmark \\
 Heavy quarks & $c$, $b$ &  &  &  & \checkmark \\
 Quarkonia & $J/\Psi$, $\psi(2S)$, $\Upsilon(nS)$ &  &  &  & \checkmark \\
 Weak bosons & $W^+$, $Z$, $h$ &  &  &  & \checkmark \\
 \hline \hline
 \end{tabular}
 \caption{Spectra of Standard Model particles that are available in \texttool{FORESEE}~\cite{Kling:2021fwx} and the relevant Monte Carlo simulation tools employed in the analysis (see the text for references). The spectra can be found in the package as text files stored in the \texttt{files/hadrons} directory.}
 \label{tab:spectra}
\end{table}

In the package, the aforementioned spectra are used to generate far-forward flux and spectra of the LLPs produced in the parent meson decays. This is then used to study the sensitivity reach of far-forward experiments at the LHC. To initialize such simulation, the user needs to define the model by specifying \textsl{i)} the LLP production rates, \textsl{ii)} their lifetimes, and, optionally, \textsl{iii)} the LLP decay branching fractions. The last information is needed if only some of the decay final states can be successfully searched for in the detector. In the case of 3-body decays, $p_0 \to p_1 p_2 p_3$, with $p_3$ being the LLP, the user needs to additionally provide the differential branching fraction $d\text{BR}/(dq^2 \ d\cos\vartheta)$. Here, $q^2=(p_2+p_3)^2$ and $\vartheta$ is the angle between $p_3$ in the rest frame of $p_2+p_3$, and the direction of $p_2+p_3$ in the rest frame of $p_0$. The production of LLPs can also be due to rare decays of long-lived mesons (charged pions, charged/neutral kaons). In this case, \texttool{FORESEE} takes into account the relevant details of the far-forward LHC infrastructure and, conservatively, neglects such decays happening past the inner triplet quadrupole absorber TAS at $z=20$m.

On top of the aforementioned rare decays of SM particles, \texttool{FORESEE} currently also supports the LLP production due to mixing with SM species and direct production in scattering processes. For the former, the mixing of dark photons with the SM vector bosons is modeled following Ref.~\cite{Berlin:2018jbm}, i.e., by assuming that the LLP and SM production rates can be related via $\sigma(LLP) = \kappa^2 \times \sigma(SM)$, where $\kappa$ describes the mixing parameter which must be provided by the user. The direct dark photon production via Bremsstrahlung or the Drell-Yan process is simulated following Refs.~\cite{Feng:2017uoz,Berlin:2018jbm}. Direct production can also be taken into account for new user-defined models. This can be done by providing the full two-dimensional LLP spectra for different LLP masses is in the previously discussed format. 

When running the package, one can also define the LLP lifetime, $c\tau$, and decay branching fractions and provide them in a tabular form (text file). The lifetime can be given in a one-dimensional table for the fixed value of the coupling constant $g_*$ as a function of the LLP mass $m$. During the simulation then the lifetime is evaluated for different values of the coupling constant $g$ using $c\tau(m,g)=c\tau(m,g_*)g_*^{2}/g^2$. It is also possible to use two-dimensional parameterization. In this case, $c\tau(m,g)$ is provided as a function of both $m$ and $g$. The same options are available for the user-defined branching fractions. 

At the beginning of the simulation, \texttool{FORESEE} generates the LLP spectra in terms of $(\theta_{LLP},p_{LLP})$. The user can then specify further details relevant for the sensitivity reach estimation. These include the \textsl{i)} distance $L$ between the IP and experiment, \textsl{ii)} acceptance in terms of the LLPs momentum and position, \textsl{iii)} luminosity, \textsl{iv)} LLP production channels employed, and \textsl{v)} allowed LLP decay channels. \texttool{FORESEE} counts the number of signal events after applying the selection criteria and the reach plots are automatically generated. The final sensitivity curve is plotted for the number of BSM signal events defined by the user, which, in turn, should depend on the expected background level for a given search in the considered experiment. 

Further detailed instructions on how to use the code are provided in the package in tutorial jupyter notebooks. The package is planned to be developed in the future to add more popular LLP models, the relevant production and decay modes, as well as to explore different signatures of new physics. Currently, on top of the vanilla dark photon and dark Higgs boson scenarios, \texttool{FORESEE} also allows one to study selected other models predicting the existence of unstable LLPs, as well as the scattering signature of light mCPs produced in the far-forward region of the LHC.

The example sensitivity plot for the FASER and FASER2 experiments obtained with \texttool{FORESEE} is shown in the right panel of \cref{fig:foresee}. This has been obtained for the vanilla dark photon model characterized by the two-dimensional parameter space spanned by the dark photon mass $m_{A^\prime}$ and kinetic mixing $\epsilon$ parameters. The plot corresponds to the search for highly displaced visible decays of $A^\prime$s, mainly into pairs of oppositely-charged SM particles. Further details of the relevant modeling for this scenario can be found in Refs.~\cite{Feng:2017uoz, FASER:2018eoc}. The past bounds are shown with a gray-shaded region. These correspond to the BaBar~\cite{Lees:2014xha}, CHARM (following Ref.~\cite{Gninenko:2012eq}), E137~\cite{Bjorken:1988as}, E141~\cite{Riordan:1987aw},  LHCb~\cite{Aaij:2019bvg}, NA48/2~\cite{Batley:2015lha}, NA64~\cite{Banerjee:2018vgk}, and NuCal~\cite{Blumlein:1990ay} experiments. We similarly show complementary future sensitivity reach lines for the Belle-II~\cite{Kou:2018nap}, HPS~\cite{Battaglieri:2017aum, Solt:2020zbi}, LHCb~\cite{Ilten:2015hya, Ilten:2016tkc}, NA62~\cite{Dobrich:2018ezn}, SeaQuest~\cite{Berlin:2018pwi}, and SHiP~\cite{SHiP:2020vbd} detectors. 

To better illustrate the capabilities of \texttool{FORESEE}, we also show in the plot the expected reach of FASER2 for only the simplest $A^\prime\to e^+e^-$ decay channel. The focus on only the electron-positron pairs has no impact on the reach for the dark photon mass below the di-muon threshold, while it moderately limits the sensitivity for larger $m_{A^\prime}$. Last but not least, in the plot we also present the mild change in the sensitivity reach of a FASER2-like detector with the radius $R=1~\m$, assuming that it has been shifted by $1~\m$ off the beam axis. We note that for larger displacements, the expected sensitivity would become degraded more significantly, see Ref.~\cite{FASER:2018eoc}.

The black solid line in the plot corresponds to the thermal relic target of the light complex scalar DM coupled to the SM via the dark photon mediator. We assume here the typical benchmark value for the coupling constant between the two dark species $\alpha_D=0.1$ and the fixed mass ratio between them $m_\chi = 0.6\,m_{A^\prime}$. This choice allows one to consider visible decay signatures of $A^\prime$, since the $A^\prime$ decay into two dark matter species is kinematically forbidden in this case. In addition, by keeping the DM mass below the mediator mass, we restrict the annihilation process relevant for the relic target. In particular, the exceptionally effective secluded annihilation into two dark vectors, $\chi\chi^\ast\to A^\prime A^\prime$, is also forbidden for this benchmark.

As can be seen in the plot, both FASER and FASER2 detectors will probe an important part of the allowed region of the parameter space, in which the correct value of the DM relic density is predicted and $\chi$ DM is not thermally overproduced in the early Universe. Importantly, this remains complementary to traditional DM direct detection (DD) searches in underground detectors. In the plot, we present such current bounds with a light-gray shaded region. These correspond to null searches for DM scatterings with electrons and nuclei in the Xenon10~\cite{Angle:2011th} and Xenon1T~\cite{Aprile:2019xxb, Aprile:2019jmx} detectors. We implement the former following Ref.~\cite{Essig:2017kqs}. For the latter, the bounds are based on interactions that employ the Migdal effect~\cite{Ibe:2017yqa}. Importantly, the DM DD bounds can become weaker, e.g., in the inelastic DM scenario, without affecting the prospects for light vector searches in the FPF.

The rich BSM experimental program of the FPF benefits a lot from dedicated phenomenological and theoretical analyses of new physics models relevant for this type of search. To support these efforts, we have introduced a numerical package \textbf{FOR}ward \textbf{E}xperiment \textbf{SE}nsitivity \textbf{E}stimator, or \texttool{FORESEE}, that can be used to estimate the sensitivity reach of the far-forward experiments at the LHC and beyond, as well as it provides the user with a list of validated far-forward spectra of light mesons and many other SM species. These can be employed in separate studies. Further development of the package is planned in the future to extend its modeling capabilities to more light new physics models and additional experimental features.

\section{Long-Lived Vector Particles}
\label{sec:bsm_llp_vec}

The underlying principle of all SM interactions is the concept of symmetries. Of the three renormalisable portal interactions---the Higgs portal, neutrino portal and vector portal---only the vector portal is fundamentally based on a new symmetry. It requires a novel $U(1)_X$ symmetry, allowing for a kinetic mixing term with the SM $U(1)$ gauge fields~\cite{Okun:1982xi, Holdom:1985ag} or a coupling to the corresponding currents. The most general Lagrangian of an extra $U(1)_X$ symmetry can be written as
\begin{align}\label{eq:u1lag}
    \mathcal{L} = \mathcal{L}_\mathrm{SM} &- \frac{1}{4} X_{\alpha\beta} X^{\alpha\beta} - \frac{\epsilon_Y}{2}  B_{\alpha\beta} X^{\alpha\beta} - g_x \, j^X_\alpha \, X^\alpha - \frac{M_{X}^2}{2} X_\alpha X^\alpha \, ,
\end{align}
where $X_\mu$ denotes the new $U(1)_X$ gauge boson, $B_{\mu\nu}$ and $X_{\mu\nu}$ the hypercharge and $U(1)_X$ field strengths, $\epsilon_Y$ the kinetic mixing parameter and $g_x$ and $j^X$ the $U(1)_X$ gauge coupling and current.

In this section, we will consider several examples of light vector particles and discuss the sensitivity of the FPF experiments to search for them. This includes the dark photon discussed in \cref{sec:llp_vec_dp}, the gauged $B-L$ gauge boson discussed in \cref{sec:llp_vec_bl}, the gauged $L_i - L_j$ groups discussed in \cref{sec:llp_vec_lilj}, the gauged $B-3L_i$ groups discussed in \cref{sec:llp_vec_b3l}, and the gauged $B$ group in \cref{sec:llp_vec_b}. We then present improved estimates of the production of light vector particles via proton bremsstrahlung in \cref{sec:llp_vec_brem} and via hadronization and radiation in \cref{sec:llp_vec_prod}, as well as their decays in \cref{sec:llp_vec_decay}. 

\subsection{Dark Photon} \label{sec:llp_vec_dp}

\begin{figure*}[t]
\centering
    \includegraphics[width=0.49\textwidth]{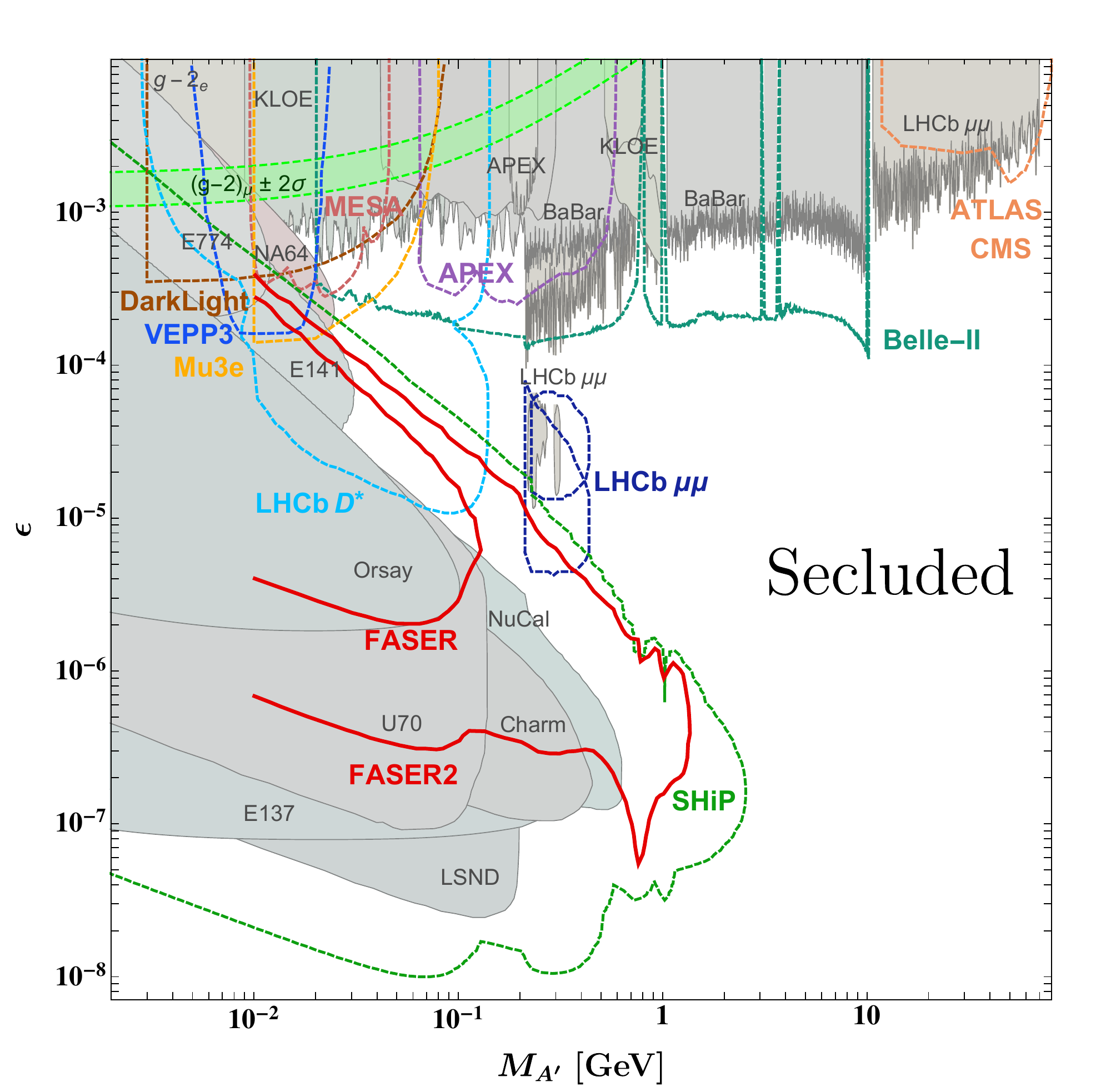}
    \includegraphics[width=0.49\textwidth]{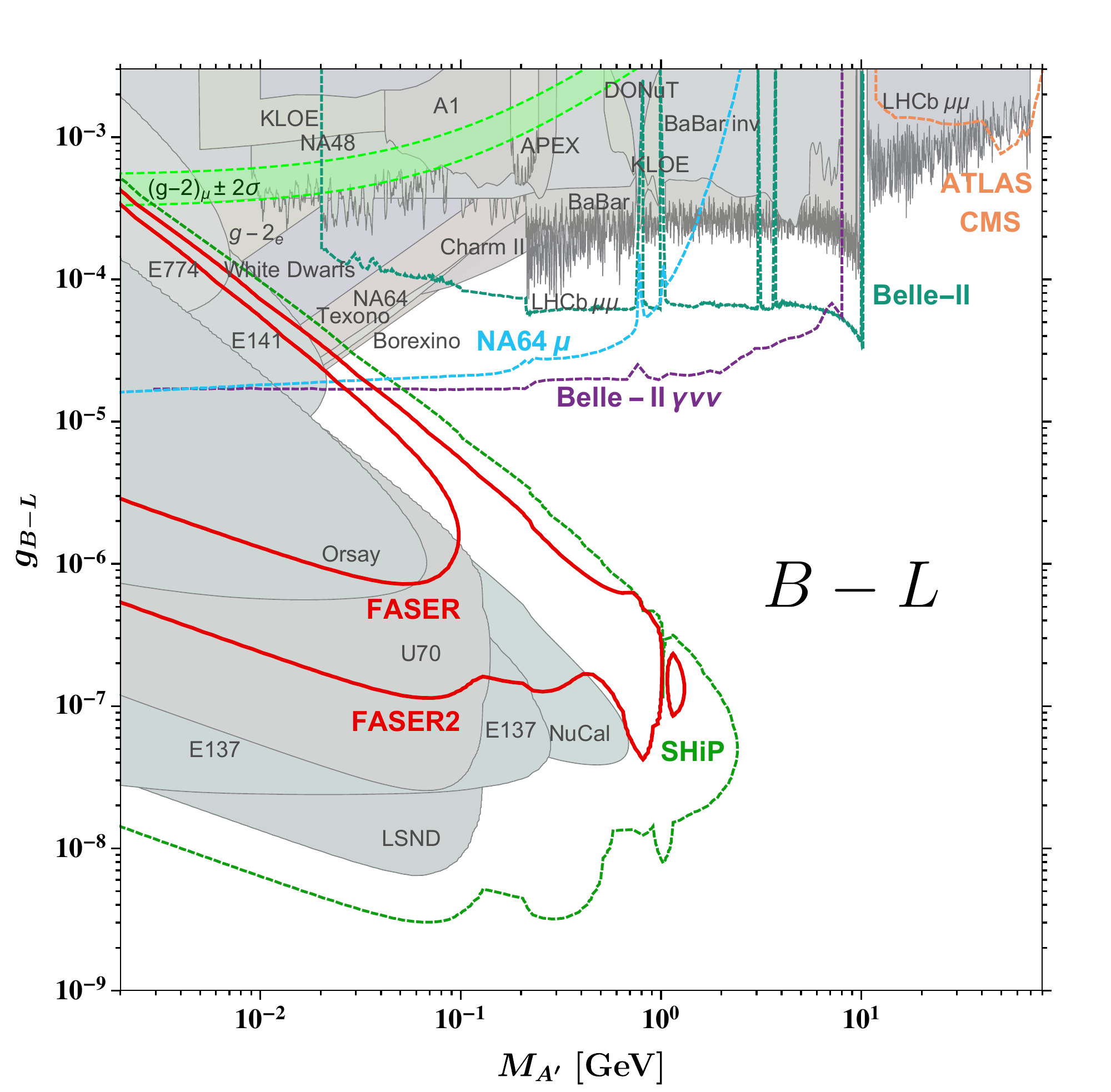}
\caption{Sensitivity reaches of FASER and FASER2 (solid red lines) to the minimal secluded hidden photon (left) and a $U(1)_{B-L}$ gauge boson (right) in the coupling-mass plane. Current existing constraints are shown as the grey shaded regions alongside projections of other future experiments and measurements shown as coloured dashed lines. See text for explanations.}
\label{fig:hp_bl}
\end{figure*}

The most minimal realisation of a novel $U(1)_X$ symmetry is one where the SM remains uncharged under the new symmetry and there are no new fermions present. This translates to $j^X_\alpha = 0$ in \cref{eq:u1lag}. Hence, the dark photon $A'$ is completely \textit{secluded} from the SM and the only interaction of the dark photon $A'$ with SM fermions proceeds through kinetic mixing (see e.g.~\cite{Essig:2013lka, Raggi:2015yfk, Fabbrichesi:2020wbt} for reviews of the secluded dark photon). The mixed kinetic terms in \cref{eq:u1lag} can be removed by a field redefinition of the neutral gauge bosons $(B_\mu$, $W^3_\mu$, $X_\mu)$. Two consecutive orthogonal rotations diagonalise the neutral gauge boson mass terms. The full transformation translating to the mass basis $(A_\mu, Z_\mu, A'_\mu)$ then reads
\begin{equation}
    \left(\begin{matrix}
     B_\mu \\
     W^3_\mu \\
     X_\mu
    \end{matrix}\right)  =
    \underbrace{
    \left(\begin{matrix}
    1 & 0 & - \frac{\epsilon_Y}{\sqrt{1-\epsilon_Y^2}} \\
    0 & 1 & 0 \\
    0 & 0 & \frac{1}{\sqrt{1-\epsilon_Y^2}}
    \end{matrix}\right)}_{G( \epsilon_Y)}
    \underbrace{
    \left(\begin{matrix}
    \cos\theta_W & -\sin\theta_W & 0 \\
    \sin\theta_W & \cos\theta_W & 0 \\
    0 & 0 & 1
    \end{matrix}\right)}_{R_1(\theta_W)}
    \underbrace{
    \left(\begin{matrix}
    1 & 0 & 0 \\
    0 & \cos\xi & -\sin\xi \\
    0 & \sin\xi & \cos\xi
    \end{matrix}\right)}_{R_2(\xi)}
    \left(\begin{matrix}
    A_\mu \\
    Z_\mu \\
    A'_\mu
    \end{matrix}\right),
\end{equation}
where $A_\mu$ denotes the SM photon, $Z_\mu$ the weak neutral gauge boson and $A'_\mu$ the mass eigenstate of the new $U(1)_X$ bosons, which we refer to as \textit{hidden} or \textit{dark photon}. Finally, couplings of the mass eigenstates of the  neutral bosons $(A_\mu, Z_\mu, A'_\mu)$ to the SM fermions are described by the interaction term of the kind,
\begin{equation}\label{eq:hp_ints}
    \mathcal{L}_\mathrm{int} = - (e\, j^\mathrm{em}_\alpha, g_Z \, j^Z_\alpha, g_x \, j^X_\alpha) \ K \ \left(\begin{matrix}
    A^\alpha \\
    Z^\alpha \\
    A'^\alpha
    \end{matrix}\right) \,,
\end{equation}
where the coupling matrix is given by~\cite{Bauer:2018onh}
\begin{align}
    K=\left[ R_2(\xi)R_1(\theta_W)G^{-1}(\epsilon_Y)R_1(\theta_W)^{-1}\right]^{-1}
    &\approx \begin{pmatrix}
    1 & 0 & -\epsilon \phantom{e}\\
    0 & 1& 0 \\
    0 & \epsilon \tan\theta_W&  1
    \end{pmatrix} \,.
\label{eq:KK}
\end{align}
Here, $\theta_W$ denotes the weak mixing angle, $e$ the electromagnetic coupling constant, the $Z$-coupling constant $g_Z=e/(\sin\theta_W \cos\theta_W)$ and $j^\mathrm{em}_\mu$ and $j^Z_\mu$ are the electromagnetic and weak neutral currents. Furthermore, we have defined the physical kinetic mixing parameter as $\epsilon=\epsilon_Y \,\cos\theta_W$.

In the domain of low-energy observables, it is sufficient to consider mixing of the new gauge boson with the photon of QED. The interaction of the mass eigenstate dark photon with the SM sector is then parametrised by~\cref{eq:hp_ints} and reads explicitly
\begin{equation}\label{eq:dp_int}
    \mathcal{L}_\mathrm{secl} =  \epsilon\,e\, j^\mathrm{em}_\mu
    A'^\mu\,.
\end{equation}
This interaction naturally suggest the name of \textit{hidden} or \textit{dark photon} for the new boson $A'$ since it couples to the  electromagnetic current $j^\mathrm{em}_\mu$ exactly analogous to the SM photon, but suppressed by the kinetic mixing parameter $\epsilon$.

Models of purely kinetically-mixed dark photons received a lot of attention in the literature when it was shown that such GeV-scale new mediators can help to explain the cosmic ray positron excess~\cite{ArkaniHamed:2008qn, ArkaniHamed:2008qp}. Especially, after it had been established that such models naturally arise from weak-scale SUSY breaking~\cite{Dienes:1996zr, Hooper:2008im, Baumgart:2009tn, Cheung:2009qd, Katz:2009qq, Morrissey:2009ur}, the secluded dark photon model rejoiced from an increased popularity in the literature~\cite{Pospelov:2008zw, Batell:2009yf, Essig:2009nc, Reece:2009un, Bjorken:2009mm}.

A consequence of the kinetic mixing interaction in~\cref{eq:dp_int} is that the dark photon can be produced and searched for in a number of electromagnetic processes where a SM photon is replaced by a dark photon $A'$. Natural candidates to search for dark photons are $e^+e^-$ and hadron colliders, where dark photons can be abundantly produced in Bremsstrahlung processes. For example, in $e^+e^-$ machines the dark photon can be produced  in radiative return ($e^+e^- \to \gamma A'$)~\cite{Essig:2009nc} or via heavy meson decays and searched for in prompt dilepeton or pion decays. These search strategies have been employed in a number of $e^+e^-$ collider experiments in the past like e.g.~BaBar \cite{Aubert:2009cp, Lees:2014xha}, Belle \cite{Abe:2010gxa,Inguglia:2016acz} and KLOE \cite{Archilli:2011zc, Babusci:2012cr, Anastasi:2016ktq, Anastasi:2015qla}. At hadron colliders like the LHC the dark photon can also be produced via Drell-Yan production ($q\bar q \to A'$) or via the decay of heavy resonances, like e.g.~$H \to Z A'$ \cite{Curtin:2014cca}, or heavy meson decays, like e.g~$D^*\to DA'$~\cite{Ilten:2015hya}. Searches for prompt decays of such produced dark photons have been conducted at ATLAS and CMS~\cite{Curtin:2014cca} and at LHCb~\cite{Aaij:2017rft, Ilten:2015hya, Ilten:2016tkc,LHCb:2019vmc}. In the left panel of~\cref{fig:hp_bl} these existing collider constraints are shown  as grey regions in the (kinetic mixing, mass) plane for a secluded dark photon. As can be seen these experiments cover the region of $\epsilon\gtrsim 10^{-3}$ over a very large range of dark photon masses.

Nevertheless, the aforementioned collider experiments lose sensitivity for light ($M_{A'}\lesssim1 $ GeV) and very weakly coupled dark photons. This region, however, can be probed by searching for displaced decays of dark photons. Abundant sources of such light and long-lived dark photons are provided by electron or proton beam dump and fixed target experiments. In these experiments beams of charged particles are dumped onto a block of material where a large number of dark photons can be produced from Bremsstrahlung or secondary meson decays. These dark photons can then be searched for at a macroscopic displacement from the target material. Such searches have been conducted in the past at electron beam dump experiments like SLAC E137 and E141~\cite{Bjorken:2009mm,Bjorken:1988as, Riordan:1987aw, Andreas:2012mt}, Fermilab E774 \cite{Bross:1989mp}, Orsay \cite{Davier:1989wz}, electron fixed target experiments like APEX~\cite{Abrahamyan:2011gv}, A1/MAMI~\cite{Merkel:2014avp,Merkel:2011ze}, HPS~\cite{Battaglieri:2014hga}, NA64~\cite{Banerjee:2016tad,NA64:2019auh}, as well as at proton beam dumps like CHARM~\cite{Bergsma:1985is}, LSND~\cite{Athanassopoulos:1997er} and U70/Nu-Cal~\cite{Blumlein:2011mv,Blumlein:2013cua}, and proton fixed-target experiments, such as SINDRUM I~\cite{MeijerDrees:1992kd} and NA48/2~\cite{Batley:2015lha}. The resulting constraints are also shown in the  of~\cref{fig:hp_bl} as grey regions. 

At the future Forward Physics Facility, experiments like FASER2 will be able to search for such long-lived dark photons in the very forward direction of LHC proton-proton interactions. As for beam dump and fixed target experiments, dark photons can be produced via proton Bremsstrahlung in the $pp$ collisions at LHC. The total number of dark photons produced can be computed from~\cite{Feng:2017uoz}
\begin{multline}\label{eq:PbeamCS}
    N=N_p\ |F_1(M_{A'}^2)|^2\int_{M_{A'}}^{E_p-m_p} dE_{A'} \frac{1}{E_{p}}\frac{\sigma_{pp}(2m_p(E_p-E_{A'}))}{\sigma_{pp}(2m_pE_p)} \int_0^{p_{\perp,\text{max}}^2}\omega_{A'p}(p_\perp^2) dp_\perp^2 \\
    \times \Theta(\Lambda_\mathrm{QCD}^2 - q^2) \, A_\mathrm{geom} \, \mathcal{P}_{A'}\,,
\end{multline}
where $N_p, E_p, m_p$ denote the total number of proton collisions, the proton energy and mass, $\sigma_{pp}$ denotes the proton-proton cross section, $F_1(p_{A'}^2)$ denotes the time-like form factor of the proton from the vector meson dominance model~\cite{Faessler:2009tn, deNiverville:2016rqh}, $A_\mathrm{geom}$ is a geometrical acceptance factor, $\omega_{A'p} (p_\perp^2)$ is a weighting function relating the $2\to3$ process $p+p\to p+p+A'$ to $p+p\to p+p$ and $\mathcal{P}_{A'} =  e^{-{L_\text{min}}/{\ell_{A'}}} - e^{-{L_\text{max}}/{\ell_{A'}}}$ is the probability that the dark photon decays within the fiducial detector volume. Details on these expressions can be found in Refs.~\cite{Blumlein:2013cua, Feng:2017uoz}.

Besides Bremsstrahlung, another important source of dark photon is provided for by secondary decays of mesons produced in the proton-proton collisions. The number of expected dark photons from meson decays can be expressed as \cite{Blumlein:2011mv}
\begin{align}
    N=\frac{N_p}{\sigma(pp\to X)}\, \int_{0}^1dx_F \int_{0}^{p_{\perp, \text{max}}^2}\,dp_\perp^2&\,\frac{d\sigma(p p\to M X) }{dx_F dp_\perp^2} \text{Br}(M \to A' \gamma)\, A_\mathrm{geom} \, \mathcal{P}_{A'}\,,
\end{align}
with $x_F$ denoting the Feynman-x variable and $d\sigma(p p\to M X) /(dx_F dp_\perp^2)$ being the differential meson production cross section for the meson $M$. Here $\text{Br}(M \to A' \gamma)$ denotes the branching ratio of the meson $M$ decaying into a photon and a dark photon, which in the case of a secluded dark photon is simply given by 
\begin{align}\label{eq:BRpi}
    \text{Br}(M\to A'\gamma)=2 \epsilon^2  \bigg(1-\frac{M_{A'}^2}{M_{M}^2}\bigg)^3 \text{Br}(M\to \gamma\gamma)\,.
\end{align}
For dark photon production at LHC the most relevant meson decays are from $\pi^0, \eta$ and $\eta'$ mesons.  

In the left panel of~\cref{fig:hp_bl} the projected sensitivities of FASER and the future FASER2 experiment located at the FPF are shown by the red solid lines. For comparison we also show the projected sensitivity of the future CERN based proton fixed target experiment SHiP~\cite{SHiP:2015vad, Alekhin:2015byh} by a dashed green line. As can be seen FASER and the future FASER2 experiments will probe regions of dark photon parameter space that has not been tested at any other beam dump or fixed target experiment. In particular, FASER2 will have a significantly increased mass reach, comparable to what can be achieved by SHiP. 

\subsection{$B-L$ Gauge Boson} \label{sec:llp_vec_bl}

The peculiar flavour structure of the SM leads to the presence of the accidental global symmetries $U(1)_B$, $U(1)_{L_e}$, $U(1)_{L_\mu}$ and $U(1)_{L_\tau}$ in the SM Lagrangian. Under the minimal extension of the SM by three right-handed neutrino fields these global symmetries can be promoted to anomaly-free gauge symmetries by combining them into either a mixed baryo-leptophilic $(i)$ or a purely leptophilic $(ii)$  two-parameter $U(1)_X$ group with~\cite{Araki:2012ip}
\begin{align}
    (i)\quad  X &= B - x_e\, L_e- x_\mu\, L_\mu - (3-x_e-x_\mu)\, L_\tau \,, \notag\\
    (ii)\quad X &= y_e\, L_e + y_\mu\, L_\mu - (y_e+y_\mu)\, L_\tau\,,
\end{align}
where $x_e,x_\mu,y_e$ and $y_\mu$ are real numbers. Amongst these groups the combination with $x_e=x_\mu=1$ plays a special role, since it leads to the flavour-universal $U(1)_{B-L}$ group under which all generations of quarks carry the same charge, as do all generations of leptons.

In contrast to the secluded dark photon, the $B-L$ gauge boson additionally couples to the $B-L$ current according to~\cref{eq:hp_ints}, with
\begin{equation}
    j^{B-L}_\mu=  \frac{1}{3}\bar Q \gamma_\mu Q 
          + \frac{1}{3}\bar u_R\gamma_\mu u_R 
          + \frac{1}{3}\bar d_R\gamma_\mu d_R
          - \bar L \gamma_\mu L 
          - \bar \ell_R\gamma_\mu \ell_R  
          - \bar \nu_R\gamma_\mu \nu_R \,.
\end{equation}
As a consequence the $B-L$ boson couples to all SM fermions proportional to the coupling strength $g_{B-L}$. Hence, as long as the kinetic mixing parameter is smaller than the new gauge coupling, $\epsilon \ll g_{B-L}$, kinetic mixing effects are irrelevant for this model. The two major differences of the phenomenology of the $B-L$ boson compared to the secluded dark photon are the different charges of the quarks under the two models, and most importantly, the fact that neutrinos are charged under $U(1)_{B-L}$ coupling them to the associated gauge boson. 

The latter  has significant consequences for the employed search strategies. First of all, the $B-L$ boson can decay invisibly into pairs of neutrinos, making searches for invisible final states a sensitive probe for this model. In particular, mono-photon (or even mono-$\Phi$) searches at BaBar~\cite{Lees:2017lec}, NA64~\cite{NA64:2019imj} and, in the future, Belle-II provide stringent constraints to this model. Furthermore, a number of neutrino scattering experiments sensitive to extra neutrino interactions provide very stringent bounds on leptonically-coupled dark photons like the $B-L$ gauge boson. Most noticeably, the evaluation~\cite{Bilmis:2015lja, Lindner:2018kjo} of reactor neutrino data from Texono~\cite{Deniz:2009mu} and the neutrino beam experiment Charm-II~\cite{Vilain:1994qy, Vilain:1993kd}, as well as the recent re-evaluation~\cite{Amaral:2020tga} of solar neutrino scattering at Borexino~\cite{Bellini:2011rx, Agostini:2017ixy} are probing new regions of parameter space in the sensitivity gap between conventional prompt collider and displaced beam dump and fixed target searches for dark photons. This is shown in the right panel of~\cref{fig:hp_bl}.

Since the coupling structure of the $B-L$ gauge boson to charged leptons and hadrons is very similar to the secluded dark photon, the constraints from visible prompt collider and displaced beam dump and fixed target experiments are also quite similar. These limits have been derived for the case of $U(1)_{B-L}$ for example in Refs.~\cite{Ilten:2018crw, Bauer:2018onh} and are shown by the grey areas in the right panel of~\cref{fig:hp_bl}. Comparable to the secluded dark photon case, the LHC forward experiments FASER and in particular FASER2 have the potential to cover large untested regions of the $B-L$ parameter space complementary to future searches of prompt invisible decays at Belle-II and NA64$\mu$~\cite{Bauer:2018onh}. The major difference of the FASER(2) projections compared to the secluded dark photon case is a slightly reduced sensitivity to a $U(1)_{B-L}$ gauge boson, since this has now a sizeable invisible branching fraction into neutrinos.

\begin{figure*}[t]
  \centering
  \includegraphics[width=0.49\textwidth]{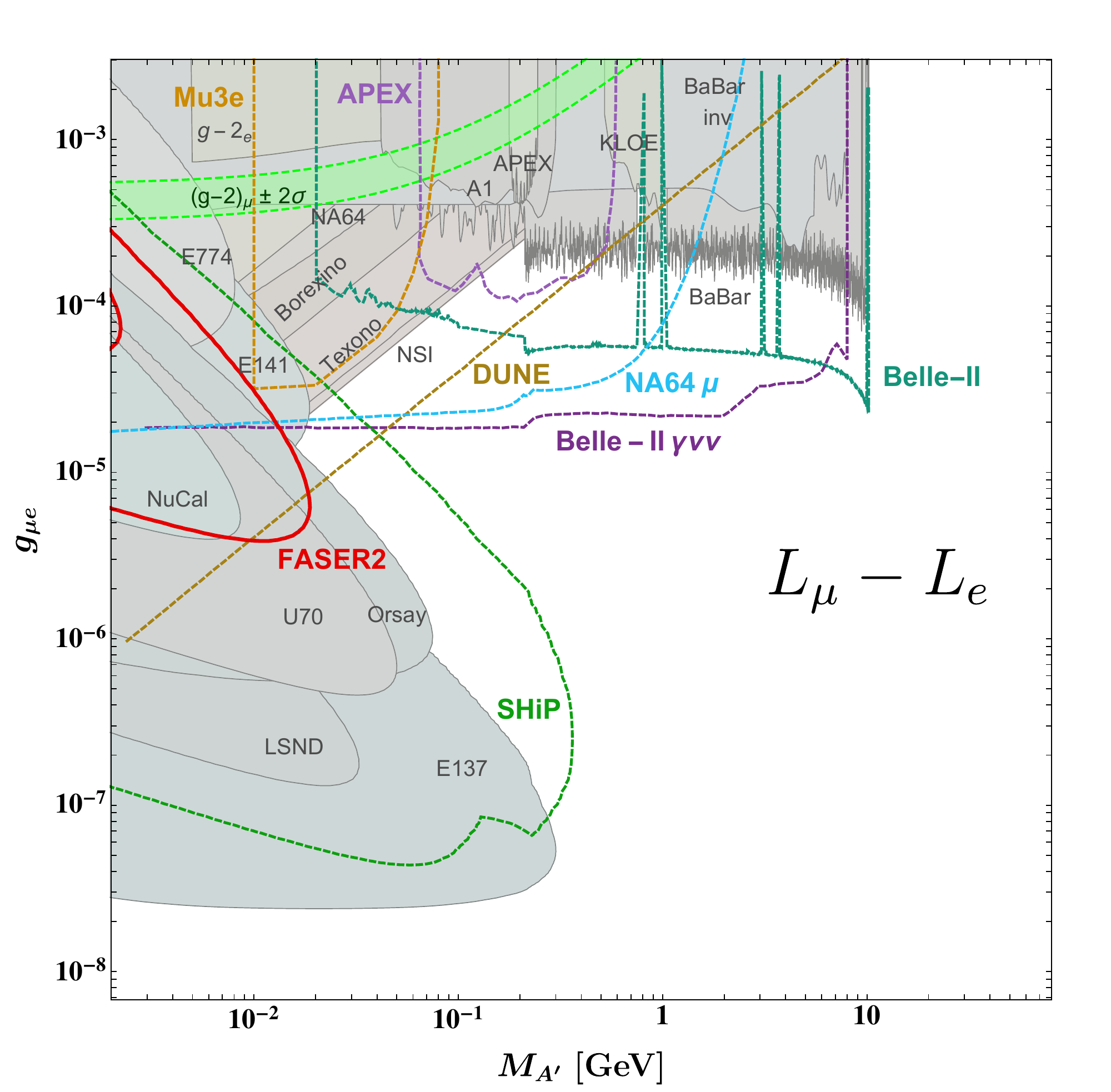}%
  \includegraphics[width=0.49\textwidth]{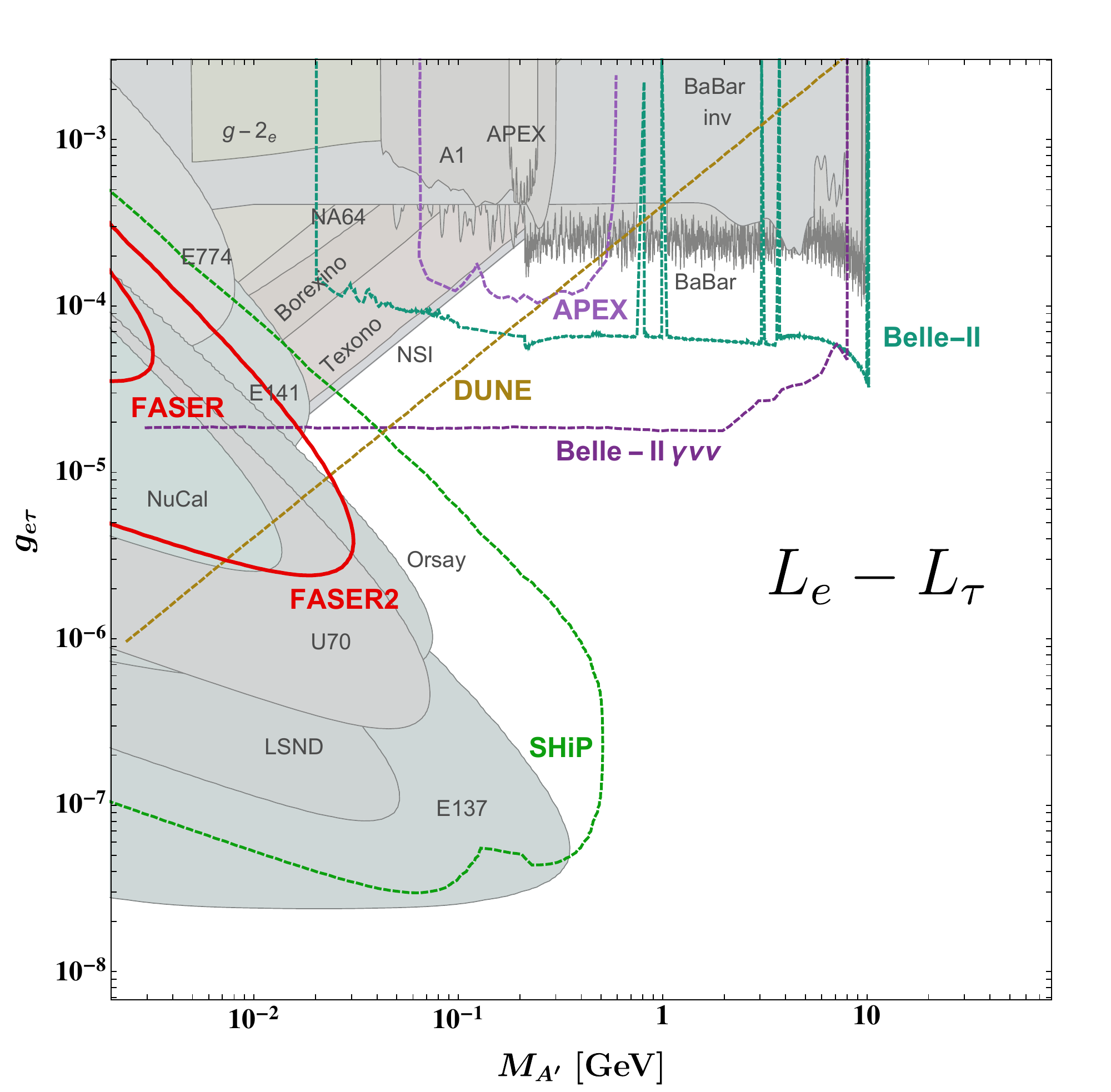}\\
  \includegraphics[width=0.49\textwidth]{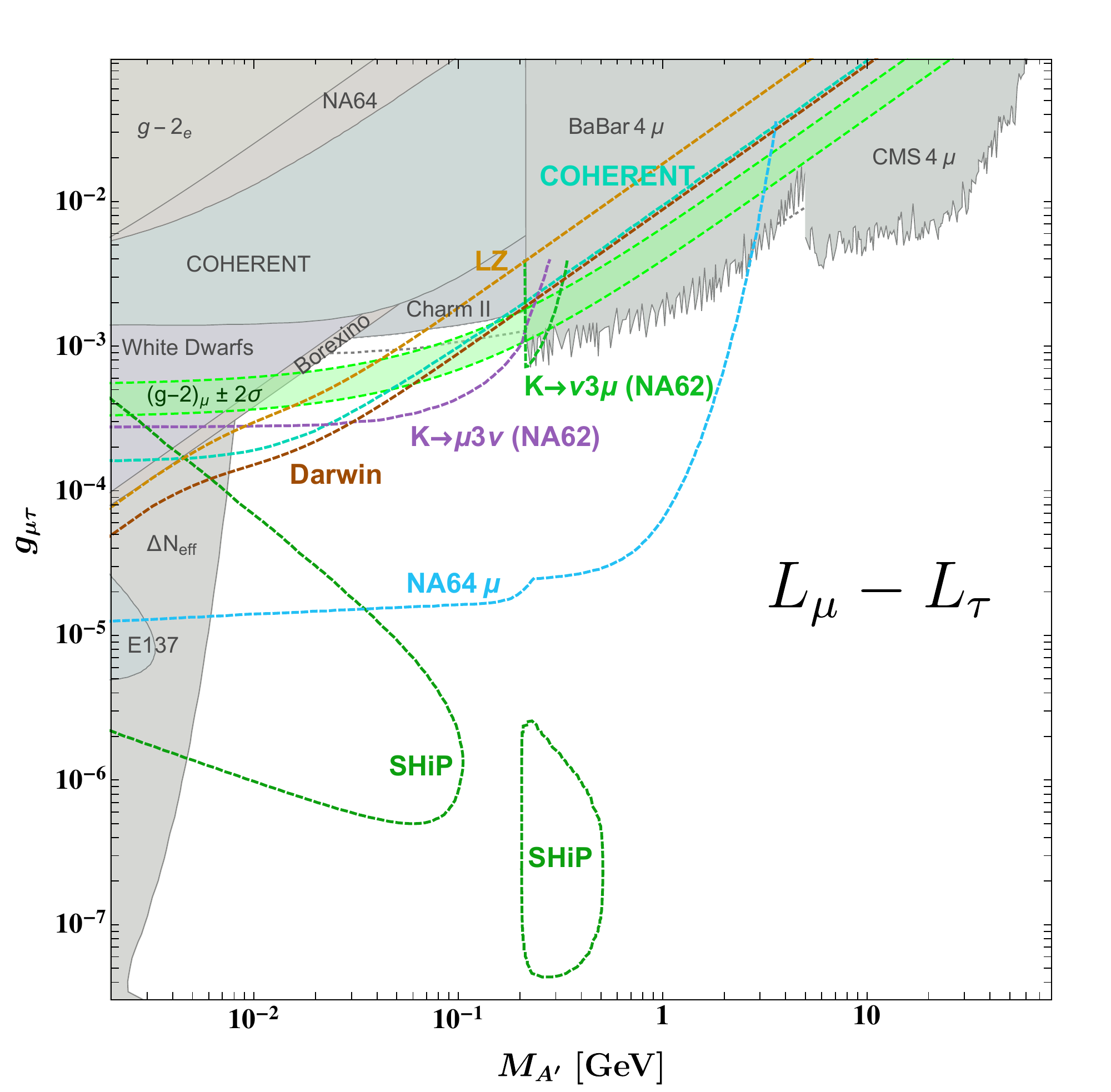}
  \caption{Current existing constraints shown as the grey shaded regions alongside projections of other future experiments and measurements shown as coloured dashed lines for a $U(1)_{L_\mu-L_e}$ (top left), $U(1)_{L_e-L_\tau}$ (top right) and $U(1)_{L_\mu-L_\tau}$ (bottom centre) gauge boson in the coupling-mass plane. Sensitivity reaches of FASER and FASER2 are shown by the solid red lines. See text for explanations.}
  \label{fig:li_lj}
\end{figure*}

\subsection{$L_i - L_j$ Gauge Bosons} \label{sec:llp_vec_lilj}

Among the purely leptophilic anomaly-free $U(1)$ extensions of the SM discussed in~\cref{sec:llp_vec_bl}, there are three special subgroups, which are anomaly-free even without the addition of right-handed neutrinos (if Majorana mass terms for the neutrinos are forbidden). These are the three groups $U(1)_{L_\mu-L_e}$, $U(1)_{L_e-L_\tau}$ and $U(1)_{L_\mu-L_\tau}$. Under these groups the associated gauge bosons couple according to~\cref{eq:hp_ints} to the gauge currents
\begin{align}
j^{\, i-j}_\mu&= \bar L_i \gamma_\mu L_i 
          + \bar \ell_i\gamma_\mu \ell_i 
          - \bar L_j \gamma_\mu L_j -\bar\ell_j\gamma_\mu \ell_j\,,
\label{eq:lilj_currents}
\end{align}
with $i\neq j=e, \mu, \tau$. The lack of any gauge interactions with hadrons sets these groups quite apart from the previously discussed secluded dark photon and $B-L$ gauge boson. 

However, the fact that part of the SM leptons are charged under these leptophilic gauge groups induces a kinetic mixing term at the one-loop level. At energies well below the weak scale, $q\ll v_\mathrm{EW}$, this mixing is to first order only with the SM photon and the loop-induced mixing parameter can be expressed 
\begin{align}\label{eq:liljmix}
\epsilon_{ij}(q^2)=\frac{e\, g_{ij}}{2\pi^2}\int_0^1 dx\,x(1-x)\ \log\left(\frac{m_i^2-x(1-x)q^2}{m_j^2-x(1-x)q^2}\right) \,,
\end{align}
where $g_{ij}$ denotes the $U(1)_{L_i-L_j}$ coupling constant and $m_i$ and $m_j$ are the masses of the charged leptons of the $i$th and $j$th generation. This irreducible loop-induced kinetic mixing is finite and effectively leads to loop suppressed interactions of the $U(1)_{L_i-L_j}$ with hadrons. When presenting results below, we assume that there is no bare tree-level kinetic mixing and the dynamics are fully-governed by the loop-induced mixing of~\cref{eq:liljmix}. Such a tree-level mixing term can be either forbidden by the underlying UV completion of the model, or it can simply be neglected if it is much smaller than the loop-induced finite mixing. 

\paragraph{$L_\mu-L_e$} The phenomenology of the $U(1)_{L_\mu-L_e}$ gauge boson is governed by its couplings to electrons and muons. Therefore, it is constrained by a large number of  $e^+e^-$ collider, as well as electron beam dump and fixed target dark photon searches, as can be seen in the top left panel of~\cref{fig:li_lj}. Similar to the case of $U(1)_{B-L}$ discussed in~\cref{sec:llp_vec_bl}, its sizeable couplings to neutrinos make it subject to bounds from invisible decays searches and neutrino scattering experiments like Borexino~\cite{Bauer:2018onh} and Texono~\cite{Bilmis:2015lja, Lindner:2018kjo}. The most stringent limit due to neutrino interactions, however, is the bound on neutrino non-standard interactions (NSI) derived from global fits to neutrino oscillation data~\cite{Coloma:2020gfv}. In the future, this bound can be even further pushed by more precise measurements of neutrino oscillation at DUNE~\cite{Wise:2018rnb}.

The major difference to the $U(1)_{B-L}$ case is the heavily suppressed couplings of the $U(1)_{L_\mu-L_e}$ boson to hadrons, which are only induced at the loop level. Consequently, experiments relying on hadronic dark photon production processes are significantly less sensitive to this boson. For example, limits from proton beam dumps like NuCal/U70 or LSND are much less sensitive.  Similarly, the limits from LHCb dimuon searches for dark photons with masses of $M_{A'}\gtrsim 10$ GeV are substantially weaker and not even visible on the plot in the top left panel of~\cref{fig:li_lj}.  Since the LHC is a proton-proton collider, the production of such a purely leptophilic gauge boson is heavily suppressed. Therefore also FASER and FASER2 will not be able to probe unconstrained parameter space of the $U(1)_{L_\mu-L_e}$ boson. Noticeably, SHiP still will have some sensitivity to untested parameter space due to its very high statistics, a suitable geometry and a high boost factor. Most progress in searches for this boson can be expected from visible and invisible searches at Belle-II, as well as  invisible searches at NA64$\mu$~\cite{Gninenko:2014pea, Gninenko:2018tlp}.

\paragraph{$L_e-L_\tau$} The phenomenology of the $U(1)_{L_e-L_\tau}$ boson is very similar to the case of $U(1)_{L_\mu-L_e}$, since both bosons have gauge interactions with electrons, but only loop-suppressed interactions with hadrons. The main difference between the two cases are the loop-suppressed interactions with muons of the $U(1)_{L_e-L_\tau}$ boson. As a consequence the limit from KLOE search for dimuon resonances is absent, and the future muon run of NA64 will not be sensitive to this scenario. 

Another difference to the $L_\mu-L_e$ case is the slightly larger loop-induced kinetic mixing parameter for $L_e-L_\tau$. The kinetic mixing parameter from~\cref{eq:liljmix} can be approximated as 
\begin{align}
   \epsilon_{ij} &\approx \frac{e \, g_{ij}}{6\pi^2}\log\left(\frac{m_i}{m_j} \right)\,,
\end{align}
yielding 
\begin{align}
   \epsilon_{\mu e}  \approx  \frac{g_{\mu e}}{37}\,, && \epsilon_{e\tau}  \approx - \frac{g_{e\tau}}{25}\,,  && \epsilon_{\mu\tau}  \approx - \frac{g_{\mu\tau}}{70}\,.
   \label{eq:eps_approx}
\end{align}
This larger kinetic mixing results in slightly more constraining limits from NuCal/U70 and LSND, and a more constraining projection for SHiP in the top right panel of~\cref{fig:li_lj}. As a consequence, also FASER and FASER2 are more sensitive to the case of $U(1)_{L_e-L_\tau}$ due to the enhanced gauge boson production in hadronic processes. FASER2 is even able to probe a small region of previously unconstrained parameter space. 

\paragraph{$L_\mu-L_\tau$} Finally, we consider the phenomenology of an extra $U(1)_{L_\mu-L_\tau}$ gauge symmetry. This differs most from all previously discussed models, since the associated gauge boson couples to electrons and hadrons only via loop-suppressed kinetic mixing. As ordinary matter is composed of electrons, protons and neutrons - and so are the experimental apparatuses -  the $U(1)_{L_\mu-L_\tau}$ boson is very hard to produce and to detect. As a consequence, we can see in the bottom panel of~\cref{fig:li_lj} that constraints from proton and electron beam dump and fixed target searches are entirely absent. By the same token there are no constraints from prompt decay searches at collider experiment, with the exception of four-muon final state searches at BaBar~\cite{TheBABAR:2016rlg} and CMS~\cite{Sirunyan:2018nnz}.

Instead, the most stringent constraints are due to the neutrino interactions of the $U(1)_{L_\mu-L_\tau}$ boson, like neutrino trident production at Charm II~\cite{Altmannshofer:2014pba, Geiregat:1990gz} and CCFR~\cite{Altmannshofer:2014pba} (which is only shown as a grey dotted line due to some dispute over additional background~\cite{Krnjaic:2019rsv}). Moreover, stringent bounds arise from solar neutrino scattering at Borexino~\cite{Bellini:2011rx, Agostini:2017ixy, Amaral:2020tga} as well as from white dwarf cooling~\cite{Bauer:2018onh}. At small masses of $M_{A'}\lesssim 10$ MeV a very strong bound arises from cosmological constraints on extra relativistic degrees of freedom, $\Delta N_\text{eff}$, in the early universe~\cite{Escudero:2019gzq}.

Most interestingly, a $U(1)_{L_\mu-L_\tau}$ group is the only of these minimal anomaly-free extra $U(1)_X$ symmetries that can accommodate a solution of the muon $(g-2)_\mu $~\cite{Bennett:2002jb, Bennett:2004pv, Bennett:2006fi, Abi:2021gix}. The preferred region of parameter space, where the $(g-2)_\mu$ anomaly is resolved is shown by the light green band in the bottom panel of~\cref{fig:li_lj}. In the future, there are a number of experimental searches that can probe this remaining parameter space. For example, one of the soonest searches to test this region will be the missing energy search at NA64$\mu$~\cite{Gninenko:2014pea, Gninenko:2018tlp} (or similarly at the proposed Fermilab M$^3$ experiment~\cite{Kahn:2018cqs}). With a similar search strategy, the $(g-2)_\mu$ region can also be tested via missing energy searches in kaon decays at NA62~\cite{Krnjaic:2019rsv}. Furthermore, future direct detection experiments like LZ and Darwin will be able to test parts of this interesting parameter space via solar neutrino scattering~\cite{Amaral:2020tga}, as will spallation source experiments like COHERENT~\cite{Abdullah:2018ykz} or CCM and ESS~\cite{Amaral:2021rzw}. However, due to the loop-suppressed hadronic couplings of the $U(1)_{L_\mu-L_\tau}$ boson, it is very hard to produce at LHC. Therefore, FASER and FASER2 will not be sensitive to this scenario. Only SHiP will be able to test large fractions of unconstrained parameter space due to its very high statistics~\cite{Bauer:2018onh}.
 
\subsection{$B-3L_i$ Gauge Bosons} \label{sec:llp_vec_b3l}

In this section we will study the groups $U(1)_{B-3L_i}$ as an example of the minimal anomaly-free baryo-leptophilic $U(1)$ extensions of the SM, discussed in~\cref{sec:llp_vec_bl}, that feature family-non-universal couplings in the lepton sector. Their phenomenology will be yet different from the previously discussed examples and searching for the associated gauge bosons requires some dedicated search strategies. To illustrate the maximal effect of  flavour-specific couplings in the lepton sector, we will set the kinetic mixing to zero, or assume that it is negligibly small, $\epsilon\approx 0$.

\begin{figure*}[t]
  \centering
  \includegraphics[width=0.49\textwidth]{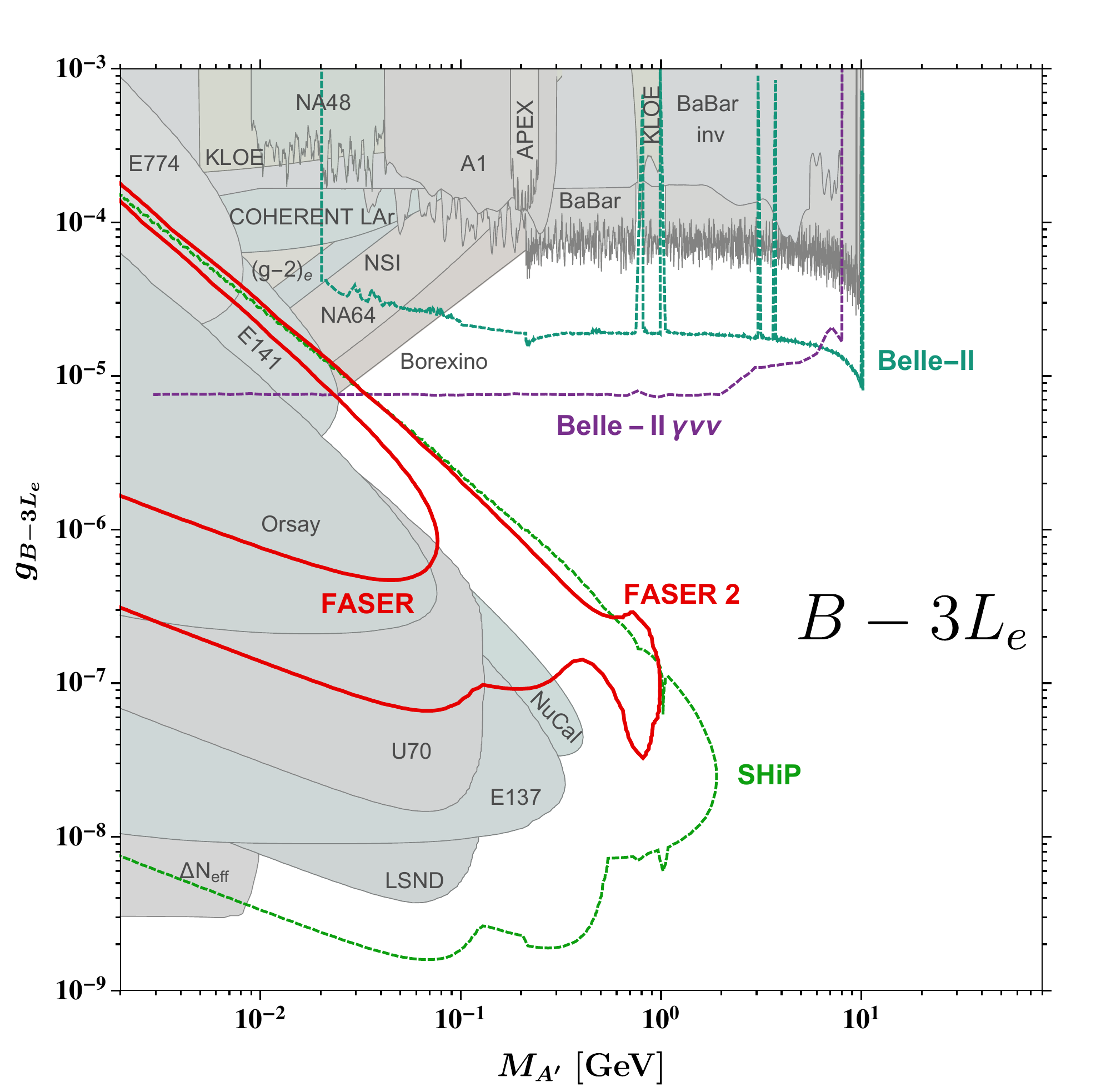}%
  \includegraphics[width=0.49\textwidth]{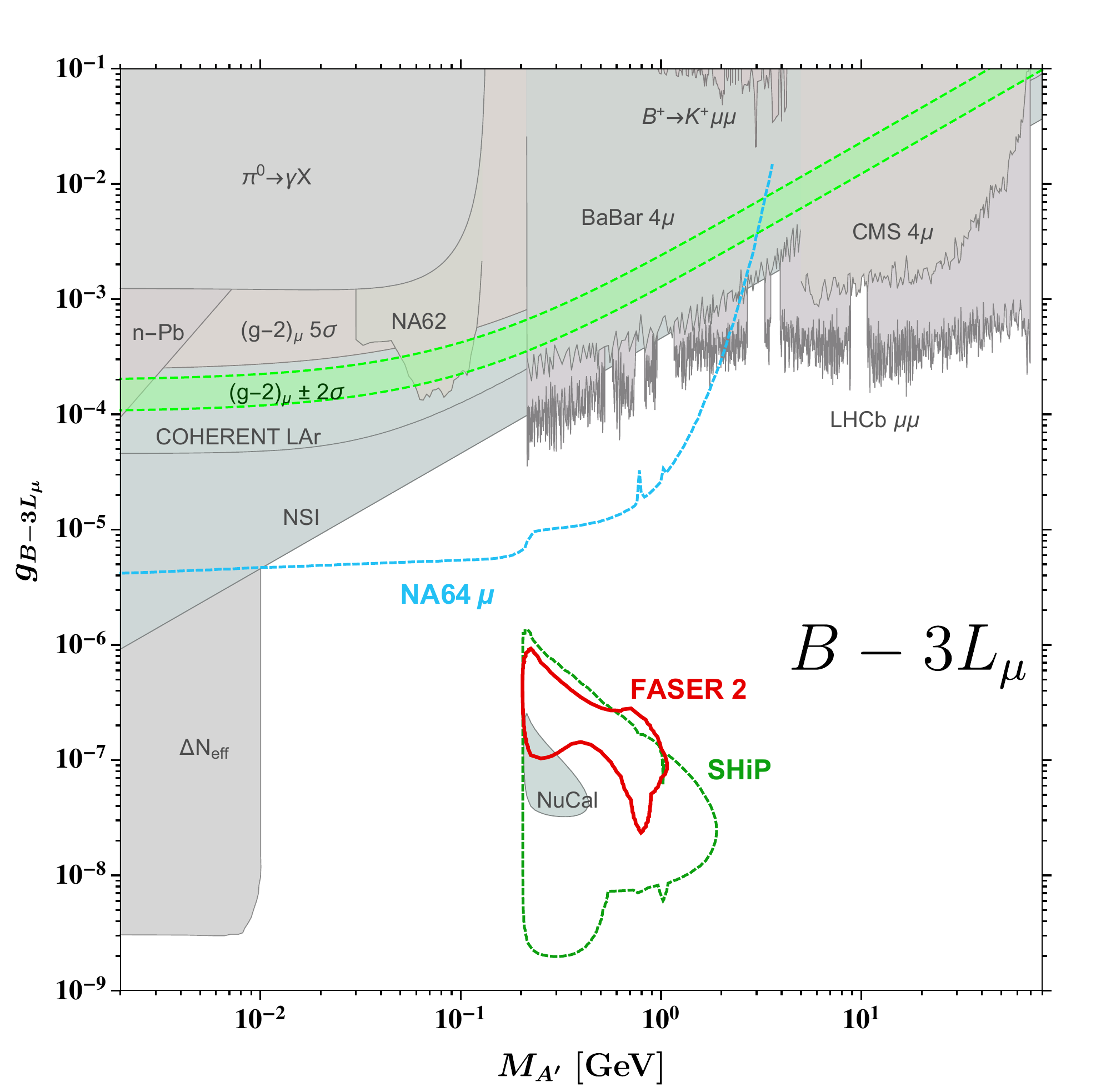}\\
  \includegraphics[width=0.49\textwidth]{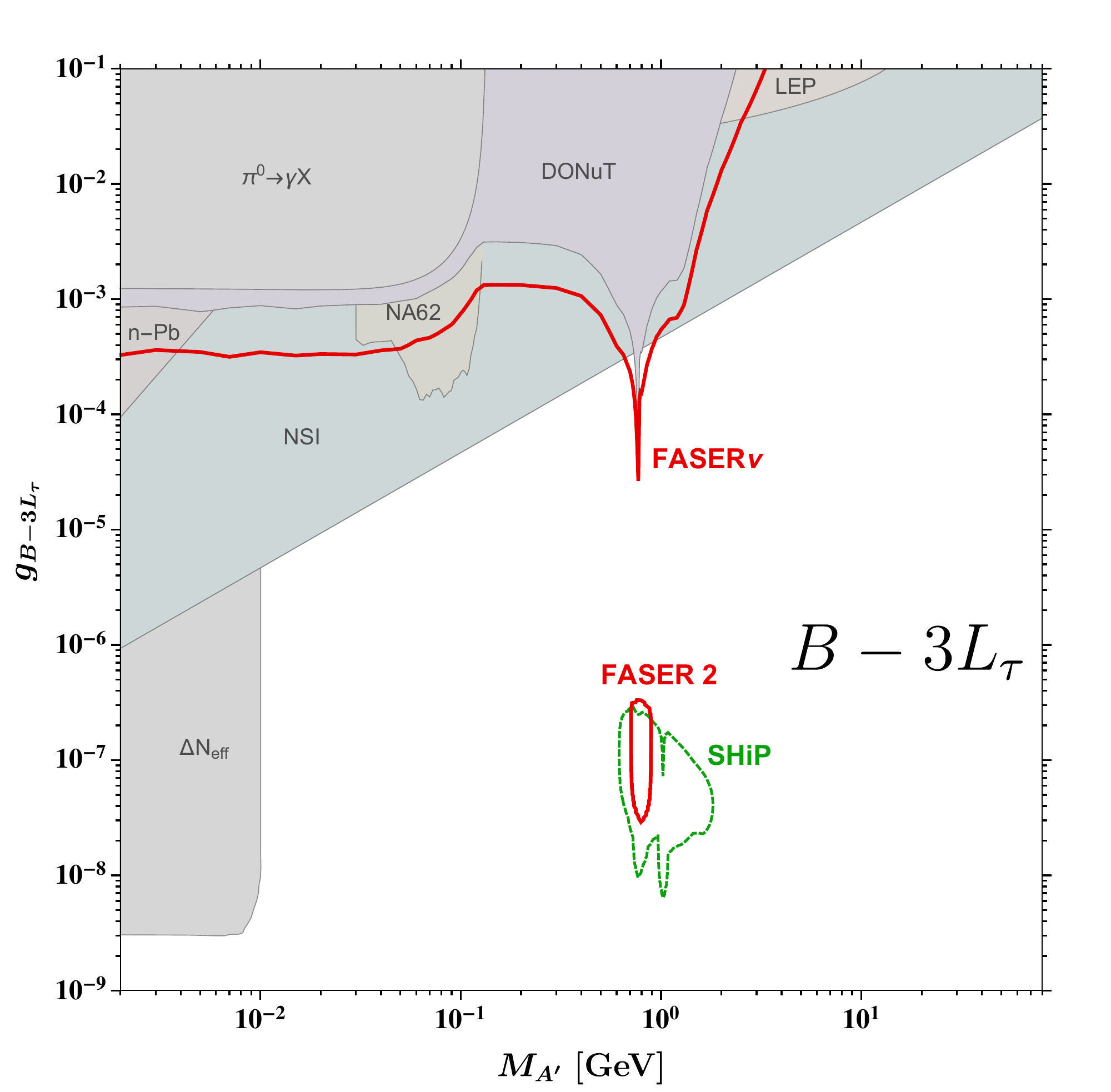}
  \caption{Current existing constraints shown as the grey shaded regions alongside projections of other future experiments and measurements shown as coloured dashed lines for a $U(1)_{B-3L_e}$ (top left), $U(1)_{B-3L_\mu}$ (top right) and $U(1)_{B-3L_\tau}$ (bottom centre) gauge boson in the coupling-mass plane. Sensitivity reaches of FASER, FASER2 and FASER$\nu$ are shown by the solid red lines. See text for explanations.}
  \label{fig:b3li}
\end{figure*} 

\paragraph{$B-3L_e$} The phenomenology of the $U(1)_{B-3L_e }$ gauge boson is still quite similar to the family-universal $U(1)_{B-L}$ case, with the major difference coming from the absence of muon couplings. Hence, e.g.~the search for dimuon resonances at LHCb is not sensitive to this scenario and consequently the LHCb limits are absent in the top left panel of~\cref{fig:b3li}. In contrast, this model is constrained from prompt and invisible searches at $e^+e^-$ colliders, like at A1~\cite{Merkel:2011ze, Merkel:2014avp}, APEX~\cite{Abrahamyan:2011gv}, BaBar~\cite{Lees:2014xha, Lees:2017lec}, KLOE~\cite{Archilli:2011zc, Babusci:2012cr, Babusci:2014sta} or the fixed target experiment NA64~\cite{NA64:2019imj}. Furthermore, very strong constraints arise from proton and electron beam dump and fixed target experiments (like e.g.~E137, E141, E774~\cite{Bjorken:2009mm}, Orsay~\cite{Andreas:2012mt}, NuCal/U70~\cite{Blumlein:2011mv, Blumlein:2013cua}, LSND~\cite{Essig:2010gu}), which have been derived in Ref.~\cite{Bauer:2020itv}. As for $B-L$, the $U(1)_{B-3L_e }$ gauge boson is also subject to constraints from neutrino experiments, like neutrino scattering at COHERENT and Borexino~\cite{Bauer:2020itv} or from searches for NSI with oscillation data~\cite{Coloma:2020gfv}.

In the future, searches for dielectron resonances and missing energy at Belle-II~\cite{Kou:2018nap, Foguel:2022ppx} will significantly push the limits from collider searches. However, due to the gauge couplings to hadrons, the $U(1)_{B-3L_e }$ gauge boson can also be abundantly produced in proton-proton collisions at LHC and searched for in electronic and hadronic final states at the FPF. Hence, both FASER and FASER2 will be able to probe previously untested regions of parameter space~\cite{Bauer:2020itv} and improve over current beam dump limits. The FASER and FASER2 reaches are shown as the red lines in the top left panel of~\cref{fig:b3li}.
Finally, also SHiP will be able to probe large areas of previously uncovered parameter space thanks to its large statistics.

\paragraph{$B-3L_\mu$} In the case of an extra $U(1)_{B-3L_\mu}$ symmetry, dark photon searches at electron beam experiments or $e^+e^-$ colliders are not sensitive since the associated boson is not coupling to electrons and hence cannot be produced in these kind of experiments. Thus, limits from electron beam dump and fixed targets are entirely absent in the top right panel of~\cref{fig:b3li}. Below the dimuon threshold, $M_{A'}< 2 m_\mu$, the $U(1)_{B-3L_\mu}$ boson can only decay invisibly into neutrinos. Therefore, in this mass range constraints mostly arise from experiments exploiting its neutrinos interactions, like searches for neutrino coherent scattering at COHERENT~\cite{Bauer:2020itv}, NSI in neutrino oscillations~\cite{Coloma:2020gfv} or extra relativistic degrees of freedom in the early universe~\cite{Escudero:2019gzq, Kamada:2015era}. Above the dimuon threshold, $M_{A'} \gtrsim 2 m_\mu$, the $U(1)_{B-3L_\mu}$ gauge boson can be searched for in dimuon resonance searches, like e.g.~at LHCb~\cite{LHCb:2019vmc}, or in four-muon final states at BaBar~\cite{TheBABAR:2016rlg} and CMS~\cite{Sirunyan:2018nnz}. In this region of parameter space, there is also a constraint coming from dark photon searches in muonic final states at the proton beam dump experiment NuCal~\cite{Bauer:2020itv}.

In the future, the muon beam experiment NA64$\mu$~\cite{Gninenko:2014pea, Gninenko:2018tlp} will be able to improve significantly over current in muon plus missing energy searches. As the $U(1)_{B-3L_\mu}$ boson has gauge interactions with hadrons it can be abundantly produced at the LHC, where it can be searched for at the FPF in displaced muonic and hadronic final states. This will allow FASER2 (and similarly the proton fixed target experiment SHiP)  to probe previously unexplored parameter space~\cite{Bauer:2020itv}, as is illustrated in the top right panel of~\cref{fig:b3li}.

\paragraph{$B-3L_\tau$} Finally, the fact that first and second generation lepton couplings are entirely absent in a $U(1)_{B-3L_\tau}$ model makes it very hard to test. In particular, it is not subject to constraints from past beam dump or fixed target experiment, since the energies have not been high enough to produce dark photons massive enough that they can decay into purely hadronic final states. The most stringent constraints on a $U(1)_{B-3L_\tau}$ boson are due to its neutrino interactions. A rather strong constraint comes from searches for the decay $\pi^0\to\gamma\, (X\to\nu\bar\nu)$ at NA62~\cite{CortinaGil:2019nuo}. Furthermore, similar to the previous case of $U(1)_{B-3L_\mu}$, the most stringent constraints arise from bounds on NSI in neutrino oscillations~\cite{Coloma:2020gfv} or extra relativistic degrees of freedom in the early universe~\cite{Escudero:2019gzq, Kamada:2015era}, as can be seen in the bottom panel of~\cref{fig:b3li}. Furthermore, a previous search for tau neutrino scattering at the DONuT~\cite{Kodama:2007aa} experiment has been used to derive limits to this model~\cite{Kling:2020iar}, which are competitive in a small region around the $\omega$-resonance.

Similarly to the DONuT search, FASER$\nu$ will be sensitive to scattering of LHC produced neutrinos. It can thus be used to search for extra tau neutrinos originating from the decay of a $U(1)_{B-3L_\tau}$ boson. The corresponding projection for the FASER$\nu$ sensitivity has been derived in Ref.~\cite{Kling:2020iar} and is illustrated by a red line in the bottom panel of~\cref{fig:b3li}. This will be able to improve over the DONuT and NSI limit in the $\omega$-resonance region. Lastly, the abundant production of $U(1)_{B-3L_\tau}$ bosons in proton collisions combined with the high energies at LHC will allow FASER2 to probe untested parameter space of this model in hadronic decays around the $\omega$-resonance~\cite{Bauer:2020itv}. Similarly, SHiP will be able to search for this boson in hadronic final states and test new regions of parameter space.

\subsection{$B$ Gauge Boson} \label{sec:llp_vec_b}

\begin{figure*}[t]
\centering
    \includegraphics[width=0.99\textwidth]{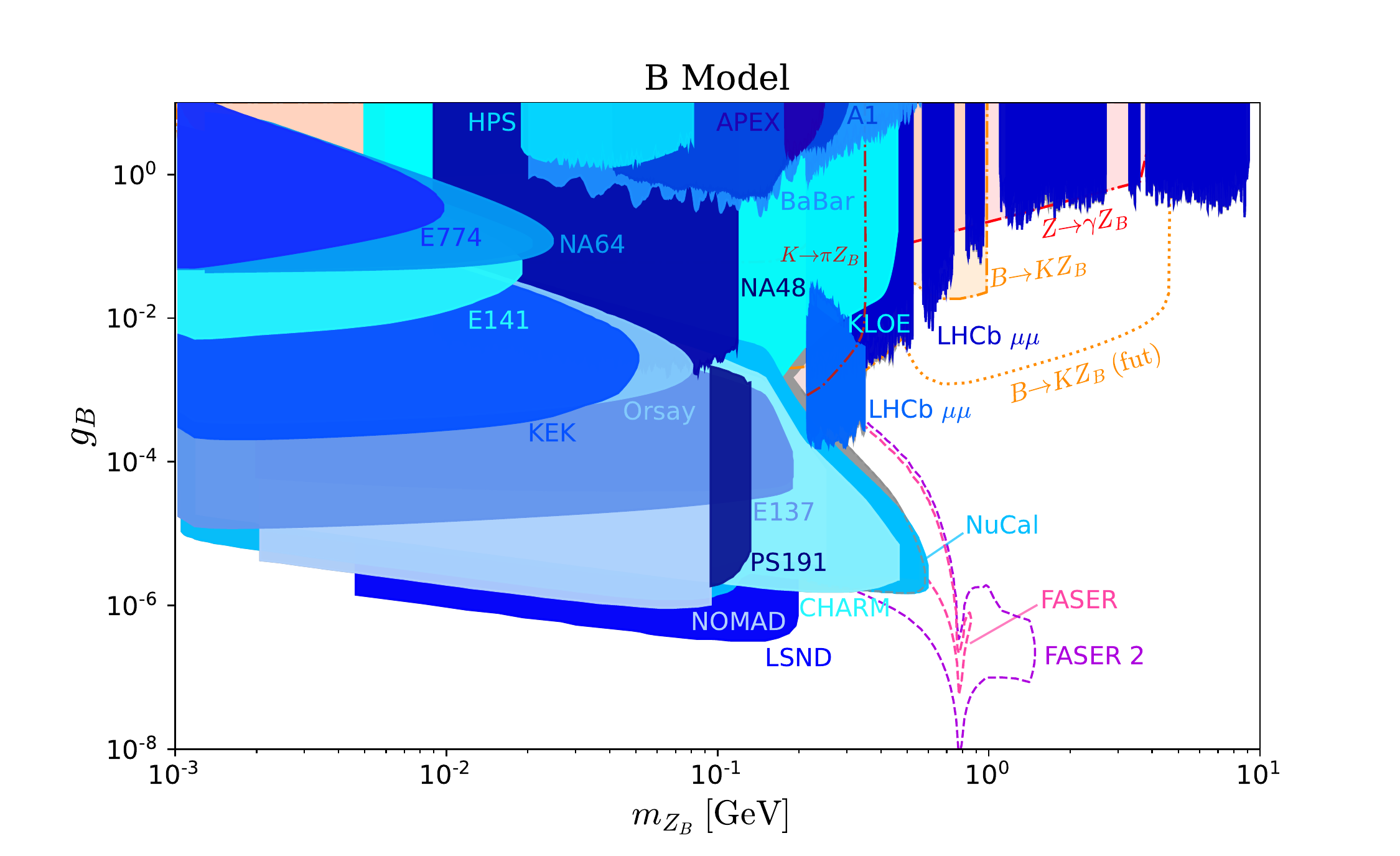}
\caption{Current bounds on the $g_B-m_B$ plane for the $Z_B$ vector mediator model. In blue we show the limits obtained for leptonic decay searches with data from the electron Bremsstrahlung experiments APEX~\cite{APEX:2011dww} and A1~\cite{A1:2011yso,Merkel:2014avp}, the proton beam dump experiments PS191~\cite{Bernardi:1985ny}, NuCal~\cite{Blumlein:1990ay, Blumlein:1991xh}, CHARM~\cite{CHARM:1985anb}, NOMAD \cite{NOMAD:2001eyx} and LSND~\cite{LSND:1997vqj, Bauer:2018onh}, the electron beam dump experiments E137~\cite{Bjorken:2009mm, Andreas:2012mt}, E141 \cite{Riordan:1987aw}, E774 \cite{Bross:1989mp}, NA64 \cite{NA64:2018lsq}, KEK \cite{Konaka:1986cb} and Orsay \cite{Davier:1989wz}, the $e^+e^-$ annihilation experiments BaBar~\cite{BaBar:2014zli} and KLOE~\cite{Anastasi:2015qla, KLOE-2:2012lii}, the fixed target experiment HPS \cite{Battaglieri:2014hga}, the LHCb experiment~\cite{LHCb:2017trq, LHCb:2019vmc} and  NA48~\cite{NA482:2015wmo}. The gray region indicate the exclusion bounds obtained in a previous study \cite{Ilten:2018crw}. The dash-dot and dotted curves are current and future limits coming from the decays $B \to K Z_B$, $Z\to \gamma Z_B$ and $K^\pm \to \pi^\pm Z_B$ that were taken from \cite{Dror:2017ehi}. Finally, the dashed curve in pink and purple show the future sensitivity predictions for the FASER and FASER 2 experiments, respectively. These were calculated using the \texttool{FORESEE} code~\cite{Kling:2021fwx} with the hadronic implementation from \cite{Foguel:2022ppx}. Modified version of Figure extracted from~\cite{Foguel:2022ppx}.}
\label{fig:limits_Bmodel}
\end{figure*}

All previously mentioned Abelian $U(1)$ mediator models are anomaly-free just by requiring a certain choice of charges and without working in some extra fermion family or embedding it into a UV-complete model. Nevertheless, we might also consider mediators that only couple to lepton or baryon number keeping in mind that it is not per se anomaly-free. In the following, we consider a $U(1)_B$ model, or $B$ model, where the mediator entertains gauge quark couplings but only couples to leptons via kinetic mixing induced by loop effects. 
The Lagrangian is given by
\begin{align}
\mathcal{L}_{Z_B,\rm int} \supset  e \epsilon J^\mu_{\rm em} Z_{B\mu} - g_B J_B^\mu Z_{B\mu} \, ,
\label{eq:LBmodelint}
\end{align}
where $J_{\rm em}^\mu$ is the SM electromagnetic current and $J_B^\mu$ a new vector current given by
\begin{align}
     J^\mu_{\rm B} =\frac13 \sum_{q} \bar{q} \gamma^\mu\, q\, .
\end{align}
The kinetic mixing parameter $\epsilon$ is usually assumed to be of typical one-loop size $\epsilon=eg_B/(4\pi)^2$ and hence, depends on $g_B$. If direct couplings with $g_B$, the loop contributions can be neglected. They only become relevant if leptonic decays are the only option since no hadronic decay channels are open. 

The $B$ model has been considered for several decades already starting with the early work~\cite{Nelson:1989fx} about MeV bosons in $\pi^0$, or $K^+$ and $\eta$ decays if the gauge boson mass is smaller than the pseudoscalar meson mass. Some of the work even considered vector masses at the order of $\mathcal{O}(10^2)$~GeV suggesting searches at collision experiments such as the LHC~\cite{Rajpoot:1989jb,He:1989mi}. But soon the focus has been shifted more towards masses below the $Z$ mass, with limits coming from $\Upsilon$ and $Z$ decays~\cite{Carone:1994aa,Bailey:1994qv}, and even further down to 1-10 GeV including rare decays of B or D mesons and quarkonia states~\cite{Aranda:1998fr}. On the theoretical side, various possibilities of model realizations have been discussed as for example in connection to the seesaw-mechanism~\cite{Foot:1989ts}, discussing the mixing with the SM photon~\cite{Carone:1995pu}, anomaly cancellations by adding a single new fermionic generation~\cite{FileviezPerez:2010gw}, and other UV-complete models as for example in the context of asymmetric DM~\cite{Graesser:2011vj}, usually containing a $U(1)_B$ breaking Higgs field~\cite{Williams:2011qb}. More recently, the $B$ model has reached the sub-GeV range again where processes have to be described within the framework of vector meson dominance (VMD)~\cite{Tulin:2014tya,Ilten:2018crw} for sub-GeV gauge bosons where several signatures have been proposed including revised descriptions of mesonic decays into vector particles or vector mediator decays into hadrons. 

Most boson searches assume leptonic couplings when searching for signatures in various experiments. It is often easier to search for a clean leptonic signature at electron facilities such as Belle-II~\cite{Belle-II:2018jsg} or NA64~\cite{Banerjee:2019dyo}, instead of hadronic signatures involving jets or mesonic decays which usually are probed at proton facilities like the LHC. If the mediator is coupling to neutrinos, a wide range of neutrino searches can further constrain parts of the parameter space. Moreover, the branching ratio into hadronic decay channels is often smaller than leptonic branching ratios as for example in $B-L$, or $B-3L_i$ models as discussed in \cref{sec:llp_vec_decay} based on~\cite{Foguel:2022ppx}. This changes drastically if we consider bosons that predominately couple to quarks as in baryon number gauging $U(1)_B$ models. We show that current and proposed far-forward experiments as proposed to be accomodated by the FPF, can enhance the potential to probe this particular model which remain beyond the reach of experiments focusing on BSM electron couplings.

\paragraph{Without DM} In \cref{fig:limits_Bmodel}, we present several constraints for the model reaching from current bounds from accelerator and collider searches, to rare anomaly-induced decays to new LLP decay signatures that can be probed in the near future. One can see that for vector mediator masses below the pion mass threshold, the only available decay channel is $Z_Q\to e^+e^-$ and hence only experiments measuring leptonic signatures (blue) can constrain the parameter space in the $g_B-m_B$ plane. In~\cite{Dror:2017ehi,Dror:2017nsg} it has been shown that the meson decays $B \to K Z_B$ and $K^\pm \to \pi^\pm Z_B$ as well as the $Z$ boson decay $Z\to \gamma Z_B$ are enhanced due to non-current conservation. 

The excluded blue regions were obtained using data from different kinds of experiments, such as the proton beam dumps NuCAL~\cite{Blumlein:1990ay, Blumlein:1991xh}, CHARM~\cite{CHARM:1985anb} and LSND~\cite{LSND:1997vqj, Bauer:2018onh}, where the $Z_B$ boson is produced via meson decays or the proton bremsstrahlung process as well as LHCb constraints with a vector mediator decaying into muons. For a more detailed description of the limits, we refer to~\cite{Foguel:2022ppx}. We also show the predicted sensitivities for the FASER  (dashed pink) and FASER2 experiments (dashed purple), that were obtained using \texttool{FORESEE} code~\cite{Kling:2021fwx} together with the hadronic branching ratios from Ref.~\cite{Foguel:2022ppx}. We can see from the figure that FASER will reach lower $g_B$ couplings and will provide constraints in a new region where the current experiments lack efficiency and sensitivity. This is further outreached by FASER2 which covers an even larger region of the parameter due to the sensitivity to three pion and kaons decays of the vector boson.

\begin{figure*}[t]
\centering
\includegraphics[width=0.85\textwidth]{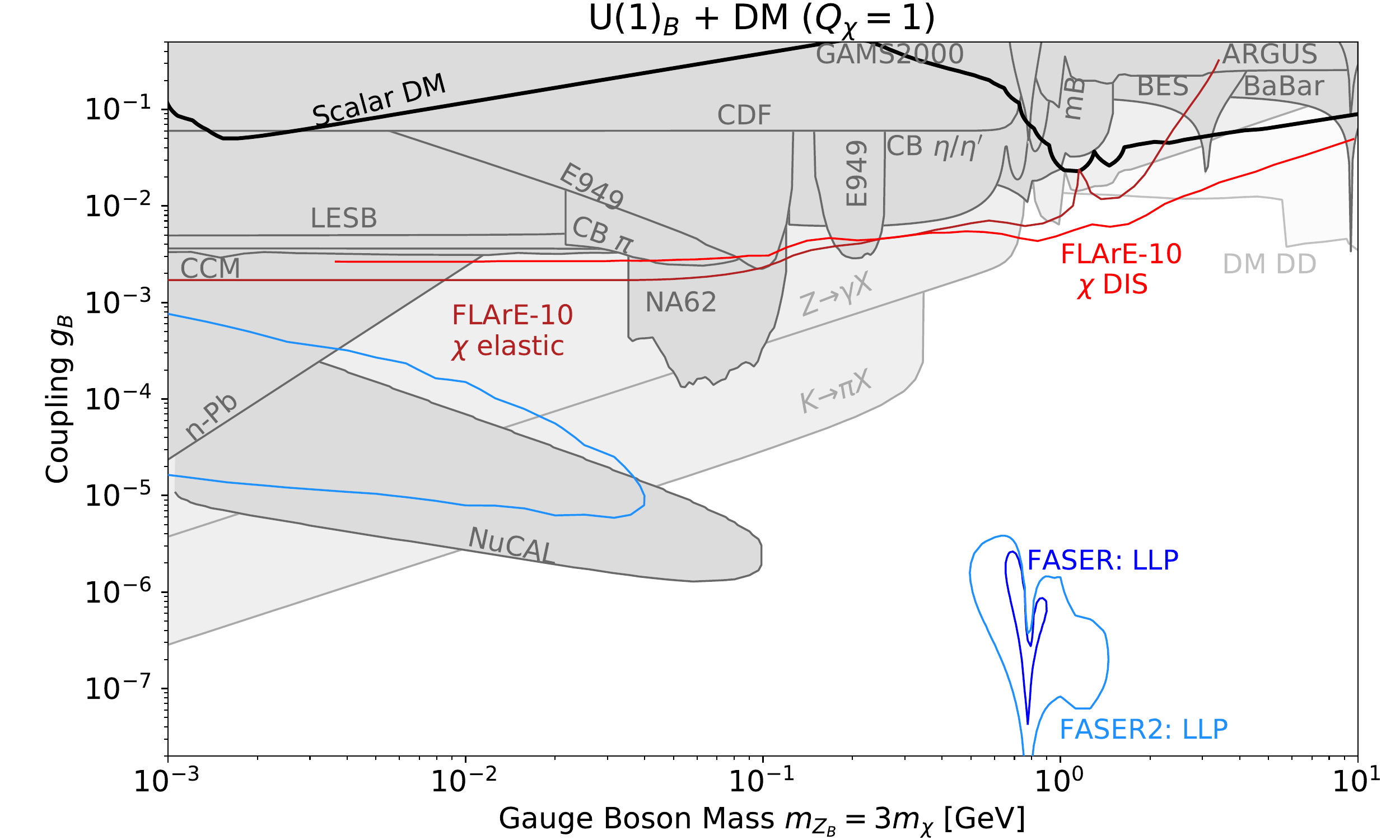}
\caption{Current bounds (gray) and FPF future sensitivities from FLArE (light and dark red), FASER (light blue) and FASER2 (dark blue) for a $Z_B$ vector mediator with couplings to a DM particle $\chi$. Modified version of figure extracted from~\cite{Batell:2021snh}.}
\label{fig:limits_BDMmodel}
\end{figure*}

\paragraph{Including DM}In ~\cite{Batell:2021snh}, the $U(1)_B$ model has been considered in context of DM. In particular, a
complex scalar with a Lagrangian
\begin{align}
    \mathcal{L}&\supset |\partial_\mu\chi|^2-m_\chi^2|\chi|^2
\end{align}
and a dark current
\begin{align}
    J_{Z_B\bar{\chi}\chi}&= g_\chi i(\partial_\mu \chi^* \chi - \chi^* \partial_\mu \chi)
\end{align}
has been presented. This model evades constraints arising from precision measurements of the CMB anisotropies~\cite{Planck:2015fie, Slatyer:2009yq} due to velocity-suppressed P-wave annihilations. The full model parameter space is specified by four parameters, $m_B, g_B, m_\chi$ and $Q_\chi$. Two choices of $Q_\chi$ values have been discussed in~\cite{Batell:2021snh}, one where DM and SM particles have comparable interactions strengths, \textit{i.e} $Q_\chi=1$, and one where $\alpha_\chi = g_B^2Q_\chi^2 / (4\pi)$ was fixed. In the following, we will only present the case of $Q_\chi=1$ while the other case will be discussed in \cref{sec:bsm_dm_hadro} in the context of scattering signatures. If we further adopt the common convention $m_B=3m_\chi$, the analysis can be reduced to the two parameters $g_B$ and $m_B$ again. 

Compared to the $U(1)_B$ model without DM, already mentioned searches are modified with changed branching ratios and are complemented by dark matter direct detection (DD), DM deep inelastic scattering (DIS), and DM elastic scattering limits. The DD bounds have to be taken with care since they do not apply for inelastic scalar DM if the mass splitting between the dark species is large enough to suppress upscattering of non-relativistic DM particles. In \cref{fig:limits_BDMmodel}, the combined results for CRESST-III~\cite{CRESST:2019jnq}, DarkSide-50~\cite{DarkSide:2018bpj}, and Xenon 1T~\cite{XENON:2018voc,XENON:2020gfr} are shown as a very light gray shaded region assuming that $\Omega_\chi h^2\simeq 0.12$~\cite{Planck:2018vyg}. Note that for parameter points that are below the thermal target line (solid black), the DM abundance has to be explained by a non-standard cosmological scenario. DM scattering events can occur when the copiously produced hadrophilic mediator decays to dark matter particles $Z_B\to \chi\chi$. Particles in this DM beam can then scatter in detectors and can be searched for in experiments like in the downstream neutrino detector of MiniBooNE~\cite{MiniBooNE:2017nqe, MiniBooNEDM:2018cxm}, or as recently with the Coherent CAPTAIN-Mills (CCM) liquid argon (LAr) detector~\cite{Aguilar-Arevalo:2021sbh}. All existing constraints are shown in dark gray.

Several of the proposed detectors under consideration for FPF have discovery prospects for the $U(1)_B$ mediators coupling to DM. The projected sensitivity is shown in \cref{fig:limits_BDMmodel}. Both FASER and FASER2 can detect decays of the long-lived vector mediator into visible final states, as shown by the blue lines. Note that this LLP signature covers a smaller but still significant part of the parameter space compared to the case of mediators that are not coupling to DM since a bigger share of the total decay width is going into invisible DM states. In addition, the proposed neutrino detectors at the FPF, FLArE-10 and FASER$\nu$2, have the capability to probe this scenario by searching for the scattering of the produced DM particles inside the detectors. The sensitivity has been estimated for DM elastic scatterings with nucleons $\chi p\to \chi p$, which lead to visible single proton tracks in the detectors (dark red), and DM DIS off nuclei $\chi N\to \chi X$, leaving a hadronic recoil with multiple charged tracks (light red)~\cite{Batell:2021snh}. More details on the scattering signature will be discussed in \cref{sec:bsm_dm_hadro}. Overall, all proposed signatures in the light of hadrophilic models highlight the rich phenomenology of FPF searches being able to both constrain couplings in the range $10^{-8}\lesssim g_B \lesssim 10^{-5}$ from LLP decays as well as above $10^{-4}\lesssim g_B$ in DIS and elastic scattering events where DM can explain the relic abundance.

\subsection{Production via Proton Bremsstrahlung} \label{sec:llp_vec_brem}

Experiments at high-luminosity facilities utilizing proton beams provide impressive sensitivity to new physics in the form of light weakly coupled degrees of freedom, namely {\it dark sectors}. The FPF at the LHC promises to provide an important new facility of this type, given its energy reach up to 14 TeV. The dominant production channels for dark sector degrees of freedom at proton beam facilities depend on the beam energy and the mass of the dark mediator. Among a variety of channels, forward bremsstrahlung of dark vectors and scalars is particularly important in the mass range from 500 MeV to a GeV, due to the possibility of enhancement via mixing with hadronic resonances with the same quantum numbers. However, computing this production rate in the forward region is a difficult task as it involves nonperturbative QCD physics associated with the forward $pp$ cross section. 

In recent work \cite{Foroughi-Abari:2021zbm} we have revisited the calculation for the rate for bremsstrahlung production of light dark vector ($A'$) and dark scalars ($S$), coupled through the corresponding renormalizable portal interactions, ${\cal L} \supset - \frac{1}{2} \epsilon F^{\mu\nu} F'_{\mu\nu} - ASH^\dagger H$. For sub-GeV mass dark sector states, the dark sector state couples with the proton coherently, and one considers the induced coupling to protons which follows directly from the portal interactions,
\begin{equation}
 {\cal L}_{\rm eff} \supset - \epsilon e A'_\mu \bar{p} \gamma^\mu p  - g_{SNN} \theta S \bar{p} p +\cdots
\end{equation}
where higher multipole couplings for $A'_\mu$ have been ignored, the $h-S$ mixing angle $\theta \simeq Av/m_h^2 \ll 1$ for low mass scalars, and $g_{SNN} = 1.2 \times 10^{-3}$ follows from QCD low energy theorems. Based on these couplings, we considered a number of different approximation schemes in calculating the production rate, 
\begin{itemize}
\item ISR and FSR in quasi-elastic scattering
\item ISR in non-single diffractive scattering via the quasi-real approximation
\item Hadronic generalization of the WW approximation
\item Modified WW approximation
\end{itemize}
This approach allows for an assessment of relative precision in the calculations, and for consideration of more direct approaches that fully model initial and final state radiation (ISR and FSR) within pomeron-mediated forward diffractive and non-diffractive proton-proton scattering, with variants of the Weizsacker-Williams (WW) approximation, developed in the 1970's and generalizing the approach used for electron beams. In particular, the modified WW approximation of \cite{Blumlein:2013cua} has been used widely to model bremsstrahlung production of dark sectors within proton beams. 

\begin{figure*}[t]
\centering
    \includegraphics[width=0.7\textwidth]{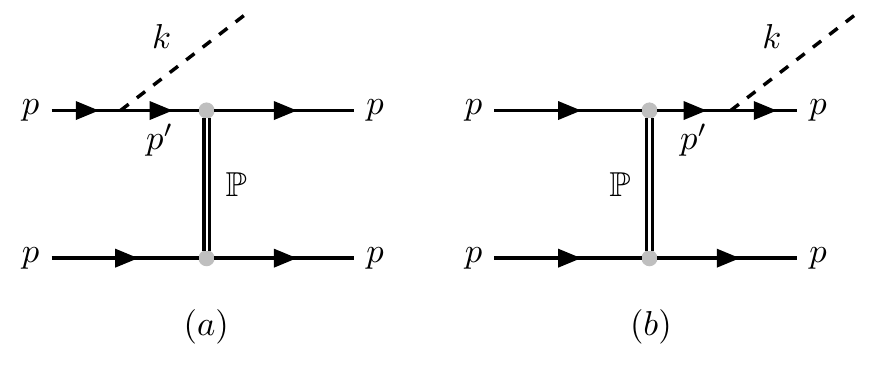}
    \caption{Dark scalar and vector radiation from (a) initial state, and (b) final state proton bremsstrahlung, with quasi-elastic scattering modeled via $t$-channel pomeron exchange. This process is subject to suppression through ISR-FSR interference, but does not account for non-single diffractive inelastic scattering, where this interference is not expected due to the variety of final states. Thus ISR alone should provide a reasonable approximation.}
    \label{fig:brem_el}
\end{figure*}

\begin{figure*}[t]
    \centering
    \includegraphics[width=0.49\textwidth]{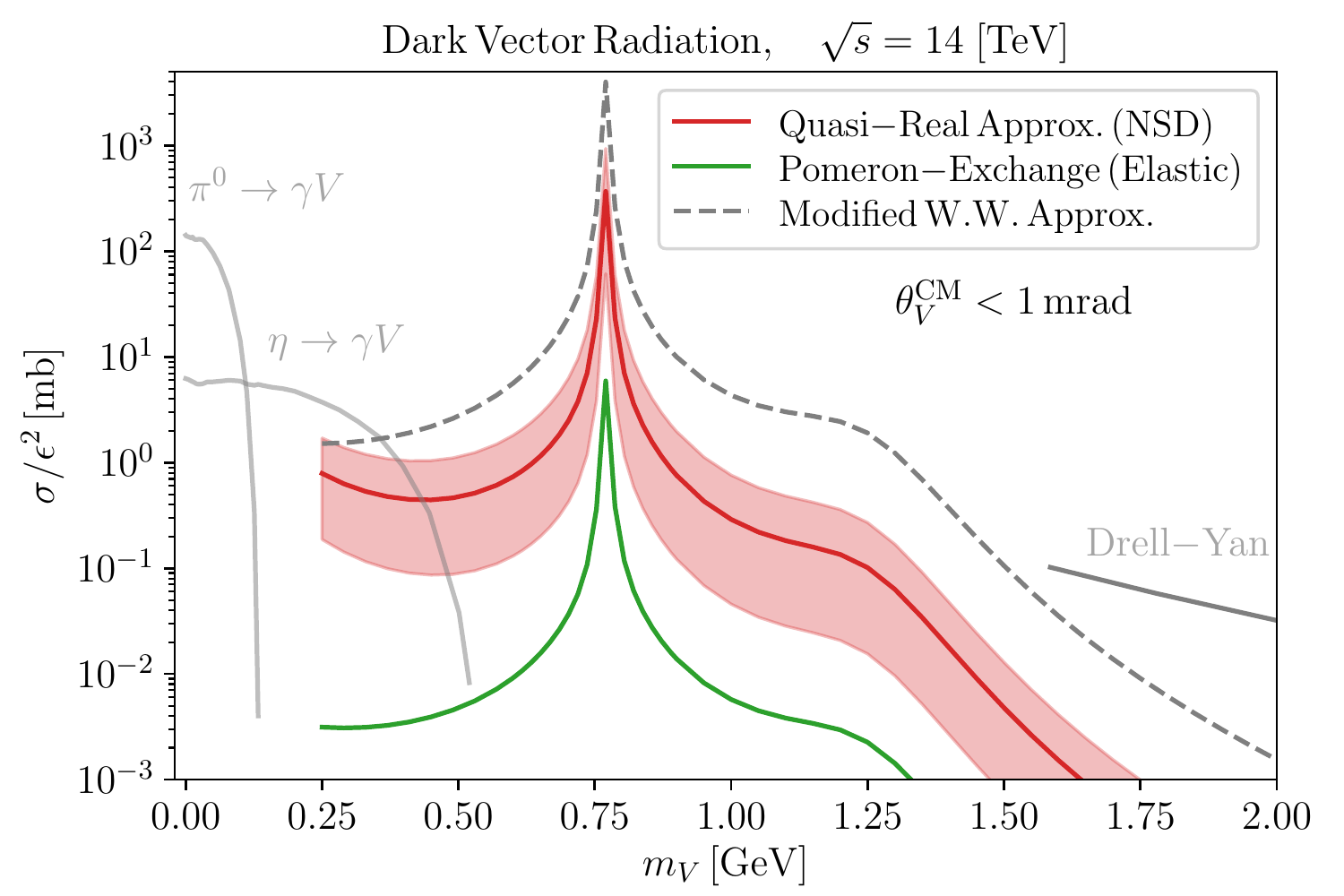} 
    \includegraphics[width=0.49\textwidth]{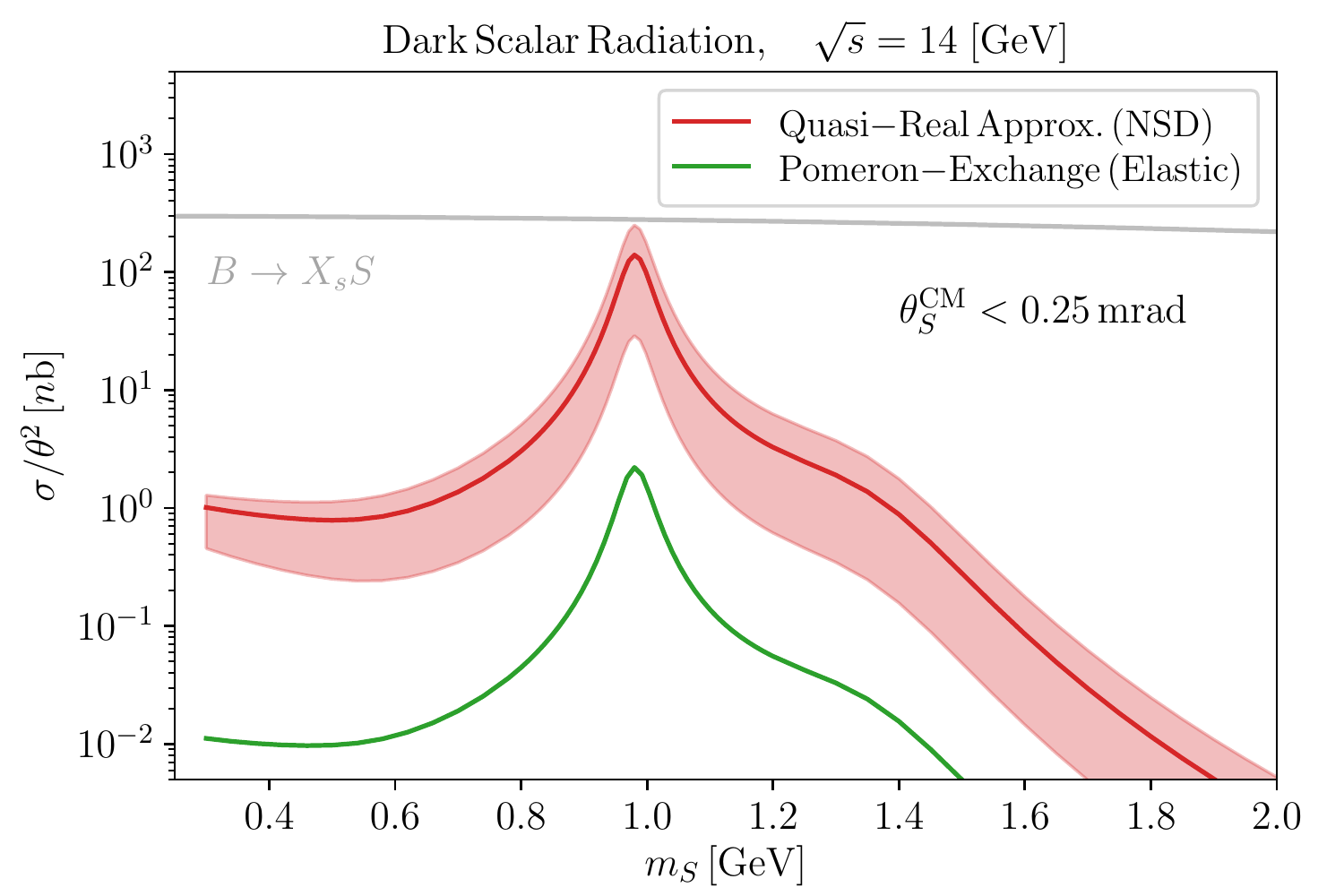} 
     \caption{The production cross sections for both dark vectors (upper) and scalars (lower) at $14$~TeV as a function of mass and within angles $\theta < 1$~mrad or $0.25$~mrad respectively of the beam axis in the centre of mass frame. The red curves denote the rates using the quasi-real approximation, or equiavelntly pomeron-mediated ISR, in non-single diffractive scattering, and the uncertainty band results from varying a cut-off scale for the intermediate off-shell proton form-factor $\Lambda_p\in [1,2]$~GeV (with central value $1.5$~GeV). The green curves show the associated rates obtained by combining ISR and FSR in quasi-elastic scattering, where interference effects cause a significant suppression. In the vector case, the dashed grey curve results from the modified WW approximation of \cite{Blumlein:2013cua} with transverse momentum $p_T< 1$~GeV, while for both plots the solid lighter grey curves indicate other hadronic production channels, e.g. from meson decay~\cite{Feng:2017uoz,Kling:2021fwx} at lower mass, and parton-level Drell-Yan~\cite{Kling:2021fwx} processes at higher mass. Note that for dark scalar radiation (lower), the bremsstrahlung channel is subleading to production via $B$ meson decays. Further details are available in \cite{Foroughi-Abari:2021zbm}.}
    \label{fig:Production14TeV}
\end{figure*}

Our final results for production rates are shown in \cref{fig:Production14TeV} for the 14 TeV LHC indicating various comparisons between different modes and caclulational schemes. These plots show the rate within an angular region that matches the expected scale of experiments within the FPF. In all cases, a timelike form-factor for coupling to the proton provides resonant enhancements.

One of the interesting conclusions from this analysis was the recognition that the combination of $t$-channel ISR plus FSR for quasielastic pomeron-mediated scattering (shown in \cref{fig:brem_el}) is suppressed via interference. This is seen explicitly in the green contour in \cref{fig:Production14TeV}. Since this cancellation is not expected to persist as scattering becomes non-single diffractive with more complex final states, ISR from the beam proton is predicted to provide a better approximation to the total rate. This rate was independently estimated using a quasi-real approximation for the incoming proton wit consistent results, and indeed leads to a rate that as shown in \cref{fig:Production14TeV} is similar in magnitude to rate obtained by evaluating a splitting function using the modified WW approximation in \cite{Blumlein:2013cua}.

\subsection{Additional Production Modes} \label{sec:llp_vec_prod}

We are planning to model dark photon production modes in meson decays, hadronization, initial state radiation (ISR), final state radiation (FSR) in \texttool{Herwig}~\cite{Bahr:2008pv,Bellm:2019zci} to study its relevance in various experimental setups. So far, it is assumed that dark photons $A'$ would mainly be produced in meson decays, e.g. $\pi^0\to \gamma A'$, Drell-Yan and in bremsstrahlung processes in proton-proton collisions and beam-dump experiments~\cite{Berlin:2018jbm}. Meson decays are described in the framework of vector meson dominance (VMD)~\cite{Sakurai:1960ju,Kroll:1967it,Lee:1967iv} to determine decays like $V\to P A'$, $P \to \gamma A'$ or via mixing $V\to A'$ where $V$ is a vector meson, $P$ a pseudoscalar meson and $\gamma$ the SM photon~\cite{Ilten:2018crw}. For dark photons heavier than the meson masses, i.e. $m_{A'}>m_{\pi,\eta,...}$, bremsstrahlung and Drell-Yan are so far the dominant production mode. The bremsstrahlung process is typically modelled by using a Fermi-Weiz\"{a}cker-Williams approximation~\cite{Blumlein:2013cua,deNiverville:2016rqh,Faessler:2009tn} which uses proton form factors including an off-shell mixing with vector mesons. This results in a relatively sharp enhancement in the production around the $\rho$ and $\omega$ mass $\sim 775$~MeV~\cite{Berlin:2018jbm}. The calculations for proton bremsstrahlung have recently been revised in~\cite{Foroughi-Abari:2021zbm} as discussed in the previous section. Above these masses, the production rate through bremsstrahlung drops sharply down to unobservably small values, and hence, additional production modes could increase the potential of dark photon searches.

In their EOI~\cite{Gligorov:2017nwh}, the CODEX-b collaboration has studied several production modes for axion-like particles (ALP) with couplings to gluons~\cite{Aielli:2019ivi} by using \texttool{Pythia~8}~\cite{Sjostrand:2014zea}. In particular, they include i) Hadron decays via mixing with the light pseudoscalar $\pi^0, \eta, \eta'$ mesons: for any hadron decay into neutral pseudoscalar mesons, there is the possibility that this hadron decays into an ALP. ii)  Additionally, an ALP can be produced in flavor-changing neutral current (FCNC) bottom and charm hadron decays as for example via $b\to s a$. iii) Axions can be radiated of any gluon in the parton shower. iv) Finally, axions can be produced during hadronization if a $q\bar{q}$ pair has pseudoscalar quantum numbers like $\pi^0, \eta, \eta'$ and hence, forms an axion through mixing into it. For the case of CODEX-b, the production in the parton shower turns out to play an important role and boosts the overall production rate by orders of magnitude. Their results could possibly also be applied to MATHUSLA~\cite{Chou:2016lxi,Curtin:2018mvb,MATHUSLA:2018bqv}, ATLAS, CMS, and LHCb. 

\begin{figure*}[t]
  \centering
  \includegraphics[width=0.9\textwidth]{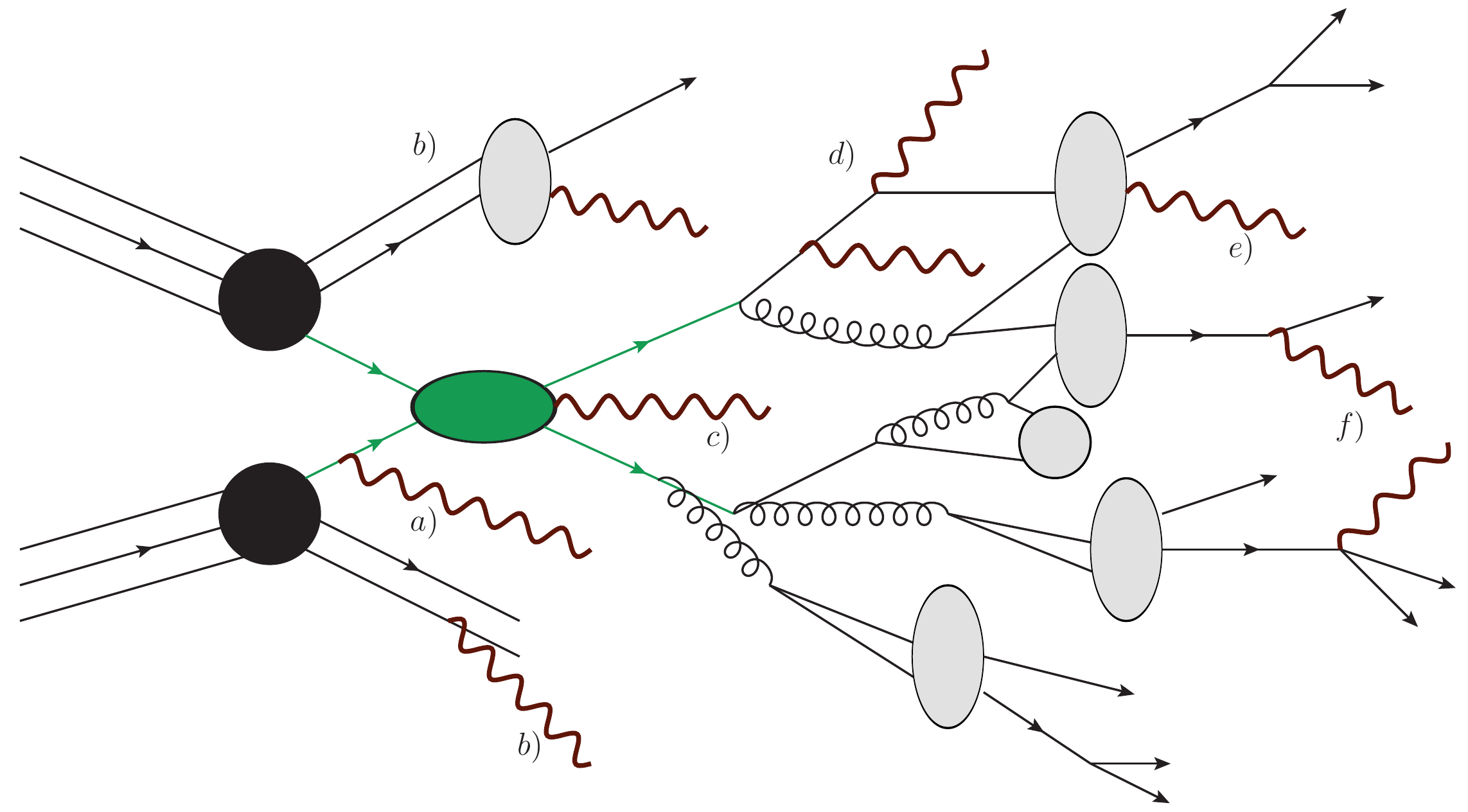}
  \caption{Dark Photon production in a) ISR, b) within the beam remnants, either by radiation or in hadronization, c) the hard process (green), d) parton shower radiation, e) hadronization, and f) decays of hadronic particles.  }
  \label{fig:DPproduction}
\end{figure*}

In the context of the FPF, the focus shifts more towards additional production modes in the forward direction. Let us take the dark photon as a model example. In \cref{fig:DPproduction}, we show several ways of producing a dark photon in proton-proton collisions, for example in the beam remnants, ISR, in the hard process, FSR, or in hadronization. 

\begin{figure*}[t]
  \centering
  \includegraphics[width=0.5\textwidth]{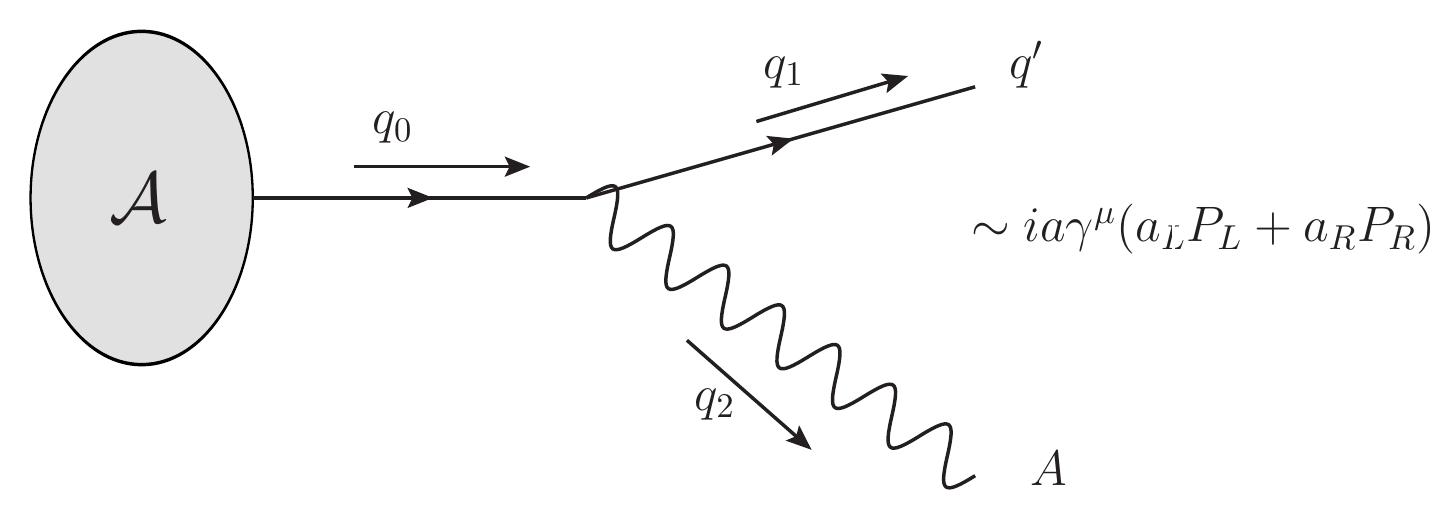}
  \caption{Diagram for a matrix element $\mathcal{A}$ with one $q \to q' A'$ splitting with left- and right-handed couplings $a_{L/R}$ and the coupling structure of a general vector splitting.  }
  \label{fig:DPsplitting}
\end{figure*}

To accurately describe these production modes, we can take advantage of the existing infrastructure available in MC simulators. In particular, we would like to use  \texttool{Herwig} to study the following dark photon production mechanisms:
\begin{itemize}
    \item ISR and FSR for parton splittings $q\to q A'$,
    \item hadronization where a quark pair $q\bar{q}$ hadronizes into $A'$, 
    \item hadron decay to $A'$.
\end{itemize}
For the implementation of ISR and FSR in \texttool{Herwig}, we can make use of the previous work~\cite{Masouminia:2021kne, Darvishi:2021het} where an angularly-ordered (AO) EW parton shower has been implemented into~\texttool{Herwig}. The general form for the $q\to q' V$ splitting function as given in~\cite{Masouminia:2021kne} and shown in \cref{fig:DPsplitting} can be easily translated into a splitting dark photon since its already been taken care of that all engaged particles have non-zero masses, especially the vector particle $V$, and additional longitudinal polarization states might be present. 

In fact, the dark photon with same couplings to left- and right-handed helicities $a_L=a_R=1$, represents a simplified version of the EW vector case. The explicit expression for the dark photon splitting function $P_{q\to qA}(z,\tilde{q})$ is then given by
\begin{align}
P_{q\to q
  A}(z,\tilde{q})=\frac{a^2}{1-z}\left[1+z^2-\frac{2m_q^2+m_A^2}{z\tilde{q}^2}\right]
\end{align}
with a light-cone momentum fraction $z$, evolution scale $\tilde{q}$ and quark and dark photon masses, $m_q$ and $m_{A'}$, respectively. The implementation of dark photon production in and after hadronization will probably follow the approach as done by the CODEX-b collaboration in~\cite{Aielli:2019ivi}. A big advantage of using \texttool{Herwig} for studying vector mediators in collision experiments is that we could not only describe the production of those vector bosons but also the decay using earlier results implemented into \texttool{Herwig}~\cite{Plehn:2019jeo}. Therein, in particular the hadronic decay of a vector particle with arbitrary couplings to quarks is described in the context of indirect detection~\cite{Bernreuther:2020koj, Reimitz:2021wcq}. But the implemented channels could easily be used in the context of dark photons in other experimental setups.

It is yet unclear how these mechanisms affect the overall production rate and as a consequence predictions for various experiments. For detectors in the forward direction of a beam-collision such as \texttool{FASER}, we expect that ISR of dark photons and the production within the hadronization of beam remnants could provide a sizable contribution. 

Our attempt to add dark photon production processes is of course not limited to vector particles produced in the forward direction. In experiments that are placed in a more transverse direction, dark photon radiation in the parton shower might yield a higher number of dark photon events. The results can be applied to various existing/proposed dark photon searches and experiments. We expect that many proton beam dump experiments, for example, would be sensitive to dark photons being produced in the shower, and hence their bounds would increase. Besides, not only vector particles could be produced. Since the project inherits the EW parton shower implementation of  Ref.~\cite{Masouminia:2021kne} including Higgs radiations, our study can, in principle, be extended to dark scalars as well.

\subsection{Decays of Light Vector Particles}
\label{sec:llp_vec_decay}

The majority of vector mediator models consider leptonic decays when searching for signatures in experiments made or used for the detection of light vector particles. For the dark photon model, the coupling strength to the SM can be constrained for a wide range of masses from a few MeV to $\mathcal{O}(100)$~GeV. In this specific model, all possible decays, and in particular the total decay width, can be extracted from $e^+e^-$ annihilations into leptons, quarks and hadrons since the dark photon kinetically mixes with the SM photon and hence, its couplings are proportional to the SM photon-fermion couplings. This yields to very accurate descriptions of dark photon decays and consequently predictions for experimental searches. Nevertheless, for the case of generic $U(1)$ gauge group models this strategy can not be conducted anymore due to different decay structures that will arise from differences in the mediator-meson couplings compared to the SM photon case. Especially in the low-energy range from MeV to $\sim 2$~GeV, hadronic decays might deviate largely from the dark photon case. In this range, the vector mediator $Z_Q^\mu$ directly mixes with the vector mesons $\rho, \omega,$ and $\phi$ through
\be
\mathcal{L}_{VZ_Q}=2\, g_{Q}  Z^\mu_Q{\rm Tr}\left[V_\mu Q^{f}\right]\, ,
\label{eq:BSMA9mixing}
\ee
with $V^\mu=T^a V^{a,\mu}$, where $T^a$ are $U(3)$ generators, $V^{a,\mu}$ the vector mesons with
\be
   \!\!\!
   \rho: \; \rho^{\mu} T_\rho =\rho^{\mu} \frac12 {\rm diag}(1,-1,0), \;\;
   \omega: \;  \omega^{\mu} T_\omega =\omega^{\mu}\frac12 {\rm diag}(1,1,0), \;\;
   \phi: \;  \phi^{\mu}T_\phi =\phi^{\mu}\frac{1}{\sqrt{2}}{\rm diag}(0,0,1).
   \!
   \label{eq:BSMA9mesons}
\ee
and $g_Q$ is the coupling constant. Whereas the dark photon with $Q^f=\text{diag}(2/3,-1/3,-1/3)$ couples to all of the mesons, most baryophilic $U(1)$ gauge bosons with $Q^f=\text{diag}(1/3,1/3,1/3)$ only couple to $\omega$ and $\phi$. The vector mesons further couple to other vector and pseudoscalar mesons which yields to a wealth of final state configurations. For example, the 2 and 4 pion channels for vector decays are purely driven by the mixing with the $\rho$ meson whereas the $\pi\gamma$ and 3 pion final states mostly come from the mixing with $\omega$ and $\phi$. A comprehensive list of final states can be found in Ref.~\cite{Foguel:2022ppx}. A change in the hadronic decays will even change the results for leptonic signatures, since limits and predictions do not depend on the partial decay width into leptons that can be calculated perturbatively, but on its relative contribution compared to other decays. Hence, to set robust constraints on these models, it is inevitable to precisely determine the branching ratio into all sorts of final states as well as the total decay width to determine the lifetime. That is the only way we can accurately predict where and how the vector boson will appear in the experimental setup.

As part the \texttool{DarkCast} tool presented in Ref.~\cite{Ilten:2018crw}, a first attempt has been made to describe hadronic channels individually, especially its splitting into $\rho$, $\omega$ and $\phi$ contributions, to not only describe dark photon decays, but general vector mediator models with arbitrary couplings to quarks. In this section, based on Ref.~\cite{Foguel:2022ppx}, we describe how we improve the description of several channels and extend the number of final state configurations compared to Ref.~\cite{Ilten:2018crw} by extracting the hadronic currents obtained in low-energy $e^+e^-$ fits. Some of the fits are taken from a previous study~\cite{Plehn:2019jeo}, and some are introduced for the first time. Decay widths, branching ratios, and other related decay quantities calculated in \cite{Foguel:2022ppx} are available in the python package \texttool{DeLiVeR}\footnote{The package is available at \href{https://github.com/preimitz/DeLiVeR}{https://github.com/preimitz/DeLiVeR} and includes a jupyter notebook tutorial.}. \medskip

\begin{figure*}[t]
\centering
\includegraphics[width=0.99\textwidth]{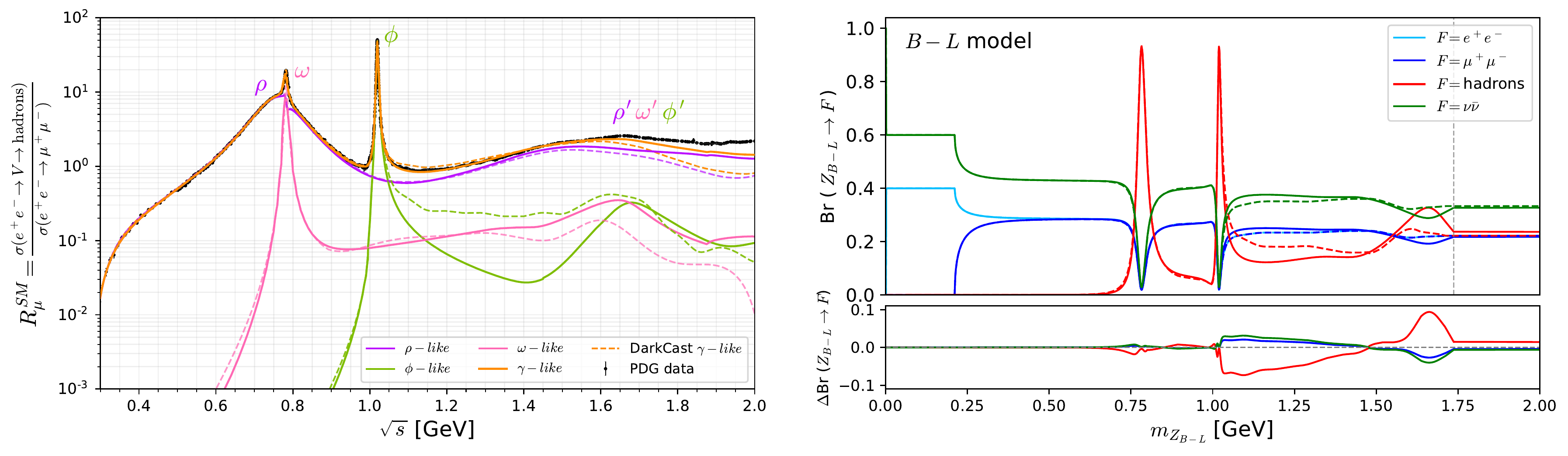}
\caption{Left: Total SM hadronic cross-section ratio $R_\mu^{SM} = \sigma(e^+ e^- \to \mathrm{hadrons})/\sigma(e^+ e^- \to \mu^+ \mu^-)$ decomposed into $\rho$ (purple), $\omega$ (pink) and $\phi$ (light green) contributions. The orange line represents the sum of these three curves and the black points are the experimental data extracted from the Particle Data Group \cite{ParticleDataGroup:2020ssz}. In both figures, the solid (dashed) lines indicate the results obtained using the hadronic calculation of this (\texttool{DarkCast}) work.  Right: In the upper panel we show the branching ratio of the $B-L$ model for $Z_{B-L}$ decaying into electrons (light blue), muons (dark blue), hadrons (red) and neutrinos (green). In the lower panel we have the deviation $\Delta$Br, i.e. the difference between the branching ratio calculation of this work minus the one from \texttool{DarkCast}. Figures extracted from~\cite{Foguel:2022ppx}.}
\label{fig:BSMA9_contributions_br}
\end{figure*}

As described in detail in Ref.~\cite{Foguel:2022ppx}, channels that have already been used in Ref.~\cite{Ilten:2018crw} are improved by considering contributions from more than one vector meson mixing with the vector mediator for each channel, by including more recent and complete data sets as well as rigorously following the vector meson dominance model (VMD)~\cite{Sakurai:1960ju, Kroll:1967it, Lee:1967iv} with only very little theoretical assumptions. As seen in the left panel of \cref{fig:BSMA9_contributions_br}, major differences arise in the $\omega$ and $\phi$ contributions when describing $e^+e^-$ annihilations to hadrons. In case of the $\omega$ meson (pink curve), this mainly comes from a different description of the $\pi\gamma$ channel for energies below the $\omega$ mass~\cite{SND:2016drm}, and because of including more processes that include the $\omega$ resonance or excited states thereof above 1.4~GeV. The big difference in the $\phi$ contribution (light green curve) between just above the $\phi$ resonance and 1.6~GeV can be traced back to a different handling of the $KK$ and $KK\pi$ channels. 

In \texttool{DarkCast}, it is assumed that $KK=2 K^+K^-$ which leads to an overestimation of the $KK$ contribution. Only attributing this channel then to the $\phi$ meson consequently results in an overestimation of the $\phi$ contribution. We fit both the charged $e^+e^- \to K^+K^-$ and $e^+e^-\to K^0\bar{K}^0$ components separately and consider contributions from $\rho$-like, $\omega$-like and $\phi$-like mesons. For the $KK\pi$ channel, \texttool{DarkCast} only considers the isoscalar subprocess $e^+e^-\to K^*(892) K$ that is responsible for $e^+e^-\to K^*(892) K \to KK\pi$. Nevertheless, besides not accurately describing the three-body phase space of $KK\pi$ by $K^*K$, the isovector contribution of $K^*K$ is missing. As before in the case of $KK$, the $KK\pi$ channel is assumed to be determined by the $\phi$ contribution only. We take into account the three components $K^0K^0\pi^0$, $K^+K^-\pi^0$ and $K^\pm K^0\pi^\mp$ for the description of the $KK\pi$ channel. In all the three processes, we have a isovector contribution that is assigned to the $\rho$ meson, and an isoscalar contribution that is described by the $\phi$ meson. 

As an example on how the different treatment of hadronic decays affects the vector mediator branching ratios, we discuss the case of the $B-L$ model. As seen from \cref{eq:BSMA9mixing}, in this model we only have a $\omega$ and $\phi$ contribution. In the right panel of \cref{fig:BSMA9_contributions_br}, we can see that slightly above the $\phi$ mass, our branching ratios are smaller due to the lower $\phi$ contribution. For mediator masses above 1.5~GeV instead, the bigger $\omega$ contribution of our calculations results in a bigger branching ratio into hadrons. Above 1.75~GeV, we do not trust VMD anymore and assume perturbative mediator decays into quarks. Overall, the difference between our branching ratios and the ones from \texttool{DarkCast} is below the 10\% level for hadrons. This difference is shared between electrons, muons and neutrinos. As a consequence, we observe that limits coming from $Z_Q\to e^+e^-, \mu^+\mu^-, \nu\bar{\nu}$ searches that reach over many order of magnitudes in the vector coupling will only change very little with differences around a few percent in the hadronic decays. Only for the case where one hadronic channel is dominating, as for example below the two-pion threshold in the $U(1)_B$ model, we can observe visible differences in the limits~\cite{Foguel:2022ppx}.

\begin{figure*}[t]
\centering
\includegraphics[width=0.99\textwidth]{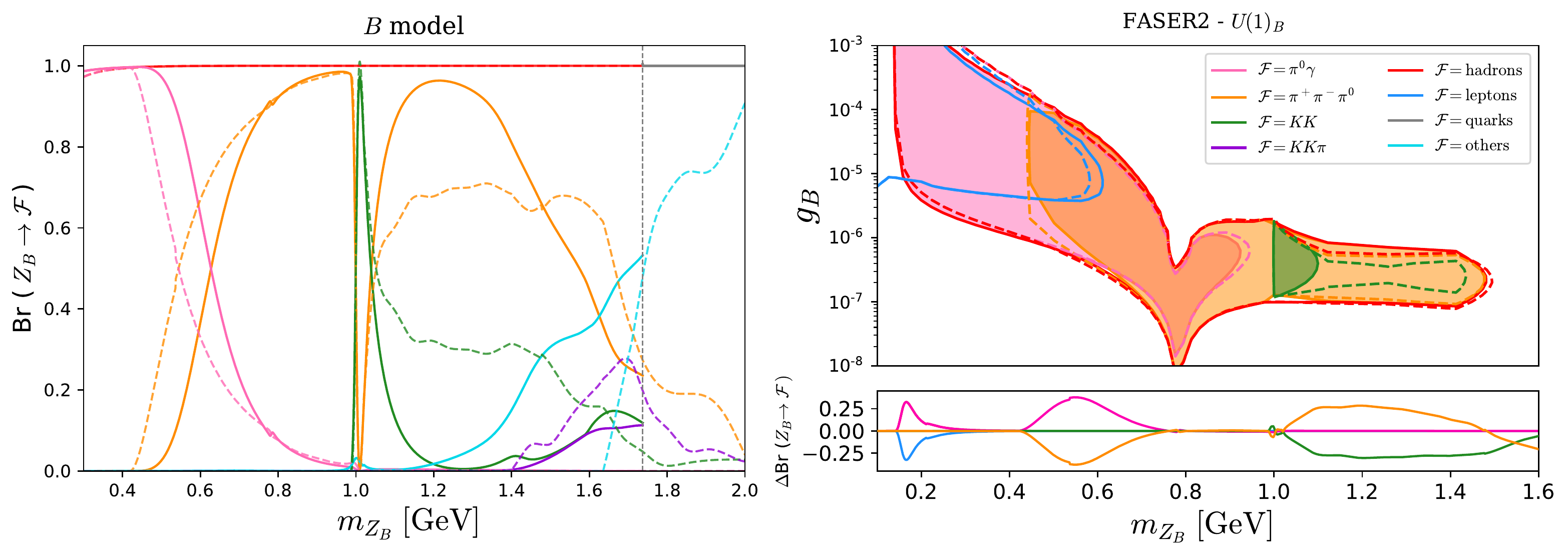}
\caption{Left: Branching ratios of the $B$ model for $Z_B$ decaying into hadrons (red) and individual hadronic states. The cyan curve represents the sum of all the other hadronic channels included in this work (see table III of Ref.~\cite{Foguel:2022ppx}). The grey dashed line close to $1.73$ GeV shows the transition between hadrons and the perturbative quark calculation. In both figures the solid (dashed) lines indicate the results obtained using the hadronic calculation of this (\texttool{DarkCast}) work. Right: In the upper panel we show the comparison between the future sensitivity of the FASER2 experiment for the $B$ model obtained by implementing our (solid lines) and \texttool{DarkCast} (dashed lines) branching ratios in the \texttool{FORESEE} code. In the lower panel we show the deviation $\Delta$Br of our branching ratios minus the ones from \texttool{DarkCast}. Figures extracted from~\cite{Foguel:2022ppx} }
\label{fig:BSMA9_br_FASER2}
\end{figure*}

Nevertheless, if we focus on signatures of a $Z_Q$ decaying into hadrons as suggested by Ref.~\cite{Batell:2021snh}, the limits might change drastically. As an example, we take the hadrophilic $U(1)_B$ model where the gauge boson couples directly to quarks but only through kinetic mixing with leptons. Hence, the mediator decays nearly a 100\% into hadrons above the $\pi\gamma$ channel threshold. Major differences in the individual hadronic channels are coming from $\pi\gamma$ and $KK$ which changes the relative contribution of all dominant channels as seen in the left panel of \cref{fig:BSMA9_br_FASER2}. With these branching ratios, we can now study the expected sensitivity for future experiments as for example FASER2. Whereas the overall sensitivity to all hadrons only changes mildly, the sensitivity to the $KK$ channel reduces only to the region around the $\phi$ mass instead of reaching from 1.0-1.4~GeV as shown in the right panel of \cref{fig:BSMA9_br_FASER2}. We can clearly see that the individual branching ratio differences of up to 30\% directly translates to different sensitivities. For deviations that cannot be explained by the branching ratios it is likely that they come from the difference of the vector mediator lifetime of both calculations.

In this section, we presented improvements in the calculation of hadronic decays of vector particles in vector mediator models. This includes providing an almost complete set of all possible hadronic and leptonic partial decay widths, branching ratios, and related vector mediator decay quantities in the python package \texttool{DeLiVeR}. Especially in decay channels related to vector mediators mixing with the $\omega$ and $\phi$ mesons, we observe significant differences between our calculations and \texttool{DarkCast}. Limits that range over several orders of magnitude in the vector coupling and mass and are based on $Z_Q$ decays into electrons, muons or neutrinos only change very little when having only a few percent differences in the corresponding branching ratios. Nevertheless, for regions that are dominated by one hadronic channel, the limits might differ. Moreover, if we study future sensitivies for hadronic decay signatures of vector mediators, we observe that, for example for the case of FASER2, the different treatment of hadronic decays will have a significant effect on the sensitivities.

\clearpage
\section{Long-Lived Scalars}
\label{sec:bsm_llp_scalar}

One of the most widely discussed examples of a renomalizable portal interaction between the SM and a dark sector is that of a scalar mediator. In this scenario, the dark sector contains a new singlet scalar $S$ which could obtain a small quartic couplings to the Higgs of the form $\epsilon S^2 H^2$. This quartic term will then induce a mixing between the singlet scalar and the SM Higgs boson. 

Let us for example consider the scenario in which both the both the SM Higgs and the singlet scalar have a quartic potential and obtain a vacuum expectatation value such that we can write $S = (v_s + s)/\sqrt{2}$ and $H = (v_h+h_{SM})/\sqrt{2}$. After diagonalization of the mass terms, the physical fields are $h$ and $\phi$ are given by
\be
 h = h_{SM} \cos\theta - s \sin\theta 
 \quad \text{and} \quad
 \phi = h_{SM} \sin\theta + s \cos\theta \ . 
\ee
In the above scenario, the mixing angle $\theta$ is given by $\theta \sim v_h/v_s$ which is constrained to be small to not violate current experimental constraints. 

Due to the mixing, the new physical scalar $\phi$ obtains Yukawa-like couplings to the SM fermions and gauge bosons, which are modified by the small mixing angle $\theta$. Its effective Lagrangian can then be written as 
\be
 \mathcal{L} \sim - m_\phi^2 \phi^2 - \sum_f \theta y_f \phi \bar f f \ . 
\ee
Due to some similarities with the dark photon case, the new scalar $\phi$ is also often called the dark Higgs. It is worth noting that the discussion presented above is not the only way to generate the mass and couplings of the dark Higgs. For a discussion of other scenarios, see Ref.~\cite{OConnell:2006rsp, Bezrukov:2009yw, Feng:2017vli}. 

\begin{figure*}[t]
\centering
\includegraphics[width=0.8\textwidth]{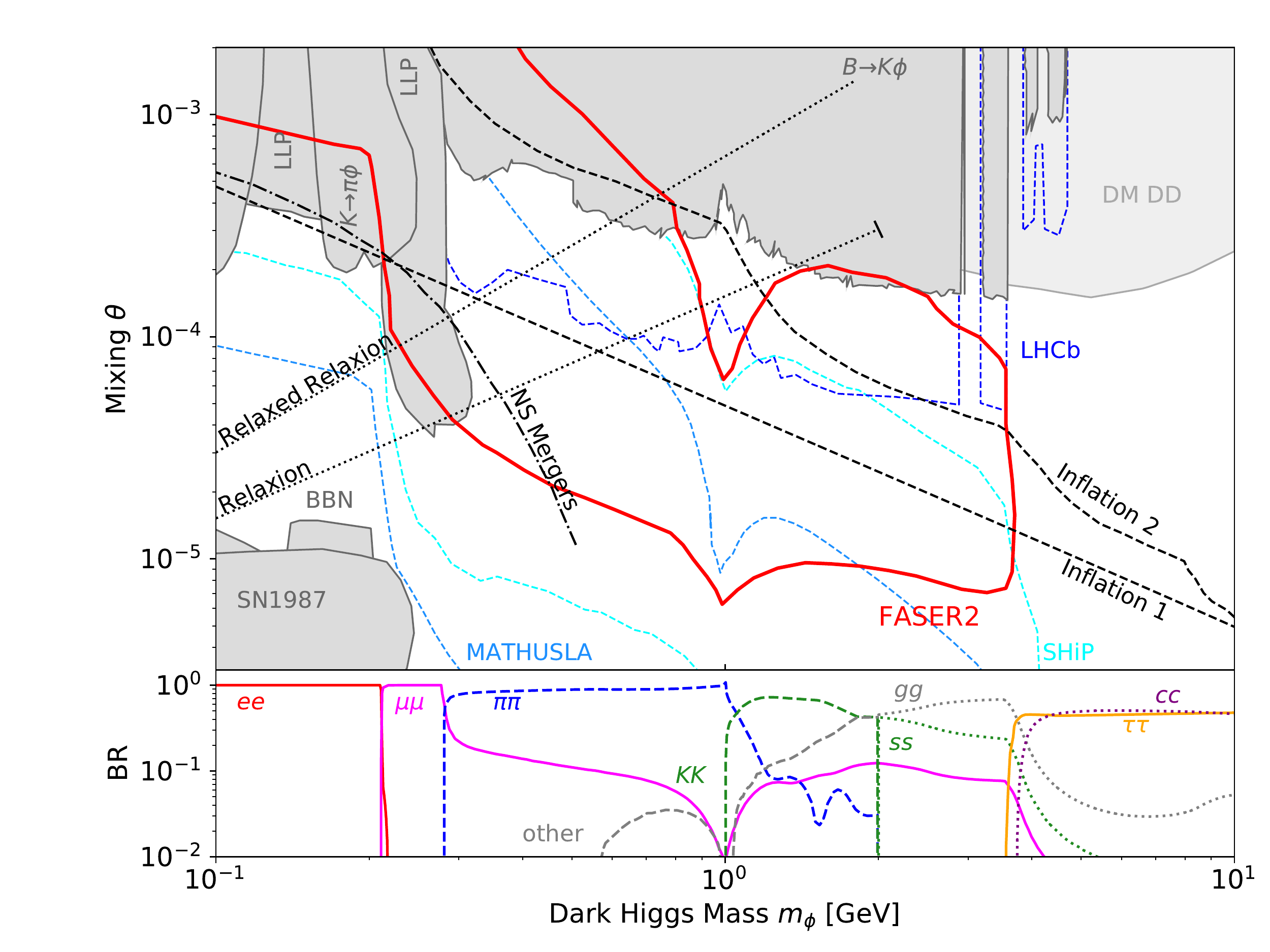}
\caption{Dark Higgs boson sensitivity in the coupling vs mass plane. The sensitivity reaches of FASER2 is shown as solid red lines alongside existing constraints (dark gray-shaded regions) and projected sensitivities of other selected proposed searches and experiments (blue dashed lines). The gray shaded region indicated regions of parameter constrained by dark matter direct detection searches in the case that the dark Higgs couples to dark matter with a coupling that reproduces the observed relic abundance.  The black lines shows target lines from scenarios in which the dark Higgs is an inflaton, a relaxion or could have observable effects on neutron star mergers. The bottom panel shows the LLP branching fractions, as obtained in Ref.~\cite{Winkler:2018qyg}. See text for details and references.}
\label{fig:bsm_scalar_dh}
\end{figure*}

The prospects of probing the dark Higgs scenario at FASER2 have been studied in Refs.~\cite{Feng:2017vli, FASER:2018eoc, Kling:2021fwx}. We present the resulting sensitivity in \cref{fig:bsm_scalar_dh}. Here the dark gray shaded regions correspond to constrains from previous searches for the dark Higgs boson. The reach of FASER2 is shown as red line. Several other dedicated experiments and searches have been proposed which would also have the opportunity to probe the dark Higgs parameter space. We present the projected sensitivities for a subset of those as dashed blue lines, while we refer for a more complete overview to Ref.~\cite{Beacham:2019nyx}.  In addition, we have included several target lines to illustrate the motivation by various theoretical models that give rise to a dark Higgs like particle
\begin{description}
 \item [Dark Matter] In the discussion above, the dark Higgs was introduced as a mediator between the visible and dark sector. Let us consider that the dark scalar couples to the DM particle $\chi$. If $m_\chi > m_\phi$ and for an appropriate choice of the scalar coupling to dark matter, secluded annihilation $\chi \chi \to \phi \phi$ can lead to the correct dark matter relic abundance via thermal freeze-out throughout the entire shown parameter space. In \cref{fig:bsm_scalar_dh} we show the current direct detection limits in this scenario assuming a fixed mass ratio $m_\chi = 3m_\phi$~\cite{DarkSide:2018kuk, CRESST:2019jnq, XENON:2018voc}. See Ref.~\cite{Feng:2017vli} for more details. 
 \item [Relaxion] In addition, a dark Higgs like scalar also arises in relaxion models, which has been introduced as a solution to the hierarchy problem. Target lines corresponding to two realizations of the relaxion are shown as dotted lines. For a more detailed discussion of this scenario, see \cref{sec:bsm_relaxion1} and \cref{sec:bsm_relaxion2} below. 
 \item [Inflation] A dark Higgs can also play the role of a light inflaton~\cite{Bezrukov:2009yw}. Several specific models have been introduced in this context, and corresponding target lines are shown as dashed lines in \cref{fig:bsm_scalar_dh}. The line labeld as \textit{Inflation~1} corresponds to model where the inflation potential exhibits classical conformal invariance, which is broken radiatively via the Coleman-Weinberg mechanism. A more detailed explanation is provided in \cref{sec:bsm_inflaton1}. A second theory, labeled \textit{Inflation~2}, considers a low-scale inflaton-curvaton model. The curve corresponds to the mass and mixing angle for an inflaton or curvaton that decays when the universe reaches a density around the electroweak scale of $\rho \sim (100~\gev)^4$~\cite{Bramante:2016yju}.
 \item [Neutron Star Mergers] If the dark Higgs is sufficiently light, it can be abundantly produced inside neutron stars. If additionally the mixing is small, the scalar has a large mean free  path and can contribute sizably to the thermal conductivity of the neutron star. Such modifications of the thermal conductivity of neutron stars could lead to observable signals in neutron star merger events recorded by gravitational wave telescopes. The dash-dotted target line in \cref{fig:bsm_scalar_dh} denotes the phase space region where the presence of the scalar contributes about 10\% to the thermal conductivity of the neutron star. A more detailed discussion of the dark Higgs is presented in \cref{sec:bsm_nsmerger}.
\end{description}

The remainder of this section is organized as follows: In \cref{sec:bsm_dh_constraints} we will summarize the existing constraints on the dark Higgs parameter space. We will then in more detail discuss several theoretical motivations: the dark Higgs as relaxion discussed in \cref{sec:bsm_relaxion1} and \cref{sec:bsm_relaxion2}, the dark Higgs as inflaton in \cref{sec:bsm_inflaton1}, and the phenomenological impact of a dark Higgs boson in neutron star mergers in \cref{sec:bsm_nsmerger}. Finally, we will then present  several other examples of light scalar particles and discuss the sensitivity of the FPF experiments to search for them. This includes a flavour specific scalar in \cref{sec:bsm_dh_flav}, a light scalar in the 2HDM in \cref{sec:bsm_dh_2hdm}, a supersymmetric sgoldstino in \cref{sec:sgoldstino}, and a dilaton in \cref{sec:bsm_dh_dilaton}.

\subsection{Existing Constraints on the Dark Higgs}
\label{sec:bsm_dh_constraints}

\begin{figure*}[t]
\centering
\includegraphics[width=0.508\textwidth]{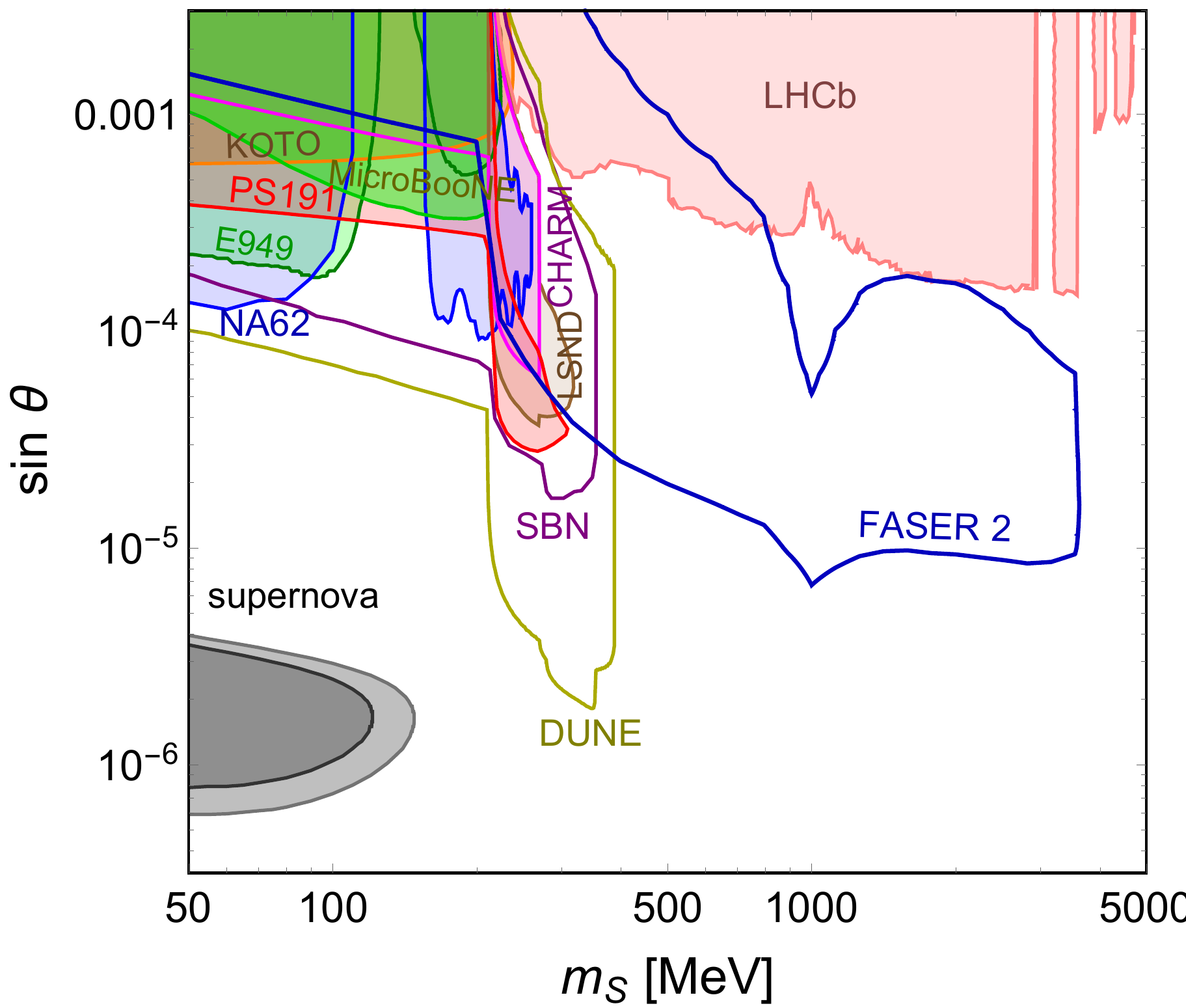}
\includegraphics[width=0.485\textwidth]{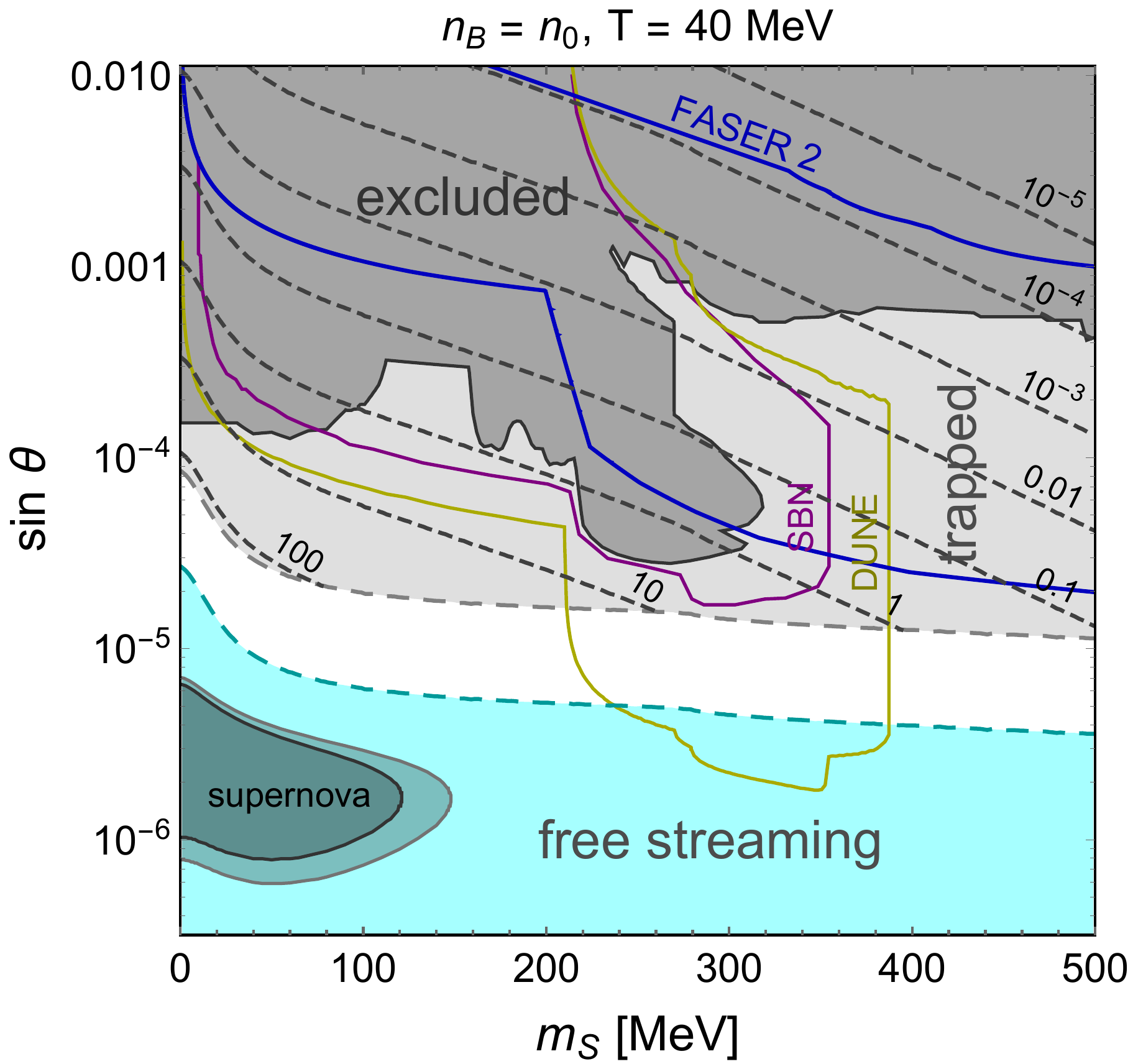}
\caption{Left: The current laboratory~\cite{Egana-Ugrinovic:2019wzj, Dev:2019hho, Foroughi-Abari:2020gju, Gorbunov:2021ccu, MicroBooNE:2021ewq, Anchordoqui:2021ghd} and astrophysical~\cite{Krnjaic:2015mbs, Dev:2020eam} constraints in the dark scalar mass-mixing plane (shaded regions). The future prospects at FPF (FASER2)~\cite{Anchordoqui:2021ghd}, as well as SBN~\cite{Batell:2019nwo} and DUNE~\cite{Berryman:2019dme} are shown as the solid purple, green, and dark blue lines, respectively. Right: Thermal conductivity ratio $\kappa_{S}/\kappa_{\nu}$ in the $m_S - \sin\theta$ plane, for the baryon density of $n_B =n_0$ and temperature $T = 40$ MeV in the NS merger. The current laboratory limits combined together are shown as the dark gray-shaded region. The future prospects at SBN, DUNE and FASER2 are shown as the solid purple, green, and light blue lines respectively. The grey and cyan shaded regions indicate the trapped and free-streaming regions inside the NS merger, corresponding to a scalar MFP less than 100 m and larger than 1 km, respectively. Figure taken from Ref.~\cite{Dev:2021kje}. \label{fig:bsm_scalar_limit}}
\end{figure*}

Before proceeding to particular physics realization of the dark Higgs and other light scalars, let us pause and review the existing constraints on the parameter space. The landscape of the dark Higgs $S$, spanned by its mass $m_S$ and mixing $\sin\theta$, is shown in the left panel of \cref{fig:bsm_scalar_limit}. As we can see, FASER2 at the FPF can detect light long-lived scalars produced in the high-energy $pp$ collisions up to a few GeV scalar masses, as shown by the FASER2 target line~\cite{FASER:2018eoc, Anchordoqui:2021ghd}. For comparison, we also show the existing limits (shaded regions) from flavor-changing neutral current (FCNC) meson decay searches and beam-dump experiments. 

The scalar $S$ can obtain FCNC couplings to the SM quarks at one-loop level via its mixing with the SM Higgs, and thus contribute to rare FCNC decays of mesons, such as $K \to \pi +X$, $B \to K + X$ and $B \to \pi + X$ with $X = e^+ e^-,\, \mu^+ \mu^-,\, \gamma\gamma$ or missing energy if $X$ decays outside the detector. There are stringent limits from $K^+ \to \pi^+ e^+ e^-$ and $\pi^+ \mu^+ \mu^-$ in NA48/2~\cite{Batley:2009aa,Batley:2011zz}; $K^+ \to \pi^+ \gamma\gamma$ and $\pi^+ \nu \bar{\nu}$ at NA62~\cite{Ceccucci:2014oza, NA62:2020fhy} and E949~\cite{Artamonov:2009sz}; $K_L \to \pi^0 e^+ e^-$, $\pi^0 \mu^+ \mu^-$ and $\pi^0 \gamma \gamma$ in KTeV~\cite{AlaviHarati:2003mr, AlaviHarati:2000hs, Alexopoulos:2004sx, Abouzaid:2008xm}; $K_L \to \pi^0 \nu \bar{\nu}$ and $\pi^0 X$ (with $X$ being a long-lived particle) at KOTO~\cite{Ahn:2018mvc}; and $B \to K +X$ with $X = e^+ e^-,\, \mu^+ \mu^-,\, \nu \bar\nu$ at LHCb~\cite{Aaij:2012vr},  BaBar~\cite{Aubert:2003cm,Lees:2013kla} and Belle~\cite{Wei:2009zv}. Similarly, the MicroBooNE collaboration has recently performed a search for light monoenergetic scalars from kaon decay at rest $K^+ \to \pi^+ + S$ with $S \to e^+ e^-$~\cite{MicroBooNE:2021ewq}. The most stringent limits coming from NA62, E949, KOTO, MicroBooNE and LHCb experiments are shown as the shaded blue, light green, orange, dark green and pink regions respectively. For the detailed calculations of the FCNC decays and the derivation of these limits, see e.g.~Refs.~\cite{Dev:2017dui, Dev:2019hho, Egana-Ugrinovic:2019wzj}.

The light scalar $S$ can also be produced in the high-intensity beam-dump experiments. The current most stringent limits come from kaon decays $K \to \pi + S$ in the CHARM experiment~\cite{CHARM:1985anb}, as shown by the magenta shaded region. At the LSND experiment, the scalar $S$ is predominately produced by the proton bremsstrahlung process, and the current LSND electron and muon data~\cite{LSND:2001aii, LSND:2003mon} have excluded the brown shaded region~\cite{Foroughi-Abari:2020gju}. The data from the fixed target experiment PS191 have also be reinterpreted in Ref.~\cite{Gorbunov:2021ccu} to set limits on the light scalar from kaon decays $K \to \pi + S$, as shown by the red shaded region. It is expected that the future high-intensity beam-dump experiments will  significantly improve these limits. For instance, the Fermilab SBN experiment, which is a combination of using the SBND and ICARUS experiments with the BNB and NuMI beams respectively, can probe a mixing angle down to $\sim 10^{-7}$~\cite{Batell:2019nwo}, as shown by the purple curve. The near detector at DUNE can further improve the prospects by over an order of magnitude~\cite{Berryman:2019dme}, as shown by the green curve.

On the astrophysical side, the scalar $S$ can be produced abundantly in the supernova cores with temperature $T \sim 30$ MeV via the nucleon bremsstrahlung process. If it escapes the supernova core, it will contribute an additional channel for energy loss; therefore, the observed neutrino luminosity ${\cal L}_\nu$ of SN1987A can be used to set limits on the mixing angle $\sin\theta$ of $S$ with the SM Higgs~\cite{Ishizuka:1989ts, Diener:2013xpa, Krnjaic:2015mbs, Lee:2018lcj, Arndt:2002yg, Dev:2020eam}. Setting conservatively ${\cal L}_\nu = 3\times 10^{53}$ erg/sec and $5\times 10^{53}$ erg/sec, the corresponding supernova limits are shown in \cref{fig:bsm_scalar_limit} respectively as the lighter and darker black shaded regions. One can also derive similar limits~\cite{Dev:2020jkh} from energy loss criteria in the Sun, white dwarfs, red giants and horizontal-branch stars; however, their core temperature is only ${\cal O}({\rm keV})$, and therefore, the corresponding limits are not relevant for the parameter space shown in \cref{fig:bsm_scalar_limit}. 

On the cosmology front, if the light $S$ stays in thermal equilibrium with the SM particles and has a lifetime $\gtrsim 1$ sec, it will contribute an extra degree of freedom and spoil the success of big bang nucleosynthesis (BBN)~\cite{Kainulainen:2015sva, Fradette:2017sdd}. However, it turns out that there is no parameter space of $m_S$ and $\sin\theta$ that satisfies both the conditions above, so the generic BBN limit is not applicable here~\cite{Dev:2020eam}. The actual BBN constraint will depend on the particular thermal history of the scalar $S$ in the early universe and can in principle be evaded, without affecting the late-time scalar phenomenology we are interested in here. Therefore, we do not show the BBN limit in \cref{fig:bsm_scalar_limit}. However, if the scalar has an additional trilinear coupling to Higgs, it can be abundantly produced via rare Higgs decays $h \to SS$ and reach thermal equilibrium. The corresponding constraints are shown in \cref{fig:bsm_scalar_dh}, where we note that the exact constraint only mildly depends on the additional Higgs-scalar coupling~\cite{Fradette:2017sdd}. 

\subsection{Dark Higgs as Relaxion}
\label{sec:bsm_relaxion1} 

Another motivation for a singlet scalar is provided by the relaxion mechanism which was introduced as a solution to the (little) hierarchy problem of the Standard Model~\cite{Graham:2015cka}. An initially large Higgs boson mass $M\gg v$ is reduced to the observed mass $m_h=\mathcal{O}(v)$ through the dynamical evolution of the relaxion $\phi$ in the early universe. The Higgs-relaxion potential reads
\begin{equation}\label{eq:relaxion}
V = (M^2 - g_1 M \phi) \,h^2 - g_2 \,M^3 \phi - \Lambda^2 \,h^2\, \cos\left(\frac{\phi}{f}\right)+\lambda\,h^4\,,
\end{equation}
where $g_{1,2}$ are dimensionless couplings of the same order and $h$ is the neutral component of the Higgs doublet. The cosine-term is generated by the coupling of the relaxion $(\phi/f)G^\prime_{\mu\nu}\tilde{G^\prime}^{\mu\nu}$ to the field strength of a new strongly coupled gauge group QCD' whose strong dynamics set the scale $\Lambda$.

During cosmic inflation, the relaxion rolls down its potential, before triggering electroweak symmetry breaking at $\phi \sim M/g_1$. The Higgs field gets displaced, turns on the periodic potential of the relaxion, and the Hubble friction then stops the relaxion in one of the resulting minima. The relaxion mass and the Higgs-relaxion mixing angle $\sin\theta$ in the minimum can be approximated as~\cite{Flacke:2016szy, Winkler:2018qyg}
\begin{align}\label{eq:higgsrelaxionmixing}
 m_\phi^2 \simeq \frac{\Lambda^2 v^2}{2 f^2} \left[\cos\left(\frac{v_\phi}{f}\right)-\frac{2\Lambda^2}{m_h^2}\sin^2\left({\frac{v_\phi}{f}}\right)\right] 
 \quad \text{and} \quad
 \sin\theta& \simeq \frac{\Lambda^2 v}{f m_h^2}\sin\left(\frac{v_\phi}{f}\right)\,,
\end{align}
where $v_\phi$ denotes the final relaxion field value. For the Hubble scale $H \gtrsim \sqrt{g_2\,M^3/f}$, the relaxion immediately stops in one of its first minima after electroweak symmetry breaking~\cite{Winkler:2018qyg} and one finds $\sin(v_\phi/f)\sim \cos(v_\phi/f)\sim 1/\sqrt{2}$ such that,
\begin{equation}\label{eq:higgsrelaxionmixing2}
 \sin\theta \simeq \frac{2^{1/4} \, m_\phi \, \Lambda }{ m_h^2}\,.
\end{equation}
Validity of the effective theory~\cref{eq:relaxion}, furthermore requires\footnote{We assumed $\Lambda\ll v$ for the derivation of~\cref{eq:higgsrelaxionmixing2}.} $f \gtrsim v \gtrsim \Lambda$ which implies $m_\phi \lesssim \Lambda$. On the other hand, $\Lambda$ below the GeV-scale is cosmologically unfavorable as the relaxion mechanism requires an enormous number of inflationary e-foldings in this case~\cite{Choi:2016luu}. The FPF will be able to probe a large part of the relaxion parameter space. In \cref{fig:bsm_scalar_dh} we used~\cref{eq:higgsrelaxionmixing2} with $\Lambda=2~\gev$ to provide an explicit FPF target line, which is labeled as \textit{Relaxion}.

\subsection{Dark Higgs as Relaxed Relaxion}
\label{sec:bsm_relaxion2} 

In Ref.~\cite{Banerjee:2020kww} it was noted that generically the relaxion would stop at the first minimum, which is characterised by a shallower potential compared with the one described by the single cosine term (of a generic axion-like models), while the mixing angle with the Higgs remains unmodified. It results in an "unnaturally" large value for the mixing angle, basically, as the relaxion is generically relaxed to an unnatural point in its parameter space, and was dubbed as the "relaxed-relaxion" with a mass-mixing-angle relation of 
\begin{align}
    m_\phi^2 \simeq n\delta \frac{\Lambda_{\rm{br}}^4}{f^2} 
    \quad \text{and} \quad 
    \sin\theta \simeq \frac{\epsilon}{\sqrt{n}\delta} \frac{m_\phi^2}{m_h^2}\,, \label{eq:relaxed_relaxion}
\end{align}
where $n$ denotes the $n$th minimum, $\epsilon = \left( \Lambda_{\rm{br}} / 4\right)^4$ and $\delta = \Lambda_{\rm{br}} / (v\Lambda)$. The upper bound of the relaxion mixing angle for a relaxion stopping in the first or a generic minimum is given by
\begin{equation}
\sin\theta\leq  (m_\phi / v)^{2/3}\,.
\end{equation}
In addition, there is a lower bound on the mixing angle, that obtained from requiring $f\gtrsim\Lambda\gtrsim \,4\pi v$:
\begin{equation}
   \sin\theta \geq \left({ m_\phi\over n^{1\over4} v}\right)^{4\over3}\,.
\end{equation}
This gives rise to a band of the natural parameter space of the relaxion. In \cref{fig:bsm_scalar_dh}, we show the lower limit as target line labeled as \textit{Relaxed Relaxion}. The upper limit would be found at larger couplings beyond those shown in the figure. As we can see, the FPF will be able to probe a the relaxion band in the mass range above $300~\mev$. The phenomenology and collider bounds of the long-lived relaxion with a mass of several GeV have been analyzed in Ref.~\cite{Fuchs:2020cmm}.

\subsection{Dark Higgs and Neutron Star Mergers}
\label{sec:bsm_nsmerger}

If the dark Higgs/singlet scalar $S$ is sufficiently light,  it can be readily produced in the hot and dense nuclear matter in the cores of stars, supernovae and neutron stars via the nuclear bremsstrahlung process $NN\to NNS$, where the $S$ can couple to any of the four external nucleon legs or to the intermediate pion mediator. Note that a crucial difference between the production of dark Higgs and that of axion/ALP is that the latter being a CP-odd particle can only be emitted from the nucleon legs but not from the mediator pions at leading order. If the scalar coupling with the SM particles (induced via its mixing $\theta$ with the SM Higgs) is sufficiently weak, it will 
free-stream out of the astrophysical bodies, thus providing an additional energy loss mechanism that would affect their evolution. On the other hand, if the coupling is sufficiently strong, the scalar particle is trapped, thereby contributing to the transport properties of the astrophysical core. 

An interestingly new astrophysical probe of the light scalar parameter space of interest comes from neutron star (NS) mergers~\cite{Dev:2021kje}.  Depending on the thermodynamic conditions in the merger, the scalar can have a wide range of mean free path (MFP). Even after taking into account all the laboratory and supernova limits on the scalar, there is still some overlapping parameter space with FPF where the scalar MFP is small enough for it to be trapped inside the merger. In this case, the scalar $S$ will form a Bose gas and contribute to thermal conductivity of the merger remnant. In the absence of any physics beyond the standard model, the neutrinos dominate the thermal transport inside the merger remnant \cite{Alford:2017rxf}.  The thermal conductivities due to the trapped scalars ($\kappa_S$) and trapped neutrinos ($\kappa_\nu$) are compared in the right panel of \cref{fig:bsm_scalar_limit} for a particular choice of the baryon density $n_B$ equal to the nuclear saturation density $n_0$ and for a merger temperature $T=40$ MeV (thermodynamic conditions which simulations reliably indicate that mergers reach \cite{Perego:2019adq}), as shown by the black dashed contours. The results for other choices of density and temperature can be found in Ref.~\cite{Dev:2021kje}. In places where the ratio $\kappa_S/\kappa_\nu$ is larger than one, fast thermal equilibration is a signature of BSM physics, as no SM mechanism yields such a high thermal conductivity. This effect can potentially lead to observable signals in future NS merger events, but this needs to be further studied with the aid of updated merger simulations including the transport of trapped scalars. It is remarkable that some of the regions with $\kappa_S / \kappa_\nu >1$ can be directly probed at the future laboratory experiments such as SBN, DUNE and FASER2, thus providing a nice complementarity between the laboratory and astrophysical probes of dark Higgs. 

\subsection{Dark Higgs as Inflaton}
\label{sec:bsm_inflaton1}

We consider the non-minimal quartic inflation in a classically conformal U(1)$_X$ extended SM~\cite{Oda:2017zul}. By imposing the conformal invariance at the classical level on the minimal U(1)$_X$ extended SM \cite{Oda:2015gna}, all mass terms in the Higgs potential are forbidden. As a result, the U(1)$_X$ gauge symmetry is radiatively broken by the Coleman-Weinberg (CW) mechanism~\cite{Coleman:1973jx}, which subsequently drives the electroweak symmetry breaking through a mixing quartic coupling between the U(1)$_X$ Higgs $\Phi$ and the SM Higgs $H$ fields~\cite{Iso:2009ss}. We identify the U(1)$_X$ Higgs field with a non-minimal gravitational coupling as inflaton. Because of the classically conformal invariance, this scenario not only leads to the inflationary predictions consistent with the Planck 2018 results~\cite{Planck:2018jri}, but also provides a direct connection between the inflationary predictions and the LHC search for the U(1)$_X$ gauge boson ($Z^\prime$) resonance. We will show that the inflaton mass and its mixing angle with the SM Higgs field lie in a suitable range, the inflaton can be searched by FASER2 with a direct connection to the inflationary predictions. Therefore, three independent experiments, namely, the inflaton search at FASER2, the $Z^\prime$ boson resonance search at the HL-LHC and the precision measurement of the inflationary predictions, are complementary to test our inflation scenario. 

\begin{figure*}[t]
\centering
   \includegraphics[width=0.42\textwidth, trim={0 -1.3cm 0 0},clip]{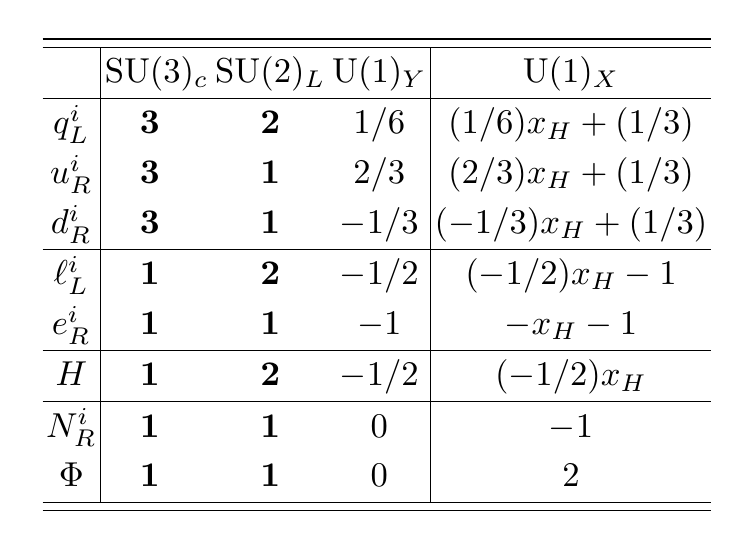} 
   \includegraphics[width=0.50\textwidth]{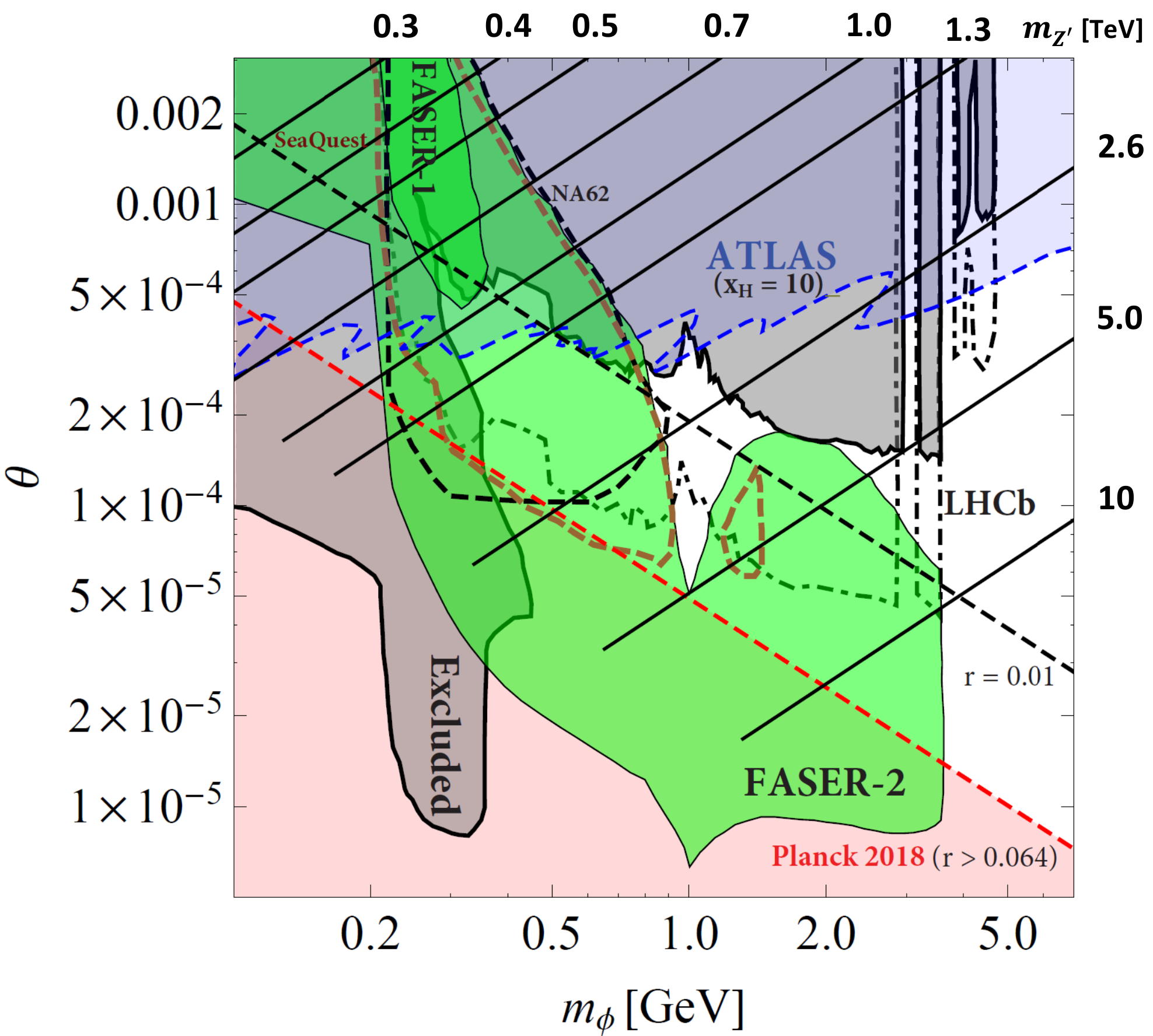}   
\caption{Left: The particle content of the minimal U(1)$_X$ model. Here, $i = 1,2,3$ is the generation index. Right: The inflaton search reach at FASER (green shaded region) and the relation with other observables. The diagonal dashed lines correspond to $\xi = 9.8 \times 10^{-3}$ ($r = 0.064$) and $\xi = 0.12$ ($r = 0.01$), respectively, from left to right. The diagonal solid lines correspond to fixed $Z^\prime$ masses. The blue shaded region (labeled ATLAS) is excluded by the ATLAS result of the $Z^\prime$ boson search for $x_H = 10$. The red shaded region is excluded by the Planck 2018 measurements.}
\label{fig:bsm_scalar_inflaton}
\end{figure*}

The particle content of the model is listed in the left panel of of~\cref{fig:bsm_scalar_inflaton}, where the U(1)$_X$ charge of a particle is defined as a linear combination of its SM hypercharge and its ${B-L}$ (Baryon minus Lepton) number. The U(1)$_X$ charges are determined by a real parameter, $x_H$, and the well-known minimal U(1)$_{B-L}$ model is realized as the limit of $x_H \to 0$. In the presence of the three right-hand neutrinos (RHNs), $N_R^{1,2,3}$, this model is free from all the gauge and the mixed gauge-gravitational anomalies. Once the U(1)$_{X}$ Higgs field ($\Phi$) develops a vacuum expectation value (VEV), $\langle \Phi \rangle = v_X/\sqrt{2}$, the U(1)$_{X}$ gauge symmetry is broken and the $Z^\prime$ boson becomes massive, $m_{Z^\prime} = 2 g_{X} v_{X}$, where $g_X$ is the U(1)$_X$ gauge coupling. The symmetry breaking also generates Majorana masses for the RHNs which play a crucial role in the type-I seesaw mechanism for generating the light neutrino masses.   

\textit{a) Radiative $U(1)_X$ Symmetry Breaking:} Imposing the classical conformal invariance, the Higgs potential of our model at the tree level is given by
\be
V = \lambda_H \!\left(  H^{\dagger}H  \right)^2
+ \lambda_{\Phi}\! \left(  \Phi^{\dagger} \Phi   \right)^2
- \lambda_{\rm mix} \!
\left(  H^{\dagger}H   \right) \!
\left(  \Phi^{\dagger} \Phi  \right), \qquad 
\label{eq:Higgs_Potential}
\ee
where we set $\lambda_{{H,\Phi},{\rm mix}} > 0$. Assuming $\lambda_{\rm mix} \ll1$, we can separately analyze the Higgs potential for $\Phi$ and $H$. The CW potential for the Higgs field $\Phi$ at the 1-loop level is given by~\cite{Coleman:1973jx}
\be
   V(\phi) =  \frac{\lambda_\Phi}{4} \phi^4 
     + \frac{\beta_\Phi}{8} \phi^4 \left(  \ln \left[ \frac{\phi^2}{v_{X}^2} \right] - \frac{25}{6} \right), 
\label{eq:CW_potential} 
\ee
where $\phi = \sqrt{2}\Re[\Phi]$, $v_{\rm X}$ is chosen as a renormalization scale, and the coefficient of the 1-loop corrections is approximately given by $16 \pi^2 \beta_\Phi \simeq 96 g_{X}^4  - 3 Y_M^4$. The stationary condition, $\left. dV/d\phi\right|_{\phi=v_{X}} = 0$, leads to  
\be
    {\overline {\lambda_\Phi}} = \frac{11}{6} {\overline {\beta_\Phi}},
\label{eq:stationary}
\ee
where the {\it barred} quantities are evaluated at $ \langle \phi \rangle = v_X$. The mass of $\phi$ is given by 
\be
  m_\phi^2 &=& \left. \frac{d^2 V}{d\phi^2}\right|_{\phi=v_{X}}    
  = {\overline {\beta_\Phi}} v_{X}^2 
  \simeq  \frac{6}{\pi} {\overline{\alpha_{X}}}m_{Z^\prime}^2, 
\label{eq:mass_phi}
\ee
where $\alpha_{X} = g_{X}^2/(4 \pi)$, and we have neglected RHN contributions. The U(1)$_X$ gauge symmetry breaking by $\langle \Phi \rangle = v_X/\sqrt{2}$ induces a negative mass squared for the SM Higgs doublet in \cref{eq:Higgs_Potential} and triggers the electroweak symmetry breaking~\cite{Iso:2009ss}. The SM(-like) Higgs boson mass ($m_h = 125~\gev$) is described as $m_h^2 = \lambda_{\rm mix} v_{X}^2 = 2 \lambda_H v_h^2$, where $v_h = 246~\gev$ is the Higgs doublet VEV. The mass matrix for the Higgs bosons, $\phi$ and $h$, is given by
\be
 {\cal L}  \supset - \frac{1}{2}
 \begin{bmatrix} h & \phi \end{bmatrix}
 \begin{bmatrix} 
    m_{h}^2 &  \lambda_{\rm mix} v_{X} v_{h} \\ 
   \lambda_{\rm mix} v_{X} v_{h} & m_{\phi}^2
 \end{bmatrix} 
 \begin{bmatrix} h \\ \phi \end{bmatrix}.  
\label{eq:massmatrix}
\ee
We diagonalize the mass matrix by 
\be
  \begin{bmatrix} h \\ \phi \end{bmatrix}
  =
  \begin{bmatrix} 
     \cos\theta &   \sin\theta \\ 
    -\sin\theta & \cos\theta  
  \end{bmatrix} 
  \begin{bmatrix} {\tilde h} \\ {\tilde \phi} \end{bmatrix}  ,
\label{eq:eigenstate}
\ee
where ${\tilde h}$ and ${\tilde \phi}$ are the mass eigenstates. For the parameter region which will be searched by FASER,  the mixing angle $\theta \simeq v_h/v_X \ll 1$ and the mass eigenvalues $m_{\tilde{\phi}, {\tilde{h}}}  \simeq m_{\phi, h}$ with $m_\phi^2  \ll m_h^2$. 

\textit{b) Non-minimal U(1)$_X$ Higgs inflaton:} Our inflation scenario is defined by the action in the Jordan frame:  
\be
 {\cal S}_J &=& \int d^4 x \sqrt{-g} 
   \left[-\frac{1}{2} (1+ 2 \xi \Phi^\dagger \Phi)  {\cal R}+ g^{\mu \nu} \left(D_\mu \Phi \right)^\dagger \left(D_\nu \Phi \right) 
 - V  \right]  , 
\label{eq:S_J}
\ee
where the non-minimal gravitational coupling constant $\xi > 0$, and the reduced Planck mass of $M_P=2.44 \times 10^{18}~\gev$ is set to be 1 (Planck unit). In our analysis, we fix the number of e-folds $N_0=50$ to solve the horizon and flatness problems. In this non-minimal quartic inflation, all the inflationary predictions are determined as a function of $\xi$. Once $\xi$ is fixed, the quartic coupling $\lambda_\Phi$ at the inflationary scale is uniquely determined. Extrapolating the $\lambda_\Phi$ to its value at $v_X$ according to the renormalization group equations and using \cref{eq:stationary}, we obtain a one-to-one correspondence between the set of two free parameters, $\{\xi, v_X\}$ and $\{m_{\phi}, \theta\}$. 

\textit{c) Searching for the inflaton at the FPF:} In the right panel of \cref{fig:bsm_scalar_inflaton}, we show our results in ($m_\phi, \theta$)-plane, together with FASER search reach, the search reach of other planned/proposed experiments (contours with the names of experiments indicated), and the current excluded region (gray shaded) from CHARM~\cite{Bergsma:1985qz}, Belle~\cite{Wei:2009zv} and LHCb~\cite{LHCb:2015nkv} experiments, as shown in Ref.~\cite{FASER:2018eoc}. Here, to ensure the readability of the figure, we have not shown the search reach of other experiments such as SHiP~\cite{Alekhin:2015byh}, MATHUSLA~\cite{Evans:2017lvd} and CODEX-b~\cite{Gligorov:2017nwh} presented in Ref.~\cite{FASER:2018eoc}. After our analysis, each point in FASER parameter space has a one-to-one correspondence with inflationary predictions and $Z^\prime$ boson search parameters. The diagonal dashed lines correspond to $\xi=9.8\times10^{-3}$ ($r=0.064$) and $\xi=0.12$ ($r=0.01$), respectively, from left to right. The light red shaded region ($r > 0.064$) is excluded by the Planck 2018 results. We find that the parameter region corresponding to the inflationary prediction $r \gtrsim 0.01$ can be searched by FASER2 in the future, a part of which is already excluded by the Planck 2018 result. The diagonal solid lines correspond to $m_{Z^\prime} [{\rm TeV}] = 0.3$, 0.4, 0.5, 0.7, 1.0, 1.3, 2.6, 5.0, and 10, from top to bottom. A point on a solid line corresponds to a fixed value of $\xi$, or equivalently, $r$. Along each line, the $\xi$ ($r$) value increases (decreases) from left to right. For example, the left (right) diagonal dashed lines denote $r = 0.0064$ and $r = 0.01$. 

In conclusion, we have considered the non-minimal quartic inflation scenario in the minimal U(1)$_X$ model with classical conformal invariance, where the inflaton is identified with the U(1)$_X$ Higgs field. FASER can search for the inflaton when its mass and mixing angle with the SM Higgs field are in the range of $0.1 \lesssim m_\phi[{\rm GeV}] \lesssim 4$ and $10^{-5} \lesssim \theta \lesssim 10^{-3}$. By virtue of the classical conformal invariance and the radiative U(1)$_X$ symmetry breaking via the Coleman-Weinberg mechanism, the inflaton search by FASER, the $Z^\prime$ boson resonance search at the LHC, and the future measurement of $r$ are complementary to test our inflationary scenario. For all the details of analysis presented here, see the original paper~\cite{Okada:2019opp}. 

\subsection{Flavor-philic Scalars}
\label{sec:bsm_dh_flav}

The minimal coupling of a singlet scalar $S$ to the SM through renormalizable interactions with the Higgs doublet, $S^{(2)} H^2$, leads to the scalar acquiring couplings proportional to those of the Higgs boson. Notably, the Yukawa coupling of the scalar to fermions is proportional to the fermion mass. In the presence of higher dimensional effective interactions involving the scalar and SM fields, however, there are additional possibilities for the scalar-fermion couplings. We consider the Lagrangian
\be
\mathcal{L} = \frac{1}{2} \partial^\mu S \partial_\mu S - \frac{1}{2} m_S^2 S^2 - \left( \frac{c_S}{M} S \bar{Q}_L u_R H_c +\ \mathrm{h.c.} \right)
\label{eq:flsplag}
\ee
where $c_S$ is a matrix containing the family structure of the $S$ coupling to the up-type quarks and $M$ is the mass scale of the new physics underlying the effective operator. While we restrict ourselves to scalar couplings to up-type quarks for simplicity, analogous theories can be written for down-type quarks and charged leptons. In particular, FPF searches for a muon-philic scalar could probe regions of parameter space consistent that could potentially explain the muon $g - 2$ anomaly~\cite{Batell:2017kty}.

Generally non-standard couplings such as those in \cref{eq:flsplag} result in flavor changing neutral currents, but certain choices of coupling structures lead to viable phenomenology. Specifically, suppose we work in the mass basis for the up quarks. If $c_S \propto \mathrm{diag}(1, 0, 0)$, $S$ couples to only the 1st generation up quark; we term this a flavor-specific hypothesis~\cite{Batell:2017kty}, and it may be thought of as a generalization of minimal flavor violation. With such a form for $c_S$, it can be shown that the radiatively generated flavor-changing couplings of the $S$ are small due to symmetry considerations. After electroweak symmetry breaking, the scalar acquires a coupling to the up quark
\be
\mathcal{L} \supset -g_u S \bar{u} u,\ g_u = \frac{c_S v}{\sqrt{2} M}
\ee
This effective coupling is sufficient to determine the production and decay of the $S$ at the FPF~\cite{Batell:2018fqo}, which can be used to calculate the projected sensitivity of long-lived particle detectors such as FASER2 to up-philic scalars.

\begin{figure*}[t]
  \centering
  \includegraphics[width=0.49\textwidth]{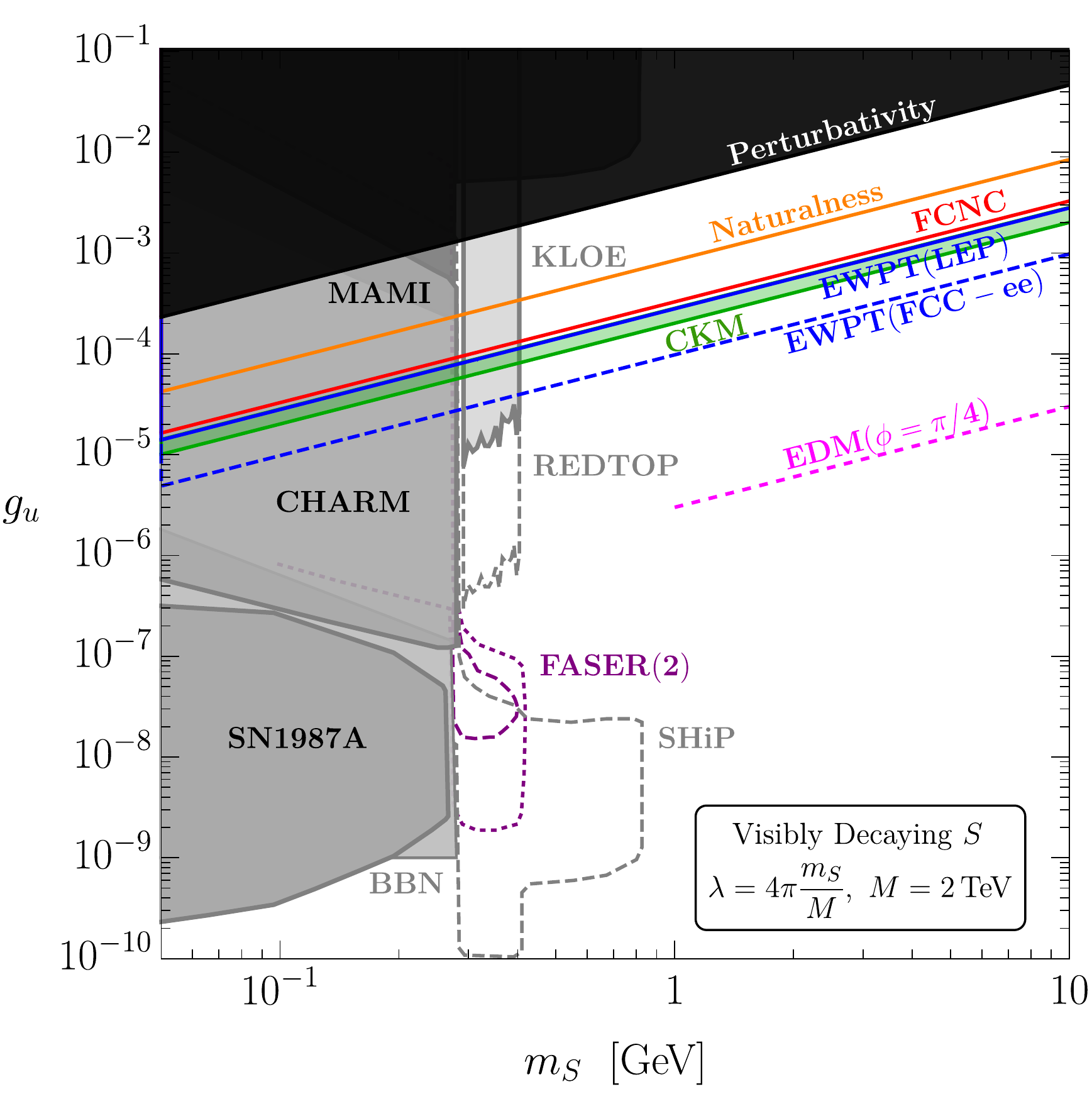}
  \includegraphics[width=0.49\textwidth]{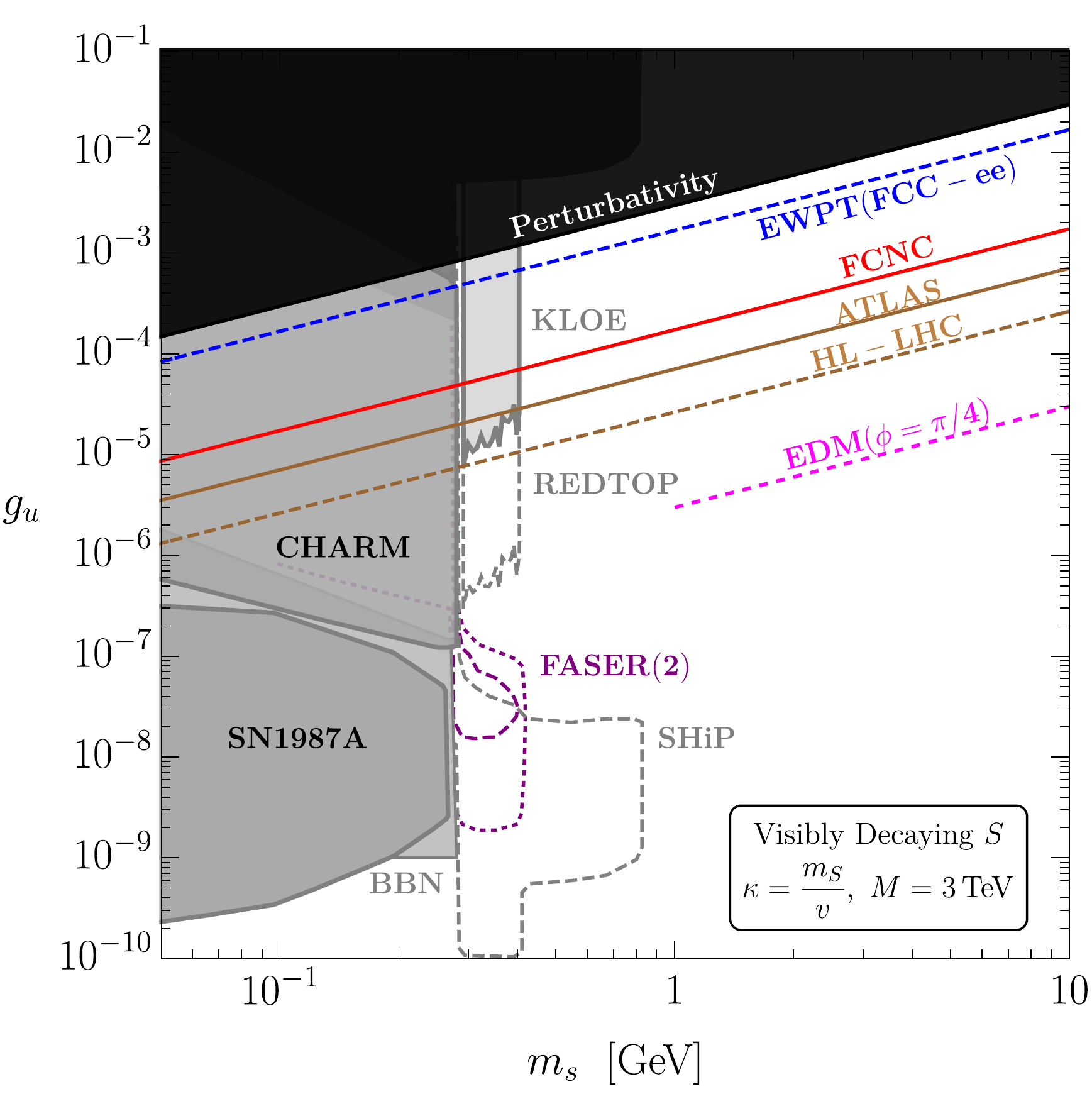}
  \caption{Current and future limits on an up-philic scalar $S$, assuming only SM decays. The left (right) plot shows constraints arising from the effective operator of \cref{eq:flsplag} as well as from its UV completion with vector-like quarks (scalars). Figures taken from Ref.~\cite{Batell:2021xsi}.}
  \label{fig:flspscalar}
\end{figure*}

Because the scalar-quark coupling is not proportional to mass as in the Higgs portal model, the experimental landscape for flavor-specific scalars is markedly different from that for minimally coupled scalars. Up-philic scalars are primarily produced at the FPF through light meson decays, notably those of the $\eta^{(')}$ and kaons; this may be compared to Higgs portal scalars, which are dominantly produced at the FPF from $B$ meson decays. Similarly, as up-philic scalars do not couple to leptons, the only possible decay mode of the $S$ is to $\gamma \gamma$ for $m_S < 2 m_\pi$. Above the pion threshold, $S$ decays hadronically. We use the estimated meson spectra at the LHC in conjunction with the $S$ decay widths in Ref.~\cite{Batell:2018fqo} to calculate the reach of FASER(2) and SHiP long-lived particle searches as a function of the scalar mass and effective coupling $g_u$. We also translate the limits on $S$ from current and future rare meson decay searches (MAMI, KLOE, BESII, REDTOP), fixed target experiments (CHARM), supernova data (SN1987A), and Big Bang Nucleosynthesis. These limits are shown in \cref{fig:flspscalar}. 

Now, the model described above may be augmented in several ways. From an effective field theory perspective, perhaps the first question is the character of the UV completion of \cref{eq:flsplag} at the heavy scale $M$. Two natural completions of the $c_S$ interaction come from vector-like quarks and additional scalars, respectively~\cite{Batell:2021xsi}. It is of interest to ask whether direct and indirect searches for these particles would be complementary to FPF searches for $S$, and we explore the interplay between these searches below. In addition, the $S$ could serve as a mediator between the SM and dark matter. Depending on the relative masses of the mediator and dark matter, it may be possible for $S$ to decay invisibly. As our focus here is on long-lived particle searches for the $S$ assuming its visible decay, we will not consider the potential couplings of $S$ to dark matter further.

Turning to the question of the UV physics underlying the effective theory in \cref{eq:flsplag}, we first write a completion with a vector-like quark $U'$ that has the same charges under $SU(3) \times SU(2) \times U(1)$ as the right-handed up quark:
\be
\mathcal{L} = \frac{1}{2} \partial^\mu S \partial_\mu S - \frac{1}{2} m_S^2 S^2 + \bar{U}' (i \cancel{D} - M) U' - \left( y \bar{Q}_L U'_R H_c + \lambda \bar{U}'_L u_R S +\ \mathrm{h.c.} \right)
\ee
The new couplings $y$ and $\lambda$ are related to the effective coupling of the scalar through $c_S = - y \lambda$, and we have identified the heavy scale in \cref{eq:flsplag} with the $U'$ mass.

The presence of the $U'$ leads to additional constraints beyond those in the effective theory. First, naturalness of radiative corrections to the $S$ masses suggest that $\lambda \ll y$, and we assume this hierarchy of couplings. Then, the $U'$ mixes with the SM $u_R$. This affects the unitarity of the effective SM CKM matrix and electroweak precision tests (EWPT). Furthermore, the $U'$ mediates box diagrams contributing to kaon mixing (FCNC). We show these limits, with future projections when available, in the left panel of \cref{fig:flspscalar}. In addition, if there is a non-trivial phase among the couplings $y$ and $\lambda$, the SM up Yukawa $y_u$, and $M$, a contribution to the neutron electric dipole moment (EDM) arises. We show this constraint assuming an order-1 phase. Lastly, there are LHC constraints from direct $U'$ searches, but these are subdominant in the parameter space of the left panel of \cref{fig:flspscalar} and are not shown.

Instead of being completed with a vector-like quark, the effective theory of an up-philic scalar may arise from integrating out a heavy new Higgs doublet $H'$ with the Lagrangian
\be
\mathcal{L} &= \frac{1}{2} \partial^\mu S \partial_\mu S - \frac{1}{2} m_S^2 S^2 + |D_\mu H'|^2 - M^2 H'^\dag H' \\ 
& \ \ \ - \left( y' \bar{Q}_L u_R H'_c + \kappa M S H^\dag H' +\ \mathrm{h.c.} \right) +\ \mathrm{quartic\ couplings}
\ee
Similarly to the vector-like quark completion, the mass of the new Higgs doublet is the heavy scale $M$, and the effective coupling above can be thought of as $c_S = - \kappa y'$.

In an analogous fashion to the $U'$ theory, we may consider additional constraints on an up-philic scalar that is completed by an additional Higgs doublet. Once again, requiring the stability of the $S$ mass under radiative corrections results in tight constraints on one of the couplings, and we will work in the regime $\kappa \ll y'$. In addition, the $H'$ contributes to neutral kaon mixing (FCNC). Then, the $H'$ and $S$ mediate one-loop diagrams involving the $Z$ couplings to quarks, which affect electroweak precision (EWPT). Also, since the $H'$ couples to quarks, it is subject to dijet searches at the LHC. As before, we show all of these limits and their future projections in the right panel of \cref{fig:flspscalar}. There is again a neutron EDM constraint if there is a phase among the new physics couplings and the SM up Yukawa, which is shown assuming that the phase is large.

From \cref{fig:flspscalar}, it is clear that searches for long-lived up-philic scalars at FASER2 can provide additional sensitivity beyond existing limits. These results would be independent of the UV completion, and complementary to direct searches for new heavy particles mediating the effective coupling of a light scalar to up quarks. Beyond the minimal renormalizable portals of hidden sectors, flavor-philic scalar theories provide novel targets for long-lived searches at the FPF.

\subsection{Two Higgs Doublet Models}
\label{sec:bsm_dh_2hdm}

Another example of a theory that could provide a scalar particle that could be probed at the FPF is the Type-I 2HDM. In the following, we will consider 2HDMs in which one of the neutral beyond the Standard Model (BSM) scalars is very light and long lived, while the others are heavy.

The Higgs sector of the 2HDMs consists of two SU(2)$_L$ scalar doublets  $\Phi_i\,(i=1,2)$ with hyper-charge $Y=1/2$. After the electroweak symmetry breaking with the neutral components of $\Phi_{1,2}$ develop vacuum expectation values of $v_{1,2}$ and $\tan\beta=v_2/v_1$,  the scalar sector of the 2HDMs~\cite{Branco:2011iw, Gu:2017ckc} consists of five physical scalars: $h,H,A,H^\pm$, with the CP-even Higgs mixing angle being $\alpha$. In this work, $h$ will be identified as the SM Higgs with $m_h=125\,\rm GeV$, while $H$ and $A$ are the BSM neutral Higgses. The alignment limit under this convention is therefore $\cos(\beta - \alpha)=0$. The effective Lagrangian for a (light) CP-even scalar $H$ and CP-odd pseudoscalar $A$ interacting with SM particles can be written as
\begin{eqnarray}
 \mathcal{L} &&=  - \sum_f  {\xi_{H}^{f}}  \frac{m_f}{v} \, H \bar{f} f   - {\xi_{H}^{W}} \frac{2m_W^2}{v} \, \phi W^{\mu+} W^-_\mu- {\xi_{H}^{Z}} \frac{m_Z^2}{v} \, H Z^{\mu} Z_\mu  +  {\xi_{H}^{ g}} \frac{\alpha_s}{12\pi v} H G_{\mu\nu}^a G^{a\mu\nu}    \nonumber \\
 &&+  {\xi_{H}^{\gamma}} \frac{\alpha}{4\pi v} H F_{\mu\nu} F^{\mu\nu} +\sum_{f=u, d, e} \xi_{A}^f \frac{im_f}{v}  \bar{f} \gamma_{5} f A+ \xi_A^g \frac{\alpha_s}{4\pi v} A G_{\mu\nu,a}\tilde G^{\mu\nu,a}+ \xi_A^\gamma \frac{\alpha}{4\pi v} A F_{\mu\nu}\tilde F^{\mu\nu}.
\label{eq:lag}
\end{eqnarray}
Expressions for the various coupling modifiers $\xi_{H,A}^{f,W,Z}$ as well as the loop induced couplings  $\xi_{H,A}^{g,\gamma}$ can be found in Refs.~\cite{Gunion:1989we, Djouadi:2005gi, Djouadi:2005gj, Domingo:2016yih}. There are four different types of 2HDMs, depending on the couplings of $\Phi_{1,2}$ with the quark and lepton sectors. Consequently, the resulting effective couplings of $H$ and $A$ with fermions, namely $\xi_{H,A}^f$, exhibit different $\tan\beta$ dependence.  

After taking into account the theoretical considerations such as unitarity, perturbativity, and vacuum stability, as well as electroweak precision measurements, the viable parameter space with a light $H$ or $A$ are:
\begin{eqnarray}
m_H\sim 0: &&\  m_{A}\sim m_{H^\pm}\lesssim 600\ {\rm GeV}, \\
m_A\sim 0: &&\  m_{H^\pm}\sim m_H\lesssim m_h,
\end{eqnarray}
with $\lambda v^2 \equiv m_H^2-m_{12}^2/ (\sin\beta\cos\beta)\approx 0$ to accommodate a large range of $\tan\beta$. 

The flavor constraints such as $K^+\rightarrow \pi^+ \nu \bar \nu$, $K^+\rightarrow \pi^+ e^+e^-$, $B\rightarrow X_s \gamma$, $B_{s,d}\rightarrow \mu^+\mu^-$, $B-\bar{B}$ mixing, decays of $B$ and $D$ baryons, impose strong constraints on the charged Higgs mass as well as the value of $\tan\beta$.   In Type-II, Type-L, and Type-F, at least one of the Higgs Yukawa couplings to quarks or lepton are unsuppressed at all regions of $\tan\beta$, which makes it hard to accommodate a very weakly coupled light neutral Higgs that has relatively long decay lifetime.  Therefore, the only viable model is the  Type-I 2HDM, in which all the Higgs Yukawa couplings are suppressed at large $\tan\beta$. 

To accommodate the current SM-like Higgs coupling measurements~\cite{ATLAS:2019nkf, CMS:2018uag}, we take the close to the alignment limit of $\cos(\beta - \alpha)\sim 0$.  For a light $H/A$ with long enough lifetime, $h\to HH/AA$ is constrained by the invisible Higgs decay.  Under the exact alignment limit of $\cos(\beta - \alpha)=0$, unlike $g_{hAA}$, which is suppressed to be only a few percent of the value of the SM trilinear Higgs coupling, $g_{hHH}$ recovers the SM value of $m_h^2/2v$, leading to an invisible decay branching fraction too big to accommodate the current bound of ${\rm Br}(h\to  {\rm invisible} )<0.24$~\cite{Khachatryan:2016whc, Aad:2015txa, Aaboud:2017bja}. In our study below for the CP-even light Higgs, we take the parameter choice of large $\tan\beta= 1/\cos({\beta-\alpha})$, which leads to simultaneous suppressions of the $g_{hHH}$  coupling as well as the effective Yukawa couplings $\xi_{H,A}^f$.     

The productions of a light CP even Higgs $H$ are mostly via the flavour changing decays of kaons and $B$ mesons. Depending on the mass of the CP-even scalar $H$, it can decay into pair of photons, leptons, and  multiple  hadrons. However, for larger values of $\tan\beta > 10$, the decay into photons dominate. This is due to our parameter choice $\cos(\beta-\alpha) = 1/\tan\beta$, which leads to suppressed couplings to fermions $\xi_H^f \sim 1/\tan^3\beta$. In contrast, the decay mode into photons arises through a $W$-boson or charged Higgs with corresponding coupling $\xi_H^\gamma \sim 1/\tan\beta$. The decay length $c\tau$ takes values between $10^3~\m$ and $10^{-3}~\m$ for light scalar masses $m_H$ in the range between 100~MeV and 10~GeV for $\tan\beta=100$.

A light CP odd scalar $A$ typically mixes with pseudo-Goldstone bosons $\pi^0$, $\eta$ and $\eta^\prime$.  Any process that produces those mesons would also produce the new pseudoscalar $A$. In addition, the scalar can be produced in the weak decays of SM mesons, in particular $K \rightarrow \pi A$ and $b \rightarrow X_s A$. At low mass, it mainly decays to diphotons and dileptons. Once the hadronic channels open up, it could decay into multiple hadrons, as well as radiative hadronic decay. Decay length $c\tau$ typically reaches between $100~\m$ and $10^{-7}~\m$ for $m_A$ in the range of 100~MeV to 10~GeV and $\tan\beta=100$.
 
\begin{figure*}[t]
    \centering
    \includegraphics[width=0.49\textwidth]{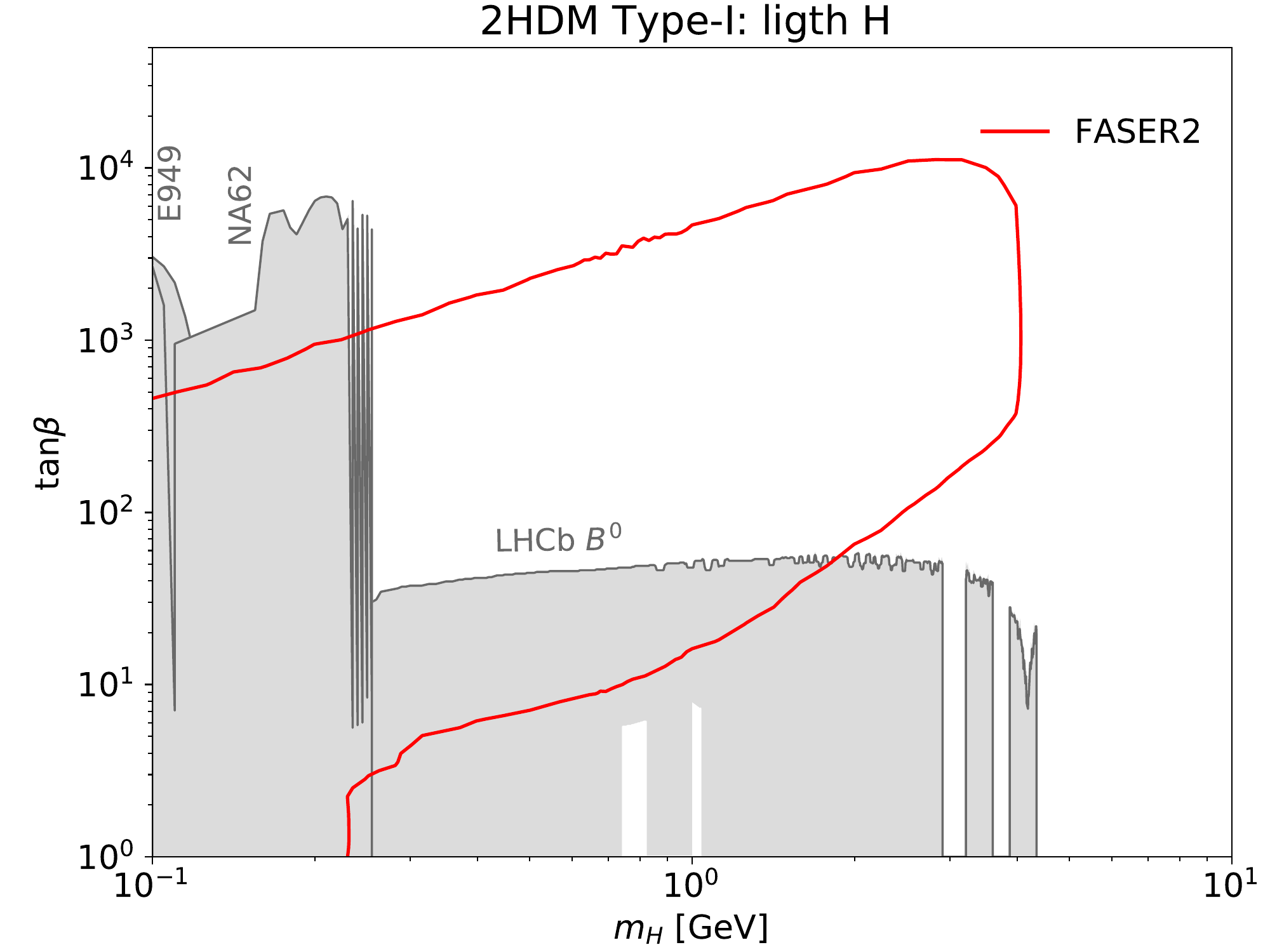}
    \includegraphics[width=0.49\textwidth]{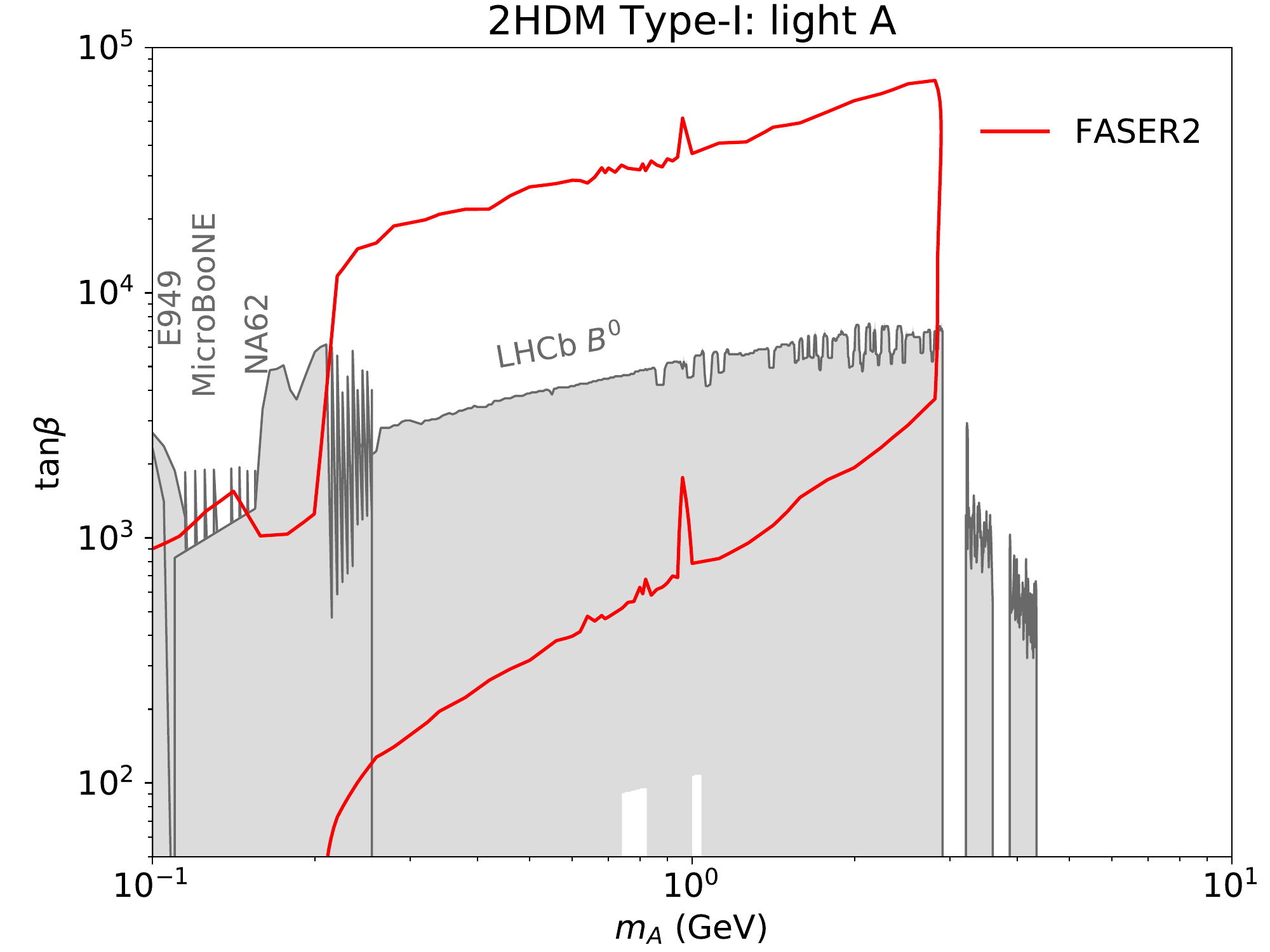}
    \caption{The FASER2 reach (red curves)  for the light CP even Higgs $H$ (left) and CP-odd Higgs $A$ (right) in the parameter space of $m_{H/A}$ vs. $\tan\beta$ plane. Various current experimental constrains, including  $B$ meson decays at LHCb~\cite{LHCb:2015nkv}, $K^+$ decays from NA62~\cite{NA62:2021zjw}, MicroBooNE~\cite{MicroBooNE:2021usw}, and E949~\cite{BNL-E949:2009dza}, are shown in grey regions. Note that for the presented slice of parameter space the scalar $H$ almost exclusively decays into photons, while decays into fermions are suppressed. 
    }
    \label{fig:2HDM}
\end{figure*}

In \cref{fig:2HDM}, we show the FASER2 reach for the light CP even Higgs $H$ (left panel) and CP odd Higgs $A$ (right panel) in the parameter space of $m_{H/A}$ vs. $\tan\beta$ plane.  Experimental constraints on the light scalars from various experiments are also shown in shaded grey regions. The benchmarks we chose are: light $H$ with $\tan\beta= 1/\cos({\beta-\alpha})$, $m_A=m_{H^\pm}=600~\gev$; light $A$ with $\cos(\beta-\alpha)=0$, $m_H=90~\gev$, $m_{H^\pm}=110~\gev$. The reach for the light CP odd $A$ has higher value for $\tan\beta$ since  $\xi_{A}^f=1/\tan\beta$, while $\xi_{H}^f \sim 1/\tan^3\beta$ for the light $H$ case. We can see that FASER2 will have the potential to probe a large region of this currently unconstrained region of 2HDM parameter space. 

\subsection{Sgoldstino}
\label{sec:sgoldstino}

Supersymmetry, if it exists in Nature, should be spontaneously broken in a hidden sector. This breaking may occur at relatively low  energy scale and it can be achieved in models with no-scale supergravity~\cite{Ellis:1984kd, Ellis:1984xe} or gauge-mediation~\cite{Giudice:1998bp, Dubovsky:1999xc}. Although the details of the hidden sector are model dependent, there should be a sector, where spontaneous supersymmetry breaking takes place, and this sector contains the Goldstone supermultiplet. Fermionic component of this supermultiplet is goldstino which becomes gravitino in supergravity extensions of the model. In the case of chiral Goldstone  supermultiplet, the superpartners of goldstino are scalar $S$ and  pseudoscalar $P$ {\it sgoldstinos}. The scale of sgoldstino masses depends on details of the hidden sector and it is phenomenologically acceptable to have it below 10~GeV. Here we discuss FASER2 reach for the light sgoldstinos~\cite{Demidov:2022ijc} and assume, in accordance with results from LHC experiments, that all the superpartners of the SM particles and particles from the hidden sector (except for gravitino which is LSP in the present scenario), to be much heavier neglecting their possible impact on sgoldstino phenomenology. In the case of light sgoldstinos their interactions with gravitino can also be safely neglected.  

Sgoldstino interactions with SM fields are determined by the soft supersymmetry breaking parameters. Corresponding low energy interactions (see, e.g.~\cite{Brignole, Gorbunov:2000th}) include couplings to photons and gluons 
\begin{eqnarray}
    {\cal L} = &-\frac{M_{\gamma\gamma}}{2\sqrt{2}F}SF_{\mu\nu} F^{\mu\nu} 
    +\frac{M_{\gamma\gamma}}{4\sqrt{2}F}PF_{\mu\nu} F_{\lambda\rho}\,\epsilon^{\mu\nu\lambda\rho} 
    -\frac{M_3}{2\sqrt{2}F}SG^a_{\mu\nu} G^{\mu\nu\;a} 
    +\frac{M_3}{4\sqrt{2}F}PG^a_{\mu\nu} G^a_{\lambda\rho}\,\epsilon^{\mu\nu\lambda\rho}
\end{eqnarray}
as well as to quarks $q_i$, $i=1,\dots,6$ and charged leptons $l_i$, $i=1,2,3$,  

\begin{equation}
    {\cal L} = -S\,\frac{v A_Q y^q_{ij}}{\sqrt{2}\,F}\bar q_i q_j 
    -P\,\frac{v A_Q y^q_{ij}}{\sqrt{2}\,F}\bar q_i \gamma_5 q_j  
   -S\,\frac{v A_l y^l_{ij}}{\sqrt{2}\,F}\bar l_i l_j 
    -P\,\frac{v A_l y^l_{ij}}{\sqrt{2}\,F}\bar l_i \gamma_5 l_j \,.  
\end{equation}
Here  $M_{\gamma\gamma}\equiv M_1\cos^2\theta_W+M_2\sin^2\theta_W$ with $M_i$, $i=1,2,3$ being the gaugino masses. The combinations $A_{Q,l} y_{ij}^{q,l}$ play the role of trilinear coupling constants. For flavour conserving couplings we assume $vy_{ii}^{q,l}=m^{q,l}_i$ for simplicity. The parameter $\sqrt{F}$ has the meaning of supersymmetry breaking scale. The model should be considered as a low energy effective field theory valid for energies $E\ll \sqrt{F}$ and in the weak coupling regime for $m_{soft} < \sqrt{F}$.  

The interactions of scalar and pseudoscalar sgoldstino to the SM fields induce their decays to the lighter SM particles. The decay pattern of sgoldstino depends on the hierarchy of the soft supersymmetry breaking parameters see e.g.~\cite{Gorbunov:2000th, Gorbunov:2000cz, Gorbunov:2002er, Demidov:2011rd, Astapov:2015otc}. In  case of soft parameters of similar size, $M_3=M_{\gamma\gamma}=A_0$, the decays of scalar sgoldstino into hadrons dominate, if allowed, while decays into photons dominate, otherwise. Let us note that for pseudoscalar sgoldstinos the hadronic modes are suppressed for $m_P\lesssim 1$~GeV, because they are three-body decays. 

Sgoldstino couplings are naturally bounded from limits on soft supersymmetry breaking terms inferred from direct searches for squarks, gluinos, etc and observations of rare processes where the superpartners might contribute to. At the same time, the models with light sgoldstinos are also constrained from testing their predictions, e.g. direct searches for light sgoldstinos, performed at kaon experiments~\cite{Tchikilev:2003ai, E391a:2008grj, KTeV:2011xrc},  $e^+e^-$~\cite{Belle:2010ryo, BESIII:2011wmh} and hadronic~\cite{LHCb:2013zjg, LHCb:2017rdd} colliders, and at beam-dump facilities~\cite{Astapov:2015otc}.    

Production mechanisms of scalar and pseudoscalar sgoldstinos in proton collisions relevant for FASER2 include direct production via gluon fusion $gg\to S(P)$ and production in decays of mesons. The latter involve flavour-conserving and flavour-violating sgoldstino couplings to quarks as well as scalar sgoldstino mixing with the Higgs  boson~\cite{Astapov:2014mea}. Examples of sgoldstino production in decays include decays of flavourless vector mesons $V\to S(P)\gamma$ and semihadronic decays of light and heavy pseudoscalar mesons $K\to \pi S$, $\eta\to \pi S$, $B\to K S(P)$. Expressions for various partial decay widths with scalar and pseudoscalar sgoldstinos in the final states can be found in Ref.~\cite{Demidov:2022ijc}.

\begin{figure}[t]  
\centering
    \includegraphics[width=0.49\textwidth]{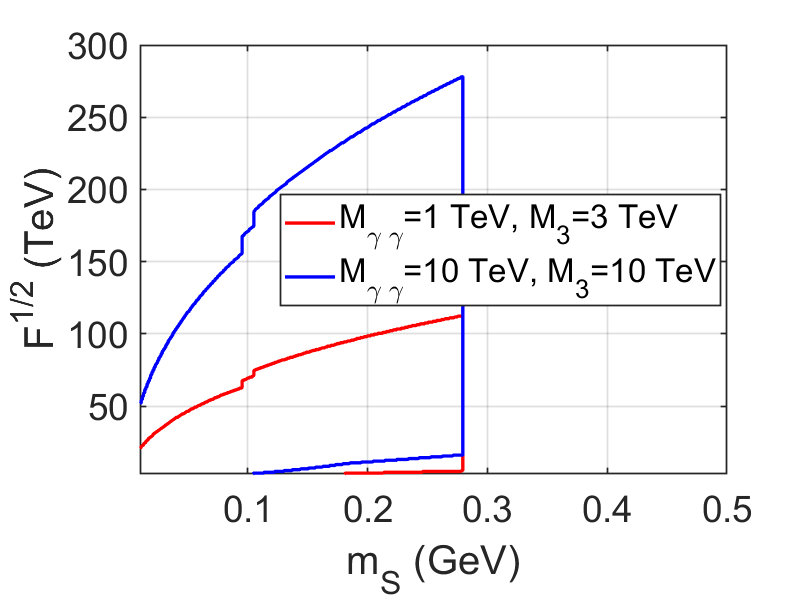}
    \includegraphics[width=0.49\textwidth]{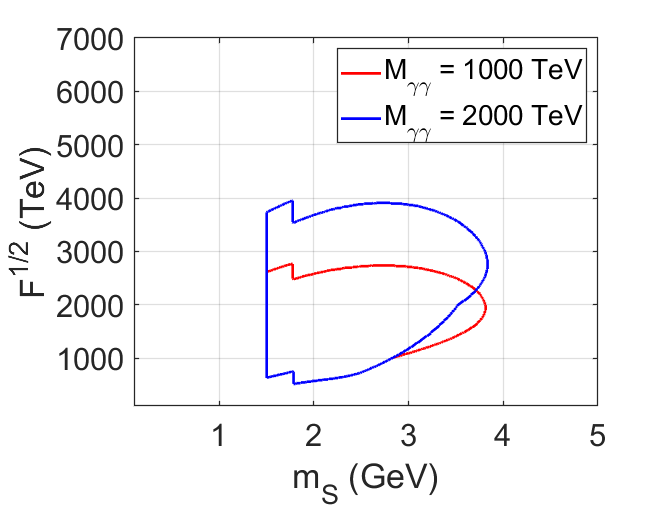}
\caption{FASER2 reach to light scalar sgoldstino decaying into a couple of photons and produced in gluon fusion (left) or in decays of $B$-mesons (right). In the latter case $A_Q = 100$~GeV, $M_3 = 3$~TeV are assumed, see Ref.~\cite{Demidov:2022ijc} for details.} 
\label{fig:sgoldstino1}
\end{figure}

\begin{figure}[t]  
\centering
    \includegraphics[width=0.49\textwidth]{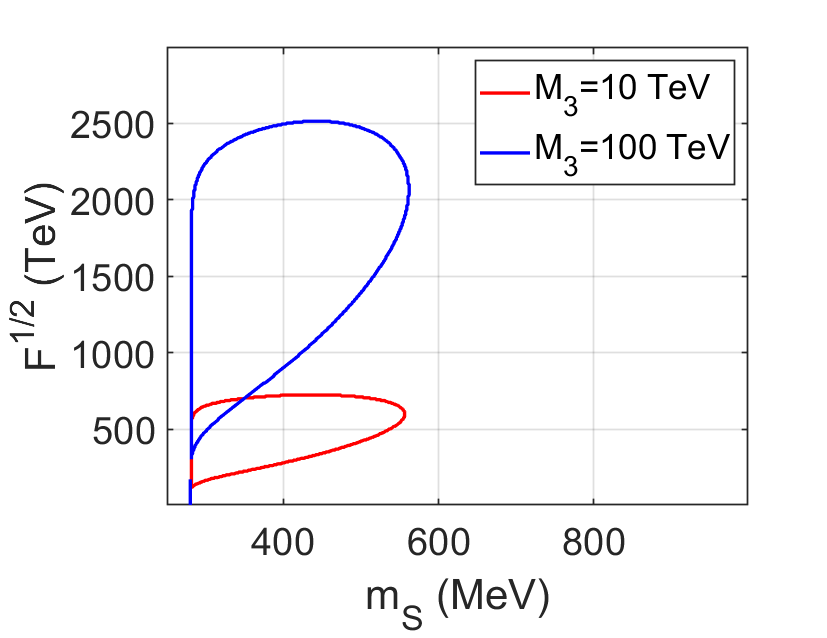} 
    \includegraphics[width=0.49\textwidth]{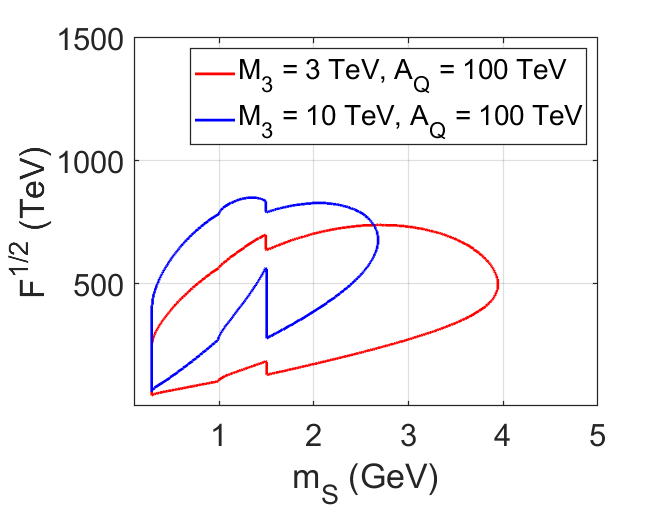} 
\caption{FASER2 reach to light scalar sgoldstinos decaying into mesons and produced directly (left) or in decays of $B$-mesons (right). In the latter case we take $A_Q = 100$\,GeV and $M_{\gamma\gamma} = 3$~TeV, see Ref.~\cite{Demidov:2022ijc} for details.} 
\label{fig:sgoldstino2}
\end{figure}

\begin{figure}[t]  
\centering
    \includegraphics[width=0.49\textwidth]{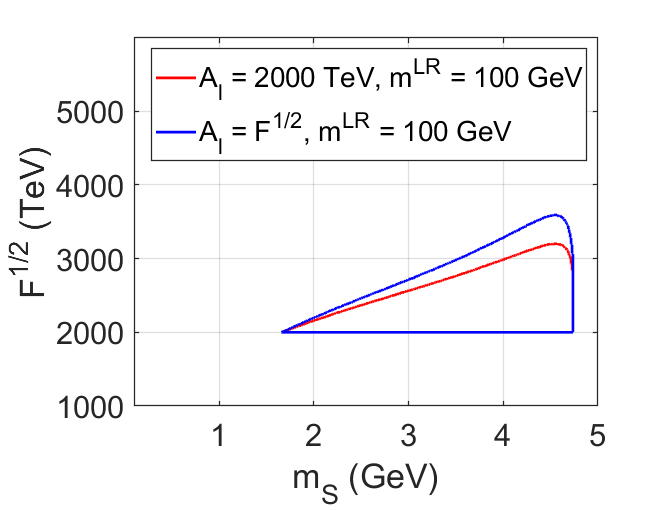} 
    \includegraphics[width=0.49\textwidth]{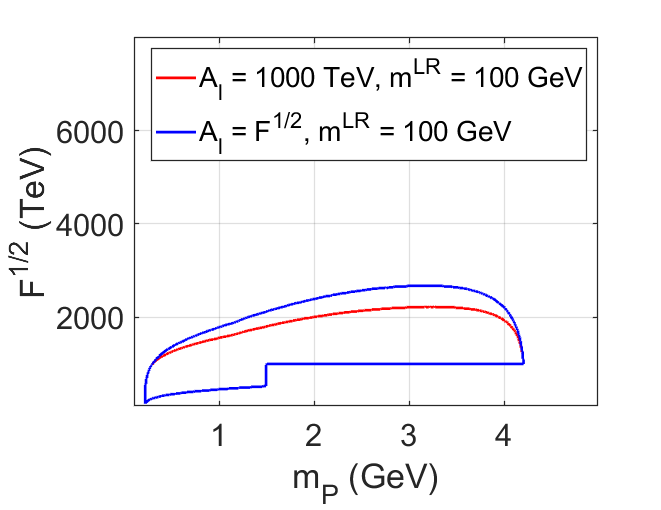} 
\caption{FASER2 reach to light scalar (left) and pseudoscalar (right) sgoldstino decaying into lepton pairs and produced in decays of $B$-mesons via flavor-violating couplings. $M_3=3$\,TeV is assumed, see Ref.~\cite{Demidov:2022ijc} for details.} 
\label{fig:sgoldstino3}
\end{figure}

On plots of \cref{fig:sgoldstino1}, \cref{fig:sgoldstino2} and \cref{fig:sgoldstino3} we show several examples of FASER2 sensitivity regions, which depend on the signal signature (pair of photons, leptons or mesons) and sgoldstino production mechanism. We estimate the FASER2 sensitivities assuming dominance of a particular decay mode. In all cases we investigate only the regions in the model parameter space which are consistent with existing bounds on  light sgoldstinos. 

The model contains several parameters and it is difficult to specify  precisely the total parameter space which can be probed by FASER2. We refer to Ref.~\cite{Demidov:2022ijc} for detailed discussion of FASER2 sensitivity to models with light sgoldstino.  To get an impression we can broadly say that FASER can probe the models with light sgoldstinos having supersymmetry breaking scale $\sqrt{F}\lesssim 1500-5000$~TeV at the first stage of experiment and about 1--2 orders of magnitude higher at the second one. 

\subsection{Crunching Dilatons}
\label{sec:bsm_dh_dilaton}

In recent years, there has been interest in cosmological solutions to the hierarchy problem of the Standard Model, wherein novel dynamics in the early universe selects a small Higgs mass~\cite{Graham:2015cka,Arkani-Hamed:2016rle,Geller:2018xvz,Cheung:2018xnu,Strumia:2020bdy,Giudice:2019iwl,Giudice:2021viw,TitoDAgnolo:2021pjo}. In contrast to traditional symmetry-based solutions to the hierarchy problem (e.g. weak-scale SUSY and composite Higgs), quadratically divergent corrections to the Higgs mass from new physics are not suppressed. These models generically require a new light, weakly-coupled particle that interacts with the Higgs~\cite{TitoDAgnolo:2021pjo}.

Here we focus on the ``crunching dilaton'' model of Ref.~\cite{Csaki:2020zqz}, which combines aspects of symmetry-based and cosmological approaches to address Higgs naturalness. The main phenomenological prediction is a new scalar (a dilaton) in the $0.1$--$10$~GeV range which mixes with the Higgs. This dilaton could potentially be observed at FASER2. It differs from a dark Higgs in that it also has a tree-level coupling to the photon, which alters its phenomenology.

We will summarize the essential points of the model, and refer the reader to Ref.~\cite{Csaki:2020zqz} for additional details. We postulate a multiverse of causally disconnected Hubble patches. Some scanning sector sets the value of the Higgs VEV in each patch, up to a UV cutoff scale. We introduce dynamics such that all patches rapidly undergo a cosmological crunch unless the Higgs VEV lies in a finite range, $v_{\rm min} < v < v_{\rm crit}$. Then the only patches that allow for a large observable universe are those in which the Higgs VEV is in this range.

This can be explicitly realized through a new CFT sector in which the conformal symmetry is spontaneously broken; the Higgs couples to the dilaton, the Goldstone boson of the broken symmetry. We employ the Goldberger-Wise mechanism~\cite{Goldberger:1999uk} to generate a stable minimum in the dilaton potential. The total vacuum energy in this minimum is always large and negative, so that any patches where the dilaton rolls down to the Goldberger-Wise minimum rapidly undergo a cosmological crunch. Interactions between the Higgs and dilaton generate a second, metastable minimum in the dilaton potential, but only when the Higgs VEV is smaller than some critical value $v_{\rm crit}$. Crucially, the conformal symmetry allows $v_{\rm crit}$ to be $\mathcal{O}({\rm TeV})$, hierarchically smaller than the UV cutoff, without fine-tuning. Thus we dynamically select a small Higgs VEV in the multiverse, generating an apparent naturalness problem.

Explicitly, the dilaton potential is given by
\begin{equation}\label{eq:crunching_dilaton_potential}
    V(\chi, v) = -\lambda \chi^4 + \lambda_{\rm GW} \frac{\chi^{4+\delta}}{k^\delta} + \lambda_2 v^2 \frac{\chi^{2+\alpha}}{k^\alpha} - \lambda_{H\epsilon} v^2 \frac{\chi^{2+\alpha+\epsilon}}{k^{\alpha+\epsilon}} - \lambda_4 v^4 \frac{\chi^{2\alpha}}{k^{2\alpha}}
\end{equation}
where $\chi$ is the dilaton, $k$ is the UV cutoff of the CFT sector, the $\lambda$'s are dimensionless couplings, and $\delta$, $\alpha$, and $\epsilon$ are related to anomalous dimensions of certain CFT operators. The first two terms correspond to the Goldberger-Wise potential, while the remaining terms depend on the value of the Higgs VEV and generate a metastable minimum when $v$ is sufficiently small. In the 5D dual description of the CFT, where the dilaton is identified with the IR brane in a slice of AdS$_5$ space, these Higgs-dependent terms can be understood as arising from IR brane-localized terms in the potential. We omit a detailed discussion of the potential here because it is not necessary to understand the phenomenological features of the model.

\begin{figure*}[t]
    \centering
    \includegraphics[width=0.99\textwidth]{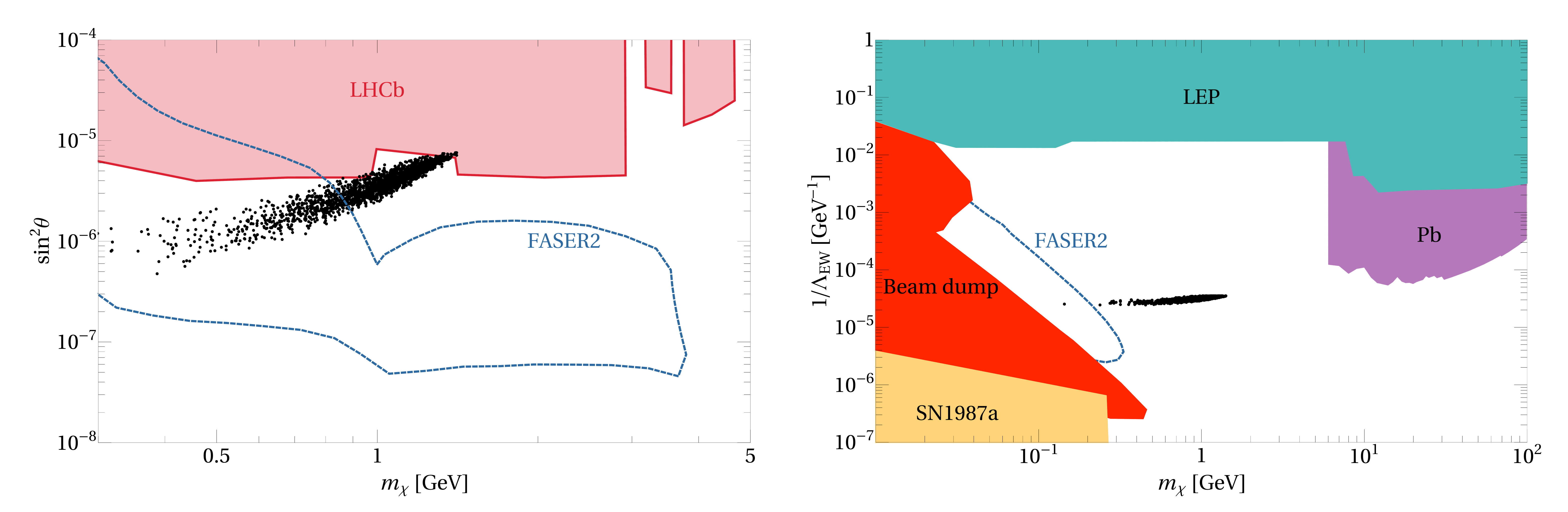}
    \caption{Experimental bounds on a random scan of the crunching dilaton parameter space, adapted from Ref.~\cite{Csaki:2020zqz}. Left: Bounds on the mixing angle with the Higgs $\sin^2 \theta$ from $B$ meson decays at LHCb~\cite{LHCb:2012juf,LHCb:2015nkv} (red), adapted from~\cite{Flacke:2016szy}, along with a projection for bounds from FASER2 (blue line). Right: Bounds on the photon coupling $1/\Lambda_{\rm EW}$, adapted from~\cite{Bauer:2017ris}, from LEP~\cite{Mimasu:2014nea,Jaeckel:2015jla} (turquoise), beam dump experiments~\cite{Bjorken:1988as}(red), lead ion collisions at the LHC~\cite{Knapen:2017ebd} (purple), and supernova SN1987a~\cite{Payez:2014xsa,Jaeckel:2017tud} (yellow), as well as a projection for FASER2 (blue line).  }
    \label{fig:crunching_dilaton_exclusion}
\end{figure*}

For the Higgs VEV to drive the dilaton potential, the dilaton must be light relative to the Higgs, typically around $m_\chi = 1$~GeV. The dilaton mixes with the Higgs and therefore inherits all the SM Higgs couplings suppressed by $\sin \theta$, where $\theta$ is the mixing angle. One can derive approximate analytical expressions for $m_\chi$ and $\sin \theta$ in terms of the potential parameters, from which it follows
\begin{equation}
    \sin \theta \propto \frac{m_\chi^2}{m_h^2} .
\end{equation}
The dilaton has negligible tree-level couplings to the SM fermions. This is most easily understood in the 5D holographic description of the CFT: the dilaton is localized towards the IR brane, while the SM fermions are localized on the UV brane, and so they do not interact significantly with the dilaton at tree level. Conversely, the Higgs propagates in the AdS$_5$ bulk, and so the electroweak gauge bosons must live in the bulk as well. Therefore the dilaton has couplings to the electroweak gauge bosons of the form~\cite{Csaki:2000zn}
\begin{equation}
    \frac{1}{4\Lambda_{\rm EW}} \chi \left( F_{\mu\nu}^2 + Z_{\mu\nu}^2 + 2 W_{\mu\nu}^2 \right) ,
\end{equation}
where $\Lambda_{\rm EW}$ is $\mathcal{O}({\rm TeV})$. The coupling to the $W$ and $Z$ turns out to be negligible relative to the coupling obtained through mixing with the Higgs, but the coupling to the photon is phenomenologically important.

In summary, there is a dilaton with mass $m_\chi = \mathcal{O}({\rm GeV})$ which mixes with the Higgs and also has a direct coupling to the photon. In the left panel of \cref{fig:crunching_dilaton_exclusion} we show the results of a random scan of the parameter space in the $m_\chi$-$\sin^2 \theta$ plane. We take $k = 10^8$~TeV, $\delta = 0.01$, $\lambda_{\rm GW} = 2 \times 10^{-6}$, $\lambda = 1.1 \lambda_{\rm GW}$, and $\alpha = 0.1$, and sample the other parameters from $\epsilon \in (0.03, 0.1)$, $\lambda_2 \in (0.5, 1) \times 10^{-2}$, $\lambda_{H\epsilon} \in (2,4) \times \lambda_2$, and $\lambda_4 \in (2,3)$, where the reader is referred to Ref.~\cite{Csaki:2020zqz} for a discussion of the naturalness of these parameter choices. The experimental bounds are similar to those on a dark Higgs, but must be rescaled to account for the photon coupling. The existing constraints on the parameter space come from searches for rare $B$ meson decays at LHCb. From \cref{fig:crunching_dilaton_exclusion}, we see that the unexplored parameter space below $m_\chi \simeq 1$~GeV could be probed at the FPF with FASER2.

One can also consider probing the photon coupling directly. The bounds on the photon coupling are shown in the right panel of \cref{fig:crunching_dilaton_exclusion}. The model is not in tension with the existing experimental constraints. However, we also see that FASER2 would lack the sensitivity to probe the region of parameter space relevant to the model. Although it is not depicted in \cref{fig:crunching_dilaton_exclusion}, we remark that this relevant parameter space could be fully explored at a future lepton collider such as the FCC-ee~\cite{FCC:2018evy,Bauer:2018uxu}.

\clearpage
\section{Long-Lived Fermions}
\label{sec:bsm_llp_fermion}

The third type of interactions between the SM sector and a new particle are fermionic portals. The only renormalisable example of such an interaction is the ``neutrino portal''. Here a new particle $N$ is introduced that couples to the left handed lepton doublet and the Higgs field, $\mathcal{L} \sim L H N$, filling the role of the right handed neutrino. Since such a particle is a singlet or neutral under all the SM gauge groups, it is also referred to as heavy neutral lepton (HNL) or a sterile neutrino. The motivation, phenomenology sensitivity projections for HNLs at the FPF are discussed in \cref{sec:LLPHNL0} and \cref{sec:LLPHNL}. A more detailed discussion of the scenario in which the HNL dominantly couples to tau leptons is presented in \cref{sec:LLPHNL2}.

In addition to HNLs, there are other examples of fermionic long lived particles. One prominent example is the neutralino in supersymmetric models. \cref{subsec:LLP-susy-neutralino-I} and \cref{subsec:LLP-susy-neutralino-II} discuss the scenario of a light long-lived neutralino in RPV SUSY models. Finally,  \cref{sec:fermion_portal} considers a higher dimensional fermion portal using effective operators. 

\subsection{Heavy Sterile Neutrinos}
\label{sec:LLPHNL0}

A sterile neutrino is a neutral lepton which is a gauge singlet with respect to the SM gauge symmetry SU(3)$_C \otimes$SU(2)$_L \otimes$U(1)$_Y$. Several extensions of the SM consider a right-chiral sterile neutrino $\nu_R$, which are CP conjugates of $\nu_L^c$, where $\nu_L$ is a left-chiral active neutrino and the superscript denotes a charge conjugate spinor: $\psi^c = C\overline{\psi}\,\! ^T$ with the charge conjugation defined as $C = i\gamma_2\gamma_0$. We consider three active and $n >0$ sterile neutrinos, \textit{i.e.} ''3+$n$'' neutrinos.

From neutrino oscillation experiments it is clear that at least two of three active neutrinos are massive. Case $n=1$ is insufficient as it yields only one massive active neutrino. Therefore the case $n=2$ is minimal, while $n=3$ is often used. Such a case complements the number of generations in the SM.

Sterile neutrinos may be either Dirac or Majorana fermions. Majorana mass term can be written directly into the Lagrangian without any need to generate it via a spontaneous symmetry breaking (SSB) mechanism. This allows practically any mass for the sterile neutrinos. In addition, sterile neutrinos may have Yukawa interactions with the SM Higgs boson: $\Delta \mathcal{L}_Y = - Y_D \varepsilon_{\alpha\beta} L_{L\alpha} \phi_\beta \overline{\nu_R}$. After SSB, they produce also a Dirac mass term. When both Dirac and Majorana mass terms are present the mass term can be written as 
\begin{equation}
    \Delta \mathcal{L} = -  \frac{1}{2}\binom{\nu_L}{\nu_R^c}^TC\begin{pmatrix}
        0 & m_D \\m_D^T & M_R
    \end{pmatrix}\binom{\nu_L}{\nu_R^c} + \text{h.c.} \ , 
\end{equation}
where the neutrino mass matrix must be diagonalized with a unitary matrix $U$ to obtain the physical mass eigenstates. Here $m_D = Y_Dv/\sqrt{2}$, where $v \approx 246$ GeV is the vacuum expectation value of the SM Higgs boson and $\operatorname{Re} [Y_D] \in \: ]0,\sqrt{4\pi}[$ is the Yukawa coupling. The case $m_D=0$ (corresponding to the pure Majorana neutrino case) produces no mixing between active and sterile neutrinos. We consider instead the type-I seesaw mechanism \cite{Schechter:1980gr,Gell-Mann:1979vob,Mohapatra:1979ia,Yanagida:1979as}, which requires $0 < m_D \ll M_R$. It gives an explanation for the lightness of active neutrino masses by requiring the sterile neutrinos to be very heavy. The light neutrino mass matrix is given by seesaw formula 
\begin{equation}\label{eq:mnu}
    m_\nu \approx -m_DM_R^{-1}m_D^T = -\frac{v^2}{2}Y_DM_R^{-1}Y_D^T,
\end{equation}
which is a valid approximation if the sterile neutrinos are heavier than $\sim 1$~eV and at most $\sim 10^{15}$~GeV. Even higher seesaw scale produces too small $m_\nu$. The unitary matrix $U$ diagonalizes the neutrino mass matrix as
\begin{equation}
    U^T\begin{pmatrix}
        \textbf{0} & m_D \\m_D^T & M_R
    \end{pmatrix}U^\dagger = \operatorname{diag}(m_1,m_2,m_3,\cdots ,m_{3+n}),
\end{equation}
where $m_1, m_2$ and $m_3$ are active neutrino masses and $m_{4},\cdots,m_{3+n}$ are the sterile neutrino masses. The mixing matrix $U$ can be parameterized in several different ways, see \textit{e.g.} Ref.~\cite{Blennow:2011vn}. We use the notation
\begin{equation}
    U = \begin{pmatrix} 
        U_\nu & \theta \\ \theta^{\prime T} & U_N  
    \end{pmatrix},
    \quad \text{and} \quad
    \begin{pmatrix}
        \nu_{L,e} & \nu_{L,\mu} & \nu_{L,\tau} & \nu_{R,1} & \cdots & \nu_{R,n}
    \end{pmatrix} = U
    \begin{pmatrix}
        \nu_{1} & \nu_{2} & \nu_{3} & N_{R,1} & \cdots & N_{R,n}
    \end{pmatrix}
\end{equation}
where $\theta \approx m_DM_R^{-1}$ is a $3 \times n$ \textit{active-sterile mixing matrix} and $U_\nu \approx (I-\frac{1}{2}\theta^\dagger \theta)U_\text{PMNS}$ is in leading order the $3 \times 3$ Pontecorvo-Maki-Nakagawa-Sakata mixing matrix $U_\text{PMNS}$. While PMNS matrix is unitary, $U_\nu$ is \textit{almost} unitary as $|\theta| \ll 1$. The $n \times n$ matrix $U_N$ is an unphysical rotation matrix, and $\theta'$ is similar to $\theta$.

\begin{figure*}[t]
\centering
    \includegraphics[width=0.75\textwidth]{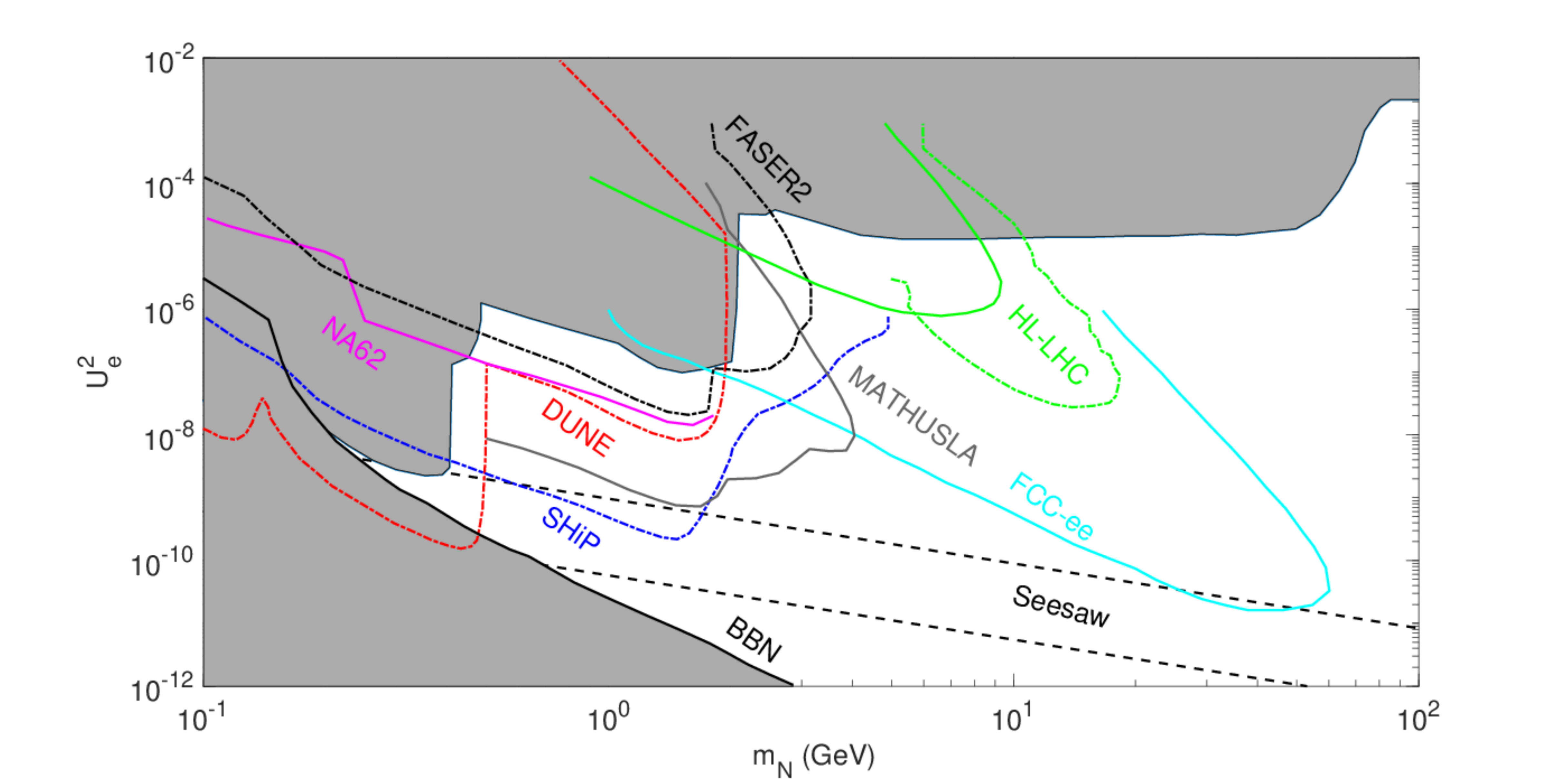}
    \includegraphics[width=0.75\textwidth]{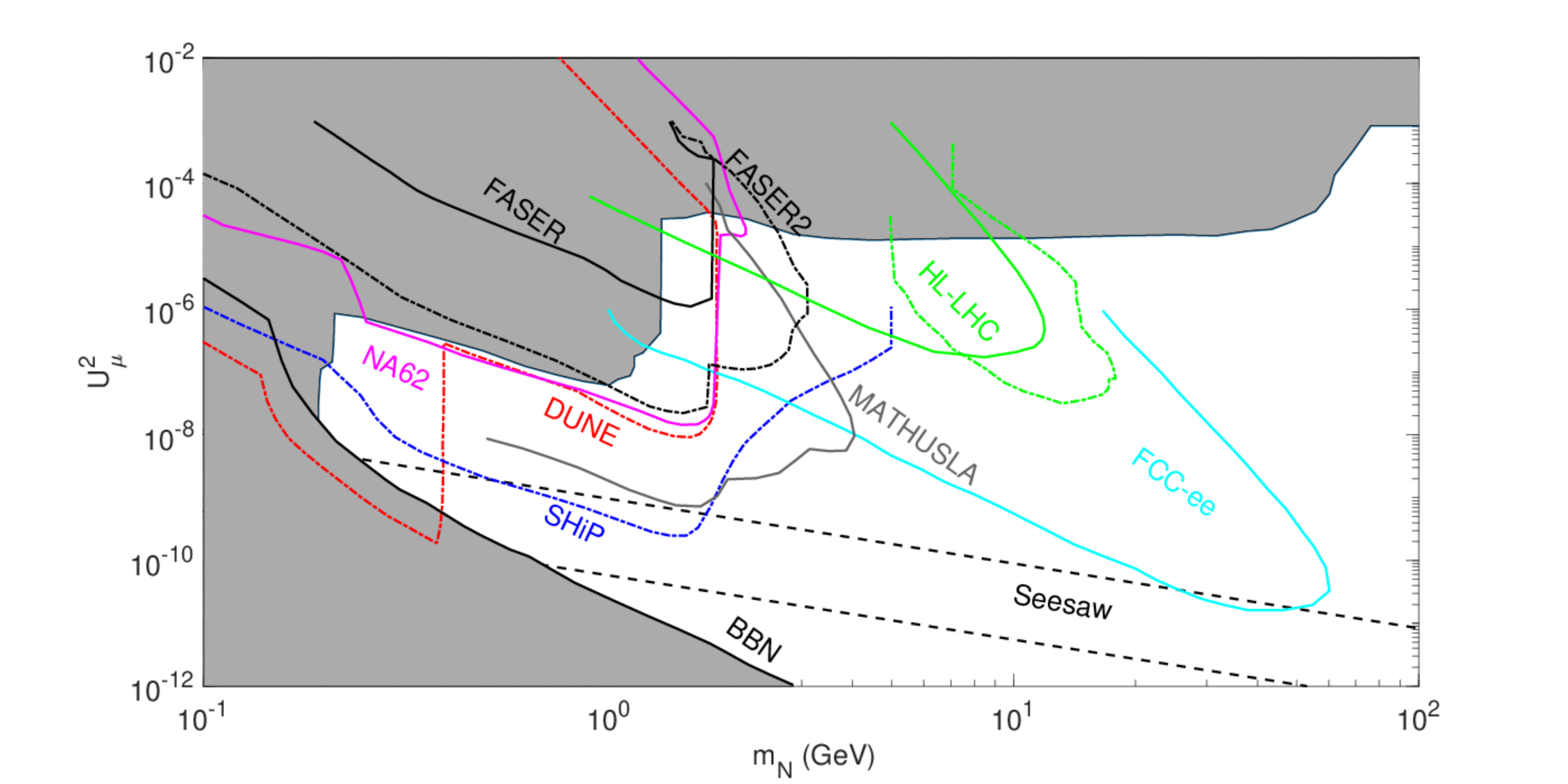}
    \includegraphics[width=0.75\textwidth]{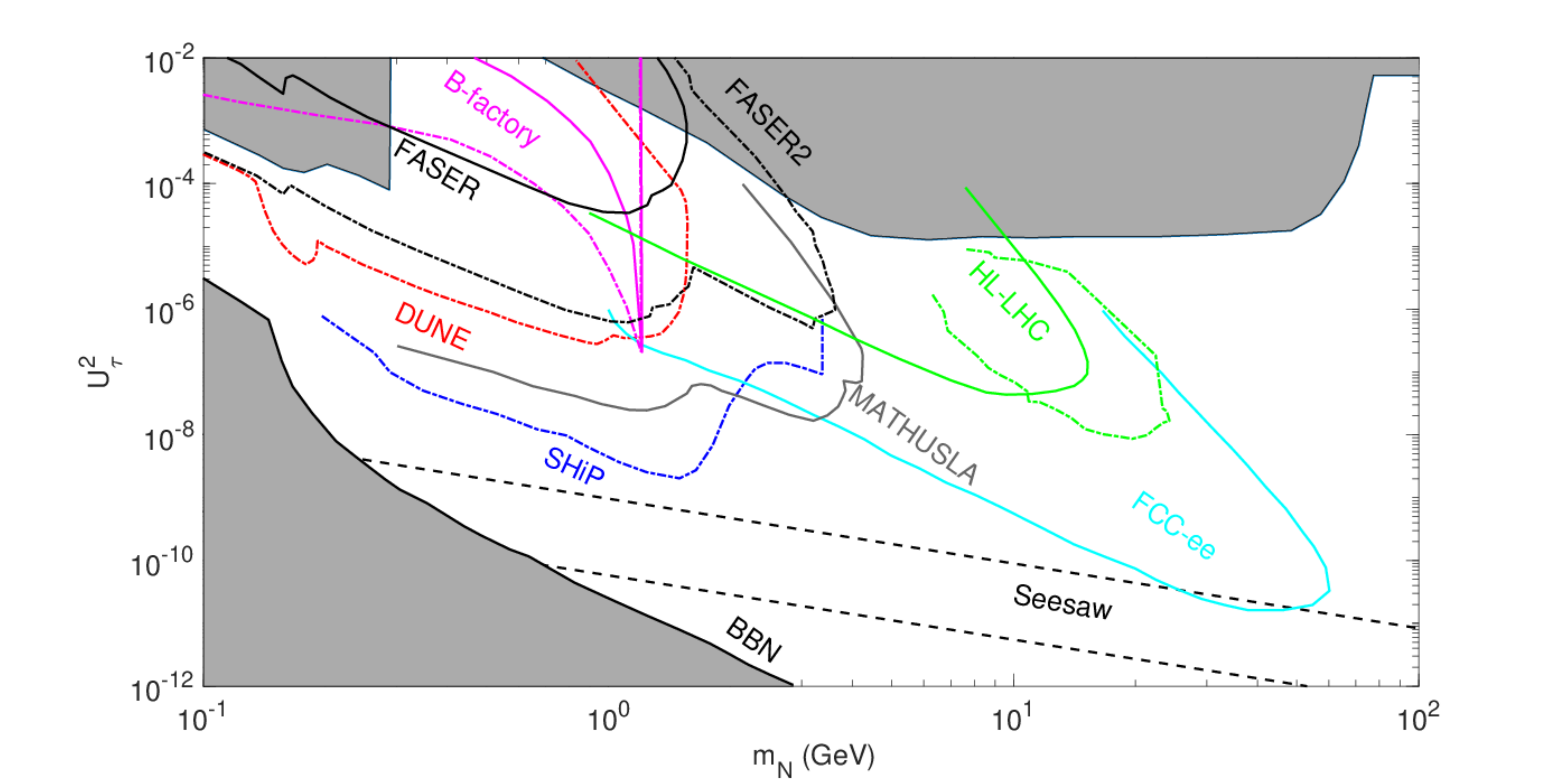}
\caption{Sensitivity of various experiments to active-sterile mixing.}
\label{fig:bsm_hnl_mixing}
\end{figure*}

Due to mixing, the physical sterile neutrinos $N_{R,i} \equiv N_i$ have couplings to Higgs, $W^\pm$ and $Z$ bosons, given by 
\begin{align}
    -\Delta \mathcal{L} &= \frac{g_L}{\sqrt{2}}\overline{N^c_{R,i}}\theta^\dagger_{i\alpha}\gamma^\mu \ell_{L\alpha}W^+_\mu - \frac{g_L}{2\cos\theta_W}\overline{N_{R,i}^c}\theta_{i\alpha}
^\dagger \gamma^\mu \nu_{L,\alpha}Z_\mu - \frac{g_L}{\sqrt{2}}\frac{m_{3+i}}{m_W}\theta_{\alpha i}h\overline{\nu_{L,\alpha}}N_{R,i} + \text{h.c.}
\end{align}
where $\theta_W$ is the Weinberg angle. The light and heavy neutrino states can be written approximately as
\begin{equation}
    \nu_i \approx (U_\nu^\dagger)_{i\alpha} \nu_{L,\alpha} - (U_\nu^\dagger \theta)_{ij} \nu_{R,j}^c 
    \quad \text{and} \quad 
    N_{R,i} \approx \nu_{R,i} + \theta_{\alpha i}\nu_{L,\alpha}^c,
\end{equation}
 The probability of active component $\nu_\alpha$ interacting in a physical sterile neutrino $N_{R,i}$ wave function is given by
\begin{equation}
    |\theta|^2 = \frac{m_\nu}{m_{3+i}} = \mathcal{O}(10^{-11}) \times \frac{\text{GeV}}{m_{3+i}},
\end{equation}
which acts as an approximate lower bound for the expected mixing. An even lower value is in conflict with the simplest seesaw mechanism, which we are considering here. Below, we have compiled the expected sensitivities of future experiments on weights of sterile components in active neutrinos $\nu_e$, $\nu_\mu$ and $\nu_\tau$, \textit{i.e.} 
\begin{equation}\label{eq:Usquared}
    U_e^2 = \sum \limits_{i=1}^n |\theta_{ei}|^2,\quad U_\mu^2 = \sum \limits_{i=1}^n |\theta_{\mu i}|^2,\quad U_\tau^2 = \sum \limits_{i=1}^n |\theta_{\tau i}|^2,
\end{equation}
respectively. The constraints corresponding to the observables in \cref{eq:Usquared} are shown in \cref{fig:bsm_hnl_mixing}. Next, we shall give details on the shown bounds. The greyed out area is excluded by experiments sensitive to peak searches, $Z$ boson decays and electroweak precision tests. The lower section labeled by ''BBN'' is excluded by big bang nucleosynthesis, which is affected by sterile neutrinos, if they are present and mix with active flavours. The effect is acceptable if a MeV scale sterile neutrino has a lifetime less than $\sim 1$~second \cite{Ruchayskiy:2012si}.

\textit{i) Peak Searches:} For MeV and GeV scale sterile neutrinos, the best constraints are given by peak searches of $\tau$ leptons and mesons. Two-body decays in parent particle's rest frame imply a particularly simple result. Decay rate of a decay $X^\pm \rightarrow \ell^\pm_\alpha N_i$ is proportional to mixing squared $|\theta_{\alpha i}|^2$ and to $m_{3+i}^2/m_\ell^2$. There is no helicity suppression for $N_i$, which has both helicities, as it is massive and not light. Sensitivity on $|\theta_{\alpha i}|^2$ actually increases with larger $m_{3+i}$ until near the cutoff mass, where it sharply decreases. Similar effect also happens with three-body decays.

At $m_{3+i} \lesssim$ 2~GeV there are considerations on meson decays $\pi^\pm \rightarrow \ell^\pm N$, $K^\pm \rightarrow \ell^\pm N$, $K^\pm \rightarrow \pi^0 \ell^\pm N$, $D^\pm \rightarrow \ell^\pm N$, $D_s^\pm \rightarrow \ell^\pm N$, decays $D^+ \rightarrow \ell^+\overline{K^0}N$ and its charge conjugate decay, and $\tau$ lepton decays $\tau \rightarrow \pi N$, $\tau \rightarrow \rho N$ and $\tau \rightarrow \ell \overline{\nu_\ell}N$. Here $\ell = e,\mu$ for all decays. For two-body decays of $D^\pm$ and $D^\pm_s$, also $\ell = \tau$ is possible. Even higher masses can be probed via $B$ meson decays $B \rightarrow X\ell N$ and $B \rightarrow \ell N$, where $X$ are mesons \cite{Belle:2013ytx}.

\textit{ii) Beam Dump Experiments:} A sterile neutrino may be considered a \textit{long-lived particle} (LLP) on a suitable mass region. At particle accelerator beam collision events, mesons are produced which may decay to sterile neutrinos and SM particles. The long-lived sterile neutrinos then propagate undisturbed away from the beam collision region and decay in a detector elsewhere. For example, the DUNE experiment is expected to further constrain the present limit on $|\theta|^2$ for $N_{R,i}$ produced in $\tau$ lepton and meson decays, which is shown as a red dot-dashed line in \cref{fig:bsm_hnl_mixing}~\cite{Coloma:2020lgy}, assuming a 5 year run with near detector and $5 \times 10^{21}$ protons on target (POT). Assuming that $L$ is the distance from sterile neutrino production point to the detector and $L_\text{det}$ is the length of its trajectory inside the detector, the probability of such a decay in the detector is
\begin{equation}
    P(E) = e^{-\Gamma L/\gamma \beta}\left( 1 - e^{-\Gamma L_\text{det}/\gamma\beta}\right),
\end{equation}
where $E$ is the sterile neutrino energy, $\beta = |\textbf{p}|/E$, \textbf{p} its momentum, $\gamma = E/m_{3+i}$ and $\Gamma$ the full decay width of $N_{R,i}$ in its rest frame \cite{Coloma:2020lgy}. Expected constraints by FASER experiment are given by black line in \cref{fig:bsm_hnl_mixing} for $U_\mu^2$ and $U_\tau^2$ (constraint for $U_e^2$ does not improve the present experimental bound and is therefore not included), assuming LHC Run 3 with integrated luminosity $\mathcal{L}_\text{int} = 150$ fb$^{-1}$, and for FASER2, assuming HL-LHC with $\mathcal{L}_\text{int} = 3$ ab$^{-1}$~\cite{FASER:2018eoc}. MATHUSLA constraint (grey line) assumes the same $\mathcal{L}_\text{int}$~\cite{Curtin:2018mvb}. NA62 constraint (magenta line) is given for LHC Run 3 with $10^{18}$ POT~\cite{Drewes:2018gkc}. SHIP constraint assumes $2\times 10^{20}$ POT during a 5 year operation~\cite{SHiP:2015vad}.

\textit{iii) Other Decays:} If the sterile neutrino is lighter than $Z$ boson, they can be produced by the decay $Z \rightarrow \nu N$. The mass region up to $\sim$ 60 GeV can be probed via a proposed FCC-ee experiment down to $|\theta|^2 = \mathcal{O}(10^{-11})$ (cyan line in \cref{fig:bsm_hnl_mixing}). Lower mass region above $B$ meson mass can be probed via HL-LHC by searching displaced vertices (DV). Solid green line corresponds to ''short DV'' strategy, where the search is performed in the inner trackers of ATLAS and CMS detectors. The dashed green line corresponds to ''long DV'' strategy, which utilizes the CMS muon tracker~\cite{Liu:2019ayx}. B-factory constraint for $U_\tau^2$ is given for both optimistic (dashed magenta) and conservative (solid magenta) estimates, assuming $10^7$ decays of $\tau$ lepton to a neutrino and three pions~\cite{Kobach:2014hea}.

\textit{iv) Seesaw Scale:} The lower bound for active neutrino masses is given by atmospheric neutrino mass splitting, $|\Delta m_{3\ell}^2|^{1/2} \approx 0.05$~eV~\cite{Esteban:2020cvm}. The upper bound of 1.1~eV is given by KATRIN experiment~\cite{KATRIN:2019yun}. However there is a more stringent bound based on cosmological data by PLANCK~\cite{Planck:2018vyg}, which set the upper limit for the sum of active neutrino masses to 0.12~eV. We do not include PLANCK limit in the figure, since it assumes a specific cosmological model which does not include sterile neutrinos. Using the KATRIN and atmospheric bounds the expected mixing can be constrained to a narrow band between the black dashed lines, shown by label ''Seesaw'' in \cref{fig:bsm_hnl_mixing}. 

\subsection{Light Long-lived Sterile Neutrinos in $\nu$SMEFT}
\label{sec:LLPHNL}

The sterile neutrino $N$, also called heavy neutral lepton (HNL), is a right-handed gauge-singlet spin-1/2 field $\nu_R$. The main motivation to include such a field is that it can couple to a left-handed lepton doublet and the Higgs field through a gauge-invariant dimension-four operator. This operator leads to a Dirac neutrino mass after electroweak symmetry breaking and can account for the observed neutrino oscillations. However, no gauge symmetry forbids a Majorana mass term for the sterile neutrino. Adding this term, in combination with the Yukawa interaction, leads to Majorana mass eigenstates and the violation of Lepton number ($L)$ by two units. Furthermore, it leads to a splitting of the neutrino spectrum into three light active neutrinos that have been observed, and additional heavier sterile neutrinos that have not been discovered. 
The masses of the heavier sterile neutrinos are not predicted and can range from the eV- to GUT-scale. Here we entertain the intriguing possibility that the neutrinos are relatively light in the MeV-GeV range. Such sterile neutrino have been linked to explanations of other problems of the SM. Light sterile neutrinos can account for dark matter \cite{Drewes:2013gca, Kusenko:2009up, Adhikari:2016bei, Boyarsky:2018tvu}, while sterile neutrinos with a broad range of masses can account for the baryon asymmetry of the Universe through leptogenesis \cite{Davidson:2008bu}. Sterile neutrinos are thus a well-motivated solution to a number of major outstanding issues in particle physics and cosmology. 

In the minimal scenario, sterile neutrinos only interact with SM fields through Yukawa interactions. In the minimal type-I seesaw model, sterile neutrinos then interact through mixing and heavy-light mixing angles scale roughly as $\sqrt{m_\nu/m_N}$ where $\m_\nu$ and $m_N$ are the masses of active and sterile neutrinos respectively. We refer to \cref{sec:LLPHNL0} and \cref{sec:LLPHNL2} for a more detailed discussion.

This is not necessarily the end of the story. In various Beyond-the-Standard-Model (BSM) models, sterile neutrinos interact through the exchange of heavy BSM fields, e.g. left-right symmetric models \cite{Mohapatra:1974gc, Pati:1974yy, Mohapatra:1980yp}, grand unified theories \cite{Bando:1998ww}, Z' models \cite{Chiang:2019ajm} and leptoquark models \cite{Dorsner:2016wpm}. While each of these models has it's own features, at relatively low energies they share many features that can be efficiently captured by the use of effective field theory (EFT). Under the assumption that BSM fields, with the exception of sterile neutrinos, have masses that lie well beyond the electroweak scale, we can integrate them out and describe them through local effective operators in the framework of the neutrino-extended Standard Model effective field theory ($\nu$SMEFT) \cite{delAguila:2008ir, Liao:2016qyd}. The relevant operators at dimension-5 are
\be
 \mathcal L^{(5)}_{\nu_L} = \epsilon_{kl}\epsilon_{mn}(L_k^T\, C^{( 5)}\,CL_m )H_l H_n\,
 \quad \text{and} \quad
 \mathcal L^{(5)}_{\nu_R}=- \bar \nu^c_{R} \,M_R^{(5)} \nu_{R} H^\dagger H\,,
\ee
which contribute to the Majorana mass of both active and sterile neutrinos after electroweak symmetry breaking.  More interesting for the present discussion are operators that appear at dimension-6. We list dimension-6 operators involving a single sterile neutrino that lead to the hadronic processes in \cref{tab:O6R}. 

\begin{table}[t]
\centering
\begin{tabular}{c|c||c|c}
	\hline 
	\hline 
	Class $1$& $\psi^2 H^3$  & Class $4$ &  $\psi^4 $\\
	\hline
	$\mathcal{O}^{(6)}_{L\nu H}$ & $(\bar{L}\nu_R)\tilde{H}(H^\dagger H)$ & $\mathcal{O}^{(6)}_{du\nu e}$ & $ (\bar{d}\gamma^\mu u)(\bar{\nu}_R \gamma_\mu e)$  \\ 
	\cline{1-2}
	Class $2$&  $\psi^2 H^2 D$ &  $\mathcal{O}^{(6)}_{Qu\nu L}$ & $(\bar{Q}u)(\bar{\nu}_RL)$  \\ 
	\cline{1-2}
	$\mathcal{O}^{(6)}_{H\nu e}$ & $(\bar{\nu }_R\gamma^\mu e)({\tilde{H}}^\dagger i D_\mu H)$ & $\mathcal{O}^{(6)}_{L\nu Qd}$ & $(\bar{L}\nu_R )\epsilon(\bar{Q}d))$ \\ 
	\cline{1-2}
	Class $3$ & $\psi^2 H^3 D$  & $\mathcal{O}^{(6)}_{LdQ\nu }$ & $(\bar{L}d)\epsilon(\bar{Q}\nu_R )$ \\ 
	\cline{1-2}
	$\mathcal{O}^{(6)}_{\nu W}$ &$(\bar{L}\sigma_{\mu\nu}\nu_R )\tau^I\tilde{H}W^{I\mu\nu}$  & &\\
	\hline
	\hline 
\end{tabular}
\caption{$\nu$SMEFT dim-6 operators \cite{Liao:2016qyd} involving one sterile neutrino field.} 
\label{tab:O6R}
\end{table}

We consider relatively light GeV-scale Majorana sterile neutrinos. Such sterile neutrinos can be produced either via direct production with parton collisions, or via rare decays of mesons that are copiously produced at the LHC interaction points~\cite{Shrock:1980vy, Shrock:1980ct}. For sterile neutrino masses below the $B$-meson threshold the primary production mode is through rare decays of mesons with subleading contributions from partonic processes, which we estimated using \texttool{MadGraph~5~3.0.2}~\cite{Alwall:2014hca} to be less than 10\%. The latter become more important and even dominant for heavier sterile neutrinos. We choose to focus on  the sterile neutrino with a mass $m_N$ below about 5~GeV and hence on the rare decays of B- and D-mesons. We do not study the production of the sterile neutrinos from the decay of lighter mesons such as $\pi^\pm$ and $K^\pm$ as their simulation in \texttool{Pythia~8}~\cite{Sjostrand:2006za, Sjostrand:2007gs} is insufficiently validated in the forward direction, which is relevant for the FASER experiments. In fact, even for $D^\pm$ and $B^\pm$ mesons, we will use \texttool{FONLL} \cite{Cacciari:1998it, Cacciari:2001td, Cacciari:2012ny, Cacciari:2015fta} to correct the behavior of \texttool{Pythia~8} in the very large pseudorapidity regime. For a more detailed discussion of forward heavy flavour production, we refer to the QCD section. Finally, we safely neglect the vector mesons decays into sterile neutrinos, as their decay width is typically many orders of magnitude larger than that of pseudoscalar mesons leading to tiny branching ratios. 

Through the mixing between active- and sterile- neutrinos and higher dimensional operators in \cref{tab:O6R}, sterile neutrinos have two-body and three-body decays. If the sterile neutrinos are relatively long-lived, their decays lead to displaced vertices that can be reconstructed in LHC detectors. We consider the LHC experiments: FASER~\cite{Feng:2017uoz, FASER:2018eoc} and  MoEDAL-MAPP~\cite{Pinfold:2019nqj, Pinfold:2019zwp}, and discuss their potential in probing active-sterile mixing angle and $\nu$SMEFT operators.  To assess and compare the sensitivities of different experiments, we show 3-event isocurves which correspond to 95\% confidence level (C.L.) with zero background and consider two scenarios: the minimal scenario and the leptoquark scenario.

\begin{figure*}[t]
\centering
	\includegraphics[width=0.49\textwidth]{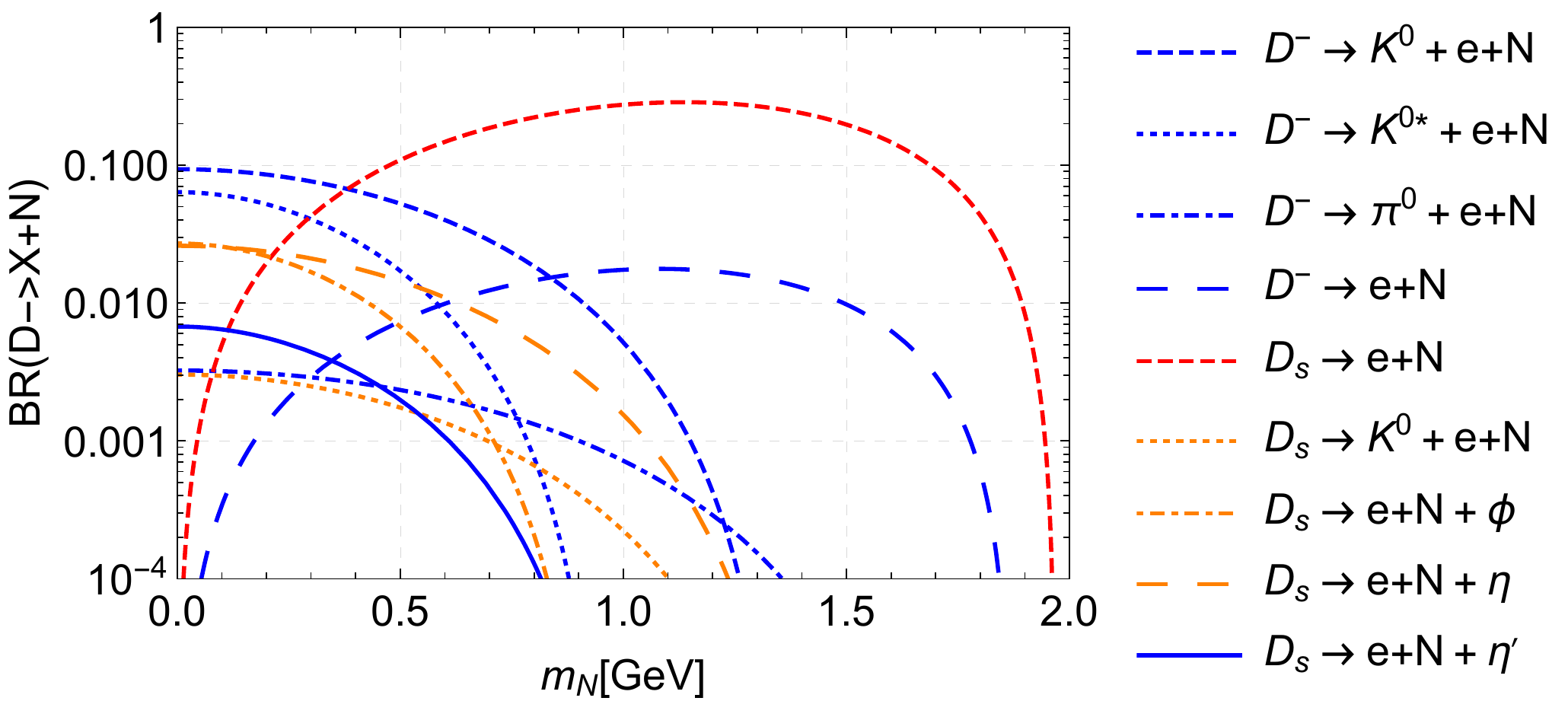}
	\includegraphics[width=0.46\textwidth]{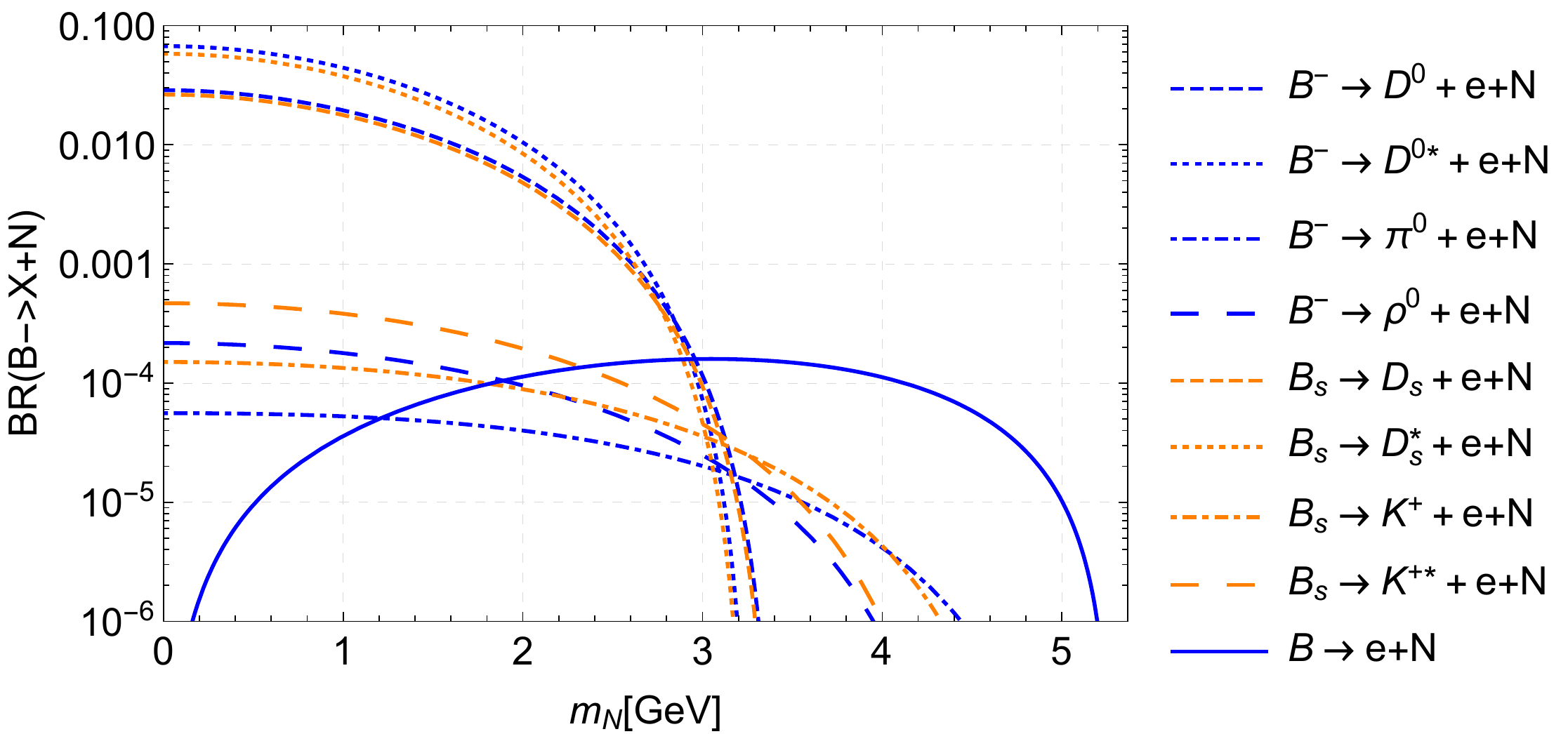}
\caption{Branching ratios of sterile neutrino production channels through $D$ (left) or $B$ (right) mesons in the minimal scenario for final-state electrons and $U_{e4} =1$.}
\label{fig:BRminimal}
\end{figure*}

\begin{figure*}[t]
\centering
	\includegraphics[width=0.50\textwidth]{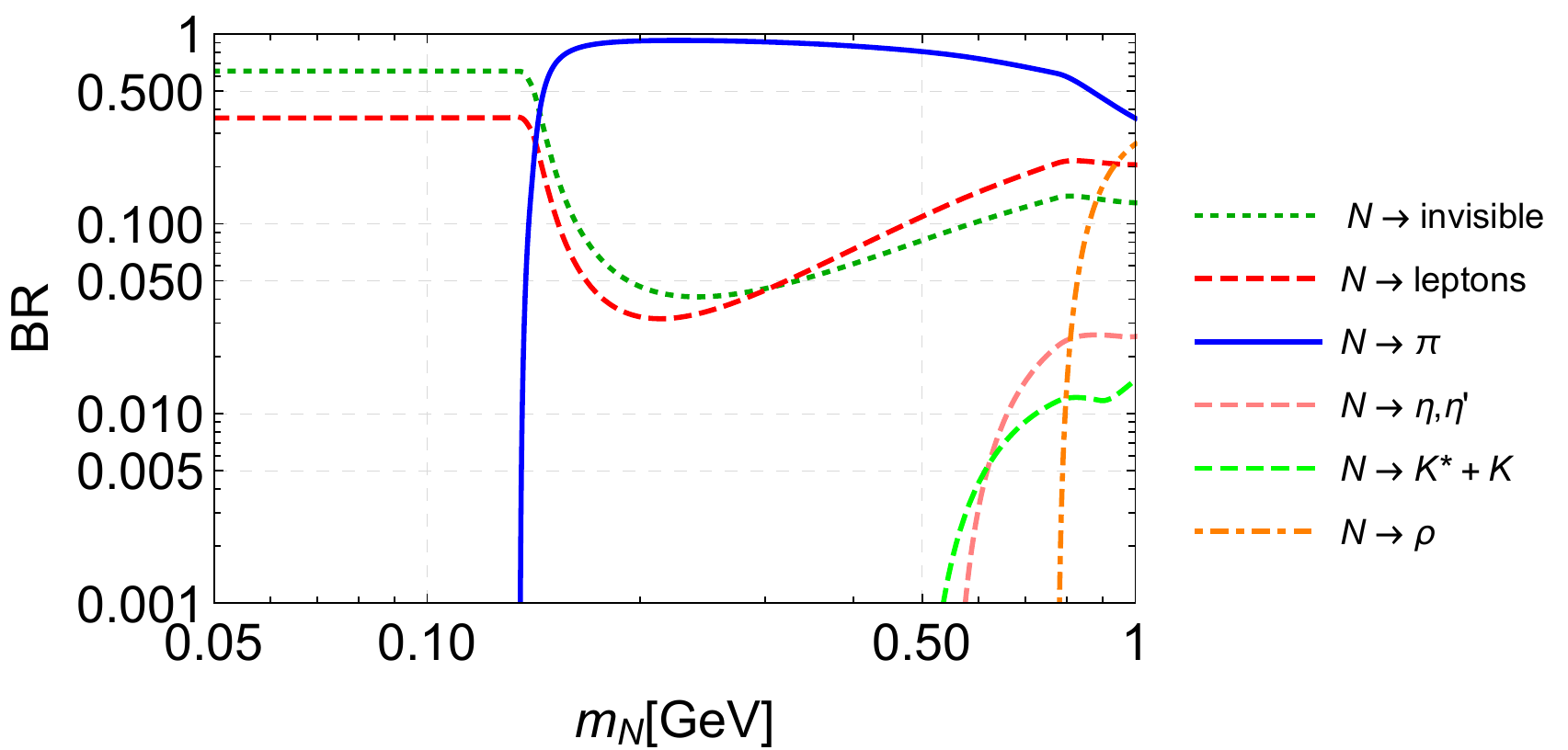}
	\includegraphics[width=0.48\textwidth]{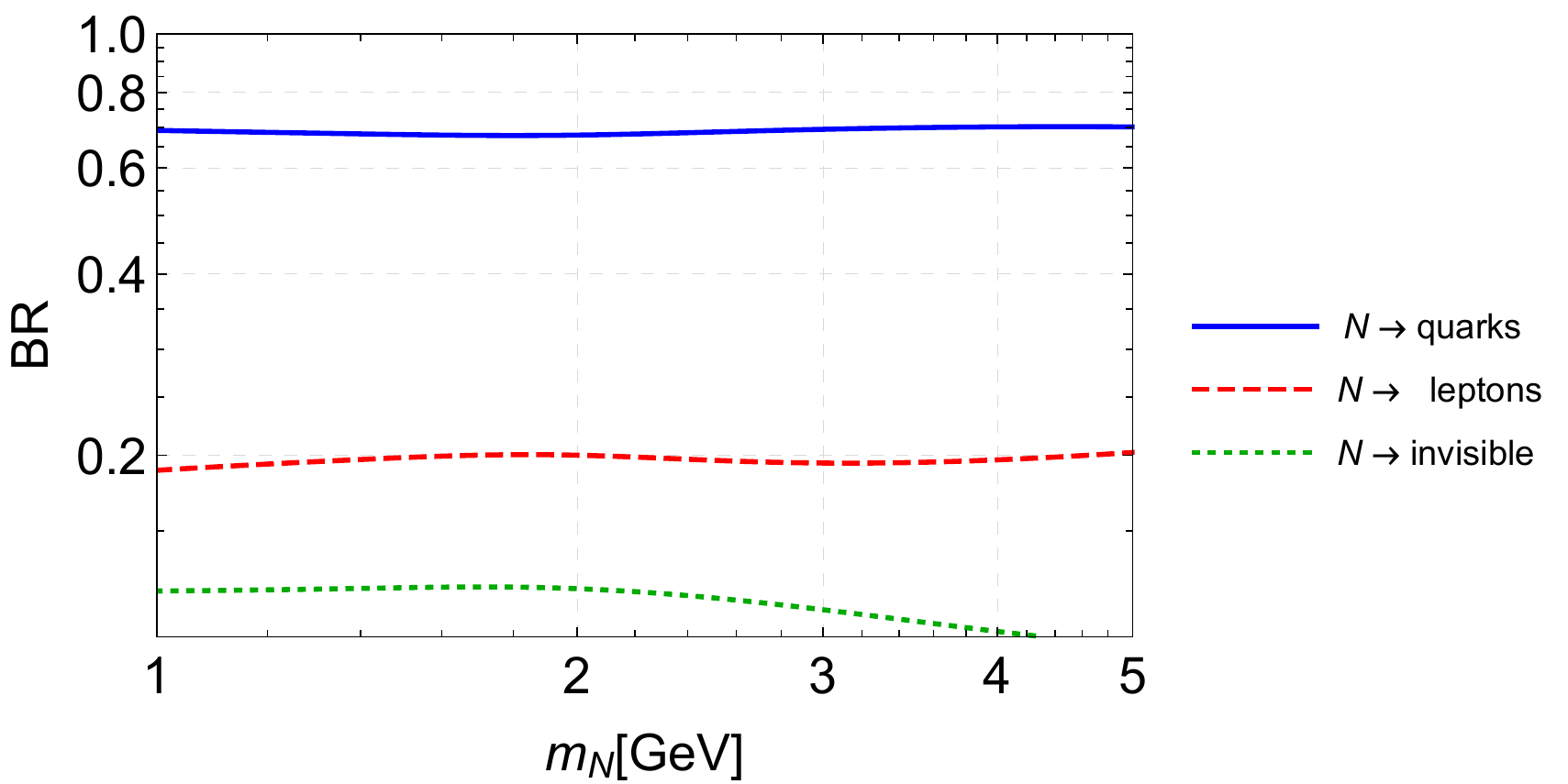}
\caption{Branching ratios in the minimal scenario for $U_{e4} =1$ and $U_{\mu4}=U_{\tau 4}=0$. Left: $m_N <1~\gev$ with decays to individual mesons. Right: $m_N > 1~\gev$ and the decays to quarks  correspond to the total hadronic branching ratio.}
\label{fig:NBR}
\end{figure*}

In the minimal scenario, we add one sterile neutrino to the SM (a 3+1 model) and it only mixes with the electron neutrino $\nu_e$ with a mixing angle $U_{e4}$. The sterile neutrino interacts with the SM fields via charged and neutral SM weak interactions. In \cref{fig:BRminimal} we show part of the branching ratios of  heavy mesons to final states including a sterile neutrino $N$ and an electron. These branching ratios have been calculated in the literature, see Refs.~\cite{Bondarenko:2018ptm, Coloma:2020lgy, DeVries:2020jbs}. We also show the branching ratios for the possible decay channels of a sterile neutrino in \cref{fig:NBR}. When the sterile neutrino mass $m_N>1~\gev$, we estimate the total hadronic  branching ratio as the branching ratio of the decays to quarks according to Ref.~\cite{Bondarenko:2018ptm}.

\begin{figure*}[t]
\centering
	\includegraphics[width=0.7\textwidth]{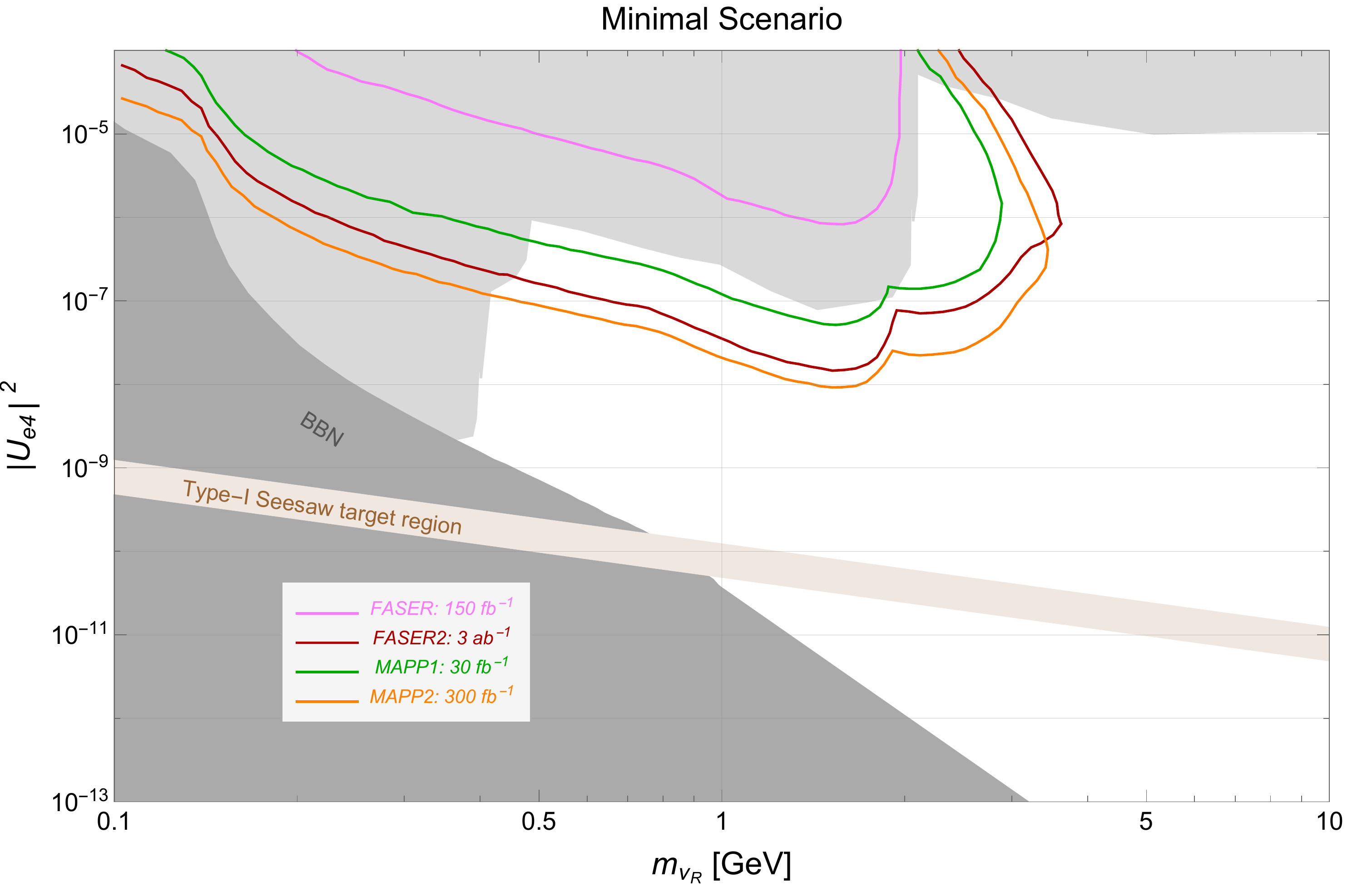}
\caption{Sensitivity for the minimal scenario with the sterile neutrino mixed solely with the electron neutrino.}
\label{fig:min_scen}
\end{figure*}

We present the results in \cref{fig:min_scen}, shown in the plane $|U_{e4}|^2$ vs. $m_{N}$. The light gray area shows the present bounds obtained by various experiments including the searches from CHARM~\cite{Bergsma:1985is}, PS191~\cite{Bernardi:1987ek}, JINR~\cite{Baranov:1992vq}, and DELPHI~\cite{Abreu:1996pa}. The dark gray area corresponds to the part excluded by big bang nucleosynthesis (BBN)~\cite{Sabti:2020yrt, Boyarsky:2020dzc}. We also show a brown band of ``Type-I Seesaw target region'' for $m_{\nu_e}$ between 0.05~eV and 0.12~eV with the relation $|U_{e4}|^2 \simeq  m_{\nu_e}/m_{N}$. These two limits are derived from neutrino oscillation and cosmological observations, respectively. The former finds that there is at least one active neutrino mass eigenstate of mass at least 0.05~eV~\cite{Canetti:2010aw} while the latter imposed an upper limit of 0.12~eV for the sum of the active neutrino masses~\cite{Aghanim:2018eyx}. FASER2 and MAPP2 extend the sensitivity reach by about an order of magnitude in the mass range between $0.5$ and $5$~GeV, but are still far away from the naive type-I seesaw predictions.

\begin{figure*}[t]
	\centering
	\includegraphics[width=0.49\textwidth]{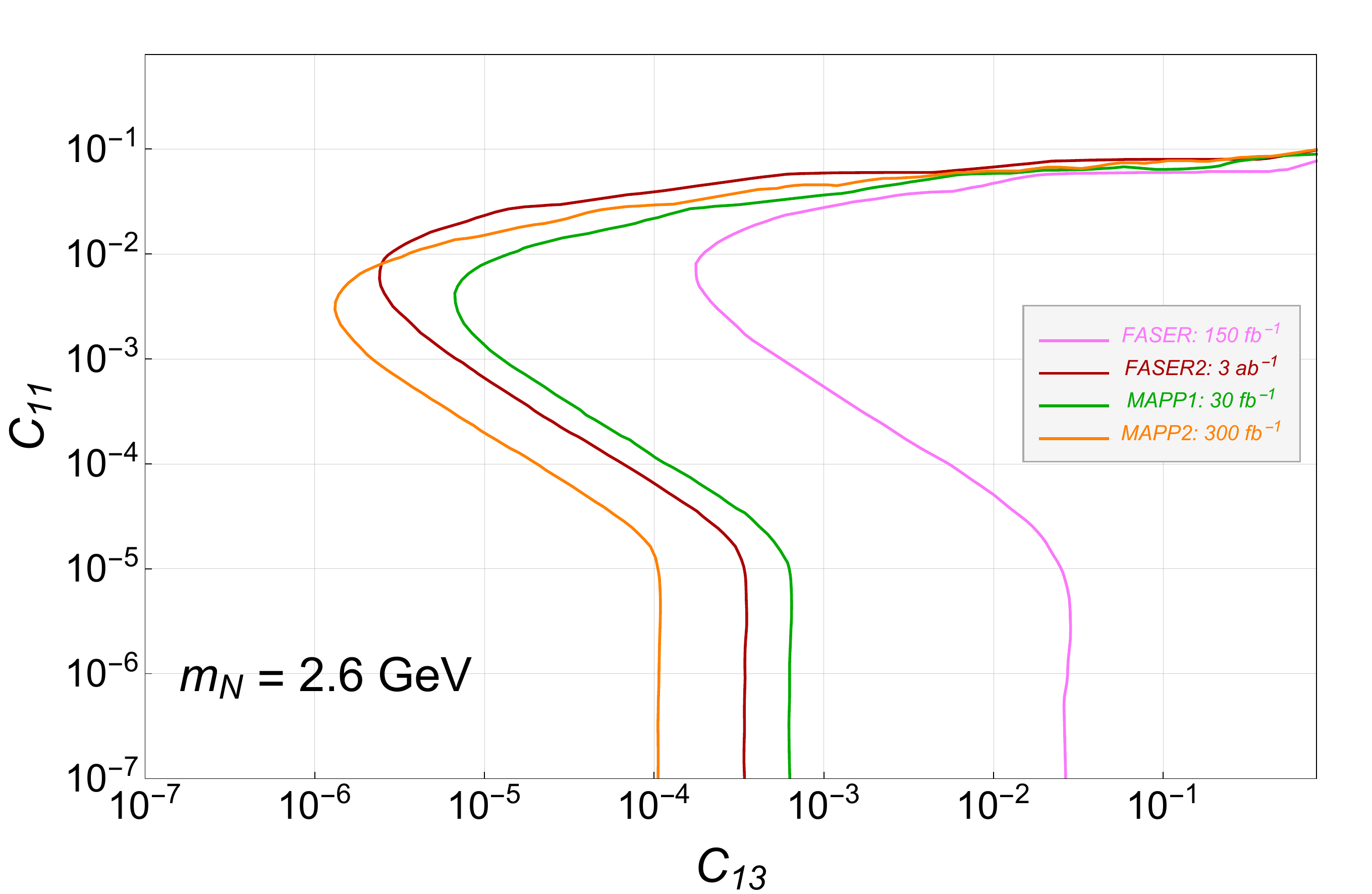}
	\includegraphics[width=0.49\textwidth]{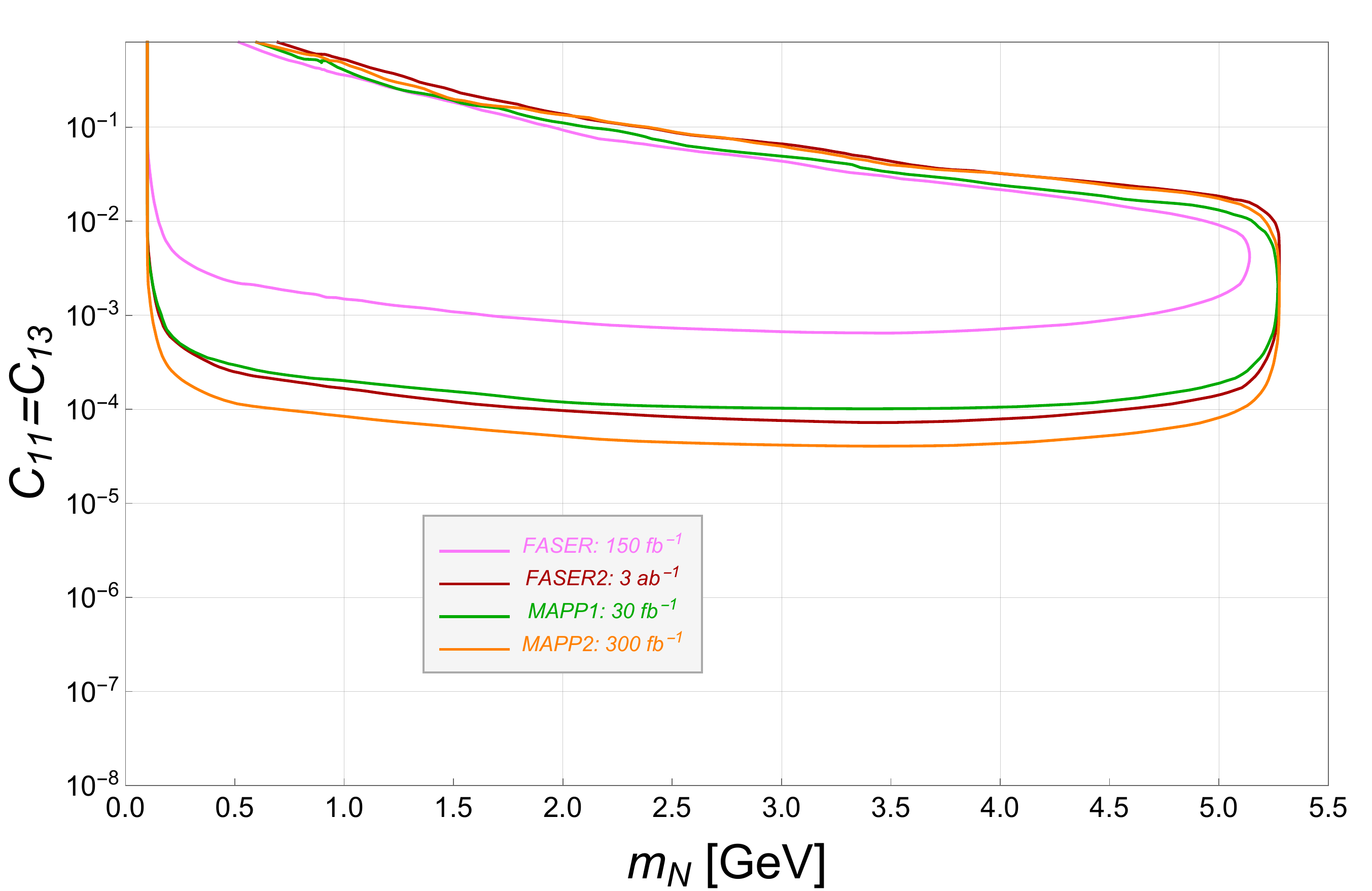}
\caption{Sensitivity for the leptoquark scenario.}
\label{fig:eft_scen}
\end{figure*}

To illustrate the EFT framework we consider a simple scenario where the minimal 3+1 model is extended with interactions generated by the exchange of leptoquarks.  We focus on the representation $\tilde R\left({\bf 3},~{\bf 2},~1/6\right)$, which can couple to the sterile neutrinos through the Lagrangian
\begin{equation}\label{eq:LQlag}
{\cal L}_{\rm LQ}=-{y}^{RL}_{jk}\bar{d}_{Rj}\tilde R^a\epsilon^{ab}L_{Lk}^{b}+y^{\overline {LR}}_{il}\bar{Q}^{a}_{Li}\tilde R^a\nu_{Rl} 
+{\rm h.c.}\,,
\end{equation}
where $a,b$ are $SU(2)$ indices and $i,j,k,l$ are flavor indices, respectively. Note that $l=1$ in the $3+1$ model. LHC constraints force the leptoquark mass, $m_{\rm LQ}$, to be above a few TeV and for low-energy purposes we can integrate it out. At tree level this leads to the effective operator 
\begin{align}
{\cal L}^{(6)}_{\nu_R}=\left(C^{(6)}_{LdQ\nu}\right)_{ijkl}\left(\bar{L}^a_k d_j \right)\epsilon^{ab}\left(\bar{Q}^b_i \nu_{Rl} \right)+{\rm h.c.}\,,
\end{align}
where 
\begin{equation}
\left(C^{(6)}_{LdQ\nu}\right)_{ijkl}=m^{-2}_{\rm LQ} \, y^{\overline{LR}}_{il}y^{RL*}_{jk}\,.
\end{equation}
To induce the production and decay of the sterile neutrino, we turn on two flavor configurations  $y^{\overline{LR}}_{11}y^{RL*}_{11}$ and $y^{\overline{LR}}_{11}y^{RL*}_{31}$ with all others being zero and set the mixing angle to the type-1 see-saw prediction. For convenience we define two dimensionless parameters $C_{11}$ and $C_{13}$ with 
\begin{equation}
C_{11}=\frac{v^2}{2}\left(C^{(6)}_{LdQ\nu}\right)_{1111} 
\quad \text{and}\quad C_{13}=\frac{v^2}{2}\left(C^{(6)}_{LdQ\nu}\right)_{1311}\, , 
\end{equation}
where $v=246$ GeV is the Higgs vacuum expectation value. These effective interactions lead to very different production and decay mechanisms of sterile neutrinos and  we refer to Ref.~\cite{DeVries:2020jbs} for a detailed analysis. We present the sensitivity plots in terms of $C_{11}$ and $C_{13}$ in \cref{fig:eft_scen}. The left panel shows results in the $C_{11}$-$C_{13}$ plane with fixed $m_{N}=2.6~\gev$. For $C_{13}<$ $10^{-6}$ no detection is possible in any of the experiments, even for large $C_{11}$. The curves become vertical for $C_{11}< 10^{-5}$ because the decay of the sterile neutrino via leptoquark interactions becomes sub-leading with respect to the contributions from minimal mixing at the new physics scale $\Lambda\sim\sqrt{v^2/C_{\mathrm{11}}} = \mathcal O(80)~\tev$. In the right panel, we fix $C_{11}=C_{13}$ and show the dependence of the sensitivity of the experiments on $C_{11}=C_{13}$ and the sterile neutrino mass $m_{N}$. MAPP2 gives the strongest sensitivity at the level of $5\times 10^{-5}$ around $m_{N}=3~\gev$, which corresponds to a new physics scale of $\mathcal{O}(20)~\tev$.

The leptoquark scenario discussed here is not by itself very interesting or compelling. We mainly presented it here to illustrate how the use of EFT methods can facilitate a quick analysis of a broad range of BSM models where sterile neutrinos only appear sterile at low energies, but in fact interact through a decoupled BSM sector. 

\subsection{Heavy Neutral Leptons with Tau Mixing in Neutrino Mass Models}
\label{sec:LLPHNL2}

\begin{figure*}[t]
\centering
	\includegraphics[width=0.49\textwidth]{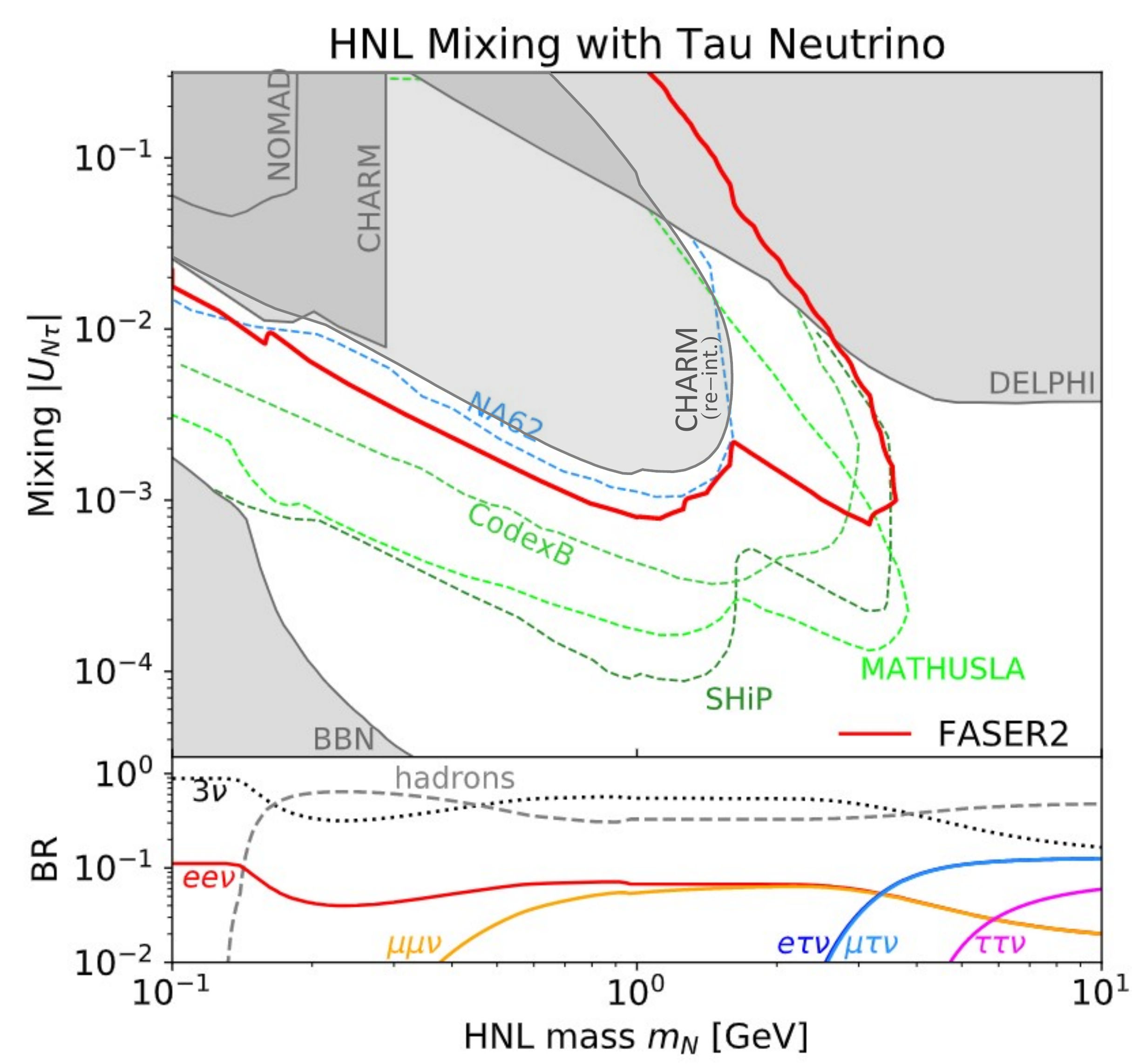}
    \caption{Sensitivity for an HNL  mixing solely with the tau neutrino in  the mass versus  mixing  plane. The  sensitivity  reaches  of  FASER2  are  shown  as  solid  red lines alongside projected sensitivities of other proposed searches and experiments, as obtained in Refs.~\cite{FASER:2018eoc, Beacham:2019nyx, Kling:2021fwx}. The gray area in the top shows the present bounds  including the searches from CHARM \cite{CHARM:1985nku} and DELPHI \cite{DELPHI:1996qcc} along with the re-interpretation of the CHARM  limits obtained in Ref.~\cite{Boiarska:2021yho}. The gray area in the bottom corresponds to the part excluded by BBN~\cite{Sabti:2020yrt, Boyarsky:2020dzc}. The bottom panels show the LLP’s branching fractions, as obtained in Ref.~\cite{FASER:2018eoc}. This figure is taken from Ref.~\cite{Anchordoqui:2021ghd}.}
\label{fig:HNLtau}
\end{figure*}

The sensitivity of FASER and FASER2 to HNLs as long-lived particles has been studied in Refs.~\cite{Helo:2018qej, FASER:2019aik, Cottin:2021lzz, Anchordoqui:2021ghd} and is also discussed in \cref{sec:LLPHNL0} and \cref{sec:LLPHNL}. As \cref{fig:HNLtau} shows, FASER2 could be sensitive to $|U_{\tau N}|^2 \sim 10^{-6}$ for HNLs masses $m_N \lesssim 5~\gev$, covering so far unexplored parts of the parameter space \cite{CHARM:1985nku, DELPHI:1996qcc} \footnote{The authors of \cite{Dib:2019tuj}  have estimated the sensitivity of HNLs coupled to taus at B-factories. Their projected limits can reach roughly $U_{\tau N}^2\sim 10^{-4}$ for HNL masses smaller than the tau mass.}. Recall, that limits from DELPHI~\cite{DELPHI:1996qcc} apply, in principle, to any SM lepton generation, $U_{\alpha N}$ with $\alpha=e,\mu,\tau$, while limits from CHARM \cite{CHARM:1985nku} apply mostly for $\alpha=e,\mu$. CHARM, however, provides very stringent limits at small HNL masses; it excludes $U_{\alpha N}^2$ down to even $U_{e/\mu N}^2 \sim 10^{-7}$ for $m_{N} \simeq 1.5~\gev$, whereas for $U_{\tau N}^2$ at most $U_{\tau N}^2 \sim 10^{-4}$ are probed. However, the authors of \cite{Boiarska:2021yho} have recently claimed that data from the CHARM experiment can be used to  obtain stronger limits on HNLs coupled to taus. Taking into account $\tau$ decays to HNLs and neutral current HNL decays to $\nu_{\tau} l^+_{\alpha} l^{-}_{\alpha}$, they rescale the published CHARM limits, to reach roughly $U_{\tau N}^2 \sim (2-3) \times 10^{-6}$ at $m_N \sim 1~\gev$. No new limits above the tau mass can be established from this re-interpretation. Given this hierarchy in limits for $U_{\alpha N}^2$ it seems very plausible that a discovery of HNLs at FASER2 would be likely to happen only if HNLs are coupled mainly to the third generation of leptons.

In this short note, we discuss the theoretical expectations for HNL decays, as motivated by different variants of the seesaw. We argue that a discovery of HNLs by FASER2 might actually shed light on the origin of neutrino masses, pointing to the underlying mechanisms being a low scale seesaw variant, such as the inverse seesaw~\cite{Mohapatra:1986bd}.

The simplest possible seesaw model is also known in the literature as the type-I seesaw. In seesaw type-I one adds three right-handed neutrinos~\footnote{Neutrino oscillation data allow for one active neutrino to be massless. In this case only two right-handed neutrinos are necessary to explain the data.} to the standard model. The well-known mass matrix for the six neutral states is given by
\begin{equation}
    \mathcal M_{\rm type-I}=\left(\begin{array}{c c} 
    0 & m_D^T \\ 
    m_D & M_R
    \end{array}\right) \, ,
\label{eq:typeImatrix}
\end{equation}
where $m_D$ is the matrix of Dirac mass terms, while $M_R$ is the Majorana mass matrix for the right-handed singlets. Note that in seesaw type-I one can always choose to work in the basis where $M_R$ is diagonal. After diagonalization of \cref{eq:typeImatrix} the light neutrino masses are given by the eigenvalues of $m_{\nu} = - m_D^T \cdot M_R^{-1}\cdot m_D$, while mixing between the light (and mostly active) and heavy (mostly sterile) neutrinos is given by
\begin{equation}
    U_{HL} = m_D^T \cdot M_R^{-1} + \dots ,
\label{eq:typeIMix}
\end{equation}
where the dots represent higher order terms, ${\cal O}(m_D^3/m_R^3)$ and higher. The Dirac mass matrix can be parametrized in terms of measured neutrino oscillation parameters (mixing matrix $U_{\nu}$ and eigenvalues ${\hat m}_{\nu}$), the eigenvalues of $M_R$ and an orthogonal matrix $R$. As was first shown in Ref.~\cite{Casas:2001sr}, the following parametrization covers all of the available parameter space for type-I seesaw:
\begin{equation}
    m_D = i \sqrt{M_R}\cdot {\cal R}\cdot \sqrt{{\hat m}_{\nu}}\cdot U_{\nu}^{\dagger}.
\label{eq:CI}
\end{equation}
The orthogonal matrix $R$ contains three complex angles, ${\cal R}= {\cal R}_3.{\cal R}_2.{\cal R}_1$. The entries in ${\cal R}_i$ can be written in terms of $s_i\equiv \sin(z_i)$, with $z_i=\kappa_i \times e^{2 i \pi \xi_i}$~\cite{Anamiati:2016uxp}. In the simplest possible choice of all $s_i=0$, the matrix $U_{HL}$ is given by:
\begin{equation}
    (U_{HL})_{ij} = (U_{\nu}^{*})_{ij} \sqrt{\frac{{\hat m}_{\nu,i}}{M_{R,i}}}.
\label{eq:typeIMixSim}
\end{equation}
For this case, (a) the mixing between active and nearly sterile heavy states is suppressed by the smallness of the light neutrino masses, typically $U_{HL}^2 \sim 5 \times 10^{-11}$, for $m_{\nu}\simeq |\Delta m^2_{\rm Atm}|^{1/2}$ and $M_R \simeq 1~\gev$. And (b) all right-handed neutrinos should, in general, decay to all three SM generations, since oscillation data has shown that no element of the neutrino mixing matrix $U_{\nu}$ can be zero. In this limit and for the best fit point of the neutrino oscillation data~\cite{deSalas:2020pgw}, one finds for the decays of the three heavy neutrinos roughly:
\begin{equation}
{\rm Br}(N_{R_i}\to l_j/\nu_{j} +X) \sim \left(\begin{array}{c c c} 
 0.66 & 0.31 & 0.022 \\
 0.09 & 0.36 & 0.55 \\
 0.24 & 0.33 & 0.42
\end{array}\right) \, .
\label{eq:Br}
\end{equation}
Here, the notation ${\rm Br}(N_{R_i}\to l_j/\nu_{j} +X)$ indicates summation over different final states $X$ (but not summing over SM generation of leptons) and we assumed $m_{\tau} \ll M_{R_i}$ for simplicity.  Clearly, the choice of $\forall s_i \equiv 0$ does not represent the most general possible case. Thus, \cref{eq:Br} is {\em not a prediction} of the type-I seesaw; it only represents a 
rough, but typical expectation.

\begin{figure*}[t]
\centering
	\includegraphics[width=0.49\textwidth]{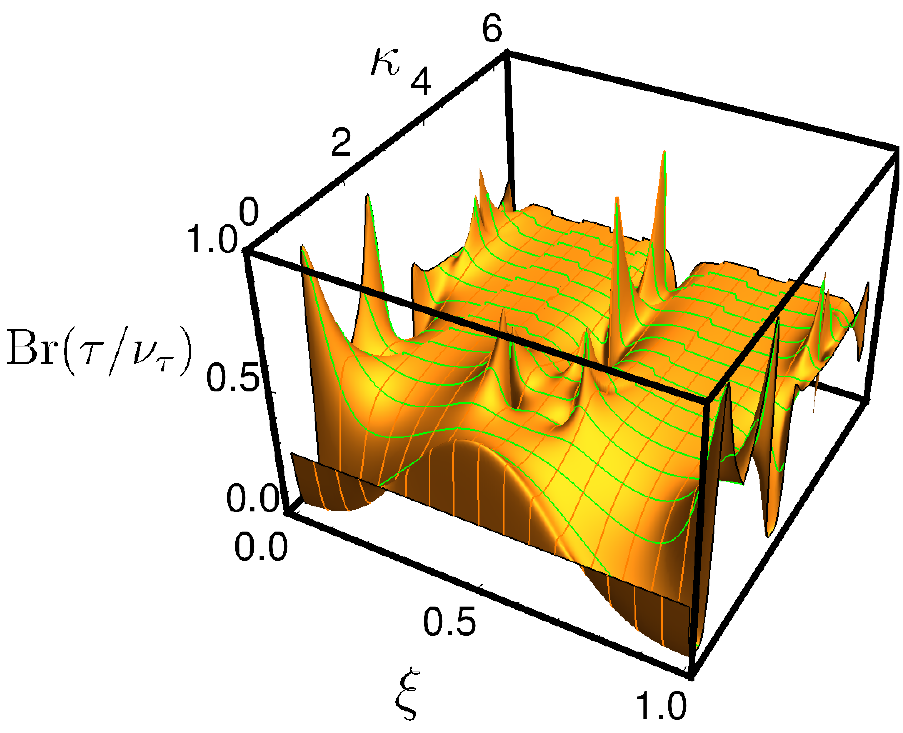}
	\includegraphics[width=0.49\textwidth]{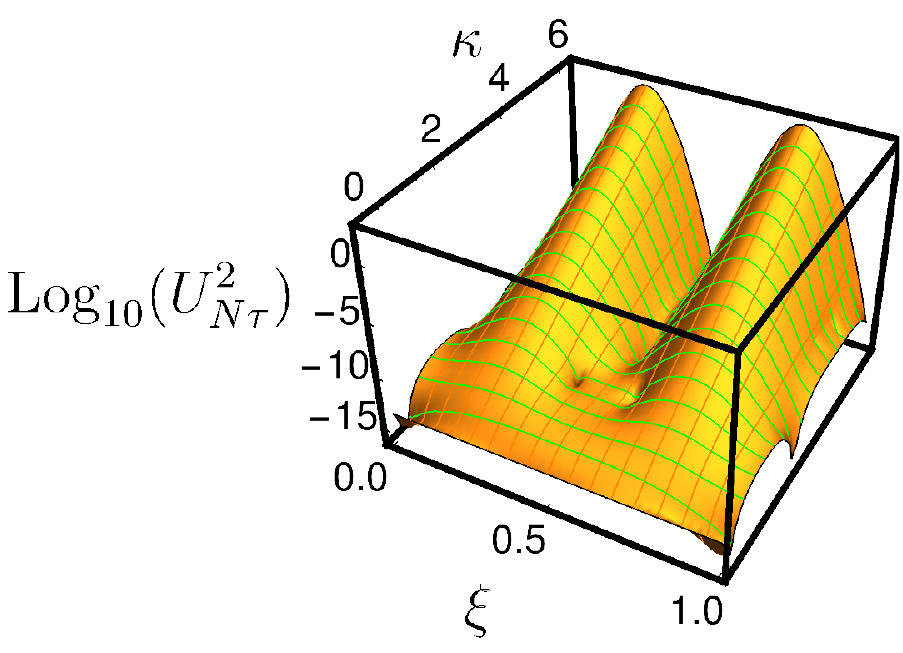}
\caption{Sum over the branching ratios of the decay of $N_1$ to final states containing either $\tau/\nu_{\tau}$ (left) and heavy-light mixing (right) as a function of the parameters $(\kappa,\xi)$, which define the Casas-Ibarra ${\cal R}$ matrix, see text.} \label{fig:mixbr}
\end{figure*}

For $s_i \ne 0$, it is possible to find particular points in parameter space, which depart strongly from the expectation expressed in \cref{eq:Br} and it is also possible to find heavy-light mixing much larger than expected from \cref{eq:typeIMixSim}. Choose, for simplicity, $(\kappa_i,\xi_i)\equiv (\kappa,\xi)$ and best fit point data for neutrino oscillations. The mixing, $U_{HL}$, of the lightest sterile neutrino, for a choice of $M_{R_1}=10~\gev$, and its branching ratio to states involving third generation leptons is then shown in \cref{fig:mixbr}. The plot to the left shows that in a very few exceptional points the branching ratio to $\tau/ \nu_{\tau}$ final states can reach values close to one. The plot to the right shows that $U_{HL}$ can be much larger than the naive expectation if the phase $2 \pi \xi$ is close to either $\pi/2$ or $3\pi/2$ {\em and} $1\ll\kappa$. We stress that, in this part of the seesaw type-I parameter space the smallness of the neutrino masses is not due to the seesaw suppression (i.e. $m_D\ll M_R$), but rather is caused by a delicate cancellation of the contributions from different right-handed neutrinos, $N_{R_i}$, to the neutrino mass matrix.  For us, it is important to note that all points, where the branching ratio to $\tau/\nu_{\tau}$ final states can approach one, lie along the lines of parameter space, where $s_i$ are real~\footnote{There are two expectional points with large branching ratios (large, but not equal to 1) to $\tau/\nu_{\tau}$ final states in this plot, for which $s$ is complex.  But these require $\kappa \sim {\cal O}(1)$ and, therefore, small $U_{HL}^2$ too.} and for real $s_i$ the absolute value of $U_{HL}^2$ is forced to be of the order ${\hat m}_{\nu,i}/M_{R,i}$, as the plot on the right of \cref{fig:mixbr} shows. In other words, for large values of the mixing $U_{HL}$, $N_1$ must decay to all SM lepton generations; points in which $N_1$ decays only to one of $e/\mu /\tau$ are not only exceptional but always occur together with (unmeasurably) small $U_{HL}^2$. While we show this here only for $N_1$, see \cref{fig:mixbr}, we have checked that qualitatively the same conclusions are reached for $N_2$ and $N_3$. We have also checked that scanning over the allowed active neutrino angles does not affect this general conclusion, although of course the exact values of $N_i$ branching ratios do depend on the numerical values of $U_{\nu}$ one chooses.

This conclusion is valid only for type-I seesaw models, very different values for $U_{HL}^2$ and $N_i$ branching ratios can be obtained in extended models. Let us discuss for example the inverse seesaw mechanism \cite{Mohapatra:1986bd} (an analogous chain of arguments can be constructed for the {\em linear} seesaw model~\cite{Akhmedov:1995ip, Akhmedov:1995vm}). In this model, three additional singlets, called $S$, are added and the ($9,9$) mass matrix can be written as:
\begin{equation}
    \mathcal M_{\rm ISS}=\left(\begin{array}{c c c} 
    0 & m_D^T & 0 \\ 
    m_D & 0 & M_R \\ 
    0 & M_R^T & \mu 
    \end{array}\right) \, .
\label{eq:ISSmatrix}
\end{equation}
This matrix has the special property that in the limit $\mu\equiv 0$ lepton number is restored and the three active neutrinos are massless. For this reason, a small value of $\mu$ is technically natural and, for $\mu \ll m_D \ll M_R$, the mass matrix for the lightest three states can be estimated via:
\begin{equation}
m_{\rm ISS} = m_D^T \, {M_R^T}^{-1} \, \mu \, M_R^{-1} \, m_D ,
\label{eq:numassISS}
\end{equation}
Heavy-light mixing is given by $U_{HL} =\frac{1}{\sqrt{2}} m_D^T \cdot M_R^{-1}$. However, the relation of $U_{HL}$ to light neutrino masses is changed, yielding the naive expectation $U_{HL} \simeq \sqrt{m_{\nu} / \mu}$. Thus, in the inverse seesaw much larger mixing between heavy and light neutrinos is naturally expected. In fact, in inverse seesaw heavy-light mixing can easily saturate existing bounds.

The inverse seesaw has more free parameters than the simplest type-I seesaw. One can formulate a parametrization of $m_D$ in terms of neutrino oscillation parameters, $M_R$ and $\mu$~\cite{Cordero-Carrion:2019qtu} in the same spirit as the Casas-Ibarra parametrization for the type-I seesaw~\cite{Casas:2001sr}. For brevity we will not show all equations here, and refer to Ref.~\cite{Cordero-Carrion:2019qtu}. However, already a superficial inspection of \cref{eq:numassISS} shows that there are a number of different limits (for $m_D,M_R,\mu$), all of which reproduce the measured neutrino angles correctly, but which give very different expectations for the decay branching ratios of the right-handed neutrinos.

Here, we will discuss only the two simplest examples. Let us (a) choose $\mu$ and $M_R$ diagonal or (b) choose $m_D$ and $M_R$ diagonal. The remaining matrix can then be fixed by the requirement of correctly reproducing neutrino oscillation data. In case (a) $m_D$ (for ${\cal R}=1$) must be proportional to $U_{\nu}^{\dagger}$. In this case the expectation for branching ratios of the heavy neutrino decays is similar to the type-I case, with the only difference being that the entries in the mixing matrix $U_{HL}$ are now a factor $f=(M_{R,i}/\mu_{i})$ larger than the type-I seesaw expectation. More interesting is case (b). In this case, each of the right-handed neutrinos will decay to only one generation of SM leptons. In particular, one of the right-handed neutrinos will decay {\em exclusively to tau leptons and tau neutrinos} (plus hadrons/mesons). While this simple discussion obviously does not cover all available the parameter space of inverse seesaw models, it serves to demonstrate that an observation of an HNL decaying to $\tau$'s at FASER2 would already provide a strong hint that the underlying neutrino mass model can not be the simplest type-I seesaw.

In conclusion, in this short note we have discussed that a search  for HNLs coupling exclusively to the third generation of leptons  at FASER2 is well motivated from the theoretical point of view.  A discovery of HNLs in the sensitivity region predicted for FASER2 would point towards low-scale seesaw models, such as the inverse or the linear seesaw, as the explanation of the observed neutrino masses and angles. 

\subsection{Tree-level Decays of GeV-Scale Neutralinos from $D$ and $B$ Mesons}
\label{subsec:LLP-susy-neutralino-I}

In the supersymmetric extension of the Standard Model (SSM)~\cite{Wess:1974tw, Nilles:1983ge, Martin:1997ns}, one additional Higgs field as well as superpartners related to each SM  field  are introduced. This symmetry is softly broken by explicit sfermion and gaugino masses, and couplings with positive mass dimensions. The resulting electroweak gauginos are not mass eigenstates. After electroweak symmetry breaking, the neutral electroweak gauginos, $\tilde{B}^0$ and $\tilde{W}^0$, mix with the neutral higgsinos, $\tilde{H}_u^0$ and $\tilde{H}_d^0$, to four Majorana mass eigenstates called neutralinos $\tilde{\chi}^0_i$ with $i=1,2,3,4$. The allowed supersymmetric particle interactions respecting gauge symmetries can be divided into pure gauge interactions, matter-gauge couplings  and solely matter interactions. In the SSM the latter are governed by the superpotential, whose most general renormalizable form can be expressed as
\be
    W&= W_{\mathrm{MSSM}} + W_{\mathrm{RPV}},\\
    W_{\mathrm{MSSM}}&= Y^u_{ij}\bar{U}_iQ_jH_u-Y^d_{ij}\bar{D}_iQ_jH_d-Y^l_{ij}\bar{E}_iL_jH_d +\mu H_uH_d,\\
    W_{\mathrm{RPV}}&= \kappa_iL_iH_u+\lambda_{ijk}L_iL_j\bar{E}_k + \lambda^\prime_{ijk}L_iQ_j\bar{D}_k + \lambda^{\prime\prime}_{ijk}\bar{U}_i\bar{D}_j\bar{D}_k\label{eq::llp_neutralinos::superpotential}
\ee
The superpotential can be further restricted to just $W_{\mathrm{MSSM}}$ by imposing for example a discrete $\mathbb{Z}_2$-symmetry called R-parity~\cite{Farrar:1978xj} (or a $\mathbb{Z}_6$ symmetry called proton hexality~\cite{Dreiner:2005rd}). If R-parity is conserved the lightest supersymmetric particle (LSP), typically the lightest neutralino $\tilde{\chi}_1^0$, must be stable, making it an attractive dark matter candidate. Currently, the lower limit on the mass of the lightest neutralino is $m_{\tilde{\chi}_1^0}>46~\gev$~\cite{ParticleDataGroup:2020ssz}. This bound is obtained via limits on the range of $\mu$ and the soft supersymmetry breaking mass parameter $M_2$ from chargino searches, and assuming the unification of gauge couplings at some larger energy scale, for example in a grand unified theory
\begin{equation}
    M_1=\frac{5}{3}\tan^2\theta_W M_2\approx \frac{1}{2}M_2\, 
\label{eq::llp_neutralinos::gut_relation}
\end{equation}
at the electroweak scale. It has been found that when \cref{eq::llp_neutralinos::gut_relation} is dropped, the neutralino can essentially be massless and still be consistent with observations from cosmology and accelerator data~\cite{Dreiner2003, Dreiner:2009ic, Dreiner:2011fp}. However, a very light neutralino can not remain stable. The Cowsik-McClelland bound~\cite{Cowsik:1972gh} and the Lee-Weinberg bound~\cite{Lee:1977ua, Barman:2017swy} imply that a neutralino in the mass range 
\begin{equation}
    \label{eq::llp_neutralinos::restricted_neutralino_mass_range}
    0.7~\text{eV} < m_{\tilde{\chi}_1^0} < 34~\gev \, ,
\end{equation}		 
must decay via R-parity violating couplings. Subsets of the couplings in $W_{\mathrm{RPV}}$ are as well-motivated as R-parity conserving interactions and should not be dismissed~\cite{Dreiner:1997uz, Dreiner:2012ae}. To conserve baryon-number and avoid bounds due to proton decays, we extend the SSM symmetry group by a discrete $\mathbb{Z}_3$ symmetry (baryon-triality)~\cite{Dreiner:2005rd, Dreiner:2011ft}, leaving only the lepton-number violating terms of $W_{\mathrm{RPV}}$. We are specifically interested in  GeV-scale neutralinos that are produced by rare meson decays and subsequently decay at a displaced vertex~\cite{Dreiner:2009er}. We focus on the $LQ\bar{D}$-operator, which together with fermion-sfermion-gaugino-interactions allows for such hadronic processes. As current sfermion mass bounds are $m_{\tilde{f}} > \mathcal{O} (500~\gev)$~\cite{ParticleDataGroup:2020ssz}, we can integrate out the heavy degrees of freedom and obtain a tree-level effective Lagrangian consisting of dimension-6 four-fermion operators, such as
\begin{align}
\label{eq:Theory_EffectiveInteractions}
    G_{ijk}^{S,l}&\times\left(\bar{\tilde{\chi}}_1^0P_Ll_i\right)\left(\bar{d}_k P_Lu_j\right)\,,
    &\hspace{5mm}&G_{ijk}^{T,\nu}\times\left(\bar{\tilde{\chi}}_1^0\sigma^{\mu\nu}P_L\nu_i\right)\left(\bar{d}_k\sigma_{\mu\nu}d_j\right)\,.
\end{align}
The new couplings $G_{ijk}$ are functions of rescaled RPV couplings $\lambda^\prime_{ijk}/m^2_{\tilde{f}}$, the electroweak gauge coupling $g_2$ and the Weinberg angle $\theta_W$. For a more detailed description we refer to Ref.~\cite{deVries:2015mfw}.
		
With the effective interactions, we can obtain the decay widths for the production of neutralinos via two-body meson decays $M^{(*)}_{jk}\rightarrow \tilde{\chi}_1^0 + l_i$, and estimate the number of produced light neutralinos over the runtime of an experiment via the decay of a meson $M$ as
\begin{equation}
    \label{eq::llp_neutralinos::prod_neutralinos}
    N_{M\tilde{\chi}_1^0}^{\text{prod}}= N_M \tau_M \Gamma (M^{(*)}_{jk}\rightarrow \tilde{\chi}_1^0 + l_i),
\end{equation}
where $N_M$ is the total number of produced mesons and $\tau_M$ is their respective lifetime. The total number of neutralinos is the sum over all possible meson contributions. As the lifetime of vector mesons is several orders of magnitudes smaller compared to pseudoscalar meson with the same quark configuration, we can neglect the contribution of vector mesons here. To estimate the number of observed neutralinos, we employ the decay widths $\Gamma(\tilde{\chi}_1^0\rightarrow M^{(*)}_{jk}+l_i )$, where both pseudoscalar and vector mesons need to be taken into account. 
		
However, it is not sufficient for the neutralino to only decay. The displaced vertex must be located inside the detector volume, the probability of which is a function of the neutralino's kinematics and the geometry of the setup. By simulating $N_{\mathrm{MC}}$ neutralino events with a suitable Monte Carlo event generator such as \texttool{Pythia~8} \cite{Sjostrand:2006za, Sjostrand:2007gs, Sjostrand:2014zea}, we can determine an average decay probability $\langle P[(\tilde{\chi}_1^0]$ in d.r.$]\rangle$. Additionally, a decay inside the detectable region does not necessarily entail that the neutralino is detectable. The final state momenta and their path through the detector need to be observable, to allow for a sufficient reconstruction of the displaced vertex. As neutral mesons in the final states are accompanied by a neutrino, the information about momentum and the path through the detector of the second particle are lost in such cases. Hence, we consider these modes ``invisible''. Decay modes involving charged mesons include charged leptons, so that these final states are detectable or ``visible''. The number of observable neutralino events is then given by
\begin{equation}
    \label{eq::llp_neutralinos::obs_neutralinos}
    N_{\tilde{\chi}_1^0}^{\text{obs}}=\langle P[(\tilde{\chi}_1^0]\text{ in d.r}]\rangle\cdot  \text{Br}(\tilde{\chi}_1^0\rightarrow \text{visible})\sum_MN_{M\tilde{\chi}_1^0}^{\text{prod}}.
\end{equation}
The process described above necessitates that at least two flavour configurations of the RPV coupling $\lambda^\prime_{ijk}$ are non-zero. As a pseudoscalar meson M decays into $\tilde{\chi}_1^0$, which subsequently decays into charged final states, including a second meson $M^\prime$, the allowed neutralino mass interval is $m_{M^\prime}+m_{l^\prime} < m_{\tilde{\chi}} < m_{M}-m_{l}$. If only a single coupling were non-zero, the mass range would either be not allowed or very restricted, which would result in low momenta and hence challenging to observe final states. Thus, we assume two non-zero $\lambda^\prime$-couplings.
		
Because the simulation of light mesons is not adequately validated in the  forward region in \texttool{Pythia} and the produced neutralinos would be very light, we do not consider them. We further compensate this insufficiency of \texttool{Pythia} in simulating the forward region for the production of heavy $D-$ and $B-$mesons by employing LHCb meson- and quark-production data \cite{Aaij:2015bpa, Aaij:2016avz} and extrapolating their findings with the help of \texttool{FONLL} \cite{Cacciari:1998it, Cacciari:2001td, Cacciari:2012ny, Cacciari:2015fta}.

\begin{table}[tb]
\centering
\begin{tabular}{l|l|l}
\hline\hline
    & charm scenario & bottom scenario \\
\hline\hline
    $\lambda^\prime_P$ production coupling & $\lambda^\prime_{122}$ & $\lambda^\prime_{131}$ 
\\
    $\tilde{\chi}_1^0$ producing meson(s) & $D_s^\pm$ & $B^0,\bar{B}^0$
\\
    $\lambda^\prime_D$ decay coupling & $\lambda^\prime_{112}$ & $\lambda^\prime_{112}$ 
\\
    visible final state(s) & $K^\pm+e^\mp$, $K^{*\pm}+e^\mp$ & $K^\pm+e^\mp$, $K^{*\pm}+e^\mp$
\\
    invisible final state(s) through $\lambda^\prime_P$ & $\big(\eta,\eta^\prime,\phi\big)+\big(\nu_e, \bar{\nu}_e\big)$ & None
\\
    invisible final state(s) through $\lambda^\prime_D$ & $\big(K_L, K_S, K^{*0}\big)+\big(\nu_e, \bar{\nu}_e\big)$ & $\big(K_L, K_S, K^{*0}\big)+\big(\nu_e, \bar{\nu}_e\big)$\\
\hline\hline
\end{tabular}
\caption{Characterizing features of the charm and bottom scenario.}\label{tab::llp_neutralino::benchmarkScenario}
\end{table}

\begin{figure*}[t]
	\centering
	\includegraphics[width=0.49\textwidth]{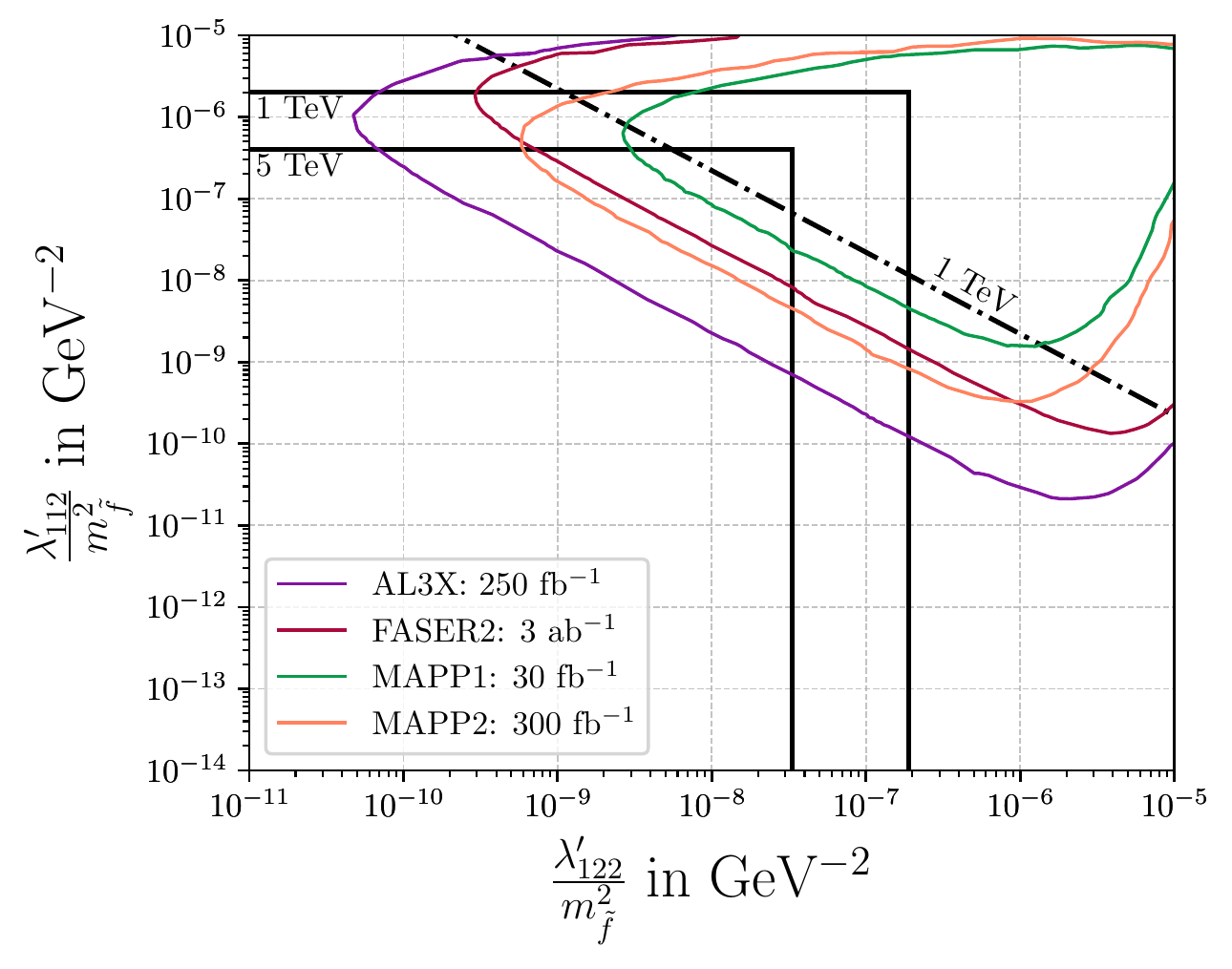}
	\includegraphics[width=0.49\textwidth]{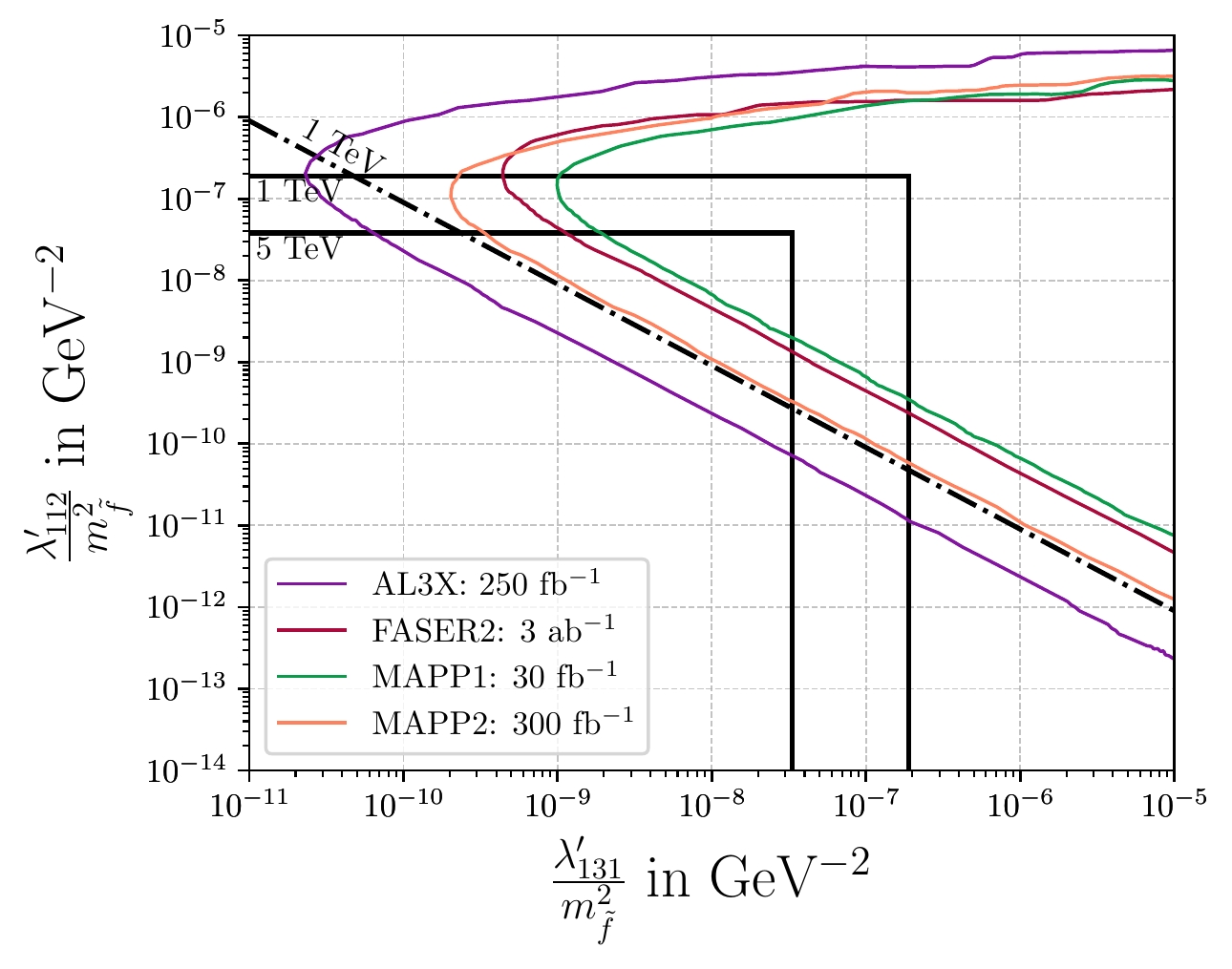}
	\caption{Sensitivity reach comparison between AL3X, FASER2, MAPP1 and MAPP2  in the $\lambda_D/m^2_{\tilde{f}}$ vs. $\lambda_P/m^2_{\tilde{f}}$ representation. Left: results for the charmed benchmark scenario with a neutralino mass of $m_{\tilde{\chi}}=1.2\gev$.  Right: results for the bottomed benchmark scenario with $m_{\tilde{\chi}}=3\gev$.}
	\label{fig::llp_neutralino::couplingvcoupling}
\end{figure*}

We display two scenarios, one considering produced charmed mesons, $D_s^\pm$, and the other bottomed mesons, $B^0,\,\bar B^0$ that can each decay into neutralinos, respectively. The specific flavour couplings, initial mesons and all final states are summarized in \cref{tab::llp_neutralino::benchmarkScenario}. The independent parameters in each scenario are the neutralino mass $m_{\tilde{\chi}}$, the production coupling strength $\lambda^\prime_P$ and decay coupling strength $\lambda^\prime_D$. For each scenario, we consider three LHC-detector concepts: AL3X~\cite{Gligorov:2018vkc}, FASER and FASER2~\cite{Feng:2017uoz, FASER:2018eoc}, and MoEDAL-MAPP \cite{Pinfold:2019nqj, Pinfold:2019zwp}, which were investigated in Refs.~\cite{Dercks:2018wum, Dercks:2018eua, Dreiner:2020qbi}. For every detector, we assume a detection efficiency of 100\%, and we do not consider the effects of background events. Our sensitivity estimates are displayed in terms of three-event isocurves that correspond to the 95\% C.L. exclusion limit with no background. Since there are three independent parameters, we either fix the relation between the couplings or fix $m_{\tilde{\chi}}$  and depict the results in terms of the remaining two parameters. For brevity, we restrict ourselves to the latter. The results shown in \cref{fig::llp_neutralino::couplingvcoupling} are for a neutralino mass $m_{\tilde{\chi}}$ in the middle of the available mass range. The solid black lines represent current limits on the individual RPV-couplings for certain sfermions masses. The dot-dashed line corresponds to an indirect bound on the product of both couplings~\cite{Allanach:1999ic}.
		
In the charm scenario all detectors can cover parameter regions beyond the current bounds, whereas only AL3X extends past the limit on the product of both couplings in the bottomed scenario by one order of magnitude. In both scenarios, AL3X can cover the largest parameter space beyond existing limits followed by the extended MAPP2 detector. 
		
\begin{figure*}[t]
	\centering
	\includegraphics[width=0.49\textwidth]{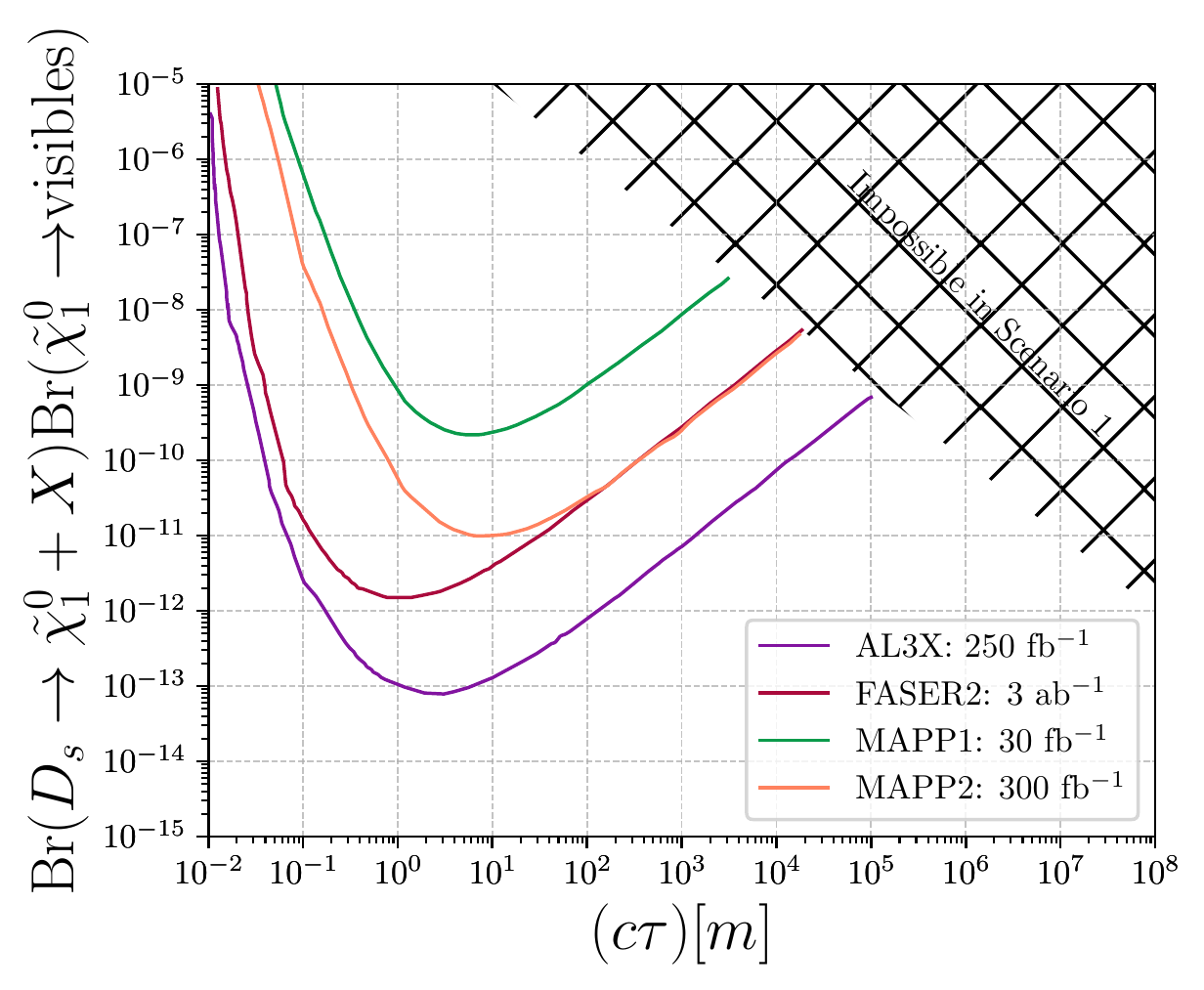}
	\includegraphics[width=0.49\textwidth]{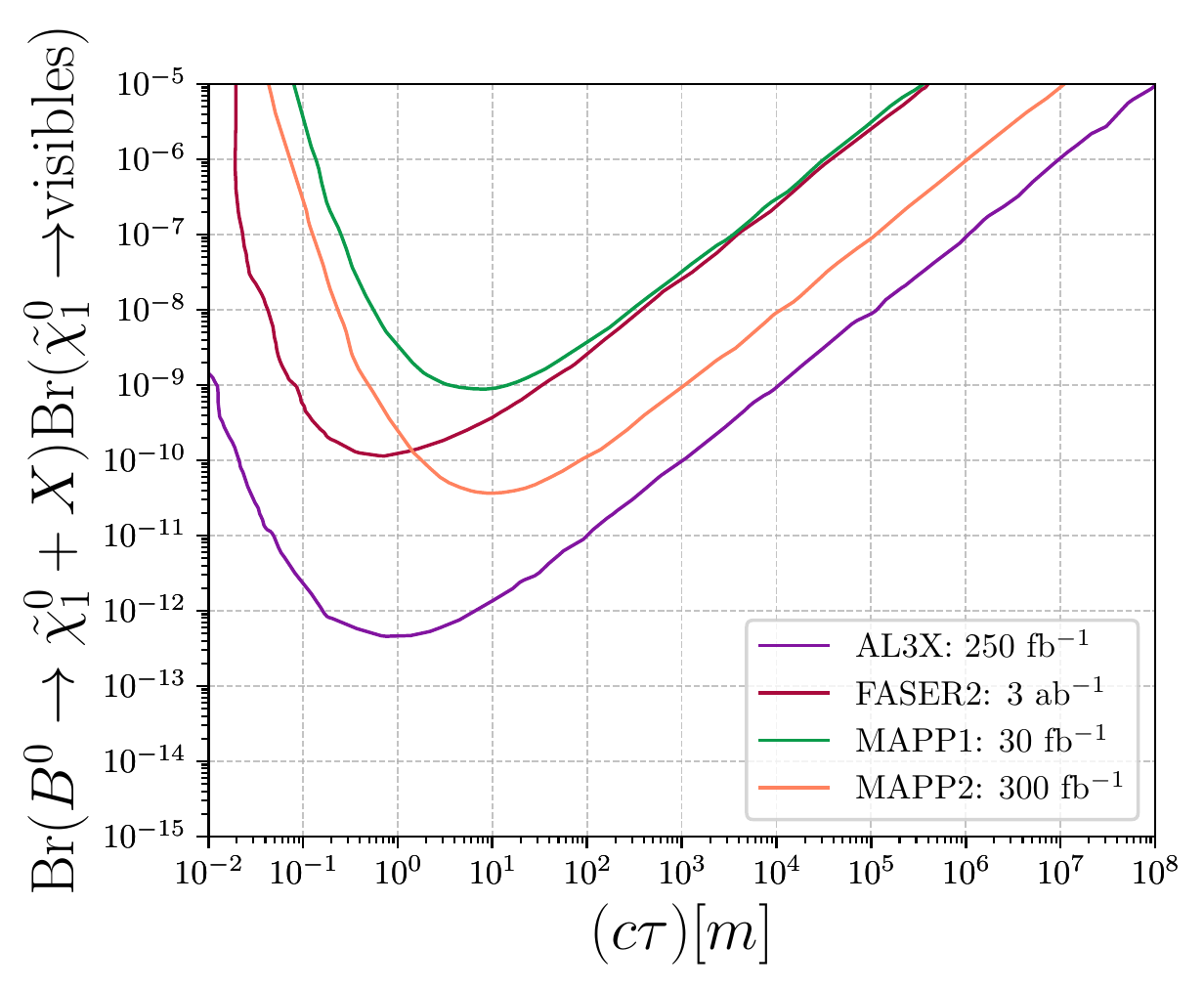}
	\caption{Comparison of Br vs. $c\tau$ estimates for AL3X, FASER2, MAPP1 and MAPP2. Left: Results for the charmed benchmark scenario with $m_{\tilde{\chi}}=1.2\gev$. The upper right region 
	of simultaneous large branching ratio and large proper decay length can not be probed in the charmed scenario (signalized 
	by a hashed grid) since the production coupling induces invisible final states. Right: Results for 
	the bottomed benchmark with $m_{\tilde{\chi}}=3~\gev$}\label{fig::llp_neutralino::brvctau}
\end{figure*}

We further present the data in terms of the proper decay length $c\tau$ and the product of production and decay branching ratio. If other theoretical models exhibit comparable topological structures, the sensitivities of such models should not vary in a qualitative sense from the one shown in \cref{fig::llp_neutralino::brvctau}. As before, AL3X can detect lower branching ratios in both scenarios followed by MAPP2 for the bottomed scenario. However, in the charmed scenario, FASER2 can probe lower products of the branching ratio than MAPP2.

\subsection{Radiative Decays of Sub-GeV Neutralinos from Light Mesons}
\label{subsec:LLP-susy-neutralino-II}

Here, we consider the case of a light neutralino, as in \cref{subsec:LLP-susy-neutralino-I}. The motivation and basic setup of the scenario are as before. However, our focus here is on even lighter neutralinos than considered there. Specifically, we will consider the case of sub-GeV neutralinos in the mass range: 
\begin{equation}
    30~\mev < M_{\tilde\chi^0_1} < 500~\mev\,.
    \label{eq:mass-range}
\end{equation}
As in \cref{subsec:LLP-susy-neutralino-I}, the neutralino is produced in hadronic collisions through rare decays of mesons~\cite{Choudhury:1999tn, Dedes:2001zia, Dreiner:2009er, deVries:2015mfw}; here we consider pions, kaons and $D_S$ mesons. The produced number is given by the master formula, \cref{eq::llp_neutralinos::prod_neutralinos} of \cref{subsec:LLP-susy-neutralino-I}. The decay of the mesons proceeds via operators of the $LQ\bar D$-type, while, for the decay of the neutralinos, we consider both $LQ\bar D$ and $LL\bar E$ couplings, \textit{cf.} the R-parity Violating (RPV) superpotential, $W_{\mathrm{RPV}}$, in \cref{eq::llp_neutralinos::superpotential}) in \cref{subsec:LLP-susy-neutralino-I}. 

\begin{figure*}[t]
\centering
	\includegraphics[width=0.49\textwidth]{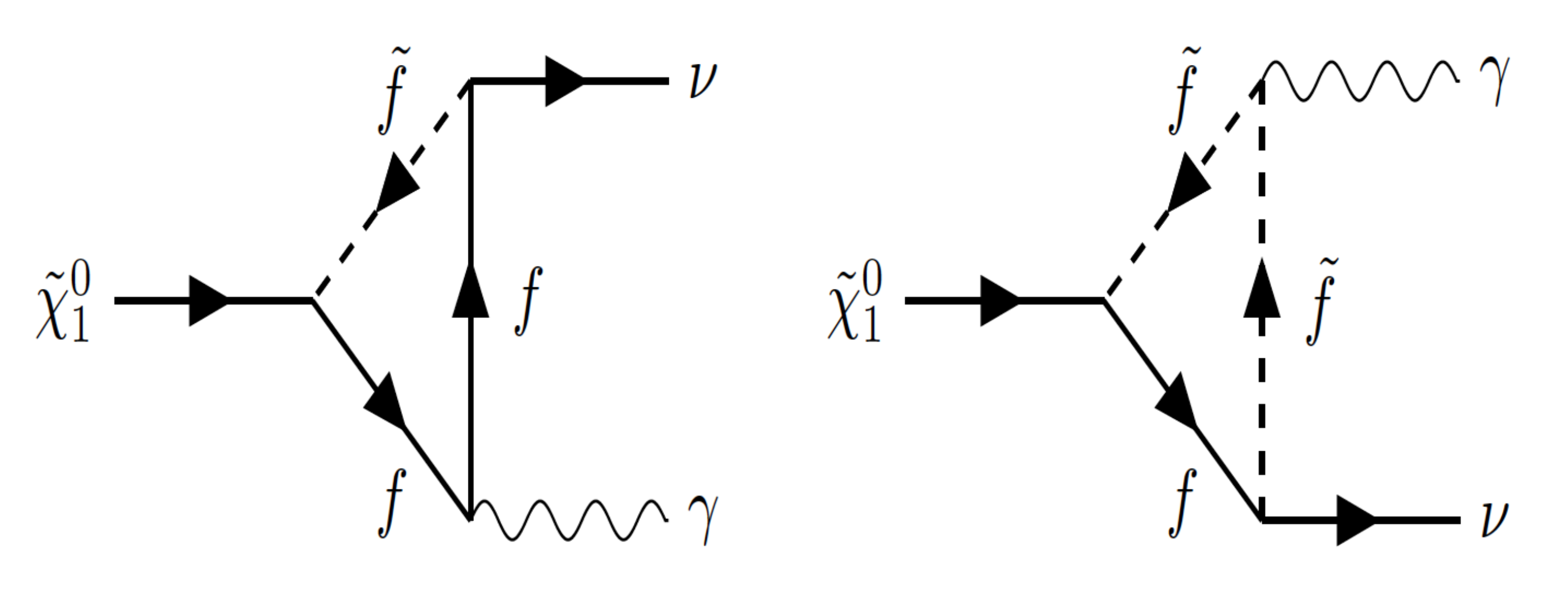}
\caption{Feynman diagrams for the radiative neutralino decay.} 
\label{fig:rad-neutralino-decay-feynman}
\end{figure*}

First, we consider the possible neutralino decays, given the mass range in \cref{eq:mass-range}. As in \cref{subsec:LLP-susy-neutralino-I}, the neutralino can decay to a meson and a lepton via an $LQ\bar D$ operator, if kinematically allowed. Similarly, it can decay as $\tilde\chi^0_1\to\ell_i^+\ell_k^-\nu_j + \text{c.c.}$ via the $LL\bar E$ operator, if kinematically allowed. For operators $L_iQ_j\bar D_j$ or $L_iL_j\bar E_j$ there is also the possibility for the loop-induced decay, 
\begin{equation}
    \tilde\chi^0_1\to \gamma +\nu_i\,,
\label{eq:rad-neutralino-decay-reaction}
\end{equation}
along with the charge conjugate state. We show example Feynman diagrams in \cref{fig:rad-neutralino-decay-feynman}. The fermions/sfermions in the loop have generation index $j$.The decay rate is given by ~\cite{Hall:1983id, Dawson:1985vr, Haber:1988px}:
\begin{equation}
    \Gamma(\tilde\chi^0_1\to \gamma +\nu)=\Gamma(\tilde\chi^0_1\to \gamma +\bar\nu) 
    = \frac{\lambda^2\alpha^2m^3_{\tilde{\chi}^0_1}} {512\pi^3\cos^2\theta_W}\left[\sum_f 
    \frac{e_f C_f m_f\left(4e_f+1\right)}{m^2_{\tilde{f}}} \left(1+\log{\frac{m^2_f} {m^2_{\tilde{f}}}}\right)\right]^2.
    \label{eq:radiative-neutralino-decay-rate}
\end{equation}
%

In the above expression, $\lambda$ is the relevant $L_iQ_j\bar D_j$ or $L_iL_j\bar E_j$ coupling, $\alpha$ is the fine structure constant, while $\theta_W$ is the electroweak mixing angle. $e_f, C_f$ and $m_f(m_{\tilde{f}})$ are the electric charge, color factor ($3$ for $LQ\bar D$, $1$ for $LL\bar E$), and the mass, respectively, of the fermion (scalar) inside the loop. The two decay widths in \cref{eq:radiative-neutralino-decay-rate} are equal due to the Majorana nature of the neutralino.

The partial width for the radiative decay mode, \cref{eq:radiative-neutralino-decay-rate}, is proportional to $\left( m^3_{\tilde{\chi}^0_1}m^2_f \right) / m^4_{\tilde f}$, compared to $m^5_{\tilde{\chi}^0_1} / m^4_{\tilde f}$ for the standard tree-level three-body decay, and can thus be important for a light neutralino. Depending on the generation indices of the dominant RPV coupling(s) and the neutralino mass, it might even be the only  kinematically allowed mode. Furthermore, given a non-zero operator $L_iQ_j\bar D_j$ or $L_iL_j\bar E_j$ the decay, \cref{eq:rad-neutralino-decay-reaction}, is always allowed, due to the light neutrino. This is unlike the tree-level three-body decays, which have a mass threshold. We make use of this fact in the choice of our benchmark scenarios below.

At the LHC, the boosted neutralino, once produced, would typically travel a macroscopic distance in the far-forward region before decaying. Thus, FASER and FASER2 are ideal candidates for probing these models~\cite{Dercks:2018eua}. In this short report, we study the sensitivity reach attainable at FASER and FASER2 by considering the three benchmark scenarios listed in \cref{tab:benchmarks-rad-neutralino}. They are each chosen so that the radiative neutralino decay is the sole kinematically allowed  mode.\footnote{We note that, in this work, we are neglecting the effects of the suppressed three-body decay that can proceed at one-loop level via an off-shell $Z$, e.g., $\tilde\chi^0_1\to 3\nu\,$. We thank Florian Domingo for a discussion on this topic.} For example, for \textbf{B2}, the tree-level decays compatible with the index structure are: $\tilde\chi^0_1\to (\nu_e K^0 + \text{c.c.}, e^\pm K^\mp)$ [$\lambda'_{112}$] and $\tilde\chi^0_1\to (\nu_\tau\mu^\pm\mu^\mp,\tau^\pm\mu^\mp\nu_\mu) + \text{c.c.}$ [$\lambda_{322}$], which are all kinematically blocked for $m_{\tilde\chi^0_1}=200\,$MeV. We note the interesting possibility in  \textbf{B3} of having a scenario with just a single non-zero RPV ($\lambda'_{222}$) coupling for both production and  decay. This is absent in the tree-level case~\cite{deVries:2015mfw}. Our signature is thus a single energetic photon.\footnote{Since we expect only few events, the accompanying neutrino will typically not be detected, even at FASER2.}  

\begin{table}[t]
\centering
\setlength{\tabcolsep}{3pt}
\begin{tabular}{c|cccc} 
    \hline\hline
    {\bf Scenario} & $\mathbf{m_{\tilde\chi^0_1}}$ & {\bf Production} $\left(\lambda^\text{P}_{ijk}\right)$ & {\bf Decay} $\left(\lambda^\text{D}_{ijj}\right)$ & {\bf Current Constraints}\\
    \hline
    {\bf B1} & $30~\mev$ & $\lambda'_{211} \left(M\!=\!\pi^\pm\!,\! \pi^0\right)$ & $\lambda'_{333}$ & $\lambda'_{211} \!<\! 0.59 \left(\frac{m_{
    \tilde{d}_R}}{\tev}\right), \lambda'_{333} \!<\! 1.04$\\
    {\bf B2} & $200~\mev$ & $\lambda'_{112}\left(M\!=\!K^\pm\!,\! K^0_{L/S}\right)$ & $\lambda_{322}$ & $\lambda'_{112} \!<\! 0.21 \left(\frac{m_{
    \tilde{s}_R}}{\tev}\right), \lambda_{322} \!<\! 0.7 \left(\frac{m_{
    \tilde{\mu}_R}}{\tev}\right)$\\
    {\bf B3} & $500~\mev$ & $\lambda'_{222}\left(M\!=\!D_S^\pm\right)$ & $\lambda'_{222}$ & $ \lambda'_{222} \!<\! 1.12$\\
    \hline\hline
\end{tabular}
\caption{Benchmark scenarios considered in this report. The neutralino is produced through the rare decay of the meson $M$ via the coupling $\lambda^\text {P}_{ijk}$: $M\to\tilde\chi^0_1+l\left(\nu\right)$. The neutralino decay is as in \cref{eq:rad-neutralino-decay-reaction} via the coupling $\lambda^\text{D}_{ijj}$. On the right we list the current best bounds on the couplings from low-energy experiments or perturbativity considerations, see for example Ref.~\cite{Allanach:1999ic}.}
\label{tab:benchmarks-rad-neutralino}
\end{table}

We now briefly discuss the relevant backgrounds\footnote{We thank Felix Kling for clarifying comments on the photon signature at FASER.}. We follow the arguments made in Ref.~\cite{Jodlowski:2020vhr}, where the same signature has been studied in a different context. At FASER, the single photons are detected as high-energy deposits in the electromagnetic (EM) calorimeter. Other objects may also cause such deposits, \textit{e.g.}, neutrinos interacting deep inside the calorimeter via charged current interactions. To differentiate the photon signal, one can use the dedicated pre-shower station in front of the calorimeter~\cite{FASER:2021cpr}, which first converts the photon, thereby identifying it. 

With single photon events identified, there can still be some background caused by muons or neutrinos, capable of traversing the $\mathcal{O} \left(100~\m\right)$ of rock in front of FASER. The former can radiate energetic photons through Bremsstrahlung. However, such muon-associated events can be rejected with very high efficiency using the veto system in front of FASER~\cite{FASER:2018bac}. Neutrinos can also produce energetic photons (among other particles) through their interactions in the detector. However, typically, such processes produce $\mathcal{O}\left(10\right)$ charged tracks, leading to a high probability that some of these would be visible in the tracker, allowing an efficient veto. Finally, we note that requiring an energy threshold for the signal can further reduce background; see Ref.~\cite{Jodlowski:2020vhr} for details. Thus, we expect negligible background and for the purposes of this study, we will neglect it completely. A more reliable estimate would require a detailed detector simulation.

We now describe our simulation procedure. We use the package \texttool{FORESEE}~\cite{Kling:2021fwx} (see also \cref{sec:foresee}) to obtain the neutralino spectrum in the far-forward region, relevant for FASER and FASER2. As mentioned, the dominant sources of the neutralinos are the rare decays of the light mesons:
\begin{equation}
    M\to\tilde\chi^0_1+l\left(\nu\right)\,.
    \label{eq:meson-decay-reaction}
\end{equation}
Direct pair production, in comparison, is expected to be several orders of magnitude lower and hence, we neglect it here. Thus, the \texttool{FORESEE} package can determine the neutralino spectrum, given the branching fraction for \cref{eq:meson-decay-reaction}. We note that, in the simulation of the long-lived mesons, e.g., charged pions and kaons, the package requires all mesons to decay before they hit any absorber material or leave the beam pipe.

The neutralino decay is also simulated by \texttool{FORESEE}, taking into account the lifetime $\tau_{\tilde \chi^0_1}$ and kinematics. We assume that every neutralino decaying inside the detector is visible; see \cref{tab:detector_simulation} for the values corresponding to the detector position and geometry we employ in our simulations for FASER and FASER2. For our sensitivity study, we require 3 radiative neutralino decays observed in the detector for an integrated luminosity at the LHC of $150~\ifb$ for FASER, and $3000~\ifb$ for FASER2. We show our results in \cref{fig::llp_neutralino::couplingvcoupling1}, \cref{fig::llp_neutralino::couplingvcoupling2}, and \cref{fig::llp_neutralino::B3} for the three benchmark scenarios \textbf{B1}, \textbf{B2}, and \textbf{B3}, respectively. 

\begin{table}[t]
\setlength{\tabcolsep}{3pt}
\centering
\begin{tabular}{c|ccccc} 
    \hline\hline
    {\bf Detector} & {\bf Luminosity} $\mathbf{\mathcal{L}}$ & {\bf Energy}  & {\bf Distance} $\mathbf{L}$ & {\bf Detector Length} $\mathbf{\Delta}$ &  {\bf Detector Radius} $\mathbf{R}$\\
    \hline
    {\bf FASER} & $150~\ifb$ & $14~\tev$ & $480~\m$ & $1.5~\m$ & $10~\cm$ \\
    {\bf FASER2} & $3000~\ifb$ & $14~\tev$ & $480~\m$ & $5~\m$ & $1~\m$ \\
    \hline\hline
\end{tabular}
\caption{Integrated luminosities and geometries of the detectors used in the simulations.}
\label{tab:detector_simulation}
\end{table}

\begin{figure*}[t]
	\centering
	\includegraphics[width=0.49\textwidth]{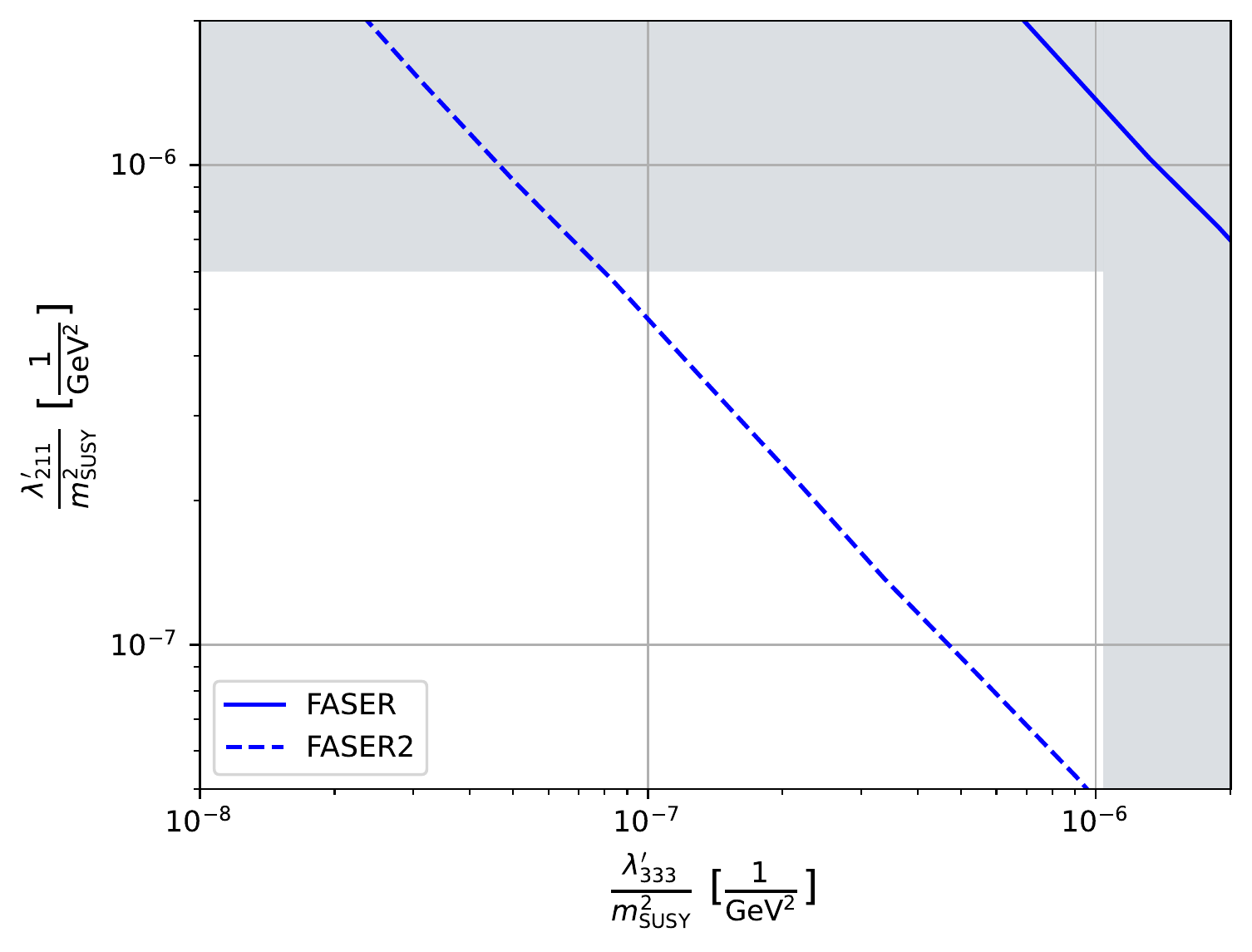}
	\includegraphics[width=0.48\textwidth, trim={0 -0.5cm 0 0}, clip]{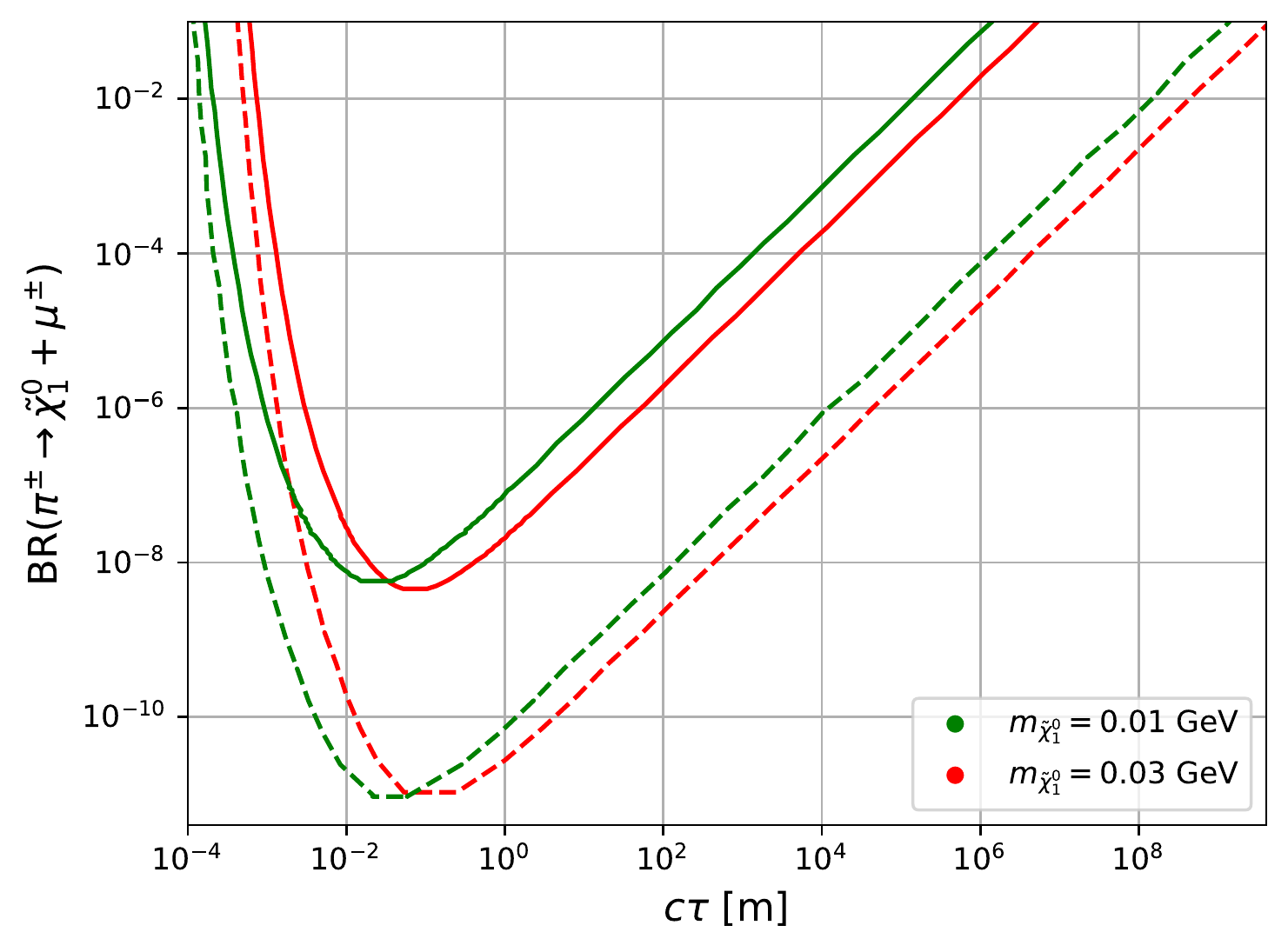}
\caption{Sensitivity reach for FASER (solid lines) and FASER2 (dashed lines) for \textbf{B1}, cf. \cref{tab:benchmarks-rad-neutralino}. Left: The production coupling sensitivity ($\lambda^{\prime}_{211} / m_\text{SUSY}$) vs. the decay ($\lambda^{\prime}_{333} / m_\text{SUSY}$). The gray areas are excluded by current bounds. Right: The sensitivity reach in BR$(\pi^\pm \to \tilde {\chi}^0_1 + \mu^\pm)$ as a function of the neutralino decay length $c\tau$ for $m_{\tilde\chi^0_1}=10,$ and $30~\mev$.}
\label{fig::llp_neutralino::couplingvcoupling1}
\end{figure*}

We see that FASER has no new sensitivity for \textbf{B1}, beyond the current bounds, whereas FASER2 can extend the reach by more than an order of magnitude in $\lambda^{\mathrm{P}}$ and $\lambda^{\mathrm{D}}$. The right plot in \cref{fig::llp_neutralino::couplingvcoupling1} is model independent, in that it is valid for any new neutral long-lived particle (LLP) produced in charged pion decays, which decays radiatively, here specifically with a mass of $10$ or $30~\mev$. The maximum sensitivity depends on the location and geometry of the detector, and also on the momentum distribution of the produced pions and, correspondingly, of the decay product neutralinos~\cite{Dercks:2018eua}. The latter point explains why the minima of the curve shifts to smaller lifetimes for lighter masses, which are more boosted. We see that FASER (FASER2) can probe charged pion decay branching ratios down to about $10^{-8}$ $\left(10^{-11}\right)$.

\cref{fig::llp_neutralino::couplingvmass1} shows the sensitivity reach of FASER2 for \textbf{B1}, however allowing the neutralino mass to vary but fixing $\lambda^P=\lambda^D$. The production and decay mode are as in \textbf{B1} of \cref{tab:benchmarks-rad-neutralino}. We see the sensitivity reach in the couplings is reduced as the neutralino becomes lighter. It is even more sharply reduced for increased neutralino mass, $m_{\tilde\chi^0_1}\gtrsim 30~\mev$. The former is because the lifetime of the neutralino becomes too large to be seen at FASER2, \textit{cf.} \cref{eq:radiative-neutralino-decay-rate}. The latter is because the charged pion decay mode is kinematically blocked. The branching fraction of the neutral pion mode, $\pi^0 \to \tilde \chi^0_1 + \nu_\mu \left( \bar{\nu}_\mu\right)$, is suppressed by its rather short lifetime.

\begin{figure*}[t]
	\centering
	\includegraphics[width=0.49\textwidth]{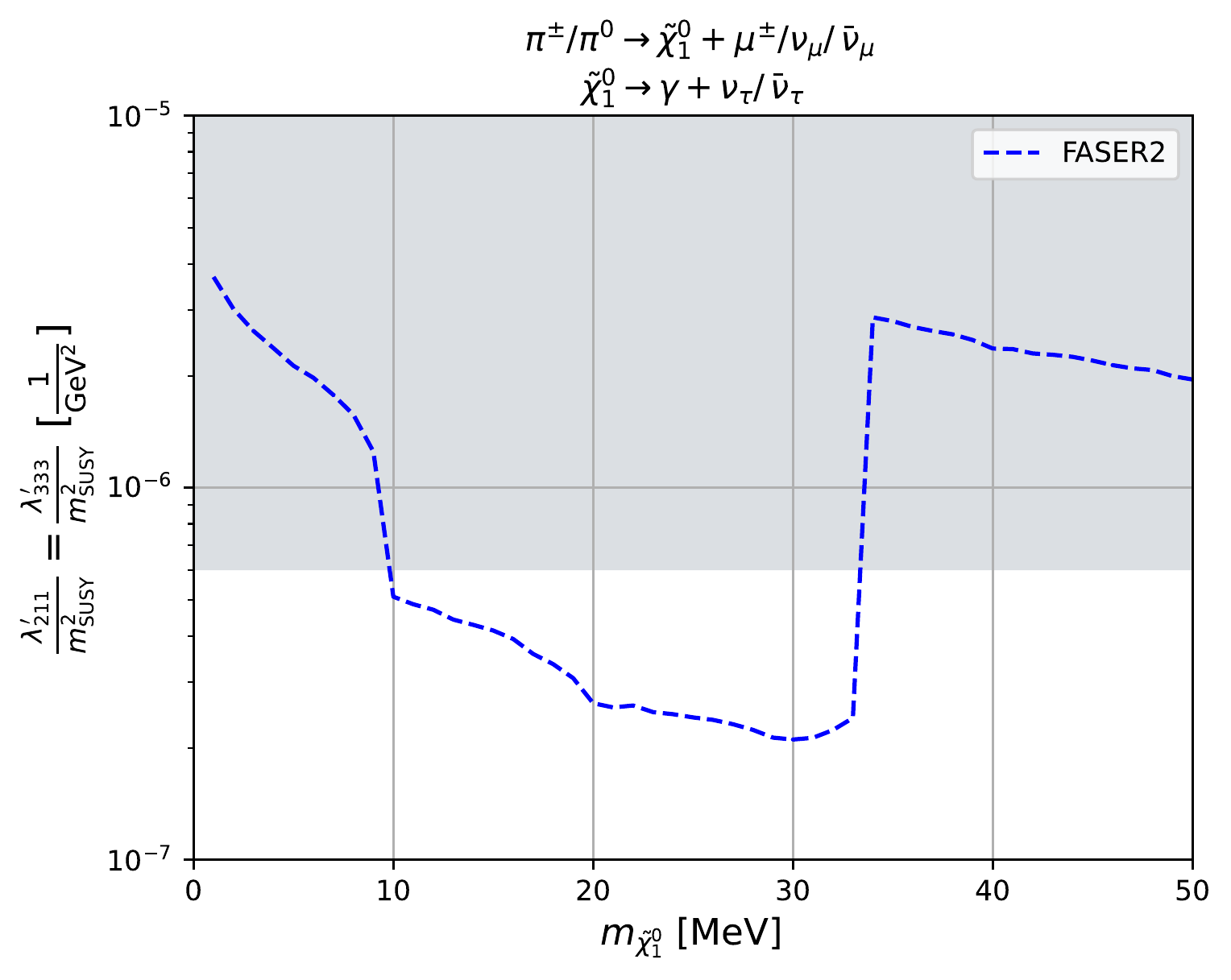}
	\includegraphics[width=0.49\textwidth]{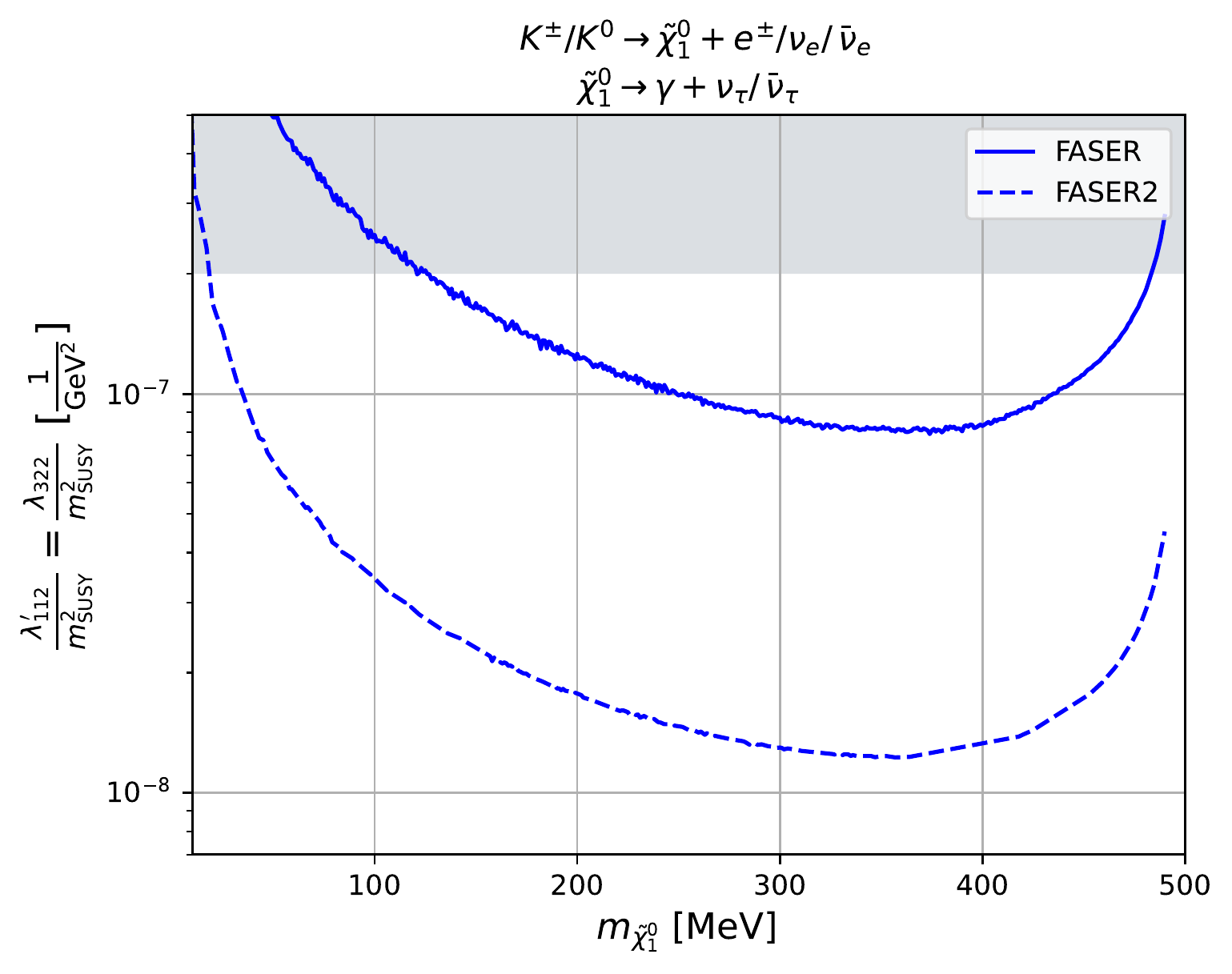}
\caption{Sensitivity reach in the coupling-mass plane for FASER2 for the same physics scenario as in \textbf{B1} (left) and \textbf{B2} (right) but with variable  neutralino mass. The production and decay couplings have been set equal. The gray areas are excluded by current bounds. The sensitivity reach corresponds to FASER (solid line) and FASER2 (dashed line).}
\label{fig::llp_neutralino::couplingvmass1}
\end{figure*}

\begin{figure*}[t]
	\centering
	\includegraphics[width=0.48\textwidth]{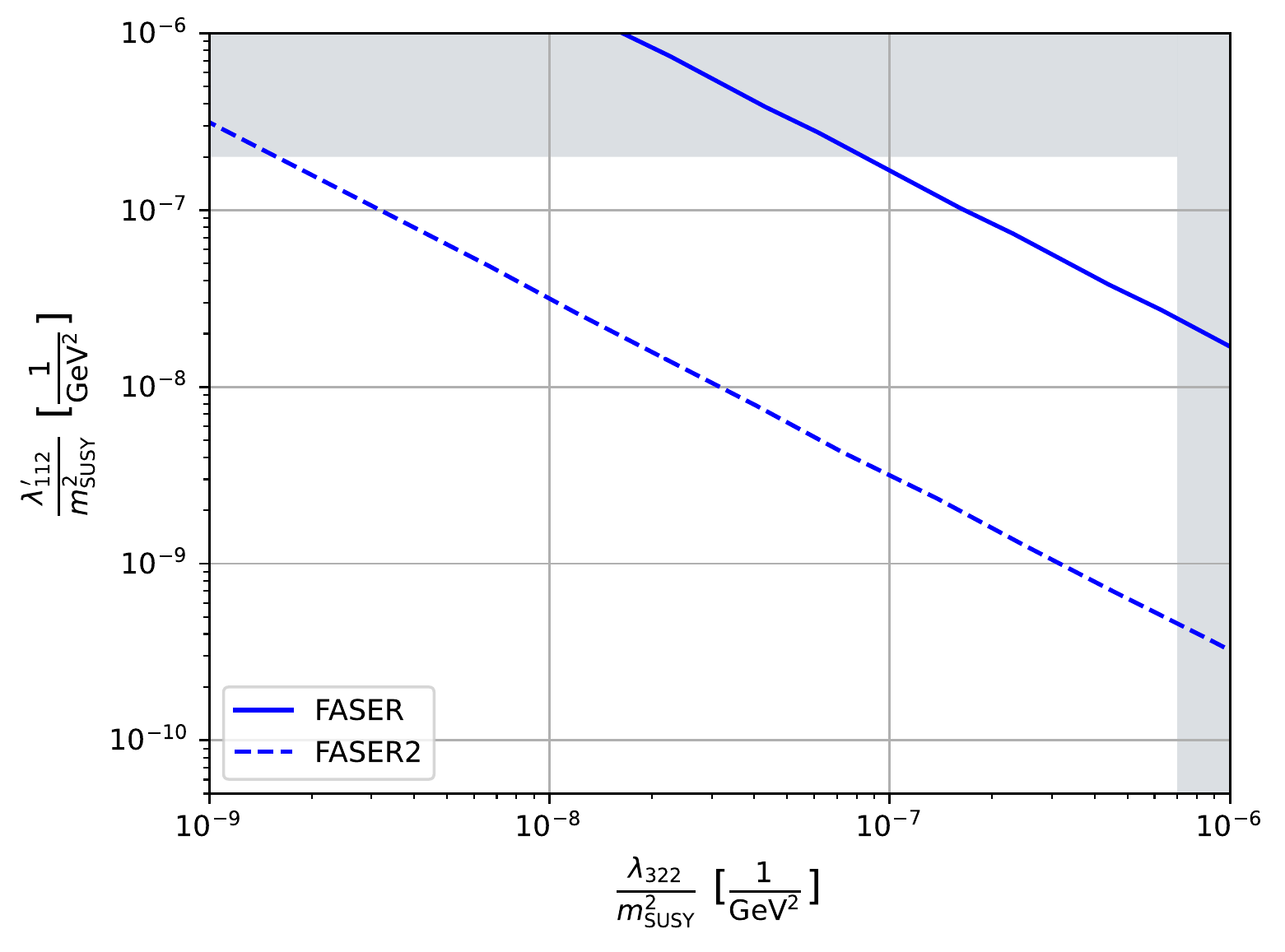}
	\includegraphics[width=0.49\textwidth]{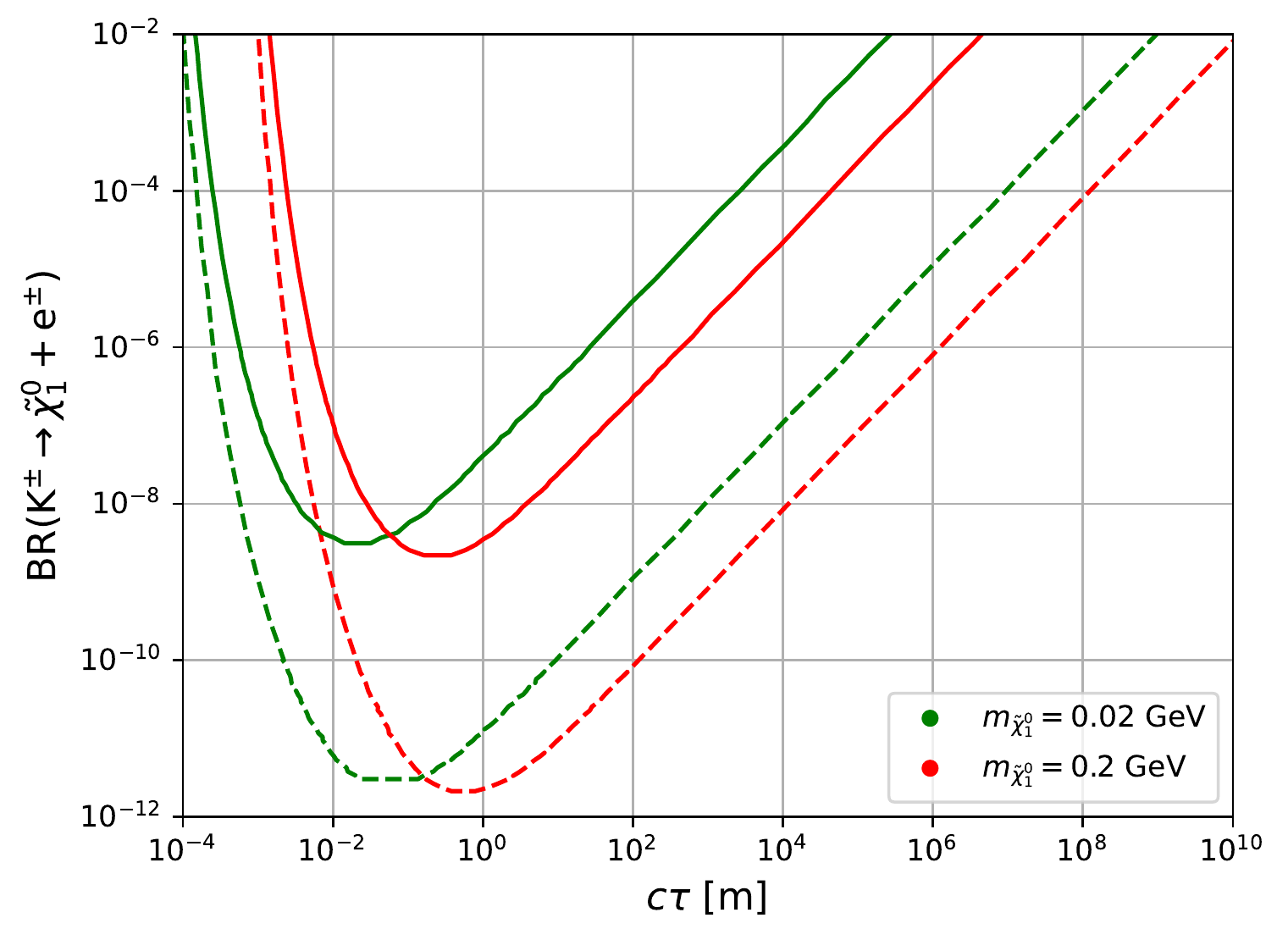}
\caption{As in \cref{fig::llp_neutralino::couplingvcoupling1} but for \textbf{B2}. The right plot shows the sensitivity reach in BR$(K^\pm \to \tilde {\chi}^0_1 + e^\pm)$ as a function of the neutralino decay length $c\tau$ for $m_{\tilde\chi^0_1}=20,$ and $200~\mev$.}
\label{fig::llp_neutralino::couplingvcoupling2}
\end{figure*}

In \cref{fig::llp_neutralino::couplingvcoupling2} and the right panel of  \cref{fig::llp_neutralino::couplingvmass1} we show the corresponding plots for benchmark scenario \textbf{B2}. The left plot of \cref{fig::llp_neutralino::couplingvcoupling2} shows that now both FASER and FASER2 have significant new reach in the couplings $\lambda^{\mathrm{P}}$ and $\lambda^{\mathrm{D}}$. The right plot looks very similar to \cref{fig::llp_neutralino::couplingvcoupling1}, but here we have now considered LLP masses: $20$ and $200~\mev$. The coupling-mass plane plot in the right panel of \cref{fig::llp_neutralino::couplingvmass1} shows that the sensitivity at FASER2is reduced for lower masses as before, but unlike the pion case is robust over the entire higher mass regime, right up to the kaon threshold. This is because: (a) The electron is much lighter than the muon in \textbf{B1} and (b) The charged and neutral decay modes have comparable branching fractions. We note that for $m_{\tilde{\chi}^0_1} \gtrsim 2m_{\mu}$ the decay mode $\tilde{\chi}^0_1\to \mu^+ \mu^-\nu_{\tau} + \text{c.c.}$ opens up, leading to additional visible events. These are not included in the right panel of \cref{fig::llp_neutralino::couplingvmass1}.

In \cref{fig::llp_neutralino::B3}, we show the sensitivity reach for \textbf{B3}, which only has one RPV coupling. The left plot shows the sensitivity reach of FASER and FASER2 in the coupling-mass plane. Here for larger masses, the neutralino has new decay modes into $\eta, \eta'$ and $\phi$ plus $\nu_\mu$ opening up at the respective mass thresholds; as before, we only count the photon events as signal. We see once again that the reach offered by FASER2 outperforms existing constraints by an order of magnitude over large areas of the accessible phase space. The reduced sensitivity in the large mass, large coupling regime is because the lifetime of the neutralino becomes too short, decaying well before reaching the FASER or FASER2 detectors. The model-independent plot on the right now corresponds to an LLP mass of $500~\mev$.

\begin{figure*}[t]
	\centering
	\includegraphics[width=0.49\textwidth]{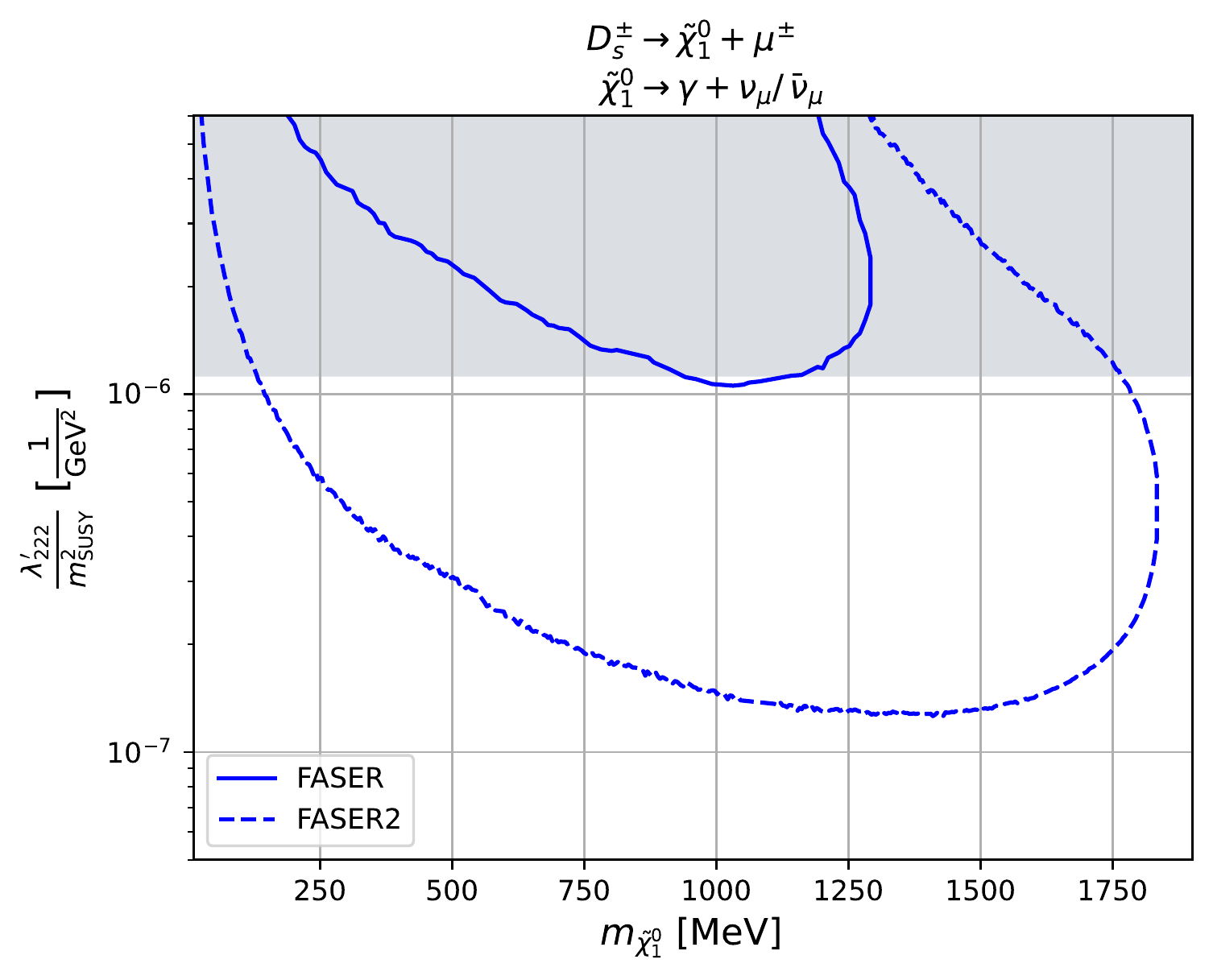}
	\includegraphics[width=0.49\textwidth]{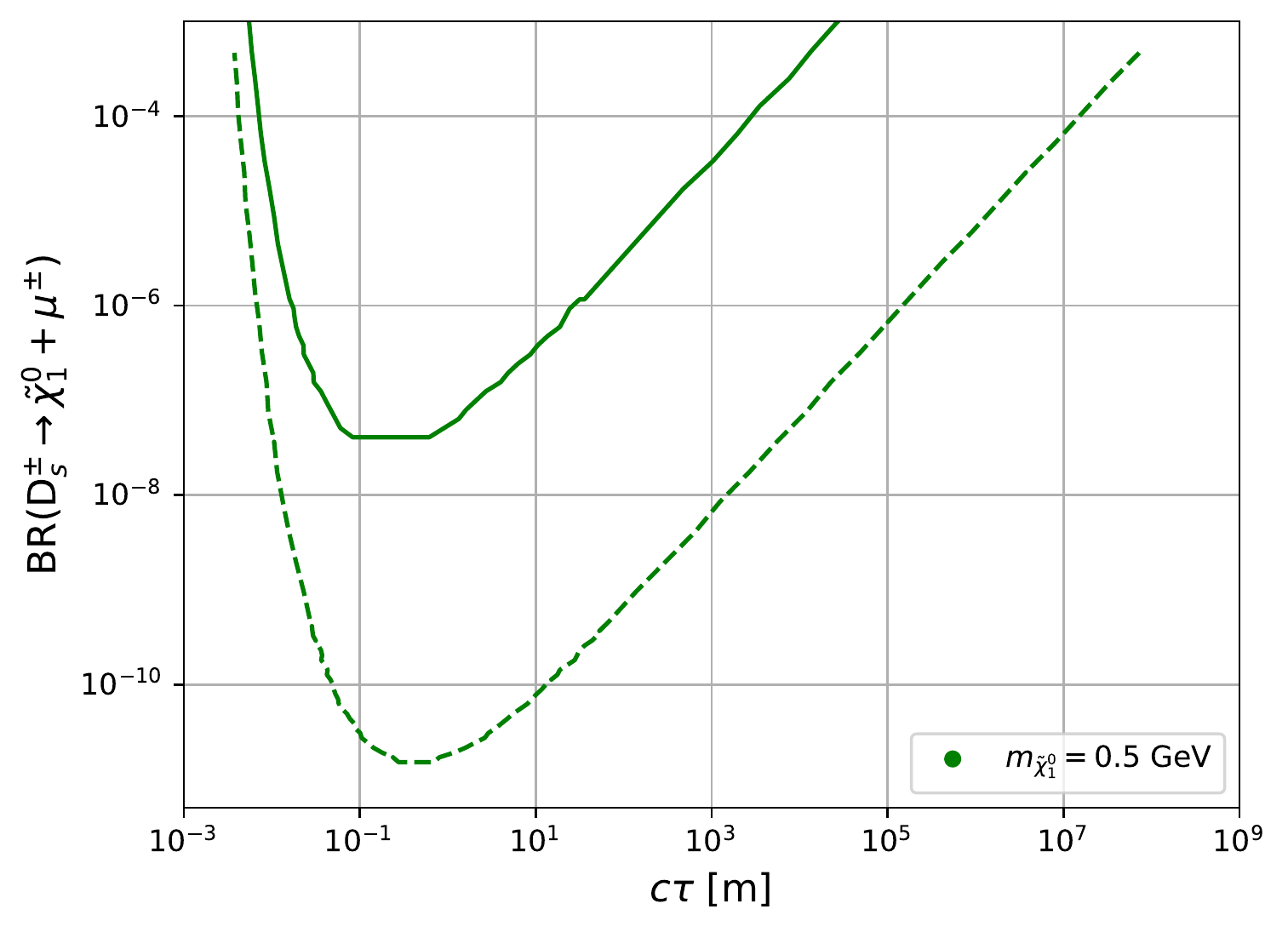}
\caption{Same as \cref{fig::llp_neutralino::couplingvmass1}, and
the right plot same as \cref{fig::llp_neutralino::couplingvcoupling1}, but now for \textbf{B3}.}
\label{fig::llp_neutralino::B3}
\end{figure*}

\subsection{Fermion Portal Effective Operators }
\label{sec:fermion_portal}

\textit{i) Introduction:} The visible sector energy density is dominated by just a few members of the zoo of states. This non-minimality could be true of a dark sector of light particles as well, and could be experimentally discovered. Dark sector states are, by definition, singlets under the Standard Model (SM) $SU(3)\times SU(2) \times U(1)$ gauge group. Interactions between the visible and dark sectors can be classified according to so-called portal operators --- singlet combinations of SM fields. The Higgs, vector, and neutrino portals are the operators of lowest mass dimension and their phenomenology has been well studied (see e.g. Refs.~\cite{Holdom:1985ag, Schabinger:2005ei, Patt:2006fw, Batell:2009di, Falkowski:2009yz}). The fermion portal, meanwhile, contains a singlet bilinear of SM fermions. Since this bilinear has mass dimension 3, the fermion portal must necessarily be a higher-dimensional operator if it forms an operator with dark sector fermions. We will focus here on the dimension-6 four-fermion operator involving a pair of visible fermions and a pair of dark sector fermions. The feeble interactions between the two sectors are then due to suppression by the scale of the heavy mediator responsible for generating the four-fermion operator at lower energies. The phenomenology of this fermion portal operator was first systematically investigated in Ref.~\cite{Darme:2020ral}. 

FASER is an experiment that is capable of probing long-lived dark sectors in this Effective Field Theory (EFT) framework. In the {\tt DarkEFT} public code of Ref.~\cite{Darme:2020ral}, a variety of constraints and projections from past, present, and future experiments were obtained on the sensitivity to fermion portal four-fermion interactions of the vector and axial-vector forms, allowing the user to specify the flavour structure of the couplings. Here we extend this to include projections for phase 2 of FASER (FASER2) in the newly proposed Forward Physics Facility (FPF). 

The nature of the four-fermion operator is critical in determining the dark sector production rates, especially for rare meson decays. In Ref.~\cite{Darme:2020ral} we consider both vector ($\Gamma_\mu = \gamma_\mu$) and axial-vector ($\Gamma_\mu = \gamma_\mu \gamma^5$) operators,
\begin{equation}
\mathcal{L} \supset \sum_{q \in u,d} \frac{g_q}{\Lambda^2}(\bar\chi_1 \Gamma_\mu \chi_2) (\bar{q} \gamma^{\mu} q) \, .
\label{eq:opV}
\end{equation}
We have included in the above the possibility that the dark sector consists of two dark sector fermions, $\chi_1$ and $\chi_2$, with $M_2 \geqslant M_1$. We then define the normalised splitting
\begin{equation}
\delta_\chi \equiv \frac{|M_2| - |M_1|}{|M_1|} \ ,
\label{eq:dChi}
\end{equation}
which can be taken to zero to recover the single state case. The case $\delta_\chi \gg 1$ thus corresponds to the limit scenario where $\chi_1$ is much lighter than $\chi_2$.

\textit{FASER2 sensitivity at FPF:} The FPF will have an excellent sensitivity to the fermion portal. We use the code \texttool{DarkEFT}~\cite{Darme:2020ral} to show in \cref{fig:final_plot} the projection for FASER2 recasted from the inelastic dark matter result from~\cite{Berlin:2018jbm} in the case where the effective couplings are aligned with electromagnetism and the lighter dark sector has a negligible mass. Current existing constraints are shown in grey, and the full HL-LHC dataset of $3 \rm ab^{-1}$ is assumed. Remarkably, FASER phase-2 can probe effective scales in the multi-TeV range for heavy state masses $M_2 \gtrsim \text{GeV}$, complementary to astrophysical constraints. The sensitivity to the axial-vector operator is qualitatively similar, and is shown in Ref.~\cite{Darme:2020ral} for different choices of $\{g_e,\,g_d,\,g_u\}$.

\begin{figure*}[t]
    \centering
    \includegraphics[width=0.7\textwidth]{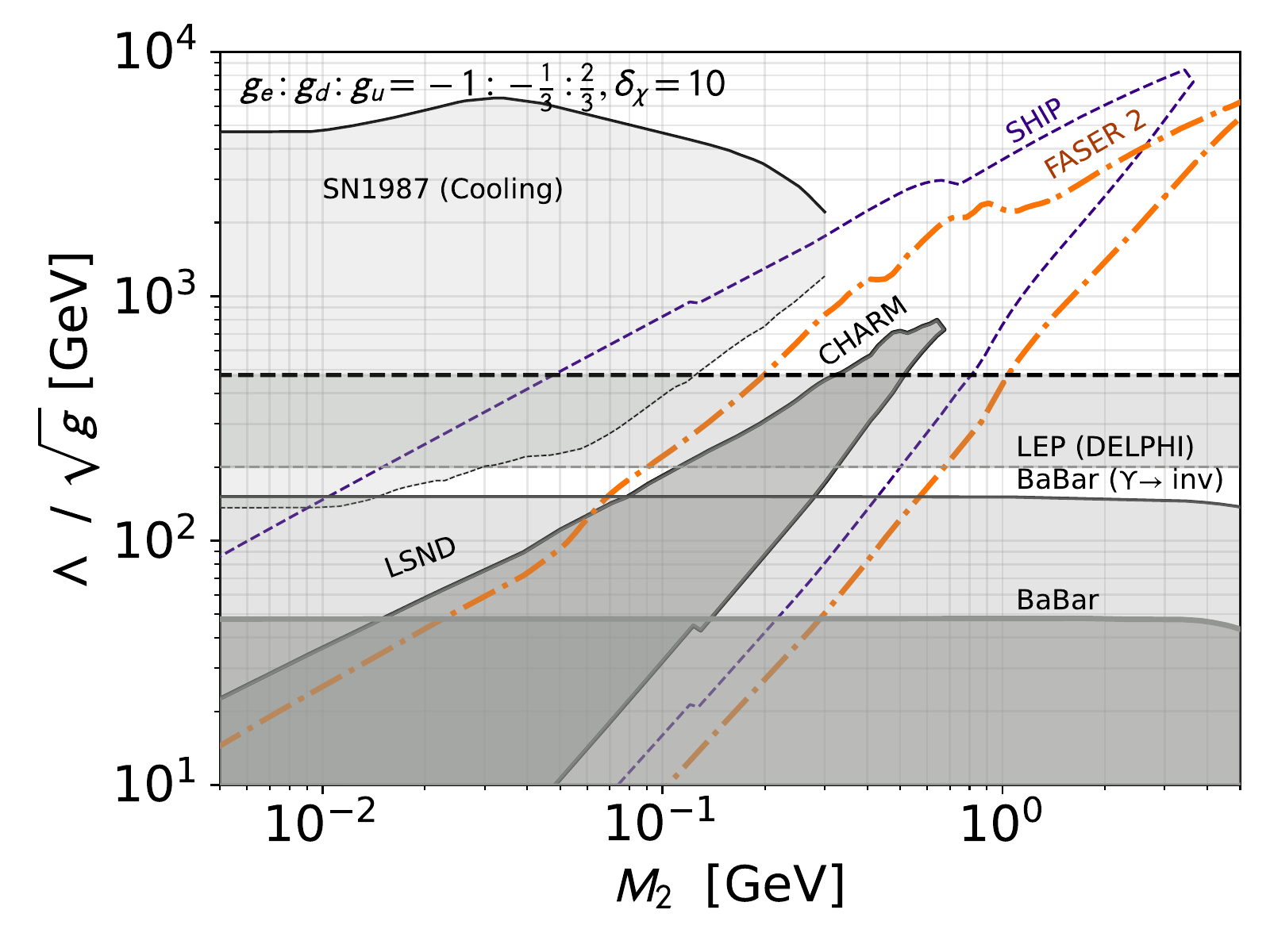}
    \caption{Limits and projected sensitivity to the vector operator effective scale $\Lambda/\sqrt{g}$ in the case of effective coupling electromagnetically-aligned as a function of $M_2$. Grey region indicates coverage from existing experiments: mono-photon at BaBar~\cite{Essig:2013vha}, limit from cooling of SN1987A~\cite{DeRocco:2019jti}, mono-$\gamma$ limit from LEP~\cite{Fox:2011fx}, limits from $\chi_2 \to \chi_1 e^+ e^-$: from LSND~\cite{Darme:2018jmx} and CHARM~\cite{deNiverville:2018dbu}. We show a projection for FASER~\cite{Berlin:2018jbm} in dot-dashed orange and for SHIP in thin dashed indigo~\cite{Darme:2020ral}.
    The normalised splitting $\delta_\chi$ is defined in \cref{eq:dChi}, and is set to $\delta_\chi = 10$ here. When estimation the FASER2 sensitivity, we use a cylindrical decay volume of radius $1$m and length $10$m as a reference~\cite{Anchordoqui:2021ghd}.
    \label{fig:final_plot}}
\end{figure*}

\begin{figure*}[t]
\centering
	\includegraphics[width=0.7\textwidth]{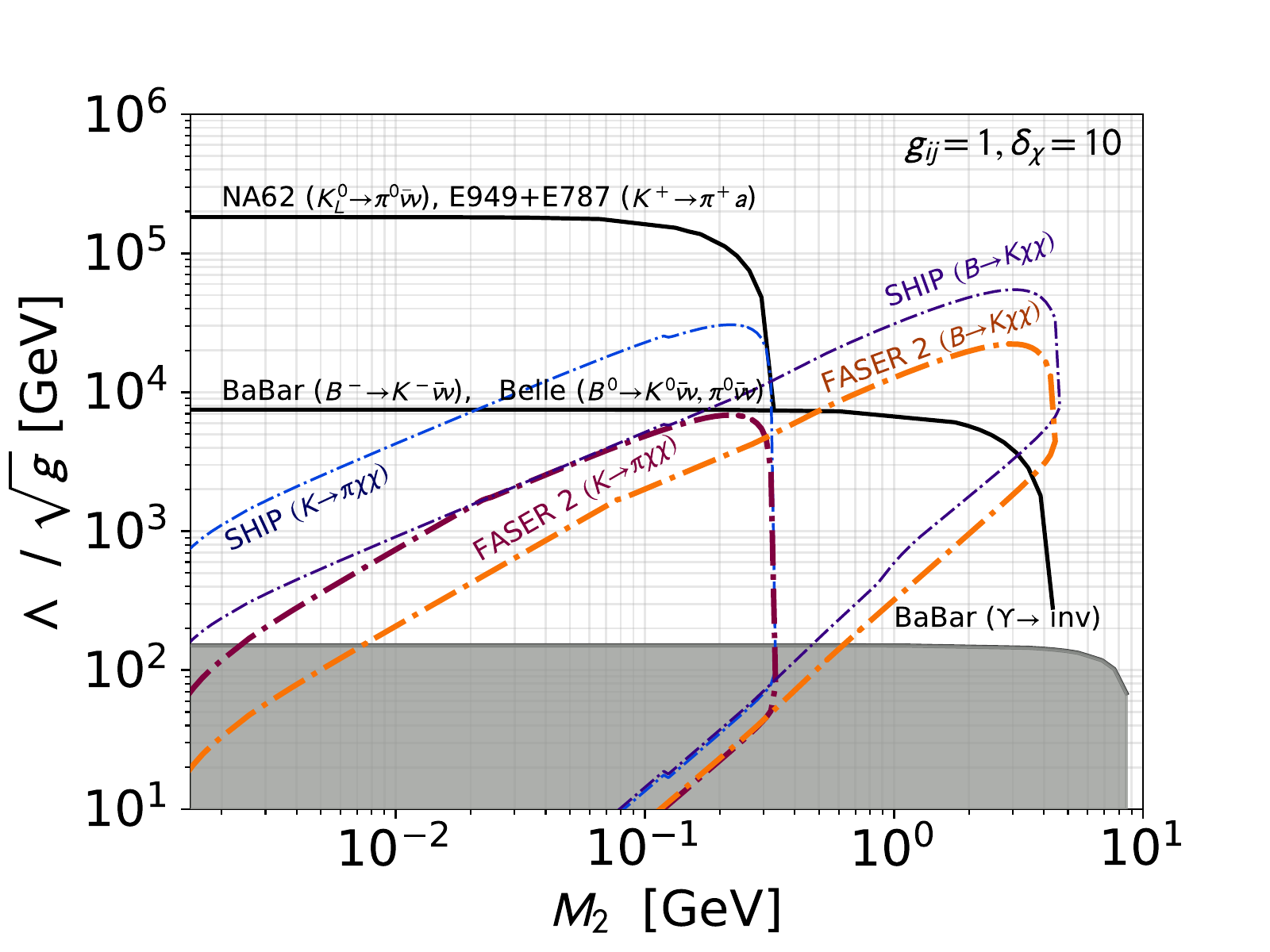}
	\vspace{-0.2cm}
	\caption{Heavy $B$ and $K$ meson limits and projected sensitivity to $\Lambda/\sqrt{g}$ for the vector-current operator as a function of $M_2 $. Regions outlined by thick dot-dashed dark red (orange) lines show the 10 event projection at FASER for $\chi_2 \to \chi_1 e^+ e^-$ decay processes, produced by  $K \to \pi \chi_1 \chi_2$ ($B \to K \chi_1 \chi_2$). Similar projection for SHiP are shown in blue and indigo thin dashed lines~\cite{Darme:2020ral}. The solid lines denote actual bounds from invisible $B$ and $K$ decays while dotted lines are future projections. The normalised splitting $\delta_\chi$ is defined in \cref{eq:dChi}, and is set to $\delta_\chi = 10$ here.}
	\label{fig:limHeavymes}
\end{figure*}

The LHC also produces a large number of flavoured mesons. In the presence of flavourful effective operators with heavy quarks, the $B\to K \chi_2 \chi_1$ and/or $K\to \pi  \chi_2 \chi_1$ transitions can dominate the production rates. Assuming that the process $\chi_2 \to \chi_1 e^+ e^-$ is allowed, this would lead to a compelling signature at the FPF. We illustrate this for the vector-operator case with $g_{sb}=g_{ee}=1$ or $g_{sd}=g_{ee}=1$ in \cref{fig:limHeavymes}. The FASER projection assumes 10 signal events based on an average boost factor of $\mathcal{O} (1) \ \rm TeV$ for the heavy states and $N =10^{13}$ $B$-mesons produced during the HL-LHC run within FASER's acceptance~\cite{Feng:2017vli, FASER:2018eoc, Anchordoqui:2021ghd}. As shown in \cref{fig:limHeavymes}, the FPF will have the potential to probe effective scales in the tens of TeV range, an order of magnitude above current B-factory limits. We have overlaid the bounds from B and K-factory experiments on  $B\to K \ \rm inv.$ and $K\to \pi \ \rm inv.$ decays. Even if invisible meson decay constraints appear to exclude a large region of the parameter space that FASER will probe, we note that this depends strongly on our assumption $g_{sd}=g_{ee}=1$. For instance, reducing the flavour-violating coupling while keeping the product fixed will boost the FPF limit compared to the K-factory one. The bounds from FPF and invisible meson decays are therefore complementary. The axial-vector heavy flavour constraints are not currently implemented in \texttool{DarkEFT}~\cite{Darme:2020ral}, but could be included in a future version of the code.

Finally, in the above we have shown results for $\delta_\chi = 10$, where the sensitivity of the FPF to $\Lambda/\sqrt{g}$ saturates. This saturation is discussed in Ref.~\cite{Darme:2020ral} in greater detail. At smaller $\delta_\chi$, the production and decay of the $\chi_i$ states depends more sensitively on the splitting, so that for FASER, at small $\delta_\chi \lesssim 0.3$, the upper limit $\Lambda/\sqrt{g} \propto \delta_\chi^{5/8}$.

\textit{Conclusion:} The fermion portal to dark sector fermions is a family of four-fermion operators parametrising interactions between visible and dark sectors. This EFT approach provides a model-independent way to capture the phenomenology of light dark sector fermions interacting via a heavy mediator, which exhibits distinct characteristics compared to the usual vector or scalar portals. In this framework the weakness of the interaction can then be related to the heavy scale of new physics. In this white paper contribution we have quantified the sensitivity of FASER2 at the FPF to the scale and couplings of a representative set of four-fermion portal operators. This sensitivity can reach the tens of TeV scale for $\mathcal{O}(1)$ couplings and a mass splitting $\delta_\chi = 10$ between the light dark sector states, providing complementary or stronger bounds to existing constraints.

\clearpage
\section{Long-Lived Axion Like Particles}
\label{sec:bsm_llp_alp}

In the previous sections, we have considered new particles that couple to the SM via renormalizable portals. Another broadly studied scenario is that of axion-like particles (or ALPs), which are light pseudo-scalar particles that couple to the SM via dimension-5 operators. The perhaps best motivated example of this class of model is the QCD axion~\cite{Peccei:1977hh, Peccei:1977ur, Wilczek:1977pj}, which was introduced as solution to the strong CP problem~\cite{Peccei:2006as}. More generally, ALPs can appear as pseudo Nambu-Goldstone bosons in theories with broken global symmetries and also generically appear in string theory~\cite{Jaeckel:2010ni}. 

The phenomenology of the ALP $a$ is characterized by its mass $m_a$ and its couplings to the SM particles~\cite{Bauer:2017ris}. In the most general scenario, the ALP can have arbitrary couplings to the SM gauge bosons and fermions. In this case, we can write down a Lagrangian
\be
\mathcal{L} = -\frac12 m_a^2 a^2 - \sum_{f} C_f \partial^\mu a \bar{f} \gamma_\mu \gamma_5 f  - \sum_{F} \frac14 C_{F} F^{a}_{\mu\nu} \tilde{F}^{a\mu\nu} 
\ee
Here we summed over the SM fermion fields $f$ as well as the field strength tensors $F$ for the electroweak gauge boson and the gluon fields. In the above expression, we have considered individual coefficients $C_f$ and $C_F$, whose connection is expected to be determined by an underlying UV completion. It is worth noting that the existence of one non-vanishing coupling at some higher scale $\Lambda_{UV}$ typically also induces non-vanishing values for the remaining couplings at loop level in the low energy theory. 

In the remainder of the section, we will first give an overview on the often considered cases of ALPs with a single coupling to photons, the electroweak gauge bosons, fermions and gluons in \cref{sec:bsm_alp}. We then discuss the special case of an with ALPs dominant couplings to charm quarks in \cref{bsm_alp_charm}. In \cref{bsm_alp_brem} we then investigate an additional ALP production mechanisms via Bremsstrahlung which will lead to a significant increase in the FPF sensitivity.

\subsection{Overview on Axion Like Particles}
\label{sec:bsm_alp}

\begin{figure*}[t]
	\centering
	\includegraphics[width=0.49\textwidth]{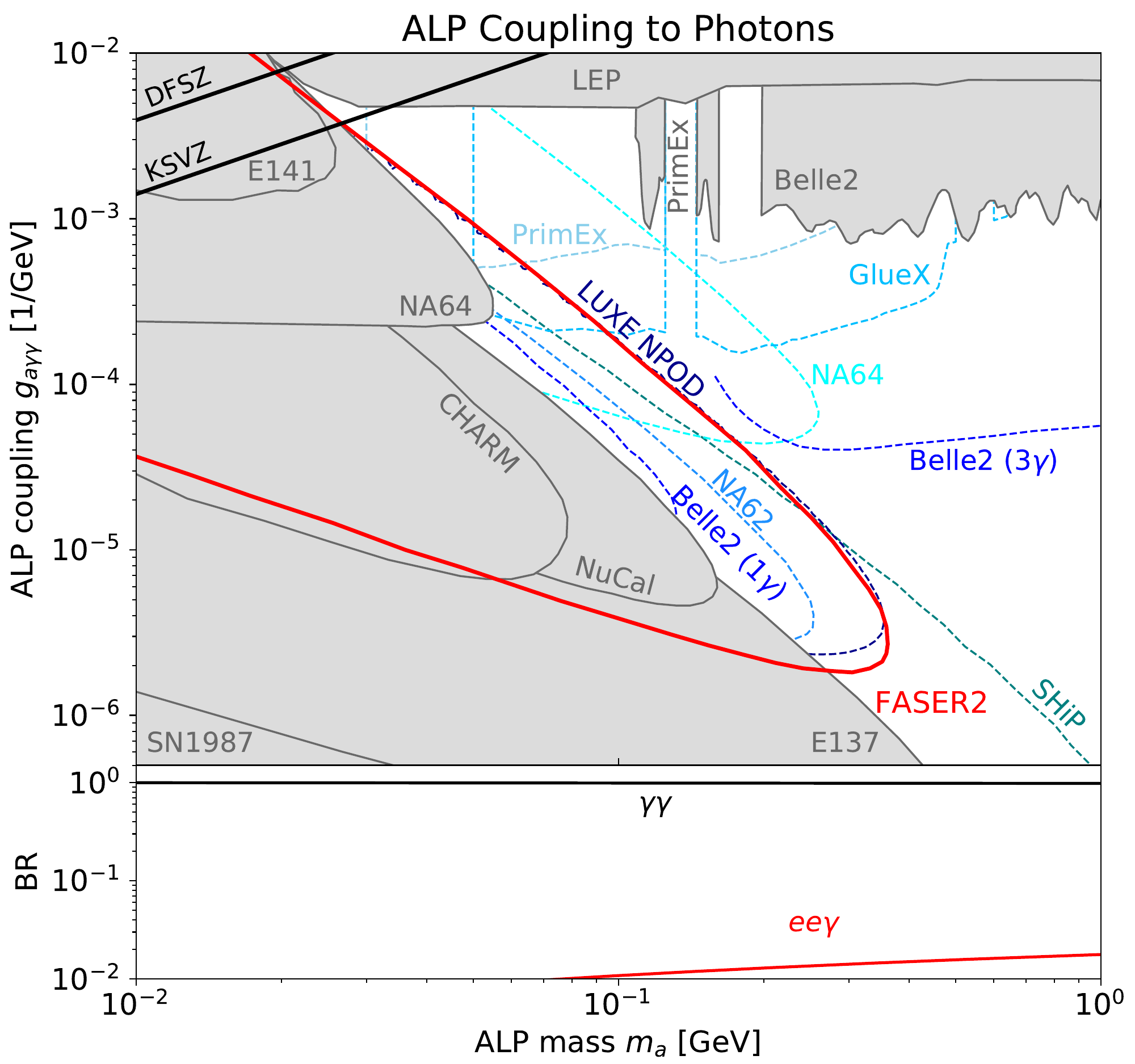}
	\includegraphics[width=0.49\textwidth]{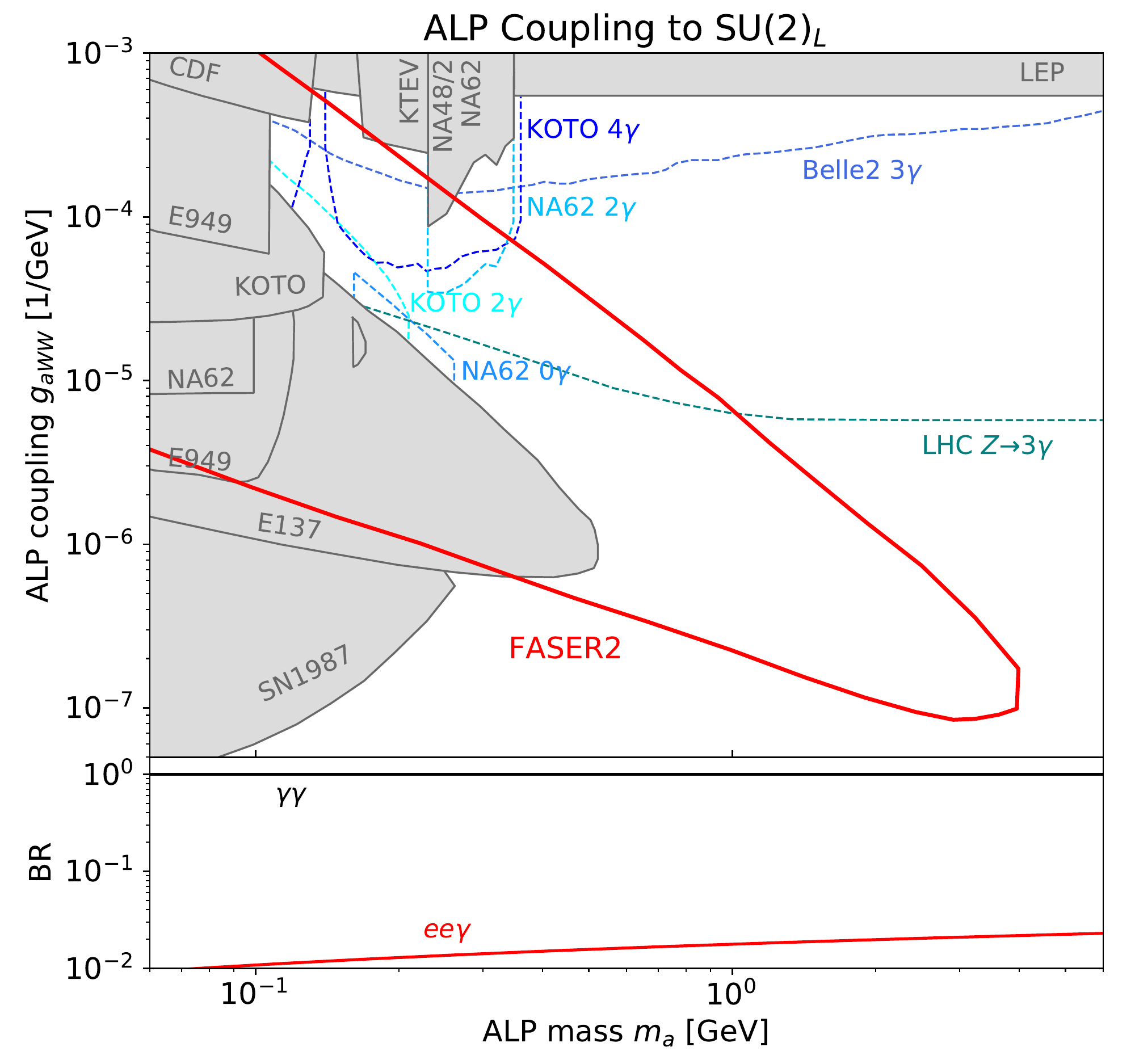}
	\includegraphics[width=0.49\textwidth]{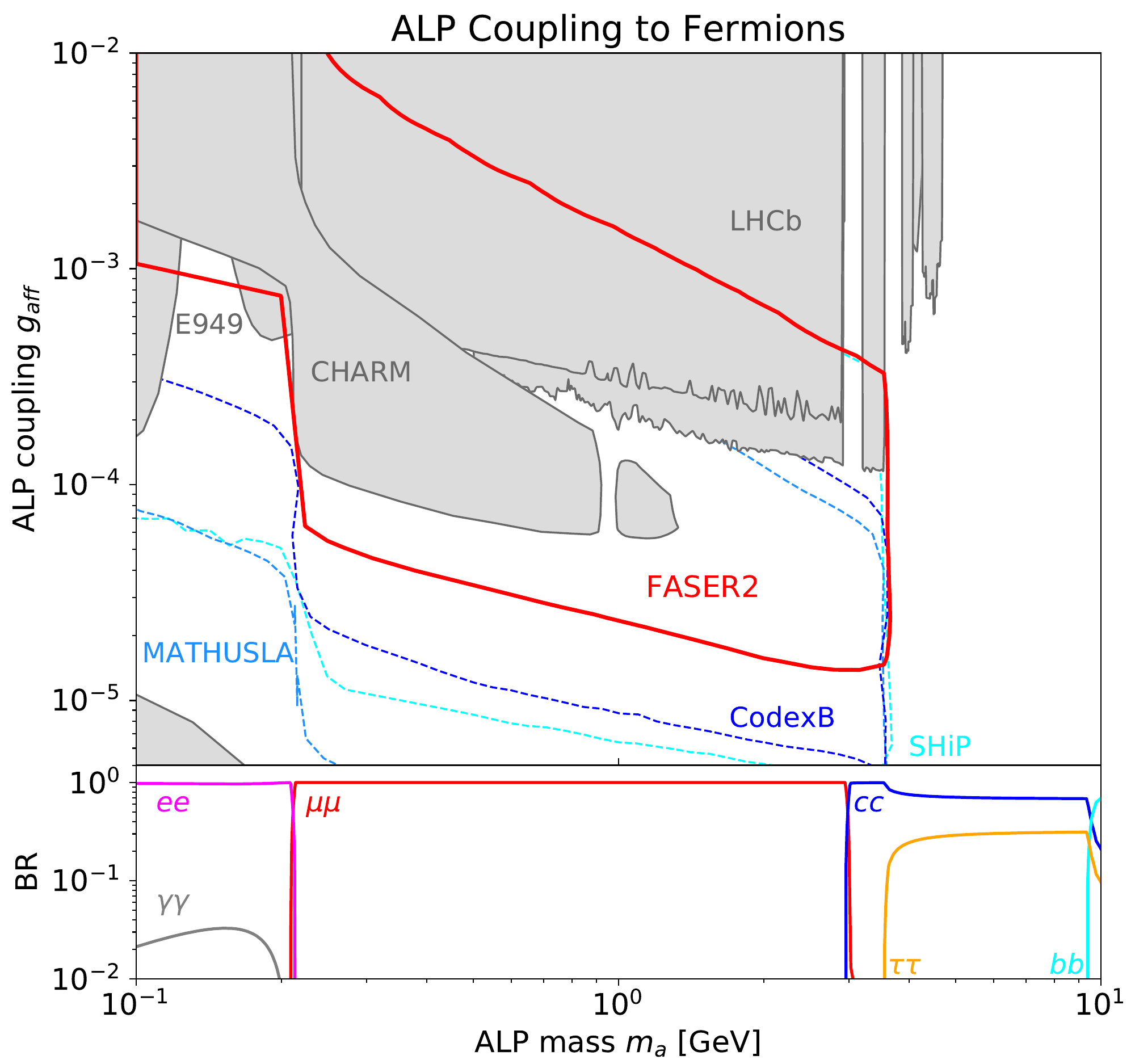}
	\includegraphics[width=0.49\textwidth]{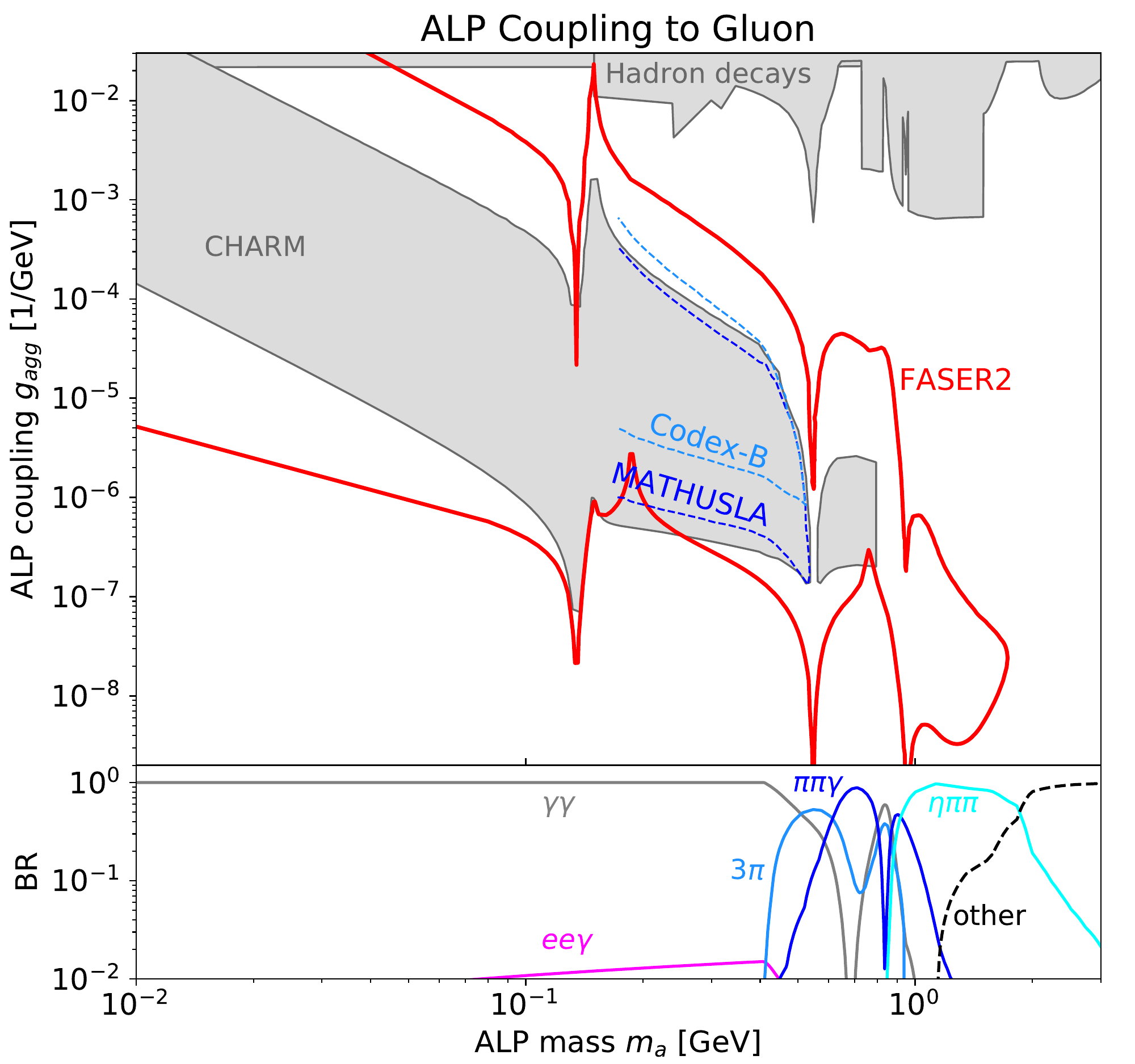}
	\caption{Sensitivity for ALP searches at the FPF. We consider four specific scenarios in which the ALP couples dominantly to the photon (upper left), $SU(2)_L$ (upper right), all fermions (lower left) and gluons (lower right). The sensitivity reaches of FASER2 is shown as solid red lines alongside existing constraints (gray shaded regions) and projected sensitivities of other selected proposed searches and experiments (blue dashed lines). The bottom part of each panel shows the ALP branching fractions.  In panel (d), the ALP reach is enhanced by a recent calculation including a consistent treatment of proton bremmstrahlung productions, for more details, see \cref{bsm_alp_brem}. See text for details and references.}
	\label{fig::bsm_alp}
\end{figure*}

Given the large number of free coupling parameters for ALPs, it is instructive to consider some special cases in which at the high-energy scale $\Lambda_{UV}$ only one of the couplings is non-vanishing. In the following we consider four commonly considered cases:
\begin{description}
  \item[Coupling to Photons] The perhaps most studied scenario considers the case in which the ALP exclusively couples to the photon field~\cite{Dobrich:2015jyk}. The corresponding effective Lagrangian for this benchmark model is $\mathcal{L} = -\frac12 m_a^2 a^2 - \frac14 g_{a\gamma\gamma} a F^{\mu\nu}\tilde{F}_{\mu\nu}$, where the dimensional coupling $g_{a\gamma\gamma}$ parameterizes the interaction with the photon field. 
  
  At the FPF, this particle is mainly produced via the Primakoff process: high energy photons originating from pion decay can convert into ALPs when hitting the TAN neutral particle absorber~\cite{Feng:2018pew}. The ALP then almost exclusively decays into a pair of photons couplings mainly decay into a pair of photons. A subleading decay channel into a photon and an electron pair has a suppressed branching fraction of order $\text{BR}(a \to \gamma e e ) \approx \text{BR}(\pi^0 \to\gamma e e ) \approx 1\%$. 
  
  The projected sensitivity reaches for FASER2 is shown in the upper left panel of \cref{fig::bsm_alp}. The gray shaded region are already excluded by previous searches from LEP~\cite{OPAL:2002vhf}, PrimEx~\cite{Aloni:2019ruo}, Belle~2~\cite{Belle-II:2020jti} E141~\cite{Dobrich:2017gcm}, NA64~\cite{NA64:2020qwq}, NuCal~\cite{Blumlein:1990ay, Dobrich:2019dxc}, CHARM~\cite{CHARM:1985anb} and E137~\cite{Bjorken:1988as}.  In addition we show the projected sensitivity for a subset of other proposed search at SHiP~\cite{Dobrich:2015jyk}, Belle~2~\cite{Dolan:2017osp}, NA62~\cite{Dobrich:2015jyk}, NA64~\cite{Gninenko:2320630}, PrimEx~\cite{Aloni:2019ruo}, GlueX with $1~\ipb$~\cite{Aloni:2019ruo}, and LUXE-NPOD (phase-1)~\cite{Abramowicz:2021zja, Bai:2021dgm}. See Ref.~\cite{Dobrich:2019dxc} and Ref.~\cite{Beacham:2019nyx} for a more complete overview on proposed searches. 
  
  The solid black lines correspond to the parameter space for the DFSZ~\cite{Dine:1981rt, Zhitnitsky:1980tq} and KSVZ~\cite{Kim:1979if, Shifman:1979if} models for the QCD axion, as presented in Ref.~\cite{NA64:2020qwq}. 
  \item[Coupling to $SU(2)_L$] The previous scenario focused on the ALP's coupling to photons. More generally, this is expected to arise if the ALP couples either to the field strength tensor of the $U(1)_Y$ or $SU(2)_L$ fields in the unbroken theory. In the following, let us concentrate on the specific case in which the axion only couples SM field strength tensor $W_{\mu\nu}$ of the $SU(2)_L$ group, which was studied in Ref.~\cite{Gori:2020xvq}. The corresponding Lagrangian is  $\mathcal{L} = -\frac12 m_a^2 a^2 - \frac14 g_{aWW} a W^{a\mu\nu}\tilde{W}^{a}_{\mu\nu}$. After electroweak symmetry breaking, the ALP obtains coupling to all SM electroweak gauge bosons, including $WW$, $ZZ$, $Z\gamma$ and $\gamma\gamma$. Notably, the coupling to photons is suppressed relative to $W$ bosons by $g_{a\gamma\gamma} = g_{aWW} \sin^2 \theta_W$. 
  
  Due to its coupling to $W$ bosons, the ALP can be produced in flavour changing hadron decays, especially $K \to \pi a$ at low masses and $B \to X_s a$ at higher masses, which are induced via the usual loop diagrams. These rare meson decays turn out to provide the dominant production mode, while the previously considered Primakoff process still contributes subdominantly. For the light ALP masses of interest for the FPF, decays into the heavy $Z$ and $W$ bosons do not play a role, and the ALP dominantly decays into two photons, $a \to \gamma\gamma$. 
  
  The projected sensitivity for FASER2 has been obtained in Ref.~\cite{Kling:2020mch} and is shown in the upper right panel of \cref{fig::bsm_alp}. The gray shaded region denotes the parameter space excluded by previous searches as obtained in Refs.~\cite{Gori:2020xvq, Izaguirre:2016dfi} and references therein. In addition, we show the projected sensitivity for searches for visible and invisible ALP decays at KOTO and NA62 as obtained in Ref.~\cite{Gori:2020xvq}, and visible ALP decays at Belle~2 and the LHC as obtained in Refs.~\cite{Dolan:2017osp, Bauer:2017ris}. We can see that FASER2 at the FPF provides very complementary constraints compared to the other proposed searches. 
  \item[Coupling to Fermions] In addition to the electroweak bosons, the ALP can also couple to the SM fermions. For simplicity, we will assume that all fermion coupling constraints are identical at the relevant low-energy scale. This is equivalent to the assumption that all the SM fermions carry the same PQ charge. In this case, the ALP obtains Yukawa-like couplings to the SM fermions. The corresponding Lagrangian can then be written as $\mathcal{L} = -\frac12 m_a^2 a^2 - i \, g_{aff}\, a \sum_f y_f \, \bar f  \gamma_5 f $, where we introduced the dimensionful coupling $ g_{aff}$. 
  
  Similar to the previous case, addition non-diagonal couplings arise loop level, such as the flavour changing $b-s-a$ vertex. This coupling then induces the rare $B$-meson decay $B \to X_s a$, which is the dominant source of ALP passing through the FPF. The relevant decay rates are discussed in Refs.~\cite{Batell:2009jf, Dolan:2014ska, Beacham:2019nyx}. Due to the Yukawa-like fermion couplings, the ALP typically decays into  pairs of the heaviest kinematically available SM fermions. Below the charm threshold, the ALP mainly decays into leptons while hadronic decays were found to be suppressed~\cite{Dolan:2014ska, Domingo:2016yih}. 
  
  The projected sensitivity for FASER2 was obtained in Ref.~\cite{FASER:2018eoc} and is presented in the lower left panel of \cref{fig::bsm_alp}. As before, the gray shaded region corresponds to already excluded parameter space, where the main constraints arise from searches for $K \to \pi a$ at E949~\cite{E949:2004uaj, BNL-E949:2009dza}, searches for $B \to K a$ at LHCb~\cite{LHCb:2015nkv, LHCb:2016awg} and LLP searches at CHARM~\cite{CHARM:1985anb}, as discussed in Ref.~\cite{Dolan:2014ska, Dobrich:2018jyi}. We also show the projected sensitivity for other proposed experiments to search for Long-Lived ALPs as blue dashed lines as presented in Ref.~\cite{Beacham:2019nyx}. 
  
  It is worth noting that this model is qualitatively similar to a light pseudoscalar with Yukawa-like coupling, as for example arising in the Type-I 2HDM discussed in \cref{sec:bsm_dh_2hdm}. However, due to the different way in which electroweak symmetry is broken, there are notable differences in the loop-induced couplings. Two relevant examples are the decay width into photons, and the size of flavor-changing $b - s - a$ coupling. See Ref.~\cite{Dolan:2014ska} for a more detailed discussion and Ref.~\cite{FASER:2018eoc} for the FASER2 reach for a pseudoscalar with Yukawa-like coupling. 
  \item[Coupling to Gluons] In the last scenario, we consider an ALP that primarily couples to the gluon field strength tensor. The corresponding Lagrangian defined at some scale $\Lambda$ then reads $\mathcal{L} = -\frac12 m_a^2 a^2 - \frac{g_s^2}{8} g_{agg} a \text{Tr}G^{\mu\nu}\tilde{G}_{\mu\nu}$. At lower scales, the ALP will obtains diagonally couplings to quarks at one loop, and even further suppressed non-diagonally couplings at two-loop. The latter will again induce flavour changing heavy meson decays, such as $B \to X_s a$.  
  
  Another interesting feature of this scenario is that the ALP can mix with the neutral pseudo-scalar mesons, especially the $\pi^0$, the $\eta$ and the $\eta'$ mesons (see Ref.~\cite{FASER:2018eoc, Beacham:2019nyx} for details). As a consequence, the ALP can be  produced in any process that produces these pesudo-scalar mesons, for example the hadronization of a hadronic shower. In addition, the ALP can also be produced in flavor changing meson decays like $B \to X_s a$, similar to the previous cases. Depending on the mass, the ALP will either decay into photons or hadrons. Here we use the decay width obtained in Ref.~\cite{FASER:2018eoc, Beacham:2019nyx}. The presented branching fractions for the ALP with gluon couplings in \cref{fig::bsm_alp} were adopted from Ref.~\cite{Aloni:2018vki}. 
  
  The projected sensitivity for the ALP with gluon couplings for FASER2 is presented in the lower right panel of \cref{fig::bsm_alp}. Here we include recent work considering the ALP emission from the proton, which is discussed in more detail in \cref{bsm_alp_brem}. The existing constraints from flavour physics, as obtained in Ref.~\cite{Aloni:2018vki}, and LLP searches at CHARM, as presented in Ref.~\cite{FASER:2018eoc}, are shown as gray shaded area. In addition, we also show the projected sensitivity for MATHUSLA and Codex-B as presented in Ref.~\cite{Beacham:2019nyx}. 
  
\end{description}

\subsection{Charming ALPs}
\label{bsm_alp_charm}

Strongly coupled dark sectors are a particularly interesting class of dark matter models as they can inherit the SM flavour structure via a flavoured portal. One likely feature of these models is the emergence of pseudo Nambu-Goldstone bosons which couple dominantly to either up or down quarks and can be long-lived. This happens for instance in  models featuring a QCD-like dark sector where a heavy scalar mediator, charged both under the SM and a dark colour group $SU(N)_D$, connects the dark quarks with either right-handed SM up-type or down-type quarks~\cite{Bai:2013xga, Blanke:2017tnb, Blanke:2020bsf, Cheng:2019yai, Jubb:2017rhm, Mies:2020mzw, Renner:2018fhh, Schwaller:2015gea, Strassler:2006im, Carmona:2021seb}, depending on the hypercharge of the mediator. A similar outcome occurs e.g. in Froggatt-Nielsen (FN) models~\cite{Froggatt:1978nt} where only the up- or down-type quarks are charged under the additional global symmetry. Here, based on~\cite{Carmona:2021seb}, we focus on light new states that are mainly coupled to up-type quarks and thus produced in $D$ meson decays instead of from $B$ mesons or Kaons. 
 
For concreteness, we consider an ALP $a$ coupling only to right-handed up-type quarks at tree-level, described by the following EFT Lagrangian~\cite{Georgi:1986df, Choi:1986zw}:
\be
	\mathcal{L}&=\frac{1}{2}(\partial_{\mu}a)(\partial^{\mu}a)-\frac{m_a^2}{2}a^2+\frac{\partial_{\mu} a}{f_a} \left[(c_{u_R})_{ij}\bar{u}_{Ri} \gamma^{\mu} u_{Rj}+c_{H}H^{\dagger} i\overleftrightarrow{D_{\mu}} H\right] 
	\\
	&\ \ -\frac{a}{f_a}\left[c_{g}\frac{g_3^2}{32\pi^2}G_{\mu\nu}^{a}\tilde{G}^{\mu\nu a}+c_{W}\frac{g_2^2}{32\pi^2}W_{\mu\nu}^I\tilde{W}^{\mu\nu I} +c_{B}\frac{g_1^2}{32\pi^2}B_{\mu\nu}\tilde{B}^{\mu\nu}\right].
	\label{eq:alag}
\ee
The gauge couplings of $U(1)_Y$, $SU(2)_L$ and $SU(3)_C$ are denoted $g_1$, $g_2$ and $g_3$, and $B_{\mu\nu}$, $W_{\mu\nu}^I$, $I = 1,2,3$ and $G_{\mu\nu}^{a}$, $a = 1,...,8$ their respective field strength tensors with $\tilde{B}^{\mu\nu} = \frac{1}{2} \epsilon_{\mu\nu\alpha\beta} B^{\alpha\beta}$. $H$ is the SM Higgs field and the Wilson coefficients $c_g$, $c_W$, $c_B$, $c_H \in \mathbb{R}$, while $c_{u_R}$ is a hermitian $3\times3$ matrix. We do not consider the QCD axion case, so the ALP mass $m_a$ and decay constant $f_a$ are independent parameters. The operators 
\begin{align}
    &\mathcal{O}_{H}=(\partial^{\mu}a/f_a) H^{\dagger}i \overleftrightarrow{D_{\mu}} H \text{ and}\\
    &\mathcal{O}_{W}=\frac{a}{f_a}\frac{g_2^2}{32\pi^2}W_{\mu\nu}^I\tilde{W}^{\mu\nu I},
\end{align}
lead to flavour-violating couplings with the down-type quark sector at the one-loop level~\cite{Gavela:2019wzg}. However, we assume that their Wilson coefficients are small enough so that the leading flavour violating effects appear in the up sector and that $c_g=0=c_B$ at tree-level. We dub these scenarios  ``charming ALPs''. 

For ALP masses $m_a\lesssim 1$~GeV the dominant ALP decay mode is to hadrons. In this case the interactions are best described in chiral perturbation theory, see Refs.~\cite{Georgi:1986df, Choi:1986zw, Bardeen:1986yb, Krauss:1986bq} for more details. For masses $m_a\gtrsim 1$~GeV, this approach is no longer valid and we use quark-hadron duality~\cite{Poggio:1975af, Shifman:2000jv} to compute the ALP decay width into hadrons. Constraints on the parameter space arise from  astrophysics, such as supernovae cooling~\cite{Raffelt:1996wa} and red giant burst~\cite{Raffelt:2006cw, Capozzi:2020cbu}, as well as cosmology, e.g. from potential distortions of the cosmic-microwave background, modifications of the predicted big-bang nucleosynthesis or possible impact on $N_{eff}$ as discussed in Refs.~\cite{Cadamuro:2011fd, Millea:2015qra, Depta:2020wmr}.

Charming ALPs also contribute to several flavour processes and constraints on the parameter space arise from $D^0-\bar{D}^0$ mixing~\cite{HFLAV:2019otj}, flavour violating kaon~\cite{NA62:2020xlg}, $B$ meson~\cite{Ammar:2001gi} and $D$ meson~\cite{Eisenstein:2008aa, Ablikim:2019rpl} decays, with the kaon and $B$ meson decays loop-suppressed, as well as radiative $J/\psi$ decays~\cite{Insler:2010jw}. The decay width for meson decays of the form $M \to N a$ is 
\be
	\Gamma(M \to N a)&=\frac{m_M^3|\varkappa_{MN}|^2}{64\pi f_a^2}\left(1-\frac{m_{N}^2}{m_M^2}\right)^2 (f_0^{MN}(m_a^2))^2\\ 
	&\ \ \times\sqrt{\left(1-\frac{(m_{N}+m_a)^2}{m_M^2}\right)\left(1-\frac{(m_{N}-m_a)^2}{m_{M}^2}\right)},
\ee
with $m_N$ and $m_M$ being meson masses and $f_0^{MN}$ the scalar form factor. The coupling $\varkappa_{MN}$ is defined as follows
\begin{align}
	\mathcal{L}\supset \varkappa_{MN} \frac{\partial^{\mu}a}{2 f_a} \bar{q}_i \gamma_{\mu} q_j+\mathrm{h.c.},
\end{align}
with $\varkappa_{D\pi} = (c_{u_R})_{12}$ for the case of flavour violating $D$ meson decays, and  
\begin{equation}
    \varkappa_{MN} = \frac{1}{16\pi^2 v^2}V_{ri}^{\ast}(\mathcal{M}_u)_{rr} (c_{u_R})_{rs}  (\mathcal{M}_u)_{ss} V_{sj}\ln\left(\frac{f_a^2}{\mu^2}\right)
\end{equation}
for loop-mediated decays of kaons and $B$ mesons, where $i$ and $j$ denote the quark in the initial and final state mesons and $\mu$ is the relevant energy scale of the process.

\begin{figure*}[t]
\centering
\includegraphics[width=0.48\textwidth]{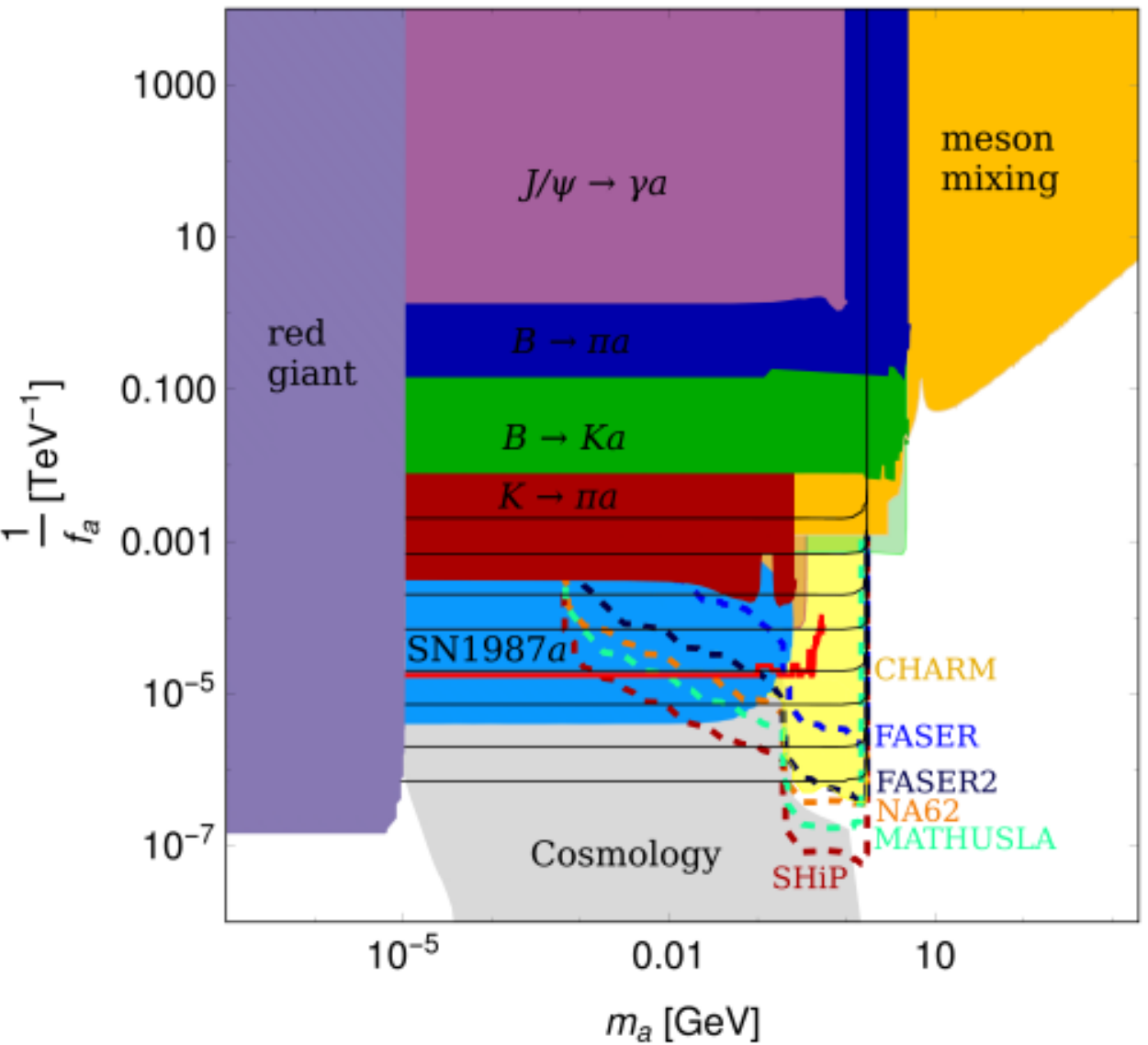}
\includegraphics[width=0.48\textwidth]{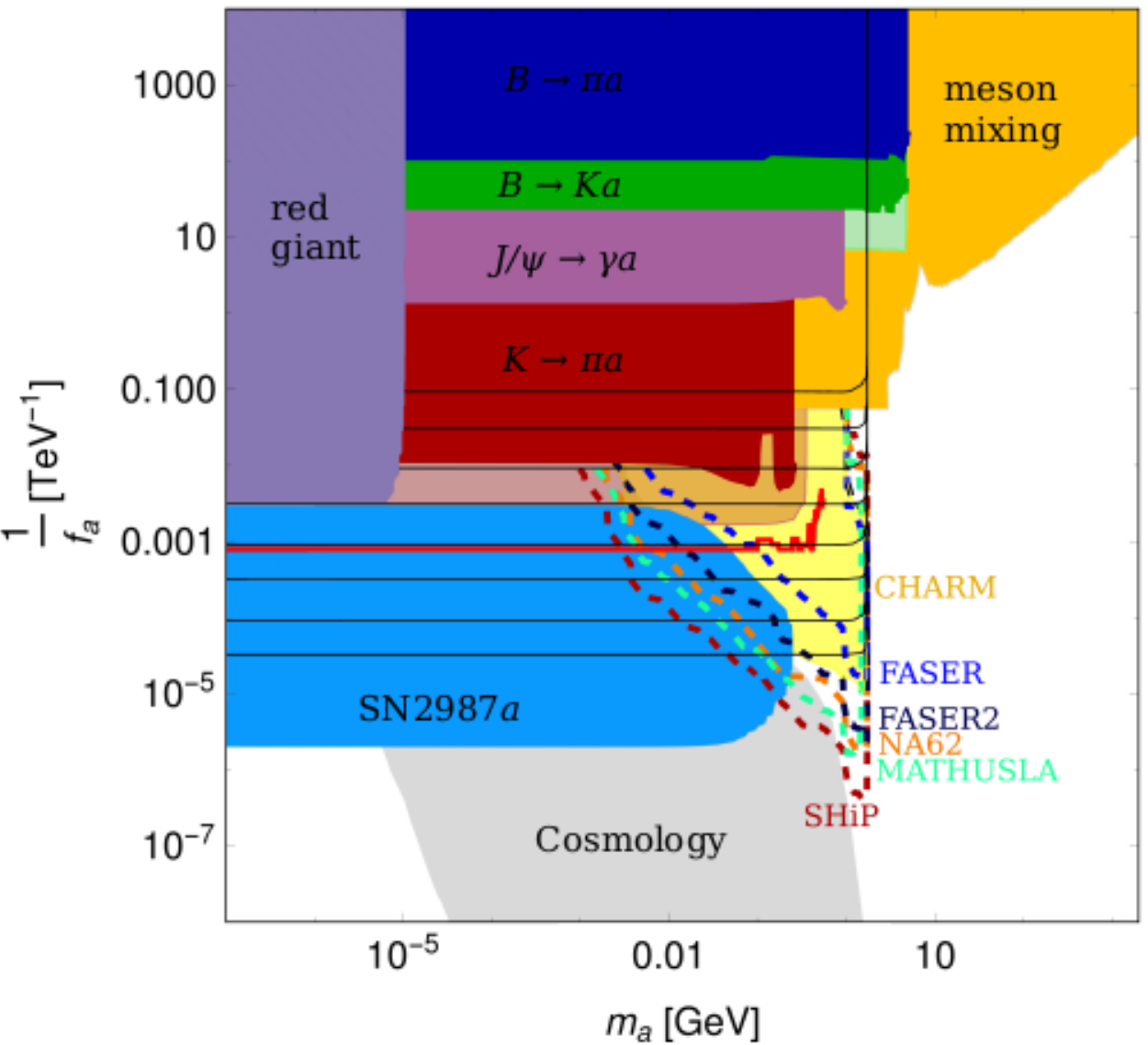}	
\includegraphics[width=0.48\textwidth]{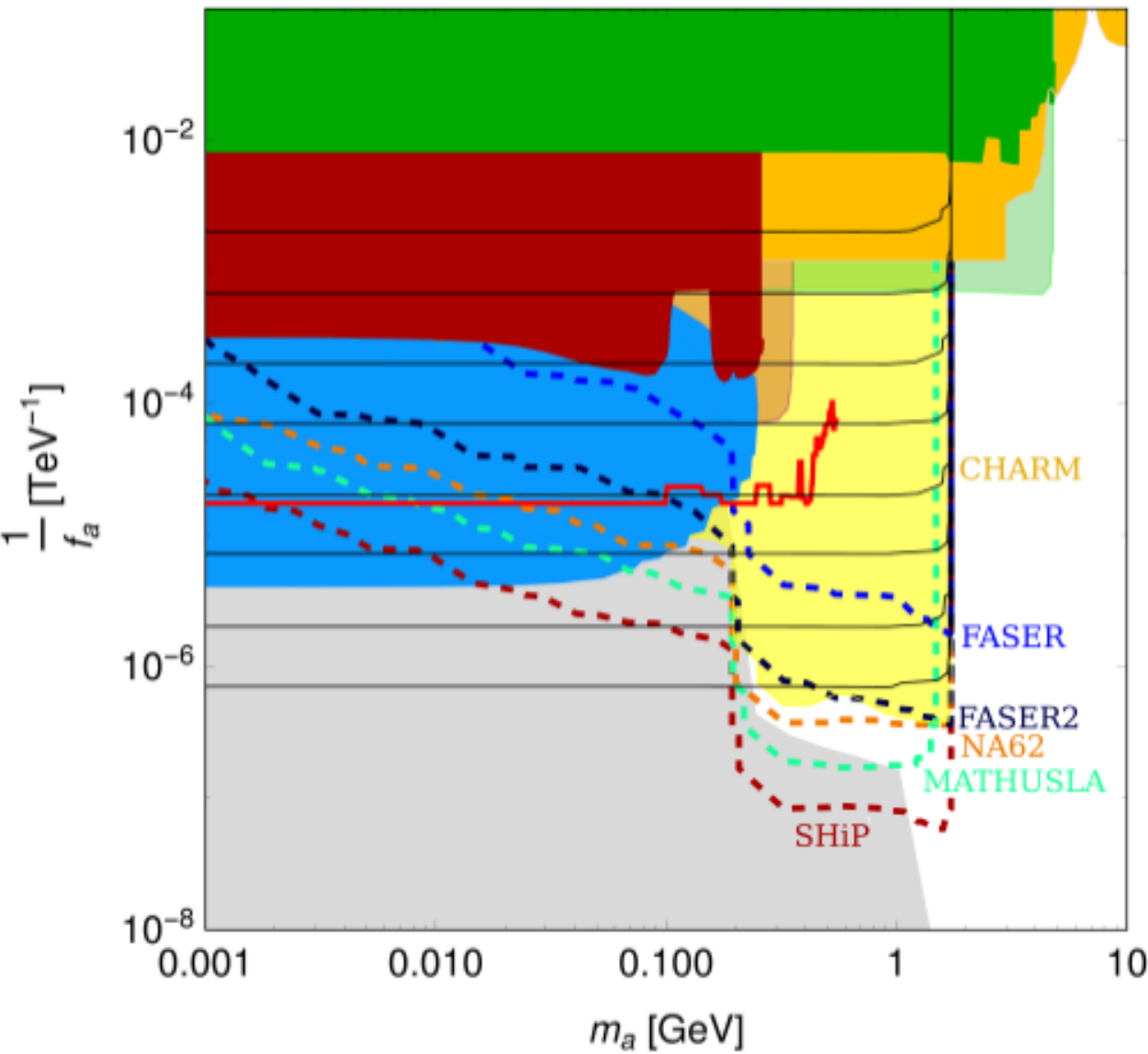}
\includegraphics[width=0.48\textwidth]{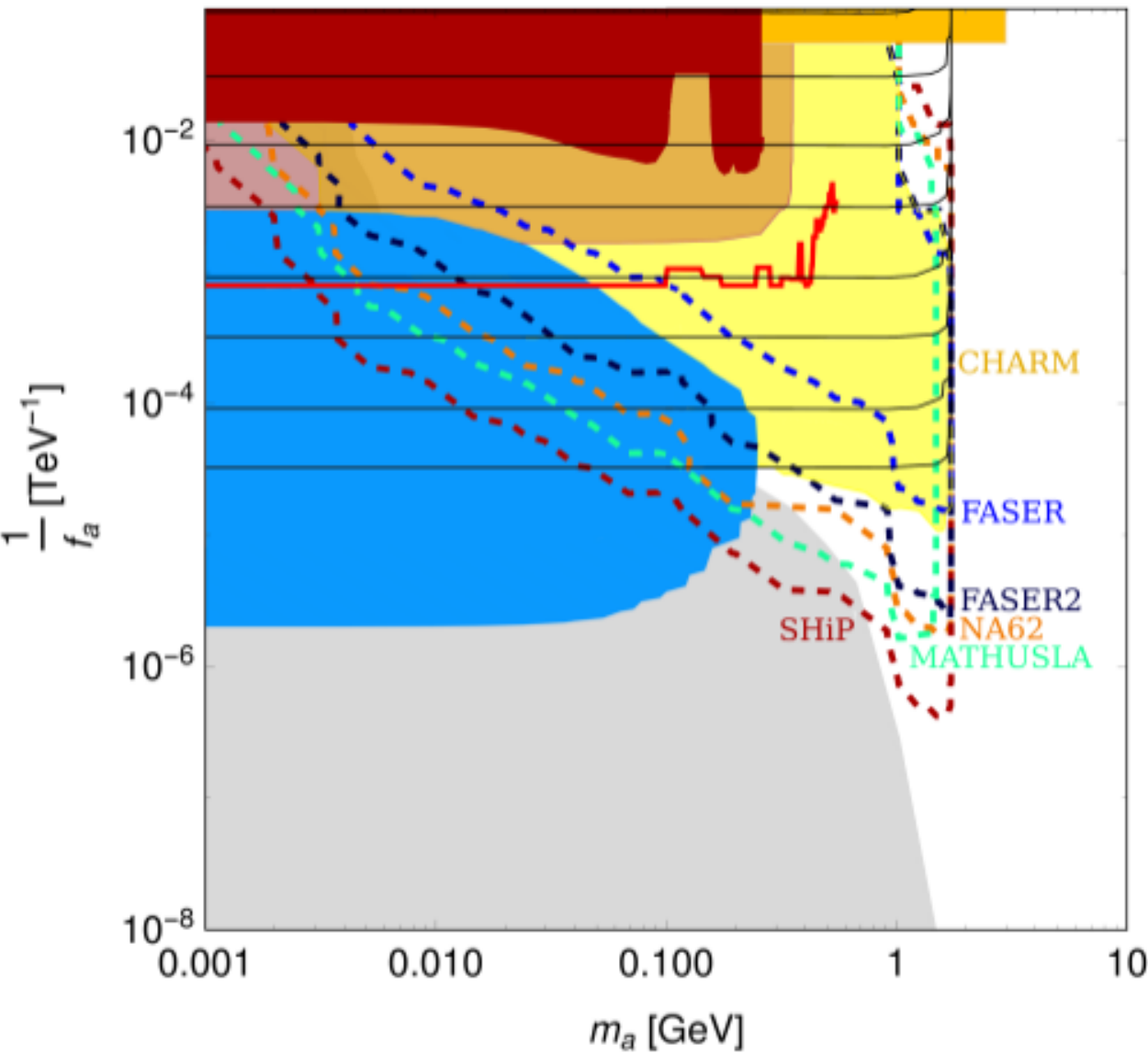}
\caption{Parameter space of charming ALPs for the anarchic (left) and FN like scenario (right) of Ref.~\cite{Carmona:2021seb} in the $1/f_a$-$m_a$ plane. The coloured regions are excluded by constraint from flavour, astrophysics and cosmology. The thick red line shows the indirect bound on $\mathrm{Br}(D\to \pi a)$ from $D\to \tau \nu \to \pi \nu\nu$ measurements, while the black lines show the model's predictions of $Br(D\to \pi a)$ for $\mathrm{Br}(D\to \pi a)$ $=$ $10^{-1}-10^{-8}$, each line one decade smaller. Above the dashed lines more than three ALPs are expected to decay inside of the respective detector's decay volume.}
\label{fig:charming_ALP_results}
\end{figure*}

The decay $D\to\pi a$ is also one of the main production modes when searching for charming ALPs in the forward direction as can be done with fixed target experiments like NA62~\cite{NA62:2017rwk} and the proposed SHiP experiment~\cite{Alekhin:2015byh}, as well as detectors in the forward direction such as FASER/FASER2 or MATHUSLA~\cite{MATHUSLA:2018bqv, MATHUSLA:2020uve}. Charming ALPs decay mostly to photons, muons and pions in the part of the parameter space these experiments can probe. For a more detailed discussion of the branching ratios see~\cite{Carmona:2021seb}. The number of charming ALPs decaying in the detector volume is
\begin{align}
	 N_{a}=N_D\cdot \mathrm{Br}(D\to\pi a)\cdot\varepsilon_{\rm geom}\cdot F_{\rm decay}\,,
 \end{align}
where $N_D$ is the number of $D$ mesons produced at the interaction point, $F_{\rm decay}$ is the fraction of ALPs decaying inside the detector volume, and the geometric acceptance $\varepsilon_{\rm geom}$ is given by the fraction of ALPs whose laboratory frame momentum falls within the detector opening angle. The $D$ meson momentum distribution for FASER/FASER2 and MATHUSLA was simulated with FONLL with CTEQ6.6~\cite{Cacciari:1998it}, for NA62 and SHiP taken from Ref.~\cite{CERN-SHiP-NOTE-2015-009} and $\varepsilon_{\rm geom}$ and $F_{\rm decay}$ were calculated as in Ref.~\cite{Renner:2018fhh}. We assume that three detected events are needed for discovery. 

In \cref{fig:charming_ALP_results}, the excluded regions of the parameter space and detection prospect of charming ALPs at FASER/FASER2 in the $1/f_a$-$m_a$ plane are shown for two benchmarks defined in Ref.~\cite{Carmona:2021seb}: the so-called anarchic (left) and FN-like (right) scenarios. The lower panels show a zoomed in look at the region where the experiments are sensitive. In addition, the prospects for NA62, SHiP and MATHUSLA, as well as the region excluded by CHARM~\cite{Bergsma:1985qz} are shown. Above the dashed lines more than three ALP decays are expected in the decay volumes of the respective detectors. The main difference between the two benchmark scenarios is the coupling of ALPs to top quarks. In the anarchic scenario all  ALP-quark couplings are $\mathcal{O}(1)$ leading to comparatively large contributions from the loop processes. On the other hand, in the FN like scenario the coupling to tops is $\mathcal{O}(10^{-5})$, while the other diagonal couplings are $\mathcal{O}(1)$ and the off-diagonal couplings $\mathcal{O}(10^{-2}-10^{-4})$. 

Comparing the right and left panels one can clearly see the impact of the ALP top coupling. In the left panel, for the anarchic scenario, the bounds from all loop processes as well as the bound from CHARM, which searched for ALPs decaying to photons, electrons and muons, are more stringent than for the FN like case in the right panel, where the ALP top coupling is small. Especially the impact of this coupling on the bounds from CHARM also affect the prospects of FASER and FASER2: for large ALP top couplings FASER and FASER2 will mostly validate the CHARM constraints. On the other hand, for smaller ALP top couplings, as shown in the right panel, FASER is still mainly testing the constraints from CHARM, but FASER2 has a significant larger reach and will be able to probe so far untested parameter space up to $1/f_a\sim 10^{-6}$~TeV$^{-1}$ in the kinematically accessible region. NA62 and MATHUSLA have a similar range, while SHiP will be able to probe an even larger part of the unexplored parameter space.

We studied the parameter space and discovery prospect of charming ALPs as an example for long-lived light new physics states coupling dominantly to the up quark sector. In the here considered case $D$ meson decays are the main ALP production mode. We showed that detectors in the forward direction, such as FASER and FASER2, but also MATHUSLA and fixed target experiments like NA62 and SHiP, will be able to probe large parts of the so far unexplored parameter space of such models, especially if the ALP top coupling is small. 

\subsection{Bremming Enhanced ALP Productions and FPF Sensisivity}
\label{bsm_alp_brem}

The searches for ALPs are strongly motivated by general dark sector new physics considerations, the strong CP puzzle, and the axion quality problem. In particular, the ALP to gluon coupling is critical in connection to the strong CP puzzle (for a recent model-building and phenomenology discussion for GeV-scale Axion, see~Refs.~\cite{Hook:2019qoh, Csaki:2019vte, Gherghetta:2020keg, Gherghetta:2020ofz} and references therein). The searches for ALPs with gluonic couplings turned on, without further suppressions compared to other operators, such as in the vanilla axion models, are hence of great importance and interest. The generic and motivated gluon coupling also poses a challenge, that the axion lifetime will be on shorter size for long-lived particle detectors such as those at FPF. One will need to rely on the large production rate at colliders to probe the ALPs, and here focus on the bremsstrahlung process for the proton-proton collision that has not been incorporated in the previous ALP studies~\cite{Feng:2018pew, Aielli:2019ivi}. As we shall see, the inclusion of the bremsstrahlung production process significantly extends the FPF coverage for ALP in a broad class of models~\cite{LiuLyu}. 

In the forward region, proton collisions can be categorized into four distinct categories, namely the elastic scattering, single dissociative scattering, double dissociative scattering, and non-diffractive scattering~\cite{Foroughi-Abari:2021zbm}. For the elastic scattering, the protons stay unbroken before and after the collision, which can be parametrized by exchanging Pomeron at the lowest order. The second refers to the single diffractive scattering in which one proton keeps intact while the other proton dissociates into partonic debris. The third refers to both protons dissociating, which can be parametrized by exchanging Pomerons. The last one is the general (deep) inelastic scattering in which partons within the ``broken" protons scatter with relatively large transverse momentum. With the four categories, we can divide the total cross-section of p-p collision into $\sigma_{\rm tot} = \sigma_{\rm el} + \sigma_{\rm SD} + \sigma_{\rm NSD}$. Note that we cannot tell apart the double and non-diffractive scattering experimentally, since there is no proton in the final states, and group them together as non single diffractive scattering. Ref.~\cite{Foroughi-Abari:2021zbm} has discussed the soft radiation of dark photon and dark CP-even scalar by the bremsstrahlung process of protons. It turned out that the initial state radiation and final state radiation largely cancel each other in the forward region for the elastic and single diffractive processes. A similar case applies to our ALP radiation. Hence we can focus on the ISR of the incoming proton in the non single diffractive processes. In principle, ALP can be radiated by either the proton or the partons. Our computations have shown that the partonic bremsstrahlung is subdominant compared with the bremsstrahlung process of the proton. Hence in the following we just show the calculation of the proton bremsstrahlung.\footnote{One shall be careful with the validity of the proton bremming calculation, given that proton is a composite particle. We restrict ourselves to low momentum transfer, characterized by the low off-shellness of the proton after radiation. For detailed discussion, see Refs.~\cite{Foroughi-Abari:2021zbm, LiuLyu}}. The proton-axion interaction can be parametrized via~\cite{ParticleDataGroup:2020ssz}
\begin{equation}
    \dfrac{c_p}{2 f_A}  \bar{p} \gamma^\mu \gamma_5 p \, \partial_\mu a
\end{equation}
To deduce the axion proton coupling, we can match the effective Lagrangian in the perturbative QCD regime to the chiral perturbation theory in the  strong dynamics confinement phase. The relevant effective axion Lagrangian is given by
\begin{equation}
\dfrac{a}{8\pi f_a} \bigg( c_3 \alpha_3 G \tilde{G} + c_2 \alpha_2 W \tilde{W} + c_1 \alpha_1 B \tilde{B} \bigg) + \sum \dfrac{c_q}{2 f_a}  \bar{q} \gamma^\mu \gamma_5 q \, \partial_\mu a
\end{equation}
which characterize the ALP coupling to the gauge field and quarks. As for the  gauge field, usually two cases are considered: the first is the Gluon Dominance with $c_3 =1, c_2 = c_1 = 0$ and the second is the Codominance with $c_3=c_2=c_1$~\cite{Kelly:2020dda}. The two cases match well respectively to the KSVZ  and DFSZ scenario of the minimal axion theory, where one needs to also include the non-suppressed fermionic coupling for the latter. Correspondingly the coupling of proton to axion in the case of axion mass far smaller than the $\pi^0$ mass is given by~\cite{GrillidiCortona:2015jxo}
\begin{equation}\label{eq:alp_p_coupling}
c_p^{\rm KSVZ} = -0.47 \quad\quad c_p^{\rm DFSZ} = -0.617 + 0.435 \sin^2 \beta \pm 0.025 
\end{equation}
Note that the root of the enhanced production considered here is from the ALP to gluon coupling, which generic ALP models share, and the lifetime is only affected by $\mathcal{O}(1)$ in the relevant regime. So we focus here on the KSVZ-like scenario with quark coupling turning off, dubbed as ``gluon-dominance" following~\cite{Kelly:2020dda}, and one can also estimate the enhanced sensitivities for other scenarios in a similar way. The coupling in \cref{eq:alp_p_coupling}  is computed in the 2-flavor scenario. For the ALP mass around $O(1)$ GeV in our interests, we need to exploit the 3-flavor ChPT and Effective Baryon Theory to extract the ALP-proton coupling. Following \cite{Blinov:2021say}, the Lagrangian terms describing the nucleon-ALP are given by
\begin{equation}
\begin{split}
\mathcal{L} \supset &\text{tr} \bar{B} \left( i \slashed{D} - m_N \right) + \dfrac{D}{2} \text{tr} \bar{B} \gamma_\mu \gamma_5 \{ u^\mu, B \} + \dfrac{F}{2} \text{tr} \bar{B} \gamma_\mu \gamma_5 [ u^\mu, B ] + \dfrac{D_s}{2} \text{tr} \bar{B} \gamma_\mu \gamma_5 B \text{tr} u^\mu
\end{split}
\end{equation}
in which $u^\mu$ is the veilbein which contains the meson field and ALP field. After expansion and combined with the ChPT Lagrangian, the physical ALP's coupling to the proton is given by
\begin{equation}
\begin{split}
g_{pa} &= \dfrac{(2 m_{\eta}^2 - 5 m_{\eta^\prime}^2 + 3m_{\pi}^2 ) \big[ 6 D (m_a^2 - m_{\eta}^2) + D_s (9 m_a^2 - 8m_{\eta}^2 - m_{\eta^\prime}^2) + 2 F (m_{\eta}^2 - m_{\eta^\prime}^2) \big] }{108 f_a (m_a^2 - m_\eta^2)( m_a^2 -m_{\eta^\prime}^2 ) }\\
&+ \dfrac{\delta_I m_{\pi}^2 (D+F)  (2 m_{\eta}^2 + m_{\eta^\prime}^2 - 3 m_{\pi}^2 ) (2 m_{\eta}^2 -5 m_{\eta^\prime}^2 + 3 m_{\pi}^2) }{108 f_a (m_a^2 - m_\pi^2) (m_\pi^2 - m_\eta^2) ( m_\pi^2 -m_{\eta^\prime}^2 )}
\end{split}
\end{equation}
in which $\delta_I = (m_d-m_u)/(m_d+m_u)$. The three coefficients are chosen  $D \approx 0.80, F \approx 0.46$ and $D_s \approx -0.41$ from experimental data fitting and Lattice QCD computations~\cite{Blinov:2021say,Alexandrou:2020okk,Close:1993mv,Borasoy:1998pe}. Here we have included the light meson mixing. It should be noted that the ALP-mesons kinetic mixing has been cancelled by the proton-ALP coupling in \cref{eq:alp_p_coupling} hence only ALP-meson mass mixing effects play the role. For $m_a \gg m_{\eta^\prime}$, $g_{pa}$ becomes smaller as $m_a$ increases due to the mixing propagator suppression.

Next we come to compute the ALP production rate at LHC interaction points (IP). In the forward region with low momentum transfer we can use the splitting function to factorize the total cross section into the product of splitting vertex  and the subprocess signal rate. The cross section for radiating one axion can be written as
\begin{equation}
\sigma_{\rm brem} = \int dz d p_T^2 \dfrac{d \mathcal{P}}{d z d p_T^2} \hat{\sigma}_{\rm NSD}(\hat{s})
\end{equation}
in which the non-single diffractive scattering can be expressed as~\cite{Foroughi-Abari:2021zbm}
\begin{equation}
\sigma_{\rm NSD}(\hat{s}) = 1.76 + 19.8 \bigg( \dfrac{s}{\text{GeV}^2}\bigg)^{0.057} \text{mb} 
\end{equation}
In the splitting function, $z$ denotes the energy fraction that is carried by the remnant off-shell proton and $p_T$ is the transverse momentum of the radiated axion. The FPF detector impose a physical cutoff of the polar angle of the axion, which translate into an integration upper limit on $p_T$ that satisfies our approximation. The differential behavior of the splitting function is given by
\begin{equation}
\dfrac{d \mathcal{P}}{dz \, dp_T^2} = \dfrac{1}{16 \pi^2} \dfrac{z}{1-z}\dfrac{1}{(p'^2-m_p^2)^2}  \bigg[ \dfrac{1}{2} \sum |\mathcal{M}|^2\bigg] \bigg[ F_H(p^\prime) F_a(m_a^2) \bigg]^2
\end{equation}
the hadronic form factor $F_H(p^\prime)$ is given by
\begin{equation}
    F_H(p') = \dfrac{\Lambda^4}{\Lambda^4 + (p^{'2} - m_p^2)^2}
\end{equation}
this factor constrain the off-shellness of the intermediate photon. If it deviates from the on-shellness too much, the proton would break up into partons. The second factor is the nucleon form factor which characterize the internal structure of the proton. In principle this factor is composed of the intrinsic part and the contribution from the meson cloud. Since we have discussed on the role of the light pseudoscalar mesons, we only focus on the axion vector part to avoid double counting. The lightest axion vector meson is $a_1(1260)$ with quantum number $I^G(J^{PC}) = 1^-(1^{++})$. We will use the generalized form factor in time-like region as in~\cite{Blinov:2021say}
\begin{equation}
\begin{split}
F_a(k^2)|_{\text{timlike}} =& \dfrac{1}{(1- \gamma e^{i\delta} k^2)^2}
\bigg[ 1 - \alpha + \alpha \dfrac{m_{a_1}^2(m_{a_1}^2 - k^2 + i m_{a_1} \Gamma_{a_1}) }{\left(m_{a_1}^2 - k^2\right)^2 + m_{a_1}^2 \Gamma_{a_1}^2}  \bigg]
\end{split}
\end{equation}
in which now an addition phase factor is introduced. It is found that $\alpha = 0.95, \gamma = 0.515 \text{GeV}^{-2}, \delta = 0.397$ in \cite{Adamuscin:2007fk,Bijker:2004yu}. For the resonance, $m_{a_1} = 1.23$ GeV and its width is $\Gamma_{a_1} = 400$ MeV.
From Ref.~\cite{LiuLyu}, it is convenient to transform from the integrated variables ($p_T,z$) to ($E_a,\sin\theta_a$) in which $E_a$ denotes the energy of the radiated ALP and $\theta_a$ is the polar angle the ALP is emitted relative to the beam direction. 
After being generated at the collision point, the axion will start to decay in propagation. The FPF facility is built at a distance $D$ away from the collision point, with the decay volume length $L$ and available diameter $d$. Eventually, the number of  ALPs decaying in the detector is given by
\begin{equation}
N_d \!=\! \int d\sin\theta_a d E_a  \dfrac{dN_0}{d E_a d\sin\theta_a} \exp\bigg(\!-\!\frac{D m_a}{\sqrt{E_a^2 - m_a^2} c \tau(m_a)} \bigg) \bigg[ 1 - \exp\bigg(\!-\!\frac{L m_a}{\sqrt{E_a^2 - m_a^2} c \tau(m_a)}  \bigg) \bigg]
\end{equation}
in which $c\tau(m_a)$ is the decay length of the ALP. It is heavily dependent on the ALP mass near 1 GeV. 
\begin{figure*}[t]
	\centering
	\includegraphics[width=0.49\textwidth]{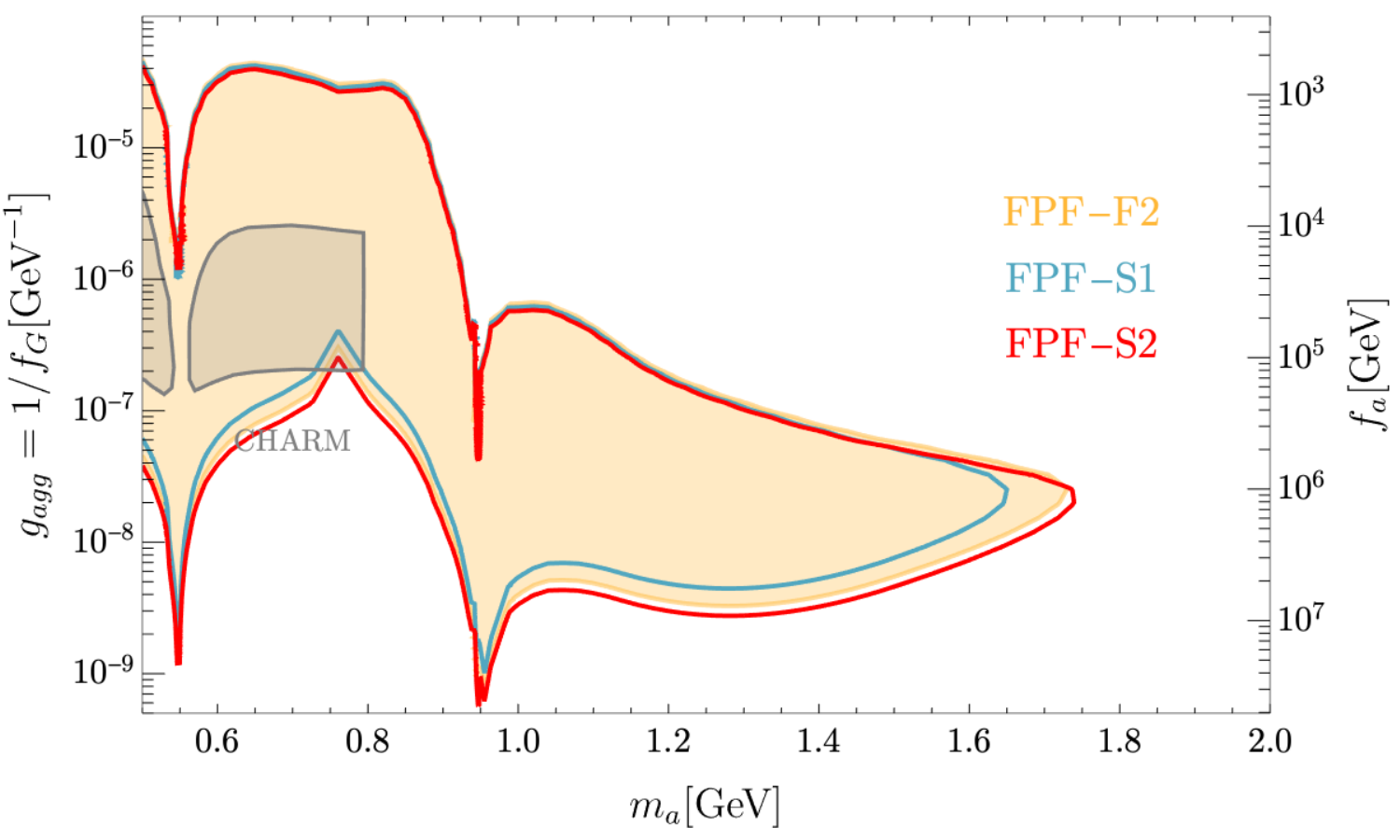}
	\includegraphics[width=0.49\textwidth]{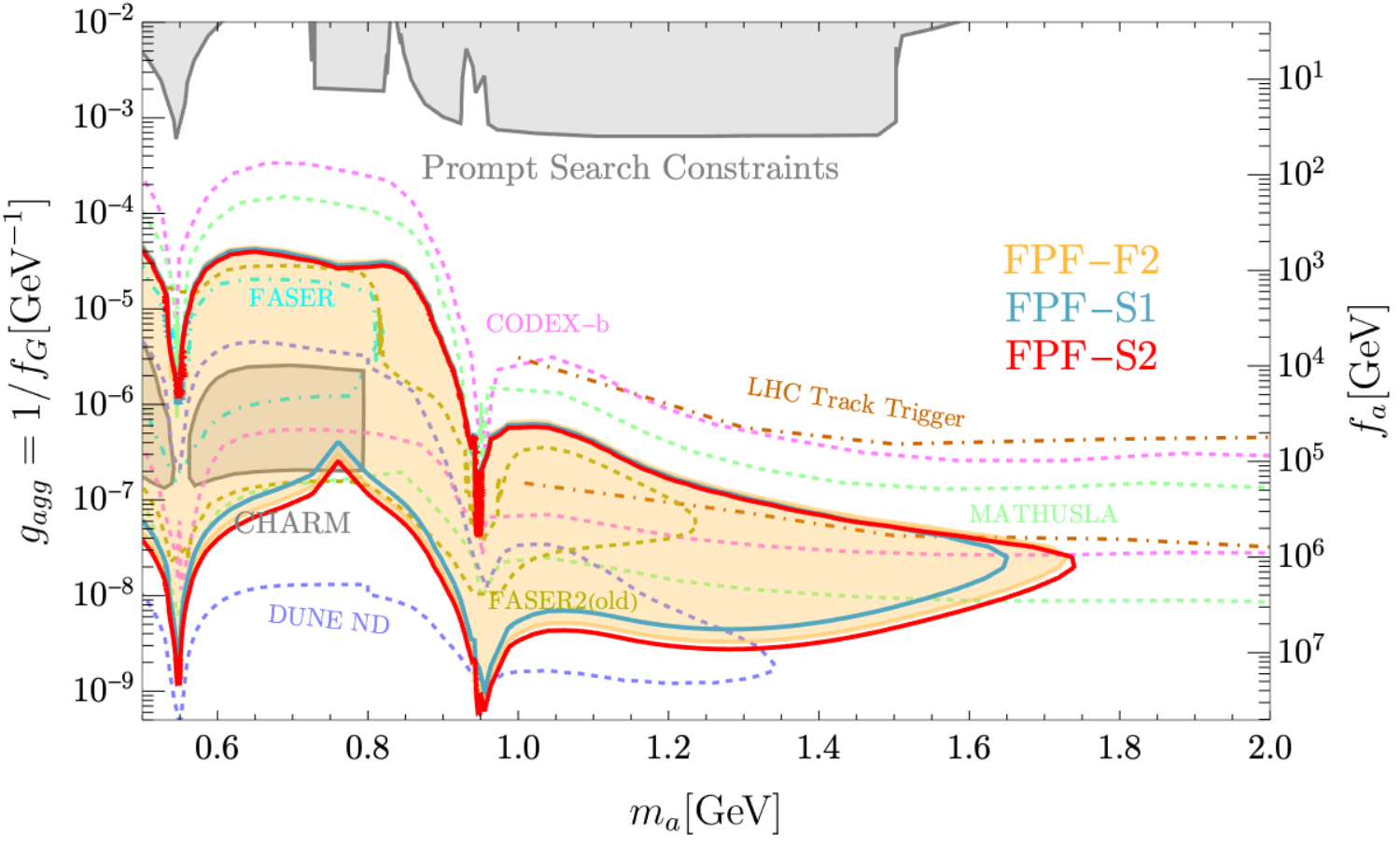}
	\caption{The sensitivity of generic ALPs at FPF using the proton bremming production. Left: Current constraints and FPF projections in the relevant mass-coupling regime. Right: Current constraints, FPF projections, and other future experiment projections in extended mass-coupling plane. Here we choose $\Lambda=1.5$ GeV.
	\label{fig:breamALP}}
\end{figure*}

In \cref{fig:breamALP}, we show how FPF would probe ALPs after including proton-bremming production.  In the left panel, we show the projected ALP sensitivities and existing constraints. We can see the different sensitivities with proposed setups of the FASER2 experiment defined in Sec.III A of \cite{Anchordoqui:2021ghd}: S1, S2, and F2 in Cyan, Red and Orange, respectively. The $\Lambda$ is set to be $1.5$ GeV. We can probe the ALP mass up to around $1.7$ GeV. There are two obvious sharp enhancement position which is due to the resonance mixing with $\eta$ and $\eta^\prime$. For $m_a$ above 1 GeV we can probe $f_a$ as higher as $10^7$ GeV. Varying $\Lambda$ from 1 GeV to 2 GeV only brings small changes. On the right panel, we show various projections from other experiments of this scenario~\cite{Kelly:2020dda} in addition. Note that the previous projection for ALP at FASER2 does not go beyond 1.2~GeV~\cite{Feng:2018pew,Aielli:2019ivi}. Our new result can enhance the probed parametric space of the precious FASER and FASER2. We can see from the figure that proton-bremming production enhances the ALP searches at FPF and makes it more competitive and complementary to other future experimental proposals in the GeV realm.

\clearpage
\section{Long-Lived Particles in Non-Minimal Models}
\label{sec:bsm_llp_nonminmal}

We have so far discussed BSM models which could be probed by searching for a single light new physics portal particle decaying into the SM final states. While the full BSM content of such models could be complicated, it was assumed that it has a minor impact on the FPF phenomenology and, therefore, can be omitted in the analysis. Many realistic BSM scenarios, however, predict the existence of multiple new states that can manifest their existence in more complicated experimental signals to be measured in the FPF. Below, we will discuss examples of such non-minimal models; see also contribution to Snowmass 2021 \textsl{Big Idea: rich dark sectors}~\cite{RF06WP_3} for further discussion.

To this end, we begin with a popular example model of inelastic DM coupled to the SM via the dark vector portal or via the additional dark scalar particle, as discussed in \cref{sec:bsm_nonmin_iDM,sec:bsm_nonmin_iDMdarkHiggs}. In \cref{sec:DDM}, we discuss the search for inelastic DM coupled to the SM via the dipole operator which could be embedded in the framework of Dynamical Dark Matter. We then explore other non-minimal scenarios with a dark gauge boson that is coupled to a dark scalar, cf. \cref{sec:bsm_nonmin_darkgaugeandscalar}, generates charge lepton flavor violation interactions, cf. \cref{sec:bsm_nonmin_clfv}, or can be associated with a new gauge group under which the right-handed SM fermions are charged, cf. \cref{sec:bsm_nonmin_U1T3R}. In \cref{sec:bsm_nonmin_darkaxion}, we discuss the model in which the dark vector species is coupled to the SM via the axion-like portal particle. Another prominent type of non-minimal BSM model contains dark gauge bosons coupled to right-handed neutrinos with an important impact on FPF phenomenology. We provide examples of such scenarios in \cref{sec:bsm_nonmin_BmLZprime,sec:bsm_nonmin_sterilegauge,sec:bsm_nonmin_nuRphilic}. In \cref{sec:bsm_nonmin_secondary,sec:bsm_nonmin_chaindecays} we discuss further experimental consequences of the existence of multiple light degrees of freedom due to the possible secondary LLP production and chain decay. Last but not least, we stress that in non-minimal models FPF searches could also be able to constrain heavy dark sector species with the masses of order tens or even hundreds of GeV. Possible such search for dark bound states is described in \cref{sec:bsm_nonmin_boundstates}. 

\subsection{Inelastic Dark Matter}
\label{sec:bsm_nonmin_iDM}

One example for a light dark sector model that is able to explain the observed dark matter relic abundance while avoiding strong constraints from direct and indirect detection searches is inelastic dark matter (or iDM). In these class of models, the dark sector contains both a stable particle $\chi_1$, which will take the role of dark matter, as well as a nearly degenerate excited state $\chi_2$. These two states are then assumed to couple off-diagonally to a mediator particle, such as a dark photon. This has a variety of phenomenological consequences: i) the dark matter freeze-out mainly proceeds through coannihilations with the heavier state into SM particles; ii) for sizable mass splittings, DM scattering rate at direct detection experiments is kinematically suppressed by the small DM velocity; iii) the heavier state can decay into DM plus SM final states with macroscopic lifetimes, leading to displaced vertex signatures at long-lived particle experiments. 

For concreteness, Let us consider a specific scenario consisting of a Dirac pair of two-component Weyl fermions, $\xi_1$ an $\xi_2$, which are oppositely charged under a broken $U(1)_D$ symmetry. Let us further assume that the associated gauge boson, the dark photon $A'$, couples to the SM photon field via the usual kinetic mixing term. The symmetries of this model allow is to write down a Dirac mass $m_D$. In addition, we introduce a small Majorana mass $m_M \ll m_D$ that softly breaks the $U(1)_D$ symmetry. We can write for the mass term
\be
  - \mathcal{L} \supset   m_D \xi_1 \xi_2 + \frac{1}{2} m_M \xi_1^2  + \frac{1}{2} m_M \xi_2^2 
\ee
where we for simplicity assumed both Majorana masses to be the same. In the limit of small Majorana masses, $m_M \ll m_D$, the mass eigenstates re given by the pseudo-Dirac pair
\be
\chi_{1} = \frac{i}{\sqrt{2}} (\xi_1 - \xi_2)  
\quad \text{and} \quad
\chi_{2} = \frac{1}{\sqrt{2}} (\xi_1 + \xi_2)  
\ee
with mass $m_{1,2} \approx m_D \mp m_M$. Conventionally, one often introduces the relative mass splitting $\Delta = (m2-m1)/m1 = m_M/m_D$. These mass eigenstates now couple off-diagonally to the dark photon, $\mathcal{L}\supset i e_D A'_\mu \bar{\chi}_1 \gamma^\mu \chi_2$. 

\begin{figure*}[t]
\centering
    \includegraphics[width=0.99\textwidth]{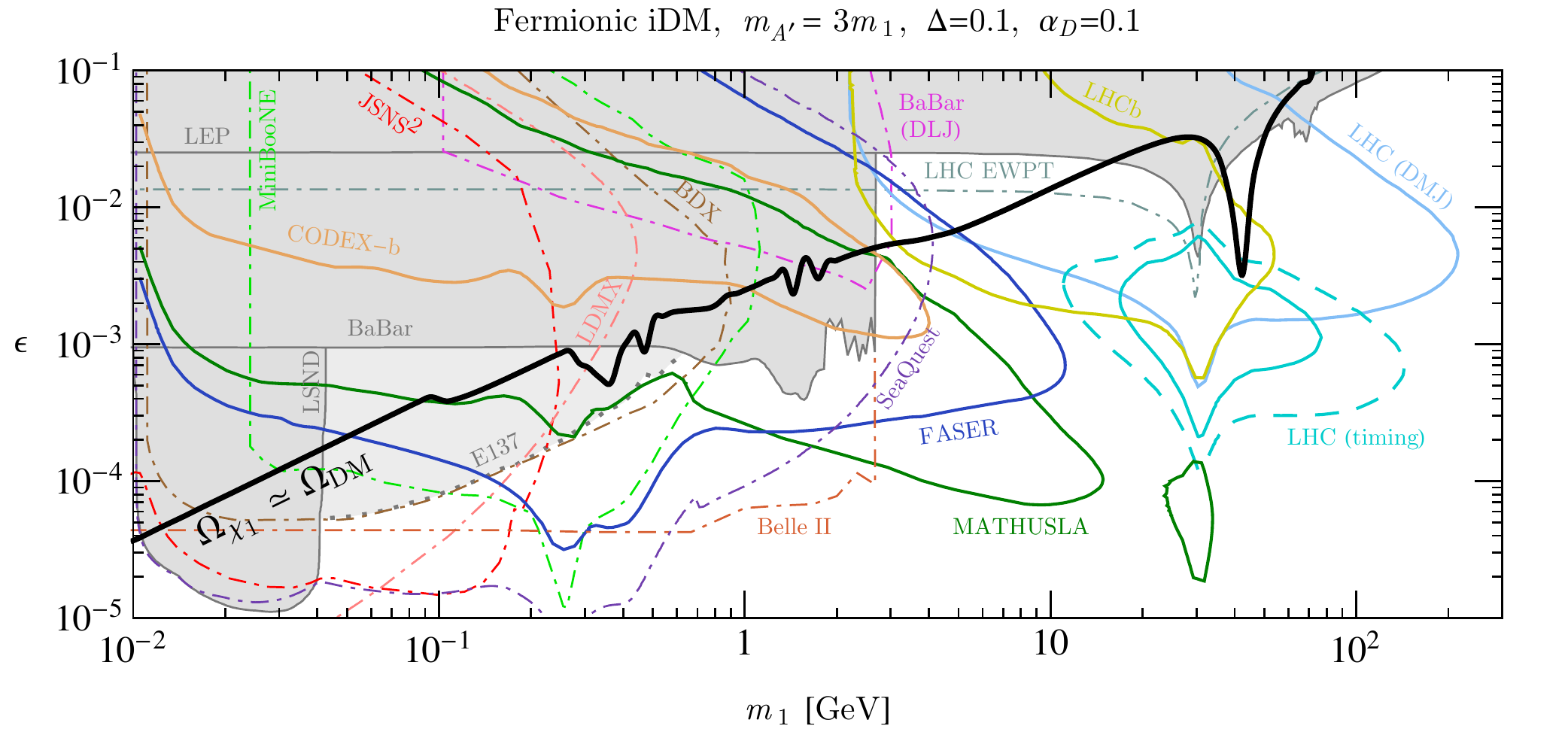}
    \includegraphics[width=0.485\textwidth]{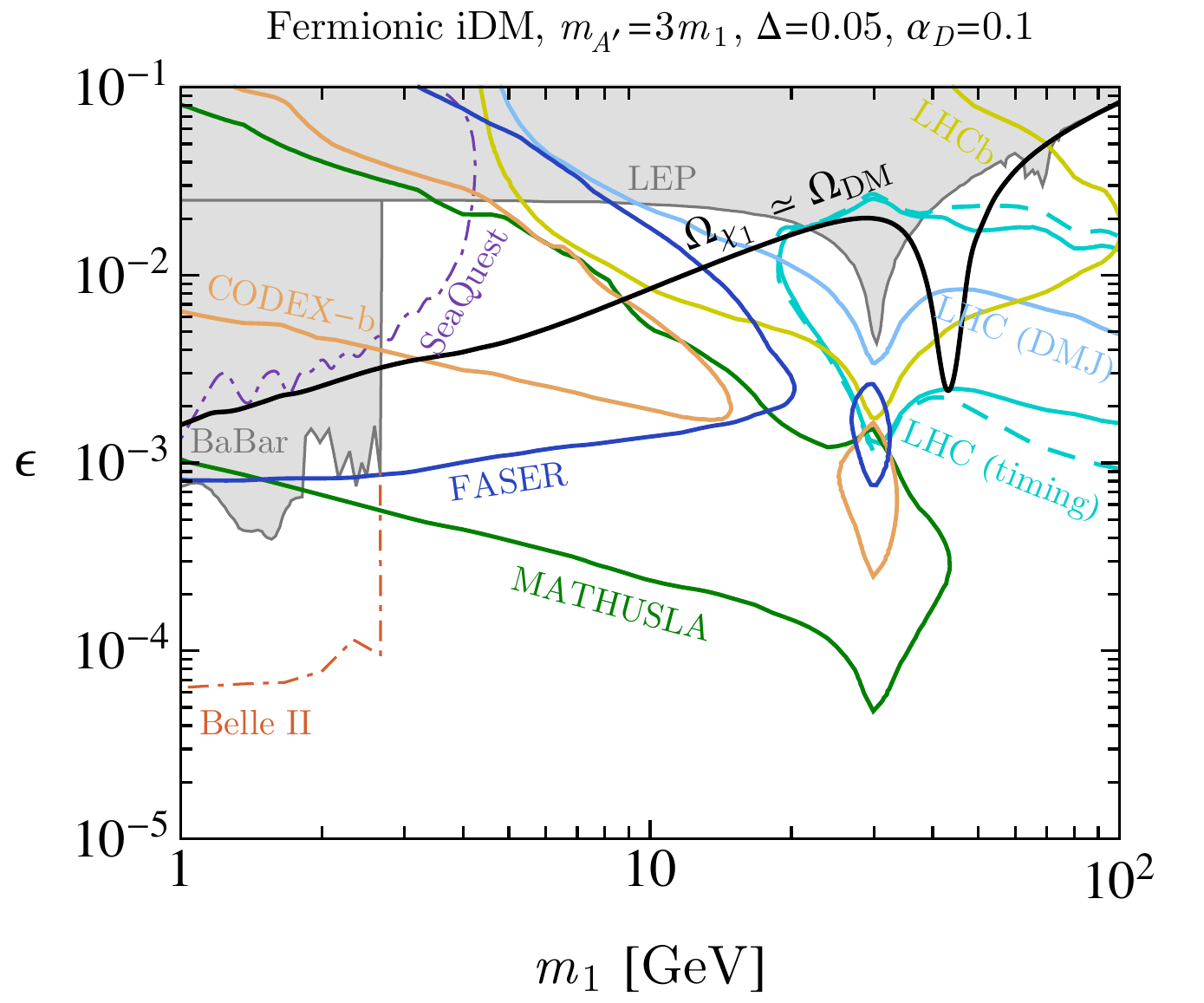}
    \includegraphics[width=0.485\textwidth]{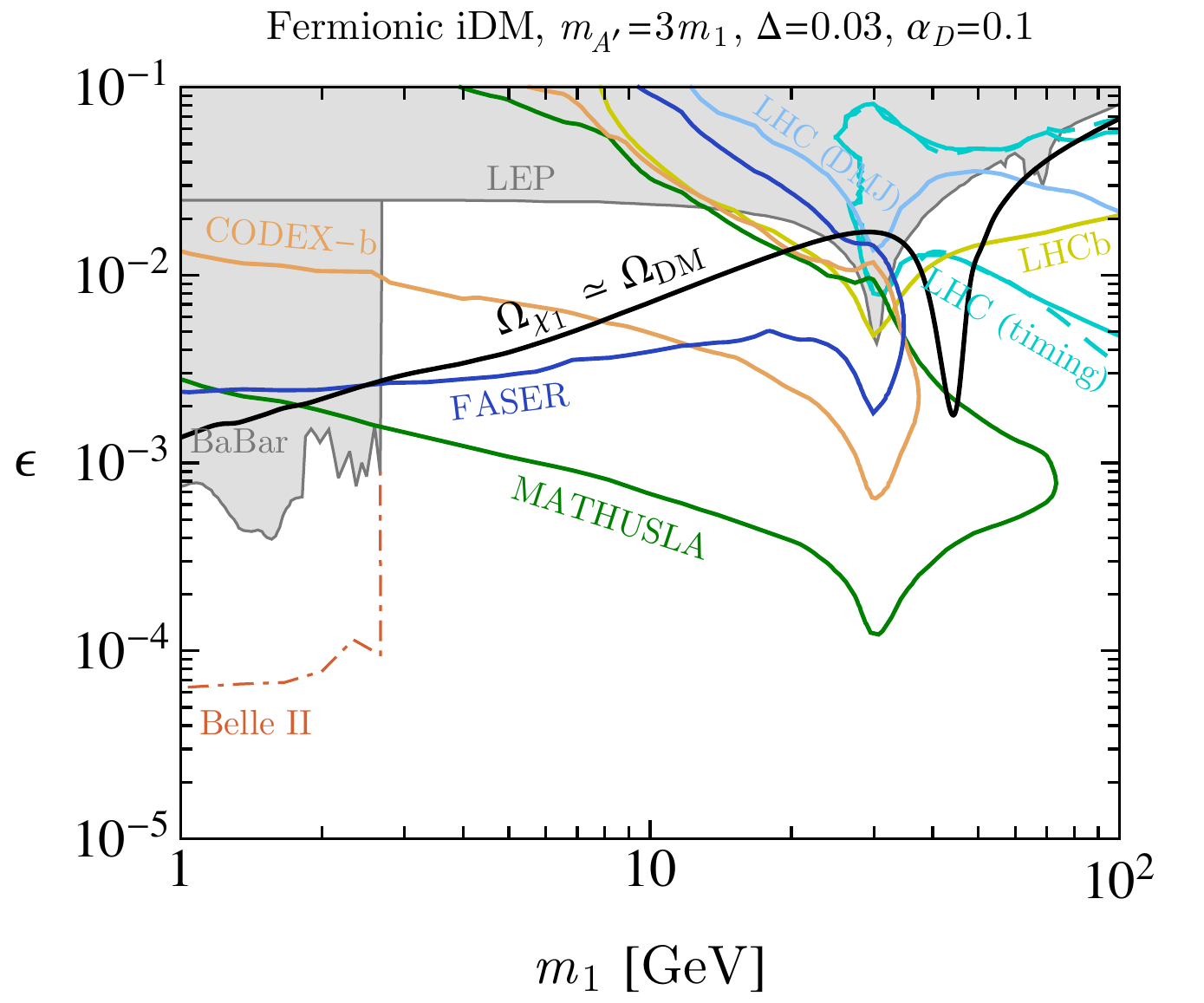}
\caption{Sensitivity reaches of FASER2 (blue line) to inelastic dark matter model for a fixed mass ratio $m_1 = 3 m_{A'}$ and fixed $\alpha_D=0.1$. The upper panel considers the case $\Delta=0.1$ over the a large mass range. Existing constraints are shown as the grey shaded regions alongside projections of other future experiments. The solid black curve indicated the relic target line for this model. The lower panel show similar scenarios for $\Delta=0.05$ and $0.03$, focussing on the high-mass ragion. Figures taken from Ref.~\cite{Berlin:2018jbm}.}
\label{fig:bsm_idm}
\end{figure*}

In the following, we will consider the case of $m_{A'} > m_1+m_2$, in which case $\chi_1$ and $\chi_2$ are be produced via dark photons decay. At the LHC, the dark photon could be produces in a variety of ways as discussed in detail in \cref{sec:bsm_llp_vec}. For the results presented below, we consider its production via meson decays, proton bremsstrahlung and quark antiquark fusion. If kinematically allowed, the heavier state  $\chi_2$ will then decay down into the lighter state $\chi_1$ and a pair of SM leptons via an offshell dark photon. The associated decay width scales as $\Gamma \sim \epsilon^2 e^2 e_D^2 \Delta^5$, and leads to microscope lifetimes even for relatively large values of $\epsilon$.

The projected sensitivity reaches for FASER2 is shown in \cref{fig:bsm_idm} as blue line. The gray shaded regions are excluded from LEP~\cite{Hook:2010tw, Curtin:2014cca}, BaBar~\cite{Lees:2017lec}, LSND~\cite{Auerbach:2001wg, deNiverville:2011it}, E137~\cite{Bjorken:1988as, Batell:2014mga}. The solid colored contours show the projected sensitivity of several other proposed LLP searches and experiments at the LHC, as obtained in Ref.~\cite{Berlin:2018jbm}: displaced vertex searches at ATLAS and CMS~\cite{Izaguirre:2015zva}, precision timing search at CMS~\cite{Liu:2018wte}, displaced vertex searches at LHCb~\cite{Aaij:2017rft, Ilten:2016tkc, Pierce:2017taw}, CODEX-b)~\cite{Gligorov:2017nwh} and MATHUSLA~\cite{Chou:2016lxi}. Also shown as dot-dashed lines are proposed searches for LLPs at Belle II~\cite{Kou:2018nap}, SeaQuest~\cite{Berlin:2018pwi},  BaBar~\cite{Izaguirre:2015zva}, MiniBooNE~\cite{Izaguirre:2017bqb}, JSNS$^2$~\cite{Jordan:2018gcd}, BDX~\cite{Izaguirre:2017bqb}, and LDMX~\cite{Berlin:2018pwi, Berlin:2018bsc, LDMX:2018cma}, as well as  the future sensitivity from electroweak precision tests at the LHC ~\cite{Curtin:2014cca}. The solid black lines correspond to the parameter space where the abundance of $\chi_1$ matches the observed dark matter energy density. It is worth noting that the probed mass range at FASER2 extends up to tens of GeV for the DM mass, which corresponds to about 100~GeV for the dark photon mass, making it one of the scenarios with the highest LLP mass that can be probed at the FPF. See Ref.~\cite{Berlin:2018jbm} for more details.

An important feature of this model is the typically smaller energy of the visible final states $E_\text{vis}$. In most minimal models, such as the dark photon or dark Higgs, the visible final states inherit the full energy of the LLPs: $E_\text{vis} = E_{LLP}$. In contrast, in the iDM scenario most of the energy of the long-lived heavy state $\chi_2$ is transferred to the DM state $\chi_1$ and hence the visible energy is suppressed, and roughly given by $E_\text{vis} \sim \Delta E_{LLP}$. This implies that iDM searches at FASER2 require a reasonable acceptance even for small final state energies.  

\subsection{Inelastic Dark Matter from Dark Higgs Boson Decays\label{sec:bsm_nonmin_iDMdarkHiggs}}

\paragraph{Introduction} Inelastic DM (iDM) is one of the compelling candidates for sub-GeV thermal dark matter \cite{Tucker-Smith:2001myb, Tucker-Smith:2004mxa}, which was originally motivated by the annual modulation reported from the DAMA/LIBRA experiment \cite{DAMA:2008jlt, Bernabei:2013xsa, DAMA:2010gpn, Bernabei:2018yyw}. In the iDM scenario, there exist two dark particles with different masses i.e. the lighter dark matter state and heavier excited state. Elastic interactions of both states are assumed to be absent or much suppressed, and inelastic one mainly occurs in scatterings. Then the DM state inelastically scatters off the Standard Model particles through a mediator and is converted to the excited state, or vice versa. From this interaction nature, iDM can evade constraints from direct detection experiment and residual DM annihilations from the Cosmic Microwave Background. In this section, we consider fermionic and scalar iDM models with dark photon and dark Higgs which is the origin of the dark photon mass. Similarly to the previous section, we take into account the decays of the dark Higgs, and show the sensitivity of the search for these dark matter particles at the FASER2 experiment. Details of this section can be found in Ref.~\cite{Li:2021rzt}.  

\paragraph{Models} We consider two iDM models for a fermion and scalar DM, respectively, with the dark photon of local $U(1)_X$ symmetry. Each dark matter candidate is SM singlet, and denoted as {\it Dirac fermion $\chi$} and {\it complex scalar $S = \frac{1}{\sqrt{2}} (s + i a)$}. Both $\chi$ and $S$ have $U(1)_X$ charge $+\frac{1}{2}$. All the SM particles are assumed to be neutral under the $U(1)_X$ symmetry. The $U(1)_X$ symmetry is spontaneously broken by a SM singlet scalar field $\varphi$ with $U(1)_X$ charge $+1$. Then, the dark photon acquires a mass, and the models have a remnant $Z_2$ symmetry where $\chi/S$ is odd and the other particles are even under it. Furthermore, the dark matter candidates split into two mass eigenstates due to the interaction terms. The Lagrangian of dark sector part is given by
\begin{align}
\mathcal{L}  = \mathcal{L}^{\chi(S)}_{DM} - \frac{1}{4} X^{\mu \nu} X_{\mu \nu} - \frac{\epsilon}{2} B_{\mu \nu} X^{\mu \nu}  + (D^\mu \varphi)^\ast (D_\mu \varphi)  - V, 
\end{align}
where $\mathcal{L}^{\chi(S)}_{DM}$ is the Lagrangian for our fermion(scalar) iDM scenario shown below. The gauge fields of $U(1)_X$ and $U(1)_Y$ are denoted by $X$ and $B$. The scalar potential $V$ for the SM Higgs $H$ and $\varphi$ is given by
\begin{align}
V =  -\mu_H^2 H^\dagger H - \mu^2_\varphi \varphi^\ast \varphi + \frac{\lambda_H}{2} (H^\dagger H)^2 + \frac{\lambda_\varphi}{2} (\varphi^\ast \varphi)^2 + \lambda_{H \varphi} (H^\dagger H)(\varphi^\ast \varphi).
\label{eq:scalar-potential}
\end{align}
In the following discussion, we assume that $\mu_H^2$ and $\mu_\varphi^2$ are positive. The Lagrangians for the fermion and scalar DM candidate are given by
\be
\mathcal{L}^\chi_{DM} &=  \bar{\chi} ( i \slash{D} -M_\chi) \chi + \left(y_L \overline{\chi_L^c} \chi_L \varphi + y_R \overline{\chi_R^c} \chi_R \varphi + h.c.\right), \\
\mathcal{L}^S_{DM} &= (D^\mu S)^\ast (D_\mu S) - M^2_S S^\ast S - \mu (\varphi S^\ast S^\ast + c.c.) \nonumber \\
& \quad - \lambda_S (S^\ast S)^2 - \lambda_{HS} (S^\ast S)(H^\dagger H) 
 - \lambda_{\varphi S} (\varphi^* \varphi)(S^\ast S),
\ee
where the superscript $c$ denotes charge conjugation of a field.

\textit{i) Scalar Boson}: After $H$ and $\varphi$ develop a vacuum expectation value (VEV), $v/\sqrt{2}$ and $v_\varphi/\sqrt{2}$, respectively, the $U(1)_X$ and electroweak symmetries are spontaneously broken. Then, two physical CP-even scalar bosons remain in the spectrum as a mixture of the real parts of $H$ and $\varphi$. Denoting the real parts as $\tilde{h}$ and $\tilde{\phi}$, the CP-even scalar bosons in the mass eigenstate,  $h$ and $\phi$, are expressed as
\begin{align}
\begin{pmatrix} h \\ \phi \end{pmatrix}
&= U \begin{pmatrix} \tilde{h} \\ \tilde{\phi} \end{pmatrix}, 
\label{eq:scalar-eigenstates} 
\end{align}
where the diagonalization matrix $U$ and mixing angle $\alpha$ are defined by
\be
U &=
\begin{pmatrix}
\cos\alpha & -\sin\alpha \\
\sin\alpha & \cos\alpha
\end{pmatrix}
\quad \text{with} \quad 
\tan 2\alpha &= \frac{2 \lambda_{H \varphi} v v_\varphi}{\lambda_H v^2 - \lambda_\varphi v_\varphi^2}. \label{eq:scalar-mixing}
\ee
Note that $h$ is defined to be the SM Higgs boson in the limit of $\alpha \to 0$. The scalar boson $\phi$ can interact with the SM fermions and weak gauge bosons through the mixing. The interaction Lagrangian is 
given by
\begin{align}
\label{eq:int-f-phi}
\mathcal{L}^{\mathrm{SM}}_{\phi\mathrm{-int}} &= \sum_{f} \frac{m_f}{v} \sin \alpha \phi \bar f f
+ \frac{2 m_W^2}{v} \sin \alpha \phi W^+_\mu W^{-\mu}+ \frac{m_Z^2}{v} \sin \alpha \phi Z_\mu Z^\mu ,
\end{align}
where $f$ runs over the SM fermions. The interactions of $\phi$ with the dark matter are given in the following.

\textit{ii) Dark Photon}: After the symmetry breaking, the electrically neutral components of the gauge bosons mix each other through off-diagonal masses and the kinetic mixing while the charged ones remain the same as those of the SM. Assuming $\epsilon \ll 1$, the new gauge field $X$ is approximately identified as mass eigenstate and we denote it as dark photon $A'$ hereafter. The gauge interaction of the dark photon with the SM particles and $\phi$ is given by 
\begin{align}
\label{eq:int-gauge}
\mathcal{L}^{\mathrm{SM}}_{A'\mathrm{-int}} &=
 e \epsilon \cos \theta_W J_{\rm EM}^\mu A'_\mu  + g_X m_{A'} \cos \alpha \phi A'_\mu A'^\mu,
\end{align}
where $\theta_W$ is the Weinberg angle, and $e$ and $J_{\rm EM}^\mu$ are the elementary charge and electromagnetic currents of the SM.

\textit{iii) Dark Matter}: The Dirac fermion $\chi$ splits into two mass eigenstates $\chi_1$ and $\chi_2$ after the symmetry breaking. The mass eigenvalues are given 
\begin{equation}
m_{\chi_1,\chi_2} = \frac{m_L + m_R}{2} \pm \frac12 \sqrt{(m_L - m_R)^2 + 4 M_\chi^2},
\end{equation}
where $m_{L(R)} \equiv \sqrt{2} y_{L(R)} v_\varphi$. In the following, we chose $m_{\chi_1} < m_{\chi_2}$ as convention. Mass eigenstates $\chi_1$ and $\chi_2$ are also given by
\be
\begin{pmatrix} \chi_1 \\ \chi_2 \end{pmatrix} = 
\begin{pmatrix} \cos \theta_\chi & - \sin \theta_\chi \\ \sin \theta_\chi & \cos \theta_\chi \end{pmatrix}
\begin{pmatrix} \chi_L \\ \chi_R^c \end{pmatrix}
\quad \text{with} \quad
\tan 2 \theta_\chi = \frac{2 M_\chi}{m_L - m_R}.
\ee
The gauge interactions among the mass eigenstates and $A'$ can be written by
\begin{align}
\mathcal{L}^{\chi}_{A'\mathrm{-int}}
=  g_X A'_\mu \left[ \cos 2 \theta_\chi (\bar \chi_1 \gamma^\mu \chi_1 - \bar \chi_2 \gamma^\mu \chi_2) +  \sin 2 \theta_\chi  (\bar \chi_1 \gamma^\mu \chi_2 + \bar \chi_2 \gamma^\mu \chi_1) \right]. 
\end{align}
Then $\theta_\chi \simeq \pi/4$, corresponding to $y_L \simeq y_R$, the gauge interactions of the fermions become inelastic. The interaction to $\phi$ and $h$ are 
\begin{align}
\mathcal{L}^{\chi}_{\phi\mathrm{-int}}
= & \frac{1}{\sqrt{2}} y_L (c_\alpha \phi - s_\alpha h)(c_\chi^2 \overline{\chi^c_1} \chi_1 +  c_\chi s_\chi (\overline{\chi^c_1} \chi_2 + \overline{ \chi_1} \chi^c_2) + s^2_\chi \overline{\chi^c_2} \chi_2) \nonumber \\
& + \frac{1}{\sqrt{2}} y_R (c_\alpha \phi - s_\alpha h)(s_\chi^2 \overline{ \chi^c_1} \chi_1 -  c_\chi s_\chi (\overline{\chi^c_1} \chi_2 + \overline{\chi_1} \chi^c_2) + c^2_\chi \overline{\chi^c_2} \chi_2) + h.c.,
\end{align}
where $s_\chi(c_\chi)$ stands for $\sin \theta_{\chi} (\cos \theta_{\chi})$.

In the case of the scalar iDM, the complex scalar $S$ splits into CP-even state $s$ and CP-odd state $a$. These states have different masses given by
\begin{equation}
m^2_{s,a} = \tilde M^2_S \pm \sqrt{2} \mu v_\varphi,
\end{equation}
where $+$ and $-$ in RHS stand for $s$ and $a$ respectively. 
The interaction terms among $A'$ and $s,~a$ are written by
\begin{equation}
\mathcal{L}^{S}_{A'\mathrm{-int}} = \frac{1}{2} g_X A'_\mu (s \partial^\mu a - a \partial^\mu s  ) 
+ \frac{1}{8} g_X^2 A'_\mu A'^\mu (s^2 + a^2).
\end{equation}
The inelatic interaction naturally emerges from the gauge interaction. 
In addition the interactions to the scalar bosons are given from the potential as  
\begin{equation}
\mathcal{L}^{S}_{\phi\mathrm{-int}} =  -\frac{\mu}{\sqrt2} \tilde \phi (s^2 - a^2) - \frac{\lambda_{HS}}{4} (\tilde h^2 + 2 v \tilde h)(s^2+a^2) - \frac{\lambda_{\varphi S}}{4} (\tilde \phi^2 + 2 v_\varphi \tilde \phi)(s^2 +a^2),
\end{equation}
where $\tilde \phi$ and $\tilde h$ are written by the mass eigenstate by \cref{eq:scalar-eigenstates}.

\paragraph{Results} We show the sensitivity plots of the iDM decays at the FASER2 experiment. To illustrate the production of the scalar decays, we consider the following spectra as benchmark point.
\be
1.&~~m_{\chi_1(s)} : m_{\chi_2(a)} : m_{A'} = 1 : 1.2 : 2.1, \\
2.&~~m_{\chi_1(s)} : m_{\chi_2(a)} : m_{A'} = 1 : 1.4(1.3) : 2.3(2.2) ,
\label{eq:spectrum}
\ee
where we chose $m_{\chi_1(s)} + m_{\chi_2(a)} \sim m_{A'}$ to enhance the (co)annihilation cross section. In these spectra, $A'$ dominantly decays into the dark matter $\chi_1$ and $s$, and hence is invisible. It should be noticed that the excited states cannot be produced from the on-shell $A'$ decay in the above two spectra. Therefore the scalar boson decay is the main source of $\chi_2$ and $a$ in the above spectra. \cref{fig:process} shows the sketch of the process for the dark matter signal. 

\begin{figure*}[t]
\centering
\includegraphics[width=0.9\textwidth]{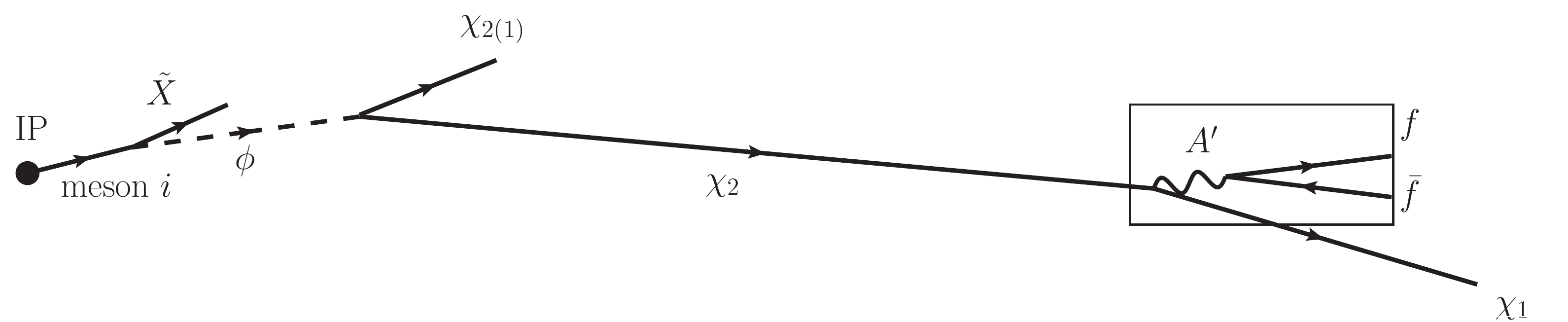}
\caption{Sketch of the production of $\chi_2$ from the decay of the scalar boson, and the subsequent decay into $\chi_1$ and $f$-$\bar{f}$ pair. IP denotes the Interaction Point at the ATLAS detector. }
\label{fig:process}
\end{figure*}

To examine the sensitivity from the scalar boson decay, we adopt the mass relation for $m_{\chi_1(s)}$ and $\phi$, and the scalar mixing angle 
\begin{align}
    m_{\chi_1(s)} : m_\phi = 1: 4
    \quad \text{and} \quad
    \alpha = 10^{-4}.
\end{align}
With these parameters, the expected number of the signal event $\chi_2 \to \chi_1 f \bar{f}$ or $a \to s f \bar{f}$ is calculated by Eq.(33) or (34) of Ref.\cite{Li:2021rzt}, respectively. For the FASER setup, we do not find viable sensitivity region and therefore only show the results for the FASER2 setup. 

\begin{figure*}[t]
\centering
\includegraphics[width=0.48\textwidth]{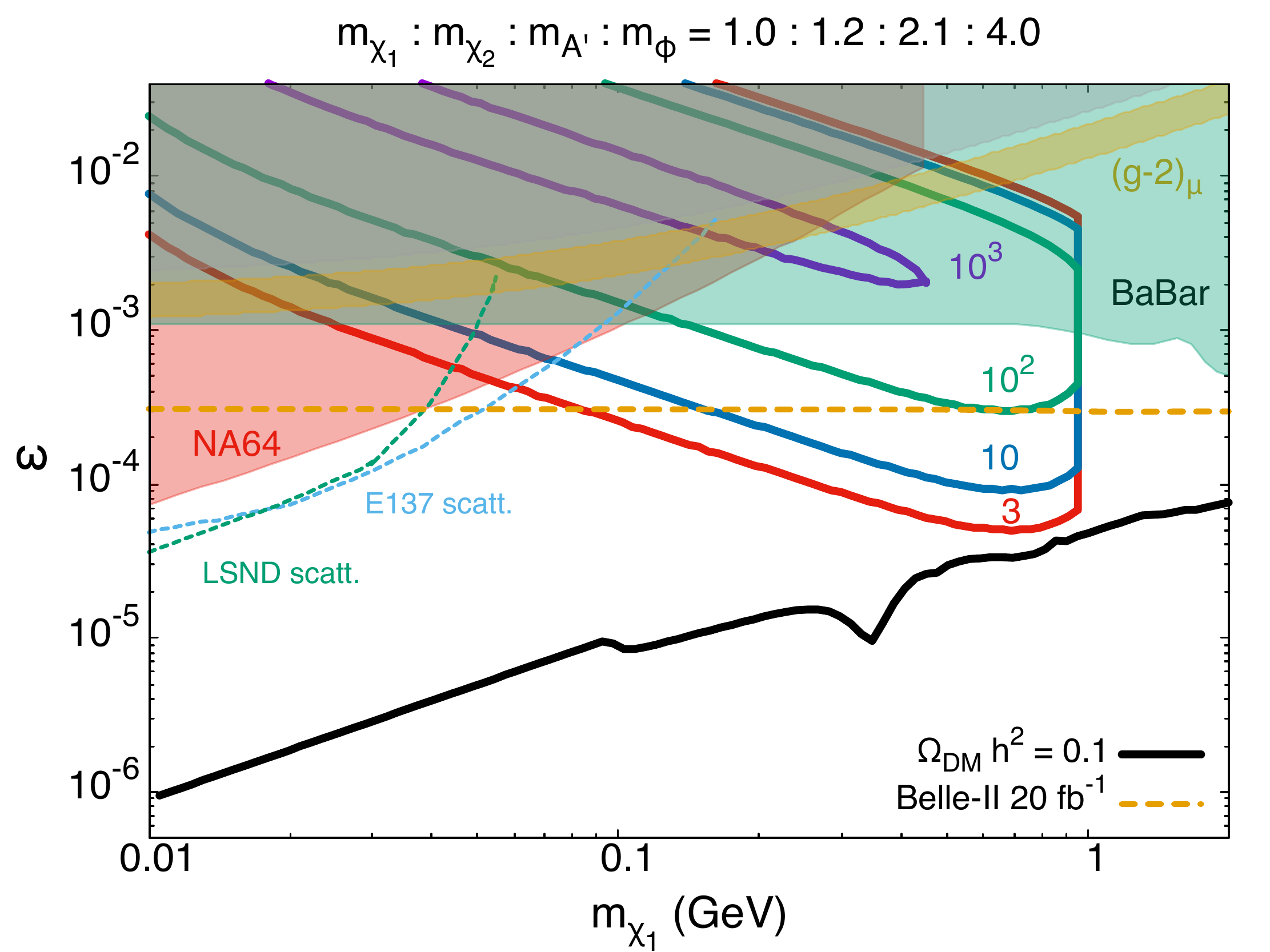} \quad
\includegraphics[width=0.48\textwidth]{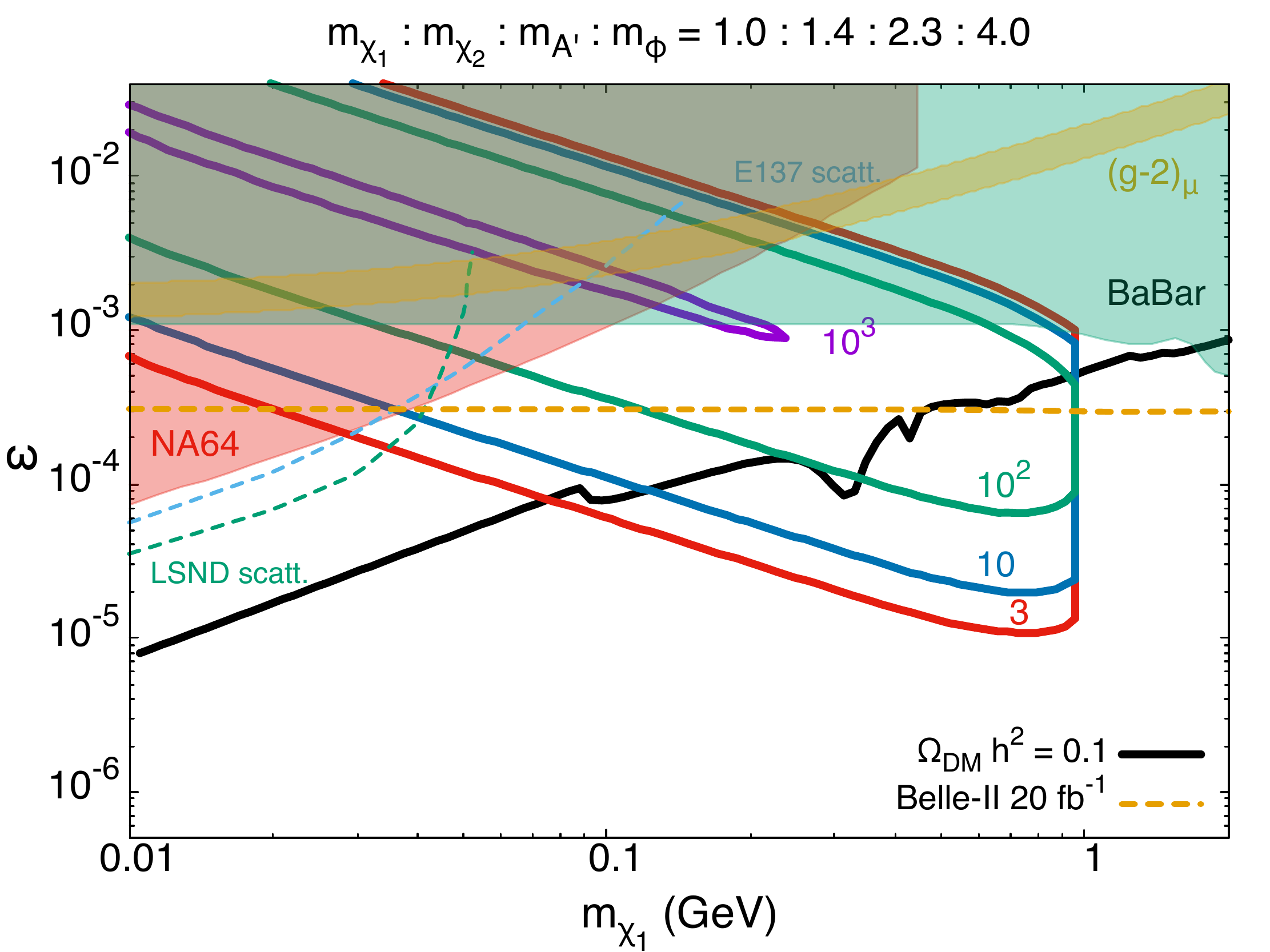}
\caption{
Sensitivity region of the fermion inelastic dark matter decay. See text for the details of curves and colored regions.
}
\label{fig:f-idm}
\end{figure*}

\cref{fig:f-idm} shows contour plots of the sensitivity region at FASER2 for the fermion inelastic dark matter with the mass spectrum 1 (left) and 2 (right) given in \cref{eq:spectrum}. Red, blue, green and purple contours correspond to the expected number of the signal events $3$ (95\% C.L.),$~10,~10^2$ and $10^3$, respectively, and the black one to the relic abundance of the dark matter $\Omega_{\mathrm{DM}} h^2=0.1$~\cite{ParticleDataGroup:2020ssz}. In this case we find that annihilation process $\chi_1 \chi_1 \to A' \to \bar f f$ plays dominant role in relic density calculation. Note that region above(below) black curve correspond to $\Omega_{\mathrm{DM}} h^2 <(>) 0.1$. The filled color regions are excluded by the invisible decay search of the dark photon by NA64 (red) \cite{Banerjee:2019pds} and BaBar (green) \cite{BaBar:2017tiz}, which we rescaled according to our sample spectra, and dashed light blue and green curves are the limit from the E137 \cite{Batell:2014mga} and LSND \cite{LSND:2001akn} for reference\footnote{We rescale this curves from Fig.~6 of Ref.~\cite{Izaguirre:2017bqb}. The spectrum is different from but similar to our spectrum.}. Yellow dashed line is the projection of the sensitivity at Belle-II \cite{Belle-II:2018jsg}. Orange band is the favored region of muon anomalous magnetic moment within $2\sigma$. In both panels, one can see that the small kinetic mixing below the BaBar exclusion region can be explored by the FASER2 experiment. The sensitivity regions at $95$\% C.L. (red curve) reach to $\epsilon \sim \mathcal{O}(10^{-4})$ in case 1 and $\epsilon \sim  \mathcal{O}(10^{-5})$ in case 2. The sensitivity region at $95$\% C.L. (red curve) covers the smaller kinetic mixing below the projection of Belle-II sensitivity. For case 2, the parameter region where $\chi_1$ satisfies the relic dark matter abundance can be examined. Larger kinetic mixing is required to satisfy the observed value of the relic abundance in case 2 than in case 1. This is simply because the number of $\chi_2$ at the freeze-out time of $\chi_1$ is much smaller in case 2 and hence the coannihilation mechanism does not work.

\begin{figure*}[t]
\centering
\includegraphics[width=0.48\textwidth]{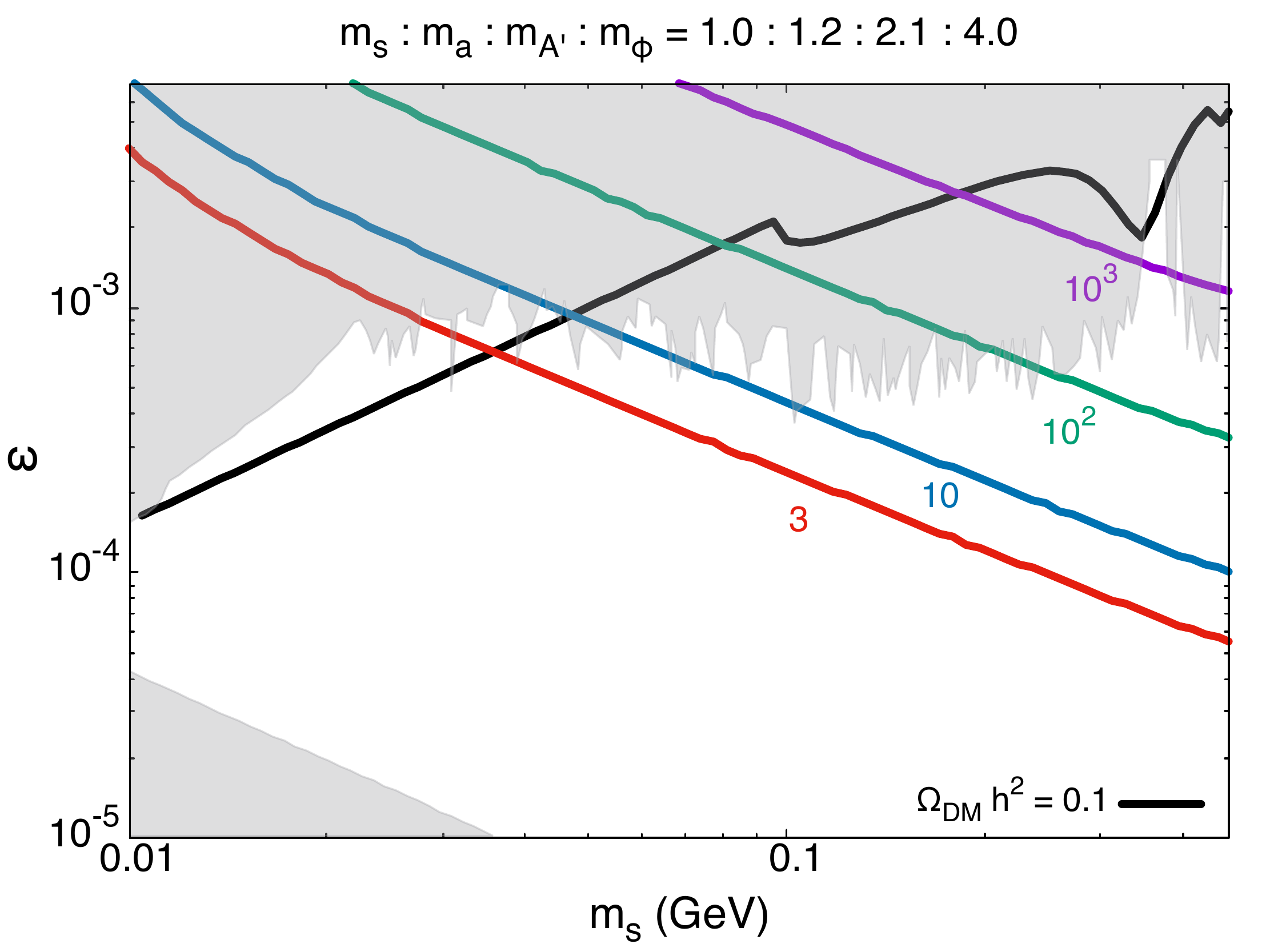} \quad
\includegraphics[width=0.48\textwidth]{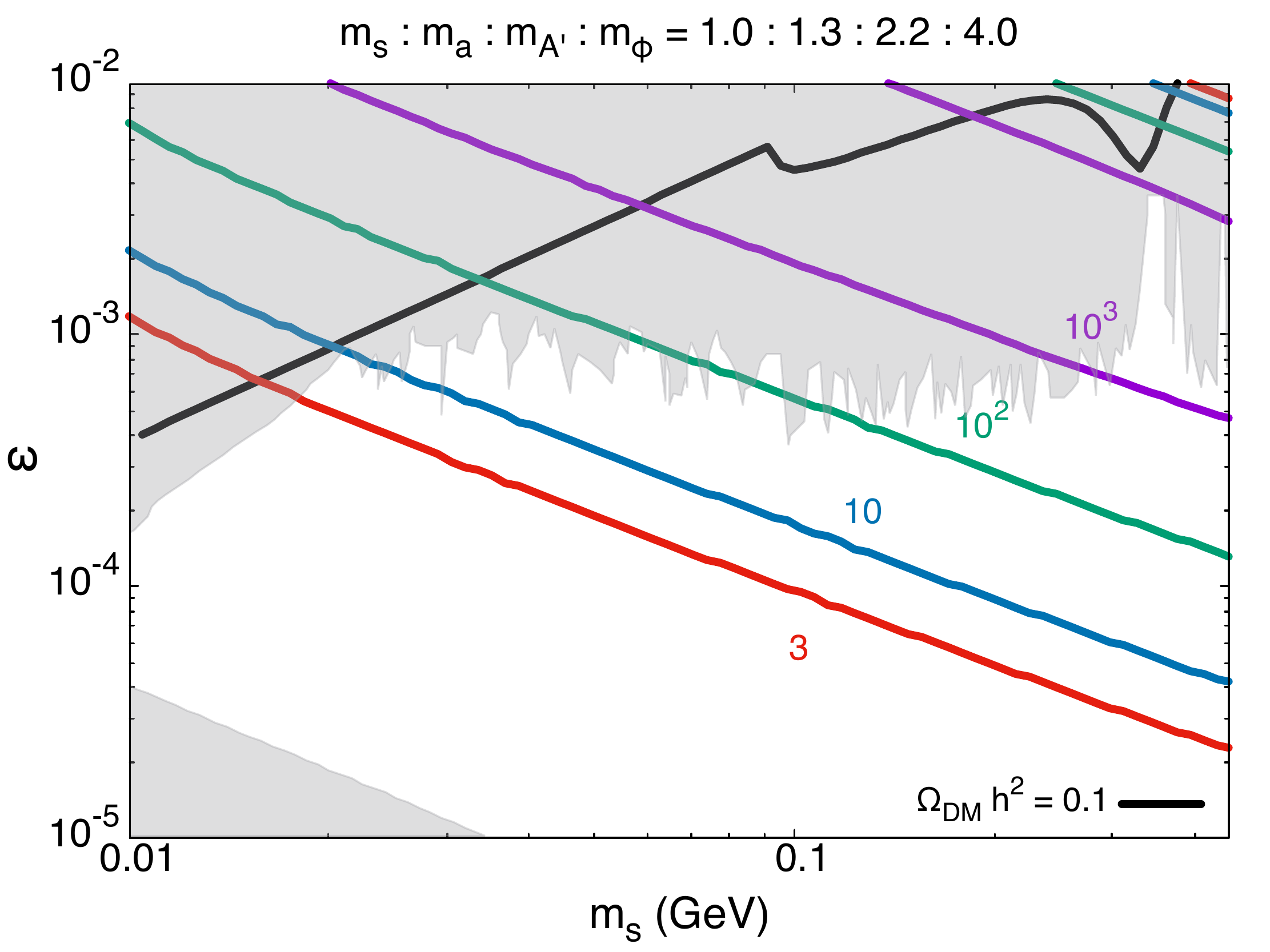}
\caption{The same figure for the scalar DM. The gray regions are taken from \cite{Jodlowski:2019ycu}.
}
\label{fig:s-idm}
\end{figure*}

\cref{fig:s-idm} shows the same plots for the scalar dark matter case. The gray region is exclusion region taken from \cite{Jodlowski:2019ycu}. One can see that most of the parameter region satisfying $\Omega_a h^2 < 0.1$ is already excluded or results in a few signal events. In the scalar inelastic dark matter case, $s$ can annihilate only through the coannihilation mechanism. With the spectrum 1 and 2, the coannihilation mechanism is less efficient, and requires to large kinetic mixing. The curve for $\Omega h^2 =0.1$ can be shifted to lower $\epsilon$ region if we chose mass parameters that is close to resonance $m_{A'} = m_{a} + m_{s}$ and/or smaller $m_a -m_s$. Note that the contours for number of events will shift in larger $\epsilon$ region when we make $m_a-m_s$ smaller since the lifetime of $\chi_2$ becomes longer.

\paragraph{Conclusion} Employing the sample spectra in which the excited dark particles are mainly produced from the decay of the scalar boson, we have analyzed the sensitivity of the signal events at the FASER experiment for the fermion and scalar inelastic dark matter. We found that the scalar boson decays provide a sizable number of the dark particles. We showed that the FASER2 experiment is able to explore unconstrained parameter space in the fermion inelastic dark matter scenario. On the other hand, in the scalar inelastic dark matter scenario, we found that most of the parameter space consistent with the dark matter relic abundance is excluded by existing experiments for our choice of the mass spectra, while FASER2 can still probe unconstrained regions of the parameter space in this scenario.

\subsection{Dynamical Dark Matter\label{sec:DDM}}

The recently constructed FASER~\cite{FASER:2018eoc} experiment and its proposed successor FASER2~\cite{Anchordoqui:2021ghd} at the FPF are well-suited to study LLPs. While many classes of theories can give rise to LLPs, some of the most intriguing are those in which the LLPs exist within a dark sector.  Here we discuss how FASER2 can be useful for probing Dynamical Dark Matter (DDM), a framework for non-minimal dark sectors that naturally includes LLPs and which has already been shown to give rise to a plethora of new signals at colliders. We first discuss the DDM framework, and then present results from a preliminary study of a simplified case to understand the potential of the FPF to probe DDM.

\paragraph{Dynamical Dark Matter: Definition and General Features}

Many models of decaying DM transcend the canonical WIMP or axion frameworks and populate new regions of the DM parameter space.  However, perhaps none do so as dramatically as those that arise within the DDM framework~\cite{Dienes:2011ja}, which posits that the dark matter in the Universe comprises a vast {\it ensemble}\/ of interacting fields with a variety of different masses, lifetimes, and cosmological abundances. Moreover, rather than imposing the stability for each field individually, the DDM framework recognizes that the decay of a DM component in the early Universe is not excluded if the cosmological abundance of that component is sufficiently small at the time of its decay.  The DDM framework therefore posits that those ensemble states with larger masses and SM decay widths have correspondingly smaller relic abundances, and vice versa.   {\it In other words, DM stability is not an absolute requirement in the DDM framework, but is replaced by a balancing of lifetimes against cosmological abundances across the entire ensemble.}\/ For this reason, individual constituents of the DDM ensemble are decaying {\it throughout}\/ the evolution of the Universe, from early times until late times and even today. In general, these decay products can involve SM states as well as other, lighter ensemble components.  DDM is thus a highly dynamical scenario in which cosmological quantities, such as the total DM abundance $\Omega_{\rm DM}$, experience a non-trivial time dependence beyond those normally associated with cosmological expansion.  Moreover, because the DDM ensemble cannot be characterized in terms of a single well-defined mass, decay width, or interaction cross section, the DDM framework gives rise to many striking experimental and observational signatures which transcend those usually associated with dark matter and which ultimately reflect the collective behavior of the entire DDM ensemble.

The DDM framework was originally introduced in Ref.~\cite{Dienes:2011ja}, while in Refs.~\cite{Dienes:2011sa, Dienes:2012jb} explicit models within this framework were constructed which satisfy all known collider, astrophysical, and cosmological constraints. Since then, there has been considerable work in fleshing out this framework and exploring its consequences.  One major direction of research consists of analyzing the various signatures by which the DDM framework might be experimentally tested and constrained. These include unique DDM signatures at direct-detection experiments~\cite{Dienes:2012cf}, at indirect-detection experiments~\cite{Dienes:2013lxa, Boddy:2016fds, Boddy:2016hbp}, and at colliders~\cite{Dienes:2012yz, Dienes:2014bka, Curtin:2018ees, Dienes:2019krh, Dienes:2021cxr}. DDM scenarios can also leave observable imprints across the cosmological timeline, stretching from structure formation~\cite{Dienes:2020bmn, Dienes:2021itb} all the way to late-time supernova recession data~\cite{Desai:2019pvs} and unexpected implications for evaluating Ly-$\alpha$ constraints~\cite{Dienes:2021cxp}. Such dark sectors also give rise to new theoretical possibilities for stable mixed-component cosmological eras~\cite{Dienes:2021woi}. DDM scenarios also give rise to enhanced complementarity relations~\cite{Dienes:2014via, Dienes:2017ylr} between different types of experimental probes.

\paragraph{DDM at Colliders} Models within the DDM framework can give rise to a variety of  distinctive signatures at colliders.  Intra-ensemble decays---decays in which one DDM ensemble constituent decays into a final state involving one or more lighter ensemble constituents---can play an important role in the collider phenomenology of DDM.  Since there may be a large number of transitions between the ensemble of DDM states, there may be a variety of lifetimes. For those with proper decay lengths between 100 m and $10^7$ m, which would typically appear only as $\met$ within the main collider detector wherein they were initially produced, potentially spectacular signals at dedicated LLP detectors such as MATHUSLA~\cite{Chou:2016lxi, Curtin:2018mvb}, FASER~\cite{Feng:2017uoz}, and Codex-b~\cite{Gligorov:2017nwh, Aielli:2019ivi} are possible.  MATHUSLA, for example, is capable of probing regions of DDM parameter space inaccessible to the ATLAS and CMS detectors themselves~\cite{Curtin:2018mvb, Curtin:2018ees}. Moreover, correlating information obtained from LLP searches at MATHUSLA  with information obtained from a variety of searches at the main CMS detector has been shown to yield further insight into the structure of a DDM ensemble and the properties of its constituents. 

\paragraph{Case Study: Inelastic Dipole Interactions} To estimate how well DDM could be probed at experiments at the FPF like FASER2, we start with the simplified benchmark of two dark states $\chi_0$ and $\chi_1$, with a mass splitting $\Delta \equiv m_1 - m_0$. Although there are many inelastic DM interactions one can explore, one interesting possibility is the dimension-5 magnetic dipole operator:
\begin{equation}
{\cal{L}} \supset \mu_{i,j}{\overline{\chi}_i}\sigma^{\mu \nu} \chi_j F_{\mu \nu} + \rm{h.c} \ . 
\label{DDM_inelasticDipoleOperator}
\end{equation}
where $i, j = 0, 1$. One interesting feature of such a minimal model is that diagonal interactions with $i=j$ vanish for Majorana fermions.  These models are therefore very hard to probe in direct detection experiments for sufficient $\Delta$, while for $i \ne j$, the interaction type implies the interesting signature of a single photon. At the LHC, this minimal model gives rise to production and decay processes shown in \cref{fig:DDM_cartoon}.  Large numbers of $\chi_1 \chi_0$ pairs can be produced in the forward direction from $\pi^0$ and $\eta$ decays~\cite{Feng:2017uoz}. Decays $\chi_1 \to \chi_0 \gamma$ then lead to mono-photon signals. Inelastic scattering of $\chi_{0,1}$ off the detector material is also potentially observable~\cite{Izaguirre:2017bqb, Izaguirre:2015zva}. We will focus on the mono-photon signal, and since we focus on production from light meson decays, we study cases where $m_{0,1}< 100$ MeV. A further contribution to the mono-photon signal is the up-scattering of a $\chi_0$ near the detector with subsequent $\chi_1$ decay inside the decay volume~\cite{Jodlowski:2019ycu}. Although we don’t report the rates from this secondary production here, it is expected that it will probe shorter lifetimes, and hence larger $\mu_{0,1}$, than that of prompt $\chi_1$ production.

\begin{figure}[t]
    \centering
    \includegraphics[width=0.95\textwidth]{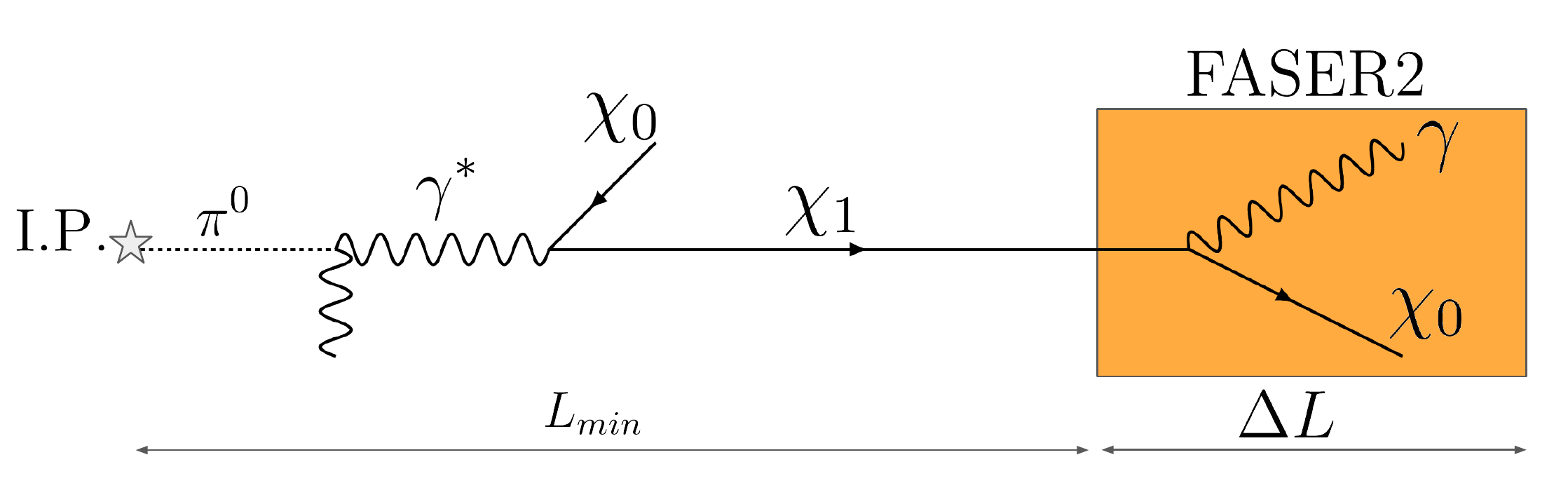}
    \caption{A mono-photon signal inspired by the DDM framework.  A neutral pion is produced and promptly decays through $\pi^0 \to \gamma \overline{\chi}_0 \chi_1 $. The heavier state DDM $\chi_1$ may then free stream a distance $L_{\text{min}}$, and decay inside the decay volume of FASER2 (with length $\Delta L$), producing a photon which can be detected.}
    \label{fig:DDM_cartoon}
\end{figure}

The decay length of the heavier $\chi_1$ state is
\begin{equation}
    {\overline{d}}=\frac{|\vec{p_2}|}{m_1}~\frac{1}{\Gamma_{\chi_1}} \sim \frac{|\vec{p_2}|}{m_1} ~ \frac{1}{\mu_{0,1}^2 ~ \Delta^3} \ ,
    \label{DDM_dbar}
\end{equation}
where $\vec{p_1}$ and $m_1$ are the 3-momenta and mass of $\chi_1$. For $m_1\approx 10$ MeV, and $\Delta\approx 10$ MeV, a decay inside FASER's decay volume implies $\mu_{0,1} \approx 10^{-4}$ GeV$^{-1}$. 

To determine the reach of FASER and FASER2 for this DDM-inspired model, we use the forward $\pi^0$ spectra, which can be found in Ref.~\cite{Kling:2021fwx} (we use the \texttool{EPOS-LHC} configuration), decay the $\pi^0$'s isotropically to produce a spectra of $\chi_1$ and make a cut for $\chi_1$ to intercept FASER2's decay volume. The probability for the heavier $\chi_1$ to decay in the detector is then
\begin{equation}
P_{\rm{decay}}=e^{-L_{\text{min}}\slash \bar{d}} - e^{-(L_{\text{min}}+\Delta L)\slash\bar{d}} \ ,
\end{equation}
where $L_{\text{min}}=620$ m (480 m), and $\Delta L=5$ m (1.5 m) for FASER2 (FASER). We convolute this probability with the $\chi_1$ momentum distribution to determine the number of decays, $N_{\rm{decays}}$, inside FASER2 (FASER) during the HL-LHC era with ${\cal{L}}=3~\rm{ab}^{-1}$ integrated luminosity (Run 3, ${\cal{L}}=150~\rm{fb}^{-1}$). In \cref{fig:DDM_results}, we show our projections at FASER and FASER2 for a fixed $m_0=10 ~\rm{MeV}$. The results show that FASER can probe values of the magnetic dipole operator mass scale as large as 10 TeV, and FASER2 can probe even larger regions of parameter space, given its larger decay volume and the greater luminosity at the HL-LHC.  

\begin{figure}[t]
    \centering
    \includegraphics[width=0.49\textwidth]{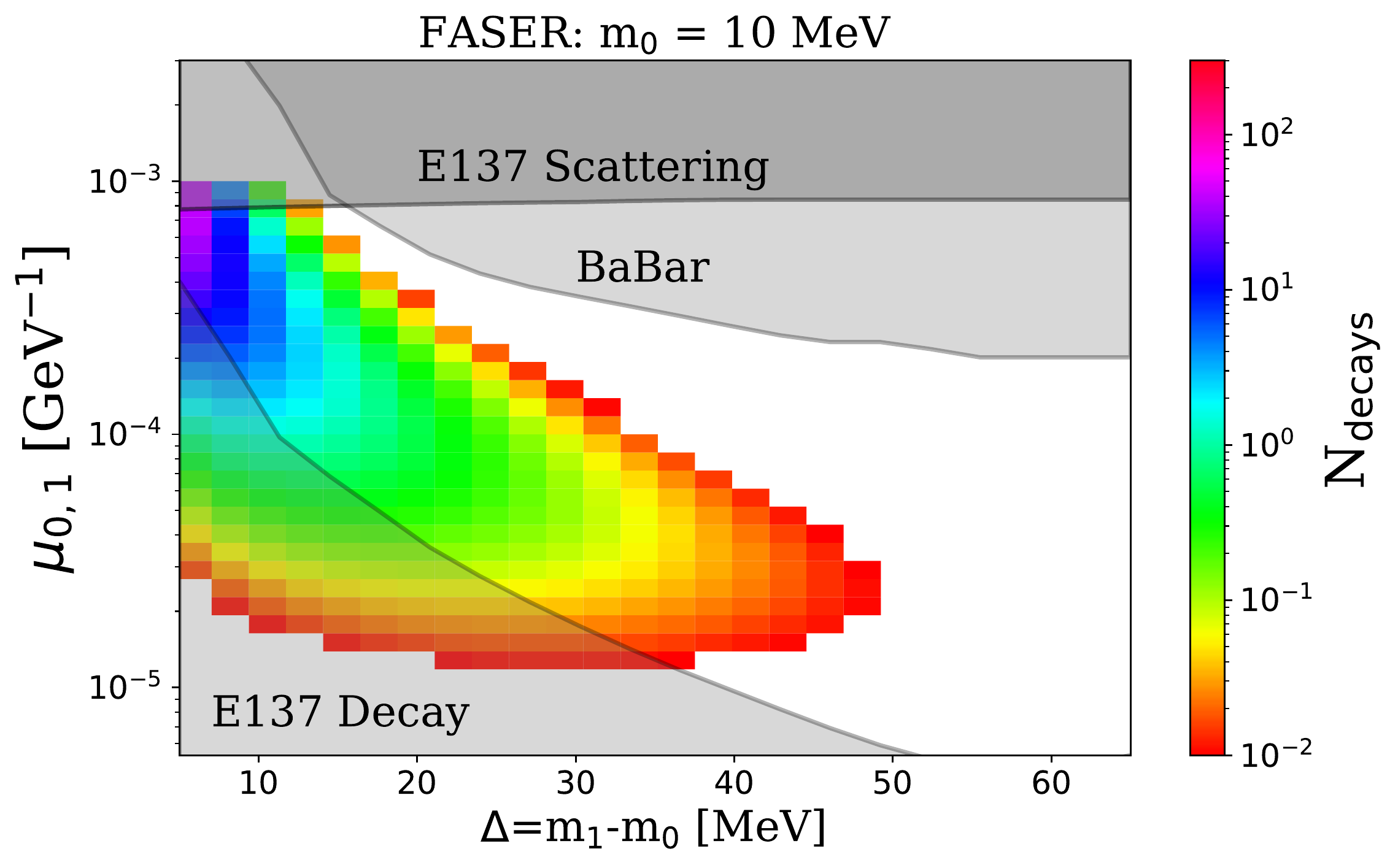}
    \includegraphics[width=0.49\textwidth]{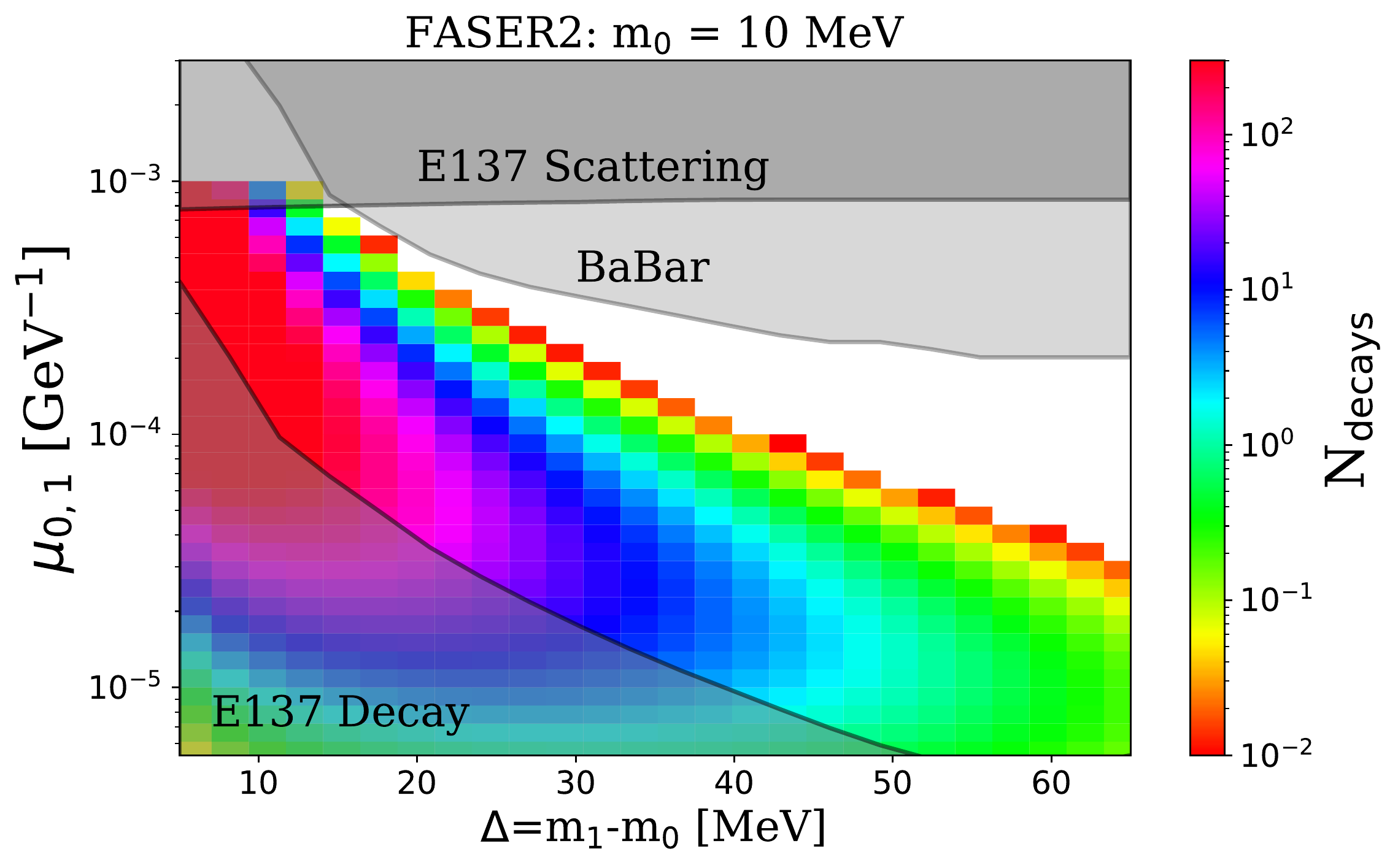}
    \caption{Projections at FASER (left, ${\cal{L}}=150~\rm{fb}^{-1}$) and FASER2 (right, ${\cal{L}}=3~\rm{ab}^{-1}$) for fixed $m_0 = 10~\rm{MeV}$ in coupling $\mu_{0,1}$ and mass splitting $\Delta$ parameter space. Here, 
    $N_{decays}$ is the number of predicted $\chi_1$ decays inside each experiment. Also shown are bounds for E137 (scattering and decay) and BaBar (decay).}
    \label{fig:DDM_results}
\end{figure}

Of course, the reach of FASER and FASER2 in parameter space must be compared to existing bounds.  Beam dump experiments, lepton colliders, indirect detection, and direct detection all impose constraints on the parameter space. The leading current constraints are also shown in \cref{fig:DDM_results}. Here we briefly discuss a few of them.  

Proton beam dump facilities such as LSND~\cite{LSND:2001akn} and MiniBooNE~\cite{Katori:2007uf} produce large pion fluxes, and thus can produce $\chi$ pairs in large numbers \cite{Chu:2020ysb}. LSND and MiniBooNE produced a total of $10^{22}~\rm{and}~10^{20}~ \pi^0$'s. Using this $\pi^0$ flux, we can determine the $\chi$ flux by calculating the branching ratio. For $m_0,\Delta \sim {\cal{O}}(10~\rm{MeV})$, we find that the branching ratio for $\pi^0$ decay is given by
\begin{equation}
{\rm{BR(}}\pi^0\rightarrow \gamma {\overline{\chi}_0} \chi_1{\rm{)}}=\frac{\mu_{0,1}^2}{2\pi ~}\frac{(m_1^2-m_0^2)^3}{m_1^3} \frac{1}{\Gamma_{\pi^0}} \sim \frac{\mu_{0,1}^2}{\rm{GeV}^{-2}}\times 10^{-4} \ ,
\label{DDM_pi0BR}
\end{equation}
where $\Gamma_{\pi^0}\approx 7.8~\rm{eV}$ is the total decay with of the pion. For decay signals, both LSND and MiniBooNE probe $\mu_{0,1}$ values that are an order of magnitude below FASER2's region of interest. This can be seen by looking at the decay length at each facility. The couplings probed scale as $\mu_{0,1}^2\propto \frac{E}{L}$, but the $\chi$ states produced at the LHC have energies $\sim 10^3$ ($\sim 10^2$) times larger than those at LSND (MiniBooNE), while the decay lengths probed by FASER2 are only $\sim 10$ ($\sim 1$) times longer. Scattering signals on the other hand constrain $\mu_{0,1} \approx 10^{-3}$. The resulting bonds are not shown in \cref{fig:DDM_results}, as they are sub-leading compared to other current bounds that we now discuss.

Electron beam dump experiments can probe these models through $e^- ~N \to e^- N\gamma^*\to e^- N {\overline{\chi}_0} \chi_1$, where $N$ is the nuclear target~\cite{Izaguirre:2017bqb, Batell:2014mga}. One of the leading electron beam dump experiments constraining sub-GeV DM is E137~\cite{Bjorken:1988as}, where the possible signals include DM scattering off the detector or the decay $\chi_1 \to \chi_0 \gamma $ producing a photon that can be seen in the detector.   Following Ref.~\cite{Batell:2014mga} and implementing this model in \texttool{Madgraph~5}~\cite{Alwall:2011uj} using \texttool{FeynRules}~\cite{Alloul:2013bka}, one can obtain the $\chi$ production rate for scattering off an aluminum target and use the null results of E137 to place a bound on $\mu_{0,1}$. While the decay length at E137 is of the same order as at FASER2, the typical $\chi$ energies are a factor of 100 lower, and so E137 constrains smaller $\mu_{0,1}$. 

Lepton colliders provide a low-background environment to constrain DM. In our model, $\chi$ pairs can be produced via $s$- or $t$-channel $e^+ e^-$ annihilation and can be probed at high-luminosity $B$ factories. In Ref.~\cite{Izaguirre:2015zva}, the authors use the BaBar experiment \cite{BaBar:1995bns} to constrain inelastic GeV-scale DM with magnetic dipole interactions. Here the authors perform a missing energy search using the monophoton trigger which was implemented at BaBar for a subset of the dataset ($\approx 60~\rm{fb}^{-1}$)~\cite{BaBar:2015jvu}. We follow their analysis and obtain the $\chi_1$ momentum distribution, and select events that would pass the monophoton trigger ($E_{\gamma}>2~\gev$). We find that BaBar constrains larger $\mu_{0,1}$ than those probed by FASER2. While BaBar does not cover our region of interest, there may be other lepton colliders such as Belle II which will be relevant for our parameter space~\cite{Chu:2018qrm}.

This model is also constrained by astroparticle searches.  In particular, $\chi_0$ pair annihilations into monochromatic photons can be observed. The most stringent are line searches at gamma-ray telescopes such as Fermi-LAT~\cite{Fermi-LAT:2009ihh}. To determine the bounds from indirect detection, we estimate the thermally-averaged cross section as
\begin{equation}
\sigma v \approx \mu_{0,1}^4 \, m_{\chi}^2 \, v \ .
\label{DDM_ID}
\end{equation}
For DM velocities today, and FASER2's region of interest ($\mu_{0,1}=10^{-4} ~\gev^{-1}$), \cref{DDM_ID} implies cross sections of $10^{-41}~\rm{cm}^3~\rm{s^{-1}}$, approximately 7 orders of magnitude below the existing bounds found in Ref.~\cite{Bartels:2017dpb}. Bounds on higher-dimensional operators from indirect searches have also been analyzed in Ref.~\cite{Kavanagh:2018xeh} and their results lead to similar conclusions.

Since we are focusing on the sub-GeV regime, direct detection experiments via nuclear recoils are also not effective due to small recoil energies. Instead, electron recoils are often exploited in direct-detection experiments (for a review, see, for example, Ref.~\cite{Lin:2019uvt} and references therein) as the energy threshold is much lower. For the magnetic dipole interaction in \cref{DDM_inelasticDipoleOperator}, relevant current bounds in the limit $\Delta \rightarrow 0$ can be found in Refs.~\cite{Kavanagh:2018xeh, Catena:2019gfa, Catena:2021qsr}. However, in the case where $\chi_i$ are Majorana fermions, the interaction in \cref{DDM_inelasticDipoleOperator} is strictly off-diagonal, which means $\Delta \neq 0$, and electron recoils can only occur if the incoming ground-state particle $\chi_0$ is energetic enough to upscatter and create a $\chi_1$. 

Since the kinetic energy of $\chi_0$ has to be at least larger than $\Delta$ for an electron recoil to occur, a quick estimate yields 
\begin{equation}
    E_{k,1}\approx \frac{1}{2}m_0v^2=5 \times \left(\frac{m_0}{10~\rm MeV}\right)\left(\frac{v}{10^{-3}}\right)^2 ~{\rm eV}.
\end{equation}
Thus, for typical dark-matter velocities in our galaxy, $v\sim 10^{-3}$, or even the maximum possible velocity in the lab frame where the escape velocity combined with the motion of the Earth and the Sun gives $v_{\rm max}\sim 10^{-2}$, the kinetic energy of $\chi_0$ is only $\mathcal{O}(5-50)~\rm eV$. Therefore, as long as $\Delta$ is a few or a few tens of $\rm eV$, constraints from current direct-detection experiments can be significantly weakened. 

Given all of these considerations, the leading competitive bounds are those shown in \cref{fig:DDM_results}, with decay (scattering) signals at E137 constraining smaller (larger) couplings and BaBar constraining larger couplings.  Bounds from decays at LSND and MiniBooNE were found to be subleading to the E137 decay bounds and are not shown, while the scattering bounds from the proton beam dumps were found to be comparable to E137.

In summary, although there is still work to do, FASER and FASER2 appear to probe new parameter space in this model, and the promising reach of FASER and FASER2 in this simplified 2-state inelastic DM model demonstrates that a full DDM framework with a tower of states may be probed at the FPF. While we do not consider other experiments proposed at the FPF in this section, it is expected that FLArE will be competitive with FASER2 for our model. Furthermore, there are other proposed and upcoming experiments (LDMX, Belle II, FORMOSA, etc.) that merit further investigation.

\subsection{Light Dark Scalars through $Z'$ and EFTs}
\label{sec:bsm_nonmin_darkgaugeandscalar}

\paragraph{Introduction} Dark sectors, consisting of new light particles at the GeV scale or lower and interacting feebly with the Standard Model (SM), have gained considerable attention in the recent years. They could contain dark matter candidates, and they might be connected to the solutions of some open questions of particle physics. An intense and diverse experimental programme is underway to target these scenarios. In this section, based on Ref.~\cite{Bertuzzo:2020rzo}, we will consider the phenomenology of a dark sector containing two non-degenerate light real scalar particles denoted by $\phi_1$ and $\phi_2$. The mass splitting will be quantified by the relative difference
\be
\delta = \frac{m_2 - m_1}{m_1},
\ee
where $m_2$ and $m_1$ are, respectively, the masses of the heaviest and lightest state. We can always imagine such non-degenerate states to emerge from a complex scalar $\phi = (\phi_1 + i \phi_2)/\sqrt{2}$ once a $B \phi^2 + h.c$ term is added to the Lagrangian. Since this term breaks the $U(1)$ symmetry associated with $\phi$ phase rotations, it is technically natural to have small values for $\delta$. The heaviest scalar $\phi_2$ can thus be long-lived, and can be searched for at beam dump or LHC experiments far away from the interaction point.

The light scalars under consideration will in general have Higgs-portal interactions. Since such interactions have been thoroughly studied in the literature (see e.g.~\cite{Arcadi:2019lka}), we will assume that they are negligible, and we focus instead on a situation in which the portal between the dark and visible sectors is given by some heavy particle\footnote{The phenomenology of light mediators has also been extensively studied in the literature, see for instance~\cite{Curtin:2018mvb, Feng:2017uoz, SHiP:2020noy, Berlin:2020uwy, Berlin:2018jbm, Berlin:2018bsc, Berlin:2018pwi}.}. More precisely, we will focus on two cases:
\begin{itemize}
	\item The dimension-6 Effective Field Theory (EFT) Lagrangian
	\be
	\mathcal{L}_{EFT} = \frac{1}{\Lambda^2} \bigg( (\partial^\mu \phi_2) \phi_1 - \phi_2 (\partial^\mu \phi_1) \bigg) \left( \sum_{f_L} c_{f_L} \bar{f}_L \gamma_\mu f_L + \sum_{f_R} c_{f_R} \bar{f}_R \gamma_\mu f_R \right)+ \dots
	\label{eq:EBMT_EFT}
	\ee 
	where $f_L$ and $f_R$ are the left handed and right handed SM fermions and the dots represent higher order operators. These effective operators can be generated integrating out a heavy vector boson, or heavy vector-like fermions~\cite{Bertuzzo:2020rzo}. In our analysis we will consider, for definiteness, the case in which $c_{f_L} = c_{f_R}$ in \cref{eq:EBMT_EFT}, i.e. the case of a purely vector current without axial component. The only exception will be given by neutrinos, that will interact via the usual V-A current;
	\item A simple UV completion for the EFT defined in \cref{eq:EBMT_EFT} in terms of a $Z^{\prime}$ mediator.
	More concretely we will focus on the case of a dark photon, 
	interacting with the SM  via the kinetic mixing~\cite{Holdom:1985ag,Galison:1983pa,Foot:1991kb}:
    \be
    \label{eq:EBMT_DP}
    \mathcal{L} \supset \frac{\epsilon}{2\,c_w} F^{\prime}_{\mu\nu} B_{\mu\nu} \ee 
    In the previous formula, $B_{\mu\nu}$ and $F^{\prime}_{\mu\nu}$ are the field strengths of the hypercharge boson and dark photon, respectively, while $c_w$ is the cosine of the weak angle. The dimensionless parameter $\epsilon$ is the kinetic mixing parameter. After transforming into the physical basis, $\epsilon$ controls the interactions between the dark photon and the SM states. Keeping only the leading contributions in $\epsilon \ll 1$ and $M_{Z'}/M_Z$ (where $M_{Z'}$ and $M_Z$ are the dark photon and $Z$ boson masses, respectively), and integrating out the $Z^{\prime}$, the model matches the EFT in \cref{eq:EBMT_EFT} with $\Lambda=M_{Z'}$ and
    \be
    c_{f_L} = c_{f_R} = g_{\phi}\, \epsilon\, e\, Q_f ,
    \ee
    with $e Q_f$ the electric charge of the fermion, and $g_{\phi}$ the coupling of the $Z^{\prime}$ to the dark current $ (\partial^\mu \phi_2) \phi_1 - \phi_2 (\partial^\mu \phi_1)$. 
\end{itemize}

Our focus are past experiments and future proposals which can look for the decays of the long-lived particle $\phi_2.$ More specifically, we consider the past beam-dump experiment CHARM and the future facility SHiP, both characterized by a center-of-mass energy of $\sqrt{s} \simeq 28$ GeV, and the future FASER and MATHUSLA  detectors, to be placed near the LHC interactions points ($\sqrt{s} = 14$ TeV).In all cases, the processes of interests are
\be
p + {\rm target} \to \phi_1 + \phi_2~~~~~ {\rm followed~by} ~~~~~~ \phi_2 \to \phi_1 + {\rm SM} ,
\ee
with the decay happening inside the detector. For all experiments, the number of signal events can schematically be determined as the product between the total number of $\phi_2$ particles produced in a given experiment, $N_{\phi_2}$, and the fraction of those events which decay inside the detector (taking also into account the experimental efficiency for the reconstruction of the signal), $f_{\rm dec}$. For $N_{\phi_2}$ we consider two main contributions: the production of $\phi_1\phi_2$ pairs from meson decays and from parton level processes. The first is obtained simulating $pp$ collisions using \texttool{EPOS-LHC}~\cite{Pierog:2013ria} and \texttool{Pythia}~\cite{Sjostrand:2007gs} to produce a sample $\pi_0$, $\eta$, $\eta'$, $\rho$, $\omega$, J/$\psi$ and $\Upsilon$ mesons, which are then decayed into channels containing $\phi_1 \phi_2$ pairs. For the second production mechanism, we employ \texttool{MadGraph5\_aMC@NLO}~\cite{Alwall:2014hca}. The fraction of $\phi_2$ states which decay inside the detector, $f_{\rm dec}$, can be computed from the sample of $\phi_2$ events produced in the simulations mentioned above. For this purpose, we compute the $\phi_2$ lifetime and we model the geometry of the different detectors considered in the analysis to estimate the probability that $\phi_2$ states decay in the detector volume. Finally, the sensitivity reach of the different experiments is computed under the assumption, justified by present studies, that backgrounds can be reduced to a negligible level. More details can be found in Ref.~\cite{Bertuzzo:2020rzo}.

Before discussing the main results of our analysis, we shall mention an important aspect related to validity of the EFT. An EFT properly describes only those physical processes occurring at energy scales smaller than its cut-off $M_{\rm cut}.$ Therefore, to ensure that the EFT in \cref{eq:EBMT_EFT} is used inside its domain of validity, in our analysis we require $\sqrt{\hat{s}} \leq M_{\rm cut}$, with $\hat{s}$ the center of mass energy of the partonic event. In practice, from the sample of events simulated with \texttool{MadGraph5\_aMC@NLO}, we select only those satisfying this condition. The cut-off can be written as $M_{\rm cut} = g_* \Lambda$, where $g_*$ is a combination of couplings of the UV theory, that we take to be $g_* = 1$. Of course, this procedure is not needed when we consider the dark photon model. Finally, we do not consider direct parton production in the context of the EFT in \cref{eq:EBMT_EFT} for experiments with small angular acceptance, i.e. CHARM, SHiP and FASER. This is because, in this context, the production of $\phi_1\phi_2$ corresponds to events with small momentum transfer, for which the description in perturbative QCD is affected by considerable uncertainties. Instead, the situation is different when the mediator of the interaction can be produced resonantly, as for the case of the $Z^{\prime}$ model. Therefore, in the case of FASER and for the dark photon model,  we include the $\phi_1\phi_2$ production from parton level processes, extending the results of~\cite{Bertuzzo:2020rzo}.

\begin{figure*}[t]
\centering
	\includegraphics[width=0.485\textwidth]{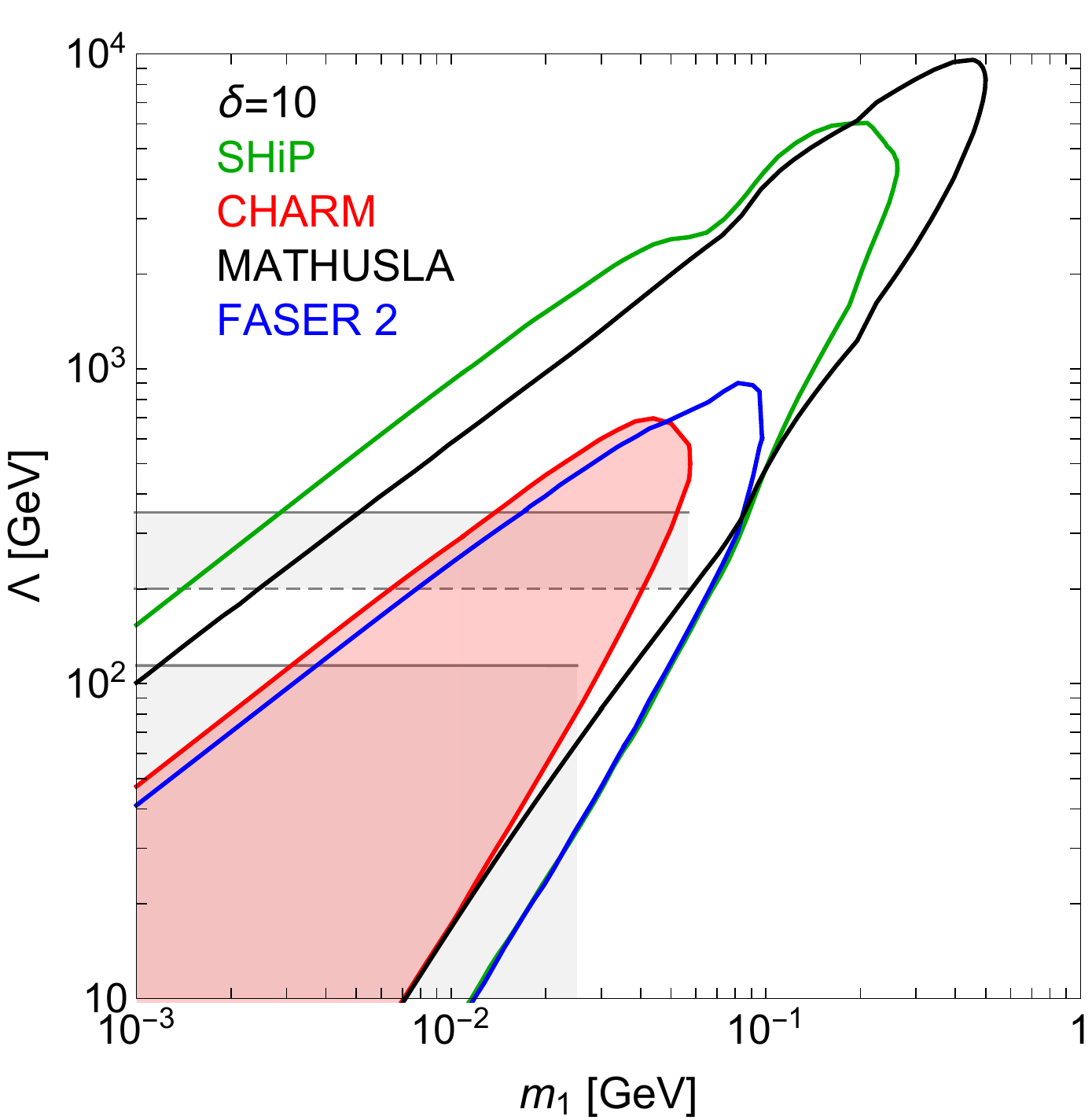}
	\includegraphics[width=0.49\textwidth]{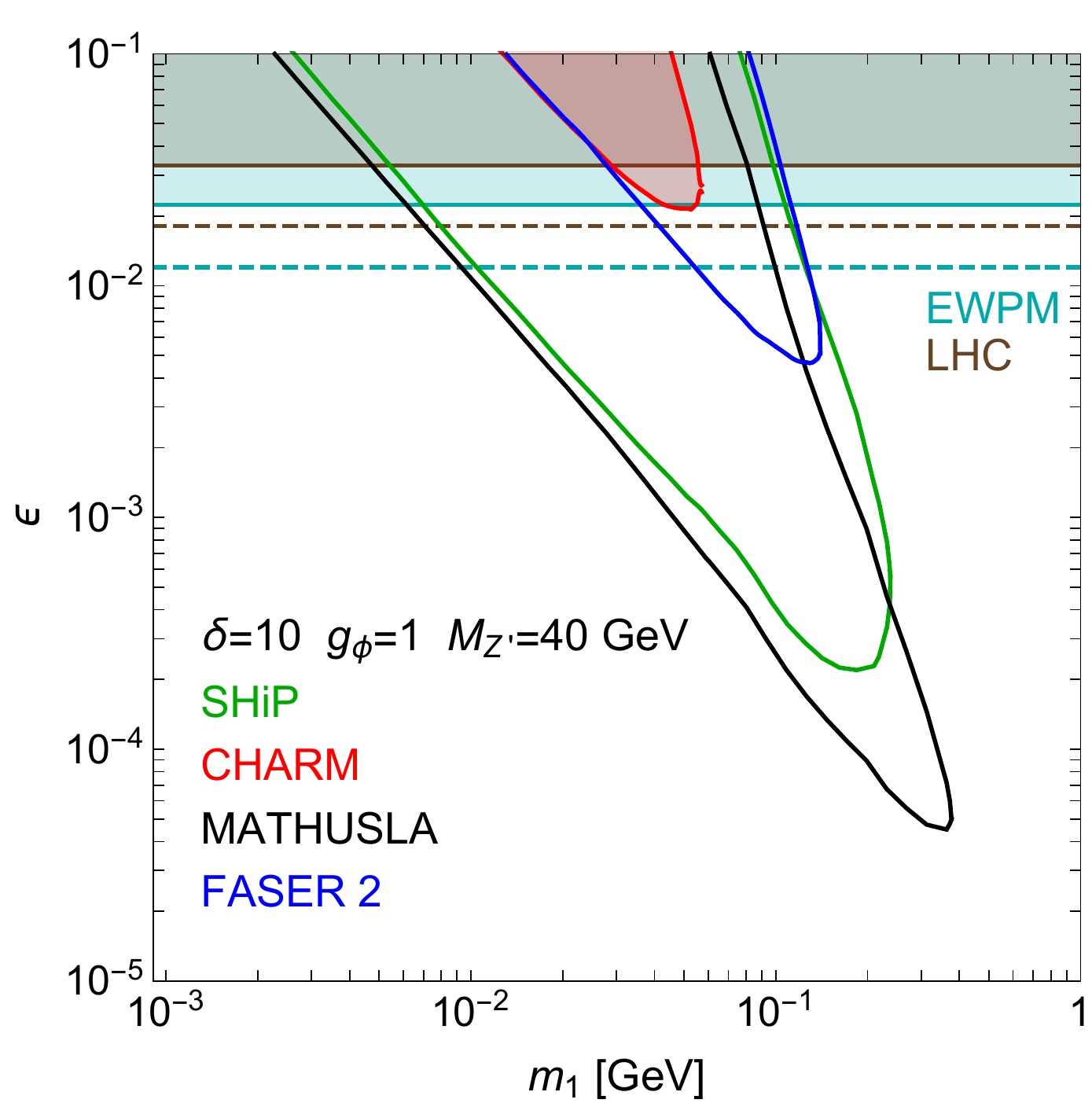}
\caption{Exclusion limit (CHARM) and projected sensitivities (SHiP, MATHUSLA and FASER) on the scalar dark sector under consideration. Left: We focus on the EFT operators in \cref{eq:EBMT_EFT}. Gray regions are excluded by the constraints derived from LEP and BaBar data. Right: We consider the scenario with a dark photon mediator, \cref{eq:EBMT_DP}. Cyan and gray regions are excluded by EWPM and LHC searches. A mass splitting $\delta=10$ is adopted for both panels. Taken from Ref.~\cite{Bertuzzo:2020rzo}.}
\label{fig:EBMT_figure1}
\end{figure*}

\paragraph{Results} In \cref{fig:EBMT_figure1} (left panel) we present an analysis of the effective operators in \cref{eq:EBMT_EFT}. We show our results in the ($m_1$,\,$\Lambda$) plane for a representative value of the mass splitting $\delta=10$, and for a democratic coupling to all the SM fermions, i.e. $c_{f_L} = c_{f_R}=1.$ The gray regions are excluded by searches of mono-photon events at LEP, and searches of invisible decays of heavy quarkonium states performed by BaBar. For the former constraints, we recast the results of~\cite{Fox:2011fx}, where LEP results have been interpreted in terms of an EFT describing the interactions of a fermionic DM candidate with the SM fermions. In addition, following the previous discussion, to ensure the validity of the EFT we impose that its cut-off, which we identify with $\Lambda$, is larger than the center of mass energy at LEP, i.e. $E_{\rm cm}\simeq 200\,{\rm GeV}$. This leads to the lower limit shown with a dashed gray line in \cref{fig:EBMT_figure1}. Concerning Babar, we exploit the upper limits of the invisible decay of the $\Upsilon(1s)$ resonance reported in~\cite{BaBar:2009gco}. We found that less-stringent constraints are obtained from the the invisible decay of the $J/\psi$, and mono-photon searches at Babar. The bounds from searches of missing energy at BaBar and LEP apply only when the $\phi_2$ particles decay outside the detectors. Imposing this requirement leads to a cut of the constraints in \cref{fig:EBMT_figure1} : this is because for large enough masses the proper decay length of $\phi_2$  significantly reduces. 

We shall now discuss the sensitivities of the experiments discussed in the previous paragraph to the long-lived particle present in our scenario. As shown in \cref{fig:EBMT_figure1}, the results of the CHARM experiment extend the  bounds from LEP at larger values of $\Lambda$. The future facilities FASER (2), SHiP and MATHUSLA will significantly improve current constraints, probing $\Lambda$ up to the TeV and multi-TeV domain. Let us comment that the feature visible in \cref{fig:EBMT_figure1} in the sensitivity curve of SHiP (and less visible for the other experiments) corresponds to a kinematical threshold where the dominant production of $\phi_1\phi_2$ pairs transits from decays of $\omega$ to $J/\psi$ mesons. Predictions for values of $\delta$ different than the one considered in \cref{fig:EBMT_figure1} can be found in\,\cite{Bertuzzo:2020rzo}. For instance, for $\delta=0.1,$ similar results are obtained at qualitative level, but the range of $\phi_1$ masses which can be probed shifts to larger values, due to the compressed mass spectrum under consideration.

Finally, additional relevant constraints could be inferred from LHC searches. However, the issue of a proper interpretation in terms of an EFT framework becomes particularly important, due to the high-energy scales probed by the LHC. Conservative but consistent bounds can be obtained following the strategy discussed before, i.e. restricting to signal events with a center of mass energy below the cut-off of the EFT. Following this strategy,~\cite{Bertuzzo:2017lwt} found that mono-jet searches at LHC can probe the parameter space of the EFT of a fermionic dark matter singlet for $g_*=4\pi,$ while no bounds can be obtained for $g_*=1.$ Instead of repeating a similar analysis for our scenario and using current LHC data, we prefer to resort to a UV completion of the EFT in \cref{eq:EBMT_EFT}. This allows a thorough comparison with LHC constraints, including the possibility to produce on-shell the mediator of the interaction among the dark states and the SM. For this purpose, we consider the dark photon model in \cref{eq:EBMT_DP}. Results are reported in \cref{fig:EBMT_figure1} (right panel), for a relatively light vector boson, with a mass of 40 GeV, and adopting a mass splitting $\delta=10.$ Relevant constraints on this scenario are from electroweak precision measurements (EWPM)~\cite{Curtin:2014cca}, and searches of a light $Z^{\prime}$ resonance at LHCb~\cite{LHCb:2019vmc} and CMS~\cite{CMS:2019buh}. These limits are shown in \cref{eq:EBMT_DP}, as well as future prospects. In the same plot we present the sensitivities for CHARM, SHiP, FASER and MATHUSLA. As evident, future experiments are able to improve current LHC limits, even probing regions of the parameter space beyond the reach of the High-Luminosity LHC. Finally, let us mention that the case of a TeV scale $Z^{\prime}$ is discussed in\,\cite{Bertuzzo:2020rzo}.   

\paragraph{Conclusions} We have analyzed the prospects for detection of long-lived particles with the future experiments SHiP, FASER and MATHUSLA, in the context of a dark sector containing a pair of non-degenerate scalars. We have shown that these experiments are complementary to searches at large-scale detectors at the LHC and they are important in testing the dark sector under consideration. Finally, we shall mention that the lightest scalar of the model, $\phi_1,$ could potentially play the role of a dark matter candidate. Some considerations along these lines can be found in\,\cite{Bertuzzo:2020rzo}.   

\subsection{Beyond the Minimal Dark Photon Model: Lepton Flavor Violation \label{sec:bsm_nonmin_clfv}}

\paragraph{Introduction} Charge lepton flavor violation (CLFV) interactions are generally predicted in the generation of neutrino masses and mixing. The most well-studied interaction in this regard is the one with a new scalar boson responsible for the neutrino masses. The scalar-type CLFV interactions originate from Yukawa interactions with the new scalar bosons in extensions of the Standard Model (SM), such as two Higgs doublet models~\cite{Kubo:2006yx} and type-II seesaw models~\cite{Ma:2001mr, Chun:2003ej, Kakizaki:2003jk, Abada:2007ux, Akeroyd:2009nu, Fukuyama:2009xk, Fukuyama:2010mz, Primulando:2019evb}. After diagonalizing mass matrix of charged lepton, the misalignment of Yukawa couplings between the SM Higgs boson and the new scalar boson results in the CLFV interactions.

Another type of CLFV interaction is the one with a new gauge boson. The vector-type CLFV interactions can appear in flavor or family gauge symmetric models, such as gauged $L_\alpha - L_\beta$ models where$\alpha,\beta= e,\mu,\tau$. When charged leptons are non-universally charged under the new gauge symmetry, and have non-diagonal Yukawa couplings, the CLFV interactions emerge in the gauge sector after the symmetry breaking~\cite{Foot:1994vd, Heeck:2016xkh, Altmannshofer:2016brv, Iguro:2020rby, Cheng:2021okr}. Furthermore, when we consider dimension $5$ operators, the dipole-type CLFV interactions are possible even in flavor universal gauge symmetric models, such as $B-L$ models and dark photon models. Such dipole-type interactions are generated by integrating out heavy scalars and/or fermions propagating in loop diagrams~\cite{Nomura:2020azp}. On the other hand, the new gauge boson has a mass. One of the possible origins of the mass is spontaneously symmetry breaking. In \cite{Araki:2020wkq}, it was shown that a new Higgs boson, which breaks the extra gauge symmetry, can be a new source of the dark photon and the sensitivity of the dark photon search at FASER can be improved. In this section, we consider the CLFV decays of the light and long-lived new bosons with scalar-, vector- and dipole-type interactions, and discuss the sensitivity to CLFV couplings at the FASER2 experiment. For analysis of the new gauge boson, we take into account new production process from the new Higgs boson which gives the origin of the gauge boson mass.

\paragraph{Interaction Lagrangian} We study the CLFV decays of light bosons for three types of interactions and refer to them as the scalar-, the vector-, and the dipole-type interaction. The FASER2 detector will be able to identify an electron and a muon, whereas the identification of a tau is difficult. Therefore, we restrict our analyses to the CLFV interactions only in the electron-muon sector. Some of the interaction Lagrangians shown below were recently studied in the context of constraints from the E137 electron beam dump experiment in Ref.~\cite{Araki:2021vhy}, in which their possible origins were also discussed based on a multi Higgs doublet model, an ALP model, a gauged $L_\mu - L_\tau$ model, and a loop-induced dark photon model. 

The scalar-type CLFV interaction is given by
\begin{align}
    \mathcal{L}_{\rm scalar} = 
    \frac{\theta_{h\phi}}{v} \sum_{f} m_f \overline{f} \phi_l f
    +\left( y_{e\mu}\overline{e_L} \phi_l \mu_R
    +y_{\mu e}\overline{\mu_L} \phi_l e_R +h.c. \right)~,
\label{eq:Lscl}
\end{align}
where $\phi_l$ stands for the CLFV dark-higgs boson, the angle $\theta_{h\phi}$ represents the mixing between the SM Higgs boson $h$ and $\phi_l$, the subscript $f$ runs over all the SM fermions with $m_f$ being its mass, the VEV of the SM Higgs boson is denoted as $v=176$ GeV, and $y_{e\mu}$ and $y_{\mu e}$ are CLFV coupling constants. Left-handed and right-handed fermions are denoted as $f_L$ and $f_R$, respectively.

With \cref{eq:Lscl}, the total decay width of $\phi_l$ is given by 
\begin{align}
    \Gamma_{\mathrm{total}} = 
      \Gamma(\phi_l \rightarrow {\rm hadrons})
      + \sum_{\ell = e,\mu,\tau}\Gamma(\phi_l \rightarrow \ell\bar{\ell})
      + \Gamma(\phi_l \rightarrow e\bar{\mu})
      + \Gamma(\phi_l \rightarrow \mu\bar{e})~.
\end{align}
The partial decay width into the charged leptons is written as
\begin{align}
    \Gamma(\phi_l \rightarrow \ell\bar{\ell'})
    = \frac{1}{16\pi}m_\phi~
    \lambda\left( \frac{m_\ell^2}{m_\phi^2}, \frac{m_{\ell'}^2}{m_\phi^2} \right) \left[ 
    S_1 \left( 1-\frac{m_\ell^2+m_{\ell'}^2}{m_\phi^2} \right)
    -4 S_2 \frac{m_\ell m_{\ell'}}{m_\phi^2} \right]~,
\end{align}
where $m_\phi$ and $m_{\ell(\ell')}$ stand for a mass of $\phi_l$ and that of $\ell(\ell')$ charged lepton, respectively, and the function $\lambda$ is the Kallen function defined as follows:
\begin{align}
\lambda(a,b) = \sqrt{1 + a^2 + b^2 - 2a - 2b - 2 ab}~.
\label{eq:kallen}
\end{align}
The constants $S_1$ and $S_2$ are defined as $S_1 = 2 S_2= 2(\theta_{h\phi} m_\ell)^2/v^2$ for the charged lepton flavor conserving (CLFC) decays, while $S_1 = |y_{e\mu}|^2 + |y_{\mu e}|^2$ and $S_2 = {\rm Re}(y_{e\mu} y_{\mu e})$ for the CLFV decays. As for the hadronic decays, we use the decay widths given in Refs.~\cite{Bezrukov:2013fca, Feng:2017vli}

For the vector-type CLFV interaction, we consider the following Lagrangian,
\begin{align}
    \mathcal{L}_{\mathrm{vector}} &= g' Z'_\rho (
    s^2~ \overline{e} \gamma^\rho e 
    + c^2~ \overline{\mu} \gamma^\rho \mu
    + sc~ \overline{\mu} \gamma^\rho e 
    + sc~ \overline{e} \gamma^\rho \mu) \nonumber \\
    & \qquad +g' Z'_\rho (- \overline{\tau} \gamma^\rho \tau
    + \overline{\nu_\mu} \gamma^\rho \nu_\mu
    - \overline{\nu_\tau} \gamma^\rho \nu_\tau )~,
\label{eq:Lvec}
\end{align}
where $Z'$ and $g'$ are the new gauge boson and the gauge coupling constant, respectively, while $s=\sin\theta_{e\mu}$ and $c=\cos\theta_{e\mu}$. Here, $\nu_\mu$ and $\nu_\tau$ are left-handed muon and tau neutrinos. For simplicity, we omit the kinetic mixing throughout this paper by assuming it is negligibly small. In \cref{eq:Lvec}, the U(1)$_{L_\mu - L_\tau}$ symmetry is restored in the limit of $\theta_{e\mu} \to 0$.

From the Lagrangian in \cref{eq:Lvec}, the total decay width of $Z'$ is obtained as
\begin{align}
    \Gamma_{\mathrm{total}} = 
    \Gamma(Z' \rightarrow \nu\bar{\nu}) 
    + \sum_{\ell=e, \mu,\tau}\Gamma(Z' \rightarrow \ell\bar{\ell})
    + \Gamma(Z' \rightarrow e\bar{\mu})
    + \Gamma(Z' \rightarrow \mu\bar{e})~,
\end{align}
where neutrinos are assumed to be massless Dirac particle. The partial decay width into the charged leptons is given by 
\begin{align}
    \Gamma(Z' \rightarrow \ell\bar{\ell'}) 
    = \frac{V^2}{24 \pi} m_{Z'}~\lambda\left(
    1, \frac{m_\ell^2}{m_{Z'}^2}, \frac{m_{\ell'}^2}{m_{Z'}^2}\right) 
    \left[ 2
    - \frac{m_\ell^2 - 6 m_\ell m_{\ell'} + m_{\ell'}^2}{m_{Z'}^2}
    - \frac{(m_\ell^2 - m_{\ell'}^2)^2}{m_{Z'}^4}
    \right]~,
\end{align}
where $V=g' s^2$ or $g' c^2$ for the CLFC decay into $ee$ or $\mu\mu$, while $V=g' s c$ for the CLFV decays into $e\bar{\mu}$ and $\bar{e}\mu$. The function $\lambda$ is defined in \cref{eq:kallen}.

The dipole-type interaction is given by
\begin{align}
    \mathcal{L}_{\mathrm{dipole}} &=
    \frac{1}{2} \sum_{\ell=e,\mu,\tau} \mu_\ell \overline{\ell} \sigma^{\rho \sigma} \ell A'_{\rho \sigma}
    + \frac{\mu'}{2} 
      \left( 
          \overline{\mu} \sigma^{\rho \sigma}e 
        + \overline{e} \sigma^{\rho \sigma}\mu 
      \right) A'_{\rho \sigma}~,
\label{eq:Ldpl}
\end{align}
where $\mu_\ell$ and $\mu^\prime$ are CLFC and CLFV dipole couplings, respectively, and $A_{\rho\sigma}^\prime$ stands for the field strength of $A^\prime$. Here the dipole couplings are assumed to be real. Electromagnetic CLFV interactions similar to \cref{eq:Ldpl} can be obtained by replacing $A'$ with a photon. However, such dangerous CLFV interactions could be suppressed when electrically neutral CP-even and odd scalar propagate in loop 
as discussed in Ref.~\cite{Araki:2021vhy}.

Given the Lagrangian in \cref{eq:Ldpl}, the total decay width of $A'$ is given by
\begin{align}
    \Gamma_{\mathrm{total}} = 
    \sum_{\ell=e,\mu,\tau}\Gamma(A' \rightarrow \ell\bar{\ell})
    + \Gamma(A' \rightarrow e\bar{\mu})
    + \Gamma(A' \rightarrow \mu\bar{e})~.
\end{align}
The partial decay width into the charged leptons is written as
\begin{align}
    \Gamma(A' \rightarrow \ell\bar{\ell'}) 
    = \frac{D^2}{12 \pi} m_{A'}^3 ~\lambda\left(\frac{m_\ell^2}{m_{A'}^2}, \frac{m_{\ell'}^2}{m_{A'}^2}\right) 
    \left[ \frac{1}{2} 
    + \frac{1}{2}\frac{m_\ell^2 + 6m_\ell m_{\ell'} + m_{\ell'}^2}{m_{A'}^2}
    - \frac{(m_\ell^2 - m_{\ell'}^2)^2}{m_{A'}^4}
    \right],
\end{align}
where $D = \mu_\ell$ or $\mu'$ for CLFC or CLFV decays, respectively, and the function $\lambda$ is given in \cref{eq:kallen}.

\paragraph{Production and Signal of Light Bosons} Here, we discuss the production of the light bosons and the number of CLFV events at the FASER2 detector. The production mechanisms considered in this paper are different for each type of the light boson.

In the case of the scalar-type interaction, the CLFV dark-higgs boson, $\phi_l$, is produced by meson decays through the SM Higgs-$\phi_l$ mixing, and dominant production processes are $B$ and $K$ meson decays~\cite{Feng:2017vli}. On one hand, $B$ mesons are short lived and can be assumed to decay into $\phi_l$ at the IP. On the other hand, $K$ mesons travel macroscopic distances, and quite a few $K$ mesons are absorbed by the TAN neutral particle absorber for $K_{L,S}$, or deflected by the superconducting quadrupole magnets for $K^\pm$, before decaying into $\phi_l$. Because of this reduction, the production from $K$ mesons is subdominant in comparison with that from $B$ mesons.Given this fact, in this work, we only consider the production from $B$ mesons and use the branching ration of
\begin{align}
\label{eq:br_Btophi}
    {\rm Br}(B \to X_s \phi) \simeq 5.7 \left( 1 - \frac{m_\phi^2}{m_b^2} \right)^2 \theta_{h\phi}^2~, 
\end{align}
which is obtained in Ref.~\cite{Feng:2017vli} in the limit of $\theta_{h\phi} \ll 1$, where $m_b$ is the b-quark mass and $\theta_{h\phi}$ denotes the SM Higgs-$\phi_l$ mixing angle defined in \cref{eq:Lscl}. It should be noted that we have checked that the sensitivity region remains almost the same even if the production from $K$ mesons are included.

For the cases of the vector- and the dipole-type interaction, the gauge bosons, $Z'$ and $A'$, cannot directly be produced from meson decays since they do not interact with quarks. However, given the fact that the gauge bosons are massive, it is natural to expect the existence of a new scalar boson which spontaneously breaks the gauge symmetry and gives non-zero mass to the gauge bosons. Moreover, the new scalar boson is presumed to mix with the SM Higgs boson and have interactions with the SM fermions, like the dark-Higgs boson. Based on these considerations, for the vector- and the dipole-type interaction, we further introduce the following interaction Lagrangian
\begin{align}
\label{eq:laglangian-X}
    \mathcal{L}_{\phi_g} = 
    g' m_G \phi_g G_\mu G^\mu + \frac{\theta_{h\phi}}{v} \sum_{f} m_f \bar{f} \phi_g f~,
\end{align}
where $G = Z'$ or $A'$, $\phi_g$ stands for the symmetry breaking scalar boson, and we refer to $\phi_g$ as the gauged dark-higgs boson in what follows. In the second term, $\theta_{h\phi}$ denotes the mixing angle between the SM Higgs boson and $\phi_g$, similarly to the CLFV dark-higgs boson. With \cref{eq:laglangian-X}, the gauge bosons can be produced from meson decays followed by $\phi_g \rightarrow GG$, as shown in Ref.~\cite{Araki:2020wkq}. The decay width of $\phi_g$ into a pair of the extra gauge boson is enhanced by the factor of $m_\phi^2/m_G^2$ for $m_\phi \gg m_G$. Note that we only consider the production from $B$ meson decays in analogy with the scalar-type interaction. Note also that we use the common symbols $m_\phi$ and $\theta_{h\phi}$ for both the gauged and the CLFV dark-higgs boson throughout this paper.

To calculate the number of the signal events, we use the probability of the CLFV decays of the light boson inside the FASER2 detector is given in \cite{Araki:2020wkq}. With this probability, the total number of events of the gauge boson decays inside the FASER2 detector is given by
\begin{align}
\label{eq:num-of-event_gauge}
   N 
   &= \mathcal{L} \sum_{j=1,2} \int dp_B d\theta_B 
   \int dp_{G} \int dp_\phi
   \frac{d\sigma_{pp \to B}}{dp_B d\theta_B } {\rm Br}(B \to X_s \phi) {\rm Br}(\phi_g \to G_1 G_2) \nonumber \\   
   &\qquad \times {\rm Br}(G \to e \mu)
   \mathcal{P}_{G_j}^{\rm det}(p_{G}, p_\phi)~,
\end{align}
where $p_B$ and $\theta_B$ denote the momentum and the angle of $B$ mesons, and the expected integrated luminosity is written as $\mathcal{L}$ .

\begin{figure*}[t]
\centering
\includegraphics[width=0.49\textwidth]{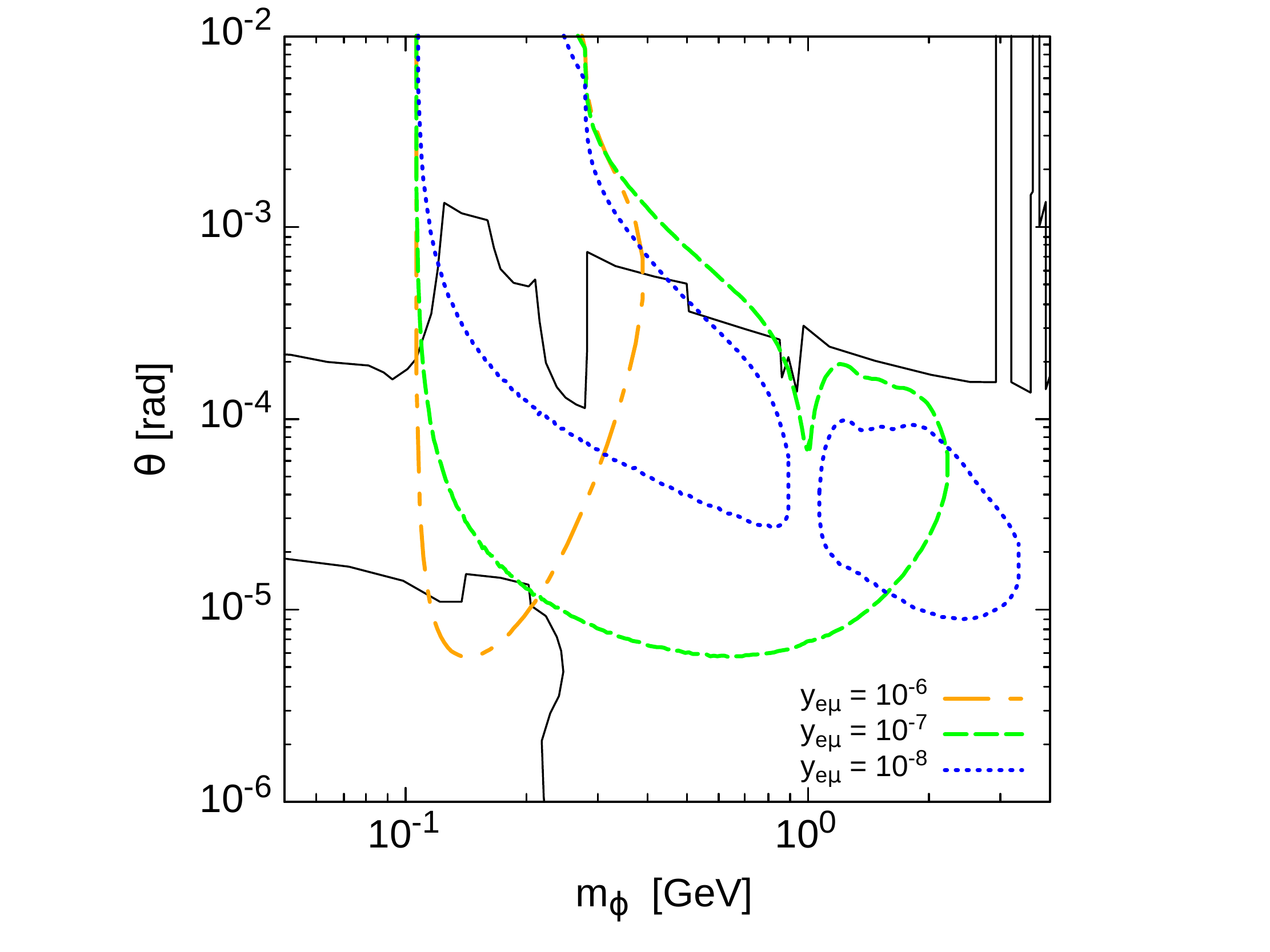}
\caption{The contour plots of 95 \% C.L. sensitivity regions for the scalar-type interaction. The inside of black curves display the current exclusion regions obtained in Ref.~\cite{Winkler:2018qyg}.}
\label{fig:dh95}
\end{figure*}

\paragraph{Results} We here show our results of numerical calculations for the scalar-type interaction for the FASER2 setup following \cite{FASER:2018eoc}. For simplicity, we assume $y_{e\mu}=y_{\mu e}$ and that all the coupling constants are real. In \cref{fig:dh95}, we present the 95\% C.L. sensitivity regions for the scalar-type interaction in the $m_\phi - \theta_{h\phi}$ plane. The shaded regions display the current exclusion regions summarized in Ref.~\cite{Winkler:2018qyg}. As $y_{e\mu}$ increases, the sensitivity region becomes narrow toward a smaller $m_\phi$ region, since the decay length is too short for a larger $m_\phi$. Also, as $y_{e\mu}$ decreases, the sensitivity region gets small because the branching ration of the signal process, $\phi_l \rightarrow e\mu$, becomes small.

\begin{figure*}[t]
\centering
\includegraphics[width=0.49\textwidth, trim={2cm 0 2cm 0}, clip]{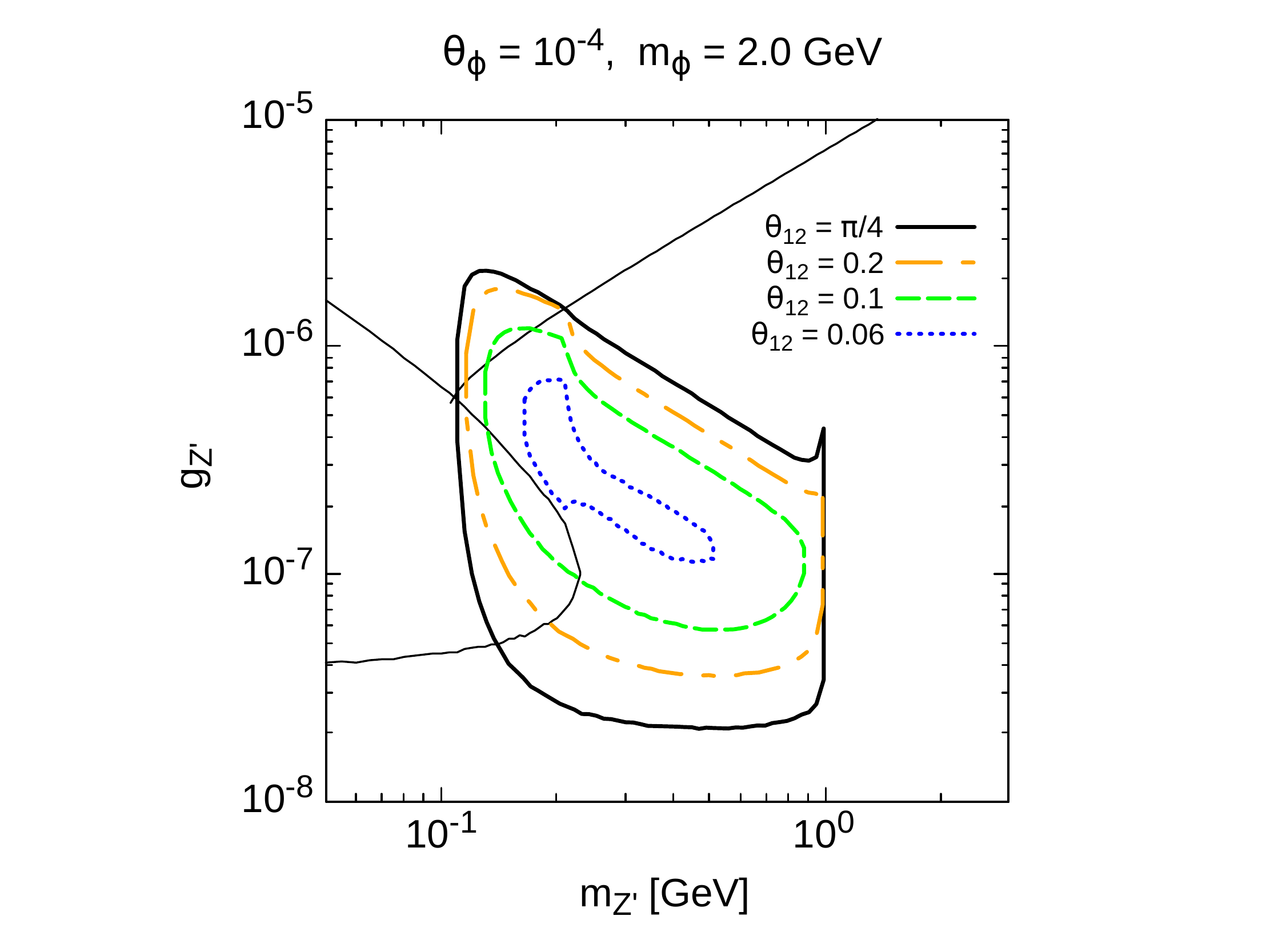}
\includegraphics[width=0.49\textwidth, trim={2cm 0 2cm 0}, clip]{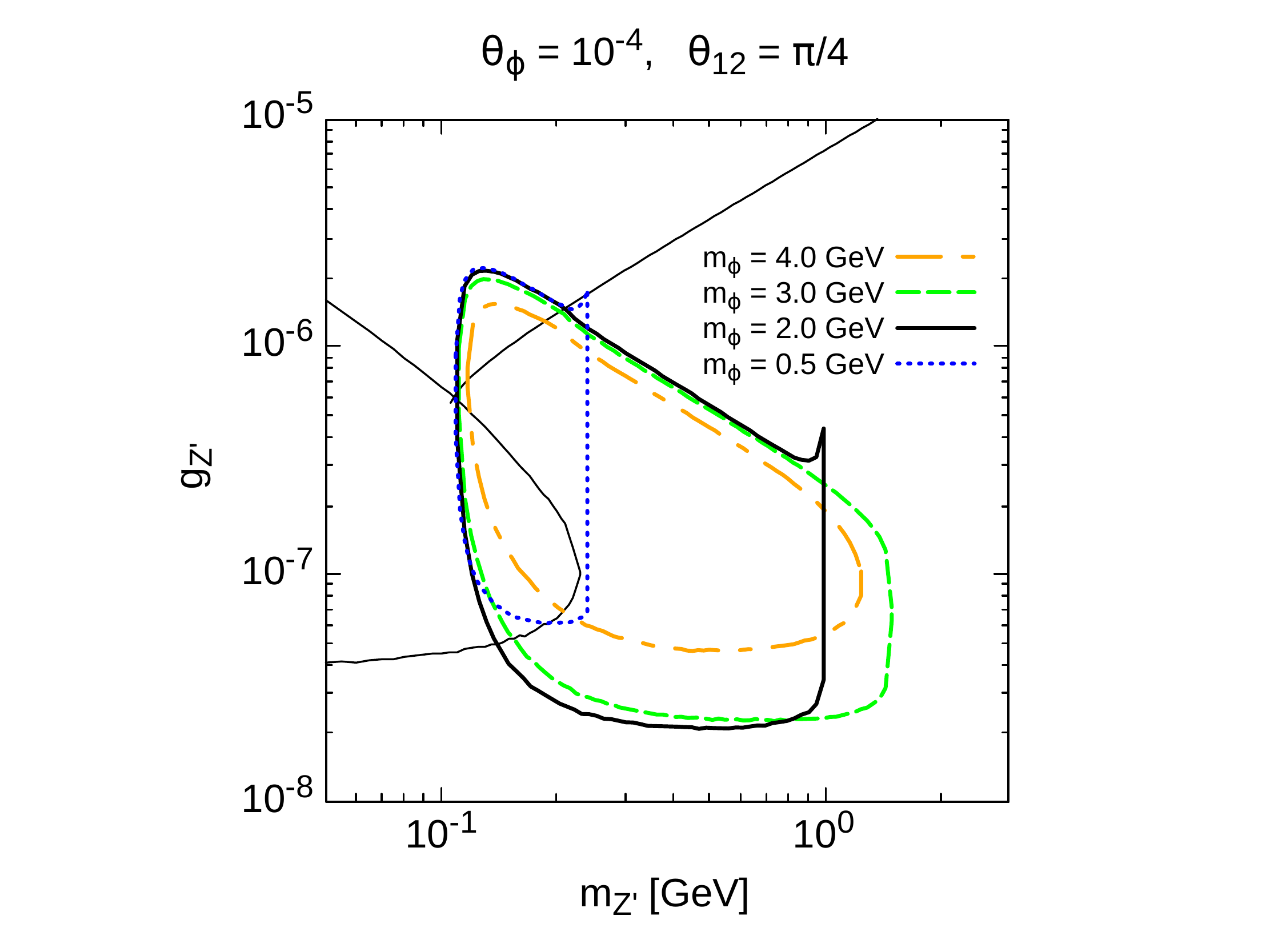}
\caption{The contour plots of 95\% C.L. sensitivity regions for the vector interaction. The black curves represent the constraints from $\mu\rightarrow eee$ and E137 for $\theta=\pi/4$.}
\label{fig:mt95}
\end{figure*}

In \cref{fig:mt95}, we show the contour plots of 95\% C.L. sensitivity regions for the vector-type interaction in the $m_{Z'}-g'$ plane. In the left panel, we vary the mixing angle $\theta_{e\mu}$ as 0.06 (blue dotted), 0.1 (green dashed), 0.2 (orange dashed-dotted), and $\pi/4$ (black solid). As the mixing angle decreases, the sensitivity region becomes narrow due to the decreasing of the branching ratio of the signal process $Z' \rightarrow e\mu$. 
In the right panel, we vary $m_\phi$ as 0.5~GeV (blue dotted), 2.0~GeV (black solid), 3.0~GeV (green dashed), and 4.0~GeV (orange dashed-dotted). The sensitivity region broadens as $m_\phi$ increases, because the range of $m_{Z'}$ satisfying $m_{Z'} < 2m_\phi$ widens. For $m_\phi > 3$ GeV, however, the sensitivity region turns to narrow since the decay length becomes too short to reach the detector. In the figures, the small spikes, for instance around $m_{Z'} = 1$ GeV and $g' = 4\times 10^{-7}$ in the left panel, arise due to the rapid increase of the decay length of $\phi_g$. We find that $N>3$ events are expected for the range of $0.07 < \theta < 1.5$, and the contour for $\theta = \pi /4 + x$ is almost the same as that for $\theta = \pi /4 - x$.

\begin{figure*}[t]
\centering
\includegraphics[width=0.49\textwidth, trim={2cm 0 2cm 0}, clip]{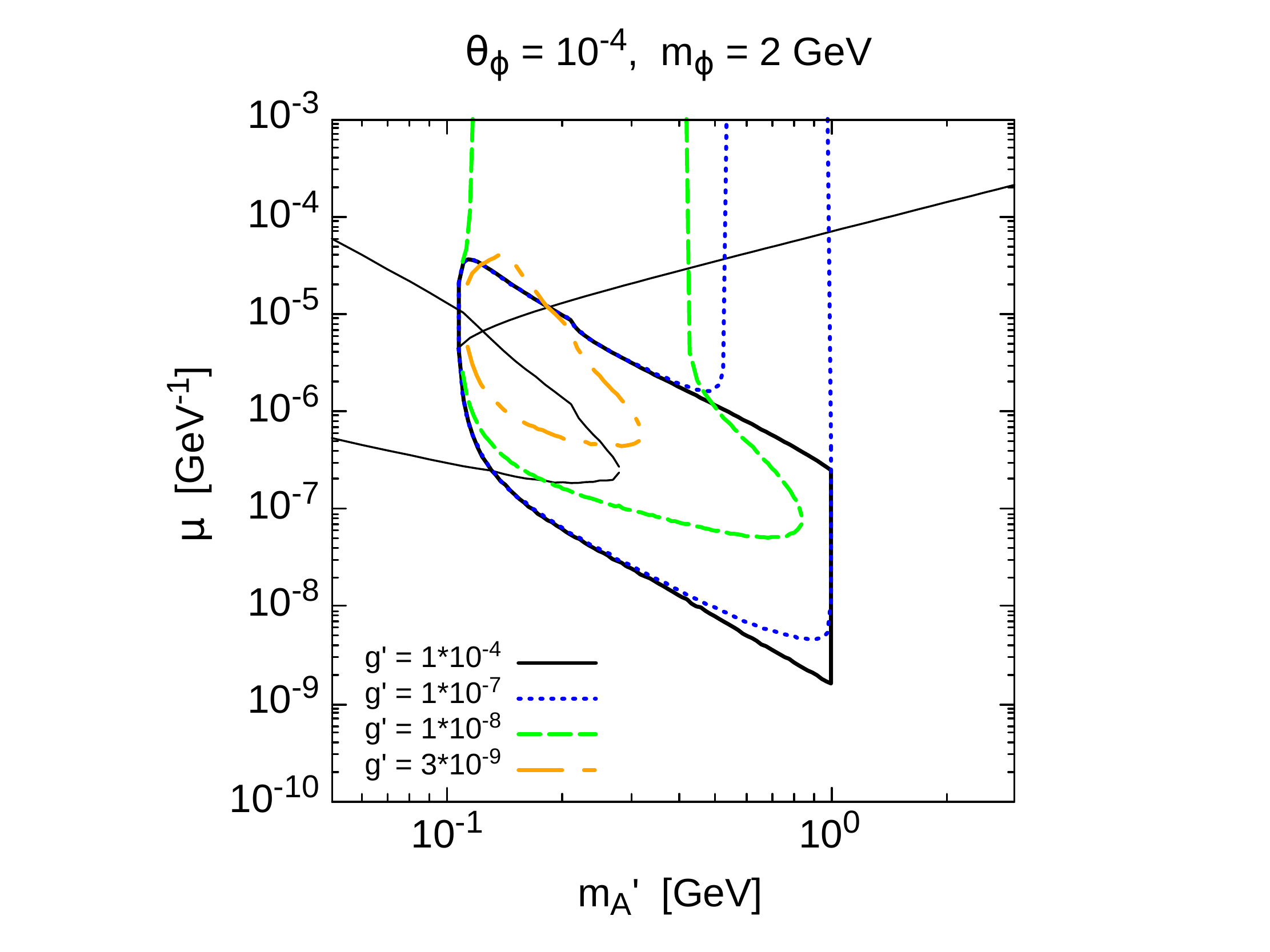}
\includegraphics[width=0.49\textwidth, trim={2cm 0 2cm 0}, clip]{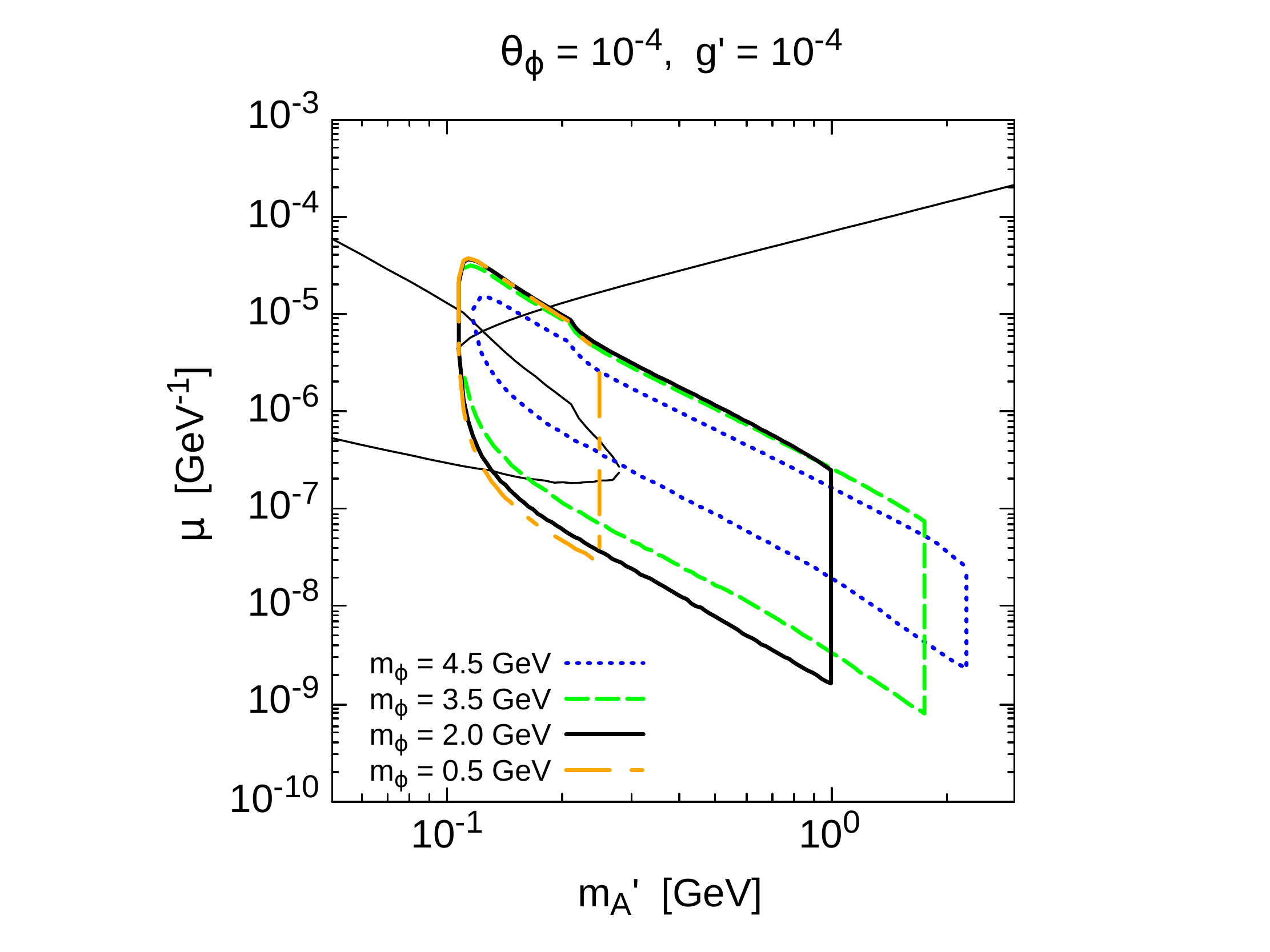}
\caption{The contour plots of 95\% C.L. sensitivity regions for the dipole interaction. The black curves represent the constraints from $\mu\rightarrow eee$ and E137.}
\label{fig:dp95}
\end{figure*}

In \cref{fig:dp95}, we show the 95\% C.L. sensitivity regions of the dipole-type interaction in the $m_{A'} - \mu$ plane for $g'=10^{-4}$-$3 \times 10^{-9}$ in the left panel, while for $m_{\phi}=0.5$-$4.0$ GeV in the right panel. In the left panel, the sensitivity region does not change much for $g' = 10^{-6} - 10^{-4}$ with which the decay length is dominated by that of $A'$. For $g' < 10^{-7}$, $\phi_g$ can travel macroscopic distances so that some of them can reach the detector. In this case, the decay length of $A'$ can be extremely short, which means the dipole coupling $\mu$ can be large.  This is the reason why the sensitivity regions rise up toward a larger $\mu$ region for $g' < 10^{-7}$. On the other hand, as $g'$ becomes small, production of the gauge boson via $\phi_g \rightarrow A'A'$ decreases; Especially, the decreasing is much more significant for larger $m_{A'}$, because the enhancement from the longitudinal mode is weak. This behavior can be seen by the case of $g' = 10^{-8}$ (blue dotted curve): although the decay length is long enough to reach the detector for a wide range of $m_{A'}$, the production of the gauge boson is not enough to yield three events in the large $m_{A'}$ region. Eventually, for $g' < 10^{-9}$, the sensitivity region starts to shrink. In the right panel, the sensitivity region for $m_\phi=4.5$~GeV is narrower in comparison with the others, since it is close to the kinematical threshold of $B \rightarrow X_s \phi$.

\paragraph{Conclusion} We have shown that the FASER2 experiment will be able to explore charged lepton flavor violation in the decays of the scalar and vector bosons. We found that the flavor violating couplings below the present bounds can be searched for the scalar-, vector- and dipole-interactions. The results for another type of CLFV interactions, i.e., pseudo-scalar or ALP type, will be presented in the future study. 

\subsection{$U(1)_{T3R}$ Gauge Boson\label{sec:bsm_nonmin_U1T3R}}

There are a variety of scenarios in which the existence of new gauge groups lead to interactions between Standard Model particles and new sub-GeV particles.  When new gauge interactions are introduced, it is necessary for gauge and gravitational anmoalies to be cancelled. As outlined above, there are several well-studied examples in which this occurs, including the $U(1)_{B-L}$ group, the $U(1)_{L_i - L_j}$ group, and a secluded $U(1)_X$ group (under which all Standard Model particles are neutral)~\cite{Foot:1990mn, He:1990pn, He:1991qd, Borah:2020swo, Costa:2020krs}. Another well-studied example is $U(1)_{T3R}$ group~\cite{Dutta:2019fxn, Dutta:2020enk, Dutta:2020jsy, Dutta:2021afo}, under which one or more full generations of right-handed Standard Model fermions are charged (including right-handed neutrinos), with up- and down-type fermions having opposite charge.

$U(1)_{T3R}$ was originally introduced in the context of left-right models, see for example Refs.~\cite{Pati:1974yy, Mohapatra:1974gc, Senjanovic:1975rk}. In that context, the Standard Model Higgs was also charged under $U(1)_{T3R}$, tying the symmetry-breaking scale of $U(1)_{T3R}$ to the TeV-scale. Recent interest has focused on scenarios in which the $U(1)_{T3R}$ is broken by the condensation of a dark Higgs, which provides a new independent dimensionful scale which is decoupled from the TeV-scale.  Since only right-handed fermions are charged under $U(1)_{T3R}$, their masses are also protected by this symmetry, and are thus proportional to this new symmetry-breaking scale.  Recent interest has thus focused on the case in which the new symmetry-breaking scale is $V \sim {\cal O}(10)\gev$, and in which only first- or second-generation fermions are charged under $U(1)_{T3R}$~\cite{Dutta:2019fxn}.  In this scenario, the Yukawa couplings of the low energy effective field theory (defined below the electroweak scale) need not be particularly small, thus providing an explanation for the Standard Model flavor hierarchy.  Moreover, this new symmetry-breaking scale naturally sets the mass scale of dark sector particles which are only charged under $U(1)_{T3R}$, while being singlets under Standard Model gauge groups.  This scenario is particularly interesting, then, because it theoretically motivates the appearance of new sub-GeV particles.

This scenario also presents an opportunity for instruments at the FPF to search unexplored parameter space.  There are a few key theoretical features which are worth noting:
\begin{itemize}
\item Because $U(1)_{T3R}$ protects the masses of Standard Model fermions, the dark Higgs whose vev breaks $U(1)_{T3R}$ must couple to Standard Model fermions, in addition to the dark photon.  This scenario thus inherently contains two mediators which interact with SM fermions, unlike other examples of new gauge groups.
\item Because the dark photon has chiral couplings to Standard Model fermions, the longitudinal polarization does not decouple.  Because this mode has its origin as a Goldstone boson, the couplings of the dark photon are thus related to those of dark Higgs.
\item Because the SM fermions which couple to $U(1)_{T3R}$ have masses which are not much smaller than the symmetry-breaking scale, the Yukawa couplings of the low-energy effective field theory are not very small.  Thus, the dark Higgs (and, necessarily, the dark photon) must have relatively large couplings to the Standard Model.
\end{itemize}

The fact that these couplings are actually reasonably large creates a window of opportunity for experiments at the FPF. Many current and upcoming experiments are designed to search for the visible decays of long-lived particles at displaced detectors, but are generally focused on the scenario in which the couplings are very small, and particle decays occur very far from production.  Models in which couplings are relatively large have been subject to less focus, because they would generally lead to large corrections to the magnetic moment of the muon.  But because constraints on $g-2$ are indirect constraints which sum corrections from all mediators, this conclusion is only generally applicable to models with a single mediator.  The $U(1)_{T3R}$ scenario necessarily contains two mediators, a scalar and a vector, which yield opposite contributions to the muon magnetic moment, because the dark photon has an axial coupling.  Moreover, because the coupling of the dark photon is tied to that of the dark scalar, one cannot decouple their magnitudes --  one generally expects to find regions of parameter space in which the scalar and vector contributions cancel. With no constraint from $g-2$, the best opportunity for constraining these models is with displaced detectors which are nevertheless close enough to interaction point that the dark mediators can reach the detector before decaying.  The FPF utilizes interactions at the LHC, which produce among the mostly highly boosted particles of any facility searching for long-lived particles, thus ensuring that even particles with a relatively short liftime may still have a reasonably long decay length.  Moreover, as the FPF is a new facility, some opportunity exists to locate a detector as close to the interaction point as possible, thus potentially improving sensitivity.

\paragraph{The model} The details of this scenario are explained in Refs.~\cite{Dutta:2019fxn,Dutta:2020jsy}, but we will briefly review the salient points. To ensure that all gauge and gravitational anomalies are cancelled, we will assume that one right-handed up-type quark, down-type quark, charged lepton and neutrino are charged under $U(1)_{T3R}$ with $Q= \pm 2$, and with up-type and down-type fermions having opposite sign.  Note that, although these Standard Model fermions constitute a full generation, they need not all be in the same generation.  It is technically natural for the charged lepton and either the up-type or down-type quark charged under $U(1)_{T3R}$ to be a mass eigenstate~\cite{Batell:2017kty}.  For simplicity, we will assume that all fermions charged under $U(1)_{T3R}$ are mass eigenstates, 
$U(1)_{T3R}$ will be broken to a parity by the condensation of a complex scalar field $\phi$ with charge $Q_\phi = 2$.  We may then express $\phi$ as $\phi = V + (1/\sqrt{2}) (\phi' + \imath \sigma)$, where $V$ is taken to be real.  The real scalar fields $\phi'$ and $\sigma$ are the dark Higgs and the Goldstone boson, respectively.

We assume that $U(1)_{T3R}$ is broken well below the electroweak symmetry-breaking scale.  In the low energy effective field theory defined below electroweak symmetry breaking, the mass and Yukawa coupling of a fermion $f$ charged under $U(1)_{T3R}$ arise from a term
\begin{eqnarray}
{\mathcal{L}}_{Yuk.} 
=-\lambda_f \phi^{(*)} \bar f P_R f + h.c. ,
=- m_f \bar f f - \frac{m_f}{\sqrt{2} V} 
\phi' \bar f f - \imath \frac{m_f}{\sqrt{2}V} 
\sigma \bar f \gamma^5 f ,
\end{eqnarray}
where $m_f = \lambda_f V$.  We thus see that if $V$ is only slightly above the mass scale of the fermions, the Yukawa coupling $\lambda_f$ need not be unnaturally small.

The dark photon, the gauge boson of $U(1)_{T3R}$, is denoted by $A'$, and has a mass given by $m_{A'}^2 = 2 g_{T3R}^2 Q_\phi^2 V^2$, where $g_{T3R}$ is the gauge coupling of $U(1)_{T3R}$.  We consider the case in which $\phi$ has a quartic potential which can be written as 
\begin{eqnarray}
V_\phi &=& \mu_\phi^2 \phi \phi^*  
+ \lambda_\phi (\phi \phi^*)^2 ,
\end{eqnarray}
in which case we find $V = (-\mu_\phi / 2\lambda_\phi)^{1/2}$, $m_{\phi'}^2 = -\mu_\phi^2 = 2\lambda_\phi V^2$.  We thus see that for perturbative couplings the masses of the dark photon and dark Higgs are below the symmetry-breaking scale $V$.  

A dark matter candidate naturally arises in this scenario.  If there is a Dirac fermion $\eta$ which is charged only under $U(1)_{T3R}$ with charge $Q_\eta =1$, then its Lagrangian generally has non-derivative quadratic terms of the form
\begin{eqnarray}
{\mathcal{L}}_\eta &=& ... -m_D \bar \eta_R \eta_L 
-\frac{1}{2} \lambda_L \phi \bar \eta_L^c \eta_L 
-\frac{1}{2} \lambda_R \phi^* \bar \eta_R^c \eta_R .
\end{eqnarray}
$\eta$  now has a Dirac mass as well as  Majorana mass terms which are proportional to $V$.  If the Dirac mass terms are small, we are left with two dark sector Majorana fermion mass eigenstates $\eta_{1,2}$, with mass $\propto V$ and with a small mass splitting.  $U(1)_{T3R}$ is broken to a parity under which all Standard Model particles are even, but $\eta_{1,2}$ are odd.  The lightest is thus stable, and is a dark matter candidate.

We consider the case in which the Standard Model fermions which are charged under $U(1)_{T3R}$ are $u$, $d$ and $\mu$.  This case is interesting because it avoids tight constraints which arise from atomic parity violation experiments and cosmological observables (if the dark photon couples to electrons) as well as constraints on the anomalous kaon decay (if the dark photon couples to second-generation quarks).  

As we have seen, we have a scenario in which we we have two new mediators (along with a dark matter candidate) whose masses are all $\lesssim V$.  As a benchmark, we will take the symmetry-breaking scale $V = 10~\gev$.  In this case, the dark Higgs coupling to muons is $\sim m_\mu / V \sim 10^{-2}$.  We will then find that the most interesting case is $m_{A',\phi'} < 2 m_\mu$, as otherwise the mediators would decay promptly to muons, a scenario which is already tightly constrained by data from $B$-factories.

\paragraph{A UV completion} Although the fermions masses arise from a renormalizable operator in the effective field theory defined below the EWSB scale, in the field theory defined above this scale this same term must arise from the non-renormalizable operator
\begin{eqnarray}
{\mathcal{L}}_{Yuk.} &=& -\frac{1}{\Lambda_f} 
H \phi \bar f P_R f ,
\end{eqnarray}
where $\lambda_f = \langle H \rangle / \Lambda_f$. This operator can be arise from renormalizable operators in a UV-completion utilizing the universal seesaw mechanism~\cite{Berezhiani:1983hm, Chang:1986bp, Davidson:1987mh, DePace:1987iu, Rajpoot:1987fca, Babu:1988mw, Babu:1989rb, Babu:2018vrl} 
if we add a new set of vector-like heavy fermions $Q_f$, which are neutral under $U(1)_{T3R}$, and have the SM gauge charges of a right-handed fermion. We may then write a mass term 
\begin{eqnarray}
{\mathcal{L}} &=& -\tilde m_{\chi_f} \bar \chi_f \chi_f 
- \lambda_{Lf} H^{(*)} \bar \chi_{fR} f_L 
- \lambda_{Rf} \phi^{(*)} \bar \chi_{fL} f_R +h.~c.,
\end{eqnarray}
where $f_R$ is a right-handed fermion charged under $U(1)_{T3R}$, and $f_L$ is the corresponding $SU(2)_L$ doublet containing the left-handed fermion.  Integrating out the heavy fermion $\chi_f$ yields the effective Lagrangian we have given, with $1/\Lambda_f \sim \lambda_{Lf}\lambda_{Rf} / \tilde m_{\chi}$, and $m_f \sim \lambda_{Lf}\lambda_{Rf} \langle H \rangle \langle \phi \rangle / \tilde m_{\chi}$. For $\lambda_{Lf,Rf} \sim {\cal O}(1)$, we would need $\tilde m_{\chi_\mu} \sim {\cal O}(10\tev)$, which is beyond the range of the LHC, but potentially within reach of the next generation of energy-frontier colliders.  

\paragraph{Anomalous magnetic moment of muon} The correction to $a_\mu \equiv (g_\mu -2)/2$ in this model is given  by~\cite{Dutta:2021afo}
\begin{eqnarray}
\delta a_\mu = 6.98 \times 10^{-7} \times ( V / 10\gev )^{-2} \times (C_{\phi'} - C_{A'} ) ,
\end{eqnarray}
where 
\begin{eqnarray}
C_{\phi'} =
\int_0^1 dx \frac{(1-x)^2(1+x)}{(1-x)^2+x r_{\phi'}^2} 
\quad \text{and}\quad 
C_{A'} =  \int_0^1 dx \frac{x(1-x)(2-x) r_{A'}^2 +x^3}{x^2+(1-x)r_{A^\prime}^2} ,
\end{eqnarray}
are the contributions from one-loop diagrams with the $\phi'$ and $A'$ in the loop, respectively,and $r_{\phi'} \equiv m_{\phi'}/m_\mu$, $r_{A'} \equiv m_{A'}/m_\mu$.  Note that the contributions of $\phi'$ and $A'$ are necessarily of opposite sign, because the $A'$ has both vector and axial couplings to the muon.  

Interestingly, the contribution of the $A'$ diagram to $\delta a_\mu$ is nearly universal, with $C_{A'}$ confined to lie between $1/2$ and $2/3$.  In particular, even at small coupling ($m_{A'} / V \ll 1$), although the transverse polarizations decouple, the contribution of the longitudinal mode remains unsuppressed; it becomes essentially the Goldstone mode, as expected from the Goldstone Equivalence theorem.  Moreover, if $m_{\phi'} \lesssim m_\mu$, then $C_{\phi'}$ is also an ${\cal O}(1)$ number.  This also is a result of the Goldstone Equivalence theorem.  $C_{A'}$ varies only slightly, but for small $m_{A'}$, $C_{A'}$ receives contributions only from the Goldstone mode $\sigma$.  Since the $\sigma$ and $\phi'$ have the same coupling, $C_{A'}$ and $C_{\phi'}$ must be comparable in magnitude and opposite in sign. If they cancel to within ${\cal O}(1\%)$, then this model is consistent with measurements of $g_\mu -2$.  Interestingly, this cancellation occurs in a region of parameter space which is not excluded by current experiments, but which can be probed by experiments at the FPF.

\paragraph{Current Constraints and Future Opportunities from Visible Decays} We discuss various constraints that are applicable for the scenario. We will focus here on the case in which the mediators $A'$ and $\phi'$ decay dominantly through the visible channels $A' \rightarrow e^+ e^-$ (the $\gamma \gamma$ channel is forbidden by the Landau-Yang theorem) or $\phi' \rightarrow \gamma \gamma$.  The decay widths for these processes are 
\begin{eqnarray}
\Gamma_{A' \rightarrow e^+ e^-} &=& 
\frac{\epsilon \alpha_{em} m_{A'} }{3} 
\left(1-\frac{4m_e^2}{m_{A'}^2} \right)^{-1/2} 
\left(1+\frac{2m_e^2}{m_{A'}^2} \right),
\\
\Gamma_{\phi' \rightarrow \gamma \gamma} &=& 
\frac{\alpha_{em}^2 m_\mu^4}{4\pi^3 V^2 m_{\phi'}} 
\left[1+\left(1-\frac{4m_\mu^2}{m_{\phi'}^2} \right)
\left(\sin^{-1} \frac{m_{\phi'}}{2m_\mu} \right)^2\right]^2,
\end{eqnarray}
where $\epsilon$ parameterizes the kinetic mixing between $U(1)_{T3R}$ and $U(1)_{em}$. In general, $\epsilon$ is a free parameter.We will take $\epsilon = g_{T3R} \sqrt{\alpha_{em} / 4\pi^3}$, which is the magnitude of the contribution one would get from a one-loop diagram involving a right-handed fermion.

Note that fixing $m_{A'}$, $m_{\phi'}$ and $V$ is sufficient to specify the mediator couplings, production cross sections, and decay rates.  In \cref{fig:VisibleSensitivity}, we plot bounds on this scenario in the $(m_{A'}, m_{\phi'})$-plane, setting $V = 10~\gev$ as a benchmark.  We see that for $m_{A', \phi'} > 2m_\mu$, this scenario is tightly constrained by searches at BaBar~\cite{Aubert:2009cp, Lees:2014xha, Bauer:2018onh} for prompt decays of the mediators to muons.  

But we see that for $m_{A',\phi'} < 2m_\mu$, there is open parameter space.~\footnote{There is also open parameter space when $\phi'$ is heavy enough that production at $B$-factories is suppressed.}  This region of parameter space lies above the ``ceiling" of current displaced detector searches. The reason, essentially, is that the lifetime of the mediators decreases with increasing mediator mass. Below the threshold for tree-level decay, the lifetime of the mediators is long enough for them to escape near detectors, but still short enough that they decay before reaching displaced detectors.  This window remains open until the decay lengths become long enough for the particles to reach existing displaced detectors, with leading current bounds being set by U70/NuCal~\cite{Gninenko:2014pea, Davier:1989wz, Bauer:2018onh} (in the case of $A'$) and E137~\cite{Riordan:1987aw, Bjorken:1988as, Bjorken:2009mm} (in the case of $\phi'$). Importantly, this open window includes the region in which constraints on $g_\mu -2$ are also satisfied. New instruments at FPF can probe this open window.

We focus here on sensitivity to the dark photon. An estimated sensitivity of experiments at FPF to this scenario can be extrapolated from a sensitivity to that of a secluded $U(1)$ as described in Ref.~\cite{Dutta:2020enk}, These sensitivities are plotted in \cref{fig:VisibleSensitivity}.

\begin{figure*}[t]
\centering
\includegraphics[width=0.8\textwidth, trim={0 0 0 1.5cm},clip]{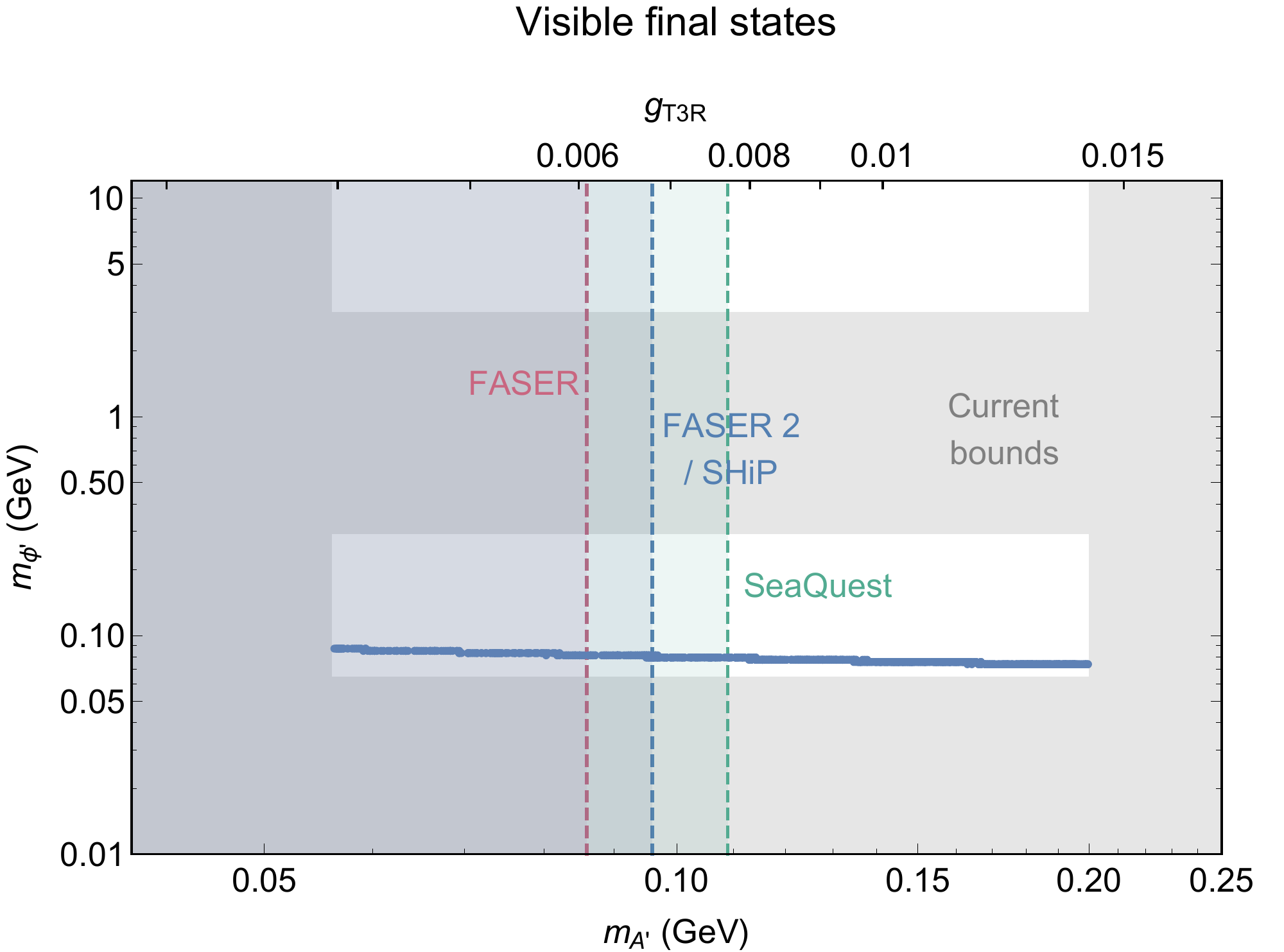}
\caption{\label{fig:VisibleSensitivity} Plot from Ref.~\cite{Dutta:2021afo} of the region in the $(m_{A'}, m_{\phi'})$-plane which is consistent with current measurements of $g_\mu-2$ (blue), along with current exclusion bounds 
(grey) from U70/NuCal~\cite{Gninenko:2014pea, Davier:1989wz, Bauer:2018onh}, E137~\cite{Riordan:1987aw,Bjorken:1988as, Bjorken:2009mm}, and Babar~\cite{Aubert:2009cp, Lees:2014xha, Bauer:2018onh}, and the future sensitivity of 
FASER~\cite{Feng:2017uoz, FASER:2018ceo, FASER:2018bac, FASER:2018eoc, FASER:2019aik} (red transparent), FASER2/SHiP~\cite{SHiP:2015vad, Alekhin:2015byh} (blue transparent) and SeaQuest~\cite{Berlin:2018pwi, Aidala:2017ofy} (green 
transparent).  $g_{T3R}$ is shown on the top axis.}
\end{figure*}

\paragraph{Interesting Features of this Scenario} We have focused on aspects of this scenario which are directly relevant to instruments at the FPF.  But it is worthwhile to consider the other interesting features of this scenario, relevant to other experimental approaches.

\begin{itemize}
 \item {\it Qualitatively new constraints.}
There is a large literature discussing laboratory, astrophysical, and cosmological constraints on models with a light dark photon or dark Higgs.  But if the new gauge group is $U(1)_{T3R}$, then there are some qualitatively different constraints~\cite{Dutta:2020jsy}.  Because the dark photon couples to right-handed SM fermions, the longitudinal polarization does not decouple from tree-level processes.  This yields an enhanced cross section for any process in which a hard dark photon is produced from a tree-level process.  As an example, we can consider the scenario where the muon is the only charged lepton coupling to the dark photon. This scenario is typically subject to much weaker constraints. But it has  been shown that, if the Universe reheats to a sufficiently high temperature ($\gtrsim 0.1\gev$), the coupling of the dark photon to right-handed muons would lead to enhanced production of the dark photon in the early Universe; constraints on $\Delta N_{eff}$ thus rule out such scenarios for $m_{A'} \lesssim 1\mev$ for {\it arbitrarily small coupling} unless the symmetry-breaking scale is $> {\cal O}(10^6)\gev$.  Recent astrophysical constraints on ALPs coupling to muons in  supernovae~\cite{Bollig:2020xdr,Croon:2020lrf} can easily be repurposed as constraints on the longitudinal polarization of the dark photon (equivalently, the Goldstone mode), and these constraints are comparable.  This constraints together place tight bounds on scenarios with $m_{A'} \lesssim 1\mev$.

\item {\it Direct detection.}  Direct detection experiments can play an especially interesting role.  Since the dark photon couples to up-type and down-type fermions with opposite charge, it leads to isospin-violating~\cite{Chang:2010yk, Feng:2011vu, Feng:2013vod} spin-independent scattering.  Moreover, since the dark matter candidate(s) are Majorana fermions, one necessarily has inelastic scattering.  Indeed, scattering via a dark photon is necessarily inelastic if the dark matter is charged only under spontaneously-broken continuous symmetries; in that case, the dark matter is generically a real degree of freedom, which cannot couple through a diagonal vector current. A variety of new techniques for probing low-mass dark matter are being studied, but few with the inelastic or isospin-violating scattering in mind.  Since this is a generic phenomenon, it would be good to study these prospects (for some recent work, see~\cite{Bell:2021zkr,Bell:2021xff}).

\item {\it Relic density.}
Despite tight constraints on low-mass dark matter annihilation from Planck~\cite{Planck:2015bpv, Planck:2018vyg}, there are regions of parameter space in which the dark matter candidate can have thermal relic density set by the freeze-out mechanism.  Bounds from Planck can be evaded by $p$-wave annihilation, and by co-annihilation, both of which can be realized in this scenario~\cite{Dutta:2019fxn}.  But it worthwhile to determine if other mechanisms can be found for generating the correct relic density, which can expand the viable parameter space.

\item {\it Flavor Anomalies.} 
Note, we have only considered a simple UV-completion in which we have added heavy fermions which couple to the fermions charged under $U(1)_{T3R}$.  But more generally, one could add more heavy fermions which couple to $b$ and $s$ in a similar manner.  This UV completion can introduce new processes contributing to $B_s \rightarrow \bar \mu \mu$ and to $B \rightarrow K^{(*)} \ell^+ \ell^-$, which could in turn explain anomalies seen in measurements of $R_{K^{(*)}}$~\cite{RKstar,Belle:2019oag, Aaij:2021vac}. This model successfully explains both $R_{K^{(*)}}$ and $g_\mu-2$ in allowed region of the parameter space~\cite{Dutta:2021afo}.

\item {\it Other FPF detection possibilities.} The new scalar and gauge boson can be produced from charged pion 3-body decays, i.e., $\pi \rightarrow \ell \nu_\ell A'(\phi')$. The three-body decay mode is not helicity-suppressed, unlike the two-body decay mode, and hence can be large. The couplings of $A'$ and $\phi$ must obey experimental constraints on various charged pion decays. Recently, the three body decay of charged pion has been utilized to explain the MiniBooNE excess~\cite{Dutta:2021cip} which can potentially be accommodated in this model. The $A'$ also can be produced from the neutral meson decays. The FPFs  can explore both the light scalar and the gauge boson of his model.

We have mostly focused on the case in which the mediators decay visibly.  But FPF experiments an also probe the scenario in which the mediators decay invisibly through $A' \rightarrow \eta_1 \eta_2, \nu_R \nu_R$, $\phi' \rightarrow \eta_1 \eta_1, \nu_L \nu_R$.  In this case, the decay products may scatter against a distant target, producing events in FPF detectors such as FASER-$\nu$.
\end{itemize}

\begin{figure*}[t]
\centering
\includegraphics[width=0.8\textwidth]{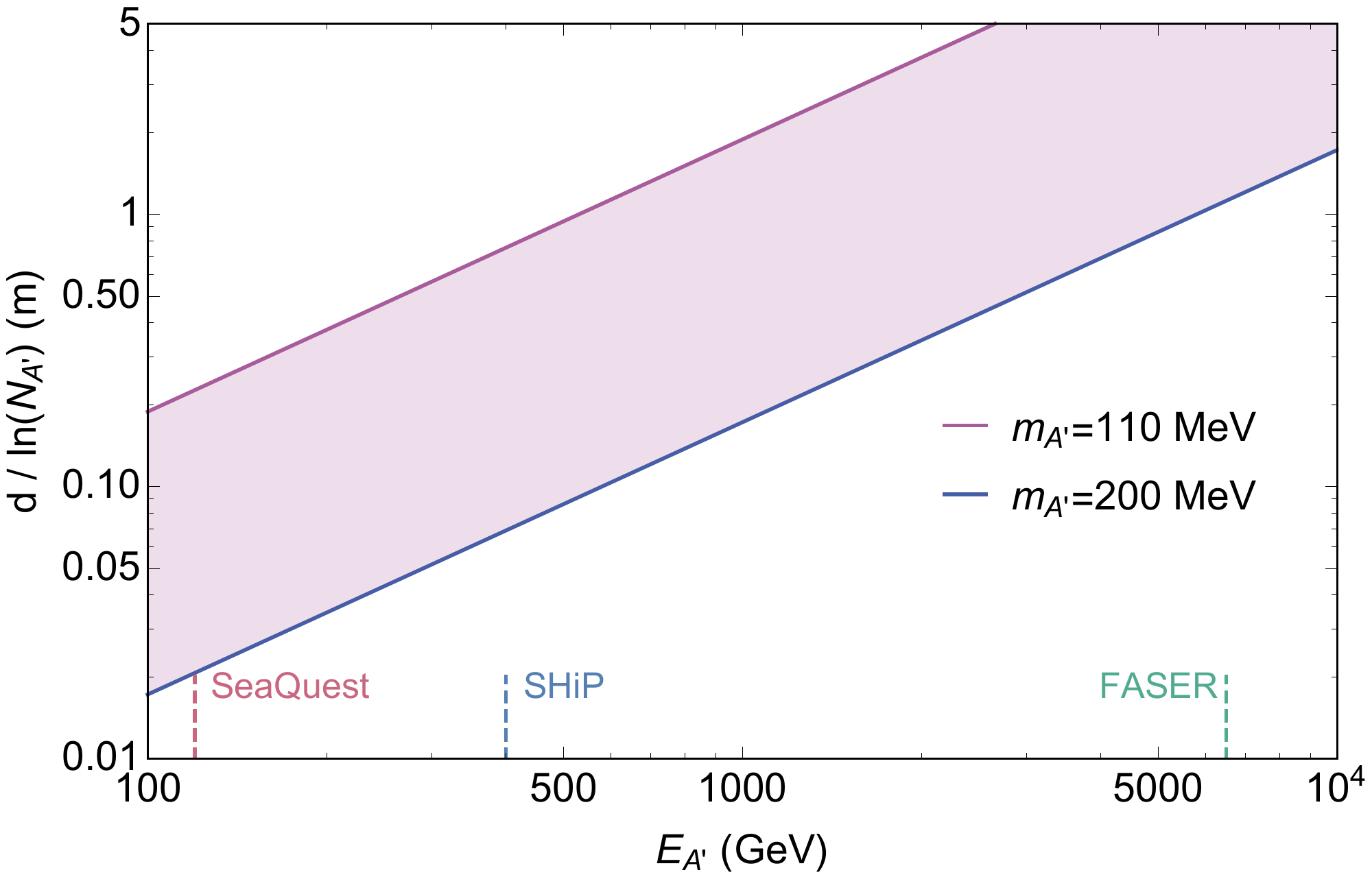}
\caption{\label{fig:dvsEAplot} A rough estimate of maximum $d/\ln(N_{A'})$ necessary for an experiment to be able to probe this scenario for $m_{A'} \in [110\mev, 200\mev]$, as a function of the maximum $A'$ energy produced by the experiment~\cite{Dutta:2021afo}. $d$ is the displacement of the detector from the beam dump, and $N_{A'}$ is the number of $A'$ at energy $E_{A'}$ produced in a beam aimed at the detector.  The maximum $A'$ energies of FASER, SHiP and SeaQuest are also shown.}
\end{figure*}

\paragraph{Conclusion} Note that, for $m_{A'} \sim 100~\mev$, a dark photon with energy $E \sim 5000~\gev$ would have a decay length of ${\cal O}(10\m)$.  Thus, one could see a significant variation of the $A'$ flux within the FPF itself, and it would be beneficial to place the detector as close to the LHC interaction point as possible.  We have assumed here that the FPF is located at the UJ12 cavern, with the detector being $480\m$ from the interaction point.  If the FPF were instead located at a proposed purpose-built facility located $\sim 620\m$ from the interaction point, the $A'$ flux, and resulting sensitivity, would be substantially reduced. The FPF can  probe both light mediators of this model,  $A'$ and $\phi'$.

To set a context for future searches for the $A'$ in this scenario, we consider an experiment which can produce $N_{A'}$ dark photons of energy $E_{A'}$ in the direction of a detector a distance $d$ away.  The number of dark photons which actually reach the detector is $\sim N_{A'} \exp [-d/d_{dec}(E_{A'})]$,  where $d_{dec}(E_{A'})$ is the decay length of a dark photon of energy $E_{A'}$, and is determined entirely by the model.  We consider the regime in which a large fraction of such dark photons decay within the detector and are observed, whereas the background is negligible.  In this case, a evidence for new physics can be found if an ${\cal O}(1)$ number of photons reach the detector, implying $d_{dec} \sim d / \ln (N_{A'})$, where the right-hand side of this relation is determined entirely by the experiment. \cref{fig:dvsEAplot}, we plot the range of $d / \ln (N_{A'})$ which would be needed, as a function of $E_{A'}$ to probe the window which would still be left open by experiments such as FASER2, SHiP and SeaQuest.

\subsection{Dark Axion Portal\label{sec:bsm_nonmin_darkaxion}}

\paragraph{Dark axion portal} Among various portals that connect the standard model sector and the dark sector are those that involve the photon. The axion portal and the vector portal provide some of the most popular connections between a possible dark sector and the Standard Model. When the axion ($a$) and dark photon ($\gamma'$) coexist, together with a photon ($\gamma$), they can form the dark axion portal as~\cite{Kaneta:2016wvf}
\begin{equation}
    \mathcal{L}_\text{dark axion portal} = \frac{G_{a\gamma\gamma'}}{2}aF_{\mu\nu}\tilde X^{\mu\nu} + \frac{G_{a\gamma'\gamma'}}{4}aX_{\mu\nu}\tilde X^{\mu\nu},
\end{equation}
where $X^{\mu\nu}$ is the field strength of the dark photon. Though the second term ($a$-$\gamma'$-$\gamma'$ vertex) is not a portal relating the Standard Model sector and the dark sector, once the first term ($a$-$\gamma$-$\gamma'$ vertex) is introduced, the second term follows naturally. Note that these are not simply a combination of the vector portal and the axion portal, but rather exploit the dark gauge couplings \cite{Kaneta:2016wvf}.

To take a model-independent approach, we take $m_a \ll m_{\gamma^\prime}$, and assume no kinetic mixing ($\varepsilon = 0$). We also assume that the effect of the axion portal ($G_{a\gamma\gamma}$) is negligibly small in our experimental probes. It is important to note that the dark axion portal is not closed even if the vector portal is because the dark axion portal does not rely on kinetic mixing.

\begin{figure*}[t]
\centering
  \includegraphics[width=0.5\textwidth]{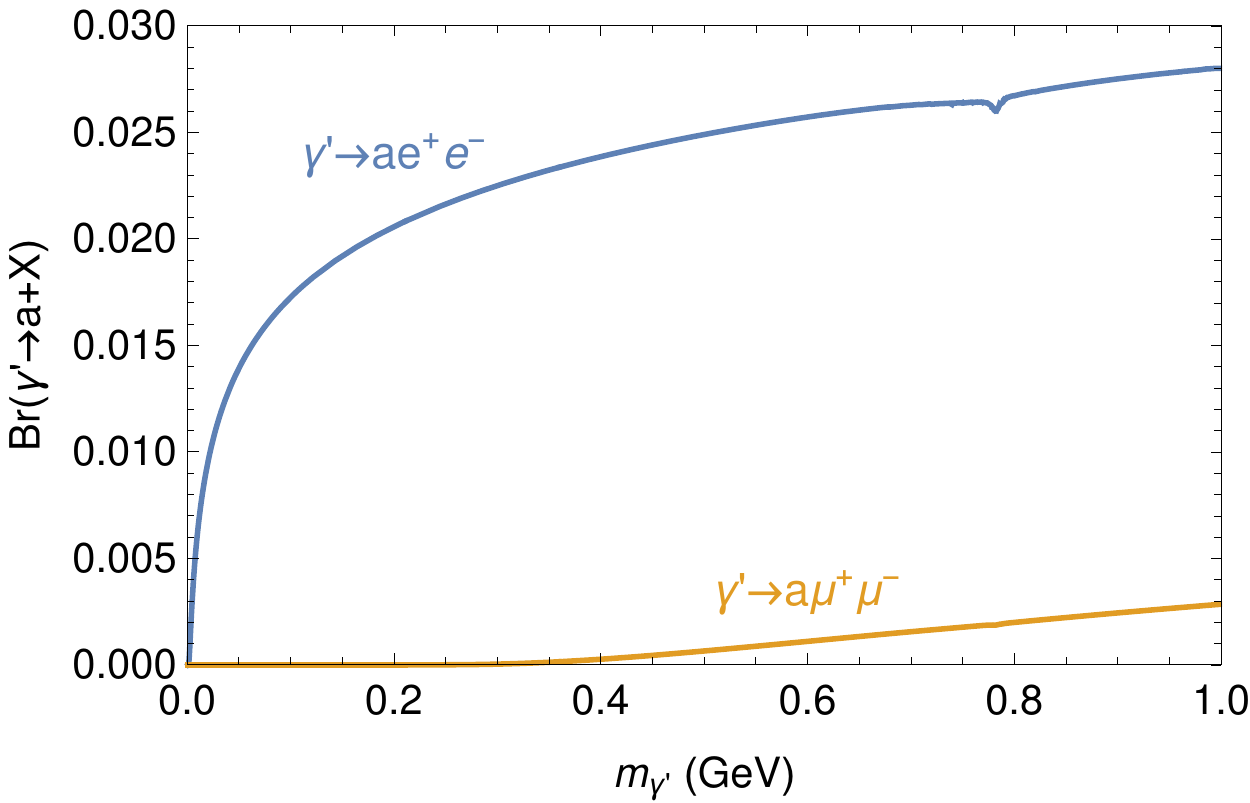}
  \includegraphics[width=0.4\textwidth, trim={0 -1cm 0 0}, clip,]{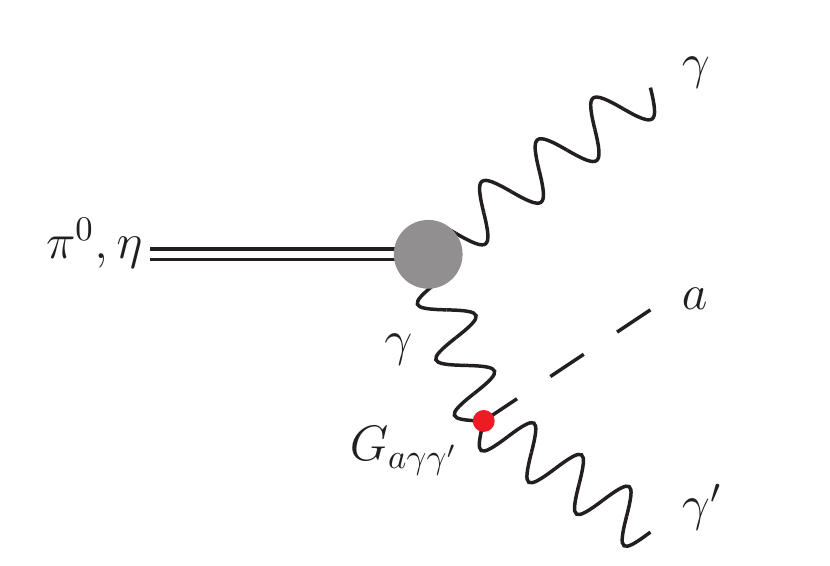}
\caption{Left: The branching ratio of $\gamma^\prime \to a e^+ e^-$ and $\gamma^\prime \to a \mu^+ \mu^-$ through the $a$-$\gamma$-$\gamma'$ vertex in the $m_a \ll m_{\gamma'}$ limit. The slight dip near 0.8\,GeV is due to the hadronic decay channels. Figure taken from Ref.~\cite{deNiverville:2019xsx}. Right: Decay of the pseudoscalar mesons $\pi^0$ and $\eta$ to $\gamma a \gamma^\prime$ through the $a$-$\gamma$-$\gamma'$ vertex. Figure taken from Ref.~\cite{deNiverville:2018hrc}.}
 \label{fig:A_to_aee}
\end{figure*}

The dark photon may decay through a number of different channels. The two-body decay $\gamma^\prime \to a\gamma$ is the dominant channel with a decay width
\begin{equation}
 \Gamma(\gamma^\prime\to\gamma a) = \frac{G_{a\gamma \gamma^\prime}^2}{96\pi} m_{\gamma^\prime}^3\left(1-\frac{m_a^2}{m_{\gamma^\prime}^2} \right)^3.
\end{equation}
The three-body decay processes $\gamma^\prime \to e^+ e^- a$ and $\gamma^\prime \to \mu^+ \mu^- a$ are also possible, with a branching fraction of a few percent as shown in the left panel of \cref{fig:A_to_aee}. Although subdominant, these channels provide a signature very similar to the $\ell^+\ell^-$ signal used to search for the dark photons through the vector portal, and those searches can be repurposed as probes of the dark axion portal.

\paragraph{Production and detection} The $a$ and the $\gamma^\prime$ could be produced in radiative decays of the pseudoscalar mesons $\pi^0$ and $\eta$ through the off-shell photon and the dark axion portal ($G_{a\gamma\gamma'}$) as shown in the right panel of \cref{fig:A_to_aee}. They provide an important source of dark photons whose visible decay products could be detected in the FASER detector. The partial decay width of the decay $\pi^0 \to a \gamma \gamma^\prime$ is given by \cite{deNiverville:2018hrc}
\begin{equation}
 \frac{d^2\Gamma}{dm_{12}^2 dm_{23}^2} = \frac{1}{(2\pi)^3}\frac{1}{32m_{\pi^0}^3}\overline{|\mathcal{M}|^2},
\end{equation}
where $m_{ij}^2=(p_i+p_j)^2$ for $i,j=1,2,3$, where particle 1 corresponds to the $\gamma$, particle 2 is the $a$ and particle 3 is the $\gamma^\prime$, and the amplitude is
\begin{align*}
 \overline{|\mathcal{M}|^2} &= \frac{e^4 G_{a\gamma\gamma^\prime}^2}{64 \pi^4 f_\pi^2 m_{23}^2} \bigg[m_{23}^2 \left(m_{12}^2+m_{23}^2-m_a^2-m_{\pi^0}^2 \right)^2-m_{23}^2\left(m_{23}^2-m_{\pi^0}^2\right) \left(m_{23}^2-m_a^2+m_{\gamma^\prime}^2\right)\\
&\times\left(m_{12}^2+m_{23}^2-m_a^2-m_{\pi^0}^2\right)+\frac{1}{2}\left(m_{23}^2-m_{\pi^0}^2\right)\left(m_{23}^2-m_a^2+m_{\gamma^\prime}^2\right)^2-m_{23}^2 m_{\gamma^\prime}^2\left(m_{23}^2-m_{\pi^0}^2\right)^2\bigg] .
\end{align*}
The same expression holds for $\eta$, but with $m_{\pi^0}$ replaced by $m_{\eta}$.

The detection search was performed by generating a list of $\eta$ and $\pi^0$ 4 momenta and production locations with some external code (the choice for FASER is described in the next section). As the lifetime of the $\pi^0$ and $\eta$ are extremely short at $\mathcal{O}(10^{-17}\,\mathrm{s})$ or less \cite{ParticleDataGroup:2020ssz}, they do not propagate a significant distance before decaying. We simulate the three-body decay $\pi^0,\eta\to a\gamma\gamma^\prime$ as described in Ref.~\cite{deNiverville:2018hrc}, discarding the $\gamma$ and $a$ 4-momenta as they do not contribute to the signal. The resulting list of $\gamma^\prime$ 4-momenta is used to calculate the expected dark axion portal signal.

The probability that a $\gamma^\prime$ with label $i$ decays inside the decay pipe through an observable channel is given by
\begin{equation}
\label{eq:dec_prob}
 P_\mathrm{decay,i} = \mathrm{Br}_X \left[\exp\left(-\frac{L_{1,i} E_i}{c\tau m_{\gamma^\prime}}\right) - \exp\left(-\frac{L_{2,i} E_i}{c\tau m_{\gamma^\prime}}\right)\right],
\end{equation}
where ${Br}_X=Br(\gamma^\prime\to a \ell^+ \ell^-)$ is the probability that the dark sector particle decays visibly to a pair of leptons (specifically electrons in this case); $E_i$ is the energy of the $\gamma^\prime$; $\tau$ is the lifetime of a $\gamma^\prime$ with mass $m_{\gamma^\prime}$ and coupling strength $G_{a\gamma\gamma^\prime}$; $L_{1,i}$ is the distance the $\gamma^\prime$ propagates before entering the decay volume; and $L_{2,i}$ is the total distance traveled before exiting the decay volume.
    
Each $\gamma^\prime$ is decayed into an $a e^+ e^-$ final state, while one could also search for muon final states in an identical manner. For this state to be accepted, both leptons must intersect with the end cap of the decay volume to enter the calorimeters. The total event rate expected from the decays of meson $j$ can be calculated as
\begin{equation}
 \label{eq:decay_event_rate}
 N_{\mathrm{event},j} = \frac{N_j \epsilon_\mathrm{eff}}{N_\mathrm{trials}} \mathrm{Br}(j\to a \gamma \gamma^\prime) \sum_i P_\mathrm{decay,i} \theta(p_{e^+,i},p_{e^-,i}),
\end{equation}
where $j=\pi^0,\eta$; $P_\mathrm{decay,i}=0$ if the $\gamma^\prime$ does not intersect the detector; $\theta(p_{e^+,i},p_{e^-,i})=1$ if the end-state leptons satisfy the cuts mentioned above and 0 otherwise; $\epsilon_\mathrm{eff}=1$ is the detection efficiency; and $N_\mathrm{trials}$ is the total number of $\gamma^\prime$ trajectories generated. The total event rate is found by summing over all generated events $j$.

\begin{figure*}[t]
  \centering
  \includegraphics[width=0.7\textwidth]{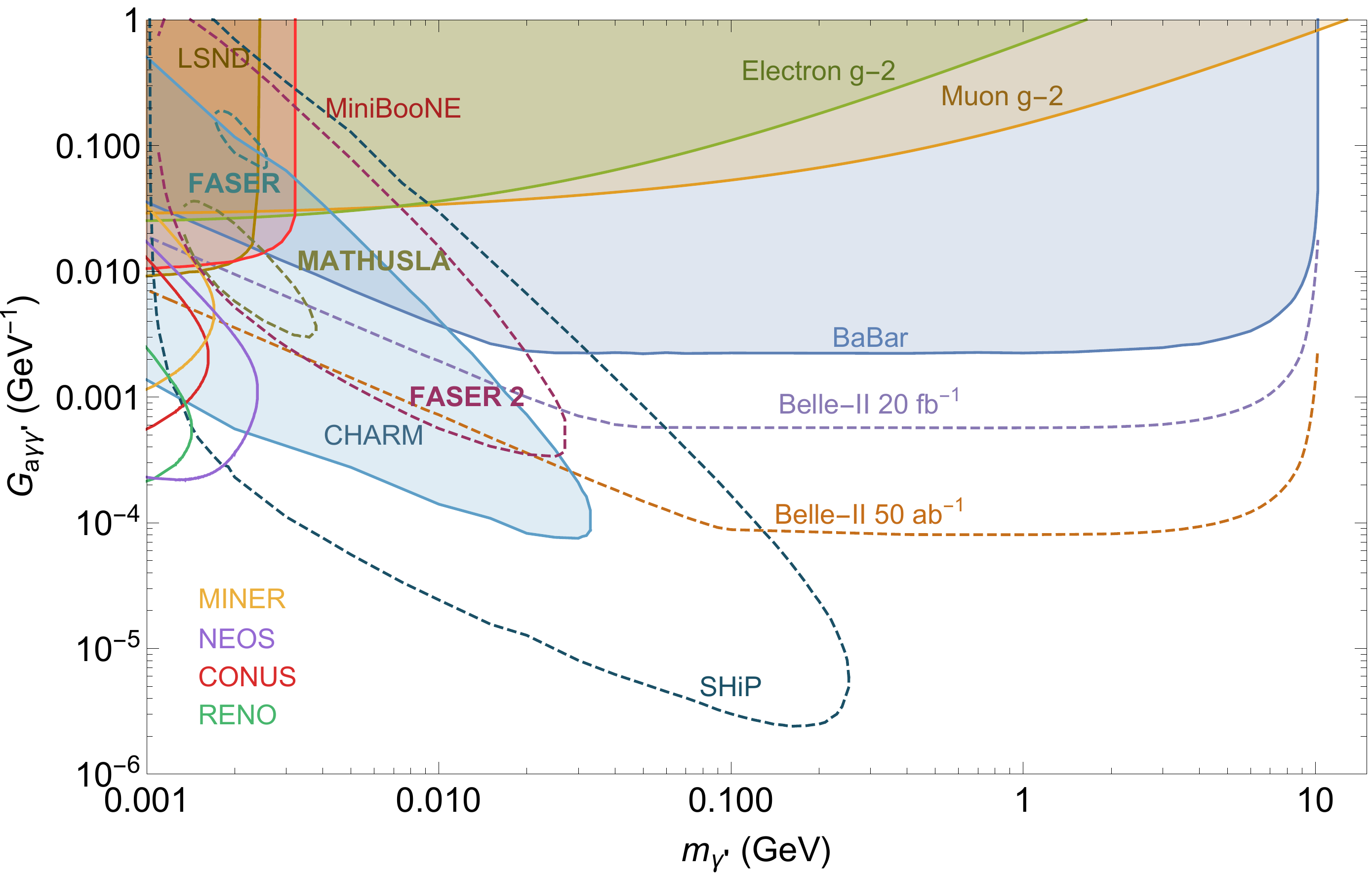}
  \caption{(Plot taken from \cite{Deniverville:2020rbv}.) Limits on the dark axion portal $G_{a\gamma\gamma^\prime}$ from high intensity experiments and the projected sensitivity of the ongoing or future experiments, for $m_a \ll m_{\gamma^\prime}$ with $\varepsilon = 0$.
  The FASER, FASER2 and SHiP projections denote excesses of greater than three $e^+ e^-$ events produced through the decay $\gamma^\prime \to a e^+ e^-$ reaching the end of their respective decay volumes. The limits from LSND and MiniBooNE come from excess neutral current-like elastic scattering events off electrons. The CHARM constraint reflects sensitivity in monophoton production through $\gamma^\prime \to a\gamma$ decays in the CHARM fine-grain detector. The electron and muon $g-2$ lines indicate where the scenario would significantly degrade the agreement between theory and experiment, though the muon line was generated before the release of the Muon g-2 2021 measurement \cite{Muong-2:2021ojo}. The BaBar and Belle-II lines all represent exclusions from the monophoton excesses through the annihilation process $e^+ e^- \to a (\gamma^\prime \to a \gamma)$. Expected sensitivities of the reactor experiments at CONUS, MINER, RENO, and NEOS at 95\% C.L. for one year of data are also shown. These limits were taken from Refs.~\cite{deNiverville:2018hrc,deNiverville:2019xsx,Deniverville:2020rbv} where further details on the limits are provided.}
\label{fig:DAP}
\end{figure*}

\paragraph{Sensitivity at FASER} The FASER collaboration has performed a preliminary analysis of its sensitivity to visible dark photon decays in Refs.~\cite{Feng:2017uoz,FASER:2018eoc}, and we have adapted these searches to the dark axion portal. The inelastic proton-proton cross section was measured to be $\sigma_\mathrm{inelastic} \sim 75\,\mathrm{mb}$ during the 13 TeV LHC run, and is not expected to differ greatly for the 14 TeV collisions of LHC run 3. The total number of inelastic collisions is therefore expected to be $N_\mathrm{inelastic} \approx 1.1\times 10^{16}$ for 150\,fb$^{-1}$ in LHC run 3.

As described in the previous subsection, we expect $\pi^0$ and $\eta$ decays to provide the primary source of dark axion portal particles. The $\pi^0$ and $\eta$ production rates and distributions were simulated with \texttool{EPOS-LHC} \cite{Pierog:2013ria} through the \texttool{CRMC~v1.7} framework \cite{CRMC}. Note that these rates were also compared with \texttool{SIBYLL~v2.3} and found to be consistent at the small angles required by FASER \cite{Ahn:2009wx, Riehn:2015oba}. The total number of $\pi^0$s ($\eta$s) produced per interaction in one hemisphere of the interaction point was calculated to be 19 (2.1). The production estimate is conservative, as there should be secondary meson production from other collision products impacting on material between the interaction point and the FASER detector, though this is likely to be less energetic than the primary products.

For this analysis, FASER is assumed to be a 1.5\,m long cylindrical decay region with 10\,cm radius located 480\,m from the interaction point operating during LHC run 3. We also considered the sensitivity of a hypothetical FASER2 detector, a 5\,m cylinder with 1\,m radius located 480\,m downstream from the interaction point. FASER2 would take data during the high luminosity LHC runs with an expected luminosity twenty times larger than that of LHC run 3.

FASER is sensitive to the dark axion portal decay $\gamma^\prime \to a e^+ e^-$, as it possesses a signature very similar to that of the kinetically mixed dark photon. We impose the same cuts as the FASER dark photon analysis: The dark photon must decay in the decay volume, both the electron and positron must cross through the downstream face of the decay volume, and the total visible energy must satisfy $E_{e^+}+E_{e^-}>100\,\mathrm{GeV}$. If we only consider mesons with energies greater than $100\,\mathrm{GeV}$, the number of $\pi^0$s ($\eta$s) per POT drops to 2.43 (0.43) in the hemisphere facing FASER.

Assuming negligible background, we exclude parameter space predicted to generate more than three events. We show the resulting contours in  \cref{fig:DAP}, where the small inner contour represents the sensitivity of FASER and the larger outer contour that of FASER2. The comparatively low luminosity and decay volume hampers the ability of FASER to probe the scenario compared to beam dump experiments. FASER2, with its much larger volume, is capable of excluding new parameter space. Improving on this search with a monophoton analysis would be challenging, as CHARM already excludes the region where the most improvement is expected.

We did not consider the $a \mu^+ \mu^-$ final state, nor the possibility of bremsstrahlung $\gamma^\prime$ production through the dark axion portal. Their inclusion could extend the sensitivity to slightly larger values of $m_{\gamma^\prime}$. One could also consider the combined vector and dark axion portal scenario such as the production of dark photons through the vector portal and decay to a monophoton through $\gamma^\prime \to a \gamma$.

\subsection{Heavy Neutrino Production via a $B-L$ Gauge Boson\label{sec:bsm_nonmin_BmLZprime}}

\paragraph{Introduction} Within the $B-L$ model \cite{Mohapatra:1980qe}, we consider the possibility of a pair of heavy RH neutrinos $N$ to be produced via the associated $Z'$ gauge boson~\cite{Deppisch:2019kvs}. The model, based on the gauge symmetry $SU(3)\times SU(2)_L \times U(1)_Y \times U(1)_{B-L}$, is a natural extension of the SM incorporating the spontaneous breaking of lepton number and thus generating heavy Majorana neutrino masses $m_N$ analogous to the SM Higgs mechanism. In turn, the active neutrinos acquire light masses $m_\nu \lesssim \mathcal{O}(0.1~\text{eV})$ via the Seesaw mechanism. We focus on $Z'$ and $N$ that are lighter than $\approx 20$~GeV such that the produced final states can be detected at the FPF.

\begin{figure*}[t]
	\centering
	\includegraphics[width=0.60\textwidth]{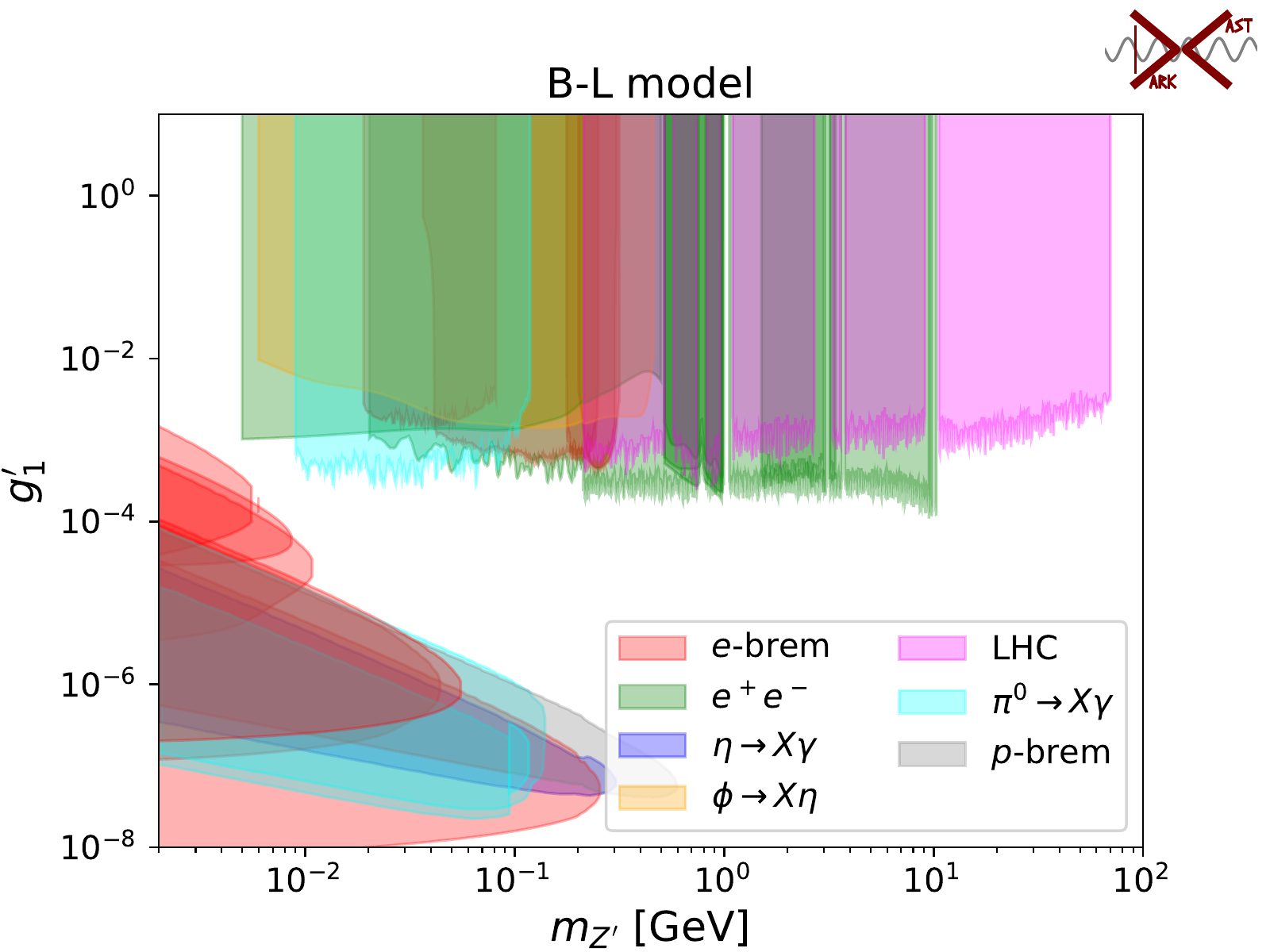}
	\caption{Limits on the $B-L$ gauge coupling $g_{B-L}$ as a function of the associated gauge boson mass $m_{Z'}$, derived from decays to SM final states for a $B-L$ model containing three generations of heavy neutrinos. Taken from \cite{Deppisch:2019kvs}, using the tool \texttool{ DarkCast}~\cite{Ilten:2018crw}.}
	\label{fig:b-l-zprime-limits}
\end{figure*}

The $Z'$ gauge boson couples to the SM fermions and the RH neutrinos according to their $B-L$ charges $Y^f_{B-L}$,
\begin{align}
    \mathcal{L} \supset -g_{B-L} Z'_\mu \sum_f Y^f_{B-L} \bar{f}\gamma^\mu f.
\end{align}
Such a $Z'$ gauge boson has been widely searched for and the current limits on the associated gauge coupling $g_{B-L}$ are shown in \cref{fig:b-l-zprime-limits}. Above $m_{Z'} \gtrsim 10$~GeV, existing limits from LHC searches are of the order $g_{B-L} = 10^{-3}$. The $B-L$ model includes three generations of heavy neutrinos $N_i$ with active-sterile mixing strengths $V_{\ell N_i}$ to the SM lepton states $\ell$. For simplicity we assume that only one of the sterile neutrinos is accessible and it couples dominantly to muon flavour via $V_{\mu N}$. If $m_N < m_{Z'}/2$, heavy neutrinos can be produced in the decay $Z'\to N N$. As benchmark we choose a fixed ratio $m_N/m_{Z'} = 0.3$ which makes the heavy neutrinos sufficiently light to avoid threshold effects but keeps them at a similar scale as the $Z'$. The heavy neutrino in this scenario decays dominantly through three-body final states such as $N\to \mu q\bar{q}$ and $N\to \mu\mu\nu_\mu$ via off-shell SM $W$ and $Z$. The branching ratios of the $Z'$ and $N$ are discussed in Ref.~\cite{Deppisch:2018eth}. The heavy neutrino is naturally expected to be long-lived with an approximate proper decay length of
\begin{align}
\label{lengthapproxi}
	L_N \approx 25~\text{m}
	\cdot \left(\frac{10^{-5}}{|V_{\mu N}|}\right)^2 
	\cdot \left(\frac{10~\text{GeV}}{m_N}\right)^5,
\end{align} 
for $m_N \lesssim m_Z$. The required masses and active-sterile mixing to get relevant decay lengths are thus naturally of the order expected to generate the light neutrino masses in a canonical Seesaw mechanism, $m_\nu \sim |V_{\mu N}|^2 m_N$.

\paragraph{Detection of long-lived heavy neutrinos} The signal under consideration consists of the process $pp \to Z' \to NN$ where one of the heavy neutrinos decays to a visible final state. To estimate the sensitivity of the the proposed displaced vertex detectors CODEX-b~\cite{Berlin:2018jbm}, FASER~\cite{FASER:2018eoc}, MAPP~\cite{Pinfold:2019nqj} and MATHUSLA~\cite{Chou:2016lxi}, we assume that every heavy neutrino decay is detected, except those that are fully invisible. We also compare the sensitivity with that of displaced vertex searches at the LHC detectors LHCb and CMS where we instead consider the exclusive decays $N\to \mu q\bar{q}$ and $N\to \mu\mu\nu_\mu$, respectively. We include the exponential decay probability based on the proper decays length of the heavy neutrino and its boost according to the production process. Simulated events are passed through the kinematic and geometric selection criteria corresponding to the different detectors.

As an example, the FASER detector is placed 480 meters from the interaction point in the forward direction. Its two-phase design proposes two different detector geometries. The FASER experiment operating during LHC Run 3 will not have sufficient sensitivity towards heavy neutrinos in our model and we thus consider FASER2 with a detector volume modelled as a cylinder with radius of 1~m and a length of 5~m along the forward direction. We assume that it runs in conjunction with the high luminosity LHC with an integrated luminosity of 3000~fb$^{-1}$. We use the same luminosity for CMS and MATHUSLA, and a luminosity of 300~fb$^{-1}$ for LHCb, MAPP and CODEX-b. The geometries of the detectors are described in detail in Ref.~\cite{Deppisch:2019kvs}. We take an optimistic view assuming no backgrounds for the displaced vertex signatures. For detectors located relatively far away from the interaction point, i.e., FASER2, MAPP, MATHUSLA and CODEX-b, this choice is better justified. 

\begin{figure*}[t]
\centering
\includegraphics[width=0.49\textwidth]{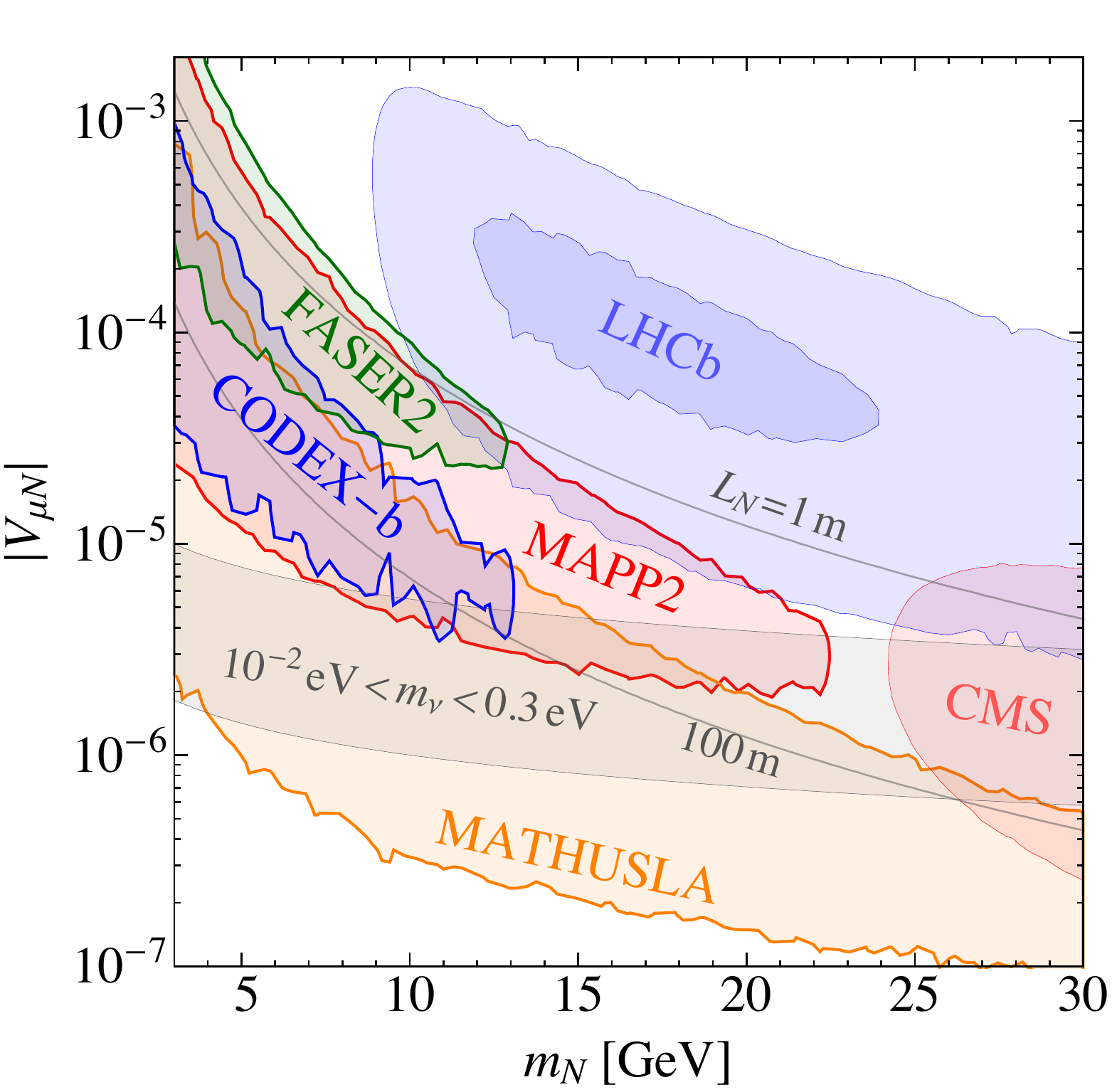}
\caption{Projected sensitivity of displaced vertex detectors at $95\%$~CL towards long-lived heavy neutrinos with mass $m_N$ and active-sterile mixing $V_{\mu N}$ in the $B-L$ model. The corresponding gauge coupling strength $g_{B-L}$ and $Z'$ mass are chosen as $g_{B-L} = 10^{-3}$ and $m_{Z'} = 3.33\times m_N$, respectively. The displaced vertex detectors are as indicated, with events produced in 14~TeV $pp$ collisions and using projected luminosities detailed in the text. We assume no background to displaced vertex searches, except for the darker LHCb region which applies for a pessimistic background estimate from data \cite{Deppisch:2019kvs}. The grey curves denote proper decay lengths of the heavy neutrino $N$ and the shaded band indicates the parameter region where the light neutrinos acquire a mass between $10^{-2}$~eV and 0.3~eV in a canonical Seesaw mechanism.}
\label{fig:b-l-n-sensitivities}
\end{figure*}

\paragraph{Sensitivity to active-sterile mixing} To estimate the sensitivity on the active-sterile mixing $V_{\mu N}$ as a function of the heavy neutrino mass $m_N$, we choose $g_{B-L} = 10^{-3}$ and $m_{Z'} = 3.33\times m_N$ as motivated above. Following the Poisson distribution, the non-observation of any signal event excludes the parameter space with an expected number of events above 3.09 at 95~$\%$ CL. The resulting sensitivities are shown in \cref{fig:b-l-n-sensitivities}. The horizontal band represents the preferred parameter space where the light neutrinos acquire a mass scale $m_\nu = |V_{\mu N}|^2 m_N$ in the range $10^{-2}~\text{eV} < m_\nu < 0.3$~eV. The lower limit is motivated by the observed mass splitting in oscillation experiments and the upper limit is due to constraints by direct neutrino mass measurements and cosmological observations. This band thus corresponds to the successful generation of light neutrino masses; above it, the light neutrinos will be too heavy, though this can be avoided in more involved scenarios with quasi-Dirac heavy neutrinos; below the band, the light neutrinos will be too light. In any case, the band should only be taken as an indication as fitting all light neutrino masses and oscillation mixing angles requires an analysis with three generations of heavy neutrinos.

It is nevertheless worth noting that the different proposed facilities can probe interesting parameter regions close to or overlapping with this preferred band, in a complementary way. Due to their geometries and distances to the interaction point, CODEX-b, FASER2 and MAPP are sensitive to small heavy neutrino masses $m_N \lesssim 20$~GeV and larger mixing $|V_{\mu N}| \gtrsim 10^{-5}$. Mainly due to its large distance and size, MATHUSLA can probe smaller mixing and larger masses, well motivated by light neutrinos. As can be seen, these dedicated displaced vertex detectors provide a complementary coverage compared to displaced searches at the LHCb and CMS detectors.

\paragraph{Discussion} This shows that displaced vertex searches in general, and dedicated facilities specifically, can probe the origin of neutrino masses and discover the underlying New Physics states. We have here concentrated on a scenario based on a $U(1)_{B-L}$ extension of the SM which provides a simple yet well motivated mechanism generating light neutrino masses. It exploits the natural expectation that the mediators of this mechanism, i.e., the heavy Majorana neutrinos, are long-lived due to the suppressed active-sterile mixing required to explain the light neutrino masses $\lesssim \mathcal{O}(0.1~\text{eV})$. Besides being well motivated, the $B-L$ model has the phenomenological advantage that the heavy neutrinos are not sterile but are charged under the $U(1)_{B-L}$. Hence, they can be produced with much larger rates than expected from the small active-sterile mixing if they were completely sterile, where it is challenging to probe the preferred parameter region~\cite{Bolton:2019pcu}. Searches for heavy neutrinos in the $B-L$ model should be considered in conjunction with $Z'$ searches~\cite{Ilten:2018crw, Amrith:2018yfb} that limit the exotic gauge coupling and thus the production rate. We finally note that the $B-L$ model also includes an exotic Higgs state, associated with the breaking the $B-L$ symmetry, that can facilitate the production of long-lived heavy neutrinos~\cite{Deppisch:2018eth} and even $Z'$~\cite{Deppisch:2019ldi}.

\subsection{Search for Sterile Neutrino with Light Gauge Interactions\label{sec:bsm_nonmin_sterilegauge}}

\paragraph{Model} A sterile neutrino ($\nu_s$) is assumed to be a SM gauge singlet, but could have originated from a dark fermion coupled to dark photon \cite{Ko:2014bka}.  Let us consider a dark sector with $U(1)_s$ dark gauge symmetry, fermion dark matter (DM) $\chi$ and a dark fermion  $\psi$. We can assign different $U(1)_s$ charges to $\chi$ and $\psi$ in such a way that $\chi$ is stable while $\psi$ mixes with the SM active neutrinos $\nu_\alpha$ and the right-handed neutrino $N_R$ through Yukawa couplings (see Eq. \cref{eq:ko_Yukawa} below). $U(1)_s$ gauge boson $X$ gets massive from the nonzero VEV of $\phi_X$ (dark Higgs field). Some problems ({\it e.g. core-cusp problem, etc.}) of the vanilla cold dark matter paradigm can be resolved  using the DM self-scattering $\chi\bar{\chi}\rightarrow \chi\bar{\chi}$ through light $X$ exchange, if the dark photon $X$ mass is $\sim O(1)$MeV.

We choose dark charges of $\chi, \psi, \phi_X$ in such a way that only the 
following Yukawa coulings are allowed ($Q_\psi = Q_{\phi_X} = 1 \neq Q_\chi$):
\begin{equation}
    \mathcal{L} \supset - y_{i \alpha} \overline{N_{R,i}} 
    \tilde{H}^\dagger L_\alpha 
    - f_i \phi_X^\dagger \overline{N_i^C} \psi + g_i \phi_X \bar{\psi} N_i 
    + H.c.  
    \label{eq:ko_Yukawa}
\end{equation}
Then there will be mixings among $\nu_\alpha$ (SM active neutrinos), $N_R$ (sterile neutrinos) and $\psi$ (dark fermion) after the electroweak and dark gauge symmetry breaking. And the physical sterile neutrino $\nu_s$ is a mixture of these three objects, and can couple to the dark photon through its $\psi$ component. The model can address the LSND $\sim O(1)$eV sterile neutrino and also contributes some amount of $\Delta N_{\rm eff}$ that could relieve the Hubble tension~\cite{Ko:2014bka}. One can also address the MiniBooNE anomaly within the same model for different mass scales of dark photon and sterile neutrino~\cite{Bertuzzo:2018itn, Ballett:2018ynz}.

For phenomenological study at FASER etc., we can simplify the gauge sector of the above model to the following `$\nu_4$+$X$' model. $U(1)_s$ gauge boson $X$ couples to the active neutrinos as well as the charged SM fermions via active-sterile mixing $U_{\ell4}$ and the hypercharge-$U(1)_s$ kinetic mixing $\epsilon$, respectively. The effective interaction at low energies is given by
\begin{eqnarray}
\mathcal{L} & \supset & - g_X U_{\ell 4} \bar{\nu}_\ell \gamma^\mu \nu_4 X_\mu - g_X \epsilon \cos \theta_W Q_f \bar{f} \gamma^\mu f X_\mu, \ \ \ \label{modelB_nu_int}
\end{eqnarray}
where $\ell=\mu, \tau$ and $\nu_4$ is a nearly sterile neutrino in the mass eigenbasis. This `$\nu_4$+$X$' models can be probed by various searches at FASER, IceCube using double-bang topology in atmospheric and astrophysical neutrino data, and previous fixed-target searches and rare meson decays.

\paragraph{LLP searches} The expected number of events $N_{\rm sig}$ is
\begin{eqnarray}
N_{\rm sig.} & = & \int_{E_{\min}}^{E_{\max}} dE_{\nu_4} \left [ \frac{dN_{\nu_4}(E_{\nu_4})}{dE_{\nu_4}} \times \left ( e^{- \frac{L-\Delta}{d} } - e^{- \frac{L}{d} } \right ) \times \text{Br}(X /\nu_4 \to \text{visible}) \times A_{\rm eff} (E_{\nu_4}) \frac{}{} \right ],
\end{eqnarray}
where $dN_{\nu_4}/dE_{\nu_4}$ is the flux of sterile neutrinos entering into decay region and $d$ is the decay length of $\nu_4$. The total decay length is $d = \gamma_{\nu_4} c \tau_{\nu_4} + \gamma_X c \tau_X$ when $m_4 > m_X$, and $d = \gamma_{\nu_4} c \tau_{\nu_4}$ when $m_4 < m_X$, where $\gamma_\alpha$ is the Lorentz factor. The detection efficiencies $A_{\rm eff}(E_{\nu_4})$ for each experiment are described in \cref{Table_LongLivedParticleSearches}. 

\begin{table}[t]
\centering
\setlength{\tabcolsep}{1pt}
\begin{tabular}{  c || c | c | c | c | c | c | c }
\hline\hline
Experiment &  $N_{\rm POT}$ or $\mathcal{L}_\text{int}$  & $\sqrt{s}$ &  $E_{p \text{ beam}}$  &  $L$  &  $\Delta$  &  $\langle E_{\nu_4} \rangle$  &  95\% C.L.    \\
\hline
\hline
FASER (LHC run 3)& $\mathcal{L}_\text{int} \!=\! $ $150$ fb$^{-1}$ & \multirow{2}{*}{\ 14 TeV \ } & \multirow{2}{*}{\ \ \ 7 TeV \ \ } &  \ \multirow{2}{*}{480m} \ & \ \multirow{2}{*}{1.5m} \ & \ \multirow{2}{*}{$ \sim 1 \text{ TeV} $} \ & \ \ \multirow{5}{*}{$N_{\rm sig} \geq 3$ } \\
\cline{1-2}
FASER2 (HL-LHC) & $\mathcal{L}_\text{int} \!=\! $ $3$ ab$^{-1}$ & & & & & & \\
\cline{1-7} 
SHiP & $N_{\rm POT} \!=\! 2 \!\times\! 10^{20}$ & \multirow{2}{*}{27.4 GeV} & \multirow{2}{*}{400 GeV} & 110m & 50m & &  \\
\cline{1-2} 
\cline{5-6} 
CHARM &$N_{\rm POT} \!=\! 2.4 \!\times\! 10^{18}$ & & & 515m & 35m & \ $\sim 50$ GeV \ &  \\
\cline{1-6} 
NOMAD & $N_{\rm POT} \!=\! 4.1 \!\times\! 10^{19}$ & 29 GeV & 450 GeV & 835m & 290m & & \\
\hline\hline
\end{tabular}
\caption{The fiducial region, collision energies and the number of protons on target (POT) of long-lived particle searches in fixed target and head-on collision experiments. Table from Ref.~\cite{Jho:2020jfz}}
\label{Table_LongLivedParticleSearches}   
\end{table}

The sterile neutrinos are produced through the mixing with the active neutrino so that its flux is  given as 
\begin{eqnarray}
\frac{dN_{\nu_4}}{dE_{\nu_4}} & \approx & \frac{dN_{\nu_\ell}}{dE_{\nu_\ell}}  \times | U_{\ell 4} |^2 \times (\text{phase/helicity space suppresion factor}),
\end{eqnarray}
where the phase space and helicity suppression factors are from the nonzero sterile neutrino masses~\cite{Orloff:2002de}. The most relevant current limits (CHARM/CHARM-II) and future searches (FASER/
FASER2/SHiP) are as follows:
\begin{itemize}
    \item CHARM (using $400$ GeV$/c$ proton beam and copper target) \cite{Orloff:2002de} has provided the exclusion limits on tau active-sterile mixing ($U_{\tau 4}$) for $m_4 = 10-290$ MeV.
    \item CHARM-II collaboration studied double vertex events. The most promising decay mode is from $\nu_4 \to \nu_\mu \mu^+ \mu^-$ with the required minimum vertex separation is $l=1.5$m. The expected number of events is given by $N = N_{\nu_4} \cdot {\rm Br}(\nu_4 \to \mu^+ \mu^- \nu_\mu ) \cdot \epsilon\left( m, \vert U \vert^2 \right)$ where $N_{\nu_4}$ is the number of sterile neutrinos produced in a neutral current interaction, and $\epsilon$ is efficiency \cite{CHARMII:1994jjr}, giving a constraints on muon active-sterile mixing ($U_{\mu 4}$) for $m_4 = 0.2-2.5$ GeV.
    \item FASER has an outstanding projection to active-sterile neutrino mixings $| U_{\ell 4} |^2$ thanks to its huge number of neutrinos produced from the decays of mesons \cite{Abreu:2019yak}. We explore the sensitivity at FASER ($L=150$ fb$^{-1}$) and FASER2 ($L=3$ ab$^{-1}$) for projected limits of our model.
    \item SHiP is proposed to use the high-intensity proton beam from CERN SPS, with the planned protons on target (POT) about $2 \times 10^{20}$ which is much larger than the previous searches such as CHARM and NOMAD. 
\end{itemize}

\paragraph{Neutrino telescope} IceCube-DeepCore (inside the IceCube detector volume) has been designed to detect neutrinos with $E_\nu = 1-100$ GeV \cite{Aartsen:2016psd}. From the 2015-2016 data, about $N_{\rm NC}^{\nu_\tau} = 1.4 \times 10^4$ NC tau neutrino events ($\nu_\tau N \to \nu_\tau N'$) in $E_\nu = 5.6 - 56$ GeV \cite{Aartsen:2019tjl} have been analyzed. Due to the $U_{\ell 4}$ mixing, there are $\nu_\ell N \to \nu_4 N'$ events. Once produced, $\nu_4$ will travel about $20$ m then leave a double-cascade signature. The event number is estimated as
\begin{eqnarray}
N_{\rm sig.} & \simeq & \int dE_\nu \Bigl [ \frac{dN_{\rm NC}^{\nu_\ell}}{dE_\nu}  \times \left ( e^{- \frac{L-\Delta}{d} } - e^{- \frac{L}{d} } \right ) \times \text{Br}(X/\nu_4 \to \text{visible}) \Bigr ],
\end{eqnarray}
where $L=300$m is the fiducial vertical length of the DeepCore and $L-\Delta = 20$m is the minimum length to distinguish the double-cascade event from the background events.  A similar analysis without the $X$ boson was done in Ref.~\cite{Coloma:2017ppo}. We set the limit as $N_{\rm sig.} \geq 10$.

\begin{figure*}[t]
\centering
\includegraphics[width=0.43\textwidth]{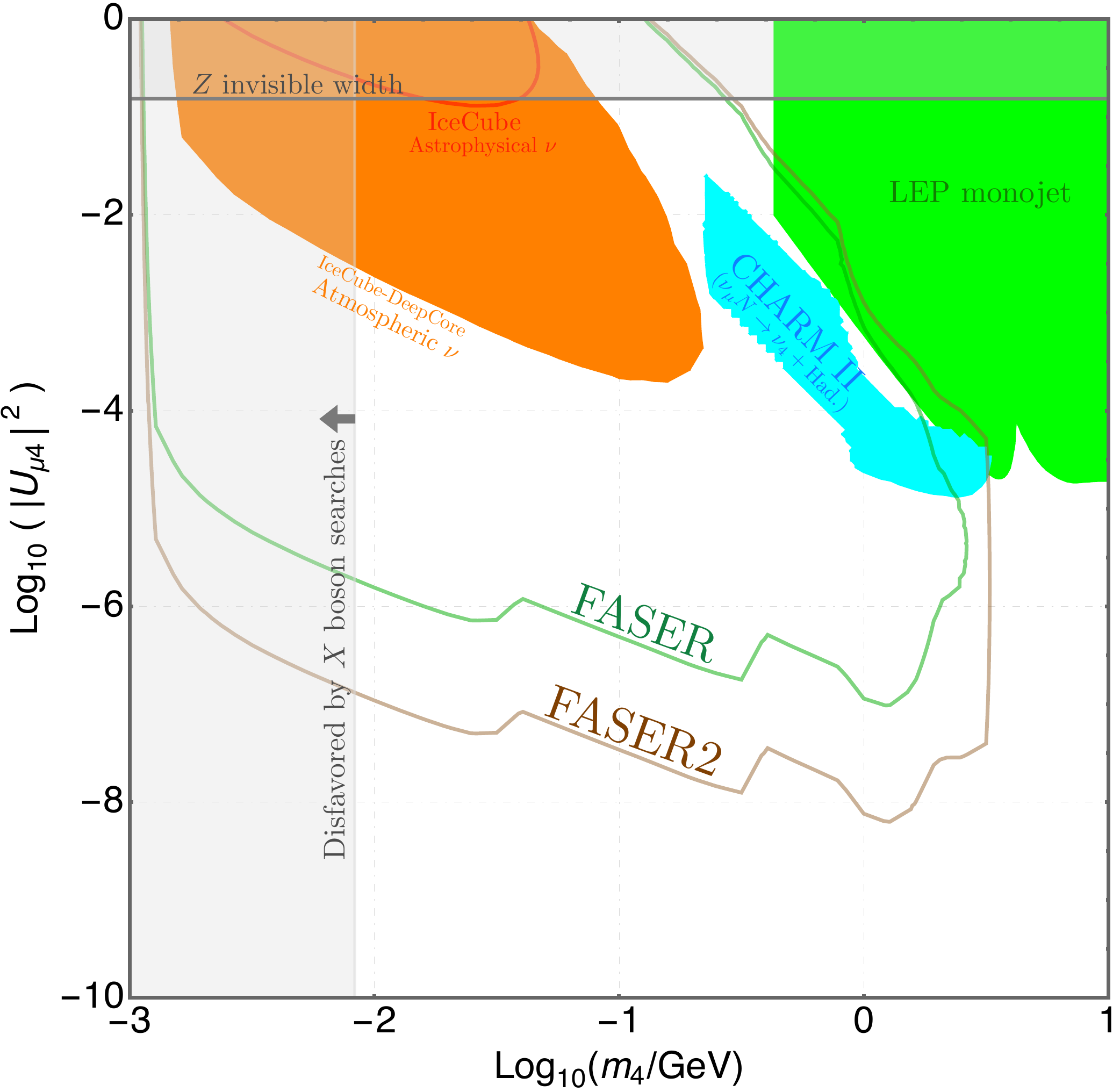} \quad \quad
\includegraphics[width=0.43\textwidth]{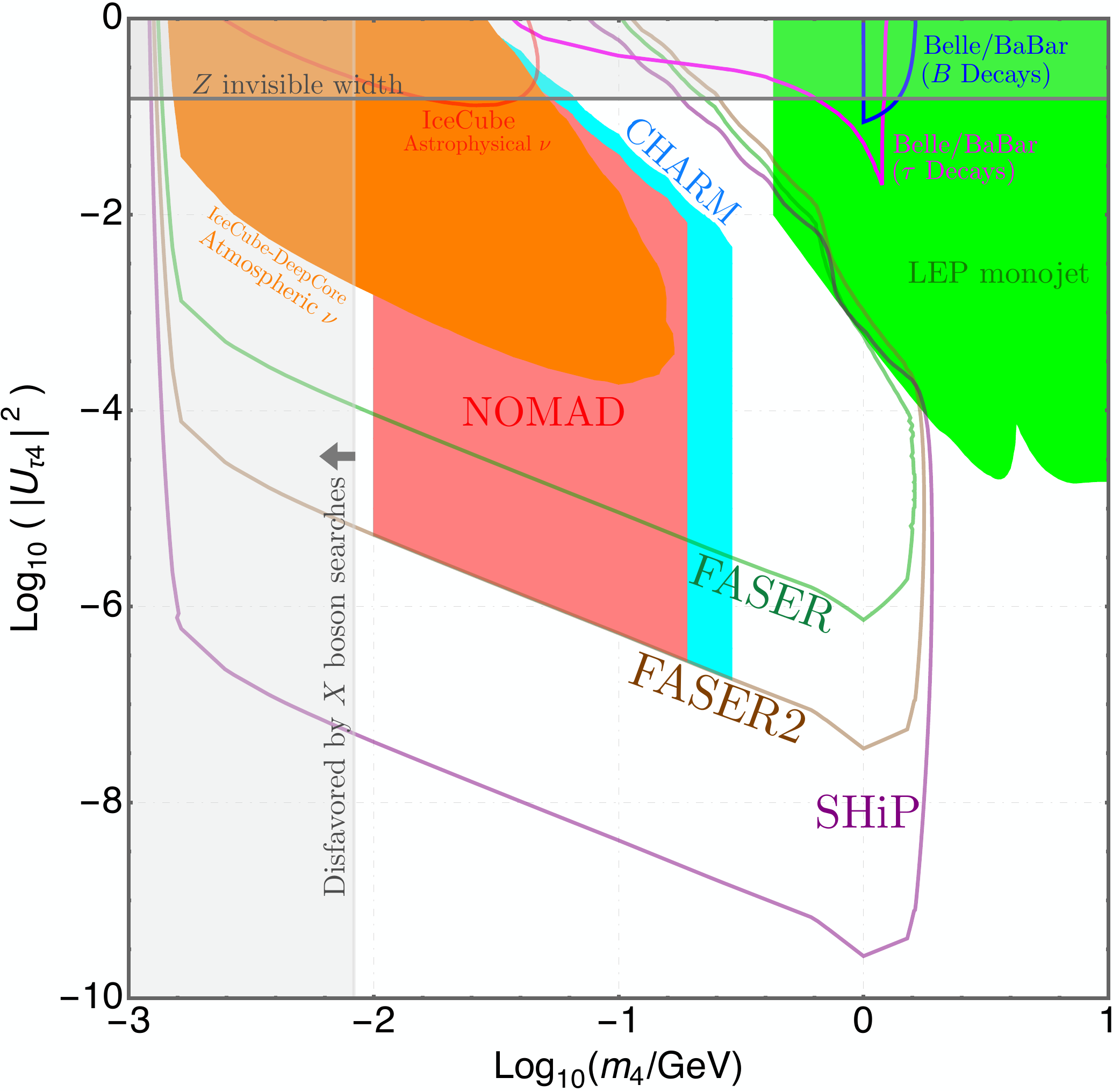}
\caption{Current and future limits on the active-sterile mixings $| U_{\mu 4}|^2$ (left) and $| U_{\tau4}|^2$ (right), with $\nu_s$-philic gauge boson parameters $m_X/m_4 = 1.2$, $\epsilon_{\gamma X} = 3\times 10^{-4}$. Figures from Ref.~\cite{Jho:2020jfz}.}
\label{fig_BP1_model}
\end{figure*}

\paragraph{$N_{\rm eff}$/monojet bound at Z-pole} The invisible and monojet decay searches at Z-pole severely constrain $O(1)$ active-sterile mixing $U_{\ell 4}$ and relatively heavy mass of $\nu_4$ ($m_4 > 6$ GeV).
\begin{itemize}
\item The upper limit of active-sterile mixing $U_{\ell 4}$ ($\ell=e,\mu,\tau$) are given by 
\begin{eqnarray}
| U_{l 4} |^2 & < & \frac{1}{\text{Br}(X \to \text{invisible})}  \cdot \left ( \frac{\Gamma_{Z \to \text{invisible}}^{\rm Exp.}}{\Gamma_{Z \to \text{invisible}}^{\rm SM}} - 1 \right ), \ \ \ \ \ \ 
\end{eqnarray}
where experimental observation of invisible $Z$ width at LEP and its SM prediction are~\cite{ALEPH:2005ab, Voutsinas:2019hwu, Janot:2019oyi}
\begin{eqnarray}
\Gamma_{Z \to \text{invisible}}^{\rm Exp.}  =  499.0 \pm 1.5 \text{ MeV} 
\quad\text{and}\quad
\Gamma_{Z \to \text{invisible}}^{\rm SM}  =  501.69 \pm 0.06 \text{ MeV}.
\end{eqnarray} 
\item DELPHI reported the weak isosinglet neutral heavy lepton ($\nu_4$) search with $3.3\times 10^6$ $Z$ bosons at LEP-I experiment. Several separate searches have been performed e.g., for promptly decaying $\nu_4 \to \text{monojet}$ ($m_4 \geq 6$ GeV), and for long-lived $\nu_4$ giving secondary vertices (\mbox{$1 \text{ GeV} \leq m_4 \leq 6$ GeV})  \cite{Abreu:1996pa}. It provides a limit on $Z$ decay width into sterile neutrinos as
\begin{eqnarray}
\text{Br}(Z \to \nu \nu_4) & < & 1.3 \times 10^{-6}\,\,(\text{$95\%$ C.L.)}
\end{eqnarray}
in a wide window $m_4 = 3.5 - 50$ GeV.
\end{itemize}

\paragraph{Conclusions} Sterile neutrinos might couple to new $U(1)$ gauge boson, $X$. We explore phenomenology for various experiments in sterile neutrino-specific $U(1)_s$ model. Concentrating on $U_{\ell 4} (\ell=\mu,\tau)$, we show all relevant results from various collider experiments, IceCube, and fixed target experiments. Additionally, we investigate the future sensitivities for LLP searches such as FASER, SHiP and CHARM. Our main results are summarized in \cref{fig_BP1_model}. As can be seen, FASER2 in the FPF has very good prospects for detecting the relevant HNL decay signature.

\subsection{The $\nu_R$-philic Dark Photon\label{sec:bsm_nonmin_nuRphilic}}

Right-handed neutrinos ($\nu_{R}$) are often considered as one of the most well-motivated solutions of known problems of the SM such as neutrino masses~\cite{Minkowski:1977sc, yanagida1979proceedings, Gell-Mann:1979vob, glashow1979future, Mohapatra:1979ia}, dark matter~\cite{Asaka:2005pn, Asaka:2005an, Ma:2006km}, and baryon asymmetry of the universe~\cite{Fukugita:1986hr}. As an intrinsically dark sector, $\nu_{R}$ might be charged under a hidden gauge symmetry~\cite{Batell:2010bp, Chang:2011kv, Ma:2013yga, Lindner:2013awa, Ballett:2019pyw, Berbig:2020wve, Abdullahi:2020nyr, Jodlowski:2020vhr}. The new gauge boson arising from this symmetry does not directly couple to other fermions except for $\nu_{R}$ at tree level. We refer to such a new gauge boson as the $\nu_{R}$-philic dark photon. 

\begin{figure*}[t]
\centering
\includegraphics[width=0.7\textwidth]{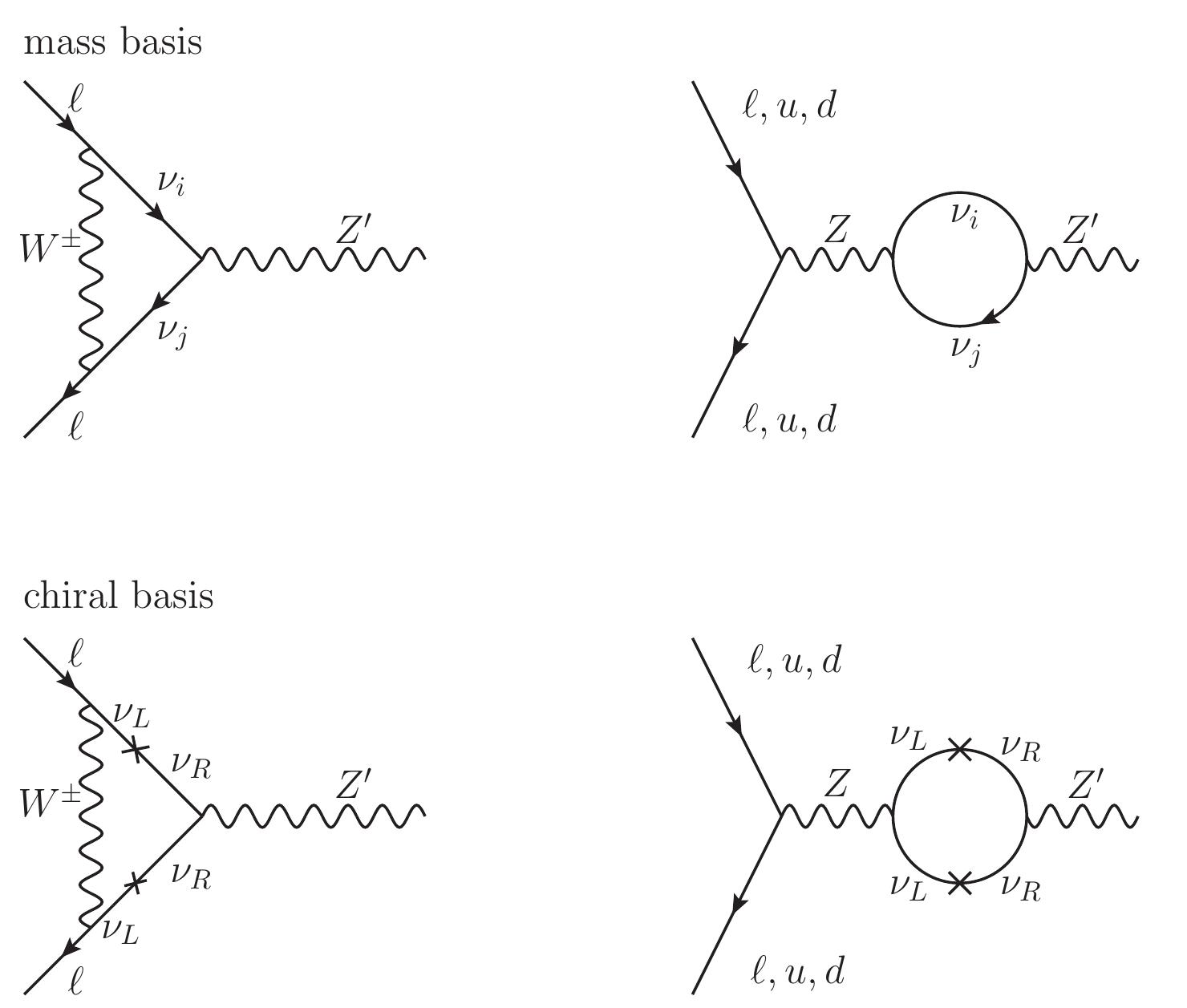}
\caption{ Loop-induced couplings of the $\nu_{R}$-philic dark photon to charged fermions in the mass basis (upper panels) and in the chiral basis (lower panels). Figure adapted from Ref.~\cite{Chauhan:2020mgv}.}
\label{fig:loop}
\end{figure*}

At the one-loop level, the $\nu_{R}$-philic dark photon interacts with normal matter via the loop diagrams shown in \cref{fig:loop}, provided that $\nu_{R}$ mixes with the left-handed neutrinos $\nu_{L}$. In Ref.~\cite{Chauhan:2020mgv}, we computed these loop diagrams to address the question of how dark the $\nu_{R}$-philic dark photon could be. 

We find that the loop-induced couplings are UV finite as a consequence of the orthogonality between the SM gauge-neutrino couplings and the new ones. Compared to our previous study on loop-induced $\nu_{R}$-philic scalar interactions~\cite{Xu:2020qek}, the couplings in the vector case are not suppressed by light neutrino masses. Instead, the loop-induced effective couplings are roughly of the order of
\begin{equation}
g_{{\rm eff}}\sim\frac{G_{F}m_{D}^{2}}{16\pi^{2}}\thinspace.\label{eq:loop-eff}
\end{equation}
Specific results of the loop-induced couplings can be found in Ref.~\cite{Chauhan:2020mgv}. 

\begin{figure*}[t]
\centering
    \includegraphics[width=0.85\textwidth]{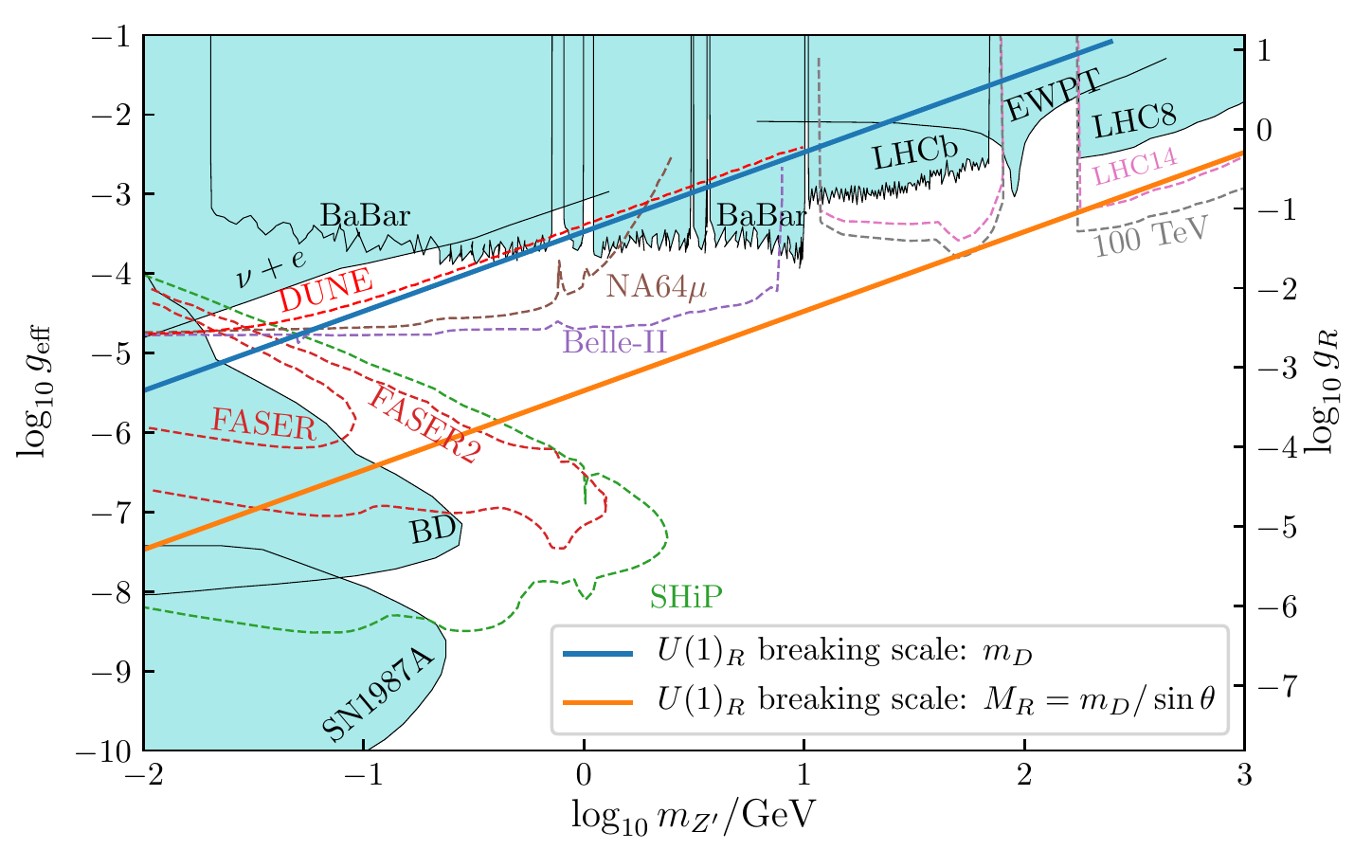}
\caption{Sensitivity of future experiments (SHiP, FASER, Belle-II) on the $\nu_{R}$-philic dark photon. Here $g_{{\rm eff}}$ is the loop-induced coupling of $Z'$ to electrons. The quark couplings are of the same order of magnitude as $g_{{\rm eff}}$ and we have ignored the difference between them when recasting constraints on quark couplings. The theoretically favored values of $g_{{\rm eff}}$ is below the solid blue or orange lines, assuming $U(1)_{R}$ breaks at the different scales. Taken from Ref.~\cite{Chauhan:2020mgv}.
\label{fig:vR-philic-large-mass}}
\end{figure*}

The theoretically favored values of the loop-induced couplings are confronted with experimental constraints and prospects in \cref{fig:vR-philic-large-mass} (see also Ref.~\cite{Chauhan:2020mgv} for the results for a wider mass range). We find that the magnitude of loop-induced couplings allows current experiments to put noteworthy constraints on it. Future beam dump experiments like SHiP and LLP searches in the proposed FPF, together with upgraded searches at large-scale collider experiments, will substantially improve sensitivity on such a dark photon. 

Hence, as the answer to the question of how dark the $\nu_{R}$-philic dark photon could be, we conclude that the $\nu_{R}$-philic dark photon might not be inaccessibly dark and could be of importance to a variety of experiments, especially to the future FASER2 detector at the FPF.

\subsection{Secondary Production in BSM and Neutrino Interactions\label{sec:bsm_nonmin_secondary}}

The search for highly displaced LLP decays in the far-forward region of the LHC is one of the most important physics goals of the proposed FPF. It benefits from i) the spatial separation of the detector from the primary interaction point, and ii) very well shielding, which allows one to reject SM backgrounds and perform a clean search for two high-energy oppositely-charged tracks coming from the LLP decays. This search then remains a very promising avenue to potentially uncover BSM physics, as the observation of only a few such events could be sufficient evidence supporting new discoveries. 

However, such searches are limited by the lifetimes of LLPs, which must travel the entire distance from the LHC interaction point to the new facility. Notably, however, typically the most phenomenologically appealing regions in the parameter space that generate experimental anomalies in other searches correspond to relatively large coupling constants to the SM that predict too small LLP lifetimes for these particles to be able to reach the far-forward detectors. 

This can be circumvented in models predicting more rich dark sectors. In particular, in the presence of at least two light new physics species and large dark coupling constants between them, the LLP with a small lifetime can be efficiently generated in front of the detector, i.e., at a much closer distance, in interactions of the other dark particle. A similar \textsl{secondary production} mechanism~\cite{Jodlowski:2019ycu} can also take place in BSM neutrino scatterings in front of the FPF or even inside the detectors~\cite{Jodlowski:2020vhr}. The newly produced LLP can subsequently visibly decay inside the decay volume of FASER2. This allows one to study LLP lifetimes much smaller than usual, while still searching at essentially zero background. We schematically illustrate this in the upper panel of \cref{fig:LLP_idea}. In the plot, at the top, we show the main idea of secondary production occurring by upscattering of the lighter species - $\textrm{LLP}_1$, denoted by red - in interactions with nuclei or electrons in the material in front of the detector. This produces a heavier unstable particle - $\textrm{LLP}_2$, which we mark with blue - which then decays inside the detector.

\begin{figure*}[t]
\centering
\includegraphics[width=0.9\textwidth]{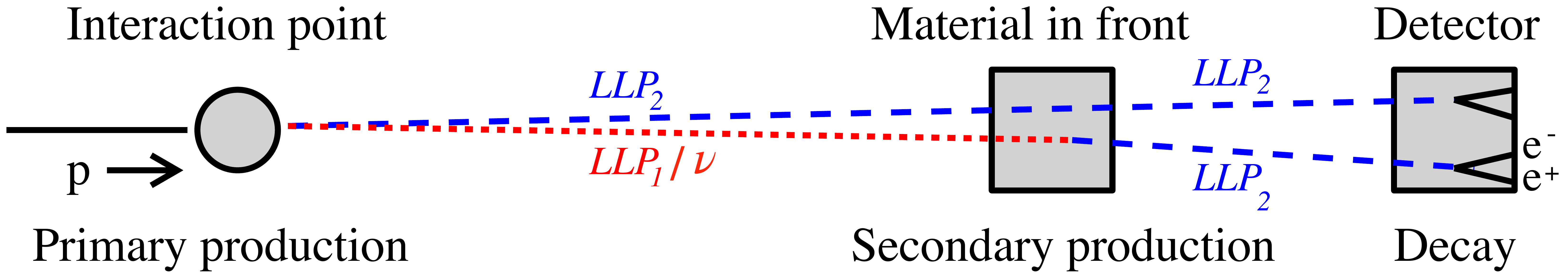}
\includegraphics[width=0.85\textwidth]{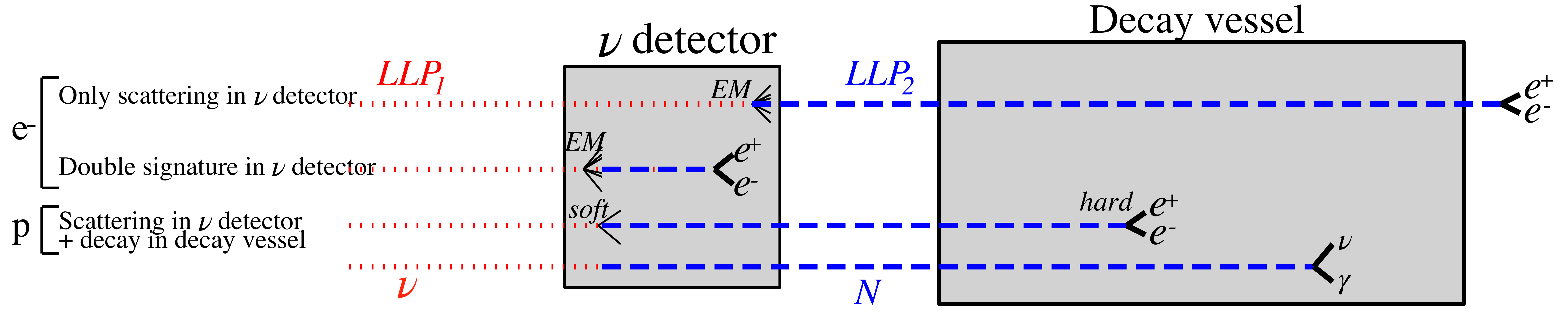}
\caption{
Illustration of secondary LLPs production in extended light new physics models. LLPs are produced in front of the detector, and their main signatures in the detectors are also shown. Adapted from Ref.~\cite{Jodlowski:2019ycu}.}
\label{fig:LLP_idea}
\end{figure*}

The bottom panel of \cref{fig:LLP_idea} also shows other possible signatures of new physics that can employ the secondary production mechanism. Besides the aforementioned displaced LLP decays inside the decay vessel of FASER2, we also illustrate there the idea of the search based on the scattering of the BSM species or neutrinos with electrons or protons in neutrino detectors placed in front of FASER2. Last but not least, the BSM physics can also be searched for in double scattering events, in which the detectable scattering is followed by the time-coincident, collinear, and spatially-separated decay inside the neutrino detector or the decay vessel of FASER2.

\begin{figure*}[t]
\centering
\includegraphics[width=0.51\textwidth]{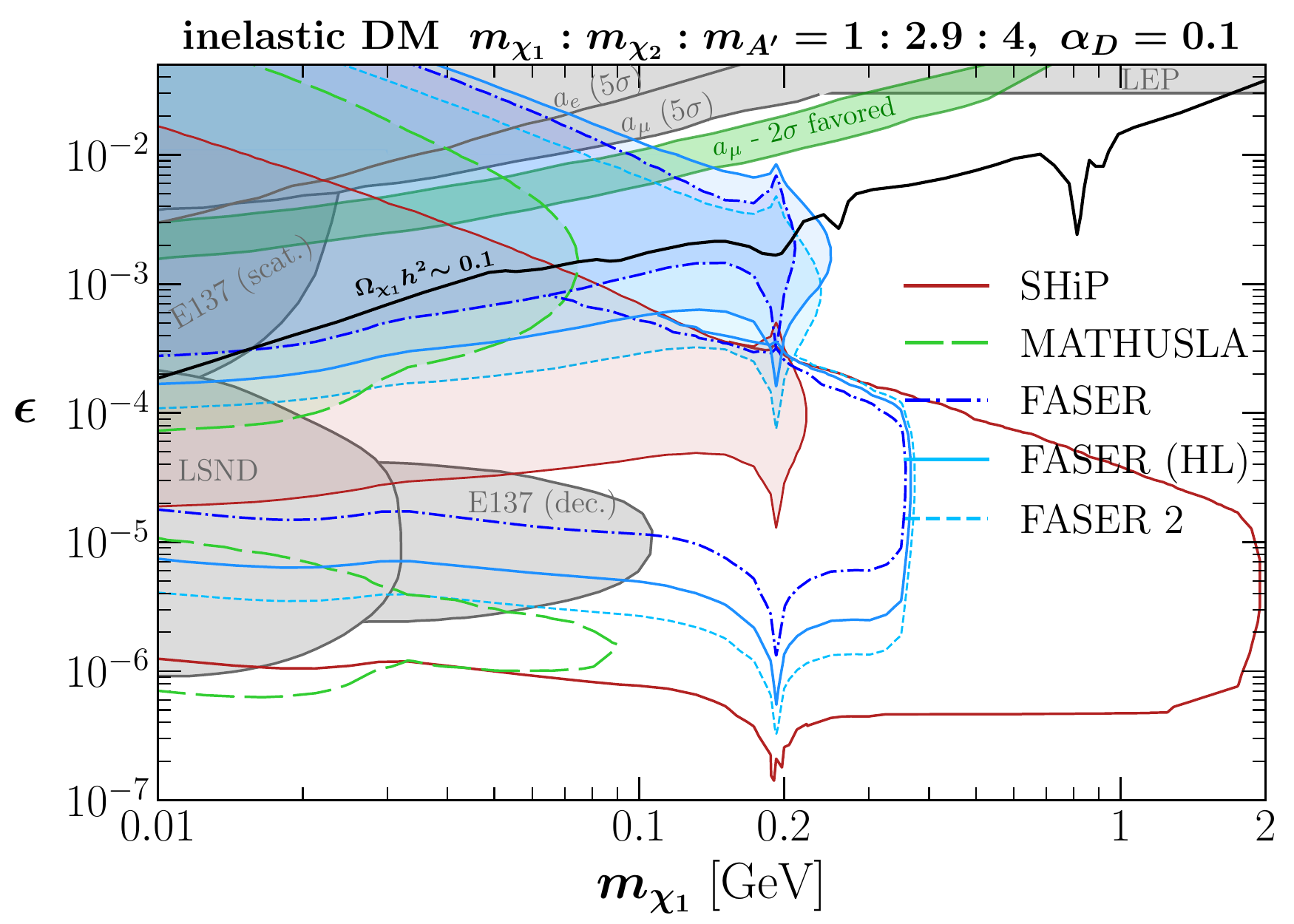}
\hspace*{0.2cm}
\includegraphics[width=0.44\textwidth]{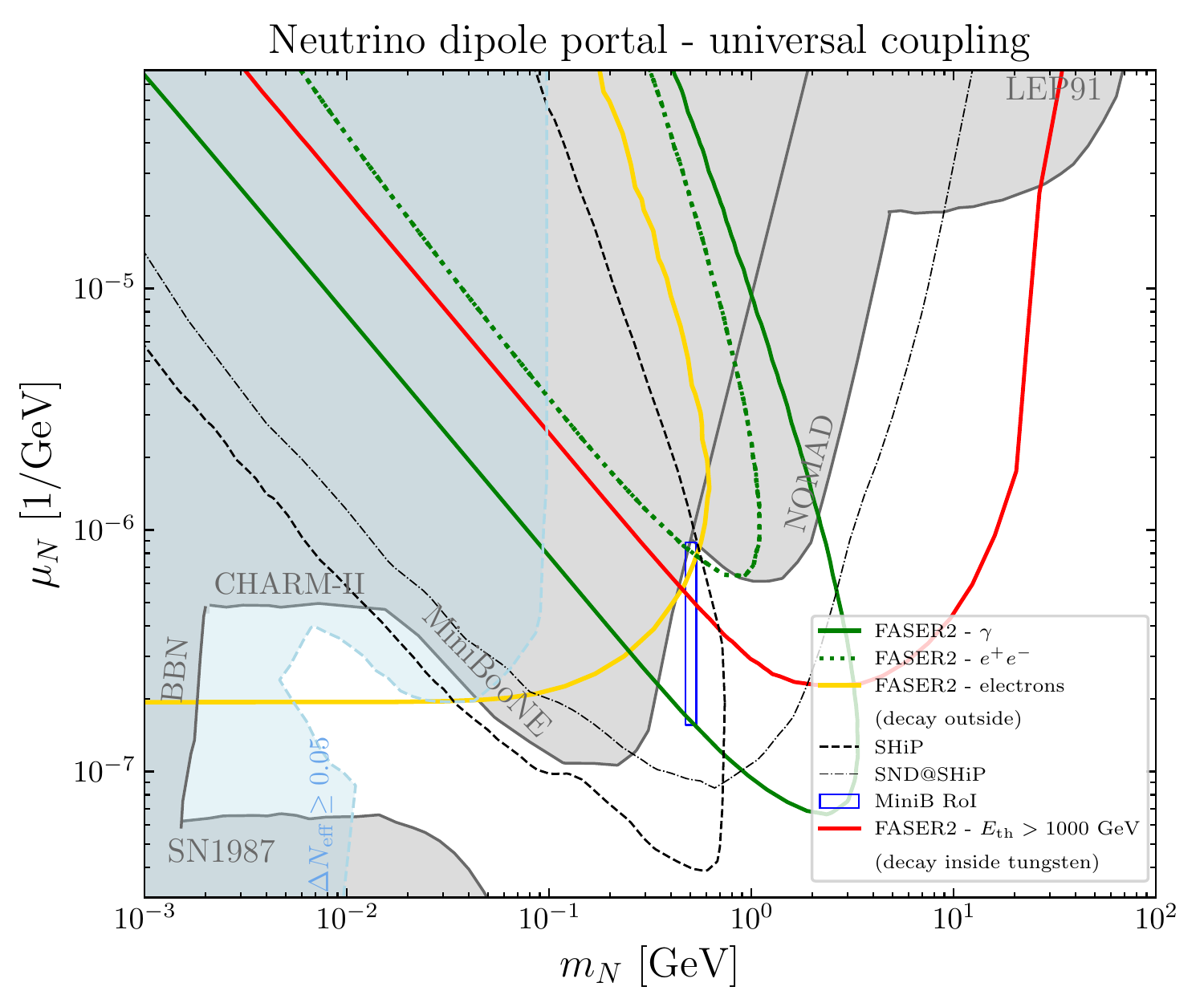}
\caption{Sensitivity reach plots with extended coverage of the parameter space corresponding to the large kinetic mixing parameter thanks to secondary LLP production for the inelastic DM model (left), and the neutrino dipole portal with a universal coupling to all neutrino flavors (right); see text for details. In the left, the colorful shaded regions correspond to the sensitivity induced by the secondary production mechanism. This can be compared with the expected sensitivity without such production, which is shown with colorful lines without shading. On the right, different colors correspond to distinct experimental signatures in the FPF, as discussed in the text. Taken from Refs.~\cite{Jodlowski:2019ycu, Jodlowski:2020vhr}.
}
\label{fig:LLP_results_2}
\end{figure*}

Below, we present the expected improvements in the sensitivity reach of the FASER2 experiment due to the secondary production mechanism for two cases: the inelastic DM scenario coupled to the SM via the dark photon portal, and the neutrino dipole portal. We also compare the FASER2 reach in these cases with the proposed MATHUSLA~\cite{Chou:2016lxi,MATHUSLA:2018bqv,Curtin:2018mvb} and SHiP~\cite{Bonivento:2013jag,SHiP:2015vad,Alekhin:2015byh} experiments.

In the inelastic DM model, we introduce two fermionic matter fields with split masses coupled to a dark photon by the dominant non-diagonal coupling~\cite{Tucker-Smith:2001myb}. This can lead to the efficient upscattering transition from the lighter, stable state into the heavier, unstable one. The Lagrangian of the model takes the following form:
\begin{equation}
\mathcal{L}\supset \left(g_{12}\,\bar{\chi}_2\gamma^\mu\chi_1\,A^
\prime_\mu+\textrm{h.c.}\right),
\end{equation}
where $A^\prime_\mu$ is a massive dark photon, $\chi_{1,2}$ are dark fermions, and $g_{12}$ is the dominant, non-diagonal coupling. In this scenario, the upscattering of $\chi_1$ to $\chi_2$ in interactions with electrons or nuclei in front of the detector is followed by the $\chi_2\to \chi_1 e^+e^-$ decay in the decay vessel, which proceeds via the intermediate off-shell dark gauge boson $A^{\prime\ast}$. 

We present the resulting reach of both FASER and FASER2 detectors as color-filled contours with various shades of blue on the left panel of \cref{fig:LLP_results_2}. As can be seen, the secondary LLP production allows one to cover large parts of the parameter space in which the inelastic DM obtains the correct thermal relic density~\cite{Berlin:2018jbm}, as well as the region where loop contributions coming from the dark photon can explain~\cite{Pospelov:2008zw, Mohlabeng:2019vrz} the persistent discrepancy between the SM predictions and the measurements of the muon anomalous magnetic moment~\cite{Bennett:2006fi, Muong-2:2021ojo}. The latter is shown as a green band in the plot.

As mentioned above, the secondary production of light new physics species can also lead to similar experimental signatures appearing due to BSM neutrino interactions~\cite{Jodlowski:2020vhr}. We illustrate this in the right panel of \cref{fig:LLP_results_2} for the dipole portal between the SM neutrinos and sub-$\gev$ heavy neutral leptons, cf. e.g. Refs.~\cite{Alekhin:2015byh}. This induces an effective neutrino-right handed neutrino-photon coupling described by the following higher-dimensional operator
\begin{equation}
\mathcal{L} \supset \mu_N\,\bar{\nu}_L \sigma_{\mu \nu}N_R F^{\mu \nu}+\textrm{h.c.},
\label{eq:Lagdipole}
\end{equation}
where $\mu_N$ is a dimensionful coupling constant, $\nu$ is a light SM neutrino, $\sigma^{\mu \nu}=\frac i2 [\gamma^\mu, \gamma^\nu]$, $N_R$ represents the right-handed neutrino, and $F^{\mu \nu}$ is the field strength tensor of electromagnetic field. The results shown in the plot correspond both to the FASER2 and FASER$\nu$2 experiments. We assume the universal coupling strength $\mu_N$ to all neutrino flavors.

In the plot, we present the expected sensitivity of the FPF experiments to several complementary signatures: i) the upscattering of light SM neutrino into the right handed neutrino $N$ followed by its decay into a light neutrino and a photon inside the decay vessel of FASER2 (shown with green lines), ii) the same upscattering process but followed by a prompt decay inside the neutrino emulsion detector producing very high-energy photons with $E_\gamma> 1$ or $3~\tev$ (red lines), and iii) the scattering-only signature with the electrons in the emulsion detector (gold line). We have also highlighted two particularly interesting regions of parameter space: the thin blue rectangle marks the region that has been invoked~\cite{Gninenko:2009ks, Magill:2018jla} to explain the MiniBooNE anomaly~\cite{AguilarArevalo:2007it, MiniBooNE:2020pnu}, while the light blue shaded region marks the parameter space where $N$ increases the effective number of relativistic degrees of freedom in the early Universe. It is known~\cite{Brdar:2020quo} that such a contribution, as long as it is small enough not to violate the BBN bounds, can relax the Hubble tension~\cite{Schoneberg:2021qvd}.

Light BSM physics often invokes unstable states with a range of decay lifetimes. The usual searches in fixed target or beam-dump experiments mainly cover the regime of lifetimes that are large enough to allow the new particle to reach the detector before decaying. This lifetime cannot be too small due to a typical long distance between the LLP production point and their decays, similar is true for the FPF. The secondary production of LLPs occurring upstream of the main detectors can overcome such difficulties and extend the sensitivity of the FPF and other intensity frontier experiments towards the smaller lifetimes.

\subsection{Light Dark Sector Going Through Chain Decay\label{sec:bsm_nonmin_chaindecays}}

For years, Weakly Interacting Massive Particles (WIMPs) with a mass of 10~GeV to few TeV held special place of honor among dark matter candidates, promising to be discovered soon via three distinct alternative methods (direct and indirect dark matter search experiments and/or direct production and detection at CMS and ATLAS.) Null results from all these searches have caused a shift of paradigm to consider different mass ranges for dark matter candidates and the possibility of having a rich dark matter sector with  multiple dark particles interacting with each other. 

Forward physics experiments will provide an ideal setup to look for dark sector of GeV mass scale. Let us suppose that there is a new particle of a GeV mass with a small coupling to the partons. As a working example, Ref.~\cite{Bakhti:2020vfq} takes it as a GeV scale pseudoscalar, $X'$ with coupling to gluons of form $X' G_{\mu\nu}^i G_{\alpha \beta}^i \epsilon^{\mu \nu \alpha \beta}/\Lambda$. Then, $X'$ particles can be abundantly produced via interaction of two colliding partons with Bjorken variables of $x_1\sim 0.1$ and $x_2 \sim 10^{-6}$ at IP despite small couplings. The large disparity between $x_1$ and $x_2$ means that the produced $X'$ will be highly boosted in the forward direction. The same couplings that lead to the production of $X'$ at the IP can also impede it traversing the rock and concrete before the detector. If, however, $X'$  promptly decays into dark sector ($X$) with tiny or zero coupling with the matter fields, a significant flux of dark particles can arrive at the forward detector. If the $X$ particles decay inside the detector into hadrons or charged leptons, they can be detected. Non observation of any signal from $X$ decay can then constrain the parameters of the model.

As was pointed out by Lee and Weinberg in 1977 \cite{Lee:1977ua}, the dark matter particles, $Y$, with mass ($m_Y$) less than 10~GeV and weak coupling strength to the SM cannot be thermally produced with the correct relic density because, in this case, $G_F^2 m_Y^2/(4\pi)\ll$pb. One of well-known tricks to obtain thermally produced GeV scale  dark matter is to introduce intermediate light neutral particles into which dark matter pair can annihilate, $\langle v\sigma(Y+\bar{Y}\to \eta+ \bar{\eta})\rangle \sim pb$ where $\eta$ and $\bar{\eta}$ eventually decay into SM particles. As discussed in \cite{Bakhti:2020vfq}, such GeV scale dark sector involves chain decays that lead to spectacular phenomenology at forward experiments such as FASER$\nu$ and SND@LHC which enjoy high spatial resolution. In the specific model in \cite{Bakhti:2020vfq}, the annihilation $Y+\bar{Y}\to \eta +\bar{\eta}$ in the early universe takes place via coupling to intermediate $X$ particles. Let the $X$ particles be coupled to the aforementioned $X'$ particles which in turn is coupled to the SM fermions. At the LHC, the $X$ particles can be produced via the $X'$ decay and pass through the rock and concrete, leading to a signal at the scattering detectors in the FPF (FASER$\nu$2, FLArE, Advanced SND@LHC) as schematically shown in \cref{fig:chain_decay}. The DM $Y$ particles will show up as missing transverse momentum. By constructing the four momenta of the final leptons and their production vertex, the mass and lifetime of the intermediate particles can be reconstructed \cite{Bakhti:2020vfq}.  Such a distinct signal will be background free, especially in the detectors allowing for a dynamical observation of the events. For a given lifetime of $X$, the null signal can be interpreted as a bound on the coupling of $X'$ to the partons \cite{Bakhti:2020vfq}.
 
\begin{figure*}[t]
\centering
	 \includegraphics[width=0.4\textwidth]{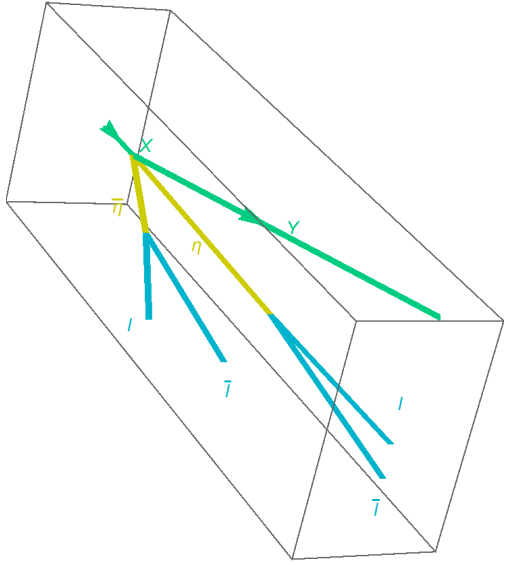}
	 \caption{ Typical signal of GeV scale dark sector at the FPF scattering detectors for the chain decay scenario. $X$ enters the detector from the direction of IP (forward direction). $Y$ is dark matter particle which appears as missing energy-momentum. By measuring the transverse momentum of the final charged particle, the emission of $Y$ along with them can be tested. The decay vertices of $\eta$ and $\bar{\eta}$ can be located with high precision. Moreover, by measuring the momenta of the final leptons, denoted by $l$ and $\bar{l}$ the directions of the tracks of $\eta$ and $\bar{\eta}$  and therefore the $X$ decay vertex can be reconstructed. The masses of $\eta$ and $\bar{\eta}$ can extracted by measuring the four-momenta of  the final leptons. Moreover, if statistics is large enough, the decay rate of $\eta$ can be derived by measuring the distances between $X$ and $\eta$ decay vertices. More information can be found in \cite{Bakhti:2020vfq}.   }
\label{fig:chain_decay}
\end{figure*}

\subsection{Bound State Formation and Long-Lived Particles\label{sec:bsm_nonmin_boundstates}}

Colored co-annihilation within a rich dark sector can lead to rich phenomenology that can evade current experimental bounds both from LHC and direct detection experiments. A class of such models being currently investigated by the DM community is dubbed by 't-channel' simplified models.  If these dark sector particles interact via a light mediator, long range effects like Sommerfeld enhancements can enhance (diminish) the velocity averaged annihilation cross section if it is an attractive (repulsive) potential \cite{Sommerfeld:1931qaf, Drees:2009gt, Beneke:2019qaa, Harz:2018csl}. Additionally, bound states can form via the emission of mediators acting as an additional annihilation channel. We analyze the impact of these effects both in terms of relic density calculations as well as collider searches in our forthcoming paper \cite{BSFLHC}. A key feature of this analysis is the emergence of bound states at colliders. While CMS/ATLAS can probe a part of the parameter space relevant to bound states depending on the lifetime, future alternative experiments can shed light on the full scope of these processes. 

For the purposes of this note, we summarize our findings and comment on the prospects of probing bound states at FASER during LHC Run 3 and FASER2 for the HL-LHC era. Although FASER2 is normally associated with light new particles, probing bound states is a significant additional physics motivation. We describe the salient features of our analysis and speculate on the possibility of FASER2 to probe a part of the parameter space of interest. 
We focus on t-channel simplified models \cite{Mohan:2019zrk} with a single Majorana dark sector fermion $\chi$, as well as three color-triplet complex scalar fields $X_{i}$ interacting with each other via Yukawa coupling $g_{DM}$. The scalars are charged under the SM gauge group $(SU(3)\times SU(2))_Y$ as:
\begin{equation}
    (3,1)_{2/3},\quad(3,1)_{-1/3},\quad(3,2)_{-1/6}.
\end{equation}
We impose a $\mathbb{Z}_2$ symmetry in the dark sector such that $\chi$ is the lightest stable particle and a DM candidate. The interaction Lagrangian is  given by:
\begin{equation}
    \mathcal{L}\supset\sum_{i}(D_\mu X_i)^\dagger(D^\mu X_i)+g_{\text{DM},ij}X_i^\dagger \Bar{\chi}P_R q_j + g_{\text{DM},ij}^* X_i\Bar{q}_j P_L \chi,
    \label{eq:Lagr}
\end{equation}
where $D_\mu$ is the covariant derivative. The index $i$ runs over the quark and mediator flavours of the model considered (up-type right-handed quarks, down-type right-handed quarks, left-handed quarks). Moreover, we set the Yukawa couplings to be real, flavour-diagonal and flavour-universal. The processes that determine the overall contribution to the DM density are direct annihilations of the DM-candidate particles, $\chi\chi\rightarrow \text{SM SM}$, as well as co-annihilation and colored annihilations processes into SM particles involving $\chi-X$, $\chi-X^\dagger$, $X-X^\dagger$ and $X-X$ as initial scattering states. A careful treatment of the relic density calculation will be accounted for in \cite{BSFLHC}.  We quote the effective annihilation cross section as, 
\begin{equation}
    \langle\sigma_{\text{eff}} v_{\text{rel}}\rangle=\sum_{ij}\langle\sigma_{ij}v_{ij}\rangle \dfrac{Y_i^{\text{eq}}}{\tilde{Y}^{\text{eq}}}\dfrac{Y_j^{\text{eq}}}{\tilde{Y}^{\text{eq}}},
    \label{eq:effective_sigmav_coannih}
\end{equation}
where $\langle\sigma_{ij}v_{ij}\rangle$ comprises all the annihilation cross-sections of two co-annihilating species $i$ and $j$. 

Processes involving two massive colored particles in the initial state are subject to the long-range behaviour of the gluonic Coulomb potential generated by the strong force exerted between them. This influence distorts their scattering wave functions in such a way that they can experience a further attraction or repulsion, depending on the color charge they carry, that is to say on their color representation, which directly influences the gauge coupling related to the gluon potential. The static `Coulomb' potential for two incoming non-relativistic colored annihilating particles in representations $\textbf{R}_\mathbf{1}$ and $\textbf{R}_\mathbf{2}$, can be written as
\be
    V_{[\hat{\textbf{R}}]}(r)=-\alpha_g^{[\hat{\textbf{R}}]}/r,
    \quad \text{where} \quad
    \alpha^{[\hat{\textbf{R}}]}_g=\alpha_s\times\frac{1}{2}[C_2 (\textbf{R}_\mathbf{1})+C_2 (\textbf{R}_\mathbf{1})-C_2 (\hat{\textbf{R}})]\equiv\alpha_s\times k_{[\hat{\textbf{R}}]},
    \label{eq:alpha_g}
\ee
with $C_2(\textbf{R})$ being the quadratic Casimir invariant of the given representation \textbf{R}.

For the scalars (antiscalars) that belong to the \textbf{3} of SU(3) (the antiscalar fields to the conjugate one, $\bar{\mathbf{3}}$), the possible potentials we obtain from $\mathbf{3} \otimes \Bar{\mathbf{3}} = \mathbf{1} \oplus \mathbf{8}$ and $\mathbf{3} \otimes \mathbf{3} = \mathbf{\bar{3}} \oplus \mathbf{6}$:
\begin{equation}
  V(r)_{\mathbf{3}\otimes\Bar{\mathbf{3}}}=
    \begin{cases}
    -\dfrac{4}{3}\dfrac{\alpha_s}{r}\quad[\mathbf{1}]\\[8pt]
    +\dfrac{1}{6}\dfrac{\alpha_s}{r}\quad[\mathbf{8}]
    \end{cases}; \quad   V(r)_{\mathbf{3}\otimes\mathbf{3}}=
    \begin{cases}
    -\dfrac{2}{3}\dfrac{\alpha_s}{r}\quad[\mathbf{\bar{3}}]\\[8pt]
    +\dfrac{1}{3}\dfrac{\alpha_s}{r}\quad[\mathbf{6}]
    \end{cases}. \label{eq:ColorPotential}
\end{equation}

Observe that the singlet state accounts for the most attractive potential. This results in the Sommerfeld enhancement of the the annihilation cross-section,
\be
    \sigma_{\text{SE},[\mathbf{R}]}v_\text{rel}=c_{[\textbf{R}]} S_{0,[\textbf{R}]}\,\sigma_0,
    \quad \text{where} \quad
    S_{0,[\textbf{R}]}=S_0\left(k_{[\textbf{R}]}\dfrac{\alpha_s}{v_{\text{rel}}}\right)
    \label{eq:somfact_color}
\ee
is the $s$-wave Sommerfeld factor and where $c_{[\textbf{R}]}$ is a factor coming from the color decomposition of the amplitude and marks the relative contribution of the $\hat{\textbf{R}}$ initial state to the total cross-section. Furthermore, in the regime where the Sommerfeld effect is important, color charged particles can also form unstable bound states via the emission of a gluon:
\begin{equation}
    X_1 + X_2 \rightarrow \mathcal{B}(X_1 X_2) + g.
\end{equation}
As an example, a scalar-antiscalar pair transforming in the fundamental representation with degenerate masses $m_X$, the cross-section for the bound-states is \cite{Harz:2018csl}:
\begin{equation}
        \sigma_{\{100\}}^{[\mathbf{8}]\rightarrow[\mathbf{1}]} v_\text{rel}=\dfrac{2^7\,17^2}{3^5}\dfrac{\pi\alpha_{s,[\mathbf{1}]}^{\text{BSF}}\alpha_{s,[\mathbf{1}]}^B}{m_X^2}\;S_{\text{BSF}}(\zeta_S,\zeta_B).
    \label{eq:BSF_sigmav_singlet}
\end{equation}
where, $\alpha_{s,[\mathbf{1}]}^{\text{BSF}}$ represents the strong coupling constant from the gluon emission vertex, while $\alpha_{s,[\mathbf{1}]}^B$ arises from the ladder diagrams involving the bound state wavefunction. Once the bound state is formed, it eventually decays at a characteristic time scale and opens up an additional annihilation channel for the DM to annihilate. These effects are taken into account in our analysis in the coupled Boltzmann equation that controls the evolution of the number densities of the dark sector. 

\begin{figure*}[t]
\centering
    \includegraphics[width=0.6\textwidth]{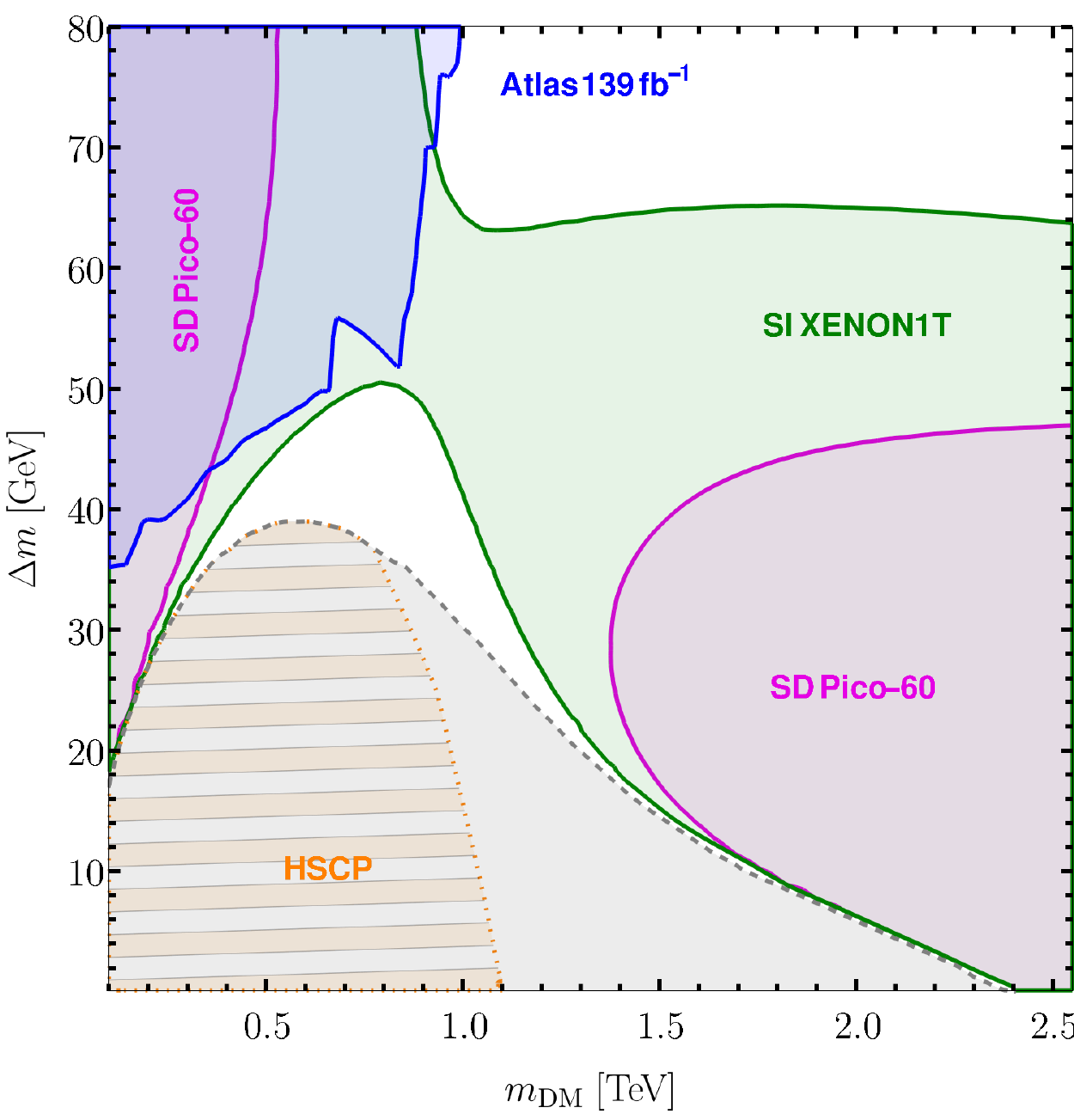}
\caption{Exclusion limits from various experiments in the co-annihilating region.  The limits from spin-independent DD (spin-dependent DD, colliders, unitarity) are denoted in green (magenta, blue, black). Additionally,  the orange shaded region provides an estimate for the potential reach of LLP searches constraining the area of the parameter space where the correct relic density can be reproduced only in the conversion-driven freeze-out regime. The region enclosed by a dotted line estimates the size of the coupling where conversion-driven freeze-out is relevant. The region enclosed by orange dashed lines assumes an earlier onset of conversion-driven freeze-out.}
\label{fig:Summary_uR_zoom}
\end{figure*}

Taking into account all of the above considerations on the relic density, we identify regions in this particular simplified model where Sommerfeld enhancement and bound state formation is important. For large mass gaps and large mass splittings between the mediator and the dark matter one needs a large value of the Yukawa coupling $g_{DM}$, and therefore bound state formation is disfavored in this region. The strongest constraints here originate from mono and multi-jet + missing energy searches at the LHC (which we recast using the MadAnalysis5 Public Analysis database~\cite{Dumont:2014tja}) as well as direct detection from Xenon1T and PICO60~\cite{Mohan:2019zrk}, and constrain mass gaps of upwards of 100~GeV for the region of parameter space that do not overproduce dark matter. In the small mass gap region, for $M_{X}, M_{DM} < 1000$ GeV and a mass gap less than 35 GeV, Sommerfeld enhancement and bound state formation enhance the thermally averaged annihilation cross section significantly. However, dark matter is overabundant in this region with conventional freeze-out mechanism. However the conversion driven freeze-out mechanism is relevant here and can provide the correct relic density, as it suppresses the total annihilation cross section. 

In this region the strongest constraints come from the search for heavy stable charged particles (HSCP) at the LHC, which can constrain mass gaps upwards of 5~GeV and $M_{DM}\simeq$ 1-50~GeV for couplings that reproduce the correct relic density. Note that bound state formation is extremely relevant here. For smaller DM masses and mass gaps, the produced bound states are highly energetic in the longitudinal direction and boosted. Depending on the lifetime of these particles, they can decay within the fiducial region of the FASER and FASER2 detector. An account of this will be provided in our future work \cite{BSFLHC}. For the purposes of this note, we summarize the preliminary findings of this paper in \cref{fig:Summary_uR_zoom}, where we plot the current constraints on the `so-called' $u_{R}$ model.  

We observe that the large mass gap `low' $m_{DM}$ region  is primarily constrained by ATLAS mono and multi-jet + missing energy searches, while the `large' $m_{DM}$ region is constrained by direct detection searches. For `low' $m_{DM}$ and $\Delta m$, the constraints come primarily from HSCP searches at the LHC. For longer lifetimes, including smaller couplings and low mass gaps, we envisage FASER/FASER2 to significantly enhance the reach for bound states in regions of parameter space satisfying cosmological constraints. 

%% file: sec_bsm2.tex
\contributors{Ahmed Ismail, Felix Kling, Sebastian Trojanowski, Yu-Dai Tsai (conveners), Brian Batell, Alexey Boyarsky, Matthew Citron, Jonathan L. Feng, Max Fieg, Saeid Foroughi-Abari, Jinmian Li, Jui-Lin Kuo, Roshan Mammen Abraham, Alex Mikulenko, Maksym Ovchynnikov, Junle Pei, Subir Sarkar, Lesya Shchutska}\\

In addition to the decay signatures considered in the previous section, a wide range of BSM theories predicts new stable particles which can scatter. This is most common in models of light DM. Typically, a collimated, high energy DM beam is created through meson decay, bremsstrahlung, or Drell-Yan production. The DM beam subsequently undergoes scattering in dense FPF detectors, including FASER$\nu$2, FLArE, and Advanced SND@LHC. Potential signatures arise from both electron and nuclear scattering. Unlike the decay signatures of the previous sections, there are typically non-negligible backgrounds arising from SM neutrino and muon scattering. These can be reduced through a combination of experimental vetoes and kinematic techniques. As will be shown, in many cases, DM models which can produce the observed relic density through thermal freeze-out can be newly tested at the FPF. Nevertheless, the contributions below also consider DM scattering signatures independently of any assumptions on cosmological history.

\begin{figure}[th]
\centering
\includegraphics[width=0.99\textwidth]{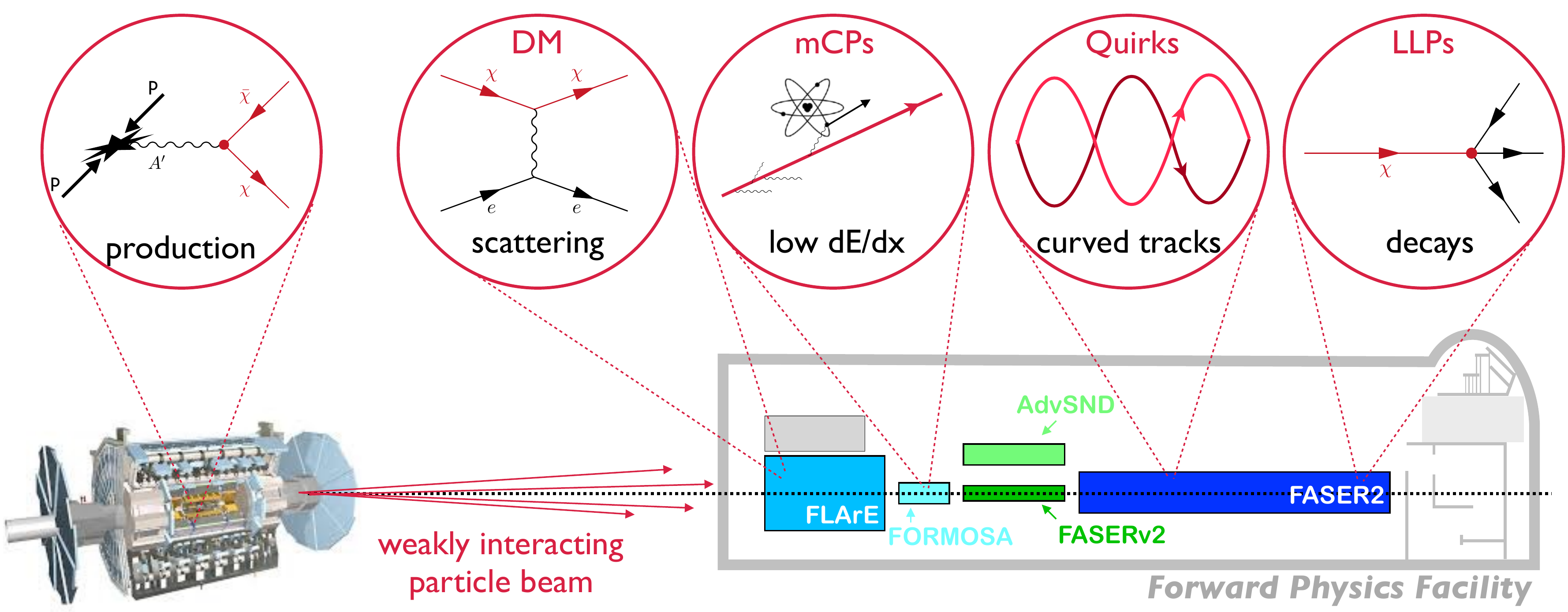}
\caption{Schematic illustration of the most important experimental signatures to be used in the FPF searches for BSM physics. On the left, we show the production of a dark sector state $\chi$ in the proton-proton scattering at the LHC Interaction Point, leading to a strongly collimated beam of these particles pointed at the FPF. On the right, we show the FPF cavern with its experiments and the BSM signatures that they can probe: i) scattering of DM in the neutrino detectors FLArE, FASER$\nu$2 and AdvSND@LHC detectors; ii) non-standard energy deposition of mCPs characterized in the FORMOSA and FLArE detectors; iii) and helical tracks in quirk models at FASER2; and iv) decays of long-lived particles to be studied in FASER2.}
\label{fig:bsm_scattering_sum}
\end{figure}

Because standard thermal freeze-out requires new mediators for DM at or below the GeV scale, theories with light DM can feature multiple signatures associated with new particles besides the DM itself. There is often interesting complementarity between the scattering signatures considered in this section and potential signatures in other sections. For instance, the mediators connecting DM to the SM can decay to SM final states. If the mediator couples to neutrinos, neutrino cross-sections and kinematic distributions at the FPF can also be affected. This section focuses on models with significant new sensitivity arising from DM scattering signatures.

Finally, some BSM theories contain particles that behave unlike traditional SM objects inside a detector. One of the most well-studied examples is mCPs, which interact far more weakly than typical SM charged particles in detectors. The unusual pattern of energy deposition arising from a mCP could be observed using the proposed FORMOSA detector at the FPF in analogy with existing collider searches. We also describe the potential signatures of quirks, which leave unique tracks owing to their long-range confining interactions. These examples both highlight the diversity of potential BSM signatures at the FPF.

In the rest of this section, we first describe the scattering of light DM at the FPF in \cref{sec:bsm_dm_scatter}, considering a range of models from the standard dark photon scenario to more exotic interactions. Then, the FPF sensitivity to mCPs is studied in \cref{sec:mCPs}. \cref{sec:quirks} evaluates the ability of long-lived particle detectors at the FPF to search for quirks. We illustrate the relevant signatures along with the example FPF detectors that could probe them in \cref{fig:bsm_scattering_sum}.

\section{Dark Matter Scattering}
\label{sec:bsm_dm_scatter}

On top of the rich experimental program to search for highly-displaced decays of unstable LLPs that can, i.a., mediate the interactions between the DM and SM sectors, the FPF experiments will also be well-suited to directly detect the scattering of DM species produced in the forward direction at the LHC. The direct search for sub-GeV DM in the FPF will be highly complementary to more traditional missing-energy searches at the LHC that target heavier DM particles, cf. Ref.~\cite{Kahlhoefer:2017dnp} for recent review. Importantly, such light DM particles can be efficiently thermally produced in the early Universe, in a way that is similar to weakly interacting massive particles (WIMPs)~\cite{Boehm:2003hm, Pospelov:2007mp, Feng:2008ya}. This has prompted extensive discussions about the discovery prospects of such light DM species, cf. Ref.~\cite{Alexander:2016aln, Battaglieri:2017aum, Beacham:2019nyx} for reviews; see also contribution to Snowmass 2021 \textsl{Big Idea: dark matter production}~\cite{RF06WP_1} and \textsl{RF6 - Overview of Facilities and Experiments}~\cite{RF06WP_4}. The direct search for DM particles at the FPF contributes significantly to these efforts, bringing to bear the unique capabilities of far-forward searches at the LHC and complementing experimental proposals based on the missing energy/momentum signatures. 

In addition, DM searches at the FPF will offer distinct discovery prospects with respect to standard DM direct detection experiments placed in deep underground facilities, cf. Ref.~\cite{Billard:2021uyg} for recent review. This is first due to the large DM energies accessible at the LHC. Notably, in this case, DM particles with masses even much below the $\gev$ scale will generate recoil energies of the SM targets of order $E_r\gtrsim 100~\mev$ up to tens of $\gev$. This is far above the typical energies induced in the scatterings of non-relativistic light DM, which currently limit traditional direct detection searches. The large energies characteristic of the LHC also enable probes of DM interactions in the relativistic regime, opening new discovery prospects in models predicting suppressed non-relativistic scattering rates. In addition, DM searches at the FPF rely on the highly-collimated far-forward flux of DM particles produced at the LHC. As a result, even a relatively small detector like FLArE, with fiducial mass of order $10$ ton concentrated within a $1~\m$ radius along the beam collision axis, can successfully probe thermal relic targets of popular theoretical scenarios predicting the existence of light DM species~\cite{Batell:2021blf}. 

Notably, the expected DM event rate in the FPF depends on the DM-SM coupling constant $g_{\textrm{DM-SM}}$ in a different manner to the traditional direct detection searches that are sensitive to the local DM number density. In the former case, the increasing value of such coupling generates both the larger DM flux and the growing scattering rate. The expected number of events in the detector then scales as $g_{\textrm{DM-SM}}^4$. In the latter searches, the growth of the scattering probability is compensated for by the decreased thermal relic density of DM, resulting in the predicted constant event rate independent of $g_{\textrm{DM-SM}}$. Simultaneous observation of the DM signal in both types of searches could then shed more light on the specific nature of the DM coupling to the SM. 

The DM search in the FPF will rely on the possibility to discriminate the BSM signal rates from the muon- and neutrino-induced backgrounds. In particular, the muon-induced backgrounds can significantly limit the DM discovery potential of the FPF experiments, if it is not actively vetoed. However, these can be greatly suppressed by detecting the through-going muon and collecting information about the time of the event, with minimal impact on the DM signal rate. In the following, we will assume that such a vetoing occurs, which remains straightforward for the FLArE detector, while it would require additional electronic detectors to provide the time information about the event in the emulsion detector FASER$\nu$2, cf. Ref.~\cite{Batell:2021blf} for further discussion. We also assume that muon-induced backgrounds can be rejected in the DM search at the Advanced SND@LHC detector. 

Instead, rejecting the neutrino-induced backgrounds requires employing specific kinematic cuts, as will be illustrated below. This allows for probing DM species below the specific (LHC-driven) ``neutrino floor'' present in DM direct detection searches in the far-forward region of the LHC. Below, we first present in more detail the prospects for the direct search for relativistic DM particles at the FPF for the popular dark photon mediator models and we highlight possible complementarities between different experimental signatures. We then discuss the expected exclusion bounds for the hadrophilic DM models coupled to the SM via the dark vector mediator. 

\subsection{Dark Photon Mediator Models} 
\label{sec:DMdarkphoton}

We first investigate a well-motivated and commonly studied class of light DM models based on a massive dark photon mediator $A'_\mu$ of a spontaneously broken U(1)$_D$  gauge symmetry, that couples to the SM via kinetic mixing. For the (sub-)GeV $A'$ masses of interest here, the Lagrangian in the physical basis is given by
\be
\mathcal{L} \supset -\frac{1}{4} F'_{\mu\nu}F^{'\mu\nu} +\frac{1}{2} m_{A'}^2 A'_\mu A^{'\mu} + A'_\mu (\varepsilon \, e \, J_{EM}^\mu + g_D \, J_\chi^\mu),
\ee
where $m_{A'}$ is the dark photon mass, $\varepsilon$ is the kinetic mixing parameter, $g_D \equiv \sqrt{4 \pi \alpha_D}$ is the U(1)$_D$ gauge coupling, $J_{EM}^\mu$ is the SM electromagnetic current, and  $J_\chi^\mu$ is the current for the DM particle $\chi$, with mass denoted by $m_\chi$.
For our DM candidates, we will study two cases: (1) complex scalar DM and (2) Majorana fermion DM, with respective currents given by
\be
\label{eq:JD}
J_\chi^\mu = 
\begin{cases}
\displaystyle{\ i \chi^* \overset{\text{\footnotesize$\leftrightarrow$}}{\partial_\mu} \chi} \quad \text{(complex scalar DM)}  \\
\displaystyle{\ \frac{1}{2} \overline \chi \gamma^\mu \gamma^5 \chi} \quad \text{(Majorana fermion DM)} \ .
\end{cases}
\ee

Both scalar and Majorana DM can be produced in the early universe through simple thermal freeze-out with the correct relic density. Notably, for $m_{A'} > 2 m_\chi$, DM annihilates directly to SM fermions through $s$-channel dark photon exchange, $\chi \chi \to A^{'(*)} \to f \bar f$, with an annihilation cross section given by (for $m_{A'} \gg m_\chi$)
\be
\label{eq:sigv}
\sigma v \propto \alpha \, v^2 \, \frac{\varepsilon^2 \, \alpha_D \, m_\chi^2}{m_{A'}^4}  = \alpha \, v^2 \, \frac{y}{m_\chi^2} \ ,
\ee
where $\alpha$ is the SM electromagnetic fine structure constant and $y \equiv \varepsilon^2 \alpha_D (m_\chi/m_{A'})^4$, following Ref.~\cite{Izaguirre:2015yja}. Below we will present our sensitivity estimates for DM scattering searches at the FPF detectors in the $(m_\chi, y)$ plane for the well-studied benchmark choices $\alpha_D = 0.5$ and $m_{A'}  = 3 m_\chi$. We will also show in this plane the thermal targets, corresponding to regions of parameter space explaining the observed DM relic abundance~\cite{Berlin:2018bsc}. 

Before turning to a discussion of the scattering signatures, it is worth highlighting the velocity suppression present in \cref{eq:sigv}, corresponding to $P$-wave annihilation. This implies the consistency of these models with bounds derived from observations of the cosmic microwave background anisotropies. We also remark that direct detection bounds are naturally evaded in the Majorana DM case due to the rate suppression from momentum-dependent scattering and can also be evaded in a simple way in the scalar DM case by introducing a small mass splitting, causing the scattering reaction to proceed inelastically~\cite{Tucker-Smith:2001myb}.

The direct search for the DM scattering in the FPF will rely on several distinct detection modes. In the case of the positive signal, the complementarity between them will provide additional information about the DM couplings. Here, we briefly present these signatures based on the discussion in Refs.~\cite{Batell:2021blf, Batell:2021aja}, in which the DM search in the FLArE and FASER$\nu$2 detectors in the FPF has been proposed. 

As mentioned above, the direct search for the DM scattering in the FPF will rely on the highly-collimated flux of DM particles produced in the far-forward region of the LHC and will employ several distinct detection modes. In our modeling, we employ the \texttool{FORESEE} code~\cite{Kling:2021fwx}, cf. \cref{sec:foresee} for further details, and rely on Ref.~\cite{Kling:2021gos} for the forward neutrino flux and spectrum. The complementarity between different scattering signatures will provide additional information about the DM couplings. Here, we briefly present these signatures based on the discussion in Refs.~\cite{Batell:2021blf, Batell:2021aja}, in which the DM search in the FLArE and FASER$\nu$2 detectors in the FPF has been proposed. 

The most important feature of sub-GeV DM scatterings mediated by the light dark photon, which allows the discrimination of the DM signal events from neutrino-induced backgrounds, is the typical low momentum exchange present in the DM interactions. This is in contrast to neutral current (NC) neutrino scatterings mediated by the heavy $Z$ bosons that favor larger visible energy depositions in the detector. This can be seen from the approximate expression for the differential electron scattering cross section valid in the limit of the large incident DM energy, $E_\chi\gg m_\chi$, and large electron recoil energy, $E_e\gg m_e$,
\begin{equation}
    \frac{d\sigma}{dE_e} \approx \frac{ 8\pi\, \epsilon^2\, \alpha\, \alpha_D\,m_e} {(m_{A^\prime}^2 + 2m_eE_e)^2}\ .
\end{equation}
The relevant scattering rate is dominated by events characterized by $2m_eE_e\lesssim m_{A^\prime}^2$. For $m_{A^\prime}\lesssim 100~\mev$, this corresponds to $E_e\lesssim 10~\gev$. This allows for a sensitive search for DM through its \textsl{electron scatterings}, $\chi e\to \chi e$, by employing a cut favoring low recoil energy events. 

Below, we present the results for a search requiring the electron recoil to be within the range $30~\mev (300~\mev) < E_e < 20~\gev$, where the lower bound is driven by the FLArE (FASER$\nu$2) detector capabilities to study low-energy EM signal, while the upper bound is set to reduce backgrounds from neutrino interactions. The typical electron recoil energy depends on the dark photon mediator mass and varies between $E_e\sim 100~\mev$ and $E_e\sim 10~\gev$ in the sensitivity reach plots below. When obtaining the sensitivity curves, additional bounds on the electron recoil angle $\theta_e$ have been imposed, as discussed in Ref.~\cite{Batell:2021blf}. These, however, play only an auxiliary role in the analysis. Importantly, though, the detection of only a very small electron recoil angle with respect to the beam collision axis will facilitate the identification of the events related to DM particles coming from the $pp$ collisions at the distant LHC IP. Thanks to the cuts on $E_e$ and $\theta_e$, the neutrino-induced backgrounds in this search can be reduced to $\mathcal{O}(10)$ events for the $10$-tonne detector during the entire HL-LHC era.

A complementary way of probing light DM in the FPF is to search for their \textsl{elastic scatterings off protons}, $\chi p\to \chi p$. This leads to a single proton charged track in the detector that can be efficiently distinguished from the electron-induced signals. In this case, however, the larger target mass makes it more difficult to generate sufficiently large proton recoil energies when a very light dark vector is exchanged in the scattering. As a result, the relevant expected exclusion bounds are typically less constraining than the bounds derived based on the electron scattering, especially in the limit of $m_{A^\prime}\lesssim 100~\mev$. When presenting the result for this signature, we employ the upper cuts on the proton recoil momentum $p_p\lesssim 500~\mev$ ($1~\gev$) for FLArE (FASER$\nu$2). The relevant lower bounds are driven by detector capabilities, $E_{k,p}\gtrsim 20~\mev$ for FLArE and $p_p\gtrsim 300~\mev$ for FASER$\nu$2. This leads to $\mathcal{O}(100)$ expected neutrino interactions that could mimic the DM signatures in the $10$-tonne detector. We additionally employ \texttool{GENIE}~\cite{Andreopoulos:2009rq, Andreopoulos:2015wxa} and take into account the possibility that the recoiled proton will rescatter (final-state interactions, FSI) before leaving the nucleus, which could affect the identification of the event, cf. Ref.~\cite{Batell:2021aja} for further discussion.

For  dark photon masses above the QCD scale, the DM interaction can also probe the internal structure of the nucleon in the \textsl{deep inelastic scatering} (DIS) regime. This leads to events with rich hadronic activity, including potentially multiple charged tracks, and searches for these signatures can offer the best sensitivity for heavier DM species. In the following, we will present the expected bounds in a search employing cuts on the hadronic energy, $1~\gev < E_{\textrm{had}}< 15~\gev$, and the hadronic transverse momentum of the event $1~\gev < p_{T,\textrm{had}} < 1.5~\gev$. Again, by focusing on signatures with  relatively low visible energies, one can significantly suppress neutrino-induced backgrounds from more than $10^5$ events to $\mathcal{O}(10^3)$ events over the full HL-LHC era at the $10$-tonne detectors. Instead, the impact of these cuts on the DM signal rate is comparatively minor, cf. Ref.~\cite{Batell:2021aja} for further discussion. While we do not study systematic uncertainties, we note they could be reduced when searching for an excess in the ratio between the NC and CC scattering rates, $r = $NC/CC.

\begin{figure}[t]
\centering
\includegraphics[width=0.99\textwidth]{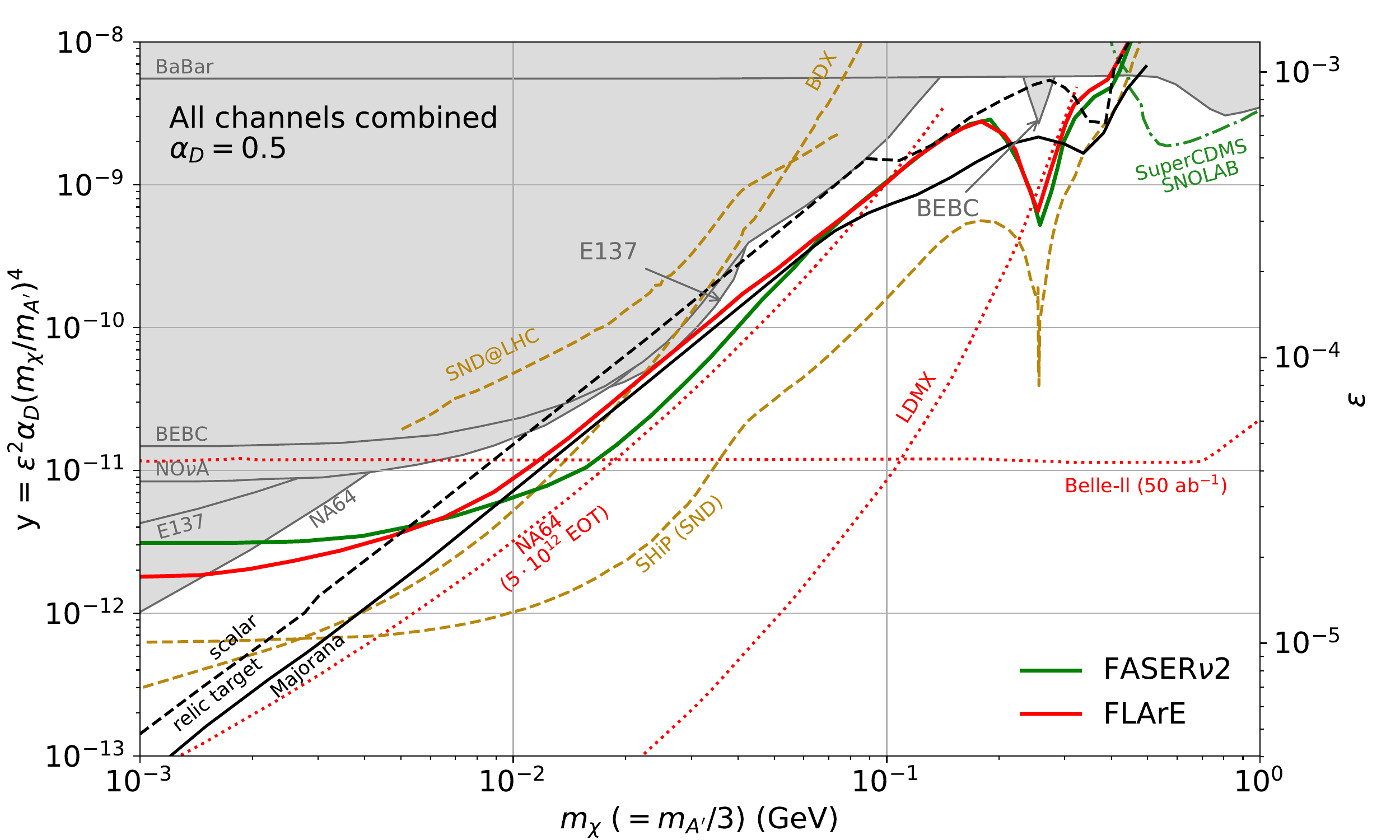}
\caption{The projected 90\% CL exclusion bounds combining all channels for the FASER$\nu$2 and FLArE-10 detectors at the HL-LHC with 3 ab$^{-1}$ of integrated luminosity. Existing constraints and projected reaches from other experiments are shown with the gray-shaded region and colorful lines, respectively, see text for details. Adapted from Ref.~\cite{Batell:2021aja}.}
\label{fig:DMdarkphotonboundsFLArEFASERnu2}
\end{figure}

In \cref{fig:DMdarkphotonboundsFLArEFASERnu2}, we present the expected DM exclusion bounds in the $(m_\chi,y)$ plane for FLArE and FASER$\nu$2 for the benchmark scenario with $m_{A^\prime} = 3m_\chi$ and $\alpha_D=0.5$. Here, all the aforementioned experimental signatures have been taken into account. In particular, the reach in the low mass regime, $m_\chi\lesssim 200~\mev$, is dominated by the search for DM-electron scattering, while for heavier masses the expected bounds from DIS become important or even the most stringent. The search based on elastic scatterings of protons provides equally strong bounds for $100~\mev \lesssim m_\chi\lesssim 300~\mev$. In the plot, the gray shaded region is currently excluded based on the data from the BaBar~\cite{BaBar:2017tiz}, BEBC~\cite{BEBCWA66:1986err}, E137~\cite{Bjorken:1988as,Batell:2014mga}, LSND~\cite{deNiverville:2011it}, MiniBooNE~\cite{MiniBooNEDM:2018cxm}, NA64~\cite{Banerjee:2019pds}, and NO$\nu$A~\cite{Wang:2017tmk} experiments following Refs.~\cite{Buonocore:2019esg, Marsicano:2018glj}. In addition, we also present future projected bounds from other searches in the BDX~\cite{BDX:2016akw}, Belle-II~\cite{Belle-II:2018jsg}, LDMX~\cite{LDMX:2018cma}, NA64~\cite{Gninenko:2019qiv}, SHiP~\cite{SHiP:2020noy}, and SND@LHC~\cite{SHiP:2020sos} detectors. We also show there the expected future bounds from the SuperCDMS experiment~\cite{Battaglieri:2017aum, Berlin:2018bsc, LDMX:2018cma}, which assume the Majorana DM particle.

We also present there the thermal relic targets for the scalar and Majorana DM. These are indicated with the dashed and solid black lines, respectively, while in the regions above these lines, the thermal DM relic density is predicted to be lower than the observed DM relic abundance. As can be seen, the FPF experiments will cover important parts of the currently allowed regions in the parameter space of these models that remain consistent with the standard cosmology.

\subsection{Hadrophilic Dark Matter Models}
\label{sec:bsm_dm_hadro}

Many of the most sensitive experimental probes of a dark photon mediator stem from its couplings to leptons. However, on general grounds one can envision a variety of phenomenologically distinct mediator coupling patterns, and it is important to consider such scenarios to obtain a clearer picture of the physics potential of proposed experiments. Since the LHC collides protons, it can be especially sensitive to mediators with primarily hadrophilic coupling patterns, i.e., sizable couplings to quarks, but suppressed couplings to leptons. 

With this motivation, we now consider scenarios of this kind featuring hadrophilic couplings of a vector mediator, here denoted as $V_\mu$; see Ref.~\cite{Batell:2021snh} for further details. As case studies, we will discuss two representative hadrophilic dark sector models based on 1) an anomalous U(1)$_B$ and 2) and anomaly-free U(1)$_{B-3L_\tau}$ gauge symmetries. In the physical basis, the Lagrangian of the vector mediator $V_\mu$ is 
\begin{equation}
\label{eq:L-V}
{\cal L} \supset -\frac{1}{4} V_{\mu\nu} V^{\mu\nu}+\frac{1}{2} m_V^2 V_\mu V^\mu + V_\mu (J^\mu_{\text{SM}} + g_V Q_\chi J^\mu_\chi) \ ,
\end{equation}
where $m_V$ is the vector mass. The current $J^\mu_{\rm SM}$ containing SM fields is given by 
\begin{equation}
J^{\mu}_{\text{SM}} = 
g_V [J^\mu_B - 3 x (\overline \tau \gamma^\mu \tau +  \overline \nu_\tau \gamma^\mu P_L \nu_\tau) ] + \varepsilon \, e \, J^\mu_{\text{EM}} \ ,
\end{equation}
where $g_V \equiv \sqrt{4 \pi \alpha_V}$ is the new U(1) gauge coupling, $J_B^\mu$ and $J^\mu_{\text{EM}}$ are the baryon number and electromagnetic currents, respectively, $\varepsilon$ is the kinetic mixing parameter, and $x= 0$ (1) for the U(1)$_B$ (U(1)$_{B-3L_\tau}$) model. As in the dark photon study above, we will again consider both complex scalar and Majorana fermion DM candidates, with current given by \cref{eq:JD}, while $Q_\chi$ is the charge of the DM under the new gauge symmetry. 

In our sensitivity studies below, we will assume that the kinetic mixing is generated radiatively, with characteristic one-loop size $\varepsilon = e  g_V/(16 \pi^2)$. As in the dark photon models above, we will further fix $m_V = 3 m_\chi$. Our results will be presented in the $(m_V, g_V)$ plane under two qualitatively distinct assumptions for the DM charge $Q_{\chi}$. In one scenario, we will assume DM and SM particles have comparable interaction strengths with the mediator $V$, fixing
\begin{equation}
\label{eq:Qchi}
Q_\chi =  
\begin{cases}
1 \, , \  \text{U(1)$_B$ models} \\
3 \, , \  \text{U(1)$_{B-3L_\tau}$ models} \ . 
\end{cases}
\end{equation}
In the $B-3L_\tau$ model, we have fixed the DM charge to be opposite that of the $\nu_\tau$, $Q_\chi = -Q_\tau$. As a second, qualitatively distinct, scenario, we will assume DM couples to $V$ with a fixed strength, 
\begin{equation}
\alpha_\chi \equiv \frac{g_V^2 Q_\chi^2}{4 \pi} = 0.01 
 \ . 
\label{eq:largeQ}
\end{equation}

The production rates and distributions of the mediator $V$ in the far-forward region of the LHC is obtained using the \texttool{FORESEE} package~\cite{Kling:2021fwx}. We consider several production channels, including light meson decays, proton bremsstrahlung~\cite{Blumlein:2013cua, deNiverville:2016rqh} (See also Ref.~\cite{Foroughi-Abari:2021zbm}), and Drell-Yan. We then consider the decays of the mediator to dark matter and SM final states. For the latter, we employ the the \texttool{DarkCast} package~\cite{Ilten:2018crw}, which uses a data-driven approach to estimate the hadronic widths (see also Ref.~\cite{Plehn:2019jeo}). 

As in the dark photon mediator models, the correct DM relic abundance can be obtained through thermal freezeout of DM annihilation to SM particles, and below we will present the thermal targets in the $(m_V, g_V)$ plane. In the U(1)$_{B-3L_\tau}$ models, due to the open channel to tau neutrinos,  DM annihilates efficiently throughout the entire mass range. Instead, in the U(1)$_B$ model, annihilation is only efficient for masses  $m_V = 3 m_\chi \gtrsim {\cal O}( 1~{\rm GeV})$, due to the lack of unsuppressed kinematically open channels at lower masses. Under the assumptions described above, the annihilation cross section displays the following parametric dependence 
\begin{equation}
\label{eq:sigv-parametric}
\sigma v 
\sim v^2 \frac{ g_V^4 Q_\chi^2}{ m_V^2} 
\sim v^2 \frac{ g_V^2 \alpha_\chi}{m_V^2}  \ .  
\end{equation}
For our two scenarios with either fixed $Q_\chi$ or $\alpha_\chi$, the thermal targets exhibit clear trends in the $(m_V, g_V)$ plane that can be understood from \cref{eq:sigv-parametric}. We also note that similar remarks to those made in the dark photon model regarding $P$-wave suppressed annihilation 
(see \cref{eq:sigv-parametric}) and the lack of CMB constraints, as well the model dependence of direct detection constraints, apply to the hadrophilic models considered here. 

\begin{figure}[t]
\centering
\includegraphics[width=0.48\textwidth]{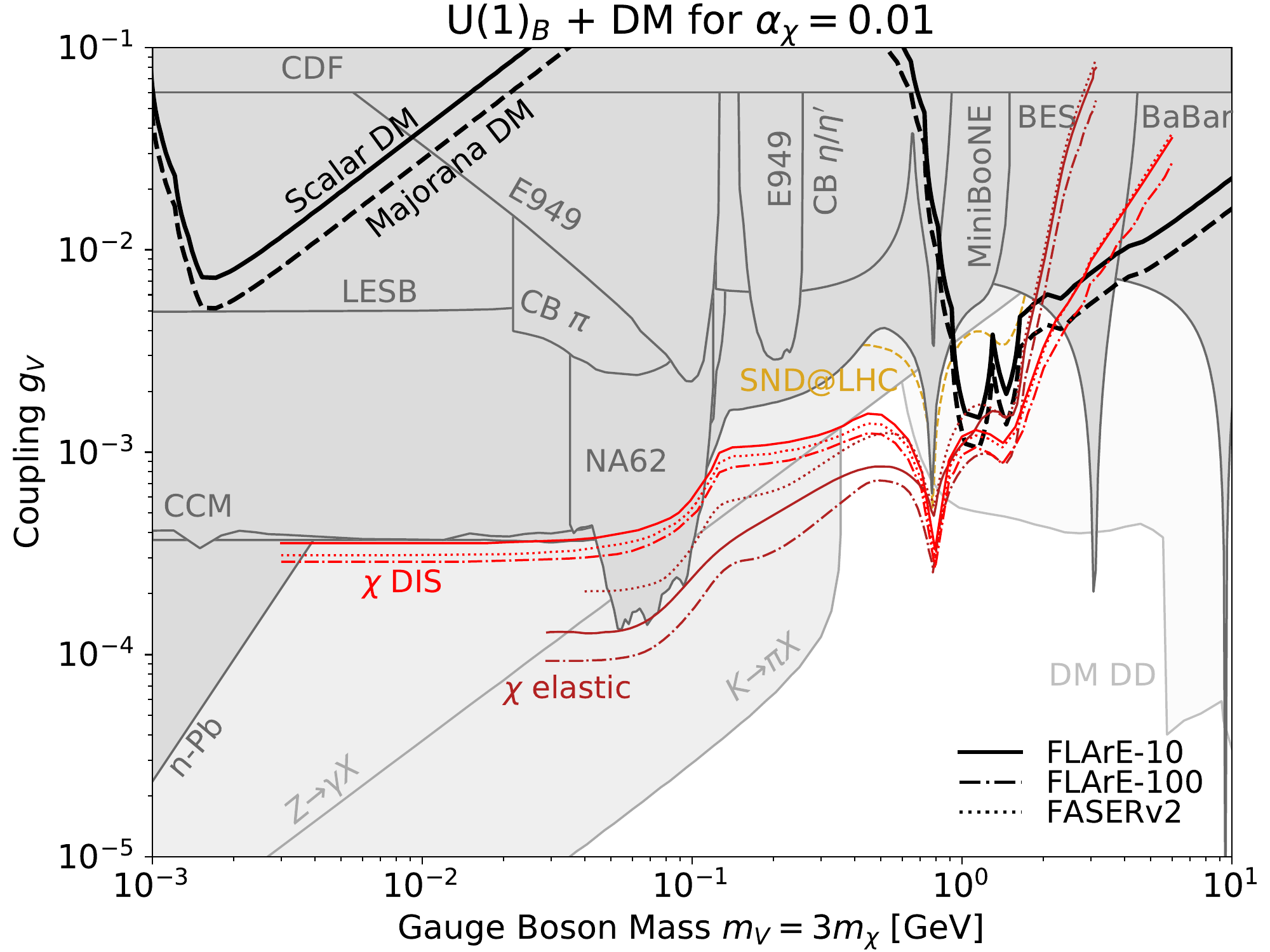}
\includegraphics[width=0.48\textwidth]{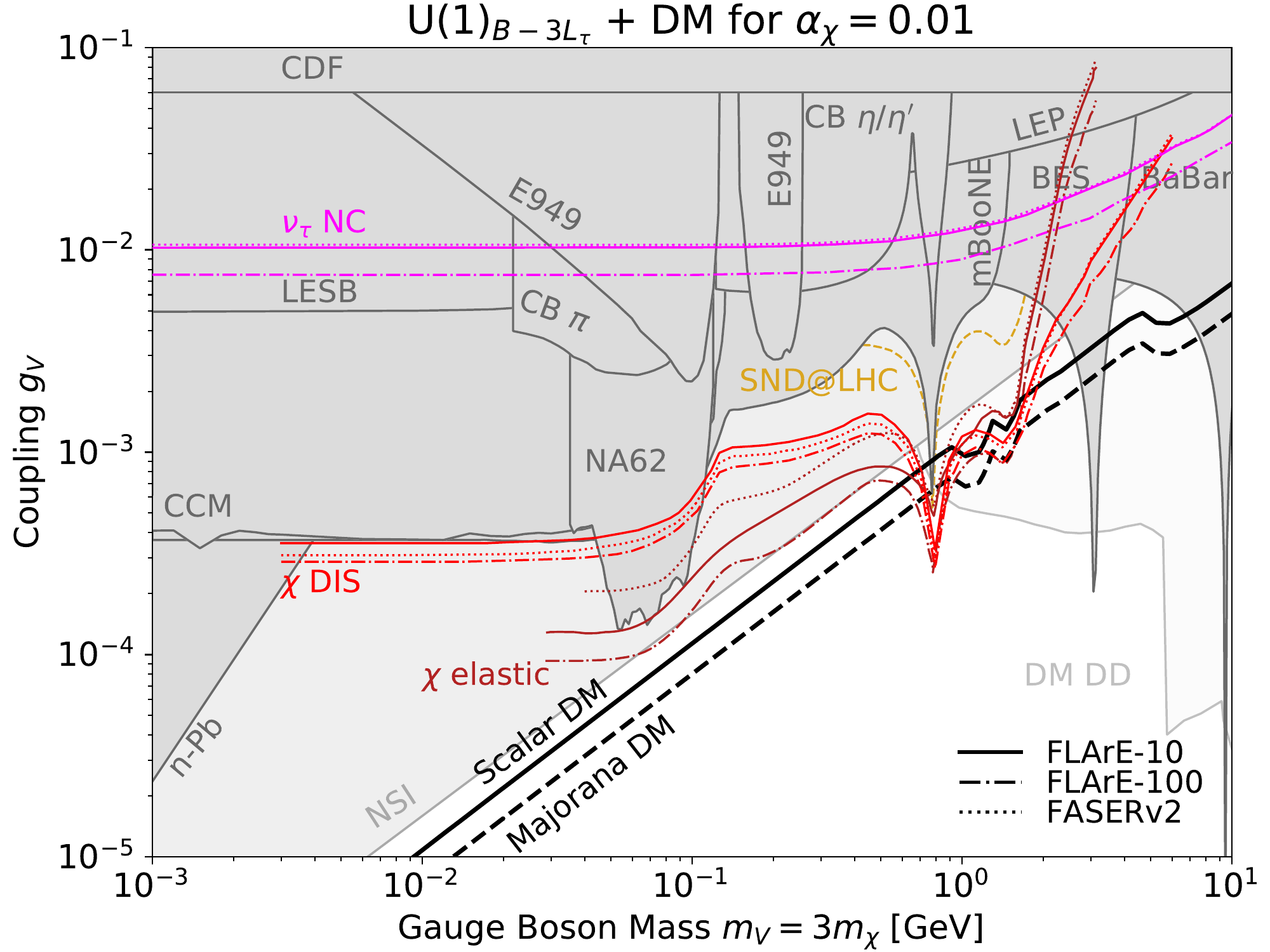}
\caption{The expected exclusion bounds in the search for the scattering of the hadrophilic DM with $\alpha_\chi=0.01$ and $m_V = 3\,m_\chi$ is shown for the $U(1)_B$ (left) and $U(1)_{B-3\tau}$ (right) models with invisibly decaying dark vector mediator. The light (dark) red lines correspond to the DIS (elastic) scattering signatures. The results obtained for the FLArE and FASER$\nu$2 detectors is shown with the solid and dotted lines, respectively. In the left, we also show with the purple line the projected bounds obtained based on the search of excessive neutral-current neutrino interactions in FLArE. The gray-shaded regions are excluded by current bounds (see text). The black solid and dashed lines correspond to the relic targets for scalar and Majorana DM, respectively. Taken from Ref.~\cite{Batell:2021aja}.}
\label{fig:hadrophilicDM}
\end{figure}

The expected exclusion bounds on the hadrophilic models for both the FLArE and FASER$\nu$2 detectors are shown in \cref{fig:hadrophilicDM} following Ref.~\cite{Batell:2021snh}. In the figure, we focus on the case with the fixed dark coupling constant, $\alpha_\chi = 0.01$ and present the results for both the $U(1)_B$ and $U(1)_{B-3\tau}$ scenarios in the left and right panels, respectively. As can be seen, in both cases FLArE and FASER$\nu$2 detectors have capabilities to probe parts of the relic target lines for the DM mass $m_\chi\sim\gev$, up to a few $\gev$, which remain unexcluded. Interestingly, one expects excesses over the neutrino-induced backgrounds in both the elastic DM-proton scattering and  DM-DIS channels, cf. \cref{sec:DMdarkphoton} for details  about these signatures.

The current bounds on the hadrophilic models are shown with the dark gray-shaded regions and are based on null searches in ARGUS~\cite{ARGUS:1986nzm}, BaBar~\cite{BaBar:2009gco}, BES~\cite{BES:2007sxr}, CDF~\cite{CDF:2012tfi, Shoemaker:2011vi}, Coherent CAPTAIN-Mills (CCM)~\cite{Aguilar-Arevalo:2021sbh}, Crystal Ball~\cite{CrystalBarrel:1996xfs}, E949~\cite{BNL-E949:2009dza}, GAMS-2000~\cite{Serpukhov-Brussels-LosAlamos-AnnecyLAPP:1987kiw}, LESB~\cite{Atiya:1992sm}, MiniBooNE~\cite{MiniBooNE:2017nqe, MiniBooNEDM:2018cxm}, NA62~\cite{NA62:2019meo}, NuCal~\cite{Blumlein:1991xh} experiments, as well as on the measurements of the low-energy neutron-lead scattering cross section~\cite{Barbieri:1975xy, Barger:2010aj} (see also Refs.~\cite{Pospelov:2008zw, Batell:2014yra}). Instead, the light gray-shaded region corresponds to the gauge anomaly-induced bounds from the rare $Z$ boson and flavor-changing meson decays~\cite{Dror:2017ehi, Dror:2017nsg} in the $U(1)_B$ case, and to neutrino non-standard interactions (NSI) constraints~\cite{Han:2019zkz} for the U(1)$_{B-3\tau}$ model. In both cases, the are further bounds on the scenarios from DM direct detection searches in CRESST-III~\cite{CRESST:2019jnq}, DarkSide~\cite{DarkSide:2018bpj}, Xenon-1T~\cite{XENON:2018voc, XENON:2020gfr}. These, however, apply only to the complex scalar DM case. Even in this case, the bounds can be significantly weakened if the scattering is rendered inelastic through the introduction of a small DM mass splitting.

It is worth highlighting that beyond the search for DM scatterings these hadrophilic models can lead to additional striking signatures at the FPF experiments. In particular, in the right panel of \cref{fig:hadrophilicDM} we present additional expected exclusion bounds that can be obtained from searches for BSM-induced $\tau$ neutrino scatterings in FLArE that lead to an excess of neutrino NC interactions. The complementarity between different searches is also evident in the results presented in \cref{fig:limits_BDMmodel} (see \cref{sec:llp_vec_b}), which has been obtained for the complex scalar DM coupled to the SM via the $U(1)_B$ gauge boson with the dark charge set to $Q_\chi = 1$. For this natural choice of the dark charge, one expects the dark vector to decay both visibly, either into lighter hadronic states or $e^+e^-$ pairs, and invisibly into the DM particles. Such visible decays will be successfully probed in FASER2, which extends the reach of the FPF experiments in this model towards small values of the coupling constant, $g_V\sim (10^{-8}-10^{-5})$ for $m_\chi$ of order several hundred $\mev$. 

Similar complementary bounds can be obtained in the $U(1)_{B-3\tau}$ case, assuming order one values of $Q_\chi$. In this case, additional bounds can be derived from the search for the enhanced production of  $\tau$ neutrinos in the far-forward region of the LHC due to  dark vector decays~\cite{Kling:2020iar}. These can then be studied via the measurement of the $\nu_\tau$ charge current (CC) scatterings in the FPF, cf. Ref.~\cite{Batell:2021snh} for further discussion.

\subsection{Dark Matter Search in the Advanced SND@LHC Detector}

The successor of the SND@LHC experiment during HL-LHC, Advanced SND is planned to be made of two detectors to perform QCD measurements with improved accuracy and neutrino physics, including cross section measurements. The first detector (\sndfar) to be located at the FPF in the pseudorapidity range $\eta\in[7.2,8.4]$ aims to continue the physics program of the SND@LHC, increasing the number of detected neutrino interactions. The second detector (\sndnear) will be at $55$ m from the interaction point and cover larger angles $\eta\in[4, 5]$, allowing to reduce the systematic uncertainty in charm hadron production by using the charm yield in the same angular range from LHCb. Both AdvSND layouts follow the design of the SND@LHC, consisting of a target region for the vertex reconstruction and EM energy measurement and a muon ID system acting also as HAD calorimeter. In addition, the new detectors will be equipped with a magnet for the measurement of the muon charge and momentum. The preliminary parameters of the detectors are given in \cref{tab:AdvSNDparameters}.

\begin{table}[t]
\centering
\begin{tabular}{c|c|c|c|c}
     \hline\hline
     &target cross section& distance $l_{\text{min}}$ & $\eta$&target mass \\
     \hline
     SND@LHC & $39\times 39 \, \text{cm}^2$ & \multirow{2}{*}{480 m} & [7.2,8.6] & 800 kg  \\
     \cline{1-2} \cline{4-5}
     \sndfar& $100\times 40\, \text{cm}^2$ & & $\approx$[7.2,8.4] & \multirow{2}{*}{5 tons}\\
     \cline{1-4}
     \sndnear & $120\times 120  \, \text{cm}^2$ & $55$ m & $\approx$[4,5] &\\
     \hline \hline
\end{tabular}
\caption{Parameters of SND@LHC detector, and the two detectors of the Advanced SND experiment used in this estimate.}
\label{tab:AdvSNDparameters}
\end{table}

We consider the sensitivity of the Advanced SND configurations to various BSM models. New feebly interacting particles (FIPs) may be produced in the $pp$ scatterings at the LHC interaction point, propagate to the detector and decay or scatter inside it. Similarly to the previous work~\cite{Boyarsky:2021moj}, we consider here scatterings of light dark matter particles (LDM) via leptophobic $U_{B}(1)$ mediator, as well as decays of Heavy Neutral Leptons (HNLs), dark scalars and dark photons. 

The number of events with FIPs in the AdvSND detectors are estimated as
\begin{equation}
    N_{\text{events}} = N_{\text{prod}}\times \epsilon \times P
    \label{eq:Nevents}
\end{equation}
where $N_\text{prod}$ is the total number of FIPs produced during the operation time of the experiment (HL phase of the LHC), $\epsilon$ is the geometrical acceptance -- the fraction of particles that move towards the detector, and $P$ is the average probability to decay/scatter inside the detector for these FIPs
\begin{align}
    \label{eq:p-scatt}
    &P_\text{scatter} = l^\text{scatter}_\text{det} \langle \sigma \rangle n \\
    & P_\text{decay} = \left\langle e^{-\frac{l_{\text{min}}}{c\tau \gamma v}} -e^{-\frac{l_{\text{min}}+l^\text{decay}_\text{det}}{c\tau \gamma v}} \right\rangle
\end{align}
with $\sigma$ being the scattering cross section, $n$ being the number density of target particles, $\tau$ being the lifetime, and $l_\text{det}$ being the effective length of the detector (the length of the target material for scatterings and the actual target length for decays). The average $\langle \dots\rangle$ is computed over the distribution of particles passing through the detector.

\paragraph{Scatterings} We consider here two scattering signatures: elastic -- an excess of neutrino-like elastic scatterings over the SM yield due to $\chi+p$ process, and inelastic -- an excess of the ratio of neutral-to-charged current-like events, $r = N_{\text{NC}}/N_{\text{CC}}$, over the SM prediction $r \approx 0.31$ due to $\chi+\text{nucleus}$ deep inelastic scattering (DIS). SND@LHC allows measuring the $r$ ratio with the accuracy $(\Delta r/r)_{\text{SND@LHC}} = 10\%$. We assume that the accuracy of the advanced configurations will be improved, and consider $(\Delta r/r)_{\text{AdvSND}} = 1\%$ as the reference value.

The model we study is the leptophobic portal with a mediator $V_{\mu}$ and scalar LDM particles $\chi$. The Lagrangian is
\begin{equation}
    \mathcal{L}\supset g_{B}V_{\mu}\times \frac{1}{3}\sum_{q}\bar{q}\gamma^{\mu}q + |(\partial_{\mu}-ig_{B}q_{\chi}V_{\mu})\chi|^{2},
    \label{eq:leptophobic-portal}
\end{equation}
where the sum $\sum_{q}$ goes over all quarks, $g_{B}$ is the leptophobic coupling, and $q_{\chi}$ is the leptophobic charge. The overview of constraints and phenomenology of the portal is described e.g. in~\cite{Boyarsky:2021moj}.

\begin{figure}[t]
\centering
    \includegraphics[width = 0.49\textwidth]{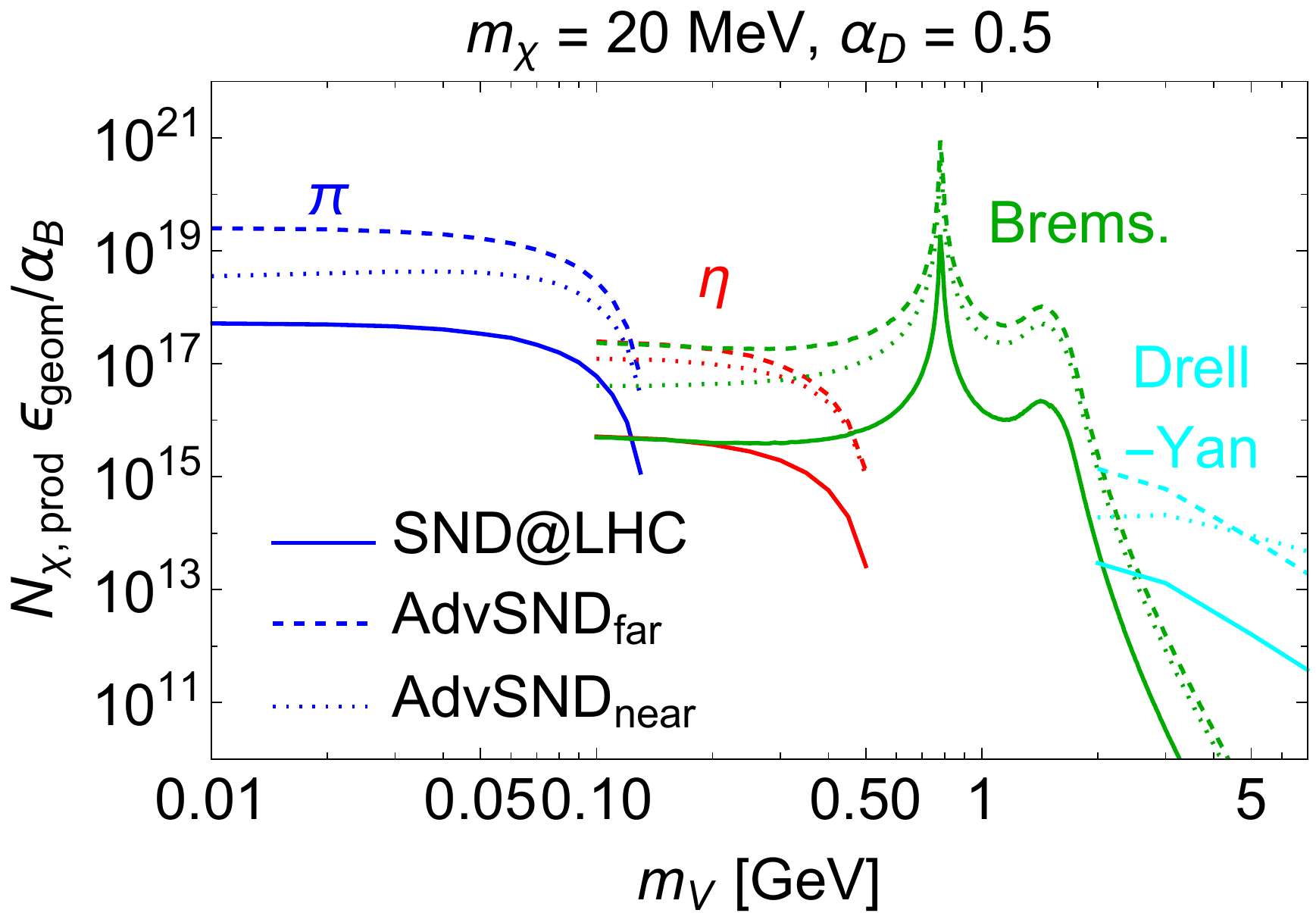}~
    \includegraphics[width = 0.51\textwidth]{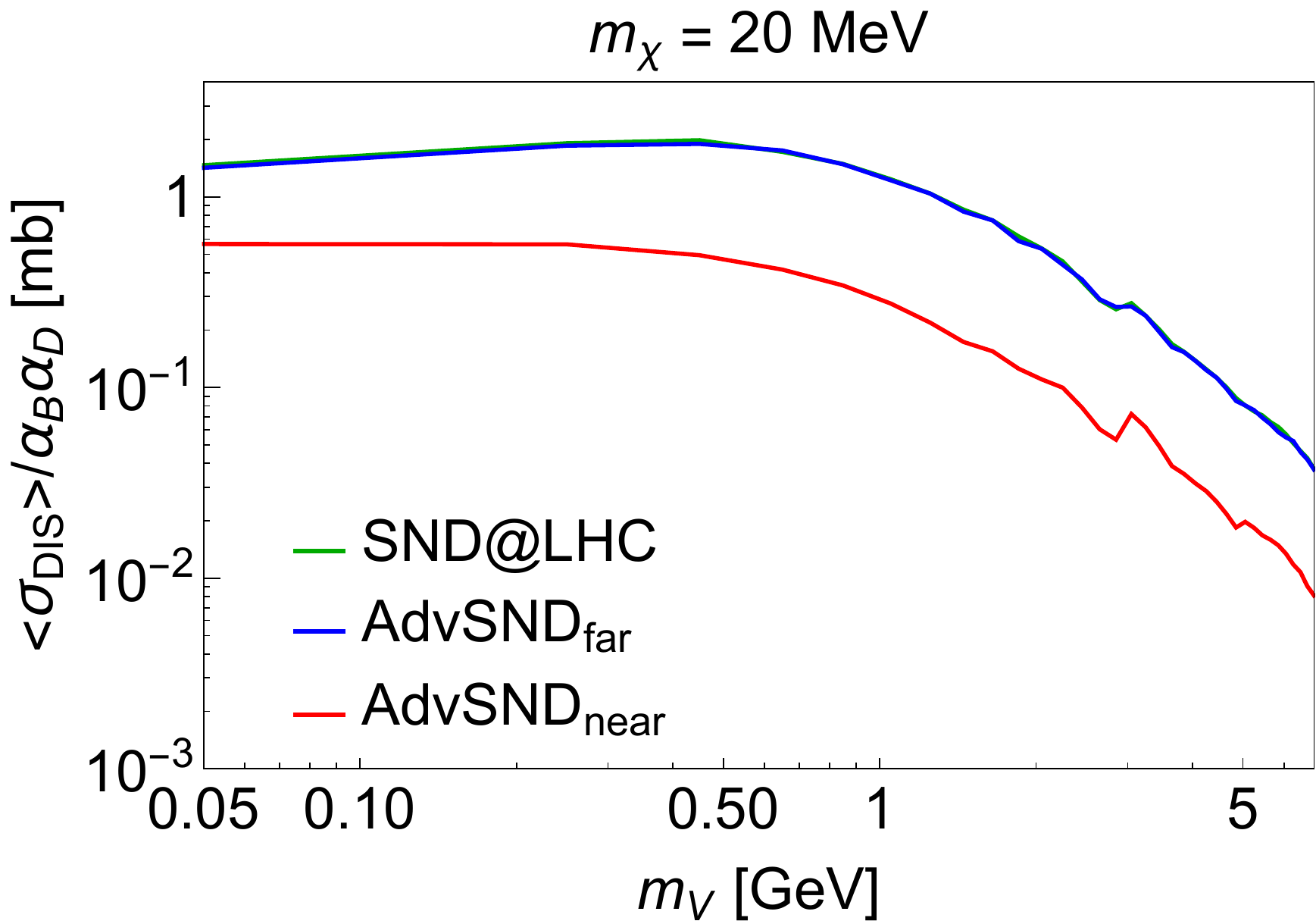}
\caption{Left: The number of $\chi$ particles produced from $\pi$, $\eta$, bremsstrahlung, and Drell-Yan process in the direction of the SND@LHC, $\sndnear$, and $\sndfar$. Right: The cross section of deep inelastic scattering $\chi$ off a nucleon averaged over the energy of $\chi$ particles produced in the direction of the SND@LHC, $\sndnear$, and $\sndfar$.}
\label{fig:LeptophobProductionAtSND}
\end{figure}

When calculating the number of events, we follow~\cite{Boyarsky:2021moj}. $\chi$ particles are produced in decays of $V$, whose dominant production channels are decays of $\pi$, $\eta$ for masses $m_V\lesssim m_\eta$, proton bremsstrahlung for $m_\eta\lesssim m_V \lesssim 3\text{ GeV}$, and the Drell-Yan process for $m_V\gtrsim 3~\gev$, see \cref{fig:LeptophobProductionAtSND}. These processes produce $V$ (and hence $\chi$) in the far-forward direction, and therefore the geometric acceptance for $\chi$ particles flying into $\sndfar$, which is located off-axis, is much lower than for those pointed to SND@LHC and $\sndnear$. 

The target particle density $n$ in~\cref{eq:p-scatt} is equal to the detector's atomic number density (the tungsten material is considered) times atomic $Z$ (elastic) or mass $A$ (NC/CC) numbers of the target material. The mass dependence of the averaged DIS cross section $\langle \sigma_{\text{DIS}}\rangle$ is shown in \cref{fig:LeptophobProductionAtSND}. The cross section grows with the mean energy of $\chi$ particles, which explains why the cross section at $\sndnear$ (where off-axis particles have relatively low energies) is suppressed as compared to the DIS cross section at $\sndfar$/SND@LHC (where particles travel in the far-forward direction and have large $\gamma$ factors).

The number of scattering events in~\cref{eq:Nevents} scales as
\begin{equation}
    N_{\text{events}} \propto \alpha_{B}\cdot \text{Br}(V\to \chi \chi)\cdot \alpha_{B}^{2}q_{\chi}^{2}, \quad \alpha_{B} = g_{B}^{2}/4\pi,
\end{equation}
where $\alpha_{B}\cdot \text{Br}(V\to \chi \chi)$ comes from the $\chi$ production, while $\alpha_{B}^{2}q_{\chi}^{2}$ from the $\chi$ scattering. To represent the reach of AdvSND, we marginalize over $q_{\chi}$ and $m_{\chi}$, choosing the $\alpha_{D} \equiv \alpha_{B}q_{\chi}^{2} = 0.5$ and $m_{\chi} = 20\text{ MeV}$. In this case, $\text{Br}(V\to \chi\chi) = 1$, and the number of events behaves as $N_{\text{events}}\propto \alpha_{B}^{2}$. In addition, for such small $\chi$ masses there is no constraints coming from the dark matter direct detection experiments (which are currently sensitive to the DM masses $m_{\chi} \gtrsim 160\text{ MeV}$~\cite{CRESST:2019jnq}, and otherwise impose very strong bounds).

Let us now discuss backgrounds and signal reconstruction for the scattering signatures (see Ref.~\cite{Boyarsky:2021moj}). The LDM elastic scattering produces an isolated low-energy proton track. In order to be recognized as a proton from the elastic scattering event, this track must be isolated from wide cascades of particles produced by neutrino DIS events. SND@LHC is equipped with the emulsion detectors that have limited capabilities to disentangle single proton tracks from the DIS events. This may be not the case of the AdvSND configurations proposed to be equipped with pixel detectors. The reconstruction efficiency and backgrounds in this case is a subject of detailed simulations of neutrino DIS interactions including secondary interactions and detector response, and goes beyond the scope of the present estimate. Within this study, we show iso-contours corresponding to the elastic signature by requiring 10 elastic events with $\chi$.

For the LDM DIS scattering signature, the background are neutrino DIS events. We estimate the number of these events with the help of the approach presented in~\cite{Kling:2021gos}, where we generated the primary flux of mesons using the \texttool{EPOS-LHC}~\cite{Pierog:2013ria} event generator  (for light mesons $\pi,K$), and \texttool{Sibyll~2.3d}~\cite{Riehn:2019jet} (for $D$ mesons) as a part of the \texttool{CRMC} package~\cite{CRMC}. The amounts of obtained CC DIS scatterings for the experiments are $N_{\text{CC}}\simeq 6\cdot 10^{4}$ for $\sndfar$ and $\simeq 2400$ for v. In order to see an excess in the NC/CC ratio over the precision measurement of SND detectors at $2\sigma$, we require the number of events with LDM scatterings to be
\begin{equation}
N_{\text{events}}>2\sqrt{N_{\text{NC}}+(0.01\cdot N_{\text{NC}})^{2}},
\label{eq:scattering-nevents-sensitivity}
\end{equation}
which is $\simeq 450$ for $\sndfar$ and $\simeq 100$ for $\sndnear$.

Let us now discuss the comparison of the reach of SND@LHC, $\sndnear$ and $\sndfar$. Given closely the same $\eta$ range covered by SND@LHC and $\sndfar$, and the same DIS cross sections, the sensitivity of $\sndfar$ may be estimated using a simple rescale:
\begin{equation}
\alpha^{\sndfar}_{B,\text{lower}}\sim\alpha^{\snd}_{B,\text{lower}} \left(\frac{\mathcal{L}_{\text{Run 3}}}{\mathcal{L}_{\text{HL}}}\times \frac{\epsilon_{\text{azimuthal}}^{\snd}}{\epsilon_{\text{azimuthal}}^{\sndfar}}\right)^{1/4}\simeq 0.3\cdot \alpha^{\snd}_{B,\text{lower}},
\end{equation}
with $\mathcal L$ being the luminosity. Here, the scaling $1/4$ comes from the scaling of the signal and background $N_{\text{events}} / \alpha_{B}^{2}, N_{\text{NC}} \propto \epsilon_{\text{azimuthal}}\mathcal{L}$ in \cref{eq:scattering-nevents-sensitivity}. The number of events at $\sndnear$ is lower due to the geometric acceptance and the cross section, which is only partially compensated by $\simeq 10$ times lower number of background events. Therefore, $\sndnear$ has worse sensitivity.

\begin{figure}[t]
    \centering
    \includegraphics[width=0.7\textwidth]{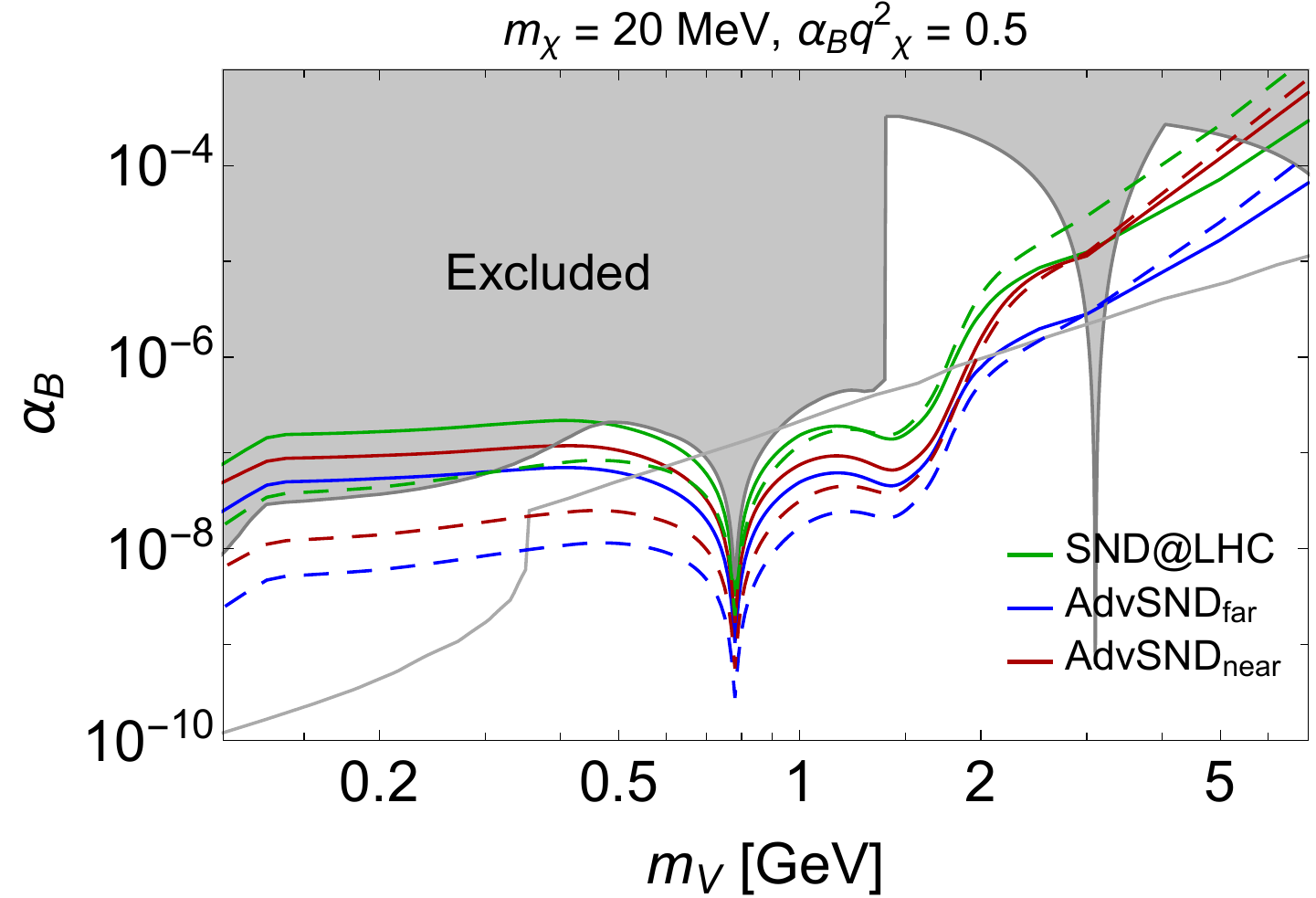}
    \caption{The potential of SND@LHC and its advanced configurations, $\sndnear$ and $\sndfar$, to probe the parameter space of the model \cref{eq:leptophobic-portal} of LDM particles interacting with SM via the leptophobic mediator. We show the sensitivity of the NC/CC signature (the solid lines) as well as the iso-contour corresponding to $10$ elastic signature (dashed lines), see text for details. The constraints coming from past experiment (the solid gray domain), as well as the model-dependent constraint from the anomaly-driven $B$ decays (the gray line) are described in~\cite{Boyarsky:2021moj}.}
    \label{fig:sensitivity-ldm-sndadv}
\end{figure}

We show the sensitivity curves of $\sndnear$, $\sndfar$, and the sensitivity of the SND@LHC experiment from~\cite{Boyarsky:2021moj} for the model~\cref{eq:leptophobic-portal} in \cref{fig:sensitivity-ldm-sndadv}. We see that $\sndfar$ may improve the sensitivity of SND@LHC by a factor of $3$.

\paragraph{Decays} Similarly to the case of scatterings, we estimate the number of decays of dark scalars $S$, HNLs $N$ with dominant muon or tau mixings, and dark photons $V$, following~\cite{Boyarsky:2021moj}. We assume that the ratio of the effective detector lengths for decays and scatterings for the advanced detectors is the same as for SND@LHC, i.e. $l^{\text{decay}}_{\text{det}}/l^{\text{scatt}}_{\text{det}} = 5/3$.

In the mass region of interest $m_{\text{FIP}} \sim $ GeV, $S$s are mainly produced in decays of $B$ mesons, $N$s are produced in decays of $B$ mesons at $m_N\gtrsim 2\text{ GeV}$, while at lower masses - in decays of $D$ mesons (the muon mixing) or in $D\to \tau\to N$ decay chains (the tau mixing). $V$s are produced similarly to the leptophobic mediator - in decays of $\pi$, $\eta$ mesons, bremsstrahlung, and by the Drell-Yan process.

\begin{figure}[t]
    \centering
\includegraphics[width = 0.49\textwidth]{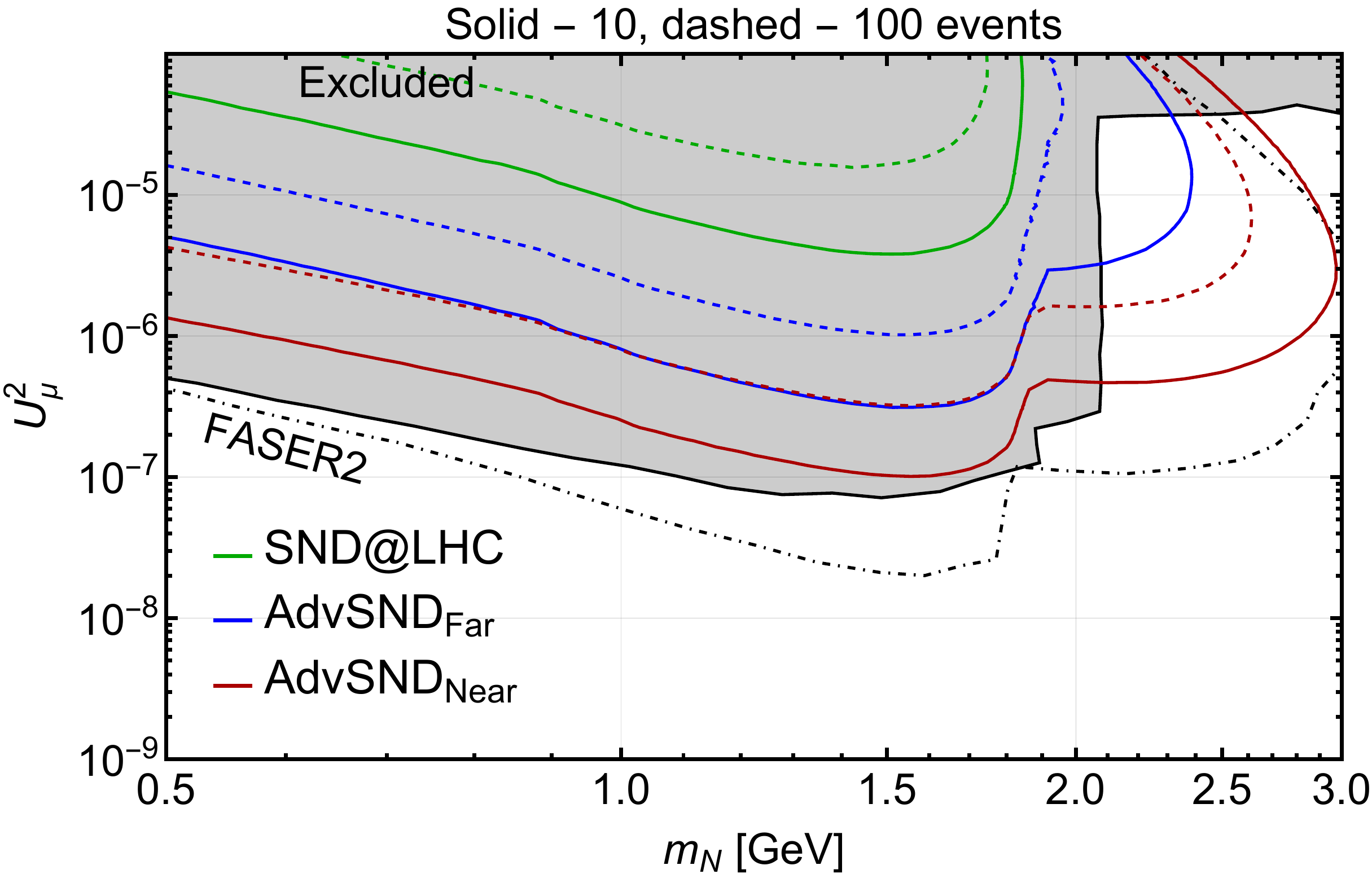}
\includegraphics[width = 0.49\textwidth]{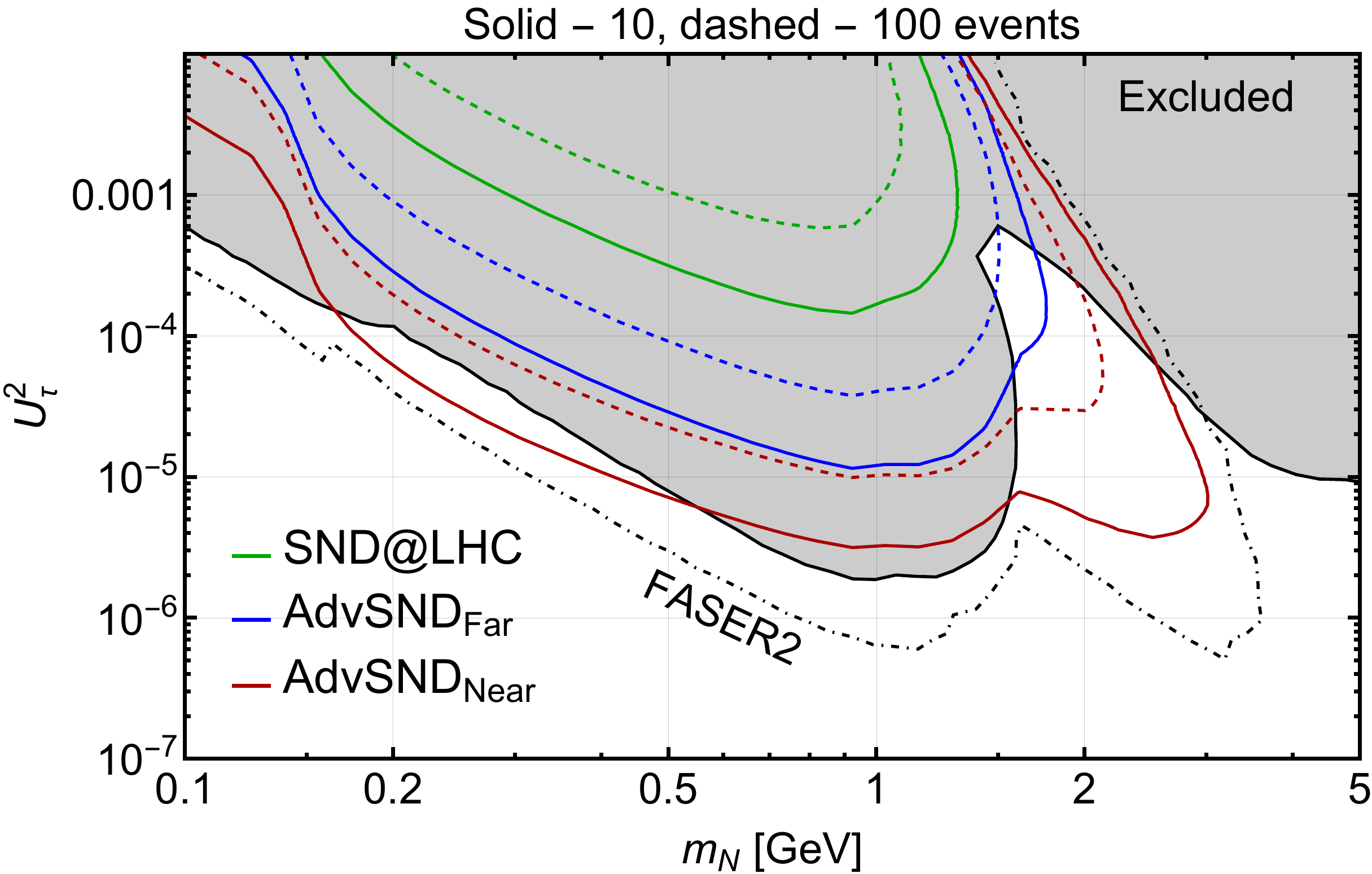}\\
\includegraphics[width = 0.49\textwidth]{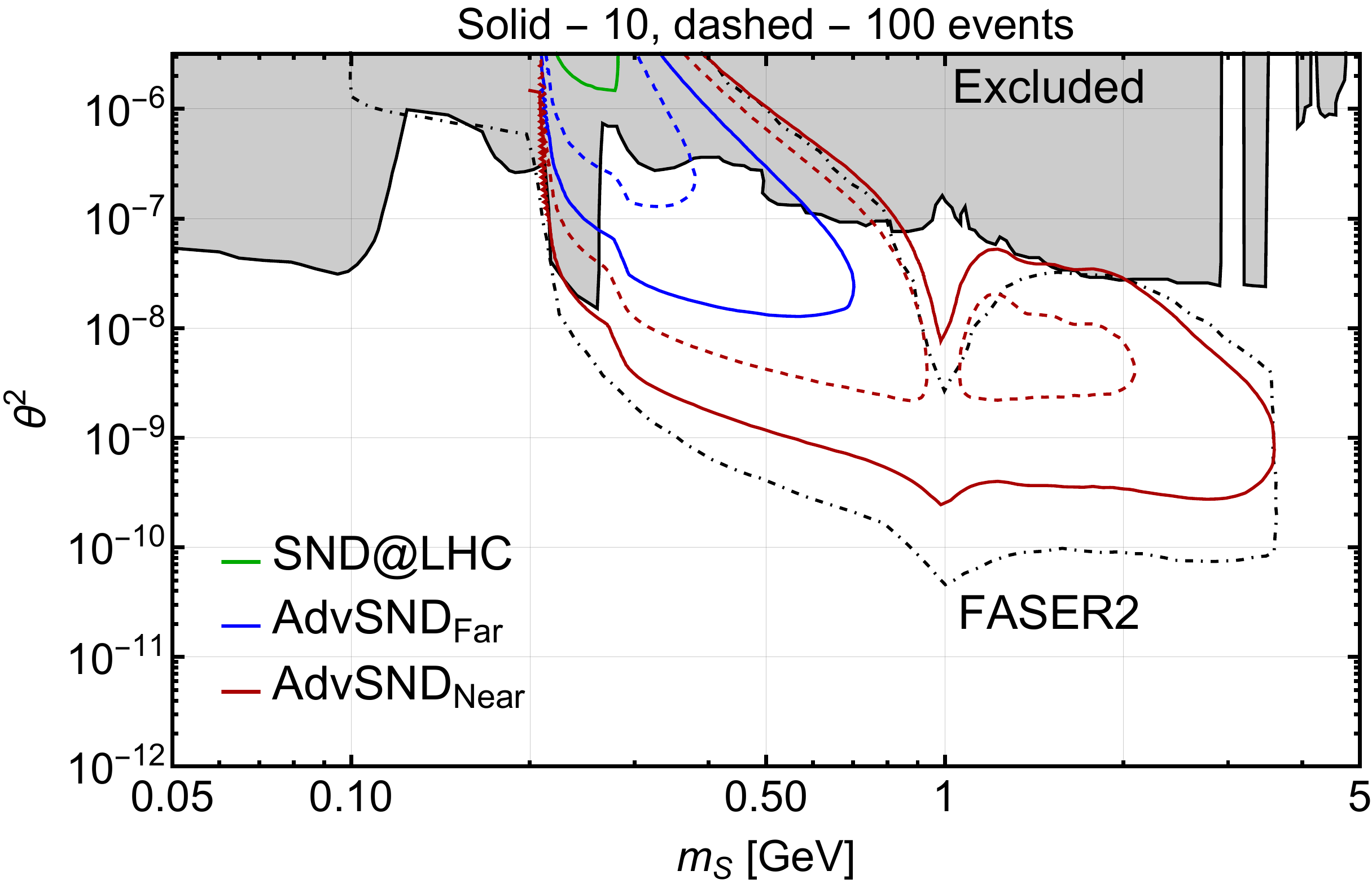}
\includegraphics[width = 0.49\textwidth]{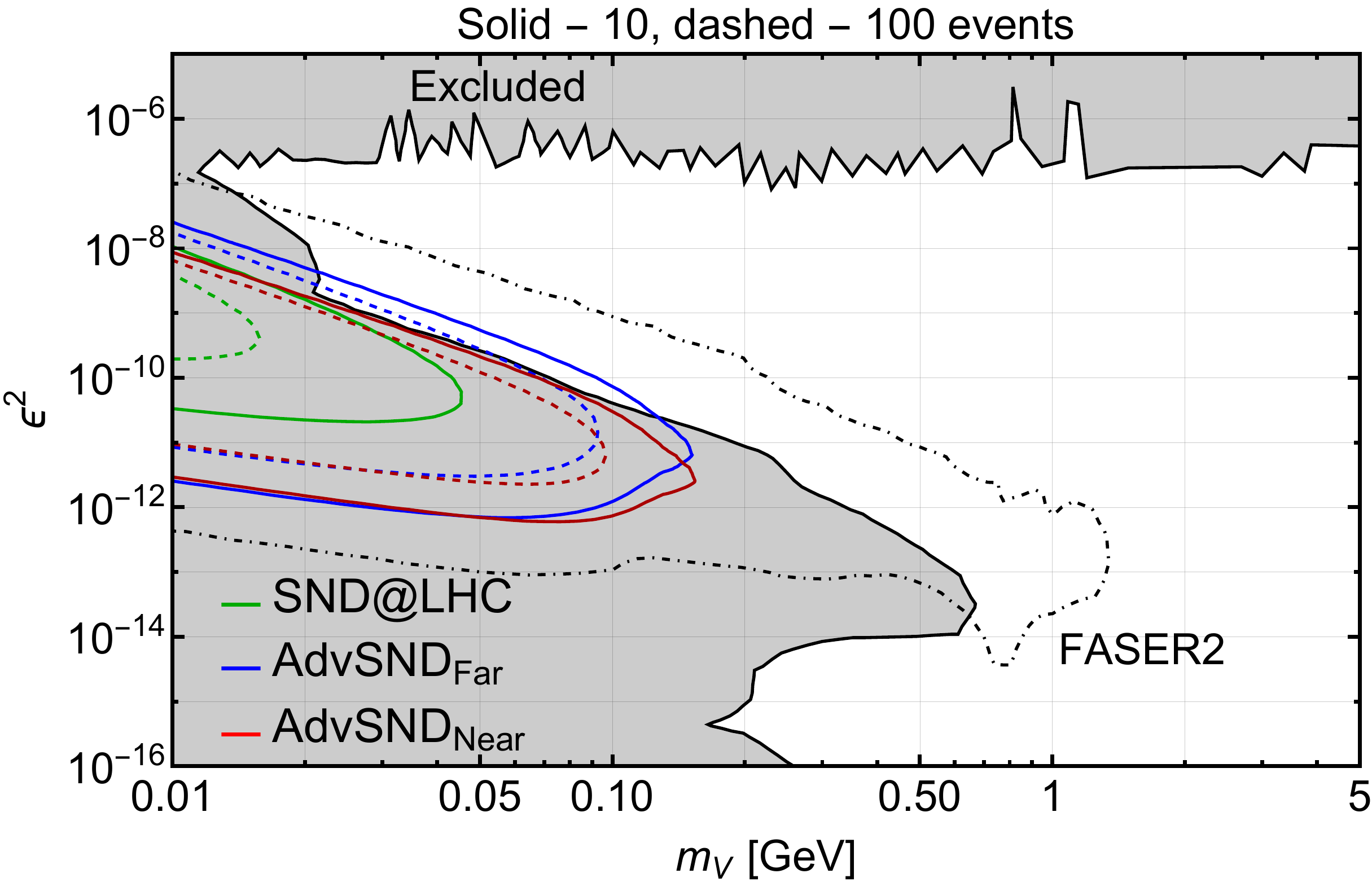}
\caption{Event contours of 10 (solid) and 100 (dashed) events for HNLs with dominant muon (top left) and tau mixing (top right), dark scalar (top left) and dark photons (top right). Event reconstruction efficiency is not included in this estimate. Sensitivities of the FASER2 experiment and of the previous experiments are reproduced from~\cite{Beacham:2019nyx,Boiarska:2021yho}.}
\label{fig:advsnd-decays-sensitivity}
\end{figure}

Decays produce either a pair of charged tracks or monophotons, in dependence on the final states. In order to estimate the reconstruction efficiency, selection efficiency and the number of background events, detailed simulations are required. Therefore, in \cref{fig:advsnd-decays-sensitivity} we show iso-contours instead of the sensitivity.

Let us now make a qualitative comparison of the sensitivities of these experiments requiring the same number of events, mainly following the approach of~\cite{Bondarenko:2019yob}. At the lower bound of sensitivity, the number of events scales with the coupling parameter of a model $\theta$ and the parameters of the detector as 
\begin{equation}
\label{eq:naivelower}
    N_{\text{events}} \propto \theta^4  \mathcal L \times \epsilon_{\text{geom}} \times \frac{l^\text{decay}_{\text{det}}}{\langle p_{\text{FIP}} \rangle}
\end{equation}
with $\mathcal L$ being the luminosity of the experiment. The value of $\theta^2$ corresponding to fixed number of events scales as $\theta_{\text{min}}^2 \propto \left(\mathcal L \epsilon_{\text{geom}}l^\text{decay}_{\text{det}}/\langle p_{\text{FIP}}  \rangle\right)^{-1/2}$.

At the upper bound, the number of particles is exponentially suppressed by the probability to survive before reaching the detector $e^{-l_{\text{min}}/c \tau}$. The exponential factor determines the minimal possible lifetime and hence maximal coupling $1/\tau \propto \theta^2$, leading to the following scaling for the sensitivity to a fixed number of events:
\begin{equation}
\label{eq:naiveupper}
    \theta^2_{\text{max}} \propto \frac{l_{\text{min}}}{\langle p_\text{FIP}\rangle}
\end{equation}
The ratios of $\theta^2_{\text{min}}$, $\theta^2_{\text{max}}$ are given in \cref{tab:thetaminratios}. Note that these estimates are not valid near the endpoint of the iso-contours. 

Similarly to the case of leptophobic mediators, the distribution of dark photons is localized on-axis and quickly drops at larger angles. This implies that the sensitivity of the off-axis $\sndfar$ is worse than that of $\sndnear$. On the other hand, dark scalars and HNLs are produced in decays of heavy mesons, leading to higher transverse momentum of FIPs, such that the distribution is smeared over larger angles, see discussion in Sec.~4 of~\cite{Boyarsky:2021moj}. In this case, $\sndnear$ benefits from the larger solid angle and has better sensitivity than $\sndfar$.  

\begin{table}[t]
\centering
\begin{tabular}{c|c|c|c}
    \hline\hline
    & $S$ & $N$ &$V$  \\
    \hline
    $(\theta^2_{\sndfar}/\theta^2_{\text{SND}})_\text{min}$ & 0.1 & 0.1 & 0.1 \\
    \hline
    $(\theta^2_{\sndnear}/\theta^2_{\text{SND}})_\text{min}$ & 0.02 & 0.02 & 0.1 \\
    \hline
    $(\theta^2_{\sndfar}/\theta^2_{\text{SND}})_\text{max}$ & \multicolumn{3}{c}{1}\\
    \hline
    $(\theta^2_{\sndnear}/\theta^2_{\text{SND}})_\text{max}$ &0.8 & 1 & $\lesssim 0.3$  \\
    \hline\hline
\end{tabular}
\caption{The ratio of $\theta^2_{\text{min/max}}$ for fixed number of events from the naive estimates \cref{eq:naivelower,eq:naiveupper} for dark scalars $S$, HNLs $N$, and dark photons $V$. Here, $N$ represents both HNLs with $\mu$ or $\tau$ mixing.}
\label{tab:thetaminratios}
\end{table}

\subsection{Dark States with Electromagnetic Form Factors}

\paragraph{Interactions between dark states and SM photons} To quantify how ``dark" is the dark sector, electromagnetic (EM) properties of dark states, including not only its EM charge but also multipole expansion of EM charge such as dipole moment, needs to be examined carefully. As an example, we consider Dirac dark states $\chi$, that are electromagnetic (EM) neutral, yet they can still couple to Standard Model (SM) photons through higher-dimensional effective operators such as mass-dimension 5 electric/magnetic dipole moment (EDM/MDM) and mass-dimension 6 anapole moment/charge radius (AM/CR); see \cref{sec:mCPs} for discussion on millicharged dark states. We note, however, that EM form factors can also be generalized to scalar or vector dark states; see~\cite{Hisano:2020qkq} for example.  

The interactions between $\chi$ and photon field strength $F_{\mu\nu}$ can be described by~\cite{Chu:2018qrm}
\begin{align}
\label{eq:EMDM_Lagrangian}
\mathcal{L}_\chi \supset \dfrac{1}{2} \mu_\chi \bar{\chi} \sigma^{\mu\nu} \chi F_{\mu\nu} + \dfrac{i}{2} d_\chi \bar{\chi} \sigma^{\mu\nu} \gamma^5 \chi F_{\mu\nu} - a_\chi \bar{\chi} \gamma^\mu \gamma^5 \chi \partial^\nu F_{\mu\nu} + b_\chi \bar{\chi} \gamma^\mu \chi \partial^\nu F_{\mu\nu}\,.
\end{align}
Here $\mu_\chi$ and $d_\chi$ are mass-dimension $-1$ coefficients of MDM and EDM, $a_\chi$ and $b_\chi$ are mass-dimension $-2$ coefficients of AM and CR, and $\sigma^{\mu\nu} \equiv i[\gamma^\mu, \gamma^\nu]/2$, respectively. The signs in front of each terms in \cref{eq:EMDM_Lagrangian} is chosen such that the Lagrangian in the non-relativistic limit reduces to the interaction Hamiltonian in classical regime. Each EM form factor interaction has different properties under discrete Lorentz transformations, resulting in distinct phenomenology; see discussion in~\cite{Chu:2018qrm}. The vertex function for Feynman-diagrammatic calculations reads~\cite{Chu:2018qrm}
\begin{align}
\Gamma_\chi^\mu (k) = i\sigma^{\mu\nu} k_\nu (\mu_\chi + i d_\chi \gamma^5 )+ (k^2 \gamma^\mu - k^\mu {\slashed k} ) ( -a_\chi \gamma^5 + b_\chi)\,,
\end{align}
where $k$ is the $4$-momentum goes into $\chi$-pair.

This \textit{photon portal} is entertained in both its signal in laboratory~\cite{Sigurdson:2004zp, Kopp:2014tsa, Chu:2018qrm, Chu:2020ysb} and in direct/indirect search~\cite{Pospelov:2000bq, Sigurdson:2004zp, Schmidt:2012yg, Ho:2012bg, Kopp:2014tsa, Ibarra:2015fqa, Sandick:2016zut, Kavanagh:2018xeh, Chu:2018qrm, Trickle:2019ovy} as well as its astrophysical and cosmological implications~\cite{Sigurdson:2004zp, Ho:2012bg, Chu:2019rok, Chang:2019xva, Kuo:2021mtp}. Possible UV-completion scenarios include, \textit{e.g.}, composite dark states~\cite{Foadi:2008qv, Bagnasco:1993st, Antipin:2015xia} or radiative generation through new charged states at UV scale~\cite{Raby:1987ga, Pospelov:2008qx}. Here we remain agnostic of the UV physics by assuming the UV scale being higher than the center-of-mass (CM) energy of considered experiments\footnote{For some parameter space, \textit{e.g.}, $\mu_\chi^{-1} \sim E_{\rm CM}$ with $E_{\rm CM}$ being the CM energy of considered experiments, this assumption is no longer valid. To obtain consistent result, one needs to take UV physics into account.}, such that effective description of the EM form factor interactions is valid. 

In general, to probe the photon portal, the CM energy enters the interaction is of utmost importance considering the energy-dependence of the effective operators. Given that FPF utilizes energetic particle flux produced from the LHC with a CM energy at TeV scale, we can envision FPF, even though with less intense flux, can yield competitive sensitivity for dark states mass below $\mathcal{O}({\rm TeV})$, when compared to previous intensity frontier experiments such as electron/proton fixed-target/beam-dumped experiments which typically have CM energy around $\mathcal{O}(10\,{\rm GeV})$. In the following, we discuss the production and detection of such dark states and demonstrate projected sensitivity of FLArE.

\begin{figure}[t]
\centering
  \includegraphics[width=0.49\textwidth]{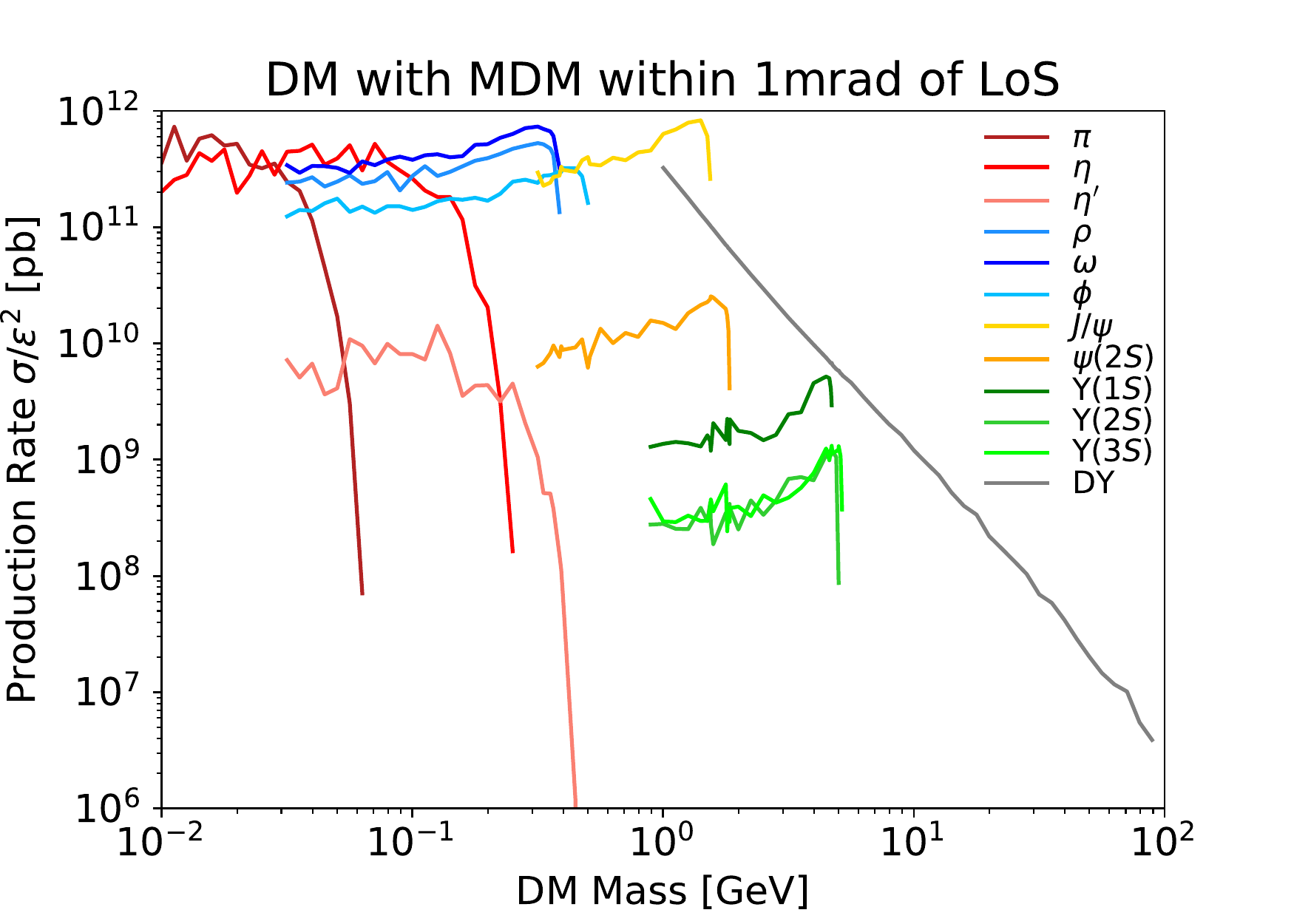}
  \includegraphics[width=0.49\textwidth]{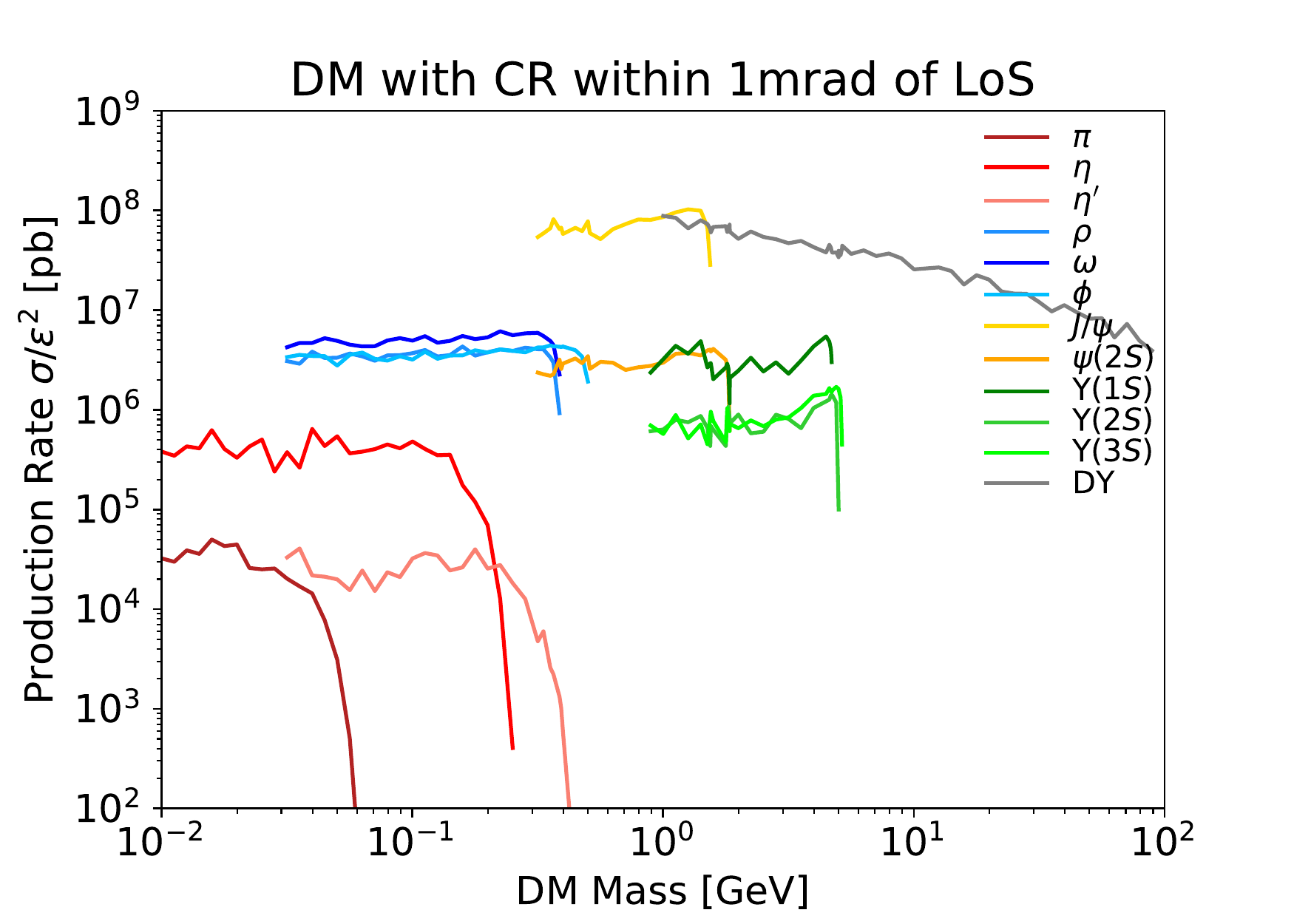}
  \caption{Left: Production rates for mass-dimension $5$ operators with MDM interaction as representative. For sub-GeV mass range, $\chi$-production is dominated by decay of $\pi, \eta, \rho, \omega$ and $J/\psi$. The DY channel extends to multi-GeV regime. Right: Production rates for mass-dimension $6$ operators with CR interaction  as representative. Recalling that ${\rm BR} \propto M^4$ for AM and CR, only $J/\psi$ decay is important for $m_\chi \lesssim \mathcal{O}({\rm GeV})$. For $\chi$ with a mass above GeV, only DY production is relevant.}
  \label{fig:prod}
\end{figure}

\paragraph{Production and detection in FPF} Similar to dark state production in proton-beam experiments~\cite{Chu:2020ysb}, in FPF dark states $\chi$ can source from Drell-Yan (DY) processes at the beam collision point, decay of pseudoscalar/vector meson decay, proton-proton bremsstrahlung, and pion capture processes.
\begin{itemize}
\item Drell-Yan processes: $\chi$ can be pair-produced through quark-antiquark annihilation. Event generators such as \texttool{MadGraph}~\cite{Alwall:2014hca} and \texttool{Pythia}~\cite{Sjostrand:2014zea} are adopted for correct estimation of the resulting energy spectrum and angular distribution of $\chi$. Drell-Yan production is essential when going beyond the kinematic endpoint of meson decay and become more important for higher-dimensional operators.
\item Meson decay: as long as the process is kinematically allowed, pseudoscalar/vector mesons produced from primary interaction can decay into $\chi$-pair through an off-shell photon. According to the spin of meson, pseudoscalar meson such as pions can have $3$-body decay into $\chi$-pair plus a photon while vector meson such as $\rho$ meson can decay into $\chi$-pair directly. For higher-dimensional operators, production of $\chi$ from heavier meson decay become more important due to the mass scaling in the branching ratio ${\rm BR}\propto M^a$, where $a = 2 (4)$ for mass-dimension $5 (6)$ operators.
\item Proton-proton bremsstrahlung: $\chi$-pair can also be emitted via an off-shell  photon in proton-proton interaction. The production rate can be estimated from the Fermi-Weizsacker-Williams method with a simplified phase space~\cite{Fermi1924, vonWeizsacker:1934nji, Williams:1934aa}. For the EM form factor interactions considered here, we find the production rate is dominated by vector meson resonance, \textit{i.e.} when $s_{\chi\bar\chi} \simeq m_{\rm V}^2$ with $m_V$ being vector meson mass; see also~\cite{Faessler:2009tn}. Since vector meson resonance is already included in production from meson decay, we do not consider proton-nucleus bremsstrahlung production to avoid double counting.
\item Pion capture: the capture of pion by proton (nucleus) can produce dark state pair via $p \pi \rightarrow n \gamma^* \rightarrow n\chi \bar{\chi}$. However, this process mostly result in  dark states with energy smaller than  $\mathcal{O}({\rm MeV})$ which cannot result in a strong enough recoil signal to be detected; thus, pion capture is not relevant for the reach of experiments considered here. 
\end{itemize}

In summary, we take Drell-Yan processes and meson decay as main production channels and derive corresponding energy spectrum and angular distribution of dark states $d^2 N_\chi/(dE_\chi d\cos\theta_\chi)$ with $\theta_\chi$ being the angle between $\chi$'s momentum and beam axis. For Drell-Yan processes, we adopt \texttool{MadGraph} to generate $\chi$-flux produced at the IP. On the other hand, the numerical package \texttool{FORESEE}~\cite{Kling:2021fwx}, in which energy spectrum and angular distribution of mesons produced from $pp$-collision are incorporated, is utilized for meson decay production with relevant formulas given in~\cite{FLArE_EM}. The production rate for each channel is summarized in \cref{fig:prod}, where an angular cut $\theta_\chi < 1\,{\rm mrad}$ from the detector geometry. It is evident that production from heavier meson decay and DY process becomes increasingly important for higher dimensional operators, consistent with the observation in~\cite{Chu:2020ysb}. For mass-dimension $5$ operators, $J/\psi$ decay is comparable with pion decay, while for mass-dimension $6$ operators, pion decay is much suppressed compared to $J/\psi$ decay.  The DY production is the only production channel when meson decays are kinematically forbidden, \textit{i.e.}, $M < 2 m_\chi$. The total production rate, used later in estimating the event rate, is given by the sum of production rate of individual channel. 

For the detection of $\chi$, we focus on signals of electron recoil. In the regime that produced dark states are highly boosted such that $E_\chi$ is much larger than $m_\chi$, electron mass $m_e$ and electron recoil energy $E_R$, the differential scattering cross section $d\sigma_{\chi e}/dE_R$ reads, \textit{e.g.}, for MDM and CR
\begin{align}
\label{eq:diff_xsec}
\dfrac{d\sigma_{\chi e}^{\rm dim\text{-}5}}{dE_R} \simeq \dfrac{\alpha \mu_\chi^2 }{E_R}\,, \quad \dfrac{d\sigma_{\chi e}^{\rm dim\text{-}6}}{dE_R} \simeq 2 \alpha b_\chi^2 m_e\,,
\end{align}
where $\alpha$ is the fine-structure constant. Note that \cref{eq:diff_xsec} also applies to EDM and AM with the replacement of the form factor coupling. The $E_R$-scaling of differential scattering cross section is different from that of the millicharged case $d\sigma_{\chi e}^{\rm dim\text{-}4}/dE_R \simeq (2\pi\alpha^2 \epsilon^2)/(E_R^2 m_e)$ with $\epsilon$ being the charge fraction, implying the event rate of EM form factors being less IR-biased. Full formulas for differential cross section can be found in~\cite{Chu:2018qrm}, which is implemented in \texttool{FORESEE}. The number of of $e$-recoil event can be obtained via 
\begin{align}
N_{\rm event} = n_e L_{\rm det}\int dE_R \,\epsilon (E_R) \int dE_\chi \int d\cos\theta_\chi \, \dfrac{d^2 N_\chi}{dE_\chi d\cos\theta_\chi} \dfrac{d\sigma_{\chi e}}{dE_R}\,,
\end{align}
where $n_e$ is electron density in the detector, $L_{\rm det}$ is the length of the detector along the beam axis, and $\epsilon (E_R)$ is the detection efficiency. Assuming the target electron is at rest, the maximal $E_R$ from certain $E_\chi$ reads
\begin{align}
   E_R^{\rm max}  = \dfrac{2m_e (E_\chi^2 - m_\chi^2)}{m_e (2E_\chi + m_e) + m_\chi^2}\,,
\end{align}
such that the upper boundary of $E_R$-integration is $\text{min}(E_R^{\rm max}, E_R^{\rm cut})$ where $E_R^{\rm cut}$ is the large-$E_R$ cut of the experiment. 

\begin{figure}[t]
  \centering
  \includegraphics[width=0.49\textwidth]{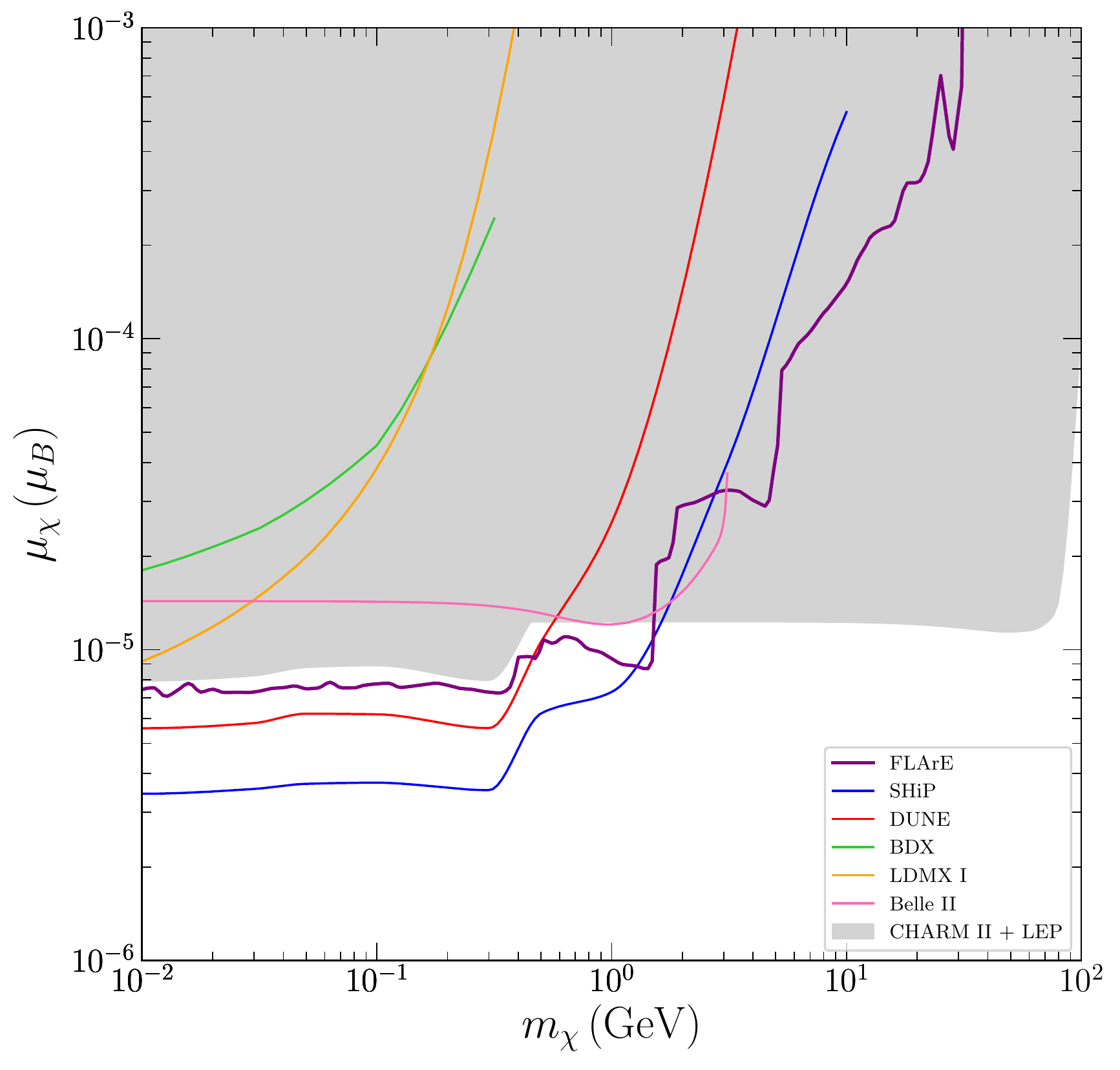}
  \includegraphics[width=0.49\textwidth]{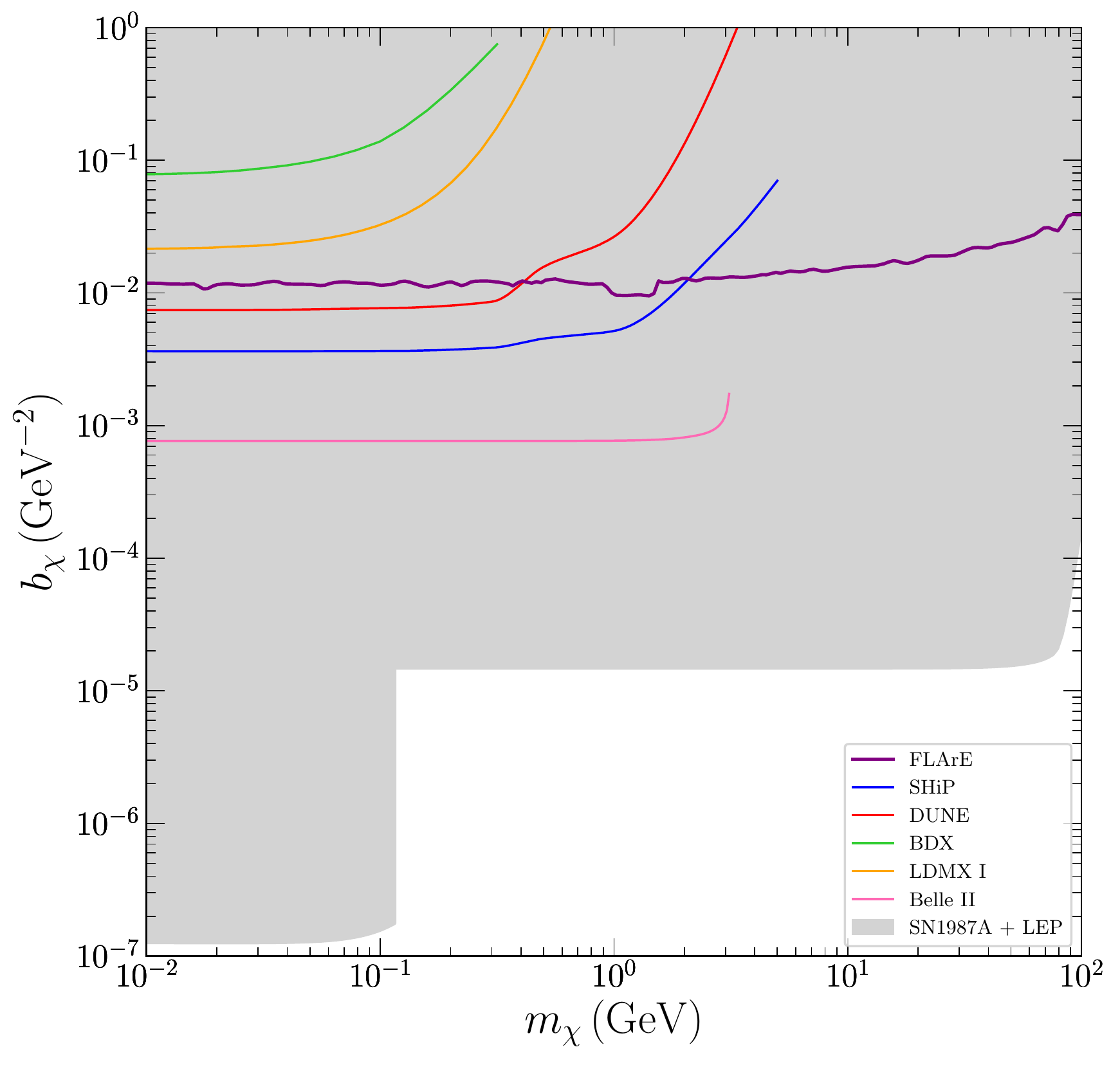}
  \caption{Left: Projected sensitivity for mass-dimension $5$ operators with MDM interaction as representative. FLArE can improve the sensitivity on sub-GeV $\chi$ compared to existing constraints and compete with proposed intensity frontier experiments~\cite{FLArE_EM}.  Right: Projected sensitivity for mass-dimension $6$ operators with CR interaction as representative. FLArE yields similar sensitivity with proposed intensity frontier experiments, but they are all  weaker than bounds from SN1987A and LEP.}
  \label{fig:sensitivity}
\end{figure}

\paragraph{Projected sensitivity of FLArE} The detector of FLArE is placed $620\,{\rm m}$ downstream of the ATLAS IP. We assume the default FLArE-10 detector geometry with $L_{\rm det} = 7\,{\rm m}$ and a transverse surface area of $1\,{\rm m}^2$. From the detector geometry, we place a cut $\theta_\chi^{\rm max} = 1\,{\rm mrad}$ such that $\chi$'s trajectory passes through the detector. The detection efficiency is assumed to be 100\% with threshold and cutoff being $E_R^{\rm th} = 30\,{\rm MeV}$ and $E_R^{\rm cut} = 3 \,{\rm GeV}$ to reduce neutrino-induced background. We have checked that extending $E_R$-range does not improve the sensitivity notably. 

In \cref{fig:sensitivity}, we show the projected sensitivity based on electron recoil together with existing constraints and projection from other proposed experiments\footnote{Note that here we do not assume $\chi$ constitutes cosmological DM. Therefore, we do not show thermal relic curves and bounds from direct detection and indirect searches in \cref{fig:sensitivity}. We refer readers to~\cite{Chu:2018qrm} for the case that $\chi$ plays the role of DM.}. Current strongest bounds (grey shaded region in \cref{fig:sensitivity}) on EM form factors are inferred from CHARM-II~\cite{Chu:2020ysb} and LEP missing-energy search~\cite{Chu:2018qrm} for mass-dimension $5$ operators, and from energy loss in SN1987A~\cite{Chu:2019rok} and LEP missing-energy search~\cite{Chu:2018qrm} for mass-dimension $6$ operators. Projected sensitivities of proposed proton-beam and electron-beam experiments~\cite{Chu:2018qrm,Chu:2020ysb} are shown in solid curves for comparison. Based on fiducial detector geometry of FLArE and energy cuts, we find that for MDM and EDM $e$-recoil at FLArE can go beyond current constraints and compete with other proposed intensity frontier experiments in $m_\chi \leq \mathcal{O}({\rm GeV})$~\cite{FLArE_EM}. On the other hand, for mass-dimension 6 operators, sensitivity of FLArE together with other past/future intensity frontier experiments are suppressed compared to bounds from LEP missing-energy search. This relative difference of constraining power of FLArE and LEP on mass-dimension 5 and 6 operators can be understood as follows: First, LEP bound for mass-dimension $6$ operators is much stronger than that for mass-dimension $5$ operators as $\chi$ production cross section from electron-positron annihilation is proportional to the CM energy. Second, the production rate of dark states with mass-dimension $6$ operators at FPF is smaller than that with mass-dimension $5$ operators since it relies on heavy meson decay which comes with a smaller flux; see \cref{fig:prod} for comparison. In addition, the scattering cross section of mass-dimension $6$ operators has no preference for low-$E_R$ in the detector; therefore, resulting signals cannot be fully disentangled from the neutrino-induced background.  We expect FLArE-100 can improve the sensitivity on mass-dimension $5$ operators, and the parameter space of mass-dimension $6$ operators can be further probed by missing-energy search in future colliders~\cite{Kadota:2014mea, Primulando:2015lfa, Alves:2017uls}.

\section{Millicharged Particles}
\label{sec:mCPs}

Millicharged particles (mCPs) are connected to many interesting topics in particle physics and cosmology and are considered as an important generic possibility by the Physics Beyond Colliders initiative~\cite{Agrawal:2021dbo}. Fractional charges arise naturally in considerations of the quantization of electric (and magnetic) charge~\cite{Dirac:1931kp}, and in grand unified and string theories~\cite{Pati:1973uk, Georgi:1974my, Kors:2004dx, Shiu:2013wxa}. mCPs are a low-energy consequence of well-motivated dark-sector models~\cite{Holdom:1985ag}, and even neutrinos may carry  millicharges~\cite{Vogel:1989iv, Giunti:2014ixa}. mCPs have been proposed as a dark matter candidate~\cite{Brahm:1989jh, Feldman:2007wj, Feng:2009mn, Cline:2012is} and were recently discussed as a solution to the 21-cm absorption spectrum anomaly reported by the EDGES collaboration~\cite{Bowman:2018yin, Barkana:2018lgd, Munoz:2018pzp, Berlin:2018sjs, Slatyer:2018aqg, Liu:2019knx, Aboubrahim:2021ohe}. We consider a mCP, labeled $\chi$, with electric charge $Q_\chi$, and define $\epsilon \equiv Q_\chi/e$. This can arise if $\chi$ has a small charge directly under the Standard Model $U(1)$ hypercharge, and also if $\chi$ is coupled to a massless dark photon via kinetic mixing~\cite{Holdom:1985ag} or to a massive dark photon via Stueckelberg mass mixing~\cite{Kors:2004dx}.

mCPs have been looked for in many experiments~\cite{Dobroliubov:1989mr, Prinz:1998ua, Davidson:2000hf, Golowich1987, Babu:1993yh, Gninenko:2006fi, Agnese:2014vxh, Haas:2014dda, Ball:2016zrp, Hu:2016xas, Alvis:2018yte, Magill:2018tbb, Kelly:2018brz, Arguelles:2019xgp}, and their signatures in astrophysical/cosmological observations have also been considered. mCPs can be produced either directly, or through the decay of secondary mesons. The main experimental signatures are: tracking ($dE/dx$), hard scattering (detecting the recoil electron) and missing momentum/energy. The electron-scattering signature is the best studied for mCPs. Since there is a $1/E$ enhancement in the scattering cross-section (where $E$ is the electron-recoil energy), experiments with sensitivity to low-energy recoil or scintillation signatures are preferred as mCP probes~\cite{Magill:2018tbb}.

\begin{figure}[t]
\centering
\includegraphics[trim={.5cm 0 0.4cm 0}, clip, width=0.9\textwidth]{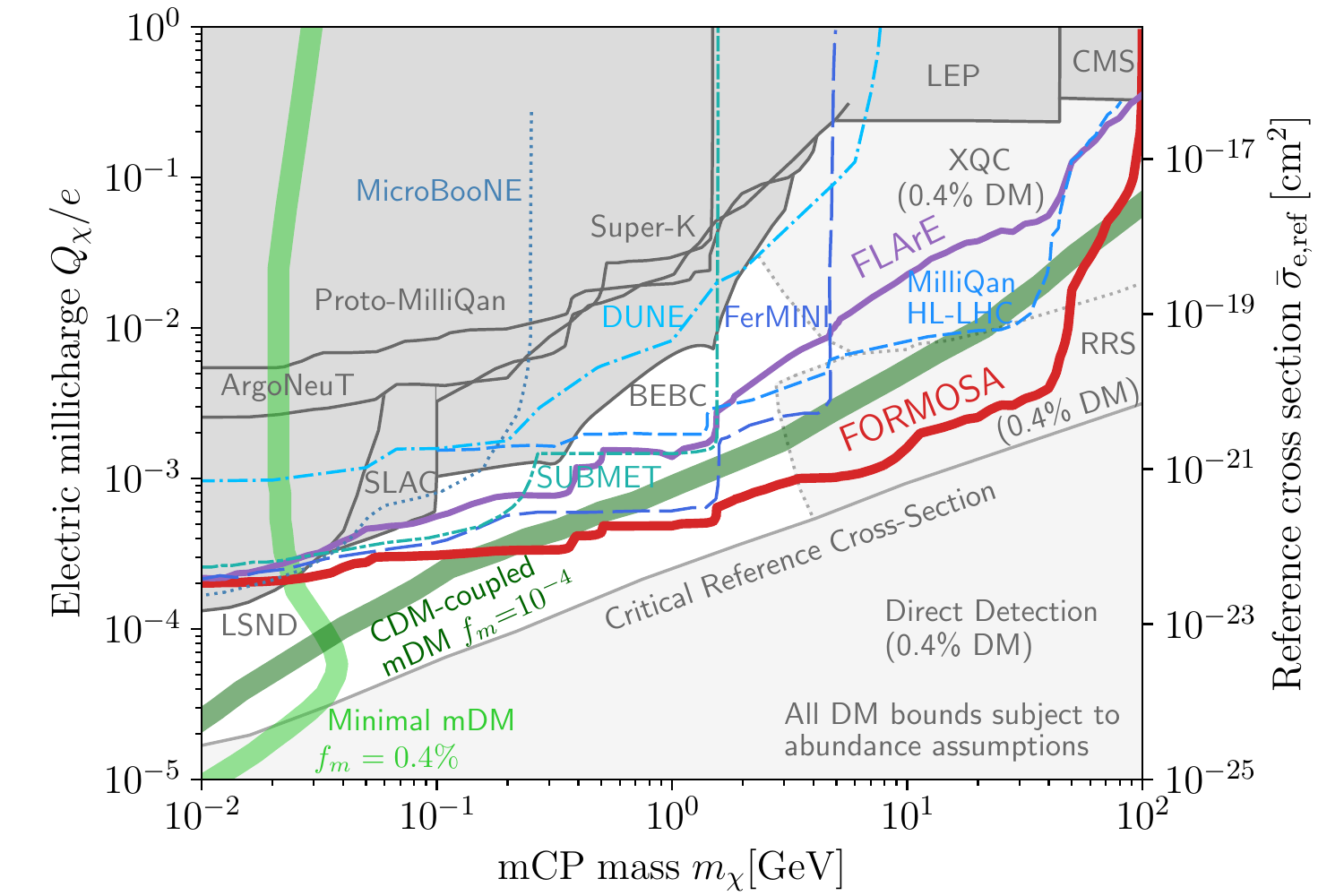}
\caption{Bounds on mCPs from previous searches (all in grey) include SLAC~\cite{Prinz:1998ua}, LEP~\cite{Davidson:2000hf, Badertscher:2006fm}, CMS~\cite{CMS:2012xi, Jaeckel:2012yz}, LSND~\cite{Magill:2018tbb}, ArgoNeuT~\cite{Acciarri:2019jly}, BEBC~\cite{Marocco:2020dqu}, Super-K limit on the diffuse supernova neutrino background~\cite{Plestid:2020kdm}, and the recent search by milliQan~\cite{Ball:2020dnx}. The sensitivities for FORMOSA~\cite{Foroughi-Abari:2020qar} and FLArE~\cite{FLArE_EM} are shown, respectively, as thick red and purple lines. Projections for milliQan AT HE HL-LHC~\cite{Haas:2014dda}, FerMINI~\cite{Kelly:2018brz} and SUBMET~\cite{Choi:2020mbk} are indicated as dashed curves.} 
\label{fig:mCP}
\end{figure}

At the FPF, one can also do a tracking search for mCPs by exploiting their scintillation signature. The FORward MicrOcharge SeArch (FORMOSA)~\cite{Foroughi-Abari:2020qar}, was proposed for this purpose. The FORMOSA detector is a large plastic scintillator array with each scintillator volume coupled to a PMT capable of detecting single photons. This will enable sensitivity to the small scintillation deposits produced by low charge particles. The use of multiple layers of scintillator provides powerful background suppression, while a large path length of active material allows sensitivity to charges as low as $10^{-4}$e. This will allow FORMOSA to provide the best probe of mCPs with mass between 0.1 and 100 GeV. A more complete description can be found in \cref{sec:FORMOSA}.

The FPF also provides the opportunity to search for mCPs via hard scattering, with the scattered electron detected by FLArE (which is extensively discussed in \cref{sec:FLArE}. The search strategy for mCPs is essentially the same as the search for LDM electron scattering and has the advantage of strong 1/E enhancement, corresponding to LDM with ultralight mediators. We show the estimate of the sensitivity in \cref{fig:mCP}, while experimental details may be found in \cref{sec:FLArE}. In \cref{fig:mCP}, we show the FORMOSA and FLArE sensitivities, along with existing constraints and other future probes.

\section{Quirks}
\label{sec:quirks}

Many interesting models beyond the Standard Model (BSM) predict the existence of quirks, which are long-lived fermionic or scalar particles charged under not only the Standard Model (SM) gauge group but also a new confining gauge group. Color neutral quirks arise from models with neutral naturalness~\cite{Curtin:2015bka}, such as folded supersymmetry~\cite{Burdman:2006tz, Burdman:2008ek}, twin Higgs~\cite{Chacko:2005pe, Craig:2015pha, Serra:2019omd, Ahmed:2020hiw}, quirky little Higgs~\cite{Cai:2008au}, minimal neutral naturalness model~\cite{Xu:2018ofw} and so on, which are built to address the little hierarchy problem~\cite{BasteroGil:2000bw, Bazzocchi:2012de}. Quirks can also be colored in more general cases. The confinement scale ($\Lambda$) of the new gauge group is so smaller compared to the mass of the lightest quirk that two quirks produced in a pair will be connected by a macroscopic gauge flux tube~\cite{Kang:2008ea}. The phenomenology of quirks is very different from that of SM particles because of the extra long-range gauge interaction. The oscillation amplitude ($L$) of two quirks in a pair is dominated by $\Lambda$ and will be macroscopic ($\sim \text{mm}-\text{m}$) when $\Lambda \sim 100~\text{eV}-\text{keV}$. Due to the macroscopic $L$ in this case, hits of quirks in the detector can not be reconstructed as helical tracks, and will be dropped in the conventional event reconstruction at the LHC.

\begin{figure}[t]
\centering
	\includegraphics[width=0.85\textwidth]{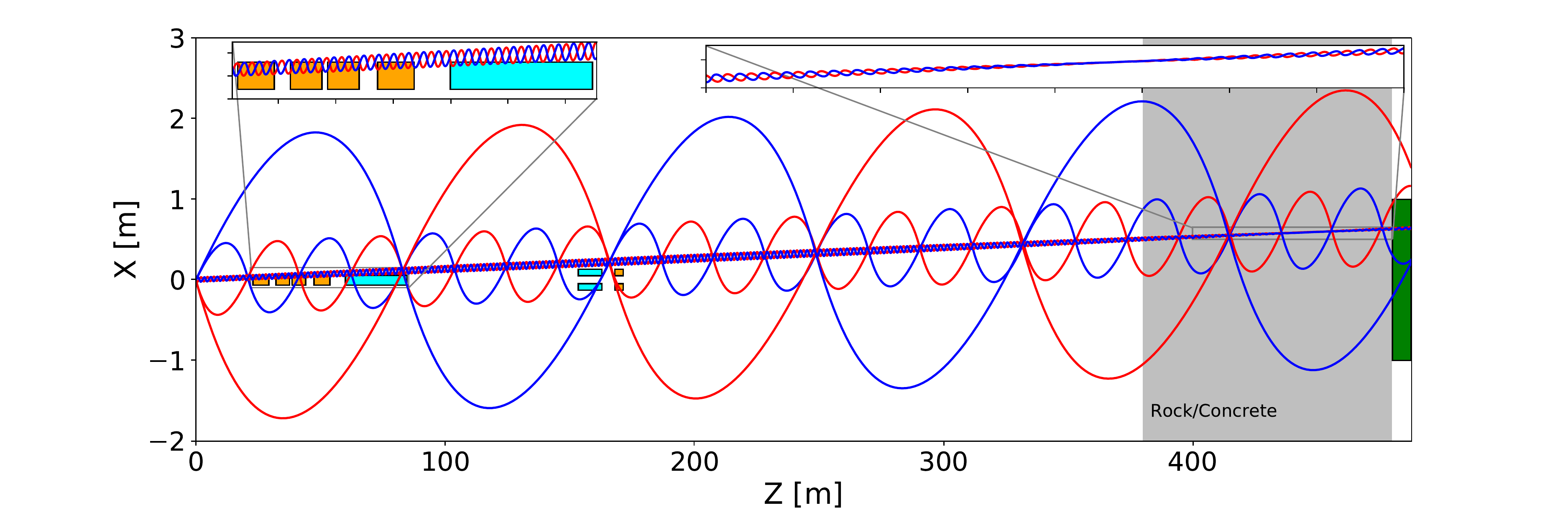}
\caption{The quirk trajectories and forward infrastructures from ATLAS IP to the FASER (2) detector. Two fragments of quirk trajectory are magnified for clearer visibility. The quirk initial momenta are $\vec{p}_1 =(-132.146, 121.085, 1167.35)$ GeV and $\vec{p}_2 =(136.381, -123.865, 2061.56)$ GeV and the quirk mass is 800 GeV. And three different confinement scales $\Lambda=50~\text{eV}, 100~\text{eV}, 400~\text{eV}$ are considered for illustration. The orange, cyan, grey, and green regions indicate the regions with quadrupole magnetic field, dipole magnetic field, rock/concrete and the FASER2 detector, respectively. }
\label{fig::quirktrajectory}
\end{figure}

As FASER (2) is located 480 m downstream from the ATLAS interaction point (IP), charged quirks produced at the ATLAS IP, as shown in \cref{fig::quirktrajectory}, need to travel through all facilities between the ATLAS IP and FASER (2) before leaving visible signals in the tracker of FASER (2). The dependence of the FASER (2) sensitivity to quirks on the quirk mass, the quirk quantum numbers, and $\Lambda$ are studied.

Four kinds of quirks are chosen for study, which under the $SU(N_{\text{IC}}) \times SU_C(3) \times SU_L(2) \times U_Y(1)$ gauge group are
\begin{align}
\tilde{\mathcal{D}} &= \left( N_{\text{IC}}, 3, 1, -1/3 \right),~~ \label{eq::qn1}\\
\tilde{\mathcal{E}} &= \left( N_{\text{IC}}, 1, 1, -1 \right),~~\label{eq::qn2}\\
\mathcal{D} &= \left( N_{\text{IC}}, 3, 1, -1/3 \right),~~\label{eq::qn3}\\
\mathcal{E} &= \left( N_{\text{IC}}, 1, 1, -1 \right) \label{eq::qn4},
\end{align}
where $\tilde{\mathcal{D}}$ ($\mathcal{D}$) and $\tilde{\mathcal{E}}$ ($\mathcal{E}$) are scalar (fermionic) and we set $N_{\text{IC}}=2$ in this study. Because of color confinement, one can never observe a single $\tilde{\mathcal{D}}$ or $\mathcal{D}$ with an electric charge of $-\frac{1}{3}$ other than the quirk-quark bound states with an integral electric charge. Around 30\% of the quirk-quark bound states have electric charge $\pm 1$~\cite{Knapen:2017kly}. For simplicity, we will refer to these charged quirk-quark bound states as $\tilde{\mathcal{D}}$ or $\mathcal{D}$ in the following discussions as only charged final states are concerned. What is more, since quirks are much heavier than quarks, the quirk-quark bound states have equation of motions (EoM) which are almost the same as those of quirks.

\begin{figure}[t]
\centering
	\includegraphics[width=0.32\textwidth]{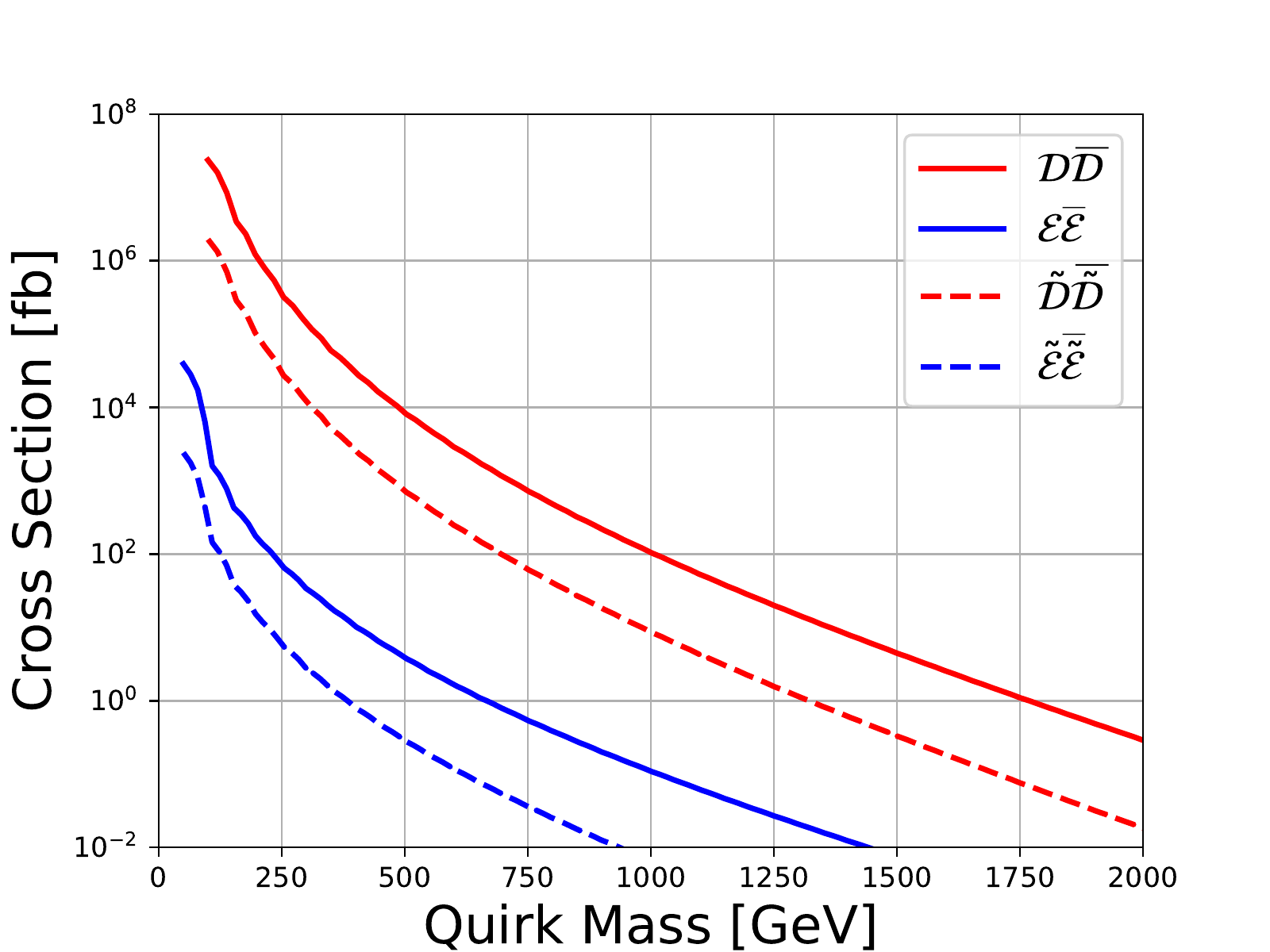}
	\includegraphics[width=0.32\textwidth]{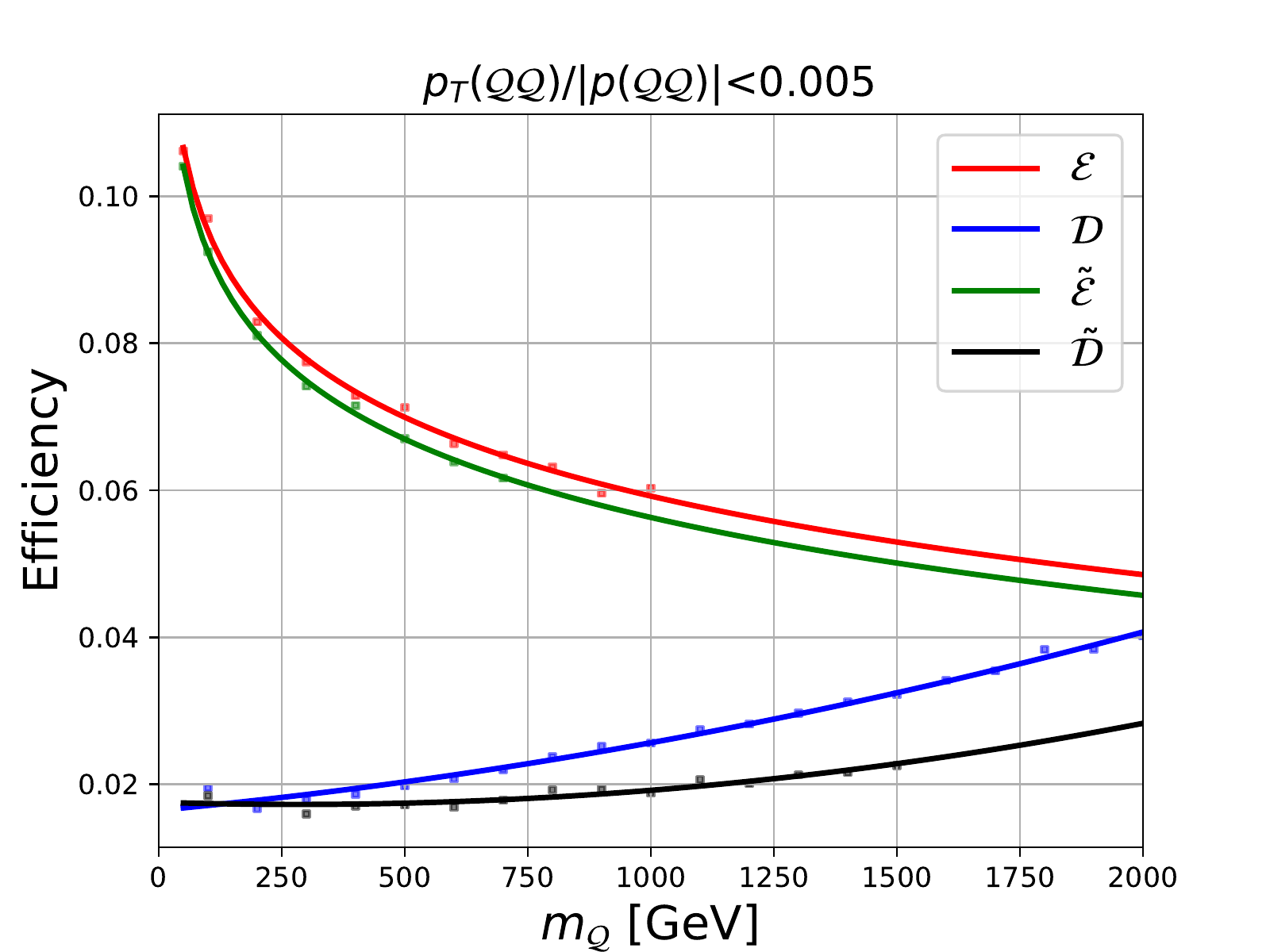}
	\includegraphics[width=0.32\textwidth]{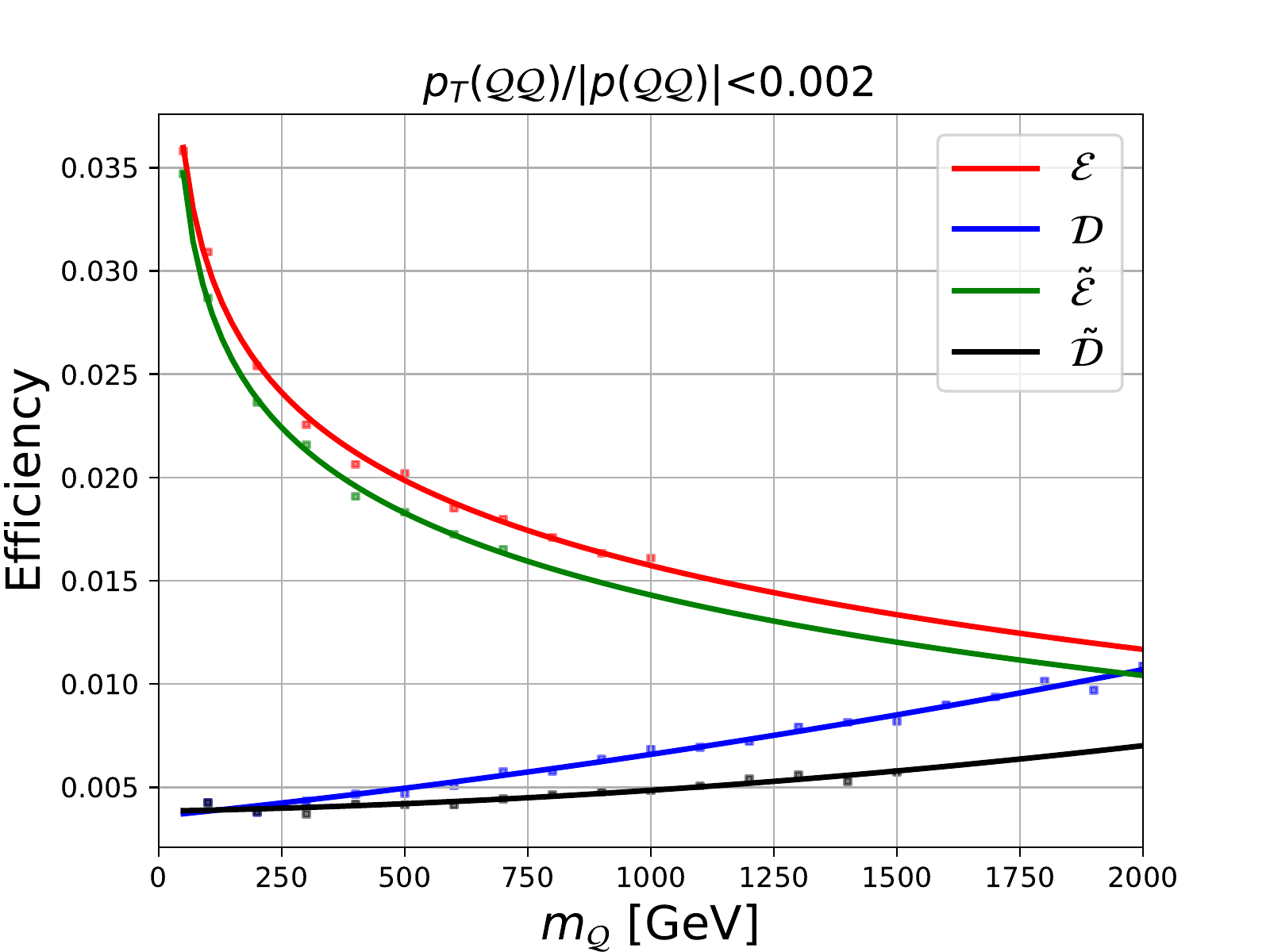}
\caption{The leading order production cross sections for the quirk pair production at the 13 TeV LHC (left). The fraction of events that have $p_T(\mathcal{Q}\mathcal{Q})/ |p(\mathcal{Q}\mathcal{Q})|< 0.005$ (middle) and $0.002$ (right) for different quirks and their masses. }
\label{fig::xsecs}
\end{figure}

The left panel of \cref{fig::xsecs} shows the leading order production cross section (calculated by MG5\_aMC@NLO~\cite{Alwall:2014hca}) of each kind of quirk. With the effects of initial state radiation (ISR), final state radiation (FSR), and the hadronization of colored final state included in event generation by \texttool{Pythia~8}~\cite{Sjostrand:2007gs}, many quirk production events can not reach the FASER (2) detector because the direction of the quirk-pair system is deflected from the beam axis. The fractions of events that have $p_T(\mathcal{Q}\mathcal{Q})/ |p(\mathcal{Q}\mathcal{Q})|< 0.005$ and 0.002 for different kinds of quirks are shown in the middle and right panels of \cref{fig::xsecs}, where $|p(\mathcal{Q}\mathcal{Q})|$ and $p_T(\mathcal{Q}\mathcal{Q})$ are the momentum size and the transverse momentum of the quirk-pair system, respectively.

The motion for each quirk moving through materials is controlled by~\cite{Kang:2008ea}
\be
\frac{\partial ({m} \gamma \vec{v})}{\partial t} =\vec{F}_{s}+\vec{F}_{\text{ion}}~, 
\quad 
\vec{F}_{s}=-\Lambda^2\sqrt{1-\vec{v}_{\perp}^{2}} \hat{s}-\Lambda^2 \frac{v_{ \|} \vec{v}_{\perp}}{\sqrt{1-\vec{v}_{\perp}^{2}}}~,
\quad
\vec{F}_{\text{ion}}= \frac{dE}{dx}\hat{v}
\ee
where $\gamma=1/\sqrt{1-\vec{v}^2}$, $v_{ \|}=\vec{v}\cdot\hat{s}$ and $\vec{v}_{\perp}=\vec{v}-v_{ \|}\hat{s}$ with $\hat{s}$ being a unit vector along the string pointing outward at the endpoint. $\vec{F}_s$ and $\vec{F}_{\text{ion}}$ correspond to the infracolor force described by the Nambu-Goto action and the force arising from the effects of ionization energy loss for charged quirk propagating through materials, respectively. The other energy loss effects such as infracolor glueball and photon radiations are not included since their effects are negligible \cite{Li:2021tsy}. The methods of numerically solving the EoM by slowly increasing the time with small steps can be found in Ref.~\cite{Li:2019wce}.

\begin{figure}[t]
\centering
	\includegraphics[width=0.50\textwidth]{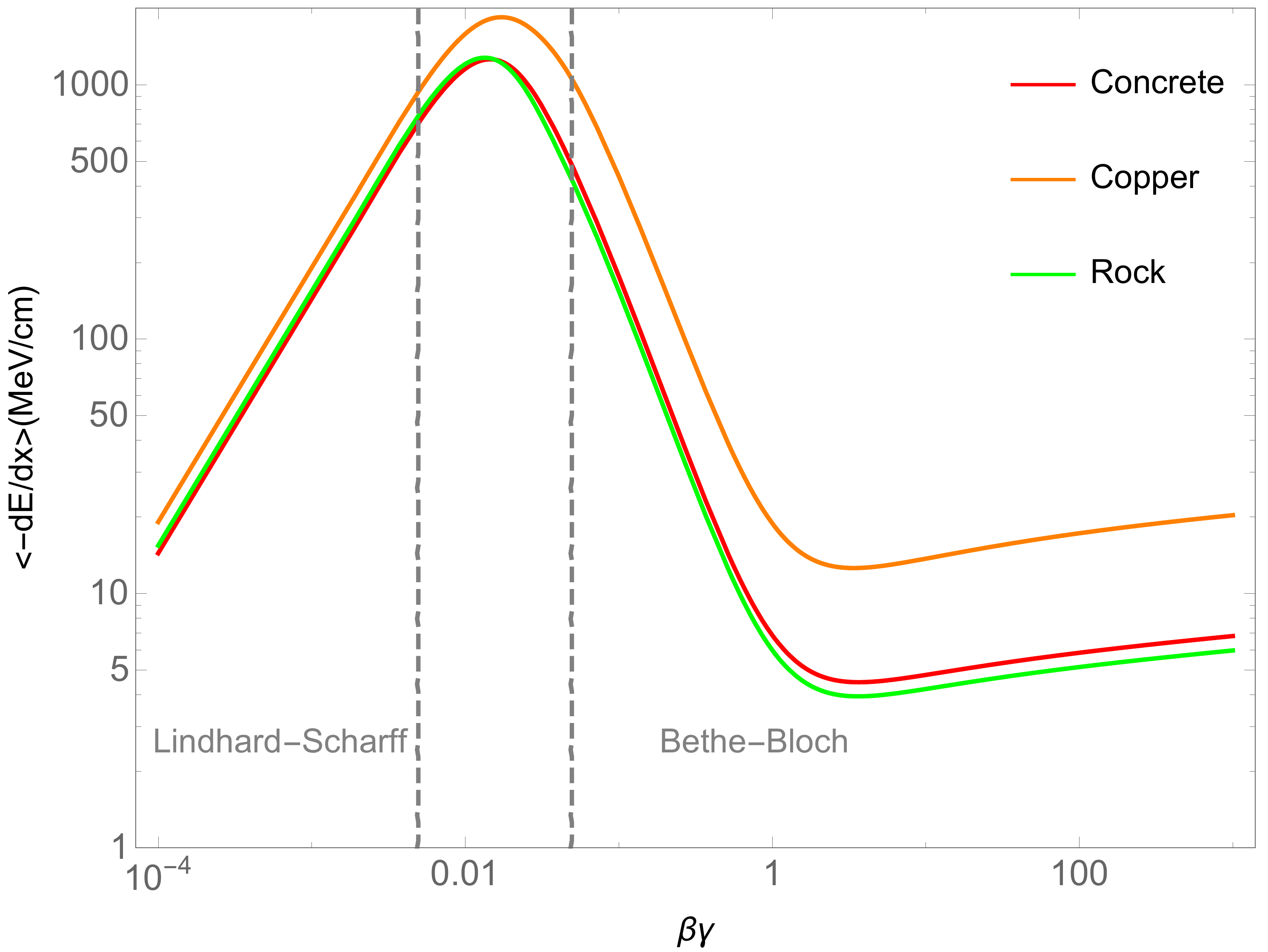}
\caption{\label{deodx} The mean rates of energy loss for charged particle traveling through concrete, copper, and rock, supposing $z=1$ and $m_e/M\ll1$.}
\end{figure}

In \cref{material}, configurations of the main infrastructures between the ATLAS IP and FASER (2) used in the simulation are listed \cite{LHCf:2006kzv}. The mean rates of energy loss when a charged particle with mass $M$ and electric charge $z=1$ moving through concrete, copper, and rock are plotted in \cref{deodx}, where values of $\langle -dE/dx \rangle$ between the Lindhard-Scharff (LS) and the Bethe-Bloch (BB) regions are obtained by interpolation. The $\langle -dE/dx \rangle$ is independent of $M$ since $m_e/M\ll 1$ is assumed. It is noted that  
the real ionization energy loss in the BB region for a charged particle travelling a long distance in the material fluctuates according to a Gaussian distribution. At each time grid of our simulation, the $-dE/dx$ is randomly generated based on this Gaussian distribution since quirks travel through macroscopic region of materials with $v/c \sim 1$ in our case.

\begin{table}[htb]
	\centering
	\begin{tabular}{c||c|c}  
		\hline\hline
		Component  & $x,y,R$[m] & $z$[m] \\
		\hline 
		TAS (Copper) & $R>0.017$ & 19$-$20.8 \\
		D1 (3.5 T) & $R<0.06$  &  59.92$-$84.65 \\
		TAN (Copper) &  $|x|<0.047$,  $-0.538<y<0.067$ &  140$-$141 \\
		D2 (3.5 T) & $(x\pm 0.093)^2+y^2<0.04^2$  &  153.48$-$162.93 \\
		Concrete & $R>0$  &  380$-$390 \\
		Rock &  $R>0$ &  390$-$480\\
		Tracker of FASER & $|x|<0.16$,  $|y|<0.16$  &  $|z-481.6/482.8/484.0|<0.041$ \\
		Tracker of FASER2 & $|x|<1$,  $|y|<1$  &  $|z-485.1/486.3/487.5|<0.041$ \\
		\hline\hline
	\end{tabular}
	\caption{\label{material} The configurations of infrastructures between the ATLAS IP and the FASER (2) detector. The ATLAS IP is the ordinate origin and the transverse distance is $R=\sqrt{x^2+y^2}$.}
\end{table}

\begin{figure}[t]
\centering
\includegraphics[width=0.24\textwidth]{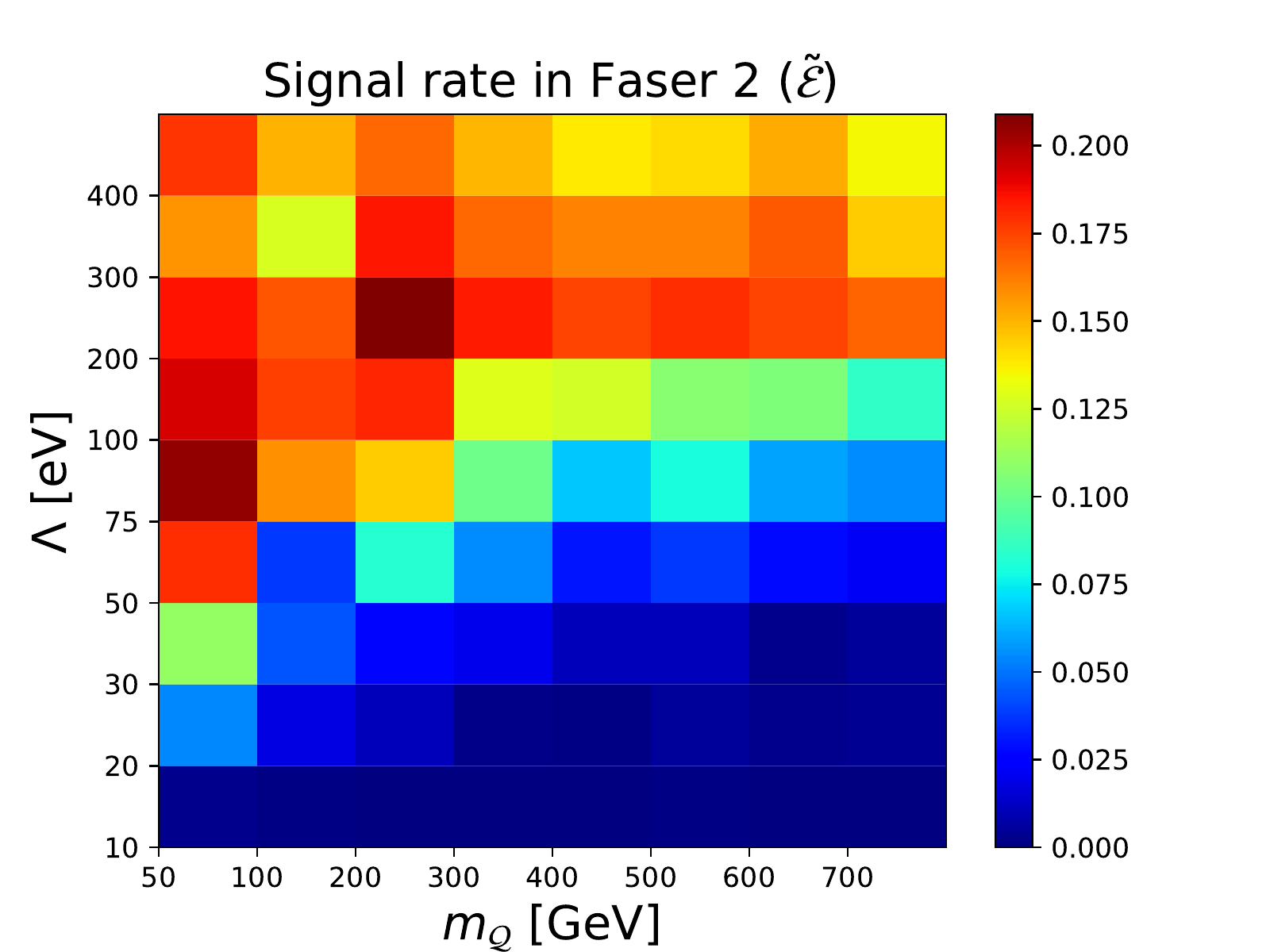}
\includegraphics[width=0.24\textwidth]{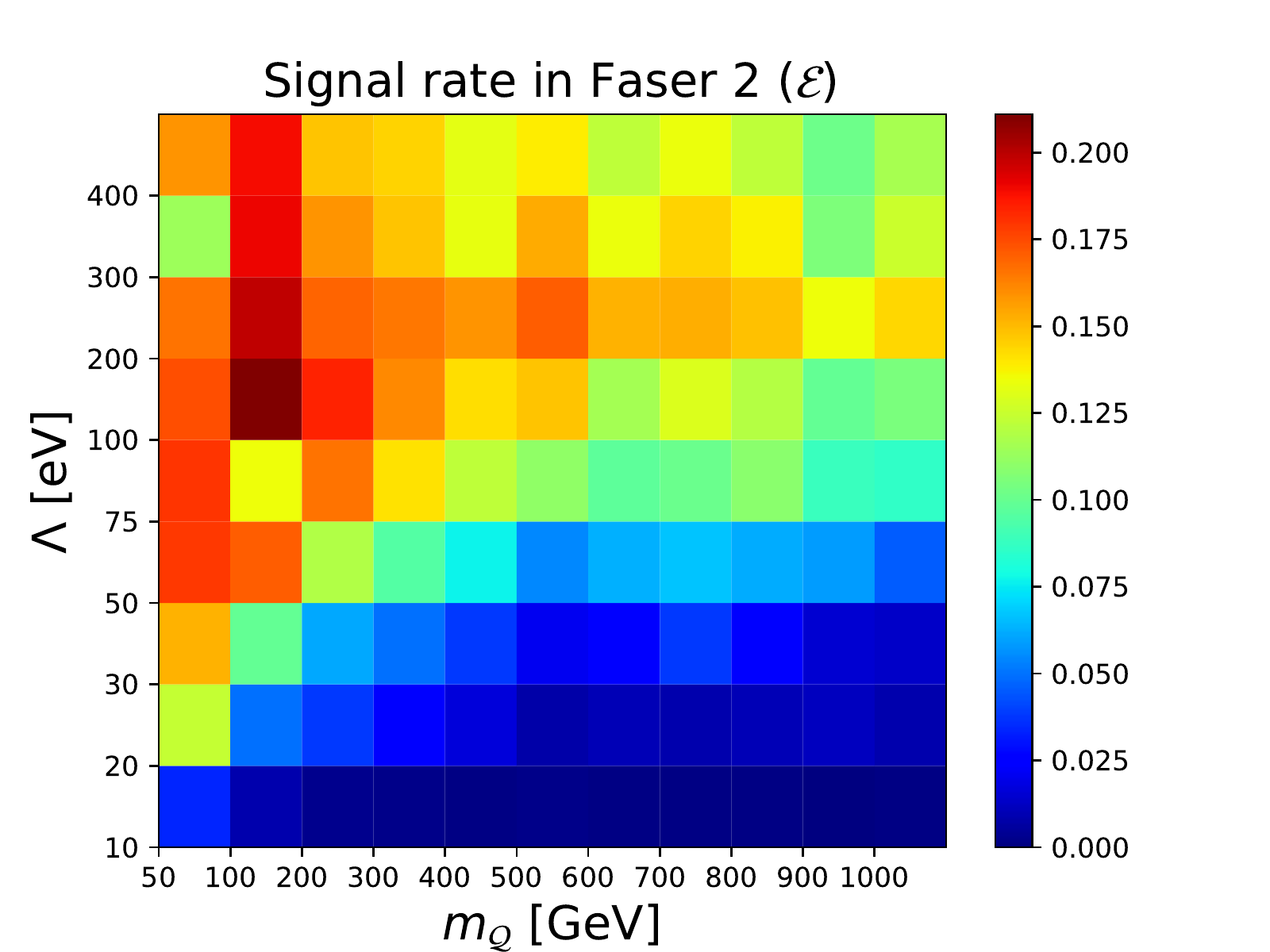} 
\includegraphics[width=0.24\textwidth]{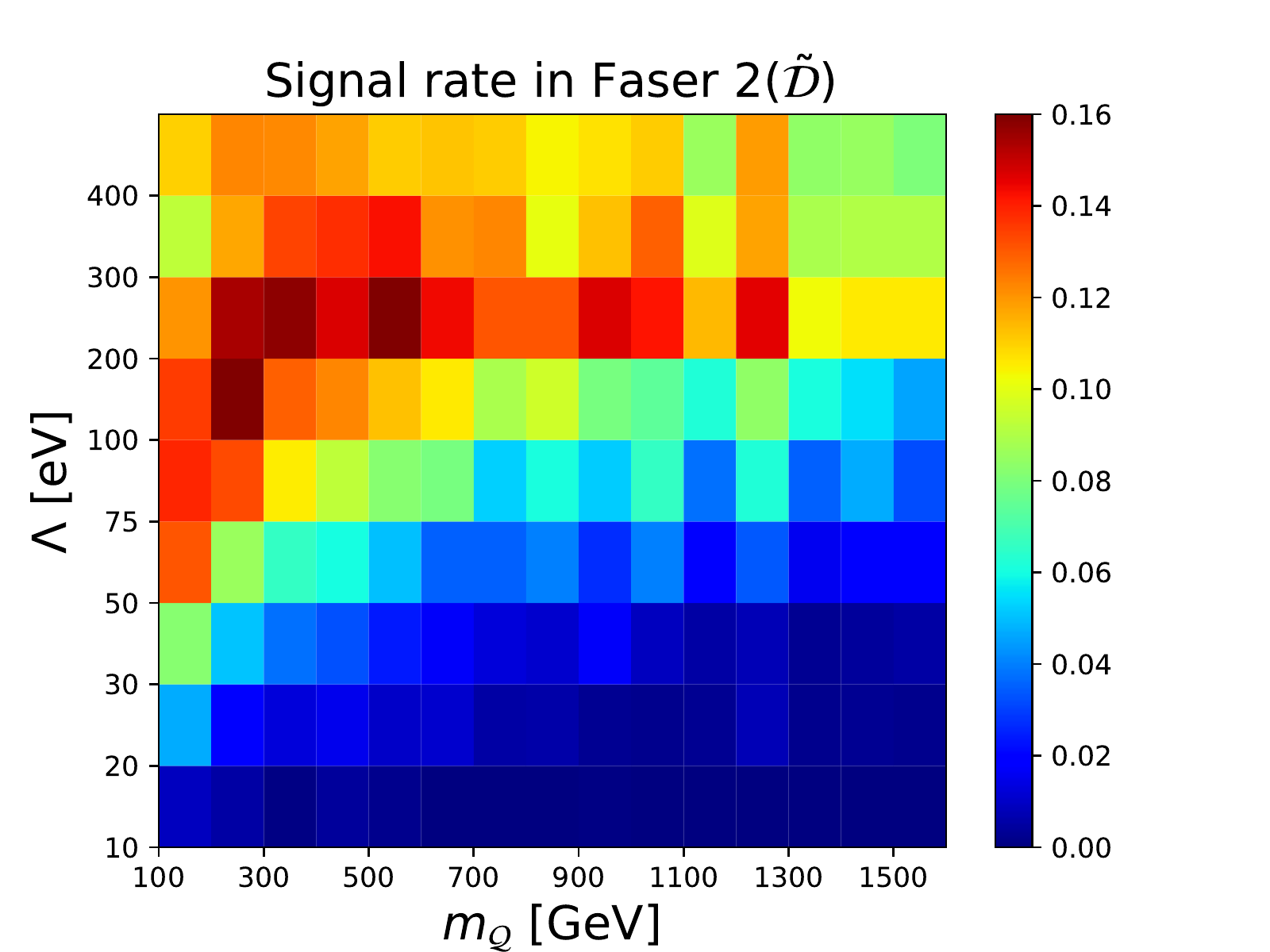}
\includegraphics[width=0.24\textwidth]{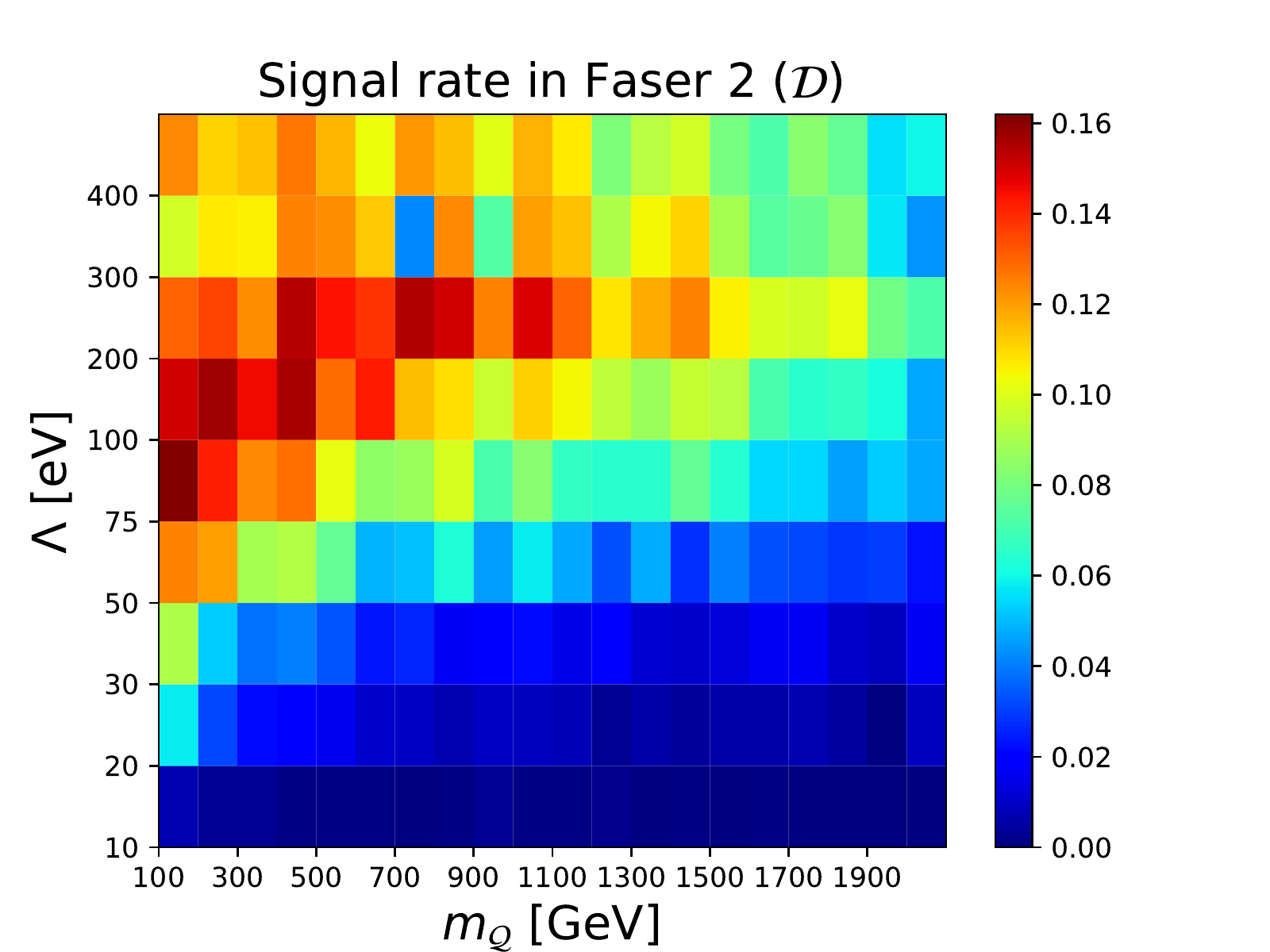}\\
\includegraphics[width=0.24\textwidth]{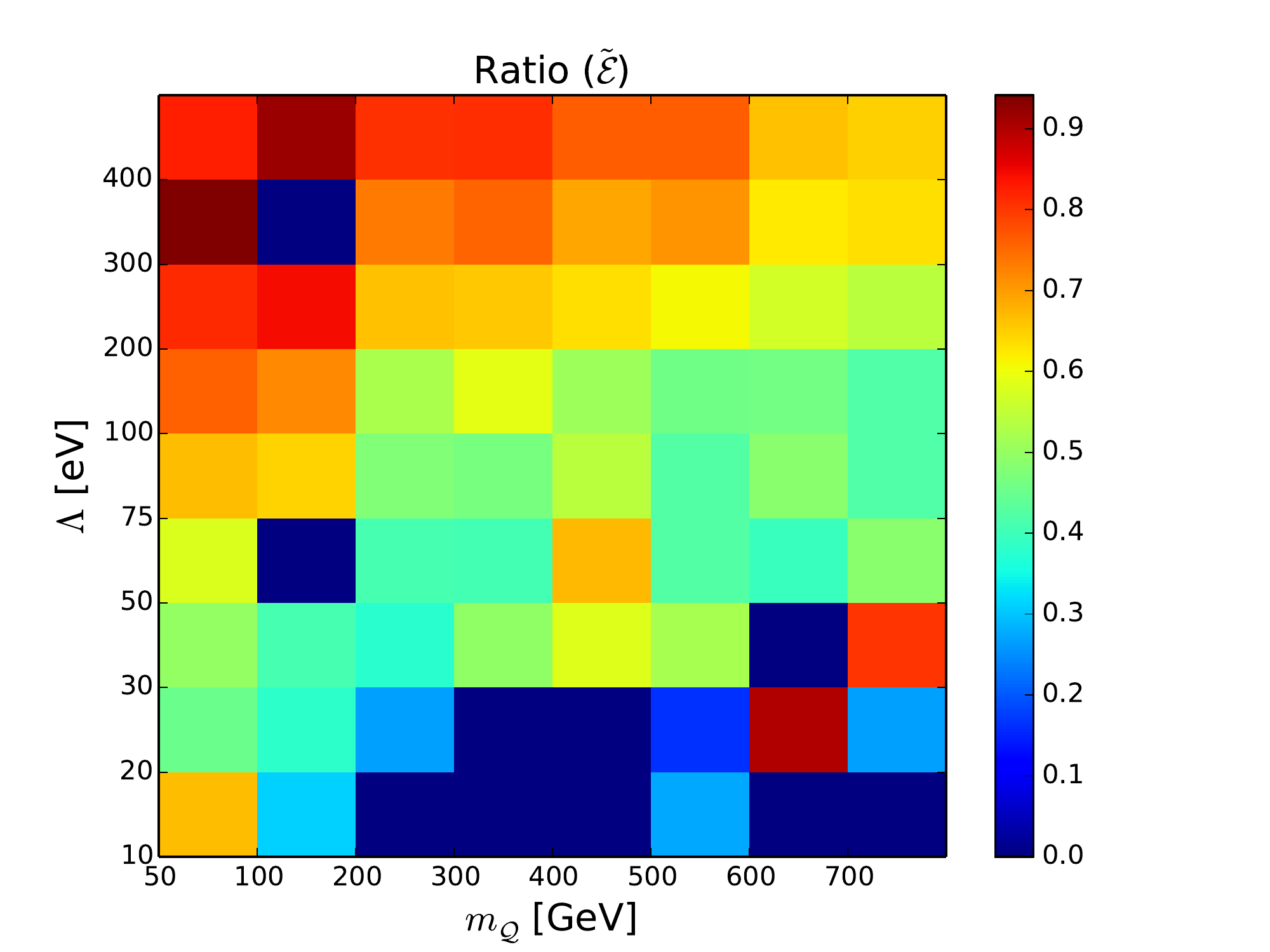}
\includegraphics[width=0.24\textwidth]{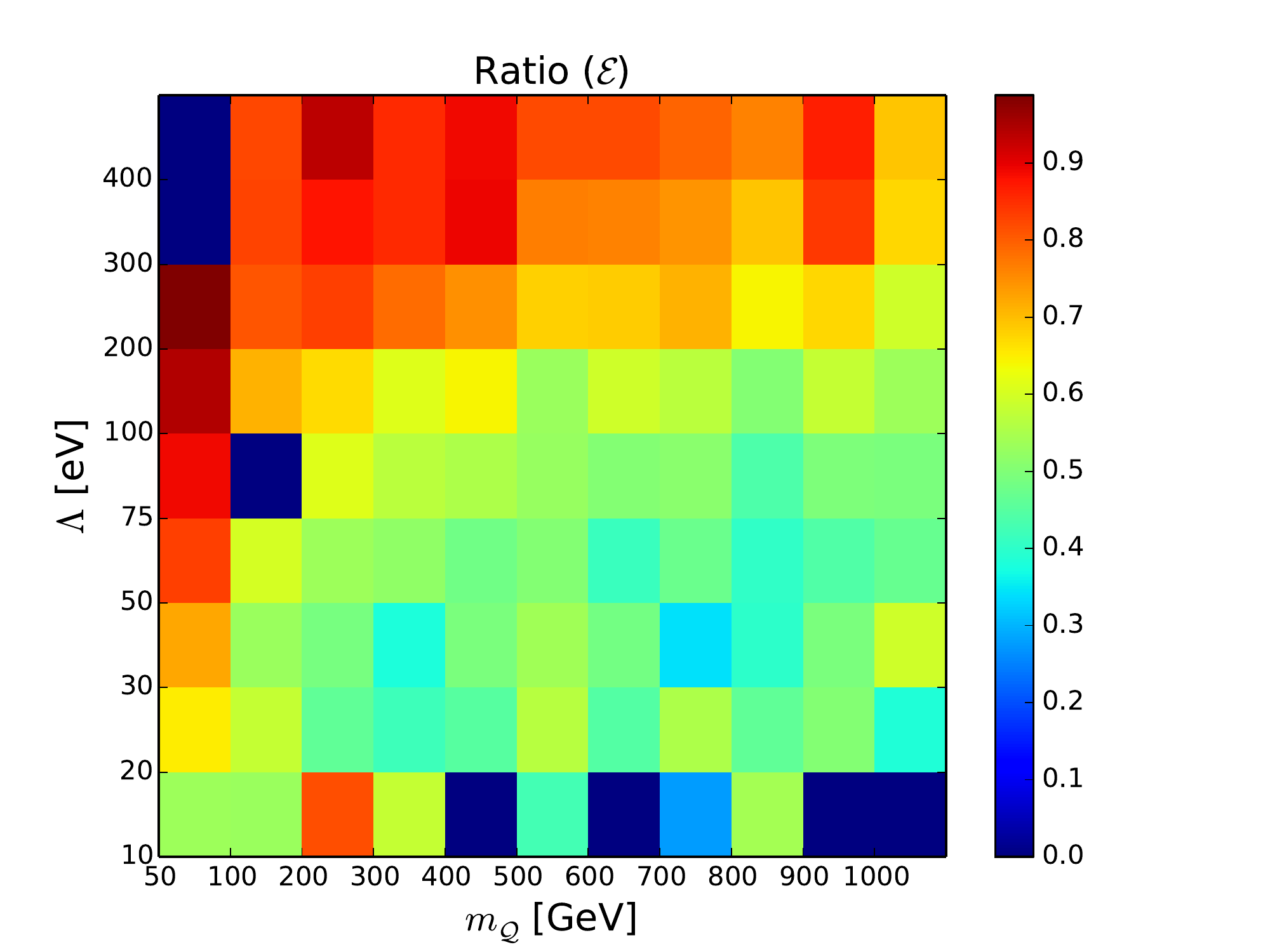} 
\includegraphics[width=0.24\textwidth]{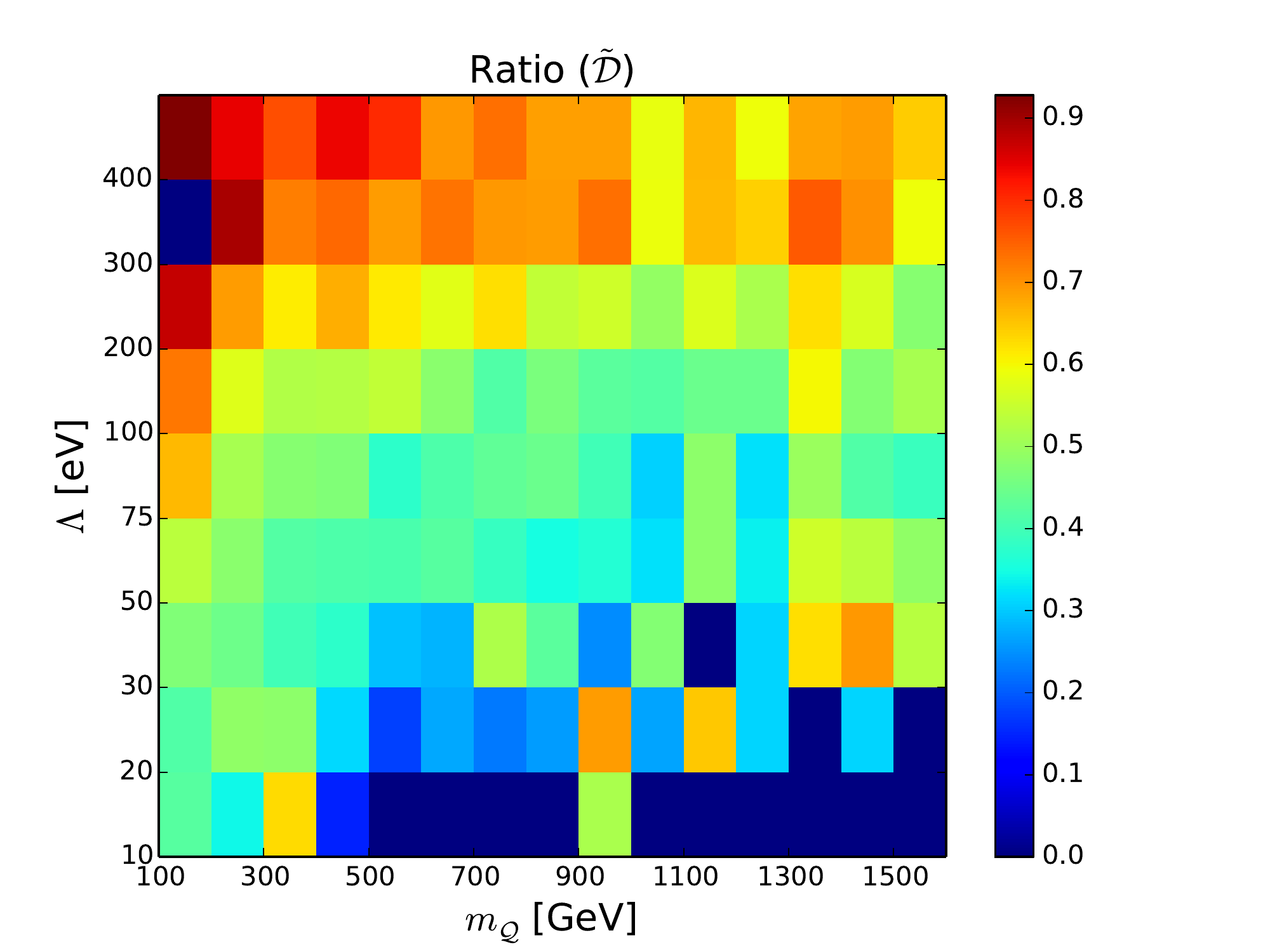}
\includegraphics[width=0.24\textwidth]{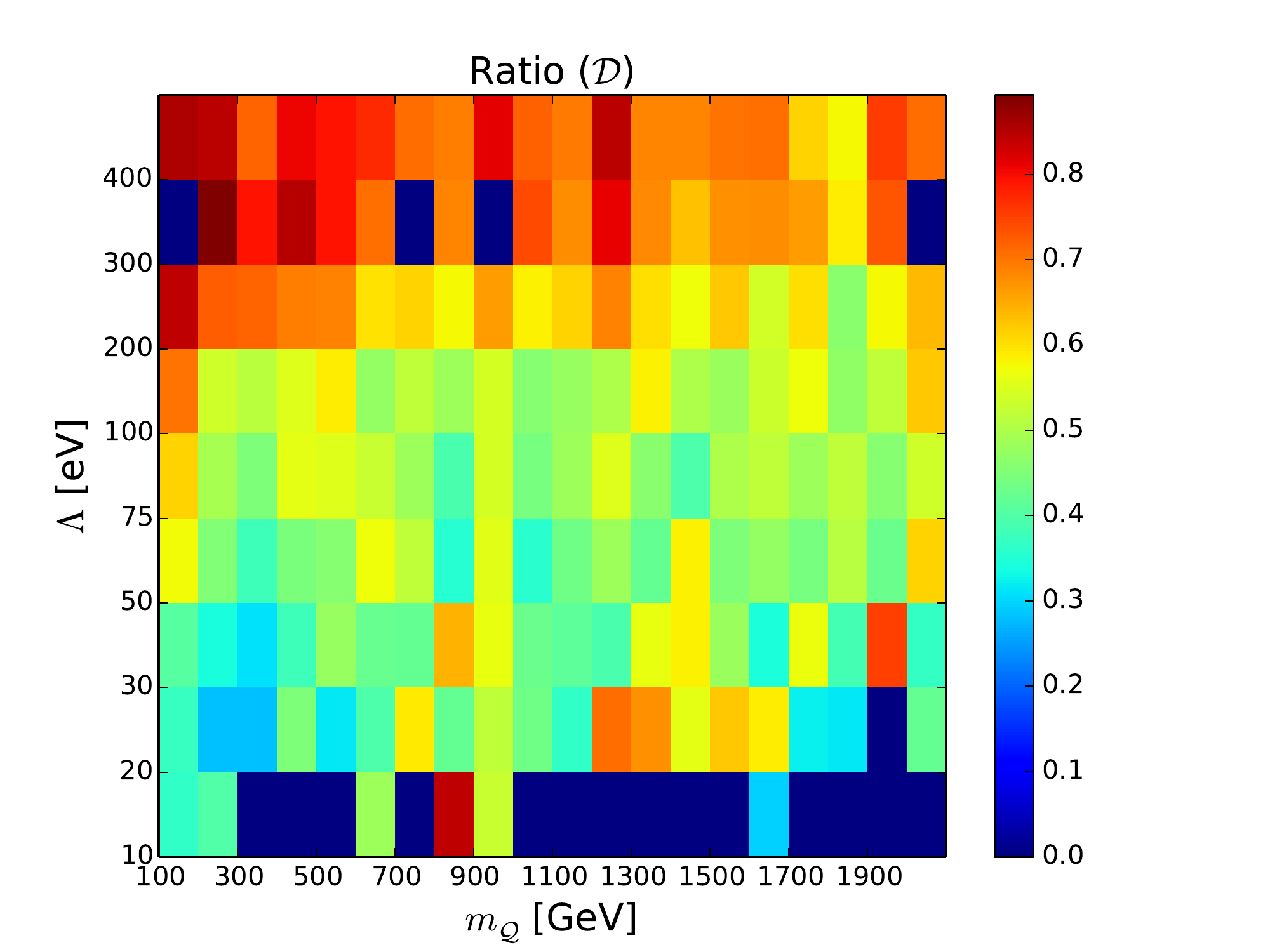}
\caption{ Upper panels: the fractions of quirk events (in event sample with $p_T(\mathcal{Q}\mathcal{Q})/ |p(\mathcal{Q}\mathcal{Q})|< 0.005)$ that have at least one quirk entering the FASER2 tracker. Lower panels: among the events which can enter the FASER2 tracker, the ratio between the number of events with $p_T(\mathcal{Q}\mathcal{Q})/ |p(\mathcal{Q}\mathcal{Q})|< 0.002$ and $p_T(\mathcal{Q}\mathcal{Q})/ |p(\mathcal{Q}\mathcal{Q})|< 0.005$ in initial state. Quirks with four different quantum numbers as given in \cref{eq::qn1} - \cref{eq::qn4} are considered.}
\label{fig::tag0052}
\end{figure}
We solve the EoM of a quirk pair with given initial momenta to see whether the two quirks in this pair could enter the tracker of FASER (2) or not. Among the events satisfying $p_T(\mathcal{Q}\mathcal{Q})/ |p(\mathcal{Q}\mathcal{Q})|< 0.005$, the fractions that have at least one quirk entering the FASER2 tracker are shown on the $m_{\mathcal{Q}} - \Lambda$ planes of the upper panels of \cref{fig::tag0052}.   In the lower panels of \cref{fig::tag0052}, the ratio between the number of events with initial $p_T(\mathcal{Q}\mathcal{Q})/ |p(\mathcal{Q}\mathcal{Q})|< 0.002$ and that with initial $p_T(\mathcal{Q}\mathcal{Q})/ |p(\mathcal{Q}\mathcal{Q})|< 0.005$ among the events that can reach the FASER2 tracker are shown. It is noted that the results for $\Lambda \le 30$ eV can not be trusted since only a few events  can reach the FASER2 tracker in this case, leading to a huge fluctuation in our simulation.

\begin{figure}[t]
\centering
	\includegraphics[width=0.45\textwidth]{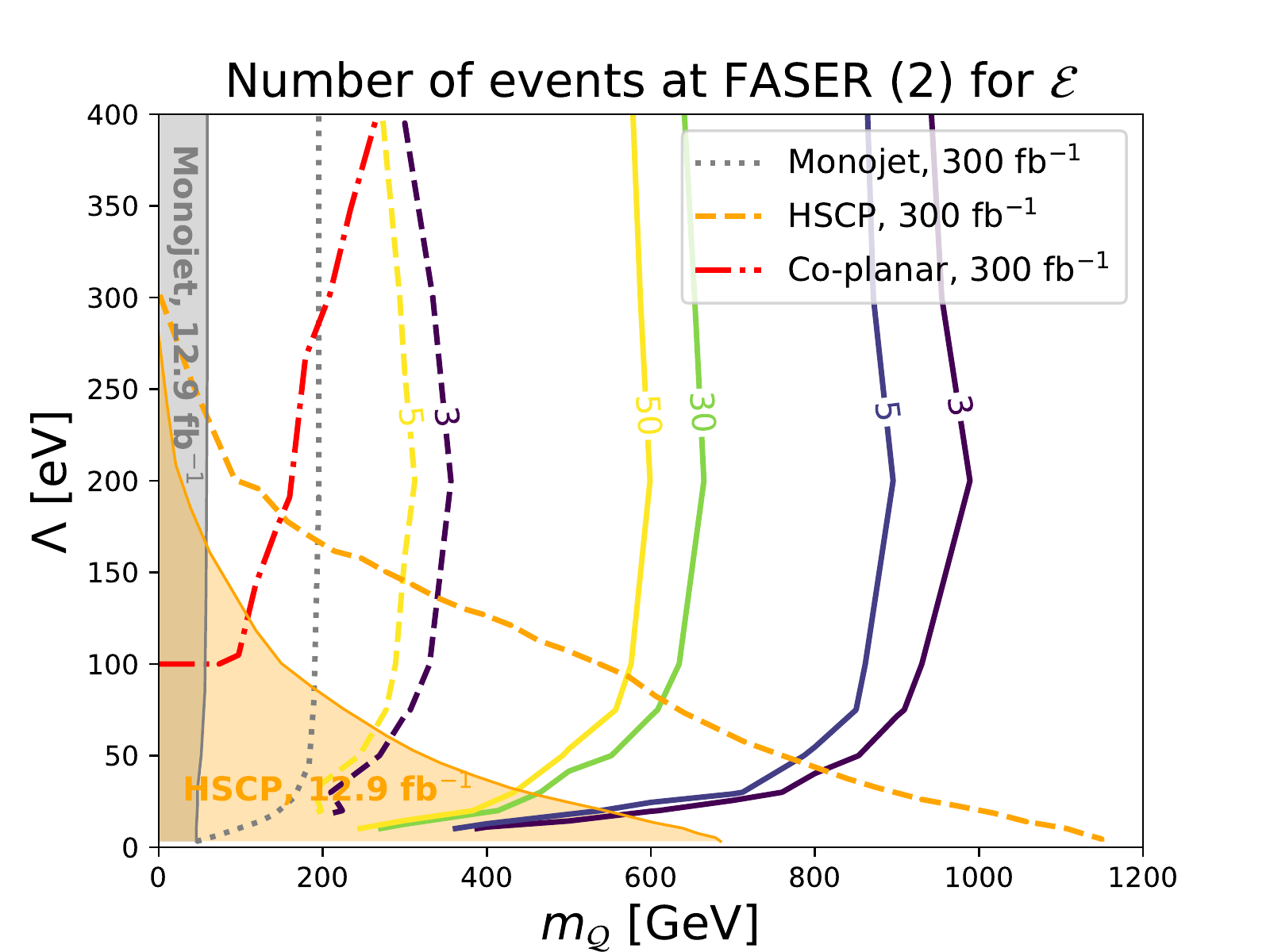}
	\includegraphics[width=0.45\textwidth]{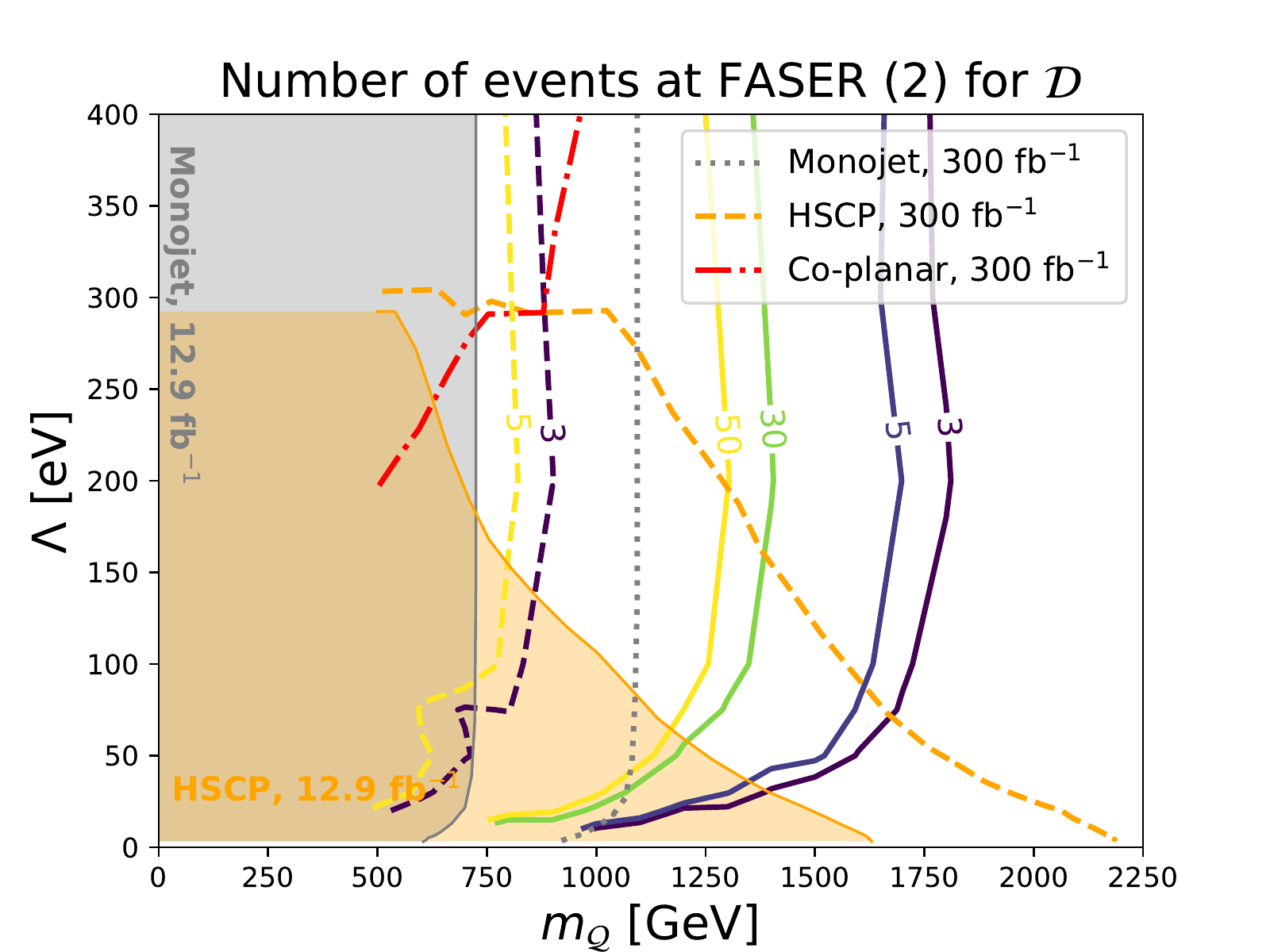} \\
	\includegraphics[width=0.45\textwidth]{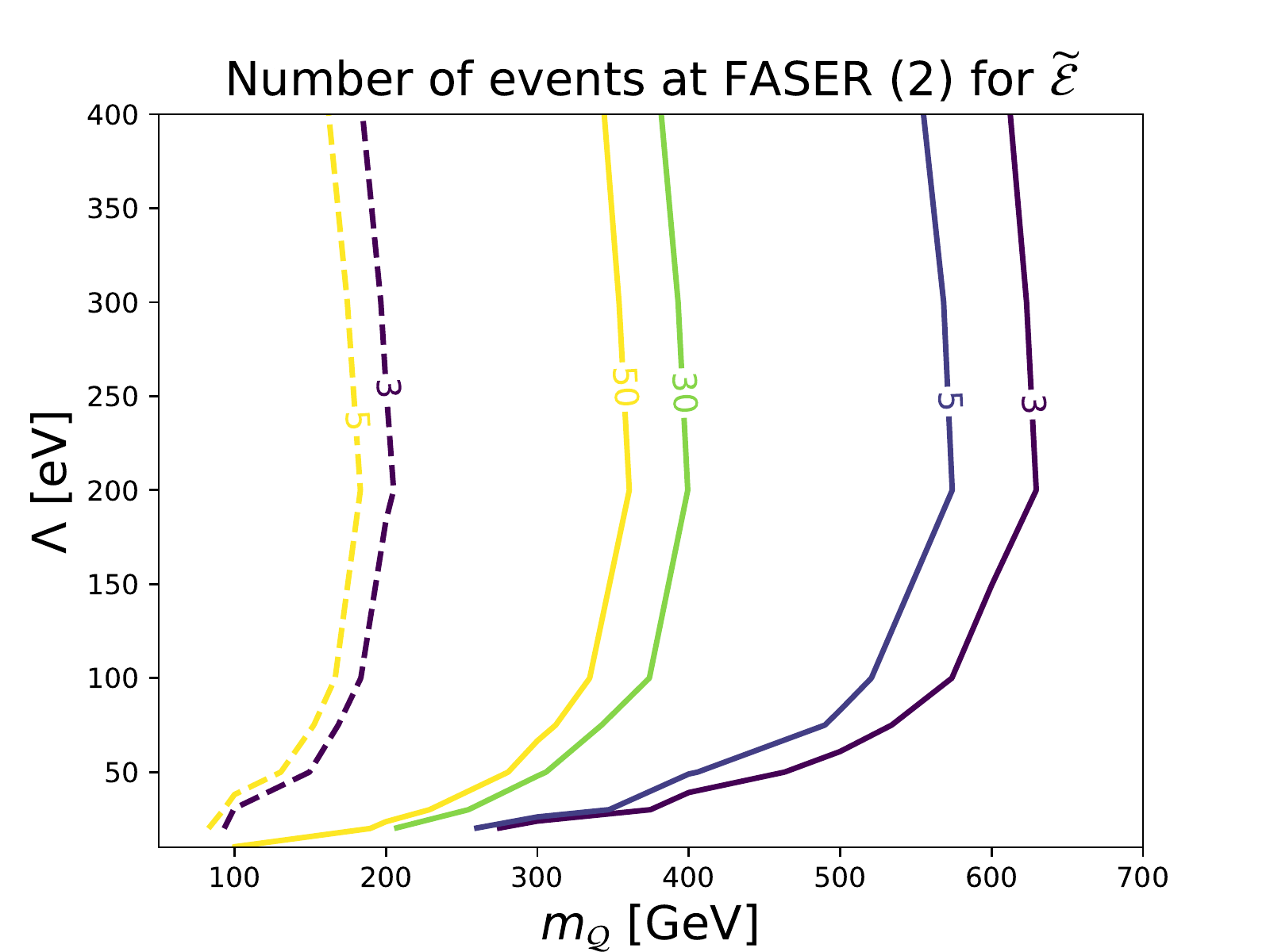}
	\includegraphics[width=0.45\textwidth]{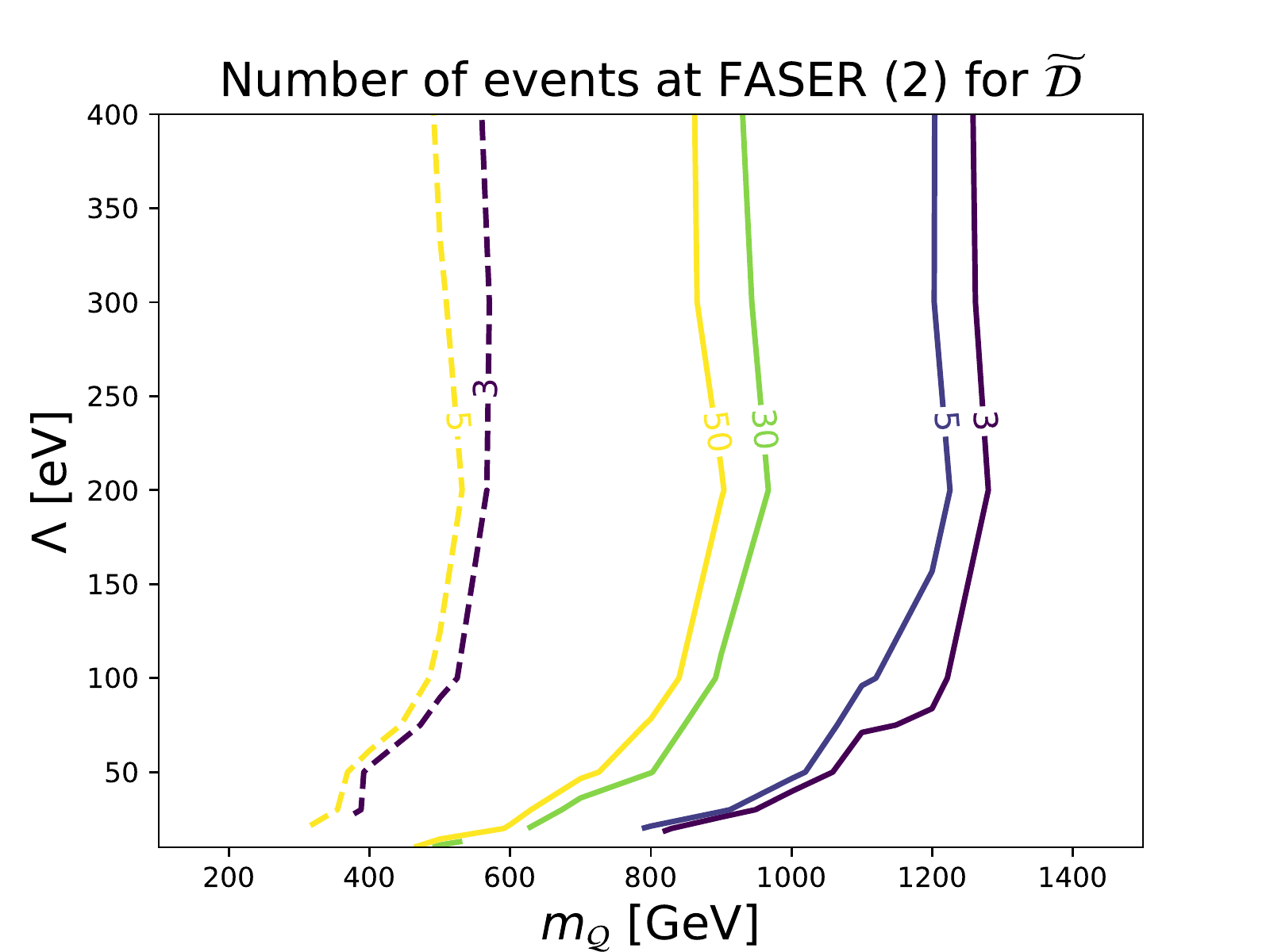} 
\caption{Contours of the number of quirk events that can reach the FASER (2) tracker in the $m_{\mathcal{Q}}$ versus $\Lambda$ plane for an integrated luminosity of 150 (3000) fb$^{-1}$. The dashed (solid) ones correspond to the event numbers at FASER (2). For two fermionic quirks ($\mathcal{E}$ and $\mathcal{D}$), the projected bounds from Heavy Stable Charged Particle (HSCP) search~\cite{Farina:2017cts}, mono-jet search~\cite{Farina:2017cts}, and coplanar search~\cite{Knapen:2017kly} (the exclusion limits are taken) are shown by orange dashed line, grey dotted curves, and red dash-dotted line, respectively. Moreover, the existing bounds from the CMS HSCP search~\cite{CMS-PAS-EXO-16-036} and ATLAS monojet search~\cite{CMS:2016xbb} are shown by grey and orange shaded regions. Those bounds are extracted from Ref.~\cite{Knapen:2017kly}. }
\label{fig::exclusion}
\end{figure}

The total number of quirk events observed by FASER (2) can be calculated by
\begin{align}
N_{\text{sig}} = \sigma \times \epsilon_{\rm fid} \times \epsilon_{0.005} \times \mathcal{L},
\end{align}
where $\sigma$~\footnote{The colored quirks $\tilde{\mathcal{D}}$ and $\mathcal{D}$ will hadronize into quirk-quark bound states, and the probability for those final states to have $\pm 1$ charges is roughly 30\%. So, the factor 0.3 is multiplied on their cross sections when estimating the number of signal events.} and $\epsilon_{\rm fid}$ stand for the quirk production cross section and the efficiency of selecting events with $p_T(\mathcal{Q}\mathcal{Q})/ |p(\mathcal{Q}\mathcal{Q})|< 0.005$ in quirk pair production, respectively, which are illustrated in \cref{fig::xsecs}. Events with $p_T(\mathcal{Q}\mathcal{Q})/ |p(\mathcal{Q}\mathcal{Q})|> 0.005$ and reaching FASER (2) are not counted due to the quite low efficiency in this kinematic region and the huge time cost of solving the EoM for them. $\epsilon_{0.005}$ is the signal efficiency shown in the upper panels of \cref{fig::tag0052}. Besides, we take the integrated luminosity $\mathcal{L}$ as $150$ and $3000$ fb$^{-1}$ for FASER and FASER2, respectively.

With an integrated luminosity of 150 (3000) fb$^{-1}$, the number of quirk events at FASER (2) is shown in \cref{fig::exclusion}. The numbers are not very sensitive to $\Lambda$ when it is larger than $\sim$ 100 eV.  According to the discussions in Ref.~\cite{Li:2021tsy}, the background in the FASER (2) detector can be highly suppressed by using the unique features of quirk tracks. With a negligible background, the contours of 3 events stand for 2-$\sigma$ exclusion limits, and the contours of 5 events imply the discovery prospects. According to \cref{fig::exclusion}, it is concluded that FASER2 (FASER) will be able to exclude the $\mathcal{E}$, $\mathcal{D}$, $\tilde{\mathcal{E}}$ and $\tilde{\mathcal{D}}$ quirks with mass below 990 (360) GeV, 1800 (900) GeV, 630 (200) GeV and 1280 (570) GeV, respectively, with an integrated luminosity of 3000 (150) fb$^{-1}$ when $\Lambda \gtrsim \mathcal{O}(100)$ eV. The bounds on fermionic quirks are much stronger than those on scalar quirks when the gauge representations are the same due to their larger production rates and higher signal efficiencies. FASER2 is much more sensitive to the quirk signal than FASER because of the larger tracking plane and the increased integrated luminosity. The projected bounds from the mono-jet search, the Heavy Stable Charged Particle (HSCP) search, and the coplanar search as provided in Ref.~\cite{Knapen:2017kly} for fermionic quirks are shown as well for comparison.  The HSCP search is most sensitive for $\Lambda \lesssim 50$ eV. When $\Lambda \gtrsim \mathcal{O}(100)$ eV, FASER2 is much more sensitive than other searches.  FASER behaves better than other searches when $\Lambda \gtrsim 150$ eV for the color neutral quirk $\mathcal{E}$.

%% file: sec_qcd.tex
\contributors{Lucian Harland-Lang,
  Juan Rojo (conveners),
  Alessandro Bacchetta,
  Atri Bhattacharya,
  Marco Bonvini,
  Victor P. Goncalves,
  Francesco Giovanni Celiberto,
  Grigorios Chachamis,
  Pit Duwentaster,
  Max Fieg,
  Rhorry Gauld,
  Francesco Giuli,
  Marco Guzzi,
  Timothy J. Hobbs,
  Steffan Hoeche,
  Tomas Jezo,
  Cynthia Keppel,
  Michael Klasen,
  Felix Kling, K. Kovarik,
  Frank Krauss,
  Aleksander Kusina,
  Chiara Le Roux,
  Rafal Maciula,
  Jorge G. Morfın,
  K.F. Muzakka,
  Pavel Nadolsky,
  Emanuele R. Nocera,
  Fred Olness,
  Richard  Ruiz,
  Ingo Schienbein,
  Holger Schulz,
  Federico Silvetti,
  Antoni Szczurek,
  Keping Xie,
  J. Y. Yu,
  Korinna Zapp,
  Michael Fucilla,
  Mohammed M. A. Mohammed,
  Alessandro Papa,
  Ina Sarcevic,
  Torbjorn Sjostrand,
  Agustin Sabio Vera,
  and Anna Stasto
}

\begin{figure}[h]
\centering
\includegraphics[width=0.95\textwidth]{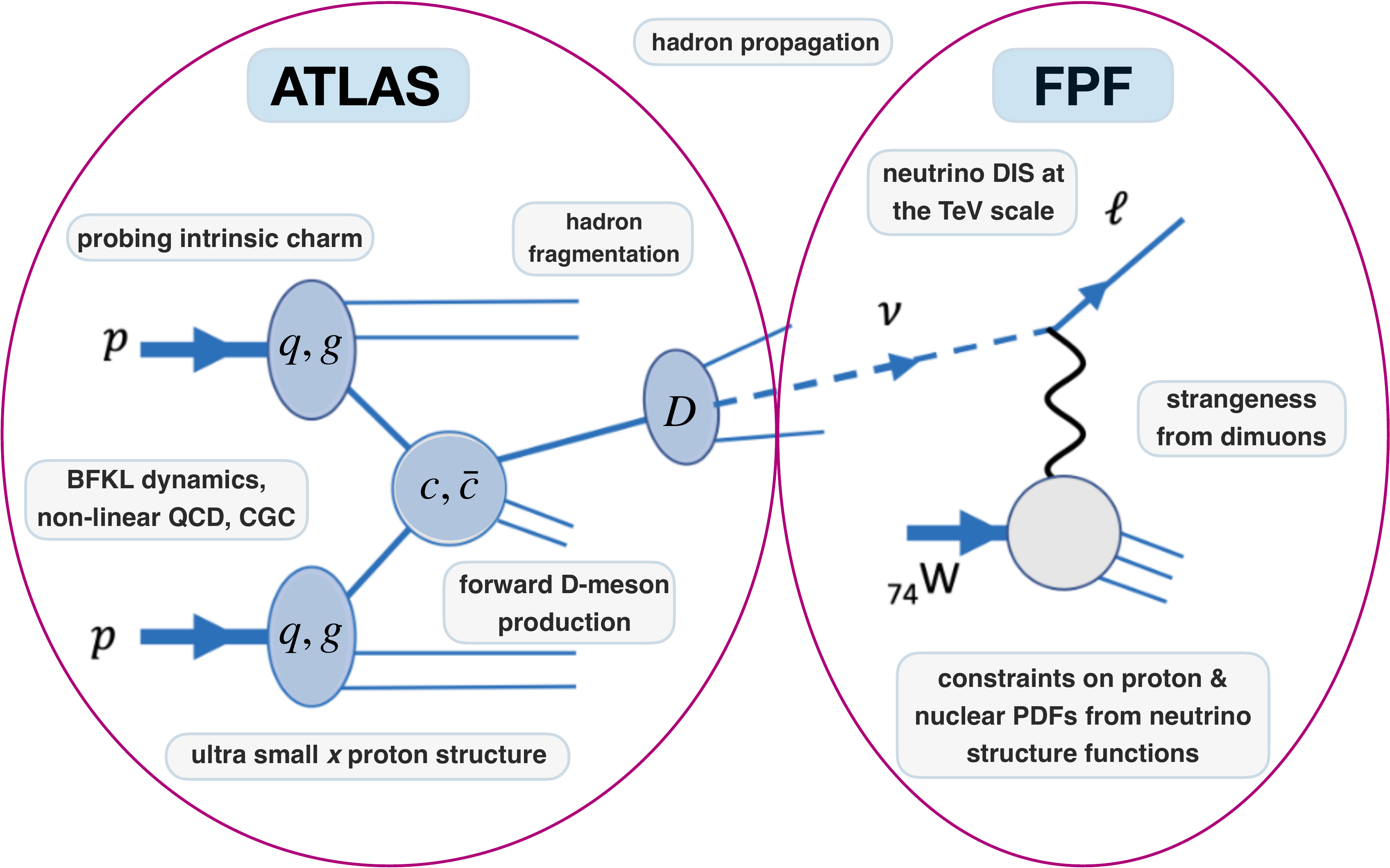}
\caption{Schematic representation of a typical QCD
process, in this case $D$-meson production, taking place at the FPF.
First of all, particles such as light hadrons
or charmed mesons are produced in the very forward region 
in proton-proton collisions at the ATLAS interaction point.
Some of these particles will decay into
neutrinos, which will travel unaffected until
they hit the FPF detectors.
The FPF operates hence as a neutrino-induced
deep-inelastic scattering experiment with
TeV-scale neutrino beams.
}
\label{fig:FPF_QCD_overview}
\end{figure}

In this chapter we explore the rich potential
for QCD studies that the Forward Physics Facility would offer.
Schematically, QCD studies at the FPF can
be classified as being related to 
forward particle production in proton-proton collisions
and neutrino deep-inelastic scattering on the target
detector, as indicated in \cref{fig:FPF_QCD_overview}.
First, one has to consider the very forward production of particles
in proton-proton collisions, such as light hadrons
or charmed mesons,  taking place
at the ATLAS interaction point.
The kinematics of such processes provide
access to the very  low-$x$ region of the 
colliding protons, as well as to novel
QCD production mechanisms such as BFKL or non-linear dynamics.

Then, some of these particles will propagate and
eventually decay into
neutrinos, which will  travel unaffected until
they hit the FPF detectors.
Detection of these neutrinos implies that the FPF effectively operates as a neutrino-induced
deep-inelastic scattering experiment with TeV-scale neutrino beams.
Measurements of the resulting DIS structure functions
provide a valuable handle on the partonic
structure of both nucleons and nuclei, in particular
concerning quark flavour separation.
The FPF will therefore continue the extremely successful
CERN program of deep-inelastic scattering with neutrino beams, which
has been instrumental in the  understanding of both the neutrino sector
of the Standard Model as well as of nucleon and nuclear structure~\cite{Gao:2017yyd}.

\begin{figure}[ht]
\centering
\includegraphics[width=0.80\textwidth]{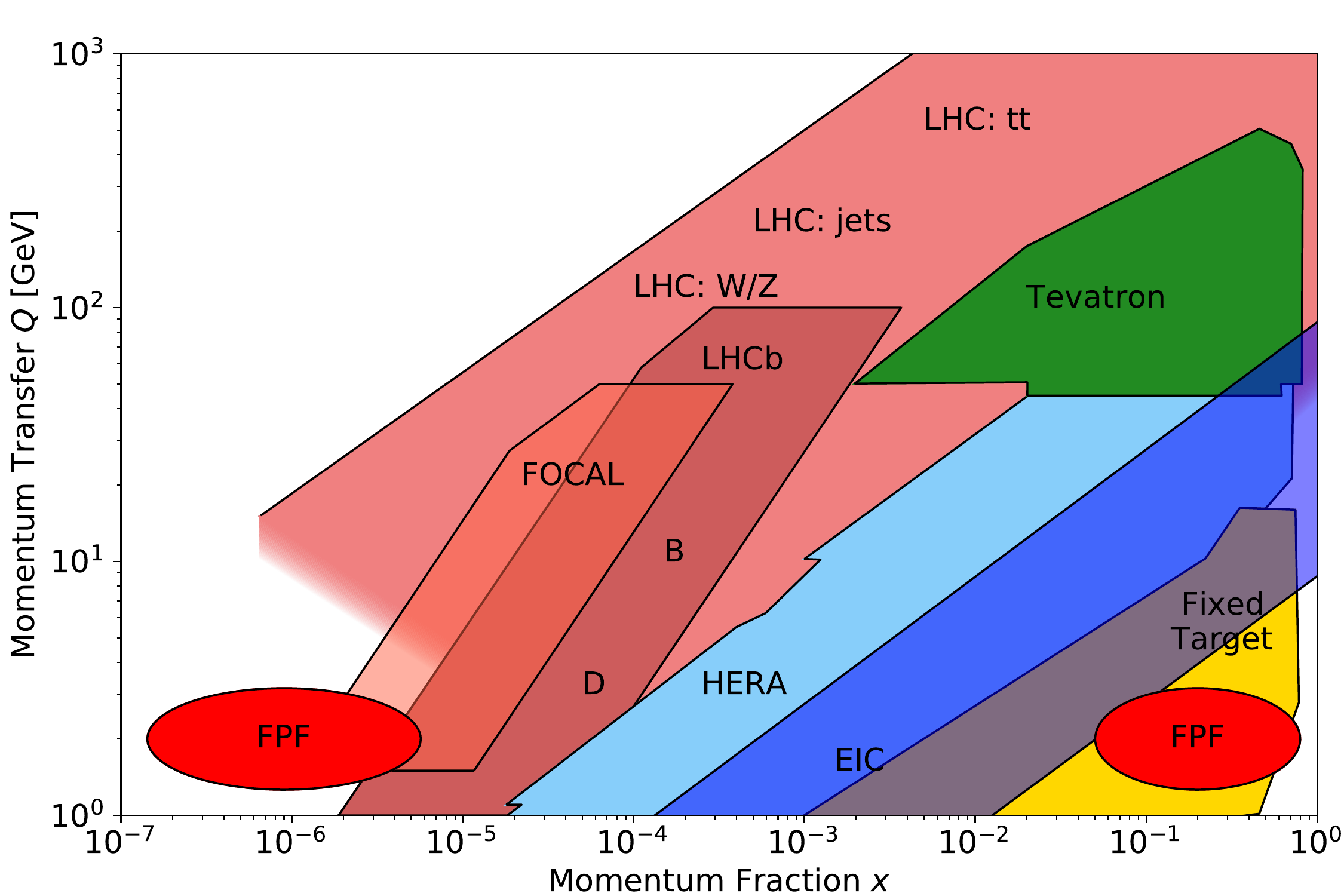}
\caption{The schematic kinematic coverage in the $(x,Q)$
plane for $D$-meson production in proton-proton
collisions at the LHC ($\sqrt{s}=14$ TeV) followed
by their decay into neutrinos falling within
the FPF acceptance.
The approximate
kinematic coverage for other experiments providing inputs
for proton  global PDF analyses, as well as that corresponding to future facilities
such as the Electron-Ion Collider and the FoCal upgrade of the ALICE experiment are indicated.
}
\label{fig:kinplot-ppcollisions}
\end{figure}

In order to illustrate the kinematic reach
in proton-proton collisions that would become available
with the FPF, \cref{fig:kinplot-ppcollisions}
displays the coverage in the $(x,Q)$
plane for $D$-meson production on proton-proton
collisions at the LHC with $\sqrt{s}=14$ TeV, followed
by the decay into neutrinos falling within
the FPF acceptance.
In the same figure we also indicate the approximate
kinematic coverage for other experiments which provide inputs
for proton  global PDF analyses, as well as that corresponding to future facilities
such as the Electron-Ion Collider (EIC)~\cite{AbdulKhalek:2021gbh}
and the Forward Calorimeter (FOCAL)~\cite{ALICE:2020mso}
upgrade of the ALICE experiment.
See also the ``Hadronic Tomography at the EIC and the Energy Frontier'' white paper~\cite{EICWhitepaper} for more details on the impact of the EIC on proton
and nuclear structure.

Inspection of \cref{fig:kinplot-ppcollisions} reveals that
 the availability of FPF measurements
would extend the coverage of LHC measurements in the
low--$x$ region by almost two orders of magnitude at
low--$Q$, reaching down to $x\simeq 10^{-7}$
depending on the specific acceptance of the FPF detectors
assumed. 
Accessing this extreme and essentially unexplored
kinematic range opens a wide range of opportunities
for QCD studies, such as charting the gluon at very
low-$x$, revealing non-standard QCD phenomena
such as BFKL dynamics, and testing our Monte Carlo models
for forward hadron production. 
In turn, this improved understanding of low-$x$
QCD and nucleon structure provides improved
predictions for key astroparticle physics processes,
such as ultra-high energy (UHE) neutrino-nucleus
and cosmic ray interaction cross-sections.
We note that the forward production of light hadrons, such as pions and kaons,
will also contribute to the overall neutrino yield at the FPF. In this case, 
a similar kinematic region to that reported in \cref{fig:kinplot-ppcollisions}
will also be probed.

It is also worth noting here that understanding small-$x$ dynamics in proton-proton collisions, which
are already important at the LHC as well as of its High-Luminosity
upgrade~\cite{Cepeda:2019klc,Azzi:2019yne}, would become crucial for any future higher-energy
proton-proton collider such as the Future Circular Collider
at 100 TeV~\cite{FCC:2018vvp,FCC:2018byv,Mangano:2016jyj,Rojo:2016kwu}.
At such extreme energies, even standard electroweak processes such as $W$ and $Z$ production
become dominated by low--$x$ dynamics, and an accurate calculation of the Higgs production cross section
requires that BFKL resummation effects be  accounted for.
Therefore, the  mapping of low--$x$ QCD dynamics that FPF measurements would allow  can provide a natural bridge between the physics program at the HL-LHC
and that of an eventual higher-energy $pp$ collider that follows it.

\cref{fig:kinplot-ppcollisions} also demonstrates that the FPF will 
be sensitive to very high--$x$ kinematics.
This region is of particular interest due to the particular sensitivity of the FPF to any intrinsic charm component of the proton~\cite{Brodsky:2015fna}.
In particular, while charm production
in $pp$ collisions is dominated  by gluon--gluon scattering, in the presence
of a non--perturbative charm PDF in the proton (known as intrinsic charm), the charm-gluon
initial state enters, and may even be dominant for forward $D$-meson
production.
Several  studies have investigated the possible existence of this intrinsic charm,
including tantalizing very recent measurements of $Z$+charm production by the
LHCb experiment~\cite{LHCb:2021stx}.
FPF measurements would  provide a complementary handle on the intrinsic charm content
of the proton, which in turn could enhance the expected flux of prompt neutrinos
arising from the decays of charm mesons produced in cosmic ray collisions in the atmosphere.
These represent a
dominant background for astrophysical neutrinos at neutrino telescopes such as IceCube and
KM3NET.

\begin{figure}[h]
\centering
\includegraphics[width=0.95\textwidth]{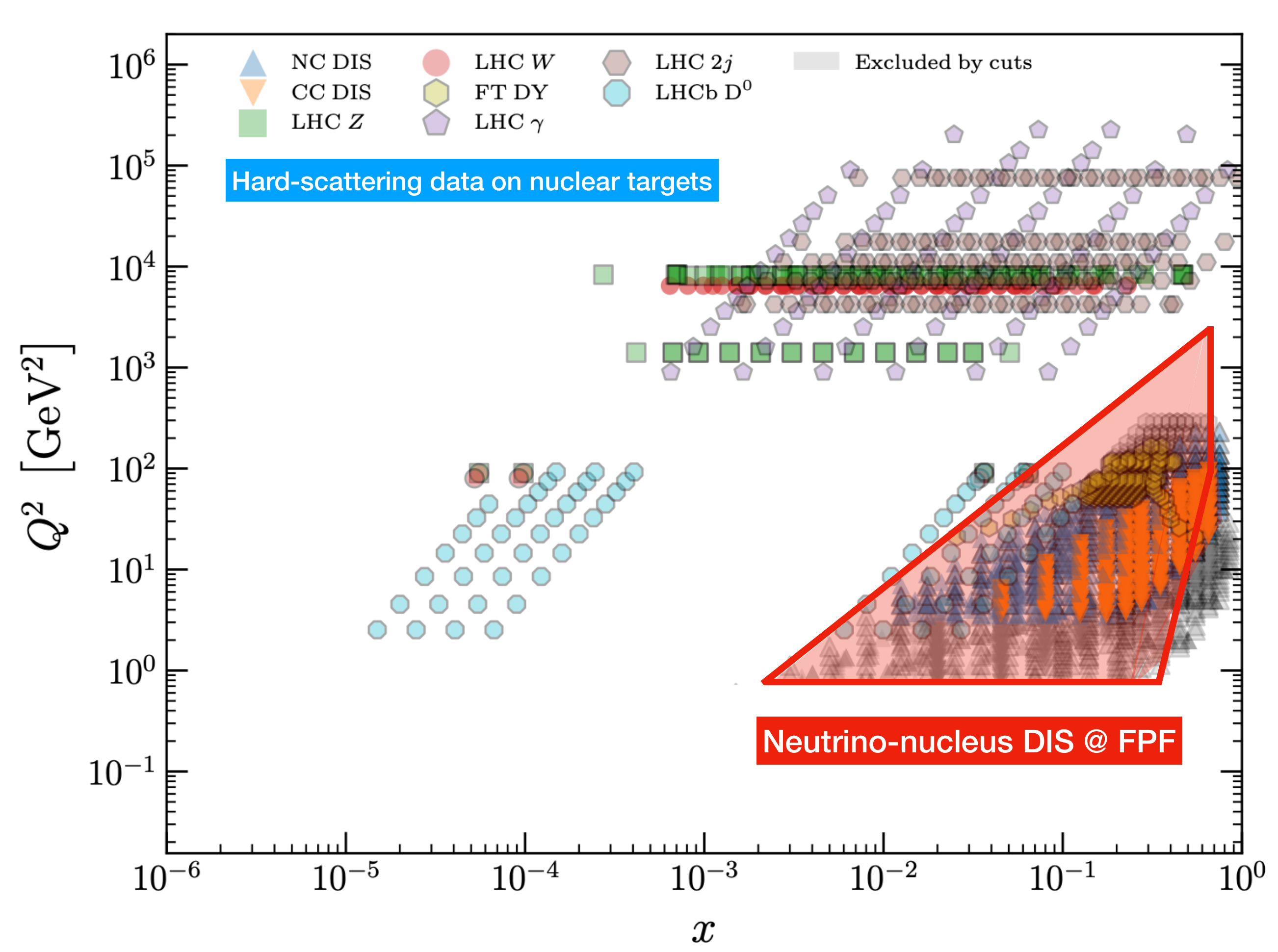}
\caption{\small The kinematic coverage in the $(x,Q^2)$ plane, assuming leading order
  kinematics, of available hard-scattering
  data on nuclear targets corresponding to the \texttool{nNNPDF3.0} global analysis~\cite{Khalek:2022zqe}
  of nuclear PDFs.
  It is compared with the expected coverage
  for neutrino-nucleus deep-inelastic scattering structure functions at the FPF.
  The coverage for these charged-current nuclear structure functions at the FPF
  would largely overlap with that for neutral-current charged-lepton structure functions expected
  at the EIC~\cite{AbdulKhalek:2021gbh,Khalek:2021ulf}.
}
\label{fig:nuclear-PDF-coverage}
\end{figure}

As indicated in the right section of \cref{fig:FPF_QCD_overview},
the FPF acts effectively as a high-energy neutrino-induced deep-inelastic scattering (DIS)
experiment, with event properties being reconstructed from the kinematics
of the outgoing charged lepton.
While in the last five decades several experiments have measured 
DIS structure functions on nuclear targets~\cite{Ethier:2020way}, the FPF beam contains neutrinos
of higher energy $E_{\nu}$  compared to these previous measurements,
 hence leading to a significant extension of the kinematic
coverage available for proton and  nuclear structure studies.

This improvement is demonstrated in \cref{fig:nuclear-PDF-coverage},
which compares the kinematic coverage in the $(x,Q^2)$ plane (assuming leading order
  kinematics) of available hard-scattering
  data on nuclear targets from the \texttool{nNNPDF3.0} global analysis
  of nuclear PDFs~\cite{Khalek:2022zqe} with the expected coverage for
   neutrino-nucleus deep-inelastic scattering structure functions at the FPF, assuming
  neutrino energies of $E_{\nu}=1$ TeV.
  We can see that FPF neutrinos extend the coverage of available NC and CC
  nuclear DIS experiments (typically affected
  by rather large uncertainties) both at low--$x$
  and at large-$Q^2$, and complement
  other datasets sensitive to quark flavour separation in nuclei such as $W,Z$ production
  in proton-lead collisions at the LHC.

Neutrino--induced CC DIS structure functions provide access to different quark flavour
combinations  compared to charged lepton DIS, and hence  FPF data
can potentially provide key information to improve global fits of proton
and nuclear PDFs in a complementary manner to other planned experiments.
For instance, the Electron Ion Collider (EIC) will measure electron-nucleus
DIS via NC scattering  in an extended kinematic range  compared
to existing measurements.
It can be shown that
the coverage for CC nuclear structure functions at the FPF
in \cref{fig:nuclear-PDF-coverage}
  broadly overlaps with that for NC charged-lepton expected
  at the EIC~\cite{AbdulKhalek:2021gbh,Khalek:2021ulf}.
Clearly, measurements of neutrino--induced DIS at the FPF would fully complement
the electron--induced DIS ones at the EIC, providing access to different quark combinations
and hence enhancing the reach of the theory interpretations
of these measurements.
Further discussion of the state of the art and prospects for PDF studies can be found in the corresponding white paper~\cite{PDFWhitepaper} (see also the white paper relating to the strong coupling~\cite{alphasWhitepaper}).

The structure of this chapter is as follows.
First of all, in \cref{sec:qcd:forward} we discuss
novel perturbative and non-perturbative QCD effects relevant for the forward particle production
that generates the neutrino flux observed at the FPF, from intrinsic charm to BFKL resummation.
Then in \cref{sec:qcd:forwardMC} the role of MC event generators in the modelling of forward particle production is considered, and a range of dedicated studies presented.
Finally, in \cref{sec:qcd:nDIS} the interaction of the neutrino beam with the FPF detector is considered, and in particular the potential for this to constrain both proton and nuclear structure via the neutrino--induced DIS process is assessed.
We also provide there
state-of-the-art predictions for neutrino-nucleus interaction cross-sections
on a tungsten nuclear target.

\section{Forward Particle Production and QCD in Novel Regimes}
\label{sec:qcd:forward}

\subsection{Introduction}\label{sec:qcd:forwardintro}

Perturbative QCD (pQCD) has proven to be a very powerful tool for describing the strong interactions at colliders.
In particular, when there are hadrons  in the initial state,
the calculation of physical observables requires the factorization of
short-distance cross sections, computable in pQCD, and of universal parton distribution functions (PDFs)
describing the long-distance internal dynamics of partons in the proton.
The collinear factorization framework~\cite{Ellis:1978sf,Ellis:1978ty, Furmanski:1980cm, Curci:1980uw} is the most widely used approach, whereby collinear logarithms in the energy scale of the process are resummed via  the DGLAP evolution equation~\cite{Gribov:1972ri,Dokshitzer:1977sg,Altarelli:1977zs}. Within this framework, partonic cross sections are usually computed at fixed order in perturbation theory.
For many processes at hadron colliders the state of the art is NNLO~\cite{Heinrich:2020ybq},
and for some processes even N$^3$LO results have been made available recently~\cite{Duhr:2020seh,Anastasiou:2015ema,Anastasiou:2016cez,Duhr:2019kwi,Chen:2019lzz,Duhr:2020kzd,Dreyer:2018qbw,Dreyer:2016oyx,Chen:2021vtu, Camarda:2021ict, Duhr:2020sdp, Duhr:2021vwj}.

However, as shown in \cref{fig:kinplot-ppcollisions}, the FPF will be sensitive to forward particle and in particular charmed meson production, via their decays to neutrinos. In this regime, the production process involves the extraction of  very low--$x$ parton from one beam and a very high--$x$ parton from the other; we can see from the figure that this will extend down to $x \gtrsim 10^{-7}$ and up to $x\lesssim 0.7$. In both of these regimes, we become sensitive to novel QCD effects that take us beyond the `standard' paradigm of collinear DGLAP--based factorization at leading twist described above.
One challenge in modelling charm production in the forward region are the large theory uncertainties
associated to missing perturbative higher orders.\footnote{See also~\cite{Brodsky:2011ig,Huang:2021hzr} for an alternative method of scale setting for fixed-order
QCD calculations.}

In the low $x$ regime in particular, we generically encounter so-called high-energy (or low-$x$) logarithms $\sim \ln 1/x$. To deal with this requires a systematic extension of our tools of pQCD, via the BFKL~\cite{Fadin:1975cb,Kuraev:1976ge,Kuraev:1977fs,Balitsky:1978ic} framework, such that these logarithmic enhancements can be systematically resummed. This can either be achieved by a suitable modification of the collinear factorization framework or by applying the alternative framework of high energy (i.e. $k_T$) factorization~\cite{Catani:1990xk,Catani:1990eg,Collins:1991ty,Catani:1993ww,Catani:1993rn,Catani:1994sq,Ball:2007ra,Caola:2010kv}. In the latter case, the production cross section is calculated in terms of matrix elements with off--shell initial--state partons, and so--called unintegrated PDFs (uPDFs), which depend on he transverse momentum of these partons. This is in contrast to former case of collinear factorization, where the matrix elements feature purely on--shell initial--state partons and the cross section is given in terms of the collinear (i.e. $k_T$ integrated) PDFs, as extracted from  global analyses. While these approaches are therefore different, they aim to describe the same low $x$ phenomena. The relationship between the collinear and $k_T$--factorized approaches, in particular in terms of their predictions for forward particle production, remains an interesting area of study.

A further novel effect we expect to encounter in this regimes is that of saturation, due to gluon recombination ($gg \rightarrow g$), which is expected to be relevant for $x \lesssim 10^{-5}$ and would tame the growth of the gluon PDF in this region. This in particular acts to restore  unitarity in the corresponding production cross sections, which might otherwise be violated. This may  also play a significant role in the low $x$ region relevant to FPF physics.

The important role of low $x$ dynamics has already been observed in various QCD analysis to inclusive HERA data~\cite{Ball:2017otu,xFitterDevelopersTeam:2018hym}. Since the kinematics of HERA is such that low-$Q^{2}$ data is also at low $x$, it has been suggested that the DGLAP resummation of $\ln Q^{2}$ terms should be augmented by $\ln(1/x)$ resummation.
It has been shown that a precise determination of the low-$x$ gluon PDF, $xg$, down to $x\sim10^{-6}$ can be obtained from LHCb charm production data in the forward region at different centre-of-mass energies $\sqrt{s} =$ 5, 7 and 13 TeV~\cite{Gauld:2016kpd,Zenaiev:2019ktw,Gauld:2015yia,PROSA:2015yid}. Similar or even stronger constraints could be expected from the corresponding forward FPF measurements, considering the aforementioned $x$ coverage.
The kinematic range of heavy-flavour production can in addition be extended by using combined and coordinated measurements from detectors at FPF and the LHC.
This is particularly interesting for events with at least two identified final-state objects, with one at central rapidity measured by ATLAS and the other one emitting a large-rapidity neutrino seen by a properly configured FPF detector.

At high $x$, intrinsic charm~\cite{Brodsky:2015fna}  contributions may strongly enhance event rate predictions for processes like $D$-meson,
$\gamma$+$D$, or $Z$+charm production~\cite{Ball:2016neh,Hou:2017khm} in comparison to calculations based on a perturbatively generated charm PDF.
The possibility that the proton wave function may contain a $\ket{uudc\bar{c}}$ component, in addition to the charm content that arises due to perturbative gluon splitting, $g\rightarrow c\bar{c}$, has been debated for decades. Light front QCD (LFQCD) calculations predict that non-perturbative intrinsic charm manifests as valence-like charm content in the PDFs of the proton~\cite{Brodsky:1980pb, PhysRevD.23.2745}; whereas, if the $c$-quark content is entirely perturbative in nature, the charm PDF resembles that of the gluon and sharply decreases at large $x$. Understanding the role that non-perturbative dynamics play inside the nucleon is a fundamental goal of nuclear physics~\cite{Brodsky:2011vgv,XQCD:2013odc,Hobbs:2013bia,Ball:2015tna, Duan:2016rkr, Sufian:2020coz}. Furthermore, the existence of intrinsic charm would have many phenomenological consequences both
in particle and in astroparticle physics.

In a recent publication by the LHCb collaboration~\cite{LHCb:2021stx}, the fraction of $Z$-boson+jet events that contain a charm jet, $\mathcal{R}_{j}^{c} = \sigma(Zc)/\sigma(Zj)$, has been measured in the different intervals of the $Z$ boson rapidity, $y_{Z}$.
The observed $\mathcal{R}_{j}^{c}$ value is consistent with the intrinsic charm hypotheses in the forward-most interval (3.5 $<|y_{Z}|<$ 4.5) for a charm momentum fraction
of $\mathcal{O}(0.5\%)$, while the measured values in the more central events 
are consistent with both the no-intrinsic charm and intrinsic charm hypotheses.
Some of these measurements can be extended to the FPF  and could further benefit from the possibility of coordination between the FPF and ATLAS detectors.
In particular, these new results would be very helpful in constraining the large-$x$ charm PDF, both in size and in shape, when incorporated into future global PDF analyses.

In the following sections, a range of studies of the novel QCD phenomena described above, and the potential impact of the FPF on our understanding of these, is presented.

\subsection{Low-$x$ Resummation at the LHC and Its Impact on the FPF Program}

As described in \cref{sec:qcd:forwardintro}, at high energies (i.e. low $x$)  logarithms $ \sim \ln 1/x$ generically appear and become large. In the $\overline{\rm MS}$ scheme, these contributions are single-logarithmically enhanced in the singlet sector, and therefore when $\alpha_s\log\frac1x\sim1$ they make the fixed-order result unreliable. When this is the case, perturbativity can be restored by resumming these contributions to all orders.
This can be achieved by using $k_T$-factorization.
In the $k_T$-factorization framework the partons are off-shell by an amount given by their transverse momentum.
Keeping the partons off-shell the $\log\frac1x$ terms do not appear in the parton-level off-shell cross section,
which can thus be safely computed at fixed order in perturbation theory.
The small-$x$ logarithms are thus ``contained'' in the off-shell uPDFs.
The latter can be linked to the ordinary (collinear) PDFs
by making use of  BFKL theory.
This procedure allows us to achieve, numerically, the resummation of small-$x$ logarithms
within the standard collinear factorization framework
(namely, the $k_T$-factorization framework is used just as a tool to resum the logarithms),
similarly to what happens for the resummation of other logarithms (e.g., the threshold logarithms).

It is very important to stress that low-$x$ logarithms also appear in the perturbative splitting functions
governing DGLAP evolution of collinear PDFs, in the singlet sector.
Their resummation is again obtainable through the link between collinear and unintegrated distributions,
and it is achieved by exploiting a duality between DGLAP and BFKL evolutions.
There are various technical issues to be addressed to obtain reliable results from this resummation,
ultimately due to a perturbative instability in the BFKL kernel itself.
It took more than a decade to finally be able to establish a reliable resummation framework for DGLAP evolution~\cite{Ball:1995vc,Ball:1997vf,Altarelli:2001ji,Altarelli:2003hk,Altarelli:2005ni,Ball:2007ra, Altarelli:2008aj}.

Recently, the machinery for low-$x$ resummation of partonic cross section and DGLAP evolution
has been exploited to describe the HERA deep inelastic scattering data~\cite{H1:2015ubc}.
In particular, this resummation has been included in a PDF fitting framework,
both by \texttool{NNPDF}~\cite{Ball:2017otu} and by \texttool{xFitter}~\cite{xFitterDevelopersTeam:2018hym, Bonvini:2019wxf}.
It turned out that HERA data below $x\sim10^{-3}$, all lying at low $Q^2$, are much better described
if low-$x$ logarithms are resummed and added to the default NNLO theory.
In particular, a well known turnover of the DIS reduced cross section measured at HERA
at $x\sim10^{-4}$ is perfectly reproduced by the resummed theory, in contrast to what happens with NNLO fits.
This result confirms that the resummation, which is required on theoretical grounds,
is fundamental for the description of experimental data at low $x$.

\begin{figure}[t]
  \centering
  \includegraphics[width=0.49\textwidth]{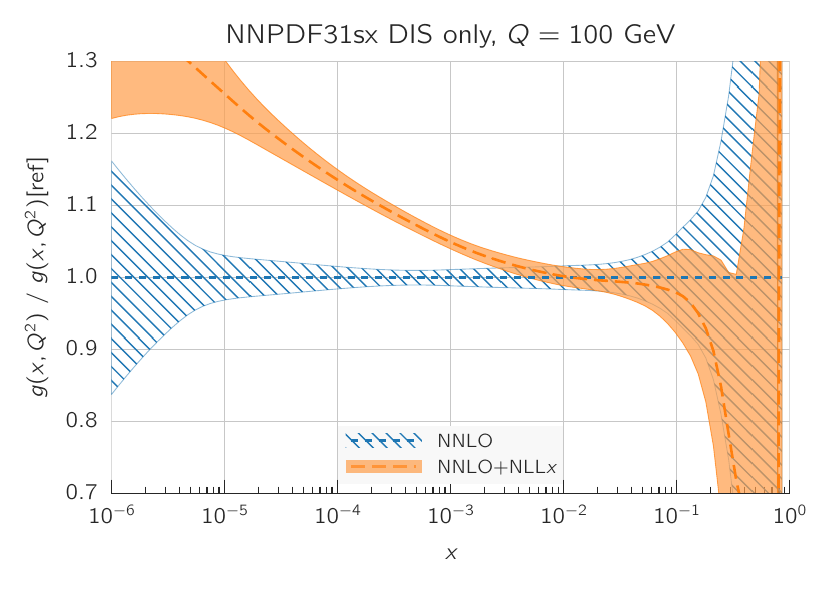}
  \includegraphics[width=0.49\textwidth]{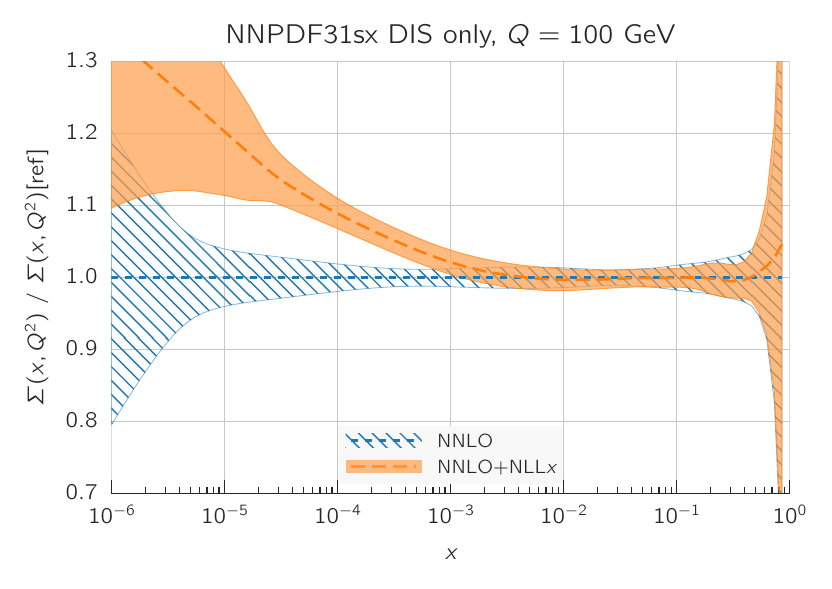}
  \caption{Comparison at $Q=100$~ GeV between the gluon PDF (left) and the total quark singlet PDF (right) determined
    with (solid orange) and without (dashed blue) low-$x$ resummation. Taken from~\cite{Ball:2017otu}.}
  \label {fig:smallxresPDFs}
\end{figure}

One of the main consequences of the resummation of low-$x$ logarithms in PDF fits is the fact
that the PDFs differ significantly from those extracted using NNLO theory.
Specifically, the quark singlet and even more the gluon PDFs are harder than their NNLO counterparts
for $x\lesssim10^{-3}$ (see \cref{fig:smallxresPDFs}).
Notably, this difference persists at all scales, and it is well outside the PDF uncertainty band. This dramatic change in the PDFs has important consequences for the physics of the LHC and, in turn, of the FPF.

\begin{figure}[t]
  \centering
  \includegraphics[width=0.49\textwidth,page=1]{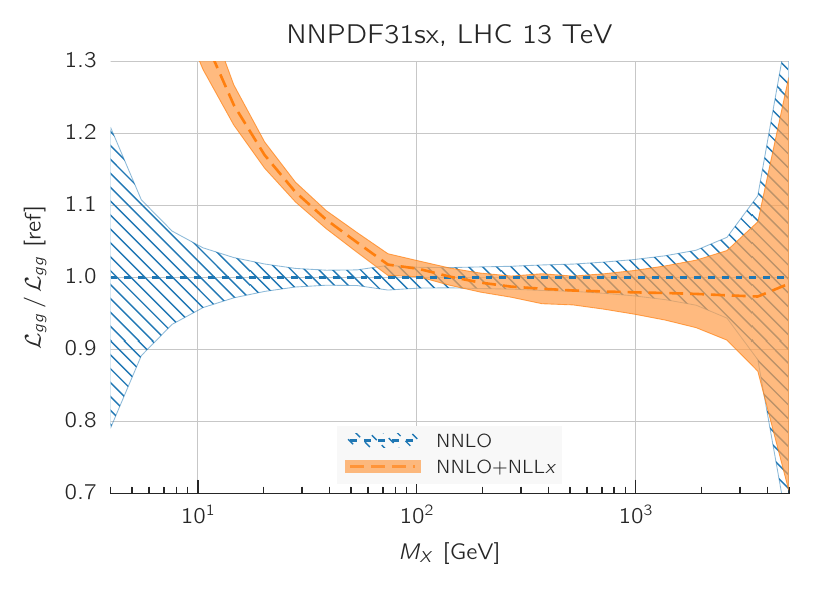}
  \includegraphics[width=0.49\textwidth,page=2]{figs_qcd/QCD_smallx_lumi_1D}\\
  \includegraphics[width=0.49\textwidth,page=1]{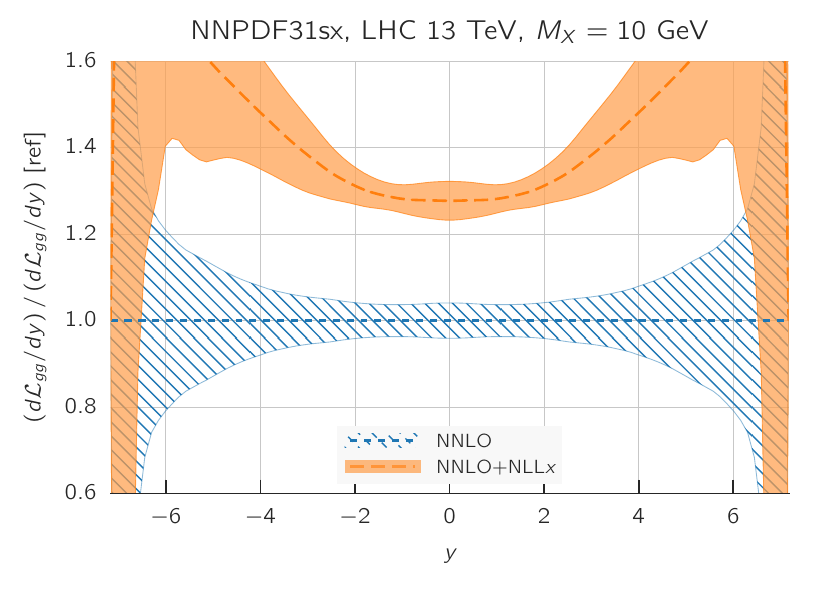}
  \includegraphics[width=0.49\textwidth,page=2]{figs_qcd/QCD_smallx_lumi_2D}
  \caption{Upper plots: Comparison between the gluon-gluon luminosity (left) and the quark-gluon luminosity (right)
    with (solid orange) and without (dashed blue) low-$x$ resummation for LHC with $\sqrt{s}=13$ TeV
    as a function of the final state invariant mass.
    Lower plots: Analogous comparison for rapidity dependent gluon-gluon luminosity,
    at low invariant mass (left) and medium invariant mass (right).
    Taken from~\cite{Ball:2017otu}.}
  \label {fig:smallxresLumi}
\end{figure}

Indeed, it is clear that any process that depends on the gluon and quark singlet PDFs at low $x$ will be affected by
the effect of resummation.
To demonstrate this, we compare in \cref{fig:smallxresLumi} some parton luminosities
computed from NNLO PDFs and PDFs obtained with the inclusion of low-$x$ resummation
(resummed PDFs henceforth).
We observe that for final state invariant masses below 100 GeV the  parton luminosities involving the gluon
differ significantly between the fixed order and resummed cases, with the latter being much larger.
The effect grows at smaller invariant masses, due to the smaller values of $x$ that contribute to the luminosities here.
For the same reason, the impact of resummation is more marked at large rapidities,
where the momentum fraction of one of the two partons is smaller.
We emphasise that the effect at low invariant masses relevant for instance for charm production,
is very important, as it can reach a few tens of percent.

\begin{figure}[t]
  \centering
  \includegraphics[width=0.5\textwidth,page=1]{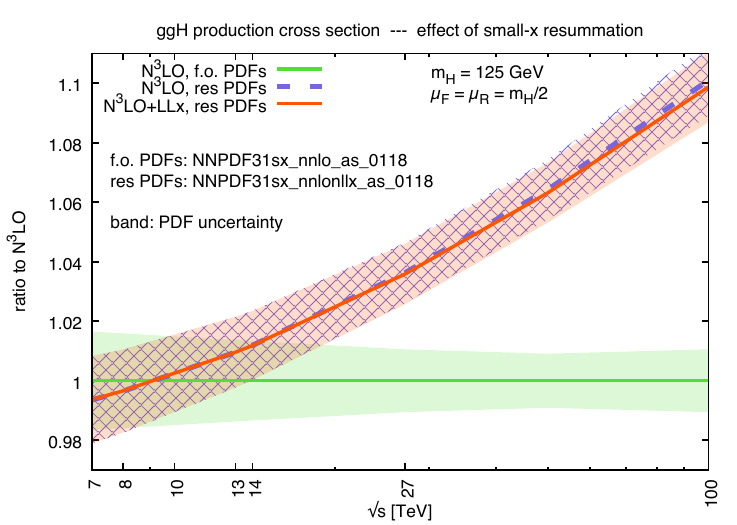}
  \caption{Comparison between the Higgs production cross section in gluon fusion
    with (red) and without (green) low-$x$ resummation as a function of the collider energy $\sqrt{s}$.
    The uncertainty band accounts for PDF uncertainty only.
    Also shown in blue the effect of resummation in PDFs but not in the partonic cross section.
    Adapted from~\cite{Bonvini:2018iwt}.}
  \label {fig:smallxresggH}
\end{figure}

Of course, a consistent prediction requires the use of resummed PDFs together with resummed partonic cross sections.
One may expect that some compensation happens between the resummation in the two objects,
thus partially mitigating the overall effect.
To understand this, we show in \cref{fig:smallxresggH} the effect of low-$x$ resummation
on the Higgs production cross section~\cite{Bonvini:2018iwt,Bonvini:2018ixe}.
While the effect is moderate (1\%) at LHC energies, it becomes substantial at higher collider energies,
reaching 10\% at the FCC energy $\sqrt{s}=100$ TeV.
In the same plot, the result obtained using resummed PDFs but without including resummation in the partonic cross section
is also shown. This result is basically identical to the consistent resummed result,
showing that, for this process, the effect of low-$x$ resummation is almost entirely driven by the resummed PDFs.

This observation may not be true in general. Indeed, in this particular case, there are two reasons
why this happens.
The first one is that the scale of the process, given by $m_H=125$~ GeV, is high enough
to suppress low-$x$ logarithms in the partonic cross section by the corresponding
small value of the strong coupling $\alpha_s(m_H)$.
Another process that probes the same values of $x$ but at smaller scale
will have enhanced logarithmic contributions due to the larger value of $\alpha_s$.
The second reason is that we are looking at the inclusive cross section,
where the sensitivity to the low-$x$ region is washed out.
Indeed, as already noticed in \cref{fig:smallxresLumi} the effect of low-$x$ logarithms
is more pronounced at large rapidities, and thus we expect to find enhanced effects
of the resummation when looking at distributions differential in rapidity.

For this reason, recent theoretical activity has been devoted to the resummation
of low-$x$ logarithms in differential distributions.
The differential version of  $k_T$-factorization has been known for some time~\cite{Caola:2010kv, Muselli:2017rjj, Muselli:2017fuf},
but no phenomenological applications were obtained because of technical complications
in the original formalism of resummation.
The recent approach of~\cite{Bonvini:2016wki, Bonvini:2017ogt, Bonvini:2018iwt},
which permitted the inclusion of resummation in PDF fits,
is particularly suitable for numerical implementation,
and its extension at the differential level has been performed~\cite{Silvetti:2019tmp}.
In particular, heavy-quark pair production has been considered as the first application of this framework~\cite{Bonvini:2022jde}.
While the theoretical setup and the analytical ingredients are all ready,
the numerical implementation is not completed yet, so a full phenomenological study
is yet to be performed.
Preliminary results show that indeed resummed logarithms are more important at large rapidities,
while they are, for instance, insensitive of the transverse momentum.

For charm production, where the hard scale is low,
the effect of low-$x$ resummation is thus expected to be important,
and one needs to include the resummation of partonic cross sections
to obtain a reliable result.
This is clearly relevant for the FPF, as the LHC production cross section of $c\bar c$
can be predicted reliably only with the inclusion of low-$x$ resummation effects.
To achieve high precision, one would also need more precise PDFs at low $x$.
Currently, the data constraining the PDFs at low $x$ used in resummed fits
are only those from HERA. 
There are for example valuable Drell-Yan data from the LHC extending to low $x$~\cite{Aad:2019wmn, Aaboud:2017ffb, Aad:2016zzw, Aad:2016izn, Aad:2014qja, Aad:2013iua, Aad:2015auj,Aad:2016naf, Sirunyan:2018owv,CMS:2014jea,Chatrchyan:2013tia,Chatrchyan:2011cm,Chatrchyan:2011wt, Aaij:2012vn,Aaij:2015gna,Aaij:2015zlq,Aaij:2016mgv},
lying at values of $Q^2$ that are higher than those probed by HERA. Including these in the
PDF fits requires  resummation, at differential level, of the
Drell-Yan process, and would also provide a strong validation of the
currently available resummed PDFs, which already  rely on the abundance of DY scattering data from the LHC~\cite{Ball:2017nwa,Harland-Lang:2014zoa,Hou:2019efy}.
A complete implementation of the resummation for the Drell-Yan process
for phenomenological applications is under development.

\subsection{Charm Production in the Forward Region within $k_T$ Factorisation}
\label{QCD:ssec:ktfac}

In this section, predictions for the neutrino flux from charm decays are obtained, as  evaluated in different QCD approaches, namely the NLO collinear and $k_T$-factorization.  We use QCD parameters, such as the choice of scales, PDFs, and fragmentation function,  determined from fitting the LHCb data for $D$ meson production. We will  show that the FPF will be able to provide valuable information about  QCD, that is complementary to IceCube measurements of prompt neutrinos. The latter is discussed in \cref{Neutrino:ssec:ktfac}. 

\begin{figure}[t]
  \centering
  \includegraphics[width=0.99\textwidth]{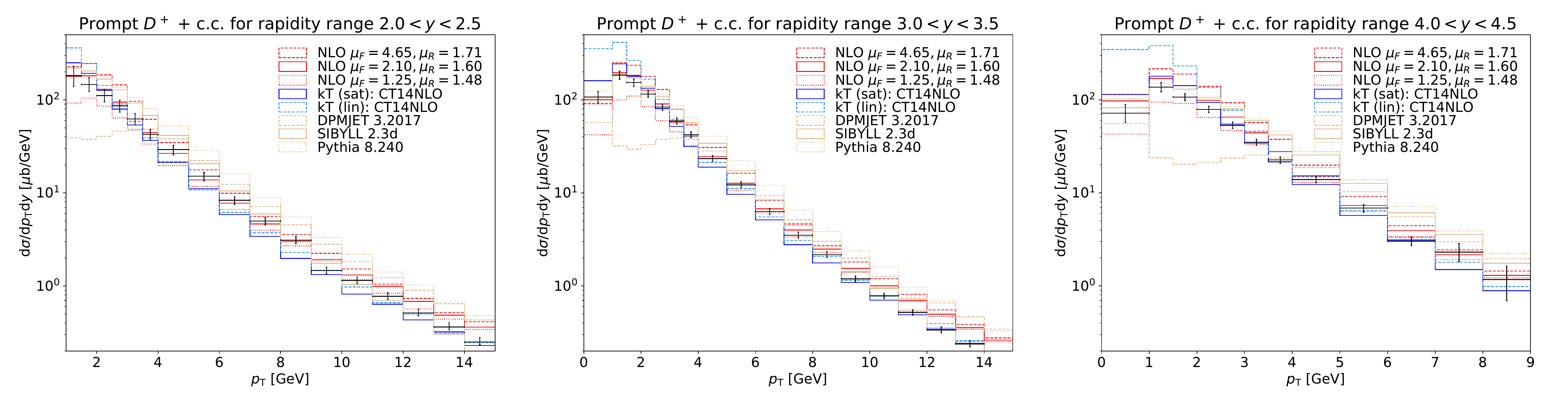}
  \caption{Predictions for the transverse momentum distribution in different rapidity intervals  at 13~TeV LHC compared to LHCb measurements. We show the predictions of different MC event generators as light brown curves. The predictions using the NLO and the $k_T$ factorization approaches are shown in red and blue, respectively.}
  \label{fig:charm_pT}
\end{figure}

\begin{figure}[t]
  \centering
  \includegraphics[width=0.99\textwidth]{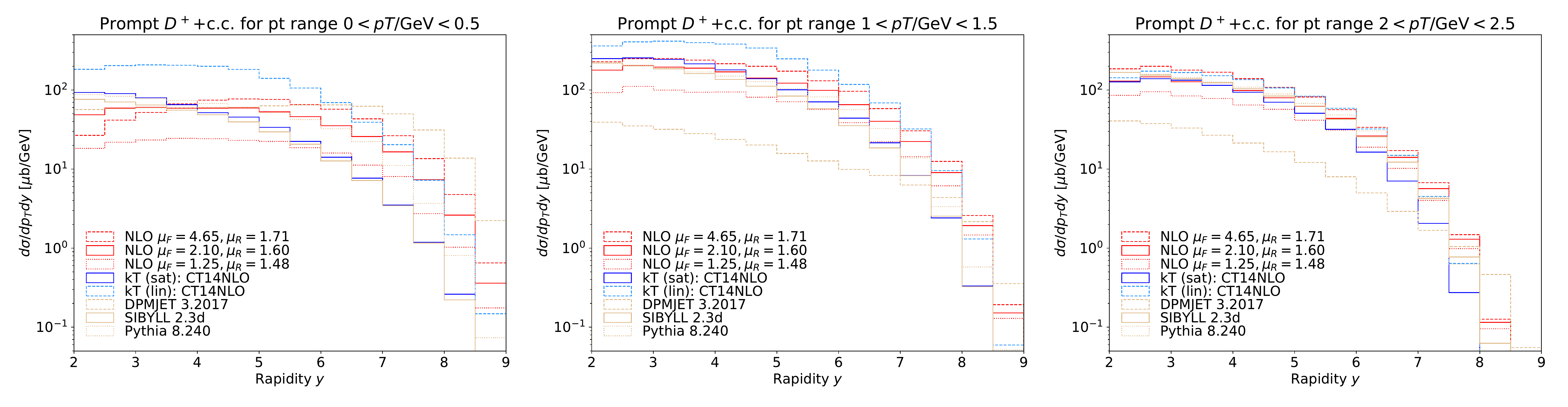}
  \caption{Predictions for the rapidity distribution in different transverse momentum intervals at 13~TeV LHC. We show the predictions of different MC event generators as light brown curves. The predictions using the NLO and the $k_T$ factorization approaches are shown in red and blue, respectively.}
  \label{fig:charm_eta}
\end{figure}

The state-of-the-art computation of charm pair production within the realm of perturbative QCD involves a NLO computation first discussed in~\cite{Nason:1987xz}. In~\cite{Nelson:2012bc} a range of factorization ($\mu_F$) and renormalization ($\mu_R$) scales were looked at, consistent with current charm production data from colliders~\cite{ALICE:2012inj,ATLAS:2011fea,LHCb:2013xam} and fixed-target experiments \cite{PHENIX:2006tli, STAR:2012nbd}. The values of $(\mu_F,\mu_R) = (2.1_{-0.85}^{+2.55}, 1.6_{-0.12}^{+0.11}) \,m_c $ were arrived at as the best-fit parameters, with $m_c = 1.27$~GeV motivated by current lattice QCD results. For simplicity, a constant fragmentation fraction, rather than a full one, is used to compute the production of charmed hadrons ($D^{\pm},\ D^0, D_s^\pm, \Lambda_c$) from charm quarks. We use these scales for our perturbative calculations, substituting $m_T = \left({m_c^2 + p_T^2}\right)^{1/2}$ for the scale multiplicity factor instead of $m_c$ for improved fits to high $p_T$ data at $\sqrt{s} = 13$ TeV from LHCb, while also using a more recent \texttool{CT14NLO}~PDF set~\cite{Dulat:2015mca}. Our benchmark results for prompt $D^{\pm}$ production, as well as the corresponding uncertainty bands, computed with these parameters using the NLO code from Refs.~\cite{Cacciari:1998it, Cacciari:2001td} are shown in \cref{fig:charm_pT} and \cref{fig:charm_eta}.

For the calculations, we use the framework from \cite{Martin:2003us}, which uses the on-shell approximation for the large $x$ gluon and keeps the low $x$ gluon off-shell. The   gluon uPDF was taken from \cite{Kutak:2012rf} which was based on the unified BFKL+DGLAP evolution supplemented with low $x$ resummation. Two sets of gluon distributions were used, based on linear evolution as well as non-linear evolution. The latter one includes the non-linear term in density, which is responsible for  saturation effects. Both sets of distributions were fitted to the data on structure functions at HERA. 
The non-linear term is important for low $x$ and low values of transverse momenta and leads to taming of the gluon distribution and therefore the resulting cross sections. 

In \cref{fig:charm_pT}, we present our predictions using perturbative QCD and $k_T$ factorisation with both linear and saturated gluons for the double differential distribution $d^2\sigma/dp_T dy$ for $D^{\pm}$ as a function of the transverse momentum $p_T$, over different rapidity ranges in the LHCb acceptance at $\surd s = 13$ TeV. We compare all these predictions against current LHCb data from~\cite{LHCb:2015swx}, as indicated by the black error bars. For the perturbative calculations, we find good agreement with data at low to moderately high $p_T \leqslant 8$~GeV, while they start to become too large at higher $p_T$. By contrast, results from the $k_T$ factorisation models, which are less dependent on undetermined parameters and therefore are more predictive, are on the higher side with respect to the data at low $p_T$, while becoming more consistent with the high-$p_T$ data. Independently of these models, we also show results from different MC generators (namely \texttool{DPMJet~3.2017}~\cite{Roesler:2000he, Fedynitch:2015kcn}, \texttool{Pythia~8.240}~\cite{Sjostrand:2014zea}, and \texttool{Sibyll~2.3d}~\cite{Riehn:2019jet}) which compute the same distribution using LO cross sections. Results from \texttool{DPMJet} are well below the data, especially at low $p_T$, while those from \texttool{Pythia} are somewhat high for all $\lbrace p_T, y\rbrace$. \texttool{Sibyll~2.3d} is specifically tuned to LHCb charm data, and its predictions are more consistent with data at low $p_T \leqslant 3$~GeV and $y \leqslant 4.5$, while still being higher  at larger $p_T$.
In general, we conclude that, for the kinematic region of most importance to detectors at FPF, i.e.\ low-$p_T$ and high $y$, current LHCb data fall slightly below predictions from the two $k_T$ factorization models, while lying within the uncertainty range of the perturbative QCD calculations discussed here.

Most neutrinos passing through the FPF originate from the decay of charmed hadrons with large rapidity and small transverse momentum. The corresponding kinematic regions are presented in \cref{fig:charm_eta}, where we show the $D$-meson rapidity distributions for three different transverse momentum intervals. Again, we show results using the NLO calculation (in red), $k_T$ factorization (in blue) and the MC event generators (in brown).  We note that our results using $k_T$ factorization with saturation predict a $D$-meson  production cross section that is almost an order of magnitude lower than that obtained using the collinear NLO approach in the kinematic region of most relevance to the FPF, i.e.\ at small $p_T$ and large rapidity.  For larger $p_T$, even at large rapidity, this difference becomes smaller.

\subsection{Forward Charm Production in $k_T$ Factorization and the Role of Intrinsic Charm}

In the case of forward charm production, a number of mechanisms may play a role.
One is related to intrinsic charm, whereby the charm
quarks/antiquarks are knocked out from the nucleon~\cite{Maciula:2020dxv}.
This mechanism was discussed both for high \cite{Maciula:2020dxv}
and recently for low \cite{Maciula:2021orz} energies.
The low energy fixed-target LHCb charm data \cite{LHCb:2018jry} suggest 
that the conventional gluon-gluon mechanism, known to be responsible 
for explanation of the LHC data (ALICE, LHCb), may be not sufficient.
The intrinsic charm contribution is in particular found to improve the description of the data. Another interesting, and slightly forgotten, phenomenon is recombination
\cite{Braaten:2001bf,Braaten:2002yt,Braaten:2001uu}. This mechanism has been discussed recently in \cite{MS2022}
for forward charm quark/meson production. The FPF opens up a new unique possibility to explore far-forward charm 
production via the observation of different kinds of neutrinos from the semileptonic
decay of charm mesons.

\begin{figure}[ht]
  \centering
  \includegraphics[width=0.60\textwidth]{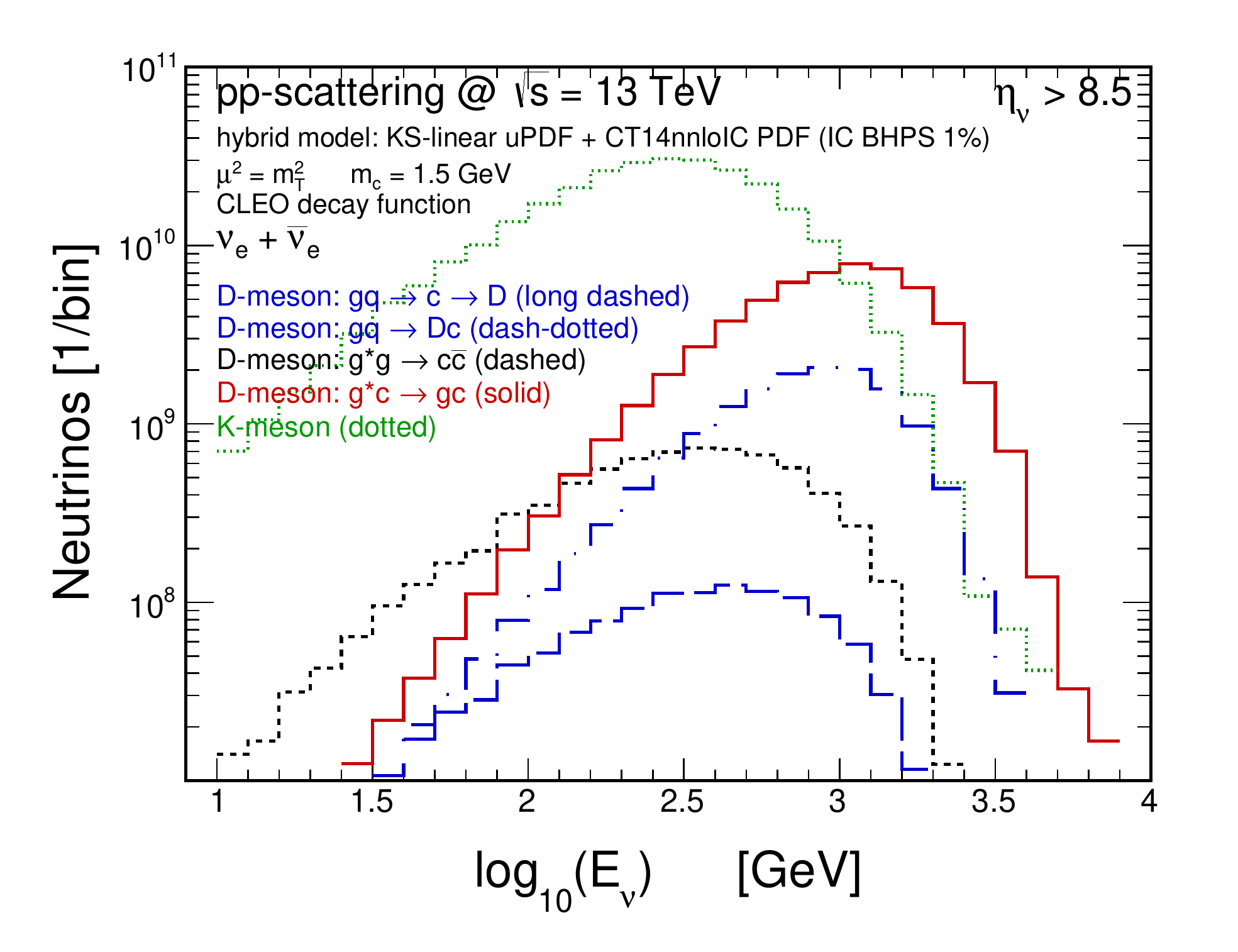}
  \caption{The energy distribution of $\nu_{e} + \bar{\nu}_{e}$ flux at 
the FPF for $\sqrt{s}$ = 13 TeV for $\eta_{\nu} > 8.5$. Here the conventional  $gg\to c\bar c$ (dashed), the intrinsic charm (solid) and the recombination (dash-dotted and long-dashed) mechanisms corresponding to charm meson component of the flux are shown separately. The $K$-meson component (dotted) is taken from ~\cite{Kling:2021gos}. An integrated luminosity $L = 150$ fb$^{-1}$ is taken.}
  \label{fig:FASER_nu_e}
\end{figure}

In \cref{fig:FASER_nu_e} we show the
$\nu_e + {\bar \nu}_e$ energy distributions from different sources, for $\eta_{\nu} >$ 8.5.
The dashed line represents the conventional $g g \to c {\bar c}$
mechanism calculated  in the $k_T$-factorization approach,
with the Kutak-Sapeta (KS) gluon uPDF; the KMR gluon uPDF gives a similar result.
We can see that this is not expected to be the dominant contribution  at the FPF.
At low neutrino energies the kaon contribution (taken here from
\cite{Kling:2021gos}) is almost two orders of magnitude larger.
At higher neutrino energies the intrinsic charm (solid line) and
recombination (dash-dotted) components start to become important.
The recombination contribution, calculated in the LO collinear
approximation, has two distinct contributions:
neutrinos from directly produced charm mesons and neutrinos from mesons
created in the hadronization of the associated charm quark/antiquark (long
dashed). The latter component is more significant in the experimental $\eta_{\nu}$ region.
At the highest neutrino energies ($\log_{10}(E_{\nu}) >$ 3) 
the direct recombination component (with $\rho = 0.1$) 
is comparable to the kaon component.

The intrinsic charm component, calculated in the hybrid factorization 
(BHPS model \cite{Brodsky:1980pb} with $P_{IC}$ = 0.01), is largest.
At present we do not know the crucial parameters ($\rho$ or $P_{IC}$) 
very well, with fixed-target data giving only an upper limit on
$P_{IC}$ of 1.65 \%, and a recent study of IC at
IceCube gives a very similar upper limit \cite{Goncalves:2021yvw}.
The FPF can therefore give a much more stringent upper limit
on this. Extraction of the precise value of
$P_{IC}$ may, however, be difficult as the shape of the recombination 
contribution is very similar to that for the intrinsic charm.
Similarly one could  extract an upper limit on the recombination contribution.
 
The situation for $\nu_{\mu} + \bar{\nu}_{\mu}$,
not discussed here, is somewhat more complicated due to large pion
contribution \cite{Kling:2021gos}.
The $\nu_{\tau} + {\bar \nu}_{\tau}$ case may also be of interest, as here
there is no light-meson background. However, the expected statistics will be
smaller, as $\tau$ neutrinos originate only from $D_s$ mesons, for
which the hadronization probability is rather small ($P_{c \to D_s} \approx$ 0.1)
and their identification is more difficult.

\subsection{Charm Production at Very Forward Rapidities in the Color Dipole Formalism}

One important open question is how to include, in a consistent way, the effect of both saturation and intrinsic charm in  heavy meson production at very forward rapidities. One possibility is the colour dipole formalism \cite{Kopeliovich:2002yv}, which allows us to take into account  nonlinear QCD effects,  higher order corrections and the contribution from intrinsic charm. 
The basic idea is that at forward rapidities, the projectile (dilute system) evolves according to 
linear DGLAP dynamics and the target (dense system) is treated using the Color Glass Condensate (CGC) formalism \cite{Gelis:2010nm}.
The $D$ meson production cross section is evaluated by including the contribution of both  the gluon and charm--initiated channels.
In the gluon--initiated case, it is assumed that before 
interacting with the hadron target, a gluon is emitted by the projectile and fluctuates into  a colour octet  $q\bar{q}$ pair. The rapidity distribution can be estimated by taking into account that the heavy quarks in the dipole as well the incident gluon (before fluctuating into 
the pair) can interact with the target. On the other hand, for the charm--initiated process, a charm quark in the projectile is assumed to interact with the dense system present in the target prior to hadronization. In \cite{Giannini:2018utr} this approach is discussed in detail and a comparison with the LHC data is presented.

In the colour dipole formalism, the $D$ meson production cross section is given in terms of the dipole--target scattering amplitude, which contains all the information about the initial state of the hadronic wave function and therefore about the non-linearities and quantum effects, which are characteristic of a system such as the 
CGC. In this analysis, we will consider the model proposed in~\cite{Boer:2007ug}, which describes several observables at HERA, RHIC and the LHC. Moreover, the cross section is also  dependent on the PDFs in the projectile. We will consider two distinct models for the description of the intrinsic component. In the BHPS model \cite{Brodsky:1980pb}, it is assumed 
that the nucleon light cone wave function has  higher Fock states,  one of them being
$|q q q c \overline{c}\rangle $. The probability of finding the nucleon in this
configuration is proportional to the inverse of the squared invariant mass of the
system. Because of the heavy charm quark mass, the probability distribution as a function 
of the quark fractional momentum, $P(x)$, is very hard in comparison to the one obtained
through  DGLAP evolution. A more dynamical approach is given by the meson cloud model (MC). In this model, the nucleon fluctuates into an intermediate state 
composed of a charmed baryon plus a charmed meson \cite{Navarra:1995rq}. The charm quark is always confined in one hadron and carries the largest part of its momentum. 
In the hadronic description we can use an effective Lagrangian to compute the charm splitting functions, which turn out to favour harder charm quarks than those due to DGLAP emission.
The main difference between the BHPS and MC  models is that the latter predicts
that  the charm and  anticharm distributions are different \cite{Cazaroto:2013wy}, since they carry 
information about the hadronic bound states in which  the quarks are found. 
In addition to these models, the \texttool{CTEQ} group has tested 
another model of intrinsic charm, denoted sea-like (SL). It consists essentially in 
assuming that at very low resolution (i.e. prior to DGLAP evolution) there is 
already some charm in the nucleon, which has a
typical sea quark momentum distribution ($\simeq 1/ \sqrt{x}$) with normalization 
to be fixed by fitting data. The main difference between these models is that the BHPS and MC models predict a large enhancement of the distribution at large  $x$  ($> 0.1$), while the SL one predicts a smaller enhancement at large  $x$, but a larger one at lower  $x$ ($< 0.2$). We follow Ref.~\cite{Pumplin:2007wg} and use the labels BHPS2, MC2 and SL2 for the versions of these models  which have the maximum amount of intrinsic charm. It is important to emphasize that  the corresponding gluon distributions are also modified by the inclusion of intrinsic charm due, e.g., to the momentum sum rule. In particular, the BHPS and MC models imply a suppression in the gluon distribution at large $x$.

\begin{figure}[t]
 \begin{center}
  {\includegraphics[width=0.475\textwidth]
{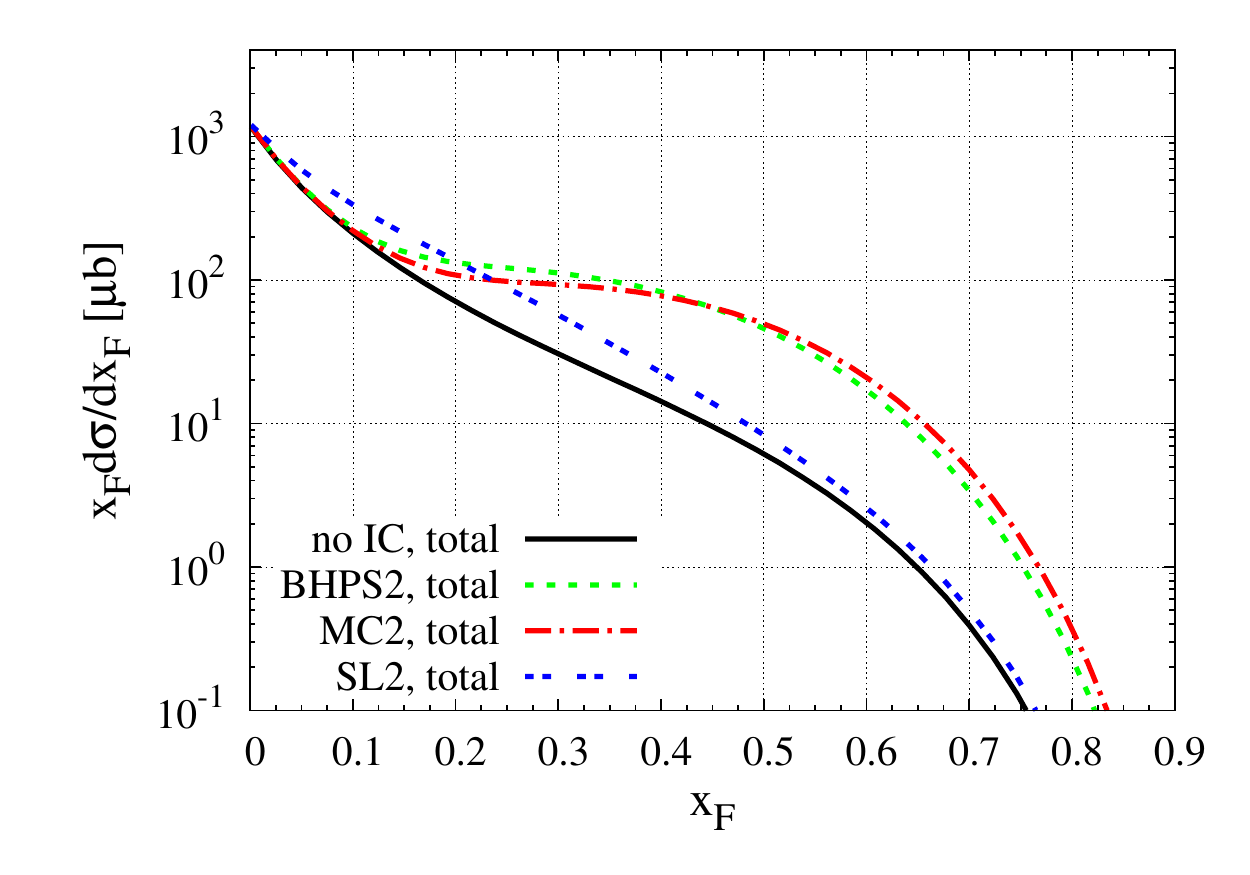}}
{\includegraphics[width=0.475\textwidth]
{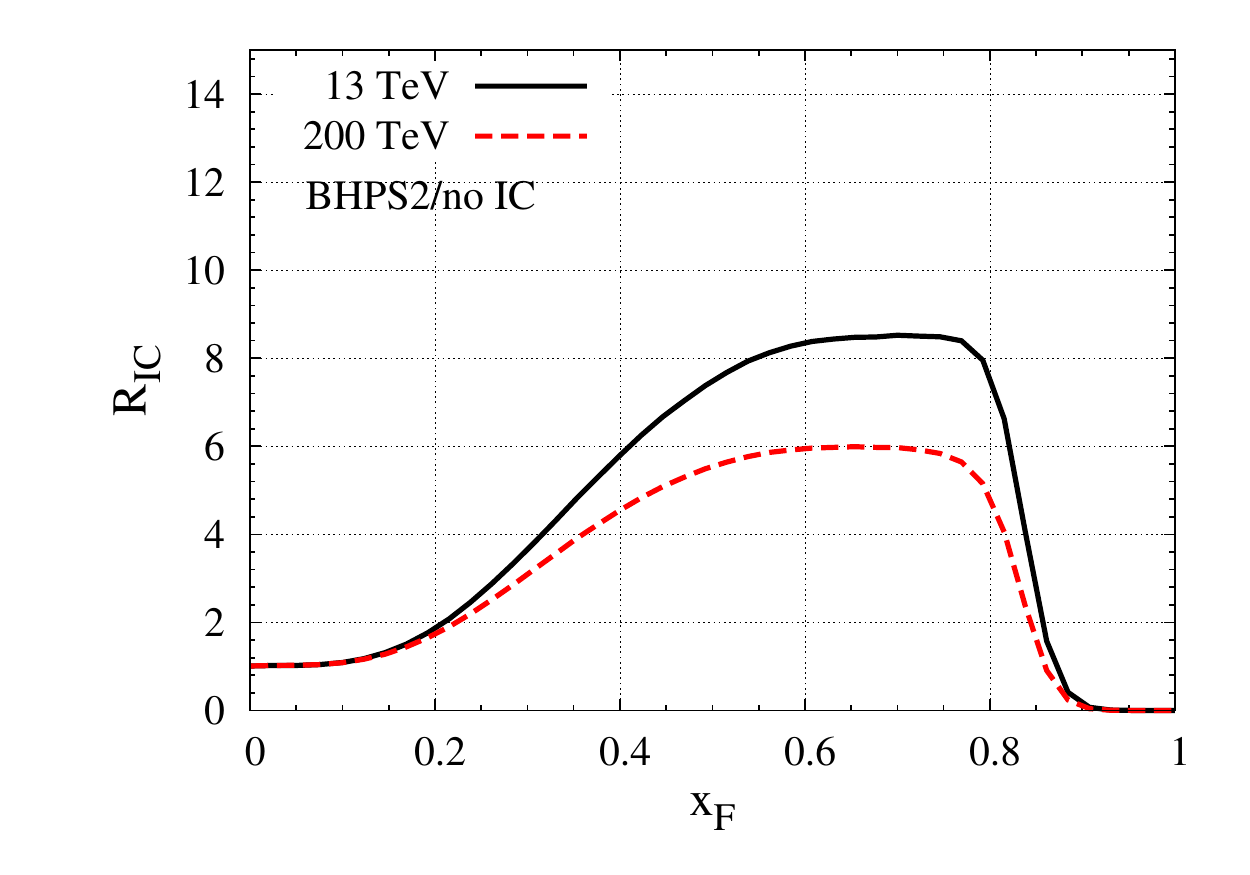}}
  \end{center}
 \vskip -0.80cm
 \caption{Left: Feynman-$x_F$ distributions  of the produced  $D^{0}+\bar{D^{0}}$ mesons in $pp$ collisions at 
 $\sqrt{s} = 13$ TeV considering different models for the intrinsic component and the contribution of both gluon -- and charm -- initiated processes. Right: Feynman-$x_F$ dependence of the ratio between the predictions calculated with the BHPS model and standard \texttool{CTEQ~6.5} parametrization. }
 \label{Fig:dsdxf_energy}
\end{figure}

The color dipole predictions for the Feynman  $x_F$ distribution of $D^{0}+\bar{D^{0}}$ mesons, produced in $pp$ collisions at $\sqrt{s} = 13$ TeV, are presented in 
\cref{Fig:dsdxf_energy} (left).  
For comparison  the standard \texttool{CTEQ~6.5} prediction, which contains no intrinsic charm, is also presented. One has that the standard \texttool{CTEQ~6.5} and the  BHPS2 and MC2 models  predict a similar 
behaviour at low $x_F$, while the SL2 model predicts a larger magnitude  associated to the enhancement of the charm distribution for $x \le 0.2$. On the other hand, the BHPS and MC models predict that the behaviour of the distribution at large values of  $x_F$ is strongly modified. 
 In order to determine the magnitude of the impact 
of intrinsic charm and the kinematical range influenced by its presence, we present in \cref{Fig:dsdxf_energy} (right panel) our predictions for the ratio between the $x_F$ 
distributions predicted by the BHPS  model and the standard \texttool{CTEQ~6.5} one, for two different values of $\sqrt{s}$. Similar results are derived using the MC model.  As expected from \cref{Fig:dsdxf_energy} (left), the BHPS model predicts an 
enhancement at intermediate $x_F$ that is a factor 
6 -- 9 in the energy ranges considered. The main  aspect that should be emphasized here is that { the enhancement occurs exactly in the $x_F$ range of the FPF experiments.} Such results indicates that a future measurement can be useful to probe the presence (or not) of the intrinsic component as well as to constrain the formalism used to describe the heavy meson production at very forward rapidities.

\subsection{Charm Production in the Forward Region and Intrinsic Charm in the CT Framework}

The presence of an intrinsic charm component in the proton would lead to an increase in the production of forward high-energy charmed hadrons. Therefore, measurements of charm hadroproduction and $Z$ + $c$ production at the LHC can constrain intrinsic charm contributions in $pp$ collisions~\cite{Hou:2017khm}.  
At large $x$, intrinsic charm contributions are generated by higher-twist processes which can in principle enhance the event rate based on the leading-power calculation. Recent global QCD analyses~\cite{NNPDF:2017mvq,Hou:2017khm} introduced 
 ``Fitted charm'', that is a phenomenological parametrization of the intrinsic charm introduced as an independent PDF functional form~\cite{Dulat:2013hea,Jimenez-Delgado:2014zga,Hou:2017khm,Ball:2016neh}. 

\begin{figure}
\centering
\includegraphics[width=0.45\textwidth]{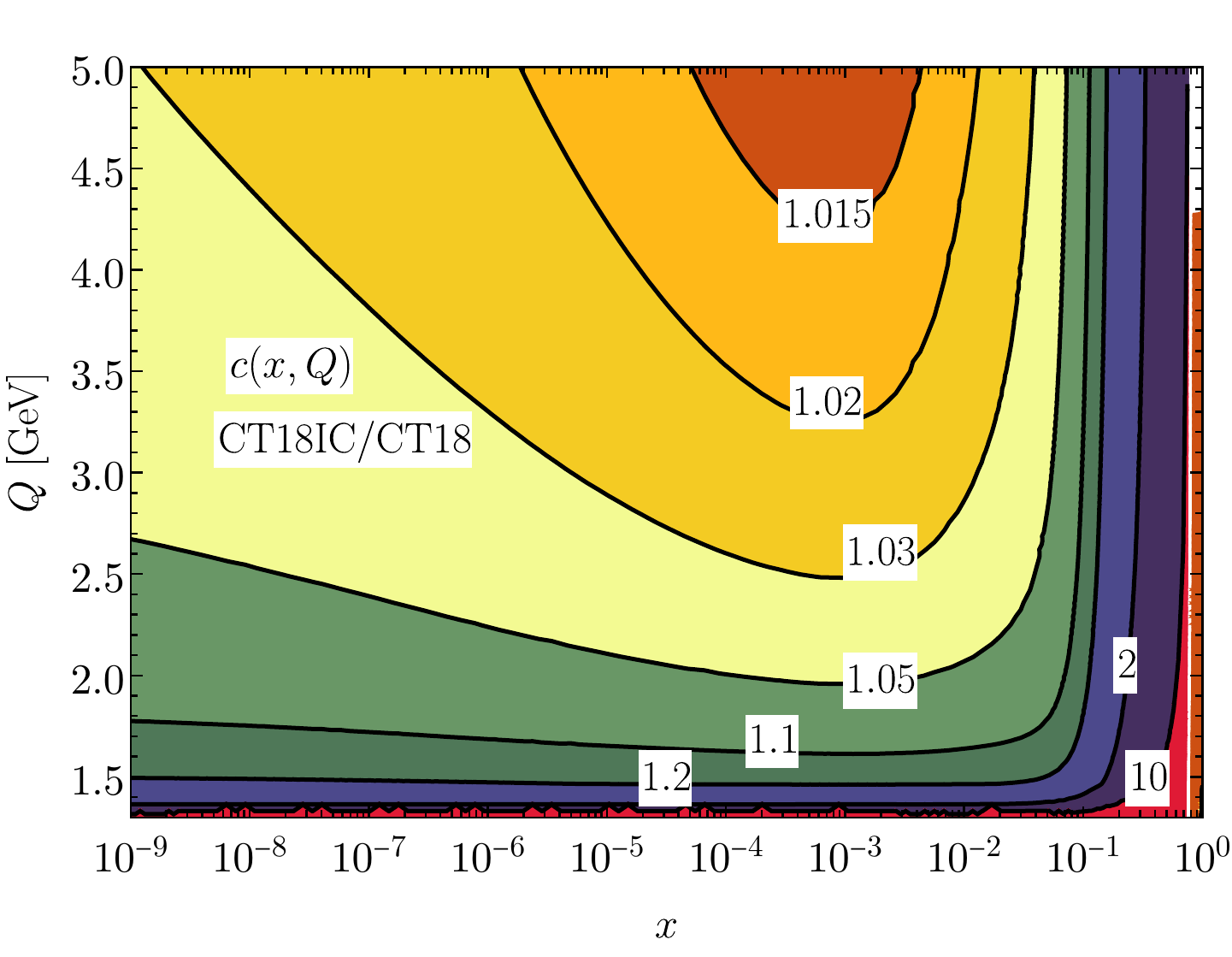}
\includegraphics[width=0.45\textwidth]{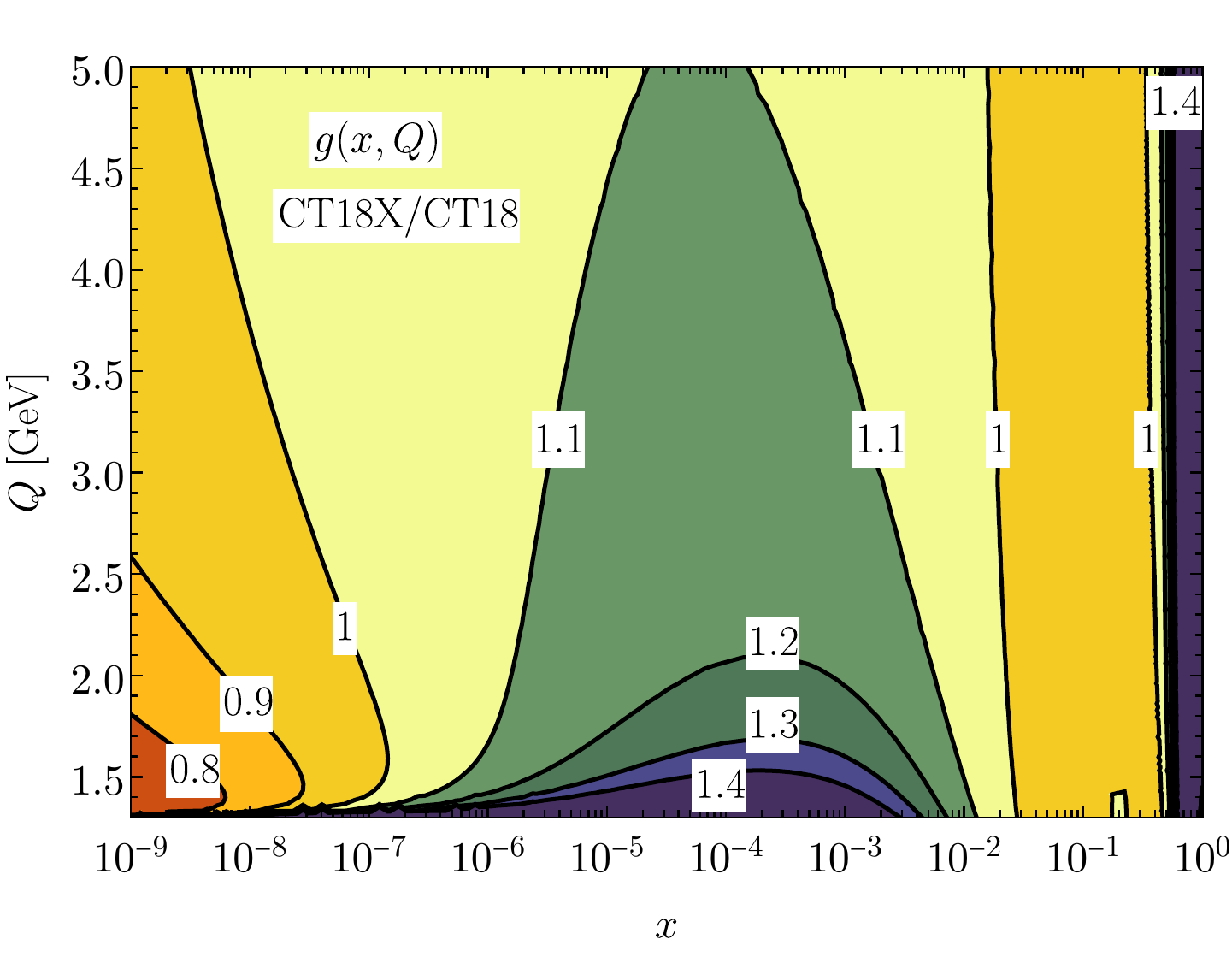}
\includegraphics[width=0.45\textwidth]{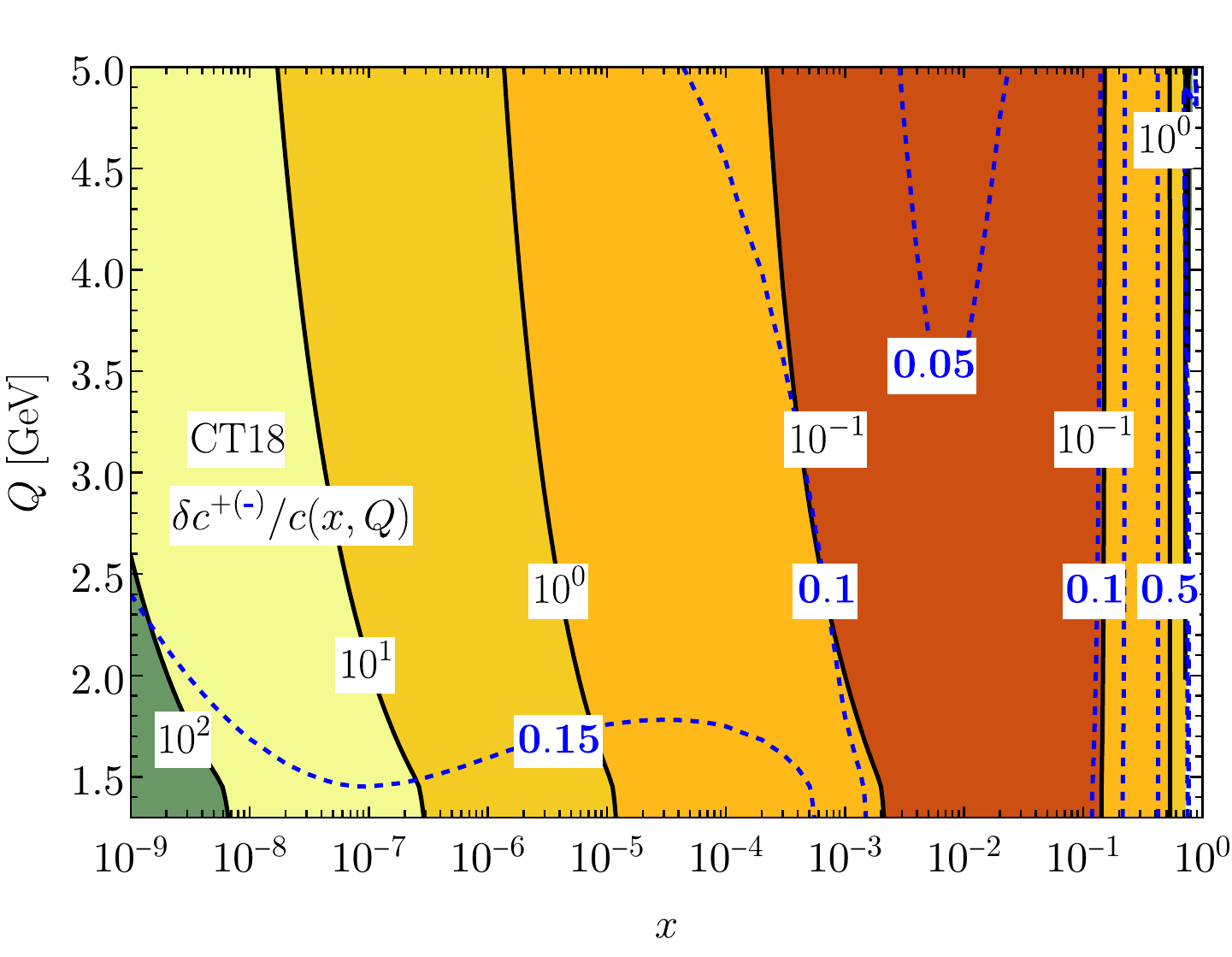}
\includegraphics[width=0.45\textwidth]{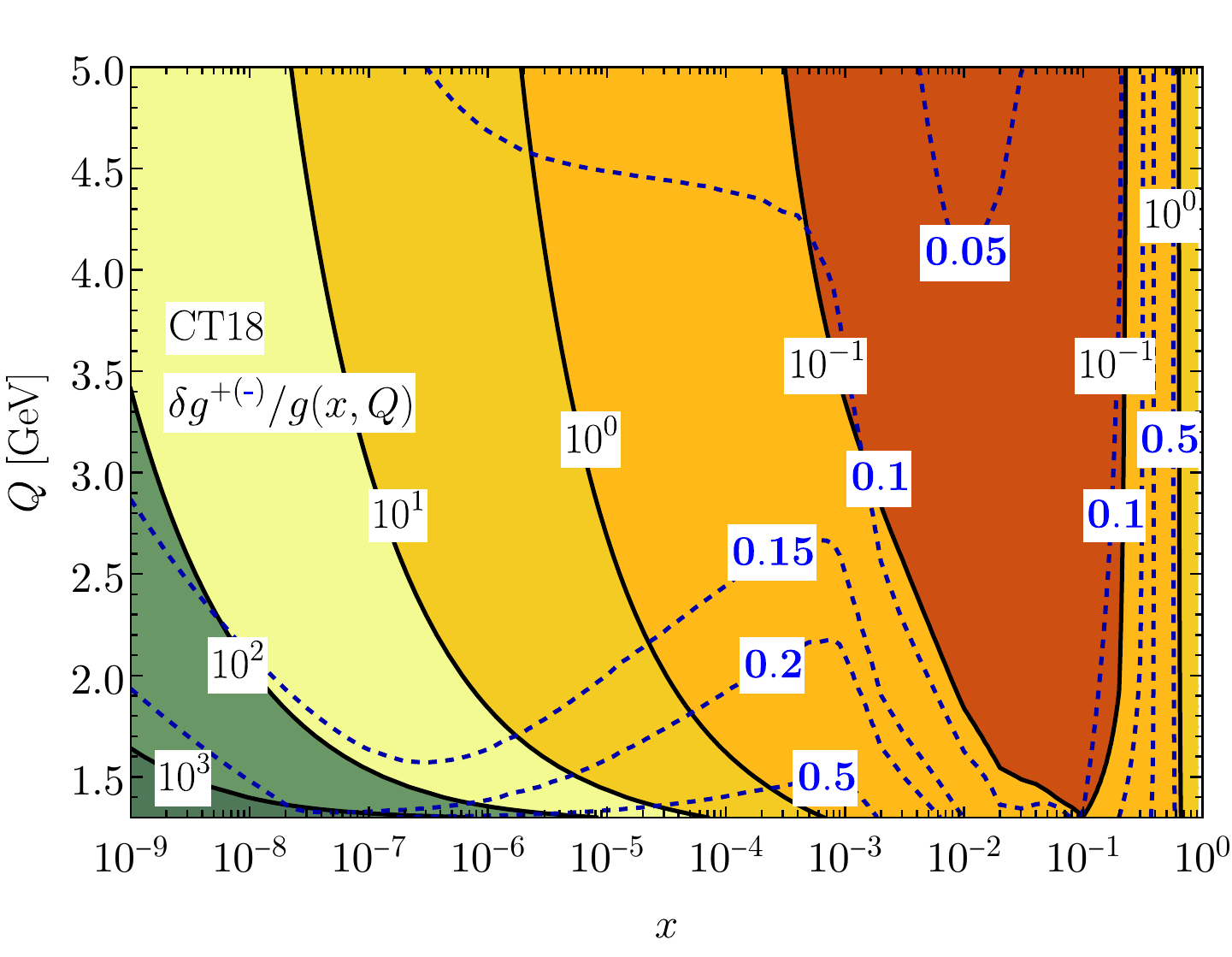}
\caption{Comparison of central (upper) and uncertainties (lower) of charm and gluon PDFs in the \texttool{CT18}, \texttool{CT18X}, and \texttool{CT18IC} analyses at NLO in QCD.
The PDF uncertainties are calculated with the asymmetric Hessian approach at 90\% C.L.~\cite{Hou:2016sho}, with positive (negative) directions denoted with black solid (blue dashed) curves.
}
\label{fig:pdfs}
\end{figure}

Low-$x$ QCD effects can also be probed by the FPF. In fact, at $x < 10^{-4}$, $\ln(1/x)$ logarithmic contributions are rapidly enhanced at factorization scales of 1 GeV and can in principle contribute to charm hadroproduction.  Therefore, the FPF has the potential to access kinematic regimes where both large-$x$ and low-$x$ QCD effects contribute to charm hadroproduction rate. 
In \cref{fig:pdfs}, the \texttool{CT18} charm and gluon PDFs at NLO are compared to a global fit with intrinsic charm (BHPS3 model) named \texttool{CT18IC}, and to the \texttool{CT18X} fit, that is described in~\cite{Hou:2019efy} and uses an $x$-dependent factorization scale to mimic effects of low-$x$ resummation. The default \texttool{CT18} fit is chosen as reference fit. 
The four insets in \cref{fig:pdfs} reflect the FPF kinematics at low $Q\approx (p_T^2+m_c^2)^{1/2}$. At both large and low $x$, 
the PDFs exhibit a large relative uncertainty due to the lack of constraints from experimental data and a significant intrinsic charm contribution is still  allowed by the current data. 
Therefore, FPF measurements will be critical to constrain this extreme kinematic regions.

\begin{figure}
\includegraphics[width=0.6\textwidth]{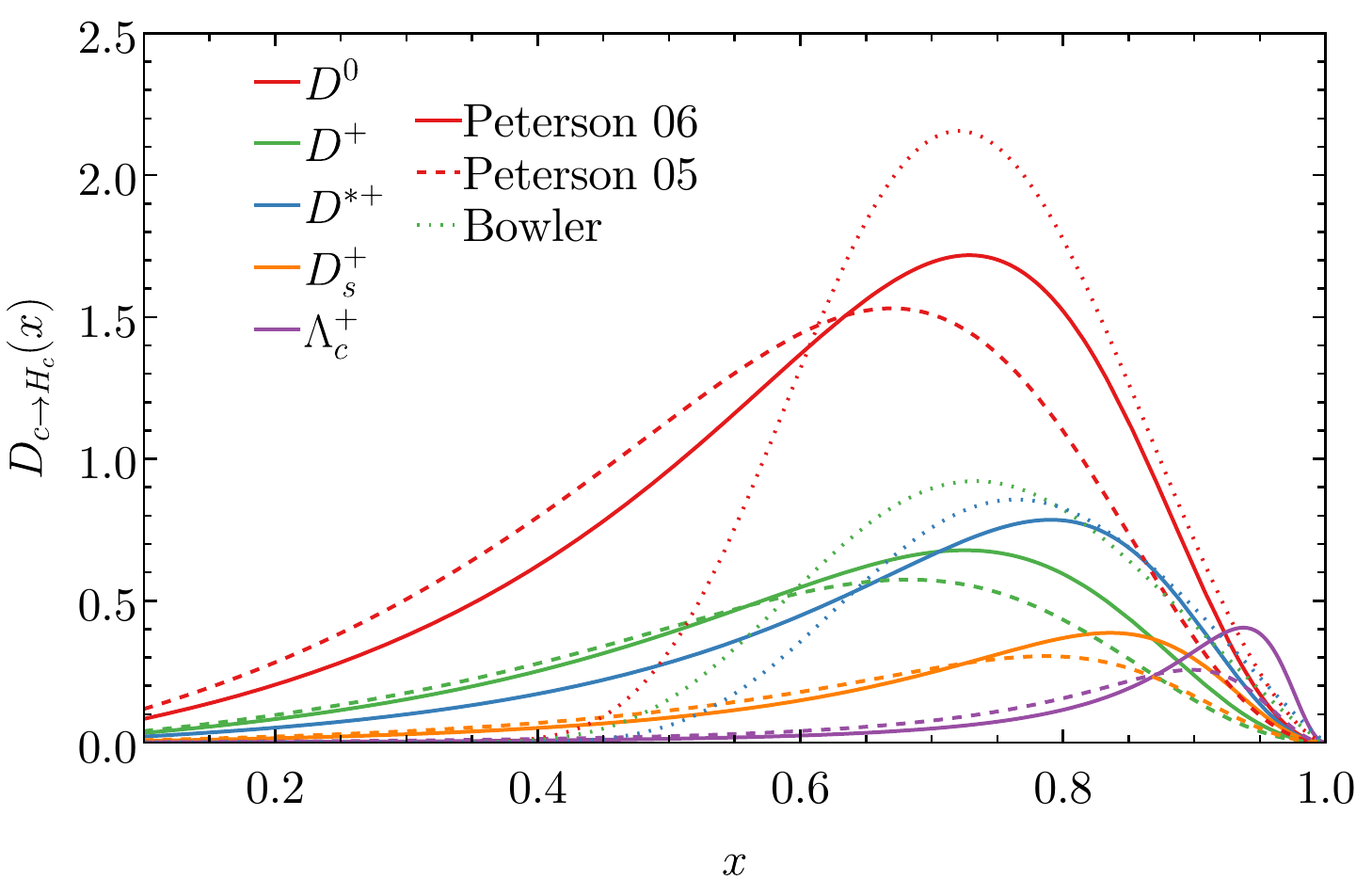}
\caption{Fragmentation functions of charm quark into hadrons in the Peterson~\cite{Peterson:1982ak} and Bowler~\cite{Bowler:1981sb} models, with parameters determined by analyses with data at LEP~\cite{Kniehl:2005de,Kniehl:2006mw,Kneesch:2007ey}.}
\label{fig:FF}
\end{figure}

In this section, we present a phenomenological study for the production of charm quark in the far forward region. 
The theory prediction for the charm production cross section calculation at NLO 
is obtained by using the recently developed S-ACOT-MPS~\cite{Xie:2019eoe} that is a general-mass variable-flavor-number scheme based on the simplified-ACOT scheme with massive phase space. S-ACOT-MPS is applied to heavy-flavor production in $pp$ collisions.  Here, we shall extend and improve the S-ACOT-MPS theory prediction by incorporating fragmentation to describe charm quark final-state hadronization. 
The hadron-level cross section for a charm quark with momentum $p_c$ which fragments into a hadron $H_c$ with momentum $\vec{p}_H=z \vec{p}_c$ can be written as
\begin{equation}
\sigma(pp\to H_cX)=\int \dd z D_{c\to H_c}(z,\mu^2)\hat{\sigma}_{pp\to cX},
\end{equation}
where $D_{c\to H_c}(z,\mu^2)$ is the fragmentation function, $\mu$ is the fragmentation scale, and $\hat{\sigma}_{pp\to cX}$ is the parton-level cross section. 
For charm production, $\hat{\sigma}_{pp\to cX}$ in the far-forward direction has recently been calculated within S-ACOT-MPS in ~\cite{Anchordoqui:2021ghd}. The fragmentation function $D_{c\to H_c}(z,\mu^2)$ can be modelled at scale $\mu=2m_c$ by using different functional forms (e.g., Peterson~\cite{Peterson:1982ak}, Bowler~\cite{Bowler:1981sb}) with parameters that are determined from analyses of LEP data~\cite{Kniehl:2005de,Kniehl:2006mw,Kneesch:2007ey}. 
Results are shown in \cref{fig:FF}.  In general, the branching fractions and average momentum fraction are defined as
\begin{equation}
B_{H_c}(\mu^2)=\int\dd z D_{c\to H_c}(x,\mu^2), ~~~~~ \langle z_{H_c}\rangle(\mu^2)=\frac{1}{B_{H_c}(\mu^2)}\int\dd z D_{c\to H_c}(z,\mu^2)
\end{equation}
respectively. We observe good agreement between the branching fractions and the average momentum fraction within the fragmentation function uncertainty. 

\begin{figure}
\includegraphics[width=0.45\textwidth]{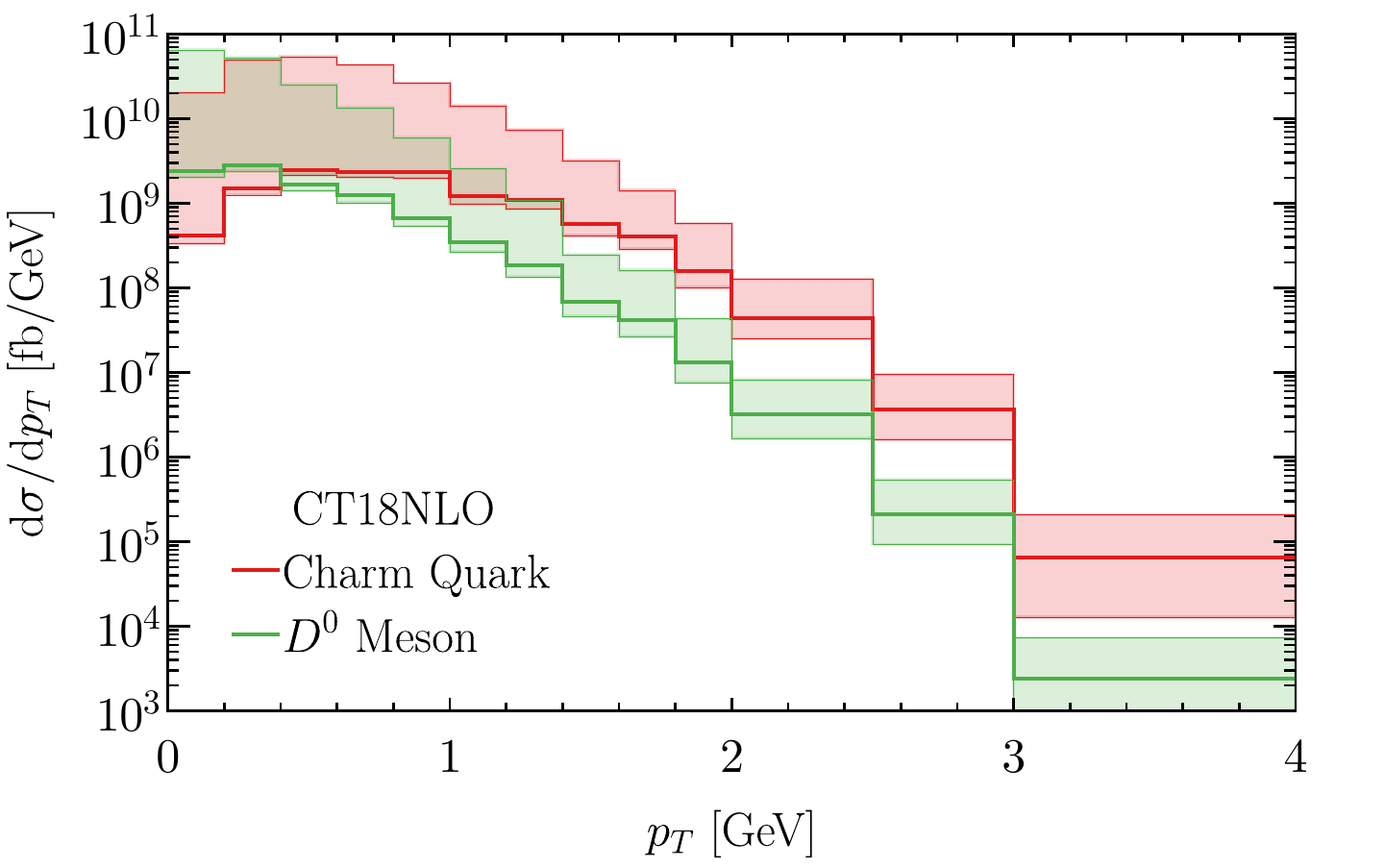}
\includegraphics[width=0.45\textwidth]{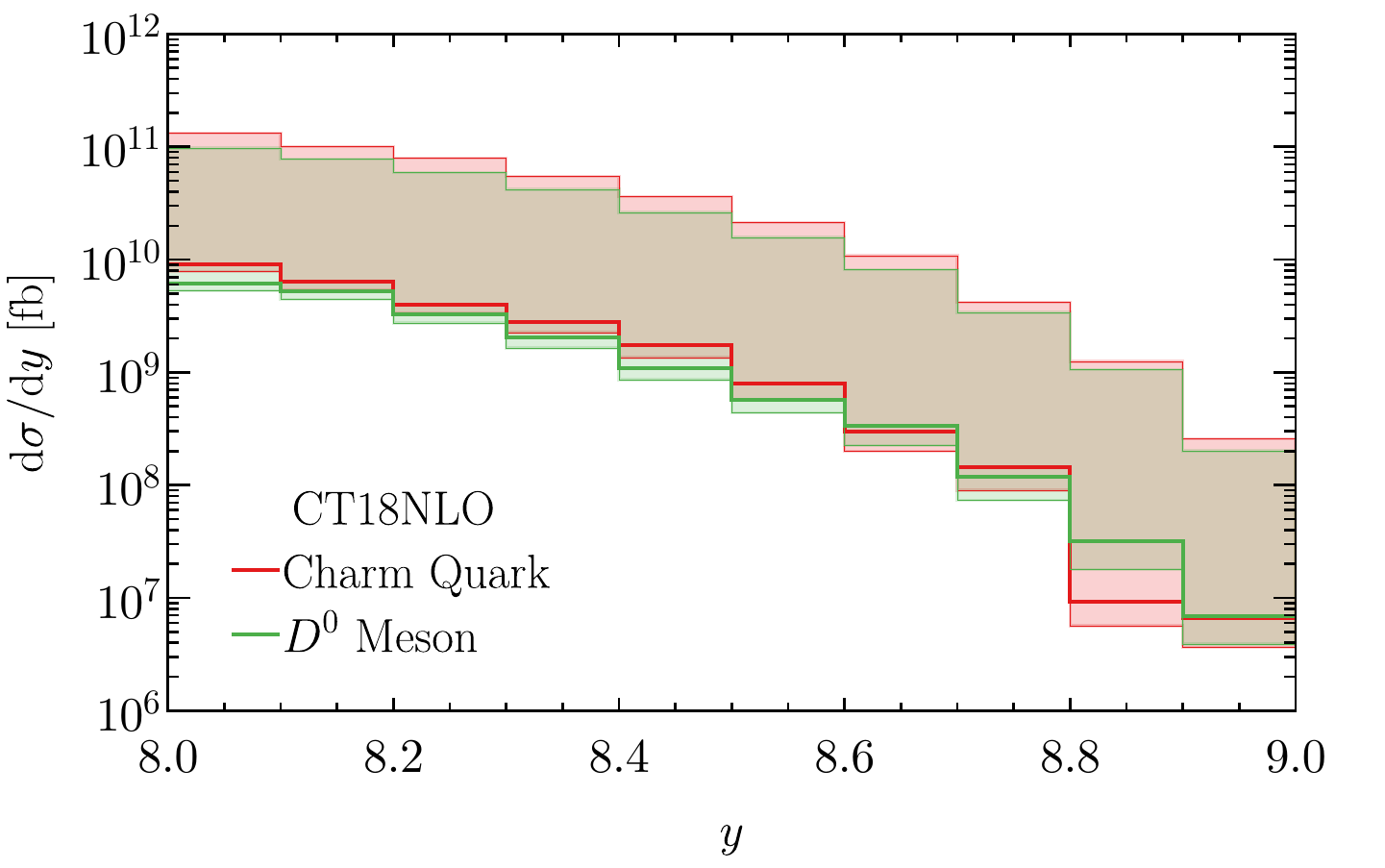}
\caption{The quark- and hadron-level of charm production in the far forward region of $pp$ collision at 13 TeV.}
\label{fig:C2D}
\end{figure}

\begin{figure}
	\includegraphics[width=0.45\textwidth]{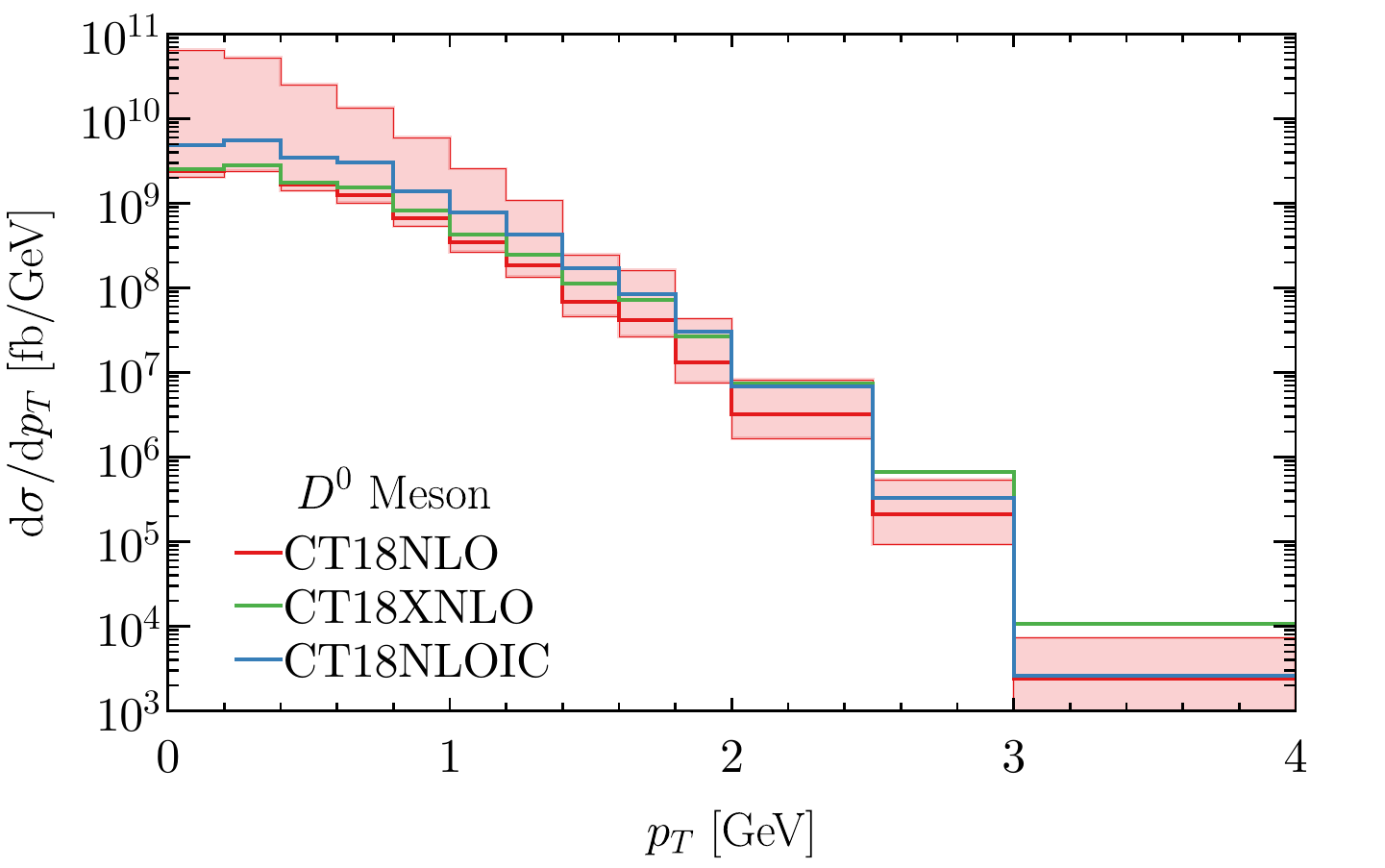}
	\includegraphics[width=0.45\textwidth]{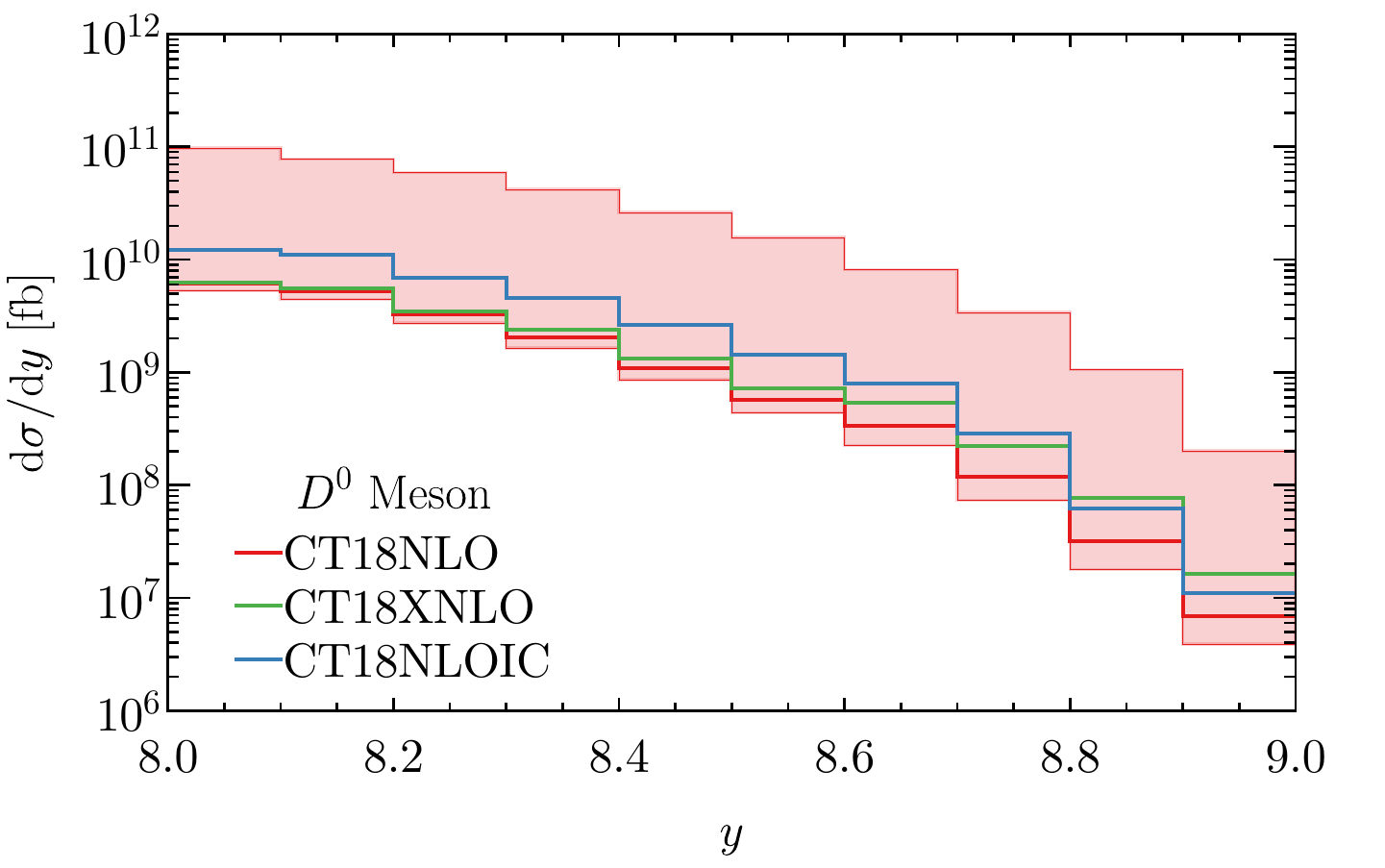}
	\caption{Differential distributions in $p_T$ (left) and rapidity (right) for $D^0$ hadroproduction in the far forward region in $pp$ collision at the LHC 13 TeV.
          The error bands in the \texttool{CT18NLO} calculation correspond to the NLO scale variation envelope.
        }
	\label{fig:D0}
\end{figure}

Charm hadroproduction in $pp$ collisions at 13 TeV in the very forward region is illustrated in \cref{fig:C2D} and \cref{fig:D0}, where we show the $p_T$ spectrum  and rapidity distribution for $D^0$ charmed meson production at very forward rapidities $y>8$. The uncertainty bands represent the PDF uncertainty at  90\% C.L.. The total rate can be well estimated by assuming a factorized branch fraction $\sigma_{H_c}\sim \sigma_{c}B_{H_c}$. The impact of fragmentation  softens the hadron $p_T$ spectrum as $\vec{p}_H=z\vec{p}_c$, while the rapidity distribution is rescaled by an overall factor as expected. PDF uncertainties are large and are dominated by charm and gluon PDF errors at low and high $x$, as shown in \cref{fig:pdfs}. 
In \cref{fig:D0} we show a comparison between the theory predictions for the $p_T$ spectrum and rapidity for $D^0$ production in $pp$ collisions at 13 TeV, obtained with \texttool{CT18}, \texttool{CT18X}, and \texttool{CT18IC} PDFs.   
The \texttool{CT18} and \texttool{CT18X} best fits give similar results for both the $p_T$ and $y$ distributions, while \texttool{CT18IC} produces an enhancement due to the intrinsic charm contribution at the large $x$, as can also be observed in \cref{fig:pdfs}. 
Future measurements at FPF will be crucial to investigate the intrinsic charm content in proton PDFs and will allow us to constrain PDFs in extreme kinematic regions at both large and low $x$. 

\subsection{Probing the Multidimensional Structure of Hadrons at the FPF}
\label{QCD:ssec:HAS_FPF}


\begin{figure}[t]
\centering
\includegraphics[width=0.30\textwidth]{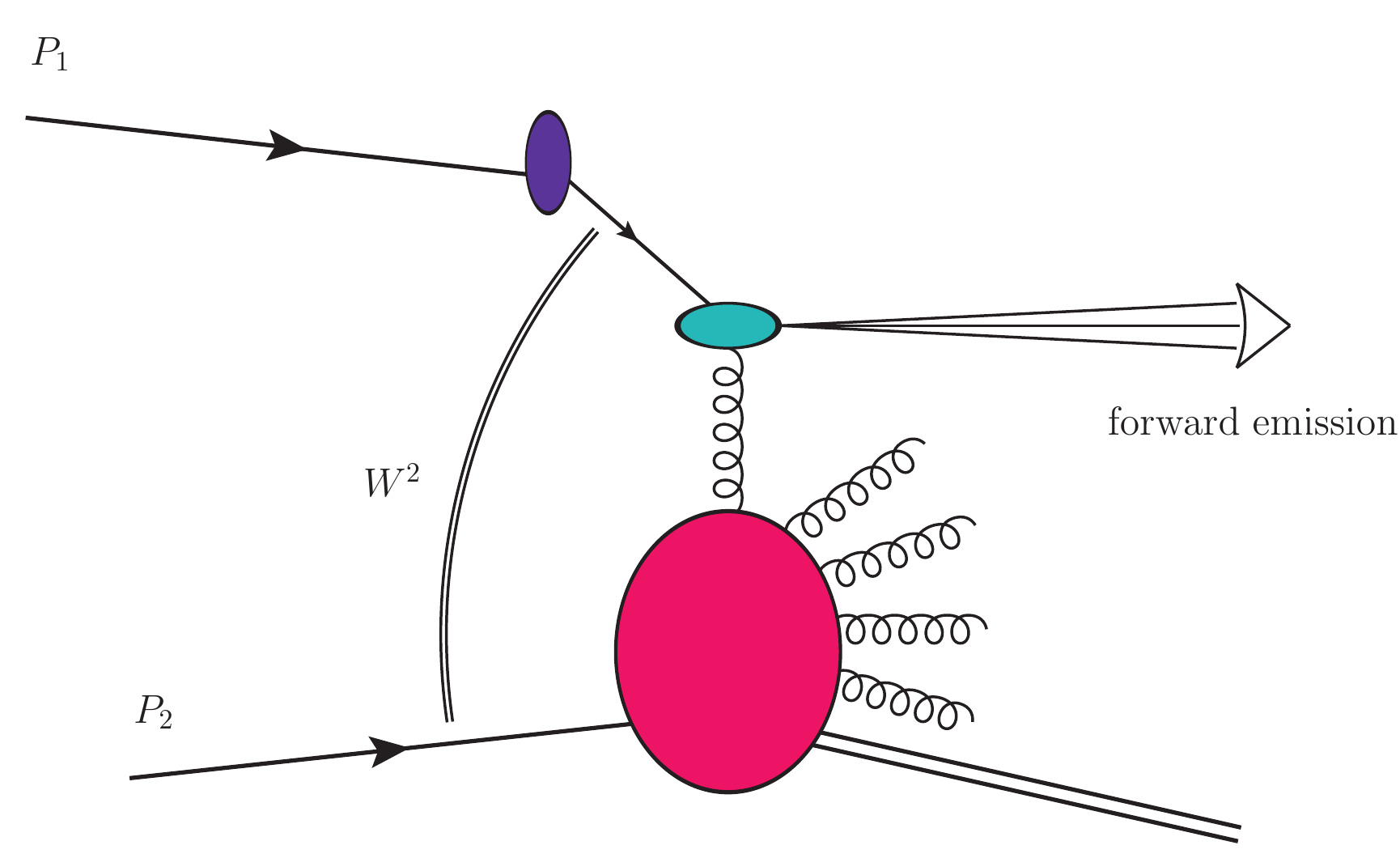}
\hspace{0.50cm}
\includegraphics[width=0.30\textwidth]{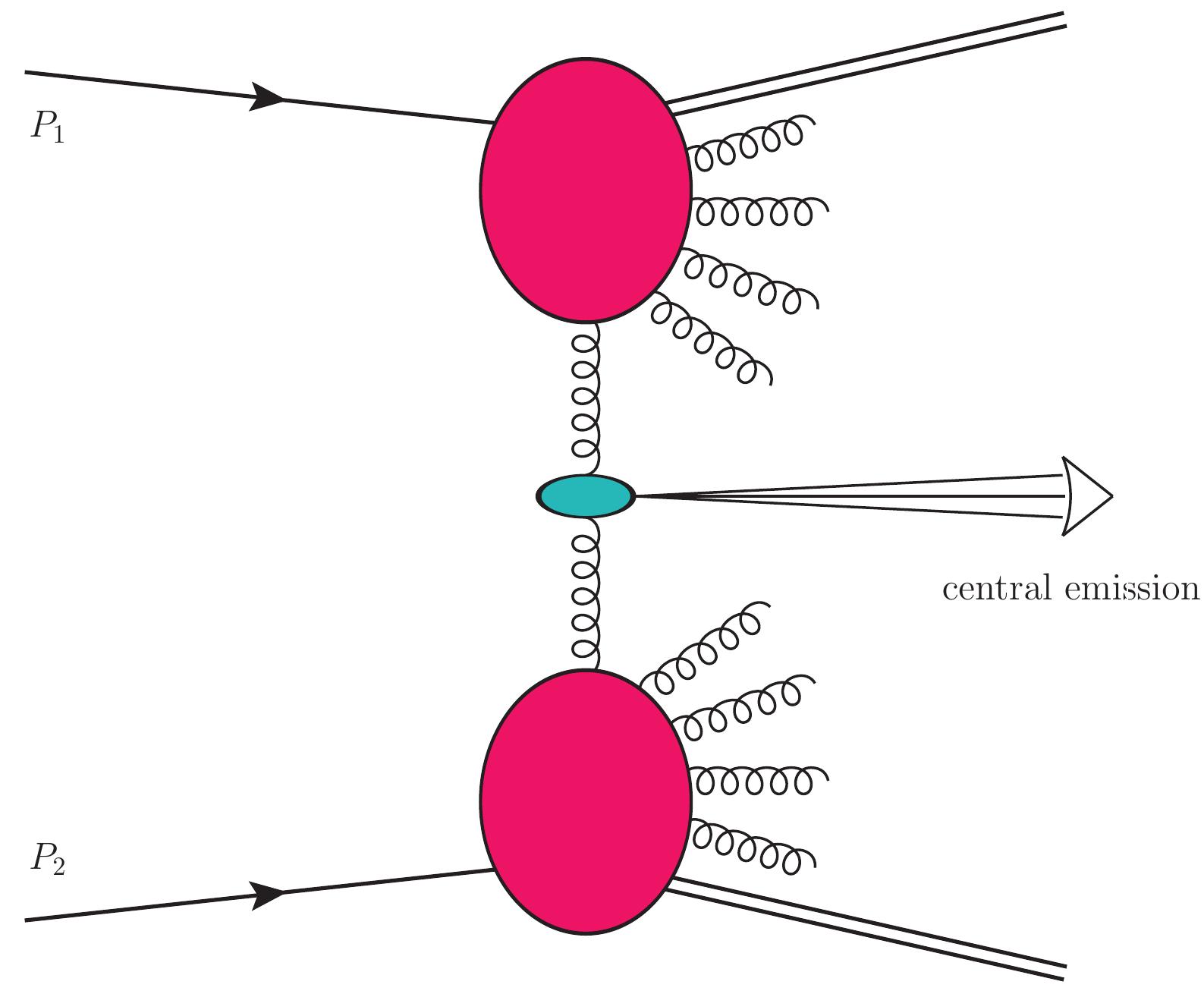}
\hspace{0.50cm}
\includegraphics[width=0.30\textwidth]{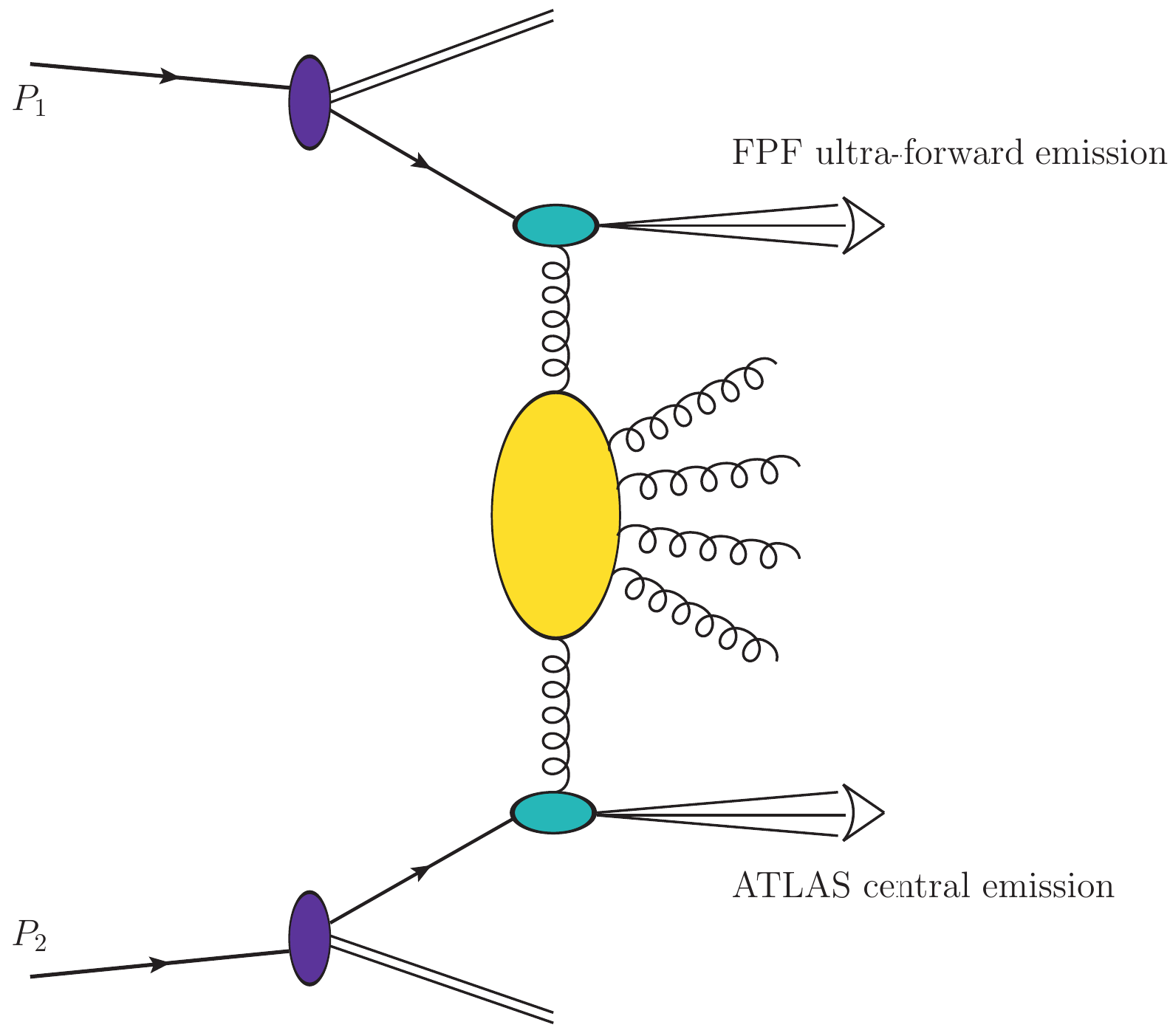}
\\ \vspace{0.25cm}
a) Single forward detection \hspace{1cm}
b) Single central detection \hspace{1cm}
c) FPF~$+$~ATLAS detection
\caption{Schematic representation of a) an inclusive single forward detection in hybrid high-energy/collinear factorization, b) an inclusive single central detection in pure high-energy factorization, c) an inclusive FPF~$+$~ATLAS detection in hybrid factorization. The violet blob in diagram a) depicts collinear PDFs, whereas the sea-green ones in both diagrams stand for the hard part of the off-shell vertex, corresponding to the emission of a generic particle in forward and/or central regions of rapidity. The gluon uPDFs, depicted in red, encode non--perturbative information about the gluon content in the proton at high energy/low-$x$. The BFKL Green’s function is represented by the yellow oval. Gluon-induced emissions from the collinear region(s) in panel a) and c), not shown here, are embodied in the sea-green blob(s).}
\label{fig:processes_HE-QCD}
\end{figure}

A key role in addressing the fundamental questions of QCD, such as the origin of proton mass and spin, is played by our capability of reconstructing the multidimensional distribution of the partons inside parent nucleons, that is going beyond standard collinear PDFs~\cite{NAP25171,Accardi:2012qut,Geesaman:2015fha,AbdulKhalek:2021gbh}. 
Different extensions of PDFs can be defined, which are based on different factorization theorems, exhibit peculiar universality properties, and obey distinct evolution equations. 

When considering the distribution of partons in momentum space, the objects of interests are transverse-momentum-dependent PDFs (TMDs), defined through TMD factorization (see e.g.~\cite{Collins:2011zzd}), and the related uPDFs, defined through high-energy factorization. The connection between the two has been investigated in several papers (see e.g.~\cite{Nefedov:2021vvy,Hentschinski:2021lsh,Celiberto:2019slj}). There is an extended literature on the theory and phenomenology of TMDs, for quarks and gluons, with or without polarization. The FPF can probe kinematic ranges  that have so far been inaccessible. It can significantly improve our knowledge of the light quark and anti-quark TMDs through the study of Drell-Yan and weak-boson production. 
Access to the gluon distributions is possible through the study of heavy-flavor production, which, in high-energy hadronic collisions, is dominantly generated by gluon-gluon interactions.
The separation of different quark flavors and gluon contributions can only be achieved by using different targets and  excellent final-state particle identification.

Currently, global fits of quark TMDs are available~\cite{Bacchetta:2017gcc,Scimemi:2017etj,Scimemi:2019cmh,Bacchetta:2019sam}. They are based on data from semi-inclusive DIS, Drell-Yan, and $Z$-production processes. The FPF can give complementary information to constrain quark TMDs in the low-$x$ and high-$x$ regions. 
Experimental information on gluon TMDs is very scarce. 
First attempts to perform phenomenological studies of the unpolarized gluon TMD have been presented in Refs.~\cite{Lansberg:2017dzg,Gutierrez-Reyes:2019rug,Scarpa:2019fol}. 
A low-$x$ improved model calculation of all leading-twist gluon TMDs, including all possible combinations of proton and gluon polarizations, was recently performed~\cite{Bacchetta:2020vty} (see also Refs.~\cite{Bacchetta:2021oht,Bacchetta:2021lvw,Bacchetta:2021twk,Bacchetta:2022esb}).
Detailed studies of gluon-dominated processes at the FPF can have a dramatic impact on our knowledge of gluon TMDs.

Apart from the unpolarized gluon TMD, the distribution of linearly polarized gluons in an unpolarized nucleon, $h_1^{\perp g}$, is particularly interesting because it gives rise to spin effects even in collisions of unpolarized hadrons~\cite{Boer:2010zf,Sun:2011iw,Boer:2011kf,Pisano:2013cya,Dunnen:2014eta,Lansberg:2017tlc} and is therefore accessible at the FPF. At high transverse momentum and at low-$x$, the unpolarized and linearly polarized gluon distributions $f^g_1$ and $h^{\perp g}_1$ are connected~\cite{Dominguez:2011wm} to the gluon uPDF, whose definition naturally comes out form the BFKL formalism. Golden channels for the study and extraction of the gluon uPDF in proton collisions correspond to the inclusive emission of a single particle over forward ranges of rapidity as well as over more central regions in gluon-induced hard scatterings (see \cref{QCD:ssec:BFKL_if}).

TMD factorization is expected to be violated in processes such as~\cite{Rogers:2010dm}
\begin{equation}
pp \rightarrow h_1+ h_2+X,~~~pp \rightarrow h+{\rm
  jet}+X,~~~pp \rightarrow h+\gamma+X,
\end{equation}
where the final-state particles belong to two different jets.
For the interpretation of hadronic collisions, it would be important to experimentally measure the size of this factorization violation. This demands precise predictions on the theory side and precise experimental measurements.   

If factorization violations turn out to be negligible, it would also be important to check the nontrivial universality properties of gluon TMDs. In fact, depending on the process, different types of gluon TMDs can be probed, i.e. the so-called Weisz\"acker--Williams and dipole TMDs~\cite{Kharzeev:2003wz,Dominguez:2010xd,Dominguez:2011wm}. The FPF can offer us a unique chance to separately probe these two structures, as well as to explore possible relations between each other at low $x$.

Finally, we remark that it is possible to define multidimensional distributions that go beyond quark and gluon TMDs, i.e. the so-called generalized transverse-momentum distributions (GTMDs)~\cite{Ji:2003ak,Belitsky:2003nz,Meissner:2009ww}. 
It is extremely challenging to find experimental observables that are sensitive to GTMDs. Some ideas have been put forward in the last years and could be investigated at the FPF, for instance exclusive double Drell--Yan~\cite{Bhattacharya:2017bvs}, ultra-peripheral pA collisions~\cite{Hagiwara:2017fye}, and diffractive forward production of two quarkonia~\cite{Boussarie:2018zwg}.

\subsection{Monte Carlo Studies of High-energy QCD Reactions at the FPF}

  As discussed in \cref{sec:qcd:forwardintro}, an important area of study within QCD
 is that of scattering in the high energy limit. These studies are to a greater or lesser extent usually based on the BFKL approach~\cite{Kuraev:1977fs,Kuraev:1976ge,Fadin:1975cb,Lipatov:1976zz,Balitsky:1978ic}. 
 The main idea in the BFKL framework is that, when the center-of-mass energy $\sqrt{s} \to \infty$, 
we have the appearance of terms of the form {$\alpha_s^n 
\log^n{\left(s\right)} \sim \alpha_s^n \left(y_A-y_B\right)^n$}, where $y_{A,B}$ are the rapidities of final state particles or jets. 
These terms need to be resummed to accurately describe 
experimental observables. In this limit,
one notes a decoupling between the transverse and longitudinal degrees of freedom which allows to cast 
the cross-sections in the factorized form: 
\begin{eqnarray}
\sigma^{\rm LL} &=& \sum_{n=0}^\infty {\cal C}_n^{\rm LL}  \alpha_s^n 
\int_{y_B}^{y_A} d y_1 \int_{y_B}^{y_1} d y_2 \dots \int_{y_B}^{y_{n-1}} d y_n \nonumber\\ 
&=& \sum_{n=0}^\infty \frac{{\cal C}_n^{\rm LL}}{n!} 
\underbrace{\alpha_s^n \left(y_A-y_B\right)^n }_{\rm LL} \,,
\label{sigmaLL}
\end{eqnarray}
where LL stands for the  leading log approximation and $y_i$ corresponds to the rapidity of the final state particles
or jets. The LL BFKL formalism allows us to calculate the coefficients ${\cal C}_n^{\rm LL}$. The next-to-leading log approximation (NLL)~\cite{Fadin:1998py,Ciafaloni:1998gs} carries large corrections whereas the NLL BFKL kernel 
 is sensitive to the running of the strong coupling as well as to the choice of the energy scale in the logarithms
 and eventually resums terms of the form $\alpha \left(\alpha_s^n \log^n{\left(s\right)} \right)$. This becomes evident if we parametrize the freedom in the choice of these two scales, respectively, by introducing the constants 
${\cal A}$ and ${\cal B}$ in the previous expression: 
\begin{eqnarray}
\sigma^{LL+NLL} &=& 
\sum_{n=1}^\infty \frac{{\cal C}_n^{\rm LL} }{n!}  \left(\alpha_s- {\cal A} \alpha_s^2\right)^n \left(y_A-y_B - {\cal B}\right)^n \nonumber\\
&&\hspace{-2.4cm}= \sigma^{\rm LL}  - \sum_{n=1}^\infty   \frac{\left({\cal B}  \, {\cal C}_n^{\rm LL} +  (n-1) \, {\cal A} 
\, {\cal C}_{n-1}^{\rm LL} \right)}{(n-1)!}  \underbrace{ \alpha_s^n 
\left(y_A-y_B\right)^{n-1}}_{\rm NLL} + \dots \nonumber
\end{eqnarray}
such that we see that at NLL a power in $\log \left(s\right)$ is lost with respect to the power of the strong coupling. 

Within the BFKL framework, we can then calculate partonic cross-sections using the factorization formula (with $\Delta Y\simeq \ln\left(s\right)$)
 \begin{eqnarray}
\sigma (Q_1,Q_2,\Delta Y) = \int d^2 \vec{k}_A d^2 \vec{k}_B \, \underbrace{\phi_A(Q_1,\vec{k}_A) \, 
\phi_B(Q_2,\vec{k}_B)}_{\rm PROCESS-DEPENDENT} \, \underbrace{f (\vec{k}_A,\vec{k}_B,Y)}_{\rm UNIVERSAL}, \nonumber
\end{eqnarray}
where $\phi_{A,B}$ are process-dependent impact factors which are functions of some external scale, $Q_{1,2}$, and some internal momentum for reggeized gluons, $\vec{k}_{A,B}$.  The gluon Green's function $f$ is universal,
it corresponds to the solution of the BFKL equation
and it depends on $\vec{k}_{A,B}$ and the energy of the process $\sim e^{\Delta Y/2}$.
The impact factors and the gluon Green's function allow for the computation of the cross section for various processes that are relevant
to the FPF. 
However, these computations are based on using analytic expressions for the Green's function
and do not allow for the study of more exclusive quantities. One could in principle obtain more information on final state differential 
distributions using the analytic expressions for the gluon Green's function,
(see for example~\cite{Caporale:2015vya,Caporale:2015int,Caporale:2016soq,Caporale:2016xku,Caporale:2016zkc}) but 
this approach quickly becomes impossible  to use as the final state multiplicity grows above $N=4$.

An alternative approach is to use stochastic methods for the computation of $f$. The first step is to 
write the solution of the BFKL equation in an iterative form~\cite{Schmidt:1996fg} in transverse momentum representation, which
at LL reads
\begin{eqnarray}
f &=& e^{\omega \left(\vec{k}_A\right) \Delta Y}  \Bigg\{\delta^{(2)} \left(\vec{k}_A-\vec{k}_B\right) + \sum_{n=1}^\infty \prod_{i=1}^n \frac{\alpha_s N_c}{\pi}  \int d^2 \vec{k}_i  
\frac{\theta\left(k_i^2-\lambda^2\right)}{\pi k_i^2} \nonumber\\
&&\hspace{-1.2cm}\int_0^{y_{i-1}} \hspace{-.3cm}d y_i e^{\left(\omega \left(\vec{k}_A+\sum_{l=1}^i \vec{k}_l\right) -\omega \left(\vec{k}_A+\sum_{l=1}^{i-1} \vec{k}_l\right)\right) y_i} \delta^{(2)} 
\left(\vec{k}_A+ \sum_{l=1}^n \vec{k}_l - \vec{k}_B\right)\Bigg\}, 
\label{eq:iterBFKL}
 \end{eqnarray}
where the gluon Regge trajectory is defined to be
\begin{eqnarray}
\omega \left(\vec{q}\right) &=& - \frac{\alpha_s N_c}{\pi} \log{\frac{q^2}{\lambda^2}} 
\end{eqnarray}
and $\lambda$ is a regulator of infrared divergences. 
The next step is to compute $f$ in \cref{eq:iterBFKL} numerically by employing Monte Carlo techniques. 
This solution has been studied at length in a series of works and it
serves as the basis to construct the Monte Carlo code \texttool{BFKLex}~\cite{Chachamis:2011rw,Chachamis:2011nz,Chachamis:2012fk,Chachamis:2012qw,Chachamis:2015zzp,Chachamis:2015ico,deLeon:2020myv,deLeon:2021ecb}.  

\texttool{BFKLex} is quite flexible and one can implement
the convolution with the relevant impact factors, all in momentum space, such that it
can be used to compute the cross-sections for either single light quark forward jet production~\cite{Chachamis:2013bwa,Chachamis:2015ona}
or for DIS processes in the low-$x$ limit.
The advantage is that one can also compute extra differential information, such as single rapidity
distributions of the final state jets and jet-jet rapidity correlation functions which allows for
a more direct comparison against the usual multipurpose Monte Carlo event generators and experimental data.
Moreover, it is straightforward to use \texttool{BFKLex} within the usual collinear factorization scheme
by convoluting the partonic cross-section with PDFs
as well as in setups where both the collinear factorization and the
$k_T$-factorization schemes need to be employed (light quark forward production).

A well known example of the former case is the study of Mueller-Navelet jets~\cite{Mueller:1986ey},
 where
the differential partonic cross-section for the production of two widely separated in
rapidity jets and transverse momenta $\vec{p}_{i=1,2}$ is given by
 \begin{eqnarray}
\frac{d \hat{\sigma}}{d^2 \vec{p}_1 d^2 \vec{p}_2} &=& 
\frac{\pi^2 {\bar \alpha}_s^2}{2} \frac{f(\vec{p}_1^{~2}, \vec{p}_2^{~2},\Delta Y)}{\vec{p}_1^{~2} \vec{p}_2^{~2}} \,.
\end{eqnarray}
The longitudinal momentum fractions of the colliding partons are assumed to be $x_{i=1,2}$,
 the rapidity difference between the two jets $\Delta Y \sim \ln{x_1 x_2 s / \sqrt{\vec{p}_1^{~2} \vec{p}_2^{~2}}}$ 
 and the impact factors a simple combination of color factors and the strong coupling.
In collinear factorization the cross-section for this process reads
\begin{eqnarray}
\frac{d \sigma^{2-\text { jet }}}{d p_{1} d y_{1} d \theta_{1} d p_{2} d y_{2} d \theta_{2} } = 
\sum_{r, s=q, \bar{q}, g} \int_{0}^{1} d x_{1} \int_{0}^{1} d x_{2} f_{r}\left(x_{1}, \mu_{F}\right) f_{s}\left(x_{2}, \mu_{F}\right) d \hat{\sigma}_{r, s}\left(\hat{s}, \mu_{F}\right) \,,
\label{totcross}
\end{eqnarray}
and it can be computed with \texttool{BFKLex} while the differential information of the final state events is preserved and
can readily be compared against experimental data.

In the case of a light quark forward jet production (~\cite{Chachamis:2013bwa,Chachamis:2015ona}), one needs first to combine the gluon Green's function
and the proton impact factor to obtain an expression for an gluon uPDF:
\begin{equation}
G\left(x, q_{1}\right)=\int \frac{d \boldsymbol{q}_{2}^{2}}{\boldsymbol{q}_{2}^{2}}{\mathcal{F}}^{\mathrm{DIS}}\left(x, q_{1}, q_{2}\right) \Phi_{p}\left(q_{2}, Q_{0}^{2}\right)\,,
\label{eq:lightQ}
\end{equation} 
where ${\mathcal{F}}^{\mathrm{DIS}}$ here is the gluon Green's function adapted for DIS-like kinematics. The total
$p p \rightarrow  Q + X$ cross-section then reads 
\begin{equation}
\sigma_{p p \rightarrow Q+X}=\int_{0}^{1} d x_{Q} \int_{0}^{1} \frac{d x_{g}}{x_{g}} \int \frac{d^{2} \boldsymbol{k}}{\pi} \hat{\sigma} \cdot\left[f_{Q}\left(x_{Q}, \mu_{f}\right)+f_{\bar{Q}}\left(x_{Q}, \mu_{f}\right)\right] G\left(x_{g}, \boldsymbol{k}_{T}, Q\right)\,,
\end{equation}
where we have added the extra argument $\boldsymbol{k}_{T}$ (transverse momentum of the light quark) in $G$ and
$ \hat{\sigma} $ is the partonic cross-section for quark-virtual gluon scattering, $Q + g^* \rightarrow Q'$.

The implementation within \texttool{BFKLex} of the relevant impact factors is under way and we believe that this
Monte Carlo approach for studies in the BFKL framework is an important 
step towards a better understanding of the high energy limit of QCD and a more robust treatment
of the uncertainties involved.


\subsection{High-energy QCD via a FPF+ATLAS Timing Coincidence}\label{QCD:ssec:HEF_hh}

\begin{figure}
   \includegraphics[scale=0.38,clip]{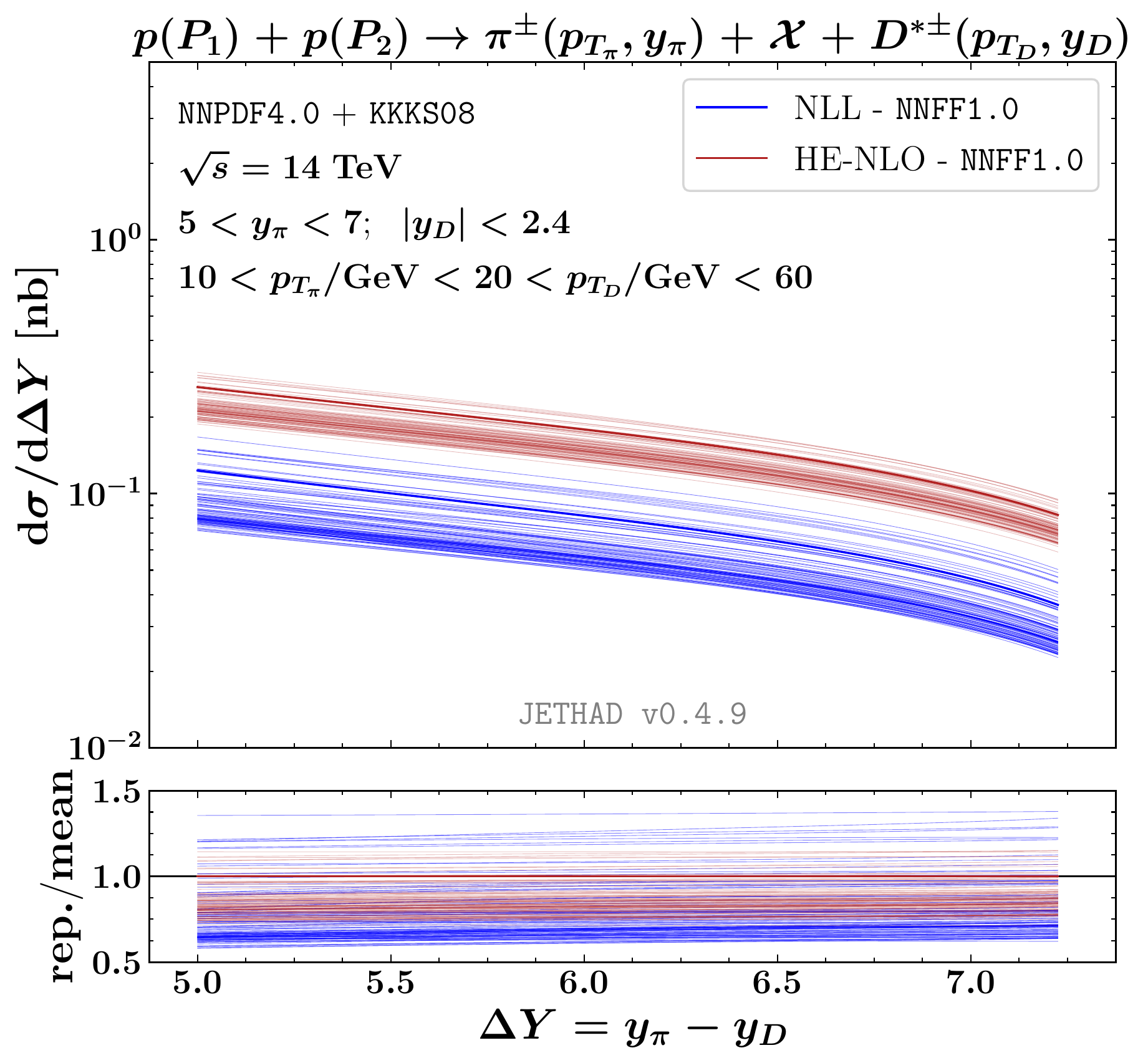}
   \includegraphics[scale=0.38,clip]{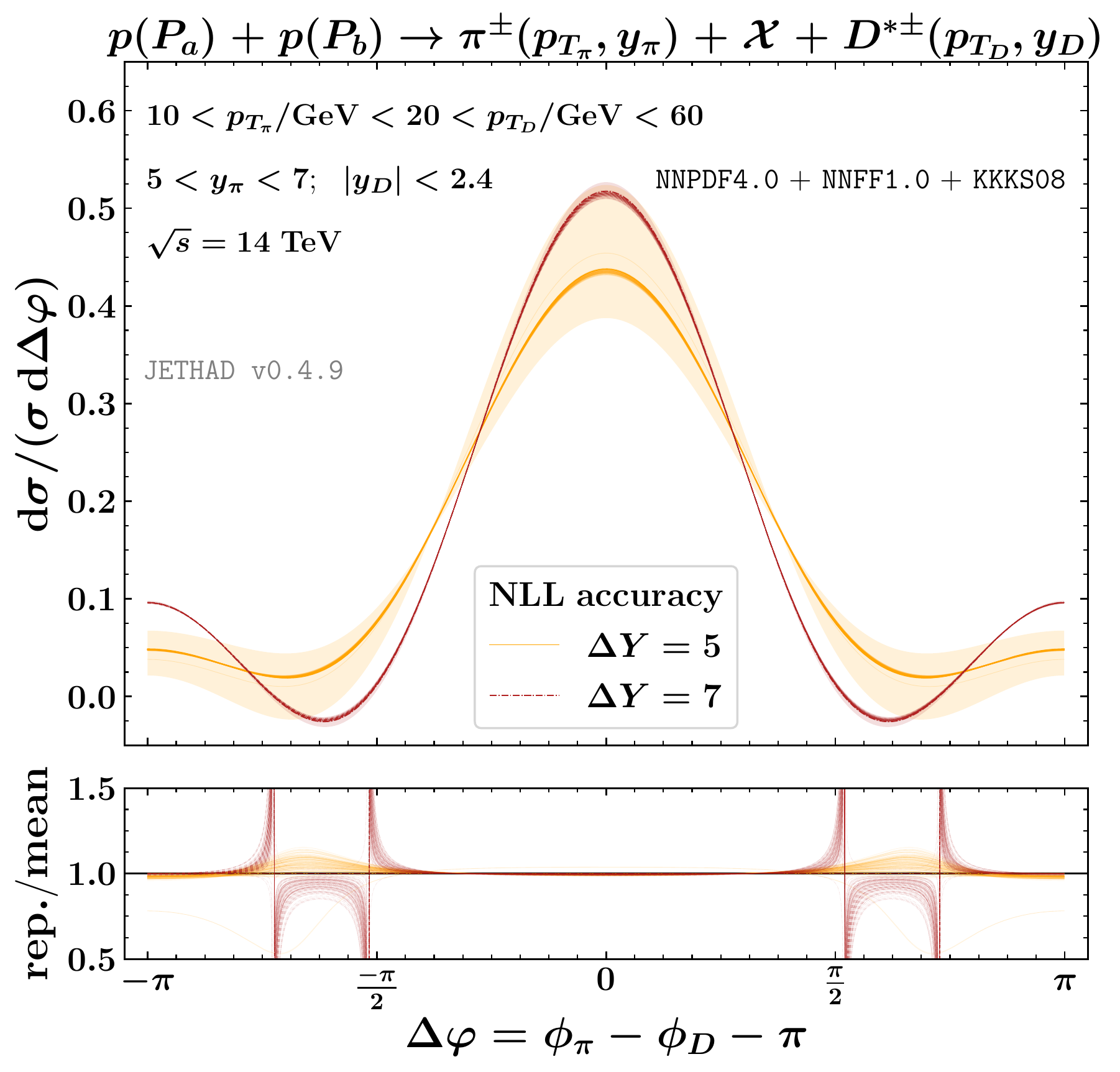}
   \includegraphics[scale=0.38,clip]{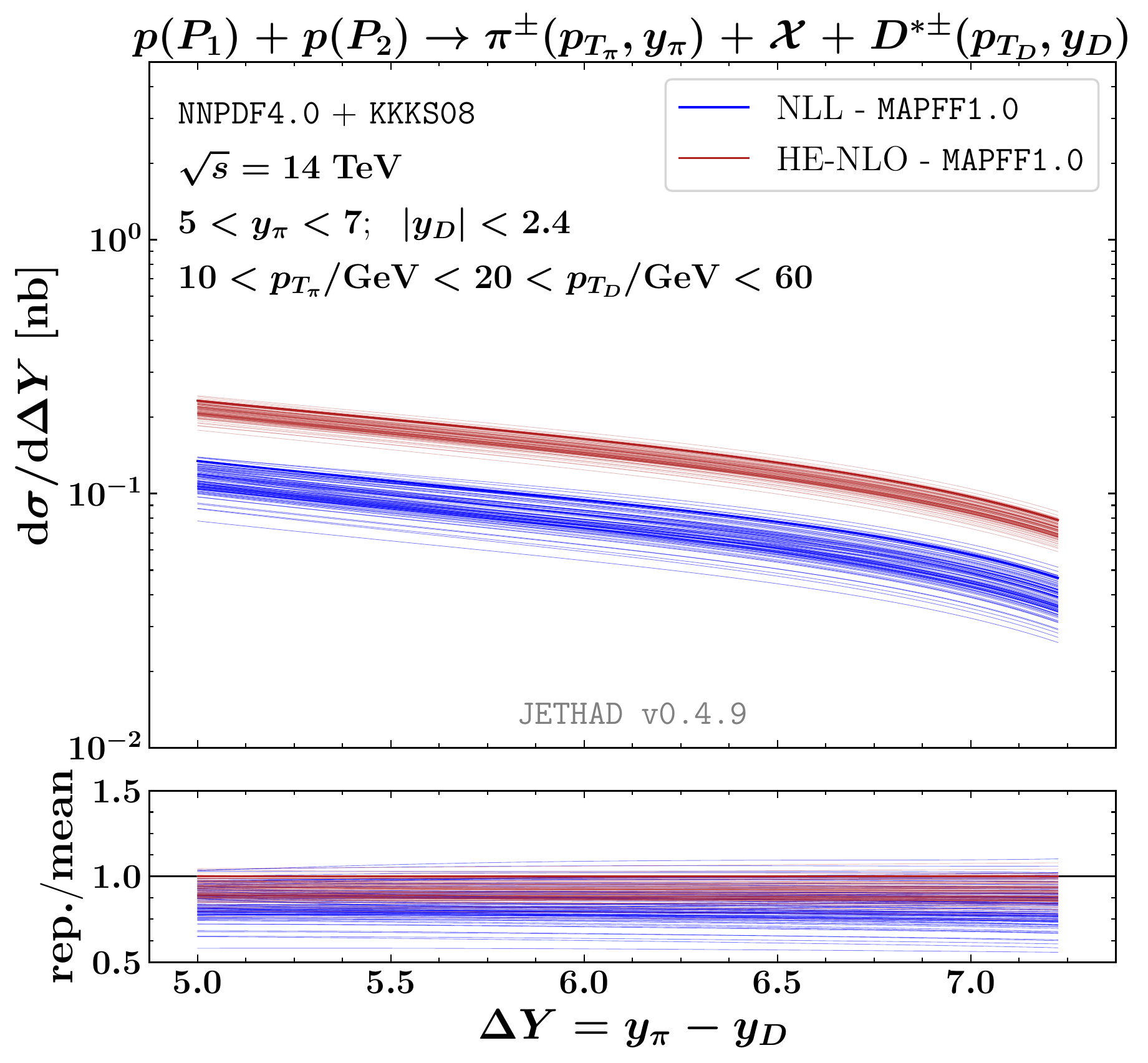}
   \includegraphics[scale=0.38,clip]{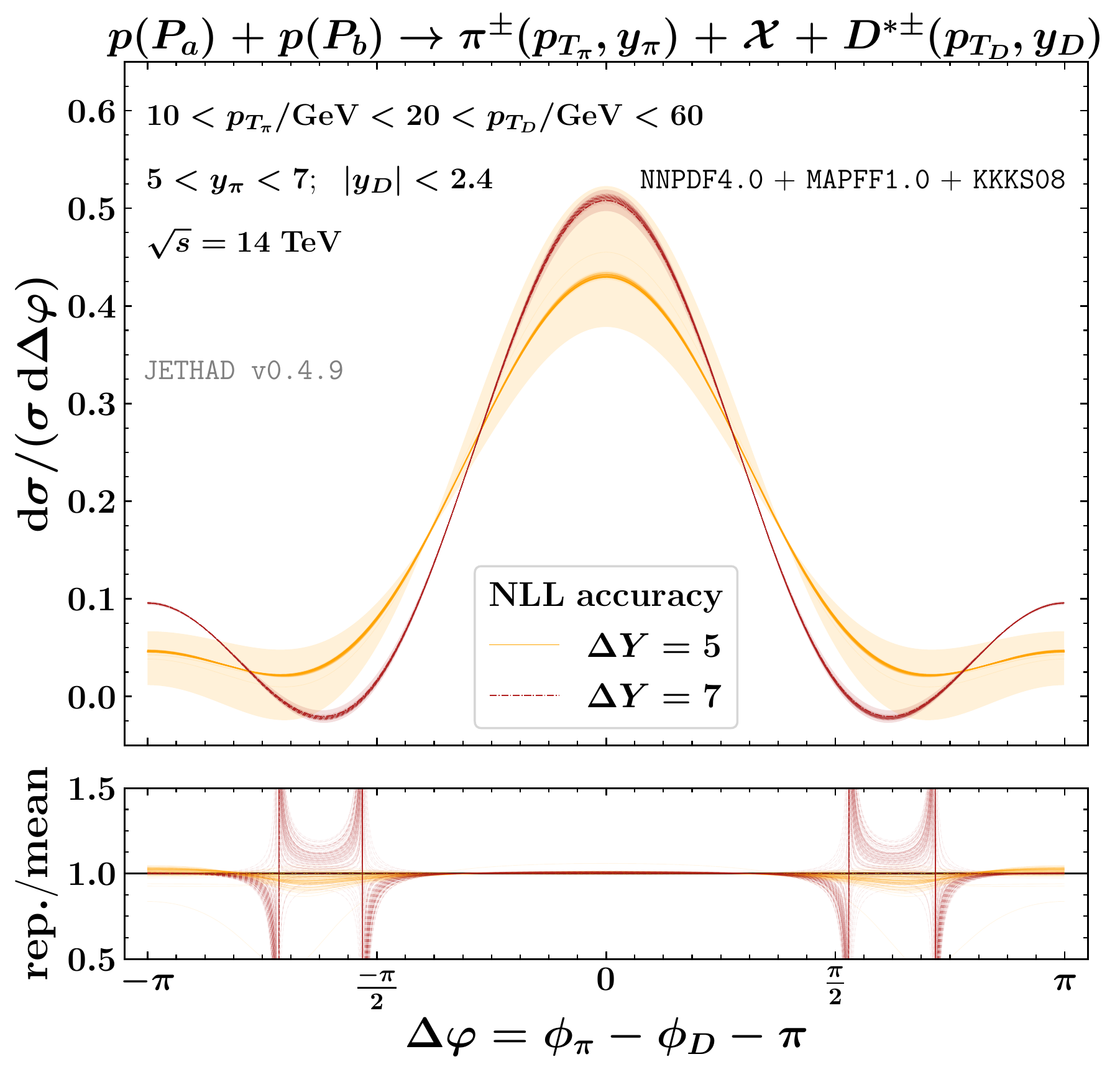}
   \includegraphics[scale=0.38,clip]{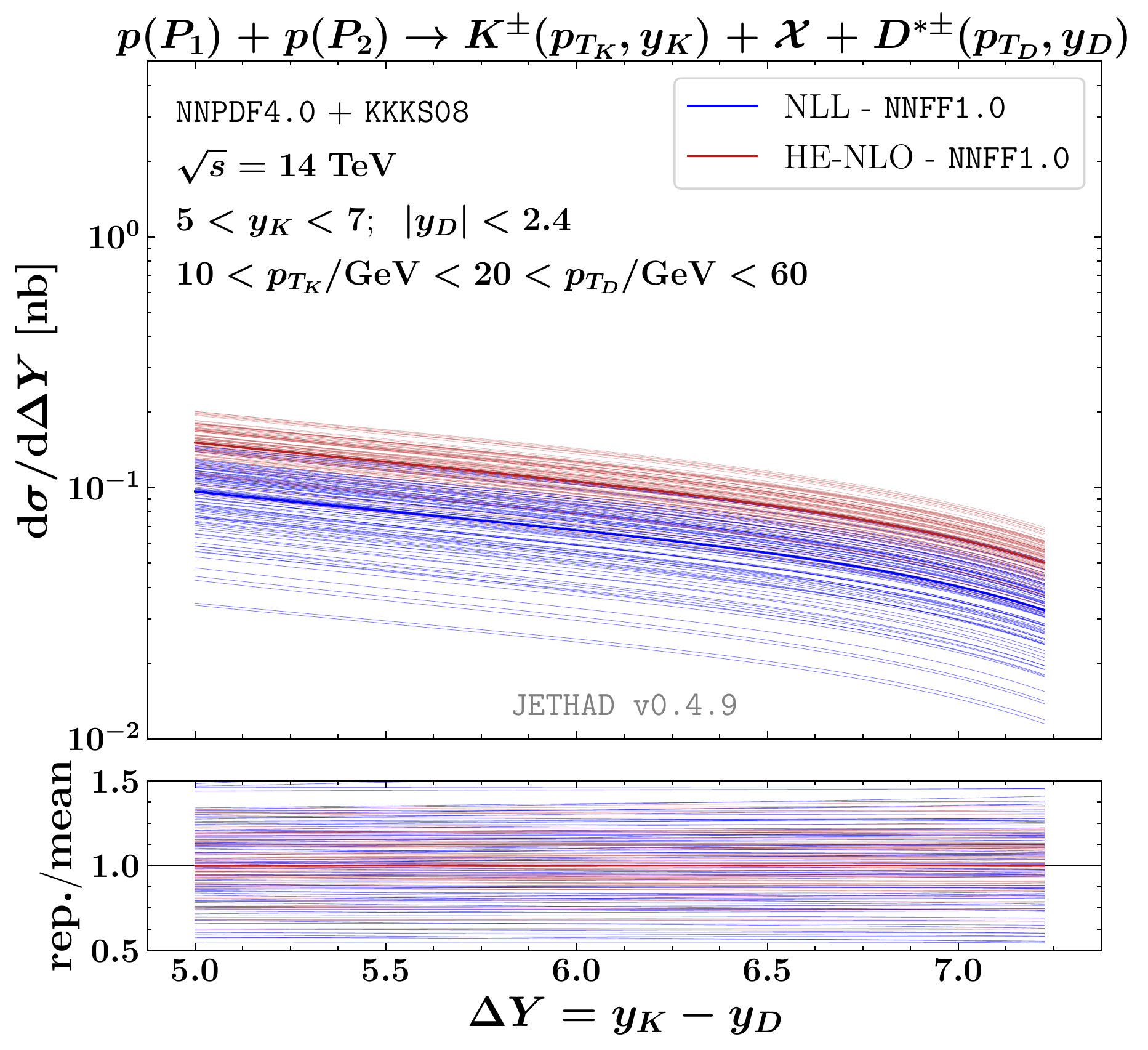}\qquad\qquad\qquad
   \includegraphics[scale=0.38,clip]{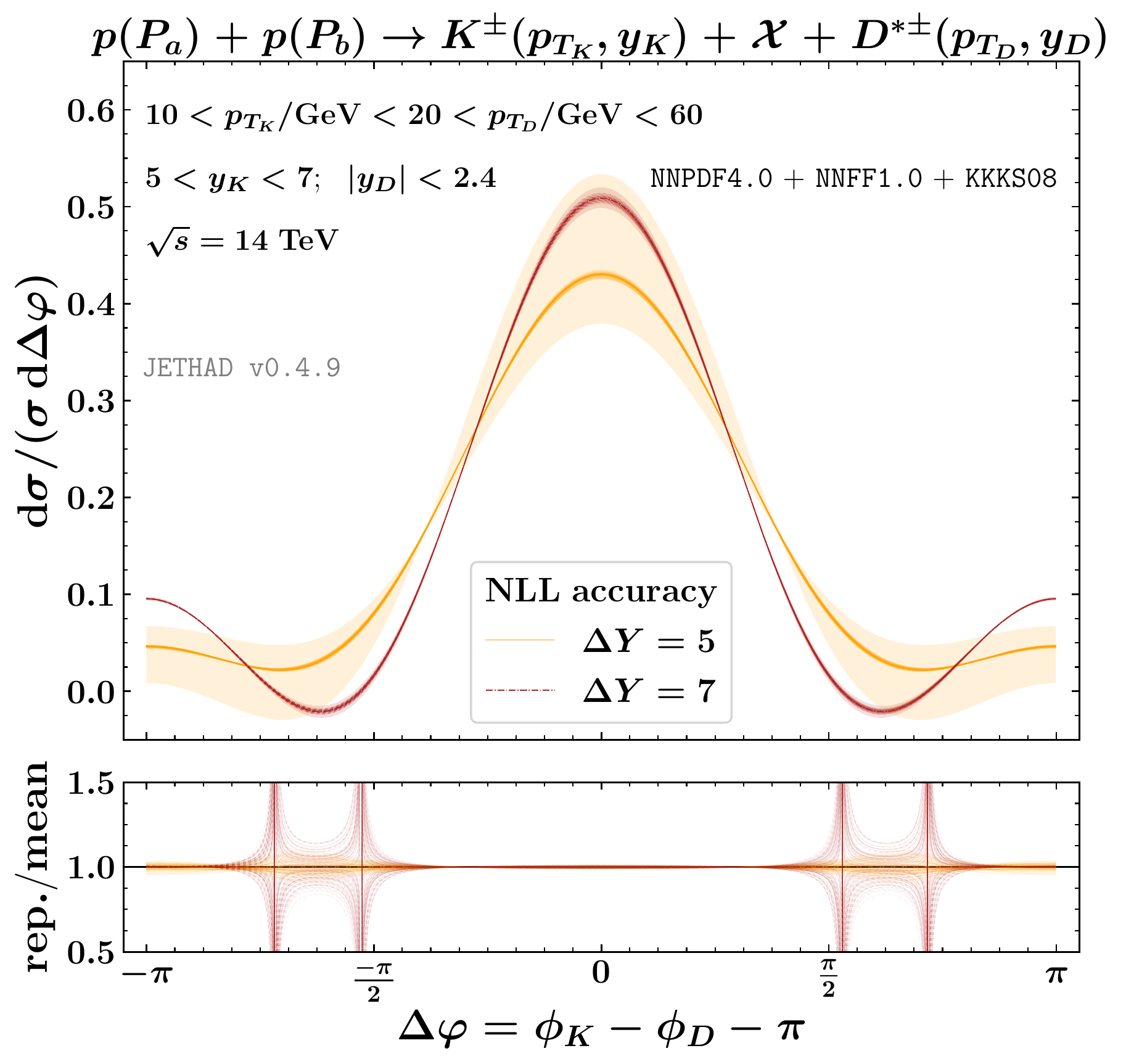}
\caption{$\Delta Y$-distribution (left) and azimuthal distribution (right) for inclusive $(\pi$~$+$~$D^{*})$ and $(K$~$+$~$D^{*})$ reactions in a tight timing coincidence setup for the FPF~$+$~ATLAS configuration, and for $\sqrt{s} = 14$ TeV.
\texttool{NNFF1.0} and \texttool{MAPFF1.0} collinear FFs are employed in the description of pion emissions, whereas only the \texttool{NNFF1.0} set is used for kaon emission.
The envelope of main results is built in term of a replica-driven study on light-hadron collinear FFs.
Ancillary panels below primary plots show the 
envelope of replicas’ predictions divided by their mean value.}
\label{fig:Y_Yphi}
\end{figure}

In the so--called semi--hard regime~\cite{Gribov:1984tu},  the scale hierarchy, $\Lambda_{\rm QCD} \ll \{ Q \} \ll \sqrt{s}$ holds, where $\Lambda_{\rm QCD}$ the QCD scale, $\{ Q \}$ a (set of) hard scale(s) typical of the process and $\sqrt{s}$ is the centre--of--mass energy. In this case, as described  in \cref{sec:qcd:forwardintro}, high energy logarithms enter which are amenable to resummation in the BFKL framework, either in the leading approximation (LL), including of all terms proportional to $\alpha_s^n \ln (s)^n$, or next-to-leading approximation (NLL), including all terms proportional to $\alpha_s^{n+1} \ln (s)^n$. It is in particular the case that large final-state rapidities (or rapidity distances) corresponding to single forward emissions (or double forward/backward emissions)  increase the impact of such high energy logarithms and hence the relevance of BFKL effects.

Over the last decade, predictions for a large number of semi-hard observables in unpolarized hadronic collisions have been obtained.
Among them, azimuthal correlations between two jets emitted with high transverse momenta and large separation in rapidity (the Mueller--Navelet dijet channel~\cite{Mueller:1986ey}) have been identified as promising observables to discriminate between BFKL-resummed and fixed-order-inspired calculations~\cite{Celiberto:2015yba,Celiberto:2015mpa}.
Several phenomenological studies have been conducted so far~\cite{Marquet:2007xx,Colferai:2010wu,Caporale:2012ih,Ducloue:2013hia,Ducloue:2013bva,Caporale:2013uva,Caporale:2014gpa,Caporale:2015uva,Mueller:2015ael,Celiberto:2016ygs,Caporale:2018qnm}, which are in fair agreement with the only set of data available, i.e. the one collected by the CMS collaboration for \emph{symmetric} ranges of the jet transverse momenta~\cite{Khachatryan:2016udy}.
In~\cite{Celiberto:2020wpk} (see also~\cite{Bolognino:2018oth,Bolognino:2019yqj,Bolognino:2019cac,Celiberto:2020rxb,Celiberto:2021xpm})  clear evidence was provided that the high-energy resummed dynamics can be sharply disentangled from the fixed-order pattern at LHC energies when asymmetric cuts for transverse momenta are imposed both in dijet and in jet plus light-hadron final states.
The BFKL resummation was then tested via a  variety of inclusive hadronic semi-hard reactions. An incomplete list includes: dihadron correlations~\cite{Celiberto:2016hae,Celiberto:2016zgb,Celiberto:2017ptm}, multi-jet emissions~\cite{Caporale:2015vya,Caporale:2015int,Caporale:2016soq,Caporale:2016vxt,Caporale:2016xku,Celiberto:2016vhn,Caporale:2016djm,Caporale:2016lnh,Caporale:2016zkc}, $J/\psi$-plus-jet~\cite{Boussarie:2017oae,Celiberto:2022dyf}, heavy-quark pair~\cite{Bolognino:2019yls,Bolognino:2019ouc}, and forward Drell–Yan dilepton production with backward-jet detection~\cite{Golec-Biernat:2018kem}.

One of the major issues emerging in the description of the Mueller--Navelet reaction is the size of NLL corrections, which turn out to be of the same order, but generally with opposite sign, with respect to the LL results. This leads to instabilities of the high-energy series which clearly become apparent when studies on renormalization/factorization scale variation are performed. Thus, it is not possible to perform phenomenological analyses of Mueller--Navelet cross sections and azimuthal correlations  around ``natural" scales~\cite{Ducloue:2013bva,Caporale:2014gpa,Celiberto:2020wpk}. 
Although the application of some scale optimization procedure, as the Brodsky--Lepage--Mackenzie (BLM) method~\cite{Brodsky:1996sg,Brodsky:1997sd,Brodsky:1998kn,Brodsky:2002ka} could help to quench this instability, it turns out that the optimal scale values are found to be much higher than the natural ones. This leadsd to a substantial lowering of cross sections and hampers any chance of making precision studies.

A first clear signal of stability of semi-hard BFKL observables under higher-order corrections at natural scales was discovered quite recently in processes featuring the emission of objects with a large transverse mass, such as Higgs bosons~\cite{Celiberto:2020tmb,Celiberto:2021fjf,Celiberto:2021tky} and heavy-flavored jets~\cite{Bolognino:2021mrc,Bolognino:2021hxx,Bolognino:2021zco}, studied with partial NLL accuracy. Strong stabilising effects in full NLL emerged in recent studies on inclusive emissions of $\Lambda_c$ baryons~\cite{Celiberto:2021dzy,Celiberto:2021txb} and bottom-flavored hadrons~\cite{Celiberto:2021fdp}. Here, corroborating evidence was provided that the characteristic behavior of variable-flavor-number-scheme (VFNS) collinear fragmentation functions (FFs) describing the production of those heavy-flavored bound states at large transverse momentum~\cite{Kniehl:2020szu,Kniehl:2008zza,Kramer:2018vde} acts as a fair stabilizer of high-energy dynamics.
We refer to this property, namely the existence of semi-hard reactions that can be studied in the BFKL approach without applying any optimization scheme nor artificial improvements of the analytic structure of cross section, as \emph{natural stability} of the high-energy resummation.

With the aim of shedding light on the physics potential to study  high-energy dynamics of QCD at the FPF, we present preliminary predictions for two semi-hard reactions that can be studied via simultaneous detections at the FPF and ATLAS detectors by means of the time-coincidence method.
In particular, we allow for an ultra-forward tag of a light meson ($\pi^\pm$ or $K^\pm$) at the FPF, in the rapidity window $5 < y_{\pi,K} < 7$ and in the transverse-momentum range 10~GeV~$< |\vec p_{T_{\pi,K}}| <$~20~GeV, together with the detection of a $D^{* \pm}$ meson in more central regions of ATLAS, $|y_D| < 2.4$, with its transverse momentum $|\vec p_{T_D}|$ ranging from 20~to~60~GeV (see \cref{fig:processes_HE-QCD} (c)).
The possibility of reconstructing events at FPF~$+$~ATLAS combined kinematics provides several benefits.
First, it is sensitive to the regime of large rapidity intervals between the two detected hadrons, $\Delta Y \equiv y_{\pi,K} - y_D \gg 1$, leading to significant transverse-momentum exchanges in the $t$-channel and thus to large energy logarithms that must be resummed via the BFKL approach.
Then, as mentioned before, the imposed asymmetric windows for the observed transverse momenta are expected to ease the discrimination between the high-energy and the fixed-order approach.
Finally, with the ATLAS tag of a heavy-flavored meson used to stabilize the high-energy series, an intriguing option is open to possibly constrain light-hadron collinear FFs in a kinematic sector complementary to the one typical of current analyses.
Numerical results presented in this section were obtained via the \texttool{JETHAD} multi-modular interface~\cite{Celiberto:2020wpk} aimed at the management, calculation and processing of observables calculated in different approaches.

In \cref{fig:Y_Yphi} (left) we present the behaviour of the cross section differential in the final-state rapidity distance, $\Delta Y$, for our reference reactions in the FPF~$+$~ATLAS setup.
NLL resummed predictions are compared with the high-energy limit of NLO fixed-order results (HE-NLO), taken as the truncation of the NLL series up to the ${{\cal O} (\alpha_s^3)}$ perturbative accuracy.
In both cases we observe promising statistics, 
together with a downtrend of the cross section when $\Delta Y$ increases. This is due to the interplay of two competing effects, namely  that pure BFKL dynamics would lead to the well-known growth with energy of partonic hard factors, but that its convolution with PDFs and FFs dampens the hadronic cross sections at large $\Delta Y$-values.
The envelope of our results is built in terms of 100 replicas for the FFs, the \texttool{NNFF1.0}~\cite{Bertone:2017tyb} and \texttool{MAPFF1.0}~\cite{Khalek:2021gxf} parameterizations being used for pions, and just the \texttool{NNFF1.0} one for kaons, while the \texttool{NNPDF4.0} proton PDF set~\cite{NNPDF:2021uiq,Ball:2021leu} is taken at its central value.
Emissions of $D^{*}$ mesons are depicted in terms of \texttool{KKKS08} FFs~\cite{Kneesch:2007ey}.
The factorization and renormalization scales are set to the  natural scale of the process, i.e. the sum of transverse masses of the two emitted hadrons.

As expected from the use of asymmetric $p_T$-windows, the resummed predictions are distinct from the fixed-order results, with the first being constantly below the second one. This effect turns out to be sharper in the $(\pi$~$+$~$D^{*})$ channel and milder in the $(K$~$+$~$D^{*})$ one.
The main outcome here is that  high-energy resummation plays a key role in the study of $\Delta Y$-differential cross sections and it needs to be accounted for in order to get a consistent description of these observables in the considered kinematic regimes.

In \cref{fig:Y_Yphi} (right) we show the azimuthal distribution of our reference processes, namely the normalized cross section differential in the angle distance between the light hadron and the $D^*$ meson on the azimuthal plane, $\Delta \varphi \equiv \phi_{\pi,K} - \phi_D - \pi$, and at fixed values of $\Delta Y$.
From a theory viewpoint, this distribution represents one of the most solid observables with which to search for high-energy effects, since it embodies high-energy signals coming from all the azimuthal-correlation moments. From an experimental perspective, its measurement is much easier with respect to standard azimuthal correlations typical of Mueller--Navelet analyses,
since it does not need to be investigated in the full $(-\pi, \pi)$ azimuthal-angle range.

Being \emph{de facto} a multiplicity, uncertainties coming from the selection of different FF sets as well as the ones from different replicas inside the same FF set are heavily dampened.
As shown in our plots (single-line envelops), they are much smaller than the ones related to the usual renormalization- and factorization-scale variation from 1/2 to 2 times their natural values (shaded bands).
We can see that the latter is larger at the lower reference value of $\Delta Y$, with its width visibly narrowing when going from $\Delta Y = 5$ to $7$.
This is a clear signal that the natural stability of the BFKL series gets more and more evident when $\Delta Y$ grows, as expected.

At variance with previous studies done in the context of inclusive light dijet~\cite{Ducloue:2013hia} and hadron-jet detections~\cite{Celiberto:2020wpk} at CMS, the azimuthal distribution for our process at FPF~$+$~ATLAS exhibits some unexpected and novel features. Indeed, its peak increases and its width shrinks when $\Delta Y$ increases. This seems to be counter-intuitive, since one of the main effects of the high-energy resummation is a loss of correlation between the two tagged objects due to the weight of rapidity-ordered inclusive-gluon emissions (labeled as $\cal X$ in our plots), that grows with $\Delta Y$.
A possible explanation could be that the strongly asymmetric rapidity windows where the ultra-forward FPF light meson and the central ATLAS $D^*$ are detected lead to a reduction of $y_{\pi,K}$ and $y_D$ combinations for the given $\Delta Y$ and, in turn, to a partial re-correlation in the final state. Furthermore, the considered kinematics directly translates into an asymmetry between the incoming-parton longitudinal fractions, with one of them  always large and the other one much smaller.
Here, possible threshold contaminations~\cite{Bonciani:2003nt,deFlorian:2005fzc,Muselli:2017bad} come into play and should be resummed as well.

All these observations support the statement that the combined tag of ultra-forward particles at the FPF and central ones in ATLAS via a tight timing coincidence setup brings along a high discovery potential and a concrete chance to widen our knowledge of strong interactions at high energies.
On the one hand, cross sections differential in the observed rapidity interval are able to disentangle BFKL from fixed-order calculations and can be used to assess the weight of the uncertainty coming from collinear FFs in ranges complementary to the currently accessible ones.
On the other hand, azimuthal-angle distributions are quite suitable observables with which to search for clear and novel high-energy signals, and can serve as a common basis to explore the interplay between BFKL and other resummation mechanisms.
Future studies will extend the present work to an opposite configuration, with heavy-flavored hadrons being tagged at the FPF.

\subsection{BFKL Phenomenology and Inclusive Forward Processes}\label{QCD:ssec:BFKL_if}

Single forward inclusive emissions in proton collisions see \cref{fig:processes_HE-QCD} (a), are characterised by an asymmetric configuration where a fast parton takes part in the hard subprocess always with an intermediate momentum fraction $x$, whereas the other parton is a low-$x$ gluon. In this case a hybrid high-energy/collinear factorization holds, so that the fast parton is described by a collinear PDF, whereas the low-$x$ gluon is described by a gluon uPDF. At LL, to get the hadronic cross section, the two distributions are convoluted with an off-shell ($g g^* \to g$) or ($q g^* \to q$) vertex corresponding to the emission of the forward particle.
On the other hand, gluon-induced single central inclusive emissions, see \cref{fig:processes_HE-QCD}) (b), lead to a pure high-energy factorization treatment. This stems from the fact that both the incoming gluons are extracted from the parent protons at low $x$. The cross section is then written as a convolution between two gluon uPDF and a doubly off-shell ($g^*g^*$) vertex initiating the emission of the central object.
The first analyses of the gluon uPDF were performed in the context of deep-inelastic-scattering (DIS) structure functions~\cite{Hentschinski:2012kr,Hentschinski:2013id}. Subsequently, the gluon uPDF was probed via the single exclusive leptoproduction of $\rho$ and $\phi$ mesons at HERA~\cite{Anikin:2009bf,Anikin:2011sa,Besse:2013muy,Bolognino:2018rhb,Bolognino:2018mlw,Bolognino:2019bko,Bolognino:2019pba,Celiberto:2019slj} and the EIC~\cite{Bolognino:2021niq}, the forward production of Drell--Yan pairs at LHCb~\cite{Motyka:2014lya,Brzeminski:2016lwh,Motyka:2016lta,Celiberto:2018muu}, and the exclusive production of heavy-quark pairs~\cite{Celiberto:2017nyx,Bolognino:2019yls} and quarkonia~\cite{Bautista:2016xnp,Garcia:2019tne,Hentschinski:2020yfm}. 
The connection between the gluon uPDF and the collinear gluon PDF was investigated through a high-energy factorization framework set up in~\cite{Catani:1990xk,Catani:1990eg,Collins:1991ty}, and via the Catani--Ciafaloni--Fiorani--Marchesini (CCFM) branching scheme~\cite{Ciafaloni:1987ur,Catani:1989sg,Catani:1989yc,Marchesini:1994wr,Kwiecinski:2002bx}. Then, first determinations of low-$x$ improved PDFs \emph{\`a la} Altarelli--Ball--Forte (ABF)~\cite{Ball:1995vc,Ball:1997vf,Altarelli:2001ji,Altarelli:2003hk,Altarelli:2005ni,Altarelli:2008aj,White:2006yh} were recently achieved ~\cite{Ball:2017otu,xFitterDevelopersTeam:2018hym,Bonvini:2019wxf}.

Further analyses of the gluon uPDF can be performed in the context of proton-proton collision, considering inclusive processes with forward production of (i) a light hadron, (ii) a heavy hadron, (iii) a heavy jet. The necessary vertices, written as the convolution of a hard partonic vertex and a suitable fragmentation function, are available with NLL accuracy in the case (i), and in the case (ii) when the VFNS can be adopted (see ~\cite{Ivanov:2012iv}). When this applies, a theoretical scheme must be used, where the NLL vertex is combined with a NLL gluon uPDF, which is straightforward for gluon uPDF models based on  BFKL, but is not trivial in other approaches.

The advent of the FPF will open a new window of opportunities to probe and study the hadronic structure at very low values of $x$.
Future data for reactions featuring single inclusive emissions of light as well as heavy hadrons detected in FPF ultra-forward rapidity ranges will be a key ingredient to perform stringent tests of the gluon uPDF.
The inclusion of these data in a larger set where the ones coming from other new-generation colliding facilities~\cite{AbdulKhalek:2021gbh,Chapon:2020heu,Arbuzov:2020cqg,Abazov:2021hku} are also collected is needed to trace the path towards the extraction of the gluon uPDF from a first global fit analysis.


\section{Modelling Forward Physics with Monte Carlo Event Generators}
\label{sec:qcd:forwardMC}

\subsection{Introduction}

The neutrino flux arriving at the FPF arises from the decay of a range of light (such as pions and kaons)
and heavy (in particular $D$-mesons) particles produced at very forward rapidities
in proton-proton collisions at the ATLAS interaction point.
The modelling of such particle production is not always amenable to the framework of
perturbative QCD, but rather requires a dedicated MC treatment. An accurate modelling of forward particle production by such event generators is therefore crucial to fully exploit the physics potential of the FPF, and conversely data from the FPF have the potential to constrain the  phenomenological models of non-perturbative QCD that are embedded in Monte Carlo event generators.
In the following sections dedicated studies in this direction, within the
widely used general-purpose \texttool{Pythia} and \texttool{Sherpa} MC generators, are described.
The white paper on event generators for high energy experiments~\cite{MCWhitepaper} complements some of the discussion here.
In addition,
we briefly explore the possibility to detect at the FPF neutrinos arising
from proton-lead (pPb) collisions at the LHC.

\subsection{Event Generation for Forward Particle Production with \texttool{Pythia~8} }

Here we present an updated study on forward charm and bottom
production with \texttool{Pythia~8}.
The reason for this study are two recent LHC results:
the LHCb observation of an asymmetry between forward $\Lambda_b$ and
$\overline{\Lambda}_b$ hadron production rates \cite{LHCb:2021xyh},
and the unexpectedly large $\Lambda_c/D$ ratio observed by ALICE
\cite{ALICE:2021dhb}, notably at small transverse momenta and at
high multiplicities \cite{ALICE:2021npz}.

\begin{figure}[tp]
\begin{minipage}[c]{0.49\linewidth}
\centering
\includegraphics[width=\linewidth]{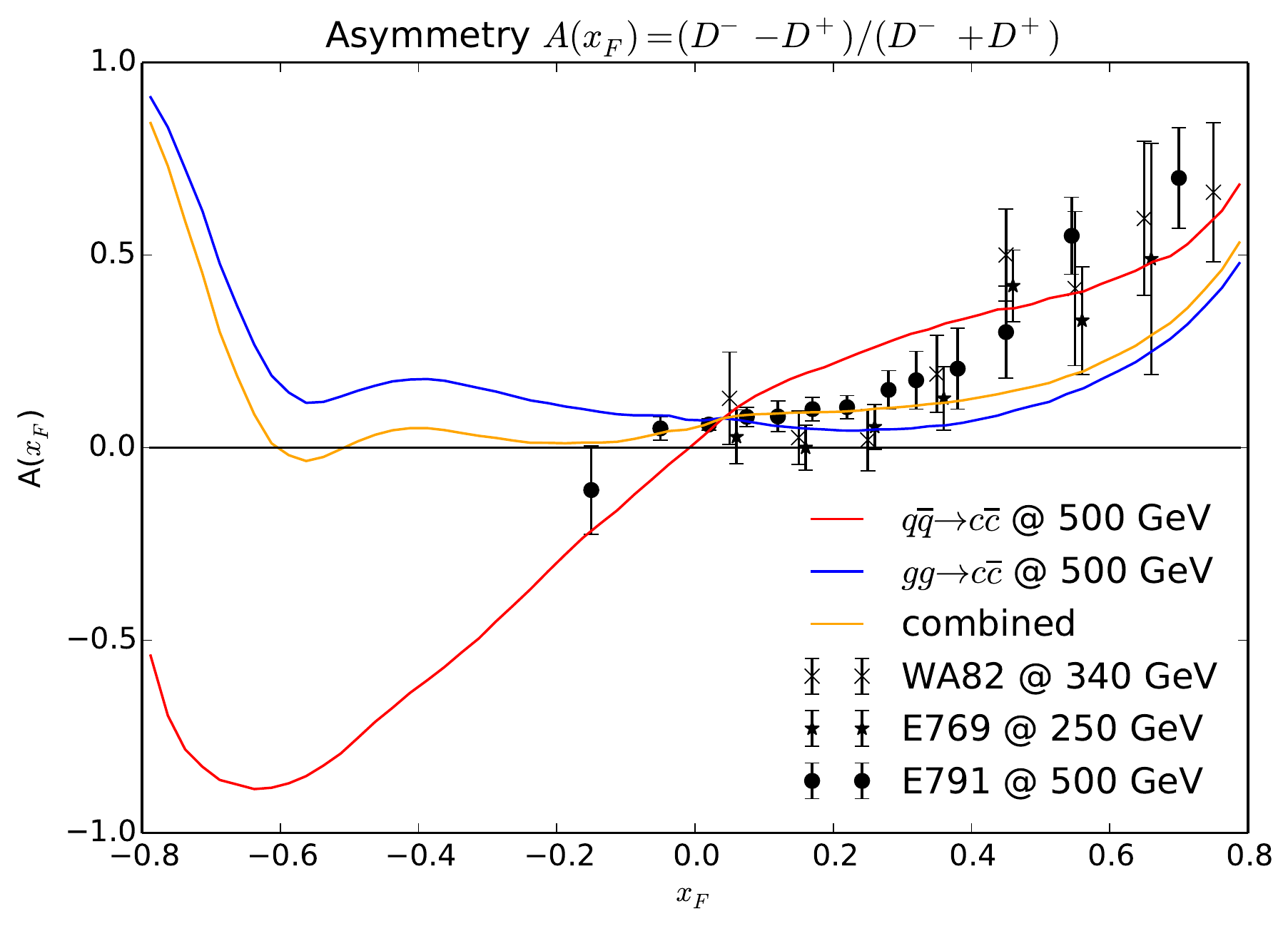}\\
(a)
\end{minipage}
\begin{minipage}[c]{0.49\linewidth}
\centering
\includegraphics[width=\linewidth]{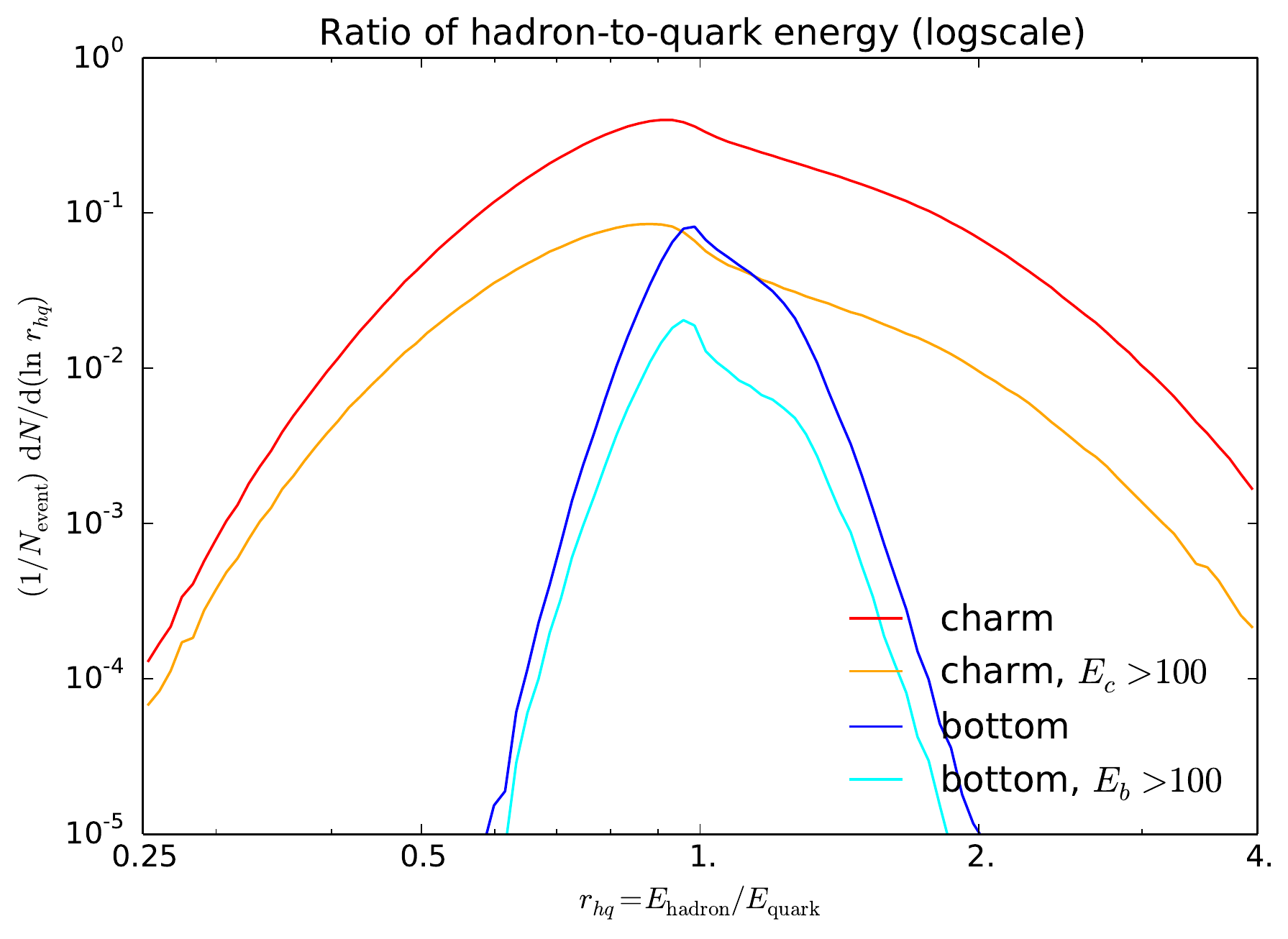}\\
(b)
\end{minipage}\\
\begin{minipage}[c]{0.49\linewidth}
\centering
\includegraphics[width=\linewidth]{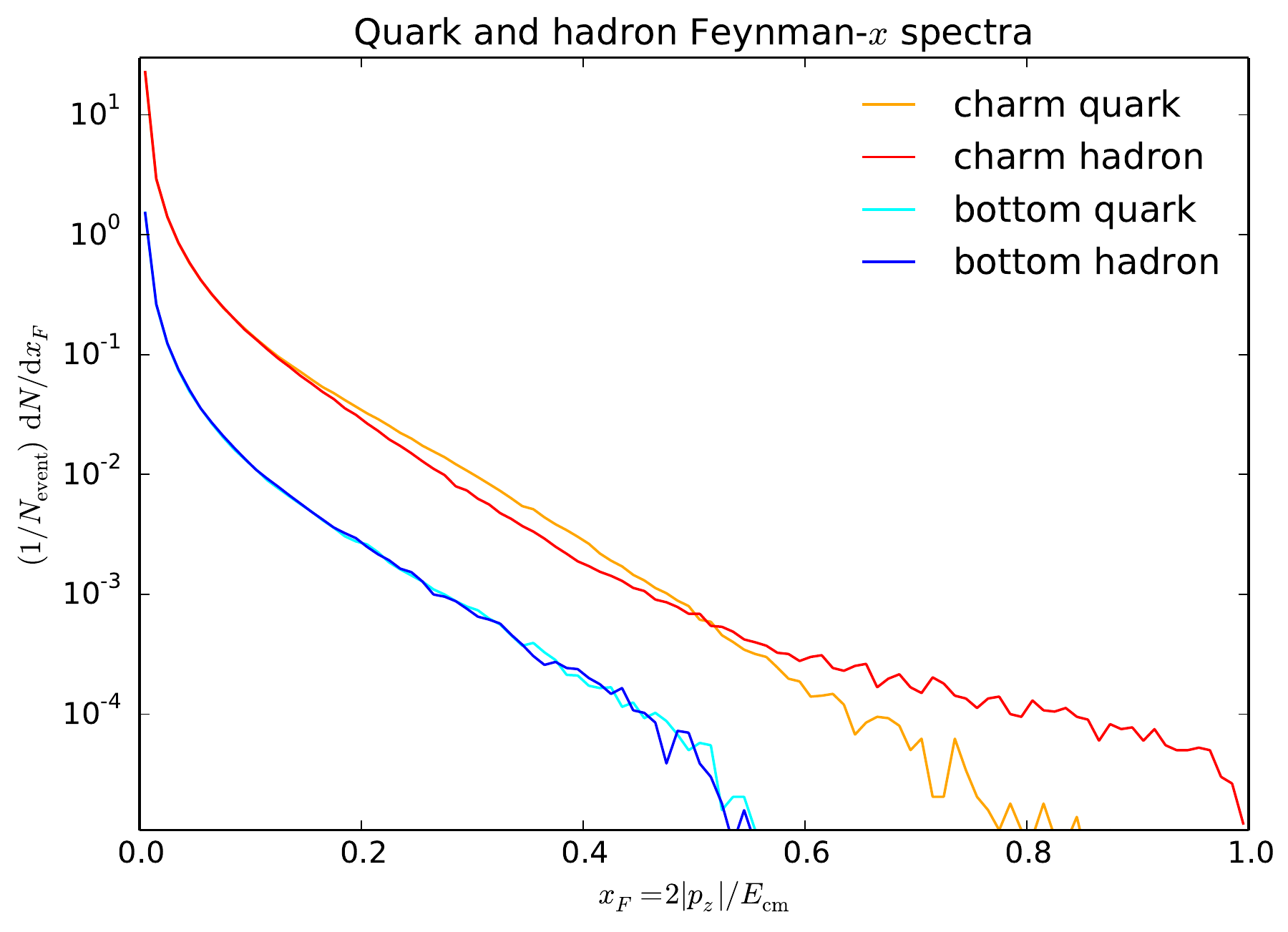}\\
(c)
\end{minipage}
\begin{minipage}[c]{0.49\linewidth}
\centering
\includegraphics[width=\linewidth]{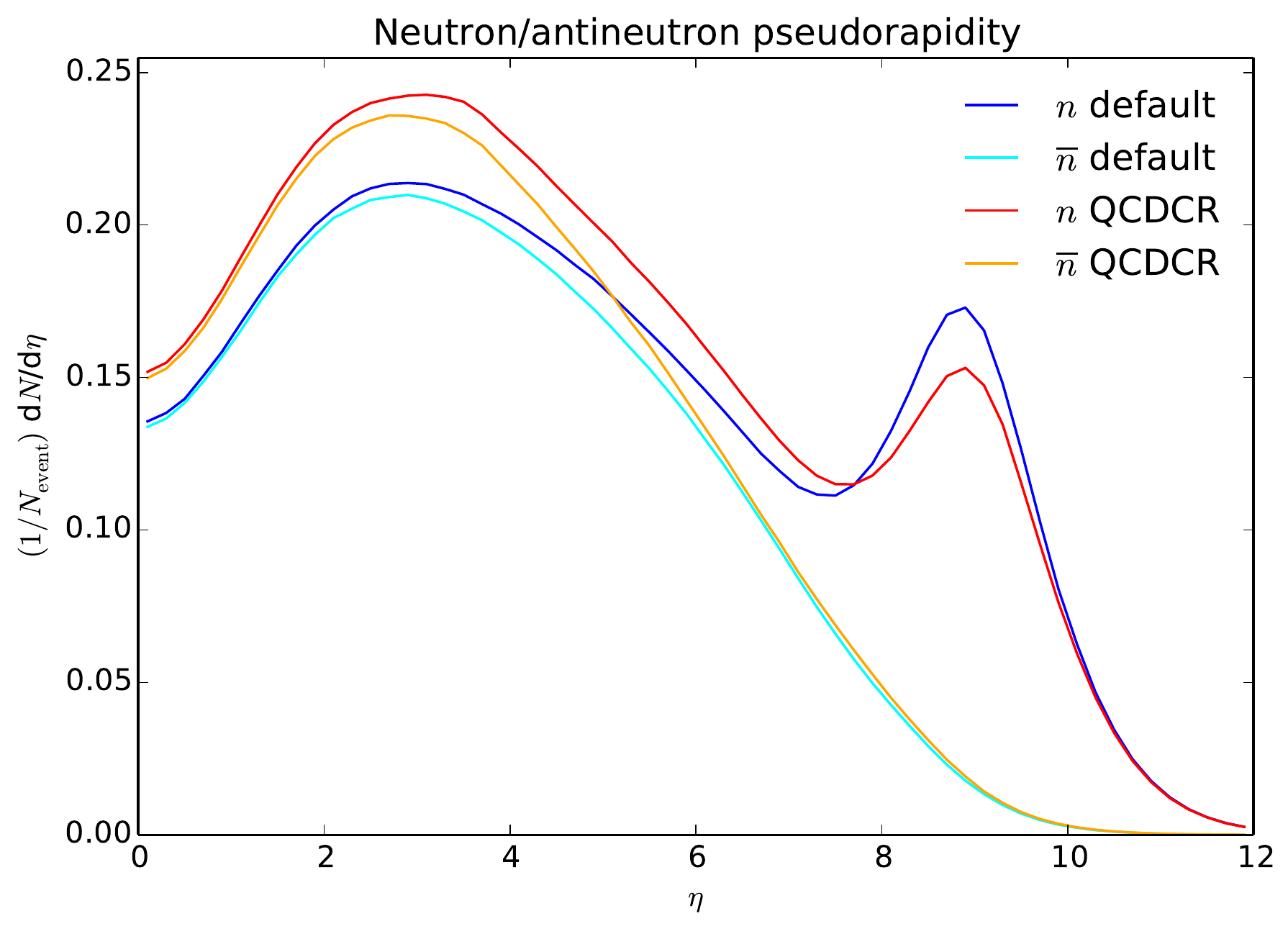}\\
(d)
\end{minipage}\\
\begin{minipage}[c]{0.49\linewidth}
\centering
\includegraphics[width=\linewidth]{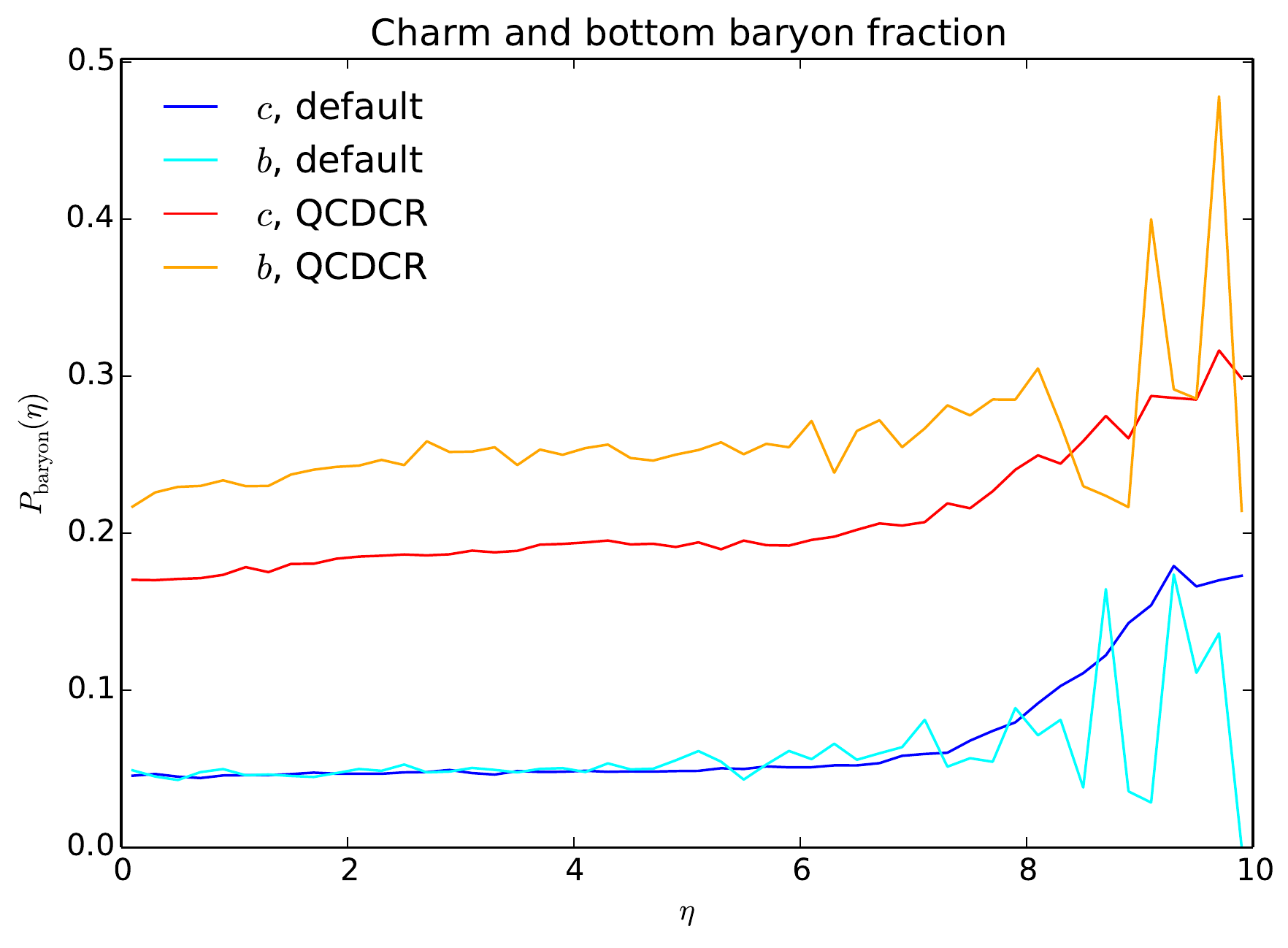}\\
(e)
\end{minipage}
\begin{minipage}[c]{0.49\linewidth}
\centering
\includegraphics[width=\linewidth]{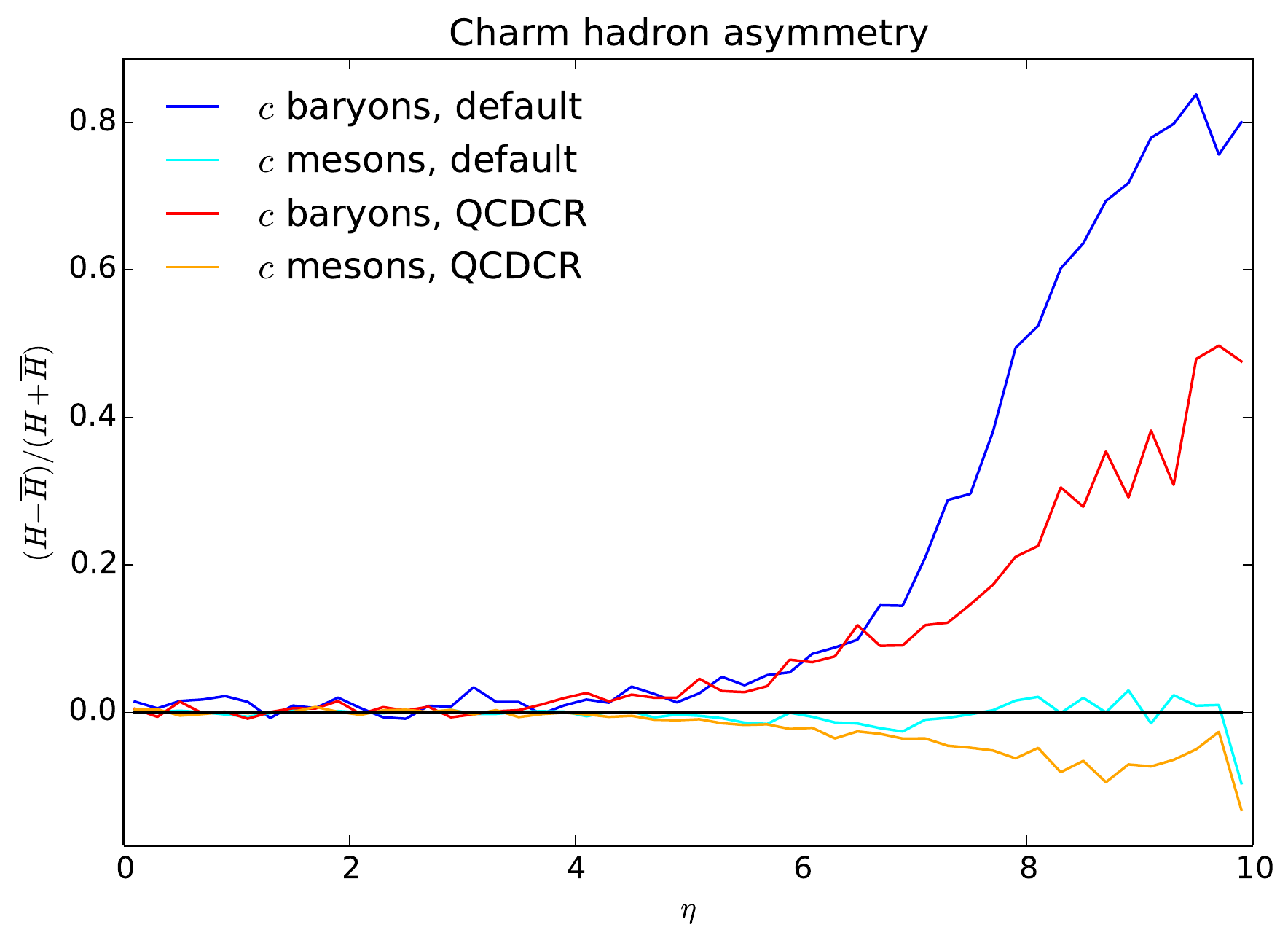}\\
(f)
\end{minipage}\\
\caption{(a) Charm hadron asymmetry in fixed-target $\pi^- p$ collisions.
Data from \cite{WA82:1993ghz,E769:1993hmj,E791:1996htn}.
(b) Ratio of the heavy-flavour hadron energy to its mother
heavy-quark energy, for all and only for those with an initial energy 
above 100 GeV.
(c) Feynman-$x$ spectra for $c/b$ quarks and hadrons.
(d) Neutron and anti-neutron pseudorapidity distributions.
(e) Charm and bottom baryon fractions.  
(f) Charm baryon and meson production asymmetries.
The latter five frames are all for an inclusive sample of inelastic
$pp$ events at 13 TeV.}
\label{figTS:pythiaRes}
\end{figure}

Production asymmetries were observed at fixed-target energies in the
90s \cite{WA82:1993ghz,E769:1993hmj,E791:1996htn}, and were
explained in terms of the colour string topologies in the relevant
processes \cite{Norrbin:1998bw}. An update of this study
is shown in \cref{figTS:pythiaRes}a. In $\pi^- A$ collisions,
where $A$ is a nuclear target, a $\overline{u}u \to c\overline{c}$
valence-quark annihilation gives a string stretched from the $\pi^-$
remnant $d$ quark to the $\overline{c}$, while the $c$ is connected to
a $p/n$ remnant $ud/uu$. This means that the string tension will pull the
$\overline{c}$ forwards and the $c$ backwards. One may even have cases
where a string system is so low-mass that it collapses to a single hadron,
notably $\overline{c} + d \to D^-$. This then gives an excess of $D^-$
over $D^+$ in the forward hemisphere. In the alternative process
$gg \to c\overline{c}$, which dominates in the combined sample, either
the $c$ or the $\overline{c}$ are pulled forwards, but in a collapse
the $c$ would give a $D^0$, so also in this case $D^->  D^+$. The current
\texttool{Pythia~8} default \cite{Sjostrand:2014zea} describes data somewhat
worse than \texttool{Pythia~6} \cite{Sjostrand:2006za} did, as minor changes
have been made over the years without any regard for these data. The old
studies also led to predicted charm and bottom asymmetries at the per cent
level for LHC \cite{Norrbin:2000zc}.   

String effects are not always taken into account for heavy flavour modelling.
At $e^+e^-$ colliders a  $c/b$ quark usually is the leading parton 
of its jet (but not always, which is important to remember in a detailed
analysis of $e^+e^-$). Thus the string tension pulls back on it, such that
the  resulting heavy hadron has a smaller energy than the mother heavy
quark. This behaviour is often parameterized in terms of a  fragmentation 
function $f(z)$, where $z$ is the ratio of the heavy hadron to the 
original heavy quark energy (or momentum), $0 \leq z \leq 1$. 
Typical values are $\langle z_c \rangle \approx 0.6$ 
and $\langle z_b \rangle \approx 0.8$, somewhat depending on the scale 
at which the perturbative process is matched to the nonperturbative 
hadronization \cite{ParticleDataGroup:2020ssz}. But, as we have already
seen, at hadron colliders the colour strings almost as often connect
$c/\overline{c}/b/\overline{b}$ to the beam remnants ``ahead'' of them,
and then these quarks are pulled forwards by the string tension, rather
than backwards. In \cref{figTS:pythiaRes}b, we show that the
hadron-to-quark energy ratio peaks slightly below unity, but with broad
wings, notably extending well above unity. The pull/push effect is seen
to be larger for the lighter $c$ than for the heavier $b$, as could be
expected. The subset with quark energy above 100 GeV demonstrates that
this is not a phenomenon specific to lower energies.  The outcome of this
double-sided smearing is that the quark and hadron longitudinal momentum
spectra almost coincide, \cref{figTS:pythiaRes}c. (More precisely, the 
comparison applies to partons at the end of the parton-shower evolution
and primary hadrons before any subsequent decays.) For charm we may
even note a hadron excess at the largest $x_F$, likely induced by the
above-mentioned collapse with a beam-remnant parton. Thus one concludes that 
LEP-style fragmentation functions are not applicable at hadron colliders.

The other new result, the enhancement of charm baryon production, was
not predicted as such, but ALICE noted that it is consistent with
what comes out of a non-default \texttool{Pythia} option. This option is
the so-called QCD-inspired colour reconnection (QCDCR) model of
Christiansen and Skands \cite{Christiansen:2015yqa}. The key point,
relative to the default colour reconnection modelling, is that it allows
the formation of junctions and antijunctions, as follows.

In its simplest form, think of a Y-shaped string topology, with
a quark at each end and a junction in the middle, where the three
strings come together. It is the junction that carries the net baryon
number of the system. Junctions can form notably when several strings
are drawn out along the collision axis in a $pp$ event. If you imagine
two strings, each with a quark end moving out in the $+z$ direction
and an antiquark along the $-z$ one, then the two strings can collapse
into one over most of the distance between the endpoints. There the
colour now flows in the opposite direction according to the simple
rule-of-thumb that red + green = antiblue, and the energy is reduced
by only having to draw out one string instead of two. Near the $qq$
end a junction is formed, which gives a baryon, and in the other end
an antijunction gives an antibaryon. This mechanism adds to the normal
baryon production one, where a single string breaks by the production
of a diquark--antidiquark pair. The junction mechanism increases in
importance when there are many nearly parallel strings, which means
at low transverse momenta in high-multiplicity events, consistent
with data.

The most obvious consequence of junction formation is that baryon
production is enhanced over the rate obtained by normal string breaks. 
The example of (anti)neutron production is shown in
\cref{figTS:pythiaRes}d, where the QCDCR is compared with the
default colour reconnection scenario. One may note that the enhancement
mainly is at central pseudorapidities $\eta$, not so much at forward ones.
Further studies could here be useful, e.g.\ to bin neutrons both in  
energy and $\eta$, to allow comparisons with LHCf data \cite{LHCf:2015nel}.
Furthermore the model may need further development and tuning in the
extreme forward region, as has recently been done for the default
scenario, \cref{subsection:improvedMC}. (Such improvements are 
not used here, to allow a fair comparison.)

The big ALICE surprise then is the significant fraction of all
(anti)charm quarks that end up in an (anti) baryon. This fraction
is shown in \cref{figTS:pythiaRes}e, for the default and QCDCR
models, where the former is consistent with an extrapolation from
$e^+e^-$ data but then inconsistent with ALICE, while the QCDCR is
about a factor three higher for charm and four for bottom. In both
scenarios the baryon fraction is increasing in the forward direction,
presumably as a consequence of the aforementioned string pull and
collapse mechanisms. (The results for bottom are marred by tiny statistics
for $\eta > 8$, but there is no reason to expect a qualitatively
different behaviour than for charm.) This is confirmed by the
hadron--antihadron asymmetries, \cref{figTS:pythiaRes}f,
which shows a significant excess of charm baryons over antibaryons,
consistent with a collapse with a remnant diquark. The asymmetry
is smaller for QCDCR, where the junction topologies would not involve
a remnant diquark, but still possibly a single remnant quark.
The asymmetries for baryons are compensated by opposite ones in the
meson sector, though less so in relative terms since there are more
of them. Asymmetries for bottom hadrons resemble those for charm,
but are somewhat higher, within the limited statistics at disposal.
We recall that the observed LHCb $\Lambda_b - \overline{\Lambda}_b$
asymmetries \cite{LHCb:2021xyh} are consistent with the QCDCR option,
but well below the default one.

In summary, it is important to recognize that perturbation theory
alone may not a good predictor of forward charm and bottom hadron
production, but must be combined with nonperturbative modelling.
This modelling involves uncertainties, that can be mitigated
by future studies, in a combination of experiment and theory.
Thus again improving our understanding of QCD is paramount in order to improve the accuracy
of (anti-)neutrino flux predictions at the FPF. 


\subsection{Event Generation for Forward Particle Production with \texttool{Sherpa} }
\label{sec:forward_generators_sherpa}

\texttool{Sherpa} is a general-purpose event generator developed initially for the Large Electron Positron collider (LEP), 
and subsequently extended to the physics case of HERA and the Large Hadron Collider (LHC)~\cite{Gleisberg:2003xi,Gleisberg:2008ta,Sherpa:2019gpd}.
The event generation framework is centered around two independent matrix element generators, \texttool{AMEGIC}~\cite{Krauss:2001iv} and \texttool{COMIX}~\cite{Gleisberg:2008fv}, and two independent parton showers, \texttool{CSShower}~\cite{Schumann:2007mg} and \texttool{DIRE}~\cite{Hoche:2015sya}, contains an implementation of the Sj{\"o}strand-Zijl multiple scattering model~\cite{Alekhin:2005dx}, a cluster hadronization model~\cite{Winter:2003tt}, a hadron and $\tau$ decay module, and an implementation of the Yennie-Frautschi-Suura algorithm for soft-photon resummation~\cite{Schonherr:2008av}. 

\texttool{AMEGIC} and \texttool{COMIX} can be employed to compute tree-level scattering processes fully differentially in nearly arbitrary physics models~\cite{Christensen:2009jx,Hoche:2014kca,Krauss:2016ely} with the help of \texttool{FEYNRULES}~\cite{Alloul:2013bka} and \texttool{UFO}~\cite{Degrande:2011ua}.
They can also be employed to perform fixed-order next-to-leading order calculations in QCD~\cite{Gleisberg:2007md} and electroweak theory~\cite{Schonherr:2017qcj} using the Catani-Seymour dipole subtraction scheme~\cite{Catani:1996vz,Catani:2002hc}, and its extension to QED~\cite{Dittmaier:1999mb} and EW. 
Some of the most challenging high-multiplicity NLO QCD and EW calculations have been computed with the help of this framework~\cite{Berger:2010zx,Bern:2013gka,Badger:2013yda,Hoche:2016elu}.

The parton showers employed in \texttool{Sherpa} are based on the Catani-Seymour dipole factorization approach~\cite{Catani:1996vz}, can be merged with higher-order tree-level matrix element calculations using the CKKW(L) techniques~\cite{Catani:2001cc,Lonnblad:2001iq,Hoeche:2009rj}, and have been matched to NLO QCD calculations through both the \texttool{MC@NLO}~\cite{Frixione:2002ik} and \texttool{POWHEG}~\cite{Nason:2004rx} techniques~\cite{Hoeche:2011fd,Hoche:2010pf}.
For a small selection of important reactions, it has also been matched to NNLO QCD calculations~\cite{Hoche:2014uhw,Hoche:2018gti}.
Merging with matched next-to-leading order calculations is supported through the \texttool{MEPS@NLO} method~\cite{Gehrmann:2012yg,Hoeche:2012yf}, which has been successfully applied to a range of important physics processes~\cite{Hoeche:2014lxa,Cascioli:2013gfa,Hoeche:2014qda,Bellm:2019yyh,Buckley:2021gfw}.

Hadronization is performed in \texttool{Sherpa} using the cluster fragmentation model~\cite{Webber:1983if}, which has been
extended to include color reconnection effects~\cite{Winter:2003tt,Krauss:2022bb}. 
Hadron decays are simulated through a dedicated hadron decay module, which also handles $\tau$ decays. 
Spin correlation effects and $B$-mixing effects are included in the decay simulation.

The radiation of soft photons, either in the hard reaction and the subsequent decay chain, or in hadron decays,
is simulated through an implementation of the Yennie-Frautschi-Suura approach to soft-gluon resummation.
Complete next-to-leading-order corrections are included for the most relevant cases of hard decays~\cite{Krauss:2018djz},
and for some selected hadron decays~\cite{Bernlochner:2010fc}.
\begin{figure}[t]
  \begin{minipage}[t]{0.575\textwidth}
  \includegraphics[width=\linewidth]{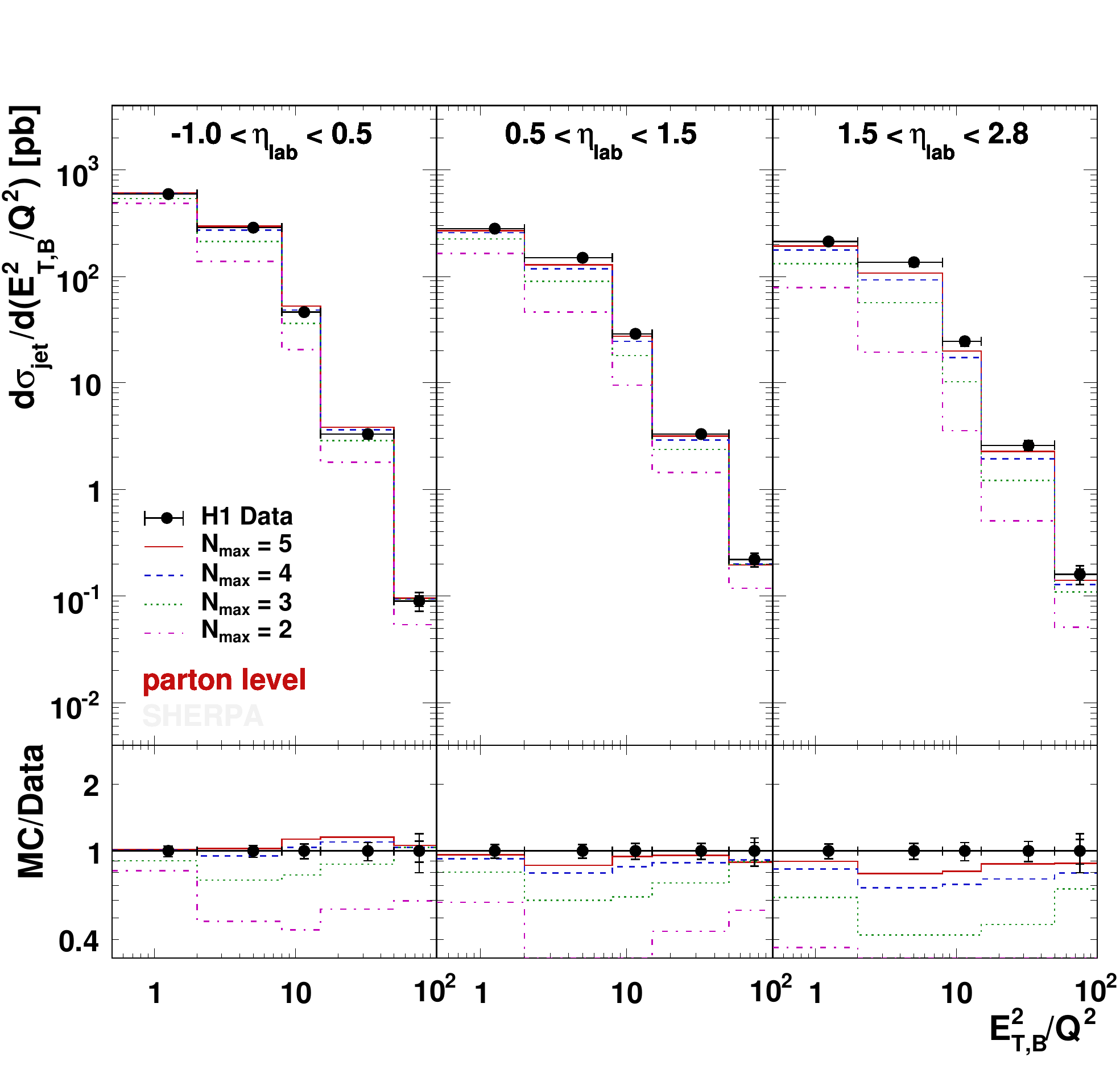}
  \end{minipage}\hfill
  \begin{minipage}[t]{0.375\textwidth}
  \includegraphics[width=\linewidth]{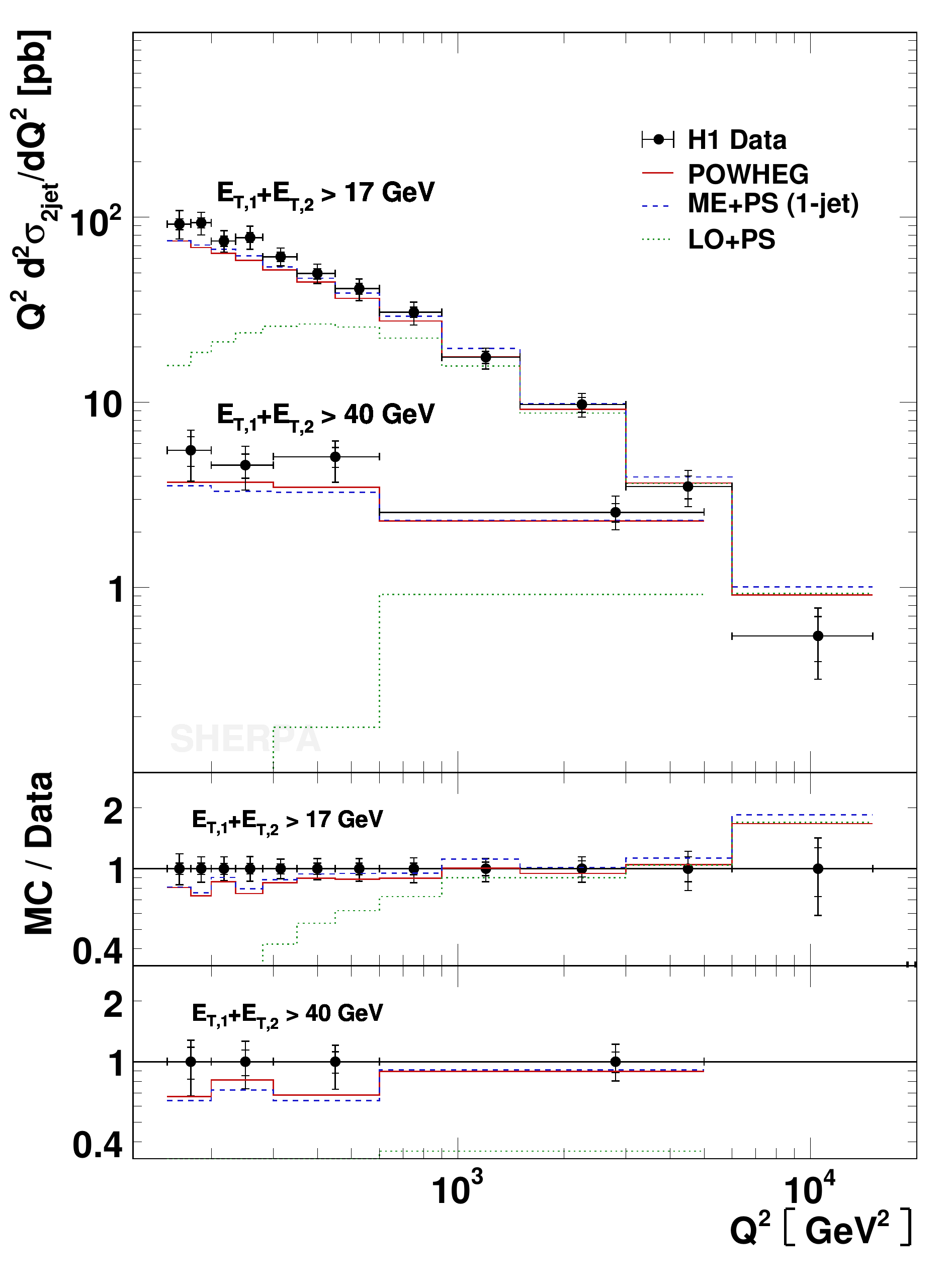}
  \end{minipage}
\caption{Left: The inclusive jet cross section as a function of $E_{T,B}^2/Q^2$ in bins                        
  of $\eta_{lab}$, measured by the H1 collaboration~\protect\cite{Adloff:2002ew}.
  Right: The di-jet cross section as a function of $Q^2$ in bins of $E_{T,1}+E_{T,2}$,
  measured by the H1 collaboration~\protect\cite{Adloff:2000tq}.
  \label{fig:sherpa:dis}}
\end{figure}
In \cref{fig:sherpa:dis} we show two examples of \texttool{Sherpa}'s physics performance in a process
relevant to the FPF: Deep-inelastic scattering. 
The left panel compares multi-jet merged predictions from~\cite{Carli:2009cg} to inclusive jet cross sections measurements from the H1 collaboration~\cite{Adloff:2002ew}.
The data are given as a function of the ratio of the jet transverse energy in the Breit frame and $Q^2$, and in different $\eta_j$ bins.
In the region $E_{T,B}^2/Q^2>1$ the reaction probes the regime where the QCD real-emission corrections cannot be approximated by a resummed calculation as provided by the parton shower, because the resummation scale $Q^2$ is too low to allow for the production of a measurable jet.
This situation is remedied by explicitly including higher-multiplicity tree-level calculations.
It is noteworthy that the corrections obtained through multi-jet merging only saturate at a jet multiplicity of four to five. 
The right panel of \cref{fig:sherpa:dis} shows predictions from \texttool{POWHEG} matching in \texttool{Sherpa} and from multi-jet merging for the di-jet cross section as a function of $Q^2$ in bins of  $E_{T,1}+E_{T,2}$, compared to measurements by the H1 collaboration~\protect\cite{Adloff:2000tq}.
It is interesting to observe the large discrepancy with the leading-order plus parton shower result, which is again due to the fact that a parton shower with resummation scale $Q^2$
is not able to populate the complete phase space relevant to di-jet production, particularly at low $Q^2$.

\begin{figure}
   \includegraphics[scale=0.32,angle=270,clip]{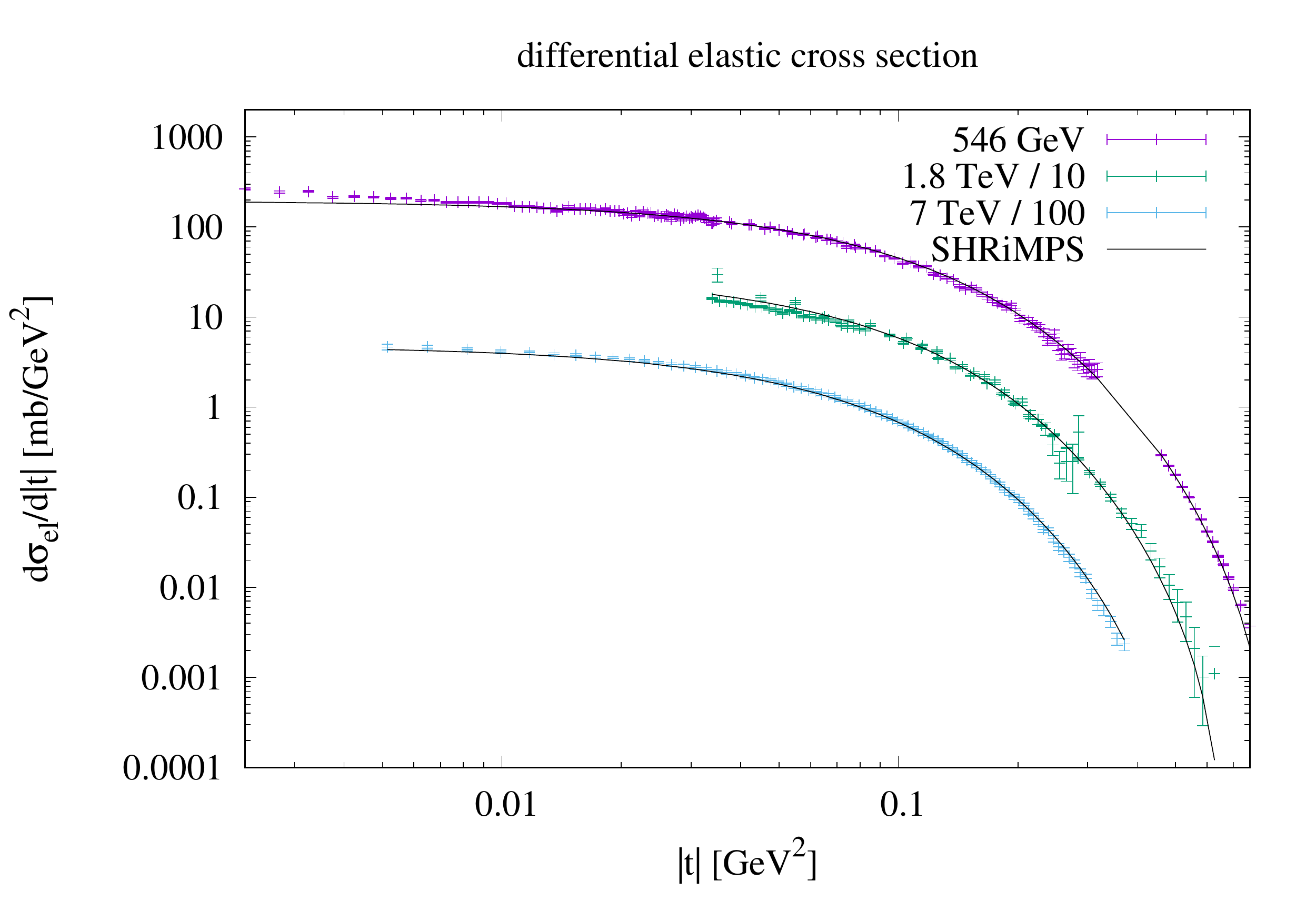}
   \includegraphics[scale=0.32,angle=270,clip]{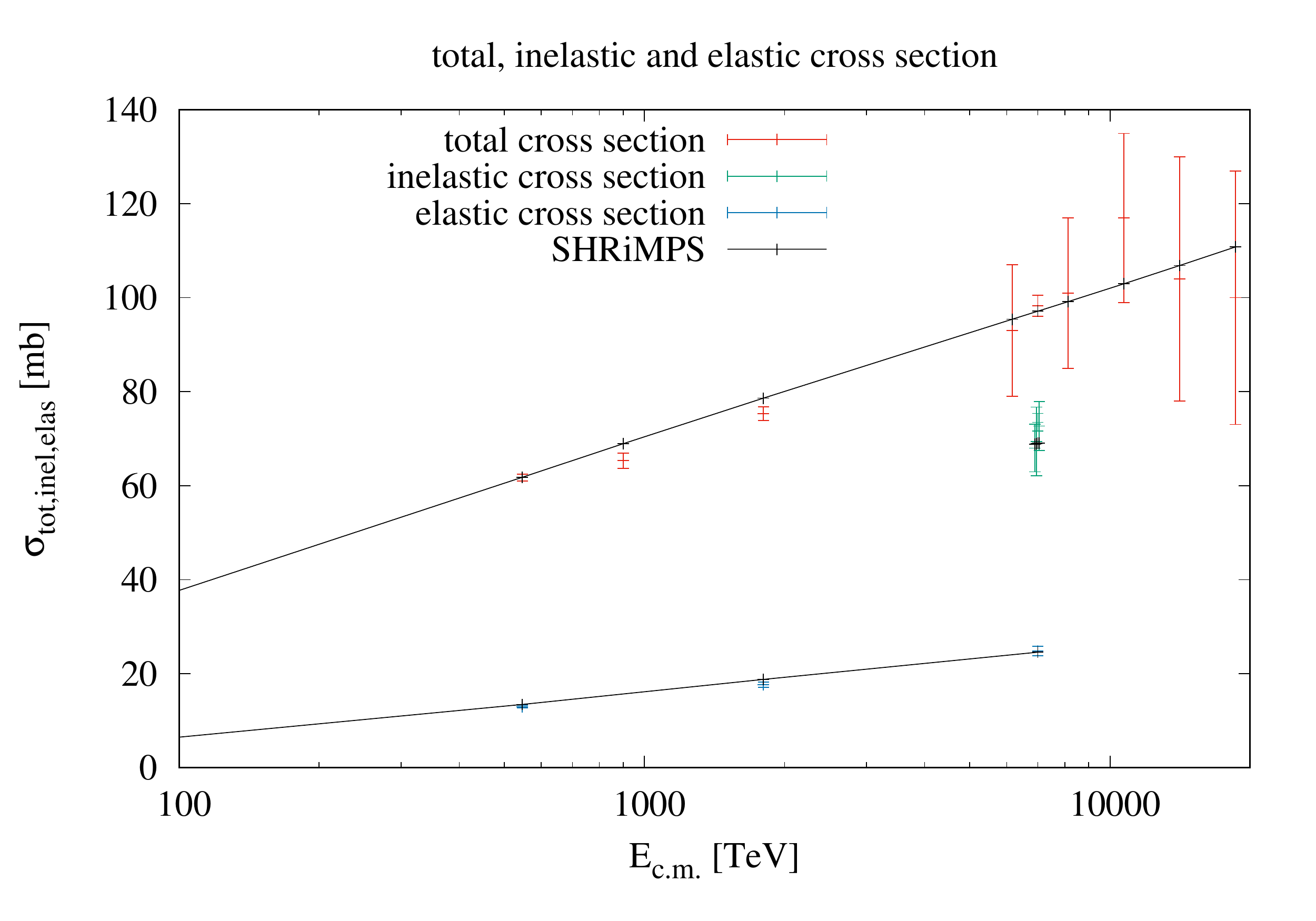}
\caption{Integrated cross sections (left) and differential elastic cross section (right) for different center-of-mass energies.}
\label{fig:shrimps_el}
\end{figure}

Finally, \texttool{Sherpa} includes a simulation of inclusive QCD scattering, \texttool{SHRiMPS}~\cite{Krauss:2022aa}, based on the Khoze--\-Martin--\-Ryskin (KMR) model~\cite{Ryskin:2009tj} and an extension of its inclusive picture to the creation of exclusive final states.
The KMR model is constructed from a simplified partonic picture of the Pomeron, together with multi-Pomeron interactions described by an effective triple Pomeron vertex. Using the notion of Good-Walker states and associated parton densities which evolve in rapidity, it successfully describes total, elastic, and diffractive cross sections in $pp$ and $p\bar{p}$ collisions at high energies.
The KMR model is an eikonal model in which the eikonals $\Omega_{ik}(y, B_\perp)$ between states $i$ and $k$ in mixed rapidity--\-impact--\-parameter space are related to the two parton densities of the incoming hadrons through
\begin{equation}\label{Eq:SingleChannel_WithoutKT}
    \Omega_{ik}(Y,\,B_\perp) \;=\; 
    \frac{1}{\beta_0^2}\,
    \int{\rm d}^2 b_\perp^{(1)}{\rm d}^2 b_\perp^{(2)}
    \delta^2\left(\vec B_\perp-\vec b_\perp^{(1)}+\vec b_\perp^{(2)}\right)
    \Omega_{i(k)}\left(y,\,b_\perp^{(1)}\right)
    \Omega_{(i)k}\left(y,\,b_\perp^{(2)}\right)\,.
\end{equation}
The parton distributions of hadron $i$ in the presence of the other hadron $k$ evolve as
\begin{equation}\label{Eq:DEQs_WithoutKTDep}
  \begin{split}
    \frac{{\rm d}\Omega_{i(k)}(y,\,b_\perp^{(1)},\,b_\perp^{(2)})}{{\rm d} y} \;=\;& 
    +\mathcal{W}_{\text{abs}}^{(ik)}(y,\,b_\perp^{(1)},\,b_\perp^{(2)})
    \;\cdot\Delta\cdot\;\Omega_{i(k)}(y,\,b_\perp^{(1)},\,b_\perp^{(2)})\\
    \frac{{\rm d}\Omega_{(i)k}(y,\,b_\perp^{(1)},\,b_\perp^{(2)})}{{\rm d} y} \;=\;& 
    -\mathcal{W}_{\text{abs}}^{(ik)}(y,\,b_\perp^{(1)},\,b_\perp^{(2)})
    \;\cdot\Delta\cdot\;\Omega_{(i)k}(y,\,b_\perp^{(1)},\,b_\perp^{(2)})\,,
  \end{split}
\end{equation}
with a parton absorption contribution given by a term like 
\begin{equation}\label{Eq:Wabs}
  \begin{split}
    \mathcal{W}_{\text{abs}}^{(ik)}(y,\,b_\perp^{(1)},\,b_\perp^{(2)}) \;=\;&
    \left\{\frac
        {1-\exp\left[-\frac{\lambda}{2}
            \Omega_{i(k)}(y,\,b_\perp^{(1)},\,b_\perp^{(2)})\right]}
        {\frac{\lambda}{2}\Omega_{i(k)}(y,\,b_\perp^{(1)},\,b_\perp^{(2)})}
        \right\}
    \left\{\frac
        {1-\exp\left[-\frac{\lambda}{2}
            \Omega_{(i)k}(y,\,b_\perp^{(1)},\,b_\perp^{(2)})\right]}
        {\frac{\lambda}{2}\Omega_{(i)k}(y,\,b_\perp^{(1)},\,b_\perp^{(2)})}
        \right\}\,.
  \end{split}
\end{equation}
The boundary conditions at maximal rapidities are given by form factors as
\begin{equation}\label{Eq:BoundaryConditionsWithoutKT}
  \begin{split}
    \Omega_{i(k)}(-Y/2,\,b_\perp^{(1)}) \;=\;& F_i(b_\perp^{(1)})\\
    \Omega_{(i)k}(+Y/2,\,b_\perp^{(2)}) \;=\;& F_k(b_\perp^{(2)})\,.
  \end{split}
\end{equation}
The parameters $\Delta$ and $\lambda$ are related to the pomeron intercept and triple-pomeron coupling, respectively, and assumed to be constant.  
While the original KMR model included three Good-Walker states and their interactions through pomerons and reggeons, its current \texttool{SHRiMPS} implementation is based on two such states and pomeron interactions only.  
Naively, one would identify the two Good-Walker states -- diffractive eigenstates -- with the nucleon $N = \{p,\,n\}$ and its first resonance $N^* = N(1440)$; however, in the \texttool{SHRiMPS} version this second eigenstates is modelled as a linear combination of $N(1440)$, $N(1700)$, and an exponentially falling distribution of a diffractive continuum.  
The latter is described by disintegrating the nucleon into a quark--\-di-quark state with a suitable mass, which is subsequently hadronized, thereby creating a spray of mostly forward hadrons. The integrated cross sections and differential elastic cross sections obtained with SHRiMPS are shown in Figure~\ref{fig:shrimps_el} for different center-of-mass energies.

The KMR model also describes inelastic cross sections, {\it i.e.}\ the inelastic production of hadronic final states, in an inclusive way.
To translate these inclusive cross sections into exclusive final states, the \texttool{SHRiMPS} model employs the notion of "cut pomerons" in a naive way.
They are represented as "ladders" with $t$-channel propagators connecting gluon "rungs".  
In each inelastic event at impact parameter $B_\perp$, selected according to the corresponding differential probability, the number of ladders exchanged between the two hadrons is selected according to a Poisson distribution of the eikonal, and their positions are taken from the integrand of \cref{Eq:SingleChannel_WithoutKT}. 
The density of emitted gluons (the "rungs") in rapidity is determined by the evolution equation \cref{Eq:DEQs_WithoutKTDep}. 
As a consequence of the triple-pomeron vertex the cut ladders can have sections that consists of un-cut pomerons which results in the $t$-channel propagators in the ladder being either in a colour singlet or an octet state, with probabilities for the different colour states determined by the eikonals. 
The transverse momenta of either the propagators or the emitted gluons are chosen from a Regge-inspired form, the gluons' momenta are then determined by momentum conservation. The emission of quark--anti-quark pairs from the ladders is also included.

\subsection{Improved MC Generation of Forward Particle Production }
\label{subsection:improvedMC}

The FPF has the potential to probe precision Standard Model (SM), as well as Beyond Standard Model (BSM) physics. To adequately test these theories at the FPF, a precise understanding of the SM's predictions in the forward direction is needed, e.g. the incoming neutrino or dark photon flux toward the FPF. One of the most commonly used Monte-Carlo (MC) event generators for predictions at the Large Hadron Collider (LHC) is \texttool{Pythia}~\cite{Sjostrand:2014zea}. Although most of the physics driving \texttool{Pythia} is computed from quantum field theory, effects like hadronization, intrinsic transverse momentum of partons, and the fate of the proton remnants after a collision must be simulated with effective models due to the breakdown of perturbative methods. The models inevitably introduce parameters, the numerical values of which need to be adjusted ("tuned") in order to yield predictions that are in agreement with experimentally observed data. The current default tune, called  "Monash"~\cite{Skands:2014pea}, has been shown to produce excellent agreement with central measurements. A drawback of Monash (and other common tunes) is that it has not been designed to agree with forward ($|\eta|>$ 5) physics data, leading to forward predictions of lower fidelity compared to their central counterpart. The forward physics data that has been obtained at the LHC, notably from the LHCf collaboration, is in strong disagreement with \texttool{Pythia}'s predictions~\cite{Anchordoqui:2021ghd, Ene:2019drk}.
This problem is not exclusive to \texttool{Pythia} and applies to other MC generators, too~\cite{Kling:2021gos, Ene:2019drk}. Resolving these issues is therefore paramount for the FPF to probe BSM and SM physics adequately. We present an ongoing effort to tune \texttool{Pythia} with the forward measurements that are currently available while trying the retain the excellent quality of \texttool{Pythia}'s predictions for central physics. Below, we will discuss the data that is being used to obtain the tune, the parameters that we fit to reproduce forward data, our methods, and some preliminary results. 

The LHCf collaboration has probed the most forward physics regime (largest $|\eta|$) at the LHC to date, producing measurements of neutron and pion production for $|\eta|>10.76$. In addition to LHCf there are the TOTEM~\cite{Aspell:2012ux,TOTEM:2013pio} and CASTOR~\cite{Chatrchyan:2013gfi} collaborations which have measured the pseudorapidity densities of charged particles and energy flows at $5.3<|\eta|<6.5$ and $5.2<|\eta|<6.6$, respectively. Finally, central measurements from ATLAS~\cite{Aad:2011eu, ATLAS:2012djz}, CMS~\cite{Khachatryan:2015gka}, and ALICE~\cite{Abelev:2012sea}, sensitive to diffractive and minimum bias physics are used to verify the compatibility of our forward tune with central physics. In \cref{table:tuninganalyses}, we summarize the LHCf analyses, their energies, references, and \texttool{RIVET}~\cite{Buckley:2010ar} modules. 
\begin{table}[h!]
  \centering
  \begin{tabular}{l||c|c|c}
  \toprule
  \hline
  \hline
	{\bf Analysis} 
	& {\bf $\sqrt{s}$}~[\tev] 
	& {\bf  Refs. }           & {\bf RIVET}  \\ 
    \hline
    \hline
    forward $\pi^0$ or $\gamma$
    & $7$ &  \cite{LHCf:2012mtr}  & {\tiny LHCF\_2012\_I1115479} 
    \\
    &$2.76$, $7$& \cite{LHCf:2015rcj}  & {\tiny LHCF\_2016\_I1385877} 
      \\
    & $13$ &  \cite{LHCf:2017fnw} & {\tiny LHCF\_2018\_I1518782}
    \\
    \hline
    forward neutrons
    & $7$ &  \cite{LHCf:2015nel} & {\tiny LHCF\_2015\_I1351909}\\
    &$13$ &  \cite{LHCf:2018gbv} & {\tiny LHCF\_2018\_I1692008}
    \\
  \hline
  \hline
  \end{tabular}
  \caption{LHCf analyses, covering $\eta>8.81$, their COM energies, references, and \texttool{RIVET} module.
  } 
\label{table:tuninganalyses}
\end{table}

The Monash tune produces too few hard neutrons and too many hard pions in the forward region. Our approach to resolving this has been to disable \texttool{Pythia}'s popcorn mechanism for beam remnants, effectively preventing remnant diquarks from hadronizing into a meson, and simultaneously tuning the fragmentation function to yield harder diquarks. As a result, more hard neutrons and fewer pions are produced, and giving much better agreement with LHCf data. Tuning these parameters affects the small $|\eta|$ predictions only marginally. It has further been noted \cite{Anchordoqui:2021ghd,Sjostrand:2021dal} that the simulation of transverse momentum, $p_{T}$, is  important as well as a larger (smaller) $p_{T}$ will lead to fewer (more) hadrons in the forward direction. Primordial $p_{T}$, which is defined as the $p_{T}$ of the partons due to their motion within an interacting hadron or beam remnant, is a main source of transverse momentum and has been useful in fitting forward measurements.  

The tuning requires sampling \texttool{Pythia} in the multidimensional parameter space of the above mentioned parameters. The MC events from \texttool{Pythia}, are analysed with \texttool{RIVET}~\cite{Buckley:2010ar} modules for a given analysis, mimicking the analysis logic applied by the experiment, yielding MC histograms that can immediately be compared with experimental data. The tuning software \texttool{Apprentice}~\cite{Krishnamoorthy:2021nwv} allows to compute high-fidelity surrogates of the MC histograms by means of polynomial or rational approximations. The surrogates can be numerically evaluated at any point in the parameter space of interest in microseconds, thus enabling numerical optimisation of a least-squares measure defined with experimentally observed and surrogate MC histograms. The best-fit point in the parameter space constitutes a tuning. 

\begin{figure}[t]
    \centering
    \includegraphics[width=0.32\textwidth]{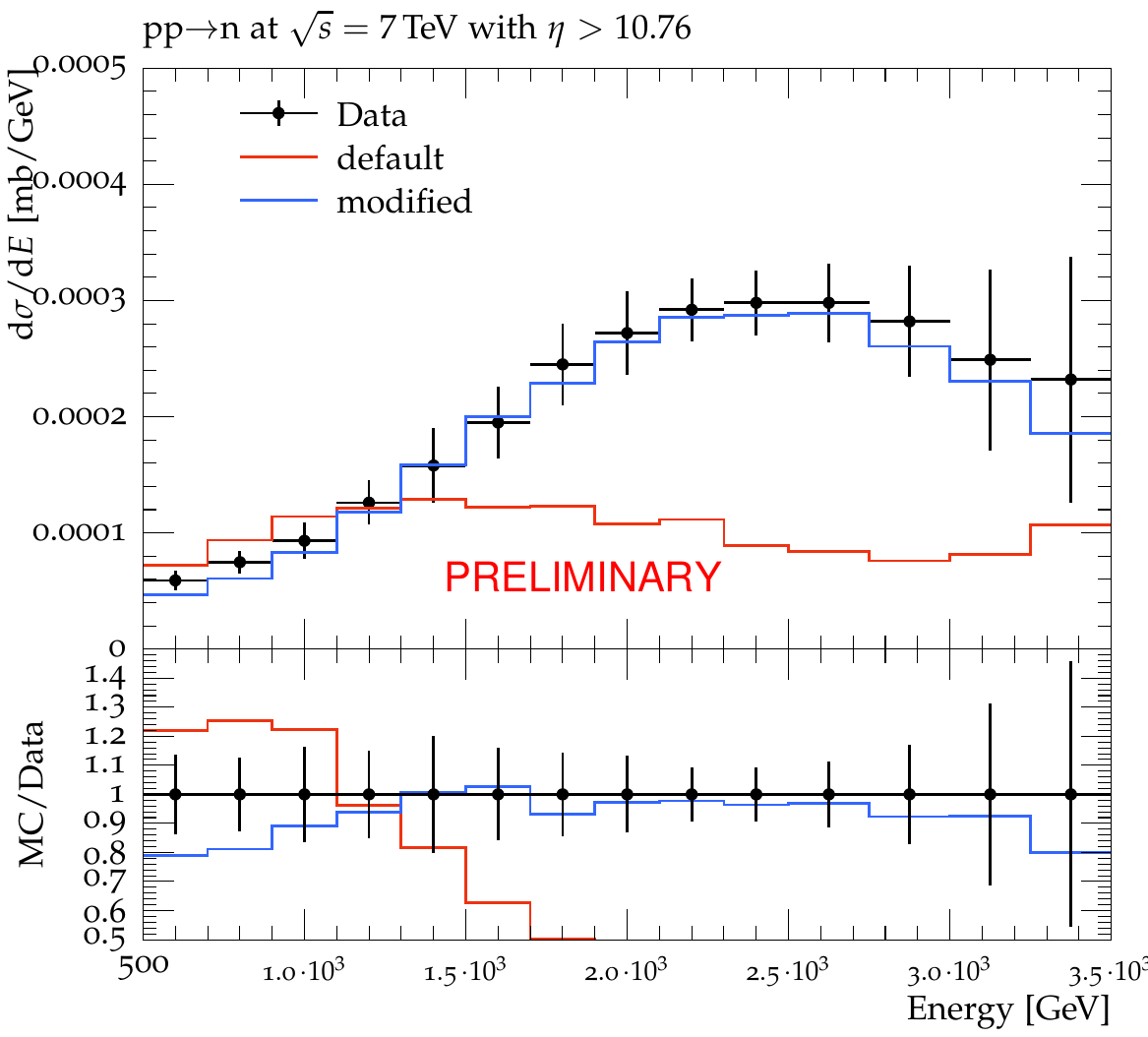}
    \includegraphics[width=0.32\textwidth]{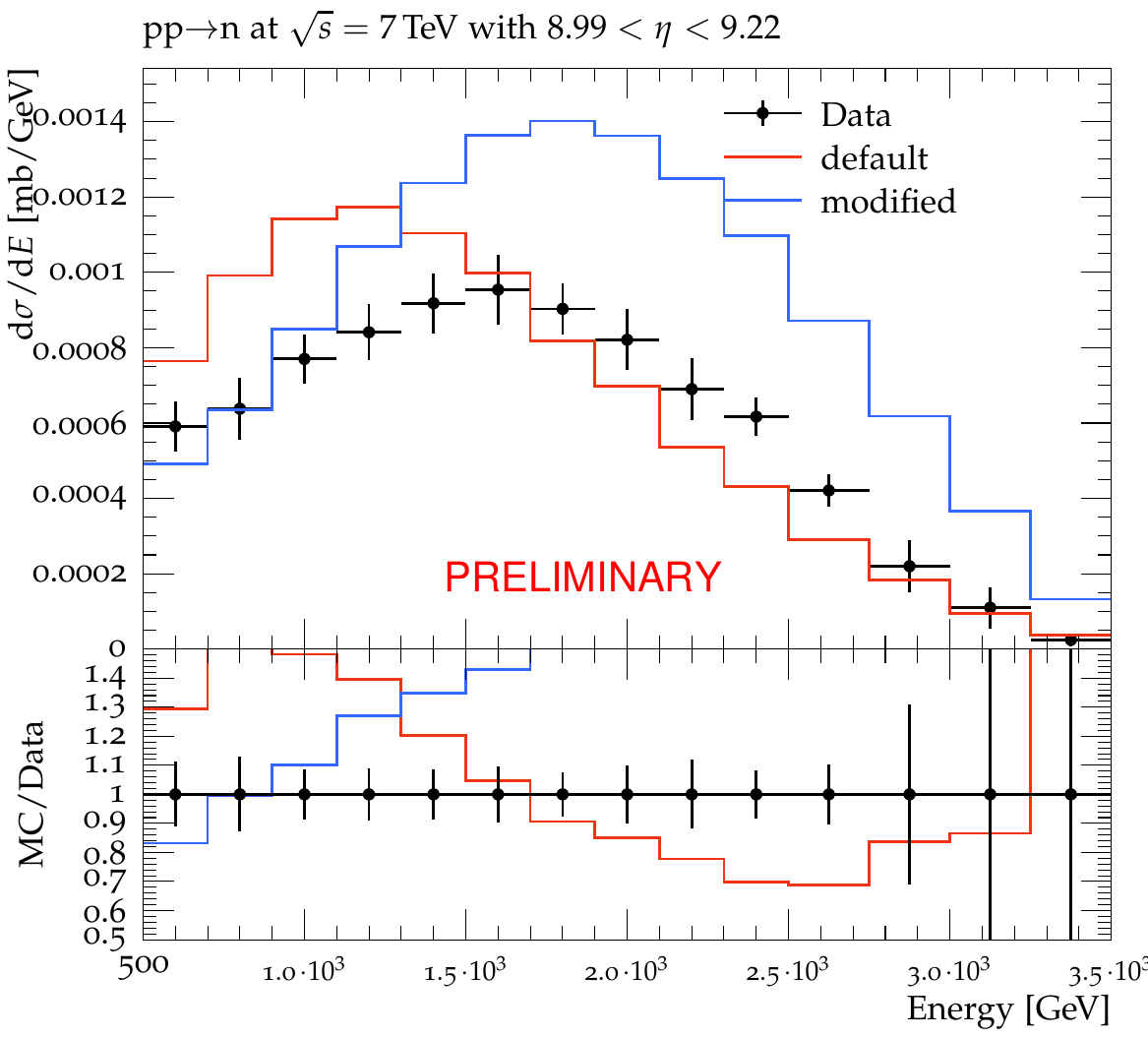}
    \includegraphics[width=0.32\textwidth]{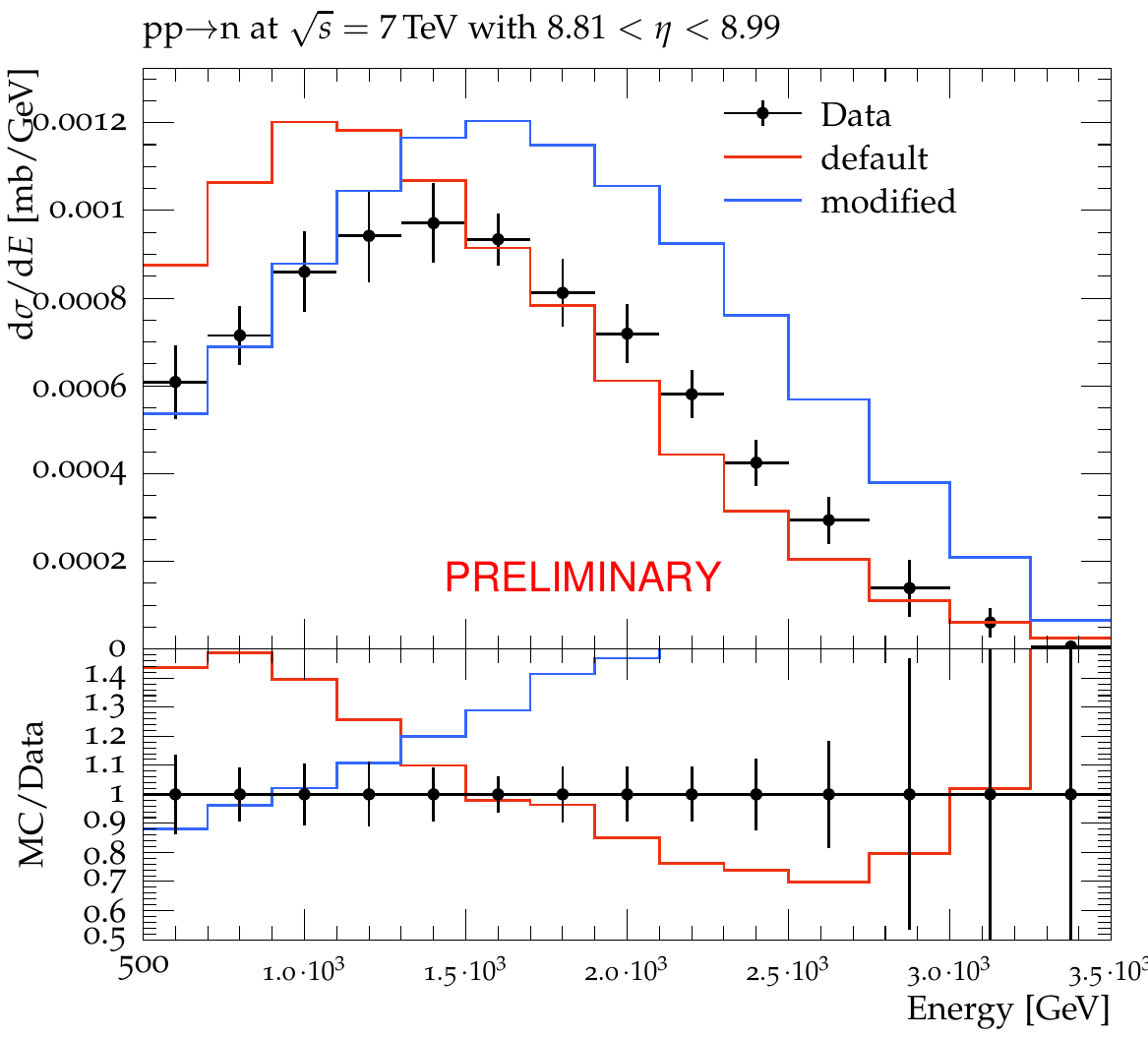}
    \includegraphics[width=0.32\textwidth]{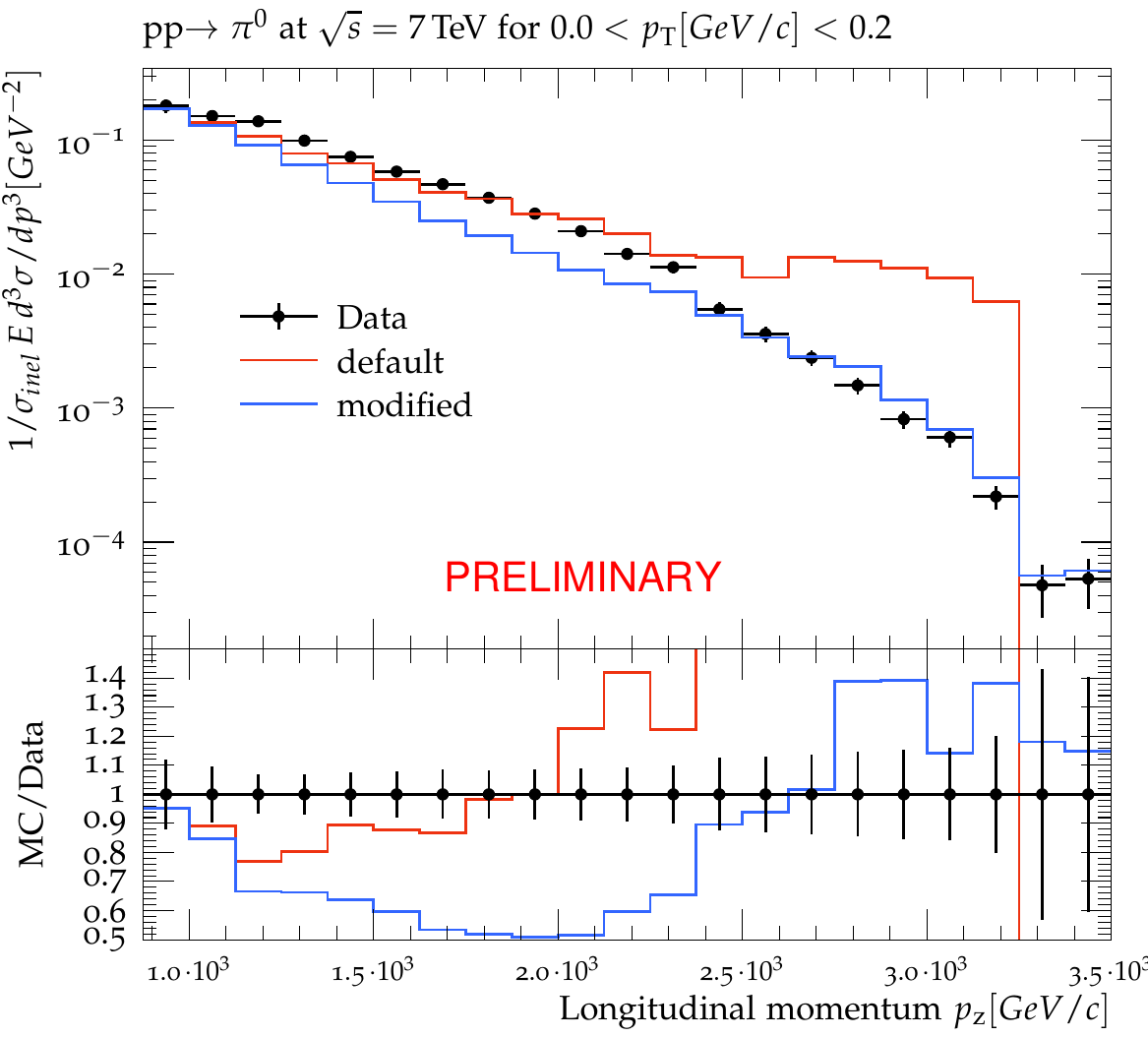}
    \includegraphics[width=0.32\textwidth]{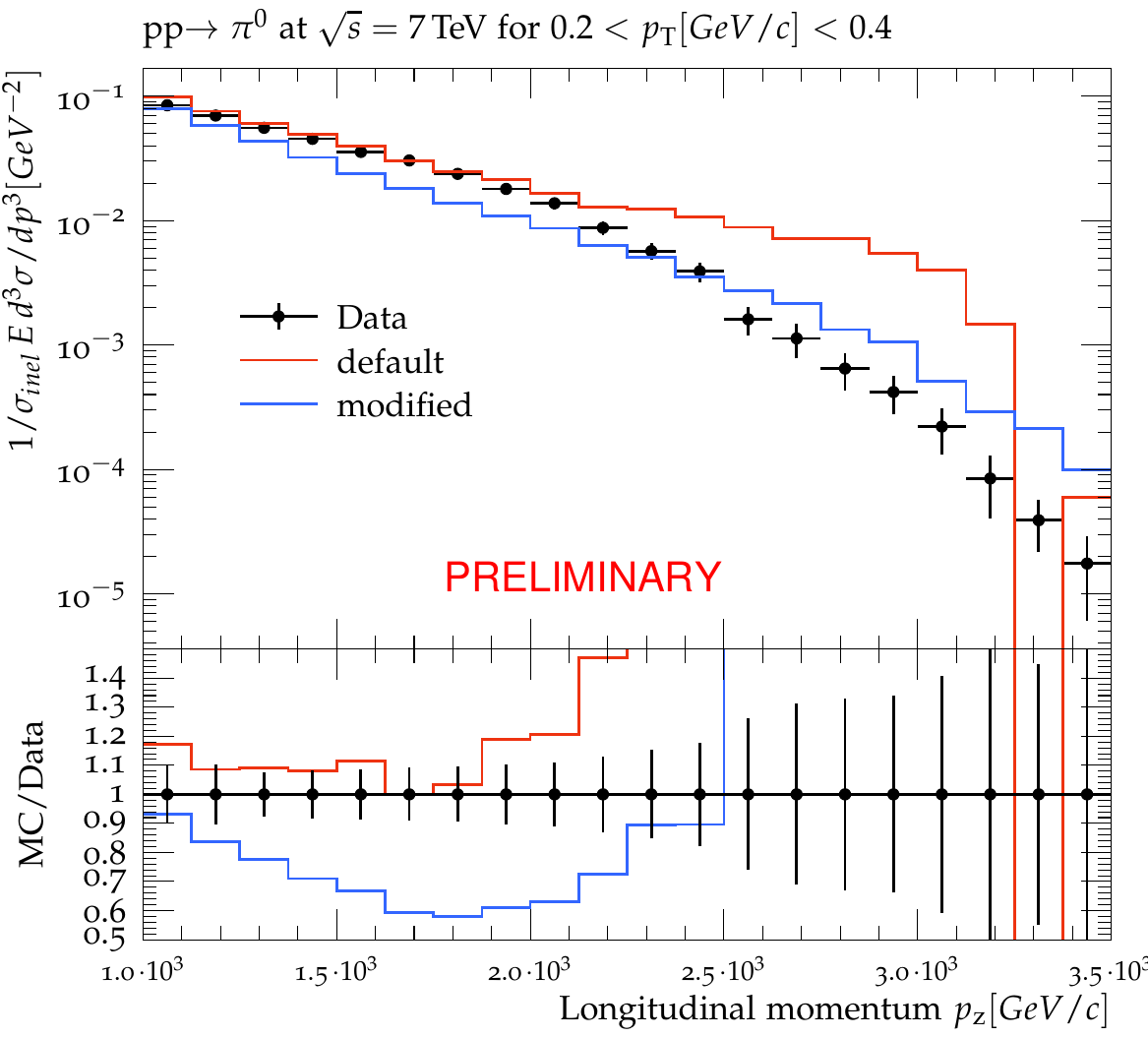}
    \includegraphics[width=0.32\textwidth]{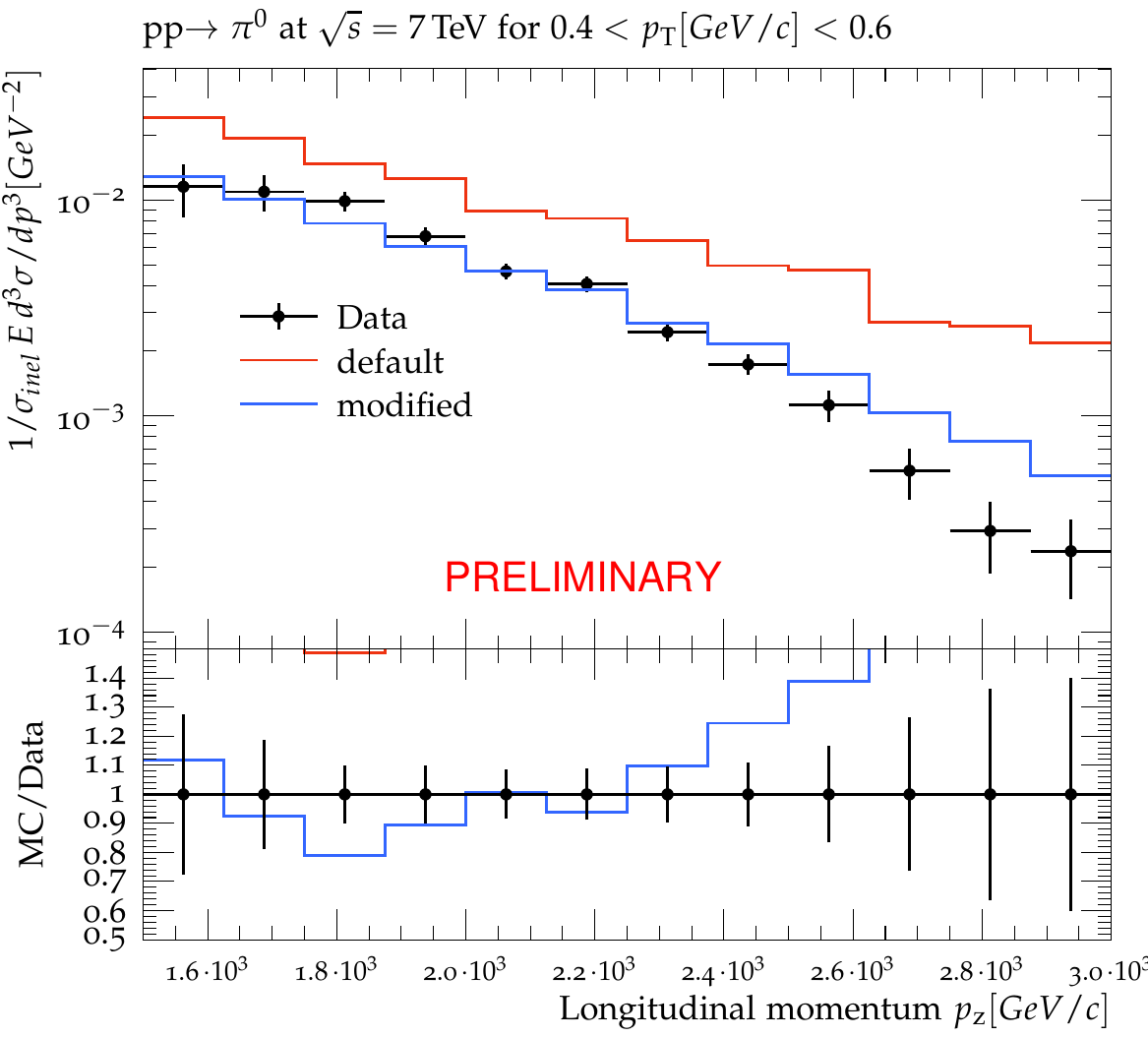}\\
    \includegraphics[width=0.32\textwidth]{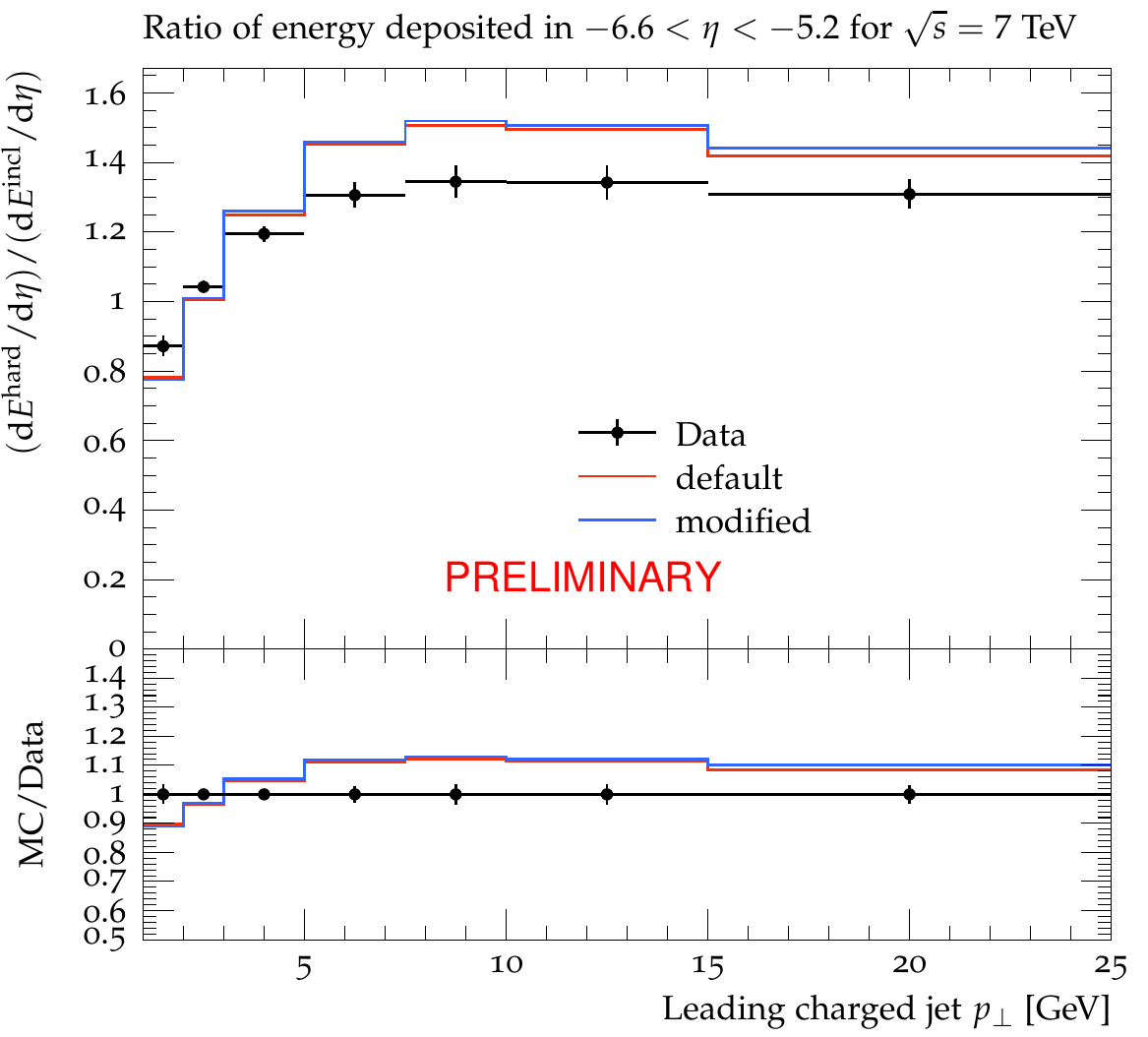}
    \includegraphics[width=0.32\textwidth]{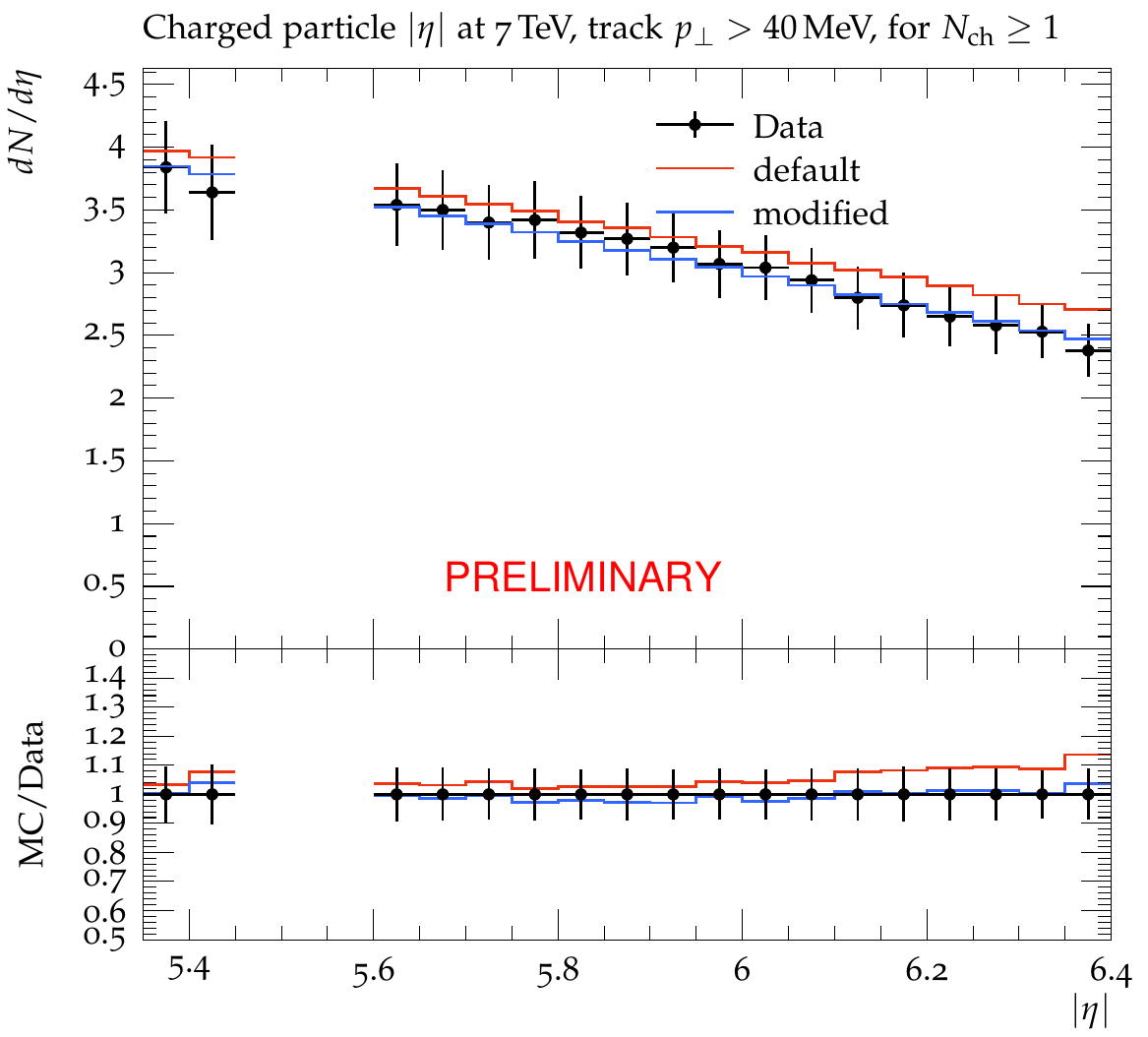}
    \includegraphics[width=0.32\textwidth]{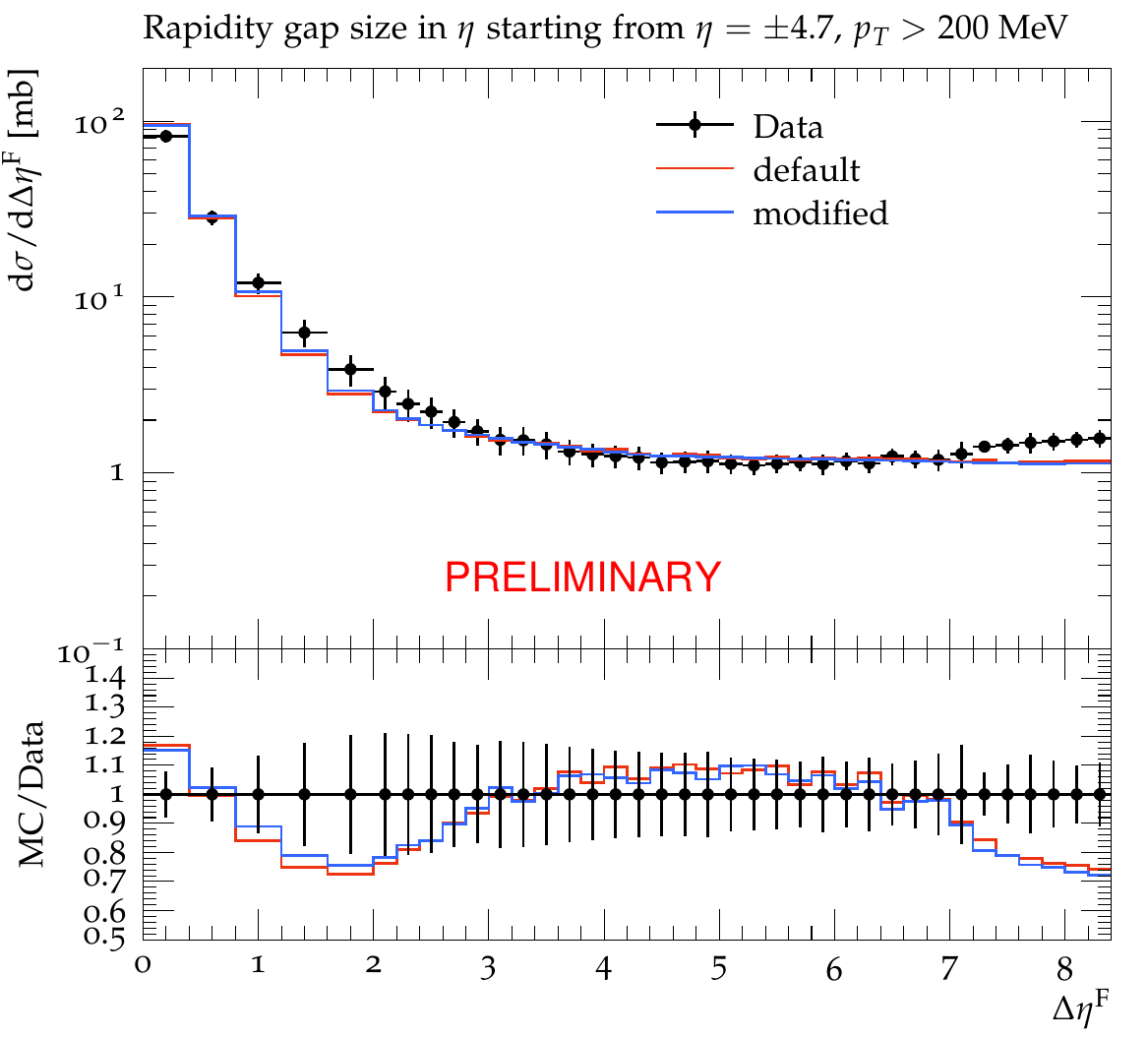}\\
    \caption{Comparing the default Monash tune (red line) with our tune (blue line) with experimental data (black data points). The first two rows are measurements of the neutron and pion spectra,  respectively, with the left column being the most forward measurements. The third row, from left to right, shows measurements by the CASTOR, TOTEM, and ATLAS collaborations. For each analysis, information on the process can be found at the top of each panel.}
    \label{fig:modified_default}
\end{figure}

In \cref{fig:modified_default} we show our results which compare the Monash tune (red line) and our preliminary tune (blue line) with experimental data (black points). The LHCf collaboration has produced neutron and pion production rates in $\eta$ and $p_T$ bins, respectively. We show a selection of representative histograms in the first two rows. The most forward measurements from LHCf, (first column, first two rows) are most relevant for forward physics facility measurements and show a good fit. For the less forward neutron spectra ($\eta<$ 9.22), our tune indicates an overproduction of neutrons. Although the default tune does not perform much better, we aim to describe these analyses well. The less forward pion spectra ($p_T>0.2~\rm{GeV/c}$), however, show a good fit as compared to the default tune. The third row compares each tune against a few representative measurements from the more central CASTOR, TOTEM, and ATLAS analyses. These plots are evidence that our tuning procedure affects these predictions only marginally.

\begin{figure}[t]
    \centering
    \includegraphics[width=0.99\textwidth]{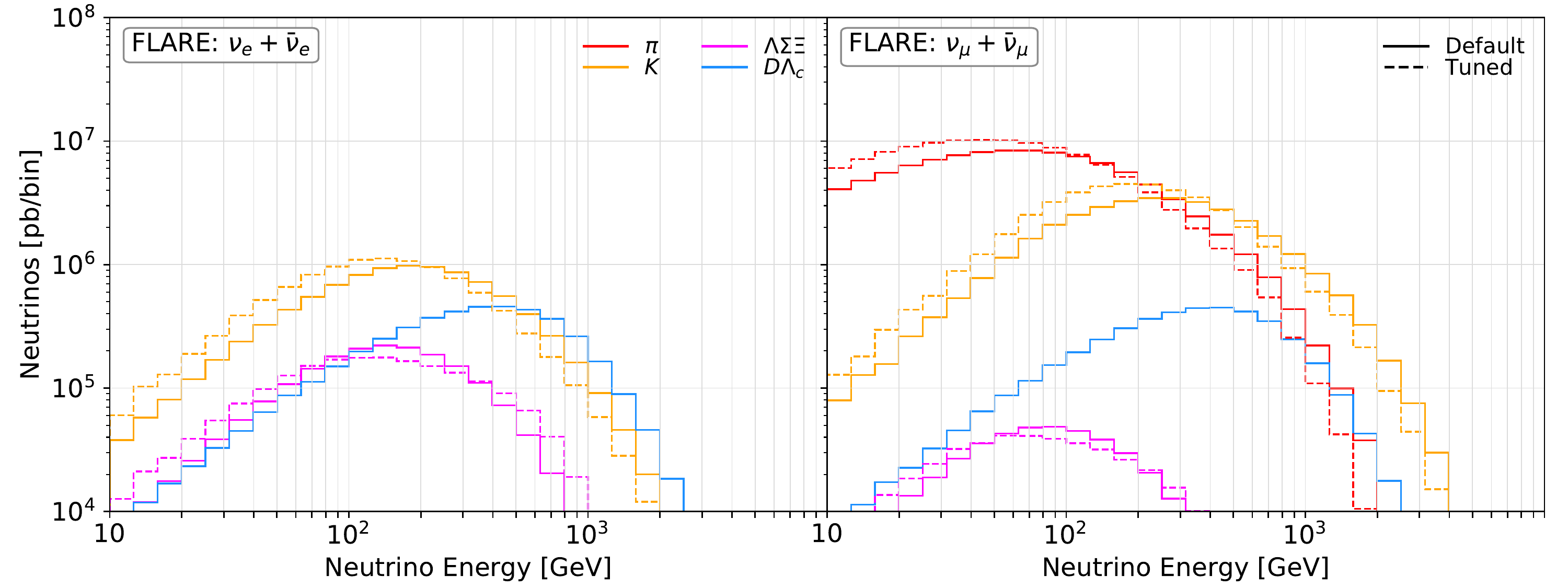}\\
    \caption{Neutrino flux through the cross sectional area of the FLArE detector for electron neutrinos (left) and muon neutrinos (right) using the Monash tune (solid) and our tune (dashed). The different colors correspond to different neutrino production modes.}
    \label{fig:modified_flux}
\end{figure}

We can use the modified tune to make updated predictions of the neutrino flux at the FPF experiments. This is shown in \cref{fig:modified_flux}, where we show the neutrino flux for electron and muon neutrinos going through the FLArE detector for both the Monash tune (solid line) and our updated tune (dashed line). Here we have used the simulation introduced in ~\cite{Kling:2021gos} to simulate the propagation of long-lived SM hadrons and their decay into neutrinos. The different line colors correspond to the neutrino production mode: pion decays (red), kaon decays (orange), hyperon decays (magenta) and charm decays (blue). We can see that the modified tune leads to a lower number of high-energy neutrinos from pion and kaon decays but a slightly larger number of high energy neutrinos from hyperon decay. 

A fundamental problem of the tuning procedure that has so far not been addressed to full satisfaction is that of assigning reasonable tuning uncertainties. Due to the unknown distribution function of the goodness-of-fit measure that we minimise, standard procedures such as confidence belt construction fail. This is in part due to the lack of information published on correlations between bins present in the experimental data histograms. Bootstrapping methods are hence of limited use only. We are therefore forced to pursue more pragmatic avenues that are partially motivated by techniques employed by PDF-fit groups. We use the hessian of the goodness-of-fit measure obtained at the best-fit point to construct a system of principal directions. This is similar to a confidence ellipsis construction but we cannot interpret the volume contained therein probabilistically. We can, however, use the principal axes and search along those for points that fulfil criteria that are compatible with a robust estimate for tuning uncertainties. One of the approaches we are currently studying requires to find points on the principal axes such that the MC histograms evaluated at those points  envelop at least two-thirds of all data bins entering the goodness-of-fit measure.



\subsection{Neutrinos at the FPF from Proton-Lead Collisions}

In addition to its usual runs with protons, the HL-LHC is also expected to collect data for proton-lead and lead-lead collisions. These heavy ion collisions will produce a large number of hadrons, sourcing a neutrino flux that can in principle be observed at the FPF experiments. From a theoretical point of view, such a measurement would be very appealing, for example to study effects of hadron propagation through nuclear matter or the quark gluon plasma in different kinematic regime.  In addition, the charm production in  heavy ion collisions would provide an opportunity to measure nuclear PDFs for the initial state gluons and test the gluon saturation, which is expected to be present at higher momentum fraction compared to the proton case. 

\begin{figure}[h]
  \centering
  \includegraphics[width=0.99\textwidth]{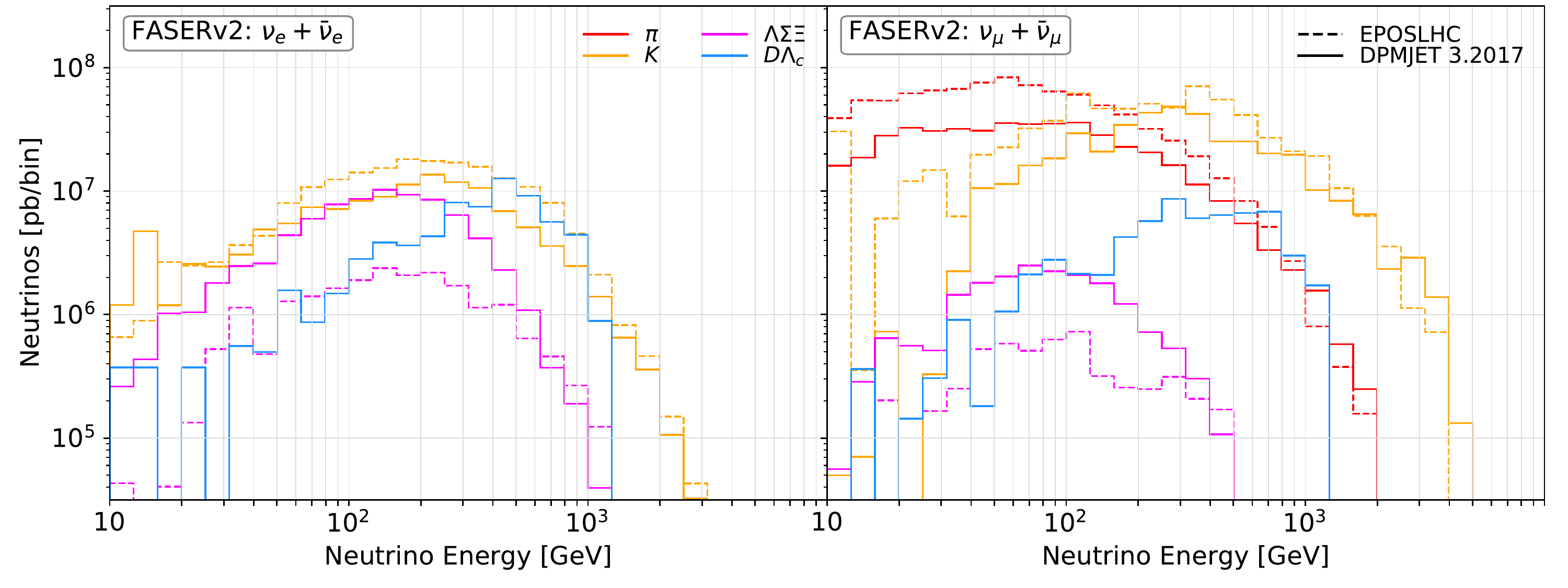}
  \caption{Predicted energy distribution of neutrinos
    produced in pPb collisions at 13 TeV LHC passing through the FLArE experiment for electron neutrinos (left) and muon neutrinos (right).
    The vertical axis shows the number of neutrinos per energy bin that  go through the considered cross sectional area for an integrated luminosity of $1~\ipb$. The different production modes are indicated by the line color: pion decays (red), kaon decays (orange), hyperon decays (magenta), and charm decays (blue). The different line styles correspond to predictions obtained from \texttool{DPMJet 3.2017} (solid) and \texttool{EPOS-LHC} (dashed).}
  \label{fig:fnfs_hi1}
\end{figure}

In the following, we present a first estimate of the expected neutrino flux from proton-lead collisions, where the proton is assumed to go towards to FPF. The results were obtained using the fast neutrino flux simulation introduced in Ref~.\cite{Kling:2021gos}. In \cref{fig:fnfs_hi1}, we show the fluxes of electron neutrinos (left) and muon neutrinos (right) going through the FASER$\nu$2 detector as function of the neutrino energy. We present results for two different MC event generators, \texttool{EPOS-LHC}~\cite{Pierog:2013ria} and \texttool{DPMJet~3.2017}~\cite{Roesler:2000he, Fedynitch:2015kcn}, which are illustrated using different line style. The different colors correspond to the neutrino production: pion decays (red), kaon decays (orange), hyperon decays (magenta) and charm decays (blue). We note that \texttool{EPOS-LHC} does currently not describe the production of charmed hadrons, and therefore only show the results from \texttool{DPMJet}. We can see a similar behaviour as for neutrinos from proton-proton collisions: pion decays are the dominant production mode for muon neutrinos at lower energies, while kaon decays provide the dominant production mode for high energy muon neutrinos and electron neutrinos. Hyperon decays mainly contribute to the anti-electron neutrino flux at lower energies, with sizable differences between the two considered models. Charmed hadron decays can become important for high energy electron neutrinos, but seem to be a bit less relevant than for proton-proton collisions. 

\begin{figure}[h]
  \centering
  \includegraphics[width=0.99\textwidth]{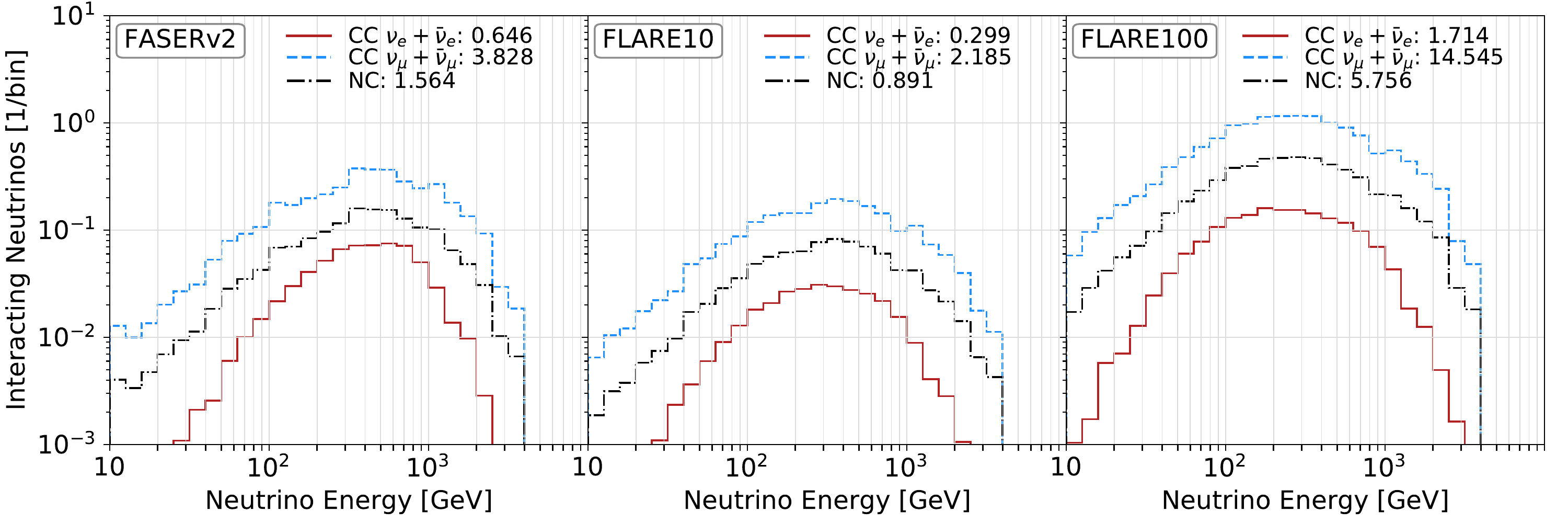}
  \caption{Number of expected neutrino interactions with the FASER$\nu$2 (left), FLArE with a target mass of 10~tons (center) and FLArE with a target mass of 100~tons (right) as function of the neutrino energy. Here we consider neutrinos were produced in pPb collisions at 13 TeV LHC and an integrated luminosity of $1~\ipb$. The red, blue and black lines correspond to charged current electron, charged current muon and neutral current interactions, respectively, and were obtained using \texttool{EPOS-LHC}.}
  \label{fig:fnfs_hi2}
\end{figure}

In \cref{fig:fnfs_hi2}, we show the expected number of neutrino interactions with the FPF neutrino detectors, again as a function of the incoming neutrino energy. Here we consider FASER$\nu$2 with a target mass of 20~tons (left), FLArE with a target mass of 10~tons (center) and a larger version of FLArE with a target mass of 100~tons with the geometry considered in ~\cite{Batell:2021blf} (right). We assume an integrated luminosity of $\mathcal{L} = 1~\ipb$, corresponding to the projected luminosity for proton-lead collisions at ATLAS during LHC Run~3 and Run~4 \cite{ATL-PHYS-PUB-2018-020}. The delivered luminosity for the whole HL-LHC could therefore be larger. The different lines show the expected number of neutrino interactions per energy bin for charged current electron neutrino interactions (red), charged current muon neutrino interactions (blue) and neutral current interactions of all neutrino flavours (black). The numbers in the legend indicate the total expected event rate, summed over the whole energy range. We can see that we expect only about one event at FASER$\nu$2 and FLArE, an about half a dozen events at a larger version of FLArE with 100~tons target mass. Unless the luminosity for heavy ion collisions is increased considerably, the prospects for neutrino measurements during the heavy ions runs are not very promising.

\section{Neutrino--induced Deep Inelastic Scattering: Constraints on Nucleon Structure}
\label{sec:qcd:nDIS}

\subsection{Introduction}

The neutrinos produced by the decays of both light and heavy-flavoured hadrons in proton--proton collisions at the ATLAS interaction point will reach the FPF, whose detectors will then permit  neutrinos and anti-neutrinos of different flavours to be distinguished.
Neutrino scattering plays an important role in the extraction of PDFs, as  neutrino-induced
structure functions provide a complementary handle towards resolving the flavour of the nucleon's constituents.
The unique ability of the weak current to probe specific quarks flavours provided
by neutrino DIS measurements significantly improves global determinations of proton
and nuclear PDFs.
Analogous measurements as the ones from previous neutrino-induced DIS experiments, such as NuTeV~\cite{NuTeV:2005wsg}, NOMAD~\cite{NOMAD:2013hbk}, CCFR~\cite{NuTeV:2001dfo} and CHORUS~\cite{CHORUS:1997wxi}, will
be obtained at the FPF providing novel constraints on future global fits for nucleon and nuclear PDFs.
As highlighted by \cref{fig:nuclear-PDF-coverage}, neutrino structure functions at the FPF
complement and extend the coverage of existing DIS measurements on nuclear targets,
while largely overlapping with the corresponding neutral-current DIS measurements in
lepton-nucleus scattering provided by the EIC~\cite{Anderle:2021wcy,AbdulKhalek:2021gbh}.

In particular, the main channel to probe the strange and the anti-strange PDFs, $xs$ and $x\bar{s}$ respectively, has historically been dimuon production in inclusive charged current  DIS data, and especially semi-inclusive charm quark production in CC DIS. These data are able to provide constraints at larger Bjorken $x$ on the strange and also the anti-strange densities trough the subprocesses $W^{+}s\rightarrow c$ and $W^{-}\bar{s}\rightarrow\bar{c}$.
The strange sea was  related to the light quark sea by a $x$-independent fraction~\cite{Jimenez-Delgado:2008orh,H1:2009pze,H1:2000muc}, such as $x\bar{s}=r_{s}x\bar{d}$, and it was often assumed that the strange sea was suppressed at the level that $r_{s}\sim0.5$.
However, the interpretation of these data is sensitive to uncertainties from charm fragmentation and nuclear corrections.
Results by the ATLAS collaboration~\cite{ATLAS:2021qnl,ATLAS:2021vod} showed that with the inclusion of more LHC data, the ratio of the strange-quark to light-quark densities, $R_{s}$, is better constrained and found to fall more steeply at high-$x$.
Along the same lines, a symmetric low-$x$ strange distribution
was reported by the \texttool{ATLASepWZ12}~\cite{PhysRevLett.109.012001} and \texttool{ATLASepWZ16}~\cite{ATLAS:2016nqi} PDF fits.
An important benefit of the FPF will be to provide measurements able to shed light on  the apparent strangeness puzzle introduced by the potential tensions between the above-described data, see also the re-analysis
of~\cite{Faura:2020oom}. 

As highlighted earlier in this document, experiments at the FPF permit charm tagging using different techniques. In particular, not only emulsion experiments, which allow  several kinds of charmed baryons and mesons to be tagged by reconstructing in detail the topology of their decays, but also experiments which allow to  charm tagging through dimuon events, will be present.
Hence the measurements of both inclusive and charm-tagged neutrino structure functions
should be feasible at the LHC.

In the following subsections, the potential impact of neutrino--induced DIS is discussed in further detail, within the context of the \texttool{nCTEQ} and (\texttool{n})\texttool{NNPDF} analyses.
We also present updated predictions for inclusive neutrino cross-sections on a tungsten nuclear
target, evaluating the associated nPDF uncertainties and the role of
including/neglecting various physics effects such as the $Q \le 1$ GeV region
and the non-isoscalarity of the target.

\subsection{Impact of Neutrino-induced DIS within the nCTEQ Framework}

Neutrino interactions have played a crucial role in characterizing 
nuclear structure in the language of parton distribution functions (PDFs)
for both protons~\cite{Hou:2019efy,Accardi:2021ysh,Accardi:2016qay,Bailey:2020ooq,Abramowicz:2015mha,Ball:2017nwa,Alekhin:2017kpj} and nuclei~\cite{Kovarik:2015cma,Kusina:2020lyz,Eskola:2016oht,AbdulKhalek:2020yuc,NNPDF:2019ubu,NNPDF:2019vjt,ATLAS:2021vod}.
The combination of separate neutrino and anti-neutrino measurements, together with parity-violating and parity-conserving structure function extractions, provide essential information necessary to disentangle the individual PDF flavors~\cite{Goncharov:2001qe, Kusina:2012vh, Faura:2020oom,SajjadAthar:2020nvy,NuSTEC:2017hzk,Eskola:2021nhw,Eskola:2022rlm}. 
A good theoretical understanding of neutrino DIS is also an important ingredient for determinations of the weak mixing angle, neutrino mass splittings, and for searches for physics beyond the Standard Model~\cite{ParticleDataGroup:2020ssz}, including at DUNE/LBNF~\cite{DUNE:2015lol,DUNE:2018tke}, which will record measurements in the
$E_\nu\! \sim$~few-GeV region wherein contributions from DIS will be significant.

In spite of ther importance, including neutrino data into global PDF fits can be challenging for several reasons. 
Due to the weak nature of neutrino-nucleus ($\nu A$) interactions, neutrino experiments typically use heavy nuclear targets such as iron or lead to obtain higher statistics~\cite{Schienbein:2009kk,Kovarik:2010uv,Kovarik:2015cma,Ball:2018twp}. Thus, there is necessarily a nuclear correction that must be included when comparing these data with measurements on the proton or other light nuclei. 
There  are in addition concerns regarding the apparent tensions between charged-current neutrino and neutral-current charged-lepton measurements. While the source of these tensions has yet to be fully understood, it will be crucial to resolve them if we are to progress toward the higher-precision analyses 
needed for the HL-LHC era.

%
The FPF will provide 
high statistics charged-current (CC) and 
neutral-current (NC) $\nu A$ measurements on a variety of nuclei.
This new data set could play an indispensable role in addressing a variety of outstanding
issues regarding the nuclear correction factors and extraction of individual partonic flavors. 
In the following, we briefly review some of the issues 
where this FPF data set can prove enlightening.

%
In~\cite{Schienbein:2007fs,Schienbein:2009kk}   initial analysis of  nuclear effects in 
deep inelastic neutrino-nucleon scattering with a focus on iron data was performed.
Iron PDFs were extracted in a global fit using the charged current (CC) neutrino--iron structure functions, 
and these results  were compared with neutral current (NC) charged-lepton--iron structure functions\cite{Kalantarians:2017mkj,Szumila-Vance:2020zpt,Arrington:2021vuu}.
As shown in Fig.~4 of~\cite{Schienbein:2009kk}, the comparison demonstrated that 
except for very high $x_{Bj}$, the 
nuclear correction factors using the CC $\nu A$ structure functions differ in both shape and magnitude from those using the NC  $\ell^\pm A$ scattering.

\begin{figure}[htb]
     \centering
     \includegraphics[width=0.45\textwidth]{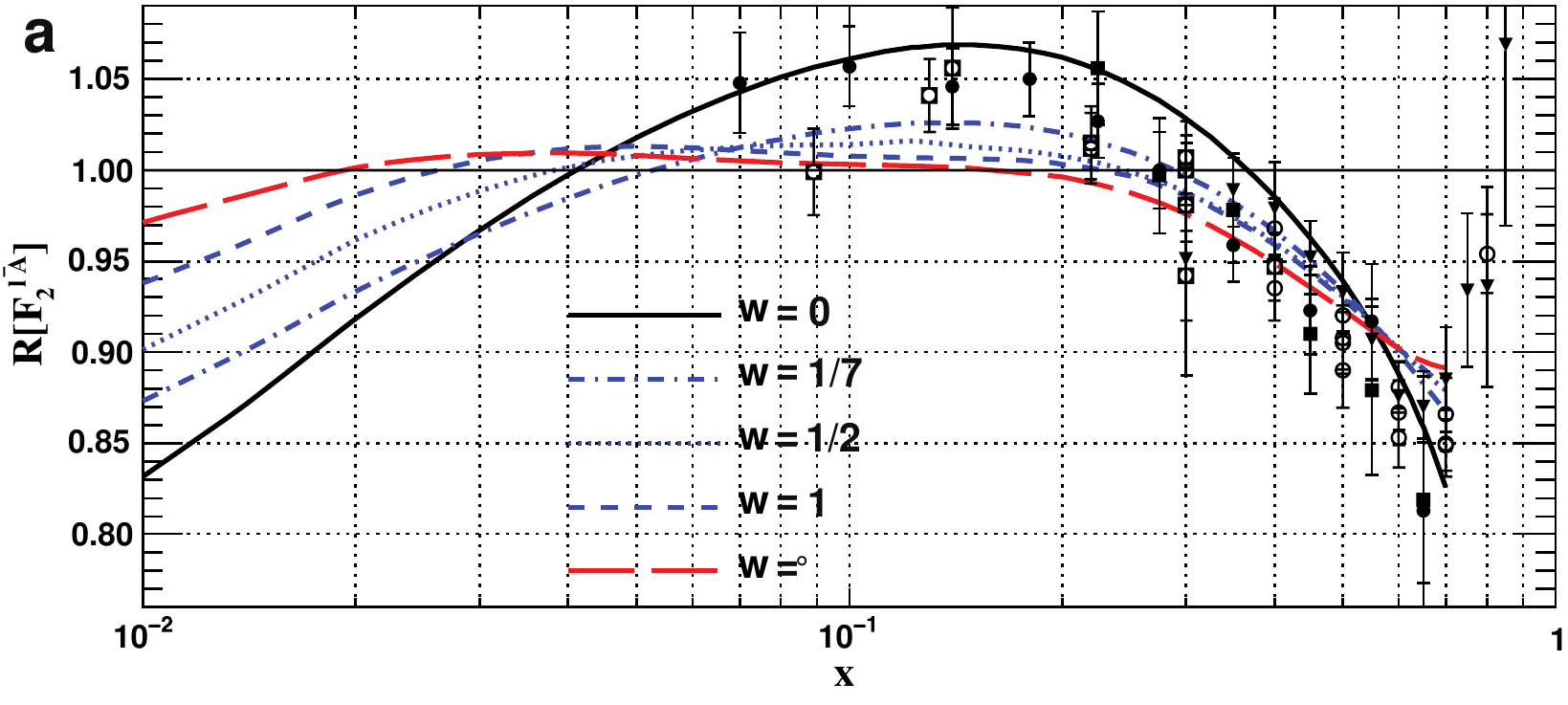}
     \hfil
     \includegraphics[width=0.45\textwidth]{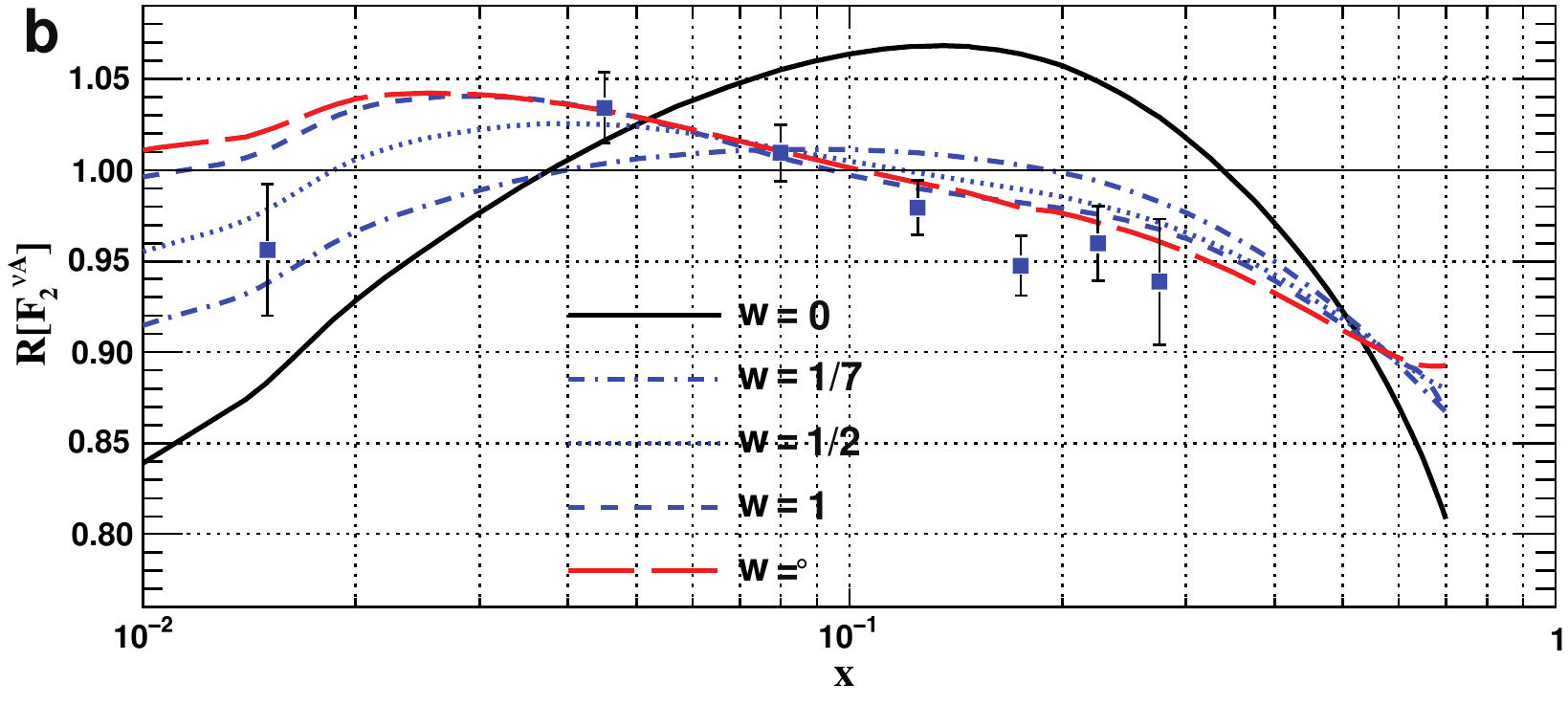}
     \caption{Fit results from ~\cite{Kovarik:2010uv}. 
     Predictions from the compromise fits for the nuclear correction factors
$R[F_{2}^{\ell Fe}]\simeq F_{2}^{\ell Fe}/F_{2}^{\ell N}$ ({\bf a})
and $R[F_{2}^{\nu Fe}]\simeq F_{2}^{\nu Fe}/F_{2}^{\nu N}$ ({\bf b})
as a function of $x$ for $Q^{2}=5\,{\rm GeV}^{2}$. The data points
displayed in ({\bf a}) are from BCDMS and SLAC experiments (for references  see \cite{Schienbein:2009kk})
and those displayed in ({\bf b}) come from the NuTeV experiment \cite{NuTeV:2005wsg}.
} 
    \label{fig:ComboFit}
\end{figure}

This apparent incompatibility of the charged-lepton ($\ell^\pm A$)  
and $\nu A$
processes in the global analysis  required  further study.
This effort was pursued in ~\cite{Kovarik:2010uv} which 
performed a combined global fit using both data sets
to determine if there might be a ``compromise'' solution compatible with both data sets. 
Both the charged-lepton and $\nu A$  
data sets were included in the fit using a relative weighting factor of $w$  such that: 
$w=0$ corresponds to no neutrino data in the fit;
$w=1$ 
corresponds to equal weight for the neutrino data and charged-lepton data in the fit;
and
$w=\infty$ corresponds to only neutrino data in the fit.
The resulting nuclear correction factors ($R=F_2^{Fe}/F_2^{N}$) are displayed in 
\cref{fig:ComboFit} along with charged-lepton data (left panel)
and neutrino data (right panel). 
As suggested by this figure, ~\cite{Kovarik:2010uv} was not able to find
a ``compromise" fit that would simultaneously satisfy  both data sets. 
It is important to note that the correlated systematic errors  were fully accounted for
in these fits; in contrast, if the errors were assumed to be uncorrelated and added
in quadrature (to obtain inflated errors) then a more satisfactory compromise fit could be constrained. 

As noted in \cref{sec:nuDIS}, 
the $\nu A$ DIS data play a key role 
in disentangling the separate partonic flavors of the proton PDF. 
This is especially the  case for the strange quark PDF which is probed by both the 
inclusive DIS ($\nu A\to eX$) and dimuon production ($\nu A\to \mu^+ \mu^- X$) data. 
Since the above nuclear correction $F_2^{A}/F_2^{N}$ is required to relate the nuclear data to the proton case, the  variation observed in \cref{fig:ComboFit} 
will necessarily influence the uncertainty of the extracted strange PDF~\cite{Schienbein:2009kk,Faura:2020oom,Alekhin:2017olj,Khalek:2022zqe,Faura:2020oom}.

At the higher energies of the LHC, $W^\pm/Z$ vector boson production
provides an alternate avenue to determine the strange PDF;
for example, 
this analysis can be directly  extract the  $R_s={(s+\bar{s})/(\bar{u}+\bar{d})}$ ratio 
with protons so that no nuclear corrections are required. 
While a number of determinations using the inclusive $W^\pm/Z$ data yielded comparably large values for $R_s\! \sim\! 1$,
other analyses of the ratio of strangeness to light quarks have obtained closer consistency with $R_s\! \sim\! 0.5$~\cite{ATLAS:2021vod}.
Achieving a simultaneous description of the $\nu A$ DIS and LHC $W^\pm /Z$ production measurements and
understanding implications for nucleon strangeness therefore remains a topic of active investigation.

%
At the LHC, 
the strange nuclear PDF (nPDF) can be obtained  using proton-lead (pPb) data
in a manner similar to the proton-proton data discussed above. 
This nuclear analysis was performed in~\cite{Kusina:2012vh,Kusina:2016fxy,Kusina:2020lyz,AbdulKhalek:2019mzd,Eskola:2022rlm},
and sample distributions for lead nPDFs are displayed in 
Fig.~18.8 of the 2021 PDF Structure Function review~\cite{ParticleDataGroup:2020ssz}, 
and also compared with other nPDFs from the literature.

There are two notable features of the extracted \texttool{nCTEQ} nuclear PDFs. First, for the strange PDF, we observe a comparatively large strange PDF in the low-$x$ region. This is similar to the proton PDF behavior discussed above~\cite{ATLAS:2021vod}. Second, we also observe an enhanced gluon distribution in the region $x {\sim} 3{\times} 10^{-2}$,
corresponding to the central-$x$ value of the $W^\pm / Z$ kinematics. 
While these observations are intriguing, it is important to determine
whether the above effects truly reflect the physical characteristics of the nucleon,
or simply an artifact of the fit exploiting the weakly constrained strange distribution. 
This is precisely the type of question which an independent, high-precision
data set from the FPF can address.

The FPF will have very broad kinematical coverage and be capable of 
probing both the low- and high-energy regimes. 
In the low-energy limit (low $Q^2$ and low $W^2$), we approach the non-perturbative region and 
encounter target mass effects and  higher twist corrections.
Recent nPDF studies~\cite{Segarra:2020gtj}  have begun to explore this region 
by relaxing the $Q^2$ and $W^2$ cuts on the  data sets~\cite{SajjadAthar:2020nvy,NuSTEC:2017hzk}. 
In particular, as we reduce the $W^2$ cut still further, we enter the shallow-inelastic scattering (SIS) and resonance regions, presenting the opportunity to study quark-hadron duality experimentally with neutrinos.  As of now only model-dependent studies have been possible and results have generally not been consistent with local quark-hadron duality in $\nu A$~interactions.
These fits obtained good agreement (in terms of $\chi^2$) with the 
neutral-current charged-lepton measurements ($e^\pm A\to e^\pm X$);
however, this analysis must still be extended to the charged current ($\nu A$) DIS data.

The high energy extreme is of equal interest as it enables us to study the high-$Q^2$ and low-$x$ limits. 
An FPF data set with a large $Q^2$ range allow us to study heavy quark production, for example, 
across the full range from the low-energy decoupling limit ($m_Q{<}Q$)
to the high-energy regime ($m_Q{\ll}  Q$) where we approach the massless limit.
Improved descriptions of heavy-quark production also help constrain the PDFs at low $x$,
and this can help constrain theoretical predictions for high-energy astrophysical phenomena, 
as demonstrated in ~\cite{Zenaiev:2019ktw}.

%
In summary, we have identified  several research areas where  neutrino DIS measurements 
from the FPF can play an indispensable role 
in resolving outstanding questions and improving the determination of nPDFs. 
The FPF can provide high statistics charged-current (CC) and 
neutral-current (NC) $\nu A$ measurements on a variety of nuclei,
and such a comprehensive data set will enormously expand our ability to 
separately determine  the nuclear corrections and the partonic flavor decomposition. 
The broad energy reach of this facility will  allow the exploration 
of extreme kinematic regimes and help us bridge the gap 
between the accelerator-based measurements and 
the ultra-high energy results from IceCube. 
The high statistics and broad phase space will also 
allow measurements  that probe into the low-$W$ transition region 
and provide an opportunity to study quark-hadron duality in the weak sector. 
Finally, the FPF has the potential to dramatically improve 
the precision of our standard model (SM) predictions (which are often limited by PDF precision),
and thus advance our search for BSM phenomena.

\subsection{Impact of Neutrino-induced DIS within the (n)NNPDF Framework}
\label{sec:nuDIS}

Dimuon production in CC neutrino--induced DIS plays a key role in the determination of the light sea quark PDFs in the proton, thanks to the properties of the weak current. Experiments that have measured either reduced cross-sections, $\sigma_{CC}^{\nu,\bar\nu}$, their ratio to the inclusive cross-section, $\mathcal{R_{\mu\mu}}$, or structure functions $F_2$, $xF_3$ (see {\it e.g.} Eqs.~(10)-(11) in~\cite{Ball:2008by} and Sect.~2.1 in~\cite{Faura:2020oom} for the definition of the observables) include CHORUS~\cite{CHORUS:2005cpn,Kayis-Topaksu:2011ols} and NOMAD~\cite{NOMAD:2013hbk} at CERN, and CCFR~\cite{Seligman:1997mc,CCFRNuTeV:2000qwc,Berge:1989hr} and NuTeV~\cite{NuTeV:2005wsg,NuTeV:2001dfo} at Fermilab. In the case of NOMAD and NuTeV, the secondary muon is tagged from the decay of a charmed meson, $\nu_\mu+N\to\mu+c+X$ with $c\to D\to\mu+x$, a fact that makes the observable specifically sensitive to strange quark and anti-quark PDFs. An accurate knowledge of these is essential to control the PDF uncertainty in weak boson mass measurements at the LHC~\cite{Nadolsky:2008zw} and to gain insights into the non-perturbative structure of the proton~\cite{Chang:2014jba}. 

The available measurements are summarized, with their references, in \cref{tab:neutrino_dataset}. Modern proton PDF sets, such as \texttool{ABMP16}~\cite{Alekhin:2017kpj}, \texttool{CT18}~\cite{Hou:2019efy}, \texttool{MSHT20}~\cite{Bailey:2020ooq} and \texttool{NNPDF4.0}~\cite{Ball:2021leu}, include a subset of all of these data sets, albeit with slight differences in the exact observable included, as also summarized in \cref{tab:neutrino_dataset}. For each PDF set, a blue tick indicates that the given dataset is included and a red cross that it is not included. A parenthesized tick denotes that a dataset was investigated but not included in the baseline fit.

\begin{table}[!t]
  \renewcommand*{\arraystretch}{1.60}
  \tiny
  \centering 
    \begin{tabularx}{\textwidth}{Xcccccccccc}
    \toprule
    &
    &
    & \multicolumn{4}{c}{Proton PDF sets}
    & \multicolumn{4}{c}{Nuclear PDF sets}\\
    \midrule
    Data set
    & 
    & Ref.
    & ABMP16
    & CT18
    & MSHT20
    & NNPDF4.0
    & EPPS21
    & nCTEQ15
    & nNNPDF3.0
    & TUJU21
    \\
    \midrule  
    CHORUS $\sigma_{CC}^{\nu,\bar\nu}$
    & Pb
    & \cite{CHORUS:2005cpn}
    & \xmark
    & \xmark
    & \cmark
    & \cmark
    & \cmark
    & \xmark
    & \cmark
    & \cmark
    \\
    CHORUS
    & Pb
    & \cite{Kayis-Topaksu:2011ols}
    & \cmark
    & \xmark
    & \xmark
    & \xmark
    & \xmark
    & \xmark
    & \xmark
    & \xmark
    \\
    NOMAD $\mathcal{R_{\mu\mu}}$
    & Fe
    & \cite{NOMAD:2013hbk}
    & \cmark
    & \xmark
    & \xmark
    & \ymark
    & \xmark
    & \xmark
    & \xmark
    & \xmark
    \\
    CCFR $x F_3^p$
    & Fe
    & \cite{Seligman:1997mc}
    & \xmark
    & \cmark
    & \xmark
    & \xmark
    & \xmark
    & \xmark
    & \xmark
    & \xmark
    \\
    CCFR $F_2^p$
    & Fe
    & \cite{CCFRNuTeV:2000qwc}
    & \xmark
    & \cmark
    & \xmark
    & \xmark
    & \xmark
    & \xmark
    & \xmark
    & \xmark
    \\
    CDSHW $F_2^p, x F_3^p$
    & Fe
    & \cite{Berge:1989hr}
    & \xmark
    & \cmark
    & \xmark
    & \xmark
    & \xmark
    & \xmark
    & \xmark
    & \cmark
    \\
    NuTeV $\sigma_{CC}^{\nu,\bar\nu}$
    & Fe
    & \cite{NuTeV:2001dfo}
    & \cmark
    & \cmark
    & \cmark
    & \cmark
    & \xmark
    & \xmark
    & \cmark
    & \xmark
    \\
    NuTeV $F_2, F_3$
    & Fe
    & \cite{NuTeV:2005wsg}
    & \xmark
    & \xmark
    & \cmark
    & \xmark
    & \xmark
    & \xmark
    & \xmark
    & \xmark
    \\
    \bottomrule
  \end{tabularx}
  \caption{The CC neutrino DIS measurements used in recent determinations
    of proton (\texttool{ABMP16}~\cite{Alekhin:2017kpj}, \texttool{CT18}~\cite{Hou:2019efy},
    \texttool{MSHT20}~\cite{Bailey:2020ooq} and \texttool{NNPDF4.0}~\cite{Ball:2021leu}) and
    nPDFs (\texttool{EPPS21}~\cite{Eskola:2021nhw}, \texttool{nCTEQ15}~\cite{Kovarik:2015cma},
    \texttool{nNNPDF3.0}~\cite{Khalek:2022zqe} and \texttool{TUJU21}~\cite{Helenius:2021tof}). For each PDF set, a blue
    tick indicates that the given dataset is included and a red cross that it is
    not included. A parenthesized tick denotes that a dataset was investigated
    but not included in the baseline fit.}
  \label{tab:neutrino_dataset}
\end{table}

\begin{figure}[!t]
  \centering
  \includegraphics[width=0.49\textwidth]{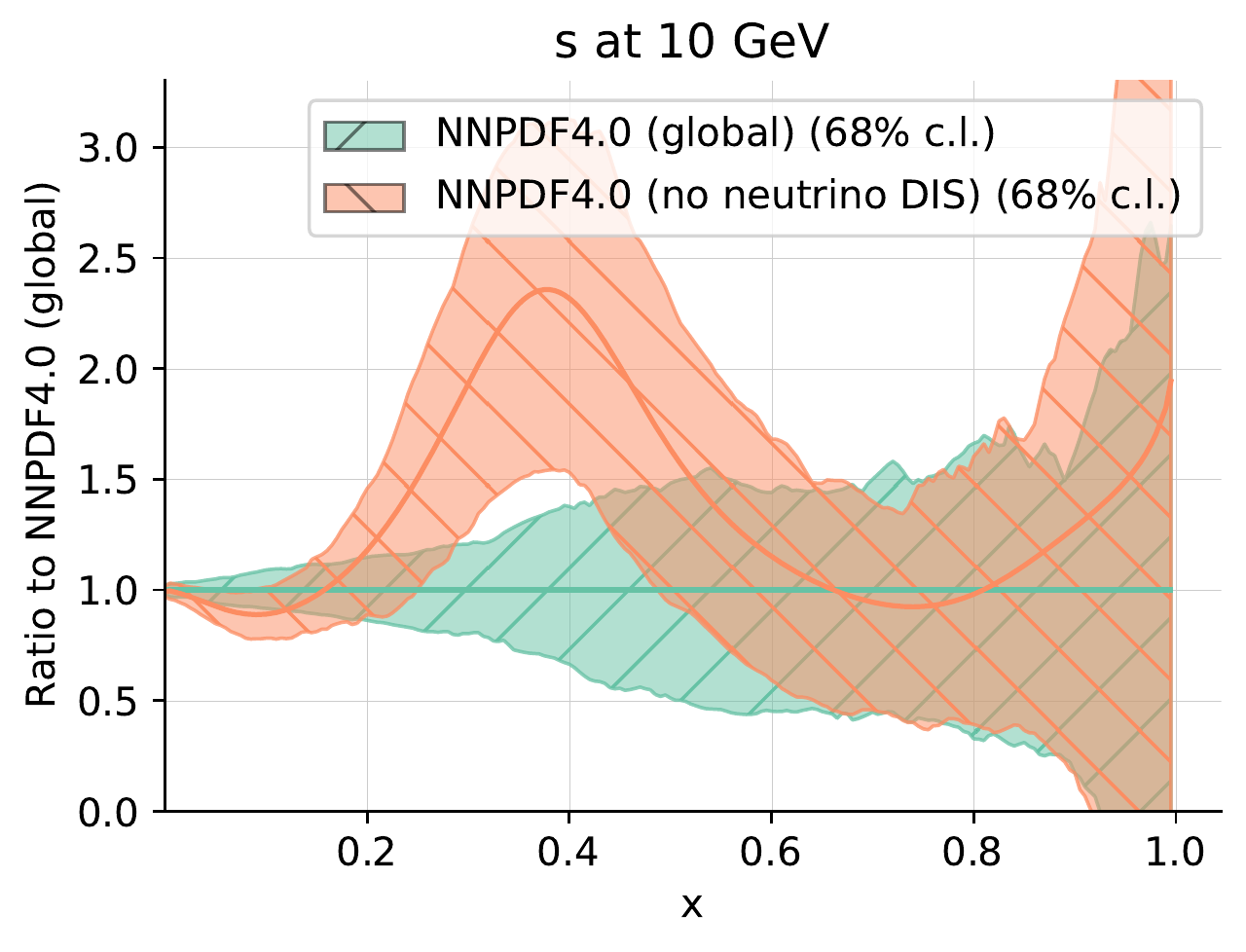}
  \includegraphics[width=0.49\textwidth]{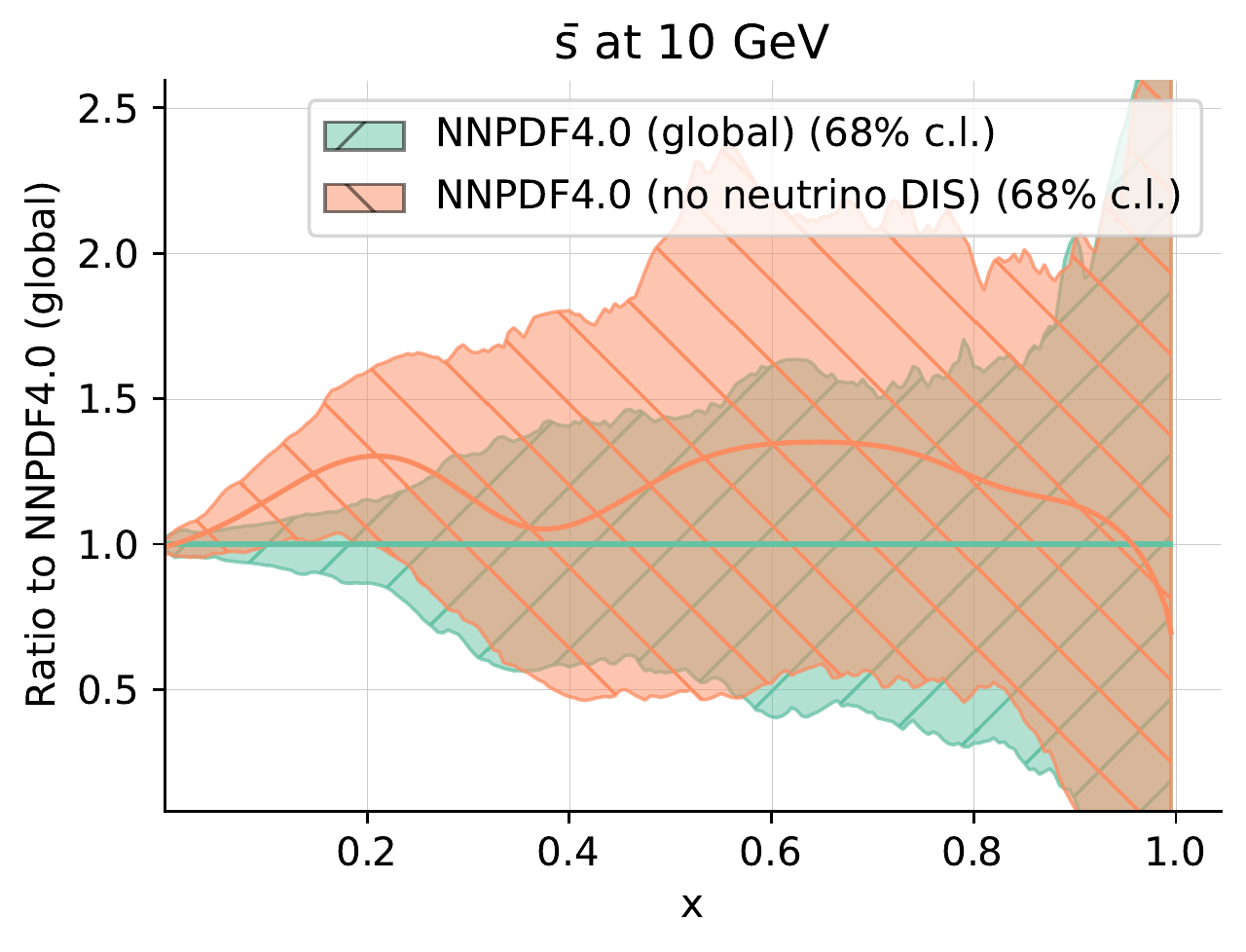}\\
  \includegraphics[width=0.49\textwidth]{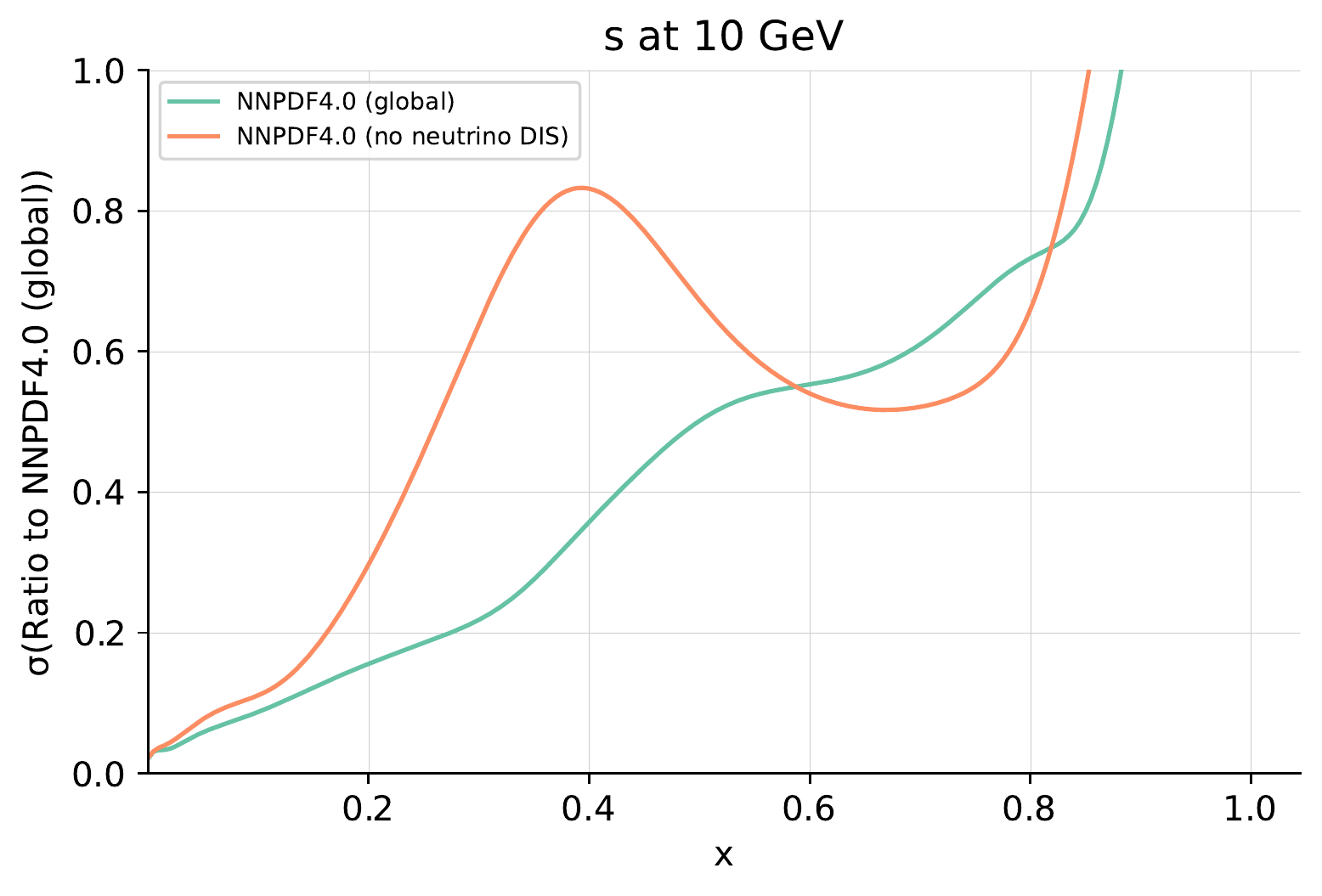}
  \includegraphics[width=0.49\textwidth]{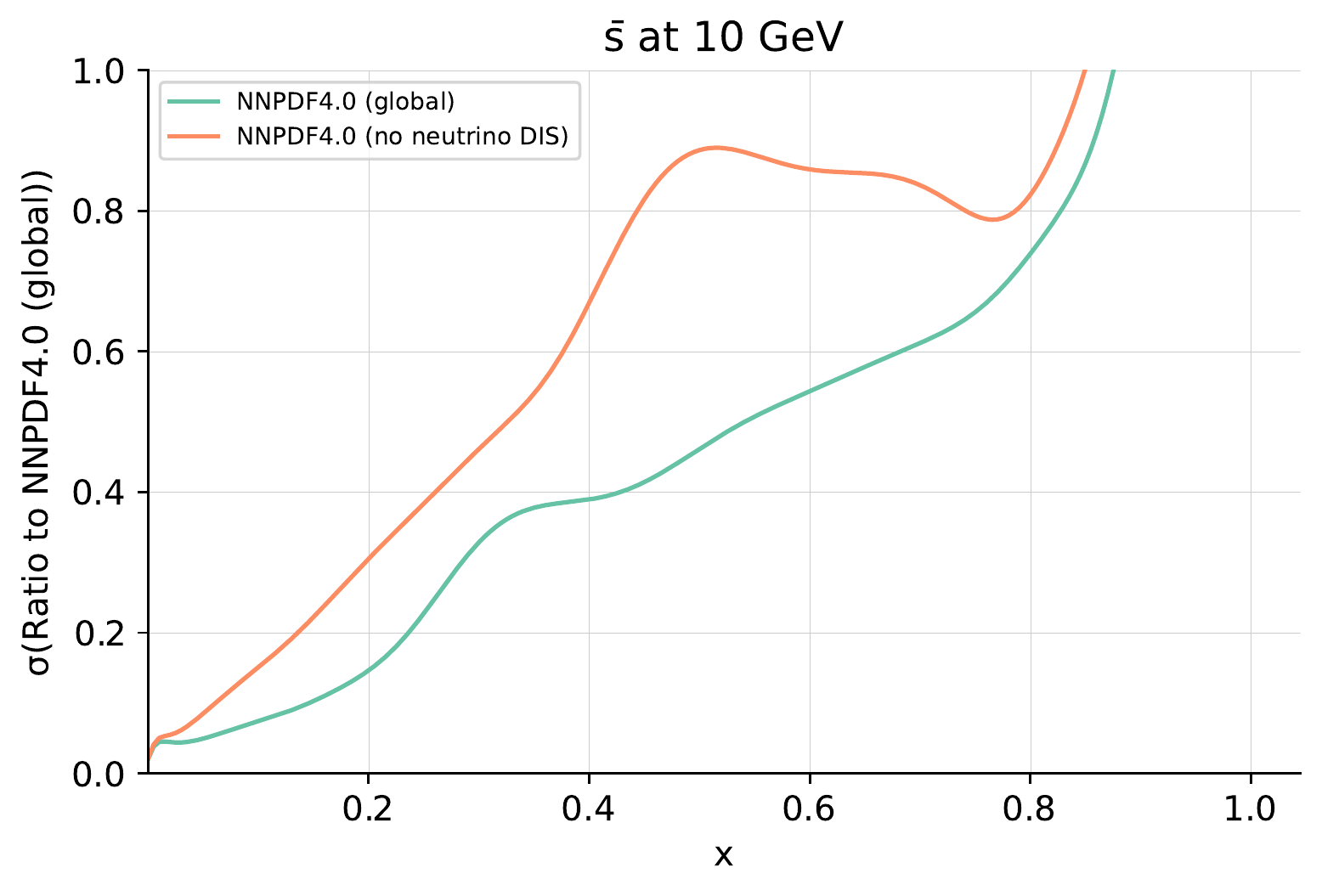}\\  
  \caption{A comparison between the strange quark (left) and anti-quark (right)
    PDFs obtained from the \texttool{NNPDF4.0} parton set and from a variant of it that does
    not include any CC neutrino DIS measurements. Results are displayed as a
    function of $x$ at $Q=10$~GeV, for the PDFs (top) and their
    relative uncertainties (bottom). In both cases, results are normalized to
    the central value of the default \texttool{NNPDF4.0} PDF set.}
  \label{fig:PDFs_neutrino}
\end{figure}

Complementary information on the strange quark and anti-quark PDFs is provided, in the proton PDF sets summarized in \cref{tab:neutrino_dataset}, by an increasing amount of complementary measurements of other processes, in particular of various production processes in LHC proton--proton collisions. These include $W$  boson production, both inclusive and in association with light jets or charm quarks. Nevertheless, the role played by CC neutrino DIS in constraining the strange quark and anti-quark distributions remains relevant in the aforementioned PDF sets. To illustrate this fact, in \cref{fig:PDFs_neutrino} we compare the strange quark and anti-quark PDFs obtained from the \texttool{NNPDF4.0} parton set and from a variant of it that does not include any CC neutrino DIS measurements. We display the strange quark and anti-quark distributions as a function of $x$ at $Q=10$~GeV, for the PDFs and their relative uncertainties. In both cases, results are normalized to the central value of the default \texttool{NNPDF4.0} PDF set. As is clear from \cref{fig:PDFs_neutrino}, the impact of CC neutrino DIS measurements is twofold: they suppress the central value of the strange quark PDF and reduce the uncertainty of the strange quark and anti-quark PDFs by about a factor of two at $x\sim 0.4$. 

\begin{figure}[!t]
  \centering
  \includegraphics[width=0.49\textwidth]{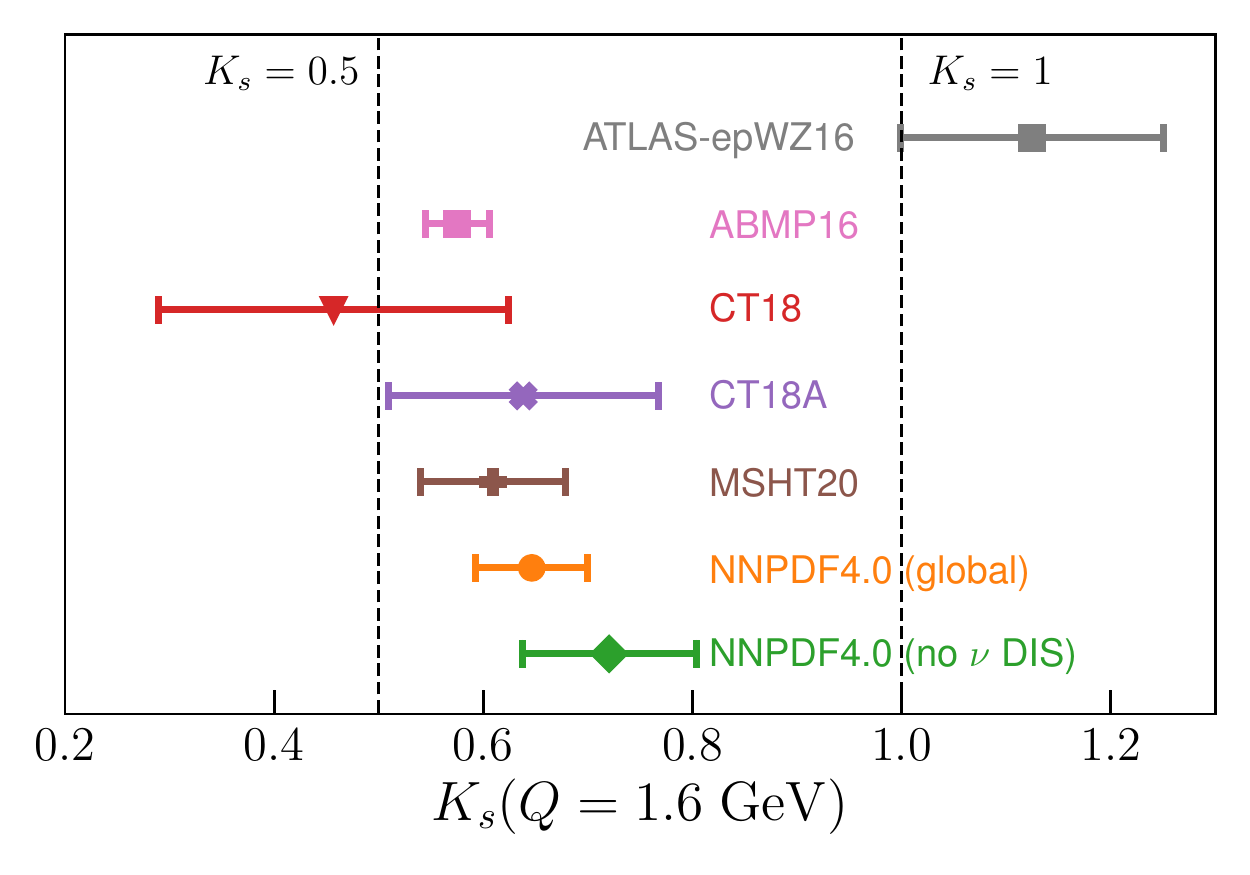}
  \includegraphics[width=0.49\textwidth]{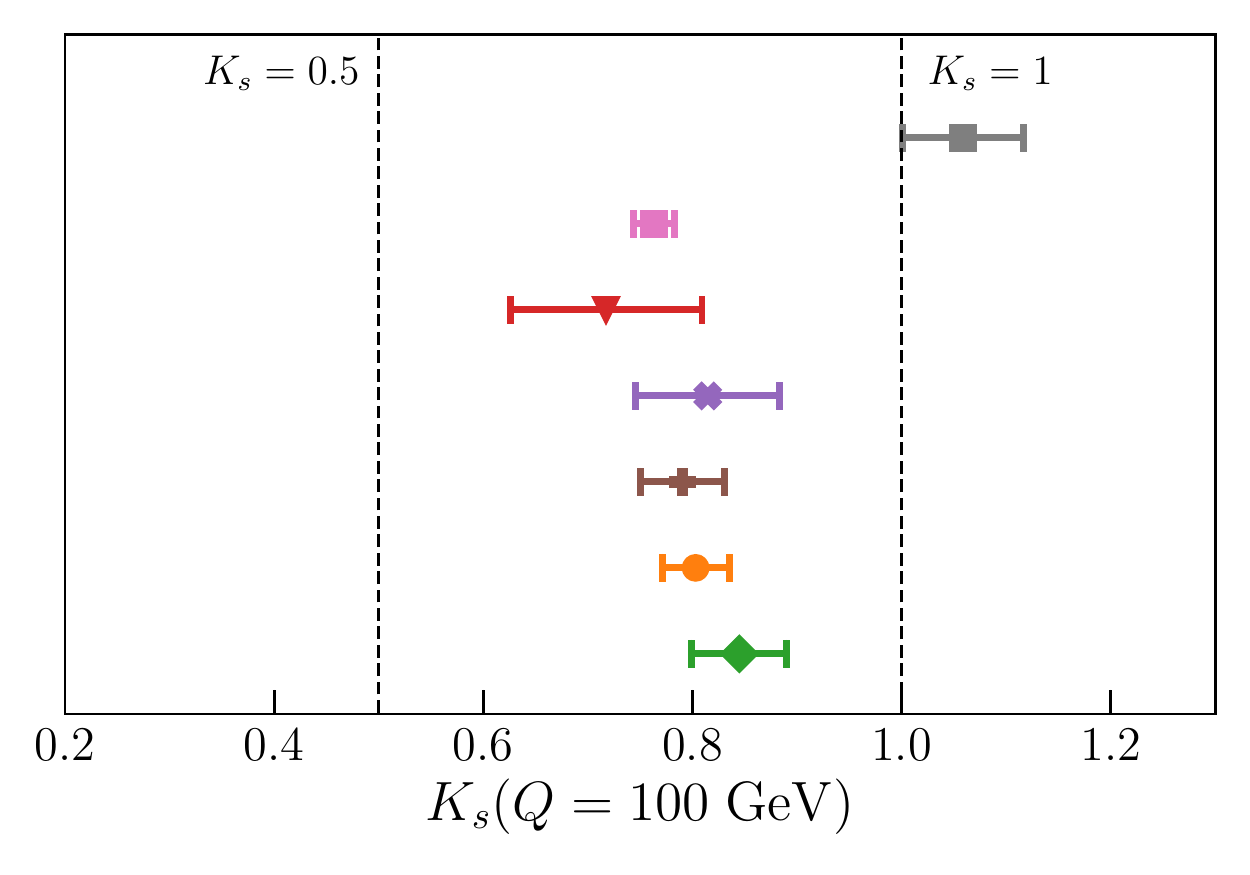}\\
  \caption{The ratio $K_s$, \cref{eq:Ks}, at $Q=1.65$~GeV (left) and
    $Q=100$~GeV (right) obtained from the following PDF sets:
    \texttool{ATLAS-epWZ16}~\cite{ATLAS:2016nqi}, \texttool{ABMP16}~\cite{Alekhin:2017kpj},
    \texttool{CT18/CT18A}~\cite{Hou:2019efy}, \texttool{MSHT20}~\cite{Bailey:2020ooq} and
    \texttool{NNPDF4.0}~\cite{Ball:2021leu} (with and without neutrino DIS data).}
  \label{fig:Ks}
\end{figure}

The compatibility of CC neutrino DIS measurements with LHC measurements has been investigated in detail. In this respect, a quantity which is usually considered is the ratio of strange to non-strange sea quark PDFs, $R_s$, possibly integrated over the $x$ range, $K_s$:
\begin{equation}
  R_s\equiv\frac{s(x,Q^2)+\bar s(x,Q)}{\bar u(x,Q) + \bar d(x,Q)}
  \qquad
  \qquad
  K_s\equiv\frac{\int_0^1dx x [s(x,Q)+\bar s(x,Q)]}{\int_0^1dx x [\bar u(x,Q) + \bar d(x,Q)]}\, .
  \label{eq:Ks}
\end{equation}
In~\cite{ATLAS:2012sjl,ATLAS:2016nqi} an analysis of inclusive gauge boson production measurements collected by the ATLAS experiment at 7~TeV suggested values of $R_s \sim 1$ when PDFs are evaluated at $x=0.023$ and $Q=1.6$~GeV. This finding is in contrast to the belief, supported by CC neutrino DIS measurements, that total quark and anti-quark strange distributions should be suppressed with respect to other light sea quarks to around $R_s\sim 0.5$ for the same values of $x$ and $Q$. Tension between CC neutrino DIS data and the ATLAS measurement~\cite{ATLAS:2016nqi} was also reported in the \texttool{CT18} global analysis~\cite{Hou:2019efy}, in which the ATLAS measurement was not included in the baseline parton set, but only in a variant set called \texttool{CT18A}. The \texttool{MSHT20}~\cite{Bailey:2020ooq} and \texttool{NNPDF4.0}~\cite{Ball:2021leu} analyses found that a larger total strange distribution, more similar to the one favored by the ATLAS measurement, also follows from CC neutrino DIS measurements if these are analysed after inclusion of recently computed NNLO charm-quark mass corrections~\cite{Berger:2016inr,Gao:2017kkx}. They also found general compatibility with other LHC measurements, namely of ATLAS and CMS $W+c$~\cite{ATLAS:2014jkm,CMS:2013wql,CMS:2018dxg} and ATLAS $W$+jet~\cite{ATLAS:2017irc} measurements, see also~\cite{Faura:2020oom} and the \texttool{ABMP16} parton set~\cite{Alekhin:2017kpj} (the only two analyses to also include NOMAD measurements). This state of affairs is summarized in \cref{fig:Ks}, where the ratio $K_s$, \cref{eq:Ks}, is displayed at $Q=1.65$~GeV and $Q=100$~GeV for the ATLAS~\cite{ATLAS:2016nqi}, \texttool{ABMP16}~\cite{Alekhin:2017kpj}, \texttool{CT18}/\texttool{CT18A}~\cite{Hou:2019efy}, \texttool{MSHT20}~\cite{Bailey:2020ooq} and \texttool{NNPDF4.0}~\cite{Ball:2021leu} (with and without neutrino DIS data) parton sets.

The FPF will provide additional measurements, in a kinematic region that extends the coverage of current CC neutrino DIS data, that may further clarify how much the strange quark and anti-quark distributions are suppressed in comparison to other light sea quark PDFs. In particular, because the FPF would probe a higher energy regime than those accessed by current data, measurements are expected to be affected by smaller theoretical uncertainties. Furthermore, the FPF may use different techniques for charm tagging, including the detailed reconstruction of the topology of the charmed meson and baryon decays achieved by emulsion experiments.

All the available CC neutrino DIS measurements make use of nuclear targets, typically Fe or Pb (see \cref{tab:neutrino_dataset}, instead of free protons. The FPF will be no exception, given the Ar or W target foreseen in LAr and emulsion experiments. This fact has two consequences. First, if the data is used to determine free-proton PDFs, nuclear corrections should be taken into account. Second, the data could instead be used to determine nuclear corrections themselves, for example by means of a determination of nuclear PDFs.

In the first case, nuclear corrections are included in global QCD analyses in various ways. In \texttool{ABMP16} and \texttool{CT18}, various nuclear models are used~\cite{Alekhin:2017kpj,Hou:2019efy}; in \texttool{MSHT20}, PDFs are corrected by means of the nuclear factors independently determined in~\cite{deFlorian:2011fp}; and in \texttool{NNPDF4.0} CC neutrino DIS data are de-weighted by a correlated uncertainty determined as the difference between the observables obtained with nuclear and free-proton PDFs~\cite{Ball:2018twp}. In this last case, the same methodology is used to determine nuclear and proton PDFs, specifically, the \texttool{nNNPDF3.0} nPDF set is used~\cite{Khalek:2022zqe}. As proton PDFs get more precise, nuclear corrections are starting to becoming of the same size of PDF uncertainties. Their inclusion in global analyses is therefore increasingly relevant to ensure the accuracy of proton PDFs.

In the second case, CC neutrino DIS data are sometimes included in modern nPDF determinations, as summarized in \cref{tab:neutrino_dataset} for the \texttool{EPPS21}~\cite{Eskola:2021nhw}, \texttool{nCTEQ15}~\cite{Kovarik:2015cma}, \texttool{nNNPDF3.0}~\cite{Khalek:2022zqe} and \texttool{TUJU21}~\cite{Helenius:2021tof} sets. Depending on the case, CC neutrino DIS measurements are not included in the proton PDF parton sets that are used as input (as is \texttool{nNNPDF3.0} and \texttool{TUJU21}), or, if they are, they are conversely not included in the nPDF determination (as for \texttool{nCTEQ15}). The impact of CC neutrino DIS measurements on nPDFs is similar to that observed for proton PDFs, if not more important given the comparatively restricted abundance of complementary LHC gauge boson measurements.
A few inconsistencies between different data sets are reported in the \texttool{EPPS21}~\cite{Paukkunen:2013grz} analysis, which  does not include NuTeV data. In the future, one may think of integrating more coherently proton and nuclear PDF determinations using one as the input to the other in a simultaneous QCD analysis. This approach may improve the overall accuracy of the determinations, especially in light of the fact that the availability of more LHC measurements will allow one to determine proton and nuclear PDFs to similar precision. If the aforementioned inconsistencies are of methodological origin, they will be likely removed.


\subsection{Neutrino DIS Cross Sections on a Tungsten Target}
\label{sec:nuDIS2}

As discussed above, the FPF will provide TeV--scale CC neutrino DIS measurements on Ar and W targets.
These will be complementary to future measurements envisioned at other facilities in terms of the energy reach and in the probed ion. Indeed, they will accompany the TeV--scale proton--Pb, Pb--Pb~\cite{Jowett:2018yqk}, and proton--O~\cite{Brewer:2021kiv} collision program carried out at the LHC, as well to the GeV--scale proton-ion collision program a the EIC~\cite{AbdulKhalek:2021gbh}. In the first respect, the FPF should help reveal whether nuclear medium effects are different in NC and CC DIS~\cite{Paukkunen:2014nqa}.
In the second respect, the availability of significant measurements for nuclei with intermediate
atomic numbers $A$ in between deuterium and Pb, for which measurements are currently the most abundant,
will allow one to test whether the commonly accepted continuous parametrization of the nPDF dependence on the atomic number $A$ is broken or not. 

\begin{figure}[h]
\centering
\includegraphics[width=0.49\textwidth]{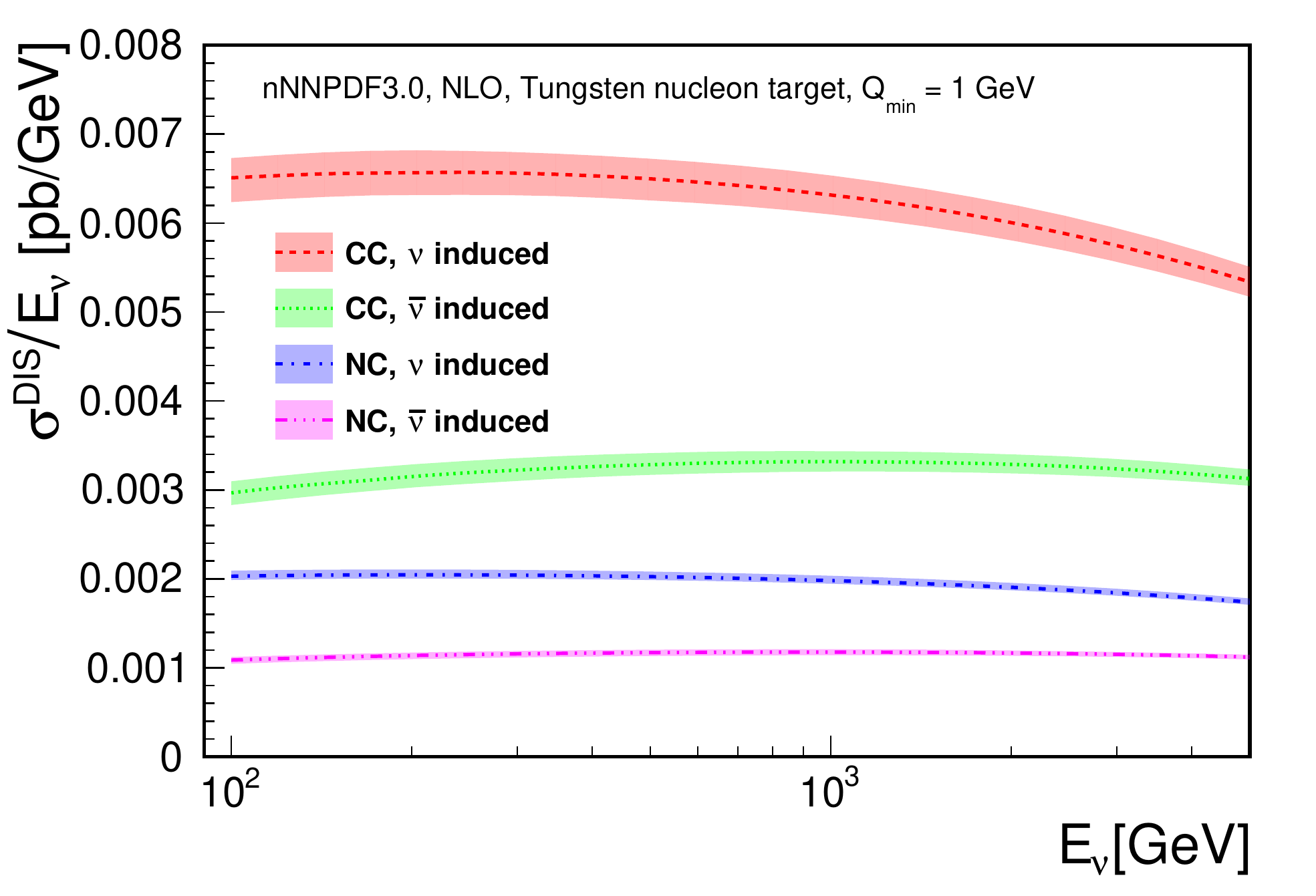}
\includegraphics[width=0.49\textwidth]{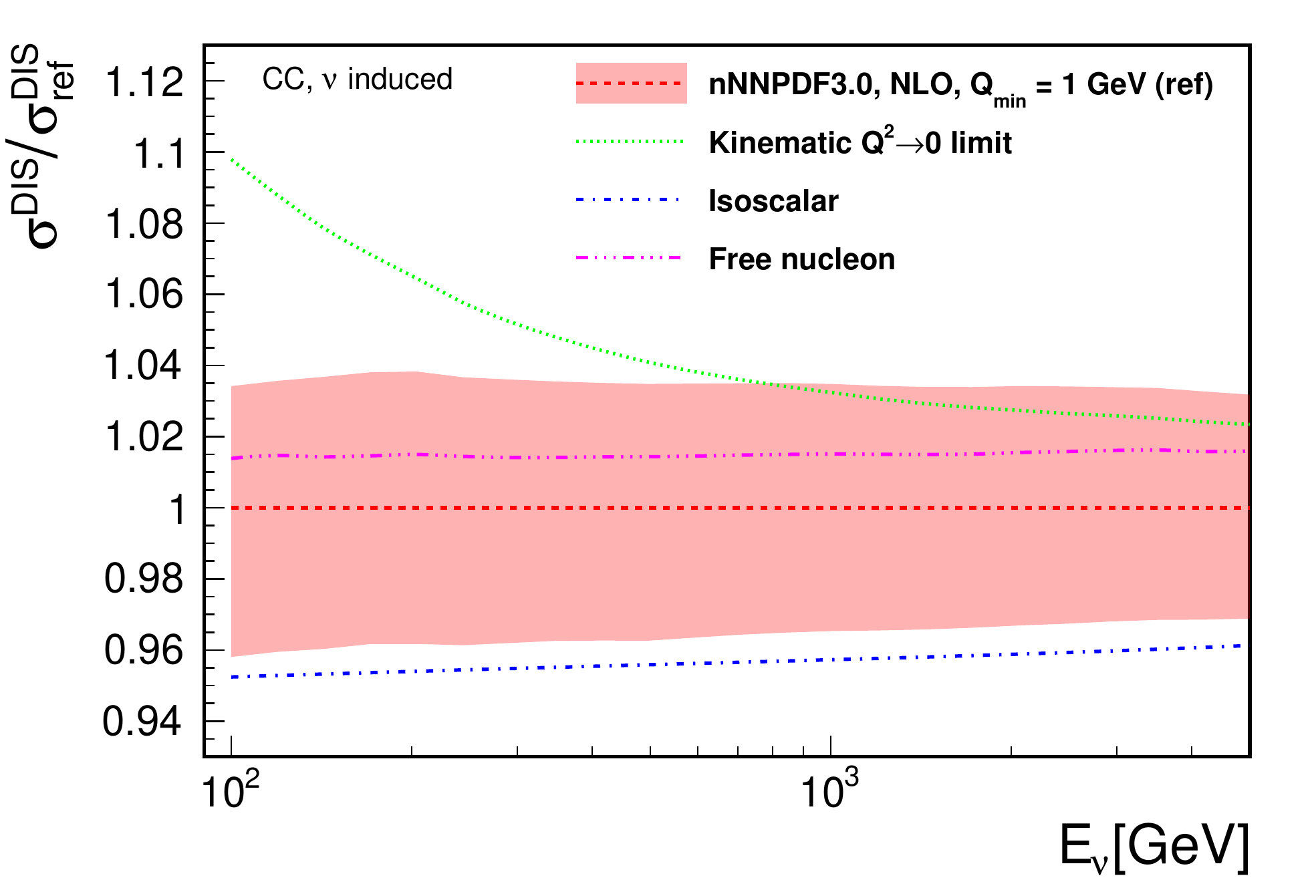}
\caption{Left: The inclusive NLO DIS cross-section as a function of the
  neutrino energy $E_{\nu}$
  for charged- and neutral-current scattering processes with either neutrino or antineutrino projectiles.
  A tungsten  target is assumed and the predictions are obtained with the \texttool{nNNPDF3.0} nPDFs.
  A cut of $Q \ge Q_{\rm min}=1$ GeV is applied to ensure the reliability of the perturbative
  QCD framework.
  Right: Normalised predictions for the neutrino-induced charged-current process
  quantifying the impact of including/neglecting various physics effects in the prediction.
  In both panels, the error bands correspond to the 68\% CL nPDF uncertainties.
}
\label{fig:nuDISinclusive}
\end{figure}

In principle, the measurement of inclusive neutrino-induced DIS cross sections can provide useful information on our knowledge of nucleon structure.
To demonstrate this, predictions for charged-current and neutral-current scattering rates are provided in \cref{fig:nuDISinclusive} (left), where the absolute DIS cross-section has been scaled by the incoming neutrino energy $E_{\nu}$. These predictions have been obtained at NLO QCD accuracy with the \texttool{nNNPDF3.0} nPDFs~\cite{Khalek:2022zqe} for a tungsten nucleon target (i.e. it is mass averaged) with \texttool{APFEL}~\cite{Bertone:2013vaa}, and follow the computational setup presented in~\cite{Bertone:2018dse}.
The four curves (for the various processes) show the uncertainty due to knowledge of the nPDFs, which is in the range of (3-4)\% in the accessible energy range.
The calculation has been carried out with a cut in the momentum transfer of $Q \geq Q_{\rm min}= 1$ GeV,
to ensure the reliability of the perturbative calculation.
The  right panel of \cref{fig:nuDISinclusive} displays then the ratio with respect to the central value is shown for the neutrino-induced charged-current process.
In addition, that plot also shows the impact of various physics effects: a kinematic extrapolation in the $Q^2\to0$ limit; the impact of assuming an isoscalar target; and the impact of neglecting nuclear corrections.

The results of \cref{fig:nuDISinclusive}
indicate that knowledge of nPDFs in the required kinematic regime is already good for the inclusive cross section, and that precise measurements will be required to further constrain the nPDFs.
However, we note that predictions based on other nPDF analyses may not be contained
by the nPDF uncertainty bands of \texttool{nNNPDF3.0} and hence FPF measurements would anyway
provide discrimination power between different nuclear PDF fits.
These results also highlight the potential impact of the $Q^2\to0$ region.
That is shown with the DIS prediction which has been obtained by freezing the lower value of $Q$ which enters the evaluation of the structure function at $Q = Q_0^{\rm PDF} = 1~{\rm GeV}$, but allowing $Q\to0$ in the kinematics of the scattering process.
This indicates that differential measurements are necessary to separate the perturbative regime, and that such measurements could also provide important information on the neutrino-induced charged-current structure functions at low-$Q^2$ values, whose current theoretical modelling relies on a number of
assumptions.

%% file: sec_neutrino.tex

\contributors{Kevin Kelly, Vishvas Pandey, Mary Hall Reno (conveners), Weidong Bai, Pouya Bakhti, Baha Balantekin, Atri Bhattacharya, Vedran Brdar, Peter Denton, Milind V. Diwan, Yong Du, Yasaman Farzan, Saeid Foroughi-Abari, Alexander Friedland, Kai Gallmeister, Alfonso Garcia, Maria V. Garzelli, Ahmed Ismail, Sudip Jana, Yu Seon Jeong, Felix Kling, Karan Kumar, Roshan Mammen Abraham, Ulrich Mosel, Laurence Nevay, Luke Pickering, Ryan Plestid, Meshkat Rajaee, Ina Sarcevic, Subir Sarkar, Jan Sobczyk, Anna Stasto, Zahra Tabrizi, Sebastian Trojanowski, Yu-Dai Tsai, Douglas Tuckler, Jiang-Hao Yu, Yue Zhang}

\begin{figure}
    \centering
    \includegraphics[width=0.95\textwidth]{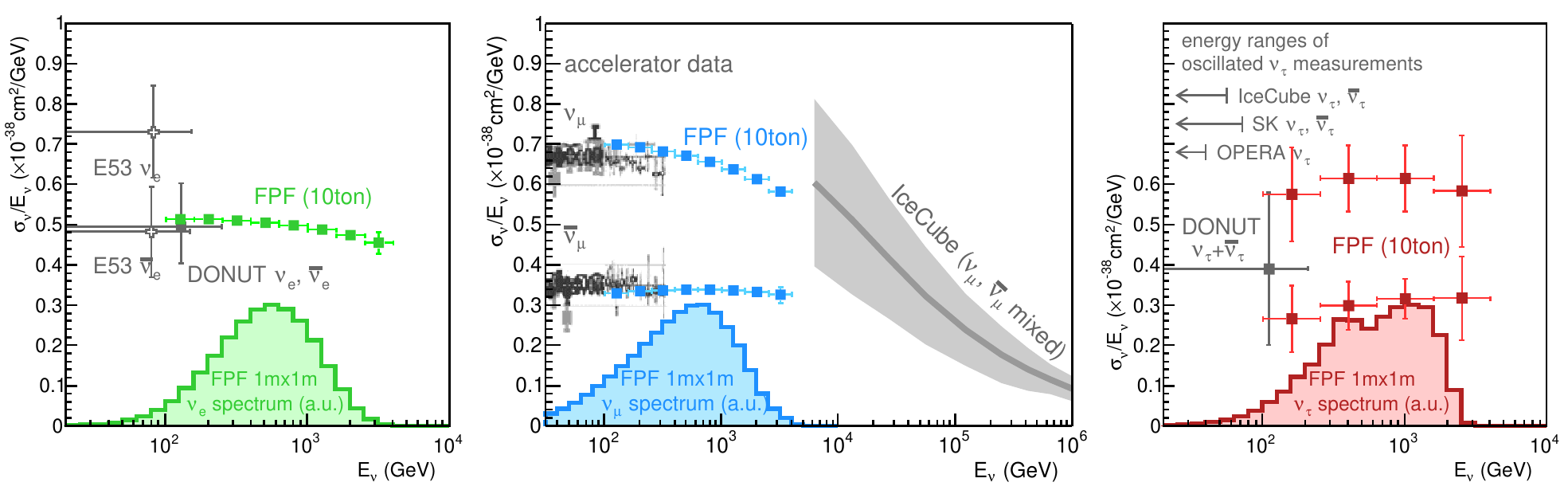}  
    \caption{The neutrino flux as a function of energy for $\nu_e+\bar\nu_e$ (left), $\nu_\mu+\bar\nu_\mu$ (middle) and $\nu_\tau+\bar\nu_\tau$ (right) for a 10 ton detector with $\eta\gtrsim 6.9$. Also shown are the expected precision of FPF measurements for neutrino plus antineutrino interactions with nucleons (left) and separate $\nu_\mu$ and $\bar{\nu}_\mu$ (middle) and $\nu_\tau$ and $\bar{\nu}_\tau$ (right) cross sections with nucleons showing statistical errors only. Data are shown from E53 \cite{Baltay:1988au}, DONUT \cite{DONuT:2007bsg}, a compilation of accelerator experiments \cite{ParticleDataGroup:2020ssz} and IceCube \cite{IceCube:2017roe}.
}
    \label{fig:nuxc-fpf}
\end{figure}

\section{Overview}

As emphasized throughout this document, measurements at the FPF offer opportunities to study QCD dynamics, neutrino interactions and BSM physics using fluxes of neutrinos of all three flavors. As  \cref{fig:nuxc-fpf} illustrates, new energy regimes unexplored by the fixed target experiments thus far will be probed by high energy, intense fluxes of $\nu_e+\bar\nu_e$, $\nu_\mu+\bar\nu_\mu$ and $\nu_\tau+\bar\nu_\tau$.  New cross section measurements at the TeV energies of the neutrino fluxes at the FPF will significantly extend accelerator measurements and open up opportunities to discover or constrain BSM physics.
Other white papers on high energy/ultrahigh energy neutrinos \cite{UHEnuWhitepaper}, on tau neutrinos \cite{nuTauWhitepaper} and on event generators for high energy experiments \cite{MCWhitepaper} complement some of the discussion here.

The fluxes of $\nu_\tau+\bar\nu_\tau$  and high energy $\nu_e+\bar\nu_e$ fluxes come from charm hadron production and decay at large rapidities. As detailed in \cref{sec:qcd}, the neutrino fluxes from charm depend on small- and large-$x$ parton distribution functions, regions uniquely probed at the FPF. Evaluations of neutrino fluxes from light and heavy mesons using Monte Carlo generators and from charm using NLO perturbative QCD with collinear factorization and $k_T$ factorization are presented in \cref{sec:NeutrinoFluxes}. As the LHC Run 3 experiments progress, the current range of predictions and assessments of uncertainties will be refined as progress is made on theoretical and experimental fronts.

Understanding theoretical predictions of the neutrino and antineutrino DIS cross sections at the few percent level requires an understanding of PDFs, quark mass effects, weak structure functions in the non-perturbative regime, and the transition region from the deep-inelastic scattering regime to resonant and quasi-elastic scattering. Augmenting measurements of the out-going lepton at the FPF will be measurements of the hadronic final state that will yield insights into hadronization in the nuclear environment. These topics are reviewed in \cref{sec:NeutrinoCrossSections}.

The development of Monte Carlo programs to model neutrino interactions with nucleons and nuclear targets has spanned decades. Much of the development has focused on the sub-GeV to few-GeV range and expanded to higher energies
as discussed in \cref{sec:tools}. Going forward, detailed comparisons of the Monte Carlo results for hundreds of GeV to TeV neutrino energies will be important for the modeling of events in detectors. Confronting the Monte Carlo results with data from the FPF will probe new kinematic regimes and reveal some of the underlying dynamics of hadronization and final state interactions/hadronic transport, complementary to the long-baseline neutrino program. 

With intense neutrino and antineutrino fluxes of all three flavors, the FPF enables tests of the both standard model and searches for signals of BSM physics at higher energies than have previously been explored. Large fluxes of neutrinos with TeV scale energies open up many possibilities for discovering new physics in the neutrino sector. Here, we consider BSM physics of a specific type: those that modify the expected neutrino flux (distributions and/or normalizations) at the FPF location and/or the neutrino interactions in the FPF detector(s). While this is a fairly simple categorization, it includes many interesting  possibilities, such as new interactions between neutrinos and other SM particles, new neutrinophilic mediators, magnetic moments of neutrinos, sterile neutrinos, and dark matter that interacts solely with neutrinos. \cref{sec:BSMNeutrinos} summarizes the prospects for these searches in the FPF detectors, focusing on complementarity between searches here and in other, contemporary experiments.

\section{Neutrino Fluxes} 
\label{sec:NeutrinoFluxes}

\subsection{Neutrino fluxes from Monte Carlo Generators}
\label{sec:flux}

The experiments at the FPF will perform a variety of measurements with LHC neutrinos. These neutrinos are produced in the weak decays of the lightest hadrons of a given flavor. In particular electron neutrinos mainly originate from the semi-leptonic decay of kaons $K \to \pi e \nu_e$ and charmed hadron decays $D \to K e \nu_e$, muon neutrinos are primarily produced in leptonic decays of charged pions $\pi^\pm \to \mu \nu_\mu$ and kaons $K^\pm \to \mu \nu_\mu$, while tau neutrinos are predominantly produced in $D_s \to \tau \nu_\tau$ decays and subsequent tau decays. Depending on the particle's lifetime, the neutrinos can either be produced promptly at the interaction point (IP) or further downstream in the LHC's vacuum beam pipe. For the neutrino physics measurements at the FPF it is important to have reliable estimates of the neutrino fluxes, which require an accurate modelling of i) the production of hadrons and ii) their propagation through the forward LHC infrastructure. 

To address the first part of the problem, the modelling of hadron production, we use several Monte Carlo event generators that are commonly used to describe forward particle production: \texttool{EPOS-LHC}~\cite{Pierog:2013ria}, \texttool{QGSJet~II-04}~\cite{Ostapchenko:2010vb}, \texttool{DPMJet 3.2017}~\cite{Roesler:2000he, Fedynitch:2015kcn}, \texttool{Sibyll~2.3d}~\cite{Ahn:2009wx, Riehn:2015oba, Riehn:2017mfm, Fedynitch:2018cbl} and \texttool{Pythia~8}~\cite{Sjostrand:2014zea} (configured with the Monash tune~\cite{Skands:2014pea}). These tools have been developed for decades, either as dedicated Monte Carlo generators for cosmic ray physics or as multi-purpose generators for collider physics, and tuned to a variety of available data sets. 

The second part of the problem regards the propagation of hadrons through the LHC infrastructure. One option to address this question is to use dedicated particle propagation tools such as \texttool{FLUKA}~\cite{FLUKA:web,FLUKA:new,Battistoni:2015epi}
or \texttool{BDSIM}~\cite{Nevay:2018zhp} (which is based on \texttool{Geant4}) described in \cref{sec:FLUKA} and \cref{sec:BDSIM}. However, these tools tend to be rather time consuming and often require special expertise or code access that is not available to the broad community. To avoid these issues, we will follow a different approach and use the fast neutrino flux simulation introduced in Ref.~\cite{Kling:2021gos}.  \medskip 

\begin{figure*}[t]
  \centering
  \includegraphics[width=0.99\textwidth]{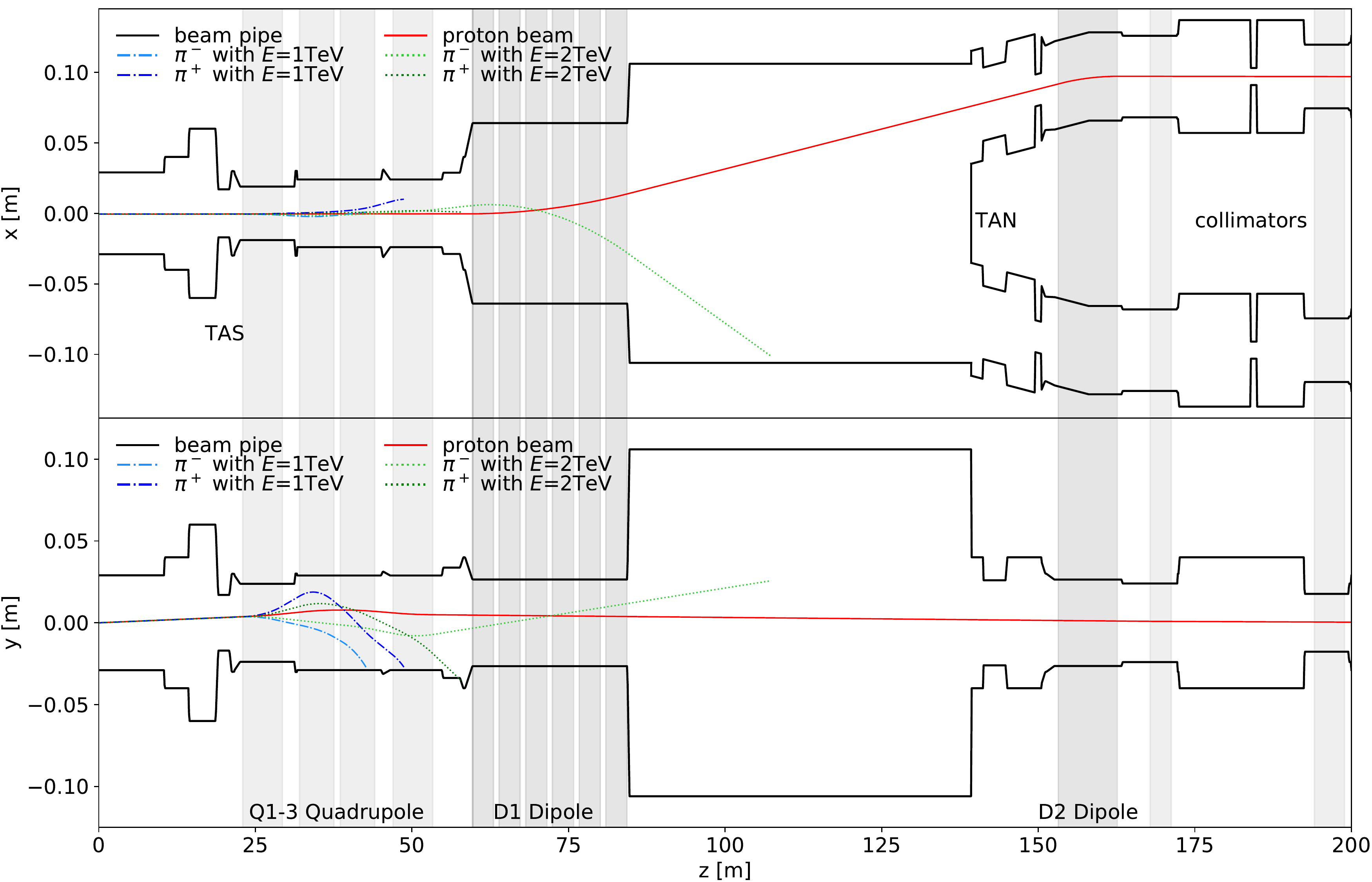}
  \caption{Geometrical model of the forward LHC infrastructure used in the fast neutrino flux simulation. The upper and lower panel show a cross sectional view in the horizontal and vertical plane, respectively. The black lines represent the boundaries of the LHC’s beam pipe and the shaded areas correspond to the quadrupole (light gray) and dipole (dark gray) magnets. 
  We additionally show the trajectory of the proton beam (red) and two oppositely charged pions with energy of 1~TeV (dot-dashed) and 2~TeV (dotted). All shown particles have an initial half beam crossing angle of 150 $\mu$rad vertically upwards.}
  \label{fig:fnfs_geometry}
\end{figure*}

This geometrical model used in the simulation is based on the \texttool{BDSIM} model described in \cref{sec:BDSIM} and corresponds to the configuration of the LHC that was used during the end of Run~2. At the collision point, it assumes a 13~TeV collision energy with a beam half-crossing angle of 150 $\mu$rad upwards in vertical directions. The layout of the forward LHC infrastructure is shown in \cref{fig:fnfs_geometry}. Here we show a cross sectional view of the beam pipe geometry in the horizontal (upper panel) and vertical (lower panel) plane. The boundaries of the vacuum beam pipe are shown as solid black lines. Additionally, the magnetized areas are highlighted as gray shaded areas.  

Placed at the about $z=20~\m$ downstream from the interaction point is the TAS front quadrupole absorber which is designed to absorbs particles with angles larger than $0.9~\mrad$ with respect to the beam axis. Located behind is a series of quadrupole magnets to focus the beam, followed by the D1 dipole magnet to separate the two proton beams. At around $z=140~\m$ the beam pipe splits into separate pipes for the individual beams. Placed at this location is also the TAN which which will absorb the forward going neutral particles. It is followed by the D2 dipole magnets, which deflects the protons beam such that they are parallel again, as well as several collimators to absorb any beam halo and debris.

As an illustration, we also show several example trajectories as colored lines in \cref{fig:fnfs_geometry}. Shown in red is the nominal proton beam which has a half crossing angle 150~$\mu$rad vertically upwards. In addition, we show the the trajectories of oppositely charged pions with energy of 1~TeV and 2~TeV as blue and green lines, produced in the same direction as the proton beam. We note that despite their large energy, even multi-TeV charged particles are significantly deflected by the quadrupole magnets. This implies that decays of charged particles occurring further downstream are  not expected to contribute much to the neutrino flux at the FPF.

We note that the LHC will undergo a variety of changes before the start of the HL-LHC era. This, for example, includes an increase of the collision energy to to 14 TeV, an upgrade of the magnets, and a relocation of the TAN of roughly $14~\m$ towards the IP~\cite{ZurbanoFernandez:2020cco}. While a geometrical model reflecting these changes will be needed to make more precise predictions, we proceed with the existing setup to obtain a preliminary estimate of the expected neutrino fluxes at the FPF.

Based on the geometrical model described above, we can now simulate the production of displaced neutrinos the LHC. To do this, we i) read an event from the MC event generator ii) propagate the long-lived hadrons through the LHC beam pipe and magnets until it hits a beam pipe boundary, iii) decay the hadrons at multiple locations along their trajectory (according to their decay distributions obtained with \texttool{Pythia~8}), and iv) store the resulting neutrino fluxes as histograms. All steps of this procedure have been implemented as a \texttool{RIVET} module and the results are saved in the \texttool{yoda} file format~\cite{Buckley:2010ar}. The produced results have been compared to a full \texttool{BDSIM} simulation and a good agreement was found, validating the performance of the fast neutrino flux simulation. 

\begin{figure*}[tp]
  \centering
  \includegraphics[width=0.96\textwidth]{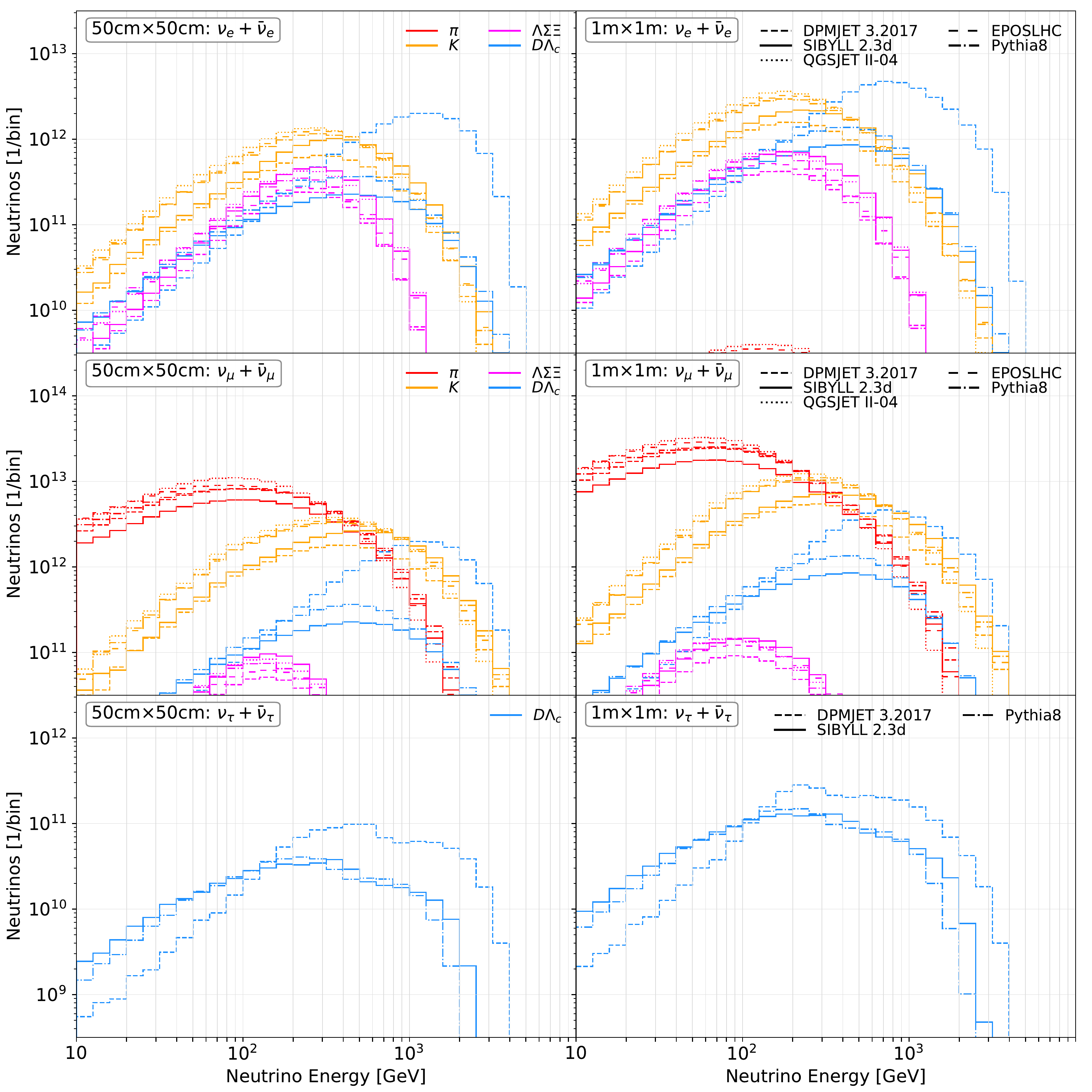}
  \caption{Predicted energy distribution of neutrinos passing through the FPF experiments. The different panels correspond to the electron (top), muon (center) and tau (bottom) neutrinos passing through a $50~\cm \times 50~\cm$ (left) and $1~\m \times 1~\m$ (right) cross sectional area at the FPF location at $z=620~\m$. The vertical axis shows the number of neutrinos per energy bin that 
  go through the considered cross sectional area for an integrated luminosity of $3000~\ifb$. The different production modes are indicated by the line color: pion decays (red), kaon decays (orange),hyperon decays (magenta), and charm decays (blue). The different line styles correspond to predictions obtained from \texttool{SIBYLL 2.3d} (solid), \texttool{DPMJet 3.2017} (short dashed), \texttool{EPOS-LHC} (long dashed), \texttool{QGSJet II-04} (dotted), and \texttool{Pythia 8.2} (dot-dashed).
  }
  \label{fig:fnfs_energy}
\end{figure*}

Before proceeding to the results, let us note that there is an additional secondary component of neutrinos originating from hadronic showers resulting from collisions of primary hadrons with the LHC infrastructure. However, the corresponding contribution to the forward neutrino flux is expected to be strongly suppressed, especially for higher energies. This is because the secondary hadrons produced in such downstream interactions both have a small probability of decay in medium before interacting again as well as a typically
broad angular spread. We have validated this statement using a full \texttool{BDSIM} simulation and found that only about 0.4~\% (2.0~\%) of the 
muon neutrinos with energy $E > 1~\tev$ ($100~\gev$) through a $0.8~\mrad \times 0.8~\mrad$ cross sectional area originate from decays in medium. For physics applications, this is an encouraging result since it allows to relate the measured LHC neutrino flux to forward hadron fluxes and hence use them as a probe of forward hadron production. We note, however, that the secondary component will become more important, and possibly even dominant, at lower energies $E \lesssim 10~\gev$. \medskip 

Let us now turn to the obtained neutrino fluxes predictions at the FPF. In \cref{fig:fnfs_energy} we show the number of neutrinos passing through a cross sectional area with dimension $50~\cm \times 50~\cm$ (left) and $1~\m \times 1~\m$ (right) at the FPF location at $z=620~\m$ downstream from the IP as function of the neutrino energy. The different rows show the energy spectrum for electron neutrinos (top), muon neutrinos (center) and tau neutrinos (bottom) where all results include both neutrinos and anti-neutrinos. 

To better understand the origin of the LHC neutrinos, the contribution corresponding to different parent particles are shown separately as indicated by the different line colors. Pion decays, shown in red, provide the dominant contribution to the muon neutrino flux at energies below a few $100~\gev$, but do not contribute to the electron neutrino flux due to the helicity suppressed branching fraction into electrons. The neutrino flux originating from kaon decays is shown in orange. Leptonic charged kaon decays provide the leading contribution to the muon neutrino flux at higher energies, while semi-leptonic neutral kaon decays are the dominant source of electron neutrinos at energies below $1~\tev$. Decays of hyperons are shown in magenta and provide a sizable contribution to the anti-electron neutrino flux at intermediate energies, mainly via the decay $\Lambda \to p e \bar\nu_e$. Finally, neutrinos from the decay of charmed hadrons, including $D$-mesons and the $\Lambda_c$ baryon, are shown in blue. Charm decays provide the dominant contribution of electron neutrinos at the highest energies. In addition, they are the main source of tau neutrinos which are produced in the both decays of $D_s$ mesons and the subsequent tau lepton decays. 

The different line styles in \cref{fig:fnfs_energy} correspond to predictions obtained with the different MC event generators. We see that the predictions for neutrinos from light hadron decay have 
small variations but larger differences of up to an order of magnitude are observed for the neutrino fluxes from charmed hadron decays. However, it is worth noting that neither \texttool{DPMJet} nor \texttool{Pythia} have been tuned to 
LHC data on charm production measurements, especially at large rapidity from LHCb. Dedicated efforts to obtain reliable predictions for the neutrino flux from charm decay are discussed in the following sections. \medskip

\begin{figure}[t]
  \centering
  \includegraphics[width=0.96\textwidth]{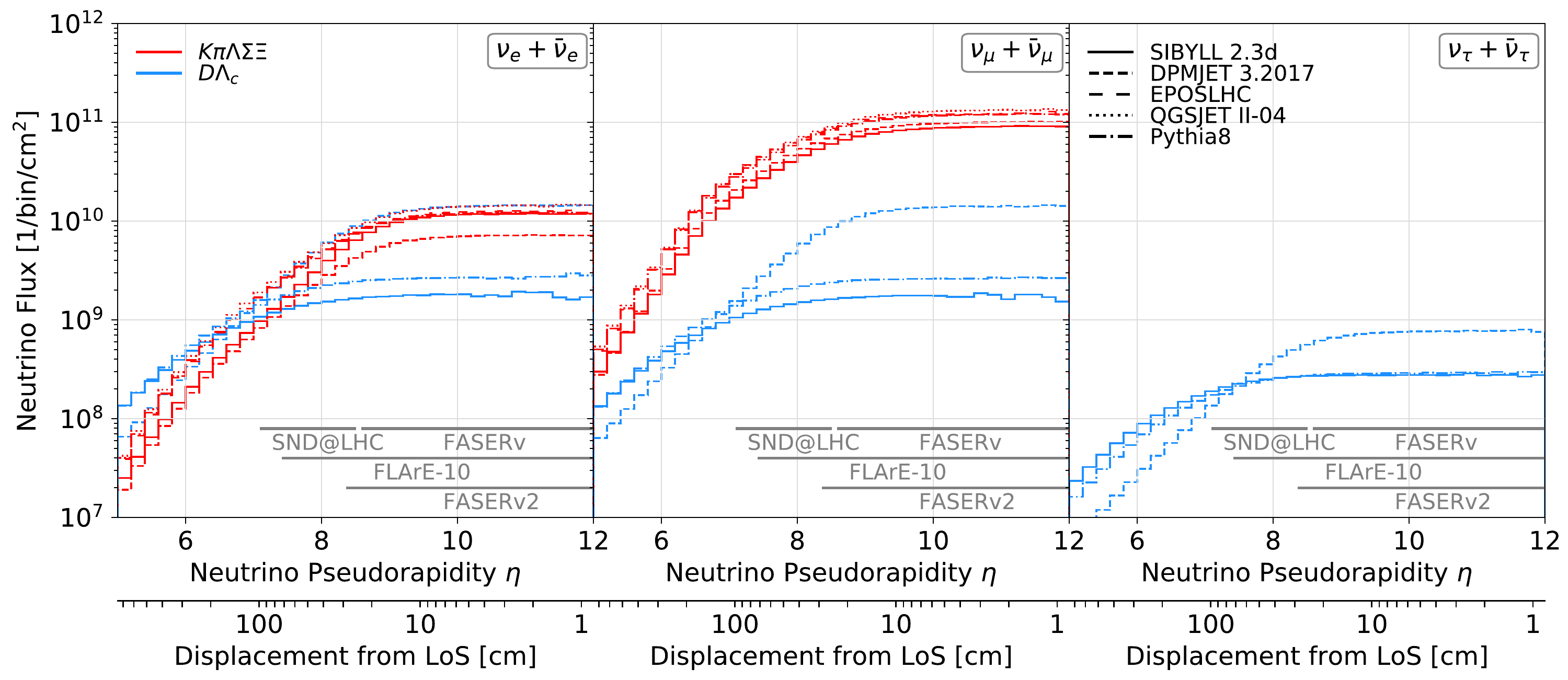}
  \caption{Predictions for the flux for electron (left), muon (center) and tau (right) neutrinos, in units of particles per area per bin at the HL-LHC with $3000~\ifb$, as function of the pseudorapidity of the neutrino $\eta$, or equivalently its radial displacement from the line of sight at $z=620~\m$. The red and blue lines correspond to the neutrino flux components from light and charmed hadron decays, respectively. The line-styles denote the different event generators. All neutrinos with energies $E > 10~\gev$ are included. Illustrated at the bottom of each panel is the angular coverage of different LHC neutrino experiments.}
  \label{fig:fnfs_rapidty}
\end{figure}

Presented in \cref{fig:fnfs_rapidty} is the rapidity dependence of the forward neutrino flux. The panels show the number of neutrinos with energy $E> 10~\gev$ per unit area and bin, as a function of the pseudorapidity. The pseudorapidity is just a
way to measure the angle with respect to the beam axis, and given the detector distance from the IP, it also corresponds to a transverse displacement from beam axis at the detector, as indicated on the additional horizontal axis in the figure. Shown at the bottom of each panel is the pseudorapidity coverage of the LHC neutrino experiments operating during LHC Run~3, FASER$\nu$ and SND~LHC, as well as the FPF neutrino experiments FASER$\nu$2 and FLArE (10 tons). 

The different panels correspond to electron neutrinos (left) muon neutrinos (center) and tau neutrinos (right). For all three flavors, the neutrino flux is maximized at the beam axis, which corresponds to $\eta = \infty$, and drops when moving away from the beam axis. The number of neutrino interaction events per detector mass is therefore maximized at the beam axis. The width of the neutrino beam depends on the neutrino production mode: while neutrinos from light hadron decays inherit a smaller transverse momentum, and are therefore more strongly collimated around the beam axis, neutrinos from heavy hadron decays typically have a higher transverse momentum and hence a larger transverse spread. This can be seen by comparing the red and blue line in \cref{fig:fnfs_rapidty}, which correspond to the neutrino flux component from light and heavy hadron decay, respectively. This means that, while the overall neutrino flux decreases when moving away from  the center of the beam, the relative fraction of neutrinos from charm decay increases. \medskip

Let us now turn to the expected neutrino event rates at the FPF. This is presented in \cref{tab:fnfs_eventrate} for both the existing neutrino detectors operating during Run~3 of the LHC, FASER$\nu$ and SND@LHC, as well as the FPF neutrino experiments, FASER$\nu$2, FLARE and AdvSND. The left part of the table summarizes the assumed detector specifications including the target mass, rapidity coverage and nominal integrated luminosity. Shown on the right are the expected number of charged current neutrino interactions occurring inside the detector volume. We show predictions for both \texttool{SIBYLL~2.3d} and \texttool{DPMJet~3.2017} which provide the maximal and minimum predictions within the set of event generators considered. 

\begin{table}[tbp]
\setlength{\tabcolsep}{3.2pt}
    \centering
    \begin{tabular}{c|c|c|c||c|c|c}
    \hline\hline
      \multicolumn{4}{c||}{Detector} & 
      \multicolumn{3}{c}{Number of CC Interactions} \\
      \hline
      Name &  Mass & Coverage & Luminosity
      & $\nu_e\!\!+\!\bar{\nu}_e$ 
      & $\nu_\mu\!\!+\!\bar{\nu}_\mu$
      & $\nu_\tau\!\!+\!\bar{\nu}_\tau$
      \\
       \hline\hline
       FASER$\nu$  
       & 1 ton & $\eta \gtrsim 8.5$ & 150~fb$^{-1}$
       & 901 / 3.4k  & 4.7k / 7.1k & 15 / 97  \\
       \hline 
       SND@LHC  
       & 800kg & $7<\eta < 8.5$ & 150~fb$^{-1}$
       & 137 / 395  & 790 / 1.0k & 7.6 / 18.6  \\
       \hline\hline
       FASER$\nu$2  
       & 20 tons & $\eta \gtrsim 8.5$ & 3~ab$^{-1}$
       & 178k / 668k  & 943k / 1.4M  & 2.3k / 20k \\
       \hline
       FLArE 
       & 10 tons & $\eta \gtrsim 7.5$ & 3~ab$^{-1}$
       & 36k / 113k & 203k / 268k & 1.5k / 4k  \\
       \hline
       AdvSND  
       & 2 tons & $7.2 \lesssim \eta \lesssim 9.2$ & 3~ab$^{-1}$
       & 6.5k / 20k & 41k / 53k & 190 / 754 \\
       \hline\hline
    \end{tabular}
    \caption{Detectors and neutrino event rates: The left side of the table summarizes the detector specifications in terms of the target mass, pseudorapidity coverage and assumed integrated luminosity for both the LHC neutrino experiments operating during Run~3 of the LHC as well as the proposed FPF neutrino experiments. On the right, we show the number of charged current neutrino interactions occurring the detector volume for all three neutrino flavors as obtained using two different event generators, \texttool{Sibyll~2.3d} and \texttool{DPMJet~3.2017}. 
    }
    \label{tab:fnfs_eventrate}
\end{table}

\begin{figure}[t]
  \centering
  \includegraphics[width=0.49\textwidth]{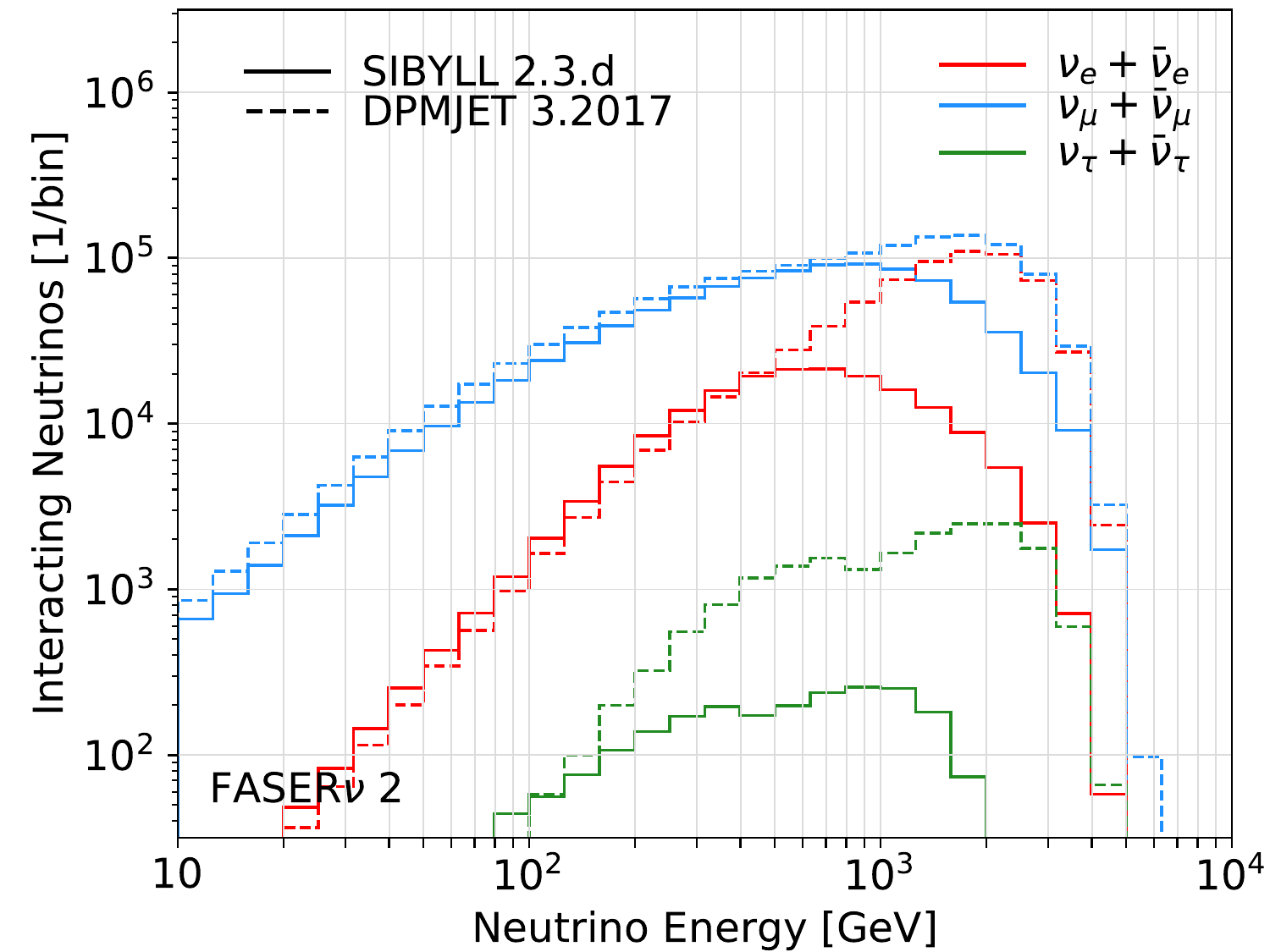}
  \includegraphics[width=0.49\textwidth]{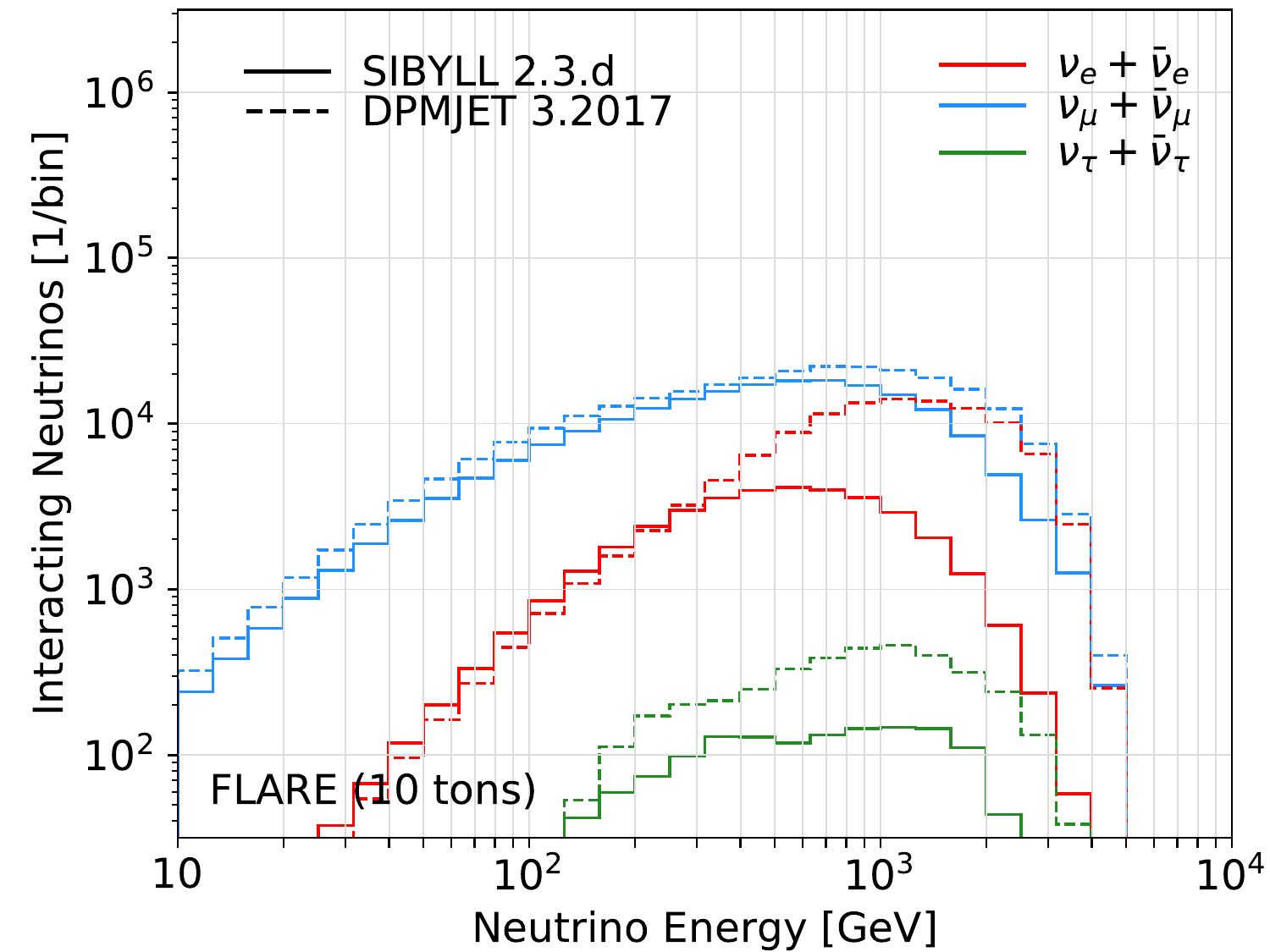}
  \caption{Number of charged current neutrino interactions with the FASER$\nu$2 (left) and FLARE (right) as function of the neutrino energy during the HL-LHC era with an integrated luminosity of $3000~\ifb$. Here we assume a target mass of 20~tons for FASER$\nu$2 and 10~tons for FLARE. The red, blue and green lines correspond to electron, muon and tau neutrinos, respectively, and were obtained using \texttool{SIBYLL~2.3d} and \texttool{DPMJet~3.2017}.}
  \label{fig:fnfs_int}
\end{figure}

The neutrino experiments at the LHC will be able to observe about a thousand electron neutrino interactions, a few thousand muon neutrino interactions and tens of tau neutrino interactions. This number will increase significantly for the FPF experiments due to both the higher integrated luminosity at the HL-LHC and bigger detectors with larger target masses. We expect about $10^5$ electron neutrino, $10^6$ muon neutrino and a few $10^3$ tau neutrino interactions to be recorded by the FPF neutrino detectors. However, these estimates currently suffer from large flux uncertainties as illustrated by the differences between the predictions by \texttool{SIBYLL~2.3d} and \texttool{DPMJet~3.2017}. In the most extreme case, the number of tau neutrinos at FASER$\nu$2, the predictions differ by about a factor 10 ranging from roughly $2\cdot 10^3$ to $20\cdot 10^3$ interactions. As already mentioned above, these large differences are  mainly associated with the neutrino flux component from charm decay and strongly motivate more detailed studies to refine the predictions and better understand the associated uncertainties. 

Finally, in \cref{fig:fnfs_int}, we show the energy spectrum of neutrinos interacting with FASER$\nu$2 (left panel) and FLARE (right panel). The different colors correspond to the three neutrino flavors, while the line styles correspond to different generators. As before, we only show \texttool{SIBYLL~2.3d} and \texttool{DPMJet~3.2017} which provide an envelope of the different generator predictions. As expected from the discussion above, the differences between the two prediction is largest where the neutrino flux component from charm decay dominates: at higher neutrino energies and for tau neutrinos.

To summarize, we have seen that in order to make reliable predictions for the neutrino flux at the LHC one needs both an accurate modelling of i) the production of hadrons and ii) their propagation through the forward LHC infrastructure. The first part can be addressed using a variety of existing MC event generators. To address the second part, we have presented a fast neutrino flux simulation, implemented as a \texttool{RIVET} module, which was made accessible to the entire FPF community\footnote{The module and presented fluxes are available at \href{https://github.com/KlingFelix/FastNeutrinoFluxSimulation}{https://github.com/KlingFelix/FastNeutrinoFluxSimulation}.} and allows LHC neutrino flux predictions to be quickly obtained. We have then used the tool to obtain the the neutrino fluxes at the FPF and presented the neutrino energy and pseudorapidity spectrum as well as the expected number of neutrino interactions.

\subsection{Neutrino Fluxes from $k_T$-Factorization}
\label{Neutrino:ssec:ktfac}

\begin{figure}[t]
  \centering
  \includegraphics[width=0.99\textwidth]{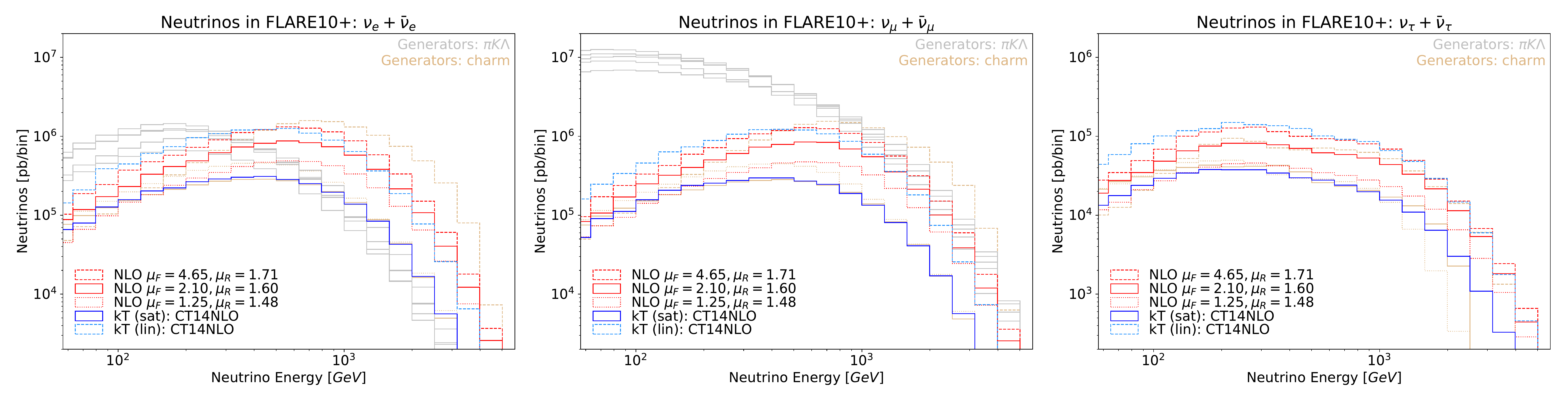}
  \caption{Predictions for the neutrino fluxes going through the $50~\cm \times 50~\cm$ cross sectional area of the FLArE detector with nominal mass of 10~tons. The panels show number of electron (left) muon (center) and tau (right) neutrinos per energy bin passing through the experiments cross section. We show the predictions of different MC event generators as light brown curves. The predictions using the NLO and the $k_T$ factorization approaches are shown in red and blue, respectively. In addition, we show the neutrino fluxes from light meson decays as gray lines.}
  \label{fig:charm_neutrinos}
\end{figure}

As outlined in \cref{QCD:ssec:ktfac}, the $k_T$ factorization formalism in QCD is well suited to describe the kinematics and dynamics of high energy hadron collisions in the forward region. 
Using the framework of Ref.~\cite{Martin:2003us} and $k_T$ dependent PDFs of Ref.~\cite{Kutak:2012rf}, 
we can make predictions for the neutrino fluxes from charmed hadron decay at the FPF. In \cref{fig:charm_neutrinos} we show the obtained fluxes for electron neutrinos (left), muon neutrinos (center) and tau neutrinos (right) going through the $1~\m \times 1~\m$ cross sectional area of the FLArE detector. We show the results using collinear factorization at NLO QCD for a range of scale choices in red, $k_T$ factorization with linear (lin) and nonlinear (sat) evolution of the gluon distribution in blue and the MC event generators in brown. For comparison, we also include the neutrino flux from light hadron decay in gray. Here, the MC predictions were obtained using \texttool{SIBYLL~2.3d}~\cite{Riehn:2019jet}, \texttool{DPMJet~3.2017}~\cite{Roesler:2000he, Fedynitch:2015kcn}, \texttool{EPOS-LHC}~\cite{Pierog:2013ria} and \texttool{QGSJet~II-04}~\cite{Ostapchenko:2010vb} to model the particle production and the fast neutrino flux simulation introduced in Ref.~\cite{Kling:2021gos} to describe their propagation and decay into neutrinos. We see that the contribution to the neutrino flux from light meson decays exceeds that from charm decays for muon neutrinos and low energy electron neutrinos. However, charm decays provide the dominant contribution for high energy electron neutrinos as well as for tau neutrinos, making them an ideal probe of forward charm production. In this regime, the different perturbative predictions of the neutrino flux differ by almost an order of magnitude, indicating that a measurement of the flux will help us to constrain the underlying modelling of the physics. This is illustrated by the $k_T$ factorization approach, which was done both in the presence (solid blue) and absence (dashed blue) of gluon saturation effects. We can see that gluon saturation leads to a suppression of the neutrino flux from charm decay by about a factor four.

\subsection{Tau Neutrino Fluxes from Heavy Flavor: PDF Uncertainties in NLO Perturbative QCD}

NLO QCD evaluations of the production of heavy-flavored hadrons followed by the decays into neutrinos and other particles can be used as a basis for predictions of  high-energy neutrino fluxes at the FPF allowing for a first estimate of the related QCD uncertainties~\cite{Bai:2020ukz, Bai:2021ira}. 
In the case of $\nu_\tau+\bar{\nu}_\tau$, the dominant contributions come from $D_s^+\to \tau^+\nu_\tau$ and $D_s^-\to\tau^-\bar\nu_\tau$, followed by the prompt decays of $\tau^\pm$. In fact, the $D^0$, $\bar{D}^0$ and $D^\pm$ have masses that are too low to permit 3-body semi-leptonic decays to taus and tau neutrinos. The 2-body decay $D^+\to \tau^+\nu_\tau$ has a small phase-space due to the small mass difference between $D^+$ and $\tau^+$ and is also Cabibbo suppressed relative to the $D_s^+\to \tau^+\nu_\tau$. Their respective branching fractions are $B(D^+\to \tau^+\nu_\tau)=(1.20\pm 0.27)\times 10^{-3}$ and $B(D_s^+\to \tau^+\nu_\tau)=(5.48\pm 0.23)\times 10^{-2}$ \cite{ParticleDataGroup:2020ssz}. 
The factor of $\sim 3$ larger fragmentation fraction of $c\to D^+$ compared to the fragmentation fraction of $c\to D_s^+$ implies that the $D^\pm$ contributions to the flux of $\nu_\tau+\bar\nu_\tau$ are  
a few percent 
of the contributions from $D_s^\pm$.  
In Ref. \cite{Bai:2020ukz} we showed that the number of $\nu_\tau+\bar{\nu}_\tau$ events for $\eta_\nu > 6.87$  from $B^0$, $\bar{B}^0$ and $B^\pm$ production and decay to taus and tau neutrinos in $pp$ collisions at $\sqrt{s}=14$ TeV 
is less than $\sim 5\%$ of the number of  events from $D_s^\pm$. In the following, we confine our discussion to the $D_s^\pm$ contributions to the flux of $\nu_\tau+\bar{\nu}_\tau$ in the forward region, with particular focus on the scale and parton distribution function (PDF) uncertainties \cite{Bai:2021ira}.

\begin{table}[t]
\setlength{\tabcolsep}{4pt}
	\begin{center}		    
		\begin{tabular}{ c||c|c|c|c|c|c||c|c|c }
			\hline\hline
            & \multicolumn{6}{c||} {UJ12 Alcoves (480 -- 521 m)}   & 
            \multicolumn{3}{c}{Purpose Built Facility (617 m)}\\                   \hline  
		     baseline  & \multicolumn{3}{c|} {480m}  & \multicolumn{3}{c||} {521 m} & \multicolumn{3}{c} {617 m} \\
		    \hline	                              
		     Experiments & FASER$\nu$2 & \multicolumn{2}{c|} {FLArE}&FASER$\nu$2 & \multicolumn{2}{c||} {FLArE}  &FASER$\nu$2 & \multicolumn{2}{c} {FLArE} \\
		    \hline	                              
		     Radius  (max) & 20 cm  & 0.5 m & 1 m &  20 cm  & 0.5 m & 1 m  & 20 cm  & 0.5 m & 1 m \\  
			\hline
		 	 $\eta_{\rm min}$   &  8.48  & 7.56 & 6.87 & 8.56 &  7.64 & 6.95 & 8.73 & 7.81 & 7.12 \\
			\hline\hline
		\end{tabular}
	\end{center}
	\caption {The minimum pseudorapidity of the FPF experiments according to the hypothetical distances from the LHC interaction point and maximum detector radii of 20 cm, 0.5 m and 1 m assumed in our computations.}
	\label{table:eta}
\end{table}

The configuration of the FPF, as well as those of the experiments which will be hosted there, are not yet determined in  detail~\cite{Anchordoqui:2021ghd}. 
We assume that all detectors are aligned with tangent to the LHC beamline 
at the interaction point (IP), and we ignore the beam crossing angle. \cref{table:eta} shows the minimum pseudorapidities for nominal baselines of 480 m, 521 m and 617 m from the LHC IP and maximum detector radii ranging from 20 cm to 1 m.  We approximate the tungsten and emulsion detector FASER$\nu$2 by using 20 tons of tungsten placed at $\eta_\nu > 8.5$. The FLArE detector concept is still under development. Here we consider $\eta_\nu>6.89$, the pseudorapidity corresponding to a 1 m radius detector positioned at 480~m from the IP. We use a detector mass of 10 tons in the case of an argon detector. For a krypton detector of the same size, the ratio of the krypton to argon density yields a detector mass of 17 tons. 

For $\eta_\nu>8.5$ and $\eta_\nu\gtrsim 6.9$, we estimate the number of neutrino events and their energy distributions for $pp$ collisions at $\sqrt{s}=14$ TeV and integrated luminosity ${\cal L}=3000$ fb$^{-1}$. In what follows, we use ``neutrino" to refer to both $\nu_\tau$ and $\bar\nu_\tau$. As discussed below, in our  QCD evaluation, 
the number of $\nu_\tau$ and $\bar\nu_\tau$ charged current (CC) events mainly differ among each other because the neutrino and
antineutrino cross sections with target nucleons are not equal.

We evaluate neutrino production in $pp$ collisions considering NLO QCD corrections for single inclusive charm production \cite{Nason:1989zy}, a phenomenological fragmentation approach and a\-na\-ly\-tical results for heavy-flavor decays. We also consider the effects \cite{Bai:2020ukz, Bai:2021ira} of a purely phenomenological Gaussian transverse momentum smearing approach, which in practice turns out to mimic the effects of higher-order corrections missing in a NLO QCD calculation and of intrinsic $k_T$ for partons confined in the proton. We assume a default value of $\langle k_T\rangle = 0.7$~GeV, which yields $D$-meson spectra that approximately correspond to those we compute by using an implementation of the
POWHEG \cite{Frixione:2007vw} NLO + Parton Shower matching formalism interfaced to \texttool{Pythia~8}
\cite{Sjostrand:2014zea,Sjostrand:2019zhc} as an alternative to the previous computation. 
Fragmentation of the charm quarks to $D$-mesons is performed in the colliding partons' center of mass frame using Peterson fragmentation functions \cite{Peterson:1982ak} to scale the charm-quark 3-momentum. Details of the evaluation of $D_s^\pm$ production can be found in Ref. \cite{Bai:2021ira}, where we also show comparisons to LHCb data for $D_s^\pm$ differential and double differential cross-sections~\cite{LHCb:2013xam,Aaij:2015bpa}. We implemented the two-body decay of $D_s^\pm$ by analytical formulas, followed by the decay of the $\tau$ as outlined in Refs. \cite{Bai:2018xum,Bai:2020ukz}.

Our default parton distribution function set is the 3-flavor \texttool{PROSA} 2019 \cite{Zenaiev:2019ktw} central PDF set, accompanied by 40 additional sets to characterize the PDF uncertainties, in the  \texttool{LHAPDF} interface \cite{Buckley:2014ana} format. The default input factorization scale $\mu_F$ in our calculation is taken to be $m_{T,2}$ where 
$   m_{T,2}^2= p_T^2+(2m_c)^2\,,$
the same scale used in the \texttool{PROSA} PDF fit when incorporating the results of heavy flavor production. The default input renormalization scale is also taken to be $\mu_R=m_{T,2}$. Alternatively, we also show predictions with $\mu_F=2\mu_R=2m_T$ for transverse mass $m_T^2=p_T^2+m_c^2$. 
The \texttool{PROSA} fit led to a charm quark mass value renormalized in the $\overline{\rm MS}$ scheme, which, when converted to the on-shell one, corresponds to approximately $m_c=1.442$ GeV. 
With $(\mu_R,\mu_F)=(1,2)m_T$ and $\langle k_T\rangle=1.2$ GeV, our central theoretical predictions for $D_s^\pm$ meson production better agree with the LHCb central data than with the default scales and transverse momentum smearing parameter~\cite{Bai:2021ira}. However, when considering the large uncertainties affecting the theory predictions, we can conclude that both sets of predictions are in agreement with LHCb data. 
We also show predictions using the default scales, $\langle k_T\rangle=0.7$ GeV and the \texttool{CT14} \cite{Dulat:2015mca}, \texttool{ABMP16} \cite{Alekhin:2018pai} and \texttool{NNPDF3.1} \cite{Ball:2014uwa} NLO PDF sets.

\begin{figure}
\centering
\includegraphics[width=.45\linewidth]{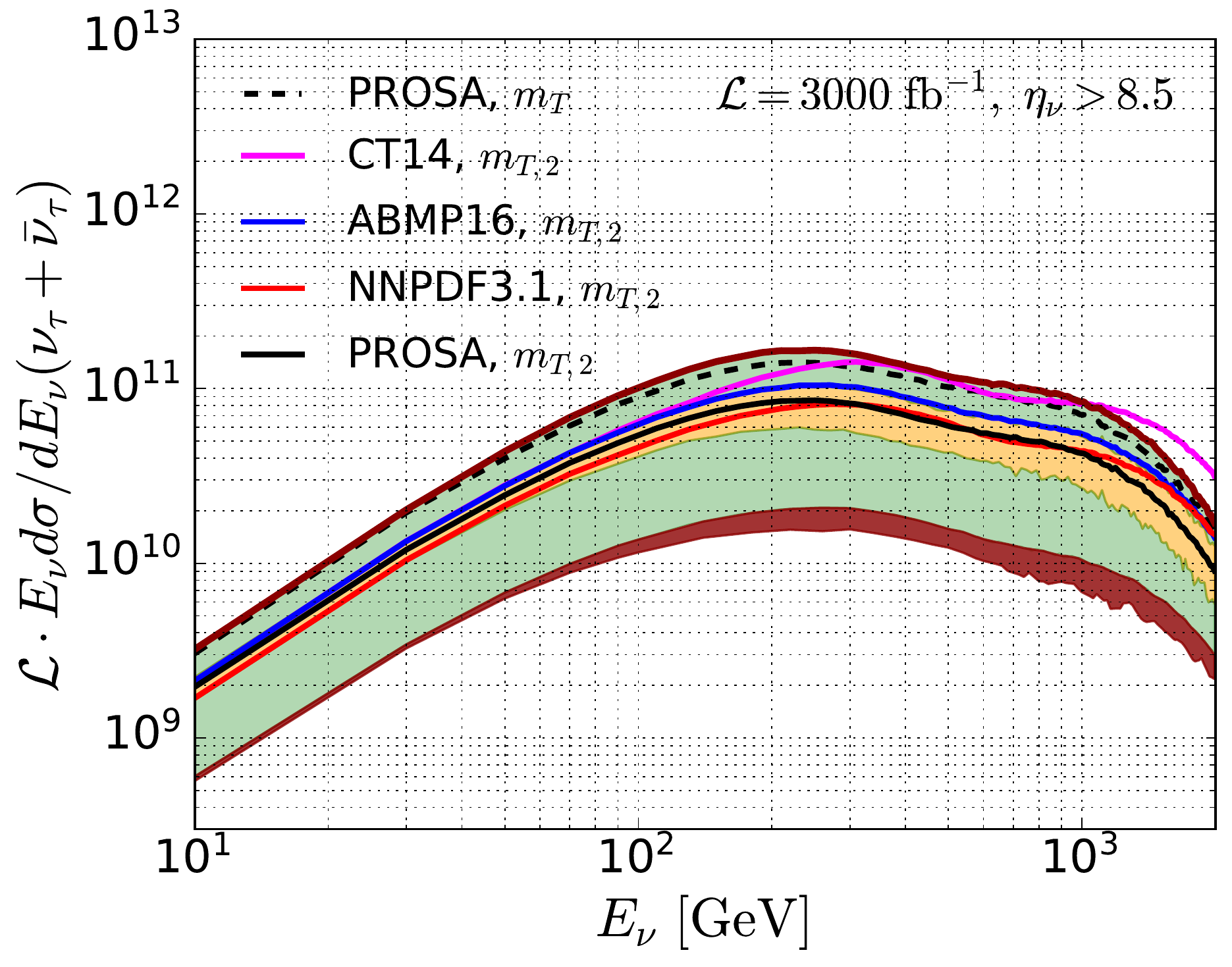}
\includegraphics[width=.45\linewidth]{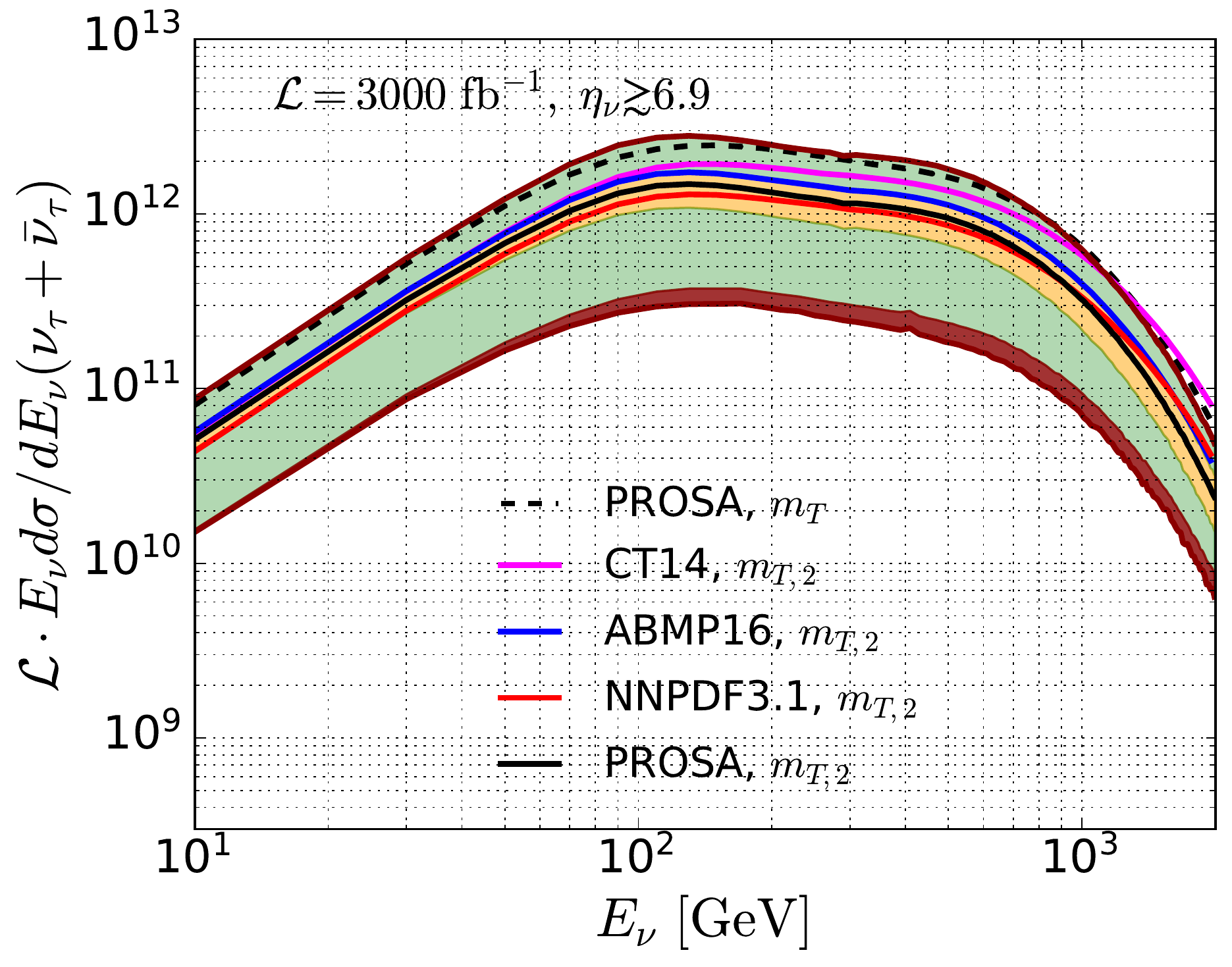}
\caption{\label{fig:fluxuncertainty} For $\sqrt{s}=14$ GeV in $pp$ collisions at the LHC, the integrated luminosity times the $E_\nu d\sigma/dE_\nu$ distribution of $\nu_\tau+\bar\nu_\tau$ for $\eta>8.5$ (left) and $\eta_\nu\gtrsim 6.9$ (right) with ${\cal L}=3000$ fb$^{-1}$ (note the change in scales of the vertical axes). The
solid curves show predictions with default renormalization and factorization scales equal to  $m_{T,2}=\sqrt{(2m_c)^2+(p_T)^2}$  and a Gaussian transverse momentum smearing amounting to $\langle k_T\rangle=0.7$~GeV for \texttool{PROSA} \cite{Zenaiev:2019ktw}, \texttool{CT14} \cite{Dulat:2015mca}, \texttool{ABMP16} \cite{Alekhin:2018pai} and \texttool{NNPDF3.1} \cite{Ball:2014uwa} NLO PDF sets. The orange and green bands show the \texttool{PROSA} PDF and scale uncertainties respectively. The dark red curves outline the combined PDF and scale uncertainty. The dashed curve shows the \texttool{PROSA} predictions for renormalization and factorization scales equal to $(\mu_R,\mu_F)=(1,2)m_T$ and$\langle k_T\rangle=1.2$ GeV. See text and Ref. \cite{Bai:2021ira} for more details. 
}
\end{figure}

Our predictions for the dimensionless quantity ${\cal L}\cdot E_\nu d\sigma/dE_\nu$ for $\nu_\tau+\bar\nu_\tau$ production
computed using the central \texttool{PROSA} PDF set 
for an integrated luminosity of ${\cal L}=3000$ fb$^{-1}$ are shown in \cref{fig:fluxuncertainty} with solid black curves, for both $\eta_\nu>8.5$ (left panel) and $\eta_\nu\gtrsim 6.9$ (right panel). The orange bands in \cref{fig:fluxuncertainty} show the associated PDF uncertainty stemming from the 40 \texttool{PROSA} PDF sets, computed following the prescriptions \texttool{PROSA} the \texttool{PROSA} collaboration.

For $\eta_\nu>8.5$, the total \texttool{PROSA} PDF uncertainty amounts to a factor of $^{+15\%}_{-25 \%}$ with respect to the central prediction in case of $E_\nu \sim 100$~GeV and increases to approximately ${\pm 40\%}$ for $E_\nu=2$~TeV. The uncertainty band for $\eta_\nu \gtrsim 6.9$ follows the same trend.
A larger uncertainty comes from the renormalization and factorization scale dependence, as shown by the green band in \cref{fig:fluxuncertainty} for the standard seven-point range of scales around $(\mu_R,\mu_F)=(1,1)m_{T,2}$.
For $E_\nu=100-1000$ GeV, the scale uncertainty in $E_\nu d\sigma/dE_\nu$, from its lower limit to its upper limit, spans a factor of almost $\sim 9$ for $\eta_\nu>8.5$ and a factor of $\sim 7$ over the same energy interval for $\eta_\nu\gtrsim 6.9$ The large scale uncertainties dominate the uncertainties in the predictions of 
$\nu_\tau + \bar{\nu}_\tau$ energy distribution. The dark red curves outline the combined PDF + scales uncertainties, obtained by adding in quadrature the scale uncertainty and the PDF uncertainty.

The three solid colored curves in the two panels of \cref{fig:fluxuncertainty} show predictions obtained by using as input the default scales and $\langle k_T\rangle$ for the \texttool{CT14} \cite{Dulat:2015mca} (magenta), \texttool{ABMP16} \cite{Alekhin:2018pai} (blue) and \texttool{NNPDF3.1} \cite{Ball:2014uwa} (red) 3-flavor NLO PDF sets with their respective charm quark pole mass values. The \texttool{CT14} prediction differs most from those of the other sets, especially at high neutrino energy and high pseudorapidity. This can be traced back to the large-$x$ behavior of these PDF sets \cite{Bai:2021ira}. Also shown in \cref{fig:fluxuncertainty} with the dashed black line is the \texttool{PROSA} central PDF prediction with the alternative set of input quantities $(\mu_R,\mu_F)=(1,2)m_T$ and $\langle k_T\rangle=1.2$ GeV. The black dashed curve lies close to the upper edge of the scale uncertainty band obtained with the default set.

\begin{figure}
    \centering
    \includegraphics[width=0.49\textwidth]{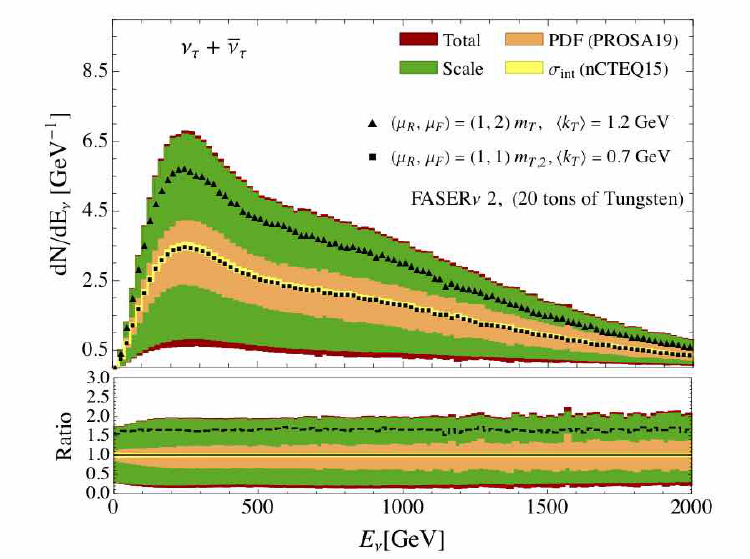}
    \includegraphics[width=0.49\textwidth]{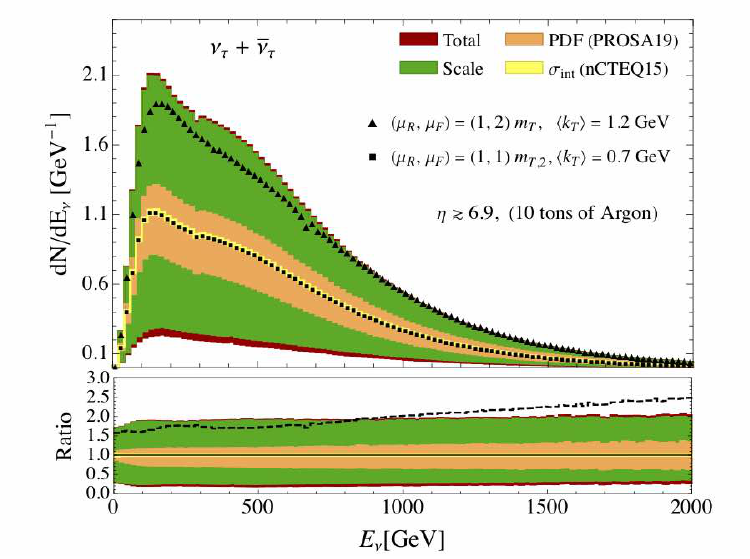}    
    \caption{Energy distribution of tau neutrino and antineutrino CC interaction events in the pseudorapidity ranges of $\eta _\nu > 8.5$ relevant for FASER$\nu$2 (left) and $\eta_\nu > 6.9$ (right) relevant for FLArE.
    The uncertainties are from the QCD scale variation (green), the 40 different sets of the \texttool{PROSA} (orange) PDF fit and the 32 sets of the nCTEQ15 (yellow) nuclear PDF fit for 20 tons of tungsten in the left panel and 10 tons of argon in the right panel, respectively.  We assume an integrated luminosity of ${\cal L}=3000$ fb$^{-1}$.
}
    \label{fig:dnde-unc}
\end{figure}

\begin{figure}
    \centering
    \includegraphics[width=0.49\textwidth]{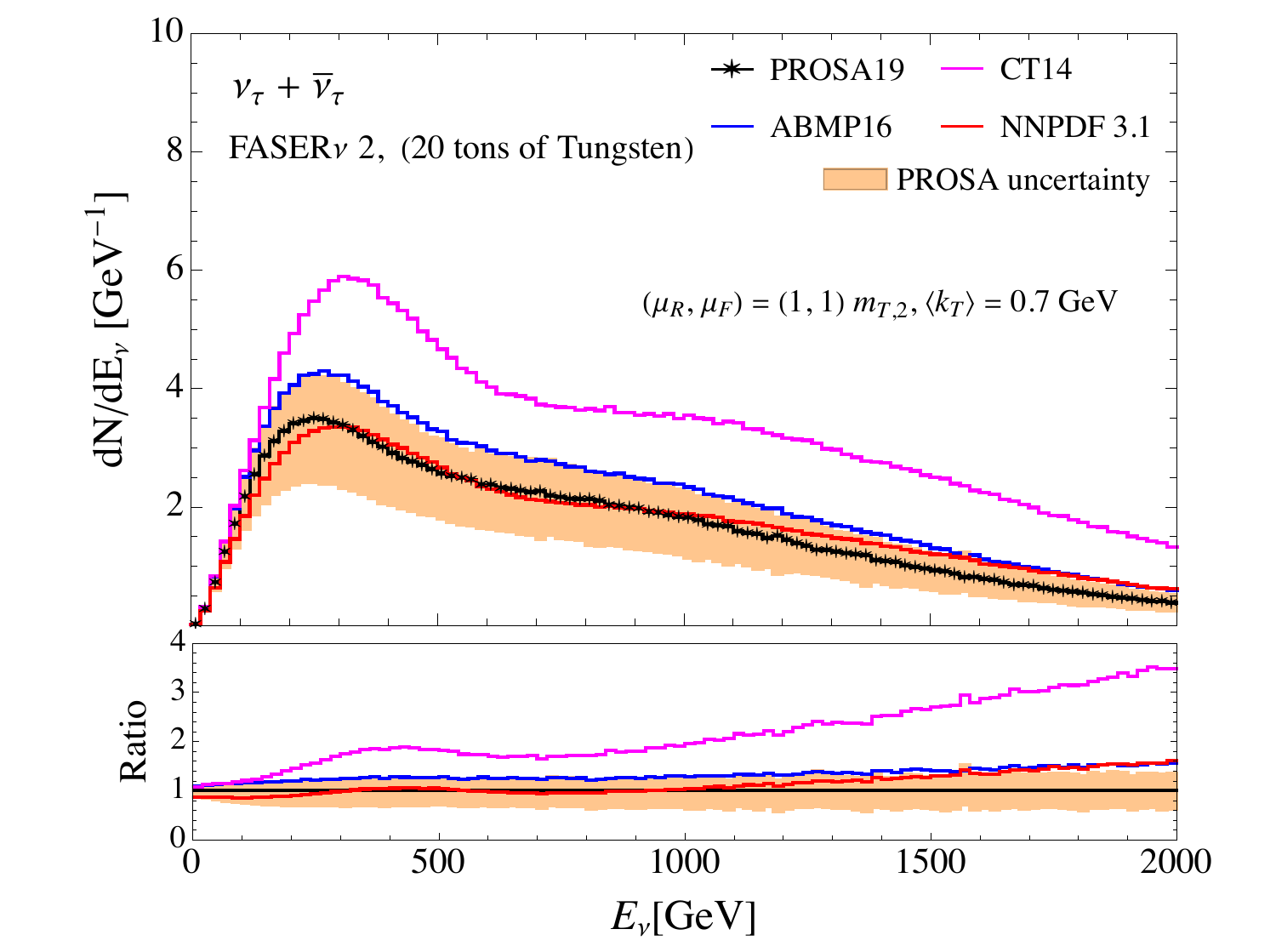}   
    \includegraphics[width=0.49\textwidth]{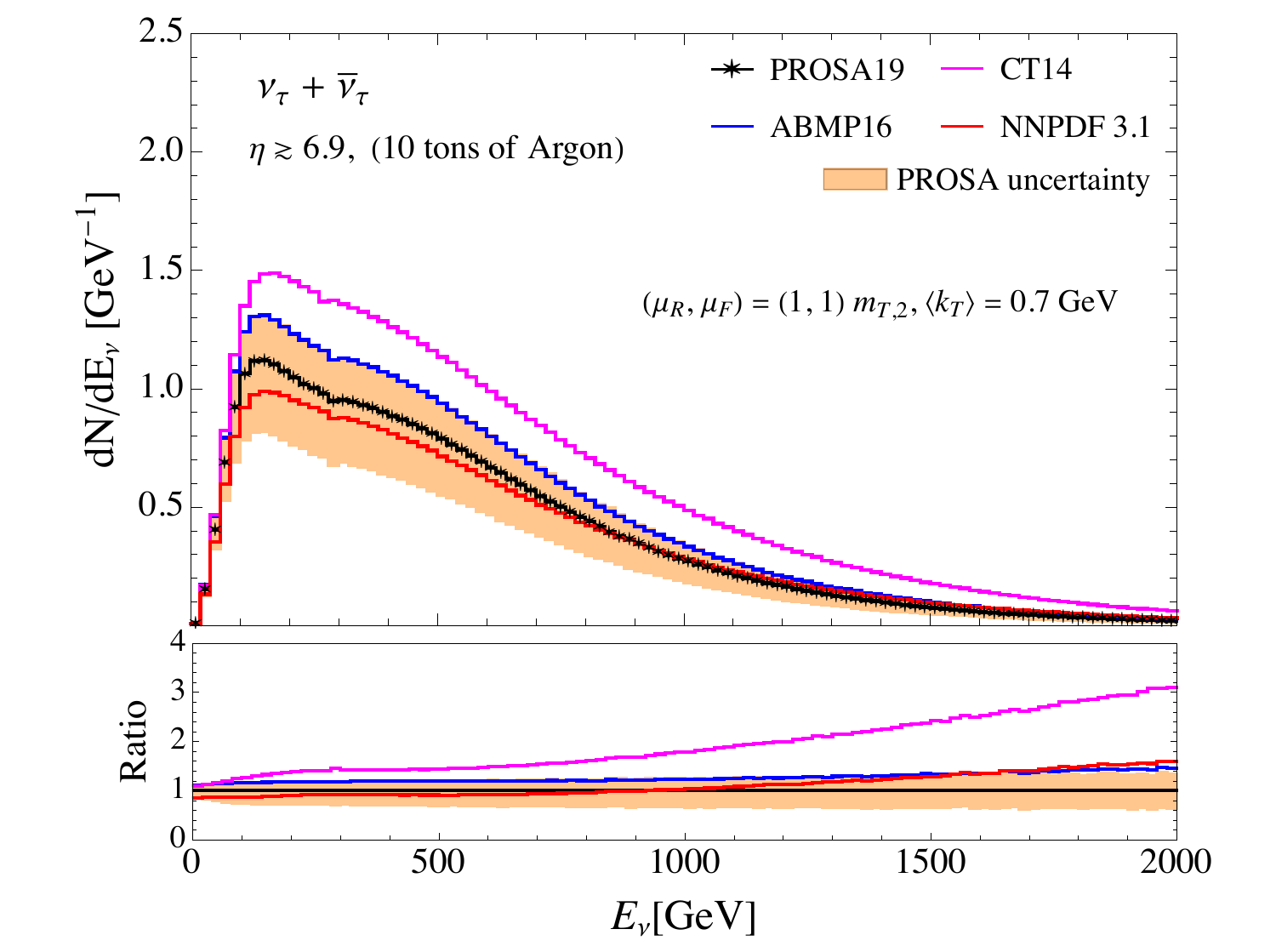}       
    \caption{Energy distribution of tau neutrino and antineutrino CC interaction events with different PDFs for pseudorapidity ranges $\eta_\nu > 8.5$ for FASER$\nu$2 with 20 tons of tungsten (left) and $\eta_\nu \gtrsim 6.9$ for 10 tons of argon (right). The integrated luminosity is ${\cal L}=3000$ fb$^{-1}$. The orange band shows the PDF uncertainty associated with the 40 different \texttool{PROSA} PDF sets.}
    \label{fig:dnde-pdf}
\end{figure}

The conversion of ${\cal L}\cdot d\sigma/dE_\nu$ to a number of charged-current events requires the density, the length of the detector (determined from the mass, density and assuming a cylindrical shape with the maximum radius in \cref{table:eta}) and the neutrino cross section.
The neutrino charged-current (CC) cross sections are evaluated using the nCTEQ15 nuclear PDFs \cite{Kovarik:2015cma} for tungsten, argon and krypton. 
\cref{fig:dnde-unc} and \cref{fig:dnde-pdf} show the energy distributions of the number of $\nu_\tau+\bar\nu_\tau$ CC events as a function of the neutrino energy for our nominal neutrino pseudorapidity cuts for both 20 tons of tungsten  ($\eta_\nu>8.5$) and 10 tons of argon ($\eta_\nu\gtrsim 6.9$). The PDF uncertainty determined from 32 nCTEQ15 nuclear PDF sets is small, less than $5\%$, shown with the yellow band in both panels of \cref{fig:dnde-unc}. Again, the \texttool{PROSA} PDF uncertainty and scale uncertainty bands are shown, and all three uncertainties are added in quadrature for the full uncertainty band. The ratio of the uncertainty bands to the central \texttool{PROSA} results with the default scales and $\langle k_T\rangle$ are shown for both rapidity ranges. The black triangles show the number of events per GeV predicted with $(\mu_R,\mu_F)=(1,2)m_T$ and $\langle k_T\rangle=1.2$ GeV.

As in \cref{fig:dnde-unc}, \cref{fig:dnde-pdf} shows the number of charged-current events per GeV. The predictions using the 
\texttool{ABMP16}, \texttool{CT14}, and \texttool{NNPDF3.1} NLO PDF sets are shown along with the central \texttool{PROSA} NLO result and the \texttool{PROSA} PDF uncertainty band. Predictions with the \texttool{ABMP16} and \texttool{NNPDF3.1} NLO PDF sets lie within the \texttool{PROSA} PDF uncertainty band, whereas those with the \texttool{CT14} PDF results do not, for the reasons already discussed when commenting \cref{fig:fluxuncertainty}. 
All energy distributions in this figure were evaluated using $(\mu_R,\mu_F)=(1,1)m_{T,2}$ and $\langle k_T\rangle=0.7$~GeV.  \cref{table:events} lists the number of $\nu_\tau$, $\bar{\nu}_\tau$ and $\nu_\tau+\bar\nu_\tau$ CC events 
for $\eta_\nu>8.5$ and 20 tons of tungsten, and for $\eta_\nu\gtrsim 6.9$ and 10 tons of argon, considering an integrated luminosity
of 3000 fb$^{-1}$.

\begin{table}[ht]
	\begin{center}		    
		\begin{tabular}{ c||c|c|c|c|c|c }
		\hline \hline
            &  $\nu_\tau$ &  $\bar{\nu}_\tau $  &   $\nu_\tau + \bar{\nu}_\tau $ &
            \multicolumn{3}{c}{$\nu_\tau + \bar{\nu}_\tau $}  \\	 
        \hline  
		      $(\mu_R, \ \mu_F)$, $\langle k_T \rangle $  & \multicolumn{6}{c} {(1, 1)  $m_{T,2}$,  0.7 GeV}  \\
		 \hline	                              
		        & \multicolumn{3}{c|}{} &
		        scale(u/l) & PDF(u/l) &$\sigma_{\rm int}$ \\
		\hline
		 	 FASER$\nu$2 & 2296 & 1088 & 3384 & +3144/-2519 & +786/-1089 &  $\pm$ 77 \\
		 	 $\eta_\nu > 8.5$, 20 tons (W)& & & & & & \\	
		\hline 
		 	$\eta_\nu > 6.9$,  10 ton (Ar) & 529  & 257 & 786 & +692/-575 & +152/-229 & $\pm$11  \\
		\hline
%
         \hline  
              $(\mu_R, \ \mu_F)$, $\langle k_T \rangle $  &
              \multicolumn{3}{c|} {(1, 2)  $m_{T}$,  1.2 GeV} & \multicolumn{3}{c} {(1, 1)  $m_{T,2}$,  0.7 GeV } \\
	     \hline
	        PDF &
            \multicolumn{3}{c|}{PROSA FFNS} &  NNPDF3.1 & CT14 & ABMP16\\
		\hline
		 	 FASER$\nu$2 & 3808 & 1804 & 5612 & 3552 & 6492 & 4338 \\
			  $\eta_\nu > 8.5$, 20 tons (W) & & & & & & \\	 	 
		\hline 
		 	$\eta_\nu > 6.9$,  10 ton (Ar) & 953  & 465 & 1418 &  748 & 1202 & 944\\
		\hline\hline
		\end{tabular}		
	\end{center}
	\caption {
	The numbers of $\nu_\tau + \bar{\nu}_\tau$ charged-current events in the FASER$\nu$2 (20 tons of tungsten, $\eta_\nu>8.5$) and 10 tons of argon ($\eta_\nu \gtrsim 6.9$) detectors for an integrated luminosity ${\cal L}=3000$ fb$^{-1}$. The separate uncertainties due to scale, PDF and $\nu_\tau+\bar\nu_\tau$ CC cross section per nucleon ($\sigma_{\rm int}$) are listed for the default scale and $\langle k_T\rangle$ evaluation.}
	\label{table:events}
\end{table}

\begin{figure}
    \centering
    \includegraphics[width=0.49\textwidth]{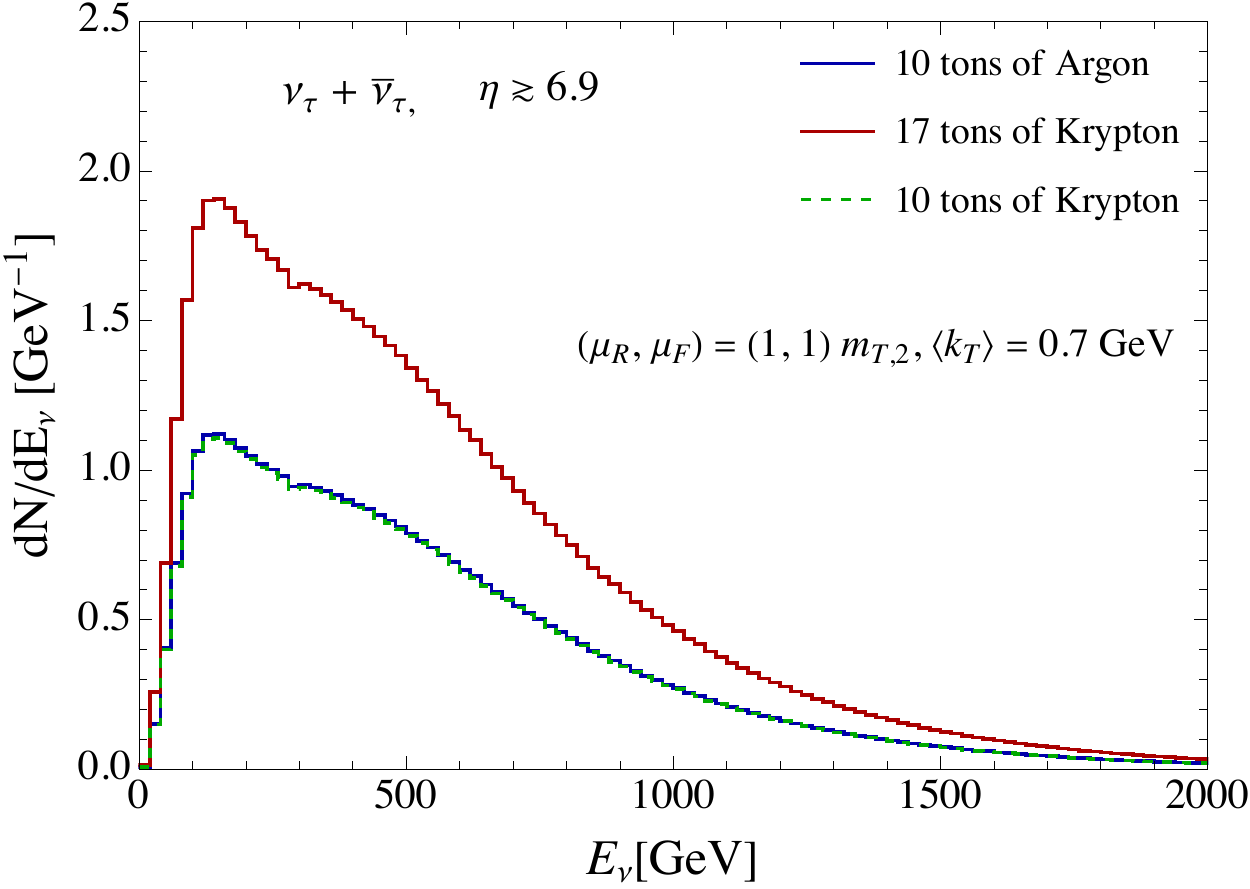}   
    \includegraphics[width=0.49\textwidth]{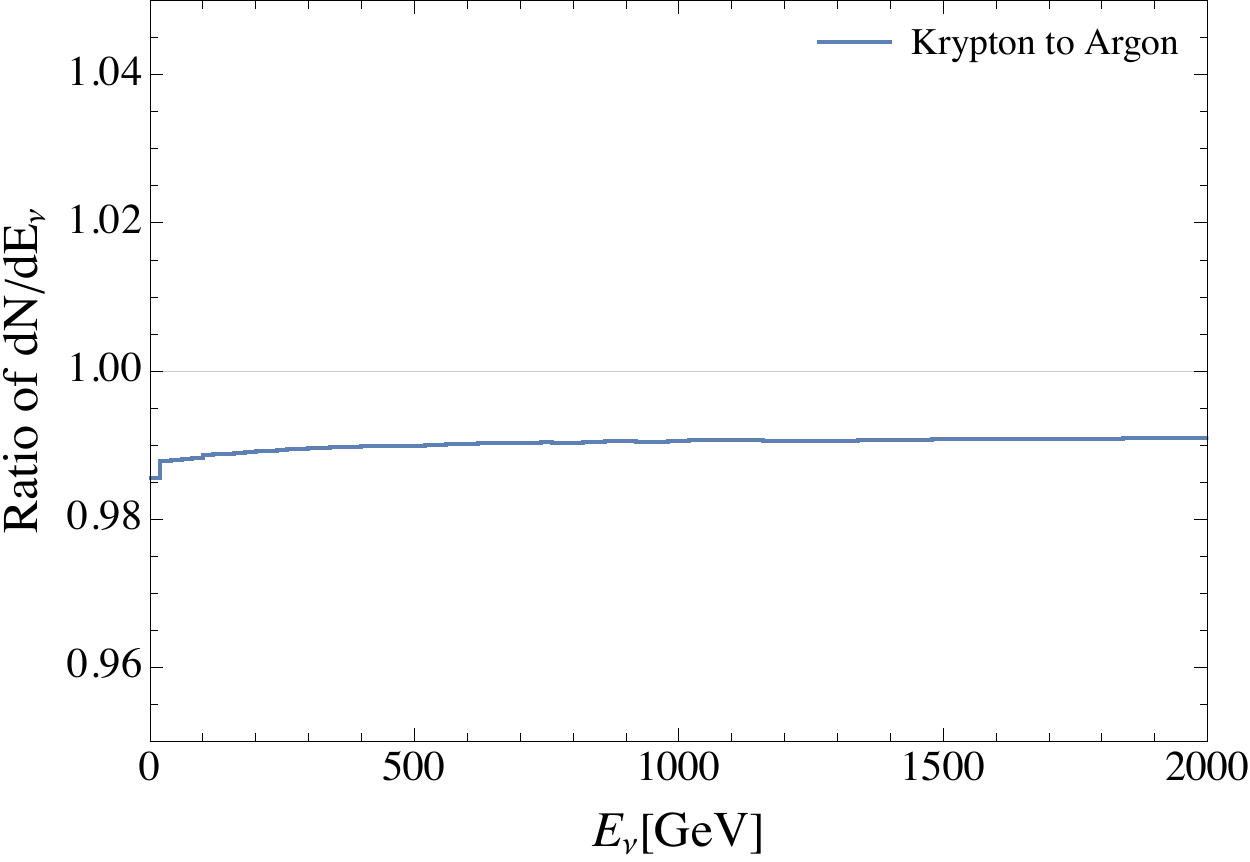}       
    \caption{Left: Energy distribution of tau neutrino and antineutrino charged-current events for $\eta_\nu \gtrsim 6.9$. These results are evaluated for 10 tons of argon, 10 tons of krypton and 17 tons of krypton.
    Right: the ratio of energy distributions of $\nu_\tau + \bar{\nu}_\tau $ events in 10 tons of krypton to 10 tons of argon. This is the ratio of the CC cross section per nucleon for krypton to argon.    }
    \label{fig:dnde-6p9}
\end{figure}

Liquid krypton, with its higher density, has advantages over liquid argon for neutrino detection. 
In \cref{fig:dnde-6p9} we show the number of events per neutrino energy for argon and krypton for $\eta_\nu\gtrsim 6.9$. A detector size of 1 m $\times$ 1 m $\times$ 7 m can contain approximately 17 tons of krypton or 10 tons of argon. 
The corresponding predictions are shown with the red and blue solid histograms, respectively. 
For 10 tons of krypton (dashed green histogram), the number of events is nearly identical to the case of 10 tons of argon. 
The ratio of the number of events for an equal mass of krypton and argon, per unit energy, is shown in the right panel of \cref{fig:dnde-6p9}. The right panel shows that the ratio of the number of events given equal masses of krypton and argon are nearly equal, so the nuclear corrections to the neutrino cross section per nucleon for krypton and argon targets are nearly identical. 

A Forward Physics Facility, with large enough neutrino detectors, in the high-luminosity LHC era, has the potential to detect hundreds to thousands $\nu_\tau+\bar\nu_\tau$ CC interaction events, arising from charm production and decay. Evaluations of heavy-flavor production including NLO QCD radiative corrections allow for a first estimate of the associated uncertainties. From our study with the \texttool{PROSA} PDF set, we find that a very large contribution to these uncertainties comes from the QCD renormalization and factorization scale dependence. Alternative PDF choices can yield predictions that lie outside the \texttool{PROSA} PDF uncertainty band and even  outside the combined scale + PDF uncertainty band. This is mainly a consequence of the different sets of data included in different PDF fits so far and of the general scarcity of data for longitudinal momentum fraction $x$ values which, although not relevant for many of the LHC analyses performed nowadays in the central interaction region, are indeed relevant for FPF calculations.  
While waiting for data from future colliders like the EIC \cite{AbdulKhalek:2021gbh} and the LHeC \cite{LHeCStudyGroup:2012zhm} capable of constraining PDFs in regions where they are currently most uncertain, the \texttool{PROSA} collaboration included LHCb open heavy-flavor production data in their fits, which helped to constrain these PDFs in the interval of $x \in$ [10$^{-6}$, 10$^{-4}$] and also impacted the large $x$ region $x > 0.1$ \cite{Zenaiev:2019ktw}. Reliable PDFs in these two regions are of utmost importance to reliably predict forward neutrino fluxes. Even though it was possible to reduce PDF uncertainties thanks to the LHCb data, as shown by subsequent PDF fits which obtained results compatible with the \texttool{PROSA} one,  
the NLO scale uncertainties are still large and dominant. As experiments and facilities develop, it is important for theoretical developments to move forward too, in order to face the new experimental challenges or match the accuracies of the data of which the new experiments, including those at the FPF, may be capable.
With this in view, 
$D$-meson production calculations including higher-order corrections are mandatory 
in order to refine the theoretical predictions of neutrino distributions for the Forward Physics Facility as well as for forthcoming $D$-meson LHCb measurements.

\section{Neutrino Cross Sections} 
\label{sec:NeutrinoCrossSections}

\begin{figure} 
\centering
\includegraphics[width=0.99\textwidth]{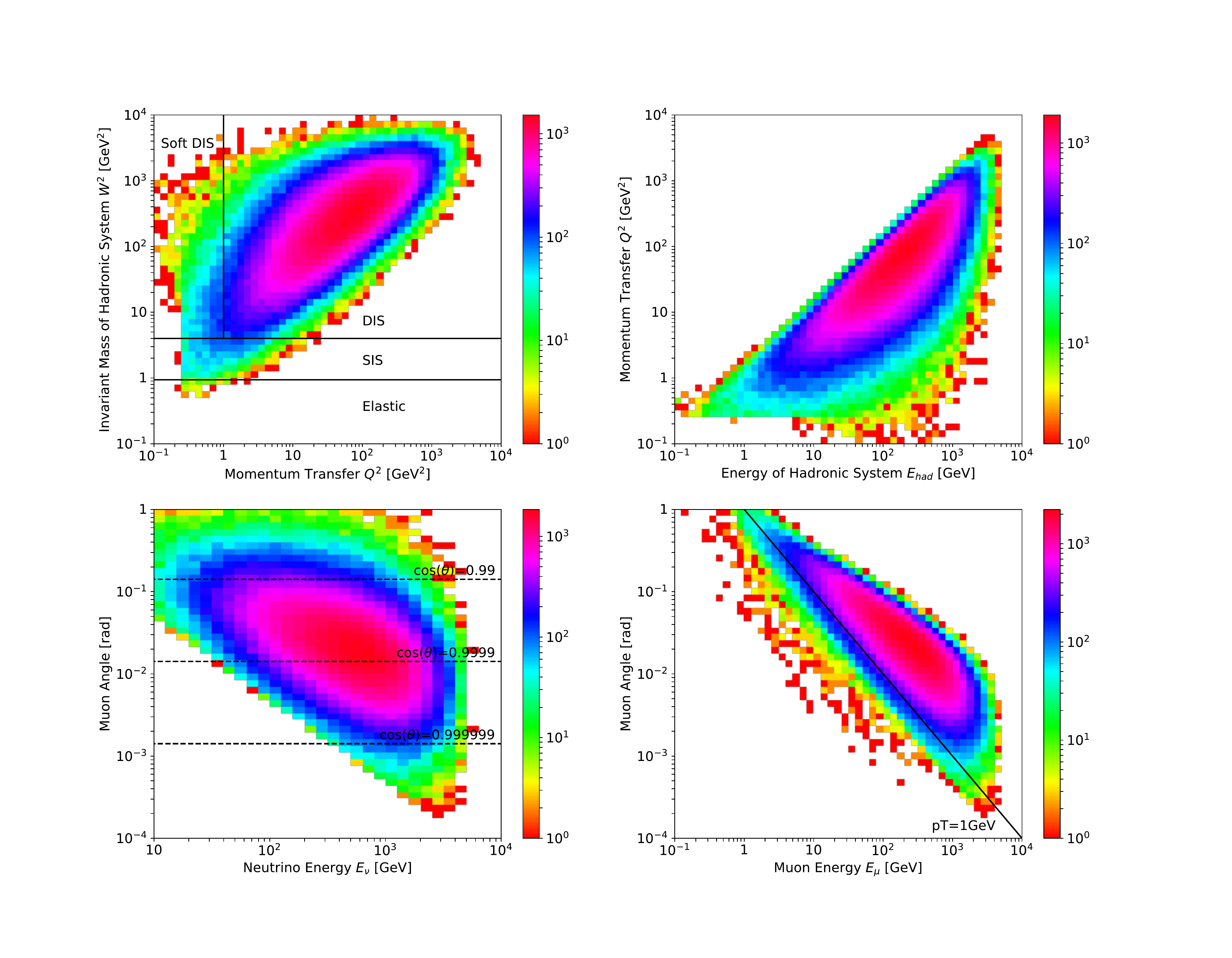}
\caption{\label{Fig:FPF_Neutrino_Kinematic} Kinematic coverage expected during HL-LHC for $\nu_\mu$ CC scattering on argon in the FLArE-10 detector. The events are generated with \texttool{Pythia~8} following the expected neutrino energy spectrum at FLArE-10.}
\end{figure}

In this section, we will discuss different neutrino interaction processes expected at the FPF. In order to demonstrate the typical kinematic coverage of neutrino interaction at FPF, in \cref{Fig:FPF_Neutrino_Kinematic}, we show examples of $\nu_\mu$ CC scattering on argon for FLArE-10 detector. The events are generated with \texttool{Pythia~8} following the expected neutrino energy spectrum at FLArE-10. The normalization on the color axis corresponds to the number of expected events per bin in the FLArE-10 detector during the HL-LHC. The top left plot shows that the large majority of events fall in the DIS region, as expected. However, there are significant events in the SIS and the soft DIS regions, 
as discussed in the next contributions. 
The bottom plots show distributions for the muon energy and scattering angle, the majority of these events have very forward going muons. 

\subsection{Deep-Inelastic Scattering}
\label{subsub:dis}

Neutrinos with energies above a few hundred GeV will primarily interact via Deep Inelastic Scattering (DIS)\footnote{The quasi-elastic and resonant cross sections  relative to DIS are less than 1\%\ at 500 GeV.}. In the DIS regime, the neutrino resolves the individual quark constituents of the nucleon via the exchange of a W or Z boson, producing a lepton and a hadronic shower in the final state. The differential cross section of this process can be described in terms of Bjorken-$x$ ($x$) and the four-momentum transfer ($Q^{2}$). The main ingredients to compute the cross section are the structure functions ($F_{i}$) which describe the underlying QCD dynamics of the nuclear target. 

In the non-perturbative regime (generally denoted as $Q<1$ GeV), the structure functions can be constructed using phenomenological models which have been tuned to data, such as the well-known Bodek-Yang model \cite{Bodek:2004pc}. This model was constructed using leading order expression for the structure functions, which were modified including a Nachtmann scaling variable \cite{Nachtmann:1973mr} and K factors with two and three free parameters respectively. These parameters account for several effects: dynamic higher twist, higher order QCD terms, transverse momentum of the initial quark, the effective masses of the initial and final quarks originating from multi-gluon interactions at low $Q^2$, and the correct form in the low-$Q^2$ photo-production limit. The parameters were extracted from a fit to inelastic charged-lepton scattering data on hydrogen and deuterium targets\footnote{Only data with an invariant final state mass $W>2 $ GeV was used.} \cite{Whitlow:1991uw,BCDMS:1989ggw,NewMuon:1996fwh,H1:2003xoe} using GRV98LO \cite{Gluck:1998xa} as input parton distribution functions (PDFs). Normalization constant were applied to have a better agreement with the data in two components: the PDFs of up- and down-quark contributions, and the K factor. In addition, $F_{1}$ was related to $F_{2}$ using a modified version of the Whitlow parametrization \cite{Whitlow:1991uw} to have a smooth transition at low $Q^2$. Finally, nuclear corrections for deuterium and iron targets were applied using a Bjorken-$x$ dependent function \cite{Seligman:1997fe}. As described in \cref{sec:tools}, this phenomenological approach is widely used by the neutrino community in Monte Carlo generators which describe few-GeV neutrino DIS interactions.

In the inelastic regime, the structure functions can be factorized in terms of coefficient functions and PDFs using perturbation theory. The PDFs are extracted from experimental data. 
The evolution of these PDFs is determined from the solutions of the DGLAP evolution equations. The coefficient functions can be computed in perturbation theory as a power expansion in the strong coupling $\alpha_s$. In the following discussion, we will focus on two NLO perturbative QCD models that use different PDFs and treatment of heavy-quark masses: CSMS \cite{Cooper-Sarkar:2011jtt} and BGR \cite{Bertone:2018dse,Garcia:2020jwr}. The CSMS calculation has been an important benchmark for the high-energy neutrino-nucleon cross section. As inputs, this calculation uses the NLO HERA1.5 PDF set \cite{Cooper-Sarkar:2010yul} and coefficient functions from QCDNUM \cite{Botje:2010ay}. In the case of the BGR model, the input PDF sets are obtained from the \texttool{NNPDF3.1sx} global analyses of collider data \cite{Ball:2017otu} and it uses the FONLL general-mass variable flavor number scheme \cite{Forte:2010ta} to account for heavy quark mass effects. 

There are no measurements of the neutrino cross sections in the energy range from 400 GeV to 10 TeV. At higher energies, the cross section has been measured using Earth absorption effects by IceCube \cite{IceCube:2017roe,IceCube:2020rnc,Bustamante:2017xuy}. However, these measurements still have large uncertainties. Neutrino beams from accelerators have been used to measure the cross section at lower energies. In particular, the NuTeV Collaboration reported the charged-current (CC) neutrino cross section up to 360 GeV \cite{NuTeV:2005wsg}. Using a segmented calorimeter followed by an iron spectrometer, the experiment collected $8.6\times10^5$ and $2.4\times10^5$ muon neutrino and antineutrino events, respectively. By selecting events with a reconstructed muon with energies above 15 GeV, they reported the relative neutrino to antineutrino cross section and that both CC cross sections are linear with energy in the 30-360 GeV range. The total CC cross section was also reported by normalizing it to the world average neutrino cross section from 30-200 GeV.

\begin{figure}[ht]
\centering
\includegraphics[width=0.7\textwidth]{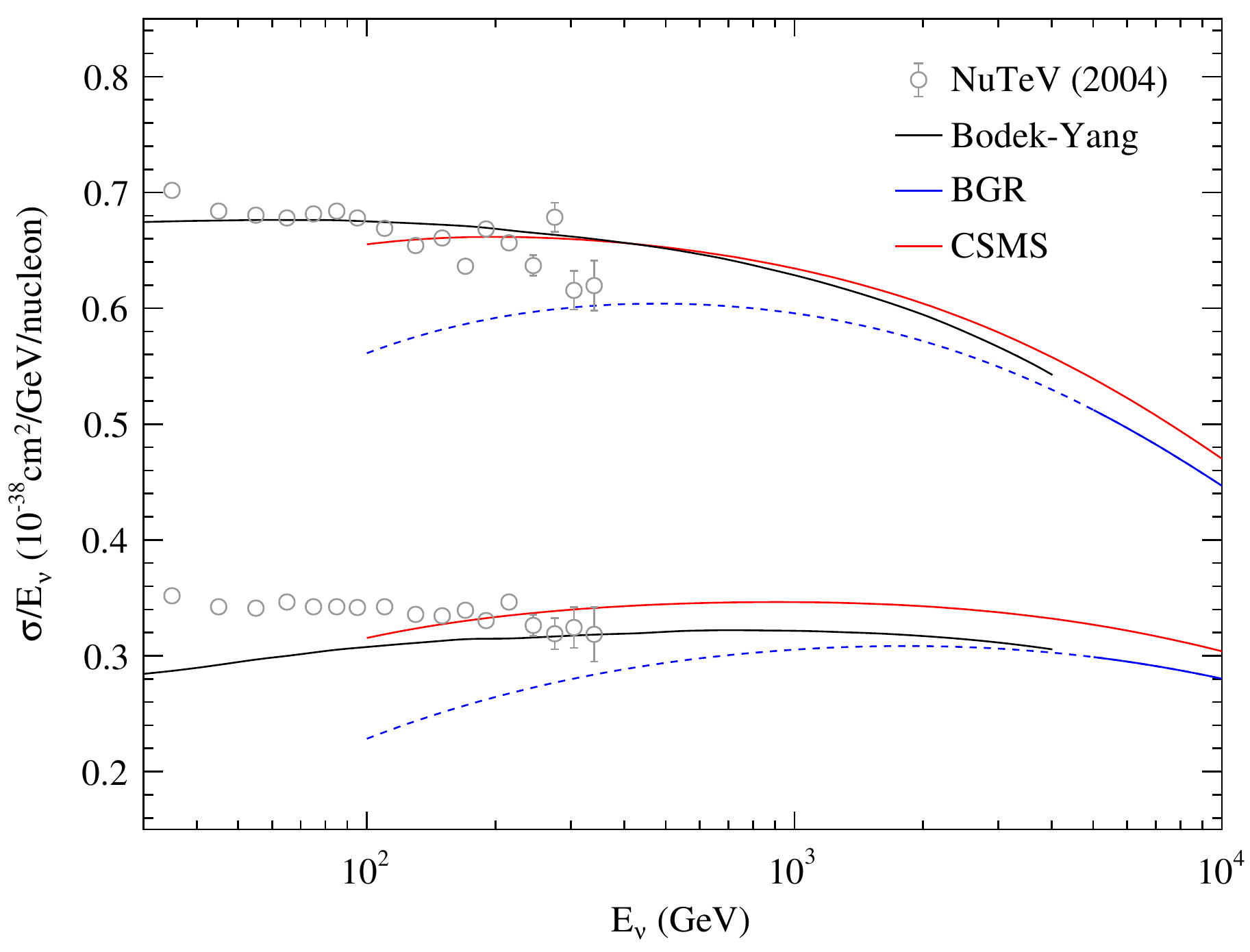}
\caption{The (anti)neutrino-iron charged-current DIS cross section as a function of the neutrino energy. We compare the results of BGR, CSMS, and Bodek-Yang calculations to the measurements from NuTeV \cite{NuTeV:2005wsg} (only statistical uncertainties are shown). The BGR calculation is extrapolated below TeV where low-$Q^2$ contributions are relevant (see text).}
\label{fig:xsec}
\end{figure}

\cref{fig:xsec} shows a comparison between the total CC cross sections measured by NuTeV and predictions from the models previously described. A good agreement is observed between data, CSMS, and Bodek-Yang model for $E>100$ GeV. At the lower energies, the DIS-only models underestimate the measured cross section since the contributions from quasi-elastic and resonant interactions start being relevant. The extrapolation of the BGR model to lower energies is also below the measurements. In the following, we will explain the main differences between these models.

\begin{figure}[ht]
\centering
\includegraphics[width=1.0\textwidth]{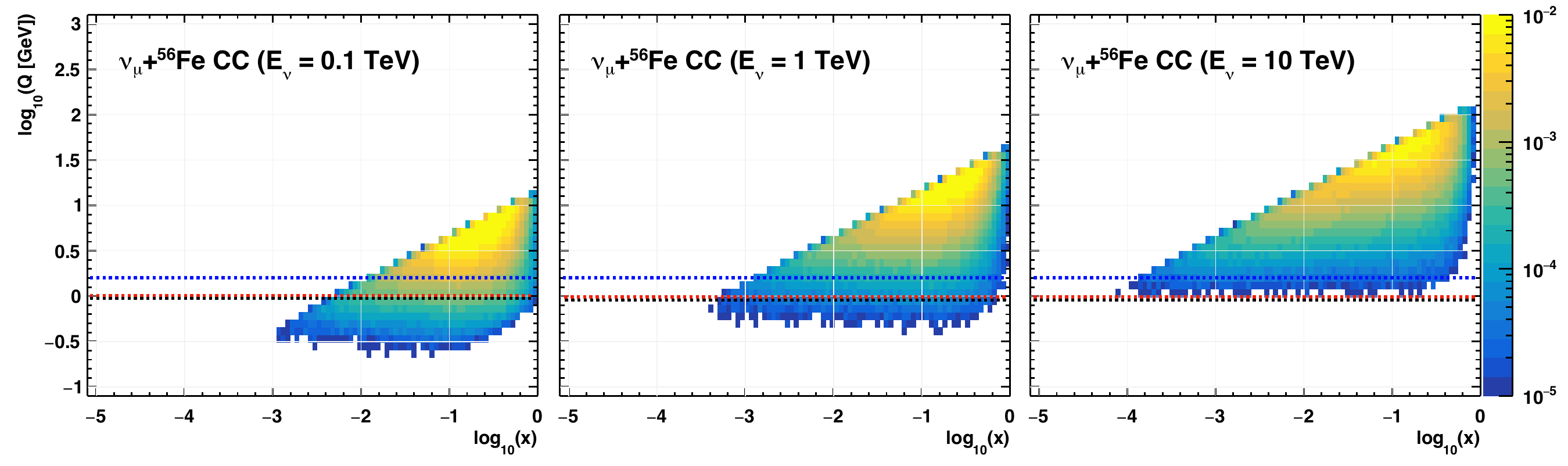}
\caption{$Q^2$ versus $x$ kinematics of neutrino-iron charged current DIS interactions. $10^{6}$ neutrino interactions were simulated using three monochromatic neutrino flux of 0.1, 1, and 10 TeV. Dashed lines represents the $Q_{0}$ value for three different models: Bodek-Yang (black), CSMS (red), and BGR (blue).}
\label{fig:dxsec}
\end{figure}

\begin{figure}[ht]
\centering
\includegraphics[width=0.5\textwidth]{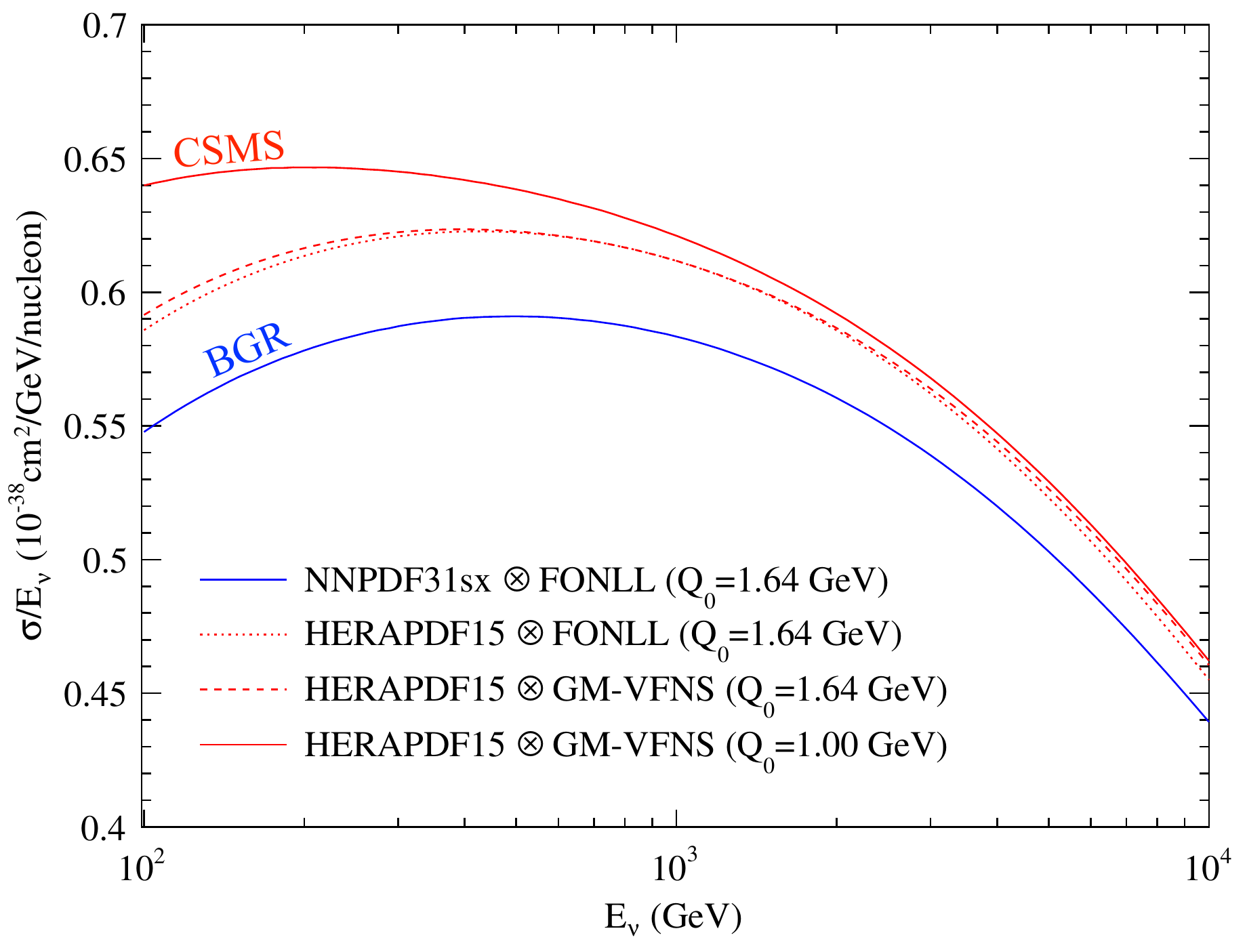}
\caption{The neutrino-iron charged-current DIS cross section as a function of the neutrino energy. We compare the results of BGR (solid blue) and CSMS (solid red) to other configurations using HERAPDF1.5 as input PDFs.}
\label{fig:xsec_q}
\end{figure}

For $E<5$ TeV, the main difference between BGR and the other models is the choice of the $Q^2$ lower bound in the cross section integration. In the pQCD models this value is determined by the $Q_{0}$ value associated with the input PDF sets: $Q_{0}^{BGR}=1.64$ GeV and $Q_{0}^{CSMS}=1$ GeV. In the case of Bodek-Yang, the $Q^2$ lower bound is set to zero but the PDFs are frozen when $Q<0.9$ GeV and K-factors are used to model $Q\rightarrow0$. \cref{fig:dxsec} illustrates the region of the $x-Q$ phase space that neutrino DIS interactions probe at different energies. For energies below 1 TeV, the contribution to the inclusive cross section from the low-momentum transfer is non-negligible. To better quantify this point, \cref{fig:xsec_q} includes, for comparison, the predictions from CSMS using both $Q_{0}=1.64$ GeV and 1.0 GeV as the integration lower limit. We conclude that a description of this low-$Q^2$ is region is required to reliably predict the cross section with percent level accuracy in the TeV regime. However, the computation of structure functions using perturbation theory starts breaking down for these values of $Q^2$. In the future, improved predictions for the low-$Q^2$ region may be assessed through a combined fit of pQCD structure function, which smoothly transition into phenomenological approaches (e.g., Bodek-Yang) for $Q<2$ GeV. 

Another major difference between the different calculations is the treatment of heavy quark mass effects, which is most relevant for the description of charm quark production in the TeV regime. In the CSMS model, the charm contribution is included using the general-mass VFN scheme including a threshold constraint ($W^{2}= Q^{2}(1/x-1)>m_{c}$), which ensures the mass of the out-going hadronic system exceeds the charm production threshold. Mass effects are instead included as part of the FONLL general-mass VFN scheme in the BGR calculation. This formalism improves the prediction obtained with the massive charm-quark structure function with higher-order terms which are resummed in the massless computation. \cref{fig:xsec_q} includes a calculation using the FONLL scheme and HERAPDF1.5 as input PDFs, which agrees with the CSMS prediction (with $Q_{0}>1.64$ GeV). Hence, the differences between CSMS and BGR models in the TeV regime are not due to the choice of the mass scheme.

\begin{figure}[ht]
\centering
\includegraphics[width=1.0\textwidth]{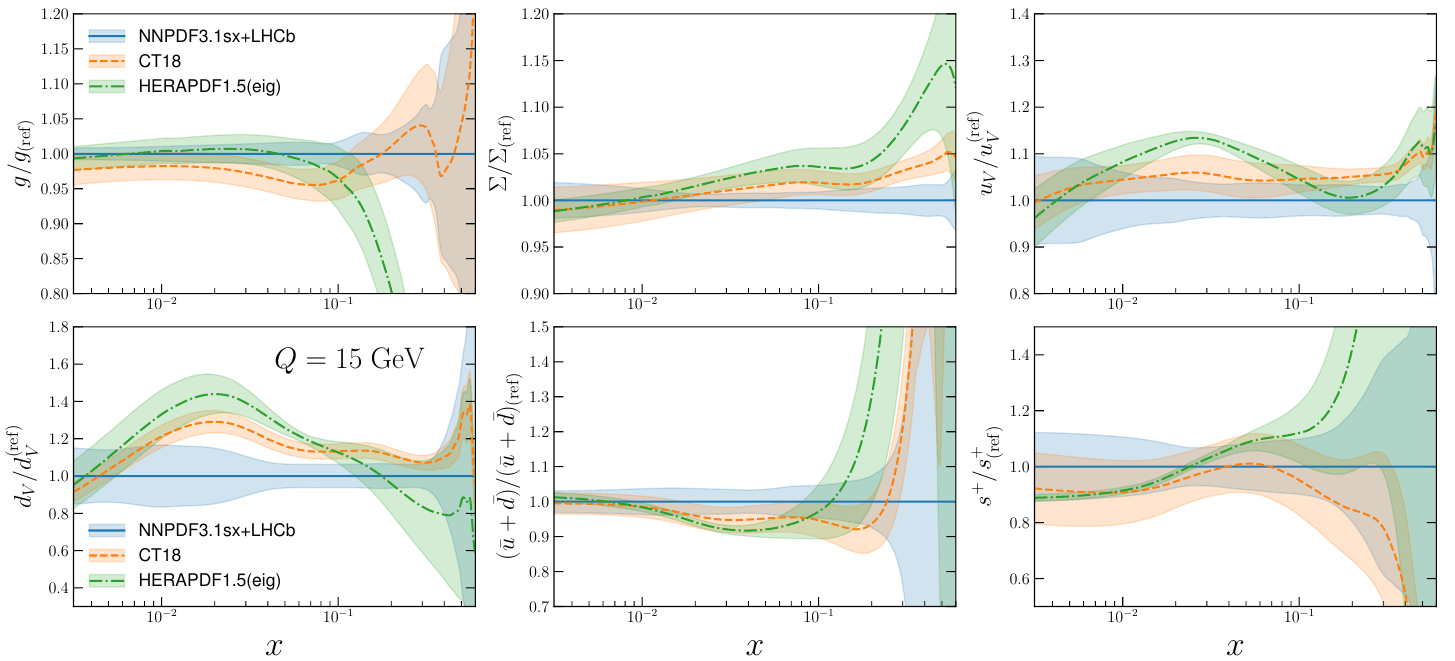}
\caption{Ratio of \texttool{CT18} and \texttool{HERAPDF1.5} to \texttool{NNPDF3.1sx+LHCb} for different PDFs combinations including singlet $\Sigma$ and $s^+=s+\bar{s}$ at $Q=15$ GeV (with the corresponding one-sigma uncertainty bands) in the $x$ region relevant for TeV neutrino-nucleus scattering. The \texttool{HERAPDF1.5} uncertainty band includes only the contribution from the experimental uncertainties and not the model and parametrization uncertainties.}
\label{fig:nu:pdfs}
\end{figure}

\cref{fig:dxsec} shows that the cross section at TeV energies is dominated by interactions with $Q\sim10$ GeV and $x\sim0.1$. A comparison between \texttool{HERAPDF1.5}, \texttool{NNPDF31sx}, and a recent \texttool{CT18} global analysis is shown in \cref{fig:nu:pdfs} for different PDFs combinations. One can observe that valence and singlet distributions for \texttool{HERAPDF1.5} and \texttool{NNPDF31sx} disagree in the range $0.01<x<0.6$. At leading order in isoscalar targets, $F_{2}$ is proportional to the quark singlet combination and $F_{3}$ to the valence content. Thus, the differences found in \cref{fig:xsec_q} can be partially attributed to the choice of the input PDFs. It is therefore essential to further study the behavior of the PDFs in this region of the $x-Q$ phase space.

\subsection{Neutral-Current Scattering}

Precise measurements of neutrino-nuclear cross sections across various energy scales are very important to test the SM as well as to probe new physics coupled to the neutrinos. Using terrestrial sources of neutrinos this interaction has been probed for energies up to $\mathcal{O}$(100)~GeV at experiments such as DONuT, NuTeV and CHARM~\cite{DONuT:2007bsg,NuTeV:2005wsg,McFarland:2005hg,CHARM:1987pwr,CHARMII:1998njb,ParticleDataGroup:2018ovx}. On the other hand, extremely high energy neutrinos, $E_\nu > \mathcal{O}$(10)~TeV, have been measured at IceCube~\cite{IceCube:2017roe,Bustamante:2017xuy}. The energy gap between these experiments (few 100~GeV to a few~TeV) can be probed at the FPF~\cite{Anchordoqui:2021ghd}, and already in Run 3 of the LHC with the FASER$\nu$ detector~\cite{FASER:2019dxq}.

\begin{figure}
    \centering
    \includegraphics[width=0.49\textwidth]{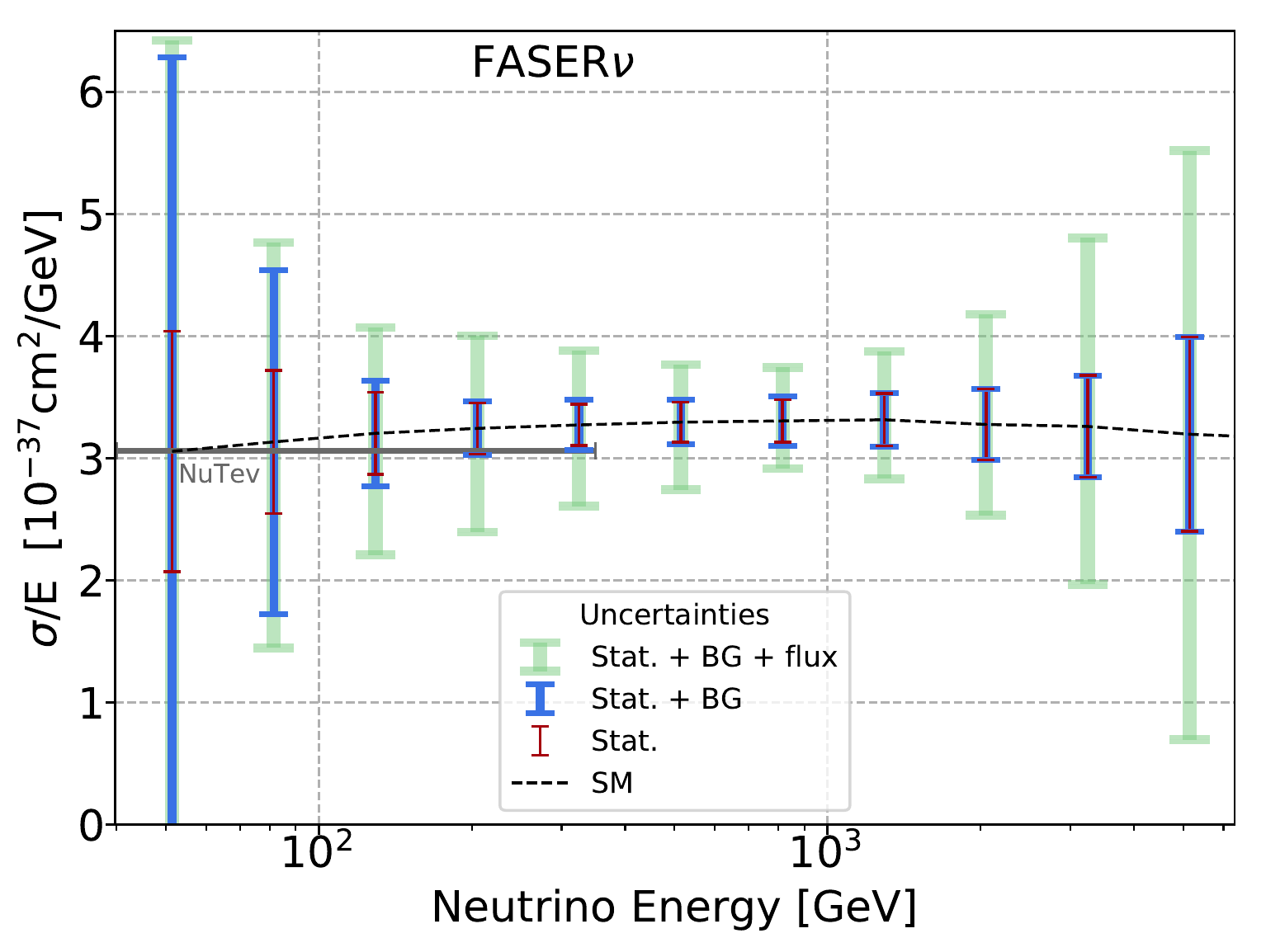}
    \includegraphics[width=0.49\textwidth]{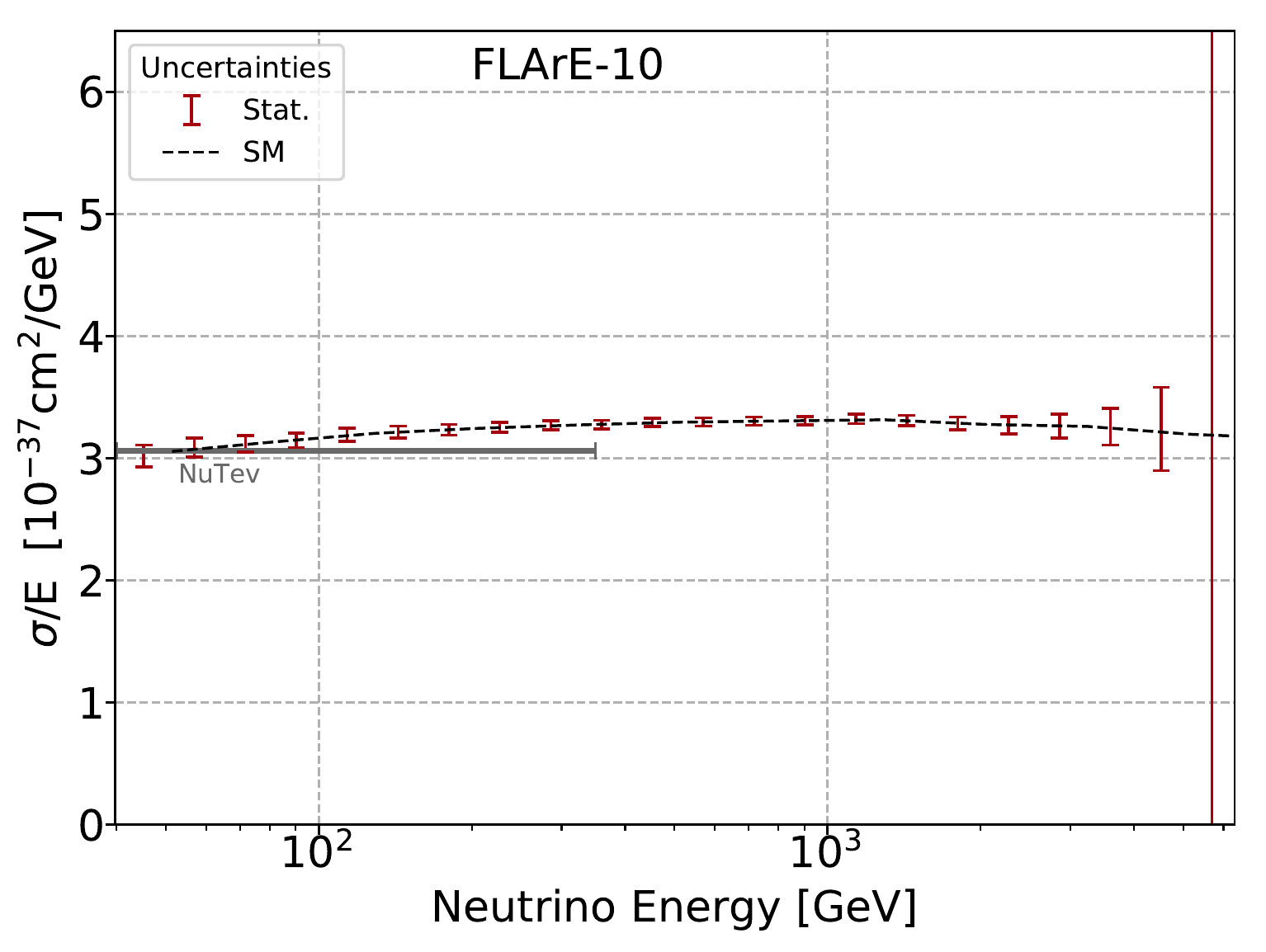}
    \caption{Estimated sensitivity to the average NC neutrino and antineutrino interactions at FASER$\nu$ (left) for $150$~fb$^{-1}$ and FLArE-10 (right) for 3~ab$^{-1}$ integrated luminosity. The black curve is the theoretical prediction for the average DIS NC cross section. Also shown are the NuTeV results~\cite{Zeller:2002he}. The left panel taken from Ref.~\cite{Ismail:2020yqc}.
    }
    \label{fig:nc_xs}
\end{figure}

Neutrinos can interact with matter via charged-current (CC) and neutral-current (NC) interactions. While CC interactions are easily identifiable by the charged lepton in the final state~\cite{FASER:2019dxq}, the situation is more complicated with NC interactions where only the recoiling target and a neutrino are present in the final state. The situation is worsened by the potentially significant backgrounds to NC interactions coming from neutral hadron interactions. These neutral hadrons could, for example, be produced in inelastic scatterings of muons coming from the LHC that pose a challenge in the absence of an active muon veto. However, emulsion detectors like those used in FASER$\nu$ have high spatial resolution which allow for precise measurement of kinematic variables like charged track multiplicities, direction and momentum of outgoing charged particles, etc. These kinematic quantities can be used to differentiate between the NC neutrino interaction signal and the neutral hadron background and estimate the neutrino energy as demonstrated in Ref.~\cite{Ismail:2020yqc}. In the first step, a classification network is trained on 9 kinematic variables to separate signal from background. The efficiency with which signal events can be identified increases with increasing neutrino energy reaching $>50\%$ for $E_\nu$ $>$ 1 TeV, while providing a good background rejection at the same time. At low energies of $\mathcal{O}$(100) GeV, backgrounds are reduced by 3 orders of magnitude and at higher energies of $\mathcal{O}$(1 TeV) backgrounds can be brought down to $\mathcal{O}$(10) events. The second step uses a regression network, again trained on the same 9 kinematic observables, to estimate the incoming energy of the neutrino. It was found that an energy resolution of about $50\%$ can be achieved in this way.

The left panel of \cref{fig:nc_xs} shows the NC cross section sensitivity thus obtained at FASER$\nu$ for LHC Run 3 with 150~fb$^{-1}$ integrated luminosity. The error bars shown include statistical uncertainties, uncertainties from background simulation, and neutrino flux uncertainties. The black dashed curve is the SM prediction for the average deep inelastic scattering (DIS) cross-section. For comparison the measurement from NuTeV~\cite{Zeller:2002he} is shown in gray. It has better statistics but is limited in the energy range is probed.

The detectors considered at the FPF, for example the FLArE liquid argon detector, have timing capabilities that allow for an efficient vetoing of neutral hadron events. Due to the higher target mass of the FPF experiments and the larger luminosity delivered in the HL-LHC era, these experiments will collect more NC events and hence have better statistics. While FASER$\nu$ is expected to collect a few thousand NC events, one can expect a few hundred thousand NC events to occur inside FLArE~\cite{Anchordoqui:2021ghd}. Considering only the statistical uncertainty and assuming perfect detector performance for simplicity, FLArE's sensitivity to NC interactions at the HL-LHC with an integrated luminosity of 3~ab$^{-1}$ is shown in the right panel of \cref{fig:nc_xs}.

\subsection{Quasi-Elastic and Resonance Regions for FPF Physics}
\label{subsub:qe}

Although the far-forward neutrinos at the LHC have typically high energies above $100~\gev$, their energy transfer to the nuclei in the FPF detectors can occasionally be much smaller. This is especially the case when they interact (quasi)elastically (QE) with the individual nucleons or in the resonant (RES) scattering regime. The capabilities to identify such events with the low hadronic activity but potentially large EM energy depositions or high-energy outgoing muons, will play an important role in better constraining the far-forward neutrino flux and spectrum at the LHC. Such measurements will allow for easier reconstruction of the incident neutrino energy. This will contribute to the BSM studies related to neutrino oscillations, cf. Ref.~\cite{Katori:2016yel} for review, as well supporting the FPF QCD physics program.

The possibility to study exclusive neutrino interaction processes of this kind could also allow for extending relevant past measurements into the high-energy regime, where only limited data have been collected in the past, cf. Ref. \cite{Formaggio:2012cpf} for review. The results obtained in the FPF experiments will substantially contribute to the efforts towards building a universal model for QE and QE-like neutrino scatterings that could be valid for a large energy range. Such studies have been identified as among the most important challenges in this field~\cite{NuSTEC:2017hzk}.

The $10$-tonne detector placed in the FPF along the beam collision axis is expected to collect of order $10^3$ charged current quasi-elastic (CCQE) events during the entire HL-LHC era, and a $2-3$ times larger number of CCRES events with a single pion in the final state~\cite{Batell:2021aja}. These will be dominated by the muon (anti)neutrino interactions, while the number of such scatterings induced by electron and tau (anti)neutrinos will be of order a few hundred and $\mathcal{O}(10)$, respectively. We summarize the relevant estimates in \cref{tab:nonDISrates}, which employs the far-forward neutrino flux and spectrum as described in Ref.~\cite{Kling:2021gos}. The mean neutrino energy in the FPF in such interactions is expected to be between $200$ and $300~\gev$. This is somewhat lower than the mean neutrino interaction energy in the DIS regime since the CCQE and CCRES scattering cross sections are not expected to grow with the energy. Therefore, the mean energy in this case directly corresponds to the average energy of the far-forward neutrinos at the LHC.

\begin{table*}[t]
\setlength{\tabcolsep}{5.2pt}
\centering
\begin{tabular}{c||c|c|c|c|c||c|c|c|c|c||c|c}
  \hline
  \hline
  &\multicolumn{5}{c||}{CCQE} & \multicolumn{5}{c||}{CCRES} & NCEL & NCRES
  \\ Detector
  & $\nu_e $    
  & $\bar\nu_e$    
  & $\nu_\mu$    
  & $\bar\nu_\mu$    
  & $\nu_\tau+\bar\nu_\tau$
  & $\nu_e $    
  & $\bar\nu_e$    
  & $\nu_\mu$    
  & $\bar\nu_\mu$    
  & $\nu_\tau+\bar\nu_\tau$
  & all
  & all
  \\
  \hline
  FASER$\nu$2
  & 60
  & 50
  & 570
  & 350
  & 3.5
  & 170
  & 180
  & 1.6k
  & 1.1k
  & 10
  & 170
  & 1.3k
  \\
  \hline
  FLArE
  & 40
  & 40
  & 420
  & 260
  & 3.5
  & 120
  & 140
  & 1.2k
  & 860
  & 10
  & 130
  & 940
  \\
  \hline
  \hline
\end{tabular}
\caption{Expected event rates for non-DIS neutrino scattering events in the FASER$\nu$2 and FLArE detectors. In the table, results for different neutrino flavors are presented for charged current quasi-elastic (CCQE), charged current resonant (CCRES), neutral current elastic (NCEL), and neutral current resonant (NCRES) interactions of neutrinos. The estimates employ  neutrino flux and spectrum predictions from Ref.~\cite{Kling:2021gos}. This table is adapted from Ref.~\cite{Batell:2021aja}.}
\label{tab:nonDISrates}
\end{table*}

\subsection{Interface of Shallow- and Deep-Inelastic Scattering}

Neutrinos and antineutrinos of all three flavors, copiously produced, offer the opportunity to probe charged-current (CC) neutrino interactions from the energy threshold of 3.5 GeV to a few TeV at the FPF. Given neutrino fluxes at the FPF, the average neutrino energy in charged-current (CC) interactions in detectors will be almost 500 GeV and higher, depending on the neutrino flavor \cite{Ahdida:2750060,FASER:2020gpr,Anchordoqui:2021ghd,Bai:2021ira,Kling:2021gos}. 
Leveraging neutrino cross section measurements at low energies, flux modeling and in-situ measurements, 
measurements of the $\nu_\tau+\bar{\nu}_\tau$ CC cross section \cite{Ahdida:2750060,FASER:2020gpr,Anchordoqui:2021ghd} should be possible. 
One is interested in how well the neutrino and antineutrino cross sections are predicted as a function of energy in the QCD-improved parton model. The low $Q^2$ contribution to the DIS cross section is noted in \cref{subsub:dis}, and exclusive processes in neutrino scattering, in \cref{subsub:qe}. Here, we focus on the $Q^2$ and the hadronic final state invariant mass dependencies of the neutrino and antineutrino DIS CC cross sections for scattering on isoscalar nucleon targets.

At the TeV neutrino energy scale $E_{\nu_\ell}$ for lepton flavor $\ell$, the nucleon target mass $m_N$, charged lepton mass $m_\ell$ and the charm quark mass are small compared to typical momentum transfers. For $E_{\nu_\ell}=10$ GeV, these mass corrections are important, as is the dependence of the CC structure functions on $Q^2 \lesssim 1$ GeV$^2$ and on how the transition from QE scattering to resonant and non-resonant production of pions to DIS is handled. In the transition, it is important to avoid double counting. If the evaluation of the neutrino cross section in the resonance region includes non-resonant diagrams, then one should not include a DIS contribution with the same final state hadronic invariant mass $W$, where
$    W^2=Q^2(1/x -1) + m_N^2\,.$
Given the $\Delta$ mass of $1.232$ GeV, the $\Delta$ resonance region is generally considered as spanning $m_N+m_\pi<W\lesssim 1.4$ GeV. Beyond the $\Delta$ resonance, additional resonances contribute out to $W\simeq 1.8-2$ GeV, where in this region, both resonant and non-resonant processes contribute.
Called shallow inelastic scattering (SIS), it is a kinematic region that is not yet clearly understood (see refs. \cite{SajjadAthar:2020nvy,NuSTEC:2019lqd} and references therein). In our evaluations of the inelastic cross section
\cite{Jeong:2022tba}, we consider three minimum hadronic final state invariant masses: $W_{\rm min}=m_N+m_\pi$, 1.4 GeV and 2 GeV. 

Even within the DIS regime, the parton model is not clearly  operative when $Q^2\to 0$. A perturbative treatment of the structure functions in terms of PDFs is not reliable for $Q^2 \lesssim 1$ GeV$^2$. Furthermore, one expects a transition of neutrino scattering from partons to neutrino scattering with hadrons as the distance scale $1/Q$ becomes the size of the nucleon. There are different approaches to handling the low $Q$ limit of the structure functions. As discussed in \cref{subsub:dis}, Bodek and Yang \cite{Yang:1998zb,Bodek:2002ps,Bodek:2004pc,Bodek:2021bde} take advantage of electromagnetic scattering data to make fits to yield effective PDFs at low $Q^2$ that use modified parton momentum fractions, multiplicative  $Q$-dependent $K$-factors and the GRV98 PDFs fixed at $Q^2=0.8$ GeV$^2$. This approach includes target mass effects through the fitted parameters, and given effective parton PDFs, the weak structure functions can be constructed. An alternative approach which is used here is to adapt phenomenological structure function parameterizations fitted to electromagnetic structure function data
\cite{Capella:1994cr,Kaidalov:1998pn,Bertini:1995cr} for $Q^2 \lesssim$ a few GeV$^2$  to neutrino scattering in the same $Q^2$ range.  

We use here the Capella et al. parameterization \cite{Capella:1994cr} (CKMT) of the electromagnetic structure function $F_2$ which has two terms with functional forms based on Pomeron and Reggeon contributions. The two terms, at large enough $Q^2$ can be interpreted as valence and sea contributions. This permits the relative normalization to be adapted to CC scattering, as outlined in Ref. \cite{Reno:2006hj}. One feature of the CKMT parameterization of $F_2$ is that $F_2(x,Q^2)/Q^2$ goes to a constant proportional to the photon-proton cross section in the $Q^2\to 0$ limit.
For $Q^2>2$ GeV$^2$, the structure functions are evaluated here with NLO QCD including target mass corrections \cite{Georgi:1976ve,Barbieri:1976rd,DeRujula:1976baf}, quark mass corrections and for tau neutrino and antineutrino scattering, tau mass effects, all summarized in refs. \cite{Kretzer:2002fr,Kretzer:2003iu}. For $Q^2 \leq 2$ GeV$^2$, the CKMT structure functions, modified for neutrino and antineutrino scattering, are used. With this approach, the effect on the cross section at low $Q^2$ can be approximated.

Here, we approximately quantify the dependence of the neutrino and antineutrino CC cross sections on the regions of
$W$ and $Q^2$ where perturbative QCD is not applicable. At the FPF, the relevant detectors will require cross sections on nuclei. PDF uncertainties in the CC neutrino cross sections with the nCTEQ15 nuclear PDFs for tungsten \cite{Kovarik:2015cma} range from 2--7\% for $E_\nu=10-1000$ GeV tau neutrinos and antineutrinos. We use 3\% as a reference PDF uncertainty, the uncertainty from the nCTEQ15 PDF sets for the $\nu_\tau$($\bar\nu_\tau$)-tungsten CC cross section for an incident energy of 100 GeV. 
The degree of impacts of the minimum W ($W_{\rm min}$) and $Q^2$ are almost the same for nucleons in tungsten and isoscalar nucleons.
For investigations of the impact of $W_{\rm min}$ and $Q^2$, we evaluate the cross sections using the nCTEQ15 PDFs \cite{Kovarik:2015cma} for isoscalar nucleons.

\begin{figure} 
\centering
\includegraphics[width=.48\textwidth]{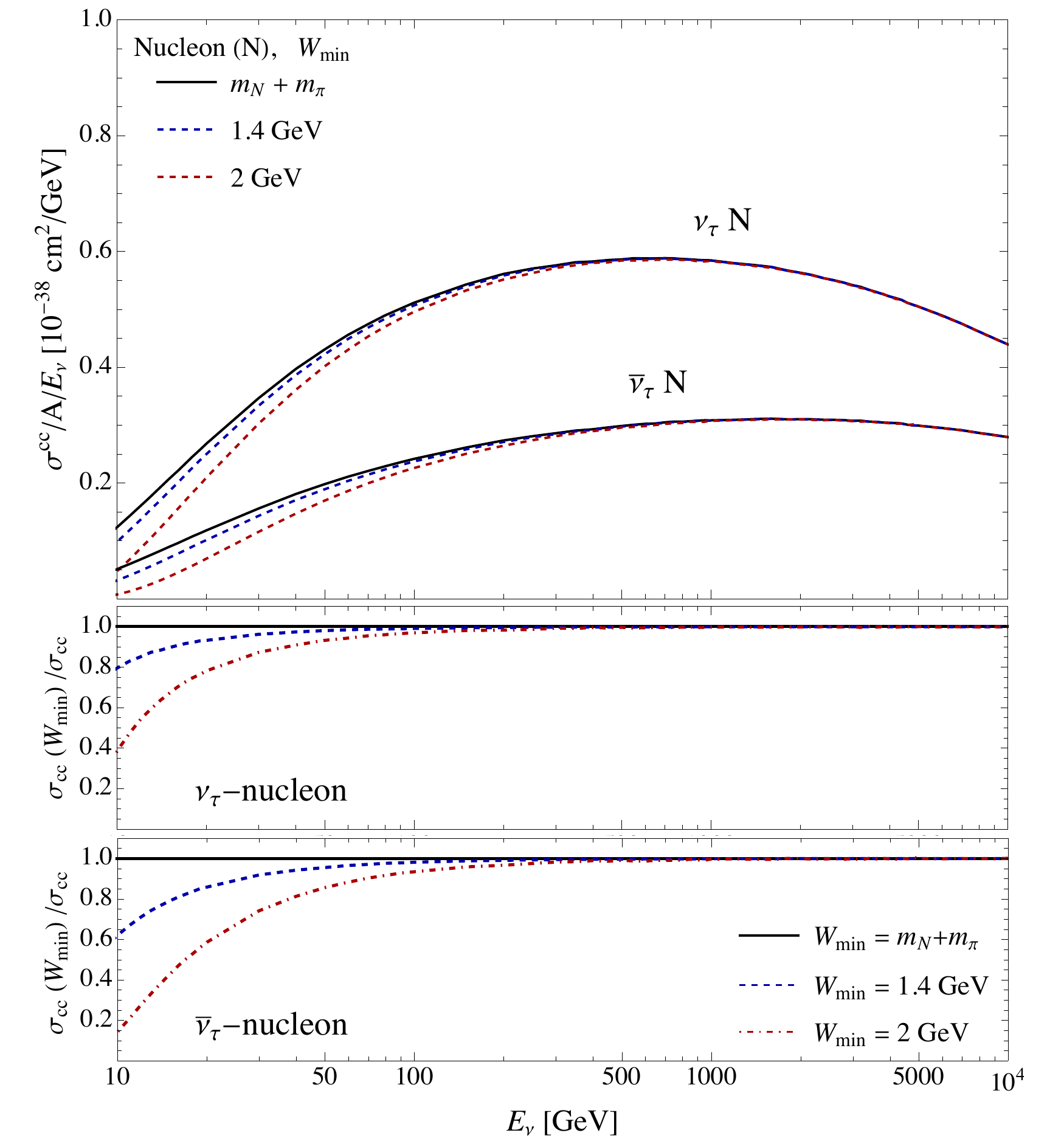}
\includegraphics[width=.48\textwidth]{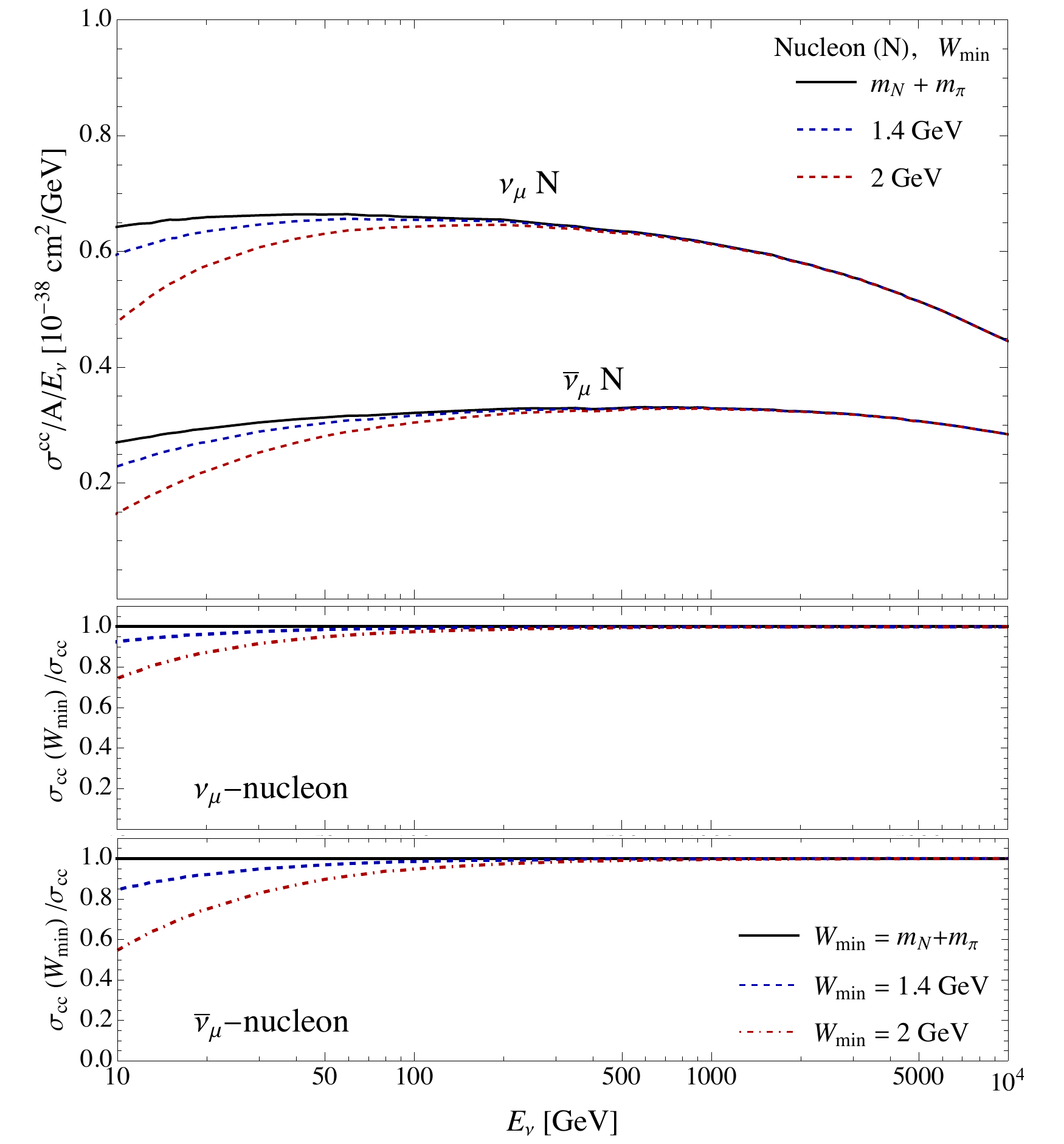}
\caption{\label{Fig:CSnutN-Wmin} 
Top row: The charged current cross section per nucleon of tau neutrino and antineutrino (left) and muon neutrino and antineutrino (right) scattering with isoscalar nucleons, for three values of the minimum hadronic final state invariant mass $W_{\rm min}$.
Bottom row: The ratios of the cross sections as a function of energy for $W_{\rm min}=1.4$ GeV and 2 GeV to the cross sections with $W_{\rm min} = m_N + m_\pi$ for tau neutrinos and antineutrinos (left) and muon neutrinos and antineutrinos (right). 
}
\end{figure}

\cref{Fig:CSnutN-Wmin} shows the $\nu_\tau$, $\bar{\nu}_\tau$, $\nu_\mu$ and $\bar{\nu}_\mu$ scattering with isoscalar nucleons.
The left (right) panel in the top row of \cref{Fig:CSnutN-Wmin} shows the charged current (CC) cross sections for tau (muon) neutrino and antineutrino interactions as function of energy for $W_{~\rm min}= m_N+m_\pi$, 1.4 GeV and 2 GeV for the full $Q^2$ range.
Compared with the muon neutrino results, the $\nu_\tau$ and $\bar{\nu}_\tau$ cross sections are suppressed below 1 TeV due to the large tau lepton mass. 
The lower panels show the ratio of the cross sections with $W_{\rm min}=1.4$ GeV and 2 GeV to the cross section with $W_{\rm min}=m_N+m_\pi$.
The effect of $W_{\rm min}$ stands out for $E_\nu \lesssim$ 100 GeV, and it is larger for antineutrinos than for neutrinos.
A focus on $E_\nu>100$ GeV shows that 
the CC cross section for $W_{\rm min}=2$ GeV is lower than the CC cross section with $W_{\rm min}=m_N+m_\pi$ by at most 3\% for neutrinos (both $\nu_\mu$ and $\nu_\tau$) and 7 (5)\% or less for $\bar{\nu}_\tau$ ($\bar{\nu}_\mu$).
To the extent that the DIS evaluation from $W=1.4-2$ GeV approximately averages over resonances beyond the $\Delta$, one can use $W_{\rm min}=1.4$ GeV. In this case, the ratios of the CC cross sections for $W_{\rm min}=1.4$ to $W_{\rm min}=m_N+m_\pi$ are greater than 99 (98)\% for both muon neutrinos and tau neutrinos (antineutrinos) for $E_\nu>100$ GeV.

\begin{figure}
\centering
\includegraphics[width=.47\linewidth]{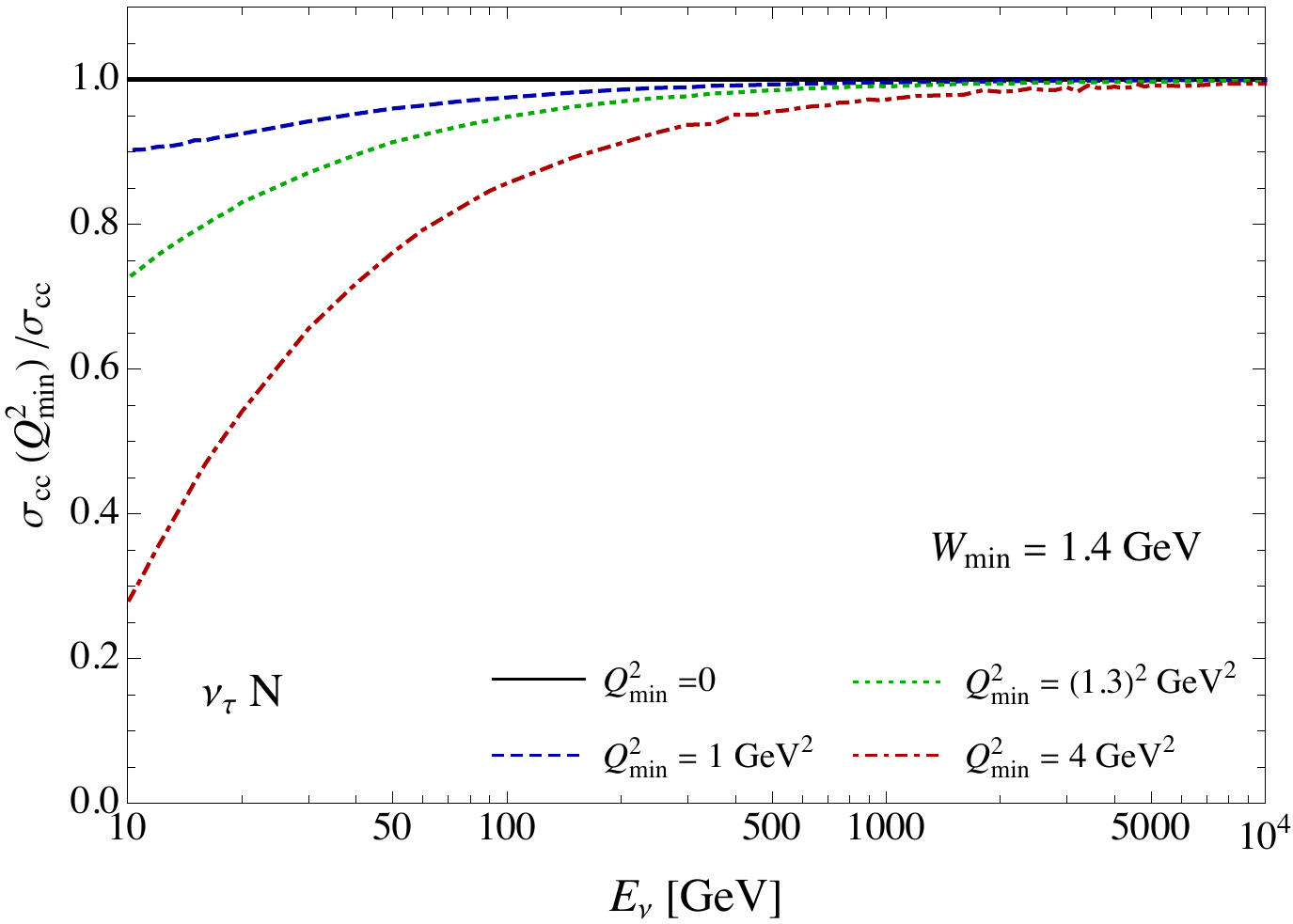}
\includegraphics[width=.47\linewidth]{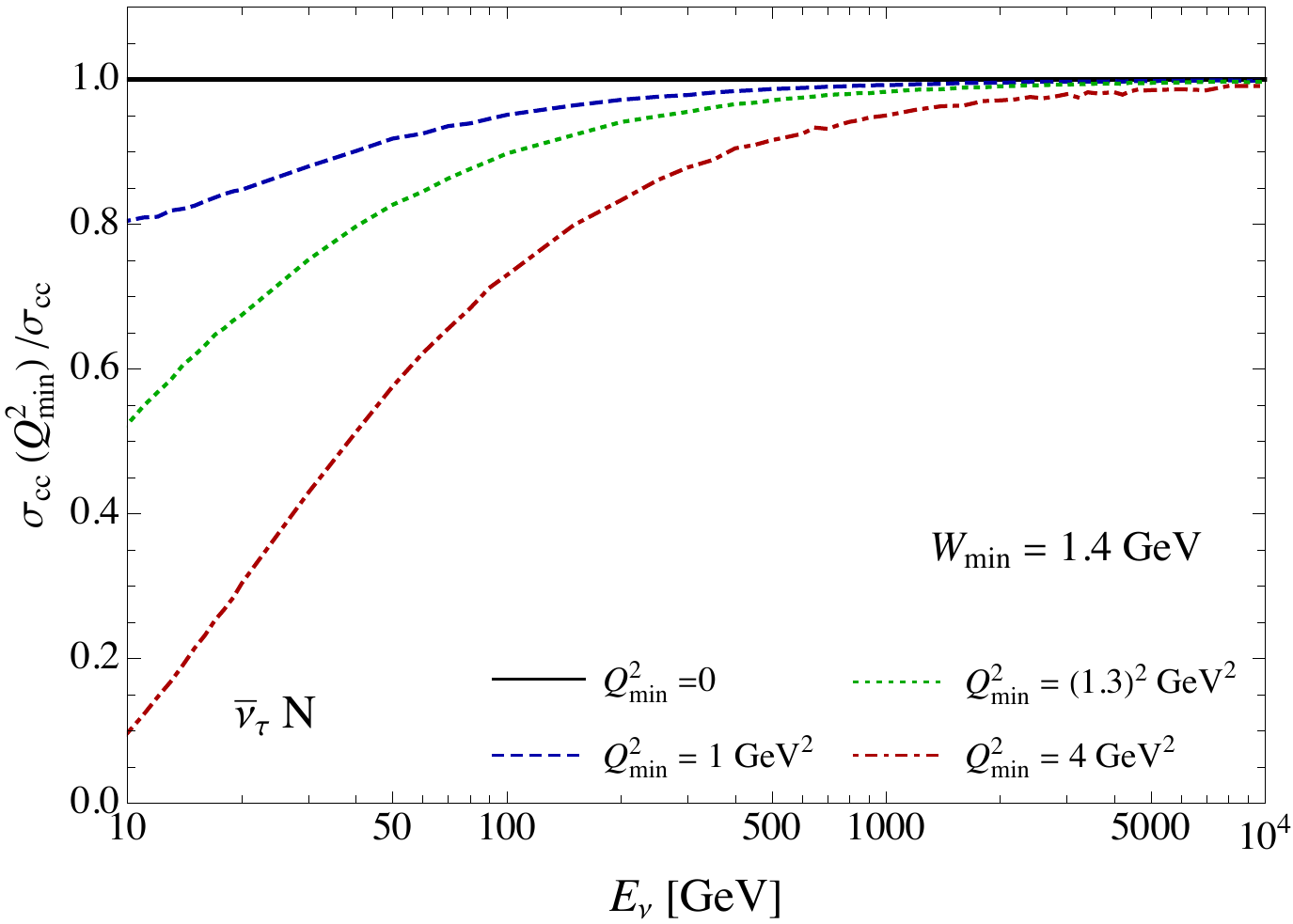}
\includegraphics[width=.47\linewidth]{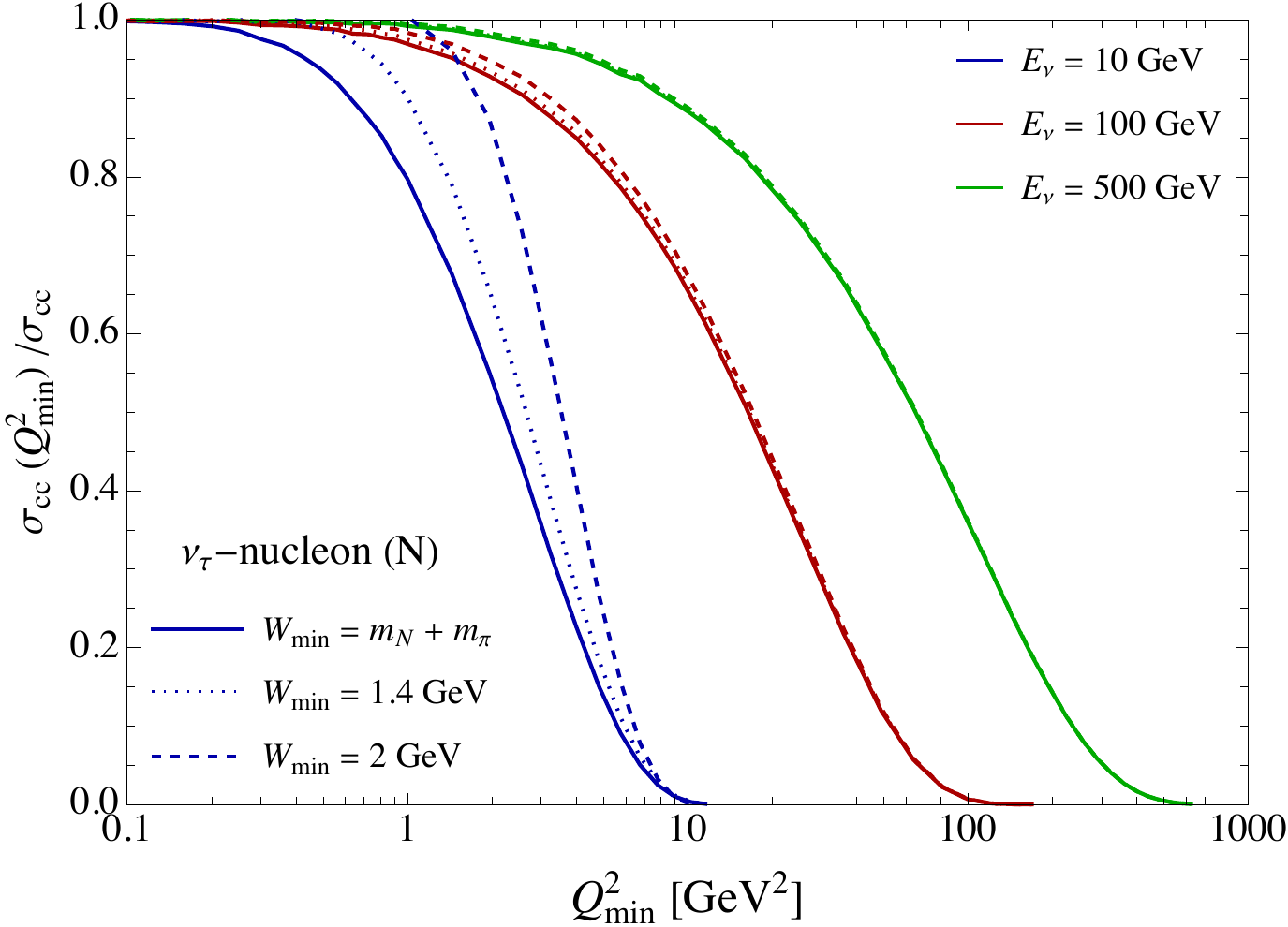}
\includegraphics[width=.47\linewidth]{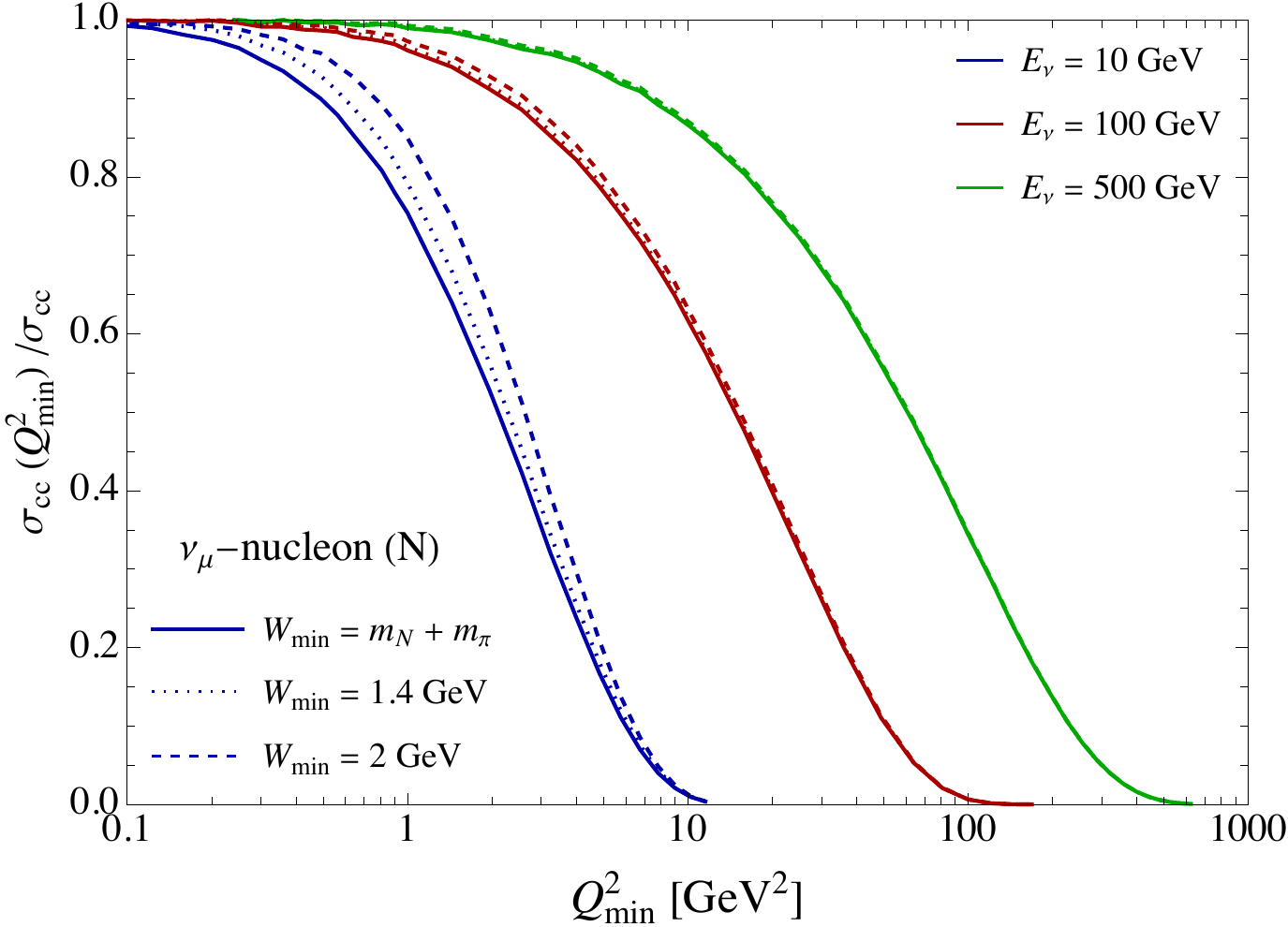}
\caption{\label{Fig:RCSnutN-Q2Wmin} 
The ratio of the charged current cross sections with the cut of $Q_{\rm min}^2$ and with all $Q^2$ for tau neutrinos (upper left) and antineutrinos (upper right) CC interaction with nucleon for $W_{\rm min}=$ 1.4 GeV. Also shown is
the ratio of the CC neutrino-isoscalar nucleon cross sections with a minimum value of $Q^2$ to the cross section with all $Q^2$, as function of $Q^2_{\rm min}$, for tau neutrinos (lower left) and muon neutrinos (lower right) for the $W_{\rm min}=m_N + m_\pi$, 1.4 and 2 GeV.
}  
\end{figure}

We now turn to the $Q^2$ dependence of the cross sections for neutrino and antineutrino CC scattering with isoscalar nucleons. We note that in order to satisfy $W>W_{\rm min}$, the region of $Q^2$ for a given parton $x$ is restricted, but for no additional restriction on $Q^2$, we denote $Q_{\rm min}^2=0$ GeV$^2$ in what follows. We consider $Q_{\rm min}=1, 1.3$ and 2 GeV in addition to $Q_{\rm min}=0$.
The impact of the $Q_{\rm min}^2$ values on the cross sections spans a wider energy range than $W_{\rm min}$ with conspicuous effects even at $E_\nu = {\cal O}(10^2)$ GeV. The upper panels of
\cref{Fig:RCSnutN-Q2Wmin} present the ratio of the CC neutrino-isoscalar nucleon cross sections for $\nu_\tau$ (upper left) and $\bar{\nu}_\tau$ (upper right) evaluated with a minimum value of $Q^2$ and those for the full $Q^2$ range when $W_{\rm min}$ = 1.4 GeV.
We checked that the corresponding results for scattering with tungsten targets are approximately the same as the ratios in the figure.

The large effect of setting $Q_{\rm min}$ between 1-2 GeV for low energy neutrino scattering reduces with energy, however, even for $E_\nu=100$ GeV, the contribution to the cross section for $Q=1-2$ GeV is large, of order $\sim 12\%$ for $W_{\rm min}=1.4$ GeV. Since the contribution from $Q$ between $1$ GeV or 1.3 GeV and 2 GeV can be calculated with evolution of the PDFs reasonably reliably, we focus on the smaller $Q^2_{\rm min}$ values, setting $W_{\rm min}=1.4$ GeV. 
The fraction of the CC cross section for tau neutrino scattering is reduced by 3\% 
at $E_\nu = $ 100 GeV for $Q_{\rm min}=1$ GeV compared to that for $Q_{\rm min}=0$. 
The effect of the cutoff on $Q_{\rm min}$ is more significant in the ratios of antineutrino cross sections,
for which 
about 5\% of the cross section comes from $Q < 1$ GeV for $E_\nu = $ 100 GeV in our evaluation.

For muon neutrinos and antineutrinos, the impact of $Q^2_{\rm min}$ is larger than for tau neutrinos and antineutrinos at low energies. In the lower panels of \cref{Fig:RCSnutN-Q2Wmin}, we show the ratio of the neutrino-nucleon CC cross sections for several $W_{\rm min}$ values, as a function of $Q^2_{\rm min}$, for $E_\nu=10,$ 100 and 500 GeV. 
At low energies, the kinematic restrictions, including the requirements for tau production and $W_{\rm min}$ are more evident in the $Q_{\rm min}^2$ dependence of the CC cross section. For $E=100$ GeV and $W_{\rm  min}=1.4$ GeV, the fraction of the neutrino CC cross section is
$3\%$ for $Q^2< 1$ GeV$^2$ with our evaluation. For antineutrinos, the corresponding fraction is 7\%.

In the framework of a phenomenological electroweak structure function extrapolation \cite{Capella:1994cr,Kaidalov:1998pn,Reno:2006hj} of the perturbative QCD evaluation of the neutrino and antineutrino CC structure functions, we find that inelastic scattering for $m_N+m_\pi<W<1.4$ GeV contributes less than 3\% of the cross section for 100 GeV incident neutrinos and antineutrinos, and less at higher energies. For $W_{\rm min}=1.4$ GeV, approximately 3--5 (7)\% of the CC cross section for incident energies of 100 GeV come from $Q<1$ GeV for $\nu_\tau$ and $\bar\nu_\tau$ ($\bar\nu_\mu$). Thus, while at very high energies, the impact of low $W$ and $Q$ regions on the neutrino and antineutrino cross sections are negligible, for energies comparable to $E_\nu\sim 100$ GeV, the neutrino
and antineutrino cross sections are not reliably predicted at the few percent level. The importance of reliable modeling of $Q<1$ GeV is emphasised also in \cref{subsub:dis} and, for example, recently in refs. \cite{Bertone:2018dse,Garcia:2020jwr}.

At lower incident neutrino and antineutrino energies, these kinematic regions of $W$ and $Q$ are even more important. 
As noted, the calculational framework for low $Q$
structure functions is not the only approach. Comparisons with Bodek-Yang \cite{Bodek:2002ps,Bodek:2004pc} and this phenomenological structure function approach in neutrino scattering have assumed that the vector and axial vector contributions to the structure functions are equal, as they are in the parton model. Recent work by Bodek and Yang  includes  axial vector structure functions not equal to the vector structure functions \cite{Bodek:2021bde}, most important at low $Q^2$.

Further investigations of the low $W$ and low $Q$ contributions for energies  $\lesssim 100$ GeV are needed.
The neutrino experiments at the FPF such as FASER$\nu$2, Advanced SND@LHC and FLArE will detect neutrinos at $E_\nu = {\cal O} (1)$ GeV -- ${\cal O} (1)$ TeV with thousands of neutrino events.
The energy range covered by the FPF and their abundant neutrino data will provide a unique opportunity to investigate neutrino interactions in the kinematic region for shallow and deep inelastic scattering and at low $Q^2$.

\subsection{Role of Final State Interactions}

An important characteristic of the planned neutrino detectors at the FPF is their precision tracking capability. This capability will make it possible to record not only the energy and direction of the outgoing lepton, but also the detailed properties of the final-state hadronic system that emerges from the large nucleus of the target (e.g., argon or tungsten). By carefully analyzing this data, one may gain unique insights into the fundamental physics of QCD in the regime of cold nuclear matter.

At first, the premise of such a study may appear paradoxical. Indeed, since the energy of the neutrino (in the TeV) range vastly exceeds the binding energy of the nucleons, one may think that it would be sufficient to consider interactions between the neutrino and an individual proton, with the rest of the nucleus being invisible to the final-state particles. This, however, is not the case. The important physical consideration here is not the nuclear binding energy, but the fact that the primary interaction vertex is surrounded by a dense nuclear medium. The newly created colored objects must be transported through this medium. How the QCD shower develops in this medium and how eventual hadronization takes place will impact the multiplicity of the final-state hadrons and their characteristic energies. As discussed in \cref{subsubsec:GiBUU}, the transport calculations by the GiBUU code predict an avalanche of particles that results in a large particle multiplicity increase in the outgoing shower.

In the case of heavy ion collisions, it is well established that propagation of energetic hadrons through the hot nuclear medium changes their properties, leading to what is known as jet quenching~\cite{Gyulassy:1990ye}. This phenomenon has been observed at RHIC~\cite{PHENIX:2001hpc,STAR:2002svs} and has been recognized as evidence for the formation of quark-gluon plasma. A quantitative investigation of the corresponding attenuation and secondary particle production effects in cold nuclear matter would be of great interest.

The possibility of probing this physics has attracted a lot of attention in the context of electron-ion collisions. In fact, investigations of this phenomenon form an important component of the physics program for the EIC. As noted in the EIC White Paper~\cite{Accardi:2012qut},  ``cold QCD matter could be an excellent femtometer-scale detector of the hadronization process from its controllable interaction with the produced quark (or gluon)''. In this sense, the neutrino detector at the FPF would be a neutrino-ion collider, which would be able to access the same physics, in a complementary way, with excellent tracking information on the final-state hadrons. Given the expected timeline of 10-15 years for the construction of the EIC, the prospects of making this study in neutrinos first are especially exciting.

Knowledge about the dynamics of the hadronization process remains limited and strongly model dependent. As an illustration, the investigation of Gallmeister and Mosel~\cite{Gallmeister:2007an} considered two models for the dependence of the FSI cross section on the post-interaction time: a linear and a quadratic. It was found that the linear model was in better agreement with the then-available data from the EMC and HERMES experiments, supporting an earlier argument by Dokshitzer {\it et al.}~\cite{Dokshitzer:1991wu}. Neutrino-scattering measurements at the FPF facility would provide crucial new data for this comparison, extending it to higher energies.

Further motivation is provided by a surprising recent measurement at Hall C of Jefferson Lab of the exclusive (e, e’p) reaction channel \cite{HallC:2020ijh}. The upgrade of the accelerator to 12 GeV made it possible to investigate recoil baryon propagation in the nuclear medium in the kinematic window where the so-called color transparency phenomenon was expected to set in (up to proton momenta of 8.5 GeV).  Yet, the experimental data does not support the onset of color transparency. Studies of this phenomenon at the much higher energies of the LHC neutrino beam will be crucial.

\subsection{Scattering with Electrons}

Neutrino--electron scattering is a useful standard candle for neutrino flux measurements but suffers from a relatively small cross section. At FASER$\nu$ energies, however, qualitatively new detection channels appear within the Standard Model, some of which have not been measured before. Of particular interest is the opportunity to measure resonant meson production off bound atomic electrons. This was recently investigated in Ref. \cite{Brdar:2021hpy} where FASER$\nu$ was identified as the most promising near-term experiment for the detection of this rare Standard Model process. 

\begin{figure} 
\centering
\includegraphics[width=.78\textwidth]{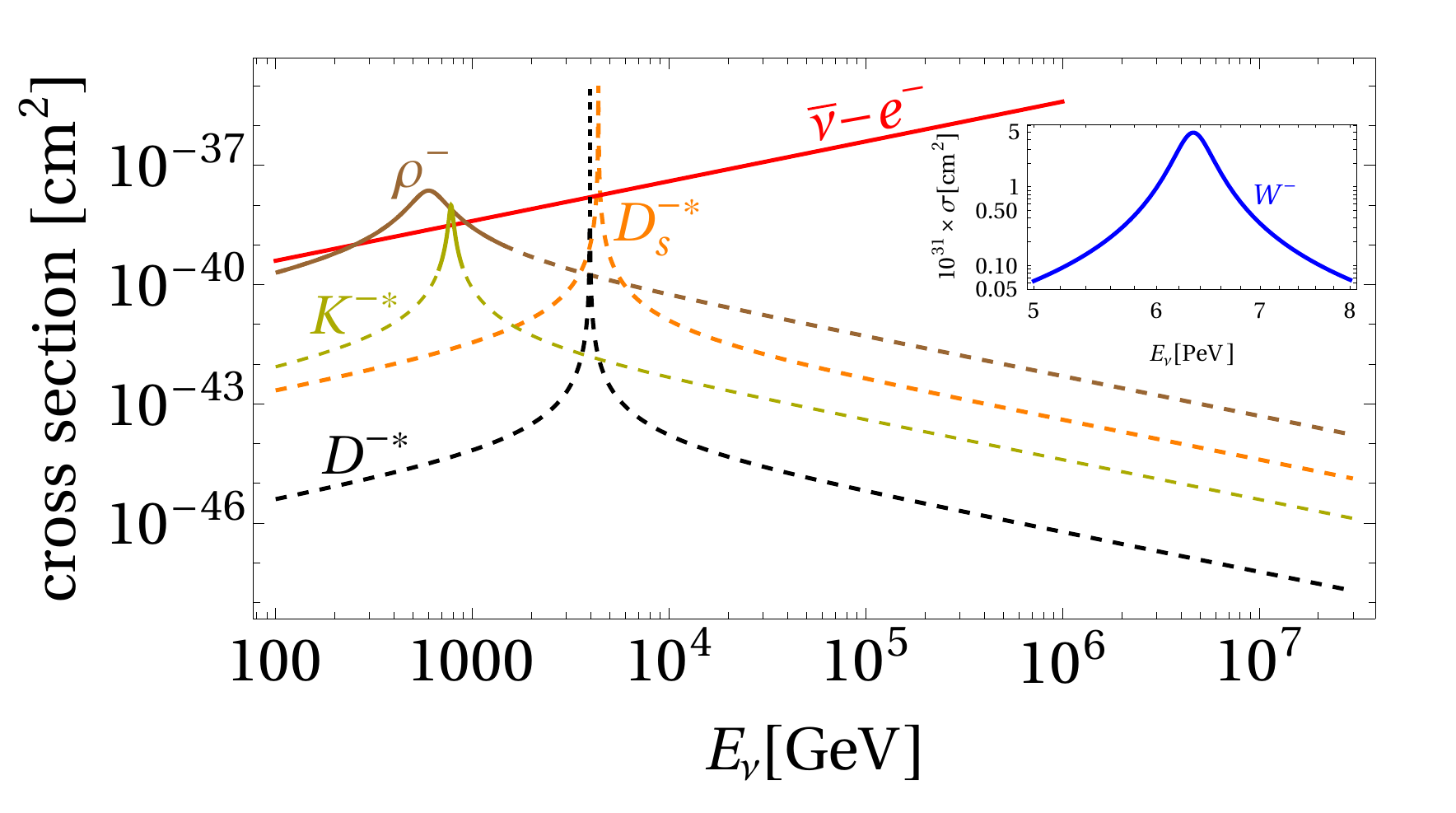}
\caption{\label{Fig:res_VR} 
Cross section for the production of vector meson resonances. For comparison, we also show elastic neutrino-electron  cross section. Figure from Ref. \cite{Brdar:2021hpy}.
}
\end{figure}

Neutrino--electron scattering can produce $s$-channel resonances, but only when neutrino energies satisfy $s=2 E_\nu m_e + m_e^2 \geq M^2$ where we have assumed that the electron is at rest, and that the $s$-channel resonance has mass $M$. For FASER$\nu$ energies, this limits the mass of a potential $s$-channel resonance to a few GeV. Pseudoscalar mesons have hopelessly small production cross sections because of the necessary chirality flip that must be supplied by the electron's mass in the initial state. The most promising detection candidates are therefore light, and necessarily charged, vector meson resonances and here the $\rho^-$ with $M\sim 770$ MeV stands out. The relevant cross sections 
are shown in \cref{Fig:res_VR}.

In Ref. \cite{Brdar:2021hpy}, the rate of $\bar{\nu}_e e^-\rightarrow \rho^- \rightarrow \pi^- \pi^0$ was estimated and found to yield a handful of events at FASER$\nu$ (a possible follow-up program with FASER$\nu$2 could yield tens of events), see \cref{table:VR}. The calculations assume an electron and rest and ignore atomic binding corrections. A sketch of potential background mitigation techniques was also discussed \cite{Brdar:2021hpy} with the conclusion being that a nuclear emulsion detector would likely be able to achieve sufficient signal to background ratios to allow for a definitive discovery with FASER$\nu$2, and possibly with FASER$\nu$. The key physics driving the signal to background ratio was the highly boosted kinematics of the $\bar{\nu}_e-e$ collision and the uncharacteristically low hadronic multiplicity of the event. Low multiplicity and small opening angle between $\pi^-$ and  $\pi^0$, together with the cut on the invariant mass of
the $\pi^- \pi^0$ pair are expected to yield signal to background ratio bigger than 1.

\begin{table}[t!]
\centering
 \begin{tabular}{c||c|c|c|c}
\hline\hline
\hspace{0.0125\linewidth} Experiment \hspace{0.0125\linewidth} &  $\rho^{-}$, $\pm\Gamma/2$ &  $\rho^{-}$, $\pm2\Gamma$ &  $K^{-*}$, $\pm\Gamma/2$ &  $K^{-*}$, $\pm2\Gamma$ \\ \hline 
FASER$\nu$ &  0.3 & 0.5  & -- & --   \\ 
FASER$\nu$2 & 23 & 37  &  0.7 & 3  \\ 
FLArE-10 &  11 & 19  &  0.3 & 2   \\
FLArE-100 & 63 & 103  &  2 & 8  \\ 
DeepCore & 3 (1) & 5 (2)  &  --  & --  \\ 
IceCube &  8 (40) & 17 (83)  &  -- & -- \\  \hline\hline
\end{tabular}
\caption{Estimated number of $\rho^-$ and $K^{-*}$ resonance-mediated events at various experiments \cite{Brdar:2021hpy}. A dash (--) indicates cases in which less than $0.1$ events are predicted.}
\label{table:VR}

\end{table}

\section{Monte Carlo Tools for Neutrino Interactions}
\label{sec:tools}

Monte-Carlo (MC) event generators are indispensable part of the experimental programs. At FPF, neutrino MC event generators will play a key role, they will be needed at each step of the experiments: in predicting neutrino interaction event rates, all final-state particles and their energies for all the target material in the detector and for all three neutrino flavors over a broad spectrum of FPF neutrino energies. Subsequently, event generators will help estimating systematic uncertainties associated with neutrino interactions that will be vital in disentangling new physics signals. Widely used neutrino generators are optimized for accelerator neutrino energies, sub-GeV to a few GeV range, but extends to higher energies relevant for FPF neutrinos. In this subsection, we briefly outline the core of the widely used neutrino event generators and present some comparison between their predictions. 

\subsection{GENIE}
\texttool{GENIE} is a ROOT-based Neutrino MC Generator, that was released in 2007~\cite{Andreopoulos:2009rq}. 
Since then, \texttool{GENIE} has been adopted by the majority of neutrino experiments, including those as the JPARC and NuMI neutrino beamlines, becoming an important physics tool for the exploitation of the world accelerator neutrino program.
It aims to become a event generator whose validity will extend to all nuclear targets and neutrino flavors over a wide spectrum of energies ranging from 1 MeV to 1 PeV.
Historically, the focus of neutrino interaction modelling in \texttool{GENIE} was the few-GeV energy range relevant for atmospheric neutrino studies, as well as for studies of accelerator-made neutrinos both at short and long baseline experiments. 
In the latest official release~\cite{GENIE:2021npt}, \texttool{GENIE} has included high energy neutrino interactions, extending its support to the field of high-energy neutrino astronomy, as well as the neutrino projects in the Forward Physics Facility at CERN.

In the TeV energy regime, where DIS dominates, two different configurations can be used to simulate neutrino interactions in \texttool{GENIE}: 

- Non-perturbative ($G00$ and $G18$ series): 
This is the default configuration in the few-GeV region. 
Differential cross sections are calculated in an effective leading order model using the modifications suggested by Bodek and Yang \cite{Bodek:2002ps}. 
It is the model used for the nonresonant processes that compete with resonances. 
The total cross section in the transition region (low multiplicity inelastic reactions) can be tuned to avoid double counting. 
This configuration also includes different neutrino charm production models \cite{Kovalenko:1990zi,Aivazis:1993kh}. 
The cross sections are computed at a fully partonic level.
The default hadronization model is AGKY~\cite{GENIE:2021wox}, which includes a phenomenological description of the low invariant mass region based on KNO scaling~\cite{Koba:1972ng}, while at higher masses it gradually switches over to the \texttool{Pythia} model~\cite{Sjostrand:2006za}. The \texttool{Pythia} configuration parameters are set to be the default values except for four parameters which were tuned \cite{Sakuda:2002rx}.
Final state interactions are also taken into account. \texttool{GENIE} includes different models: hA (data-driven code that is fully reweightable)~\cite{Andreopoulos:2015wxa}, hN (full cascade), INCL++ \cite{Mancusi:2014eia}, and \texttool{Geant4} extended Bertini \cite{Wright:2015xia}.

- Perturbative ($GHE19$ series):
This is the default configuration in the high energy regime.
Differential cross sections are calculated using the perturbative QCD formalism for $W>2$ GeV. Currently, two models are available (CSMS~\cite{Cooper-Sarkar:2011jtt} and BGR~\cite{Bertone:2018dse}). 
An interface to compute structure functions using \texttool{APFEL}~\cite{Bertone:2013vaa} is also available.
This configuration includes both charm and top production  (relevant at PeV energies). It also computes the differential cross section at parton level using the formalism described in Ref. \cite{Garcia:2019hze}, and it includes sub-leading effects, e.g., $W$-boson and trident production as in Ref. \cite{Zhou:2019vxt}.
Hadronization is based on \texttool{LEPTO}~\cite{Ingelman:1996mq} and it computes fragmentation functions and kinematics with \texttool{Pythia}.
Final state interactions can be simulated but they are turned off in the default configuration.

\subsection{NEUT}

The \texttool{NEUT} neutrino interaction simulation program library
is the interaction generator primarily used by the Super-Kamiokande and T2K neutrino oscillation experiments. It was originally written in the 1980s to predict neutrino-induced backgrounds for nucleon decay measurements made with Kamiokande. Much of the original \texttool{FORTRAN77} code is still used. Over the past 30+ years, it has been developed with a focus on sub-GeV to few-GeV neutrino scattering to fulfill the simulation needs of the Super-Kamiokande atmospheric neutrino measurements and the T2K neutrino beam measurements. 

\texttool{NEUT} is capable of simulating neutrino interactions up to a few TeV, the energy region of concern for this white paper, but none of the model components implemented are tuned for neutrino--nucleus interactions at such energies. The two most relevant components are the deep inelastic scattering (DIS) and the hadron transport (or hadron re-scattering, or FSI) models. The DIS model, which is used to produce hadronic systems with $W>2.0~\textrm{GeV}$, is based on \texttool{Pythia~v5.72} (included in \texttool{CERNLIB 2005}). The GRV98 parton distribution functions~\cite{Gluck:1998xa} are modified for low $Q^{2}$ according to the Bodek-Yang~\cite{Bodek:2005de} model. A semi-classical stepped cascade is used to transport hadrons produced in the \emph{primary} neutrino--nucleon interaction through the nuclear medium. Interaction channels are implemented for nucleons, pions, kaons, etas, and omegas. At the interface of the DIS simulation and the hadron transport, a \emph{formation zone} is implemented that shifts the positions of primary final state particles away from the interaction vertex. For more details on \texttool{NEUT} model choices and implementations, the interested reader is directed to Ref.~\cite{Hayato:2021heg}. 

\begin{figure}[ht]
    \centering
    \includegraphics[width=0.3\textwidth]{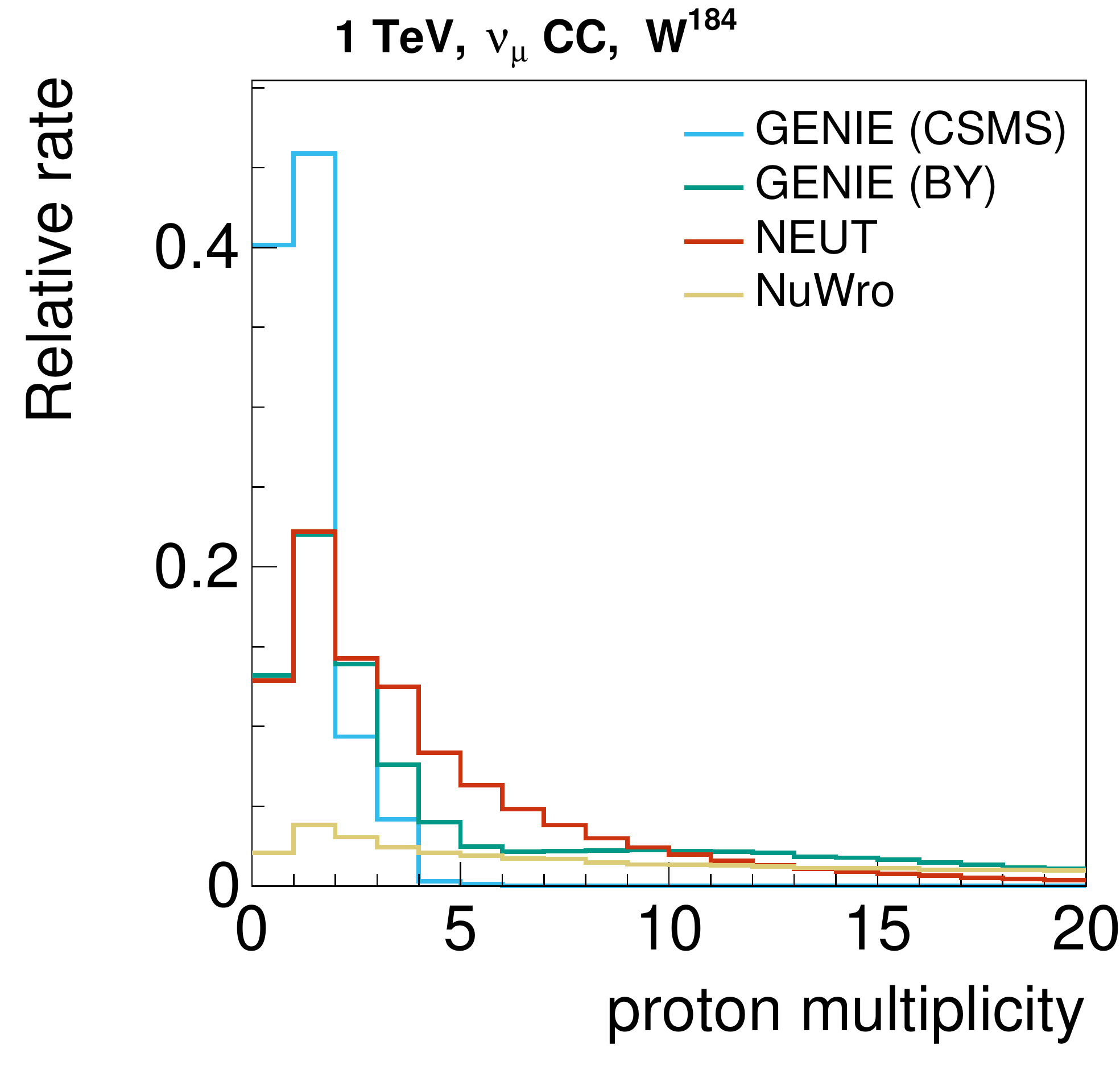}
    \includegraphics[width=0.3\textwidth]{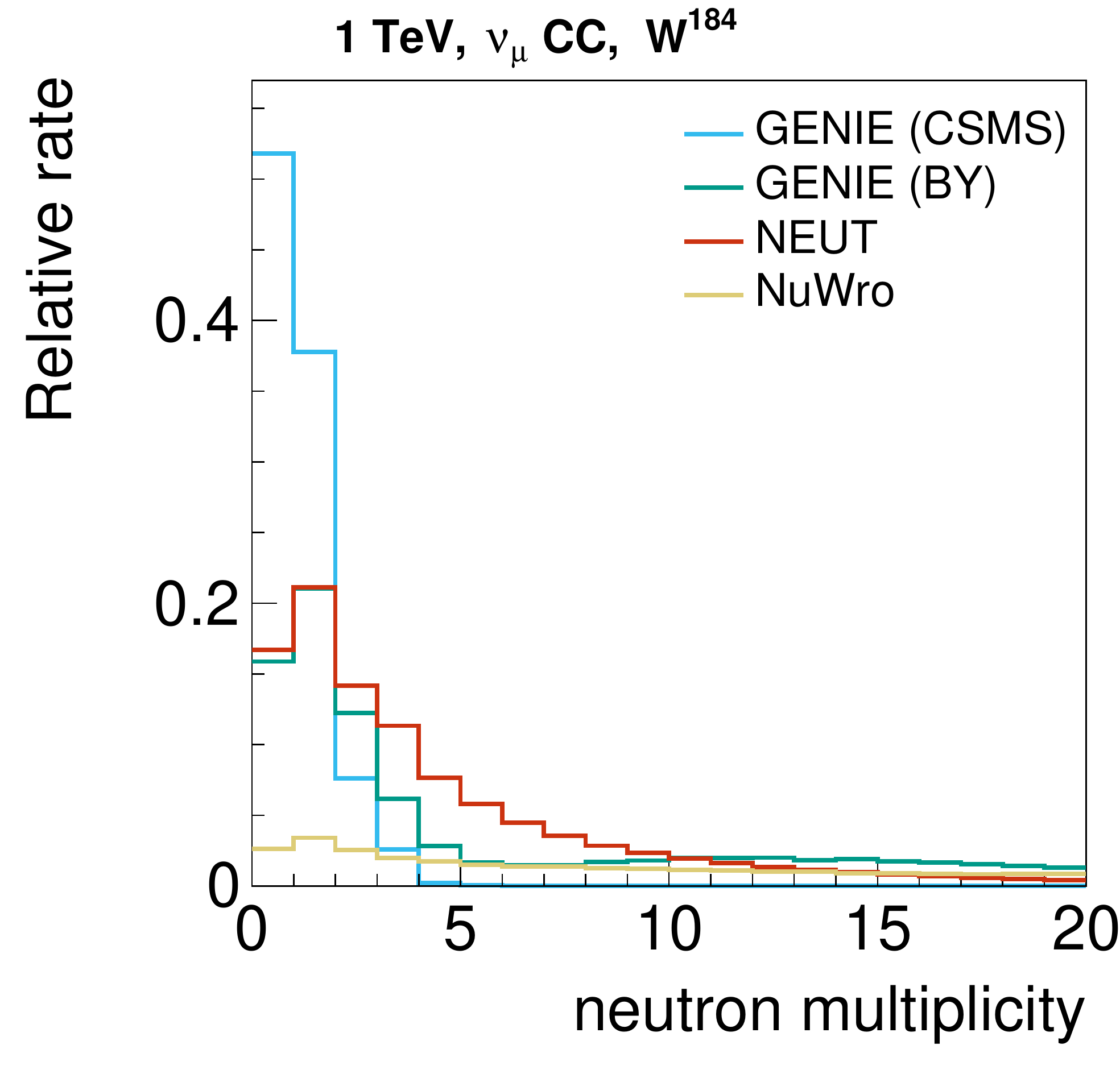}
    \includegraphics[width=0.3\textwidth]{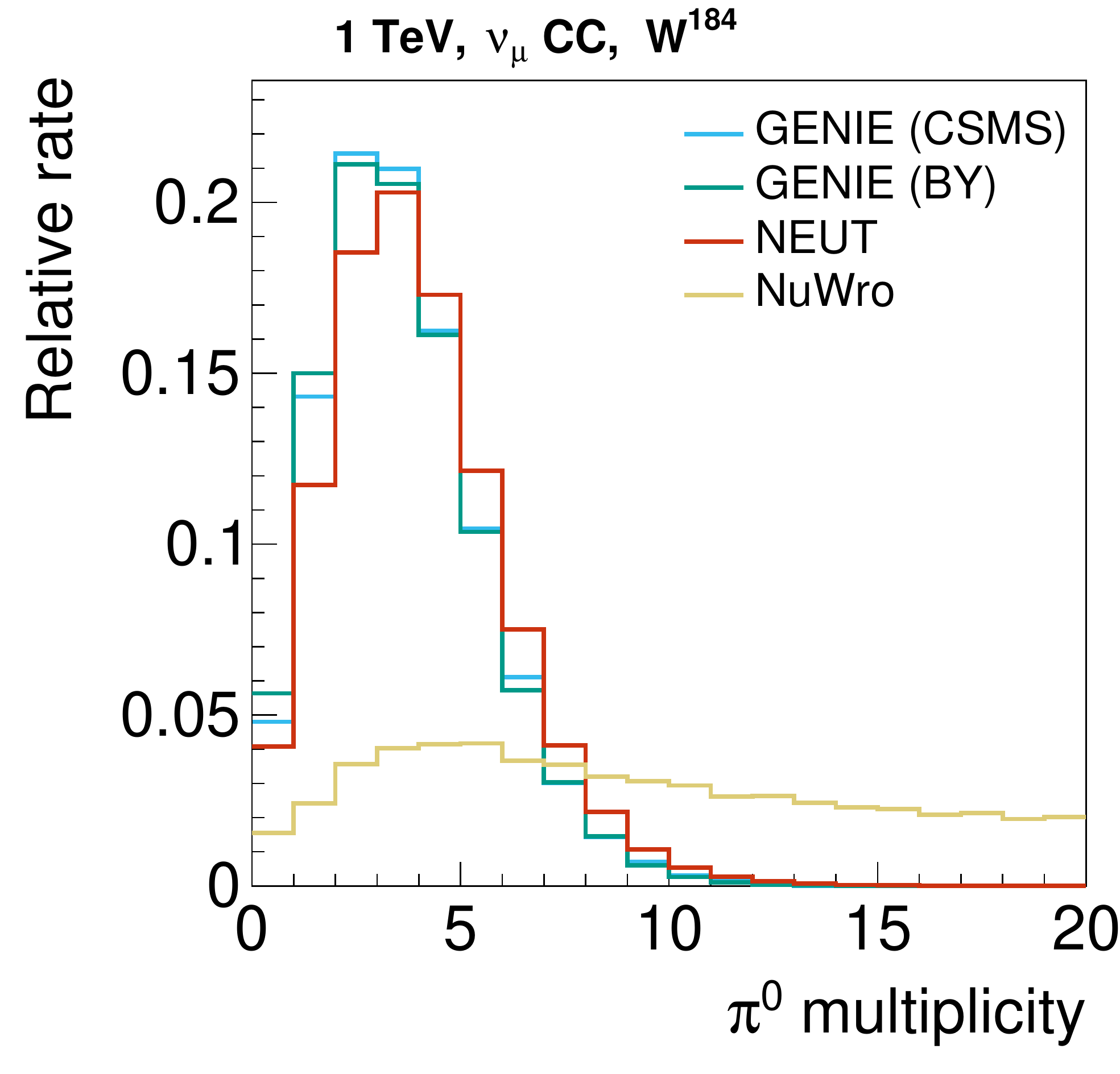}

    \includegraphics[width=0.3\textwidth]{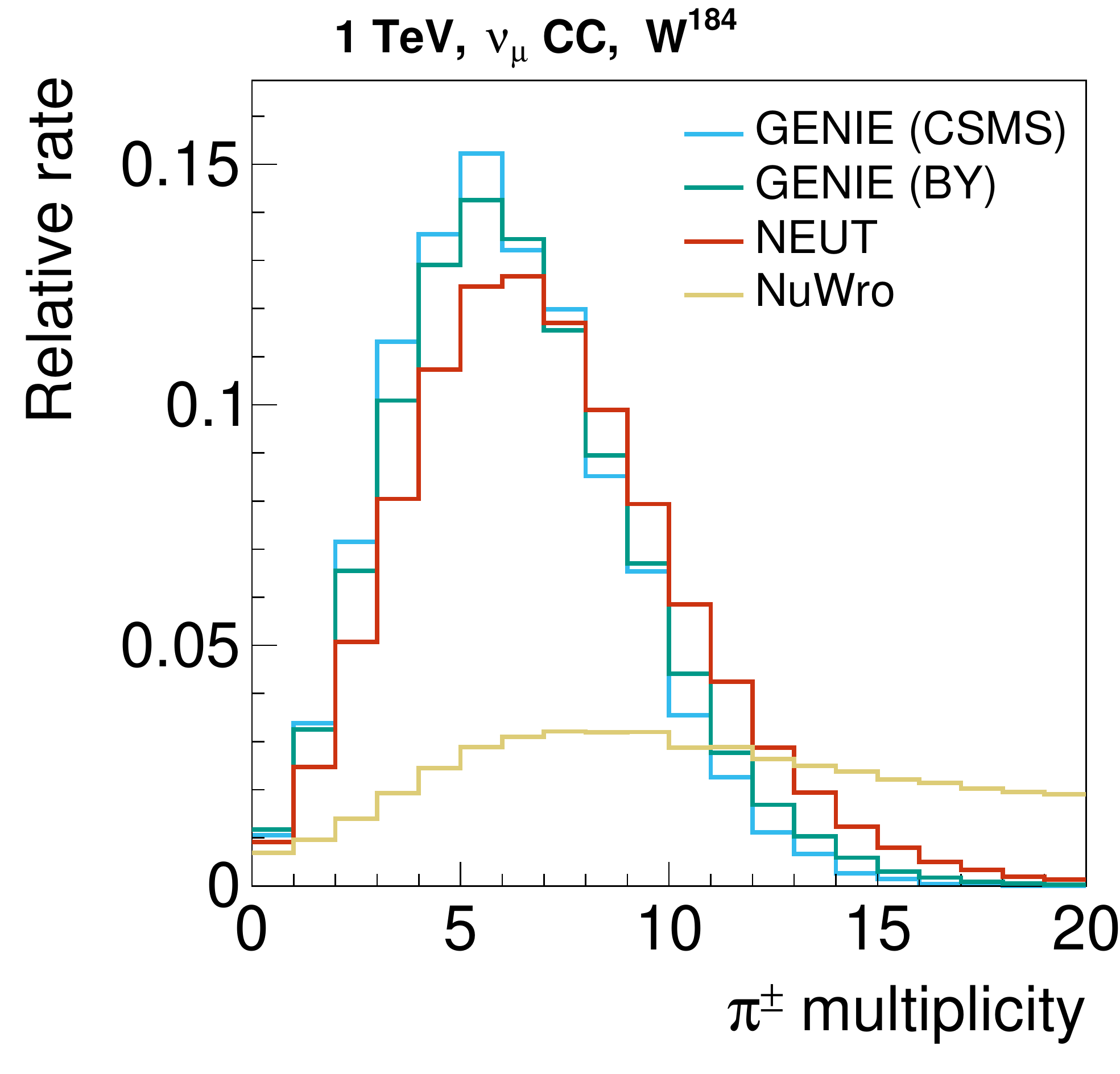}
    \includegraphics[width=0.3\textwidth]{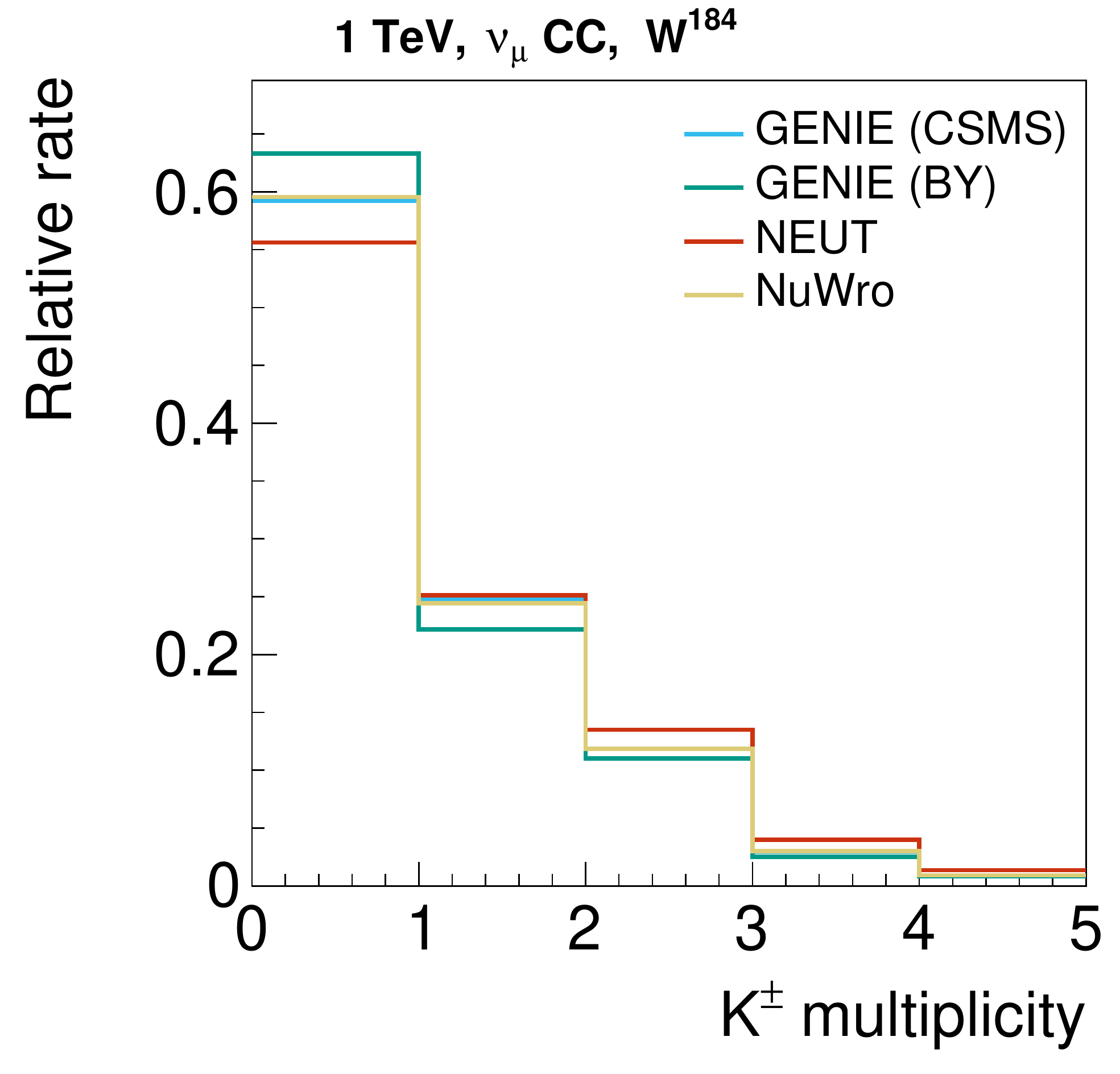}
    \includegraphics[width=0.3\textwidth]{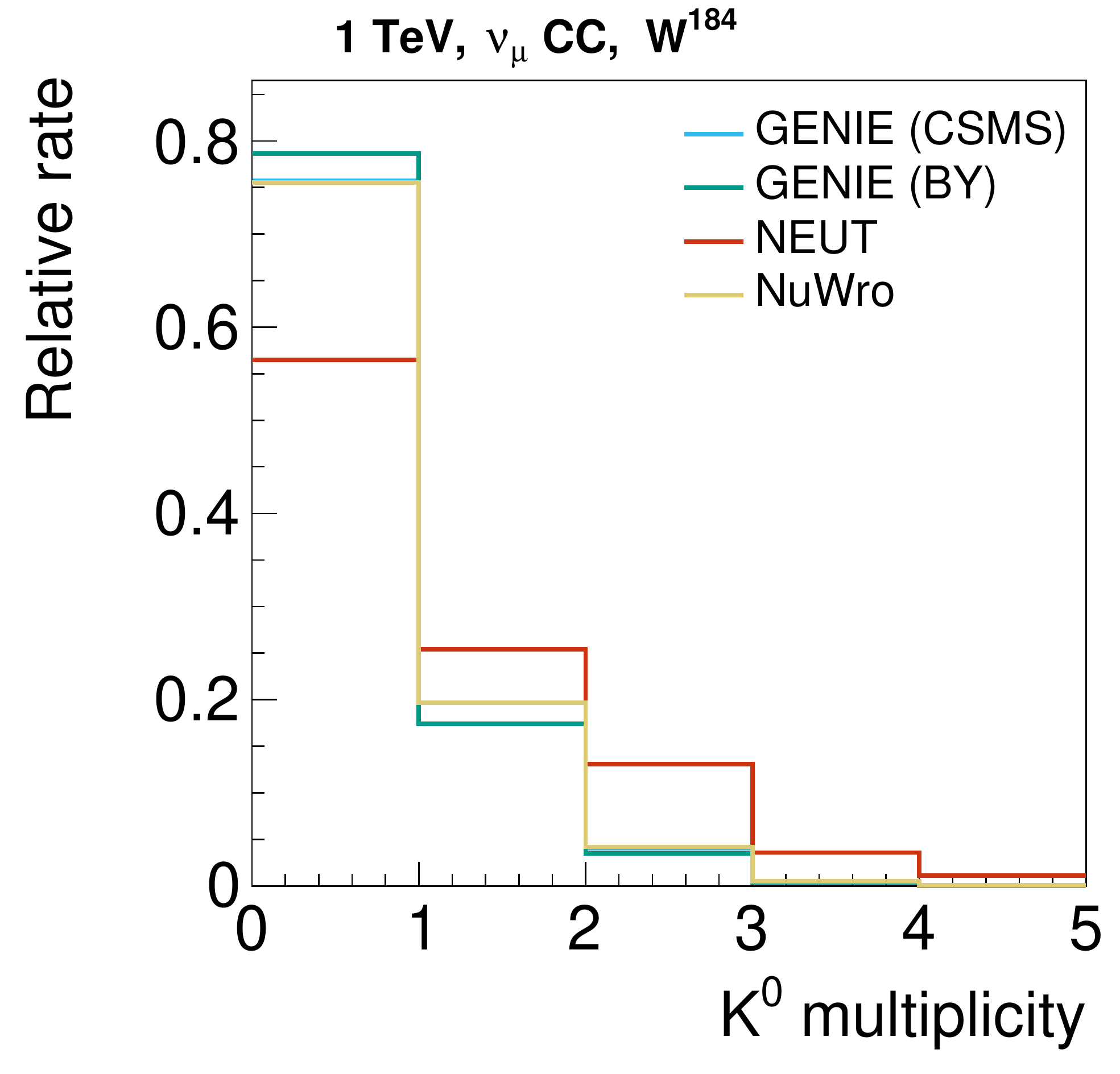}
    
    \includegraphics[width=0.3\textwidth]{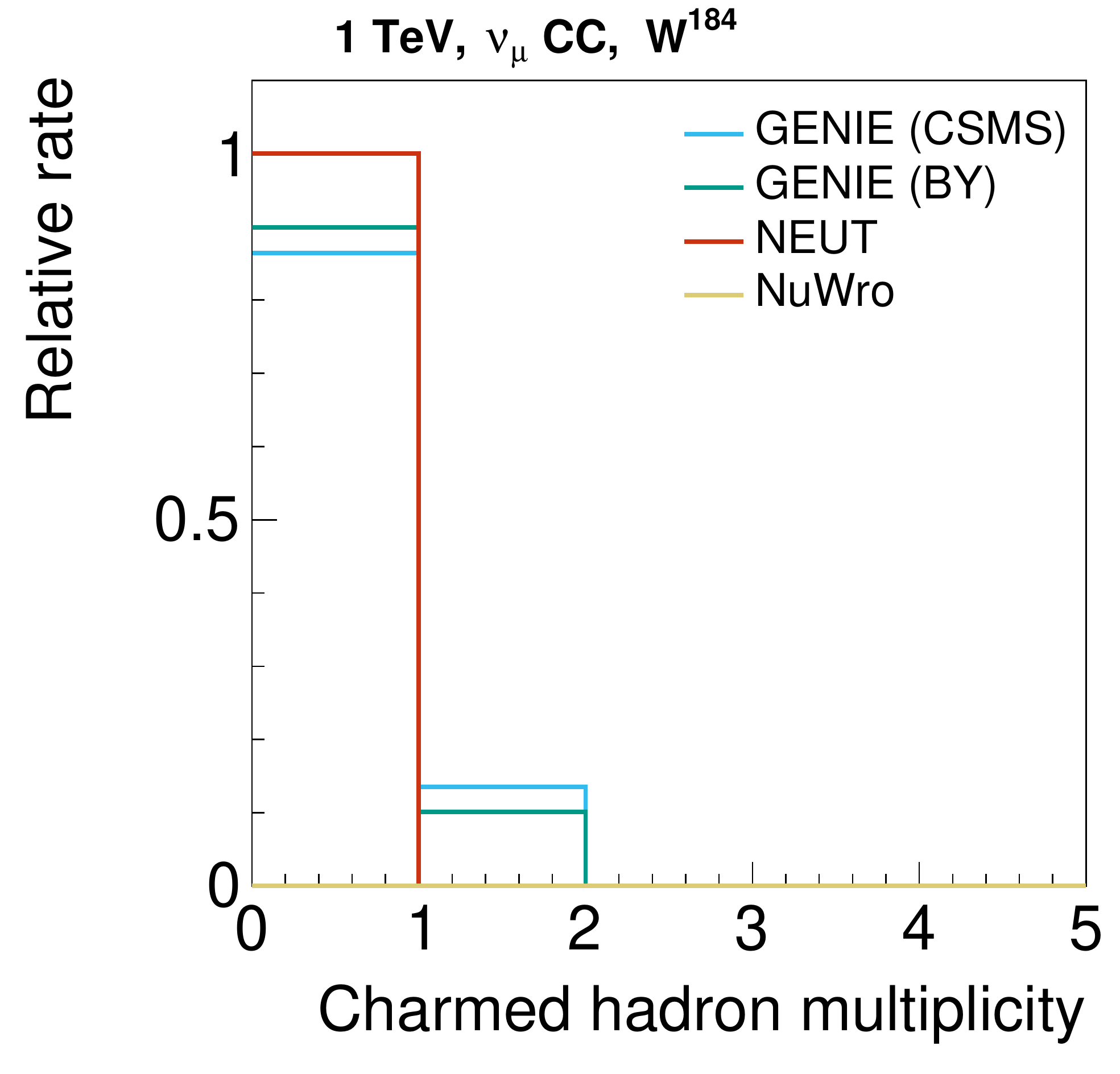}
    \includegraphics[width=0.3\textwidth]{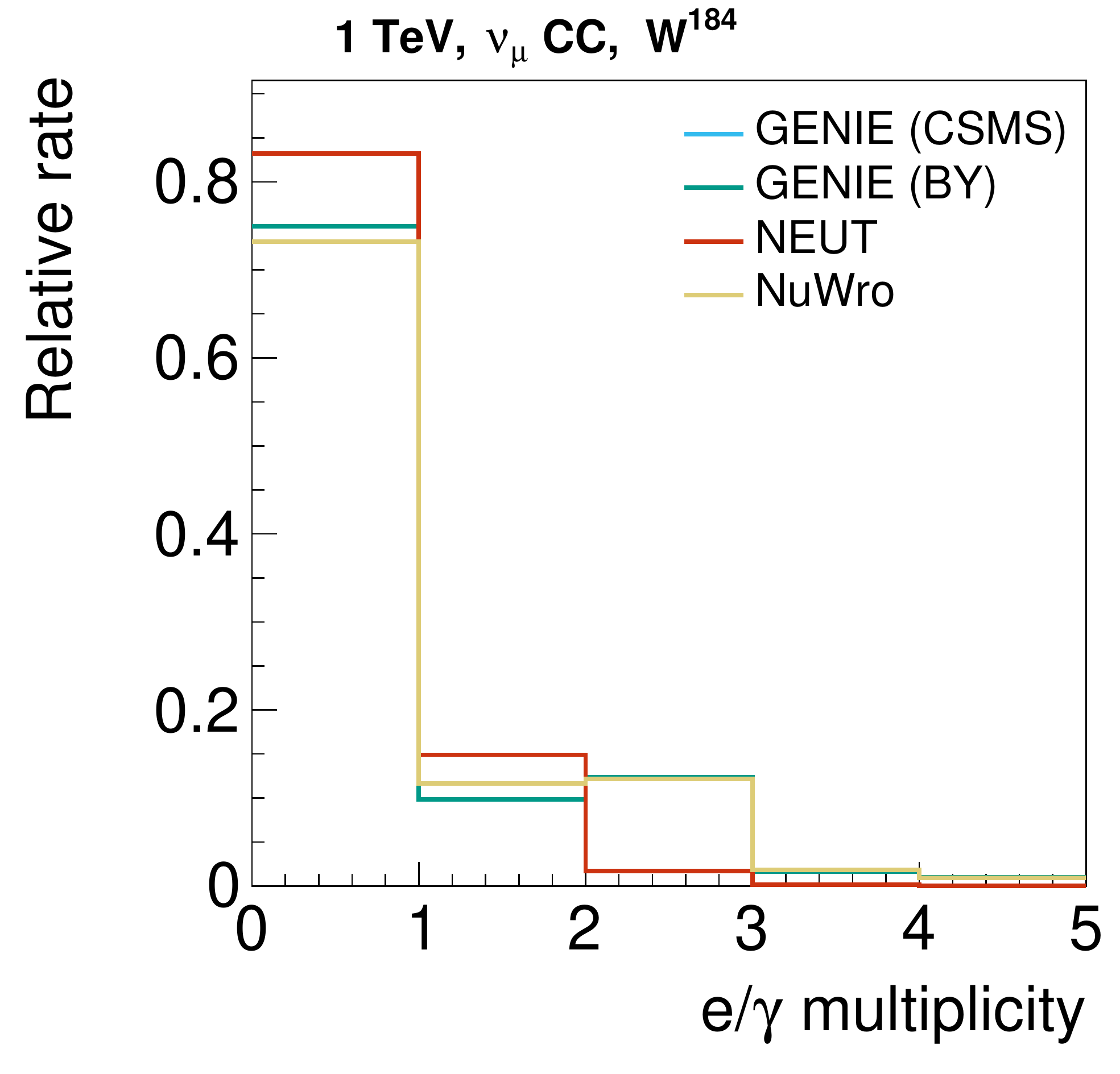}
    \includegraphics[width=0.3\textwidth]{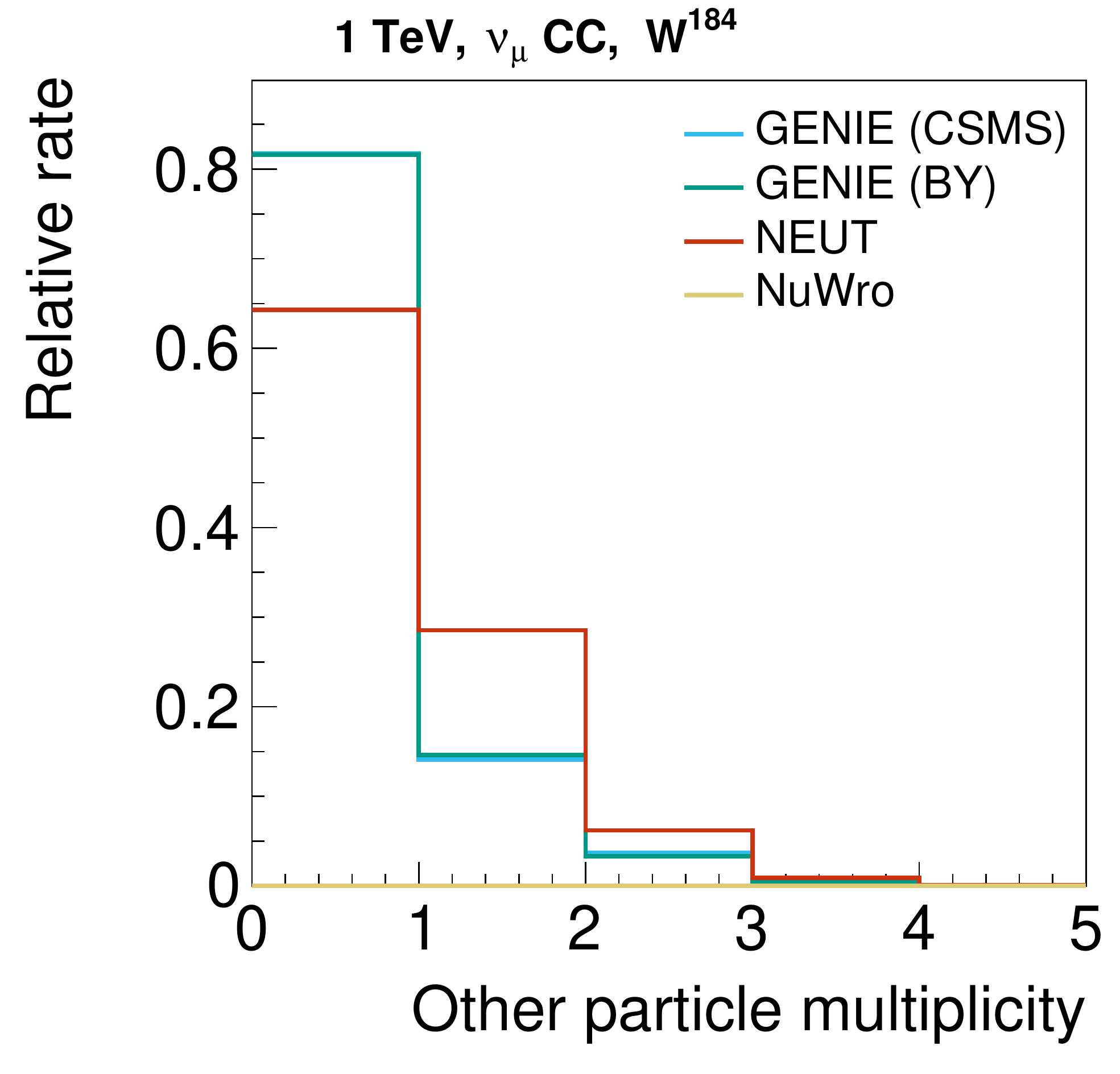}
    \caption{The particle multiplicities predicted by \texttool{NEUT}, \texttool{NuWro} and by two \texttool{GENIE} models for 1 TeV muon neutrino charged-current interactions with a tungsten target.}
    \label{fig:nugencomps::multiplicities}
\end{figure}

\begin{figure}[ht]
    \centering
    \includegraphics[width=0.3\textwidth]{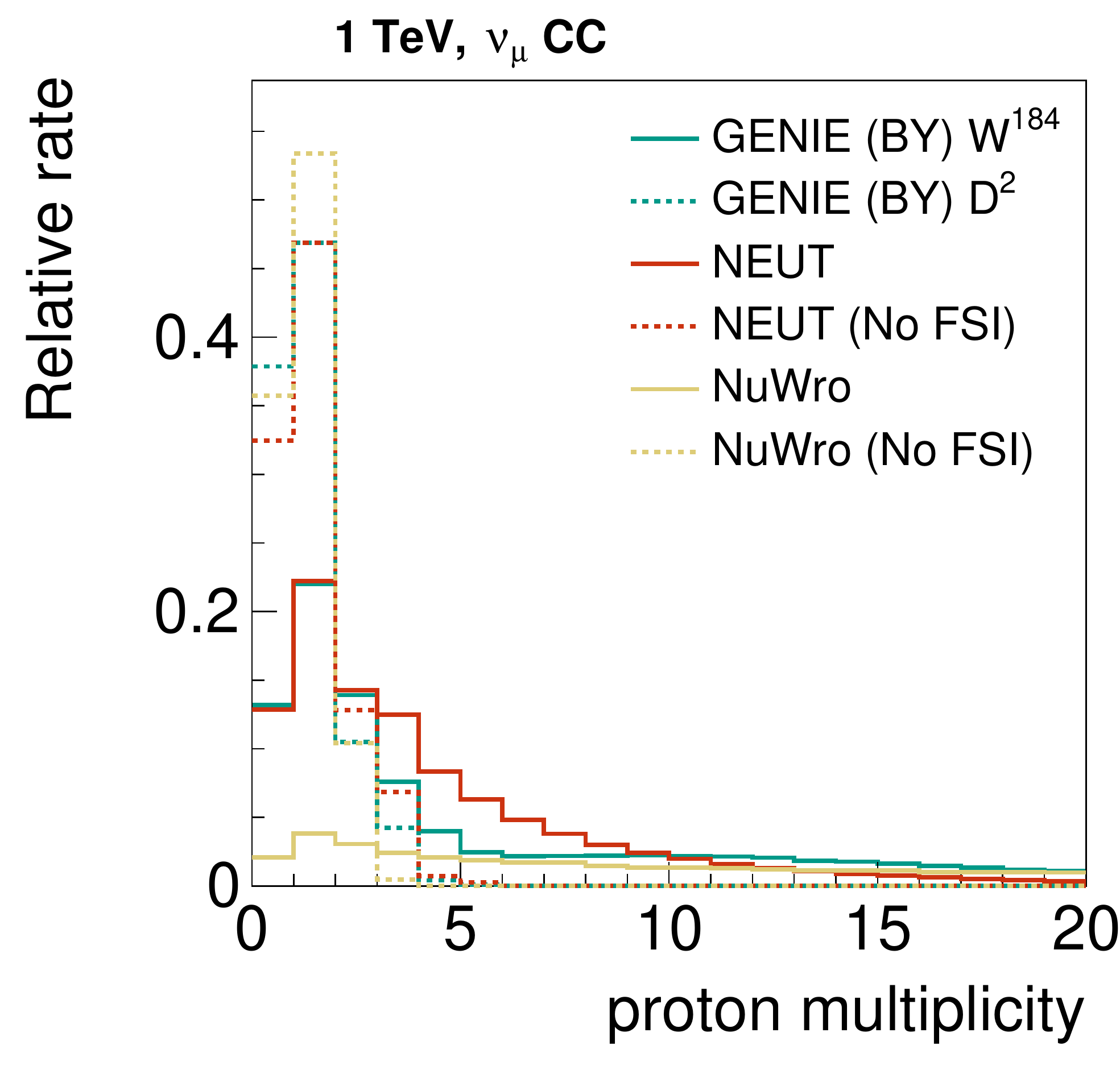}
    \includegraphics[width=0.3\textwidth]{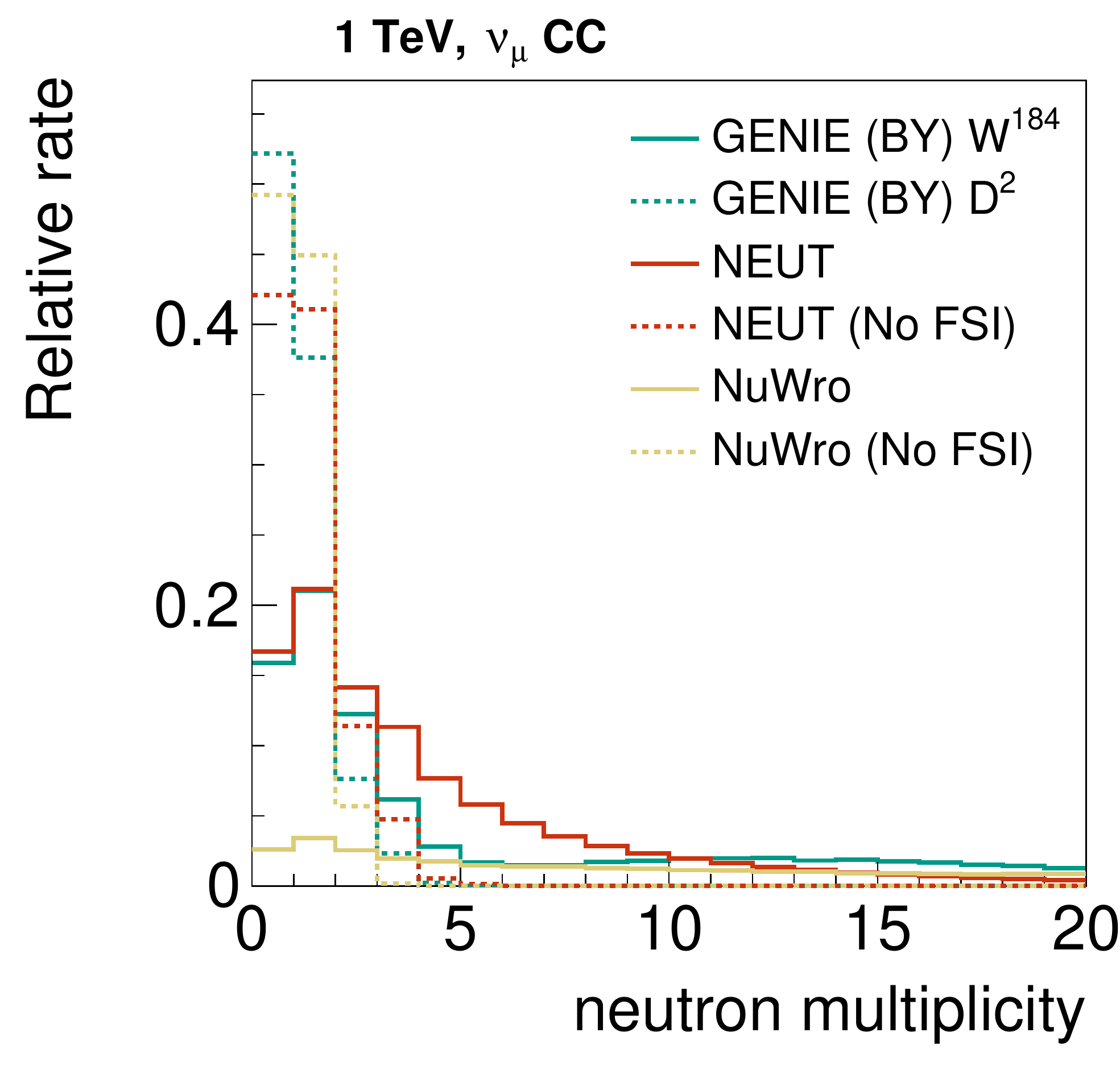} \\
    \includegraphics[width=0.3\textwidth]{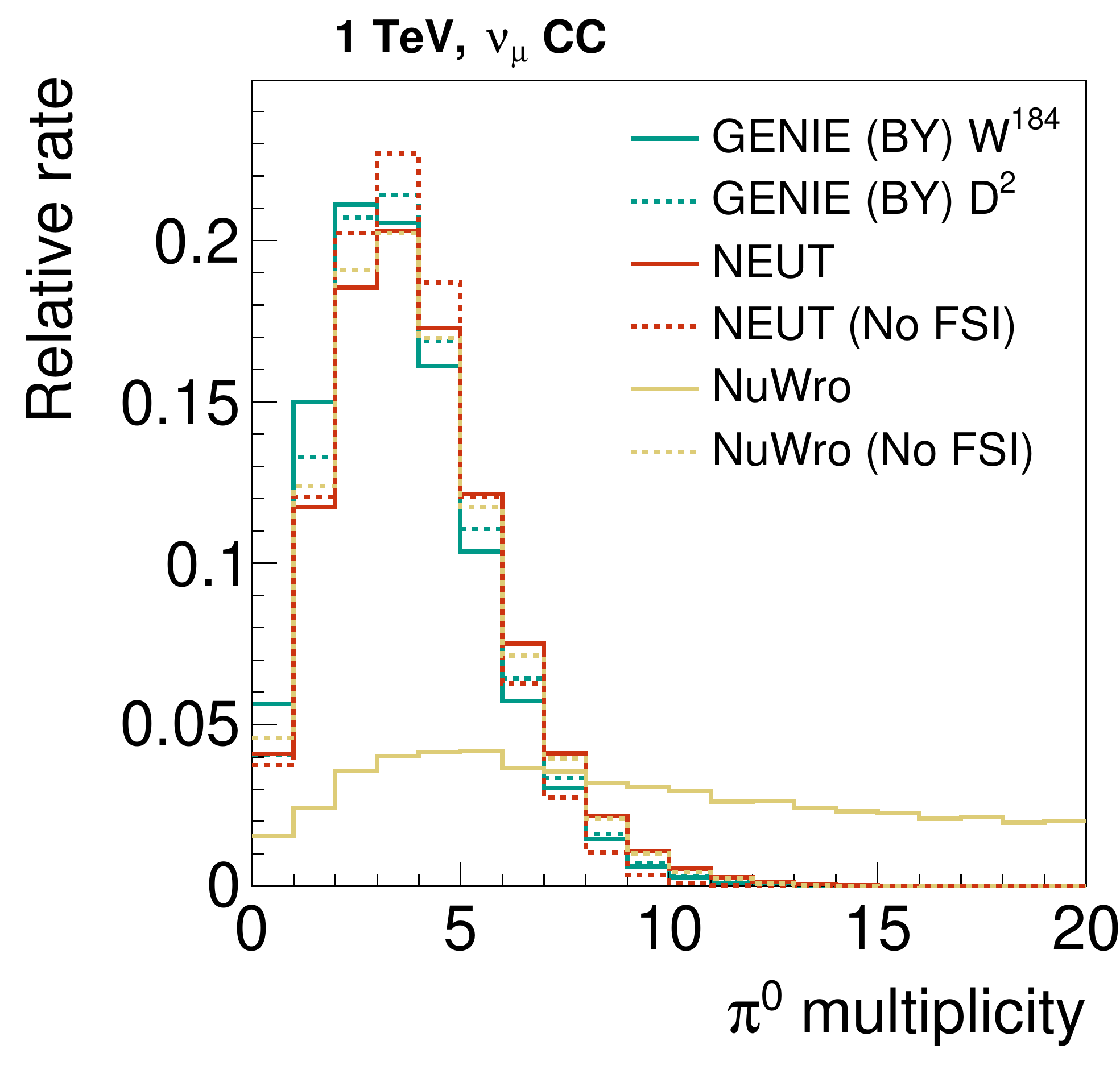}
    \includegraphics[width=0.3\textwidth]{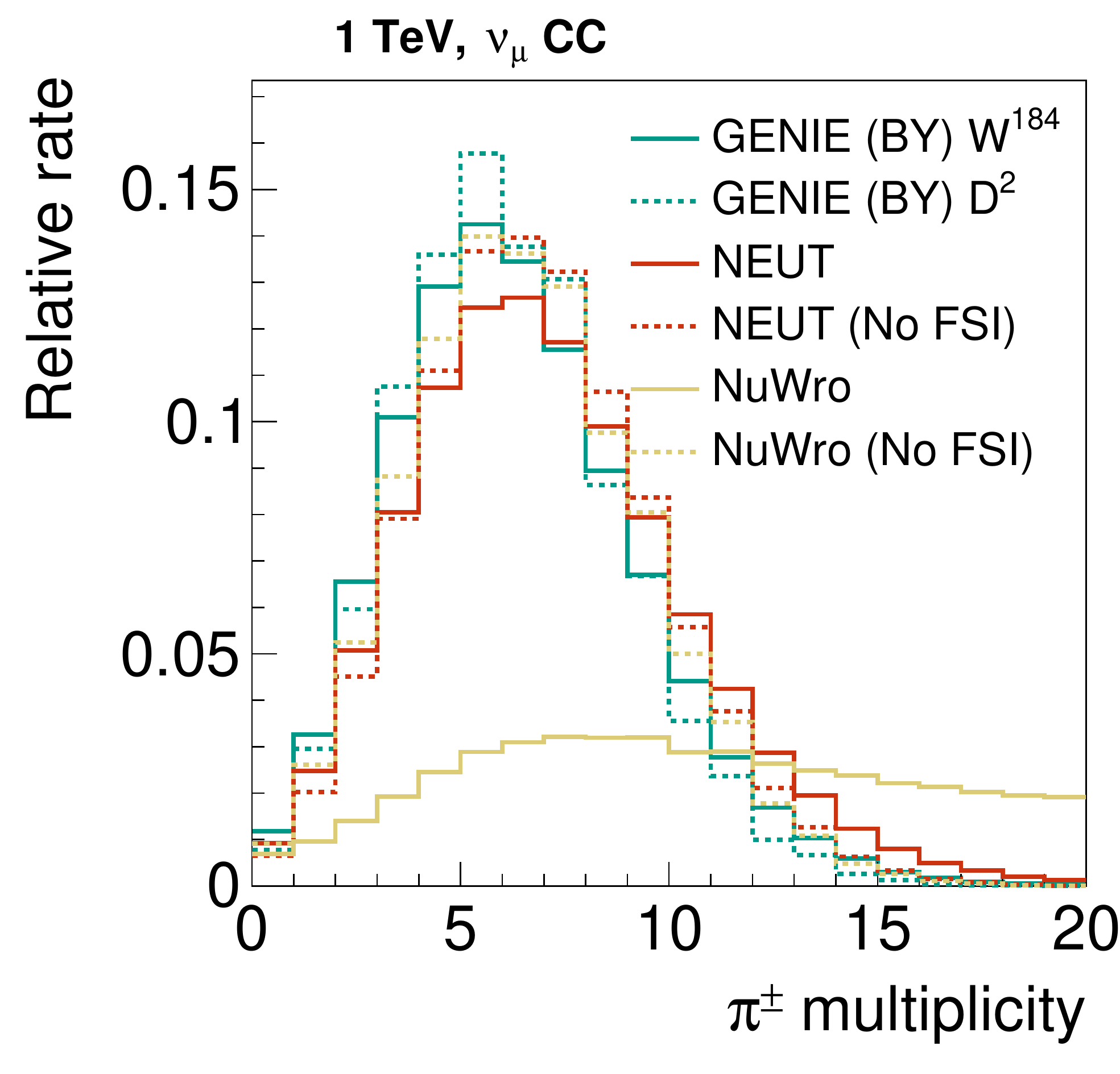}
    \caption{The impact of nuclear effects on the particle multiplicities prediction by \texttool{NEUT}, \texttool{NuWro} and by two \texttool{GENIE} models. Only distributions that were significantly affected are shown. For \texttool{GENIE}, the $D^{2}$ target predictions are used as reference, while for \texttool{NEUT} and \texttool{NuWro}, both predictions use a tungsten target, but the reference was made with the hadron transport model disabled. It is clear that the major nuclear effect in both simulations is the production of many more nucleons. Because the \texttool{GENIE} CSMS model does not implement hadron transport by default, it is not shown in these comparisons.}
    \label{fig:nugencomps::multiplicities_nuceff}
\end{figure}

\begin{figure}[ht]
    \centering
    \includegraphics[width=0.45\textwidth]{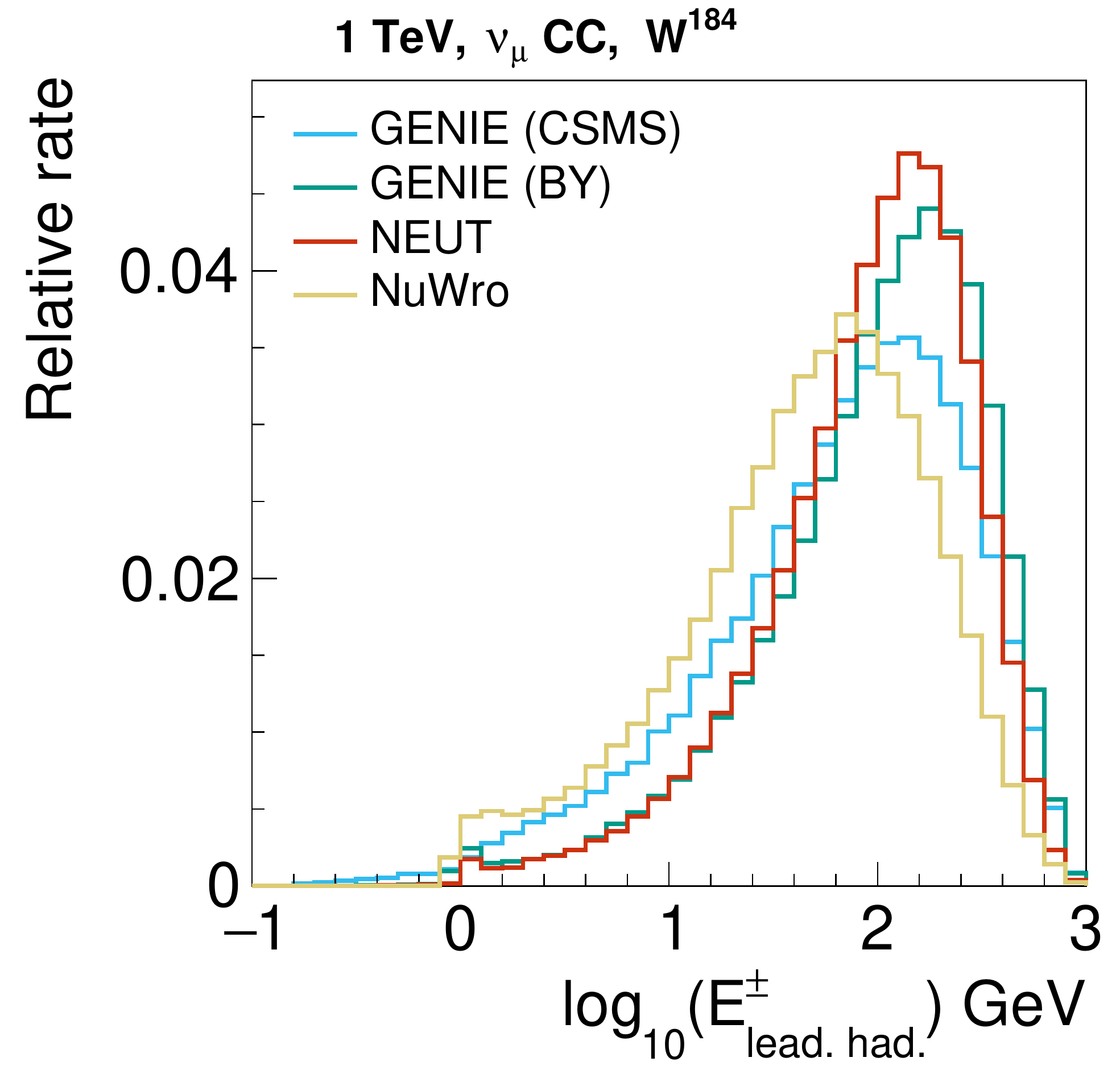}
    \includegraphics[width=0.45\textwidth]{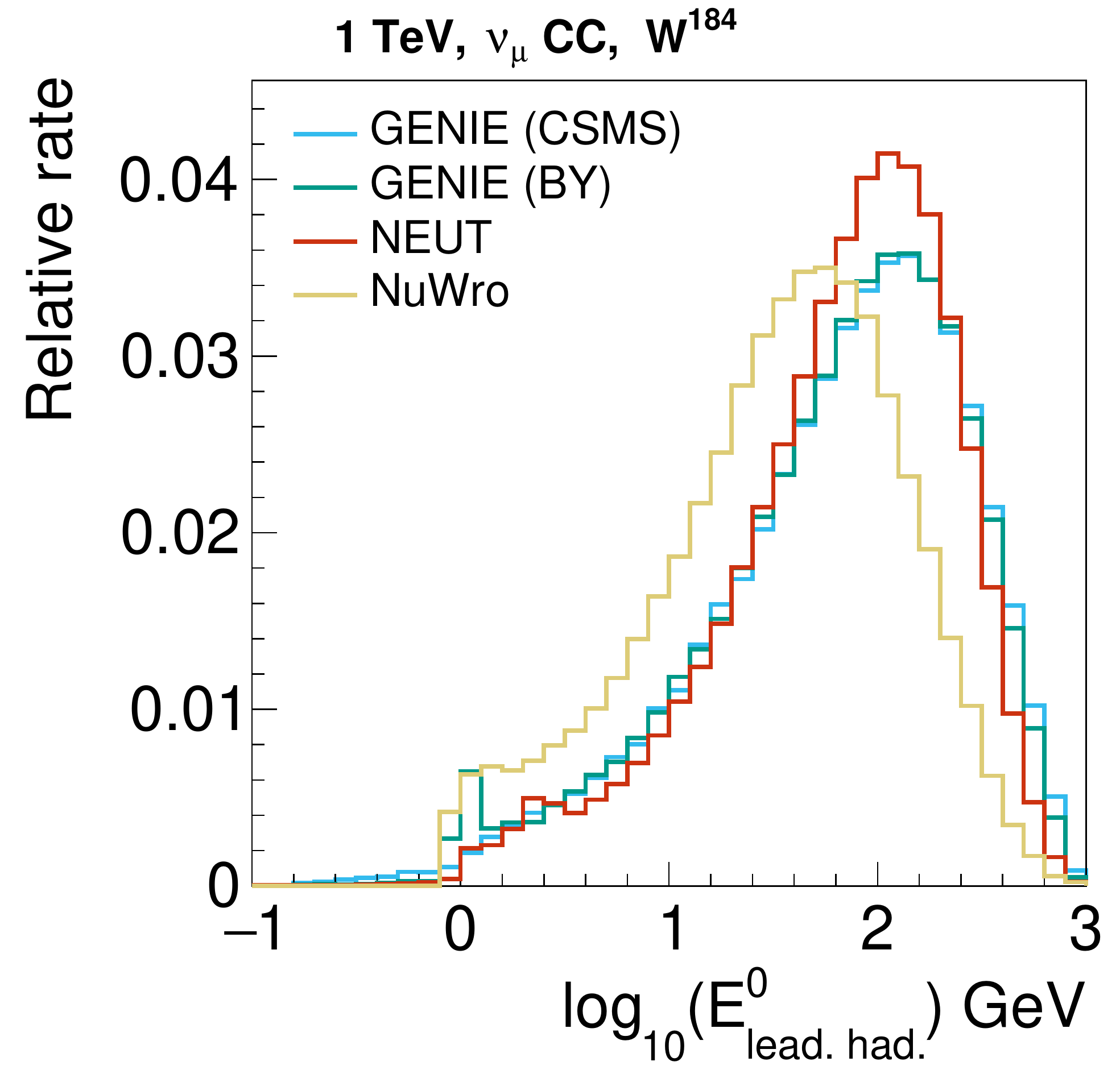}
    \caption{Leading hadron energy distributions for 1 TeV muon neutrino charged-current interactions with a tungsten target as predicted by \texttool{GENIE}, \texttool{NEUT} and \texttool{NuWro}.}
    \label{fig:nugencomps::leadinghadrons}
\end{figure}

\subsection{NuWro}
\texttool{NuWro} is a Monte Carlo generator of neutrino interactions developed at Wroc\l aw University since $\sim 2005$ \cite{Juszczak:2005zs}. It is optimized for neutrino energies of the order of $1$~GeV but it can also handle interactions at larger energies in DIS region.
In \texttool{NuWro} a definition of ``DIS" is that it includes all inelastic processes with invariant hadronic mass $W>1.6$~GeV. A general framework is based on the quark-parton model (Ref. \cite{Feynman:1969ej}). The neutrino-nucleon interactions are described as scatterings on partons (quarks or gluons). The inclusive cross section is expressed in terms of parton distribution functions $F_j$. We adopt Bodek-Yang \cite{Bodek:2002vp} parametrization of $F_j$ that correct GRV PDFs  \cite{Gluck:1994uf} based on electron scattering data.
The Bodek-Yang corrections are needed as for low values of $Q^2$ and $W$ as the perturbative QCD arguments are not valid,  and the mass of the target and higher twists are non-negligible. 

It is assumed that the cross-section is a sum of contributions from separate quarks. The cross-section for scattering on quark $q_i$ (valence or sea)
\[
\frac{{d^2 \sigma ^{q_i} }} {{dW d\nu}} \sim q_i(x) K_i
\]
where $q_i(x)$ is a parton distribution function and $K_i$ is a kinematic factor. The probability of reaction on a quark $q_i$ is defined as
\begin{equation}
P(q_i) = \frac{\displaystyle \frac{d\sigma^{q_i}}{dWd\nu}}{\displaystyle \sum \limits_i{\frac{d\sigma^{q_i}}{dWd\nu}} }
\end{equation}

As a result, the CC neutrino-proton scattering cross section becomes a sum of contributions from quark $d$, quark $s$, and anti-quark $\bar{u}$. A scheme of the algorithm for CC neutrino scattering off protons is shown in Fig. C5 in Ref.~\cite{jarekphd}.

In modelling DIS events \texttool{NuWro} uses \texttool{Pythia~6} fragmentation routines for $W$ as small as $1.6$~GeV. 
Some \texttool{Pythia~6} 
parameters are adjusted by demanding a good agreement with the charged hadron multiplicities data \cite{Nowak:2006xv, Sobczyk:2008zz}:

\begin{itemize}
 \item PARJ(32)(D=1GeV) = 0.3 is, with quark masses added, used to define the minimum allowable energy of a colour singlet parton system.
 \item PARJ(33)-PARJ(34)(D=0.8 GeV, 1.5 GeV) = 0.5 Gev, 1 GeV are, with quark masses added, used to define the remaining energy below which the fragmentation of a parton system is stopped and two final hadrons formed.
 \item PARJ(36)(D=2.0GeV) = 0.3 represents the dependence of the mass of the final quark pair for defining the stopping point of the fragmentation. Strongly correlated with PARJ(33-35)
 \item MSTJ(17) (D=2) = 3 number of attempts made to find two hadrons that have a combined mass below the cluster mass and thus allow
  a cluster to decay rather than collapse.

\end{itemize}
\subsection{Generator Comparisons}

Uncertainties associated with neutrino interactions need to be well controlled for precision neutrino and BSM measurements. Therefore, ideally, FPF would require multiple event generators and different implementations of the underlying physics assumptions to systematically analyze their impact on FPF measurements, quantify systematic uncertainties and when possible compare the predictions to data. 

In \cref{fig:nugencomps::multiplicities}, we present multiplicities of various particles produced when a 1 TeV CC $\nu_\mu$ scatters off a tungsten target. The predictions of \texttool{GENIE} (CSMS and BY models), \texttool{NEUT} and \texttool{NuWro} are compared for all multiplicity distributions. Most notably differences between \texttool{GENIE} and \texttool{NEUT} are seen in the proton and neutron multiplicities where \texttool{GENIE} CSMS model predicts much lower multiplicity compare to \texttool{GENIE} BY and \texttool{NEUT}. This difference can be associated to the absence of the FSI effects in \texttool{GENIE} CSMS model. In the presence of FSI, the primary struck nucleons re-scatter in the nucleus thereby increasing the final state multiplicity. This effect is further demonstrated in \cref{fig:nugencomps::multiplicities_nuceff}, where \texttool{GENIE} BY and \texttool{NEUT} predict similar proton and neutron multiplicity on tungsten while in the absence of FSI, \texttool{NEUT} without FSI and \texttool{GENIE} BY on deuterium, predict similar lower multiplicity. These nuclear effects do not seem to have any significant impact on the pion (and other particle’s) multiplicity. \texttool{NuWro} seem to have the strongest effects of FSI and tend to have significantly different distributions. The leading hadron energies for a 1 TeV CC $\nu_\mu$ scattering on tungsten target is shown in \cref{fig:nugencomps::leadinghadrons}. \texttool{GENIE} models and \texttool{NEUT} tend to predict similar peak energy for both leading charged and neutral hadrons. In case of leading charged hadrons, the \texttool{GENIE} CSMS model predicts more broadly distributed energy spectrum compared to other predictions, not much difference seen in the neutral hadron case. Here again, \texttool{NuWro} distributions seems to differ from the \texttool{GENIE} and \texttool{NEUT} ones and are peaked at lower energies. Understanding differences between generators, in particular the significantly different behaviours shown by \texttool{NuWro}, are left for future studies.

The differences seen between the different MC predictions will be vital in assessing the uncertainties associated with neutrino interactions and will need to be well controlled for precision neutrino and measurements to probe BSM physics. Therefore, comparison between different event generators and different implementations of the underlying physics assumptions discussed in this subsection will be vital for systematically analyzing their impact on FPF measurements and in quantifying systematic uncertainties. To this end, the data from FASER$\nu$ and SND@LHC during LHC Run 3 would be valuable to help constrain the interaction physics in generators.

\subsection{Giessen Model and GiBUU Generator}
\label{subsubsec:GiBUU}

\begin{figure}[hb]
	\label{fig:fpf-w}
	\centering
	\includegraphics[width=0.49\linewidth]{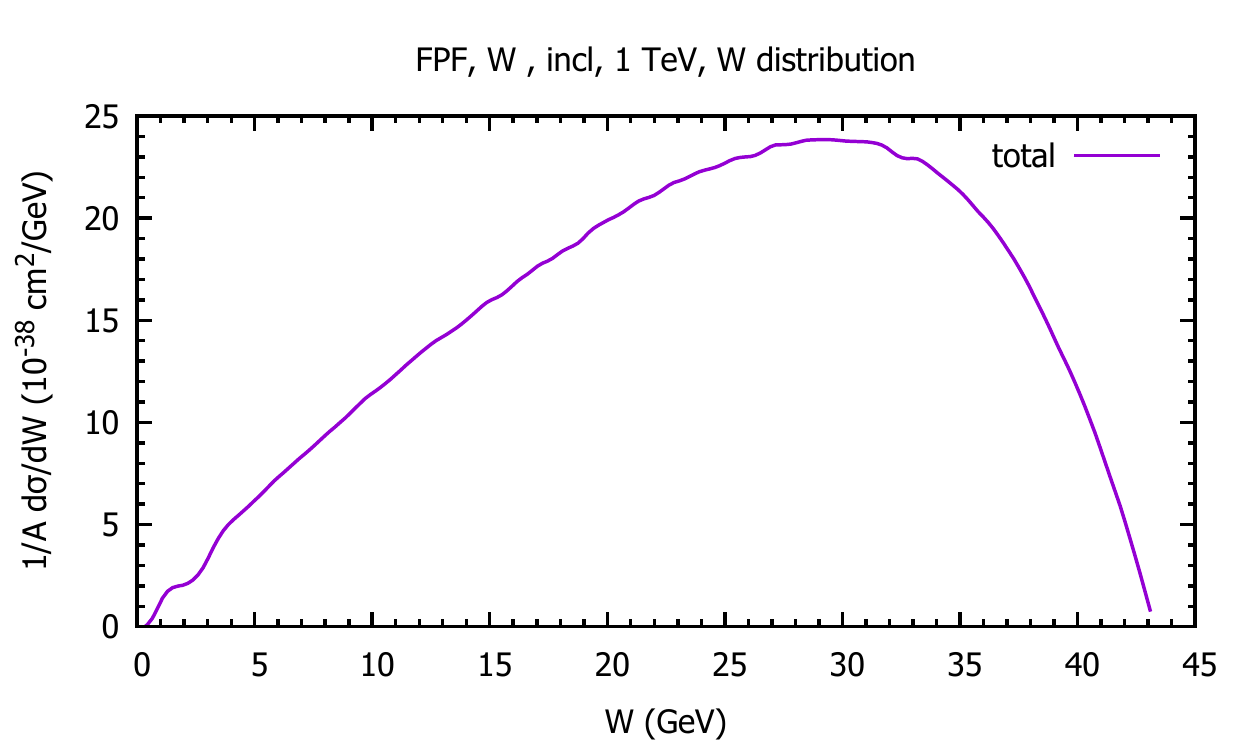}
			\includegraphics[width=0.49\linewidth]{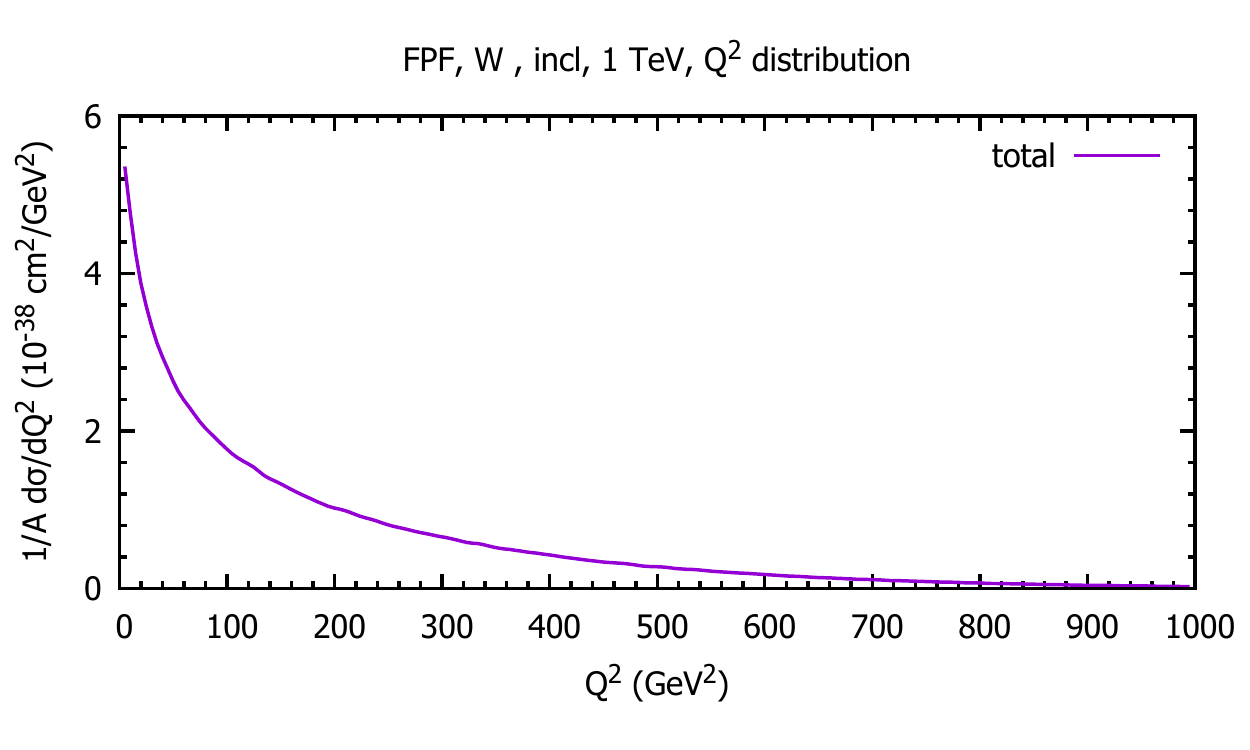}
	\caption{Distribution from \texttool{GiBUU} of invariant mass (left) and $Q^2$ distribution (right) populated in a 1 TeV neutrino reaction on a Tungsten target.}
\end{figure}

	\begin{figure}[hbt]
		\centering
		\includegraphics[width=0.49\linewidth]{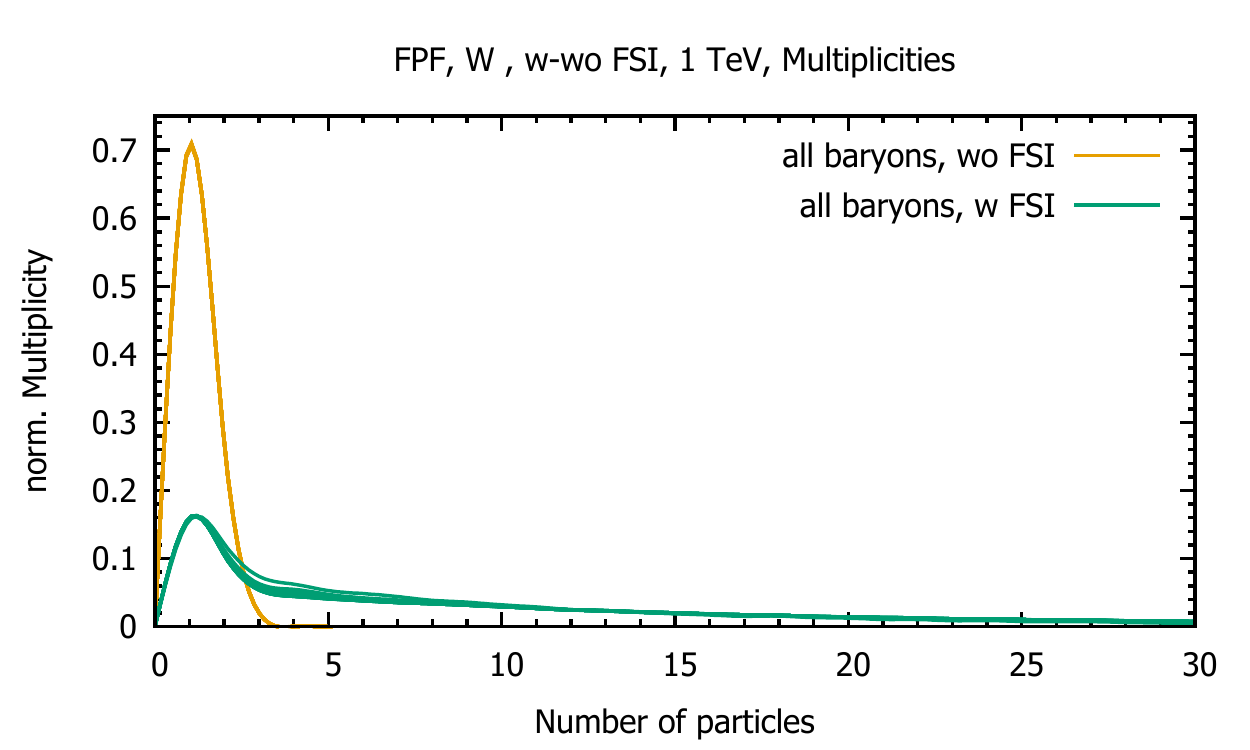}
		\includegraphics[width=0.49\linewidth]{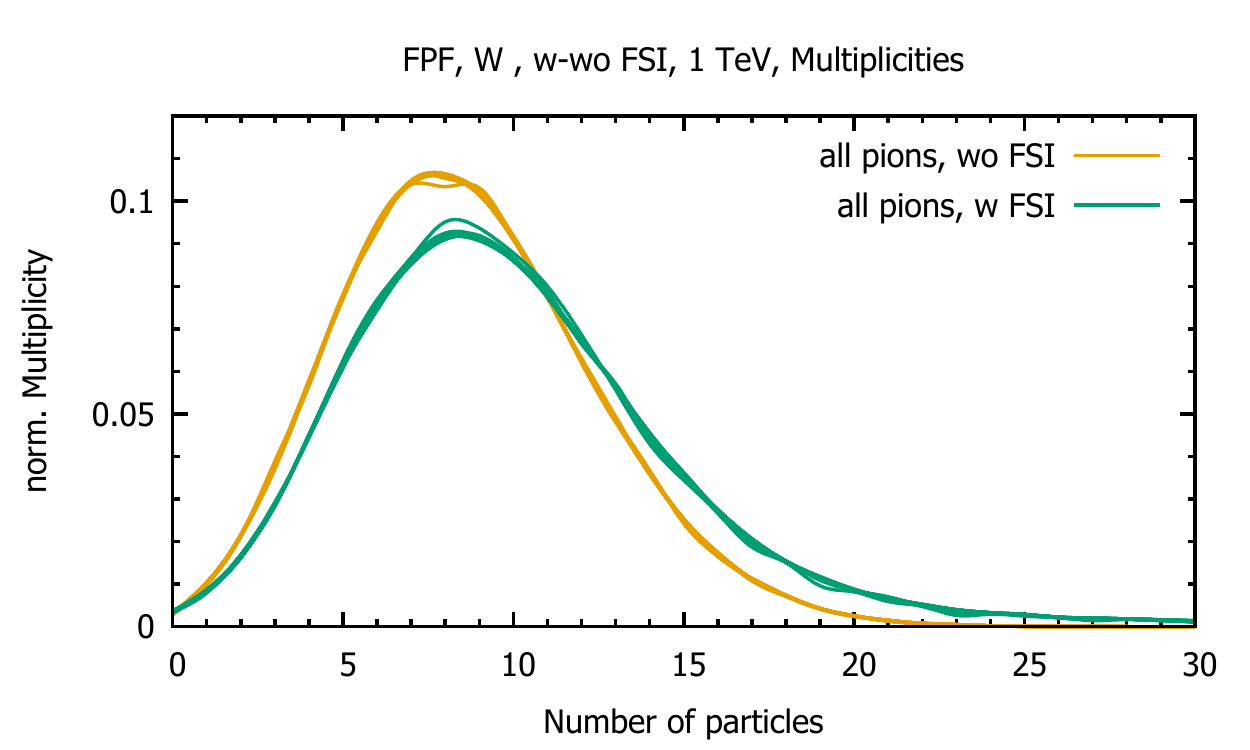}
		\caption{Multiplicity distributions from \texttool{GiBUU} of baryons (left) and pions (right) without and with FSI.}
		\label{fig:fpf-multiplfsi-compallbaryons}
	\end{figure}

The Giessen model is an extensive general theory framework to describe nuclear collisions, from relativistic heavy-ion collisions to neutrino-induced reactions on nuclear targets.  Its treatment of final state interactions (FSI) is built on quantum-kinetic transport theory \cite{Kad-Baym:1962} and as such can be used as a generator, called \texttool{GiBUU}, for the full final state of a reaction. The underlying theory is described in detail in \cite{Buss:2011mx}.
The code is being updated from year to year and its source-code is freely available from \cite{gibuu}.
We present here
some results obtained with \texttool{GiBUU} for neutrinos with an incoming energy of 1 TeV impinging on a Tungsten target are discussed. At this energy deep inelastic scattering (DIS), handled by \texttool{Pythia} inside \texttool{GiBUU}, is dominant.

In order to get a first impression of the kinematical regimes accessible by this reaction we show in the left panel of \cref{fig:fpf-w} the inclusive invariant mass distribution. The distribution peaks at about $W = 30$ GeV, i.e.\ well in the DIS regime.
This is also reflected in the four-momentum transfer distribution given in the right panel of \cref{fig:fpf-w} which reaches up to very high values.
The energy-transfer ($\omega$) distribution, on the other hand, is fairly flat with a value (per nucleon) decreasing from about $0.8 \cdot 10^{-38}$ at low $\omega$ to about $0.6 \cdot 10^{-38}$ cm$^2$/GeV at 1 TeV.

We now focus on the multiplicities of final-state particles. The left panel of \cref{fig:fpf-multiplfsi-compallbaryons} shows the multiplicity distribution of final-state baryons. The top (yellow) curve gives the result before any final state interactions. This distribution peaks at multiplicity = 1 with a tail up to about 3 - 4. This tail is caused by the production of baryon-antibaryon pairs. When FSI are turned on the multiplicity distribution changes significantly: the peak height is decreased by about a factor of 5 and a long tail reaching up to about 25 develops. This is a consequence of the so-called 'avalanche effect' in which initial nucleons collide with others on the way out of the target.	
The pion multiplicities, on the other hand show a much less effect of FSI (see right panel of \cref{fig:fpf-multiplfsi-compallbaryons}). The initial pion multiplicity distribution hardly changes.

\begin{figure}[ht]
		\centering
		\includegraphics[width=0.55\linewidth]{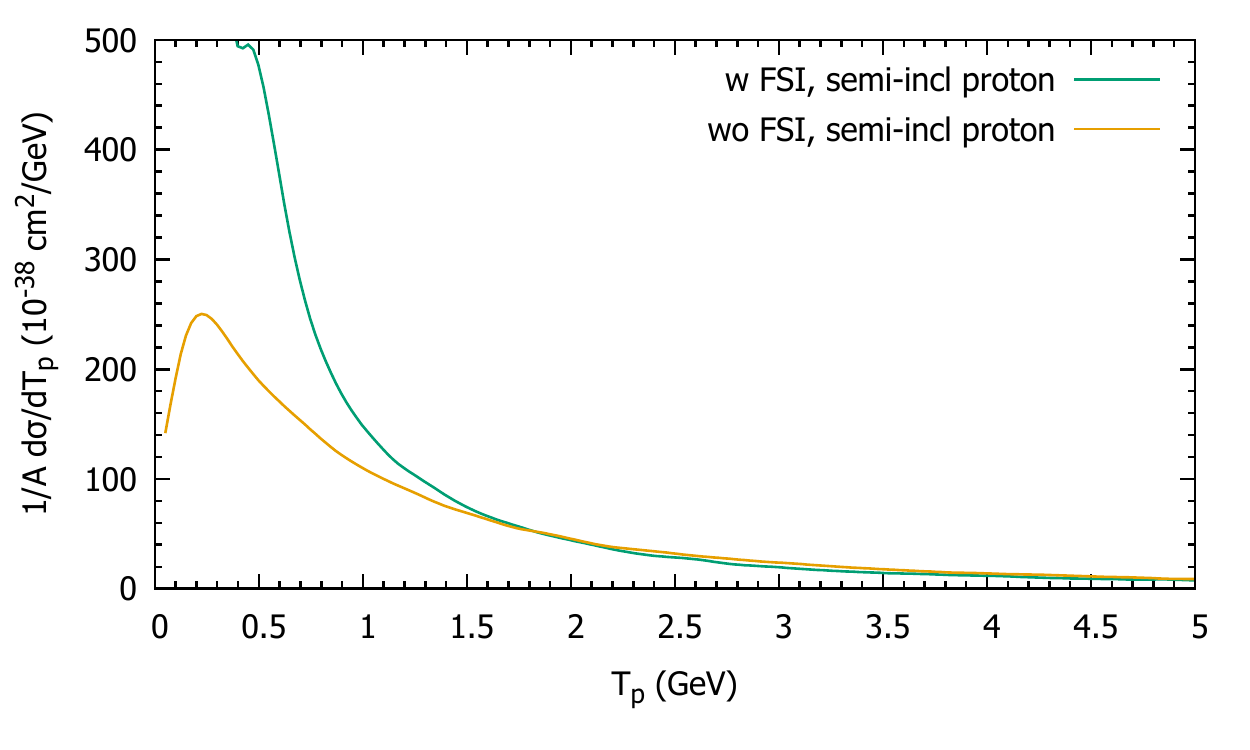}
		\caption{Distributions from GiBUU of the proton kinetic energy with and without FSI effects.}
		\label{fig:fpf-spectraexcl-incl}
	\end{figure}

The baryon avalanche effect also shows up in the kinetic energy distribution (\cref{fig:fpf-spectraexcl-incl}). While the events before FSI show a peak at about 0.3 GeV, the final state interactions change that distribution significantly. Cross sections are decreased at high $T_p > 2$ GeV and are dramatically increased at lower $T_p < 1.5$ GeV. This is again a consequence of the avalanche effect: initial high energy baryons collide with target nucleons, thus  increasing the multiplicity and consequently losing energy.
The effect of FSI on the spectra of baryons is obviously quite significant. This opens the possibility to study phenomena such as Color Transparency (CT) and hadron formation in medium \cite{Gallmeister:2007an}.

At 1 TeV bombarding energy, the DIS process is dominant. Any science program exploiting these neutrinos at the FPF should thus concentrate on this reaction mechanism. Such experiments could yield very valuable information on formation times in DIS events, color transparency (CT) effects and the tension in the EMC effect for electrons and neutrinos.
Regarding the formation time,
	earlier analyses based on HERMES and EMC electron data had shown that only hadronic FSI cross sections that increase linearly with time can describe both data sets simultaneously \cite{Gallmeister:2007an}. Data at the FPF could help to validate this result in a new kinematical regime.
	The very recent, unexpected result from JLAB that CT for baryons does not set in up to $Q^2 = 14$ GeV$^2$ presents a challenge to standard CT theory \cite{HallC:2020ijh}. It has been argued that only final state baryons that originate in a DIS event should be subject to CT \cite{Mosel:2021CT}. Experiments at the FPF where all events are dominated by DIS are ideal to investigate CT 
	further. 
It has been known for some time that the EMC effect seems to be different for electrons and neutrinos; the latter do not seem to show the strong antishadowing effect seen with electrons \cite{SajjadAthar:2020nvy}. This result presents a challenge to pQCD and thus calls  for further verification in a new kinematical regime.

\section{Beyond Standard Model Physics with Neutrinos}
\label{sec:BSMNeutrinos}

In addition to being able to study fundamental properties of neutrinos in the SM at the various FPF detectors, there is also excellent capability to search for BSM properties as well. In this section, we lay out a number of searches that can be performed, highlighting the unique capability of the FPF in this regard. The very high energies of the neutrinos involved at the LHC, in general, offer new capabilities that are not available at any other neutrino facilities in the world.

\begin{figure}[t]
  \centering
  \includegraphics[width=0.85\textwidth]{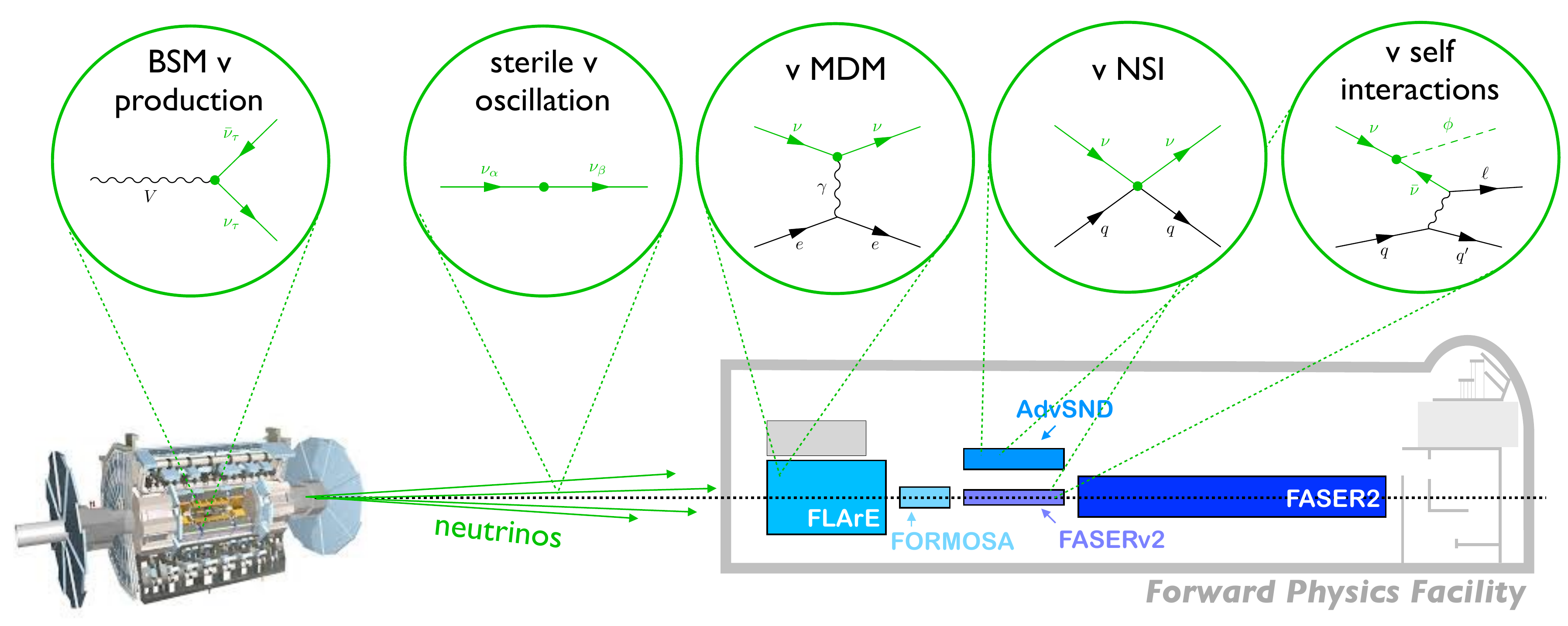}
  \caption{Schematic illustration of BSM neutrino signatures to be searched for at the FPF experiments: i) additional BSM neutrino production mechanisms; ii) sterile neutrino oscillations; iii) neutrino magnetic moments; iv) neutrino NSI; and v) neutrinophilic particles produced in neutrino interactions.   }
  \label{fig:BSMNeutrino}
\end{figure}
\cref{fig:BSMNeutrino} demonstrates a schematic view of some of these BSM possibilities, where here we focus on any BSM scenario that modifies the neutrino flux produced at/near the ATLAS interaction point (left/center) and/or the neutrino interactions in the various FPF detectors (right).

The following subsections include the following:
\begin{itemize}
    \item Non-standard Neutrino Interactions or new Effective Field Theory operators leading to new types of neutrino scattering, as presented in \cref{sec:bsm_neut_eft}, \cref{sec:bsm_neut_nsi} and \cref{sec:bsm_neut_ncnsi}.
    \item New neutrinophilic mediators which modify the predicted tau-neutrino flux at the FPF, as presented in \cref{sec:bsm_neut_lnm}, \cref{sec:bsm_neut_sni} and \cref{sec:bsm_neut_lgb}. 
    \item Neutrino magnetic moments, with or without connections to new sterile neutrinos, as presented in \cref{sec:bsm_neut_mdm} and \cref{sec:bsm_neut_dipole}. 
    \item New oscillations relevant for the energies/distances of interest here sourced by relatively heavy sterile neutrinos, as presented in \cref{sec:bsm_neut_sterile}.
    \item Emission of neutrinophilic mediators in scattering, where the new mediator can be responsible for neutrino self-interactions and/or connections to dark matter, as presented in \cref{sec:bsm_neut_neutphil}.
\end{itemize}
	
\subsection{Effective Field Theories at the FPF}
\label{sec:bsm_neut_eft}

\begin{figure}
  \centering
  \includegraphics[width=0.85\textwidth]{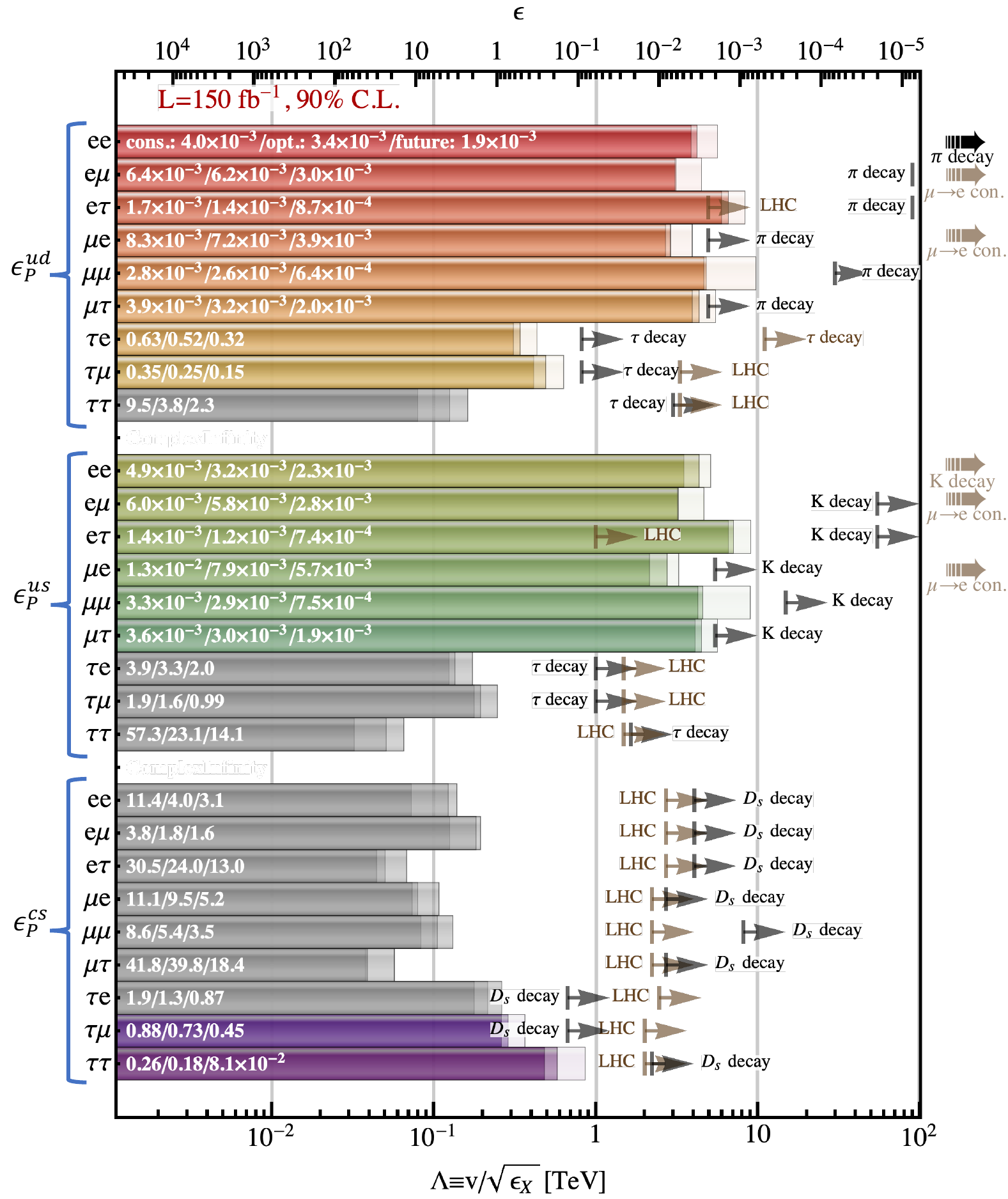}
  \caption{ Figure taken from Ref.~\cite{Falkowski:2021bkq} with the projected constraints on pseudoscalar interactions from FASER$\nu$.  See Ref. \cite{Falkowski:2021bkq} for more details.
  }
  \label{fig:EFTFASERnu}
\end{figure}

Neutrino experiments are sensitive to how neutrinos interact with matter; therefore, if as a result of physics beyond the SM there are new 4-fermion interactions between neutrinos and charged leptons or quarks, these may give observable effects at the production, propagation or detection of neutrinos. In order to systematically probe new physics beyond the neutrino masses and mixings one can use the effective field theory (EFT) language. The relevant energy at neutrino experiments is usually at a scale lower than the $Z$ boson mass, where the the most general effective Lagrangian is described by the Weak Effective Field Theory (WEFT). If the new physics is at a scale much higher than the weak scale, WEFT can be considered as the low energy part of SMEFT. The parameters of WEFT and SMEFT can be matched at the renormalization scale $\nu\sim m_W$, where the matching is shown in Ref.~\cite{Falkowski:2019xoe} for all possible lepton flavors. Hence, by measuring the parameters of WEFT at low energy experiments we can get constraints on the Wilson Coefficients of dimension-6 SMEFT operators, where these limits can be translated to several specific UV completed new physics scenarios. In this way we can indirectly have access to heavy new physics at a scale that can be out of the reach of even high energy colliders. For several dimension-6 operators neutrino experiments offer a superior sensitivity compared to what can be achieved at colliders such as LEP or LHC. As an example the authors of Ref.~ \cite{Falkowski:2018dmy} have studied the future DUNE sensitivity to dimension-6 operators in SMEFT and have shown that new physics at a scale of 30 TeV can be accessible at DUNE.  

Focusing on the charged current (CC) interactions between neutrinos, charged leptons and quarks, the most general  4-fermion interactions described by the WEFT Lagrangian reads:
\begin{align} 
\nonumber
    {\cal L}_{\rm WEFT} 
    &\supset
    - \,\frac{2 V_{jk}}{v^2} \Big\{
      [ {\bf 1} + \epsilon_L^{jk}]_{\alpha\beta}
             (\bar{u}^j \gamma^\mu P_L d^k) (\bar\ell_\alpha \gamma_\mu P_L \nu_\beta)
    \,+\, [\epsilon_R^{jk}]_{\alpha\beta} (\bar{u}^j \gamma^\mu P_R d^k)
                                          (\bar\ell_\alpha \gamma_\mu P_L \nu_\beta) \\
                                          \nonumber
    &\quad +\,
      \frac{1}{2} [\epsilon_S^{jk}]_{\alpha\beta}
             (\bar{u}^j d^k) (\bar\ell_\alpha P_L \nu_\beta)
    - \frac{1}{2} [\epsilon_P^{jk}]_{\alpha\beta}
             (\bar{u}^j \gamma_5 d^k) (\bar\ell_\alpha P_L \nu_\beta) \\
    &\quad +\,
      \frac{1}{4} [\epsilon_T^{jk}]_{\alpha\beta} (\bar{u}^j \sigma^{\mu\nu} P_L d^k)
                                      (\bar\ell_\alpha \sigma_{\mu\nu} P_L \nu_\beta)
    + {\rm h.c.} \Big \} \,
  \label{eq:EFT_lweft}
\end{align}
where $V$ is the CKM matrix, $v\simeq 246~$GeV is the vacuum expectation value of the Higgs boson, $u^j$ and $d^k$ are the up- and down-type quarks in the mass basis, $\ell_\alpha=e,\mu,\tau$ are the charged lepton fields, $P_{L/R}$ are the projection operators, and we have $\sigma_{\mu\nu}=i[\gamma_\mu,\gamma_\nu]/2$. Finally, the neutrino flavor eigenstates $\nu_\alpha$ are connected to the mass eigenstates using the PMNS matrix: $\nu_\alpha=\sum_i U_{\alpha i}\nu_i$, where $\alpha=e,\mu,\tau$ and $i=1,2,3$. In addition to the SM-like left handed interaction $(1+\epsilon_L)$, the right-handed $(\epsilon_R)$, scalar $(\epsilon_S$), pseudoscalar $(\epsilon_P)$, and tensor $(\epsilon_T)$ interactions are allowed. The systematic EFT approach to neutrino experiments was first developed in refs.
\cite{Falkowski:2019xoe,Falkowski:2019kfn}. The formalism was applied to reactor neutrino experiments Daya Bay and RENO in Ref.~\cite{Falkowski:2019xoe}, and it was shown that these experiments offer a unique way to probe tensor and scalar SMEFT operators. 

The above formalism - that can be applied to any current and future neutrino experiments - was investigated at the FASER$\nu$ detector in Ref.~\cite{Falkowski:2021bkq}. Various reasons make FASER$\nu$ an ideal place for these studies: i) Several charged and neutral hadron and meson decays contribute in the production of neutrinos, giving access to various up- and down- type quarks; ii) All different (anti)neutrino flavors can be detected at the FASER$\nu$ detector, hence all possible lepton flavors of the $\epsilon$'s in the WEFT Lagrangian can be probed; iii)  Due to the relevant range of neutrino energies the detection mechanism at FASER$\nu$ is the deep inelastic scattering, which is very well understood within the SM, and hence adding new physics on top of that is simple. The authors of Ref.~\cite{Falkowski:2021bkq}  showed that in total 81 different operators can be probed at FASER$\nu$. Particularly, it was shown that neutrinos which are produced in fully leptonic meson decays (corresponding to most of the  neutrino fluxes at FASER$\nu$) enjoy a  strong chiral enhancement for the pseudoscalar interactions. The relevant Wilson coefficients
of these pseudoscalar operators can be constrained at the per mille level $([\epsilon^{jk}_P]_{\alpha\beta}\lesssim 10^{-3})$, which 
corresponds to a new physics sensitivity at $\sim$ 10 TeV (See \cref{fig:EFTFASERnu} for the pseudoscalar constraints). Unlike other low energy (meson decays) or high energy (CMS or ATLAS) probes, because of the unique capability of FASER$\nu$ in identifying the neutrino flavors, the EFT studies at this experiment gives crucial complementary information if a new physics excess is found in the future. 

\subsection{NSI and Effective Field Theories}
\label{sec:bsm_neut_nsi}

The SM predicts massless neutrinos and the conservation of lepton flavors as a result of an accidental global ${\rm U(1)_\ell}$ symmetry. In contrast, the observed neutrino oscillations maximally violate lepton flavors and require massive neutrinos, thus directly signaling new physics beyond the SM. Experimentally, however, the agreement with SM predictions up to $\sim$1\,TeV is astonishing\,\cite{Baak:2014ora}, suggesting, on one hand, the scale of the underlying new physics is probably above the weak one, and on the other hand, the SM acts as an excellent low-energy effective theory of a larger UV completion. In other words, had we known its UV completion, we can readily obtain the full SM with the addition of some higher dimensional operators upon integrating out the heavy degrees of freedom in that UV theory. Depending on whether the SM gauge symmetry in linearly or non-linearly realized, the resulting effective field theory (EFT) can be either the SM EFT (SMEFT) or the Higgs EFT (HEFT). For a review on the SMEFT and the HEFT, see Ref.\,\cite{Brivio:2017vri}. In the following discussion, we focus specifically on the SMEFT scenario.

The aforementioned integrating out procedure can be carried out either through a diagram / amplitude matching\,\cite{Skiba:2010xn} or through the functional method combined with the covariant derivative expansion (CDE) technique\,\cite{Gaillard:1985uh,Cheyette:1987qz,Henning:2014wua}. The resulting SMEFT respects the ${\rm SU(3)}_c\otimes{\rm SU(2)}_L\otimes{\rm U(1)}_Y$ gauge symmetry of the SM and can be written as the SM Lagrangian plus a tower of higher dimensional operators. Up to dimension 6, we have
\begin{align}
\mathcal{L}_{\rm SMEFT}^{\rm dim-6} = \mathcal{L}_{\rm SM} + {C_5}\mathcal{O}^{(5)} + \sum\limits_i {C_6}\mathcal{O}_i^{(6)},
\end{align}
with the UV cutoff $\Lambda$ implicitly included in the definition of the dimensional Wilson coefficients $C_{5,6}$. The dimension-5 operator $\mathcal{O}_5=\left(\bar{\ell}_{L i}^{c} \tilde{H}^{*}\right)\left(\tilde{H}^{\dagger} {\ell}_{L j}\right)$ was firstly obtained in\,\cite{Weinberg:1979sa}, which naturally induces non-vanishing neutrino masses and also neutrino mixing when the Higgs obtains a non-zero vacuum expectation value (vev) after electroweak symmetry breaking. The dimension-6 operators have been constructed in\,\cite{Buchmuller:1985jz}, with a complete and independent set of operators written down in\,\cite{Grzadkowski:2010es}.\footnote{Recently, the complete and independent set of dimension-8 operators are also obtained in Refs.\,\cite{Li:2020gnx,Murphy:2020rsh}.}

At the weak scale, one can further integrate out $W^\pm$, $Z$, $h$ and the $t$ quark of the SMEFT to obtain the Low-energy EFT (LEFT). For the tree and one-loop matching of the SMEFT onto the LEFT, see Refs.\,\cite{Jenkins:2017jig,Dekens:2019ept}. The LEFT would then serve as an ideal framework for the study of low-energy experiments. Note however that, due to the big energy gap between the SMEFT and the LEFT, large logarithms arise and need to be properly resummed for the Wilson coefficients through the renormalization group (RG) evolution. This has been done in\,\cite{Jenkins:2013zja,Jenkins:2013wua,Alonso:2013hga}, rendering translating constraints from different energy scales onto the SMEFT or certain specific UV model above the weak scale possible. 

\begin{figure}[!thb]
\centering{
  \begin{adjustbox}{max width = \textwidth}
\begin{tabular}{ccc}
\includegraphics[width=0.8\textwidth]{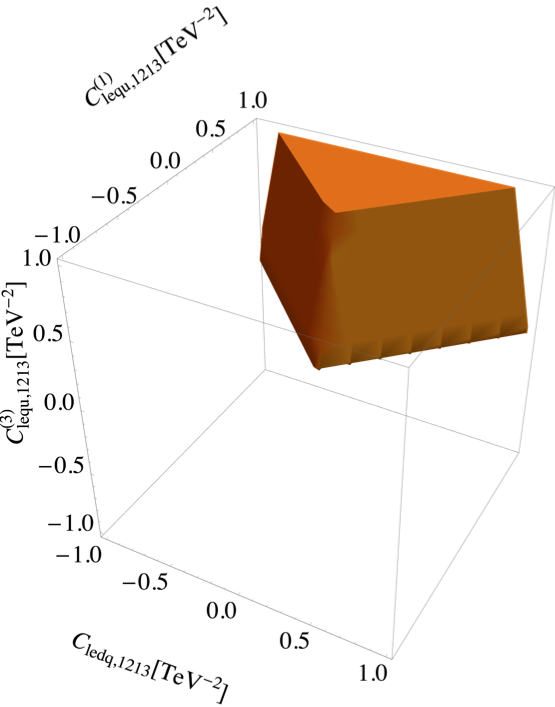} & 
\includegraphics[width=0.8\textwidth]{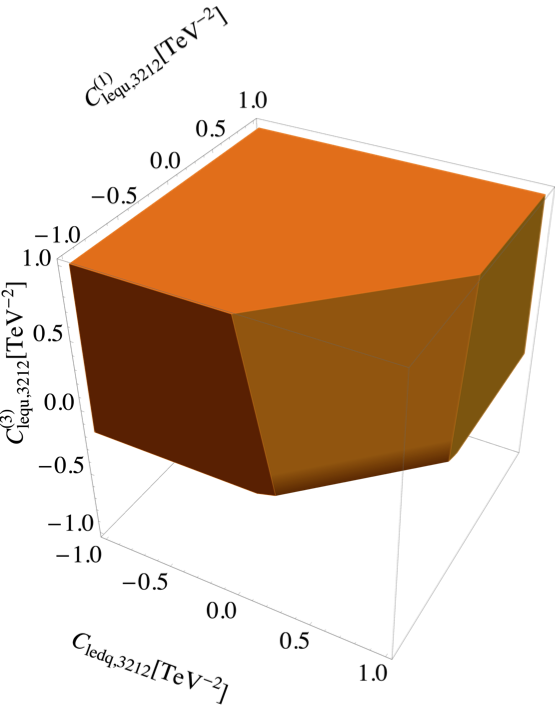} & 
\includegraphics[width=0.8\textwidth]{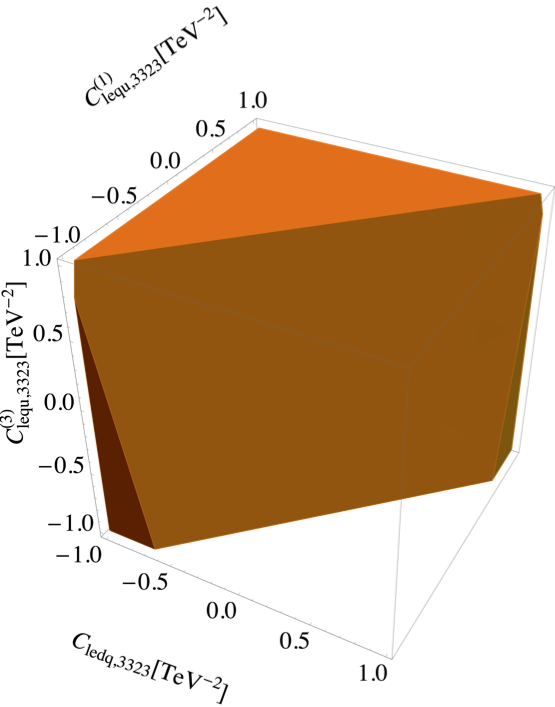}\\
\includegraphics[width=0.8\textwidth]{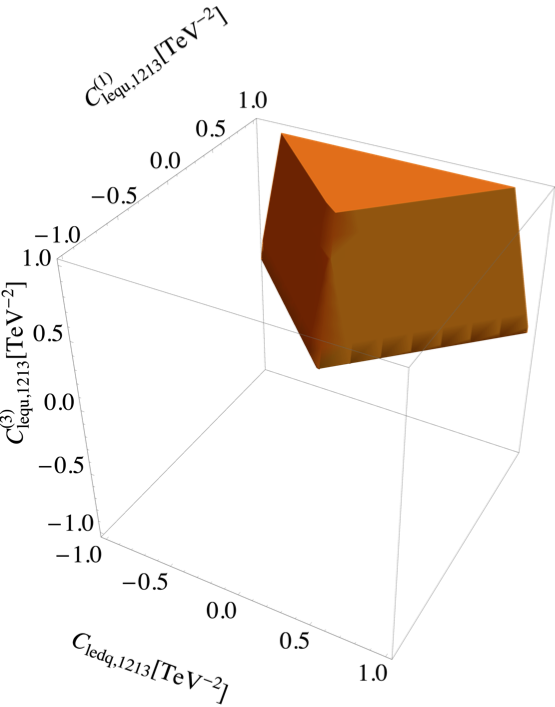} & 
\includegraphics[width=0.8\textwidth]{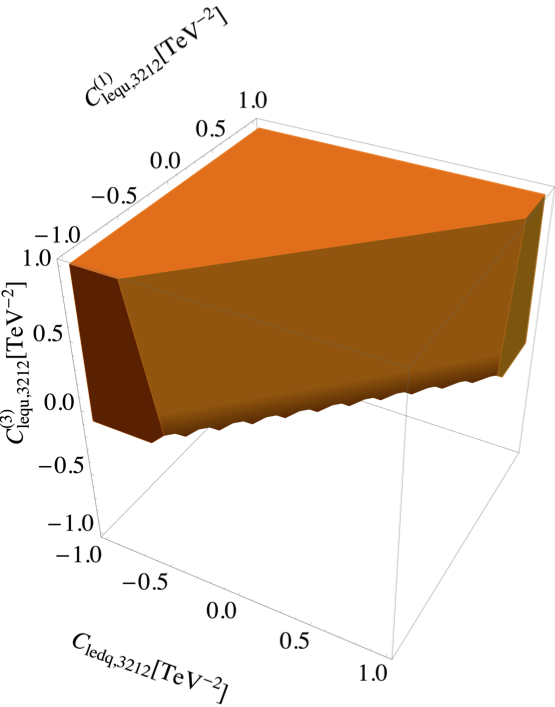} & 
\includegraphics[width=0.8\textwidth]{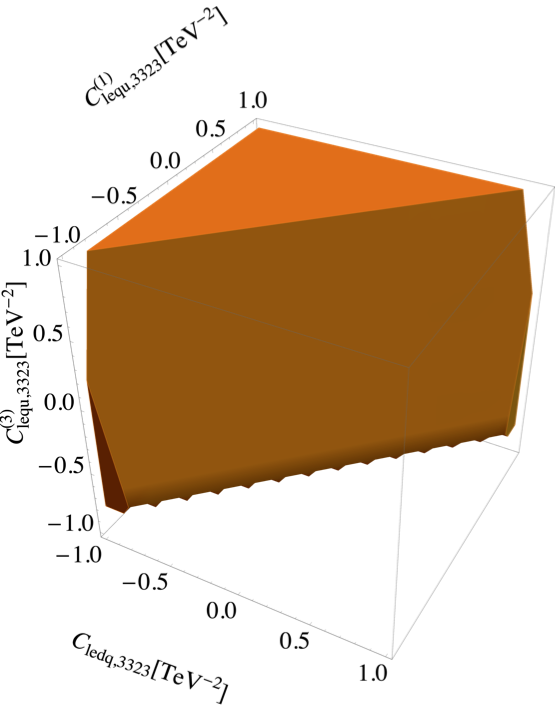}\\
\end{tabular}
\end{adjustbox}
}\caption{Constraints on the SMEFT operator $\mathcal{O}_{ledq,prst}$, $\mathcal{O}_{lequ,prst}^{(1)}$ and $\mathcal{O}_{lequ,prst}^{(3)}$ from FASER$\nu$, with $p,r,s,t$ the flavor indices. The upper row assumes a 30\%, 40\%, 50\% uncertainties in $\sigma_e$, $\sigma_\mu$ and $\sigma_\tau$, respectively\,\cite{FASER:2019dxq,Falkowski:2021bkq}. The lower row is for the High-Luminosity (HL) era of the LHC.}\label{fig:nsismeft}
\end{figure}

One application of the above picture is on neutrino non-standard interactions (NSIs). These are usually parameterized in the LEFT formalism due to the energy scale at which the corresponding experiments are performed. Depending on the currents generating these NSIs,  they are generically classified into charge-current (CC) and neutral-current (NC) ones. For neutrinos produced from $\pi^\pm$, $\mu^\pm$ and $\beta$ decay at low-energy experiments, accelerator- and reactor-type neutrino oscillation experiments for example, the presence of CC NSIs would modify the production rate of neutrinos. A similar argument applies on the detection side of neutrinos, the inverse $\beta$ decay for example. On neutrino oscillations, the influence from these CC NSIs has been recently studied in Refs.\,\cite{Du:2020dwr,Du:2021rdg} in the framework of SMEFT.

On the other hand, the NC NSIs could manifest themselves for neutrino propagation in media, usually referred to as the matter effects. The matter effects have played an essential role in explaining the solar neutrino data due to the large radius and density of the Sun. Similarly, the matter effects can neither be ignored for long-baseline terrestrial neutrino experiments like DUNE\,\cite{DUNE:2016hlj} and T2HK\,\cite{Hyper-KamiokandeProto-:2015xww}, this has also been recently investigated in Ref.\,\cite{Du:2021rdg} in the SMEFT. We comment on that not only neutrino oscillations but also the Coherent Elastic neutrino-Nucleus Scattering (CEvNS) and neutrino decoupling in the early Universe would be modified in the presence of NC NSIs. The former is a very rare process in the SM and has been finally observed in recent years with a CsI detector\,\cite{COHERENT:2017ipa} and a liquid Argon detector\,\cite{COHERENT:2020iec}, respectively. The resulting constraints on the dimension-6 SMEFT operators are presented in Ref.\,\cite{Du:2021rdg}. While for the latter, the NC NSIs could either delay or advance the decoupling time of neutrinos from the electromagnetic plasma, thus changing the prediction of $N_{\rm eff}$, the effective number of relativistic species in the early Universe. This in turn could shed some light on possible UV completions for answering the Hubble tension problem\,\cite{Aghanim:2018eyx,Riess:2019cxk}. See Refs.\,\cite{Du:2021idh,Du:2021nyb} for a model independent study on $N_{\rm eff}$ from NC NSIs up to dimension-7 in the LEFT framework.

Despite various experiments we have as mentioned above in exploring these neutrino NSIs, some of them are still very loosely or even not constrained. For example, from current and future terrestrial neutrino oscillation experiments, only $\mathcal{O}_{ledq,1211}$, $\mathcal{O}_{ledq,1212}$, $\mathcal{O}_{ledq,1213}$, $\mathcal{O}_{ledq,2211}$, $\mathcal{O}_{ledq,2212}$, and $\mathcal{O}_{ledq,1111}$ are/would be constrained\,\cite{Du:2020dwr,Du:2021rdg}, where the four digits represent the flavor indices of the fermions. Note the absence of any constraints on the third generation leptons from the current neutrino oscillation experiments due to the limited statistics for $\nu_\tau/\bar{\nu}_\tau$. More importantly, as is clear from the summary table 4 of Ref.\,\cite{Du:2021idh}, in case of 4-lepton operators, the NC NSIs are generically rather weakly constrained. Thus, one could at least ask the following question: Given the high luminosity of energetic neutrinos at FASER$\nu$ and FASER$\nu$2, what could we gain on testing these neutrino NSIs?

The answer to this question can be presented either model independently in the LEFT or the SMEFT framework, or for some specific UV models that are interesting from a phenomenological consideration. In LEFT, this has been investigated in Ref.\,\cite{Falkowski:2021bkq} by considering neutrino production from meson decay and detection from neutrino-tungsten deep-inelastic scattering. Translating these constraints onto the SMEFT has recently been done in\,\cite{YongINPREPARATION}, where the three-loop QCD and one-loop QED/electroweak running effects from\,\cite{Gonzalez-Alonso:2017iyc} have been employed, including also the threshold effects from the $t$ quark. Some representative results are shown in \cref{fig:nsismeft} for three dimension-6 SMEFT operator $\mathcal{O}_{ledq,prst}$, $\mathcal{O}_{lequ,prst}^{(1)}$ and $\mathcal{O}_{lequ,prst}^{(3)}$, which simultaneously contribute to the production and detection of neutrinos at FASER$\nu$. In each subplot, the polyhedron in orange indicates the allowed region at 90\% CL for the three Wilson coefficients indicated by the axis labels. Note that for the validity of EFTs, we require the magnitude of the Wilson coefficients to be within unity. We find that, with the High-Luminosity (HL) and energetic neutrinos at FASER$\nu$, one would indeed gain some sensitivities on the third generation leptons as seen from the last two plots in the first row of \cref{fig:nsismeft}. While in the HL-LHC era as suggested in the bottom row of \cref{fig:nsismeft}, the sensitivity to these operators would be further improved.

\begin{figure}[!thb]
\centering{
  \begin{adjustbox}{max width = \textwidth}
\begin{tabular}{ccc}
\includegraphics[width=0.8\textwidth]{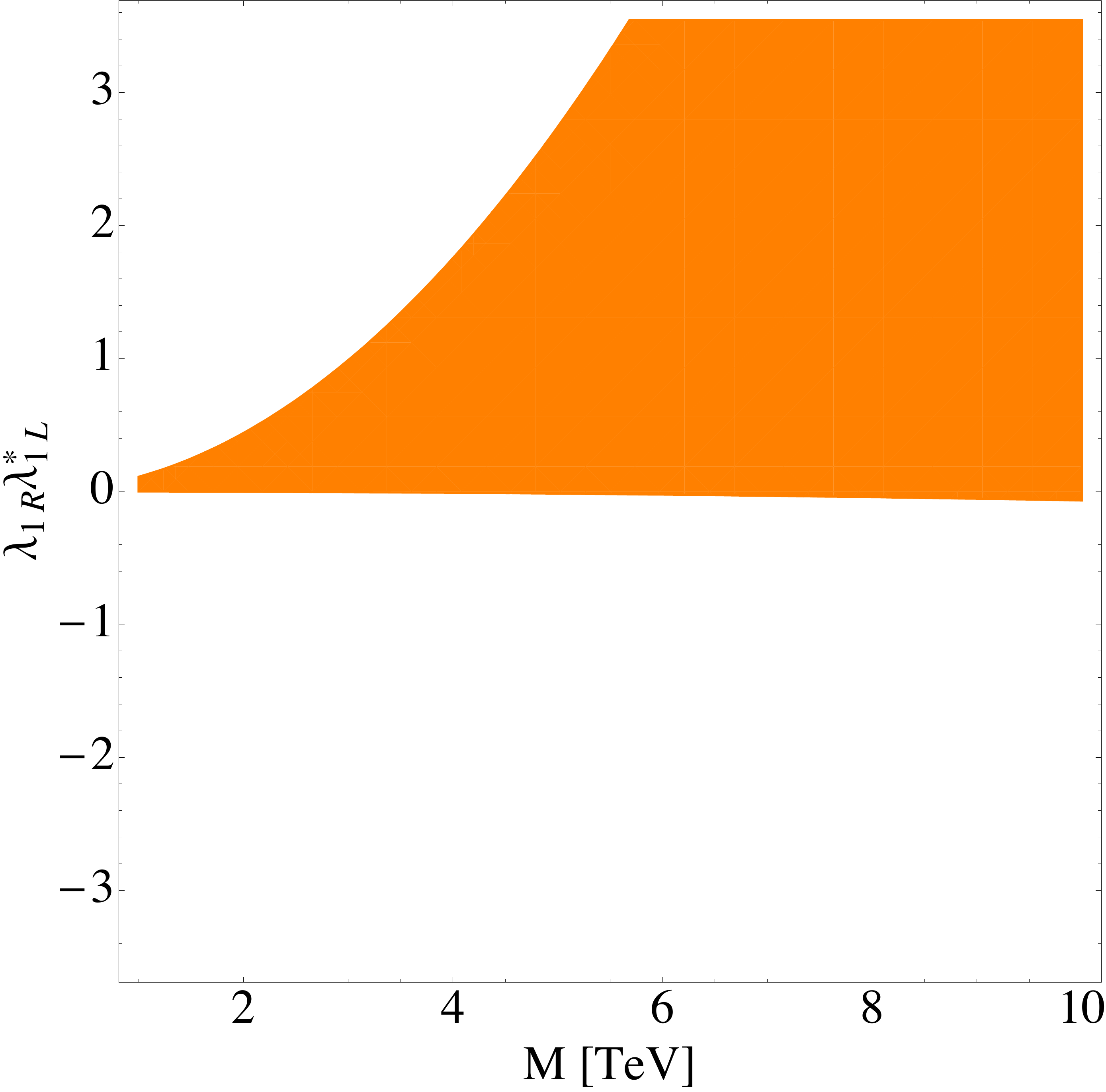} & 
\includegraphics[width=0.8\textwidth]{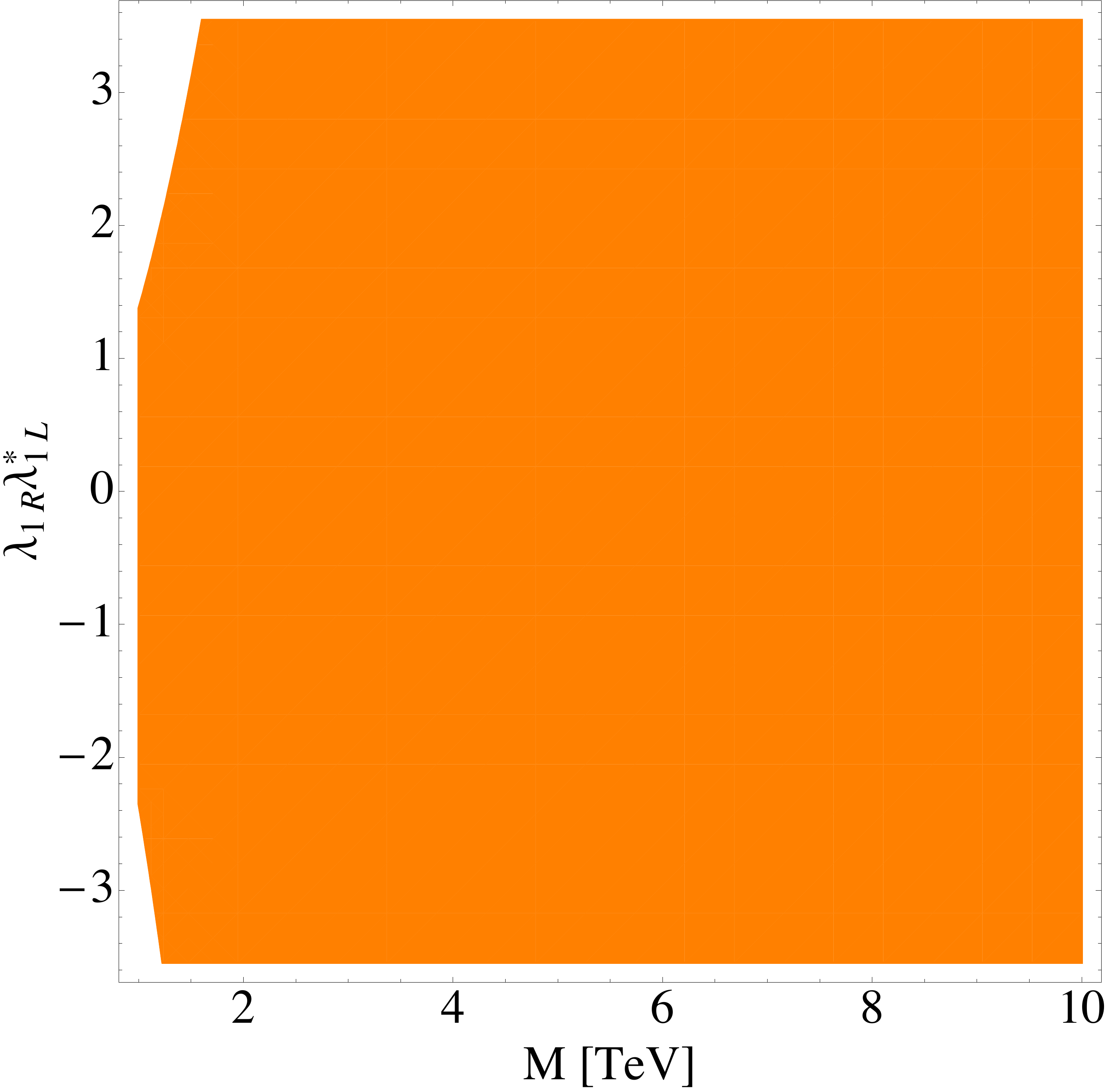} & 
\includegraphics[width=0.8\textwidth]{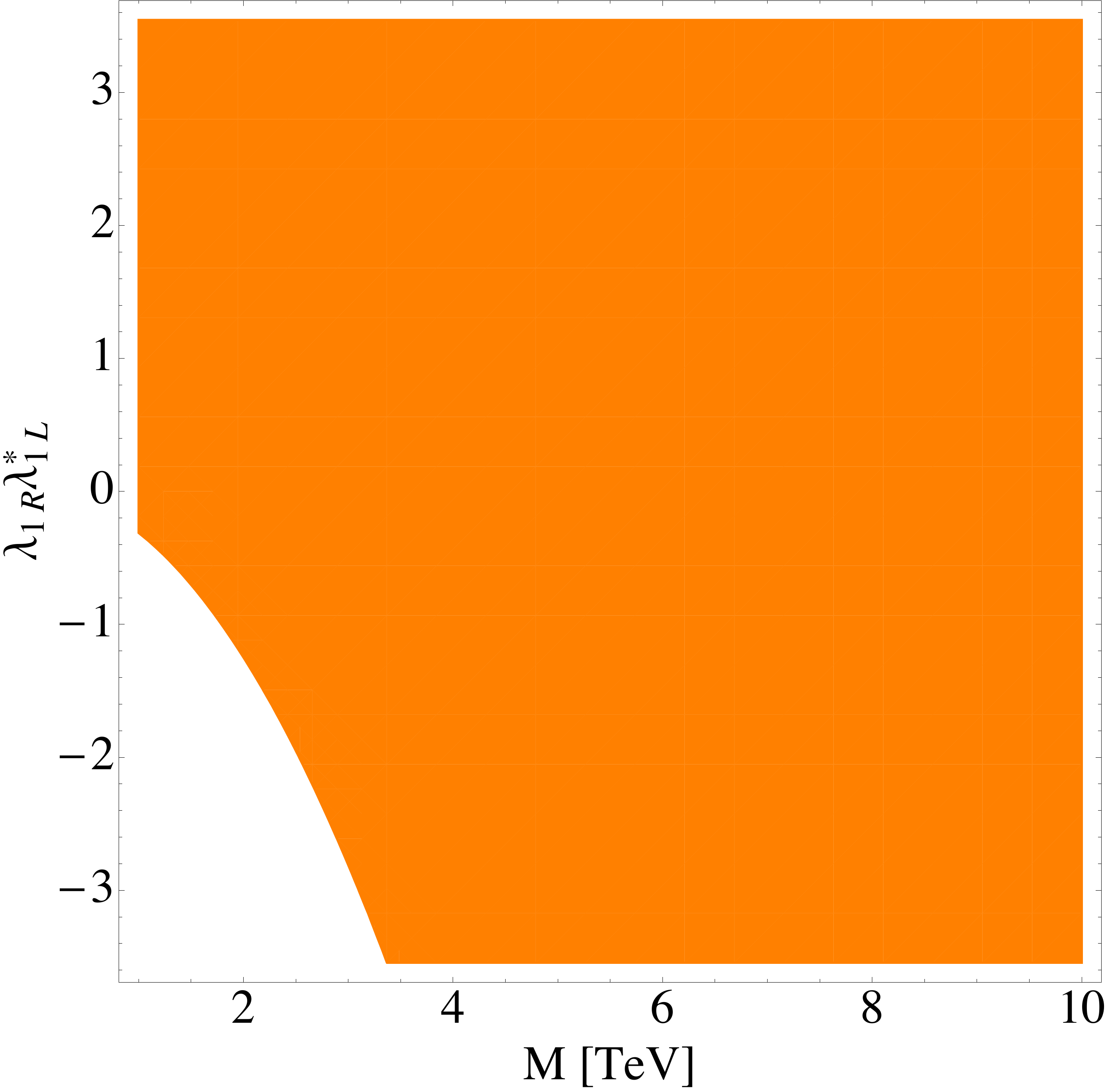}\\
\includegraphics[width=0.8\textwidth]{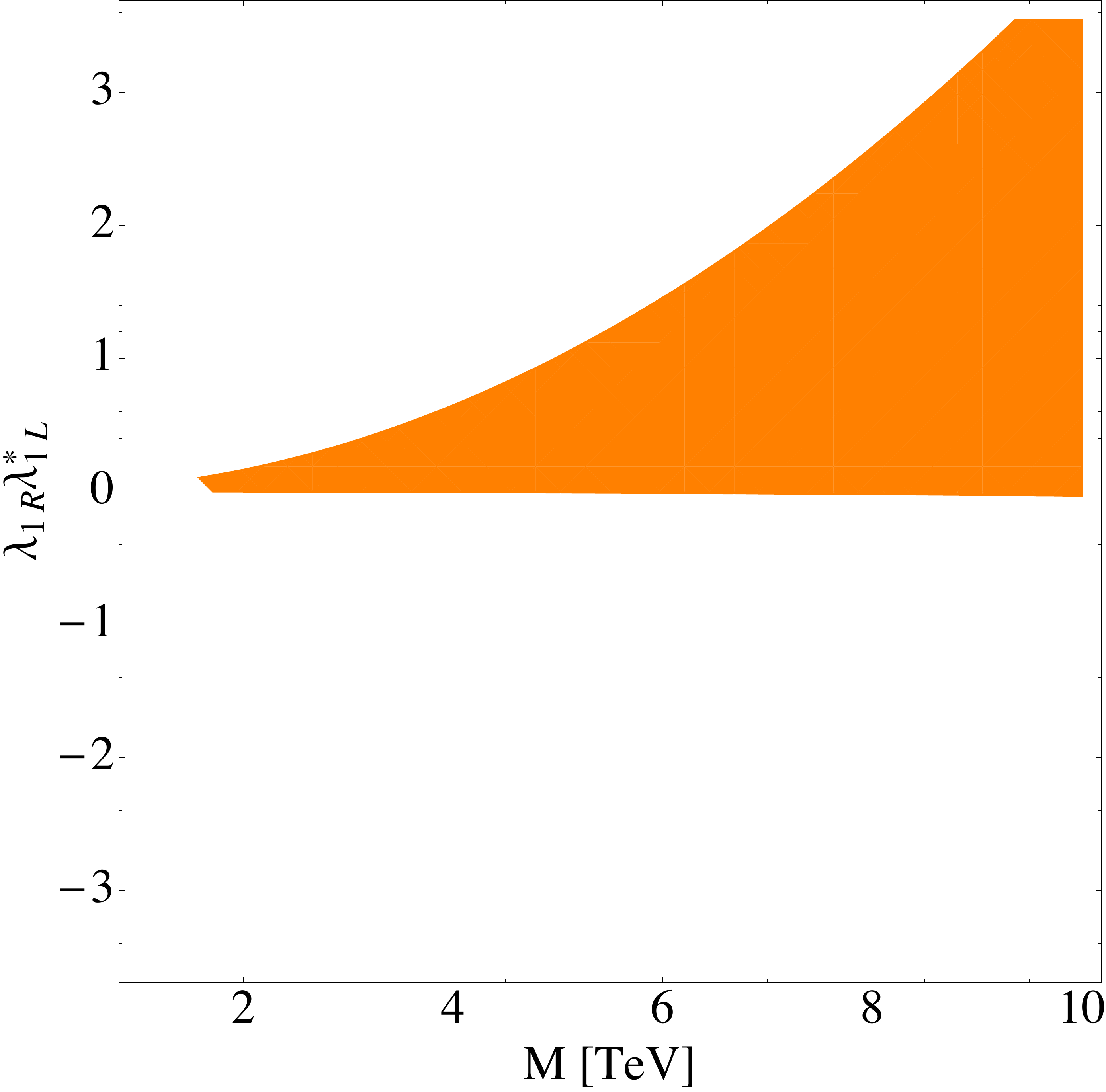} & 
\includegraphics[width=0.8\textwidth]{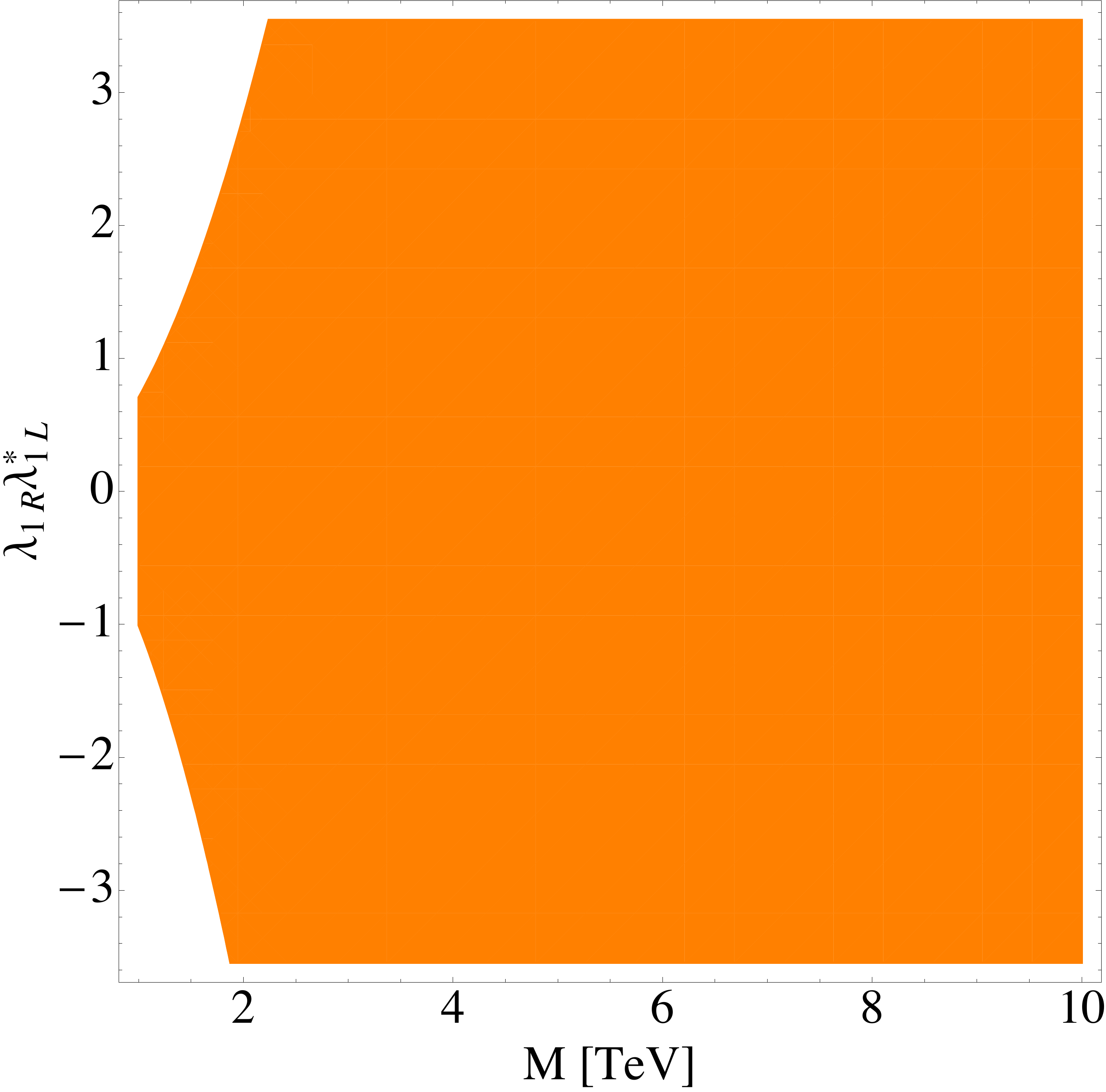} & 
\includegraphics[width=0.8\textwidth]{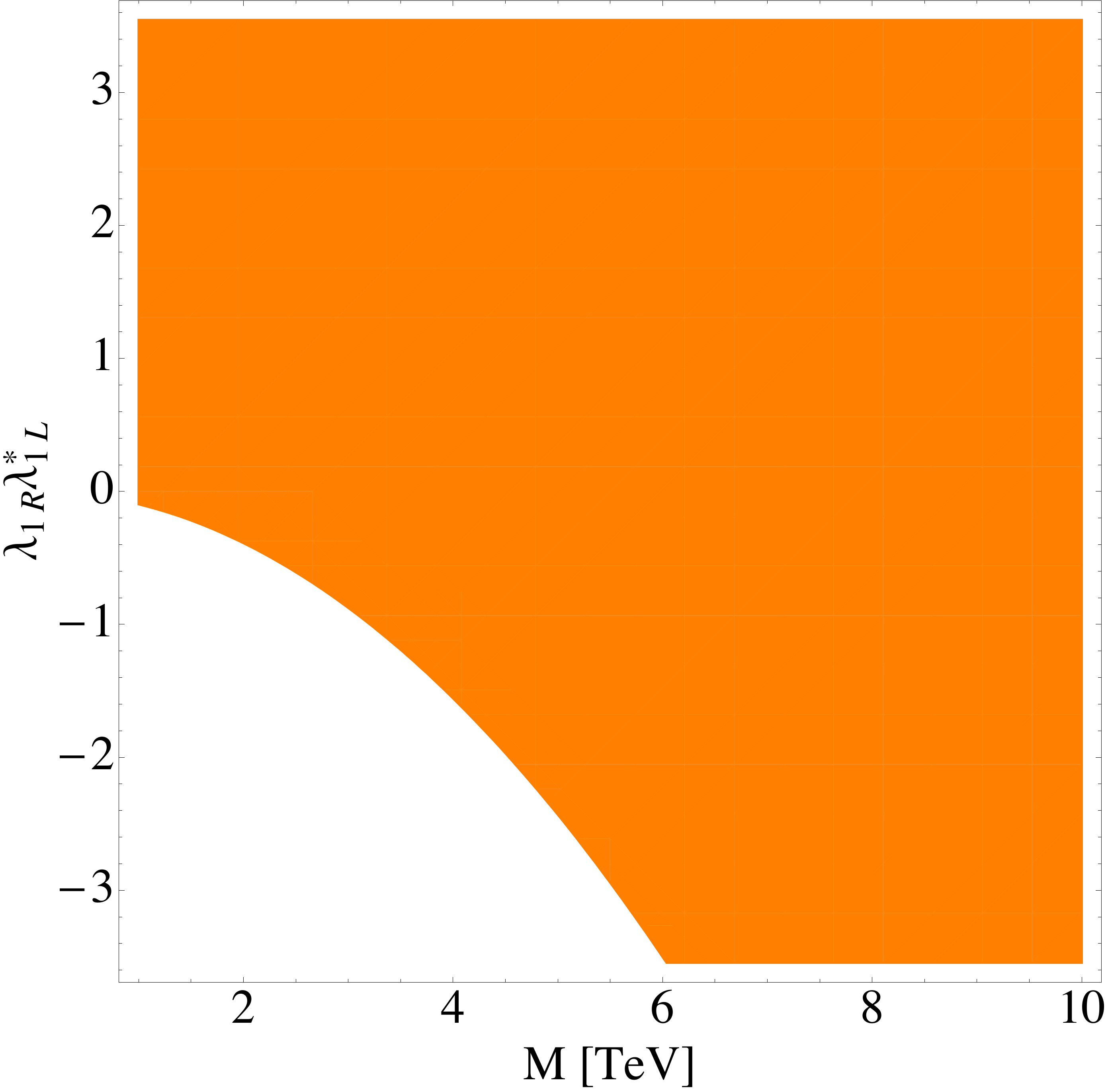}\\
\end{tabular}
\end{adjustbox}
}\caption{Constraints on the lepto-quark model parameter space with two colored scalar lepto-quarks $(\bar{\bf{3}},\bf{1})_{\frac{1}{3}}$ and $(\bar{\bf{3}},\bf{3})_{\frac{1}{3}}$. The upper (lower) row corresponds to a reinterpretation of the results in the upper (lower) row of \cref{fig:nsismeft} in this lepto-quark model.}\label{fig:nsilq}
\end{figure}

Alternatively, one can present the sensitivity to these operators as constraints on the parameter space of certain interesting UV models, the type-I, -II and -III seesaw models for example. This translation can be straightforwardly achieved once we know the matching between these Wilson coefficients and the UV parameters. For these three seesaw models, the tree and one-loop matching onto the SMEFT have been derived in\,\cite{Broncano:2002rw,Coy:2018bxr,Coy:2021hyr,Zhang:2021jdf,Ohlsson:2022hfl,Du:2022vso,Li:2022ipc}, using either the functional method or the amplitude/diagram approach mentioned above. However, due to suppression from the Yukawa couplings (for type-I and -II) or the triplet vev (for type-II) on the $\mathcal{O}_{ledq,prst}$, $\mathcal{O}_{lequ,prst}^{(1)}$ and $\mathcal{O}_{lequ,prst}^{(3)}$ operators\,\cite{Du:2022vso}, we do not expect any sensitivities to these operators from meson decay or neutrino-tungsten deep inelastic scattering discussed above\,\cite{YongINPREPARATION}. Nevertheless, we comment on that, since the type-II model can radiatively trigger electroweak symmetry breaking at a relative low scale around $\sim10^4$\,GeV\,\cite{Du:2022vso} and its distinct signatures at colliders\,\cite{Du:2018eaw}, it would be interesting to investigate the synergy of different experiments in searching for this model. By contrast, the lepto-quark models can be free of any Yukawa suppressions, the SM extension with two colored scalar lepto-quarks, $(\bar{\bf{3}},\bf{1})_{\frac{1}{3}}$ and $(\bar{\bf{3}},\bf{3})_{\frac{1}{3}}$, for example. This model could be a possible candidate simultaneously explaining the CC and NC $B$-anomalies~\cite{Buttazzo:2017ixm, Crivellin:2017zlb, Marzocca:2018wcf, Arnan:2019olv, Yan:2019hpm, Bigaran:2019bqv, Crivellin:2019dwb}, and its tree and one-loop matching have been obtained in Ref.\,\cite{Gherardi:2020det}. Using the notations of Ref.\,\cite{Gherardi:2020det} and re-interpreting the results in \cref{fig:nsismeft} for this lepto-quark model, we show our results in \cref{fig:nsilq} based on\,\cite{YongINPREPARATION}, with the orange regions allowed at 90\% CL. Note that in the SMEFT, the Wilson coefficients are only functions of $\lambda_{1R}\lambda_{1L}^*$, and we find positive $\lambda_{1R}\lambda_{1L}^*$ are generically favored from the projection of FASER$\nu$ up to $10$\,TeV.

\subsection{Neutral current cross section and non-standard interactions}
\label{sec:bsm_neut_ncnsi}

\begin{figure}
    \centering
    \includegraphics[width=0.49\textwidth]{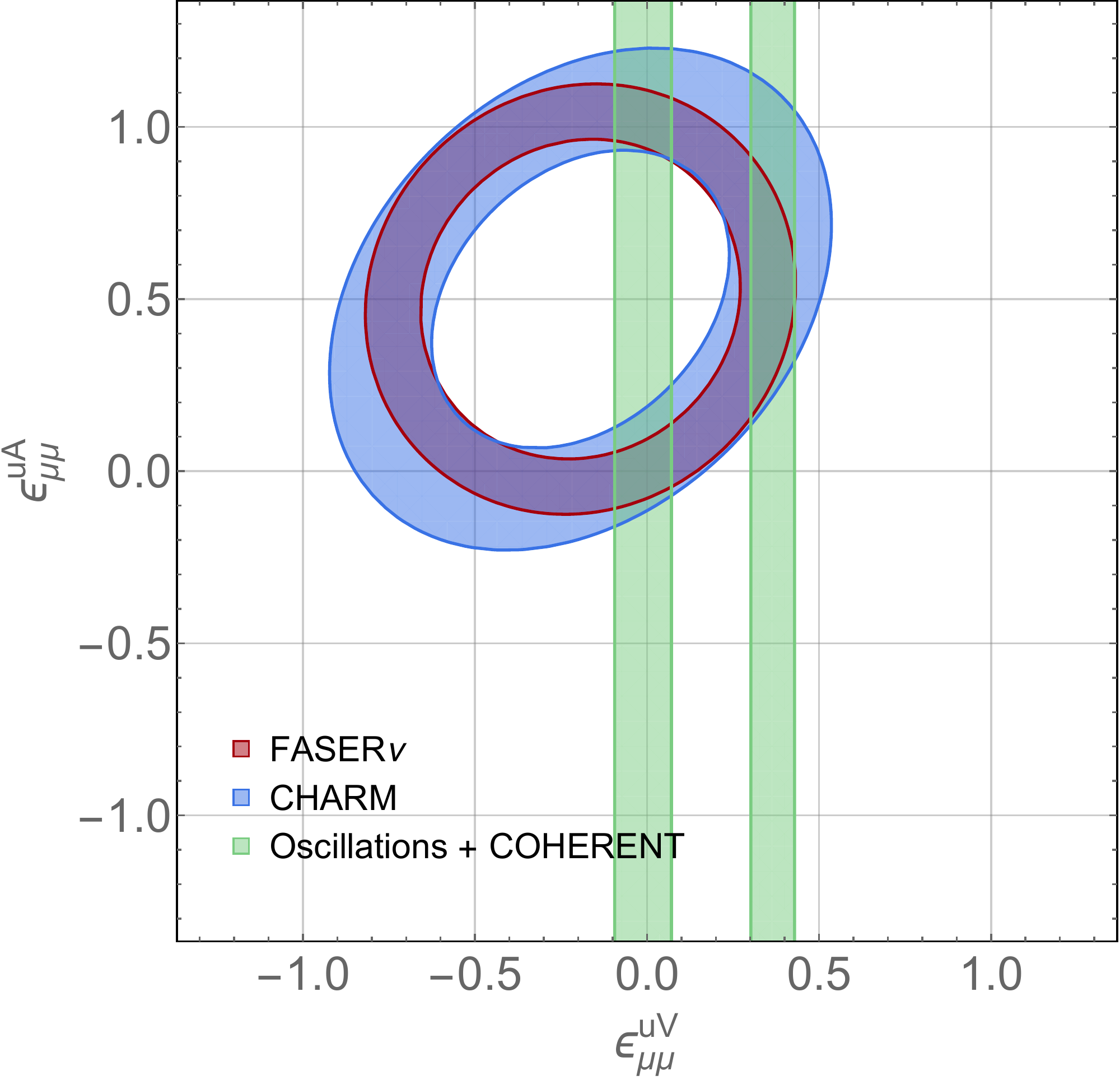}
    \includegraphics[width=0.49\textwidth]{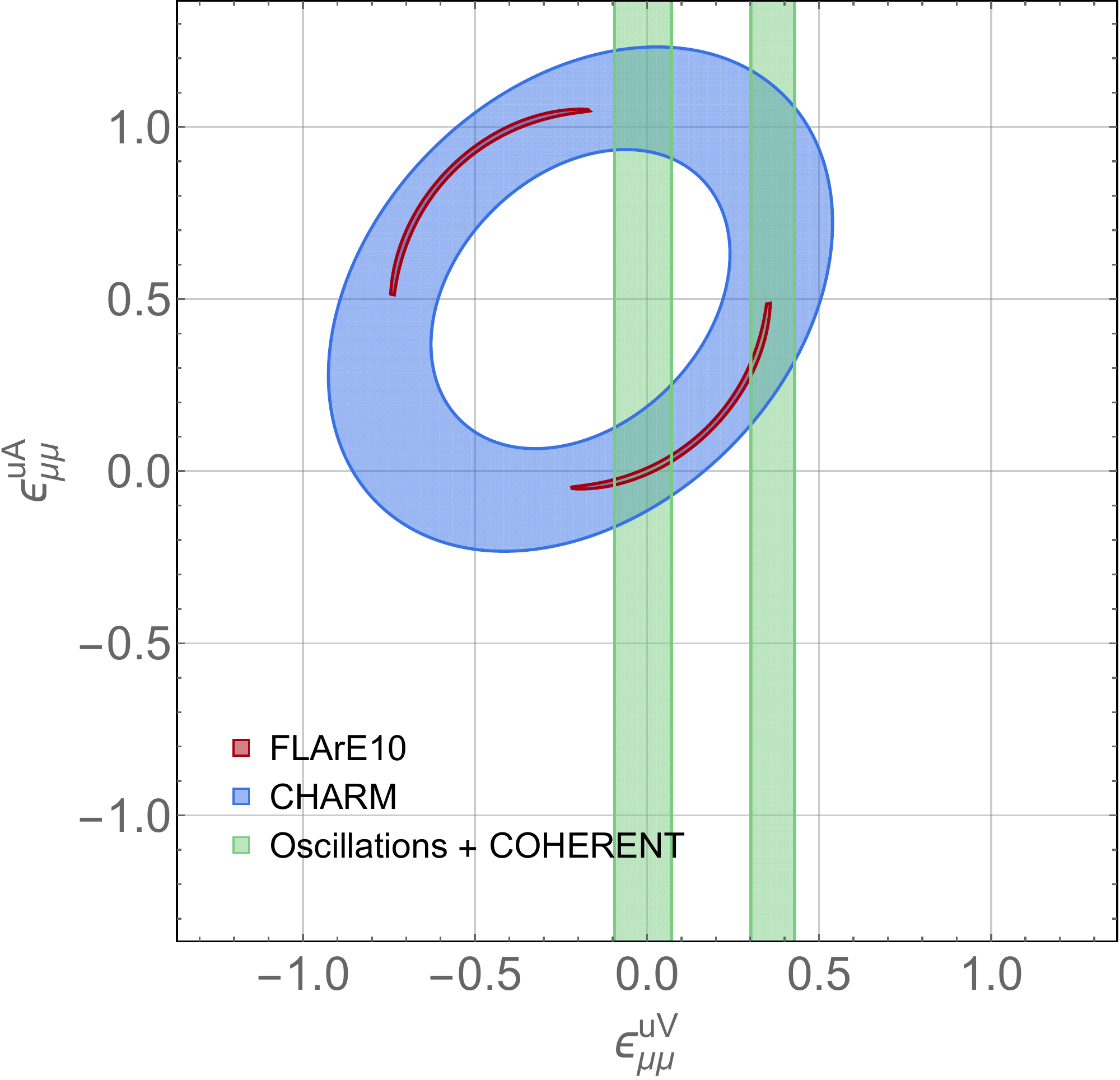}
    \caption{Estimated sensitivity to neutrino NSI parameters at FASER$\nu$ (left) for $150$~fb$^{-1}$ and FLArE-10 (right) for 3~ab$^{-1}$ integrated luminosity. Limits on NSI couplings from CHARM~\cite{CHARM:1987pwr} and on only vector NSI coupling coming from oscillation and COHERENT~\cite{Coloma:2019mbs} are also shown.}
    \label{fig:nsi}
\end{figure}

The NC to CC cross-section ratio can be used to constrain neutrino Non-Standard Interactions (NSI)~\cite{Coloma:2019mbs} at these experiments. Taking the ratio of cross-sections ensures that any uncertainty in flux estimates cancels out. Though we consider below only statistical uncertainties, it must be noted that additional experimental uncertainties are expected to contribute and will most likely dominate. Neutrino NSI can be introduced as
\begin{equation}
    \!\!\!\mathcal{L} \supset -\sqrt{2} G_F \!\!\! \sum_{f,\alpha,\beta}^{\phantom{1}}  [\bar{\nu}_\alpha  \gamma^\mu P_L \nu_\beta] [\epsilon_{\alpha\beta}^{f,V} \bar{f} \gamma_\mu  f + \epsilon_{\alpha\beta}^{f,A} \bar{f} \gamma_\mu \gamma^5  f]\!
\end{equation}
where the $\epsilon$ couplings paramaterize the NSI interactions. While neutrino oscillations~\cite{Wolfenstein:1977ue} and coherent neutrino-nucleus scattering~\cite{Barranco:2005yy} can probe only the vector couplings $\epsilon_{\alpha\beta}^{f,V}$ efficiently, this situation is alleviated in high energy experiments which can probe both axial and vector NSI couplings~\cite{Altmannshofer:2018xyo,Babu:2020nna,Liu:2020emq}. \cref{fig:nsi} shows the expected sensitivity to NSI involving up-type quarks for FASER$\nu$ (left) and FLArE (right). Here we have assumed that all the incoming neutrinos are muon (anti)neutrinos, which is the dominant flavor at FPF. The limits obtained at CHARM~\cite{CHARM:1987pwr} are also shown. Since CHARM measures only neutrinos, they probe NSIs slightly differently from FPF where both neutrinos and anti-neutrinos can be measured. This complementarity between these experiments is visible in the slightly different shapes of the region of the parameter space probed.

\subsection{BSM Interactions in Light of New Mediators}
\label{sec:bsm_neut_lnm}

Ref. \cite{Ansarifard:2021dju} has  built viable underlying UV complete model for the $(\bar{\nu}_\tau\mu)(\bar{d}u)$ as well as  $({\nu}_\tau^T c\mu)(\bar{d}u)$ effective couplings by introducing new Higgs doublets which can also be discovered at the main detectors of high luminosity LHC. Thanks to the $m_\pi^2/(m_u+m_d)^2$ enhancement relative to the SM axial vector contribution, such couplings can respectively lead to $\pi^+ \to \mu^+\nu_\tau$ and  $\pi^+ \to \mu^+\bar{\nu}_\tau$ with branching ratios of $\sim 10^{-3}$ which will be testable at forward experiments \cite{Ansarifard:2021dju}.  The energy spectrum of $\tau$ from new physics at the forward experiments will be softer than the SM prediction for the $\tau$ spectrum. This is understandable because, while the new physics predicts $\nu_\tau$ to come from the $\pi^+$ decay in the SM, $\nu_\tau$ comes from the $D$-meson decays. As a result, by reconstructing the energy spectrum of the produced $\tau$ at FASER$\nu$2, the discovery reach of this experiment can be significantly increased \cite{Ansarifard:2021dju}.

New scalar doublet coupled to quarks as well as to the second generation left-handed doublet, $L_\mu=(\nu_\mu  \ \mu_L)$ can induce the effective couplings of the following form 
\begin{equation} G_{iu} \bar{N}_{iR}\nu_\mu \bar{u}_Lu_R+ 
G_{id} \bar{N}_{iR}\nu_\mu \bar{d}_Rd_L+
G_{iL} \bar{N}_{iR}\mu_L \bar{d}_Ru_L+
G_{iR} \bar{N}_{iR}\mu_L \bar{d}_Lu_R+{\rm H.c.},\label{GuGd} \end{equation}
where $N_{iR}$ is a new right-handed neutrino. As shown in Ref. \cite{Ansarifard:2021elw}, $G$ can be as large as $10^{-5}$~GeV$^{-2}$ without violating any current constraint. With such coupling, the $\nu_\mu$ flux interacting with the forward detector can produce $N_R$ lighter than 15~GeV. On the other hand, the coupling of the new scalar to $L_\mu$ and $N_R$ can lead to a one-loop contribution to $(g-2)_\mu$ anomaly. Explaining the $(g-2)_\mu$ anomaly requires multiple $N_i$ which all can be produced by the $\nu_\mu$ interaction at FASER$\nu$ and SND@LHC and their successors. Heavier $N_i$ can chain decay into lighter ones, producing detectable $\mu \bar{\mu}$ pairs in the detector. The lightest $N_i$ decays as  $N\to \nu_\mu \bar{q}q$ or $N\to \mu \bar{q}q'$ \cite{Ansarifard:2021elw}. The corresponding displaced vertex will constitute a background free signal. As shown in \cite{Ansarifard:2021elw}, by looking for such signal,  SND@LHC,  FASER$\nu$ and FASER$\nu$2 can respectively probe $G$ with values down to {\rm few}$\times 10^{-6} $~GeV$^{-2}$, $10^{-6}$~GeV$^{-2}$ and $10^{-7}$~GeV$^{-2}$ for $m_N<15$~GeV.

Ref.~\cite{Bakhti:2020szu} shows that if the signal for new physics is two pairs of muons produced by the interaction of the neutrino beam with matter, the discovery potential of the forward experiments can be significantly increased by including the muons entering the detector from the rock before the detector. To distinguish between the signal and the accumulation of   through-going muons from IP, the arrival time of the four muon signal has to be recorded which can be done by addition of a scintillator plate in front of the detector and/or using the timing information of  the interface detector between FASER$\nu$ and FASER.

\subsection{Secret Neutrino Interaction}
\label{sec:bsm_neut_sni}

Secret neutrino gauge interaction with a
light gauge boson, $Z^{\prime}$, is motivated by several BSM scenarios. Such an interaction can affect the energy spectrum of neutrinos propagation within a supernova \cite{Das:2017iuj, Dighe:2017sur}. Moreover, if both dark matter particles and neutrinos enjoy a new gauge
interaction, Dark Mater (DM) self-interactions can alleviate the cusped-cored problem. Also, DM and neutrinos can scatter off each other via the new gauge interaction, leading to late time kinetic dark matter decoupling \cite{Chu:2015ipa, Dasgupta:2013zpn, Hooper:2007tu, vandenAarssen:2012vpm, Bringmann:2006mu, Boehm:2000gq}.
Being able to detect high-energy collider neutrinos for the first
time, in particular tau-neutrino, FASER$\nu$ will be an ideal apparatus to study the new neutrinophilic interaction and to reconstruct the flavor structure of the $Z^{\prime}$ coupling to neutrinos. In \cite{Bahraminasr:2020ssz}, we have studied the impact of the coupling of neutrinos with $Z^\prime$, with a mass of less than 500 {\rm MeV} in FASER$\nu$ experiment. This interaction term is given by
\be \sum_{\alpha ,\beta} g_{\alpha \beta}Z'_\mu\bar{\nu}_\alpha \gamma^\mu \nu_\beta \ee
where $g_{\alpha \beta}$ is the the couplings between the new
light boson $Z^{\prime}$ and neutrinos of flavor $\alpha$ and $\beta$. This interaction can lead to a new decay mode for charged mesons to a light lepton plus neutrino and $Z^{\prime}$, ($\pi^+(K^+,D_s^+)\to e^+ \nu Z^\prime$) followed by the subsequent decay of $Z^\prime$ into the pair of neutrino and anti-neutrino, ($Z^\prime \to \nu\bar{\nu}$). The produced neutrinos can be
detected at the emulsion detector of FASER$\nu$.

Let us emphasize that meson decay experiments can also search for such $Z^\prime $ \cite{Bakhti:2017jhm} but these experiments can identify the charged lepton produced in the decay of the charged meson and not the produced neutrinos. Thus, meson decay experiments are sensitive to $\sum_{\alpha \in \{ e,\mu,\tau \}} |g_{e \alpha}|^2$ and $\sum_{\alpha \in \{ e,\mu,\tau \}} |g_{\mu \alpha}|^2$. However, FASER$\nu$, being able to detect the produced neutrinos can provide complementary information on the flavor structure of $Z^\prime$ coupling.

\begin{figure*}[ht]
\centering
\includegraphics[width=0.45\textwidth]{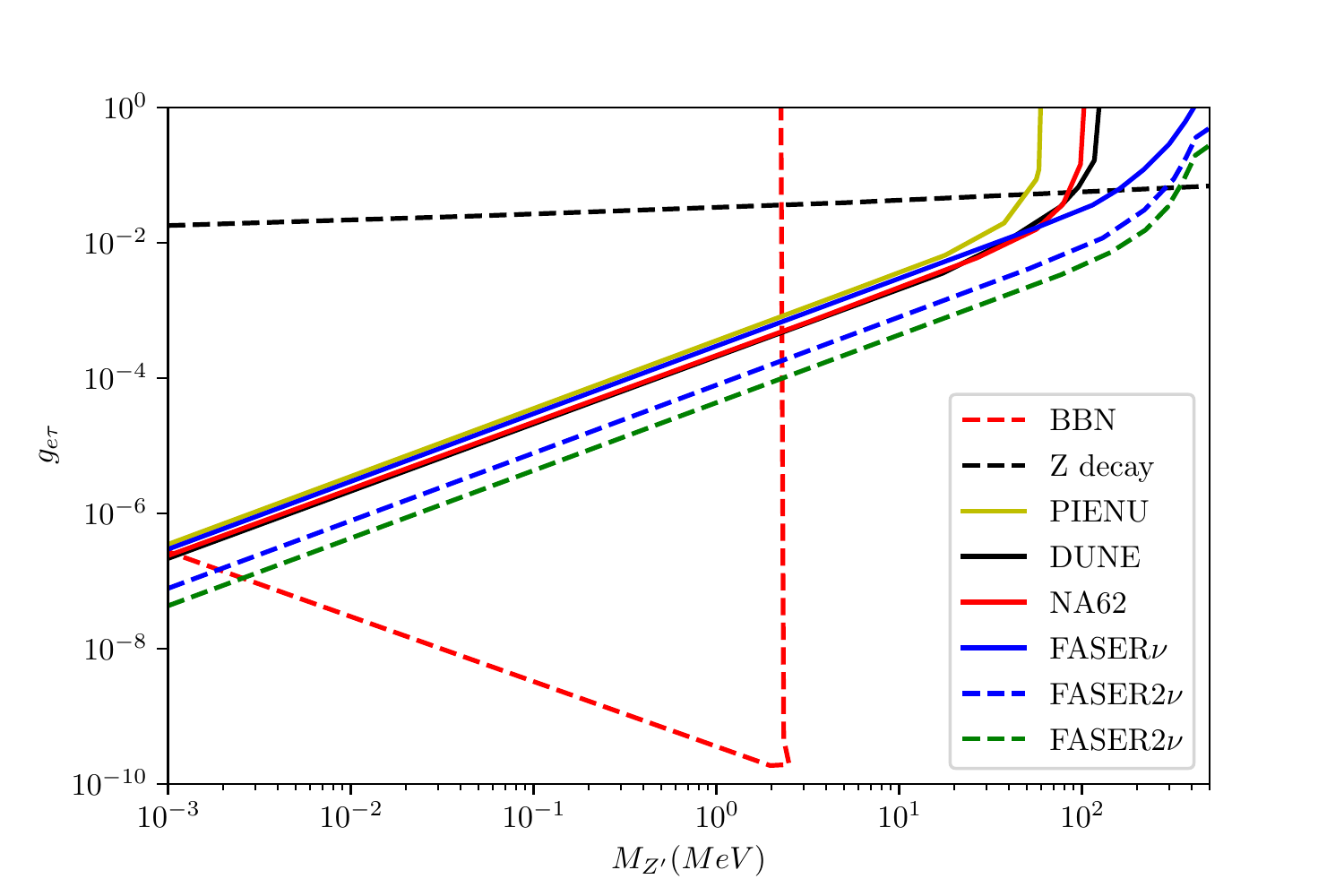}
\includegraphics[width=0.45\textwidth]{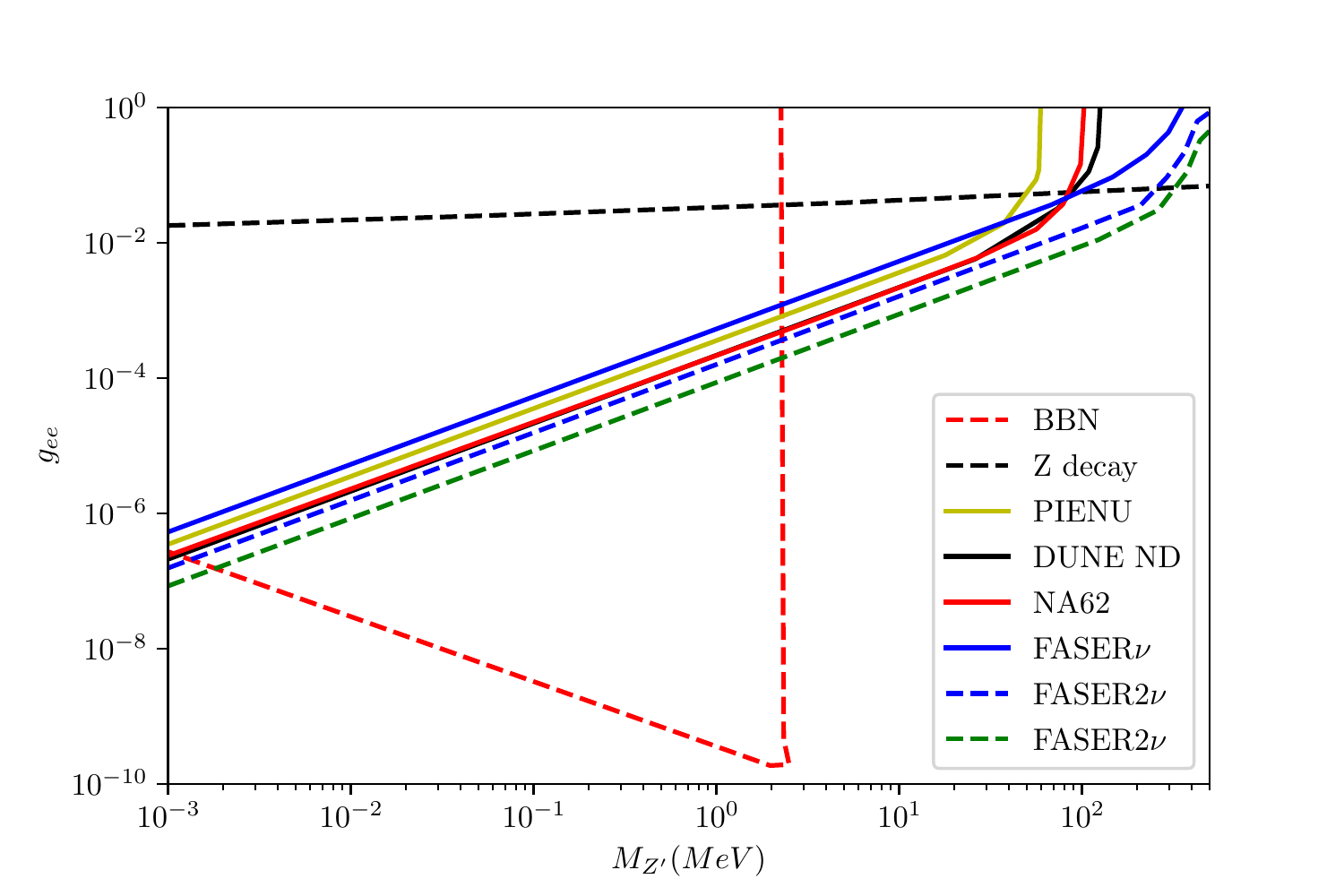}
\includegraphics[width=0.45\textwidth]{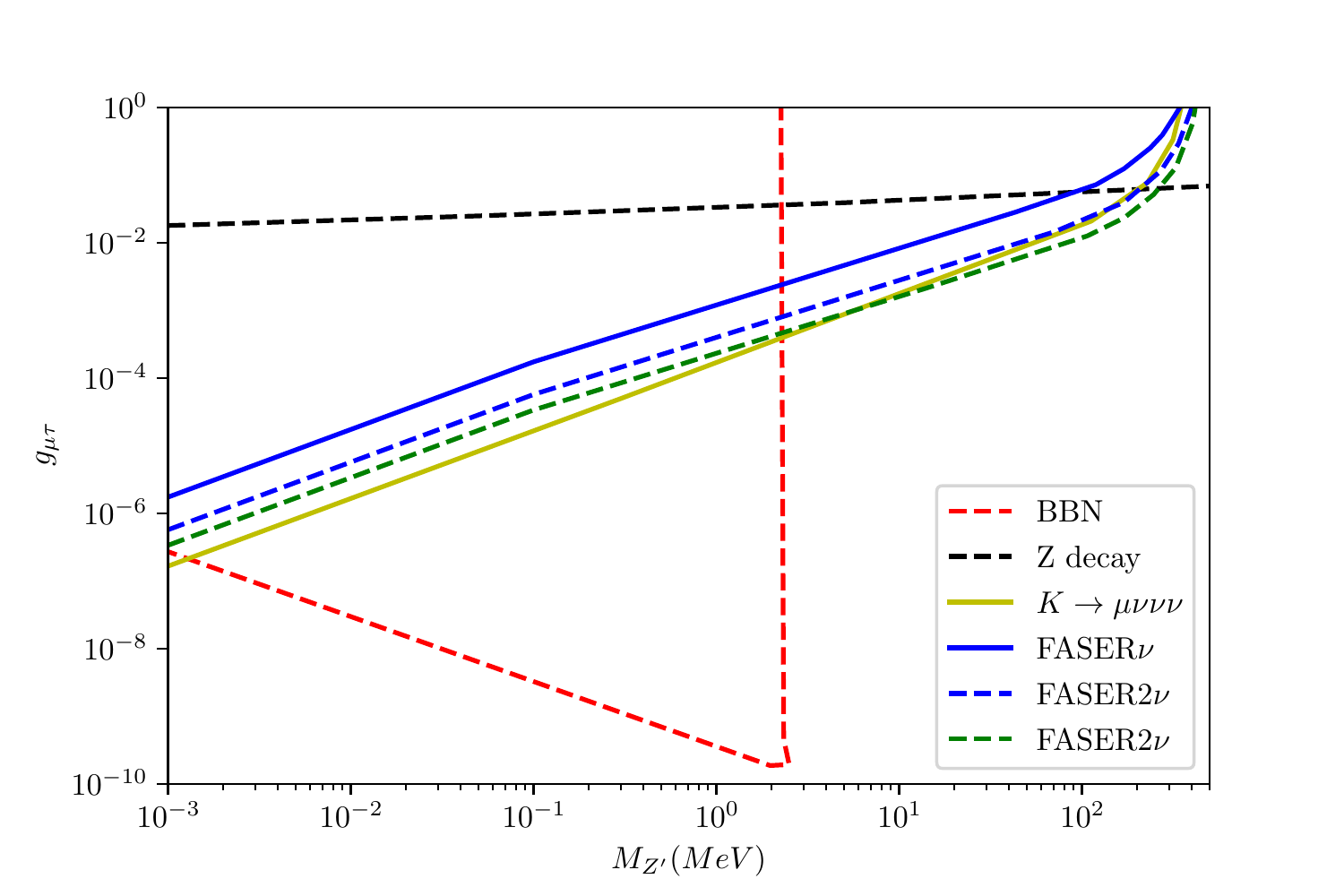}
\includegraphics[width=0.45\textwidth]{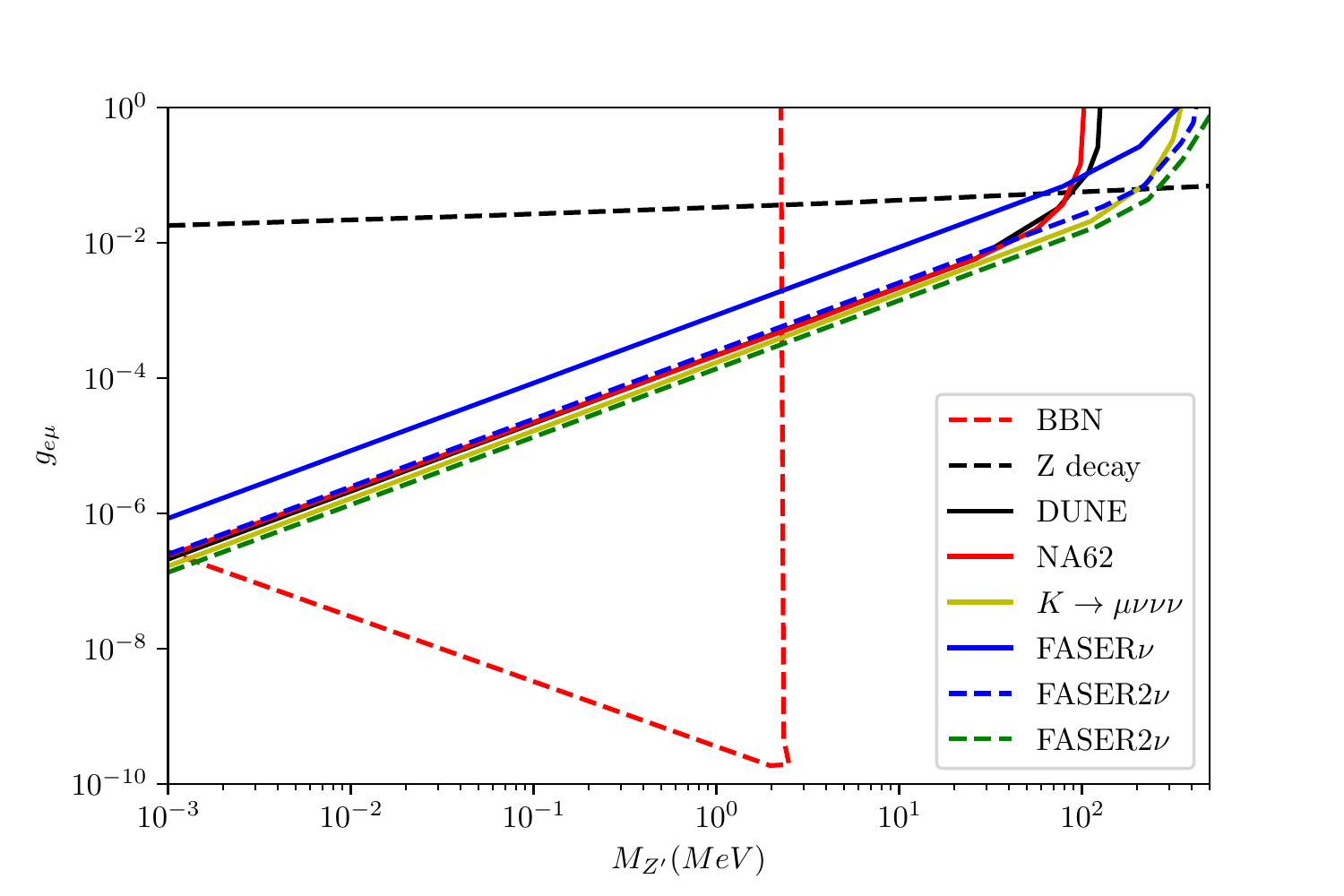}
\caption{The upper bound on $g_{e \tau}$, $g_{ee}$, $g_{\mu \tau}$ and $g_{e \mu}$ vs. $m_{Z^\prime}$ at 90$\%$ C.L.
The yellow and the blue curves shows the current bound from $K \rightarrow \mu \nu \nu \nu $ \cite{E949:2016gnh} and FASER$\nu$, respectively. The blue dashed curve and the green dashed curve indicate the constraints from FASER$2\nu$ corresponding to the assumed data of one hundred times and one thousand times larger than FASER$\nu$, respectively. We have assumed detection efficiency of $80\%$ for FASER$\nu$.
The black dashed line shows the current constraint from Z decay \cite{Laha:2013xua}. The red dashed curve shows the BBN constraint \cite{Huang:2017egl}.}\label{fig_SNI}
\end{figure*}

We consider pion and kaon leptonic decay channels, as well as the subdominant production channel $ D_s \rightarrow l \nu_\alpha Z^\prime$. Contribution from strange charm mesons is remarkable for constraining larger $Z^\prime$ mass [ $m_{Z^\prime} > 50~ $ {\rm MeV}].
For statistical inference, we used the chi-squared method. We have used the pull method to account for the systematic uncertainties.
We have considered the flux normalization uncertainty of 10$\%$. We also repeat our analysis for FASER2$\nu$ assuming it will collect 100 and 1000 times larger data than FASER$\nu$. We have assumed detection efficiency of 80$\%$ for FASER$\nu$.
Our results are shown in \cref{fig_SNI}.

As can be observed in \cref{fig_SNI}-a , FASER$\nu$ (blue curve) can constrain $g_{e \tau}$ more strongly than the current constraints and future DUNE near detector constraint ( black curve ) \cite{Bakhti:2018avv}, for $50 ~{\rm MeV} < m_{Z^\prime} < 150 ~{\rm MeV} $. However, we observed that with $100$ and $1000$ times larger than FASER$\nu$, FASER$2\nu$ can improve the limit on $g_{e \tau}$ for $ m_{Z^\prime} < 2 ~{\rm keV} $ and $3~{\rm MeV} < m_{Z^\prime}< 300~{\rm MeV}$. The current bound from PIENU \cite{PiENu:2015seu}, NA62 \cite{NA62:2012lny}, are presented by the yellow and red curves, respectively. The black
dashed line and the red dashed curve show the current constraint from Z decay \cite{Laha:2013xua} and BBN
constraint \cite{Huang:2017egl}.
As can be seen from \cref{fig_SNI}-b, FASER$2\nu$ with a data $100$ and $1000$ times larger than FASER$\nu$ data, set the strongest constraint on the $g_{ee}$ for the mass range of $ m_{Z^\prime} < 2 ~{\rm keV} $ and $3~{\rm MeV} < m_{Z^\prime}< 300~{\rm MeV}$. The results for $g_{\mu \tau}$ is indicated in \cref{fig_SNI}-c. We observed that for the mass range
$10 ~{\rm MeV} < m_{Z^{\prime}} < 300 ~{\rm MeV}$, FASER$2\nu$ with 1000 times larger data than FASER$\nu$, can set
the most stringent bound on $g_{\mu \tau}$. For $g_{e \mu}$, FASER$2\nu$ with 1000 times larger data can slightly improve the current bounds for $ m_{Z^\prime} < 2 ~{\rm keV} $ and $3~{\rm MeV} < m_{Z^\prime}< 300~{\rm MeV}$ as indicated in \cref{fig_SNI}-d.

\subsection{Probing Light Gauge Bosons via Tau Neutrinos}
\label{sec:bsm_neut_lgb}

The tau neutrinos is the perhaps least experimentally studied particle in the SM. So far, only a few handful of interactions have been identified directly at DONuT~\cite{DONuT:2007bsg} and OPERA~\cite{OPERA:2018nar}, while many other neutrino experiments have not yet observed them directly. There are several reasons for this: i) tau neutrinos are mainly produced in heavy $D_s$ and $B$ meson decays, whose production is suppressed compared to those of other lighter hadrons; ii) to observe the interactions of a tau neutrino, the neutrino beam energy needs to be sufficiently high to produce a tau lepton $E_\nu \gtrsim 3.5~\gev$; and iii) the identification of the short-lived tau lepton in the neutrino detector requires a sufficient spatial resolution, which are typically only achieved in emulsion detectors. 

This situation will change in the near future with the upcoming LHC neutrino experiments FASER$\nu$ \cite{FASER:2019dxq, FASER:2020gpr} and SND@LHC~\cite{SHiP:2020sos}, which are expected to observe tens of tau neutrino interactions. Following the same idea, the FPF neutrino experiments FASER$\nu$2, FLArE and SND@LHC would be able to detect thousands of tau neutrino interactions. This large number of tau neutrinos interactions at the FPF also it an interesting laboratory for tau-neutrinophilic BSM physics, both in the production and interaction.

In the following, let us focus on one specific example of tau-neutrinophilic BSM physics: the gauged $U(1)_{B-3L_\tau}$ group. As discussed in \cref{sec:llp_vec_b3l} this is one of the anomaly-free $U(1)$ extensions of the SM. In the minimal scenario, the associated gauge boson by $V$ couples to the SM quarks(which have charge $Q_q=1/3$), as well as the tau and neutrino (which have charge $Q_\tau=-3$). In non-minimal extensions, it might also be possible that further BSM states are charged under this symmetry. As example, we consider the case considered in \cref{sec:bsm_dm_hadro}, in which $V$ also interacts with a complex scalar dark matter state $\chi$ with charge $Q_\chi$. The Lagrangian for this model can be written as: 
\be
\mathcal{L} \supset \frac12 m_V V^\mu V_\mu +  g V^\mu \big[ 
Q_q  \sum_q  \bar{q} \gamma_\mu  q +  Q_\tau (\bar{\tau} \gamma_\mu  \tau + \bar{\nu}_\tau \gamma_\mu  \nu_\tau ) +   i Q_\chi  ( \chi \partial_\mu \chi^* - \chi^* \partial_\mu \chi ) \big]
\ee
where $m_V$ is the gauge boson mass and $g$ is the gauge coupling. Note that this scenario is an ideal example of tau-neutrinophilic new physics as it i) couples to quarks, allowing production in hadrons decays and scattering with nuclear matter, ii) strongly couples to the tau neutrino, enhancing the physics prospects, and iii) does not couple to electrons and muons, avoiding constrains from lepton beam experiments and associated precision measurements. 

\begin{figure}[t]
\centering
\includegraphics[width=0.49\textwidth]{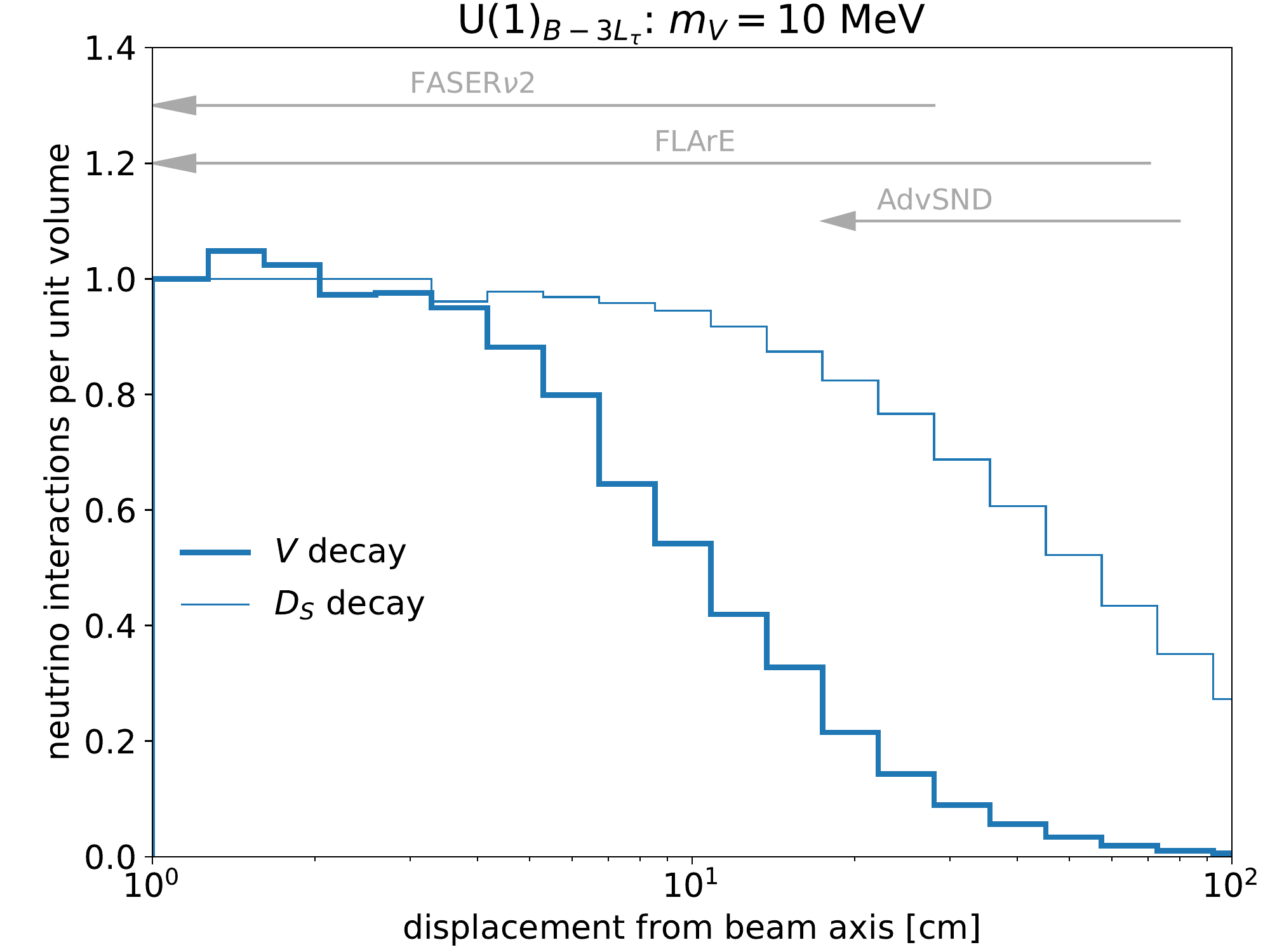}
\includegraphics[width=0.49\textwidth]{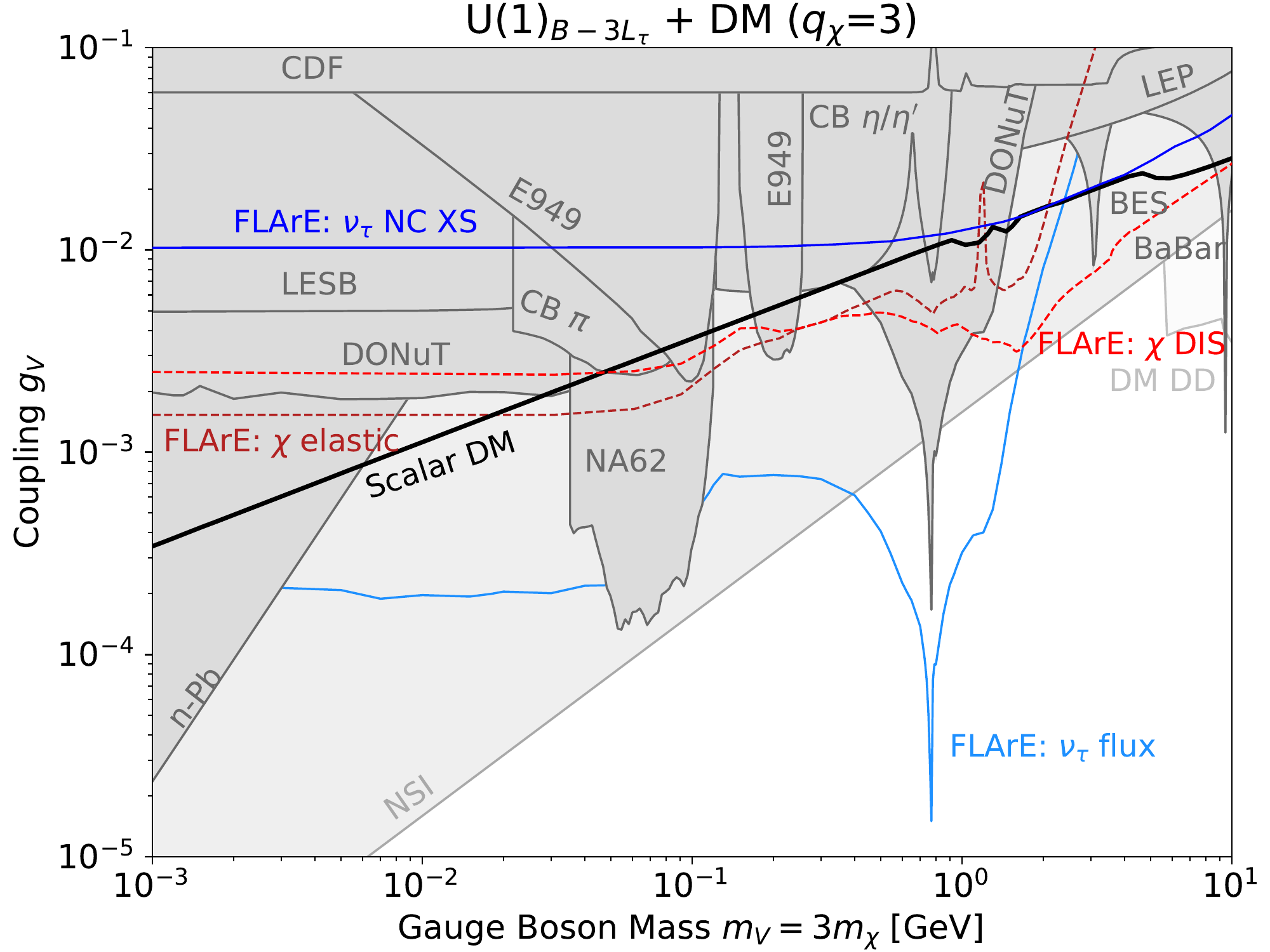}
\caption{Left: Angular distributions of tau neutrinos. We show show the rate of neutrino interactions per unit volume, normalized to its value at the beam axis, as a function of the displacement from the beam axis. The thick line corresponds to tau neutrinos from the decay of a $U(1)_{B-3L_\tau}$ gauge boson with mass 10~MeV while the thin line corresponds to tau neutrinos from $D_s$ meson decay. The gray arrows indicate the radial coverage of the FPF neutrino experiments. Modified version of figure extracted from~\cite{Kling:2020iar}. Right: Sensitivity of the FPF experiments to the vector boson of a gauged $B-3L_\tau$ symmetry. The solid blue contours show the sensitivity of FLArE using both the neutral current cross section measurement (upper line) and the tau neutrino flux measurement (lower line). In the presence of an additional dark state $\chi$ charged under the symmetry, one can also look for the corresponding scattering elastic and DIS scattering signature. Existing constraints from direct searches are shown in dark gray, while indirect constraints from scattering NSI and dark matter direct detection measurements are shown in lighter gray. Modified version of figure extracted from~\cite{Batell:2021snh}. See \cref{sec:llp_vec_b3l} and \cref{sec:bsm_dm_hadro} for more details on the model. 
}
\label{fig:neut_u1bm3ltau}
\end{figure}

In the context of the FPF, the presence of the new vector boson has several phenomenological consequences, that we can utilize to constrain this model
\begin{description}
    \item [Additional Tau Neutrino Production] In the SM, the tau neutrino flux manly originates from the decay $D_s \to \tau \nu_\tau$ and the subsequent $\tau$ lepton decay. Even at the LHC with is its enormous center of mass energy, a tau neutrino is only produced in roughly in roughly one in a hundred thousand collisions. This means that even rarely occurring new physics processes can lead to sizable contribution to the tau neutrino flux, which could be observed at the FPF through the measurement of the tau neutrino flux.
    In our case, the new gauge boson $V$ can be abundantly produced at the LHC via its coupling to the baryonic current. For sufficiently light masses it can be produced in the decay of mesons, such as $\pi^0, \eta \to \gamma V$. For heavier masses, it could be produced via bremsstrahlung of the proton beam, or in hard processes such as $ \bar{q} q \to V$. After production, the particle will quickly decay into tau neutrinos (unless $Q_\chi \gg Q_\tau$, in which it would predominantly decay into dark matter), providing an additional source of the tau neutrino flux. 
    
    Notably, the angular spectra of tau neutrinos from $V$ and $D_s$ decay can be quite different. The SM beam of tau neutrinos from $D_s$ decay has a rather broad due to the high transverse momentum gained from its decay. In contrast, the beam of tau neutrinos from a light $V$ decay can be much more narrow, reflecting the kinematics of the light parent particle. This is illustrated in the left panel of \cref{fig:neut_u1bm3ltau}. As described in Ref.~\cite{Kling:2020iar} and Ref.~\cite{Batell:2021snh}, this feature provides a useful handle that can be used to disentangle the different flux components. Here one defines a control region at higher displacements to normalize the SM tau neutrino flux, and searches for a relative excess at the center of the beam. 
    \item[Tau Neutrino Neutral Current Scattering] The presence of a new vector boson that couples both to tau neutrinos and baryons can also modify the tau neutrino neutral current scattering cross section. As discussed in Ref.~\cite{Batell:2021snh}, the additional $V$ exchange diagram leads to an enhancement of the rate for neutral current events with low recoil energies in the few GeV range. However, neutral current scattering of the larger electron and muon fluxes provide a sizable SM background to this signature. One therefore needs a rather large enhancement of the tau neutrino scattering rate in order to observe a visible effect.
    \item[DM Scattering] Similarly to the tau neutrinos, light dark matter state $\chi$ can scatter on nuclear matter via an exchange of the $V$ boson, leading to an enhancement of neutral current like events with low recoil energy. For more details, see \cref{sec:bsm_dm_hadro}. 
\end{description}

The sensitivity reach of the FLArE experiment for all three signatures is shown in the right panel of \cref{fig:neut_u1bm3ltau}. Here we following the convention of \cref{sec:bsm_dm_hadro}, and consider $Q_\chi = 3$. In this case, the branching fraction of a light gauge boson $V$ into tau neutrinos is about 90\%, while the branching fraction into DM is about 10\%. The dark gray shaded region corresponds to the direct searches for the gauge boson $V$, while the light gray region indicates constraints from NSI measurements and dark matter direct detection searches. For details on those constraints, see \cref{sec:llp_vec_b3l} and \cref{sec:bsm_dm_hadro}.  

The most sensitive signature is the modification of the tau neutrino flux through the decay $V \to \bar{\nu}_\tau \nu_\tau$, shown by the solid light blue line. It provides significant new constraints, especially around the $m_V \sim 770~\mev$ mass, where $V$ mixes resonantly with the $\omega$ meson. The tau neutrino neutral current scattering signature, shown as solid dark blue line, is only sensitive to very large couplings. It exceeds the direct constraints at masses $m > 1~\gev$. Additional constraints can also be put by searching for light DM scattering via a $V$ mediator, as shown by the red dashed lines.

\subsection{Neutrino Magnetic Moments}
\label{sec:bsm_neut_mdm}

Since it is now established that neutrinos have at least two mass eigenstates with non-zero masses, we know that they must carry magnetic dipole moments and, if additionally the time-reversal symmetry is violated, electric dipole moments. The Standard Model value of the neutrino magnetic moment is very small. Since neutrino mass differences and the mixing angles are reasonably well measured,  Ref. \cite{Balantekin:2013sda} calculates that the Dirac neutrino magnetic moment can be as low as $\sim 10^{-20} \mu_B$ for the inverted hierarchy and even lower for the normal hierarchy. However, beyond the Standard Model physics can result in much larger values of the magnetic moment. Additionally, they could explain various anomalies, such as the XENON1T excess~\cite{XENON:2020rca}, the observation of black holes in the mass gap region~\cite{Sakstein:2020axg}, and the MiniBooNE excesses~\cite{MiniBooNE:2007uho}.

\begin{figure}
    \centering
    \includegraphics[width=0.49\textwidth]{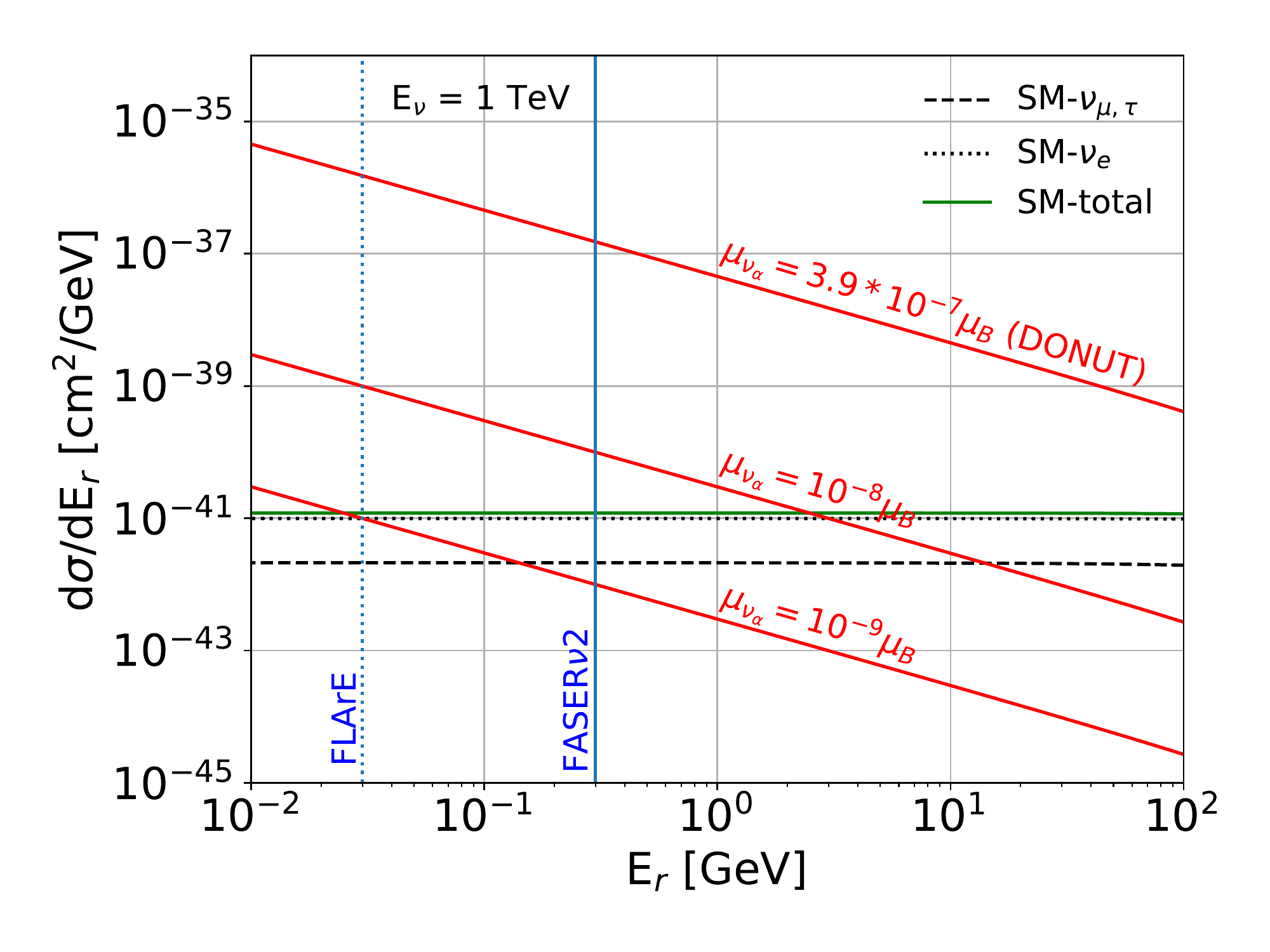}
    \includegraphics[width=0.49\textwidth]{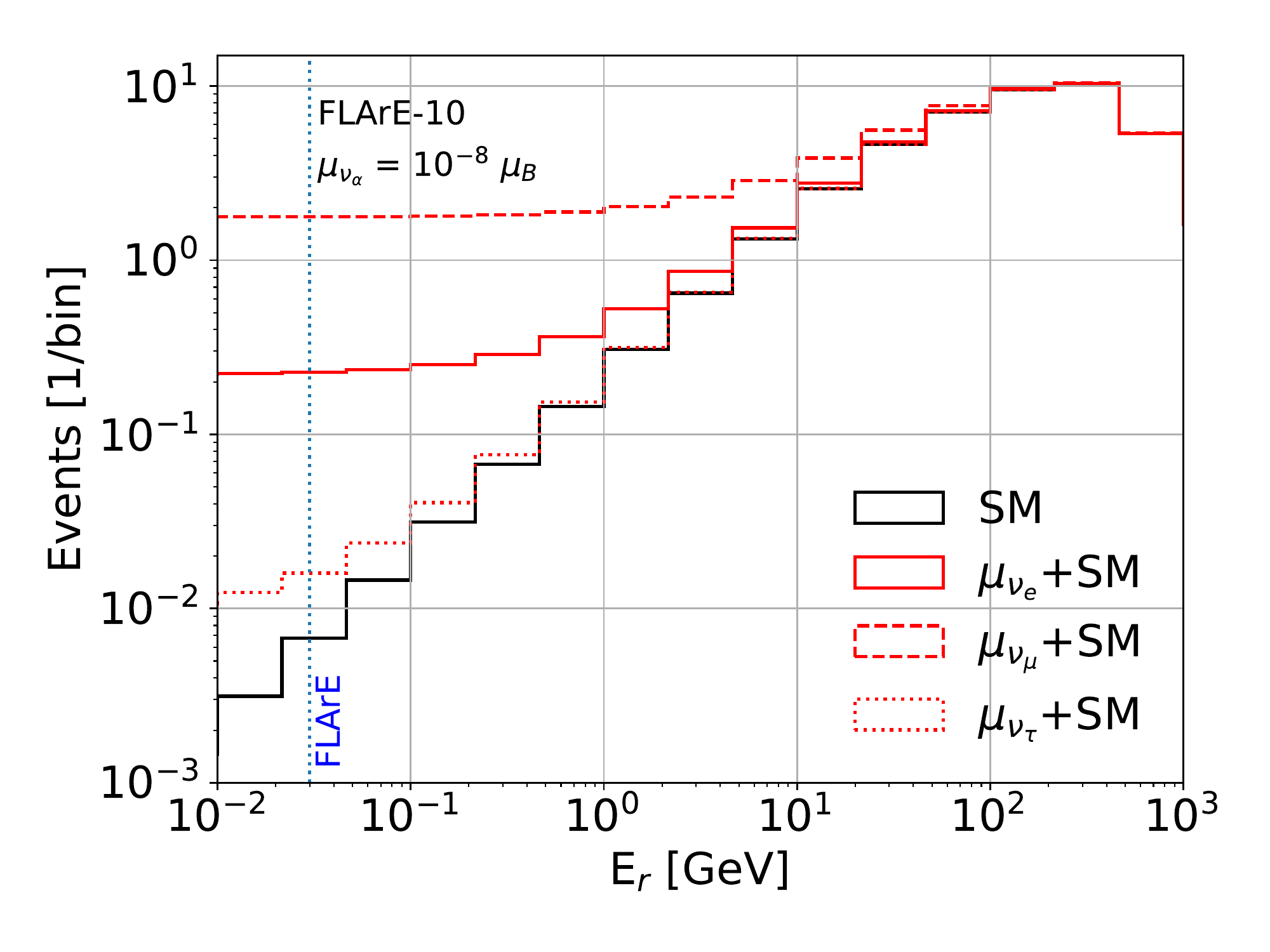}

    \caption{Left: The differential cross-section for the SM background components (black), total SM background (green), and signal (red) for an incoming neutrino with energy 1 TeV. The signal line is shown for two benchmark values of $\mu_{\nu_\alpha} = 10^{-8}\mu_B$, and $10^{-9}\mu_B$. The cross-section line for the DONUT bound of $\mu_{\nu_\tau}=3.9 \cdot 10^{-7} \mu_B$~\cite{DONUT:2001zvi} is also shown. The detector thresholds are shown for FASER$\nu$2 (dotted blue) and FLArE (solid blue).
    Right: Expected number of events at FLArE-10 detector for SM (black), and SM + BSM (red) scenario where each of the neutrino magnetic moments are set to $10^{-8}\mu_B$ one at a time. Figure taken from~\cite{AbariAbrahamTsaiKling}.
    }
    \label{fig:nmm_fpf}
\end{figure}

The cross section for the elastic scattering $\nu e^- \to \nu e^-$ of a neutrino (or antineutrino) of energy $E_{\nu}$ with an electron in the presence of a magnetic moment is given by
\begin{equation}
\frac{d\sigma_{\nu e}}{dE_{r}} = 
\bigg( \frac{d\sigma_{\nu e}}{dE_{r}}\bigg)_{\rm SM}
+ \frac{\pi \alpha ^2}{m_e ^2}  \bigg(\frac{1}{E_{r}} {-} \frac{1}{E_\nu}\bigg)\bigg(\frac{\mu_{\nu}}{\mu_{\rm B}}\bigg)^2 \ ,
\end{equation}
where $\mu_{\nu}$ is the effective neutrino magnetic moment, and $E_r$ is the recoil kinetic energy of the electron. We can see that the magnetic moment contribution to the neutrino-electron scattering cross section grows as the electron recoil energy decreases. This is also illustrated in the left panel of \cref{fig:nmm_fpf}. Hence laboratory measurements are limited by the lowest electron recoil energy that can be measured. Neither the reactor experiments nor the solar neutrino experiments observe deviations from the expected shape of the electron recoil spectrum and put a limit of $\mu_{\nu} < 2.9 \times 10^{-11} \mu_B$ \cite{Beda:2013mta,Borexino:2017fbd}. Astrophysical considerations provide somewhat better limits although they may be subject to systematic uncertainties. These studies typically invoke additional energy loss channels from stars where neutrinos are pair produced in the plasma because of the sufficiently large values of neutrino dipole moments and hence carry away energy from the star. The luminosities of the stars at the tip of the red giant branch are especially sensitive to energy losses. Comparing theoretical luminosities to the color-magnitude diagram of globular clusters the most stringent limit obtained is $\mu_{\nu} < 1.2-1.5 \times 10^{-12} \mu_B$ \cite{Capozzi:2020cbu}. Additionally if the energy losses are large enough stars cannot evolve to Cepheid stars. The existence of Cepheid stars require a magnetic moment of less than $4\times 10^{-11} \mu_B$ \cite{Mori:2020qqd}. It should be noted that neither those laboratory experiments nor the astrophysical considerations can distinguish neutrino electric dipole moment from the magnetic moment.

Some of the detectors considered at FPF have sufficiently low energy thresholds to study neutrino magnetic moments via neutrino-electron scattering~\cite{AbariAbrahamTsaiKling}. FASER$\nu$2 has an energy threshold of 300 MeV, and FLArE-10(100) has a lower threshold of 30 MeV. This makes it possible to make use of the cross-section enhancement at low electron recoil energies. Assuming the strength of the neutrino magnetic moment to be flavor dependent ($\mu_{\nu_\alpha}$), one can switch them on one by one to study them individually. The right panel of \cref{fig:nmm_fpf} shows the expected number of events at FLArE-10 for the SM and BSM cases. By placing an upper cut on the electron recoil energy of 1 GeV, one can limit $\mu_{\nu_e} < 1.4\cdot 10^{-8} \mu_B$, $\mu_{\nu_\mu} < 5.1\cdot 10^{-9} \mu_B$, and $\mu_{\nu_\tau} < 7.0\cdot 10^{-8} \mu_B$ at FLArE-10. The larger FLArE-100 can limit $\mu_{\nu_\tau} < 2.2\cdot 10^{-8} \mu_B$ which is an order of magnitude smaller than the DONUT~\cite{DONUT:2001zvi} limit.

Majorana versus Dirac nature of the neutrinos also impact their dipole moments. For Dirac neutrinos dimension 6 operators contribute both to the neutrino mass and neutrino magnetic moment. Renormalization group analysis can be used to calculate the induced radiative correction to the Dirac neutrino mass coming from the magnetic moment \cite{Bell:2005kz}. For Majorana neutrinos magnetic moments and corrections to the dimension-5 Majorana mass come from dimension 7 operators. An additional constraint comes from the fact that Majorana neutrinos cannot have diagonal magnetic moments. Hence constraints on Majorana neutrino magnetic moments are weaker as demonstrated by one-loop \cite{Davidson:2005cs} and two-loop \cite{Bell:2006wi} calculations. These results imply that if the neutrino magnetic moment is measured to be slightly below the current limits, then the neutrinos are likely to be Majorana particles. 

\subsection{Up-scattering through the Neutrino Dipole Portal}
\label{sec:bsm_neut_dipole}

The potential electromagnetic properties of neutrinos have been investigated since before the discovery of neutrinos themselves. In particular, neutrino magnetic moments have been considered recently in the context of several experimental anomalies, notably those from XENON1T~\cite{XENON:2020rca} and MiniBooNE~\cite{MiniBooNE:2018esg}. Theoretical and experimental investigations of neutrino magnetic moments are important to improve our understanding of the neutrino sector. In most extensions of the Standard Model (SM) that account for neutrino mass generation, neutrinos gain magnetic moments through loop effects \cite{Fujikawa:1980yx, Shrock:1982sc, Giunti:2014ixa, Babu:2020ivd}. Here, we consider the possibility of a dipole portal between SM neutrinos and a sterile neutrino $N_R$ arising from an interaction involving such a magnetic moment, with strength denoted by $\mu_{\nu_\alpha}$ where $\alpha$ is the flavor of the SM neutrino. The LHC provides an intense source of high energy (anti)neutrinos, making the FPF~\cite{Anchordoqui:2021ghd} well-suited to search for neutrino magnetic moments at neutrino energies far higher than previously explored. Specifically, the up-scattering of SM neutrinos to $N_R$ via the dipole portal can be studied using neutrino-electron scattering in potential FPF detectors~\cite{Ismail:2021dyp}.

To date, experiments using reactor, accelerator, and solar neutrinos have searched for evidence of neutrino magnetic moments in the process $\nu e^- \to \nu e^-$. While this reaction normally proceeds through weak boson exchange, a nonzero neutrino magnetic moment would allow for an additional photon exchange contribution. Unlike the $V - A$ weak interaction, the magnetic dipole interaction flips chirality, so the two contributions to neutrino-electron scattering do not interfere. However, the magnetic moment contribution is enhanced at low momentum transfer because the photon is massless. Consequently, detectors with low thresholds for the scattered electron can be quite sensitive to neutrino magnetic moments.

Now, magnetic moments between the active neutrinos face significant astrophysical constraints from stellar cooling~\cite{Bernstein:1963qh,Raffelt:1999tx} . We thus choose to focus on interactions between the active neutrinos and a new sterile state $N_R$, of the form
\begin{equation}
    \mathcal{L} \supset \frac{1}{2} \mu_\nu^\alpha \bar{\nu}_L^\alpha \sigma^{\mu\nu} N_R F_{\mu\nu}
\end{equation}
where $\alpha$ is a flavor index and $\mu_\nu^\alpha$ is the strength of the transition magnetic moment between the neutrino $\nu^\alpha$ and $N_R$. In principle, this interaction can arise from higher dimensional $SU(2)$-invariant operators involving the SM left-handed lepton doublets, Higgs field, the $N_R$, and an electroweak field strength tensor. However, here we remain agnostic as to the origin of the magnetic dipole interaction.

The magnetic dipole coupling leads to a new contribution to neutrino-electron scattering where a SM neutrino can up-scatter into the new sterile state~\cite{Brdar:2020quo,Shoemaker:2020kji},
\begin{equation}
\frac{d\sigma(\nu_L ^\alpha e^- \to N_R e^-)}{dE_{rec}} = \alpha \left(\mu _\nu ^\alpha\right) ^2  \big[\frac{1}{E_{rec}} - \frac{1}{E_\nu} + M_N ^2\frac{E_{rec}-2E_\nu -M_e}{4E_\nu ^2 E_{rec} M_e} + M_N ^4\frac{E_{rec}-M_e}{8E_\nu ^2 E_{rec} ^2 M_e ^2}\big] \ .
\end{equation}
where $E_\nu$ is the energy of the incoming neutrino, $E_{rec}$ is the energy of the recoiling electron and $M_N$ is the mass of the $N_R$. If the $N_R$ is long-lived, the primary signature would be a single electron track with no other visible activity. At the FPF, $E_\nu$ is typically at the TeV scale, which allows for production of sterile neutrinos up to $\mathcal{O}(1)~\mathrm{GeV}$ through the dipole interaction. By contrast, the recoiling electron can be detected down to energies of order 100~MeV. The first term in the differential cross section thus comes to dominate, leading to an excess of low-energy electrons relative to SM neutrino-electron scattering. For comparison, the analogous differential cross section for the weak scattering process $\nu e^- \to \nu e^-$ is approximately independent of $E_{rec}$ in the kinematic region of interest.

\begin{figure}
    \centering
    \includegraphics[width=0.49\textwidth]{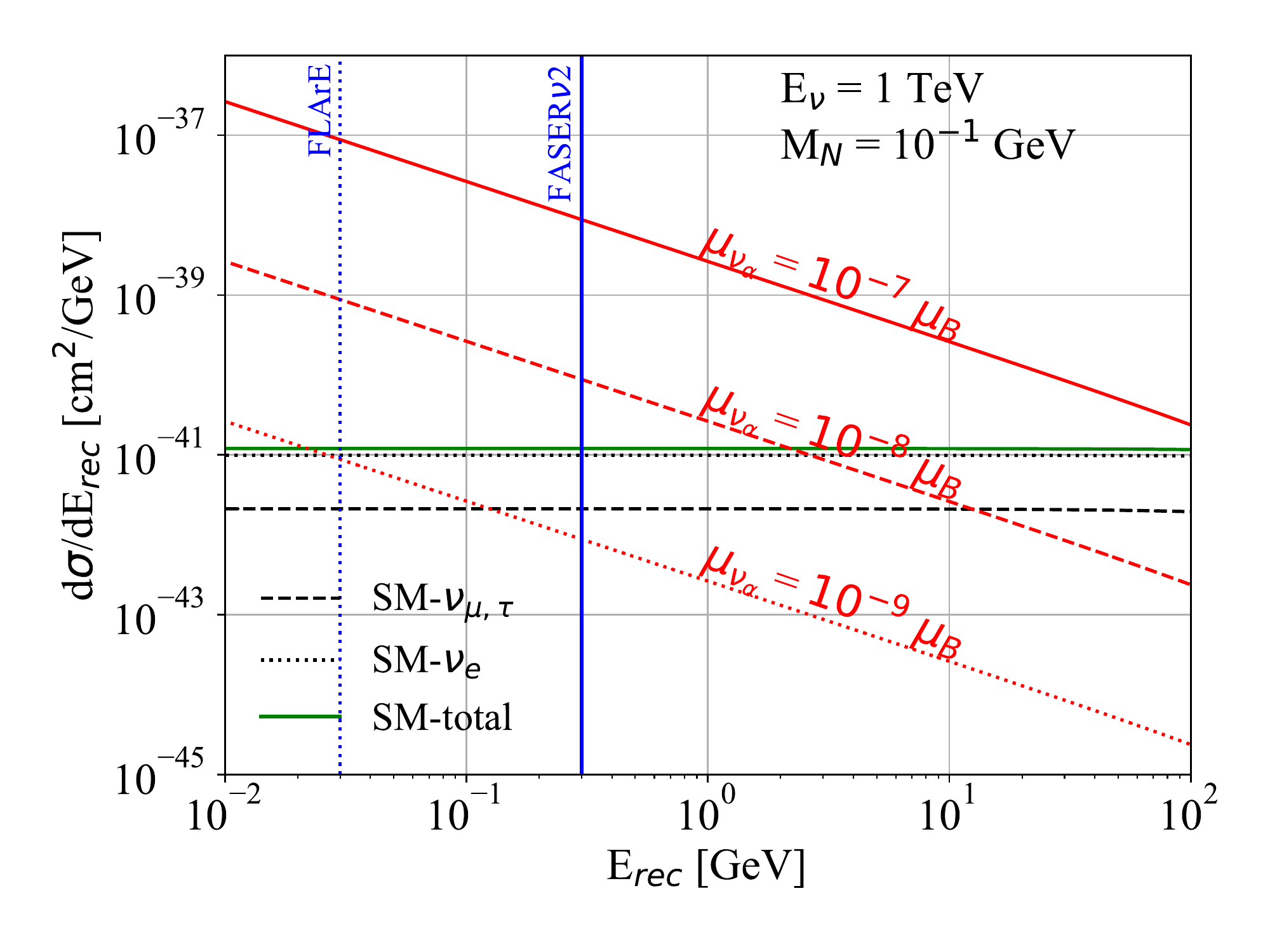}
    \includegraphics[width=0.49\textwidth, height=0.375\textwidth]{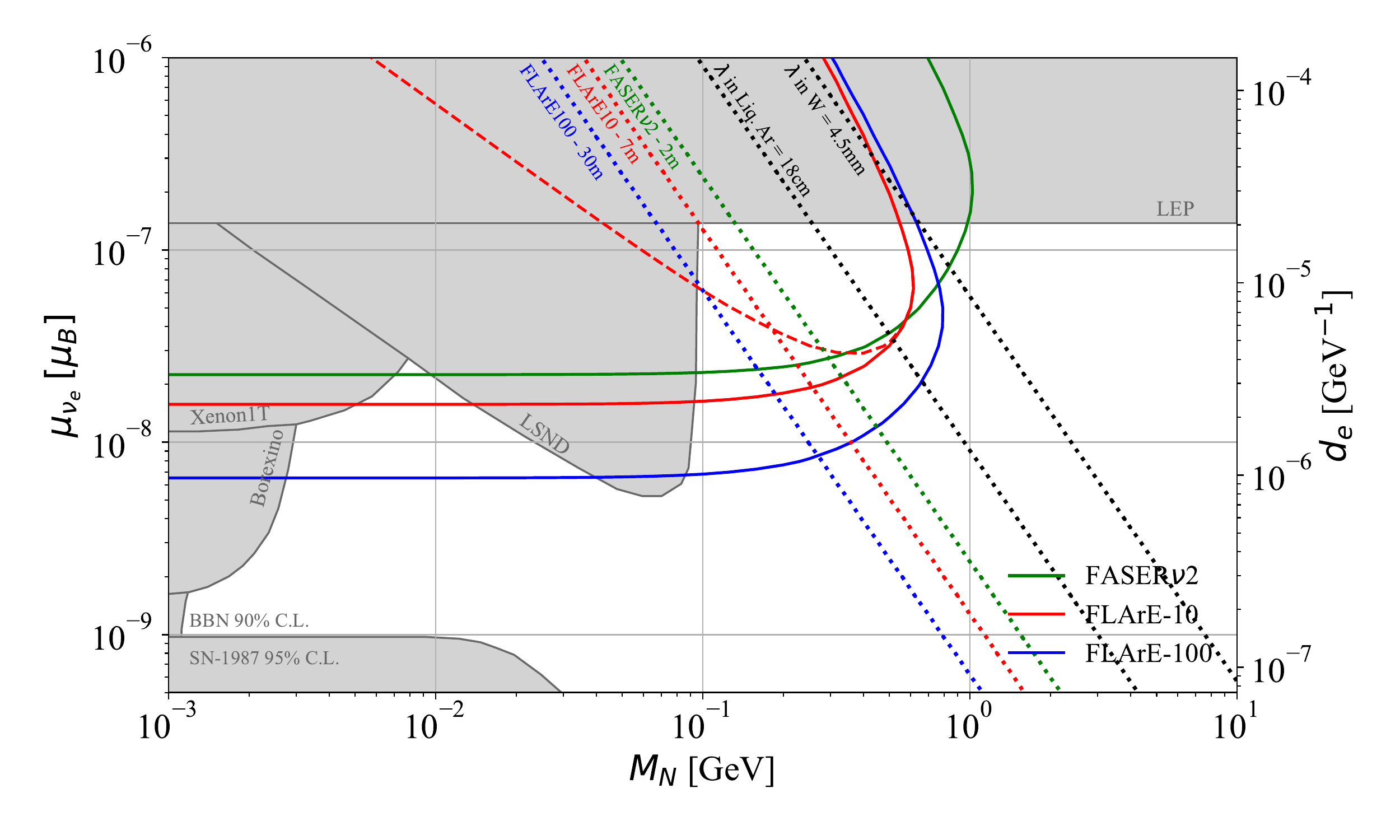}
    \caption{Left: $d\sigma / dE_{rec}$ for the SM background components (black), total SM background (green), and signal (red) with $E_\nu=1$ TeV and $M_N=10^{-1}$ GeV. Various benchmark values for $\mu_{\nu_\alpha}$ are shown. The detector thresholds at FASER$\nu$2 (FLArE) of 300 (30) MeV are shown using the solid (dotted) vertical blue line. The signal cross-section is enhanced at low recoil energies making FLArE a more promising detector with its lower energy threshold. Right: The estimated sensitivity at 90\% C.L for $\mu_{\nu_e}$ are shown at FASER$\nu$2 (solid green), FLArE-10 (solid red), and FLArE-100 (solid blue) for a total integrated luminosity of 3 ab$^{-1}$. The existing constraints, shown as the gray shaded region, are taken from~\cite{Schwetz:2020xra}. The dotted lines show the constant decay lengths of an $N_R$ with an energy of 100 GeV in the lab frame. The red dashed line shows the reach from considering only double bang events at FLArE-10, assuming its background free. Figures taken from Ref.~\cite{Ismail:2021dyp}.
    }
    \label{fig:dipolenmm_results}
\end{figure}

We convolute the cross section expression above with the neutrino fluxes expected at the FPF~\cite{Kling:2021gos}  to determine the impact of a transition magnetic moment on the electron recoil energy spectrum in neutrino-electron scattering. As the SM scattering is independent of recoil energy, we place an upper cut of $E_{rec} < 1~\mathrm{GeV}$. The minimum observable $E_{rec}$ is determined by detector considerations. We consider both an emulsion-based detector FASER$\nu$2~\cite{FASER:2019dxq} with a threshold of $E_{rec} > 300~\mathrm{MeV}$, and a liquid argon detector FLArE~\cite{Batell:2021blf} which could attain a smaller threshold $E_{rec} > 30~\mathrm{MeV}$.
As our primary signature is a single recoiling electron, the main background is SM neutrino-electron scattering, assuming that muons can be vetoed through active timing~\cite{Anchordoqui:2021ghd} . The left panel of \cref{fig:dipolenmm_results} shows the differential cross sections for both our signal and the SM background, along with our assumed detector thresholds, for incoming 1 TeV neutrinos and $M_N = 0.1~\mathrm{GeV}$. The expected number of SM events satisfying our cuts on the electron recoil energy is less than 1 with full HL-LHC luminosity~\cite{Ismail:2021dyp}. In principle, there are other backgrounds coming from $\nu_e$ charged current scattering off nuclei. Given other studies of neutrino-electron scattering at FPF detectors~\cite{Batell:2021blf}, however, we expect that with our $E_{rec}$ cut these backgrounds would be even smaller than those from electron scattering.

Using our signal and background estimates, our predictions for the FPF reach for neutrino magnetic moments are shown in the right panel of \cref{fig:dipolenmm_results}. These assume a dipole interaction between the electron neutrino and $N_R$, though analogous limits can be obtained for magnetic moments involving the SM neutrinos of other flavors~\cite{Ismail:2021dyp}. The solid lines show the improvement possible when going from a 10-ton emulsion detector, FASER$\nu$2, to an argon detector of similar mass, FLArE-10. 
The additional reach which could arise from a 100-ton version of FLArE is 
{also} shown. The FPF will be able to probe higher $N_R$ masses than other neutrino experiments because of the considerable energy of the incoming neutrinos.

Finally, we note that in most of the viable parameter space, the $N_R$ is long-lived. However, for sufficiently large $\mu_\nu$ or sterile neutrino mass $M_N$, the $N_R$ can be produced \emph{and} decay within a given FPF detector through the magnetic dipole interaction~\cite{Magill:2018jla,Jodlowski:2020vhr}. In the right panel of \cref{fig:dipolenmm_results}, we show contours of constant $N_R$ lifetime as dotted lines, assuming a 100~GeV $N_R$ energy in the lab frame. The contours correspond to possible FPF detector sizes as well as to the photon mean free paths in tungsten and argon. The latter are approximate estimates of the minimum lifetime necessary for the $N_R$ production and decay to appear displaced. The signature of the decay $N_R \to \nu \gamma$ itself would be a single photon, which would be displaced from the $N_R$ production point from which an electron track would begin. Such a ``double bang'' event would be a striking signature, which is likely to have exceedingly low background at the FPF as at other experiments such as IceCube~\cite{Coloma:2017ppo} 
and DUNE~\cite{Schwetz:2020xra,Atkinson:2021rnp}. The approximate reach from 
considering only background free double bang events 
at FLArE-10 is shown using 
the red dashed line in the right panel of \cref{fig:dipolenmm_results}.

In summary, magnetic dipole interactions between the SM active neutrino flavors and new sterile states can be probed at the FPF. The up-scattering process $\nu e^- \to N_R e^-$
via the dipole portal is enhanced at low electron recoil energy, motivating searches for the recoiling electron from $N_R$ production 
within the detector.
 The TeV neutrinos at the LHC allow for sterile states with masses up to a GeV to be produced, exceeding the kinematic reach of existing neutrino experiments. FPF detectors with low thresholds for electromagnetic showers such as FLArE would be especially well situated to test such transition magnetic moments.

\subsection{FASER/FPF Sterile Neutrino Oscillations}
\label{sec:bsm_neut_sterile}

The flux of broadband neutrinos from the LHC also provides a novel opportunity to look for neutrino oscillations using FASER$\nu$ \cite{FASER:2019dxq, FASER:2020gpr} and SND@LHC \cite{SHiP:2020sos} as well as FPF experiments.
Given a baseline of 600 m and energies in the 100 GeV to 1 TeV range, any seen oscillations would be due to a frequency $\Delta m^2_{41}\sim1000$ eV$^2$, or $m_4\sim30$ eV; thus any seen oscillations would imply a new sterile neutrino.
While existing constraints apply at this mass range in the oscillation averaged limit, there are no direct oscillation probes at this $L/E$ providing a new test of oscillations at larger $\Delta m^2$'s than are usually considered.
Neutrino oscillation searches at the LHC also benefit from the production of all three flavors of neutrinos with hierarchical production rates that are separated by $\gtrsim1$ order of magnitude each.
In addition to having all three flavors available at the source, existing and planned detectors should have full flavor discrimination capabilities allowing for, in principle, probes of all 9 oscillation channels, subject to backgrounds and flux uncertainties.

The flux uncertainties, which affect the normalization and, more importantly, the shape, see \cref{sec:flux}, represent the dominant uncertainty for neutrino oscillation searches.
The relative contribution to the neutrino flux from different particles is rather poorly understood \cite{Kling:2021gos} and these shape effects could conceivably mimic a neutrino oscillation signature \cite{Bai:2020ukz}.
Thus more theoretical work is needed to understand these fluxes to probe neutrino oscillations.

\begin{figure}
\centering
\includegraphics[width=5in]{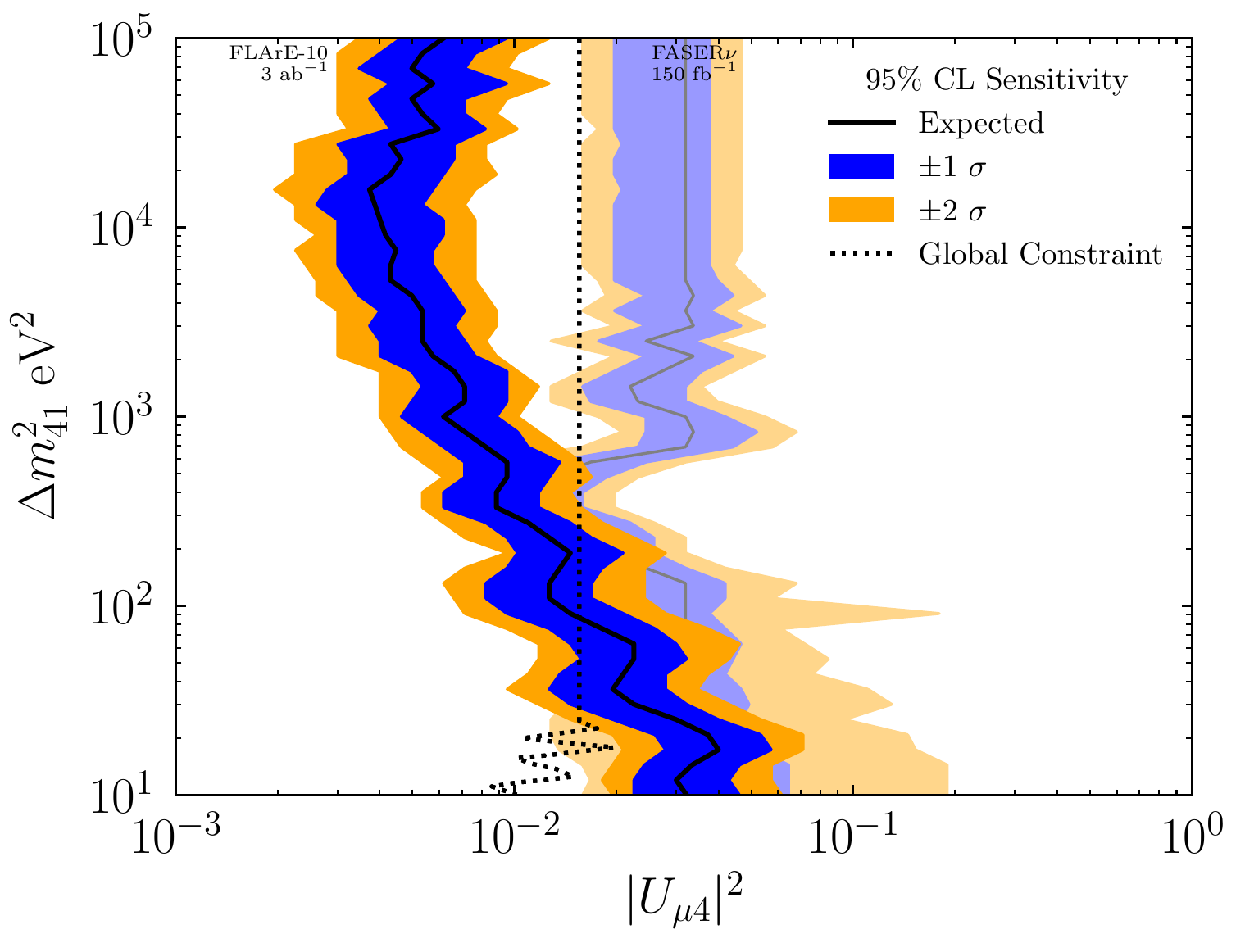}
\caption{The expected sensitivity to neutrino oscillations in the $\nu_\mu$ disappearance channel at the LHC using Feldman-Cousins and Asimov for FASER$\nu$ at the upcoming LHC run and a proposed 10-ton LAr detector in a future HL-LHC run.
The existing oscillation averaged constraints coming mostly from MINOS+ and MiniBooNE are also shown.
Figure is from \cite{Anchordoqui:2021ghd}.}
\label{fig:sterile_fpf}
\end{figure}

Nonetheless, it is possible to estimate the sensitivity to sterile neutrino oscillations at FASER$\nu$2 (see \cref{sec:fasernu2}) and FLArE (see \cref{sec:FLArE}), a proposed LAr detector in the forward direction at the LHC \cite{Anchordoqui:2021ghd}.
The most sensitive channel relative to existing constraints is the $\nu_\mu$ disappearance channel which shows sensitivity at the $\Delta m^2_{41}\sim100-1000$ eV$^2$ range down to mixings of $|U_{\mu4}|^2<10^{-2}$ which are better than existing constraints from Ref.~\cite{Dentler:2018sju} which are dominated by MINOS/MINOS+ \cite{MINOS:2017cae} and MiniBooNE \cite{MiniBooNE:2012meu}.
The sensitivity for FASER$\nu$ in the upcoming LHC run at 150 fb$^{-1}$ and FLArE-10 (10 fiducial tons of LAr) with 3 ab$^{-1}$ in the HL-LHC is shown in \cref{fig:sterile_fpf}.
The flux uncertainty is accounted for in a fairly conservative fashion by including an estimate of the impact of shape effects by varying the flux across the range of predictions from different models \cite{Kling:2021gos} with an associated $1\sigma$ pull term.
This sensitivity is then calculated using the Feldman-Cousins procedure \cite{Feldman:1997qc} at 95\% CL including the flux systematic uncertainty.
The median sensitivity the range of sensitivities is estimated with the Asimov method.

It is anticipated that LHC neutrino experiments could additionally have sensitivity to sterile oscillations for the other channels too, although this depends on the details of the flux uncertainties.

\subsection{Neutrinophilic Mediator/Dark Matter Production at the FPF}
\label{sec:bsm_neut_neutphil}

\begin{figure}[t]
\centering
    \includegraphics[width=0.65\linewidth]{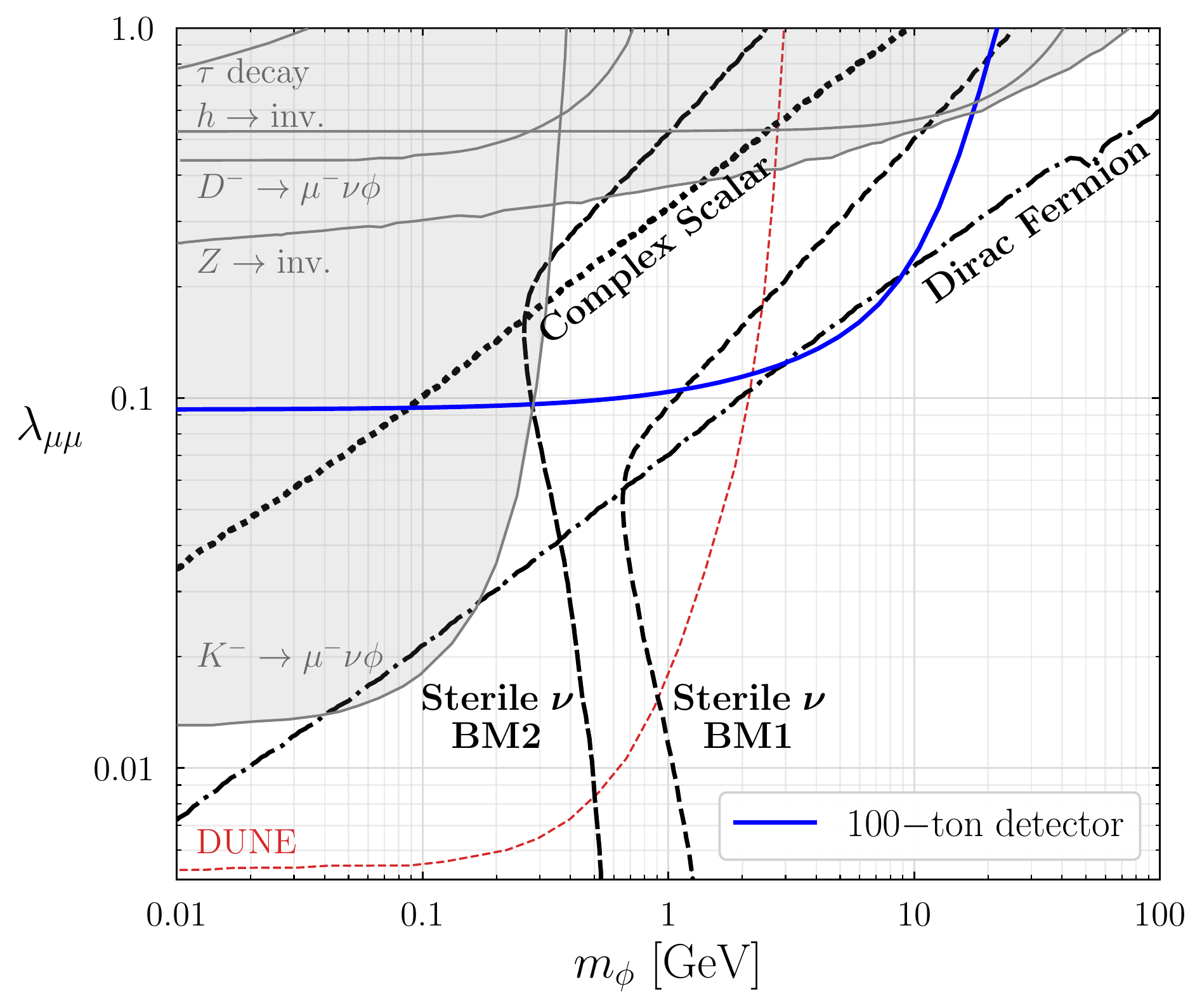}
    \caption{Constraints on $\lambda_{\mu\mu}-m_\phi$ parameter space. The gray shaded regions are from $\tau$, $D$-meson, kaon and invisible $Z$ and Higgs boson decays. The expected sensitivity of a 100 ton detector with a 15\% hadronic energy resolution (blue) and expected DUNE sensitivity (red) are shown.}
    \label{fig:DT_Bounds}
\end{figure}

New force carriers that couple predominantly to SM neutrinos are well-motivated candidates for BSM physics. These ``neutrinophilic mediators'' are predicted in extensions of the SM that are related to neutrino mass generation, and can also be connected to the DM in our universe~\cite{SelfIntNeutWhitePaper}. As a benchmark model, we consider a massive scalar $\phi$ that interacts with SM muon-neutrinos via following effective Lagrangian
\begin{equation}\label{eq:Lnuphilic}
\mathcal{L} \supset \frac{1}{2} \lambda_{\mu\mu} \nu_\mu \nu_\mu \phi + \text{h.c.},
\end{equation}
Such an operator is not gauge invariant under the SM $SU(2)_L$ but could arise from a dimension-six or higher operator. Existing constraints on the parameter space of $\phi$ that couples to muon-neutrinos is shown in \cref{fig:DT_Bounds}, where the gray shaded regions are constraints from leptonic $\tau$ decays, $D$ meson and kaon decays, and invisible decays of the $Z$ and Higgs bosons. 

The presence of the neutrinophilic force mediated by $\phi$ can also be used to address the origin of DM in our universe. First, the scalar $\phi$ can serve as a portal between the SM and a fermionic or scalar DM candidate whose relic abundance is obtained by thermal freeze out via annihilation to SM neutrinos~\cite{Kelly:2019wow}. Second, sterile neutrino DM (S$\nu$DM) can be produced via neutrino self-interactions mediated by $\phi$, as shown in~\cite{deGouvea:2019phk,Kelly:2020aks,Du:2020avz}. These DM targets are depicted by the various black curves in \cref{fig:DT_Bounds}, with the masses and couplings of the different scenarios given in \cite{Kelly:2021mcd}.

 The new neutrino self-interaction in \cref{eq:Lnuphilic} are allowed to be much larger than in the SM and can manifest themselves in neutrino experiments. The neutrinophilic mediator can be produced via bremsstrahlung off the neutrino beam when the neutrino undergoes charge-current interactions with a detector. Once produce the neutrinophilic mediator will decay invisibly, to neutrinos or DM particles, and appears invisible to the detector. However, the presence of the mediator will lead to sizable missing transverse momentum (MET) which can be inferred by measuring the visible final states. In Ref.~\cite{Kelly:2019wow,Kelly:2021mcd} this signature was dubbed the ``mono-neutrino'' signal,  in analogy to the mono-$X$ searches widely performed at various colliders to probe WIMP-like DM.

The Forward Physics Facility is in a prime position to explore the parameter space of neutrinophilic mediators by looking for the mono-neutrino signature. Compared to traditional neutrino accelerator experiments, neutrinos at the LHC have higher energies (in the hundreds of GeV to TeV range) allowing searches for heavier mediators. At the same time, the scattering of the neutrino with the detector is deep-inelastic which has smaller uncertainties compared to GeV-scale neutrino beams. 

To leverage the capabilities of the FPF to probe the scenarios mentioned above, it is necessary to distinguish between the signal process  $\nu_\mu N \to \mu^+  \phi X$ from the background from the SM charged current process $\nu_\mu N \to \mu^- X$. The following kinematic variables can be used for comparing the signal and the background:
\begin{itemize}
    \item $\slashed{p}_T$, the missing transverse momentum, reconstructed from visible final state particles,
    \item $E_{\rm vis}$, the total energy of all visible particles,
    \item $p_T^\text{max}$, the highest transverse momentum of visible final state objects.
\end{itemize}

\begin{figure}[ht]
    \includegraphics[width=1.0\linewidth]{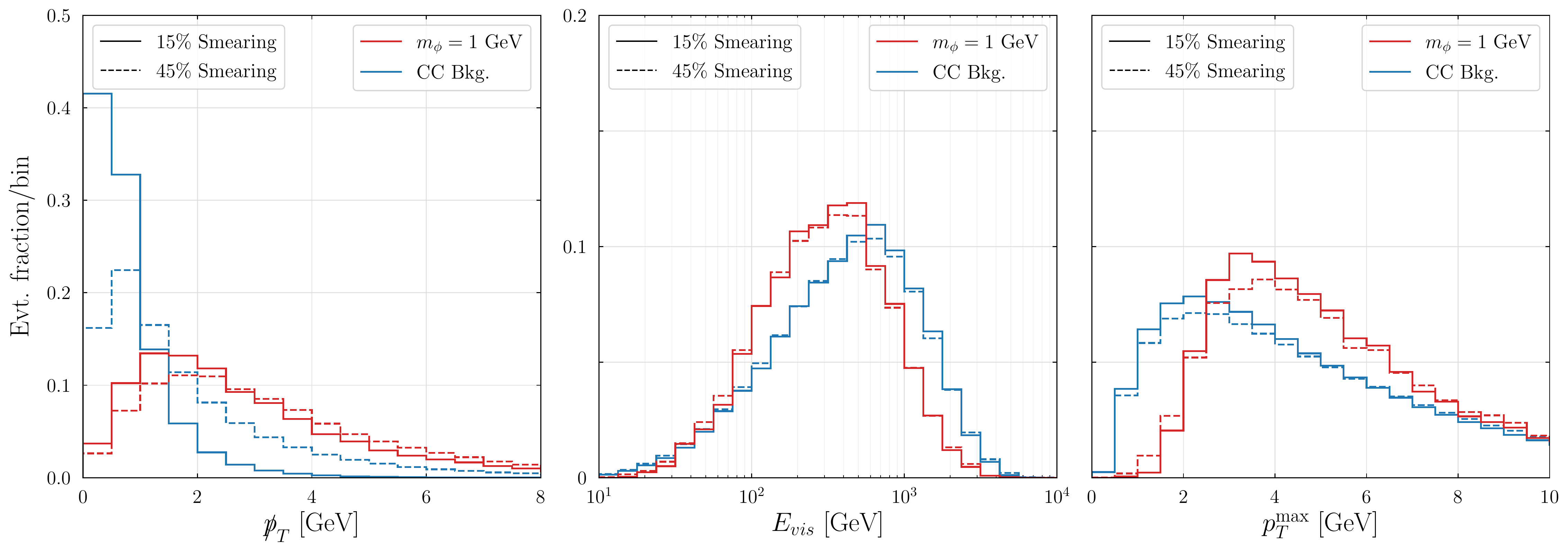}
    \caption{Distributions of missing transverse momentum, the total energy of visible particles and the highest transverse momentum of the visible final state are shown for the signal with $m_\phi=1$ GeV (red) and CC background (blue) for 15\%\ and 45\%\ hadronic energy resolution.Figure from Ref. \cite{Kelly:2021mcd}.}
    \label{fig:DT_1DDists}

\end{figure}

 \cref{fig:DT_1DDists} shows one-dimensional distributions of these kinematic variables for the signal (red, for $m_\phi=1$ GeV) and background (blue), where we assume that the FPF detector has a 15\% (solid curves) or 45\% (dashed curves) hadronic energy resolution. In both case we assume that the detector has a 5\% muon energy resolution and a perfect muon identification rate. We observe that signal has $\slashed{p}_T$ and $p_T^\text{max}$ distributions that are peaked at larger values compared to the SM background, and a $E_{\rm vis}$ distribution that is peaked at smaller values compared to the SM background.

 Using these kinematic variables, bounds on the coupling $\lambda_{\mu\mu}$ can be found by carrying out a simple cut-and-count analysis or by feeding these variables into a neural network to optimize the results. The main results of this analysis is depicted by the blue curve in \cref{fig:DT_Bounds} where we show the expected sensitivity of the FPF assuming a 100 ton detector with 15\% hadronic energy resolution. The sensitivity of of the FPF to $\lambda_{\mu\mu}$ is unmatched by existing constraints from charged meson decays above $m_\phi \approx 250$ MeV, and that the sensitivity also exceeds existing constraints from the invisible widths of the $Z$ and Higgs bosons for $m_\phi$ up to $\sim 20$ GeV. In addition, the FPF surpasses the expected DUNE sensitivity (red dashed curve) for $m_\phi \gtrsim 2$ GeV. The bounds in \cref{fig:DT_Bounds} also shows that a 100 ton detector at the FPF will have the potential to probe both the thermal freeze-out DM and S$\nu$DM targets in parameter space that is currently unconstrained by existing experiments. For more details on the simulations and analysis method used to derive the bounds we refer the reader to Ref. \cite{Kelly:2021mcd}.

%% file: sec_astro.tex
\contributors{Luis A. Anchordoqui, Dennis Soldin (conveners), Basabendu Barman, Atri Bhattacharya, Arindam Das, Hans P. Dembinski, Rikard Enberg, Carlos A. Garc\'{\i}a Canal, Anish Ghoshal, Srubabati Goswami, Andrzej Hryczuk, Yu Seon Jeong, Krzysztof Jodlowski, Spencer R. Klein, Felix Kling, Maxim Laletin, Tanmay Kumar Poddar, Mary Hall Reno, Leszek Roszkowski, Tim Ruhe, Ina Sarcevic, Sergio J. Sciutto, Jorge F. Soriano, Anna Stasto, Sebastian Trojanowski, and K.~N.~Vishnudath}

Many discoveries in the history of high-energy physics have been made through cosmic ray and cosmic neutrino observations. This includes, for example, the early landmark identification of new elementary particles, the confirmation of long-suspected neutrino oscillations, as well as measurements of cross-sections and particle interactions far beyond current collider energies. Two recent examples that demonstrate the synergies between astroparticle and high-energy physics are: 
\begin{itemize}
    \item[(i)] The measurement of the proton-proton cross-section at a center-of-mass energy of about $\sqrt{s} \sim 75~{\rm TeV}$~\cite{Collaboration:2012wt,Abbasi:2015fdr,Abbasi:2020chd}, providing evidence that the proton behaves as a black disk at asymptotically high energies~\cite{Block:2012nj,Block:2015mjw}.
    \item[(ii)] The measurements of the charged current neutrino-nucleon cross-section~\cite{Aartsen:2017kpd,IceCube:2020rnc,Bustamante:2017xuy} and the neutral to charged current cross-section ratio~\cite{Anchordoqui:2019ufu} at $\sqrt{s} \sim 1~{\rm TeV}$, which provide restrictive constraints on fundamental physics at sub-fermi distances. 
\end{itemize}

\begin{figure}[bt]
    \vspace{-1.em}
    
    \centering
    \includegraphics[width=0.82\textwidth]{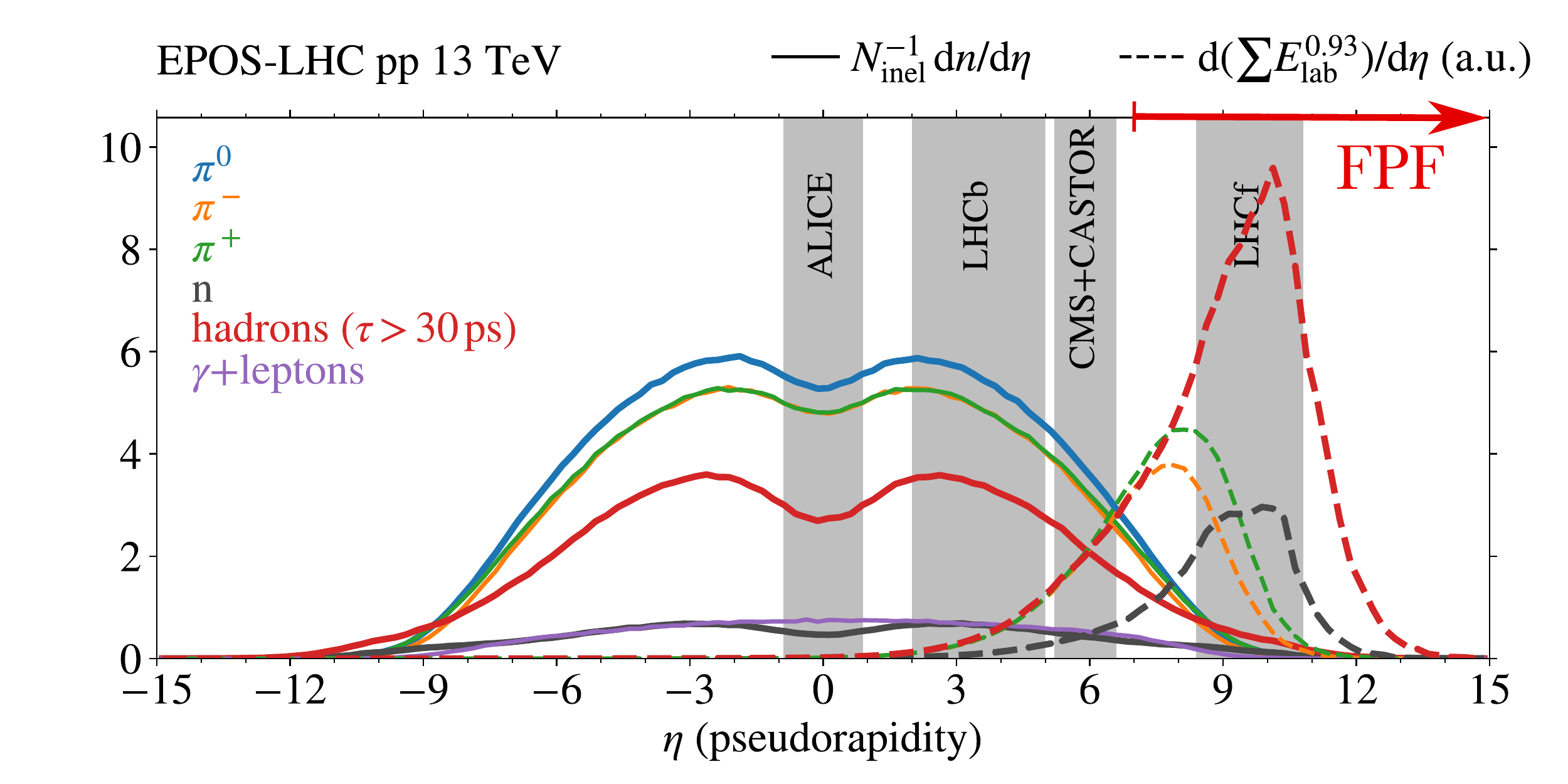}
    \vspace{-1.em}
    
    \caption{Simulated densities of particles (solid lines) in high-energy proton-proton collisions using \texttool{EPOS-LHC}. Dashed lines show the estimated number of muons produced by these particles if they were propagated through the atmosphere, assuming an equivalent energy for the fixed target collisions in the laboratory frame, $E_{\rm{lab}}$, and $N_\mu \propto E_{\rm{lab}}^{0.93}$. Figure taken from Ref.~\cite{Albrecht:2021cxw}.}
    \label{fig:eta-ranges}
    \vspace{-0.5em}
\end{figure}

Cosmic rays enter the Earth's atmosphere with energies exceeding $10^{11}\,\rm{GeV}$, where they interact with molecules in the air. These collisions produce particle cascades in the atmosphere, so-called \emph{extensive air showers} (EAS), which can be measured with large detector arrays at the ground and/or fluorescence detectors (for a detailed description, see \emph{e.g.} Refs.~\cite{Gaisser:2016uoy, UHECRWhitepaper}). \cref{fig:eta-ranges} shows simulated particle densities produced in proton-proton collisions (solid lines) compared to the pseudorapidity ranges for current LHC experiments \cite{Albrecht:2021cxw}. Also shown are the estimated number of muons, $N_\mu$, produced by these particles during propagation through the atmosphere, assuming $N_\mu \propto E_{\rm{lab}}^{0.93}$, where $E_{\rm{lab}}$ is the energy of the secondary particles in the laboratory frame (dashed lines). While the mid-rapidity ranges are only marginally relevant for the particle production in EAS, the forward region plays a crucial role and can be probed at the FPF.

The FPF will also provide key information for understanding astrophysical neutrinos~\cite{UHEnuWhitepaper} and in the context of multimessenger astronomy~\cite{MMWhitepaper}. At very high energies, above $\sim 100\,\mathrm{TeV}$, the Earth becomes opaque to neutrinos, and neutrino observatories must concentrate on downward-going neutrinos \cite{IceCube:2021aen}. Backgrounds from atmospheric muons and neutrinos, produced in EAS in the far forward region, are therefore a significant concern.

Moreover, cosmic ray measurements can potentially also lead to interesting bounds on dark matter (DM) annihilations in the Galaxy and beyond. While in the most minimal BSM scenarios with a light, sub-GeV mediator particle between the SM and DM such signals are typically suppressed in order not to violate stringent cosmological bounds, these can be avoided in more rich dark sector models. In this case, the complementarity between the LLP searches at the FPF and DM indirect detection can shed new light on the nature of the dark sector. Non-minimal DM models can also lead to interesting complementarity between searches at the FPF and DM direct detection underground experiments.

In the following, we will explore the synergistic links between astroparticle physics and the FPF. The connection between FPF and cosmic ray physics will be discussed in \cref{sec:astro_cosmic_ray}, and atmospheric neutrino fluxes will be described in detail in \cref{sec:astro_neutrinos}. The synergies between the FPF and indirect dark matter searches, considering various dark sector models, will be discussed in \cref{sec:astro_DM}.

\section{Modelling Cosmic Ray Air Showers\label{sec:astro_cosmic_ray}}

Cosmic rays have been a standard but mysterious phenomenon in astrophysics since 1912~\cite{Hess:1912srp}. After more than a century  of thorough investigation their origin and acceleration mechanism(s) remain uncertain~\cite{Blumer:2009jrd,Anchordoqui:2018qom,AlvesBatista:2019tlv}. Extragalactic cosmic rays with energies exceeding $10^{11}~{\rm GeV}$ have been observed, but their nuclear composition is still unclear (see also the contribution to Snowmass 2021 on \emph{Ultra-High Energy Cosmic Rays}~\cite{UHECRWhitepaper}). Above about $10^6~{\rm GeV}$ cosmic ray observations pivot on indirect measurements of extensive air showers. To determine the energy and nuclear composition of cosmic rays, for example, we use our understanding of particle physics to model the shower evolution and describe the main features of the atmospheric cascades~\cite{Kampert:2012mx}.  It has long been known that uncertainties in the modeling of high-energy hadronic interactions of cosmic rays with nuclei in the air propagate into the estimates of residual background rates and largely dominate the systematic uncertainties of the atmospheric cascade development~\cite{Anchordoqui:1998nq,Ulrich:2010rg}. Although several attempts have been made to describe the shower evolution correctly, significant discrepancies between experimental data and current model predictions remain. One of the main challenges is the description of multi-particle production in the forward region during the EAS development.

\subsection{The Muon Puzzle and Beyond \label{sec:astro_muon_puzzle}}

\begin{figure}[tb]

    \mbox{\hspace{-1.4em}
    \includegraphics[width=0.485\textwidth]{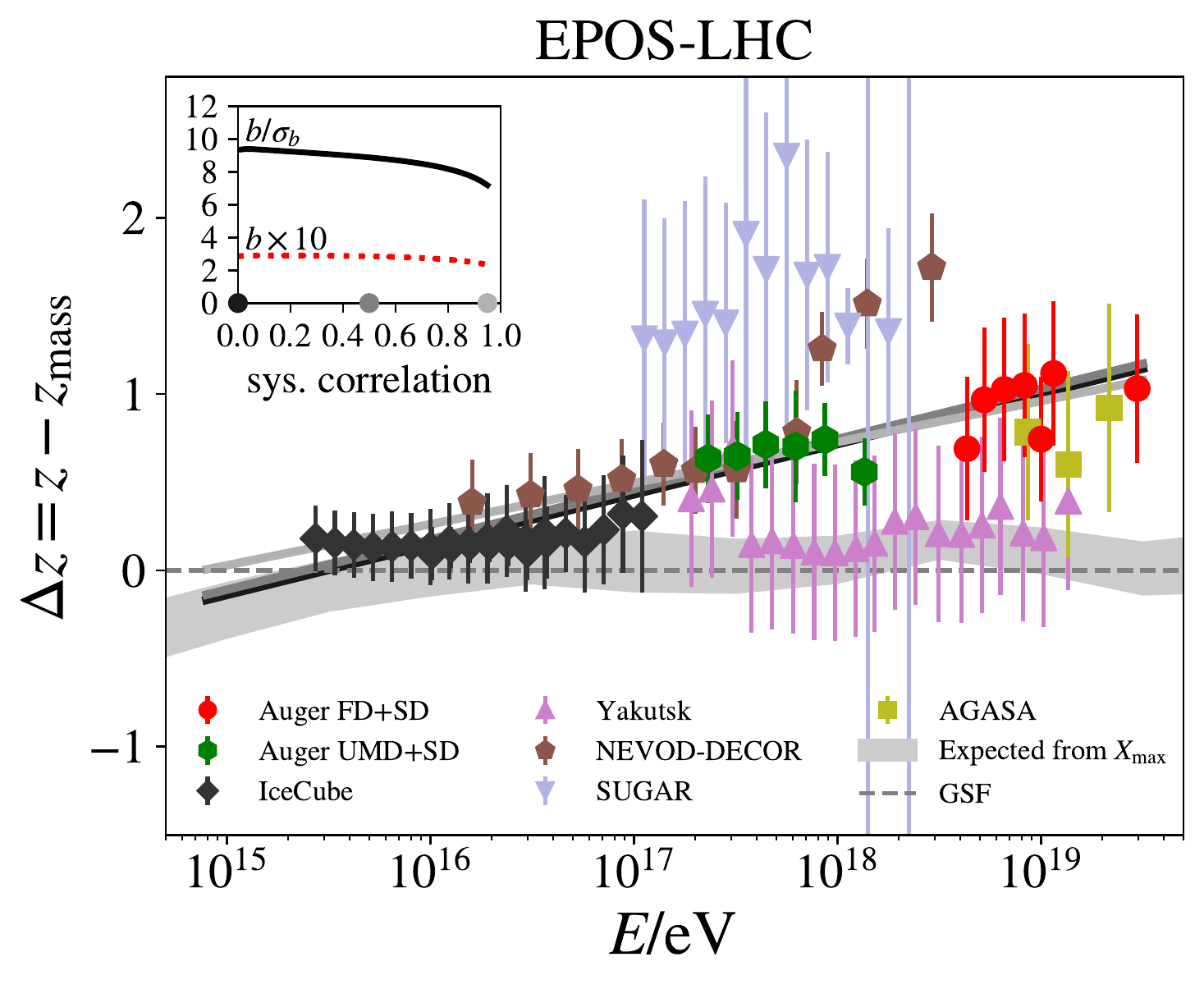}\;\;
    \includegraphics[width=0.485\textwidth]{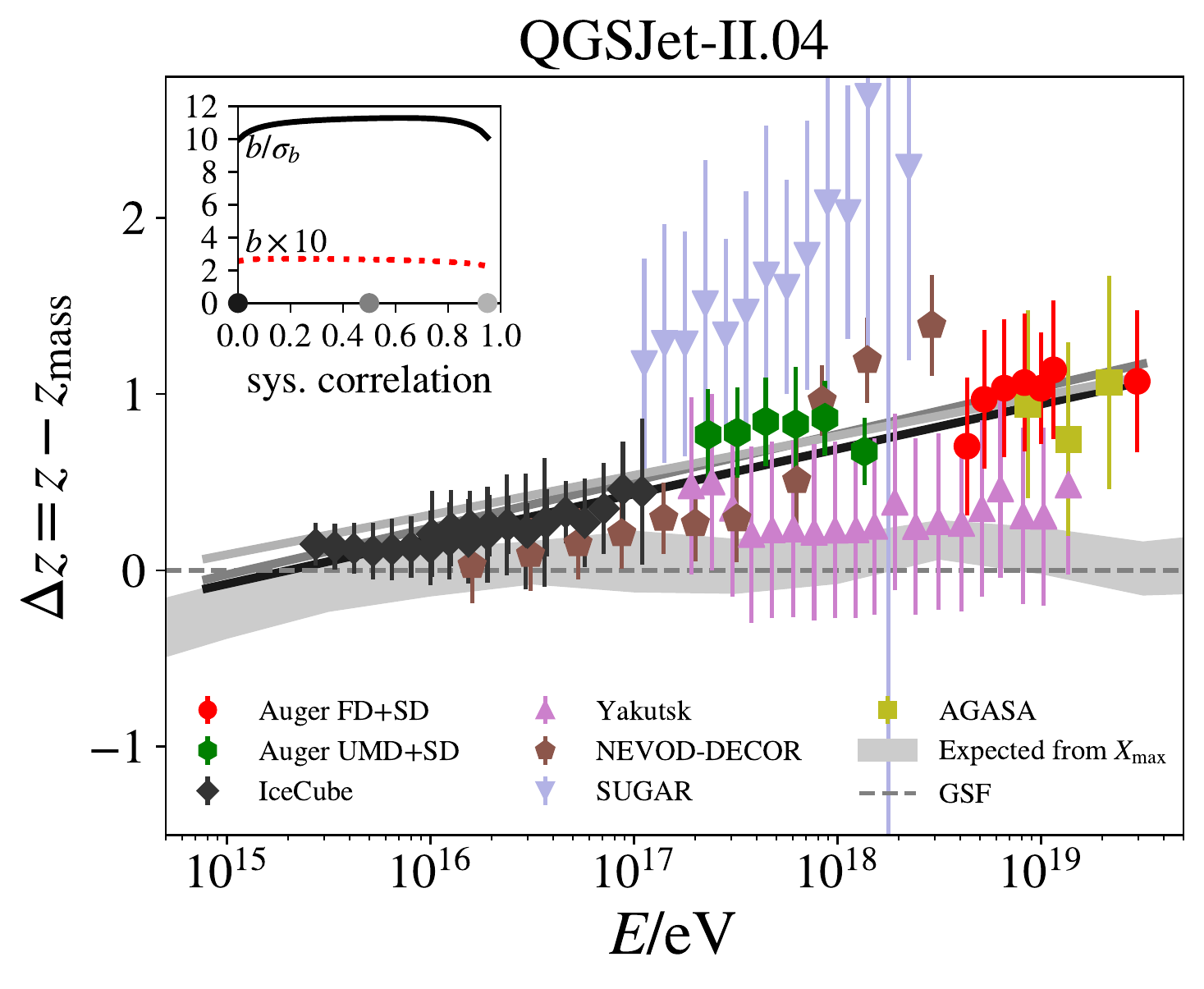}
   }
   \vspace{-1em}

    \caption{Linear fits to the $\Delta z=z-z_\mathrm{mass}$ distributions as a function of air shower energy from Ref.~\cite{Soldin:2021wyv}, where $z_\mathrm{mass}$ is the number of muons predicted by a hadronic interaction model, here \texttool{EPOS-LHC} (left) and  \texttool{QGSJet-II.04} (right), assuming a mass composition of the primaries based on experimental parameterization from Ref.~\cite{Dembinski:2017zsh} (GSF). $\Delta z$ measures the difference between the experimental data and the inferred number of muons for a given hadronic model. A positive value indicates an excess of muons in data with respect to simulations and zero indicates a perfect match. Shown in the inset are the slope $b$ and its deviation from zero in standard deviations for an assumed correlation of the point-wise uncertainties within each experiment. Examples of the fits are shown for a correlation of $0.0$, $0.5$, and $0.95$ in varying shades of gray.}
    \label{fig:muon_puzzle}
    \vspace{-0.5em}

\end{figure}

Air shower simulations reasonably reproduce many of the features in the cascade development, but there is a long-standing deficit in the number of muons produced in extensive air showers, which was first observed by the HiRes-MIA experiment more than 20 years ago~\cite{AbuZayyad:1999xa}. Since then, both simulations and experiments have made enormous progress, but the \emph{Muon Puzzle} persists~\cite{Albrecht:2021cxw}. The most unambiguous experimental evidence of the deficit was revealed in the analysis of data from the Pierre Auger Observatory~\cite{PierreAuger:2014ucz,PierreAuger:2016nfk}. A meta-analysis~\cite{Dembinski:2019uta,Cazon:2020zhx,Soldin:2021wyv} of recent muon measurements from several experiments is shown in \cref{fig:muon_puzzle}. This analysis includes recent data from the Pierre Auger Observatory (Auger)~\cite{Aab:2020frk,PierreAuger:2021qsd}, the IceCube Neutrino Observatory (IceCube)~\cite{IceCube:2021tuv,IceCube:2022yap}, the Yakutsk EAS array~\cite{Glushkov}, NEVOD-DECOR~\cite{Bogdanov:2018sfw}, SUGAR~\cite{Bellido:2018toz}, and AGASA~\cite{Gesualdi:2021yay}. In order to make these different muon measurements in air showers comparable, the $z$-scale is used,
\begin{equation}
    z=\frac{\ln\langle N_\mu\rangle-\ln \langle N_{\mu,\rm{p}}\rangle}{\ln \langle N_{\mu,\rm{Fe}}\rangle-\ln \langle N_{\mu,\rm{p}}\rangle}\:,
\end{equation}
where $\langle N_\mu\rangle$ is the average muon density estimate as observed in the detector, while $\langle N_{\mu,\rm{p}}\rangle$ and $\langle N_{\mu,\rm{Fe}}\rangle$ are the simulated average muon densities for proton and iron showers after a full detector simulation. 

After applying an energy cross-calibration, the z-scale is approximately independent of the experimental details but depends on the hadronic interaction model used in air shower simulations. In order to systematically quantify the energy-dependent trend observed in the muon measurements, the mass composition dependence also needs to be taken
into account. If the measured z-values follow $z_\mathrm{mass}$, for example obtained from the GSF flux model from Ref.~\cite{Dembinski:2017zsh}, the hadronic interaction model describes the muon density at the ground consistently. Subtracting $z_\mathrm{mass}$ will thus remove the effect of the changing mass composition (see Ref.~\cite{Dembinski:2019uta} for further details). The resulting $\Delta z=z-z_\mathrm{mass}$ distributions are shown in the figure for \texttool{EPOS-LHC} and \texttool{QGSJet-II.04}. An upward trend is observed which starts at moderate center-of-mass energies of about 10\,TeV, accessible by the LHC, followed by a linear increase with the logarithm of the shower energy. The slope of this increase deviates from zero with a significance of more than $8\sigma$~\cite{Soldin:2021wyv}, indicating shortcomings in our understanding of multi-particle production in EAS.

The muons seen by air shower experiments are of low energy (a few to tens of GeV). They are produced at the end of a cascade of hadronic interactions with up to about 10 steps on average, where the dominant process is soft forward hadron production, which can not be calculated from first principles in perturbative QCD. Effective theories are used to describe these interactions, in particular Gribov-Regge field theory. Detailed simulations \cite{Ulrich:2010rg, Baur:2019cpv} have shown that the hadron multiplicity and, in particular, the hadron species at forward pseudorapidities of $\eta \gg 2$ have the largest impact on muon production in air showers. The sensitivity to the produced hadrons is high and even small deviations of 5\,\% in the multiplicity and/or identity of the secondary hadrons have a sizeable impact on the muon production.

Proposed models that account for such deviations are based on the restoration of chiral symmetry~\cite{Farrar:2013sfa}, the production of fireballs~\cite{Anchordoqui:2016oxy}, a core-corona effect~\cite{Baur:2019cpv}, and a quark-gluon plasma~\cite{Pierog:2020ghc,Anchordoqui:2019laz}. These models have in common that the neutral particle production is suppressed with respect to the effective theories encapsulated in the current post-LHC hadronic interaction models (\emph{e.g.} \texttool{EPOS-LHC}~\cite{Pierog:2013ria}, \texttool{QGSJet-II.04}~\cite{Ostapchenko:2013pia}, \texttool{Sibyll-2.3c/d}~\cite{Riehn:2017mfm,Riehn:2019jet}, and \texttool{DPMJet-III.2017}~\cite{Roesler:2000he,Cerutti:2015lcn}). This indirectly enhances the muon content at ground without altering the remainder of the shower development. Regardless of the details of the model, generally two extremes can be distinguished: a rather strong suppression occurring in the first few interactions of the air shower -- reflecting some kind of threshold effect of exotic physics -- or a small suppression over a large range of energies where the effect on the muon content accumulates throughout the shower development. The fit shown in \cref{fig:muon_puzzle} seems to favour the latter, as $\Delta z$ is continuously increasing with shower energy. A measurement of shower-to-shower fluctuations of the muon content~\cite{PierreAuger:2021qsd} further motivates the accumulation scenario.

The amount of forward strangeness production seems of particular relevance~\cite{ALICE:2016fzo}. It is traced by the ratio of charged kaons to pions, for which the ratio of electron and muon neutrino fluxes is a proxy that will be measured by the FPF~\cite{Kling:2021gos}. Electron neutrino fluxes are a measurement of kaons, whereas both muon and electron neutrinos are produced via pion decay. However, $\nu_\mu$ and $\nu_e$ populate different  energy regions, which can help to disentangle them. In addition, neutrinos from pion decay are more concentrated around the LOS than those of kaon origin, given that $m_\pi < m_K$, and thus neutrinos from pions  obtain less additional transverse momentum than those from kaon decays. Thereby, the closeness of the neutrinos to the LOS, or equivalently their rapidity distribution, can be used to disentangle different neutrino origins to get an estimate of the pion to kaon ratio. If technically feasible, a correlation of the FPF measurements with the activity in ATLAS could also be an interesting option in order to study the charge ratio in dependence of the charged particle multiplicity in the central rapidity region~\cite{CMS:2019kap}.

In addition, it might also be possible to use the forward going muons to constrain the forward production of pions and kaons. The muon flux at FASER is estimated to be approximately 1~Hz per ${\rm cm^2}$~\cite{FASER:2018bac}.  About $2  \times 10^9$ muons will be  detected by FASER in Run 3 (2022-24).  The number at FASER2 in the FPF at HL-LHC (2027-36) is about 1000 times larger. An interesting option is to add a sweeper magnet upstream of the FPF (\emph{e.g.} where the LOS leaves the LHC beampipe) to deflect muons from the on-axis neutrino detectors. This will produce an over-density of muons at about 1-2~m off the LOS. A dedicated detector, specifically placed to record these muons, would provide complementary information to determine the ratio of charged pions to kaons. However, these measurements may be challenging because the origin of the muon flux at the FPF is currently not well understood and further investigations are needed (see for example \cref{sec:bdsim_fluxes}).  

Analogously, using the high-energy neutrino (and muon) fluxes as a proxy for pion and kaon production in the forward region, the FPF will also provide complementary data to measurements from IceCube~\cite{IceCube:2016zyt}. With its deep ice detector, IceCube measures atmospheric muons with energies from a few $100\,\mathrm{GeV}$~\cite{Abbasi:2012kza,Soldin:2015iaa,Soldin:2018vak} up to energies above $1\,\mathrm{PeV}$~\cite{IceCube:2015wro,Fuchs:2017nuo,Soldin:2018vak}, as well as the lateral separation of TeV muons~\cite{Abbasi:2012kza,Soldin:2015iaa,Soldin:2018vak}. Moreover, IceCube has measured the seasonal variations of the muon flux over several years with a statistical significance that allows to determine the pion to kaon ratio at high energies~\cite{Desiati:2011hea, Tilav:2010hj, gaisser:2013lrk, Tilav:2019xmf}. These muons are produced in the far-forward region and studies of the pion and kaon production at the FPF can provide complementary information which will help to reduce uncertainties in the modeling of the atmospheric muon fluxes. This, in turn, will reduce uncertainties for astrophysical measurements in IceCube, such as the analysis of the cosmic ray spectrum and composition~\cite{IceCube:2019hmk}, for example.

The FPF experiments will further provide complementary data on very forward hadron multiplicities. The LHCf experiment has previously measured the neutral pion and neutron production cross-sections~\cite{LHCf:2015rcj, LHCf:2015nel, LHCf:2018gbv, LHCf:2020hjf} and by using $\mu$ and $\nu_\mu$ as proxies, the FPF experiment can make complementary measurements of the charged pion production cross-section. A combination of data from FPF and LHCf will constrain the hadron composition in the very forward region. As the Muon Puzzle is assumed to be of soft-QCD origin, there is a strong connection to the QCD program of the FPF and the measurements will help to understand particle production in EAS.

\subsection{Probing Hadronic Interaction Models at the FPF \label{sec:astro_hadr_models}}

\begin{figure}
\centering
\includegraphics[width=0.478\textwidth]{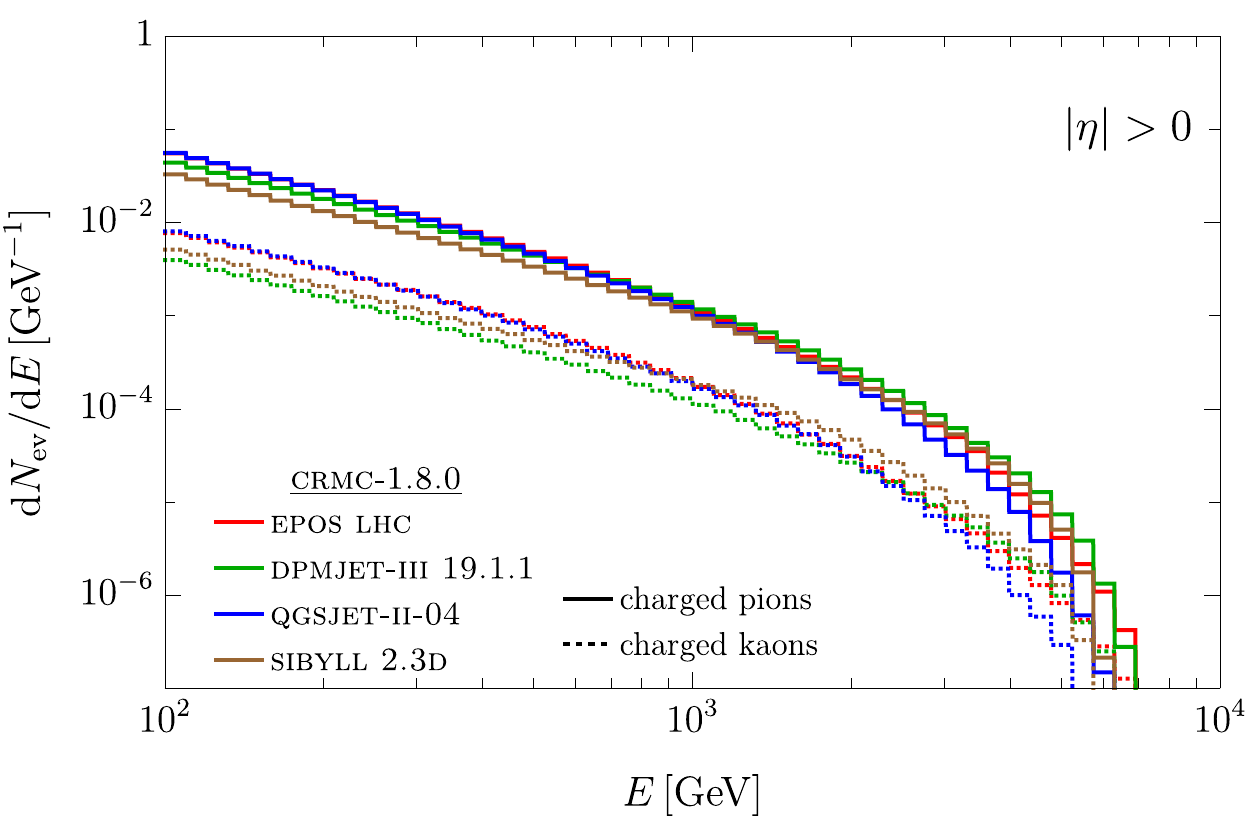}
\includegraphics[width=0.478\textwidth]{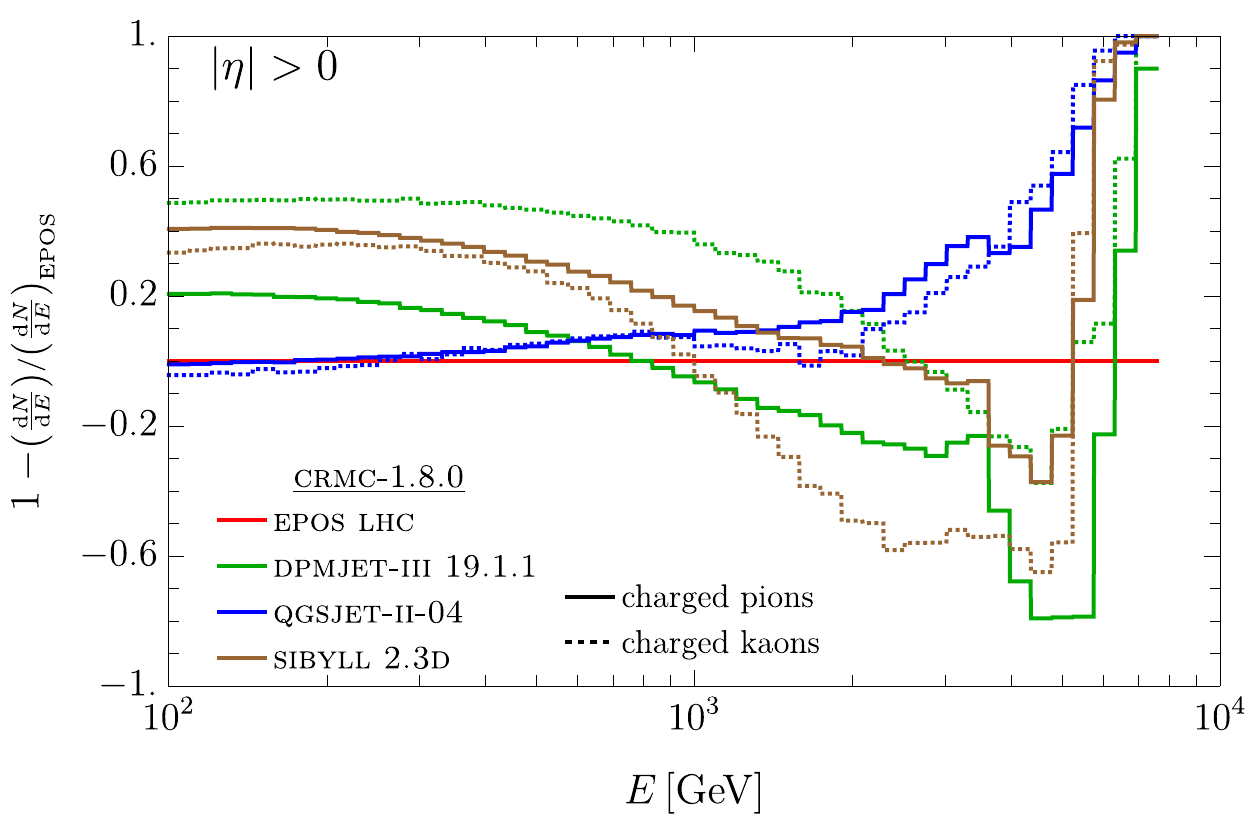}
\includegraphics[width=0.478\textwidth]{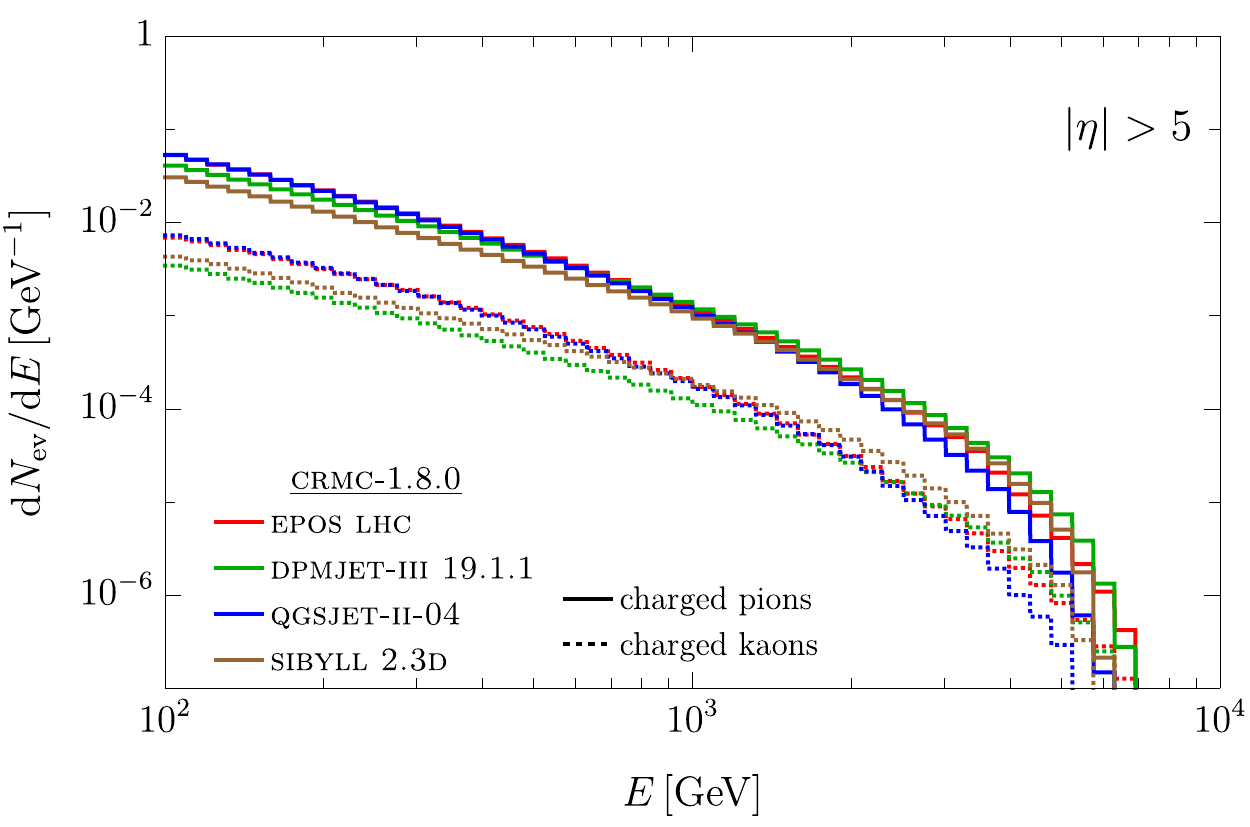}
\includegraphics[width=0.478\textwidth]{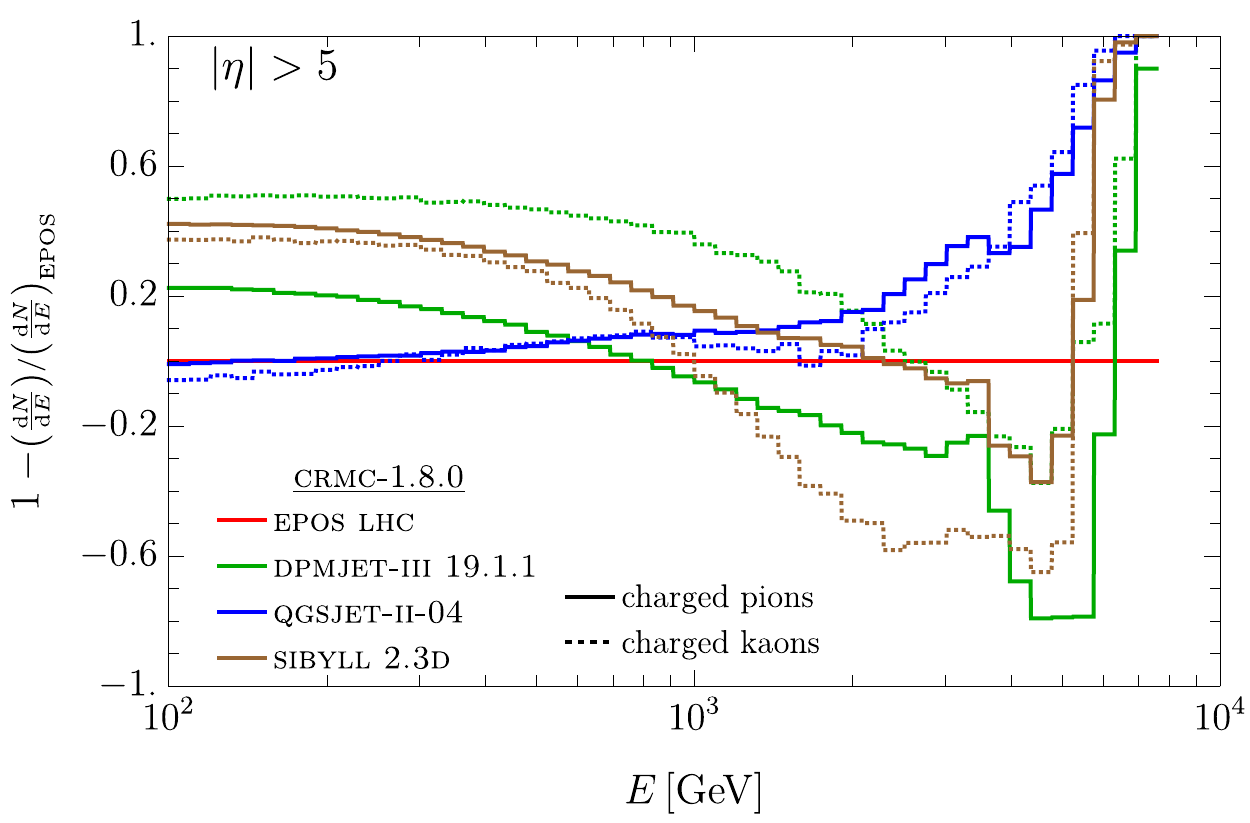}
\includegraphics[width=0.478\textwidth]{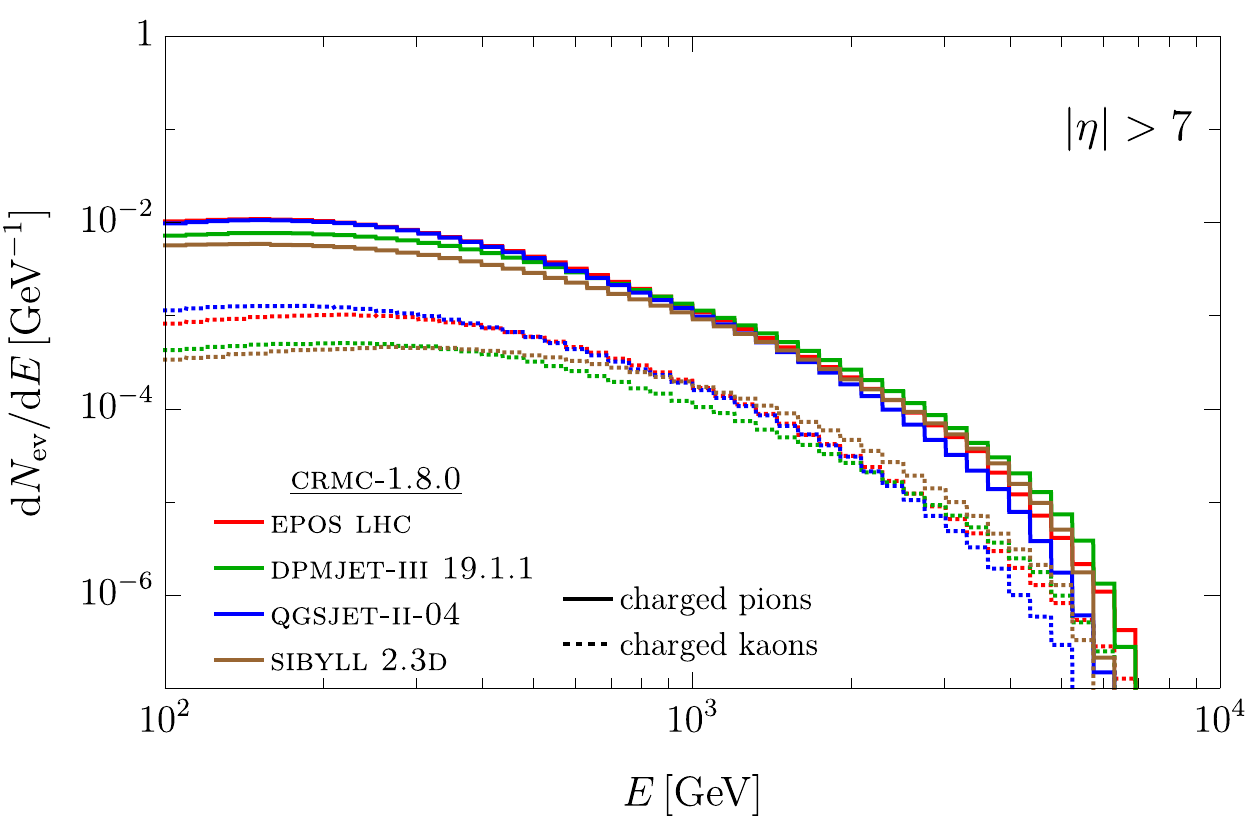}
\includegraphics[width=0.478\textwidth]{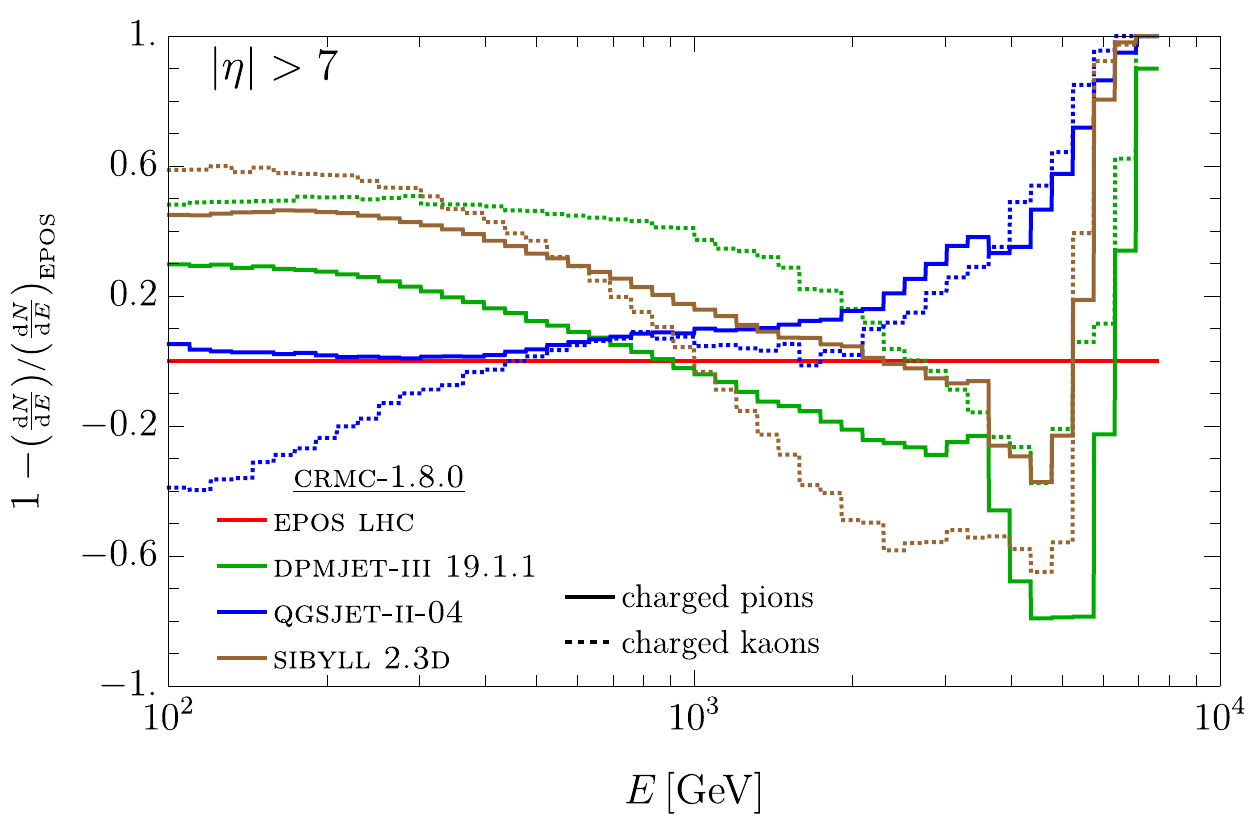}
\includegraphics[width=0.478\textwidth]{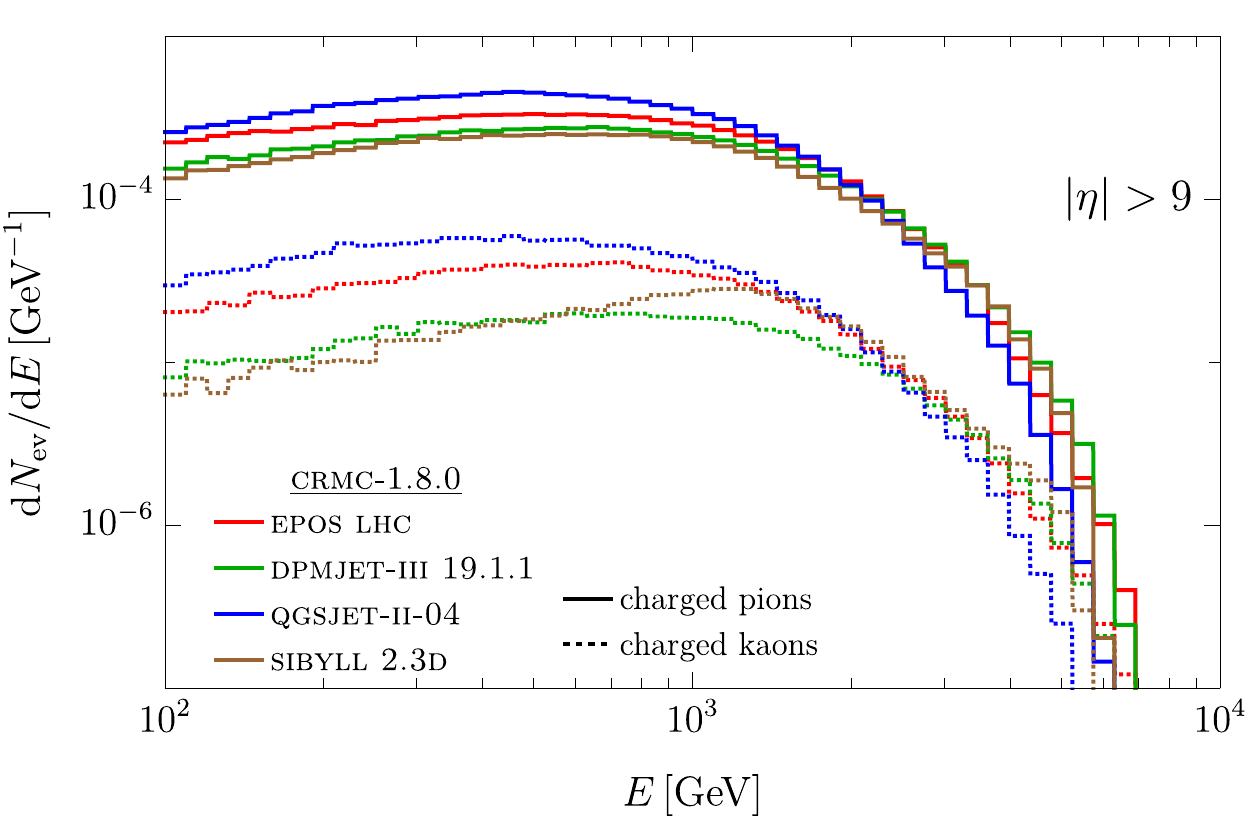}
\includegraphics[width=0.478\textwidth]{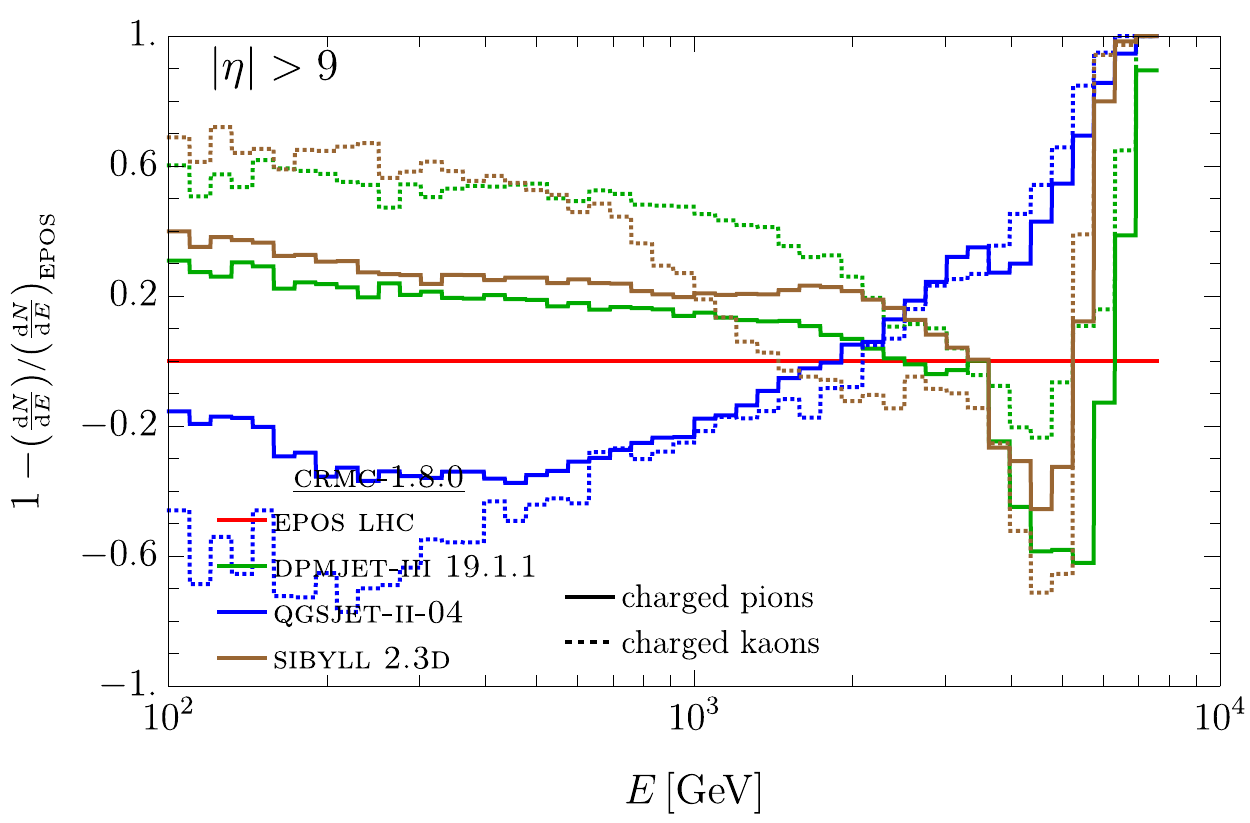}

\caption{Absolute (left) and relative to \texttool{EPOS-LHC} (right) $\pi^\pm$ and $K^\pm$ energy spectra for the four hadronic models under consideration. Obtained for different pseudorapidity ranges, increasing from top to bottom, from $10^6$ pp events at $\sqrt s=14~\mathrm{TeV}$.}
\label{fig:hadronic-models-energy}
\end{figure}

\begin{figure}
\centering
\includegraphics[width=0.474\textwidth]{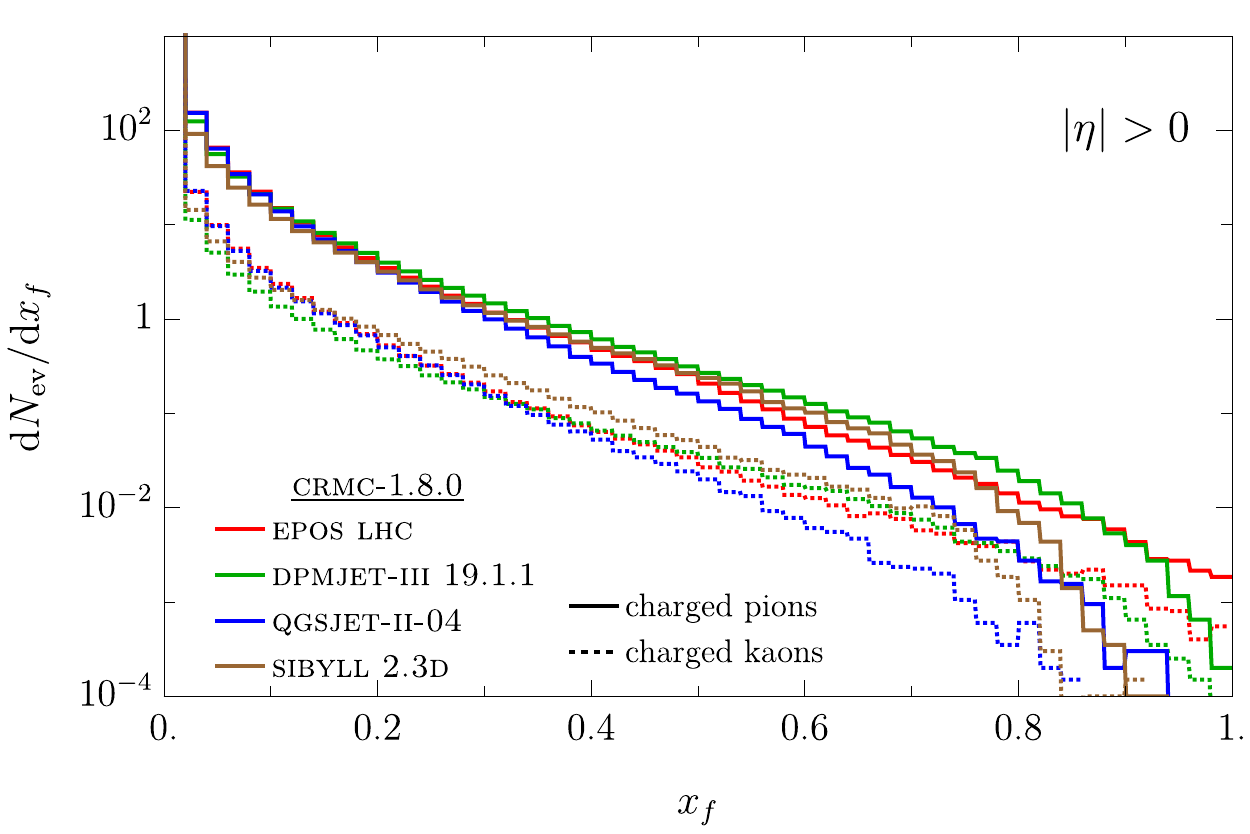}
\includegraphics[width=0.474\textwidth]{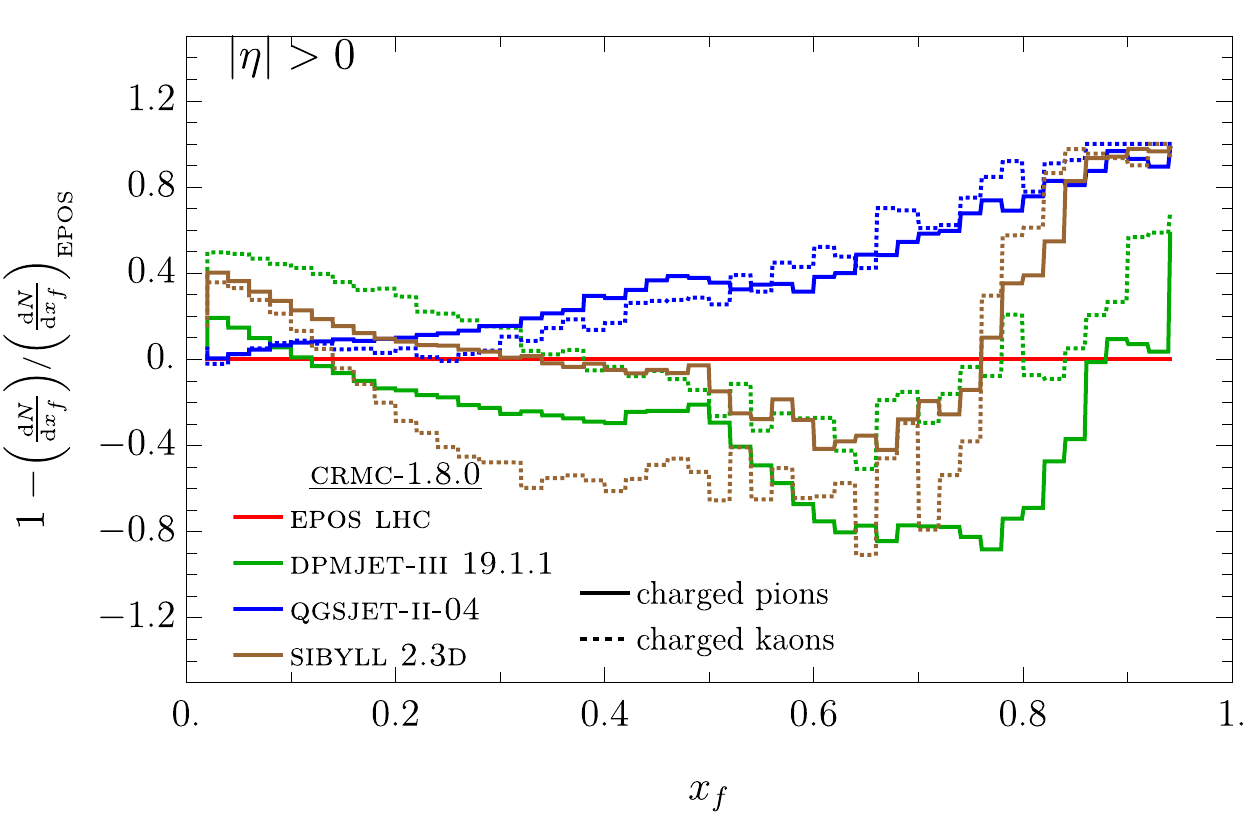}
\includegraphics[width=0.474\textwidth]{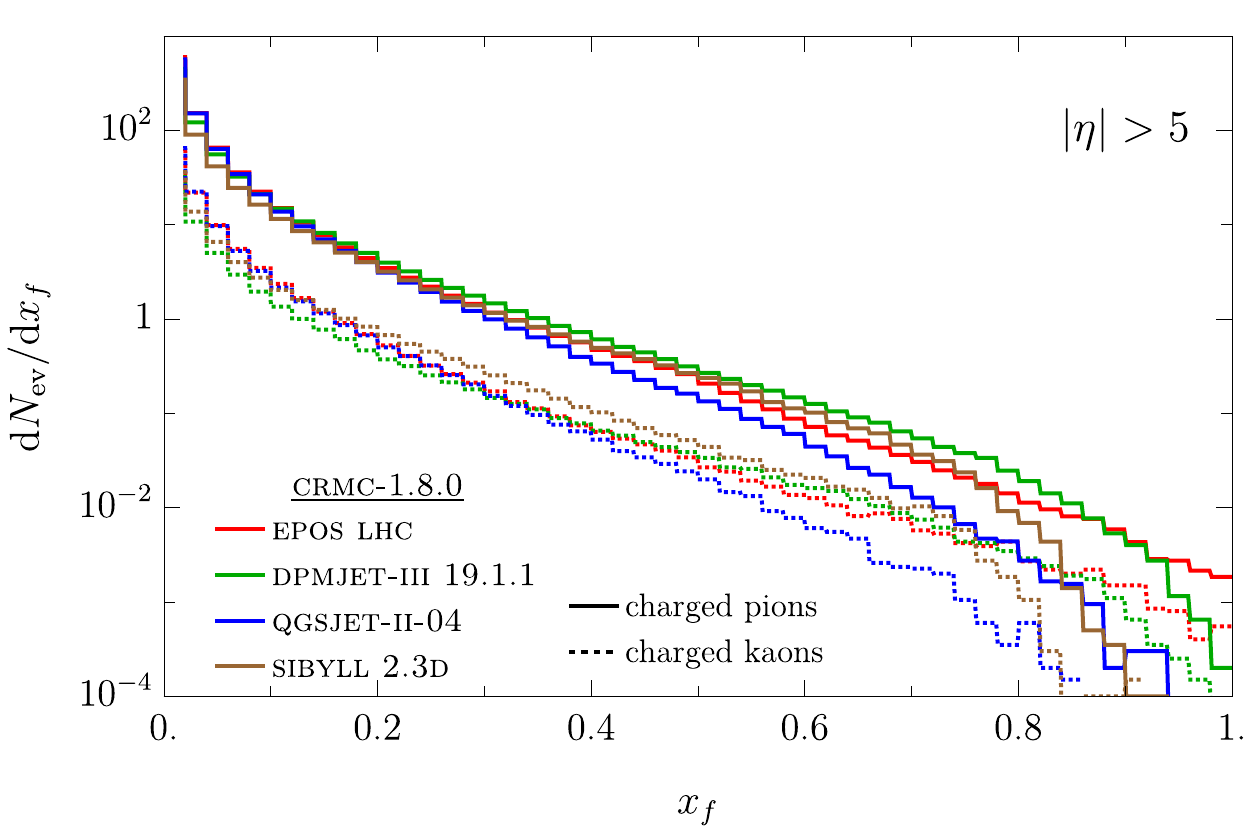}
\includegraphics[width=0.474\textwidth]{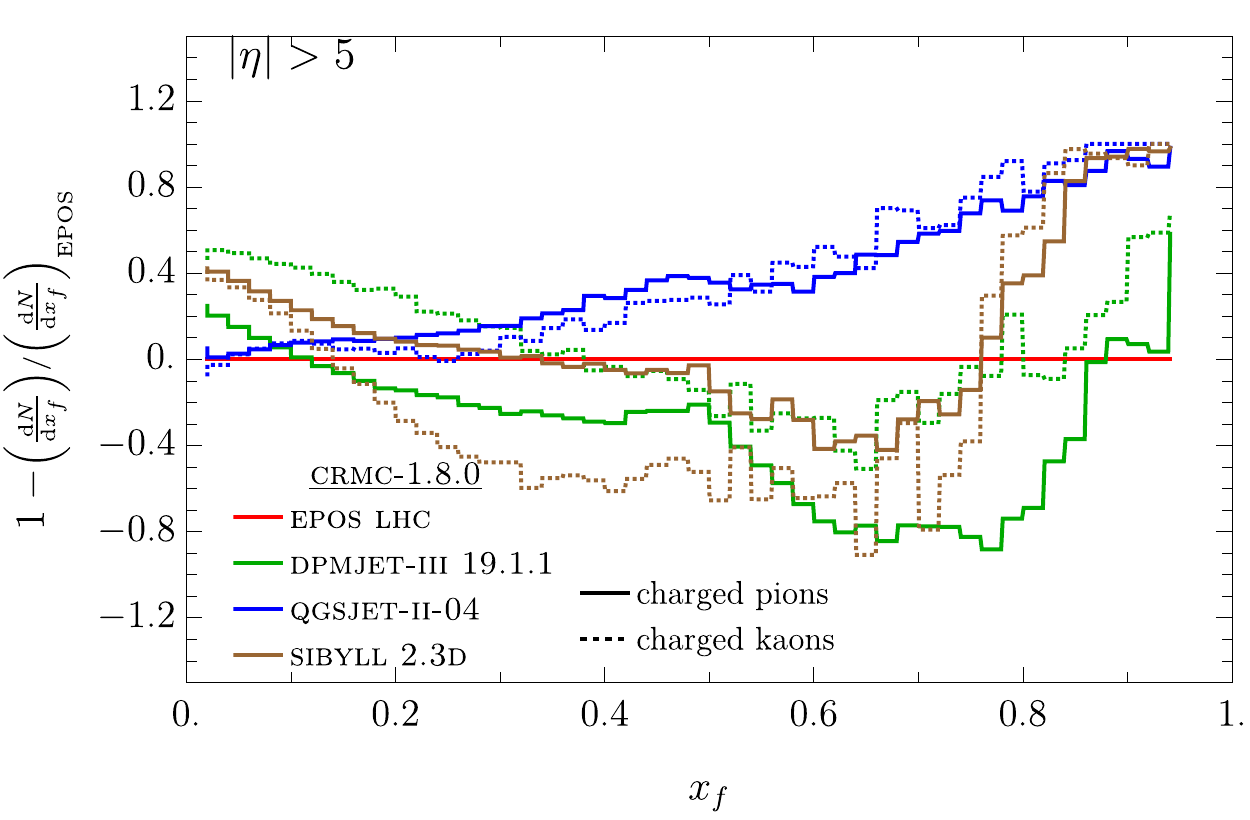}
\includegraphics[width=0.474\textwidth]{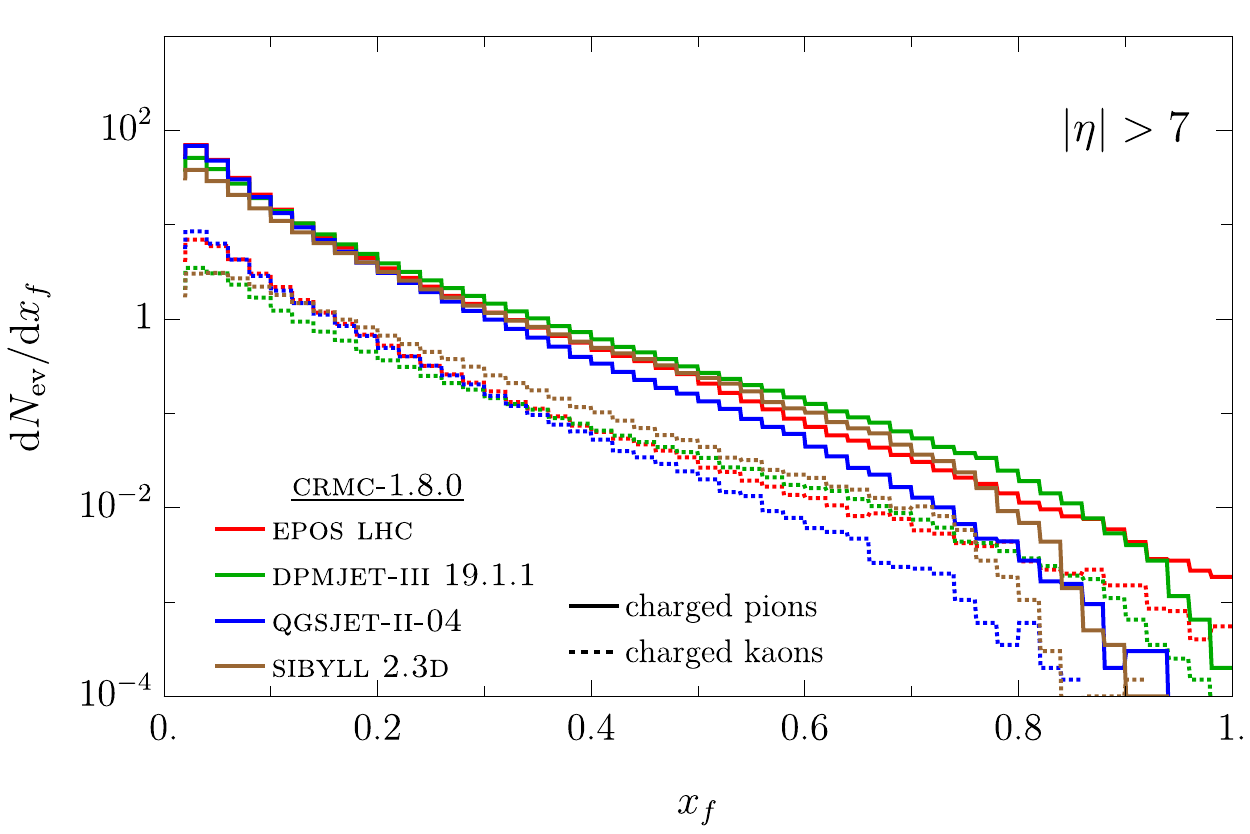}
\includegraphics[width=0.474\textwidth]{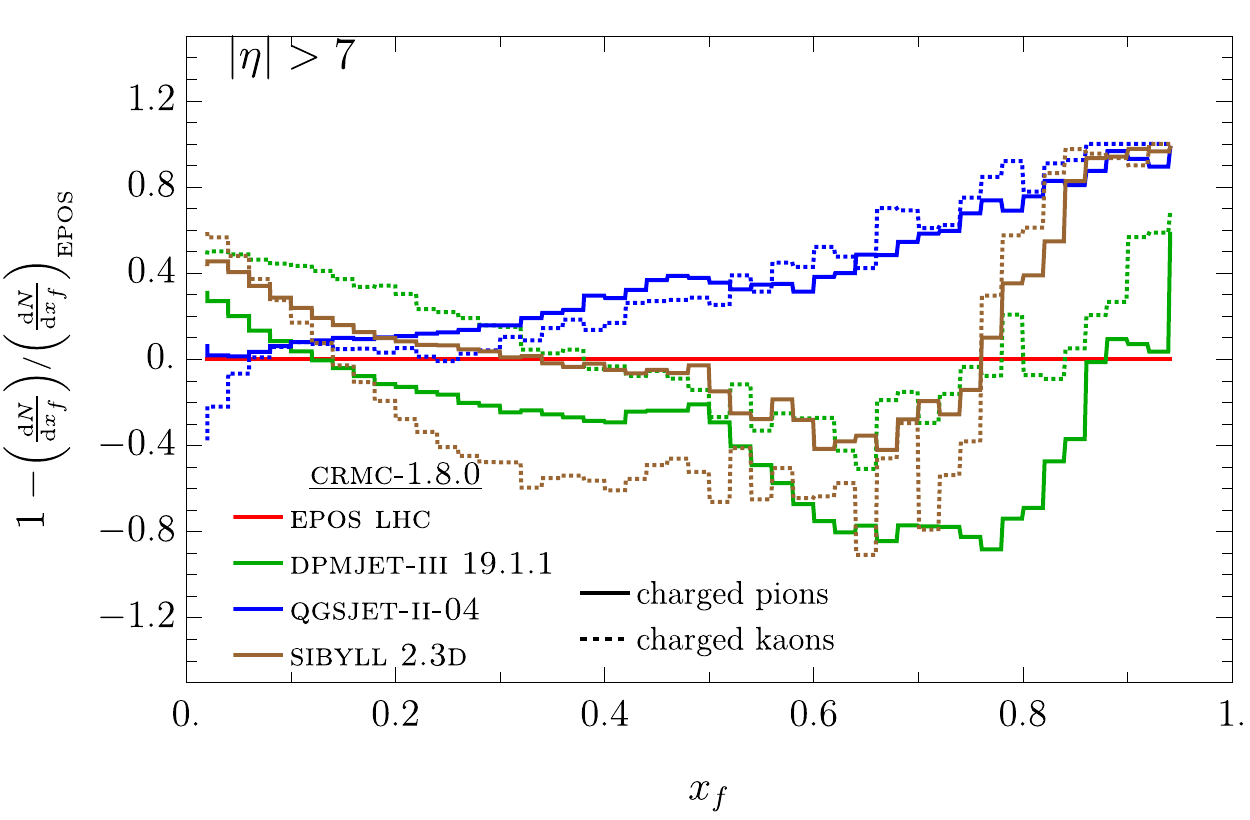}
\includegraphics[width=0.474\textwidth]{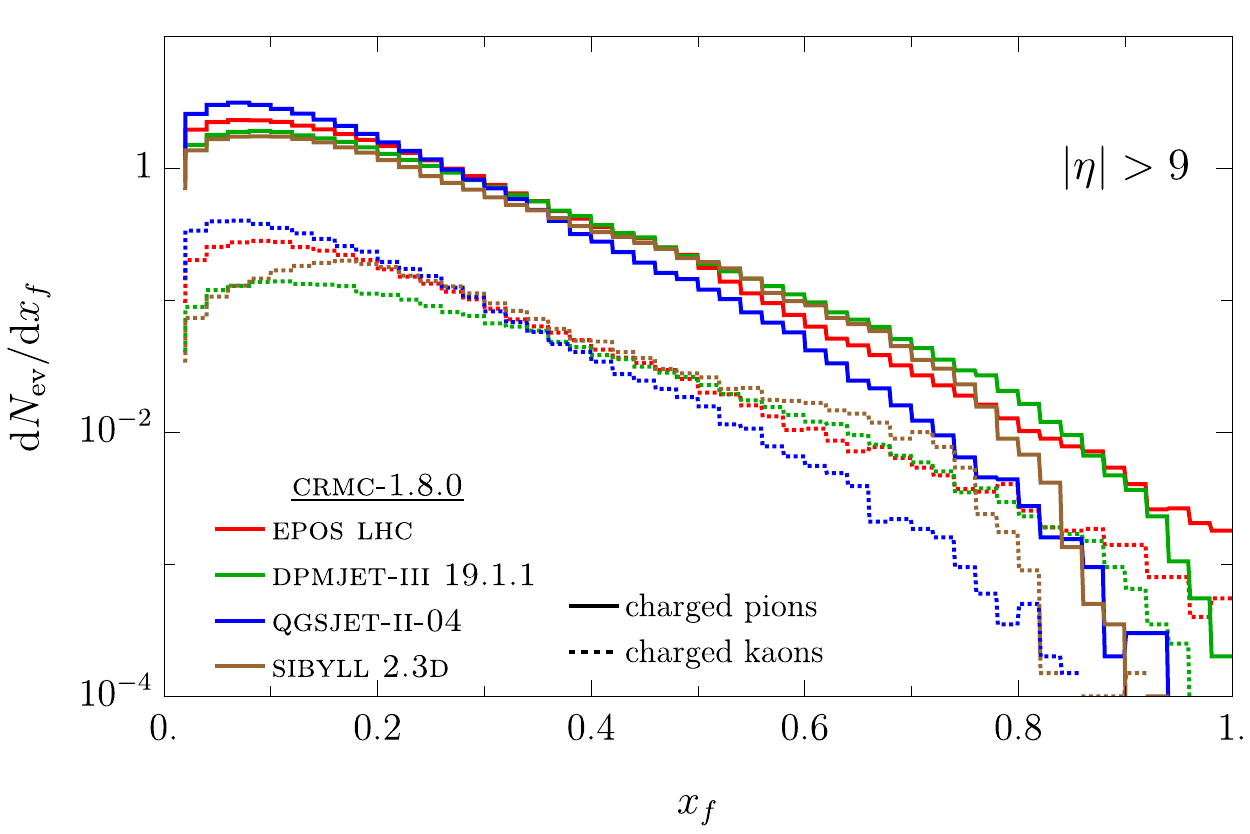}
\includegraphics[width=0.474\textwidth]{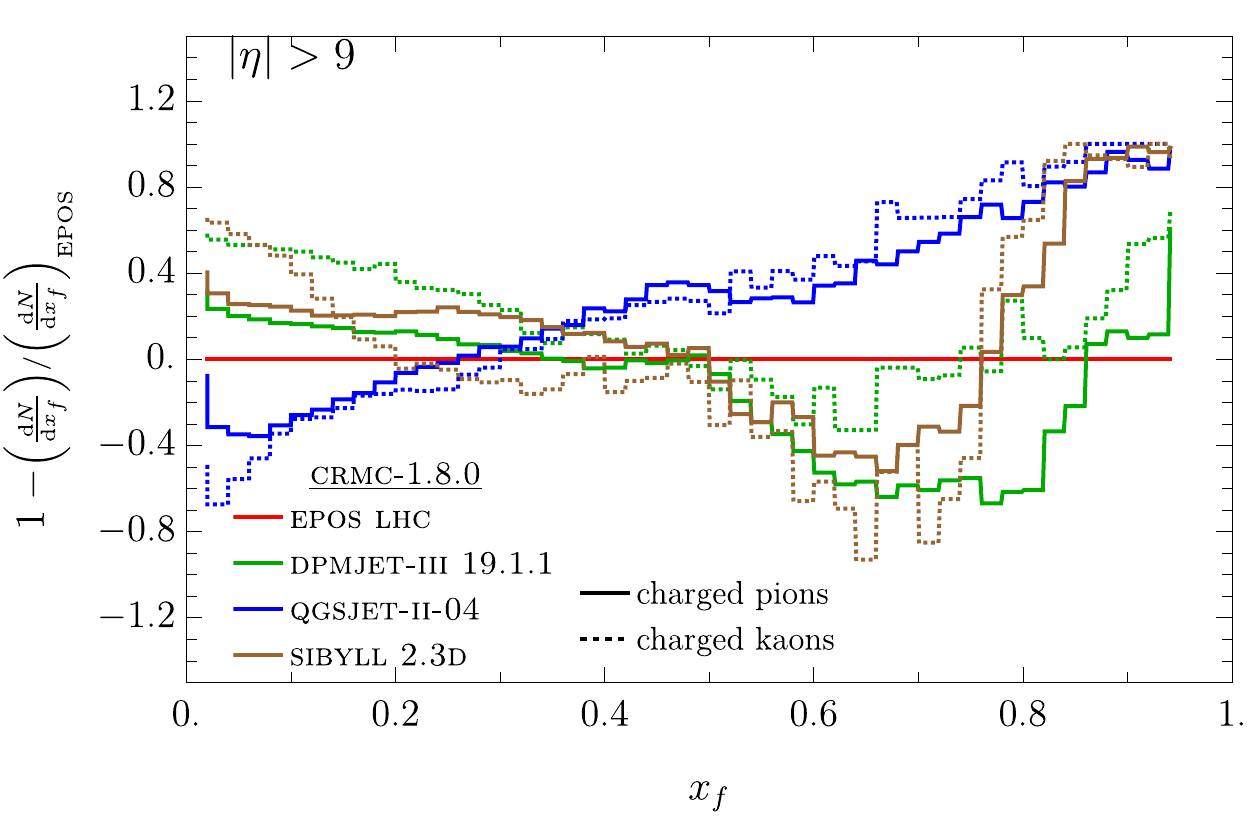}\label{fig:xratios-eta9}

\caption{Absolute (left) and relative to \texttool{EPOS-LHC} (right) $\pi^\pm$ and $K^\pm$ Feynman-$x$ spectra for the four hadronic models under consideration. Obtained for different pseudorapidity ranges, increasing from top to bottom, from $10^6$ pp events at $\sqrt s=14~\mathrm{TeV}$. }
\label{fig:hadronic-models-x}
\end{figure}

Extensive air showers are produced after a highly boosted proton or nucleus \emph{primary} undergoes a hadronic interaction with an atmospheric nucleus, which is practically at rest with respect to the Earth's surface. From there, experiments like the Pierre Auger Observatory~\cite{PierreAuger:2015eyc} or Telescope Array~\cite{TelescopeArray:2012uws,Tokuno:2012mi} observe the fluorecence produced after particles deposit their energy in $N_2$ molecules in the atmosphere, as well as the direct Cerenkov radiation of particles arriving at ground level. The development of the different components of the shower (electromagnetic and hadronic) produce signatures that are observable by such experiments. Comparison with simulations allow to constraint, among others, the hadronic models that are used to describe these primary interactions.

Since the center of mass of the colliding system is highly boosted along the speed of the primary particle, the particles produced around the center-of-mass transverse region move forward along the collision axis in the experiment frame, contributing to the shower and its observable signal. Nevertheless, unlike in circular particle accelerators, the forward region can be fully captured air shower experiments. The lack of forward region measurements from particle accelerators introduces uncertainties in what hadronic models predict for these primary interactions, as well as for the subsequent ones. These uncertainties propagate to cosmic ray shower observables, obscuring our understanding of the particle physics behind them. The FPF will address this issue by measuring and identifying the flavour of neutrinos coming from different hadrons, which will help identify the forward region hadronic spectra.

Advances in this area are of high relevance not only to provide a solid foundation for hadronic models, but to shed light on one of the unexplained observations in cosmic ray physics: the muon puzzle describe in the previous section. The ratio of pions to kaons produced in the first interaction(s) is a proxy for the ratio of the electromagnetic to hadronic energy in the shower. It is then expected that a better understanding of the hadron spectra in these interactions will provide insights on the shower development that correspond to observable quantities, rendering the aforementioned experiments as great tools to understand particle physics at the highest energies.

\begin{figure}[tb]
\centering
\includegraphics[width=0.95\textwidth]{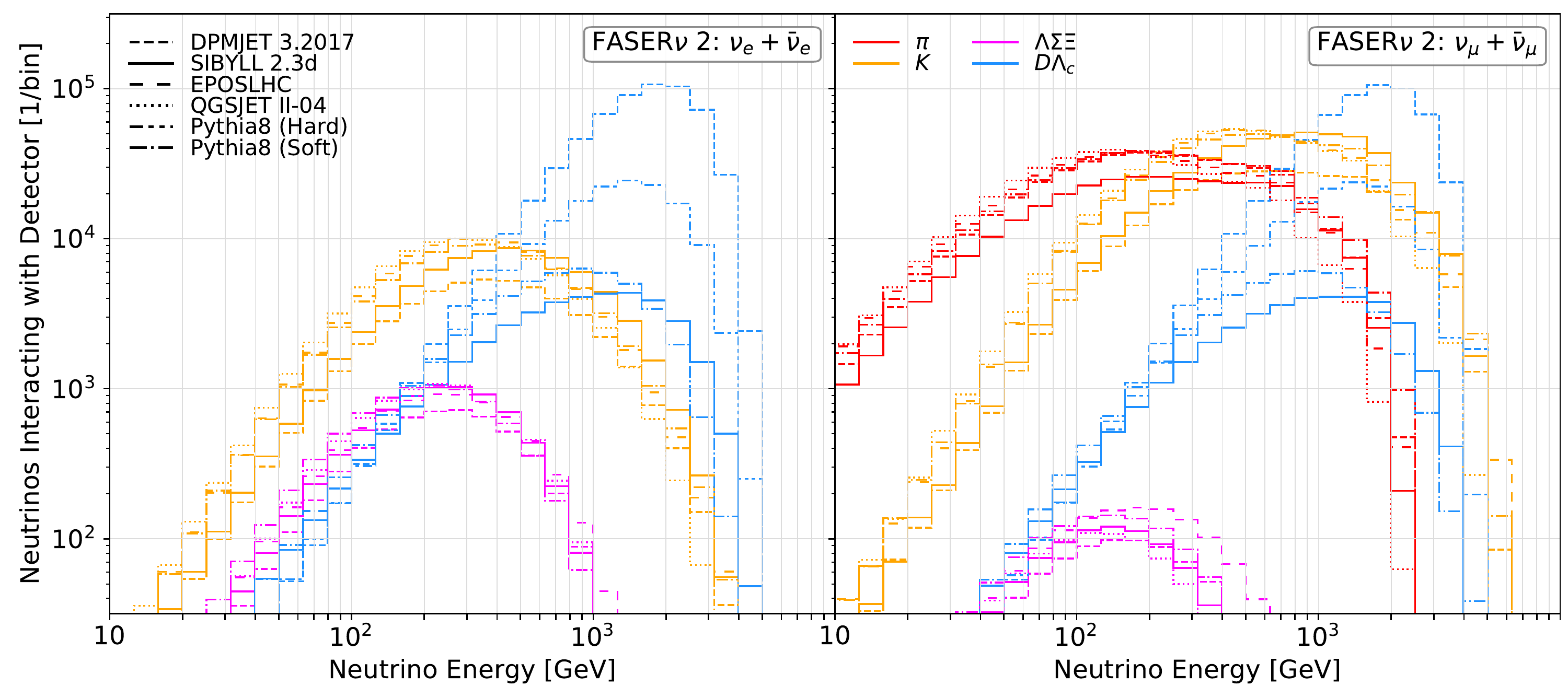}
\caption{Neutrino energy spectra for electron neutrinos (left) and muon neutrinos (right) passing through FASER$\nu$2. The vertical axis shows the number of neutrinos per energy bin that go through the detector's cross-sectional area for an integrated luminosity of $3~\iab$. The different production modes are indicated by different colors: pion decays (red), kaon decays (orange), hyperon decays (magenta), and charm decays (blue). The different line styles correspond to predictions obtained from \texttool{Sibyll-2.3d} (solid), \texttool{DPMJet-III.2017} (short dashed), \texttool{EPOS-LHC} (long dashed), \texttool{QGSJet-II.04} (dotted), and \texttool{Pythia~8.2} using soft-QCD processes (dot-dashed) and with hard-QCD processes for charm production (double-dot-dashed). Note that the predictions differ by up to a factor 2 for neutrinos from pion and kaon decays, which is much bigger than the anticipated statistical uncertainties at the FPF~\cite{Kling:2021gos}.}
\label{nu-rate}
\end{figure}

\cref{fig:hadronic-models-energy} and \cref{fig:hadronic-models-x} display the energy and Feynman-$x$ spectra for charged pions and kaons in different pseudorapidity regions after simulating $10^6$ proton-proton collisions at $\sqrt s=14~\mathrm{TeV}$. The simulations have been carried out considering standard hadronic interaction event generators: \texttool{EPOS-LHC}~\cite{Pierog:2019opp}, \texttool{DPMJet-III} (version 19.1.1)~\cite{Roesler:2000he}, \texttool{QGSJet-II.04}~\cite{Ostapchenko:2019few} and \texttool{Sibyll-2.3d}~\cite{Riehn:2019jet}. All of them are accessible via the interface \texttool{CRMC} (version 1.8.0)~\cite{CRMC}. In particular,  \cref{fig:hadronic-models-energy} (left) and \cref{fig:hadronic-models-x} (left) show, respectively, the energy and $x_f$ spectra of kaons and pions between $100\,\mathrm{GeV}$ and $10\,\mathrm{TeV}$, for different pseudorapidity cuts. On the right panels, \cref{fig:hadronic-models-energy} and \cref{fig:hadronic-models-x} show the differences between the models, relative to our choice of baseline model: \texttool{EPOS-LHC}. It is clear how all models differ substantially almost everywhere, although \texttool{Sibyll-2.3d} and \texttool{DPMJet-III} have similar features. Remarkably, as shown in \cref{nu-rate} the FPF experiments will provide complementary data on far-forward hadron production, which can be used to refine our understanding of soft hadronic processes.

\subsection{Complementary Probes of Strangeness Enhancement:  Auger Meets the FPF  \label{sec:astro_strangeness}}

\begin{figure}[tb]
\centering
\includegraphics[width=0.49\textwidth]{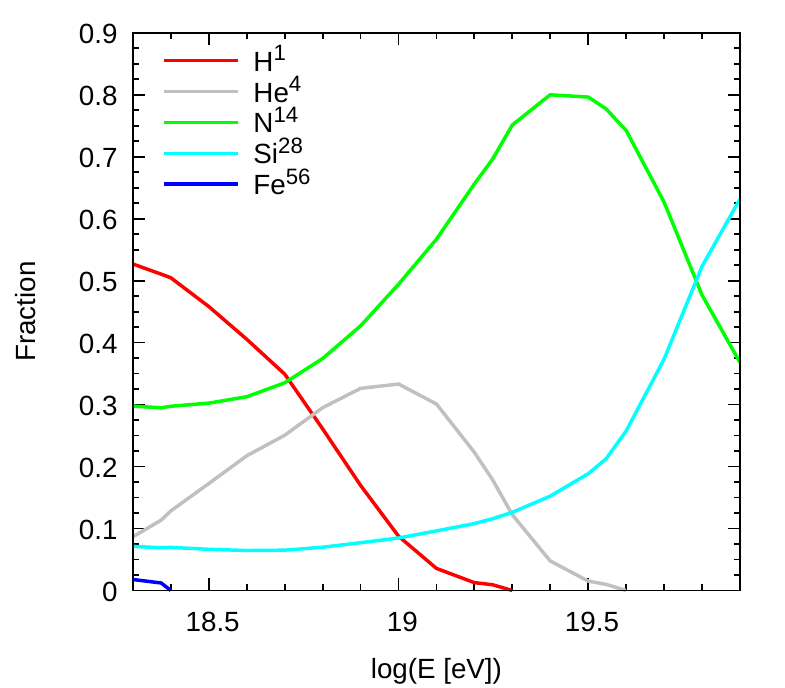}
\includegraphics[width=0.49\textwidth]{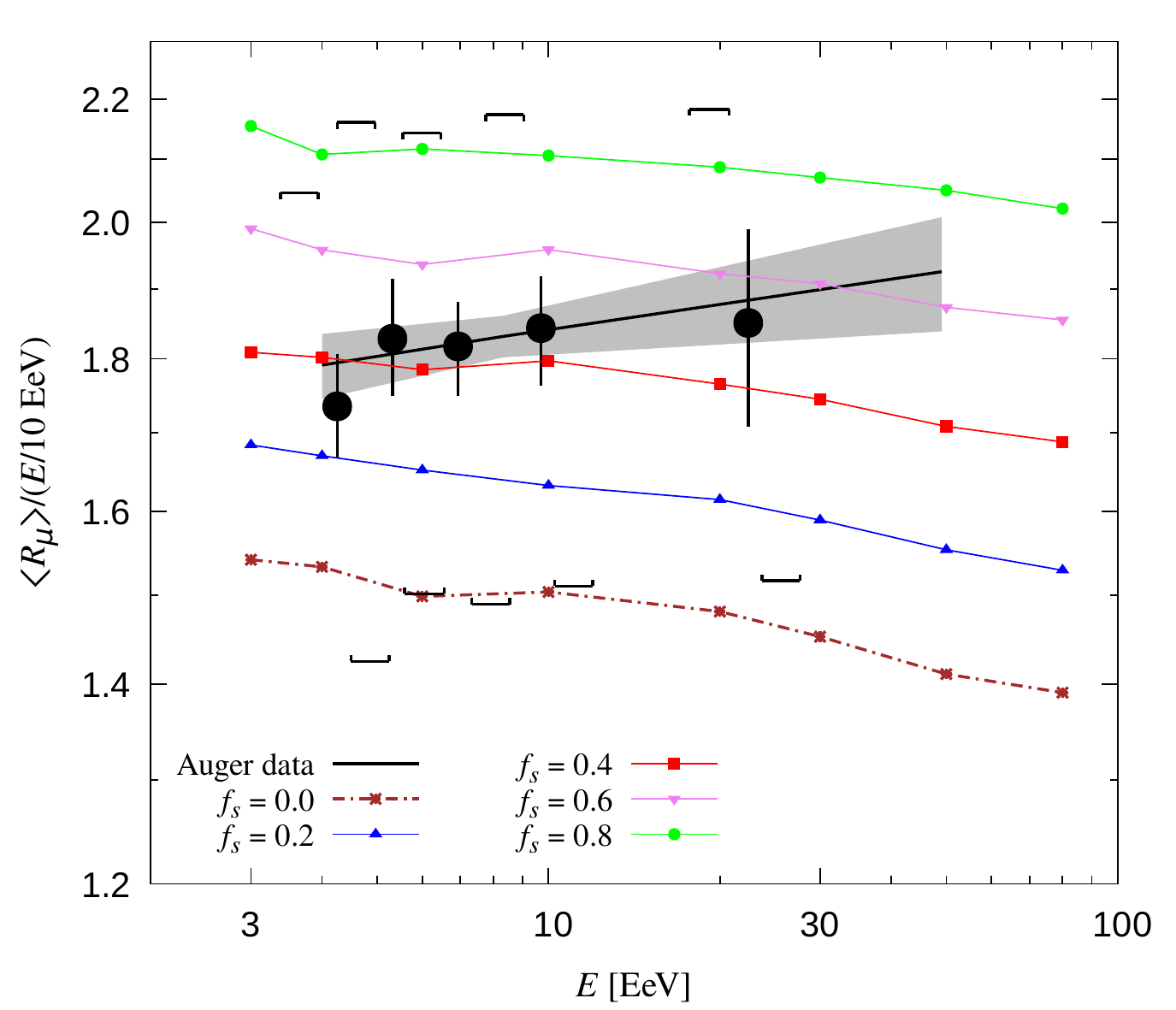}

\caption{The left panel puts on view fractions of ultra-high energy primary cosmic rays entering at the top of the Earth's atmosphere, as functions of
  the primary energy, evaluated from partial fluxes corresponding to
  the fit reported by the Pierre Auger Collaboration~\cite{PierreAuger:2016use}. The right panel exhibits estimations of the dimensionless muon content $R_\mu = N_\mu/N_{\rm ref}$ from \texttool{AIRES} simulations for different values of
  $f_s$ superimposed over Auger data with statistical
  \mbox{($\hspace{0.1em}\bullet\hspace*{-.66em}\mid\hspace{0.16em}$)}
  and systematic (\protect\rotatebox{90}{\hspace{-.075cm}[ ]})
  uncertainties~\cite{PierreAuger:2014ucz}. Here $N_\mu$ is the total number of muons (with $E_\mu >300~{\rm MeV}$) at
  ground level and $N_{\rm ref} =  1.455 \times 10^7$
  is the average number of muons in simulated proton showers at
  $10^{10}~{\rm GeV}$ (with incident angle of $67^\circ$). They  
 were obtained assuming the mixed cosmic ray composition shown in the
 left panel. Figure taken from Ref.~\cite{Anchordoqui:2022fpn}. \label{fig:uno}}
\end{figure}

In Ref.~\cite{Anchordoqui:2022fpn}, the assumption that the strangeness enhancement observed by ALICE at mid-rapidity~\cite{ALICE:2016fzo} further increases in the forward region was put into effect to study the concomitant  $\pi \to K$ swap impact on the development of extensive air showers. The study,  which was inspired on the ideas introduced in Refs.~\cite{Allen:2013hfa,Farrar:2013lra}, was carried out implementing a phenomenological toy model in \texttool{AIRES} (version 19.04.08)~\cite{Sciutto:1999jh}. The \texttool{AIRES} simulation engine provides full space-time particle propagation through the atmosphere and calls external event generators to process hadronic collisions. In the analysis of Ref.~\cite{Anchordoqui:2022fpn}  all hadronic collisions were processed using \texttool{Sibyll-2.3d}~\cite{Riehn:2019jet}, but introducing the possibility of  swapping $\pi \rightarrow K$ with a probability
\begin{equation}
  F_s (\eta)  =\left\{
\begin{array}{llc}
f_s & {\rm if} & -\infty < \eta < -4 \\
0 & {\rm if} & -4 \le \eta \le 4 \\
f_s & {\rm if} & \phantom{-!} 4 < \eta < \infty
\end{array}
\right. \,,
\label{Fs}
 \end{equation}                                       where
         $\eta$ is the pseudorapidity in the center-of-mass frame and $0<
                  f_s <1$. Particle swapping was performed in hadronic collisions whose projectile kinetic energy was larger than  $E_{\rm pmin} = 10^3~{\rm TeV}$.
This low energy threshold was selected to roughly accommodate the onset of a smooth rise of the hyperon-to-pion ratio measured at ALICE~\cite{ALICE:2013xmt}.  Secondary particles were randomly selected with probability $F_s$ and if their energies were below $E_{\rm smin} = 1~{\rm GeV}$ they are always left unchanged.  \\

In case of positive selection, the identity of particles was changed according to the following swapping criteria:
\begin{itemize}
    \item[(i)] Each $\pi^0$ was transformed onto $K^0_S$ or $K^0_L$, with 50\% chance between them.
  \item[(ii)] Each $\pi^+$ ($\pi^-$) was transformed onto $K^+$ ($K^-$).  
\end{itemize}

As shown in \cref{fig:uno}, for $0.4 <f_s< 0.6$, the toy model can partially accommodate Auger data~\cite{PierreAuger:2014ucz,PierreAuger:2016nfk}. A point worth noting at this juncture is that the shape of the best-fit curve to Auger data is driven by both strangeness enhancement and the rapid change in the nuclear composition~\cite{Sciutto:2019pqs}. Thus, nuclear effects~\cite{Anchordoqui:2016oxy} could play a conclusive role in bridging the gap  between data and simulations, hinting that  $F_s$ should also have a variation with the nucleus baryon number $A$. Along this line, a strong suppression of the production of neutral pions in pPb collisions was reported by the LHCf Collaboration after comparing to the results of pp scattering~\cite{LHCf:2014gqm}.

\begin{figure}[tb]
\centering
\includegraphics[width=0.92\textwidth]{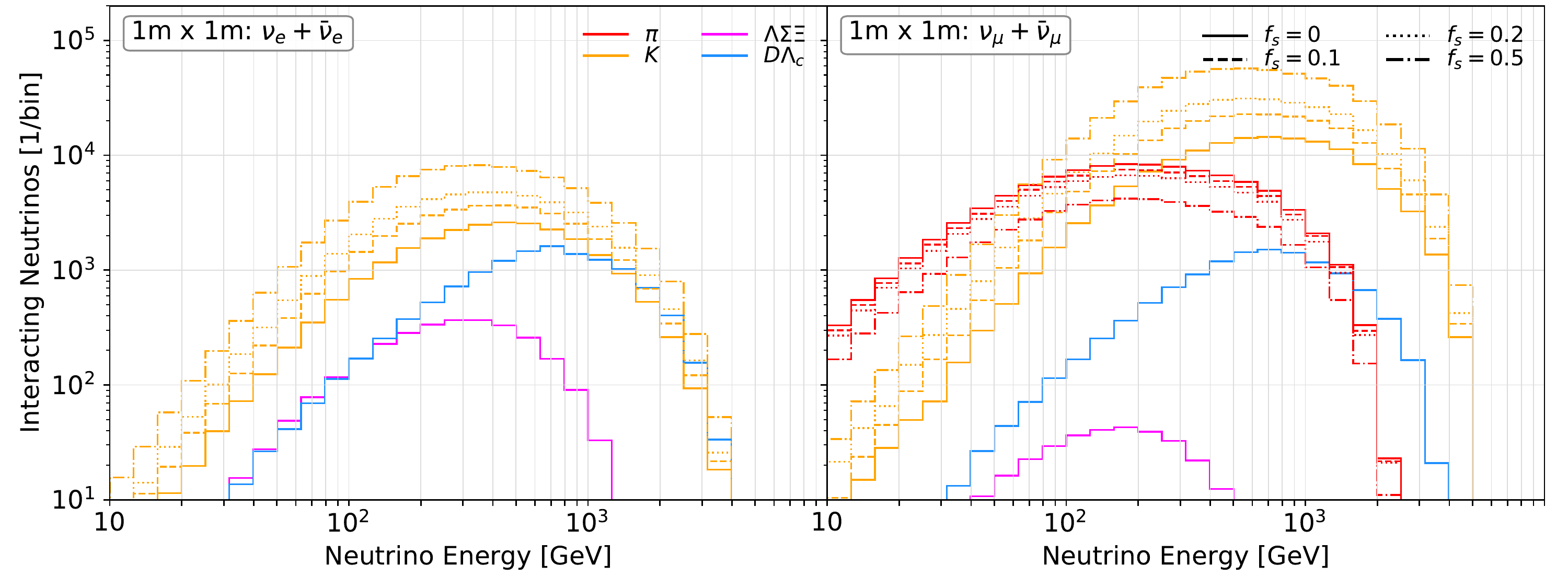}
\vspace{-1em}
\caption{Neutrino energy spectra for electron neutrinos (left) and muon neutrinos (right) passing through the FLArE detector. The vertical axis shows the number of neutrinos per energy bin that go through the detector's
cross-sectional area of $1\,\mathrm{m}^2$ for an integrated luminosity of $3~{\rm ab}^{-1}$: pion decays (red), kaon decays (orange), hyperon decays (magenta), and charm decays (blue). The different line styles correspond to predictions obtained from \texttool{Sibyll-2.3d} by varying $f_s$~\cite{Anchordoqui:2022fpn}. \label{fig:dos}}
\end{figure}

The analysis presented in Ref.~\cite{Kling:2021gos}
using the \texttool{RIVET}~\cite{Buckley:2010ar,Bierlich:2019rhm} module for fast
neutrino flux simulation has been duplicated in
\cref{fig:dos}, but considering the $\pi
\rightarrow K$ swapping
driven by $F_s$. It is remarkable that already for $f_s =0.1$ ($f_s =
0.2$) the predicted electron neutrino flux at the peak of the spectrum is a
factor of 1.6 (2.2) larger. These differences are significantly larger than the anticipated statistical uncertainties at the FPF~\cite{Kling:2021gos,Anchordoqui:2021ghd}.

Within this decade, ongoing detector upgrades of existing facilities, such as AugerPrime~\cite{PierreAuger:2016qzd}
and IceCube-Gen2~\cite{IceCube-Gen2:2020qha}, will enhance the precision of air shower measurements and reduce uncertainties in the interpretation of muon data. The FPF will provide invaluable complementary information to test models addressing the muon puzzle via  strangeness enhancement.

\section{Understanding the Atmospheric Background of Astrophysical Neutrinos\label{sec:astro_neutrinos}}

High-energy neutrinos of astrophysical origin are nowadays routinely observed by neutrino telescopes. In most cases these astrophysical neutrinos become visible as a diffuse flux~\cite{IceCube:2016umi}, but in some cases the source of the incident neutrino can be identified~\cite{IceCube:2018a,IceCube:2018b}. The next generation of neutrino telescopes like IceCube-Gen2~\cite{IceCube-Gen2:2020qha} and KM3NeT~\cite{km3net:2020a} are expected to detect one order of magnitude more cosmic neutrinos. With this increase in statistics, the number of identifiable sources is also expected to rise, as detectable sources can be five times fainter compared to IceCube. The detection of numerous neutrinos from the same source or the same class of sources, will then allow for a more detailed understanding of neutrino production mechanisms in the source and the acceleration of cosmic rays (see also the contributions to Snowmass 2021 on \emph{High Energy and Ultra-High Energy Neutrinos}~\cite{UHEnuWhitepaper} and \emph{Multimessenger Astronomy and Astrophysics}~\cite{MMWhitepaper}).

Atmospheric neutrinos, produced in the interaction of cosmic rays with nuclei in the Earth's atmosphere and the subsequent decay of mesons, are an irreducible background to searches for astrophysical neutrinos. An accurate understanding of the physics of cosmic sources therefore requires an in-depth understanding of the atmospheric neutrino flux. The FPF will provide key information for understanding astrophysical neutrinos, as described in the following.

\subsection{Atmospheric Backgrounds in Large-Scale Neutrino Telescopes\label{sec:astro_neutrinos_telescopes}}

Atmospheric neutrinos are produced by semileptonic decays of hadrons in EAS, typically $\pi^\pm$ and $K^\pm$ decays. This component of the flux is called the \emph{conventional neutrino flux} which falls steeply with increasing energy. This mainly reflects the spectral shape of the incoming cosmic ray flux, but there is an additional suppression that comes from the energy loss experienced by the mesons before they decay. The proper decay lengths of pions and kaons are ${\cal O}(1)\,\mathrm{m}$ and much longer at high energy, while their average interaction times are much shorter. They will therefore lose energy before decaying, leading to an extra factor of approximately $E_\nu^{-1}$ in the flux compared to the incoming cosmic rays. This qualitative picture is confirmed by theoretical predictions~\cite{Honda:1995hz,Gaisser:2002jj} that agree very well with measurements up to energies of roughly $10^{5}$~GeV~\cite{Schukraft:2013ya}.

On the other hand, when the energy is sufficiently high, atmospheric neutrinos also originate from the the semileptonic decay of heavy flavor hadrons, such as $D$ mesons, $B$ mesons, or $\Lambda_c$ baryons. This component is called the \emph{prompt neutrino flux}. In contrast to the pions and kaons, these heavy flavor hadrons have very short decay lengths of a few hundred \textmu{}m, and decay immediately to neutrinos after they are produced. Therefore the flux of prompt neutrinos falls off less quickly with energy than conventional neutrino flux and instead approximately reflects the energy dependence of the incoming cosmic ray flux.
As a result, the prompt neutrino flux is predicted to dominate the conventional neutrino flux above a certain energy range, roughly around $E_\nu \sim 10^5 - 10^6$~GeV.  There is also a flux of muons from the semileptonic decays of $K_S$, which have a critical energy of 120 TeV, so they are intermediate between conventional and prompt neutrinos.  At energies above this critical energy, they provide about 28\% of the neutrinos from kaons \cite{Gaisser:2014pda}.   

\begin{figure}[tb]
\centering
\mbox{\hspace{-1em}\includegraphics[width=1.0\linewidth]{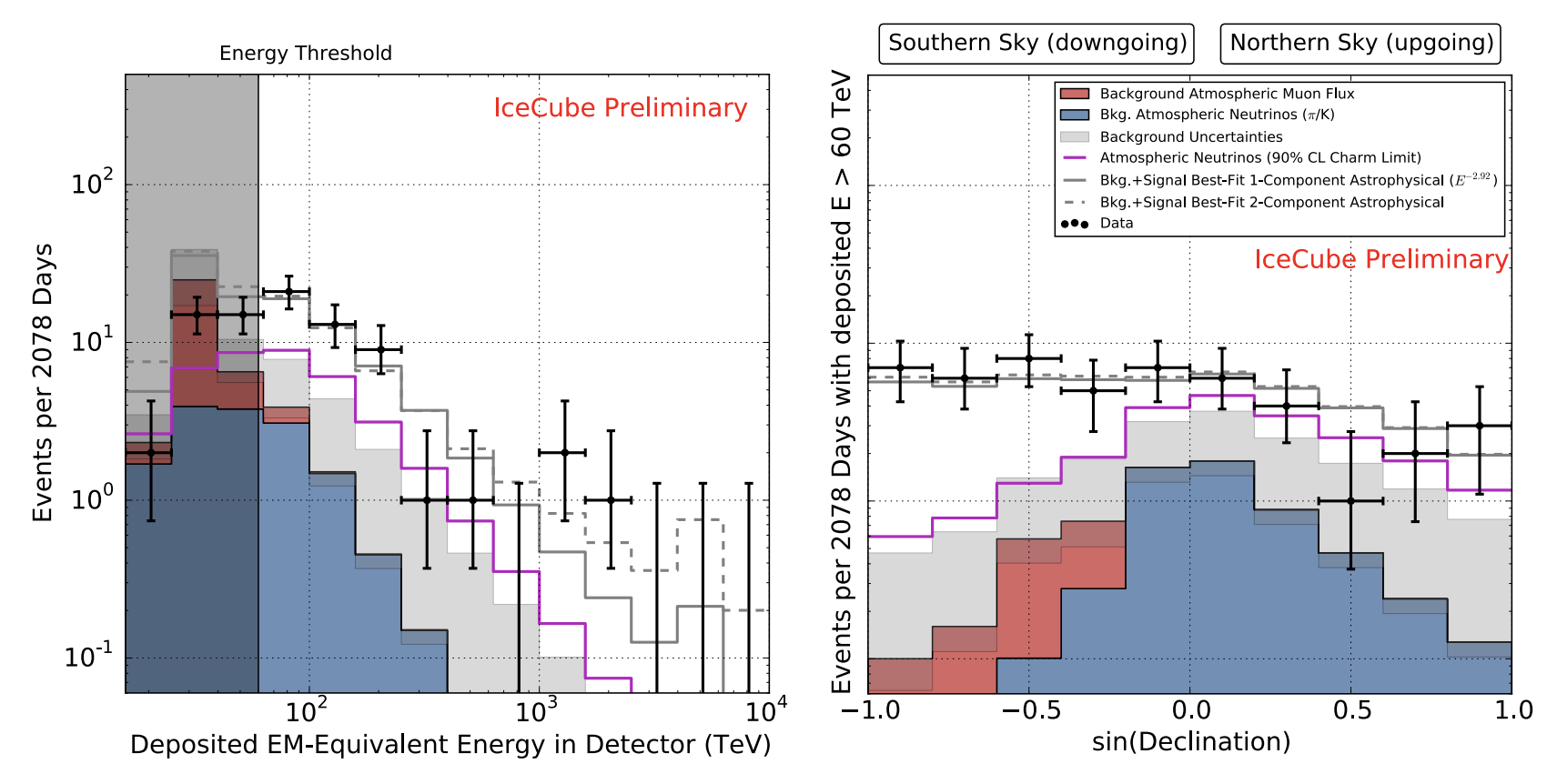}}
\caption{\label{fig:IceCubeAtmNu} 
Deposited energies and arrival directions of observed events in IceCube, as well as expected contributions from backgrounds and astrophysical neutrinos. The atmospheric muon backgrounds are estimated
from data (red), the atmospheric neutrino backgrounds (blue) are determined from simulations and include $1\sigma$ uncertainties (gray band). The $90\%$ CL upper bound on the charm neutrino flux is shown as a magenta line and the best-fit astrophysical spectra (assuming an unbroken power-law model) are shown in gray. The solid line assumes a single power-law model, whereas the dashed line assumes a two power-law model. Figure taken from Ref.~\cite{Kopper:2017zzm}.
}  
\end{figure}

At high energies, prompt atmospheric neutrinos are the primary background to astrophysical neutrinos that are hunted by high-energy neutrino observatories such as IceCube and KM3NeT. 
About a decade ago, the IceCube collaboration reported the first observation of two PeV energy neutrinos \cite{IceCube:2013cdw}, and astrophysical neutrinos have been continuously searched at energies between $1\,\mathrm{TeV}$ and $10\,\mathrm{PeV}$ \cite{IceCube:2013cdw,IceCube:2013low,IceCube:2014stg,IceCube:2015qii,IceCube:2020acn,KM3NeT:2018wnd}.
Prompt atmospheric neutrinos have not yet been observed by experiments and only upper limits have been set~\cite{IceCube:2016umi,Kopper:2017zzm,IceCube:2019lzm,IceCube:2018pgc,IceCube:2020wum}. \cref{fig:IceCubeAtmNu} shows the deposited energies and arrival directions of observed events in IceCube, as well as the expected contributions from backgrounds and astrophysical neutrinos~\cite{Kopper:2017zzm}. Also shown are the $90\%$ CL upper bound on the charm neutrino flux and the best-fit astrophysical spectra.

Ideally, one would eliminate all of the atmospheric background events, but this is experimentally not possible, instead, the remaining backgrounds must be estimated and subtracted. The acceptance for these events is typically determined using simulations and it is thus important to have experimental data to anchor these simulations and reduce the uncertainties.

At energies above $\sim 100\,\mathrm{TeV}$, the Earth becomes opaque to neutrinos, and neutrino observatories must concentrate on downward-going events \cite{IceCube:2021aen} where backgrounds from atmospheric leptons are a significant concern. Here, the main difficulty is the possibility of \emph{self-veto} which happens when the neutrino (or a muon from the neutrino) is accompanied by enough other particles from the EAS that created it so that the event is vetoed. There are generally two ways to veto atmospheric neutrino events originating from EAS in neutrino observatories:

\begin{itemize}
\item[(i)] The air shower can produce enough particles to be visible in a surface air shower array. The chance of this occurring depends on the EAS energy, and on the distance from the air shower core to the surface array. Air-Cherenkov telescopes can be used to further improve veto capabilities by measuring the EAS in the atmosphere through Cherenkov light emission \cite{Rysewyk:2019fdi,IceCube:2019yev}.
\item[(ii)] The neutrino may be accompanied by muons from the air shower.  In studies of starting events, where neutrino interacts within the detector, this muon will make the event look like a through-going muon, and so be vetoed.  In studied of through-going muons from neutrino interactions outside the detector, the additional muons will reduce the apparent stochasticity of the event, causing it to look like (and actually be) a muon bundle. There is also a very small background when a single air shower produces two neutrinos which interact in the detector \cite{vanderDrift:2013zga}. 
\end{itemize}

These self-veto probabilities must be evaluated mostly using simulations \cite{Gaisser:2014bja, Lunemann:2017blt,Anchordoqui:2021ghd,Heid:2015cfq} which usually consider neutrinos from charm and light hadrons separately.  Although charm production can, in principle, be calculated using pQCD, there are considerable uncertainties, especially in the forward direction, where one parton has a large Bjorken$-x$ and the other a very low $x$ value. Prompt atmospheric neutrino production and its uncertainties will be further discussed in \cref{sec:astro_neutrinos_charm}. 

The FPF will provide data to test and tune Monte-Carlo codes in the forward region in multiple ways. The measurements can be used to test the individual pQCD calculations and to reduce the uncertainties associated with them. In addition, the data can be used as input to calculations, which will in turn improve the available models. It is also important to take data with nuclei, especially with oxygen beams, to match the atmosphere and reduce nuclear uncertainties \cite{Klein:2009ew}.

By improving the models for prompt neutrino fluxes and by lowering the associated uncertainties, the physics program of the FPF will be highly beneficial in the context of astroparticle physics, as it will certainly lead to a better description of the major background component in searches for high-energy neutrinos from astrophysical sources. The benefit is expected to be even larger for future neutrino telescopes, like IceCube-Gen2~\cite{Lunemann:2017blt,IceCube-Gen2:2020qha}, where the measurement of neutrino spectra from certain source classes -- or even individual point sources -- comes within reach. These neutrino spectra will shed light on the acceleration of cosmic rays. In this context, it becomes clear that the background of prompt atmospheric neutrinos needs to be understood as thoroughly as possible in order to gain meaningful insights from these source spectra. 

Synergies of prompt atmospheric flux models are, however, not limited to measurements of the astrophysical neutrinos. By enabling a more detailed description of the atmospheric neutrino flux, the FPF is also expected to improve the understanding of small scale features in these spectra. Possible examples include an accurate detection of the cosmic ray knee in atmospheric neutrino spectrum, as well as a more accurate description of the impact of seasonal modulations in the atmospheric temperature on the spectra. The physics program of the proposed Forward Physics Facility is therefore crucial for an accurate measurement of the prompt atmospheric neutrino flux.

\subsection{Prompt Atmospheric Neutrino Production\label{sec:astro_neutrinos_charm}}

As described in \cref{sec:astro_neutrinos_telescopes}, the prompt atmospheric neutrino flux originates from the semileptonic decay of heavy hadrons, such as $D$ mesons, $B$ mesons, or $\Lambda_c$ baryons, and it is predicted to dominate the conventional flux above around $E_\nu \sim 10^5 - 10^6$~GeV. The atmospheric nucleon-nucleon collisions responsible for producing the prompt flux around the interesting range of neutrino energies is in fact very relevant for LHC: for a neutrino lab frame energy of $1\,\mathrm{PeV}$, the corresponding $\sqrt{s_{NN}}$ is about $8\,\mathrm{TeV}$. As we shall see, the heavy flavor hadrons are produced at very large rapidities, so we may expect to study the underlying production process at the FPF experiments.

One approach to evaluating the atmospheric neutrino flux is to use the semi-analytic $Z$-\emph{moment method}, which yields an approximate solution to the coupled cascade equations that relate the incident cosmic ray flux and the produced hadrons and leptons \cite{Gaisser:2016uoy,Lipari:1993hd}.
The general form of the cascade equation for a particle $j$ is given by %
\begin{eqnarray}
\frac{d\phi_j(E,X)}{dX}&=&-\frac{\phi_j(E,X)}{\lambda_j(E)} - \frac{\phi_j(E,X)}{\lambda^{\rm dec}_j(E,X)}
+ \sum S(k\to j)\,
\end{eqnarray}
with column depth $X$ and interaction and decay lengths $\lambda_j$ and $\lambda_j^{(\rm dec)}$. These equations thus describe the propagation of particles in the atmosphere in terms of their flux change.
The $Z$-moments are related to the source terms $S(k\to j)$ for particle generation, which involve the heavy flavor production cross-section and the decay distribution. They are given by
\be
S( k \to j ) &= \int_E^\infty dE' \frac{\phi_k(E',X)}{\lambda_k(E')}
\frac{dn(k \to j ;E',E)}{dE} \\
&\simeq 
Z_{kj}(E)\frac{\phi_k(E,X)}{\lambda_k(E)}\, .
\ee
With some approximations, \emph{e.g.} $\phi_k (E,X)\simeq \phi_k^0(E)\exp{(-X/\Lambda_k)}$, the $Z$-moments are obtained as the energy dependent factors $Z_{pp}$ for proton reproduction, $Z_{ph}$ and $Z_{hh}$ for hadron production, and $Z_{h\nu}$ for neutrino production, that contain all particle physics aspects of particle production.
The approximate solutions of the coupled cascade equations can then be expressed in terms of these $Z$-moments, and yield closed expressions for the resulting neutrino flux in terms of the $Z$-moments, decay and interaction lengths, and incoming cosmic ray flux.

The particle physics inputs are therefore the source terms $(k\to j)$, which in turn contain the differential energy distributions of heavy quarks and neutrinos. The basic quantity that must be computed is thus the differential cross-section $d\sigma/dx_q$, where $x_q$ is the longitudinal energy fraction of the produced quark. Alternatively one uses the Feynman-$x$ variable $x_F\simeq x_q$.  

Alternatively, atmospheric neutrino flux can be also evaluated numerically using, for example, \texttool{MCEq} (Matrix Cascade Equations) \cite{Fedynitch:2015zma}, which is a numerical code to solve the cascade equations. \texttool{MCEq} uses the \texttool{Sibyll-2.3} event generator to evaluate heavy flavor production cross-sections.

\begin{figure}[tb]
\centering
\includegraphics[width=.65\linewidth]{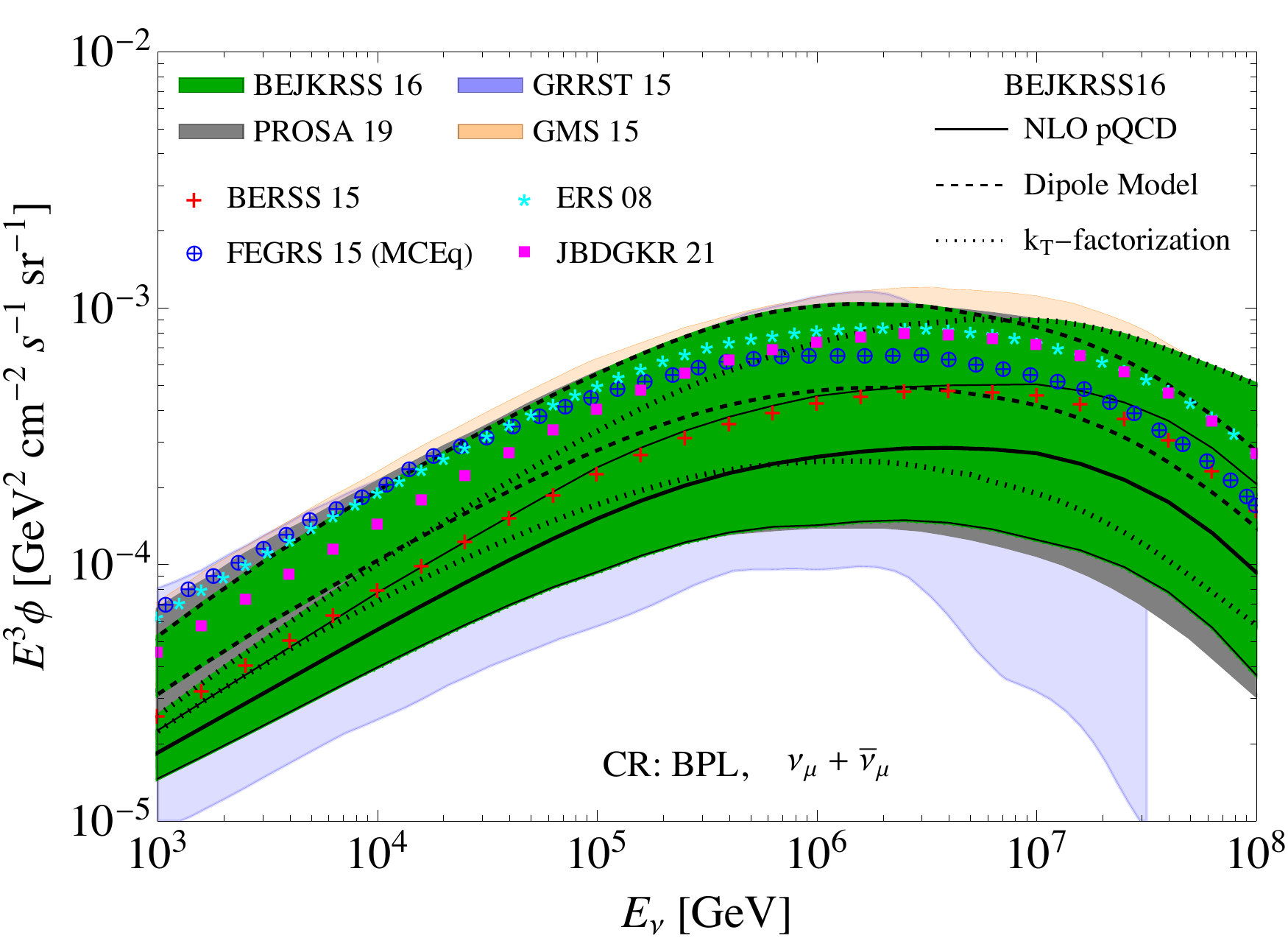}
\caption{\label{fig:PromptAtmNu} 
Comparison of the prompt atmospheric muon neutrino flux, $\phi$, from various recent flux calculations \cite{Bhattacharya:2016jce,Zenaiev:2019ktw,Gauld:2015kvh, Garzelli:2015psa,Bhattacharya:2015jpa,Enberg:2008te,Fedynitch:2015zma,Jeong:2021vqp}, as a function of the neutrino energy, $E_\nu$, following Ref.~\cite{Zenaiev:2019ktw}. The incident cosmic ray flux is approximated with a broken power law (BPL) in all predictions.
}  
\end{figure}

In \cref{fig:PromptAtmNu}, several modern predictions for the prompt $\nu_\mu + \bar{\nu}_\mu$ flux obtained by different groups are shown, using both the $Z$-moment method and the numerical method \cite{Bhattacharya:2016jce,Zenaiev:2019ktw,Gauld:2015kvh, Garzelli:2015psa,Bhattacharya:2015jpa,Enberg:2008te,Fedynitch:2015zma,Jeong:2021vqp}. These predictions use a broken power law (BPL) cosmic ray spectrum, which used to be the standard in earlier works. Although this is not a very good approximation, considering modern cosmic ray data, it allows for a transparent comparison of different particle physics calculations. However, the presented fluxes are therefore not completely realistic.

As shown in the figure, the available theoretical predictions of the prompt atmospheric neutrino flux have large uncertainties. These uncertainties come from many of the ingredients in the evaluation of the prompt fluxes, \emph{e.g.} the cross-sections for heavy flavor production, parton distribution functions (PDF) and fragmentation functions. Among these factors, the heavy flavor production cross-section brings about the most significant uncertainties. There is additional large uncertainty in the cosmic ray flux, as well in the fact that the air consists of nuclei with an average $\langle A\rangle=14.5$, so that nuclear effect may be important. We will not discuss these uncertainties here, but refer to, \emph{e.g.} Ref.~\cite{Bhattacharya:2016jce} for a discussion.

During cosmic ray interactions in the atmosphere, the heavy flavors are produced in the forward direction. In other words, the partons involved in heavy flavor production, mostly gluons, have very asymmetric longitudinal momentum fractions. The Bjorken-$x$ of the two incoming gluons are related to the final state kinematical variables through
\be
x_{1,2} = \frac{1}{2}\left( \sqrt{x_F^2+\frac{4m_{q\bar q}^2}{s}}
\pm x_F\right), 
\ee
where the invariant mass $m_{q\bar q}$ of the heavy quarks is far smaller than $\sqrt{s}$. This means that $x_1\sim x_F$ and $x_2\sim m_{q\bar q}^2/(x_F s) \ll 1$. Thus, the $x$-values are very small, and very large in the nucleons in the Air nuclei and in the cosmic rays, respectively. 

As an illustration of the $x$-values, for incoming cosmic ray energies of $E_{p}=100$~TeV or 1~PeV we get in the forward limit ($x_F\simeq 1$) that $x_2 = 3\times 10^{-5}$ or $x_2 = 3\times 10^{-6}$. These are extremely small values of $x$ -- far smaller than anything accessible at today's accelerators.

The small $x$ relevant for prompt atmospheric neutrinos are not probed by experiments at present, and available PDF parametrizations therefore are not reliable in this region. Moreover, at such small $x$, large logarithms $\ln(1/x)$ must be resummed, leading to effects described by the BFKL equation. Theoretically, one also expects parton saturation effects at very small $x$. There are some models that incorporate these effects such as the dipole model and $k_T$ factorization approach.

The standard method to calculate the heavy flavor production cross-section is to use NLO perturbative QCD with the collinear approximation, which is used in the predictions of BERSS 15 \cite{Bhattacharya:2015jpa}, BEJKRSS 16 \cite{Bhattacharya:2016jce} and JBDGKR 21 \cite{Jeong:2021vqp} in \cref{fig:PromptAtmNu}.
The models that incorporate the low $x$ effects are adopted in ERS 08 \cite{Enberg:2008te} for the dipole model, and in BEJKRSS 16 \cite{Bhattacharya:2016jce} for both dipole model and $k_T$ factorization. 
In Ref.~\cite {Bhattacharya:2016jce}, thus predictions by three different approaches for the heavy flavor production are compared, and the combined uncertainty is shown as a green band in the figure.
The other predictions are obtained from the Monte-Carlo generators. 
In particular, GMS 15 \cite{Garzelli:2015psa}, GRRST 15 \cite{Gauld:2015kvh} and PROSA19 \cite{Zenaiev:2019ktw} performed the simulations with POWHEG for the heavy quark production and \texttool{Pythia} for fragmentation, and FEGRS 15 \cite{Fedynitch:2015zma} used \texttool{Sibyll-2.3} event generator as mentioned above. 

\begin{figure}[tb]
\centering
\includegraphics[width=.65\linewidth]{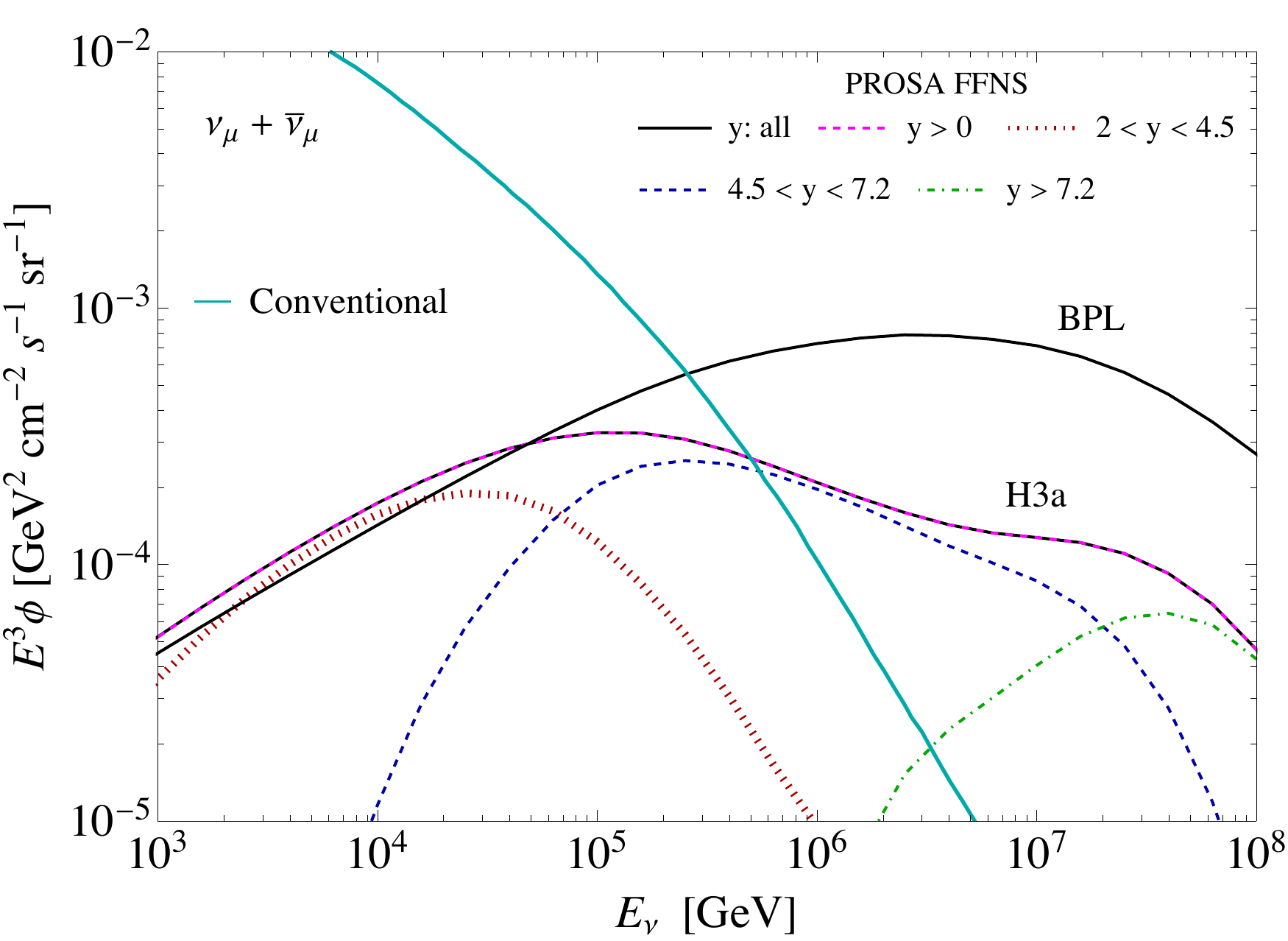}
\caption{\label{fig:PromptAtmNu-rap} 
Prompt atmospheric neutrino fluxes for $\nu_\mu + \bar{\nu}_\mu$ from the charm produced in different collider rapidity ranges~\cite{Jeong:2021vqp}. Also shown is the conventional atmospheric neutrino flux from Ref.~\cite{Honda:2006qj}. In the prompt flux evaluation, the pp charmed hadron energy distributions are scaled to account for the air target average atomic number $A = 14.5$, as described in the text. The calculation of prompt atmospheric fluxes involves pA collisions in a wide range of center-of-mass energies, including LHC energies.}  
\end{figure}

Prompt neutrinos are produced predominantly from charm meson decays. In order to evaluate the prompt atmospheric neutrino flux, one can compare the cross-sections for charm meson production with the LHC data. Charm mesons from the cosmic ray interactions in the atmosphere are produced in the forward direction, therefore it is relevant to compare in particular with the data from the LHCb experiment \cite{LHCb:2015swx}, which investigate the most forward region up to date, \emph{i.e.} $2.0<y<4.5$ for charm meson rapidities.
\cref{fig:PromptAtmNu-rap} shows the prompt atmospheric neutrino fluxes for $\nu_\mu + \bar{\nu}_\mu$ from the charm produced at different collider rapidity ranges in pp collisions \cite{Jeong:2021vqp} as well as the conventional neutrino flux. 
In splitting the rapidity ranges, we take $2.0<y<4.5$ probed by LHCb experiment, $y>7.2$ that will be investigated at the forward experiments during the Run 3 of the LHC, and the remained region between the two ranges, $4.5<y<7.2$.
In evaluating the prompt neutrino flux, the energy distributions of the cross-sections for charm production in pp collision are scaled by average atomic number for air nuclei $A = 14.5$ to approximate the $pA$ collision. 
The results in the figure are evaluated using one of more recent cosmic ray spectra than the BPL, the known as H3a, which is parameterized considering the sources and composition of cosmic rays \cite{Gaisser:2011klf} 
(for more parameterizations of such cosmic ray spectra and their application to the predicted prompt neutrino fluxes, we refer to the Refs. \cite{Gaisser:2011klf, Gaisser:2013bla, Bhattacharya:2016jce}).
For comparison, the prediction of the prompt neutrino flux with the BPL spectrum is also presented. 
As shown in the figure, the prompt atmospheric neutrinos at the energies where the prompt neutrinos are dominant are from the charm produced in the rapidity of $y > 4.5$. 

As mentioned above, the LHCb Collaboration has published charm meson production data for $2 < y < 4.5$.  
In addition, two forward experiments at the LHC, FASER$\nu$ \cite{FASER:2020gpr} and SND@LHC \cite{SHiP:2020sos}, will directly measure the prompt neutrinos during the Run 3 stage at more forward region $y > 7.2$.
The FPF during the high-luminosity era will be able to obtain data with much higher statistics and possibly with more coverage of the rapidity range, thereby providing important information to reduce the uncertainty and constrain the model for the charm production. Consequently, the FPF will play a crucial role in improving the 
predictions of the prompt atmospheric neutrino flux. 
This, in turn, will have significant impact on the searches for astrophysical neutrinos, as described in \cref{sec:astro_neutrinos_telescopes}, Moreover, an improved modelling of prompt atmospheric neutrinos might also allow for the identification of a \emph{sweet spot}, where the influence of this particular component becomes not only visible, but clearly distinguishable from all other effects. This would significantly help to measure the prompt atmospheric neutrino flux in large-scale neutrino telescopes for the first time.

\section{Dark Matter Searches and Their Impact on Astrophysics and Cosmology\label{sec:astro_DM}}

One of the primary motivations for BSM searches in the FPF is their possible connection to the efforts towards understanding the nature of DM in the Universe. In particular, as discussed in \cref{sec:bsm2}, light, sub-GeV DM particles can be abundantly produced in the far-forward region of the LHC and can be directly searched for with the scattering signatures. Further discovery prospects arise from the proposed search for light unstable species that can be mediator particles between DM and the SM, as described in \cref{sec:bsm1}. Importantly, such thermally-produced and light DM particles can additionally be strongly constrained by the possible impact of their annihilations on the cosmic microwave background (CMB) radiation (see also the contribution to Snowmass 2021 \emph{Cosmology Intertwined}~\cite{CosmologyWhitePaper}). As a result, the relevant searches in the FPF are typically competitive with current bounds in scenarios predicting suppressed DM annihilation rates, \emph{e.g.} due to a velocity dependence of the corresponding cross section.

This, however, can be circumvented in more rich dark sector scenarios, in which the DM production in the early Universe occurs differently than it would be dictated by the vanilla freeze-out mechanism. In this section, we will first illustrate possible complementarity between DM indirect detection strategies and the FPF searches for unstable light new species in example scenarios employing the freeze-in DM production mechanism in \cref{sec:astro_DM_freezein,sec:astro_DM_sterileneutrinofreezein}. In this case, the DM species remain out of thermal equilibrium with the SM particles in the early Universe. Interestingly, in these selected models, complementary search strategies employing a variety of experimental signatures can probe such scenarios well in the coming years even though the freeze-in production typically corresponds to much-suppressed couplings between DM and the visible sector. This can also extend to direct DM searches in underground detectors, as discussed in \cref{sec:astro_DM_scaleinvariance}. If DM particles thermalize in the early Universe, due to the coupling with the light unstable mediator species, the complementarity between the FPF searches and DM indirect detection can be observed in rich dark sector models in which DM is well secluded from the SM, as described in \cref{sec:astro_DM_richdarksector}.

\subsection{Dark Matter from Freeze-In Semi-Production\label{sec:astro_DM_freezein}}

An interesting alternative to thermal freeze-out is the so-called freeze-in mechanism, \emph{i.e.} a gradual production of dark matter due to an energy leakage from the SM thermal bath through a very feeble interaction. Models featuring this scenario for the origin of DM are by construction significantly harder to detect in traditional searches, both direct and indirect ones. However, a common feature of the majority of such models is the existence of a \textit{portal}, \emph{i.e.} the interaction between the dark sector (DS) and the SM sector takes place through an additional mediator field that is a singlet of the SM gauge group (see \emph{e.g.} Ref.~\cite{Bernal:2017kxu} for a review). Possible renormalizable couplings to the SM sector include mixing with the Higgs for scalar mediators, kinetic mixing with $U(1)_Y$ for vector mediators, or interactions through a connection to right-handed neutrinos, see also discussion in \cref{sec:bsm1}. It follows that even if hard to test in conventional DM searches, many freeze-in DM models can imprint signals in FPF through the presence of this mediator state.

Theories with scalar mediators connected through the Higgs portal are certainly an attractive possibility. The mixing through cubic or quartic terms is generic for any scalar and it is natural to incorporate such a mediator in the DS as well. Among recently studied ideas that possess this feature are the \textit{freeze-in from semi-production}~\cite{Hryczuk:2021qtz} (see also Ref.~\cite{Bringmann:2021tjr}) and \textit{forbidden freeze-in}~\cite{Darme:2019wpd} scenarios. These variants introduce relatively generic mechanisms for DM production and  are especially interesting phenomenologically because they require larger couplings than in the usual freeze-in. Therefore, besides the mediator signals in the forward physics experiments, these freeze-in models can be tested via common dark matter search strategies.

\begin{figure}[t]
\centering
\mbox{\hspace{-2em}\includegraphics[width=0.5\textwidth]{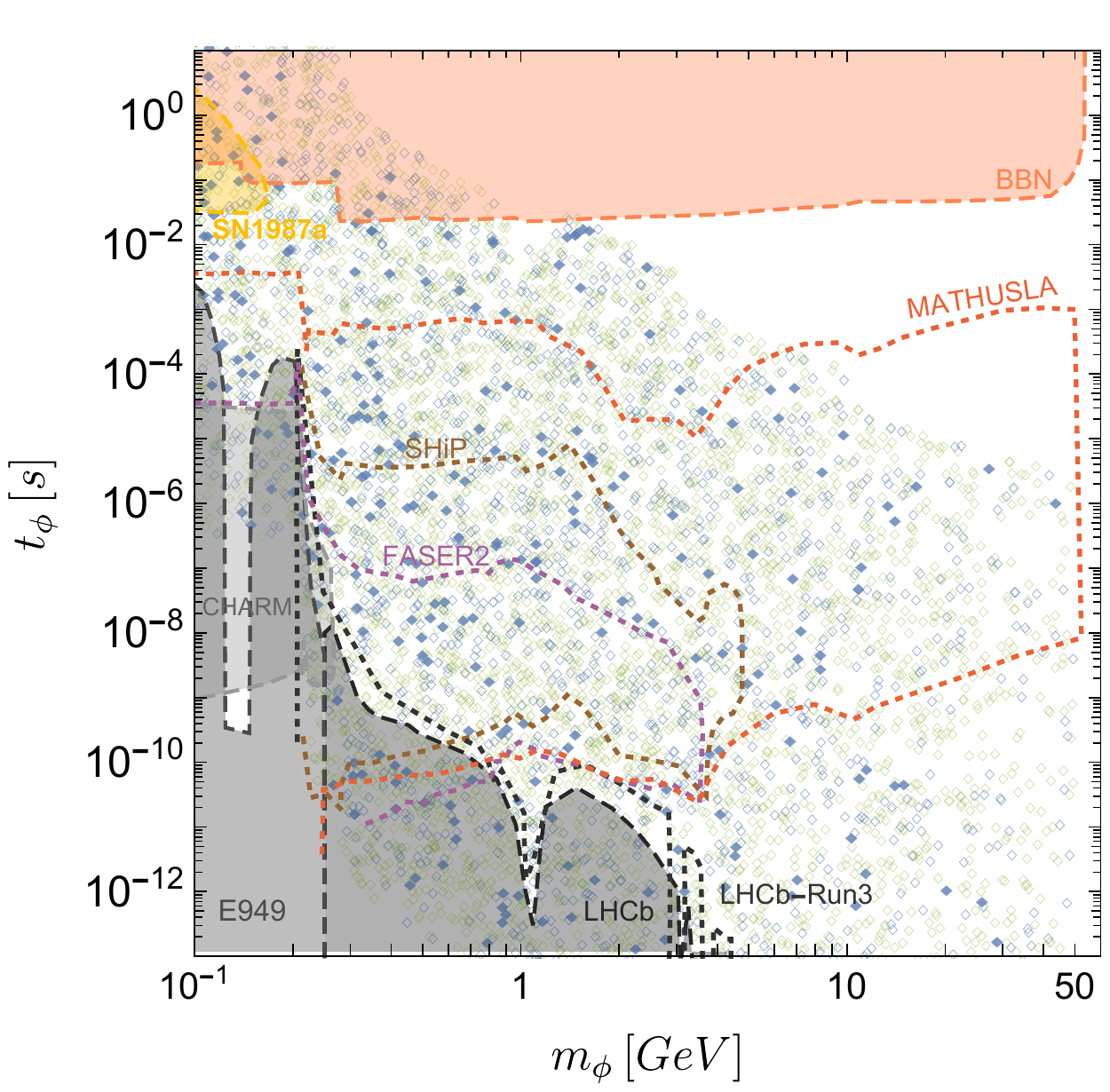}
\includegraphics[width=0.5\textwidth]{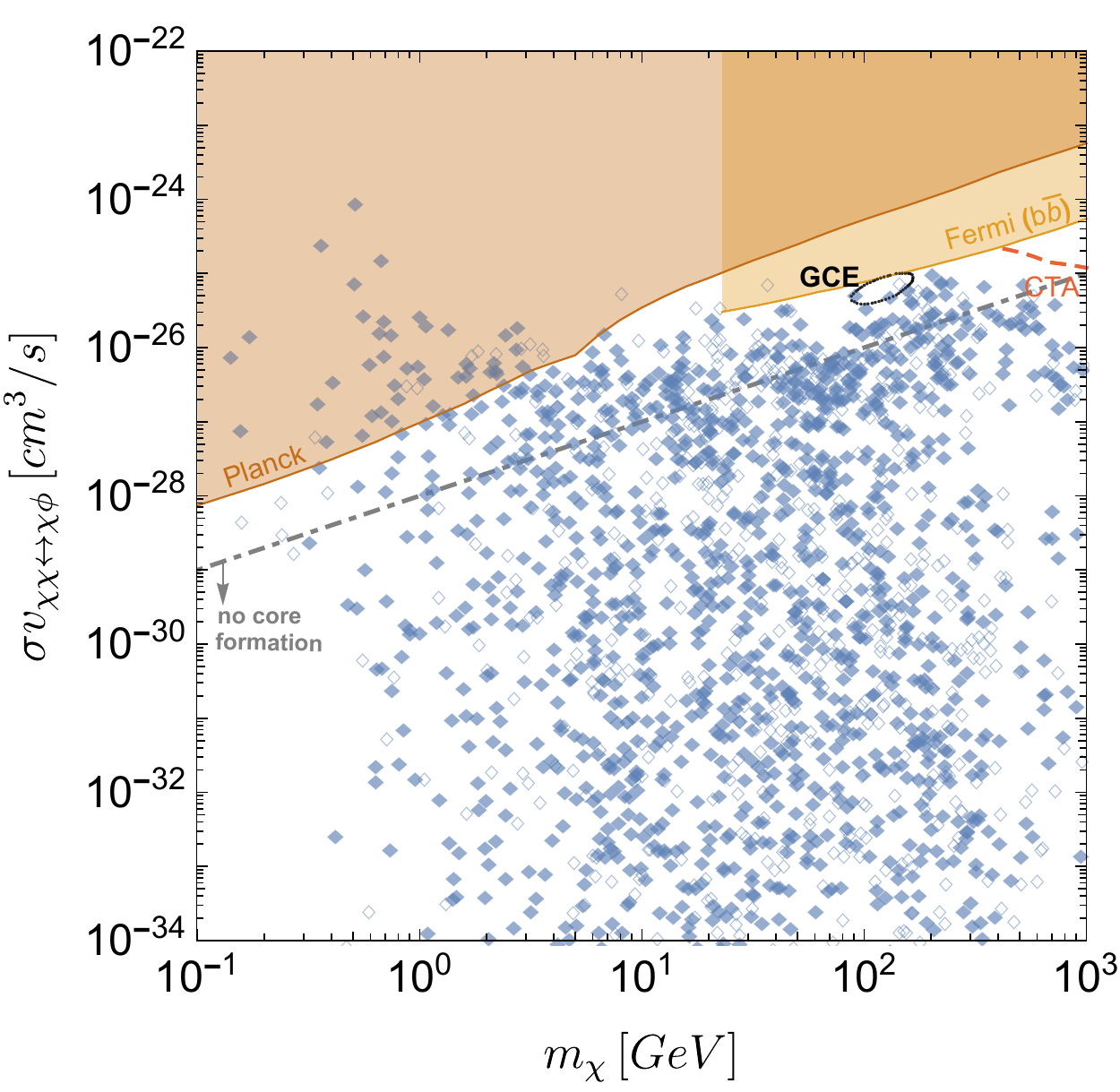}}
\caption{Left: The constraints and search prospects for the scalar mediator $\phi$ in the freeze-in mechanism via semi-production processes that decays to various SM states through the Higgs portal. The blue/green points correspond to the regime of dominant semi-annihilation/pair-annihilation and the filling of the points signify the ones that can be probed in indirect searches, while empty ones are beyond reach.  Right: Indirect detection constraints and prospects. The filled points signify the parameter sets that fall within the reach of the mediator searches, while the empty ones are beyond these prospects. For more details see Ref.~\cite{Hryczuk:2021qtz}.
}
\label{fig:scanphi}
\end{figure}

Semi-production is based on the reaction of the dark matter candidate $\chi$ with the mediator $\phi$ giving rise to two dark matter particles: 
\begin{equation}
    \chi \phi \rightarrow \chi \chi\,.
\end{equation}
Such a process is possible, and indeed can be dominant, in the models where the DM state is stabilized by symmetry that is larger than the common $Z_2$ symmetry, the simplest interesting example being $Z_3$. The reverse process, namely semi-annihilation, gives a well-known variant of the DM production in the thermal freeze-out paradigm \cite{DEramo:2010keq}. Since the rate of the $\chi \phi \rightarrow \chi \chi$ process is proportional to the DM number density, which during the early stages of freeze-in is very small, the semi-production is strongly suppressed at the beginning. Therefore, to produce substantial amounts of DM in this scenario one indeed requires larger couplings. In fact, large enough  to potentially lead to observable effects in indirect searches, especially in the case when both the mediator and the dark matter particle were never in thermal equilibrium in the early Universe. Nevertheless, most of the parameter space of the model studied in Ref.~\cite{Hryczuk:2021qtz}, still evades the current and upcoming search sensitivities in indirect detection. 

Fortunately, the prospects are much more promising in accelerator-based searches including in the far-forward region of the LHC with the FASER2 detector as can be seen by comparing the left and right panel of \cref{fig:scanphi}. Interestingly, these results also show that freeze-in from semi-production has the potential of simultaneously providing the correct relic density, cross-section in the range required to explain the Galactic Centre excess through DM (semi-)annihilation, strong enough self-heating in DM haloes to induce core formation \cite{Chu:2018nki} and, on top of this, measurable signals in the FPF. What is worth emphasizing, this mechanism is also not tied to any single model but can be realized more broadly whenever a given theory allows for a semi-annihilation process. 

Forbidden freeze-in refers to a scenario where the DM species is produced through the decay of the mediator, which at zero temperature is forbidden kinematically. In the early Universe, however, thermal effects open up this process leading to a non-zero DM production. In this scenario, the freeze-in is again suppressed, now due to the production era being limited only to a short time window at high temperatures, while still retaining independence from the unknown value of the reheating temperature. Therefore, the coupling between the mediator and the DM needs to be larger leading to a significant part of the parameter space being in a mass and lifetime range that can be probed in forward searches \cite{Darme:2019wpd}. It is worth noting that many freeze-in models relying on production from a decay may exhibit interesting, still not well studied, forbidden regimes.

\subsection{Freeze-In Sterile Neutrino Dark Matter\label{sec:astro_DM_sterileneutrinofreezein}}

A simple approach to explain the origin of  tiny neutrino mass is the seesaw mechanism, in which the suppression of the neutrino mass can be related to the presence of a  heavy particle at a high scale, which is also the scale of lepton number violation.  On the other, hand there is an alternative idea called the inverse seesaw mechanism in which the smallness of the neutrino mass can be explained by  small lepton number violating term in the Lagrangian, whereas the heavy particles can be at the TeV scale.  

This model allows for appreciable mixing between light and heavy states. We consider a general U$(1)^\prime$ extended framework \cite{Das:2016zue, Das:2021nqj, Das:2019pua} where three SM singlet right-handed neutrinos (RHNs) and three gauge singlet Majorana fermions are introduced to generate the light neutrino mass via the inverse seesaw mechanism. This model consists of an SM singlet scalar which can mix with the SM like Higgs. The presence of the three-generation of RHNs makes the model free from gauge mixed gauge-gravity anomalies \cite{Das:2019pua}. The model contains an extra $Z^\prime$ which gets mass when the $U(1)^\prime$ symmetry is broken. The cancellation of gauge and mixed gauge-gravity anomalies determines the  $U(1)^\prime$ charges of the fermions  and we find that the $Z^\prime$ interacts differently with left-handed and right-handed charged fermions. Interesting phenomenological consequences of this scenario have been discussed, for instance, in Ref.~\cite{Das:2021esm}. In this framework, we assign one pair of degenerate sterile neutrinos to be a potential dark matter candidate whose relic density can be generated by the freeze-in mechanism. We consider different mass regimes of the sterile neutrino DM $(m_{\rm DM})$ and the $Z^\prime$ boson $(M_{Z^\prime})$ which can be studied by choosing an appropriate reheating temperature $(T_R)$. We highlight the prospects for the dark vector searches in the FPF.

To study DM phenomenology from the freeze-in mechanism we consider two cases where the observed relic abundance of 0.12 can be reproduced \cite{Das:2021nqj}: 

\begin{figure}[t]
\centering
\includegraphics[width=0.47\textwidth]{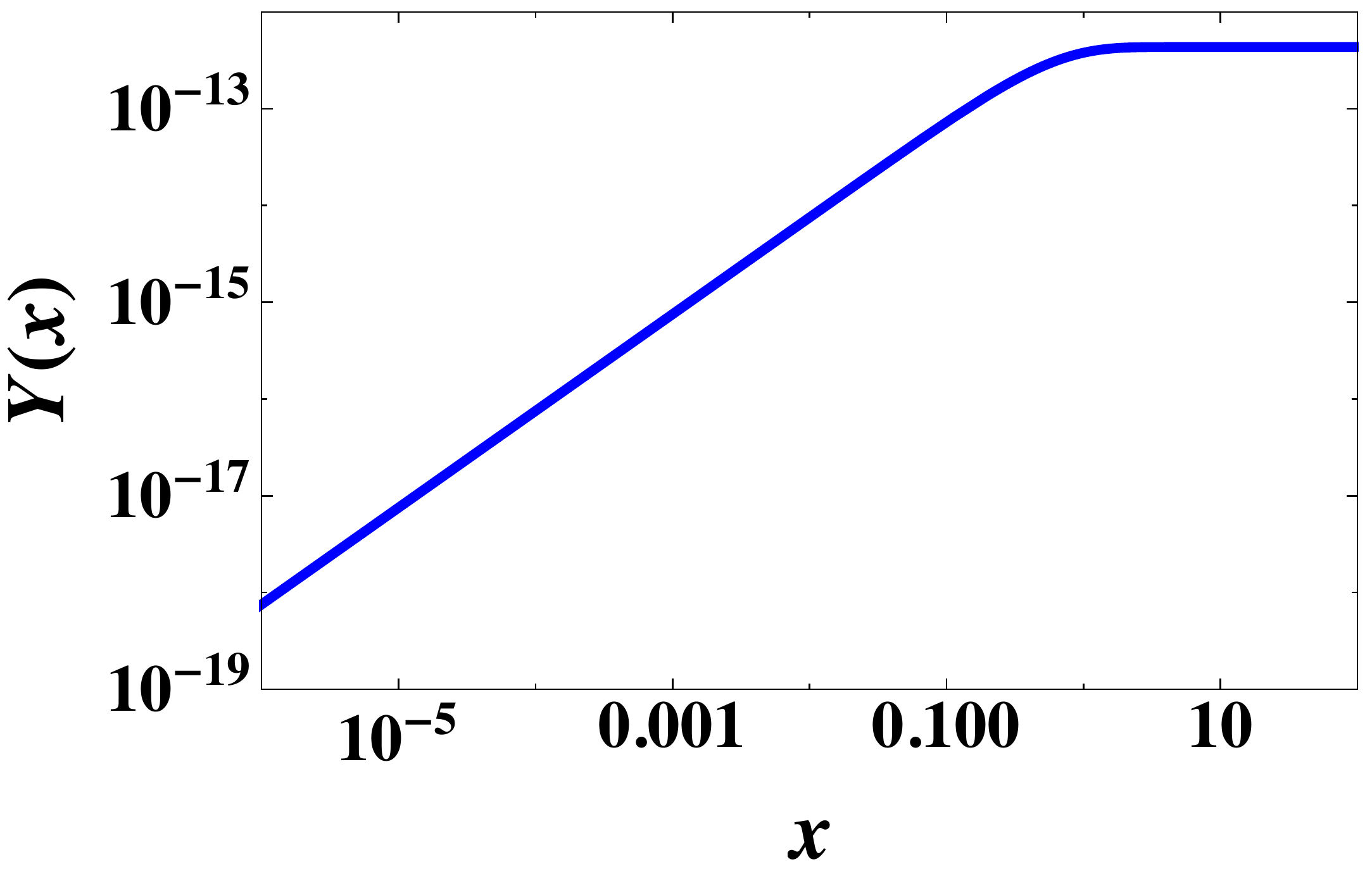}
\includegraphics[width=0.49\textwidth]{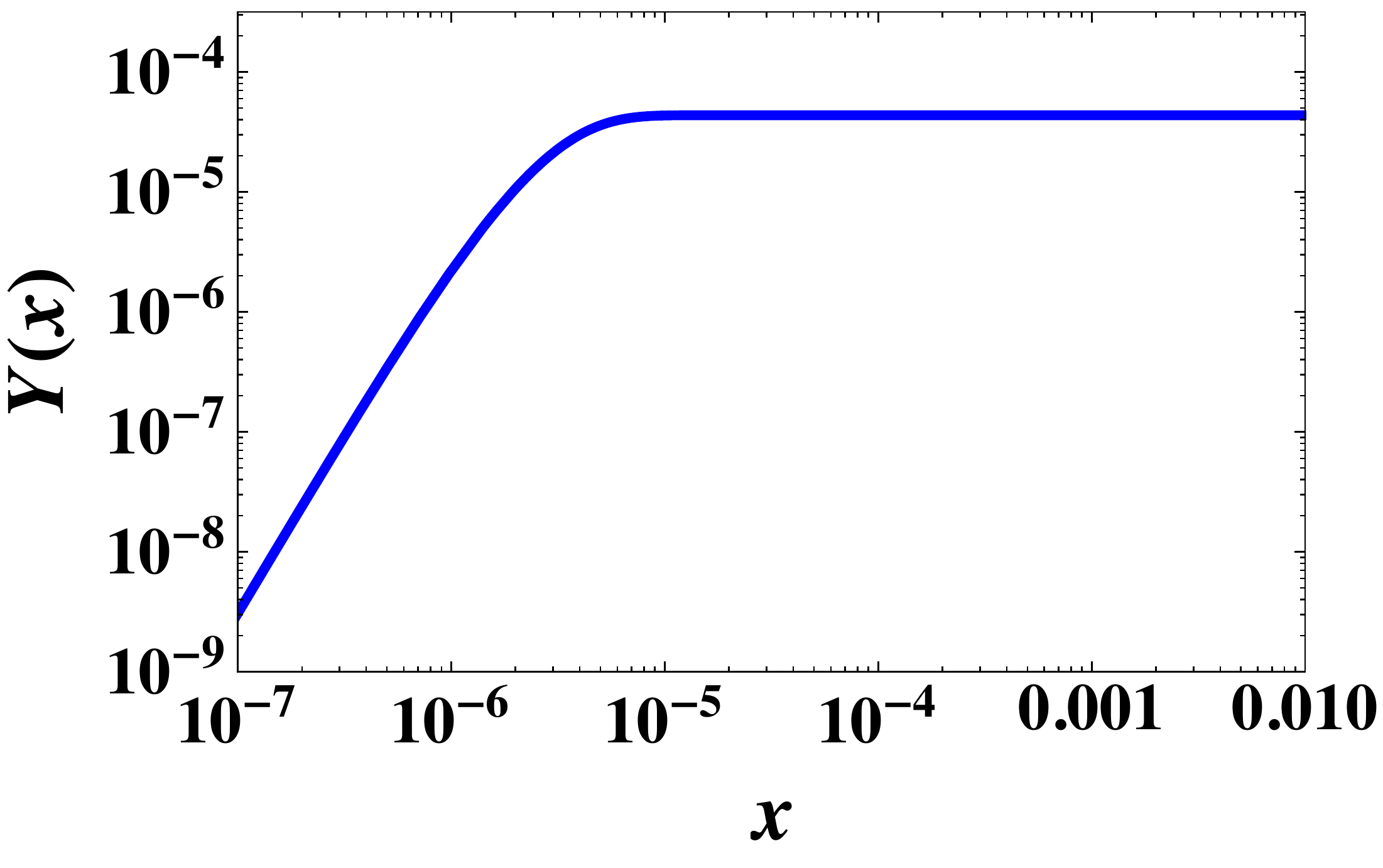}
\caption{Variation of the DM yield $Y(x)$ with $x=m_{\rm{DM}}/T_R$ for $T_R>>m_{\rm DM}>>M_{Z^\prime}$ (left) and $x$ for $T_R>>M_{Z^\prime}>>m_{\rm DM}$ (right). Figure taken from Ref.~\cite{Das:2021nqj}.}
\label{fig:m1}
\end{figure}

\begin{itemize}
    \item[(i)]In the limit of $T_R>>m_{\rm DM}>>M_{Z^\prime}$, and for $0.1$ MeV $M_{Z^\prime}\lesssim 100$ GeV the cross-section for production of the DM becomes 
    \begin{equation}
    \sigma(s)=\frac{1}{48\pi}\frac{g^{\prime4}s\Big(1-\frac{4m^2_{\rm DM}}{s}\Big)^\frac{3}{2}}{s^2}x^2_\Phi(10x^2_H+13x^2_\Phi+16 x_H x_\Phi)\;,
    \label{Eq:alpha2}
    \end{equation}
    where $x_H$ and $x_\Phi$ denote the U$(1)_X$ charges of the SM Higgs doublet and SM singlet U$(1)_X$ scalar. The DM relic density becomes
    \begin{equation}
    \Omega_{\rm DM} h^2=\frac{m_{\rm DM} Y_\infty \mathcal{S}_0}{\frac{\rho_c}{h^2}}=0.12\times \Big(\frac{106.75}{g_*}\Big)^\frac{3}{2} \Big(\frac{g^\prime}{3.04\times 10^{-6}}\Big)^4 x^2_\Phi(10x^2_H+13x^2_\Phi+16x_H x_\Phi).~~~~~
    \label{Eq:a8}
    \end{equation}
    $Y_\infty$ is inversely proportional to $m_{\rm DM}$. As a result \cref{Eq:a8} is independent of $m_{\rm DM}$. In \cref{fig:m1} (left panel) we show the variation of DM yield $Y(x)$ as a function of $x$. In this case, freeze-in occurs at $T \sim m_{\rm DM}$. We obtain $g^\prime=3.65\times 10^{-6}$ setting $x_H=1$, $x_\Phi=1$, $m_{\rm DM}=1$ TeV and $M_{Z^\prime}=0.1$ GeV to reproduce the correct relic abundance 0.12. 
    \item[(ii)]The DM can also be produced through the $Z^\prime$ boson resonance. At $s=M^2_{Z^\prime}$ the resonance occurs for the region $\sqrt{s}\geq 2 m_{\rm DM}$. Using the narrow width approximation, we can write the $Z^\prime$ propagator as
    \begin{equation}
    \frac{1}{(s-M^2_{Z^\prime})^2+M^2_{Z^\prime}\Gamma^2_{Z^\prime}}=\frac{\pi}{M_{Z^\prime}\Gamma_{Z^\prime}}\delta(s-M^2_{Z^\prime}).
    \label{Eq:q2}
    \end{equation}
    If $M_{Z^\prime} < 2 m_t$, the total decay width of $Z^\prime$ becomes,
    \begin{equation}
    \Gamma_{Z^\prime}=\frac{1}{24\pi}g^{\prime2}M_{Z^\prime} [\frac{1}{36}(241x^2_H+418x^2_\Phi+436x_H x_\Phi)+x^2_\Phi].
    \label{Eq:a15}
    \end{equation}
    Hence, the DM relic density becomes
    \be
    \Omega_{\rm DM} h^2 & \simeq  0.12 \Big(\frac{g^\prime}{6.54\times 10^{-9}}\Big)^2 \Big(\frac{m_{\rm DM}}{10~\textrm{keV}}\Big)\Big(\frac{10~\textrm{GeV}}{M_{Z^\prime}}\Big)
    \\ & \quad \times \Big[\frac{x^2_\Phi (10x^2_H+13x^2_\Phi+16x_H x_\Phi)}{\frac{1}{36}(241x^2_H + 418x^2_\Phi+ 436x_H x_\Phi)+x^2_\Phi}\Big] 
    \Big(1-\frac{4m_{\rm DM}^2}{M_{Z^\prime}^2}\Big)^{5/2}. 
    \label{comega}
    \ee
    In \cref{fig:m1} (right panel) we have show the variation of the DM yield with $x$. The DM freezes-in occurs at $x=m_{\rm DM}/M_{Z^\prime}$. We obtain $g^\prime=1.419359\times 10^{-9}$  setting $x_H=1$, $x_\Phi=1$, $M_{Z^\prime}=10~\rm{GeV}$ and $m_N=10~\rm{keV}$ to reproduce the correct relic abundance 0.12.
\end{itemize}

\begin{figure}[tb]
\centering
    \includegraphics[width=0.495\textwidth,trim={0.0cm 2.0cm 0 1cm},clip]{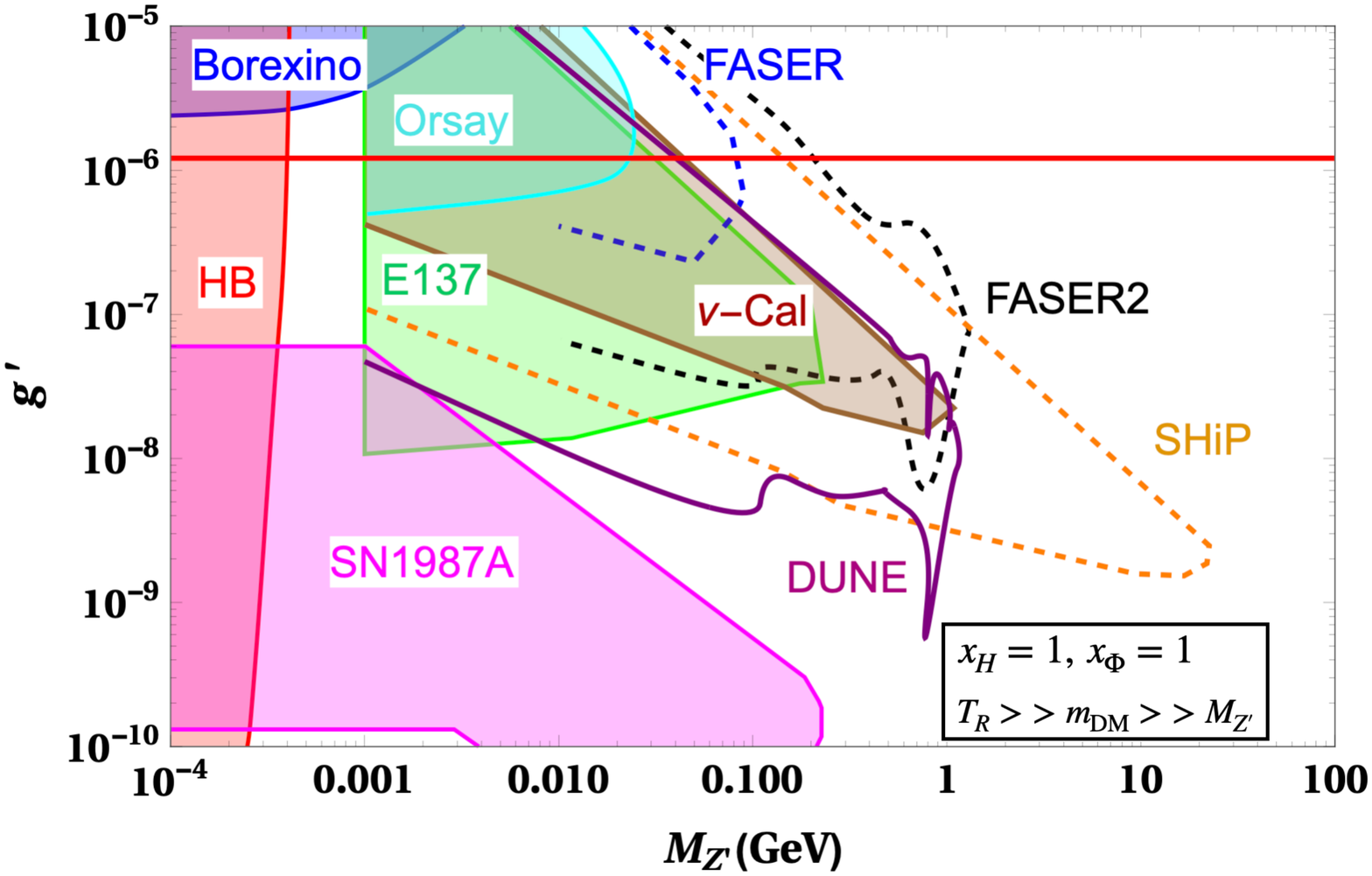}
    \includegraphics[width=0.495\textwidth,trim={0.2cm 2.0cm 0 1cm},clip]{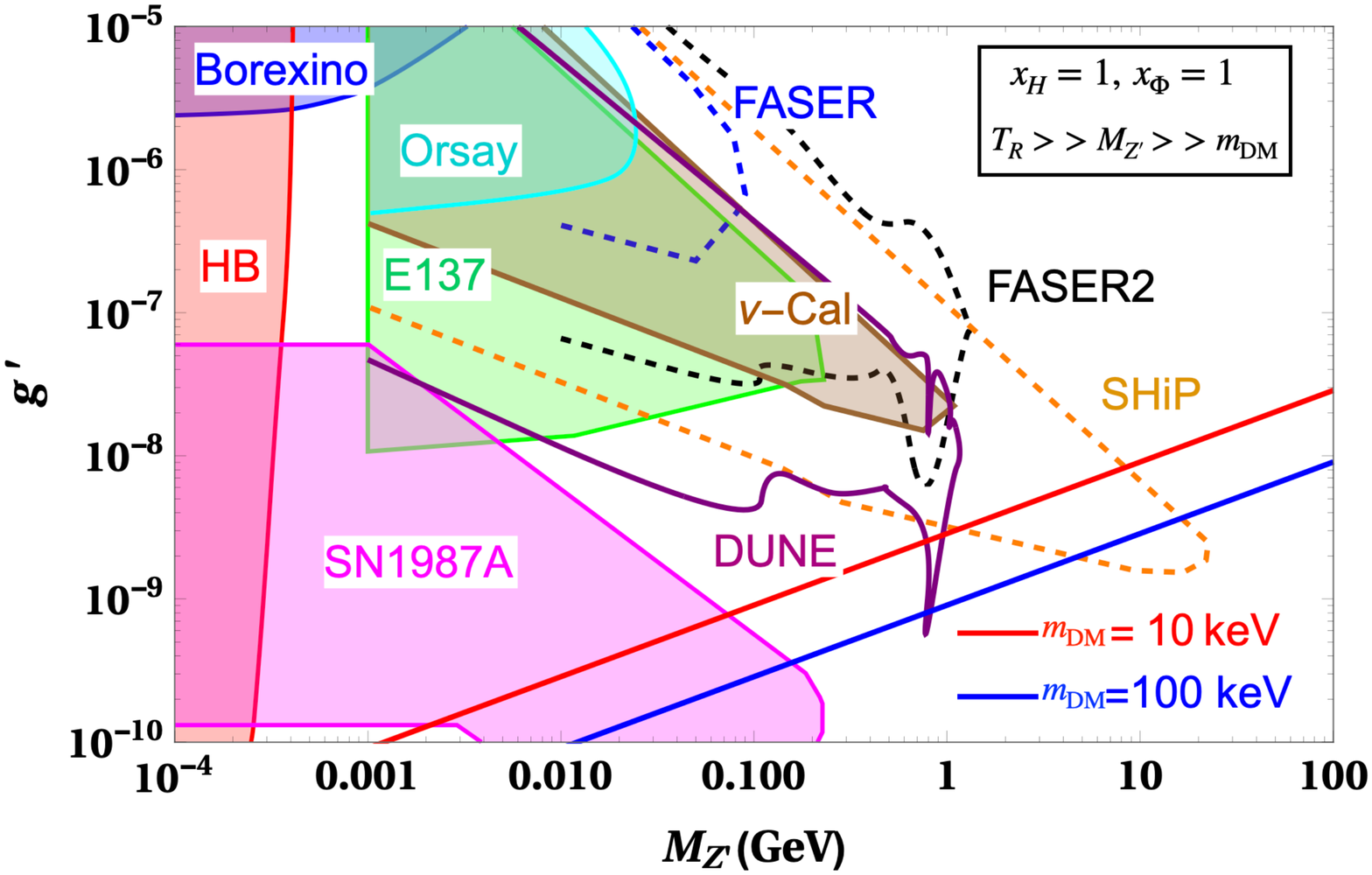}
\caption{Variation of $g^\prime$ with $M_{Z^\prime}$ considering $x_H=1$ and $x_\Phi=1$ for  $T_R>>m_{\rm DM}>>M_{Z^\prime}$ (left) and $T_R>>M_{Z^\prime}>>m_{\rm DM}$ (right) \cite{Das:2021nqj}.}
\label{fig:allplots}
\end{figure}

Requiring that the correct DM relic abundance of 0.12 is reproduced, we find bounds on the $g^\prime-M_{Z^\prime}$ plane for the model considering $x_H=1$ and $x_\Phi=1$ in \cref{fig:allplots} for two cases $T_R>>m_{\rm DM}>>M_{Z^\prime}$ (left panel)  and $T_R>>M_{Z^\prime}>>m_{\rm DM}$ (right panel). In the second case, we considered two benchmarks for the DM mass as 10 keV and 100 keV respectively. We also obtain the constraints on the $M_{Z^\prime} - g^\prime$ plane  from   neutrino-electron scattering experiments like Borexino \cite{Bellini:2011rx}, beam-dump experiments - Orsay \cite{Davier:1989wz}, E137 \cite{Bjorken:1988as}, $\nu$-cal \cite{Blumlein:2013cua, Gorbunov:2014wqa}, astrophyiscal observations like Horizontal Branch (HB) stars \cite{Raffelt:2000kp, Redondo:2013lna}, as well as lifetime frontier experiments FASER \cite{FASER:2018eoc}, FASER2 \cite{FASER:2019aik}, SHiP \cite{Alekhin:2015byh}, and also from future neutrino experiment DUNE \cite{Dev:2021qjj}. These plots are obtained from the corresponding plots for $U(1)_{B-L}$ by doing appropriate matching \cite{Kaneta:2016vkq, Croon:2020lrf, Dev:2021qjj, FASER:2018eoc}.

\begin{figure}[tb]
\centering
\includegraphics[width=0.73\textwidth]{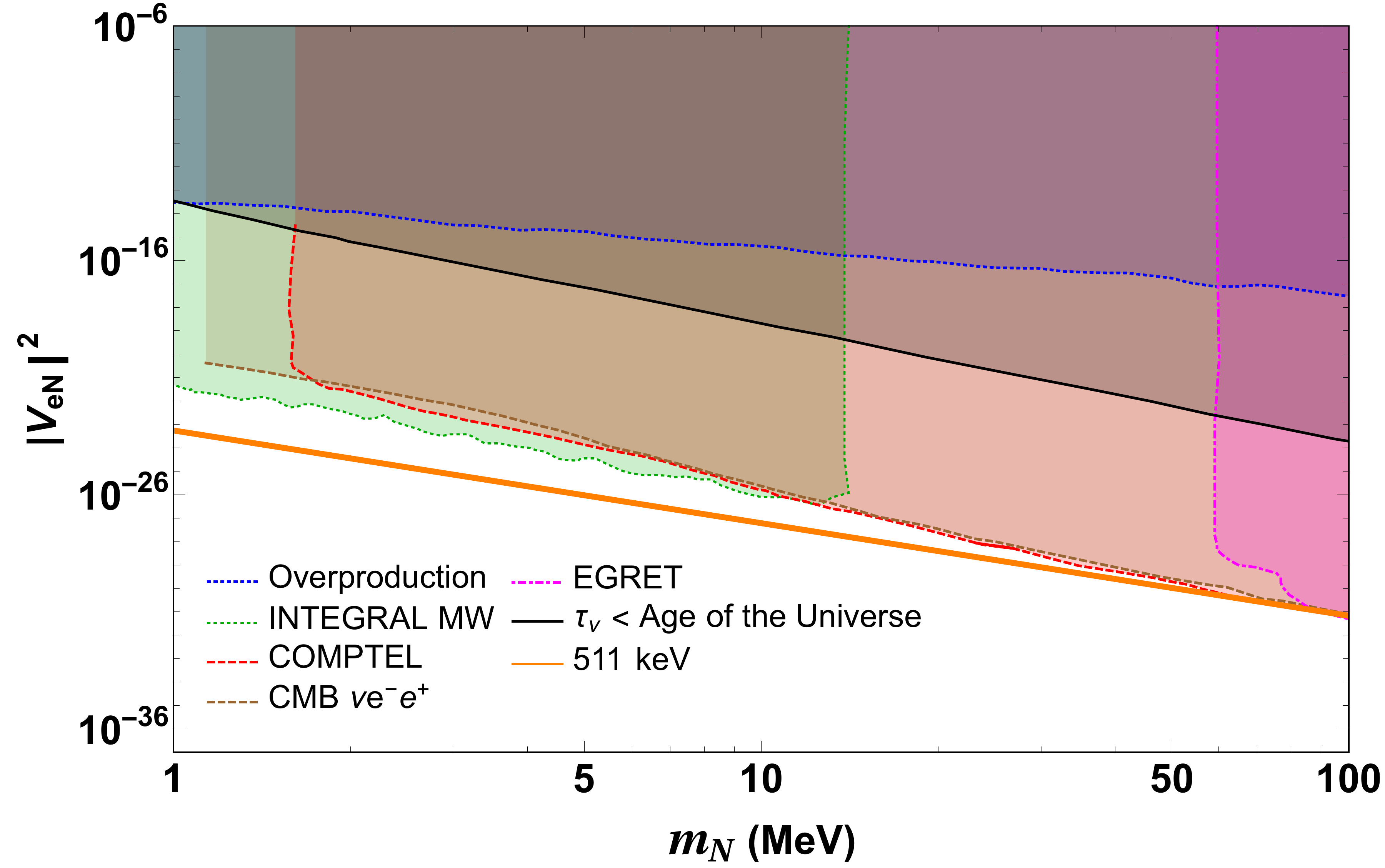}
\caption{Existing experimental bounds on the light heavy mixing squared ($|V_{eN}|^2$) and mass of the sterile neutrino DM $(m_N)$. The shaded regions are disfavored by the various experiments as indicated. 
The orange line corresponds to the values of the masses and mixing of the sterile neutrino DM that can produce the $511\,\mathrm{keV}$ line \cite{Das:2021nqj}. }
\label{mix1}
\end{figure}

We also note that if the mass of the sterile neutrino dark matter is $\geq$ 1 MeV and if the $M_{Z^\prime} > m_{\rm DM}=m_N$, the DM candidate can decay into positrons. As a result, it can explain the long-standing puzzle of the galactic 511 keV line in the Milky Way center observed by the INTEGRAL satellite \cite{Knodlseder:2003sv}. The parameter space that can give rise to the 511 keV line corresponds to  $|V_{\alpha i}|^2\sim 5.44\times 10^{-24}$. This is denoted by the orange line in \cref{mix1}.

The shaded regions in \cref{mix1} are excluded by different observations. The region above the dotted blue line corresponds to the parameter space where the sterile neutrino is overabundant (assuming production is only via mixing). In the region above the black solid line, the sterile neutrino lifetime is shorter than the age of the universe~\cite{DeRomeri:2020wng}. The DM annihilation or decay in the Milky way can give rise to $\gamma$ ray spectral lines that can be probed by various indirect detection experiments. These searches put constraints on the sterile neutrino DM mass and mixing parameter space. The regions above the dash-dotted cyan line, dashed red line and dash-dotted magenta line are disfavored by the search for $\gamma$ spectral lines by Fermi-LAT~\cite{Fermi-LAT:2015kyq}, COMPTEL~\cite{Essig:2013goa} and EGRET~\cite{Essig:2013goa, Strong:2004de} respectively. The spectral line search by INTEGRAL MW~\cite{Boyarsky:2007ge} puts bound on the active-sterile mixing for the DM of mass 0.04 MeV $<m_N<14$ MeV and is shown by the region bounded by a dotted green line. The precise measurement of the CMB anisotropies and temperature puts constraints on the sterile neutrino DM mass and mixing. The possible decay modes of sterile neutrino are $\nu\gamma$ and $\nu e^+e^-$ that result in early energy injection. The decay product $\nu e^+ e^-$ gives constraints on large sterile neutrino mass (dashed brown line) which comes from the CMB $\nu e^-e^+$ mode. Thus, it is clear from \cref{mix1} that the higher the mass of the sterile neutrino DM $(m_N)$, the lower the active-sterile mixing has to be. To achieve this in our model, the Yukawa couplings of DM candidates to the active light neutrinos and the SM Higgs have to be very small.

As can be seen, different regions of the parameter of the model under study can be probed by a variety of different experimental approaches ranging from accelerator-based searches for light unstable new physics species to DM indirect detection observations. The search for light unstable vector mediator particles in the FPF will play an important role in probing this scenario.

\subsection{Imprints of Scale Invariance and Freeze-In Dark Matter at the FPF\label{sec:astro_DM_scaleinvariance}}

Scale Invariance as a desirable UV-completion framework, in particle theory, can on one hand dynamically generate the scales of electroweak and dark sector physics~\cite{Hambye:2013dgv, Farzinnia:2013pga, Jaeckel:2012yz}, thereby ameliorating the naturalness problem of the SM Higgs mass, and on the other hand, can also incorporate a testable freeze-in~\cite{Hall:2009bx, Bernal:2017kxu} dark matter scenario. To establish this, we have given a prescription in Ref.~\cite{Barman:2021lot}, where we consider a $U(1)_X$ gauge extension of the Standard Model.  Under this new gauge symmetry the DM, which is the gauge boson $X_\mu$, becomes massive once $U(1)_X$ is spontaneously broken by the vacuum expectation value (VEV) of a new scalar $S$. This gives rise to the familiar Higgs-portal DM model, where the DM communicates with the visible sector via the portal coupling, with a strength proportional to the scalar mixing. This also opens up discovery prospects for such models in the LLP searches in the FPF.

The DM renders stability as it carries an odd $\mathbb{Z}_2$ charge, while all the SM fields are even under the $\mathbb{Z}_2$. The singlet scalar $S$, on the other hand, transforms as $S\xrightarrow[]{\mathbb{Z}_2} S^\star$, such that the stabilizing symmetry remains exact. The classically conformal gauge extension to the SM leads to several important consequences:

\begin{figure}[t]
\centering
    \includegraphics[width=0.68\textwidth]{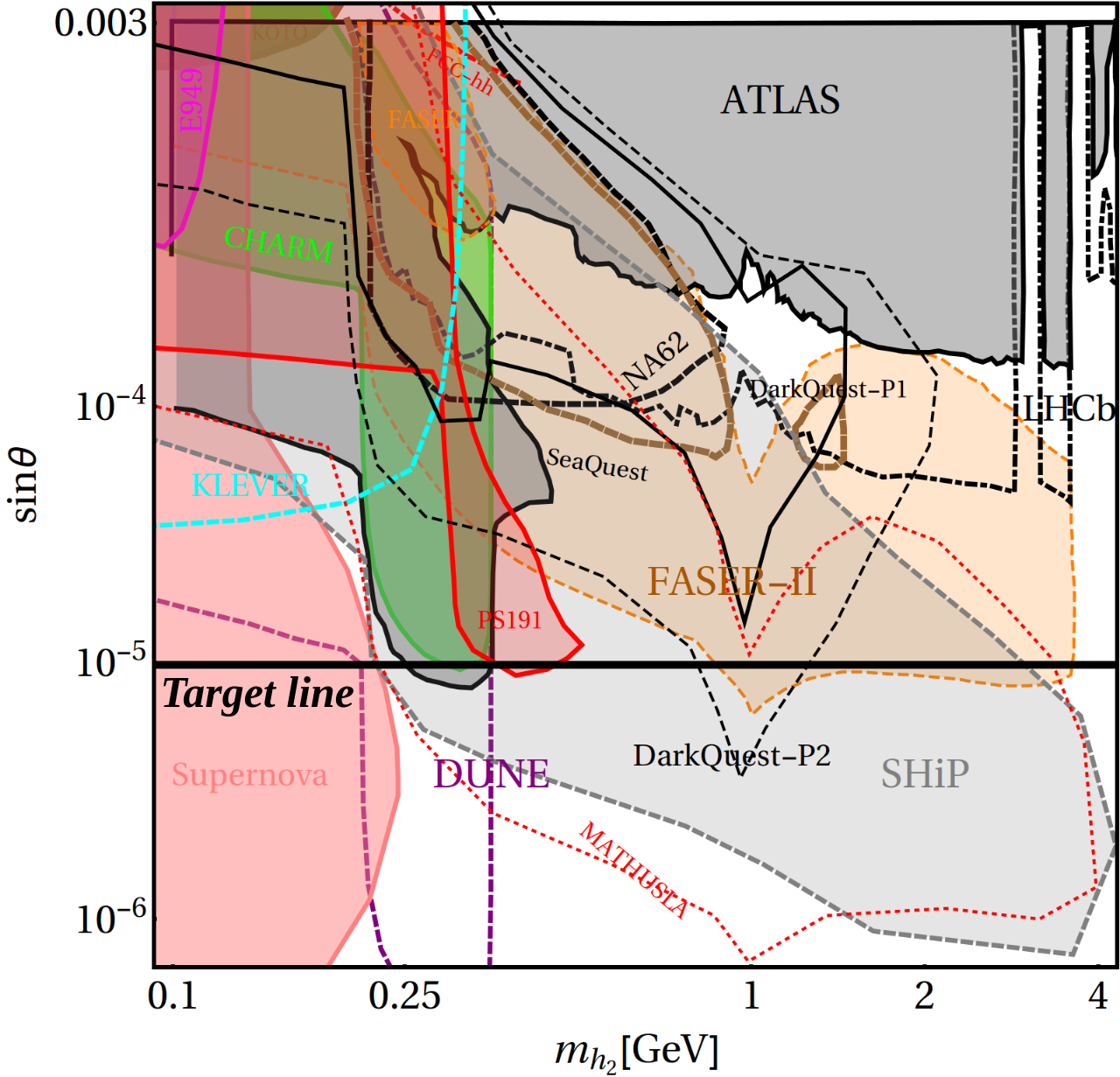}
\caption{The thick black horizontal straight line shows the DM parameter space complying with the PLANCK~\cite{Planck:2018vyg} observed relic abundance and satisfying spin-independent direct search exclusion limit in $\sin\theta-m_{h_2}$ plane. Experimental limits are shown from E949~\cite{E949:2008btt}, CHARM~\cite{CHARM:1985anb}, NA62~\cite{Kozhuharov:2014xve}, FASER and FASER2~\cite{Feng:2017vli, FASER:2018eoc, FASER:2018bac, FASER:2019aik}, FCC-hh~\cite{FCC:2018vvp, Dev:2017dui}, ATLAS~\cite{ATLAS:2012qaq,ATLAS:2014jdv, Chalons:2016jeu, Robens:2015gla}, SeaQuest~\cite{Berlin:2018pwi}, LHCb~\cite{Gligorov:2017nwh}, KLEVER~\cite{Moulson:2018mlx}, DUNE~\cite{DUNE:2015lol, Berryman:2019dme}, DarkQuest-Phase2~\cite{Batell:2020vqn}, MATHUSLA~\cite{Curtin:2018mvb}, SHiP~\cite{SHiP:2015vad}, and PS191~\cite{Bernardi:1985ny, Gorbunov:2021ccu}.}
\label{fig:exptLim1}
\end{figure}

%
\begin{itemize}
    \item[(i)] The model is extremely economical as one is left with only two independent parameters, namely the new gauge coupling $g_X$ and the DM mass $m_X$, one of which gets fixed entirely via relic density requirements. 
    The rest of the couplings follow constrained relations due to the underlying scale invariance.
    \item[(ii)] By demanding mass squared of the non-standard Higgs $m_{h_2}^2>0$, one can put a lower bound on the vector DM mass: $m_X\gtrsim 250$ GeV for $g_X\lesssim 10^{-5}$, typical coupling size needed for freeze-in production of the DM. This mass bound entirely arises from the scale-invariance of the theory. 
    \item[(iii)] For all choices of $m_X,g_X$ required for producing right DM abundance via freeze-in, we always find $m_{h_2}\ll m_X$. This provides a window for spin-independent direct detection prospects of the freeze-in DM via $t$-channel mediation of the MeV-scale non-standard Higgs.
    \item[(iv)] The requirement of DM relic abundance $\Omega_X h^2\simeq 0.12$, typically constraints the ratio $m_X/g_X$ since $\Omega_X h^2\propto\left(g_X/m_X\right)^4$, thanks to the underlying scale-invariance. This also implies, for fixed $\Omega_X h^2$, the mixing becomes constant as $\sin\theta\propto 1\Big/\sqrt{1+\left(m_X/v_h\,g_X\right)^2}$ and $g_X/m_X$ is determined from relic abundance. Thus, the scale-invariance of the theory, together with the requirement of right relic abundance, fixes $\sin\theta$ to a constant value. For all choices of $\{m_X,g_X\}$ that leaves the combination $m_X/g_X$ constant (proportional to $\Omega_X h^2$), we obtain the scalar mixing $\sin\theta\sim 10^{-5}$. We dub this as {\it Scale Invariant FIMP Miracle} (SIFM), similar to the well-known WIMP miracle.
\end{itemize}
%

\begin{figure}[tb]
\centering
\includegraphics[width=0.84\textwidth]{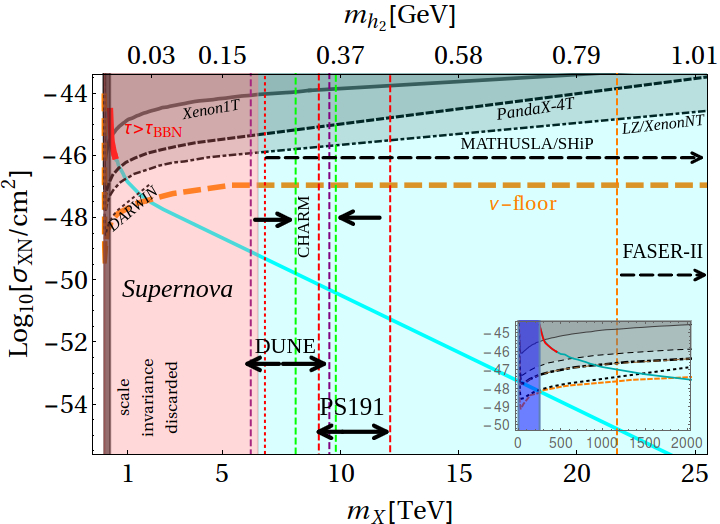}
\caption{Summary of all (experimental) bounds. The relic density allowed DM parameter space shown via the thick cyan curve. The pink region is disallowed from supernova bound. The light blue region in the background denotes the mass range that can only be probed in intensity frontier searches for light new physics. The colorful vertical dashed lines indicate the ranges in which different experimental facilities can probe the model parameter space. Inset: Sensitivity of direct detection experiments up to DM mass of 2 TeV.}
\label{fig:summaryPlt}
\end{figure}

The crucial outcome of our analysis is the experimental testability of the freeze-in DM model in intensity frontier searches for light new physics, including in the FPF, thanks to the presence of the light scalar. We find, in order to satisfy bounds from relic density, spin-independent direct detection, and big bang nucleosynthesis (BBN), the mixing within the scalar sector has to be $\sin\theta\sim\mathcal{O}(10^{-5})$. Such a mixing angle turns out to be well within the reach of experiments like DUNE~\cite{DUNE:2015lol, Berryman:2019dme}, FASER2~\cite{Feng:2017vli, FASER:2018eoc, FASER:2018bac, FASER:2019aik}, PS191~\cite{Bernardi:1985ny, Gorbunov:2021ccu}, DarkQuest-Phase2~\cite{Batell:2020vqn}, MATHUSLA~\cite{Curtin:2018mvb} and SHiP~\cite{SHiP:2015vad}. Most importantly, since the mixing $\theta$ is no more a free parameter (due to the scale invariance of the theory), it rather can be parametrized in terms of $m_X$ and $g_X$, hence, the new gauge coupling that determines the freeze-in abundance is directly being probed at a plethora of current and upcoming experimental facilities, including the FPF. 

Our main result is summarized in \cref{fig:exptLim1}. Here the thick black horizontal straight line corresponds to the right relic abundance for the vector DM, while the sensitivity reach curves for several planned and proposed experiments are shown in the $m_{h_2}-\sin\theta$ plane. We note, the light scalar mediator in the present model is within the reach of CHARM~\cite{CHARM:1985anb}, DUNE~\cite{DUNE:2015lol, Berryman:2019dme},  FASER and FASER2~\cite{Feng:2017vli, FASER:2018eoc, FASER:2018bac, FASER:2019aik}, PS191~\cite{Bernardi:1985ny, Gorbunov:2021ccu}, DarkQuest-Phase2~\cite{Batell:2020vqn}, MATHUSLA~\cite{Curtin:2018mvb} and SHiP~\cite{SHiP:2015vad} (see \emph{e.g.} Ref.~\cite{Beacham:2019nyx} for a summary on these experiments) for the allowed range of mass and mixing. The complementarity of the bounds from DM direct search and forward facility experiments is depicted in \cref{fig:summaryPlt}. Here the thick cyan curve denotes the relic density allowed parameter space for the DM. We consider exclusion limits from XENON1T~\cite{XENON:2018voc}, and projected bounds from PandaX-4T~\cite{PandaX:2018wtu}, LUX-ZEPLIN (LZ)~\cite{LUX-ZEPLIN:2018poe}, XENONnT~\cite{XENON:2020kmp} and DARWIN~\cite{DARWIN:2016hyl} experiments which provide an upper limit on the DM-nucleon scattering cross-section at 90\% C.L. As expected, for low DM mass region (below 1 TeV) these bounds are severe but become rather weak for heavier DM mass. A light scalar with mass below $\sim 250$ MeV is ruled out from supernova observations~\cite{Krnjaic:2015mbs, Dev:2020eam}. This corresponds to a DM mass $\lesssim 6.4$ TeV. Above DM mass of $\sim 1.8$ TeV the relic density allowed parameter space gets submerged into the $\nu$-floor~\cite{Billard:2013qya}, where separating DM scattering from the background neutrino scattering is rather challenging. However, in those regions FASER2 and other intensity frontier searches can provide excellent sensitivity. 

We thus conclude that a classically scale-invariant particle physics model for freeze-in production of DM is not only capable of producing the observed relic abundance in a minimal set-up, but also leaves the possibility of being probed in several light dark sector search experiments by uniquely determining the mixing in the scalar sector\footnote{The possibilities of probing freeze-in in gauge extension of the SM, beyond the realm of scale invariance, are discussed in Ref.~\cite{Barman:2021yaz}.}. Importantly, due to the classical scale invariance, complementary signals can arise in both direct DM searches\footnote{Direct detection prospects for freeze-in DM have been discussed in Refs.~\cite{Duch:2017khv, Hambye:2018dpi}.} and energy/intensity frontier experiments in probing the freeze-in parameter space of our model. This ushers in a new era where UV-completion in beyond the SM particle physics model building may lead to predictive feebly interacting DM candidates to be tested in the very near future.

\subsection{Rich Dark Sector and Complementarity with Indirect Searches\label{sec:astro_DM_richdarksector}}

On top of the examples of DM searches in the FPF discussed in \cref{sec:bsm2}, we note that light unstable new physics species can also mediate interactions between the SM and a much heavier DM particle which thermalizes in the early Universe. This, however, often leads to stringent astrophysical and cosmological bounds, although they can be alleviated if the DM particle remains secluded from the SM~\cite{Pospelov:2007mp}. In this case, visible signals related to interactions of DM are still expected to be detectable, while complementary probes of light mediator particles in the FPF are also expected to probe unconstrained regions of the parameter space of such models. The relevant signatures employ a rich dark sector structure in these scenarios and often go beyond the most direct processes. 

Interestingly, the phenomenological aspects of such models can differ from both the simple scenarios predicting the existence of LLPs, as well as from the vanilla heavy WIMP DM candidates. This is first due to expected simultaneous discovery prospects of rich dark sectors in both the intensity frontier searches for LLPs and in indirect searches for heavy DM. Further interesting features of such scenarios can be observed in the presence of even extremely long-lived particles with the lifetime of order $\tau\gtrsim 10^{11}~\textrm{s}$ that can lead to important cosmological effects in both the Big Bang Nucleosynthesis (BBN) and the CMB radiation surveys, as well as to non-standard signatures in indirect detection~\cite{Rothstein:2009pm, Chu:2017vao}.

The complementarity of such methods of searching for BSM physics with intensity frontier experiments remains largely unexplored. In Ref.~\cite{Jodlowski:2021xye}, a simplified but rich dark sector model has been studied featuring heavy DM, long-lived connectors, and a light mediator field that can illustrate the rich phenomenology of such scenarios. In the model, the LLP candidate is the light scalar $h_D$ field with a sub-$\gev$ mass, which couples to the SM via the mixing with the SM Higgs. The highly-displaced decays of $h_D$s produced in the far-forward region of the LHC can be successfully searched for in FASER2. In addition, the model includes, i.a., an additional dark vector field $A^\prime$ with the mass spanning from $\mathcal{\gev}$ to more than $100~\gev$, and the heavy complex scalar DM candidate $\chi$ with the mass fixed to $1.5~\tev$ for simplicity. The heavy DM candidate can annihilate in the dense regions of the Galaxy and beyond. This leads to detectable signatures after a cascade process, in which also a potentially very long-lived $A^\prime$ is produced that can travel to distances of order kpc before decaying. In these decays, much less long-lived, dark scalar $h_D$ appears, which then decays into the SM species. While the full structure of the model remains more sophisticated than the simplest scenarios predicting the existence of LLPs, we stress that its particle content allows one to simultaneously avoid stringent cosmological bounds, predict the proper value of the DM relic density in the Universe, and obtain novel phenomenological effects discussed below.

\begin{figure}[t]
\centering
\includegraphics[width=0.6\textwidth]{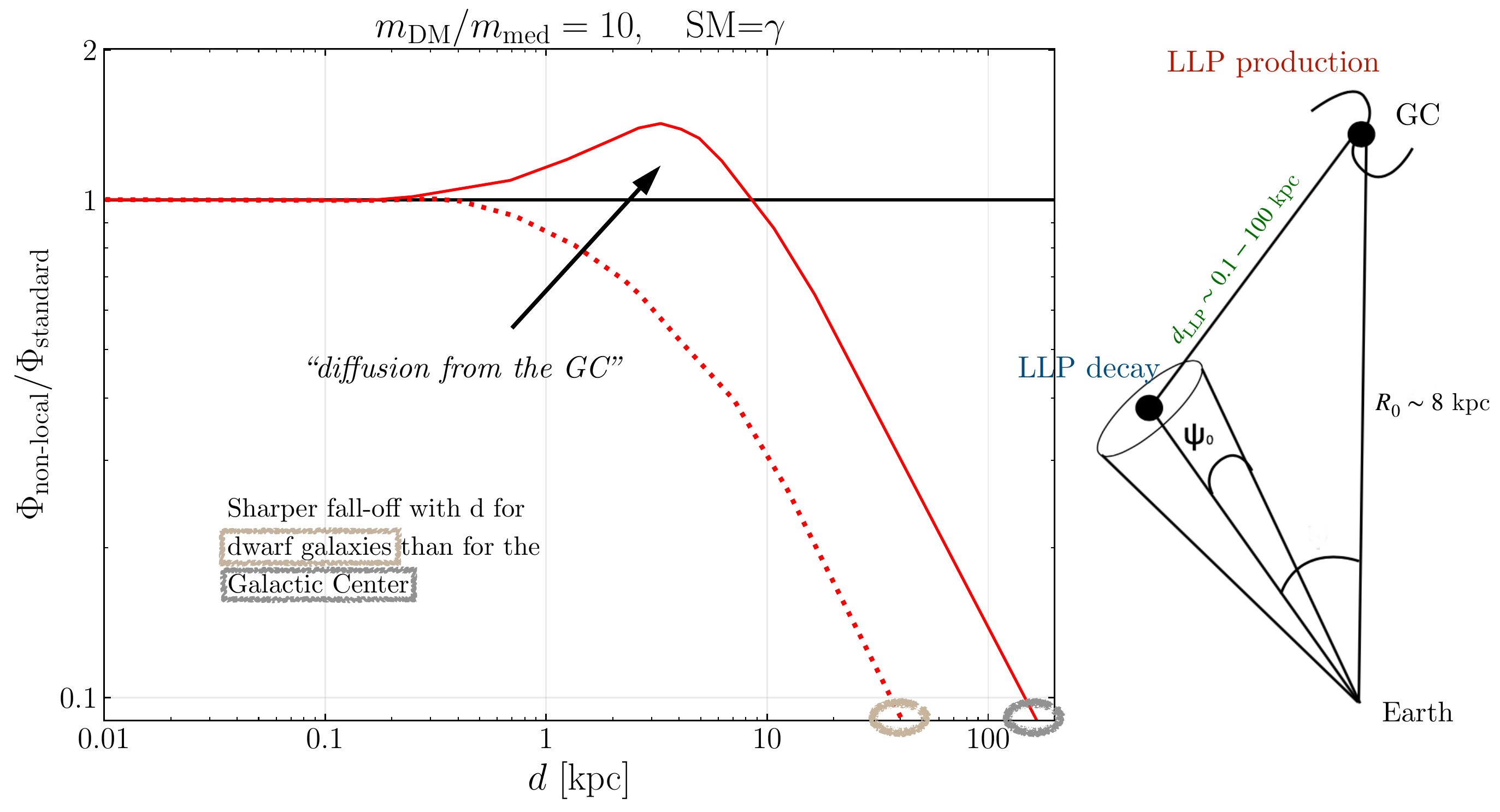}\hspace*{-0.2cm}
\includegraphics[width=0.39\textwidth]{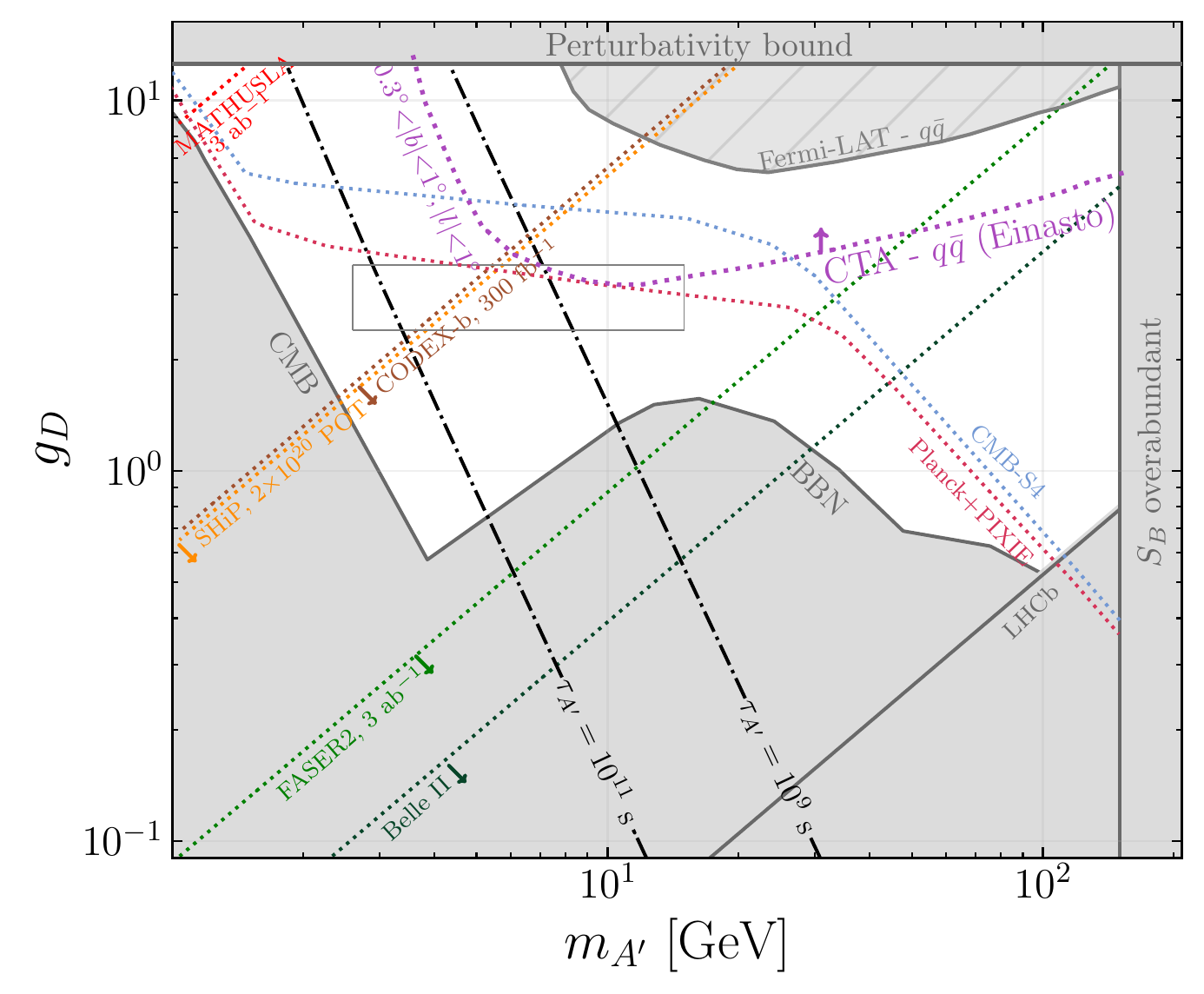}
\caption{Left: Illustration of non-local effects present in indirect detection of long-lived particles for the toy DM model with the ratio of the DM and mediator masses equal to $m_{\textrm{DM}}/m_{\textrm{med}} = 10$. Center: Schematic illustration of the separation between the DM annihilation and the LLP decay points in the Galaxy. Right: Coverage of the allowed parameter space of the model considered in Ref.~\cite{Jodlowski:2021xye} (see text), which employs secluded heavy WIMP DM, long-lived vector mediator, and a light dark higgs boson. In the plot, the complementarity between the intensity frontier searches for light new physics, DM indirect detection, and CMB surveys is illustrated. The expected sensitivity of the FASER2 experiment in the FPF is shown with the green dotted line. Taken from Ref.~\cite{Jodlowski:2021xye}.}
\label{fig:ID_LLP_results_1}
\end{figure}

As far as DM indirect detection is concerned, the resulting rich phenomenology dictated by the presence of the very long-lived $A^\prime$ species is illustrated in the left panel of \cref{fig:ID_LLP_results_1}. In the plot, the dependence of the DM-induced photon flux at Earth is presented as a function of the mediator decay length. We present this for a toy model with the DM and the mediator mass fixed to be equal to $m_{\mathrm{DM}} = 100 \mathrm{GeV}$ and $m_{\mathrm{med}} = 10 \mathrm{GeV}$, respectively. 

The effects shown in the plot contain: 
\begin{itemize}
    \item[(i)]  The \emph{Galactic Center diffusion} effect, where there is an additional contribution to the photon flux from LLPs produced near the Galactic Center (GC), which then decay along the LOF.
    \item[(ii)] A linear flux decrease in the long-lived regime due to the finite DM density support. 
    \item[(iii)] A faster decrease in flux with the LLP decay length for observations focused on small regions of interest, compared to large ones (\emph{e.g.} dwarf galaxies versus the GC). The result is the possible strong modification of the expected DM-induced photon flux due to the $h_D$ decays that are driven by the galactic-scale separation of the DM annihilation and LLP decay points. 
\end{itemize}

The discovery prospects of the model of this type can be further strengthened thanks to the intensity frontier searches for $h_D$s, as well as due to expected distortions in the CMB spectrum that can be observed in future surveys. We illustrate this in the right panel of \cref{fig:ID_LLP_results_1}. The plot shows the coverage of the allowed parameter space of the aforementioned rich dark sector model in the plane of the LLP mass, $m_{A^\prime}$, and the dark gauge coupling, $g_D$, which, i.a., dictates the coupling strength between the $A^\prime$ and $h_D$ species. We note there is a significant synergy between various proposed LHC-based intensity frontier detectors constraining the light dark Higgs boson: CODEX-b~\cite{Gligorov:2017nwh, Aielli:2019ivi}, FASER2~\cite{Feng:2017uoz, FASER:2018eoc}, MATHUSLA~\cite{Chou:2016lxi, Curtin:2018mvb} and indirect detection searches for LLP - CTA~\cite{CTAConsortium:2017dvg} - and CMB surveys: Planck~\cite{Planck:2018vyg}, PIXIE~\cite{Kogut:2011xw}, PRISM~\cite{PRISM:2013fvg}, and CMB S-4~\cite{CMB-S4:2016ple}. In particular, the future FPF search for the light dark scalar $h_D$ in the FASER2 detector will remain complementary to the CTA search for heavy WIMP DM $\chi$ and to CMB surveys sensitive to very long-lived $A^\prime$s. These can also lead to simultaneous observations in some regions of the parameter space of the model. Together with other aforementioned experiments, these searches will almost entirely probe such a scenario, employing a diverse set of experimental signatures that will differentiate this scenario from vanilla LLP and heavy DM studies.

%% file: acknowledgments_all.tex
We thank the participants of the FPF meetings and the Snowmass working groups for discussions that have contributed both directly and indirectly to this study.  We gratefully acknowledge the invaluable support of the CERN Physics Beyond Colliders study group and the work of CERN technical teams related to civil engineering studies (SCE-DOD), safety discussions (HSE-OHS, HSE-RP, EP-DI-SO), integration (EN-ACE), and discussions on services (EN-CV, EN-EL, EN-AA) and simulations (SY-STI). 

The work by J.~Alameddine, W.~Rhode, T.~Ruhe, and A.~Sandrock has been supported by the Deutsche Forschungsgemeinschaft (DFG, German Research Foundation), Collaborative Research Center SFB 876 and SFB 1491.
L.~A. Anchordoqui is supported by the US National Science Foundation (NSF) Grant PHY-2112527.
T.~Araki is supported by JP18H01210.
A.~Ariga is supported by JSPS KAKENHI Grant JP20K23373 and the European Research Council (ERC) under the European Union's Horizon 2020 Research and Innovation Programme (Grant 101002690).
T.~Ariga acknowledges support from JSPS KAKENHI Grant JP19H01909.
K.~Asai is supported by JSPS KAKENHI Grant JP19J13812 and JP21K20365.
A. Bacchetta and F. G. Celiberto acknowledge support from the INFN/NINPHA project.
P.~Bakhti and M.~Rajaee are supported by the National Research Foundation of Korea (NRF-2020R1I1A3072747).
B.~Barman received funding from the Patrimonio Autónomo - Fondo Nacional de Financiamiento para la Ciencia, la Tecnología y la Innovación Francisco José de Caldas (MinCiencias - Colombia) Grant 80740-465-2020 and the European Union's Horizon 2020 research and innovation programme under the Marie Sklodowska-Curie grant agreement No 860881-HIDDeN. 
The work of B.~Batell is supported by the US Department of Energy (DOE) Grant DE–SC0007914.
The work of A.~Berlin, T.~J.~Hobbs, S.~Hoeche, and J.~G.~Morf\'{i}n was supported by the Fermi National Accelerator Laboratory (Fermilab), a US DOE, Office of Science, HEP User Facility.
Fermilab is managed by Fermi Research Alliance, LLC (FRA), acting under Contract  DE-AC02-07CH11359.
E.~Bertuzzo acknowledges financial support from FAPESP contracts 2015/25884-4 and 2019/15149-6 and is indebted to the Theoretical Particle Physics and Cosmology group at King's College London for hospitality. 
The work of A.~Boyarsky and M.~Ovchynnikov is supported by the ERC under the European Union’s Horizon 2020 research and innovation programme (GA 694896). 
A.~Carmona acknowledges funding from the European Union’s Horizon 2020 research and innovation programme under the Marie Sk\l{}odowska-Curie grant agreement No 754446 and UGR Research and Knowledge Transfer Found - Athenea3i. A. Carmona also acknowledges partial support by the Ministry of Science and Innovation and SRA (10.13039/501100011033) Grant PID2019-106087GB-C22 and by the Junta de Andaluc\'ia Grant A-FQM-472-UGR20.
F. G. Celiberto thanks the Universit\`a degli Studi di Pavia for the warm hospitality.
The work of G. Chachamis was supported by the Funda\c{c}{\~ a}o para a Ci{\^ e}ncia e a Tecnologia (Portugal) under project CERN/FIS-PAR/0024/2019 and contract 'Investigador auxiliar FCT - Individual Call/03216/2017.'
The work of M. Citron and D. Stuart is supported by US DOE Grant DE-SC0011702.
The work of L.~Darme is supported by the European Union’s Horizon 2020 research and innovation programme under the Marie Sklodowska-Curie grant agreement 101028626.
P.~B.~Denton acknowledges support from the US DOE Grant Contract DE-SC0012704.
The work of P.~S.~Bhupal Dev was supported in part by the US DOE Grant DE-SC0017987.
The research activities of K.~R.~Dienes were supported in part by the US DOE
Grant DE-FG02-13ER41976/DE-SC0009913 and also
by the US NSF through its employee IR/D program.
M.V. Diwan acknowledges support from the US DOE Grant Contract DE-SC0012704.
Y. Du and J.-H. Yu are supported in part by National Key Research and Development Program of China Grant 2020YFC2201501, and the National Science Foundation of China (NSFC) under Grants  12022514, 11875003 and 12047503, and CAS Project for Young Scientists in Basic Research YSBR-006, and the Key Research Program of the CAS Grant XDPB15.
The work of B.~Dutta and S.~Ghosh are supported in part by the US DOE Grant DE-SC0010813. The work of S.~Ghosh is also supported in part by National Research Foundation of Korea(NRF)’s grants, grant 6N021413.
Y.~Farzan has received  financial support from Saramadan contract ISEF/M/400279 as well as from the European Union’s Horizon 2020 research and innovation programme under the Marie Skłodowska -Curie grant agreement 860881-HIDDeN.
The work of J.~L.~Feng~is supported in part by US NSF Grants PHY-1915005 and PHY-2111427, Simons Investigator Award \#376204, Simons Foundation Grant 623683, and Heising-Simons Foundation Grants 2019-1179 and 2020-1840.
M.~Fieg is supported in part by US NSF Grant PHY-1915005 and by NSF Graduate Research Fellowship Award DGE-1839285.
A.~L.~Foguel is supported by FAPESP contract 2020/00174-2.
The work of P.~Foldenauer is supported by the UKRI Future Leaders Fellowship  DARKMAP.
The work of S.~Foroughi-Abari and A. Ritz is supported in part by NSERC, Canada.
The work of J.-F.~Fortin is supported in part by NSERC, Canada.
A. Friedland is supported by the US DOE Grant DE-AC02-76SF00515. 
E.~Fuchs acknowledges support by the DFG Germany's Excellence Strategy -– EXC-2123 "QuantumFrontiers" -- 390837967. 
M. Fucilla, M. M.A. Mohammed, and A. Papa acknowledge support from the INFN/QFT@COLLIDERS project. 
A. Garcia acknowledges support from the European Union’s H2020-MSCA Grant Agreement 101025085.
C. A. Garc\'{\i}a Canal and S. J. Sciutto acknowledge support from CONICET and ANPCyT.
M. V. Garzelli acknowledges support the from German BMBF contract 05H21GUCCA.
V. P. Goncalves was partially financed by the Brazilian funding agencies CNPq, CAPES,   FAPERGS and  INCT-FNA (process number 464898/2014-5).
S. Goswami  acknowledges the J.C Bose Fellowship (JCB/2020/000011) of Science and Engineering Research Board of Department of Science and Technology, Government of India.
The work of M. Guzzi is supported by US NSF Grant PHY-2112025.
L. Harland-Lang thanks the Science and Technology Facilities Council (STFC) for support via grant award ST/L000377/1.
The work of S.~P.~Harris is supported by the US DOE Grant DE-FG02-00ER41132 as well as the US NSF Grant PHY-1430152 (JINA Center for the Evolution of the Elements). 
J.~C.~Helo  acknowledge support from Grant ANID FONDECYT-Chile 1201673 and ANID – Millennium Science Initiative Program ICN2019-044.
The work of M.~Hirsch is supported by the Spanish Grants PID2020-113775GB-I00 (AEI/10.13039/ 501100011033) and PROMETEO/2018/165 (Generalitat Valenciana).  
The work of A.~Hryczuk and M.~Laletin is supported by the National Science Centre, Poland, research grant  2018/31/D/ST2/00813.
The research activities of F.~Huang are supported by the International Postdoctoral Exchange Fellowship Program and in part by US NSF Grant PHY-1915005. 
A.~Ismail is supported by NSERC (reference number 557763) and by US NSF Grant PHY-2014071.
Y. S. Jeong acknowledges support from the National Research Foundation of Korea (NRF) grant funded by the Korea government through Ministry of Science and ICT Grant 2021R1A2C1009296.
K.~Jodlowski and L.~Roszkowski are supported by the National Science Centre, Poland, research grant 2015/18/A/ST2/00748. 
S. R. Klein is supported in part by the US NSF Grant PHY-1307472 and the US DOE contract DE-AC-76SF00098.  
F.~Kling is supported by the DFG under Germany’s Excellence Strategy -- EXC 2121 Quantum Universe -- 390833306. 
P.~Ko is supported in part by KIAS Individual Grant PG021403 and by National Research Foundation of Korea (NRF) Grant NRF-2019R1A2C3005009.
S.~Kulkarni is supported by the Austrian Science Fund Elise-Richter Grant V592-N27.
The work of J.~Kumar is supported in part by US DOE Grant DE-SC0010504.
J.-L.~Kuo is supported by US NSF Theoretical Physics Program, Grant PHY-1915005.
The work of C. Le Roux and K. Zapp is part of a project that has received funding from the ERC under the European Union’s Horizon 2020 research and 
innovation programme (Grant agreement 803183, collectiveQCD).
The work of H.-S.~Lee was supported in part by the National Research Foundation of Korea (NRF-2021R1A2C2009718).
S.~J.~Lee was supported by the Samsung Science and Technology Foundation.
Ji.~Li is supported by the National Natural Science Foundation of China Grant 11905149. 
K.-F.~Lyu and Z.~Liu are supported in part by the US DOE Grant DE-SC0022345.
The work of R. Maciula and A. Szczurek was partially supported by
the Polish National Science Centre under Grant
2018/31/B/ST2/03537. 
R.~Mammen Abraham and A. Ismail acknowledge support from the US DOE Grant DE-SC0016013. 
M.~R.~Masouminia is supported by the UK Science and Technology Facilities Council (grant ST/P001246/1).
The work of J.~McFayden was supported by the Royal Society Fellowship Grant URF R1 201519.
The work of O.~Mikulenko is supported by the NWO Physics Vrij Programme “The Hidden Universe of Weakly Inter-acting Particles” with project number 680.92.18.03 (NWO Vrije Programma), which is (partly) financed by the Dutch Research Council (NWO).
P. Nadolsky and F. Olness acknowledge support through US DOE Grant DE-SC0010129.
E. R. Nocera thanks the STFC for support by the grant awards ST/P000630/1and
ST/T000600/1. 
The work of N.~Okada is supported by the US DOE Grant DE-SC0012447. 
V. Pandey acknowledges the support from US DOE Grant DE-SC0009824.
The work of D.~Raut is supported by the US DOE Grant DE-SC0013880. 
P.~Reimitz acknowledges financial support from FAPESP contract 2020/10004-7.
M. H. Reno is supported in part by US DOE Grant DE-SC-0010113.
The work of J.~Rojo is partly supported by the Dutch Research Council (NWO).
L.~Roszkowski and S.~Trojanowski are supported by the grant ``AstroCeNT: Particle Astrophysics Science and Technology Centre'' carried out within the International Research Agendas programme of the Foundation for Polish Science financed by the European Union under the European Regional Development Fund. S.~Trojanowski is also supported in part by the Polish Ministry of Science and Higher Education through its scholarship for young and outstanding scientists (decision no 1190/E-78/STYP/14/2019). S.~Trojanowski is also supported in part from the European Union’s Horizon 2020 research and innovation programme under grant agreement No. 952480 (DarkWave
24
project).
I. Sarcevic is supported by US DOE Grant DOE DE-SC-0009913.
The work of L.~Shchutska is supported by the ERC under the European Union’s Horizon 2020 research and innovation programme (GA 758316). 
The work of T.~Shimomura is supported by JSPS KAKENHI Grant~JP18H01210,~JP18K03651, and MEXT KAKENHI Grant JP18H05543.
The work of K.~Sinha is supported in part by US DOE Grant DE-SC0009956.
The work of T. Sj\"ostrand is supported by the Swedish Research Council,
contract 2016-05996.
J. T. Sobczyk acknowledges support from NCN grant UMO-2021/41/B/ST2/02778.
D. Soldin acknowledges support from the US NSF Grant PHY-1913607.
H.~Song is supported by the International
Postdoctoral Exchange Fellowship Program.
Y.~Soreq is supported by grants from the NSF-BSF, BSF, the ISF and by the Azrieli foundation. 
A. Stasto acknowledges support from US DOE Grant DE-SC-0002145.
S.~Su is supported by the US DOE Grant DE-FG02-13ER41976/DE-SC0009913. 
W.~Su is supported by a KIAS Individual Grant (PG084201) at Korea Institute for Advanced Study.
Y.~Takubo is supported by JP20K04004.
M.~Taoso acknowledges support from the INFN grant ``LINDARK,'' the research grant ``The Dark Universe: A Synergic Multimessenger Approach 2017X7X85'' funded by MIUR, and the project ``Theoretical Astroparticle Physics (TAsP)'' funded by the INFN. 
The research activities of B.~Thomas are supported in part by US NSF Grant PHY-2014104.  
The work of Y.-D.~Tsai is supported in part by US NSF Grant PHY-1915005.
The work of A. S. Vera has been supported by the Spanish Research Agency (Agencia Estatal de Investigaci{\'o}n) through the grant IFT Centro de Excelencia Severo Ochoa SEV-2016-0597, by the Spanish Government grant FPA2016-78022-P and from the European Union's Horizon 2020 research and innovation programme under grant agreement 824093.
The work of Y.~Zhang is supported by the National Natural Science Foundation of China Grant 12175039, the 2021 Jiangsu Shuangchuang (Mass Innovation and Entrepreneurship) Talent Program JSSCBS20210144, and the ``Fundamental Research Funds for the Central Universities''. 
Y.~Zhang is supported by the Arthur B. McDonald Canadian Astroparticle Physics Research Institute.
R.~Zukanovich~Funchal is partially supported by Funda\c{c}\~ao de Amparo \`a Pesquisa do Estado de S\~ao Paulo (FAPESP) and Conselho Nacional de Ci\^encia e Tecnologia (CNPq).

The opinions and conclusions expressed herein are those of the authors and do not represent any funding agencies.